# Astrophysical Simulations and Data Analyses on the Formation, Detection, and Habitability of Moons Around Extrasolar Planets

Habilitationsschrift

eingereicht an der

Fakultät für Physik

der Georg-August-Universität Göttingen

vorgelegt von

## René Heller

aus Hoyerswerda

Göttingen, den 23. September 2019

# Contents

# List of Figures

# Abstract

While the solar system contains as many as about 20 times more moons than planets, no moon has been definitively detected around any of the thousands of extrasolar planets so far. And the question naturally arises why an exomoon detection has not yet been achieved. This cumulative habilitation thesis covers three of the key aspects related to the ongoing search for extrasolar moons: 1. the possible formation scenarios for moons around extrasolar planets; 2. new detection strategies for these moons; and 3. the potential of exomoons as hosts to extrasolar life.

Maybe one of the most important lessons that can be learned from the history of exoplanet discoveries is that theories of planet formation in the solar system are poor references for predictions of the kind of planetary systems that can be found around other stars. Or, alternatively, the solar system has not proven to be a typical planetary system. As a consequence, one could argue that extrasolar moon systems might well be very different from the moon systems around the solar system planets. It would thus not be very surprising if the first detected exomoon would be extremely different from any of the solar system moons. And indeed one of the most compelling cases for an exomoon presented in the literature claims an exomoon candidate, provisionally named Kepler-1625 b-i, which, if confirmed, would be roughly as large as the giant planet Neptune and it would likely have a mass that would be several times larger than the combined mass of the planets Mercury, Venus, Earth, and Mars. This exomoon candidate is thus a very natural object to be addressed in this thesis and we will look at it both from a formation and a detection point of view.

In fact, the original detection of Kepler-1625 b-i as a candidate was made in search for a phenomenon called the orbital sampling effect, a subtle pattern in the transit light curves of extrasolar planets with moons that was first described in a research paper that is part of this thesis. The proposed exomoon candidate Kepler-1625 b-i is just one example, though a very extreme one, for how future exomoon detections have the potential to fundamentally test our understanding of planet and moon formation. Several studies of moon formation around migrating giant planets in this thesis illustrate a connection of these models to the solar system, in which the giant planets (and possibly their primordial moon systems) have supposedly migrated as well. The synthesis of the synchronous formation of giant planets and their moons is particularly relevant for the availability of water on these moons. In the case of the four large moons around Jupiter, the so-called Galilean moons, the water contents have previously been inferred from a combined analysis of their mean densities, analyses of surface spectra, and their observed internal mass distribution as measured during spacecraft fly-bys. In this thesis, we explain these observations with simulations of the temperature properties in the dusty gas disks around young, accreting giant planets. We show evidence that the dry Galilean moons Io and Europa have formed interior to the circumjovian water ice line, where water existed in the form of vapour that could not be accreted, whereas the water-rich Ganymede and Callisto formed beyond the ice line, where they accreted solid water ice. Ultimately, water is observed to be a key ingredient for any forms of life on Earth. The detections of liquid subsurface water reservoirs on Europa, Enceladus, and potentially Titan have motivated our studies of exomoon habitability, the possibility of life on moons around extrasolar planets. In this thesis, we present the first models of exomoon habitability that take into account the irradiation from the star, the reflected and thermal light from the planet, and tidal heating inside the moon.

This work is structured as follows. Part I gives a broad introduction to the field of extrasolar moons with special attention to their formation, detection, and habitability. Part II presents the cumulative part of this thesis with a total of 16 peer-reviewed journal publications listing the author of this thesis as a lead author, and six publications with the author of this thesis as a co-author. Part III shares some insights into our ongoing research on exomoons in collaboration with master student Anina Timmermann at the Georg-August University of Göttingen and former PhD student Kai Rodenbeck at the International Max Planck Research School for Solar System Science and the University of Göttingen. The Appendix is a collection of non-peer-reviewed conference proceedings and popular science publications by the author that further disseminate our research of extrasolar moons.

# Part I

# Scientific Context of this Thesis

# Chapter 1

# Introduction

## 1.1   The Value of Moons for Planetary Sciences

The moons of the solar system planets have become invaluable tracers of the local planet formation, in particular for the Earth-Moon system (Cameron & Ward 1976; Rufu et al. 2017). Crater counts on the Moon, for example, serve as a window into the Earth's bombardment history (Öpik 1960) and give insights into the number density (or frequency) of interplanetary rocky fragments as a function of size (Hartmann 1969). Moreover, the Earth's Moon has substantially affected the Earth's spin period and the direction of its rotational axis (Laskar et al. 1993) from the very beginning. As a consequence, our natural satellite has played a major role in the evolution of the Earth's climate and therefore in the origin and evolution of life (Zahnle et al. 2007).

Beyond that, the presence of the tiny, asteroid-sized moons Phobos and Deimos around Mars sheds light on the red planet's bombardment history and early collision frequency (Rosenblatt et al. 2016). The composition of the Galilean moons (Figure 1.1) constrains the temperature distribution in the accretion disk around Jupiter 4.5 billion years ago (Pollack & Reynolds 1974; Canup & Ward 2002; Heller et al. 2015). And while the major moons of Saturn, Uranus, and Neptune might have formed from circumplanetary tidal debris disks (Crida & Charnoz 2012), Neptune's principal moon Triton has probably been captured during an encounter with a minor planet binary (Agnor & Hamilton 2006), a processed referred to as tidal disruption. The orbital alignment of the Uranian moon systems suggests a collisional tilting scenario (Morbidelli et al. 2012) and implies significant bombardment of the young Uranus. The Pluto-Charon system can be considered a planetary binary rather than a planet-moon system, since its center of mass is outside the radius of the primary, at about 1.7 Pluto radii. A giant impact origin of this system delivers important constraints on the characteristic frequency of large impacts in the Kuiper belt region (Canup 2005).

The presence of the giant planet's moons has also been used to study the orbital migration history of their respective host planets (Deienno et al. 2011; Heller et al. 2015), the properties (such as frequency and minimum encounter distances) of possible planet-planet scattering events (Deienno et al. 2014), the giant planets' bombardment histories (Levison et al. 2001), and even the characteristics of the early protoplanetary disk around the sun (Jacobson & Morbidelli 2014). On a smaller scale, planetary rings consist of relatively small sub-grain-sized to boulder-sized particles. Their own value is as indicators of moon formation and of the tidal or geophysical activity of moons, see Enceladus around Saturn (Spahn et al. 2006).

Considering all these beneficial effects of moons for the studies of planet formation in the solar system, exomoon and exo-ring discoveries can thus be expected to deliver information on exoplanet formation on a level that is fundamentally new and inaccessible through exoplanet observations alone. Despite the discovery of thousands of planets beyond the solar system within the past two decades, mostly by the space-based Kepler telescope (Borucki et al. 2010; Thompson et al. 2018), however, no moon beyond the solar system (an "exomoon") has been unambiguously detected and confirmed (Heller 2018a). And so with more than 180 moons known around just eight planets in the solar system,

Figure 1.1: The Galilean moons around Jupiter The densities of $3.5\,\mathrm{g\,cm^{-3}}$ (Io), $3.0\,\mathrm{g\,cm^{-3}}$ (Europa), $1.9\,\mathrm{g\,cm^{-3}}$ (Ganymede), and $1.8\,\mathrm{g\,cm^{-3}}$ (Callisto) and estimated water contents (see labels) suggest that Io and Europa formed interior to the circumjovian ice line of Jupiter's early accretion disk, whereas Ganymede and Callisto formed beyond the ice line.

one obvious question to ask about exomoons is: Where are they?

Chapter 2 of this thesis gives and overview of the field of moon formation with an emphasis on case studies for the formation of extrasolar moons. Chapter 5 compiles a range of research papers led by the author that tackle the question of how large, potentially detectable exomoons could form around gas giant planets.

## 1.2   An Exomoon Detection on the Horizon

So far, most of the searches for exomoons have been executed as piggy-back science on projects with a different primary objective. To give just a few examples, Brown et al. (2001) used the exquisite photometry of the *Hubble* Space Telescope (HST) to observe four transits of the hot Jupiter HD 209458 b in front of its host star. As the star has a particularly high apparent brightness and therefore delivers very high signal-to-noise transit light curves, these observations would have revealed the direct transits of slightly super-Earth-sized satellites ($\gtrsim 1.2\,R_\oplus$; $R_\oplus$ being the Earth's radius) around HD 209458 b if such a moon were present. Alternatively, the gravitational pull from any moon that is more massive than about $3\,M_\oplus$ ($M_\oplus$ being the Earth's mass) could have been detected as well. Yet, no evidence for such a large moon was found. Brown et al. (2001) also constrained the presence of rings around HD 209458 b, which must be either extremely edge-on (so they would barely affect the stellar brightness during the transit) or they must be restricted to the inner 1.8 planetary radii around HD 209458 b.

In a similar vein, Charbonneau et al. (2006) found no evidence for moons or rings around the hot Saturn HD 149026 b, Pont et al. (2007) found no moons or rings around HD 189733 b, and Santos et al. (2015) rejected the exo-ring hypothesis for 51 Peg b, the first exoplanet ever discovered around a Sun-like star (Mayor & Queloz 1995). Maciejewski et al. (2010) used ground-based observations to search for moons around WASP-3 b by studying the planet's transit timing variations (TTVs) and transit duration variations (TDVs), both of which effects we discuss in Section 3. As TDVs remained undetectable in that system, however, an exomoon scenario seems very unlikely to cause the observed TTVs of WASP-3 b. More recently, Heising et al. (2015) scanned a sample of 21 transiting planets observed with the Kepler space telescope (Borucki et al. 2010) for ring signatures and found no conclusive evidence.

The *Hunt for Exomoons with Kepler* (HEK) project (Kipping et al. 2012), the first dedicated survey targeting moons around extrasolar planets, is probably the best bet for a near-future exomoon detection. Their analysis combines TTV and TDV measurements of exoplanets with searches for direct photometric transit signatures of exomoons. The most recent summary of their Bayesian photodynamical modeling (Kipping 2011) of exomoon transits around a total of 57 Kepler Objects of Interest has been presented by Kipping et al. (2015). Using a different approach that has been developed by the author of this thesis, the so-called orbital sampling effect in the transit light curves (see Section 3.2.1 Heller 2014), the HEK team recently announced the discovery of an exomoon candidate around the transiting Jupiter-sized object Kepler-1625 b (Teachey et al. 2018). Section 1.2.2 is devoted to a more detailed discussion of this object. Other teams found unexplained TTVs in many transiting exoplanets from the Kepler mission (Szabó et al. 2013), but without additional TDVs or direct photometric transits, a robust exomoon interpretation is impossible.

### 1.2.1 Tentative Detections of Exomoons and Exo-rings

While a definite exomoon discovery remains to be announced, some tentative claims have already been presented in the literature. One of the first exomoon claims was put forward by Bennett et al. (2014) based on the microlensing event MOA-2011-BLG-262. Their statistical analysis of the microlensing light curve, however, has a degenerate solution with two possible interpretations. It turns out that an interpretation invoking a $0.11^{+0.21}_{-0.06}\,M_\odot$ star with a $17^{+28}_{-10}\,M_\oplus$ planetary companion at $0.95^{+0.53}_{-0.19}$ AU is a more reasonable explanation than the hypothetical $3.2\,M_{\rm Jup}$-mass free-floating planet with a $0.47\,M_\oplus$-mass moon at a separation of $0.13$ AU. Sadly, the sources of microlensing events cannot be followed-up. As a consequence, no additional data can possibly be collected to confirm or reject the exomoon hypothesis of MOA-2011-BLG-262.

In the same year, Ben-Jaffel & Ballester (2014) proposed that the observed asymmetry in the transit light curve of the hot Jupiter HD 189733 b might be caused by an opaque plasma torus around the planet, which could be fed by a tidally active natural companion around the planet (which is not visible in the transit light curve itself, in this scenario). But an independent validation has not been demonstrated.

Using a variation of the exoplanet transit method, Hippke (2015) presented the first evidence of an exomoon population in the Kepler data. The author used what he refers to as a superstack, a combination of light curves from thousands of transiting exoplanets and candidates, to create kind of an average transit light curve from Kepler with a very low noise-to-signal level of about 1 part per million. This superstack of a light curve exibits an additional transit-like signature to both sides of the averaged planetary transit, possibly caused by many exomoons that are hidden in the noise of the individual light curves of each exoplanet. The depth if this additional transit candidate feature corresponds to an effective moon radius of $2120^{+330}_{-370}$ km, or about 0.8 Ganymede radii. Interestingly, this signal is much more pronounced in the superstack of planets with orbital periods larger than about 35 d, whereas more close-in planets do not seem to show this exomoon-like feature. This finding is in agreement with considerations of the Hill stability of moons, which states that stellar gravitational perturbations may perturb the orbit of a moon around a close-in planet such that the moon will be ejected (Domingos et al. 2006).

Beyond the exquisite photometric data quality of the Kepler telescope, the COnvection ROtation and planetary Transits (CoRoT; Auvergne et al. 2009) space mission also delivered highly accurate space-based stellar observations. One particularly interesting candidate object is CoRoT SRc01 E2 1066, which shows a peculiar bump near the center of the transit light curve that might be induced by the mutual eclipse of a transiting binary planet system (Lewis et al. 2015), i.e., a giant planet with a very large and massive satellite. However, only one single transit of this object (or these two objects) has been observed, and so it is currently impossible to discriminate between a binary planet and a star spot crossing interpretation of the data.

There has also been one supposed observation of a transiting ring system, which has been modeled to explain the curious brightness fluctuation of the 16 Myr young K5 star 1SWASP J140747.93-394542.6 (J1407 for short) observed around 29 April 2007 (Mamajek et al. 2012). The lower panel of Figure 1.2 shows the observed stellar brightness variations, and the panel above displays the hypothesized ring system that could explain the data. This visualization nicely illustrates the connection between rings and moons, as the gaps in this proposed ring system could have been cleared by large moons that were caught in a stage of ongoing formation (Kenworthy & Mamajek 2015). The most critical aspect of this interpretation though is in the fact that the hypothesized central object has not been observed in transit. Another issue is that the orbital period of this putative ring system around J1407 can only be constrained to be between 2.33 yr and 200 yr. In other words, the periodic nature of this proposed transit event has not actually been established and it could take decades or centuries to re-observe this phenomenon, if the interpretation were valid in the first place.

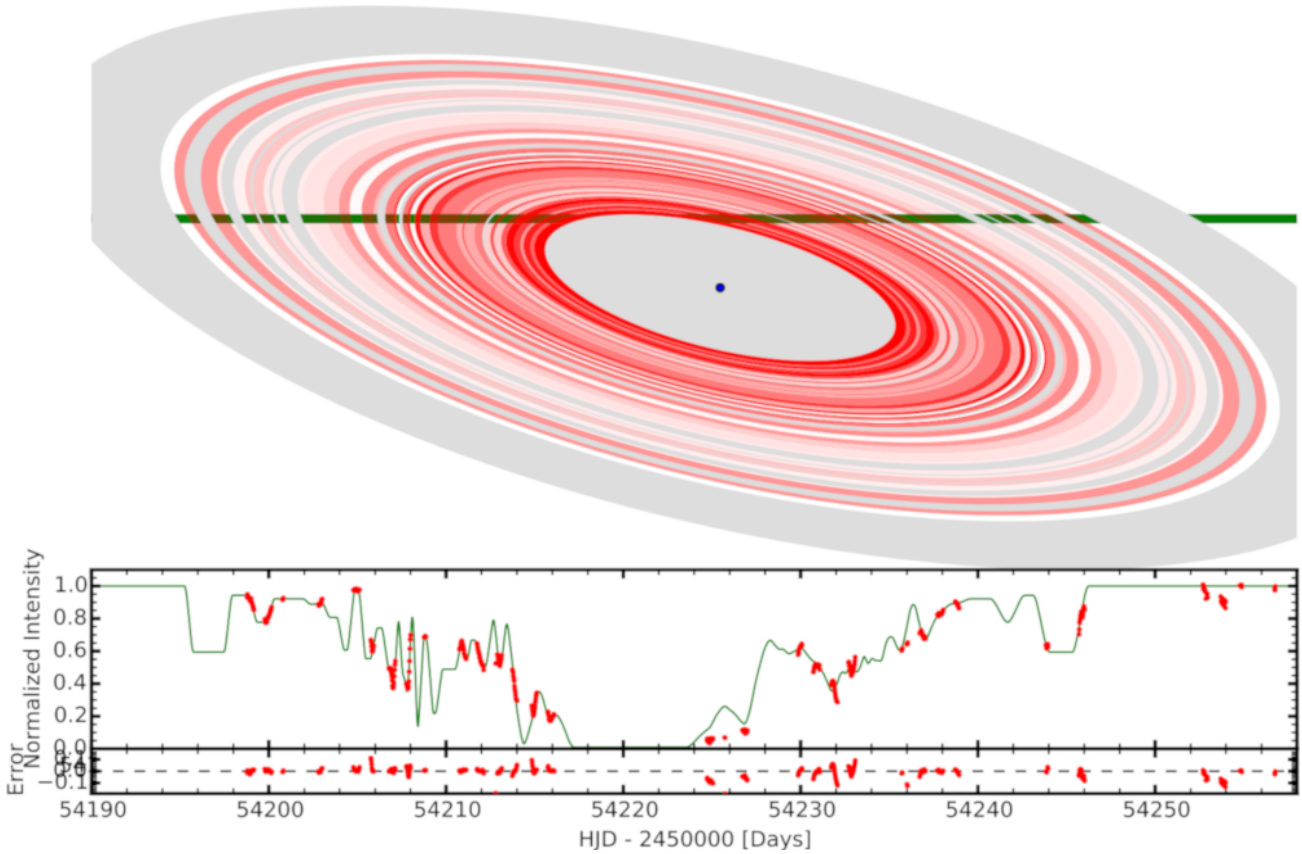

Figure 1.2: The upper panel shows a computer model of a ring system transiting the star J1407 to explain the 70 d of photometric observations from the Super Wide Angle Search for Planets (Super-WASP) in the lower panel. The thick green line in the background of the upper panel represents the path of the star relative to the rings. Gray annuli indicate regions of the possible ring system that are not constrained by the data. Gradation of the red colors symbolizes the transmissivity of each ring. Note that the hypothesized central object of the ring system (possibly a giant planet) is not transiting the star. Image credit: Kenworthy & Mamajek (2015). ©AAS.

### 1.2.2   The Exomoon Candidate Kepler-1625 b-i

One particularly interesting case of a tentative exomoon detection is the candidate Kepler-1625 b-i. In July 2017, Teachey et al. (2018) announced the detection of an exomoon-like signature in the transit light curve of the exoplanet candidate[1] Kepler-1625 b based on three transits observed with Kepler between 2009 and 2013. The candidate moon signal was found using a newly developed and efficient method to search for exomoon signals in large amounts of data that has been previously developed by the author of this thesis. In brief, this method (referred to as the orbital sampling effect, OSE) is based on a search for an additional dip in both the pre-ingress and post-egress wings of the orbital phase-folded planetary transit light curve, respectively. Geometrically speaking, in the case of a coplanar moon around a transiting planet, the moon should occasionally start its own passage of the sky-projected stellar disk prior to the planetary transits. At other times then, the planet should enter the stellar disk first and the moon would lag behind, e.g. by minutes or hours. Averaging over many transits, one can expect that the moon's sky-projected orbital position from its host planet should smear out like a ring around the planet, which reveals itself as the above-mentioned additional pre- and post-transit signature around the planetary transit. Details have been developed by Heller (2014) and Heller et al. (2016a) and can be found in Sects. 6.1 and 6.2.

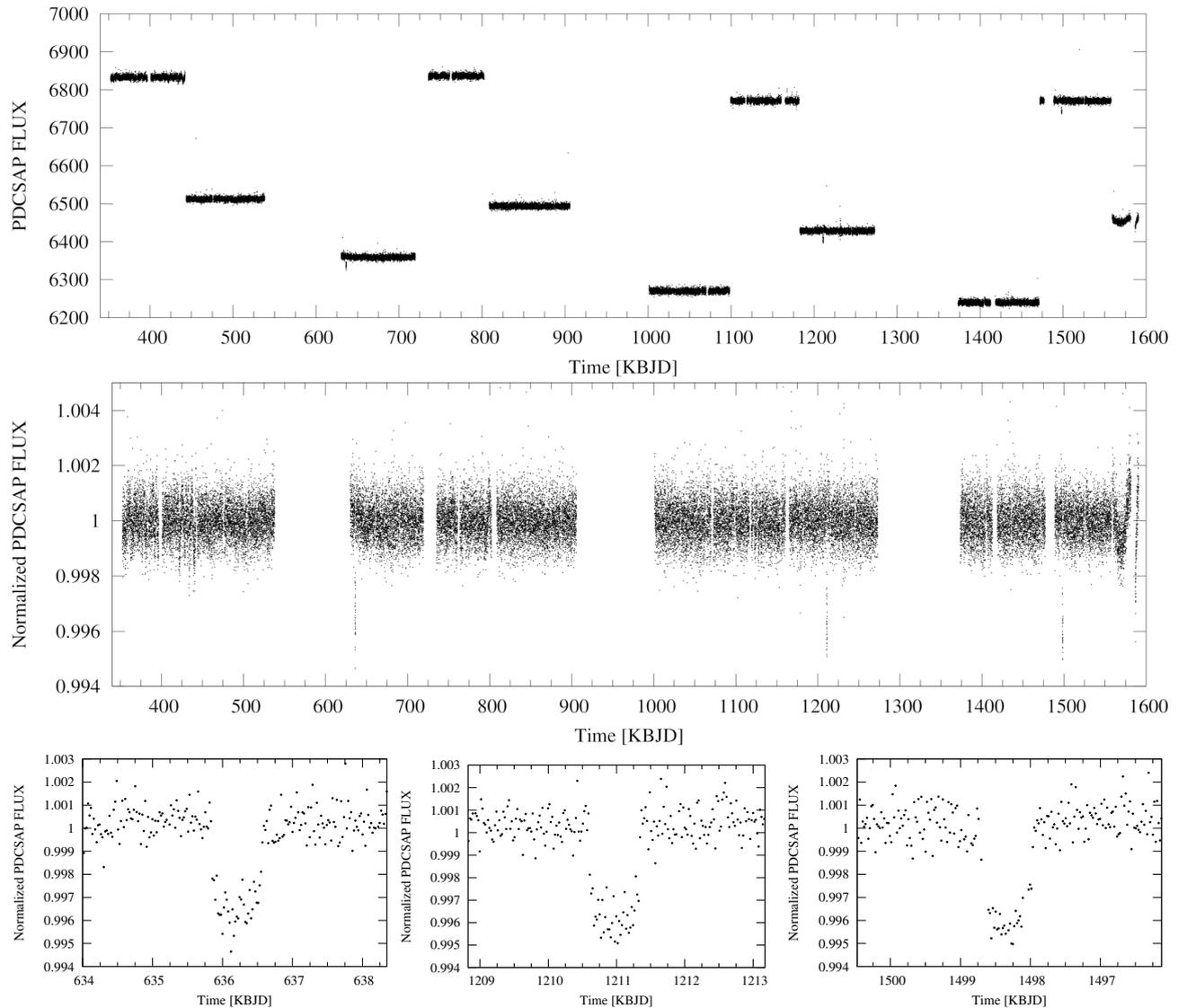

Figure 1.3: The top panel shows the Kepler light curve of the star Kepler-1625. The vertical offsets are caused by reorientations of the spacecraft every three months (called a Kepler quarter), which changed the position of the target star on the on-board CCD with different pixel sensitivities. The center panel shows the light curve with the flux per quarter normalized to 1. The three transits of the object Kepler-1625 b are now visible by eye at about 630 d, 1210 d, and 1500 d (Kepler Barycentric Julian Date, KBJD). The three panels at the bottom show zooms into these transits.

---

[1]Formally speaking, the transiting object Kepler-1625 b has not yet been confirmed as an exoplanet, which would involve observation techniques that are independent of the transit method. Such a confirmation could involve TTVs, e.g. caused by other planets in the system (Agol et al. 2005) or measurements of the stellar radial velocities (RVs), which would show periodic variations due the gravitational interaction with Kepler-1625 b. Both TTVs and RVs could be used to determine the mass of Kepler-1625 b and therefore would enable a classification of this transiting object as either a planet, or a brown dwarf, or even a low-mass star. The degeneracy of the planetary interpretation of the Kepler transit light curve alone is treated in Heller (2018c) and in Section 5.4.

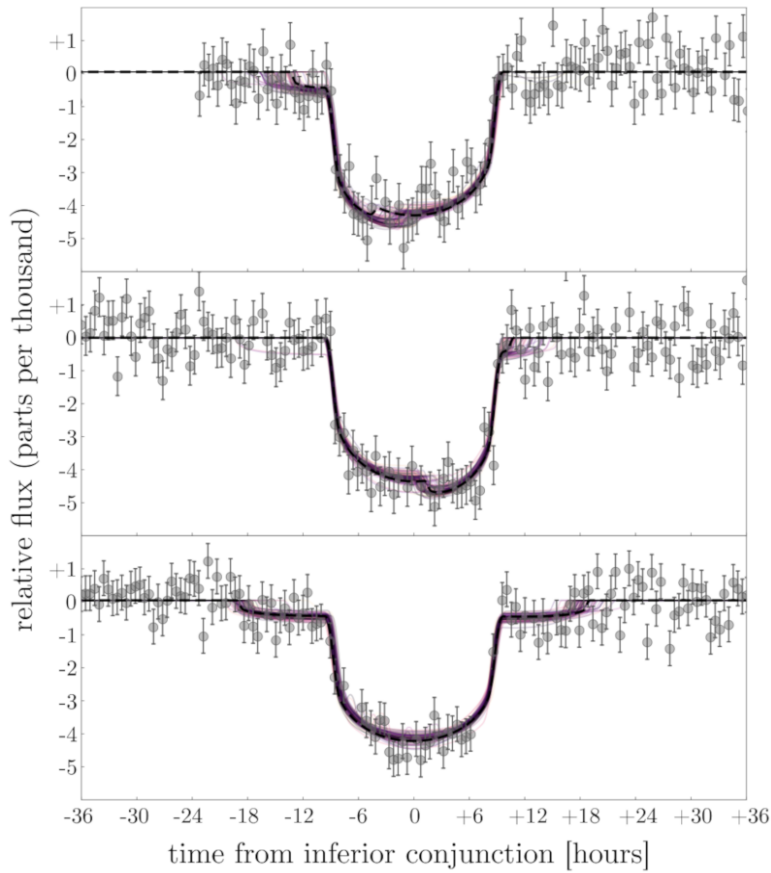

Figure 1.4: This figure is from the discovery paper of the exomoon candidate Kepler-1625 b-i by Teachey et al. (2018). Gray dots with error bars indicate Kepler data (see Figure 1.3), colored curves refer to 100 drawings from a Markov Chain Monte Carlo simulation, and the the black line refers to the best fit of a dynamical planet-moon model to the three transits. ©AAS.

Figure 1.3 illustrates the Kepler data of the host star Kepler-1625 (KIC 4760478, KOI-5084). The upper two panels show the Pre-Search Data Conditioning Simple Aperture Photometry (PDCSAP) flux, which is freely available online at the NASA Exoplanet Archive[2]. The three panels at the bottom of Figure 1.3 give an idea of the size of the planet candidate: with the normalized stellar flux dimming as much as about $d = 1 - 0.995 = 0.5\,\%$, the radius of the transiting object ($R_{\rm p}$) would be about $\sqrt{d}\,R_\star$, or about 7 % the radius of the star.

The star, located at about $2181^{+332}_{-581}$ pc from the Sun, has been classified as an evolved star with a mass of $M_\star = 1.079^{+0.100}_{-0.138}\,M_\odot$, a radius of $R_\star = 1.793^{+0.263}_{-0.488}\,R_\odot$, an effective temperature of $T_{\rm eff,\star} = 5548^{+83}_{-72}$ K (Mathur et al. 2017), and a Kepler magnitude of $K = 13.916$ (as per the NASA Exoplanet Archive). Hence, $R_{\rm p} = 1.276^{+0.187}_{-0.347}$ Jupiter radii ($R_{\rm J}$), neglecting any errors from the measurement of $d$.

The transiting object Kepler-1625 b has an orbital period of $287.377314 \pm 0.002487$ d (Morton et al. 2016), which implies an orbital semimajor axis of about 0.87 AU around the star, by means of Kepler's Third Law. The most recent transit of Kepler-1625 b occurred on 29 October 2017 at 02 : 34 : 51($\pm$ 00 : 46 : 18) UT.[3] Teachey et al. (2018) have taken observations of this transit with the *Hubble Space Telescope* and the community is eagerly awaiting their improved evaluation of the exomoon hypothesis for Kepler-1625 b-i.

Figure 1.4 shows the original light curve presented by Teachey et al. (2018), including the detrended Kepler data (gray dots) and a range of plausible model fits of a transiting planet with a moon. Even a by-eye comparison of the gray data points in Figure 1.4 with the data points shown in the bottom panels of Figure 1.3 reveals significant differences. These differences stem from an additional light curve detrending procedure applied by Teachey et al. (2018), which is supposed to remove stellar, instrumental, and other sources of time-correlated flux variability (so-called red noise) from the light curve. The aim of this procedure, sometimes referred to as "pre-whitening" of the data (Aigrain & Irwin 2004), is to remove unwanted variations in the data prior to fitting a noiseless model to the data. As we show in our follow-up study, this approach bears the risk of both removing actual signal from the data and of introducing new systematic variability (Rodenbeck et al. 2018; Sect. 6.7). In fact, a moon-like signal can be artificially injected into the data of a planet-only transit, leading to the detection of a false positive. The simultaneous fitting of the stellar, systematic, and astrophysically induced variability of the light curve, including any possible moon transit features, developed in Heller et al. (2019) is less prone to artificial injection of false positive moon signals (see Sect. 6.8). But the Bayesian information criterion (BIC) used by Teachey et al. (2018) to provide evidence for the existence of Kepler-1625 b-i was shown to be inadequate for this particular data set and given its non-Gaussian noise properties. Even more dramatic is the finding by Kreidberg et al. (2019) of an absence of the actual moon-like transit feature in the *Hubble* data, suggesting that the exomoon candidate around Kepler-1625 d really is an artifact of the data extraction procedure.

Chapter 3 of this book gives a summary of the community's efforts to search for extrasolar moons. Chapter 6 is dedicated to a number of peer-reviewed studies by the author of this thesis that present novel detection methods for exomoons, with a focus on the analysis of space-based transit photometry.

## 1.3 The Value of Moons for Astrobiology

While the interest in the possibility of an independent origin of life beyond Earth has been speculative or fictional for most of humanity's history, the advent of modern astronomy has opened the possibility of the detection and scientific study of extraterrestrial life. Among the several approaches that have been pursued in the search for life beyond Earth, the most prominent ones include the search for active or past life on Mars (McKay et al. 1996), the search for organic (that is carbon-involving) or salt chemistry in the water ice plumes of Saturn's moon Enceladus (Porco et al. 2006; Hsu et al. 2015), and the search for extraterrestrial intelligence via interstellar radio communication (Cocconi & Morrison 1959; Isaacson et al. 2017). In fact, astrophysical concepts and technological know-how have matured so rapidly over the past few decades that the in-situ detection of extraterrestrial biological activity in the solar system, the remote detection of intelligent life, or the detection of chemical biomarkers in the atmospheres of transiting exoplanets (Schwieterman et al. 2016) may all be promising on their own over the coming decades.

Most studies and concepts dedicated to the search for life beyond Earth focus on planets, which is a natural approach given the fact the only world known to harbor life is a planet, Earth. This has led planetologists and astronomers to develop the concept of the stellar habitable zone, which characterizes an exoplanet's surface habitability as a function of its distance from the star and its absorbed stellar energy flux. The oldest record of a description of a circumstellar zone suitable for life traces back to Whewell (1853, Chap. X, Section 4) who, referring to the local stellar system in a qualitative way, called this distance range the "Temperate Zone of the Solar System". The modern, much more complex version of this concept was introduced by Kasting et al. (1993). Their one-dimensional atmospheric climate model includes a parameterization of the $CO_2$ feedback and of the geological carbon-silicate cycle, both of which are key to the location of the inner and the outer edges of the HZ around the host star. The inner edge is defined by the activation of the moist or runaway greenhouse process, which desiccates the planet by evaporation of atmospheric hydrogen; the outer edge is defined by $CO_2$ freeze out from the atmosphere, which breaks down the greenhouse effect whereupon the planet transitions into a permanent snowball state.

---

[3]The transit epoch of $(2,454,833.0 + 348.83318) \pm 0.00729$ BJD corresponds to $2455181.834397 \pm 0.00729$ JD.

The principal assumption of this astronomer's concept of habitability is in the presence of liquid surface water. It has been argued that this pre-requisite is vital since, on Earth, life exists in virtually any place where there is liquid water. Some primitive forms of life, such as bacteria, archea, and unicellular eukaryotes can grow at temperatures as low as $-20°$ C and it has been possible to grow plants from seeds that had been frozen for about 32,000 yr (Yashina et al. 2012). The lower temperature limit for higher plants and invertebrates to show ongoing metabolism and to reproduce, however, is at about $-2°$ C (Clarke 2014) and liquid water has been identified as a key ingredient for the completion of their life cycles.

While extrasolar planets are natural targets for humans to search for extraterrestrial life, it has recently become clear that moons could be equally or even better suited to host life. In fact, most of the liquid water in the solar system is stored in the sub-surface oceans of Jupiter's moon Europa, which is supposed to contain as much as two to three Earth oceans worth of water, details depending on the yet unknown depth of the ocean (Squyres et al. 1983; Carr et al. 1998). Other moons, such as Ganymede (around Jupiter) as well as Titan and Enceladus (both around Saturn) also show convincing evidence for liquid subsurface water. The reason for the abundance of liquid water in the outer regions of the solar system is in the distribution of water during the formation of the planets and moons, almost 4.6 billion years ago. While the rocky planets all formed within the hot and dry regions of the protoplanetary nebula, where water only existed in the form of vapor that could hardly be incorporated into the newly forming planets[4], the giant planets and their moons formed beyond the so-called water ice line of the solar system. The ice line, thought to have been located between 2 AU and 3 AU (Hayashi 1981; Lecar et al. 2006), has been identified as a dividing line between the dry and wet (or frozen) regions of the solar system. In fact, water ice was important to rapidly form the growing cores of the soon-to-be giant planets (Lissauer et al. 2009). As a consequence, the moons of the giant planets had to form beyond the sun's water ice line as well and so sufficient amounts of water were present to also be incorporated into their structures.

Now given that water-rich moons are abundant around the giant planets of the solar system, and given that most giant planets beyond the solar system are actually found near their stellar HZs (Heller & Pudritz 2015a), one may speculate that the moons of these giant planets have taken a piggyback ride on their giant planets from their water-rich formation regions beyond the stellar ice line to their present location similar to that of the Earth around the Sun. Figure 1.5 illustrates the distribution of the knows exoplanets with mass measurements as a function of distance to their respective host stars. The abundance of Jupiter- and super-Jupiter-sized planets can readily be observed near 1 AU. It is important to keep in mind, however, that Figure 1.5 is heavily biased by the sensitivities of the different detection techniques (see figure legend). As a consequence, the observed planet populations are by no means representative of the actual exoplanet abundances, or occurrence rates (Howard et al. 2012; Dressing & Charbonneau 2015). In fact, most of the validated exoplanets have been found with Kepler via the transit method, and so they lack a firm mass estimate. These planets do not even show up in Figure 1.5. If, instead, one would plot the distance distribution of all planets with known radii (rather than masses), then this figure would be dominated by a population of super-Earths and mini-Neptunes near 0.1 AU.

---

[4]The source of water on Earth has, in fact, not been unambiguously determined. Different lines of evidence suggest either dehydration of water originally stored in Earth's internal hydrate minerals (Schmandt et al. 2014) or the delivery of water from asteroids or comets (Sarafian et al. 2014; Chan et al. 2018).

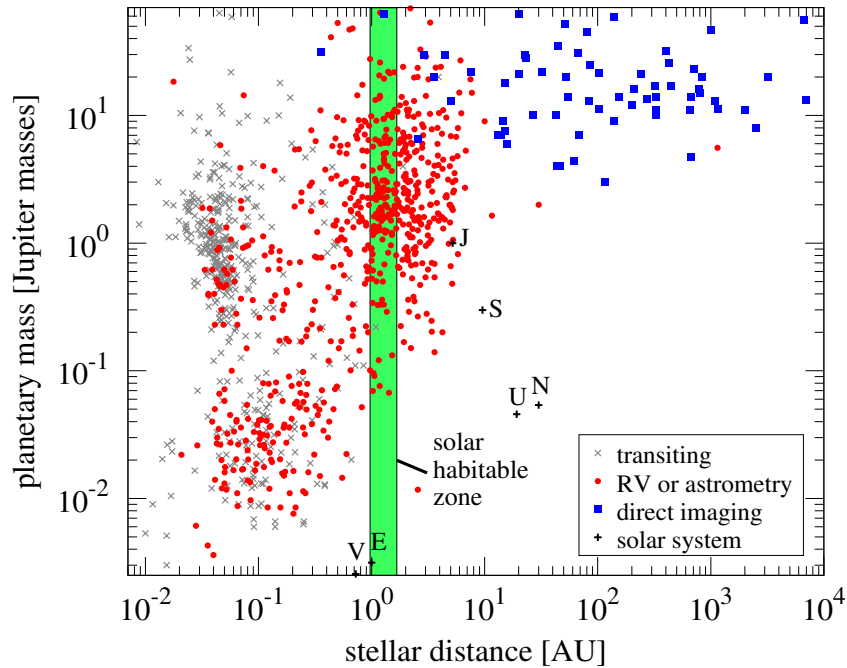

Figure 1.5: Stellar distances and planetary masses of extrasolar planets listed on www.exoplanet.eu. The discovery method of each planet is indicated by the symbol type. The locations of six planets from the solar system are shown for comparison, indicated by their initials. The cluster of Jovian and super-Jovian planets near 1 AU (along the abscissa) and between 1 and 10 $M_J$ (along the ordinate) suggests that water-rich moons in their stellar habitable zones could be highly abundant in the Milky Way. The green-shaded vertical strip denotes the solar habitable zone defined by the runaway and maximum greenhouse (Kopparapu et al. 2013). This plot is an updated version of Figure 1 from Heller & Pudritz (2015a).

The formation, dynamical (orbital) stability, and the resistance of moons against evaporative destruction during their possible circumstellar migration with their host planets has been addressed recently in the research literature. As a key outcome, Lehmer et al. (2017) showed that moons more massive than Ganymede could maintain significant amounts of surface water in their stellar habitable zones, whereas less massive moons (e.g. as light as Europa) would tend to lose their waters rather quickly through a runaway greenhouse effect and subsequent $H_2O$ photodissociation in the upper atmosphere, which would lead to the escape of atmospheric hydrogen and, hence, entire desiccation of the moon. The possibility of massive moons around Jovian and super-Jovian planets has been demonstrated through simulations of the circumplanetary accretion disks (Heller & Pudritz 2015a), and the orbital stability of these large satellites has been shown to be possible in many (though not all) cases (Namouni 2010; Spalding et al. 2016; Hong et al. 2018).

It is thus possible that moons offer a large part of the extraterrestrial habitable real estate and that large moons beyond the solar system could be habitable and detectable in the near future, as we discuss in Chapter 4. This has led the author of this thesis to develop a model for the habitable zones of moons around giant planets, which is presented in Chapter 7.

# Chapter 2

# Formation Mechanisms for Moons

Although no moon has been detected beyond the solar system, there is no reason to believe they don't exist. In fact, most astronomers were certain that planets should exist around other stars even prior to the first detections of planets beyond the solar system (Wolszczan & Frail 1992). And just a few years after the discovery of extrasolar planets around solar-type stars (Mayor & Queloz 1995), exoplanets had soon outnumbered the eight (by then formally nine) solar system planets. In the near future, thanks to the improvements of ground-based radial velocity surveys, such as the High-Accuracy Radial velocity Planetary Searcher (HARPS; Mayor et al. 2003), and thanks to the many discoveries of transiting extrasolar planets and candidates with the Kepler space telescope (Batalha et al. 2013; Rowe et al. 2014; Thompson et al. 2018), and in anticipation of thousands of new exoplanet discoveries with NASA's Transiting Exoplanet Survey Satellite (TESS; Ricker et al. 2015) and ESA's PLAnetary Transits and Oscillations of stars (PLATO; Rauer et al. 2014) satellite, we will soon know as many as a thousand times more exoplanets than planets in the solar system. Will we be able to find any moons around them? How many of these planets do actually host moons and what are their properties?

The architecture of the solar system has proven inappropriate for most extrasolar planetary systems in many regards. For one thing, the most abundant population of exoplanets turned out to be a previously unknown class of super-Earths or mini-Neptunes in relatively tight stellar orbits. The planets have radii between about 2 to 4 Earth radii ($R_\oplus$) and orbital distances of about 0.1 AU to their host stars. For comparison, the solar system has no planet with a radius between 1 $R_\oplus$ and 3.8 $R_\oplus$, the radius of Neptune. And the innermost planet, Mercury, orbits the sun at about 0.4 AU. For another thing, the first exoplanets were discovered around pulsars (Wolszczan & Frail 1992), the burnt-out remnants of rather massive progenitor stars with about 10 $M_\odot$ to 30 $M_\odot$. The survival of these planets during the late stages of stellar evolution, including the violent disruption during a supernova, remains unsolved until today. Moreover, the discovery of a hitherto unknown family of hot Jupiters (Jupiter-mass planets in very tight orbits of typically 0.05 AU) came to a big surprise of astronomers (Mayor & Queloz 1995), which triggered new theories of planet formation such as planet migration of various types (Lin et al. 1996; Trilling et al. 1998). With no firm exomoon detection available as of today, any speculation on their formation must necessarily be based on observations of the solar system moons. Hence, throughout this chapter we must keep in mind the perils of using the solar system architecture as a proxy for extrasolar moon systems.

The formation of moons in the solar system can broadly be classified into three distinct scenarios, all of which we discuss in the following subsections of this chapter: (1.) re-accretion after giant impacts on terrestrial planets; (2) accretion in the circumplanetary disks of young giant planets; and (3) capture via the tidal disruption during a close encounter between a massive planet and a planetary binary.

## 2.1 Giant Impacts on Terrestrial Planets

The formation of moons around terrestrial (that is, Earth-like, rocky) planets is arguably of highest importance to us. Several lines of evidence lead to the conclusion that the Moon formed through a giant collision of the proto-Earth with a Mars-sized object (sometimes referred to as Theia; Quarles & Lissauer 2015) about 30 Myr to 50 Myr after the formation of the primordial Earth (Hartmann & Davis 1975). Our home planet is the only rocky planet in the solar system with a substantial moon, which had a major effect on the evolution of the Earth's spin and therefore on its climate. We shall start with an overview of the key observations of the Earth-Moon system.

1. The Moon has a small iron core, possibly with a radius of about 350 km or less (Wieczorek 2006). The Moon is also depleted in volatile elements relative to the Earth (Khan et al. 2006).

2. Isotope dating of lunar rock suggests an age essentially equal to the age of the Earth.

3. The Moon's orbit carries most of the angular momentum in the Earth-Moon system (Cameron & Ward 1976).

4. The Moon is receding from Earth at a rate of about 3.8 cm per year (Walker & Zahnle 1986; Dickey et al. 1994).

5. The length of the Earth's day is increasing at a rate of about 1.8 ms per century (Stephenson et al. 2016).

The total angular momentum of the Earth-Moon binary ($L_{\leftmoon\oplus}^{\mathrm{tot}}$) is composed of the two contributions from the rotational angular momenta of the Earth ($L_{\oplus}^{\mathrm{rot}}$) and the Moon ($L_{\leftmoon}^{\mathrm{rot}}$) and of the system's orbital angular momentum ($L_{\leftmoon\oplus}^{\mathrm{orb}}$):

$$L_{\oplus}^{\mathrm{rot}} = I_{\oplus}\,\omega_{\oplus} = \frac{2}{5}M_{\oplus}\,R_{\oplus}^2\,\omega_{\oplus} = 7.1 \times 10^{33}\,\mathrm{kg\,m^2\,s^{-1}} \tag{2.1}$$

$$L_{\leftmoon}^{\mathrm{rot}} = I_{\leftmoon}\,\omega_{\leftmoon} = \frac{2}{5}M_{\leftmoon}\,R_{\leftmoon}^2\,\omega_{\leftmoon} = 2.4 \times 10^{29}\,\mathrm{kg\,m^2\,s^{-1}} \tag{2.2}$$

$$L_{\leftmoon\oplus}^{\mathrm{orb}} = |\vec{r}_{\leftmoon} \times \vec{p}_{\leftmoon}| = r_{\leftmoon}M_{\leftmoon}\,v_{\leftmoon}^{\mathrm{orb}} = 2.9 \times 10^{34}\,\mathrm{kg\,m^2\,s^{-1}} \quad, \tag{2.3}$$

where $I$ is the moment of inertia of a sphere, $\omega$ is the spin frequency, $M$ is the mass, $R$ is the radius, and $v^{\mathrm{orb}}$ is the orbital velocity of the respective object, with indices $\oplus$ and $\leftmoon$ referring the Earth and Moon, respectively. Note that in Equation (2.3) we have assumed that the Moon is in a circular orbit, which is an adequate approximation given its low orbital eccentricity of about 0.0549. Most important, Equations (2.1) to (2.3) show that most of the angular momentum is currently contained in the Moon's orbit, that is, about 80.3 %. This is a particular distinction from all other moons in the solar system, where most of the angular momentum is stored in the rotation of the host planet. Why?

The leading theory that embraces all the above observations envisions a collision between a Mars-sized impactor and the proto-Earth. This giant impact generated a circumplanetary halo of ejected material that collapsed into a disk and then into a moon within only a few years (Canup & Asphaug 2001). The satellite supposedly formed at about 3.8 $R_{\oplus}$, which is near the Earth's Roche radius, interior to which any rocky moon would be shredded by differential gravitational forces across its diameter, that is, by tidal forces. An illustration of this process is shown in Figure 2.1 that is based on Smoothed Particle Hydrodynamic (SPH) simulations from Canup (2004).

This primordial distance of the Moon must be compared to its current orbital separation of about 384,000 km, or about 60 $R_{\oplus}$. The driver for the Moon's orbital recession is found in the Earth's tides. The gravitational pull from the Moon deforms the Earth across its diameter and thereby raises two principal tidal bulges, one roughly facing the Moon, the other one on the opposite side of the Earth. The Earth's rotation period of about 24 hr is much smaller than the Moon's orbital period of about 27 d. This asynchronicity between the Earth's rotation and the Moon's orbit combined with the fact that the tidal bulges cannot instantaneously align with the line connecting the centers of mass of the Earth and the Moon due to friction inside the Earth, yields that the Earth's Moon-facing tidal bulge is constantly leading the Moon. This lag corresponds to about 600 s (Lambeck 1977; Neron de Surgy

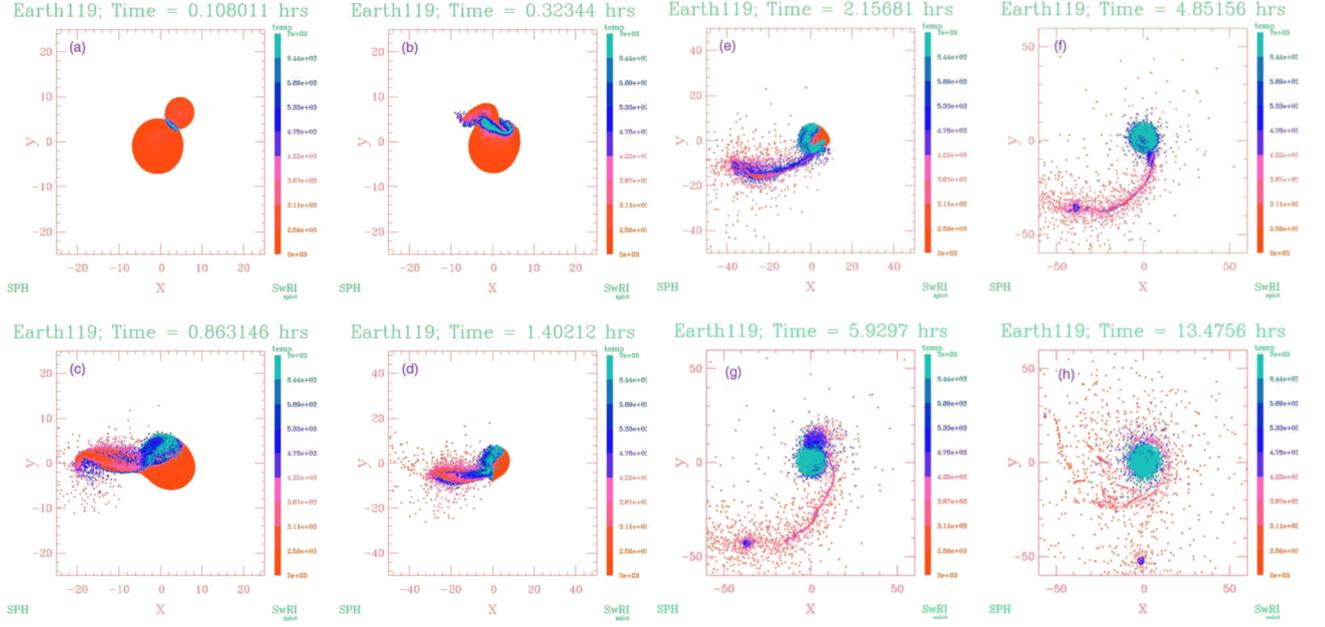

Figure 2.1: Smoothed Particle Hydrodynamic simulations of a late Moon-forming impact using 60,000 test particles. The color scale encodes temperatures ranging from 2 000 K (red) to 7 000 K (light blue). Figure adopted from Canup (2004). ©Elsevier.

& Laskar 1997). Naturally, the Moon-opposing tidal bulge on the Earth is lagging by the same time lag. However, with the Moon-facing tidal bulge being closer to the Moon, there will be a net torque of the two bulges on the Earth's rotation, which is dominated by the Moon-facing bulge. This torque is the origin of both the Moon's recession from Earth (item 4. in the list above) and the increase of the Earth's length of day (item 5.). The connection of the two processes is the transfer of angular momentum from the Earth's rotation to the Moon's orbit.

In fact, observations of the time lag of the Earth's tidal bulges can be used to estimate the recession rate and vice versa. Let us consider a tidal model for a two-body system of two deformable bodies (Hut 1981; Leconte et al. 2010), here the Earth and Moon. In this model, the change of the orbital semimajor axis $\mathrm{d}a/\mathrm{d}t$ can be calculated from first principles as (Heller et al. 2011)

$$\frac{\mathrm{d}a}{\mathrm{d}t} = \frac{2a^2}{GM_\oplus M_\mathbb{C}} \sum_{i \neq j} Z_i \left( \cos(\psi_i) \frac{f_2(e)}{\beta^{12}(e)} \frac{\omega_i}{n} - \frac{f_1(e)}{\beta^{15}(e)} \right) \ , \quad i \in \{\oplus, \mathbb{C}\} \ , \qquad (2.4)$$

where $G$ is the gravitational constant, $n$ is the orbital frequency, $\omega_i$ are the spin frequencies, $\psi_i$ are the spin-orbit angular misalignments, $\tau_i$ are the time lags, $e$ is the orbital eccentricity

$$Z_i = 3G^2 k_{2,i} M_j^2 (M_i + M_j) \frac{R_i^5}{a^9} \ \tau_i \ , \qquad (2.5)$$

$k_{2,i}$ is the potential Love number of degree 2 of the $i$th body, and the extension functions in $e$ are

$$\beta(e) = \sqrt{1 - e^2} \ ,$$
$$f_1(e) = 1 + \frac{31}{2}e^2 + \frac{255}{8}e^4 + \frac{185}{16}e^6 + \frac{25}{64}e^8 \ ,$$
$$f_2(e) = 1 + \frac{15}{2}e^2 + \frac{45}{8}e^4 \ + \ \frac{5}{16}e^6$$

following the nomenclature of Hut (1981). Note that these extensions collapse to 1 for zero eccentricities. In case of the Earth-Moon system, where the orbit is nearly circular ($e \approx 0$), the spins are misaligned by not more than a few degrees ($\psi_i \approx 0$), and the Moon's orbit is essentially synchronous ($\omega_\leftmoon = n$), Equation (2.4) simplifies to

$$\frac{\mathrm{d}a}{\mathrm{d}t} \approx \frac{2a^2}{GM_\oplus M_\leftmoon} 3G^2 k_{2,\oplus} M_\leftmoon^2 \left(M_\oplus + M_\leftmoon\right) \frac{R_\oplus^5}{a^9} \tau_\oplus \left(\frac{\omega_\oplus}{n} - 1\right) \quad . \tag{2.6}$$

With a nominal value for the Earth's second degree tidal Love number of 0.3, Equation (2.6) yields a recession rate of $\mathrm{d}a/\mathrm{d}t = 3.8\,\mathrm{cm\,yr^{-1}}$, in remarkable agreement with observations (Walker & Zahnle 1986; Dickey et al. 1994). A similar equation can be derived for the spin-down rate of the Earth, $\mathrm{d}\omega_\oplus/\mathrm{d}t$, which also yields a value close to the observed increase of the length of the day.

One major longstanding riddle of the Earth-Moon system has been in the fact that a numerical integration of $\mathrm{d}a/\mathrm{d}t$ and $\mathrm{d}\omega_\oplus/\mathrm{d}t$ backwards in time predicts a gradual infall of the Moon into the Earth with a merger at only about 1.5 billion years in the past (Bills & Ray 1999). This result is in strong disagreement with the isotopic dating of lunar rock brought back from the Apollo missions, which suggest a lunar age of about 4.5 billion years. The solution to this problem can be found in the frequency-dependence of the efficiency of tidal dissipation in the Earth's oceans, which must have been much weaker in the past (Webb 1980). In fact, the Earth currently seems to be going through a phase of enhanced tidal forcing.

Moving on to the possible formation of Moon-like natural satellites around extrasolar planets, numerical $N$-body simulations of gas-free particle disks around Earth-mass planets were able to reproduce many of the features of the Earth-Luna system (Hyodo et al. 2015). The resulting planet-to-moon mass ratios of these post-impact simulations, for example, are usually between several times $10^{-3}$ and a few times $10^{-2}$, compared to the value of about $1.2 \times 10^{-2}$ or roughly 1/81 for the Moon-to-Earth mass ratio. As an interesting side note, in about a third of these simulations by Hyodo et al. (2015), the resulting moon system consists of two moons rather than one. These results are in agreement with the findings of Elser et al. (2011), who also used $N$-body simulations to study post-impact satellite formation around Earth-like planets and who found similar mass ratios. Both studies seem to suggest, however, that the Earth's moon has become relatively large compared to the distribution of the outcomes from these simulations. As a consequence, large and possibly detectable (Luna-like) exomoons around terrestrial planets might be rare even if moon formation via giant impacts is common around Earth-like exoplanets.

## 2.2 Accretion Disks Around Young Gas Giant Planets

The most common process for moon formation in the solar system was the in-situ formation in the circumplanetary accretion disks that formed around giant planets. It has been suggested that these disks of gas ($\approx 99\,\%$ of the disk mass) and dust ($\approx 1\,\%$ of the disk mass) existed around all of the local giant planets in the vey early phase of the solar system (Crida & Charnoz 2012). Competing ideas exist about the exact physical processes that triggered moon formation. Three of the most promising theories have been reviewed by Heller et al. (2014) in the context of moon formation around planets beyond the solar system. This contribution is presented in Sect. 7.7. In brief, these models can be described and distinguished in the following manner.

### 1. Solids-enhanced minimum mass model

In what is referred to as the "solids enhanced minimum mass model" (Mosqueira & Estrada 2003a,b; Estrada et al. 2009), the formation of moons commences once sufficient gas has been removed from what started out as a massive subnebula, at which point turbulence in this new circumplanetary disk has subsided. Protomoons form from the solid materials transported to and accreted by the disk through ablation and capture of planetesimal fragments passing through the massive disk. A key ingredient to this model is the division of the disk in two density and opacity regimes. The inner, high-density part of the disk is thought to be the leftover of the gas accreted by the young planet, whereas the outer part is the result from longterm gas accretion onto the planet once it has opened up a gap in the circumstellar accretion disk. In this model, the inner Galilean moons Io, Europa, and Ganymede formed on a short time scale of between $10^3$ and $10^4$ yr in the inner disk around Jupiter, while Ganymede formed in the outer disk and on a time scale of $10^6$ yr.

## 2. Gas-starved disk model

By contrast, the "actively supplied gaseous accretion disk model" (Canup & Ward 2002, 2006, 2009) suggests that the moons around the giant planets that we observe today are the last of a number of moon generations that were constantly formed and destroyed around their respective host planets. This model of an actively supplied gaseous accretion disk assumes a low-mass, viscously evolving protosatellite disk with peak surface densities around $100\,\mathrm{g\,cm^2}$ that is continuously supplied with mass from the circumstellar protoplanetary disk. In the traditional picture of this model, the temperature profile of the circumplanetary disk is dominated by viscous heating of the gas, while the planetary luminosity has long been assumed a minor role. Protomoons are assumed to build up from dust grains that are supplied by gas infall. Once a protomoon has grown massive enough, it becomes subject to asymmetric gravitational torques from the protosatellite disk, which cause it to lose angular momentum and radially move toward the planet through a process known as type I migration (Tanaka et al. 2002). One of the key arguments in favor of the gas-starved disk model over other models is its ability to explain the universal scaling law of the observed total moon mass around the giant planets to be about $10^{-4}$ times the mass of the host planet (Canup & Ward 2006) by a balance between the moon loss rate towards the planet (via type I moon migration) and the resupply of material for new moons (from the circumstellar disk). Further refinements of the model were given by Sasaki et al. (2010), who introduced an inner magnetic cavity to the accretion disk around Jupiter and included the effect of Jupiter's gap opening in the protoplanetary disk around the sun, both of effects of which are irrelevant for Saturn. The resulting $N$-body simulations with collisions produced satellite architectures around the two gas giant planets that were ver distinct and in agreement with observations: systems of multiple large moons around Jupiter and systems dominated by a single massive moon around Saturn. The compositional aspects of the resulting moons, with a focus on their water ice contents, were investigated by Ogihara & Ida (2012).

Several papers in this thesis have contributed to the further development of the gas-starved disk model. In Heller & Pudritz (2015b), we presented the first 2D computations of the gas temperature and density distributions around Jupiter-like planets and of super-Jovian planets under the explicit inclusion of all four heating terms: (1) viscous heating in the disk, (2) accretion heating from the matter falling onto the disk, (3) the planetary luminosity, and (4) the ambient temperature of the solar nebula. An example of the resulting temperature distribution around a Jupiter-like planet at 1 AU from its young host star, about $10^6$ yr after the onset of accretion is shown in Fig. 2.2. One key result of this study is that super-Jovian planets at about 1 AU around their sun-like stars, several dozens of which have already been discovered, should regularly form massive moon systems. In fact, we found that the scaling law for the satellite-to-planet mass ratio of $M_s/M_p \approx 10^{-4}$ (Canup & Ward 2006) extends from the solar system giant planets to planets as massive as ten Jupiter masses. One crucial ingredient to this mass scaling is the location of the water ice line during the final stages of planet accretion (see the white arc in Fig. 2.2). Beyond the water ice line, ice crystals can be accreted and incorporated into the forming moons, whereas closer to the planet water vapor is too volatile to be accreted. Our results for the continued scaling law are thus the outcome of an interplay between the luminosity evolution of the planet, which determines the evolution of the radial position of the ice line, and the radial extend of the circumplanetary disk: more massive planets have their water ice lines farther out, thereby preventing the formation of massive Ganymede-like ice moons in the inner regions of their disks, but they also have larger disks. In Heller & Pudritz (2015a) we extended our computations to giant exoplanets at various distances from their host stars to study the behavior of the water line under different magnitudes of stellar irradiation. We found that the most low-mass giant planets in our test sample of simulated planets, which were as massive as Jupiter, struggle to have a water ice line in the first place if they are closer to their star than about 4.5 AU. This finding naturally led a follow-up study (Heller et al. 2015), in which we studied the evolution of the circumjovian ice line during the phase of Jupiter's orbital migration in the early solar system, a scenario known as the Grand Tack model (Walsh et al. 2011; Raymond & Morbidelli 2014). The Grand Tack scenario explains the occurrence and distribution of the present-day S and C types of asteroids and the low mass of Mars (10 % of the Earth's mass) as an outcome of Jupiter's and Saturn's coupled migration in the early solar system. In particular, Jupiter's possible formation beyond 3.5 AU and its inward migration to as few as 1.5 AU from the sun have dramatic effects on the water ice line in the accretion disk around the young Jupiter. We found that the icy moons Ganymede and Callisto cannot possibly have accreted their water contents during Jupiter's supposed Grand Tack simply because the planetary accretion disk, if still active, was too warm to build water ice in the first place. After Jupiter and Saturn got caught up in an orbital mean motion resonance, which pulled Jupiter outward to its current orbital position at 5.2 AU from the sun, any material in an accretion disk around Jupiter would have been dry due to the evaporative processes triggered during its journey to the inner, warm regions of the solar system. The fact that Ganymede and Callisto are made up of about 50 % of water by mass, respectively, means that they must have acquired large amounts of water either before or after the possible Grand Tack. If they would have received their water reservoirs by late accretion of planetesimals, however, then Io and Europa should be water-rich, too, because computer simulations have shown that planetesimal accretion on these inner two moons is very efficient (Tanigawa et al. 2014). As a consequence, our results imply that Ganymede and Callisto (and likely also Io and Europa) must have formed prior to the Grand Tack scenario, if this scenario for giant planet migration in the early solar system is actually correct. This might open the possibility of tracing the migration history of the Galilean moons through the inner solar system with the JUICE mission of ESA (Grasset et al. 2013), which is currently scheduled for launch in 2022 and arrival at Jupiter in 2029.[1]

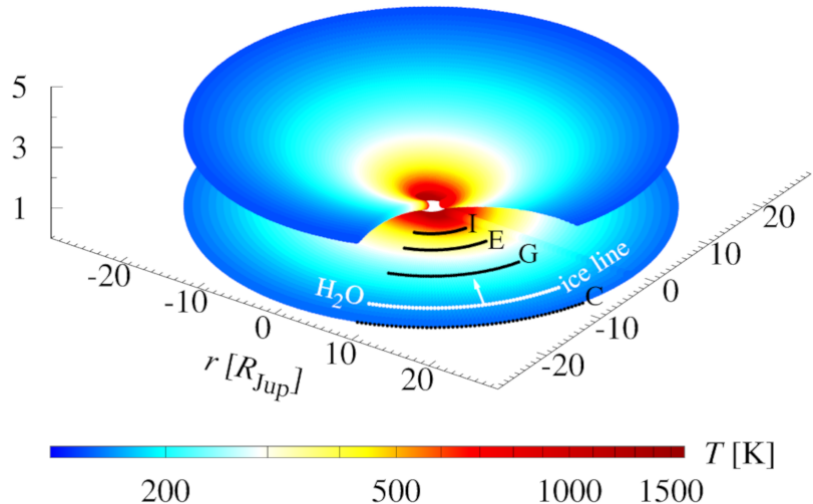

Figure 2.2: Temperature distribution in Jupiter's accretion disk about 1 Myr after then onset of planet formation. The planet (not shown) is located at the origin of the coordinate system. Distances along the axes are measured in units of Jupiter radii. The instantaneous water line, where the disk reaches 170 K, is indicated with a white arc. The contemporary orbits of the Galilean moons are shown with black arcs and labeled with initials (I: Io, E: Europa, G: Ganymede, C: Callisto). The two levels of the disk correspond to the disk mid-plane, where the pressure reaches its vertical maximum, and to the disk photosphere, where the optical depth equals 2/3. Temperatures are encoded in colors (see color bar at the bottom). Figure from Heller & Pudritz (2015b). ©AAS.

**3. Tidally spreading disk model**

Finally, Crida & Charnoz (2012) presented a model of a primordial ring system (a late, gas-free accretion disk) around a giant planet, in which satellite seeds initially form near the planet's Roche radius interior to which their aggregation is prevented due to the tidal forces of the planet. The protosatellites then move outward under the combined effects of the planetary tides and the torque from the disk. The moons then grow in mass as they move outward and clear the disk. At the same time, new material is supplied to the disk from the reservoir of debris interior to the Roche radius. This model produces small moons close to the planet and large moons far from the planet. This architecture is in agreement with the regular moon systems around Uranian, Neptunian, and – to some extent – with the Saturnian and Jovian moon systems.

## 2.3 Capture Through Tidal Disruption During Planetary Encounters

As we have seen, the Earth's moon has likely formed after a giant collision between the proto-Earth and a Mars-sized protoplanetary impactor (Sect. 2.2) whereas most of the regular massive moons around the giant planets have formed through some sort of in-situ accretion process within the dusty gas disks that fed their young host planets (Sect. 2.1). There is a third mechanism for moon formation that successfully explains the odd orbital and geological properties of Neptune's principal moon, Triton. For one thing, Triton is the only major moon in the solar system that is on a retrograde orbit around its host planet. Its orbital plane is inclined to Neptune's equator by about 157° and crater counts suggest that its surface has regions of different ages, with all regions being younger than about 50 Myr and some are as young as 6 Myr. This makes it one of the youngest planetary surfaces in the solar system (Schenk & Zahnle 2007) and it might be the result of ongoing geological activity including cryovolcanism.

Agnor & Hamilton (2006) have shown that a retrograde Triton can be naturally formed during an encounter between Neptune and a planetary binary, of which one component would be ejected from Neptune's well of gravity and of which one would be permanently captured in a stable orbit. Williams (2013) extended this model to the capture of (possibly detectable) moons around giant planets beyond the solar system. Ultimately, Heller (2018c) showed that a tidal capture scenario is the most viable mechanism to form a giant moon like the proposed Neptune-sized moon around the super-Jovian planet Kepler-1625 (see Sect. 5.4).

---

[1] https://www.cosmos.esa.int/web/juice

# Chapter 3

# Detection Methods for Exomoons[1]

About a dozen different theoretical methods have been proposed to search and characterize exomoons. For the purpose of this introduction into the field of exomoon detections, we will group them into three classes: (1.) dynamical effects of the transiting host planet; (2.) direct photometric transits of exomoons; and (3) other methods.

## 3.1 Dynamical Effects on Planetary Transits

The moons of the solar system are small compared to their planet, and so the natural satellites of exoplanets are expected to be small as well. The depth ($d$) of an exomoon's photometric transit scales with the satellite radius ($R_s$) squared: $d \propto R_s^2$. Consequently, large exomoons could be relatively easy to detect (if they exist), but small satellites would tend to be hidden in the noise of the data.

Alternatively, instead of hunting for the tiny brightness fluctuations caused by the moons themselves, it has been suggested that their presence could be derived indirectly by measuring the TTVs and TDVs of their host planets. The amplitudes of both quantities ($\Delta_{\rm TTV}$ and $\Delta_{\rm TDV}$) are linear in the mass of the satellite: $\Delta_{\rm TTV} \propto M_s \propto \Delta_{\rm TTV}$ (Sartoretti & Schneider 1999; Kipping 2009a). Hence, the dynamical effect of low-mass moons is less suppressed than the photometric effect of small-radius moons.

### 3.1.1 Transit Timing Variation

In a somewhat simplistic picture, neglecting the orbital motion of a planet and its moon around their common center of gravity during their common stellar transit, TTVs are caused by the tangential offset of the planet from the planet-moon barycenter (see upper left illustration in Figure 3.1, where "BC" denotes the barycenter). In a sequence of transits, the planet has different offsets during each individual event, assuming that it is not locked in a full integer orbital resonances with its circumstellar orbit. Hence, its transits will not be precisely periodic but rather show TTVs, approximately on the order of seconds to minutes (compared to orbital periods of days to years).

Two flavors of observable TTV effects have been discussed in the literature. One is called the barycentric TTV method (TTV$_b$; Sartoretti & Schneider 1999; Kipping 2009a), and one is referred to as the photocentric TTV method (TTV$_p$ or PTV; Szabó et al. 2006; Simon et al. 2007, 2015). A graphical representation of both methods is shown in Figure 3.1.

TTV$_b$ measurements refer to the position of the planet relative to the planet-moon barycenter. From the perspective of a light curve analysis, this corresponds to measuring the time differences of the planetary transit only, e.g. of the ingress, center and/or egress (Sartoretti & Schneider 1999).

PTV measurements, on the other hand, take into account the photometric effects of both the planet and its moon, and so the corresponding amplitudes can actually be significantly larger than TTV$_b$

---

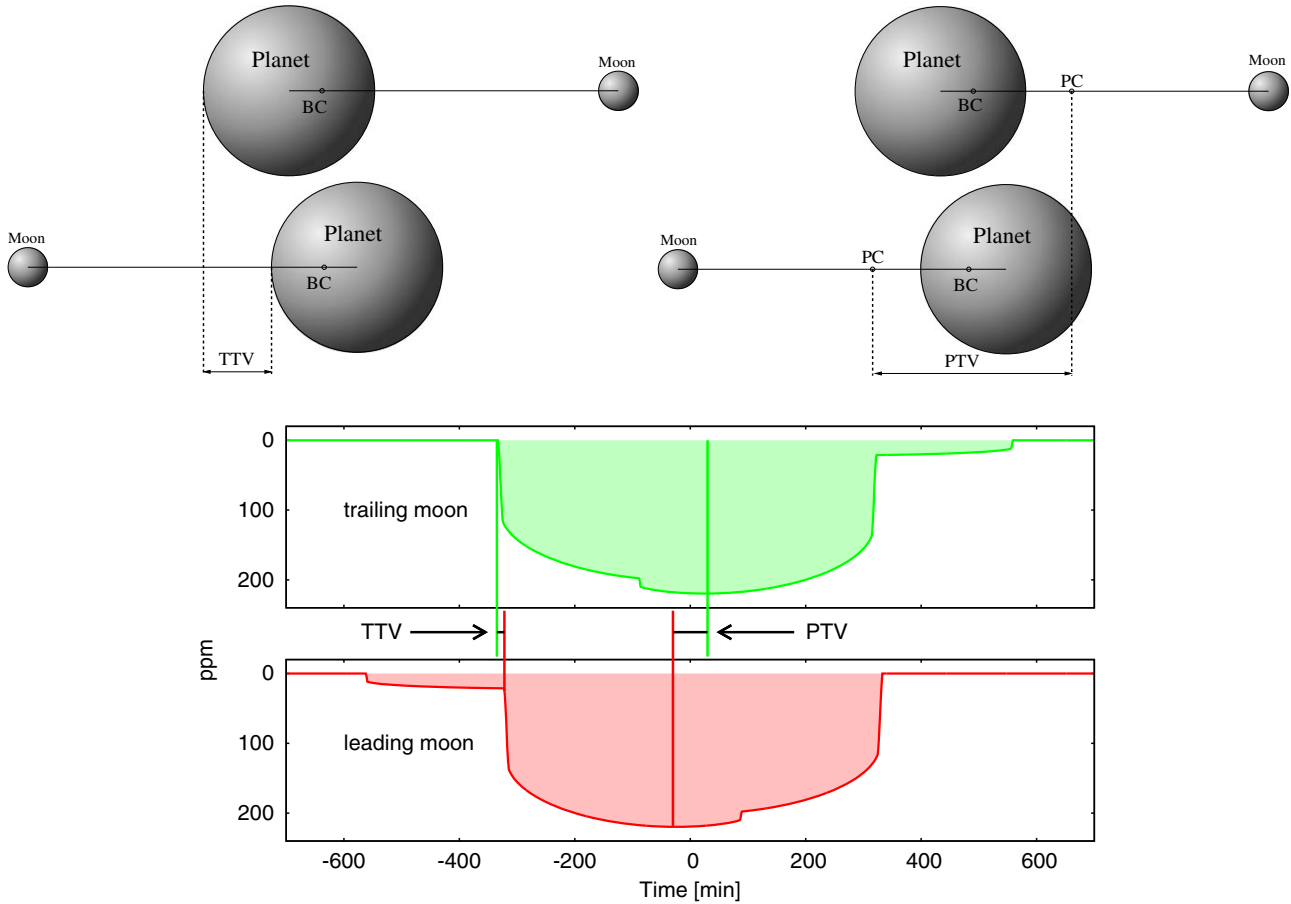

Figure 3.1: Physical explanation of the barycentric TTV (upper left) and the photocentric TTV (upper right). The two light curves at the bottom illustrate both the $TTV_b$ and the $TTV_p$ (or PTV) for a moon that is trailing the planet (upper panel, green curve) and leading the planet during the stellar transit (lower panel, red curve). Image credit: Simon et al. (2015). ©The Astronomical Society of the Pacific.

amplitudes, details depending on the actual masses and radii of both objects (Simon et al. 2015).

### 3.1.2  Transit Duration Variation

Planetary TDVs can be caused by several effects. First, they can be produced by the change of the planet's tangential velocity component around the planet-moon barycenter between successive transits (referred to as the $TDV_V$ component; Kipping 2009a). When the velocity component in the planet-moon system that is tangential to the observer's line of sight adds to the circumstellar tangential velocity during the transit, then the event is relatively short. On the other hand, if the transit catches the planet during its reverse motion in the planet-moon system, then the total tangential velocity is lower than that of the barycenter, and so the planetary transit takes somewhat longer.

TDV effects can also be introduced if the planet-moon orbital plane is inclined with respect to the circumstellar orbital plane of their mutual center of gravity. In this case, the planet's apparent minimum distance from the stellar center will be different during successive transits, in more technical terms: its transit impact parameter will change between transits or, if the moon's orbital motion around the planet is fast enough, even during the transits. This can induce a $TDV_{TIP}$ component in the transit duration measurements of the planet (Kipping 2009b).

It is important to realize that the waveforms of the TTV and TDV curves are offset by an angle of $\pi/2$ (Kipping 2009a). In a more visual picture, when the TTV is zero, i.e. when the planet is along the line of sight with the planet-moon barycenter, then the corresponding TDV measurement is either largest (for moons on obverse motion) or smallest (for moons on reverse motion), since the planet would have the largest/smallest possible tangential velocity in the planet-moon binary system. This phase difference is key to breaking the degeneracy of simultaneous $M_s$ and $a_s$ measurements ($a_s$ being the moon's semi-major axis around the planet). When plotted in a TTV-TDV diagram (Montalto et al. 2012; Awiphan & Kerins 2013), the resulting ellipse contains predictable dynamical patterns, which can help to discriminate an exomoon interpretation of the data from a planetary perturber, and it may even allow the detection of multiple moons (Heller et al. 2016b). In a related context, where close stars show mutual eclipses (rather than transits), combination of eclipse timing variations (ETVs), eclipse duration variations (EDVs), and radial velocity (RV) measurements of the stars can reveal the presence of S-type extrasolar planets, as we demonstrated in (Oshagh et al. 2017).[2]

## 3.2 Direct Transit Signatures of Exomoons

Like planets, moons could naturally imprint their own photometric transits into the stellar light curves, if they were large enough (Tusnski & Valio 2011). The lower two panels of Figure 3.1 show an exomoon's contribution to the stellar bright variation in case the moon is trailing (upper light curve) or leading (lower light curve) its planet. Note that if the moon is leading, then its transit starts prior to the planetary transit, and so the exomoon transit affects the right part of the planetary transit in the light curve. As mentioned in Section 3.1, the key challenge is in the actual detection of this tiny contribution, which has hitherto remained hidden in the noise of exoplanet light curves.

As a variation of the transit method, it has been suggested that mutual planet-moon eclipses during their common stellar transit might betray the presence of an exomoon or binary planetary companion (Sato & Asada 2009; Pál 2012). This is a particularly interesting method, since the mutual eclipses of two transiting planets have already been observed (de Wit et al. 2016). Yet, in the latter case, the two planets were known to exist prior to the observation of their common transit, whereas for a detection of an exomoon through mutual eclipses it would be necessary to test the data against a possible origin from star spot crossings of the planet (Lewis et al. 2015) and to use an independent method for validation.

### 3.2.1 Orbital Sampling Effect

One way to generate transit light curves with very high signal-to-noise ratios in order to reveal exomoons is by folding the measurements of several transits of the same object into one phase-folded transit light curve. Figure 3.2 shows a simulation of this phase-folding technique, which is referred to as the orbital sampling effect (OSE; Heller 2014; Heller et al. 2016a). The derived light curve does not effectively contain "better" data than the combination of the individual transit light curves (in fact it loses any information about the individual TTV and TDV measurements), but it enables astronomers to effectively search for moons in large data sets, as has been done by Hippke (2015) to generate superstack light curves from Kepler (see Section 1.2.1).

Ultimately, Teachey et al. (2018) used the OSE to detect the first viable exomoon candidate in the Kepler data (see Section 1.2.2).

### 3.2.2 Scatter Peak

The minimum possible noise level of photometric light curves is given by the shot noise (or Poisson noise, white noise, time-uncorrelated noise), which depends on the number of photons collected and, thus, on the apparent brightness of the star. For the amount of photons typically collected with

---

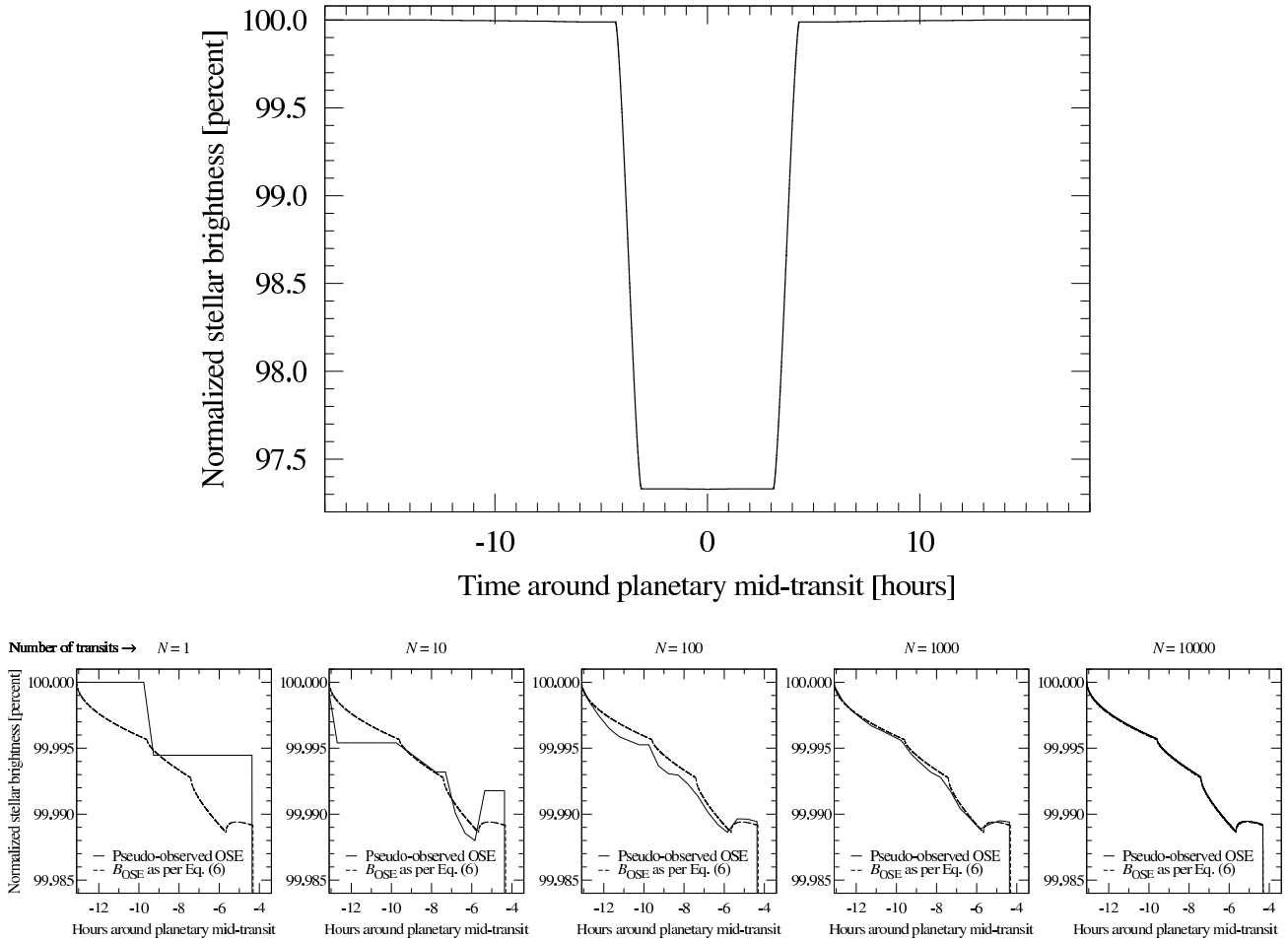

Figure 3.2: Orbital sampling effect (OSE) of a simulated transiting exoplanet with moons. The upper panel shows a model of the phase-folded transit light curve of a Jupiter-sized planet around a $0.64\,R_\odot$ K dwarf star using an arbitrarily large number of transits. The planet is accompanied by three moons of $0.86\,R_\oplus$, $0.52\,R_\oplus$, and $0.62\,R_\oplus$ in radial size, but their contribution to the phase-folded light curve is barely visible with the naked eye. The lower row of panels shows a sequence of zooms into the prior-to-ingress part of the planetary transit. The evolution of the OSEs of the three moons is shown for an increasing number of transits ($N$) used to generate the phase-folded light curves. In each panel, the solid line shows the simulated phase-folded transit and the dashed line shows an analytical model, both curves assuming a star without limb darkening. Image credit: Heller (2014). ©AAS.

space-based optical telescopes, the minimum possible signal-to-noise ratio (SNR) of a light curve can be approximated as the square root of the number of photons ($n$): $\mathrm{SNR} \propto \sqrt{n}$. And so the SNR of a phase-folded transit light curve of a given planet goes down with the square root of the number of phase-folded transits ($N$): $\mathrm{SNR_{OSE}} \propto \sqrt{N}$. In other words, for planets transiting photometrically quiet host stars, the noise-to-signal ratio (1/SNR) of the phase-folded light curve converges to zero for an increasing number of transits.

If the planet is accompanied by a moon, however, then the variable position of the moon with respect to the planet induces an additional noise component. As a consequence, and although the average light curve is converging towards analytical models (see Figure 3.2), the noise in the planetary transit is actually *in*creasing due to the moon. For large $N$, once dozens and hundreds of transits can be phase-folded, the OSE becomes visible together with a peak in the noise, the latter of which has been termed the scatter peak (Simon et al. 2012). As an aside, the superstack OSE candidate signal found by Hippke (2015) was not accompanied by any evidence of a scatter peak.

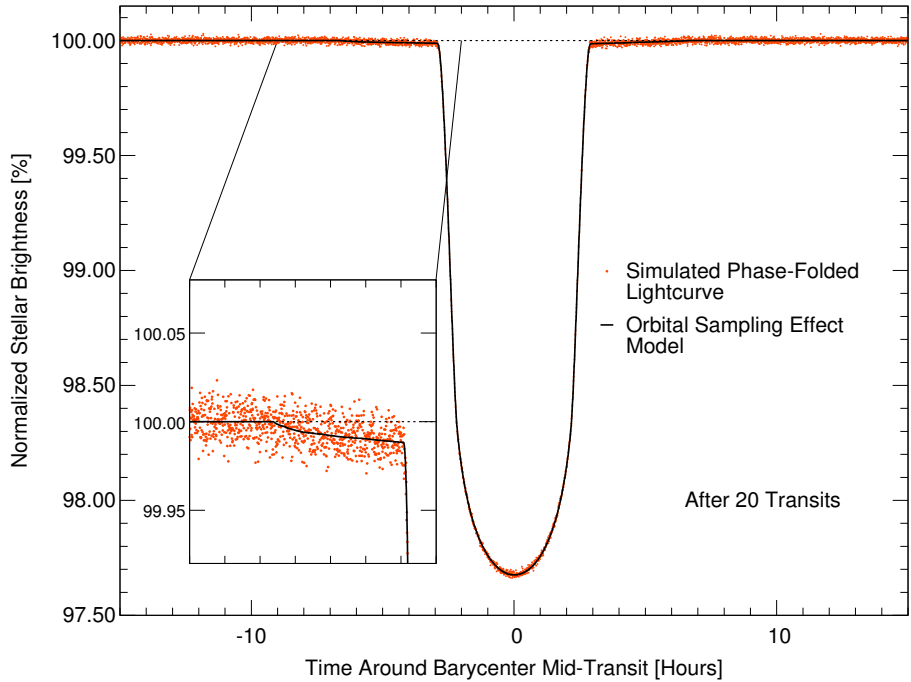

Figure 3.3: Simulated OSE of an $m_V = 11$ K dwarf star ($0.6\,R_\odot$) with a Saturn-sized planet and an Earth-sized moin in a Europa-wide orbit as it would be observed with PLATO. The distance between the star and the planet-moon system is $0.3\,\mathrm{AU}$, which yields a period of $72\,\mathrm{d}$. Only white noise is assumed, that is to say, the star is assumed to be photometrically quiet with no spots, flares or the like.

## 3.3 Other Methods for Exomoon Detection

In some cases, where the planet and its moon (or multiple moons) are sufficiently far from their host star, it could be possible to optically resolve the planet from the star. This has been achieved more than a dozen times now through a method known as direct imaging (Marois et al. 2008). Though direct imaging cannot, at the current stage of technology, deliver images of a resolved planet with individual moons, it might still be possible to detect the satellites. One could either try and detect the shadows and transits of the moons across their host planet in the integrated (i.e. unresolved) infrared light curve of the planet-moon system (Cabrera & Schneider 2007; Heller 2016), or one could search for variations in the position of the planet-moon photocenter with respect to some reference object, e.g. another star or nearby exoplanet in the same system (Cabrera & Schneider 2007; Agol et al. 2015). Fluctuations in the infrared light received from the directly imaged planet $\beta$ Pic b, as an example, could be due to an extremely tidally heated moon (Peters & Turner 2013) that is occasionally seen in transit or (not seen) during the secondary eclipse behind the planet. A related method is in the detection of a variation of the net polarization of light coming from a directly imaged planet, which might be caused by an exomoon transiting a luminous giant planet (Sengupta & Marley 2016).

It could also be possible to detect exomoons through spectral analyses, e.g. via excess emission of giant exoplanets in the spectral region between 1 and $4\,\mu m$ (Williams & Knacke 2004); enhanced infrared emission by airless moons around terrestrial planets (Moskovitz et al. 2009; Robinson 2011); and the stellar Rossiter-McLaughlin effect of a transiting planet with moons (Simon et al. 2010; Zhuang et al. 2012) or the Rossiter-McLaughlin effect of a moon crossing a directly imaged, luminous giant planet (Heller & Albrecht 2014).

Some more exotic exomoon detection methods invoke microlensing (Han & Han 2002; Bennett et al. 2014; Skowron et al. 2014), pulsar timing variations (Lewis et al. 2008), modulations of radio emission from giant planets (Noyola et al. 2014, 2016), or the generation of plasma tori around giant planets

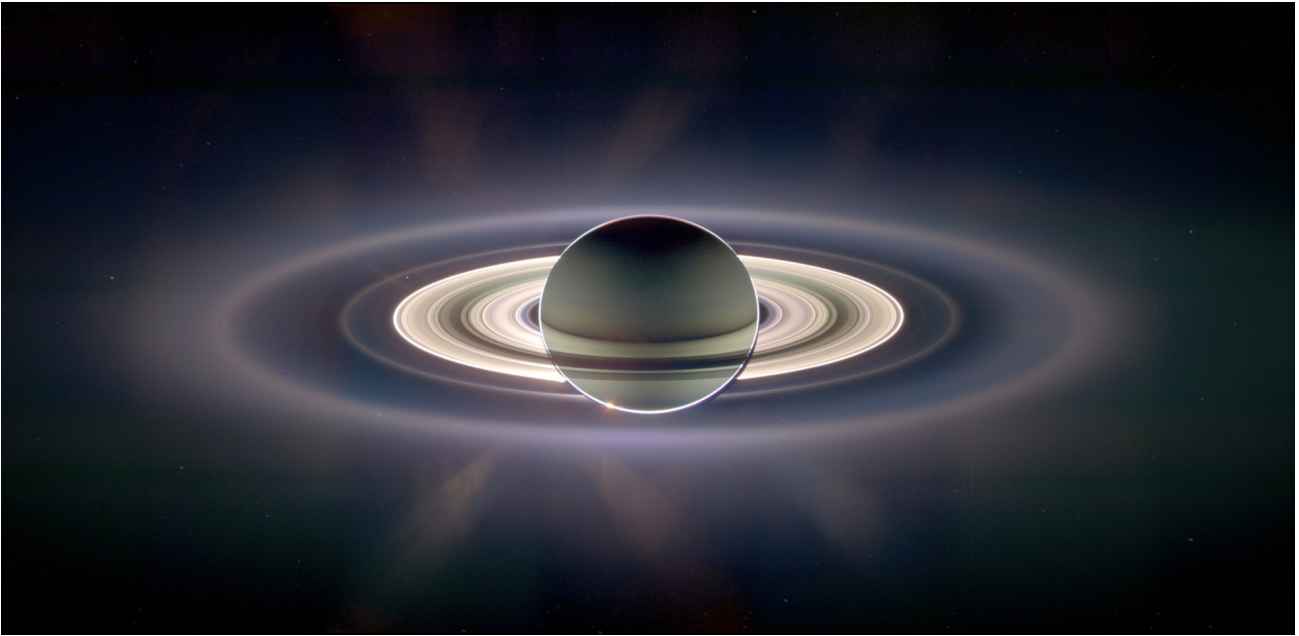

Figure 3.4: Transit of Saturn and its ring system in front of the sun as seen by the Cassini spacecraft in September 2016. Note how the rings bend the sun light around the planet, an effect knows as diffraction. Image credit: NASA/JPL/Space Science Institute.

by volcanically active moons (Ben-Jaffel & Ballester 2014).

## 3.4 Detection Methods for Exo-rings

Just like moons are very common around the solar system planets, rings appear to be a common feature as well. Naturally, the beautiful ring system around Saturn was the first to be discovered. Less obvious rings have also been detected around all other gas giants in the solar system and even around an asteroid (Braga-Ribas et al. 2014). In the advent of exoplanet detections, astronomers have thus started to develop methods for the detection of rings around planets outside the solar system.

### 3.4.1 Direct Photometric Detection

The detection of rings around exoplanets is closely related to many of the above-mentioned methods (see Section 3.2) of direct photometric transit observations of exomoons. Like moons, rings can cause additional dips in the planetary transit light curve (Tusnski & Valio 2011). But rings do not induce any dynamical effects on the planet and, hence, there will be no TTVs or TDVs. In fact, a ring system can be expected to impose virtually the same pattern on each individual transit of its host planet because rings should look the same during each transit. Moons, however, would have a different position relative to the planet during individual transits (except for the case of full-integer orbital resonances between the circumstellar and the circumplanetary orbits). This static characteristic of the expected ring signals make it susceptible to misinterpretation, e.g. by a standard fit of a planet-only model to a hypothetically observed planet-with-ring light curve: in this case, the planet radius would be slightly overestimated, while the ring could remain undetected. However, the O-C diagram (O for observed, C for calculated) could still indicate the ring signature (Barnes & Fortney 2004; Zuluaga et al. 2015)

As a consequence, rings could induce a signal into the phase-folded transit light curve, which is very similar to the OSE (Section 3.2.1) since the latter is equivalent to a smearing of the moon over its circumplanetary orbit – very much like a ring. The absence of dynamical effects like TTVs and TDVs, however, means that there would also be no scatter peak for rings (Section 3.2.2). Thus, an OSE-like signal in the phase-folded light curve without an additional scatter peak could indicate a ring rather than a moon.

One particular effect that has been predicted for light curves of transiting ring systems is diffraction, or forward-scattering (Barnes & Fortney 2004). Diffraction describes the ability of light to effectively bend around an obstacle, a property that is rooted in the waveform nature of light. In our context, the light of the host star encounters the ring particles along the line of sight, and those of which are $\mu$m- to 10 m-sized will tend to scatter light into the forward direction, that is, toward the observer. In other words, rings cannot only obscure the stellar light during transit, they can also magnify it temporarily (see Figure 3.4).

### 3.4.2   Other Detection Methods

Beyond those potential ring signals in the photometric transit data, rings might also betray their presence in stellar transit spectroscopy. The crucial effect here is similar to the Rossiter-McLaughlin effect of transiting planets: as a transiting ring proceeds over the stellar disk, it modifies the apparent, disk-integrated radial velocity of its rotating host star. This is because the ring covers varying parts of the disk, all of which have a very distinct contribution to the rotational broadening of the stellar spectral lines (Ohta et al. 2009). Qualitatively speaking, if the planet with rings transits the star in the same direction as the direction of stellar rotation, then the ring (and planet) will first cover the blue-shifted parts of the star. Hence, the stellar radial velocity will occur redshifted during about the first half of the transit – and vice versa for the second half.

Another more indirect effect can be seen in the Fourier space (i.e. in the frequency domain rather than the time domain) of the transit light curve, where the ring can potentially stand out as an additional feature in the curve of the Fourier components as a function of frequency (Samsing 2015).

## 3.5   Conclusions

In this chapter, we discussed about a dozen methods that various researchers have worked out over little more than the past decade to search for moons and rings beyond the solar system. Although some of the original studies, in which these methods have been presented, expected that moons and rings could be detectable with the past CoRoT space mission or with the still active Kepler space telescope (Sartoretti & Schneider 1999; Barnes & Fortney 2004; Kipping et al. 2009; Heller 2014), no exomoon or exo-ring has been unequivocally discovered and confirmed as of today.

This is likely not because these extrasolar objects and structures do not exist, but because they are too small to be distinguished from the noise. Alternatively, and this is a more optimistic interpretation of the situation, those features could actually be detectable *and* present in the available archival data (maybe even in the HST archival data of transiting exoplanets), but they just haven't been found yet. The absence of numerous, independent surveys for exomoons and -rings lends some credence to this latter interpretation: Out of the several thousands of exoplanets and exoplanet candidates discovered with the Kepler telescope alone, only a few dozen have been examined for moons and rings with statistical scrutiny (Heising et al. 2015; Kipping et al. 2015).

It can be expected that the Kepler data will be fully analyzed for moons and rings within the next few years. Hence, a detection might still be possible. Alternatively, an independent, targeted search for moons/rings around planets transiting apparently bright stars – e.g. using the HST, CHEOPS, or a 10 m scale ground-based telescope, might deliver the first discoveries in the next decade. If none of these searches would be proposed or proposed but not granted, then it might take more than a decade for the PLANetary Transits and Oscillations of stars (PLATO) mission (Rauer et al. 2014) to find an exomoon or exo-ring in its large space-based survey of bright stars. Either way, it can be expected that exomoon and exo-ring discoveries will allow us a much deeper understanding of planetary systems than is possibly obtainable by planet observations alone.

# Chapter 4

# Habitability of Exomoons

As explained in Sect. 1.3, astronomers regularly focus on surface habitability in their search for life beyond the solar system because this sort of life has the highest (if any) chances to be remotely detectable in the foreseeable future. Life on the surface has the observational advantage of interacting with the atmosphere, which might result in the chemical imprints of life that could be detectable through transit spectroscopy, e.g. with the James Webb Space Telescope (Beichman et al. 2014) or its successor missions HabEx or LUVOIR (Wang et al. 2018). The detection of atmospheric biomarkers on exomoons through transit spectroscopy or similar techniques using transits will be more challenging than it will be for planets, because the moon signal will almost inevitably be contaminated with the signal from its host planet. The separation of the two signals will be possible in principle by using observations of both planet and moon in transit and of only the moon in transit, but this approach will increase the amount of transit observations required (Kaltenegger 2010). Other means of remote detection of life on exomoons include observations of surface features in the photometric phase curves of planet-moon systems (Cowan et al. 2012; Forgan 2017), which also restricts this method to life on the surface. For life to exist on the surface of an exomoon, surface water is required (see Sect. 1.3). As a consequence, when referring to habitable exomoons in this thesis, we refer to exomoons that have the potential of hosting liquid water on the surface.

Although the basic concept of habitability is identical for planets and moons, there is a range of physical processes that are relevant to exomoon habitability and that might be irrelevant for exoplanets.

## 4.1 Light as a Source of Energy

The effect of planetary illumination onto an exomoon has first been described by Heller (2012) and Heller & Barnes (2013a). In this model, the moon's orbit around the planet is assumed to be coplanar with the orbit of the planet-moon barycenter around around the star, and it is assumed that the spin axis of the moon is aligned with the orbital plane normal. Planetary oblateness is neglected and all objects involved are treated as spheres. The planet is divided into a day side that reflects a certain fraction of the incoming star light (the magnitude of which is determined by the planetary Bond albedo, $\alpha_{\mathrm{p}}$) and a night side, which does not receive or reflect any stellar illumination. Both hemispheres are also emitters of thermal radiation, the amount of which is determined by their effective temperatures. A sketch of the model is shown in Fig. 4.1.

### 4.1.1 Reflected Star Light

The vast majority of planets known today, including the eight planets of the solar system and over 4000 planets and candidates known beyond, are in orbit around single stars. These planets have one principal light source, neglecting the possibility of additional light coming from their moons. Moons, however, have two principal light sources, namely, the star and the planet (Heller & Barnes 2013a).

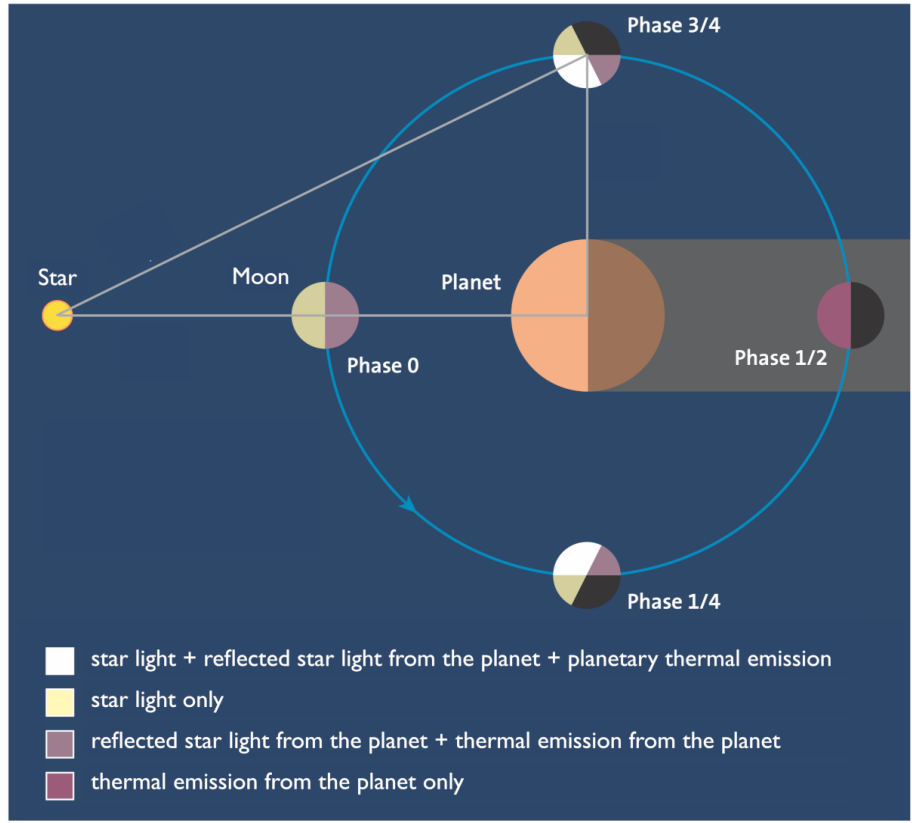

Figure 4.1: Illumination on an exomoon changes over the course of its orbital revolution around the host planet. In our model that invokes coplanar orbits around the star (of the planet-moon barycenter) and around the planet (of the moon), the moon's orbit starts an an orbital phase of 0 and then proceeds in a counter-clockwise fashion to 1. The center of the planetary eclipse occurs at a phase of 1/2. The different contributions of light falling onto the moon are explained in the legend in the lower left corner. This illustration is a modified version of a graphic shown in Heller (2013).

On Earth, the amount of reflected sun light received from the Moon is irrelevant to our climate.[1] The Earth receives about 1365 W m$^{-2}$ of photonic energy from the sun at the top of the atmosphere. For comparison, even the full moon only contributes some 0.01 W m$^{-2}$ of reflected sun light to the Earth's top-of-the-atmosphere photon flux, or just about five orders of magnitude less than the solar illumination. Exomoons around giant planets, however, can receive tens or even hundreds of W m$^{-2}$ of reflected star light from their planets, provided that the planet-moon system is close to the star and that the moon is close to its planet (Heller & Barnes 2013a).

### 4.1.2   Thermal Irradiation

In addition to the star light reflected off of the planetary surface (or atmosphere) back to an exomoon, these satellites can be subject to significant amounts of thermal illumination from their host planets (Heller & Barnes 2015; Heller 2016). This is particularly true for moons around young giant planets, which can be as hot as the most low-mass stars on the main sequence, that is, about 2000 K. Similar effective temperatures have indeed been deduced from direct imaging observations of young giant planets, e.g. of 1RXS J160929.1−210524 b (Lafrenière et al. 2008). Numerical integrations of the one-dimensional spherically symmetric structure equations of giant planets with various assumptions about their core masses show that the most massive planets can be as luminous as one millionth of the solar

---

[1] That said, it is certainly clear that the lunar cycle regulates a vast range of biological processes on Earth (Foster & Roenneberg 2008)

luminosity during the first billion years after formation (Mordasini 2013). Taking into account that moons around Jovian and super-Jovian planets supposedly form at distances of around ten planetary radii (Heller & Pudritz 2015b), or hundreds of times closer to their planet than Earth to the sun, this means that these young moons may receive up to $100 \, \mathrm{W \, m^{-2}}$ of thermal infrared radiation from their planets. As these exoplanets cool on time scales of millions to billions of years, the amount of infrared irradiation received by their potential moons also vanishes until, at some point, star light will entirely dominate the light budget on the moon (Heller & Barnes 2015).

The orbital architecture of planet-moon systems can lead to frequent stellar eclipses on the moon (Scharf 2006; Heller 2012). This may lead to the odd situation in which the subplanetary point on the moon, where the planet is in the zenith above the observer, experiences its darkest time of the day at noon (during stellar eclipse), while midnight can be quite bright with the planet acting as a giant mirror for the star light in the zenith (Heller & Barnes 2013a). These configurations are illustrated at phases 0 and 1/2, respectively, in Fig. 4.1.

## 4.2 Tidal Heating as a Source of Internal Energy

The amount of frictional heating that emanates the surface of a tidally heated body with radius $R_s$ and at a distance $a$ to its primary (with a mass $M_p$) scales roughly as $R_s^3/a^9 \times M_p^3$. For the principal moons around Jupiter (Ganymede) and Saturn (Titan), this term (in units of $\mathrm{kg^3 \, m^{-6}}$) is about $10^{20}$ or $10^{18}$, respectively. For the principal planet in the solar system, Jupiter, and taking the sun as the primary, this terms is only $10^7$. Even for the innermost planet, Mercury, we only have $10^{13}$. For Earth-sized planets in the habitable zone around M dwarf stars ($a = 0.1 \, \mathrm{AU}$), the term becomes more relevant with about $10^{17}$. But generally speaking, it can be expected that moons (habitable or not) tend to be subject to stronger tidal heating rates than planets in the stellar habitable zone (Heller et al. 2014).

The orbital evolution driven by tidal friction results in a circular orbit of the planet around the star, in which the planetary spin axis is aligned with its orbital plane normal and in which the rotation period of the planet equals its orbital period around the planet. As a consequence, tidal heating ceases typically within much less than the first $100 \, \mathrm{Myr}$ years after moon formation (Porter & Grundy 2011) in a two-body system and in the absence of continuous gravitational perturbations, e.g. by other moons or planets, or by the star. That being said, gravitational perturbations of exoplanet-exomoon system could be quite common. The most prominent example in the solar system is the 1:2:4 orbital mean motion resonance of the inner three Galilean moons Io, Europa, and Ganymede, the so-called Laplace resonance. This resonance prevents Io's orbit from being circularized by tides and currently forces its eccentricity to a value of about 0.0041. The resulting magnitude of tidal heating has been predicted to melt parts of the structure of the moon. This prediction was published by Peale et al. (1979) just shortly before the Voyager 1 probe flew by Jupiter and its moons and actually observed active volcanos on Io in 1979 (Morabito et al. 1979). A more recent view of an active volcanic region on Io, taken by the New Horizons spacecraft in 2007, is shown in Fig. 4.2.

While Io's tidal heating is maintained by gravitational interaction with its neighboring moons, moons beyond the solar system could regularly be subject to tidal heating driven by the gravitational perturbations from their host stars. As predicted by Heller (2012) and then shown by Zollinger et al. (2017) using $N$-body numerical integrations of the combined tidal and secular forces, moons in the stellar habitable zones around M dwarfs could even struggle to be habitable. With the habitable zone being so close around M dwarfs (Kasting et al. 1993; Kopparapu et al. 2013), moons around habitable zone planets will inevitably be subject to stellar perturbations. The resulting non-circularity of their orbits will lead to continuous tidal heating, possibly over billions of years. This additional internal heating would add to the warming effect of the stellar illumination in the habitable zone and therefore drive a moist or even runaway greenhouse effect on these moons, which would not allow for the presence of liquid surface water and ultimately lead to desiccation through hydrogen escape into space via photolysis in the upper atmosphere.

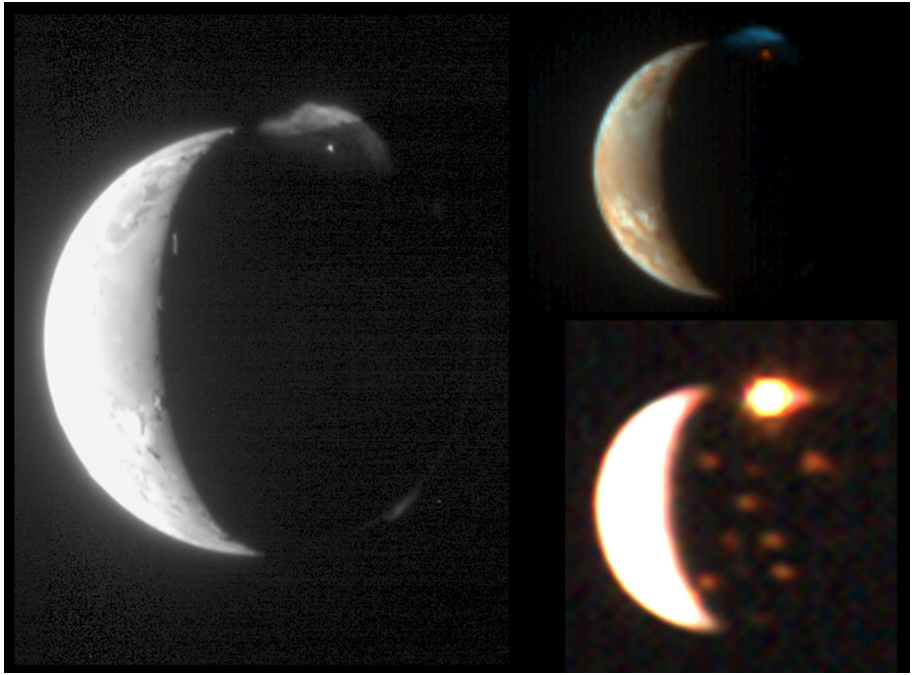

Figure 4.2: This montage shows three images of Io taken almost simultaneously by different instruments aboard the New Horizons spacecraft on 1 March 2007 (UT). *Left*: The Long Range Reconnaissance Imager presents details on Io's sunlit crescent and in the partially sunlit plume from the Tvashtar volcano. The source of the plume betrays its location as a bright nighttime glow of hot lavas. *Top right*: The Multispectral Visible Imaging Camera image shows the contrasting colors of the red lava and blue plume at Tvashtar, and the sulfur and sulfur dioxide deposits on Io's sunlit surface. *Bottom right*: The Linear Etalon Imaging Spectral Array image shows that Tvashtar's glow is even more intense at infrared wavelengths and reveals the glows of over ten fainter volcanic hot spots on Io's nightside. Image credit: NASA/Johns Hopkins University Applied Physics Laboratory/Southwest Research Institute.

## 4.3   A Concept of Exomoon Habitability Based on Energy Budged

All things from Sects. 4.1 and 4.2 combined, we can define the circumplanetary habitable space for moons. As in the scenario for planets, we consider a moon habitable if its globally averaged surface temperature is above 0° and the moon is not subject to a runaway greenhouse effect, which would ultimately free the moon of any water (Goldblatt & Watson 2012). As a principal threshold for the maximum top-of-the-atmosphere energy flux for the moon to be habitable, we estimate the runaway greenhouse limit ($F_{RG}$) from the framework of (Pierrehumbert 2010). For an Earth-sized object, this framework gives a critical flux of 295 W m$^{-2}$.

This limit needs to be compared to the total energy budget of the moon under consideration. Let us imagine a satellite with a radius $R_s$, an optical albedo $\alpha_{s,opt}$, and an infrared albedo $\alpha_{s,IR}$ in an orbit around a planet with semimajor axis $a_s$ and eccentricity $e_s$. Let us further assume that the satellite has an effective energy redistribution factor $f_s$, which accounts for the fact that the cross section of the moon is $\pi R_s^2$ while its surface area is $4\pi R_s^2$. We also consider that the moon spends a fraction $1 - x_s$ of its orbit in the shadow of the planet, neglecting the effect of ingress and egress. In other words, $x_s$ is the fraction of the satellite orbit that is *not* spent in the shadow of the planet. Moving on to the planet, we consider a radius $R_p$, a Bond albedo $\alpha_p$, an orbital semimajor axis $a_p$, and an eccentricity $e_p$ around the star[2]. Finally, the stellar luminosity is referred to as $L_\star$ and the planetary luminosity is $L_{\mathrm{p}}$. The globally averaged energy flux received by the moon from the star and from the planet can then be estimated as

$$\overline{F}_{\mathrm{s}}^{\mathrm{glob}} = \frac{L_* \left(1 - \alpha_{\mathrm{s,opt}}\right)}{4\pi a_{\mathrm{p}}^2 \sqrt{1 - e_{\mathrm{p}}^2}} \left(\frac{x_{\mathrm{s}}}{4} + \frac{\pi R_{\mathrm{p}}^2 \alpha_{\mathrm{p}}}{f_{\mathrm{s}} 2 a_{\mathrm{s}}^2}\right) + \frac{L_{\mathrm{p}} \left(1 - \alpha_{\mathrm{s,IR}}\right)}{4\pi a_{\mathrm{s}}^2 f_{\mathrm{s}} \sqrt{1 - e_{\mathrm{s}}^2}} + h_{\mathrm{s}} + W_{\mathrm{s}}\,. \qquad (4.1)$$

---

[2]Strictly speaking, the orbit of the planet around the star is defined by the three-body motions of the star-planet-moon system, in which the planet (and moon) orbit their common barycenter, which itself is in a Keplerian orbit around the

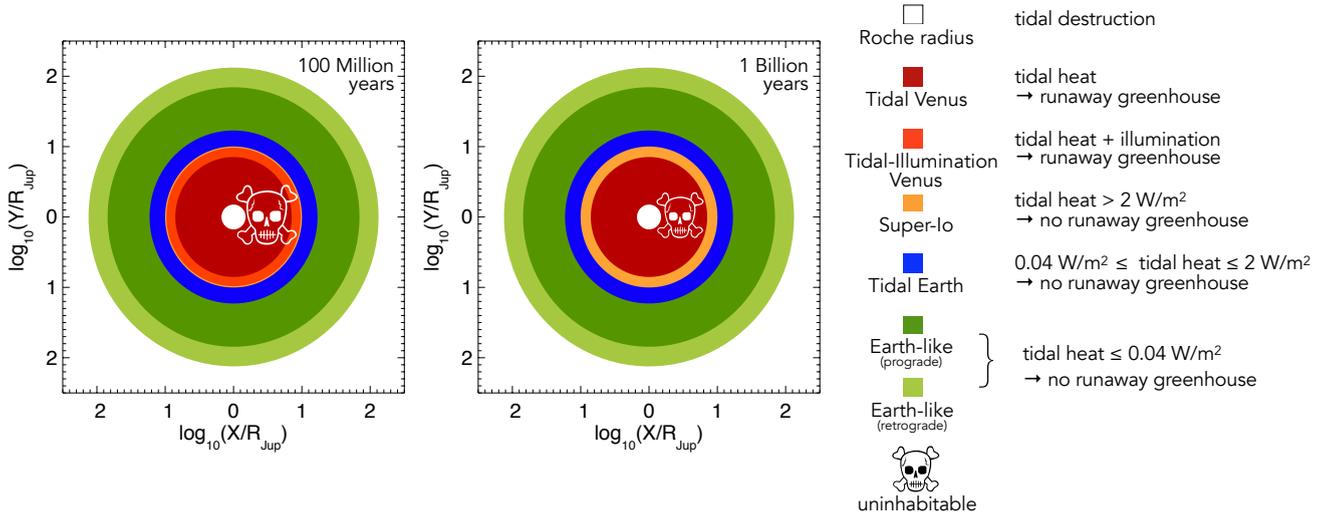

Figure 4.3: Habitability of circumplanetary orbits for an Earth-sized satellite orbiting a planet of ten Jupiter masses with an orbital eccentricity of 0.001. *Left*: At an age of 100 Myr, the planet is still relatively hot, which makes the moon uninhabitable within a distance of about $10\,R_{\mathrm{Jup}}$. *Right*: 900 Myr later, the planet has cooled substantially and the habitable edge, which defines the radial extent of the the uninhabitable space around the planet, has shrunk to about $7\,R_{\mathrm{Jup}}$. The system is modelled at 1 AU from a sun-like host star. Tidal heating contributes about $10\,\mathrm{W\,m^{-2}}$ in both panels. This plot is a modified version of Fig. 4 from Heller & Barnes (2015) that the author of this thesis presented at the astronomy seminars at Cornell University (11 May 2015) and the Center for Astrophysics at Harvard University (1 July 2015).

The first term defines the direct stellar flux received by the satellite. The terms in braces account for eclipses (e.g. $x_{\mathrm{s}} = 1$ means no eclipses) and for the amount of reflected sun light from the planet (e.g. $f_{\mathrm{s}} = 4$ means perfect redistribution of the planetary light over the moon's surface). The term invoking $L_{\mathrm{p}}$ and $1 - \alpha_{\mathrm{s,IR}}$ relates to the infrared radiation emitted by the planet and reflected by the planet. Finally, $h_{\mathrm{s}}$ is the heating of the satellite by tides and $W_{\mathrm{s}}$ is any other source of heat, e.g. heat from radioactive decay.

As an example, if we switch the roles of the moon and the planet and apply Eq. (4.1) to the Earth, on which tidal heating and other sources of heat are negligible[3] and where the effects of lunar eclipses and illumination are entirely negligible, we obtain $\overline{F}_{\mathrm{s}}^{\mathrm{glob}} = 239\,\mathrm{W\,m^{-2}}$ using an optical albedo of 70 % and zero orbital eccentricity.

A primitive version of Eq. (4.1) was first developed by Heller & Barnes (2013a), who neglected the effects of eclipses and the possibility of inefficient energy redistribution. Heller (2012) introduced the effect of eclipses and in Dobos et al. (2017) we presented the formula as shown above. In Heller & Barnes (2015) we then introduced the effects caused by the evolution of stellar and planetary

---

barycenter of the system. For $M_{\mathrm{p}} \gg M_{\mathrm{s}}$, however, our approximation is correct.

[3]The total flux of internal heat through the surface of the present-day Earth is about $0.086\,\mathrm{W\,m^{-2}}$ on average, of which approximately $0.04\,\mathrm{W\,m^{-2}}$ are due to the radiogenic decay of long-lived isotopes $^{40}$K, $^{232}$Th, $^{235}$U, and $^{238}$U (Zahnle et al. 2007). The residual heat is of primordial origin from the Earth's accretion process and from tides raised by the Moon.

luminosities to identify several classes of exomoon habitability around extrasolar planets (see Fig. 4.3). The radial extent of these classes is defined by the $a_s^{-2}$ law for the planetary illumination onto the moon and the $a_s^{-9}$ law for the effect of tidal heating in the moon. Further investigations of exomoon climates were carried out by Forgan & Kipping (2013) and Forgan & Yotov (2014), who developed a 1D latitudinal energy balance model to assess exomoon habitability under the combined stellar and planetary illumination including a geophysical carbon-silicate as a global thermostat similar to the one observed on Earth. Then Forgan & Dobos (2016) added a more sophisticated tidal model that includes the feedback mechanisms from tidal heating on the rheology of the moon (and vice versa). In Haqq-Misra & Heller (2018) we expanded exomoon climate studies into three spatial dimensions by applying an idealized 3D general circulation model with simplified hydrologic, radiative, and convective processes. More detailed investigations of the possible processes that regulate atmospheric escape on exomoons were done by Lehmer et al. (2017), so that the field of exomoon habitability and exomoon climates has evolved to a state in which the basic processes have been identified and in which further progress might only be possible using exomoon observations that will allow us to test these predictions.

# Part II

# Peer-Reviewed Journal Publications

# Chapter 5

# Formation of Moons in the Accretion Disks Around Young Giant Planets

## 5.1 Water Ice Lines and the Formation of Giant Moons Around Super-Jovian Planets (Heller & Pudritz 2015b)

Contribution:

RH did the literature research, worked out the mathematical framework, translated the math into computer code, performed the simulations, created all figures, led the writing of the manuscript, and served as a corresponding author for the journal editor and the referees.

# WATER ICE LINES AND THE FORMATION OF GIANT MOONS AROUND SUPER-JOVIAN PLANETS

René Heller[1,2] and Ralph Pudritz[1]

Origins Institute, McMaster University, Hamilton, ON L8S 4M1, Canada
rheller@physics.mcmaster.ca, pudritz@physics.mcmaster.ca
*Draft version May 6, 2015*

## ABSTRACT

Most of the exoplanets with known masses at Earth-like distances to Sun-like stars are heavier than Jupiter, which raises the question of whether such planets are accompanied by detectable, possibly habitable moons. Here we simulate the accretion disks around super-Jovian planets and find that giant moons with masses similar to Mars can form. Our results suggest that the Galilean moons formed during the final stages of accretion onto Jupiter, when the circumjovian disk was sufficiently cool. In contrast to other studies, with our assumptions, we show that Jupiter was still feeding from the circumsolar disk and that its principal moons cannot have formed after the complete photoevaporation of the circumsolar nebula. To counteract the steady loss of moons into the planet due to type I migration, we propose that the water ice line around Jupiter and super-Jovian exoplanets acted as a migration trap for moons. Heat transitions, however, cross the disk during the gap opening within ≈ $10^4$ yr, which makes them inefficient as moon traps and indicates a fundamental difference between planet and moon formation. We find that icy moons larger than the smallest known exoplanet can form at about 15 - 30 Jupiter radii around super-Jovian planets. Their size implies detectability by the *Kepler* and *PLATO* space telescopes as well as by the *European Extremely Large Telescope*. Observations of such giant exomoons would be a novel gateway to understanding planet formation, as moons carry information about the accretion history of their planets.

*Keywords:* accretion disks – planets and satellites: formation – planets and satellites: gaseous planets – planets and satellites: physical evolution – planet-disk interactions

## 1. CONTEXT

While thousands of planets and planet candidates have been found outside the solar system, some of which are as small as the Earth's moon (Barclay et al. 2013), no moon around an exoplanet has yet been observed. But if they transit their host stars, large exomoons could be detectable in the data from the *Kepler* space telescope or from the upcoming *PLATO* mission (Kipping et al. 2012; Heller 2014). Alternatively, if a large moon transits a self-luminous giant planet, the moon's planetary transit might be detectable photometrically or even spectroscopically, for example with the *European Extremely Large Telescope* (Heller & Albrecht 2014). It is therefore timely to consider models for exomoon formation.

Large moons can form in the dusty gas disks around young, accreting gas giant planets. Several models of moon formation posit that proto-satellites can be rapidly lost into the planet by type I migration (Pollack & Reynolds 1974; Canup & Ward 2002, 2006; Sasaki et al. 2010). The water ($H_2O$) condensation ice line can act as a planet migration trap that halts rapid type I migration in circumstellar disks (Kretke & Lin 2007; Hasegawa & Pudritz 2011, 2012), but this trap mechanism has not been considered in theories of moon formation so far. The position of the $H_2O$ ice line has sometimes been modeled ad hoc (Sasaki et al. 2010) to fit the $H_2O$ distribution in the Galilean moon system (Mosqueira & Estrada 2003a).

An alternative explanation for the formation of the Galilean satellites suggests that the growing Io, Europa, and Ganymede migrated within an optically thick accretion disk the size of about the contemporary orbit of Callisto and accreted material well outside their instantaneous feeding zones (Mosqueira & Estrada 2003a,b). In this picture, Callisto supposedly formed in an extended optically thin disk after Jupiter opened up a gap in the circumsolar disk. Callisto's material was initially spread out over as much as 150 Jupiter radii ($R_{Jup}$), then aggregated on a $10^6$ yr timescale, and migrated to Callisto's current orbital location. Yet another possible formation scenario suggests that proto-satellites drifted outwards as they were fed from a spreading circumplanetary ring in a mostly gas-free environment (Crida & Charnoz 2012).

We here focus on the "gas-starved" model of an actively supplied circumplanetary disk (CPD) (Makalkin et al. 1999; Canup & Ward 2002) and determine the time-dependent radial position of the $H_2O$ ice line. There are several reasons why water ice lines could play a fundamental role in the formation of giant moons. The total mass of a giant planet's moon system is sensitive to the location of the $H_2O$ ice line in the CPD, where the surface density of solids ($\Sigma_s$) increases by about factor of three (Hayashi 1981), because the mass of the fastest growing object is proportional to $\Sigma_s^{3/2}$. This suggests that the most massive moons form at or beyond the ice line. In this regard, it is interesting that the two lightest Galilean satellites, Io (at $6.1\,R_{Jup}$ from the planetary core) and Europa (at $9.7\,R_{Jup}$), are mostly rocky with bulk densities > $3\,g\,cm^{-3}$, while the massive moons Ganymede (at $15.5\,R_{Jup}$) and Callisto (at $27.2\,R_{Jup}$) have densities below $2\,g\,cm^{-3}$ and consist by about 50 % of $H_2O$ (Show-man & Malhotra 1999).[3] It has long been hypothesized that Jupiter's CPD dissipated when the ice line was between the orbits of Europa and Ganymede, at about 10 to 15 $R_{\rm Jup}$ (Pollack & Reynolds 1974). Moreover, simulations of the orbital evolution of accreting proto-satellites in viscously dominated disks around Jupiter, Saturn, and Uranus indicate a universal scaling law for the total mass of satellite systems ($M_{\rm T}$) around the giant planets in the solar system (Canup & Ward 2006; Sasaki et al. 2010), where $M_{\rm T} \approx 10^{-4}$ times the planetary mass ($M_{\rm p}$).

[1] Department of Physics and Astronomy, McMaster University
[2] Postdoctoral fellow of the Canadian Astrobiology Training Program

## 2. METHODS

In the Canup & Ward (2002) model, the accretion rate onto Jupiter was assumed to be time-independent. Canup & Ward (2006) focussed on the migration and growth of proto-moons, but they did not describe their assumptions for the temperature profile in the planetary accretion disk. Others used analytical descriptions for the temporal evolution of the accretion rates or for the movement of the $H_2O$ ice lines (Makalkin & Dorofeeva 1995; Mousis & Gautier 2004; Canup & Ward 2006; Sasaki et al. 2010; Ogihara & Ida 2012) or they did not consider all the energy inputs described above (Alibert et al. 2005).

We here construct, for the first time, a semi-analytical model for the CPDs of Jovian and super-Jovian planets that is linked to pre-computed planet evolution tracks and that contains four principal contributions to the disk heating: (i) viscous heating, (ii) accretion onto the CPD, (iii) planetary irradiation, and (iv) heating from the ambient circumstellar nebula. Compared to previous studies, this setup allows us to investigate many scenarios with comparatively low computational demands, and we naturally track the radial movement of the $H_2O$ ice line over time. This approach is necessary, because we do not know any extrasolar moons that could be used to calibrate analytical descriptions for movement of the ice line around super-Jovian planets. We focus on the large population of Jovian and super-Jovian planets at around 1 AU from Sun-like stars, several dozens of which had their masses determined through the radial velocity technique as of today.

### 2.1. Disk Model

The disk is assumed to be axially symmetric and in hydrostatic equilibrium. We adopt a standard viscous accretion disk model (Canup & Ward 2002, 2006), parameterized by a viscosity parameter $\alpha$ ($10^{-3}$ in our simulations) (Shakura & Sunyaev 1973), that is modified to include additional sources of disk heating (Makalkin & Dorofeeva 2014). We consider dusty gas disks around young ($\approx 10^6$ yr old), accreting giant planets with final masses beyond that of Jupiter ($M_{\rm Jup}$). These planets accrete gas and dust from the circumstellar disk. Their accretion becomes increasingly efficient, culminating in the so-called runaway accretion phase when their masses become similar to that of Saturn (Lissauer et al. 2009; Mordasini 2013). Once they reach about a Jovian mass

(depending on their distance to the star, amongst others), they eventually open up a gap in the circumstellar disk and their accretion rates drop rapidly. Hence, the formation of moons, which grow from the accumulation of solids in the CPD, effectively stops at this point or soon thereafter. A critical link between planet and moon formation is the combined effect of various energy sources (see the four heating terms described above) on the temperature distribution in the CPD and the radial position of the $H_2O$ ice line.

In our disk model, the radial extent of the inner, optically thick part of the CPD, where moon formation is suspected to occur, is set by the disk's centrifugal radius ($r_{\rm cf}$). At that distance to the planet, centrifugal forces on an object with specific angular momentum $j$ are balanced by the planet's gravitational force. Using 3D hydrodynamic simulations, Machida et al. (2008) calculated the circumplanetary distribution of the angular orbital momentum in the disk and demonstrated the formation of an optically thick disk within about 30 $R_{\rm Jup}$ around the planet. An analytical fit to their simulations yields (Machida et al. 2008)

$$j(t) = \begin{cases} 7.8 \times 10^{11} \left( \dfrac{M_{\rm p}(t)}{M_{\rm Jup}} \right) \left( \dfrac{a_{\star \rm p}}{1\,{\rm AU}} \right)^{7/4} {\rm m^2\,s^{-1}} \\ \qquad\qquad\qquad\qquad\quad {\rm for}\ M_{\rm p} < M_{\rm Jup} \\[2mm] 9.0 \times 10^{11} \left( \dfrac{M_{\rm p}(t)}{M_{\rm Jup}} \right)^{2/3} \left( \dfrac{a_{\star \rm p}}{1\,{\rm AU}} \right)^{7/4} {\rm m^2\,s^{-1}} \\ \qquad\qquad\qquad\qquad\quad {\rm for}\ M_{\rm p} \geq M_{\rm Jup}\ \ , \end{cases}$$

$$\text{(1)}$$

where we introduced the variable $t$ to indicate that the planetary mass ($M_{\rm p}$) evolves in time. The centrifugal radius is then given by

$$r_{\rm cf} = \frac{j^2}{GM_{\rm p}} \quad , \tag{2}$$

with $G$ as Newton's gravitational constant. For Jupiter, this yields a centrifugal radius of about 22 $R_{\rm Jup}$, which is slightly less than the orbital radius of the outermost Galilean satellite, Callisto, at roughly 27 $R_{\rm Jup}$. Part of this discrepancy is due to thermal effects that are neglected in the (Machida et al. 2008) disk model. Machida (2009) investigated these thermal effects on the centrifugal disk size by comparing isothermal with adiabatic disk models. They found that adiabatic models typically yield larger specific angular momentum at a given planetary distance, which then translate into larger centrifugal disk radii that nicely match the width of Callisto's orbit around Jupiter.[4] We thus introduce a thermal correction factor of 27/22 to the right-hand side of Equation (2) following Machida (2009), and therefore include Callisto at the outer edge of the optically thick part of our disk model. We note, though, that this slight rescaling hardly affects the general results of our simulations.

---

[3] Amalthea, although being very close to Jupiter (at 2.5 $R_{\rm Jup}$), has a very low density of about $0.86\,{\rm g\,cm^{-3}}$ (Anderson et al. 2005), which seems to be at odds with the compositional gradient in the Galilean moons. But Amalthea likely did not form at its current orbital position as is suggested by the presence of hydrous minerals on its surface (Takato et al. 2004).

[4] In particular, their isothermal model M1I, used to fit our Eq. (1), has a radial specific momentum distribution that is about 1.1 times smaller at Callisto's orbital radius than their adiabatic model M1A2. This offset means an $(1.1)^2 = 1.21$-fold increase of the centrifugal radius, which nicely fits to our correction factor of $27/22 \approx 1.23$.

Recent 3D global magnetohydrodynamical (MHD) simulations by Gressel et al. (2013, see their Sect. 6.3) produce circumplanetary surface gas densities that agree much better with the "gas-starved" model of Canup & Ward (2006), which our model is derived from, than with the "minimum mass" model of Mosqueira & Estrada (2003a).[5] The latter authors argue that Callisto formed in the low-density regions of an extended CPD with high specific angular momentum after Jupiter opened up a gap in the circumstellar disk. In their picture, the young Callisto accreted material from orbital radii as wide as 150 $R_{\rm Jup}$. In our model, however, Callisto forms in the dense, optically thick disk, where we suspect most of the solid material to pile up. Simulations by Canup & Ward (2006) and Sasaki et al. (2010) show that our assumption can well reproduce the masses and orbits of the Galilean moons.

The disk is assumed to be mostly gaseous with an initial dust-to-mass fraction $X$ (set to 0.006 in our simulations, Hasegawa & Pudritz 2013). Although we do not simulate moon formation in detail, we assume that the dust would gradually build planetesimals, either through streaming instabilities in the turbulent disk (Johansen et al. 2014) or through accumulation within vortices (Klahr & Bodenheimer 2003), to name just two possible formation mechanisms. The disk is parameterized with a fixed Planck opacity ($\kappa_{\rm P}$) in any of our simulations, but we tested various values. The fraction of the planetary light that contributes to the heating of the disk surface is parameterized by a coefficient $k_{\rm s}$, typically between 0.1 and 0.5 (Makalkin & Dorofeeva 2014). This quantity must not be confused with the disk albedo, which can take values between almost 0 and 0.9, depending on the wavelength and the grain properties (D'Alessio et al. 2001). The sound velocity in the hydrogen (H) and helium (He) disk gas usually depends on the mean molecular weight ($\mu$) and the temperature of the gas, but in the disk midplane it can be approximated (Keith & Wardle 2014) as $c_{\rm s} = 1.9\,{\rm km\,s^{-1}}\sqrt{T_{\rm m}(r)/1000\,{\rm K}}$ for midplane temperatures $T_{\rm m} \lesssim 1000\,{\rm K}$. At these temperatures, ionization can be neglected and $\mu = 2.34\,{\rm kg\,mol^{-1}}$. Further, the disk viscosity is given by $\nu = \alpha c_{\rm s}^2/\Omega_{\rm K}(r)$, with $\Omega_{\rm K}(r) = \sqrt{GM_{\rm p}/r^3}$ as the Keplerian orbital frequency.

The steady-state gas surface density ($\Sigma_{\rm g}$) in the optically thick part of the disk can be obtained by solving the continuity equation for the infalling gas at the disk's centrifugal radius (Canup & Ward 2006), which yields

$$\Sigma_{\rm g}(r) = \frac{\dot{M}}{3\pi\nu} \times \frac{\Lambda(r)}{l} \tag{3}$$

where

$$\Lambda(r) = 1 - \frac{4}{5}\sqrt{\frac{r_{\rm cf}}{r_{\rm d}}} - \frac{1}{5}\left(\frac{r}{r_c}\right)^2$$

$$l = 1 - \sqrt{\frac{R_{\rm p}}{r_{\rm d}}} \tag{4}$$

is derived from a continuity equation for the infalling material and based on the angular momentum delivered to the disk (Canup & Ward 2006; Makalkin & Dorofeeva 2014), and $\dot{M}$ is the mass accretion rate through the CPD, assumed to be equal to the mass dictated by the pre-computed planet evolution models (Mordasini 2013). We set $r_{\rm d} = R_{\rm H}/5$, which yields $r_{\rm d} \approx 154\,R_{\rm Jup}$ for Jupiter (Sasaki et al. 2010; Makalkin & Dorofeeva 2014).

The effective half-thickness of the homogeneous flared disk, or its scale height, is derived from the solution of the vertical hydrostatic balance equation as

$$h(r) = \frac{c_{\rm s}(T_{\rm m}(r))r^{3/2}}{\sqrt{GM_{\rm p}}} \quad . \tag{5}$$

We adopt the standard assumption of vertical hydrostatic balance in the disk and assume that the gas density ($\rho_{\rm g}$) in the disk decreases exponentially with distance from the midplane as per

$$\rho_{\rm g}(r) = \rho_0\,{\rm e}^{\frac{-z^2}{2h(r)^2}} \tag{6}$$

where $z$ is the vertical coordinate and $\rho_0$ the gas density in the disk midplane. The gas surface density is given by vertical integration over $\rho(r, z)$, that is,

$$\Sigma_{\rm g}(r) = \int_{-\inf}^{+\inf} dz\ \rho(r, z) \quad . \tag{7}$$

Inserting Equation (6) into Equation (7), the latter can be solved for $\rho_0$ and we obtain

$$\rho_0(r) = \sqrt{\frac{2}{\pi}}\,\frac{\Sigma_{\rm g}(r)}{2h(r)} \quad , \tag{8}$$

which only depends on the distance $r$ to the planet. At the radiative surface level of the disk, or photospheric height ($z_{\rm s}$), the gas density equals

$$\rho_{\rm s}(r) = \rho_0\,{\rm e}^{\frac{-z^2}{2h(r)^2}} \tag{9}$$

where we calculate $z_{\rm s}$ as

$$z_{\rm s}(r) = {\rm erf}^{-1}\left(1 - \frac{2}{3}\frac{2}{\Sigma_{\rm g}(r)\kappa_{\rm P}}\right)\sqrt{2}h(r) \quad . \tag{10}$$

The latter formula is derived using the definition of the disk's optical depth

$$\tau = \int_{z_{\rm s}}^{+\inf} dz\ \kappa_{\rm P}\rho(r, z) = \kappa_{\rm P}\int_{z_{\rm s}}^{+\inf} dz\ \rho(r, z) \tag{11}$$

and our knowledge of $\tau = 2/3$ at the radiating surface level of the disk, which gives

---

$$\frac{2}{3} = \kappa_P \int_{z_s}^{+\inf} dz\ \rho_g(r, z)$$

$$\Leftrightarrow z_s(r) = \mathrm{erf}^{-1}\left(1 - \frac{2}{3}\sqrt{\frac{2}{\pi}}\frac{1}{h(r)\rho_0(r)\kappa_P}\right)\sqrt{2}h(r) \quad. \tag{12}$$

Using Equation (8) for $\rho_0(r)$ in Equation (12), we obtain Equation (10).

Following the semi-analytical disk model of Makalkin & Dorofeeva (2014), the disk surface temperature is given by the energy inputs of various processes as per

$$T_s(r) = \left(\frac{1 + (2\kappa_P \Sigma_g(r))^{-1}}{\sigma_{SB}}\left(F_{vis}(r) + F_{acc}(r)\right.\right.$$
$$\left.\left. + k_s F_p(r)\right) + T_{neb}^4\right)^{1/4}, \tag{13}$$

where

$$F_{vis}(r) = \frac{3}{8\pi}\frac{\Lambda(r)}{l}\dot{M}\Omega_K(r)^2$$
$$F_{acc}(r) = \frac{X_d \chi G M_p \dot{M}}{4\pi r_{cf}^2 r}\ e^{-(r/r_{cf})^2}$$
$$F_p(r) = L_p \frac{\sin\left(\zeta(r) + \eta(r)\right)}{8\pi(r^2 + z_s^2)} \tag{14}$$

are the energy fluxes from viscous heating, accretion onto the disk, and the planetary illumination, and $T_{neb}$ denotes the background temperature of the circumstellar nebula (100 K in our simulations). The geometry of the flaring disk is expressed by the angles

$$\zeta(r) = \arctan\left(\frac{4}{3\pi}\frac{R_p}{\sqrt{r^2 + z_s^2}}\right)$$
$$\eta(r) = \arctan\left(\frac{dz_s}{dr}\right) - \arctan\left(\frac{z_s}{r}\right) \quad. \tag{15}$$

Taking into account the radiative transfer within the optically thick disk with Planck opacity $\kappa_P$, the midplane temperature can be estimated as (Makalkin & Dorofeeva 2014)

$$T_m(r)^5 - T_s(r)^4\ T_m(r) = \frac{12\mu\chi\kappa_P}{2^9\pi^2\sigma_{SB}R_g\gamma} \times \frac{\dot{M}^2}{\alpha}\Omega_K(r)^3$$
$$\times \left(\frac{\Lambda(r)}{l}\right)^2 q_s(r)^2 \quad, \tag{16}$$

where $\sigma_{SB}$ is the Stefan-Boltzmann constant, $R_{Gas}$ the ideal gas constant, $\gamma = 1.45$ the adiabatic exponent (or ratio of the heat capacities), and

$$q_s(r) = 1 - \frac{4}{3\kappa_P \Sigma_g(r)}$$

is the vertical mass coordinate at $z_s$.

### 2.2. Planet Evolution Tracks

We use a pre-computed set of planet formation models by Mordasini (2013) to feed our planet disk model with the fundamental planetary properties such as the planet's evolving radius ($R_p$), its mass, mass accretion rate ($\dot{M}_p$), and luminosity ($L_p$). Figure 1 shows the evolution of these quantities with black solid lines indicating an accreting gas giant that ends up with one Jupiter mass or about 318 Earth masses ($M_\oplus$). In total, we have seven models at our disposal, where the planets have final masses of 1, 2, 3, 5, 7, 10, and 12 $M_{Jup}$. These tracks are sensitive to the planet's core mass, which we assume to be 33 $M_\oplus$ for all planets. Jupiter's core mass is actually much lower, probably around 10 $M_\oplus$ (Guillot et al. 1997). Lower final core masses in these models translate into lower planetary luminosities at any given accretion rate. In other words, our results for the $H_2O$ ice lines around the Jupiter-mass test planet are actually upper, or outer, limits, and a more realistic evaluation of the conditions around Jupiter would shift the ice lines closer to the planet. Over the whole range of available planet tracks with core masses between 22 and 130 $M_\oplus$, we note that the planetary luminosities at shutdown are $10^{-4.11}$ and $10^{-3.82}$ solar luminosities, respectively. As the distance of the $H_2O$ ice line in a radiation-dominated disk scales with $L_p^{1/2}$, different planetary core masses would thus affect our results by less than ten percent.

The pre-computed planetary models cover the first few $10^6$ yr after the onset of accretion onto the planet. We interpolate all quantities on a discrete time line with a step size of 5,000 yr. At any given time, we feed Equations (14) with the planetary model and solve the coupled Equations (1)–(10) in an iterative framework. With $T_s$ provided by Equation (13), we finally solve the 5th order polynomial in Equation (16) numerically.

Once the planetary evolution models indicate that $\dot{M}_p$ has dropped below a critical shutdown accretion rate ($\dot{M}_{shut}$), we assume that the formation of satellites has effectively stopped. As an example, a Ganymede-sized moon can form once $\dot{M}_{shut} < M_{Gan}\,\mathrm{Myr}^{-1}$ ($M_{Gan}$ being the mass of Ganymede) and if the disk's remaining life time is $< 10^6$ yr (see Figure 1c). As $\dot{M}_p$ determines the gas surface density through Equation (3), different values for $\dot{M}_{shut}$ mean different distributions of $\Sigma_g(r)$. In particular, $\Sigma_g(r = 10 R_{Jup})$ equals $7.4 \times 10^2$, $9.7 \times 10^1$, and $7.8 \times 10^0\,\mathrm{kg\,m^{-2}}$ once $\dot{M}_{shut}$ reaches 100, 10, and 1 $M_{Gan}\,\mathrm{Myr}^{-1}$, respectively, for the planet that ends up with one Jupiter mass (see Figure 4 in Heller & Pudritz 2015).

In any single simulation run, $\kappa_P$ is assumed to be constant throughout the disk, and simulations of the planetary $H_2O$ ice lines are terminated once the planet accretes less than a given $\dot{M}_{shut}$. To obtain a realistic picture of a broad range of hypothetical exoplanetary disk properties, we ran a suite of randomized simulations, where $\kappa_P$ and $\dot{M}_{shut}$ were drawn from a lognormal probability density distribution. For $\log_{10}(\kappa_P/[\mathrm{m^2\,kg^{-1}}])$ we assumed a mean value of $-2$ and a standard variation of 1, and concerning the shutdown accretion rate we assumed a mean value of 1 for $\log_{10}(\dot{M}_{shut}/[M_{Gan}\,\mathrm{Myr}^{-1}])$ and a standard variation of 1.

To get a handle on the plausible surface absorptivi-ties of various disk, we tested different values of $k_{\rm s}$ between 0.1 and 0.5. For each of the seven test planets, we performed 120 randomized simulations of the disk evolution and then calculated the arithmetic mean distance of the final water ice line. The resulting distributions are skewed and non-Gaussian. Hence, we compute both the downside and the upside semi standard deviation (corresponding to $\sigma/2$), that is, the distance ranges that comprise $+68.3\,\%/2 = +34.15\,\%$ and $-34.15\,\%$ of the simulations around the mean. We also calculate the $2\sigma$ semi-deviation, corresponding to $+95.5\,\%/2 = +47.75\,\%$ and $-47.75\,\%$ around the mean. Downside and upside semi deviations combined deliver an impression of the asymmetric deviations from the mean, and their sum equals that of the Gaussian standard deviations.

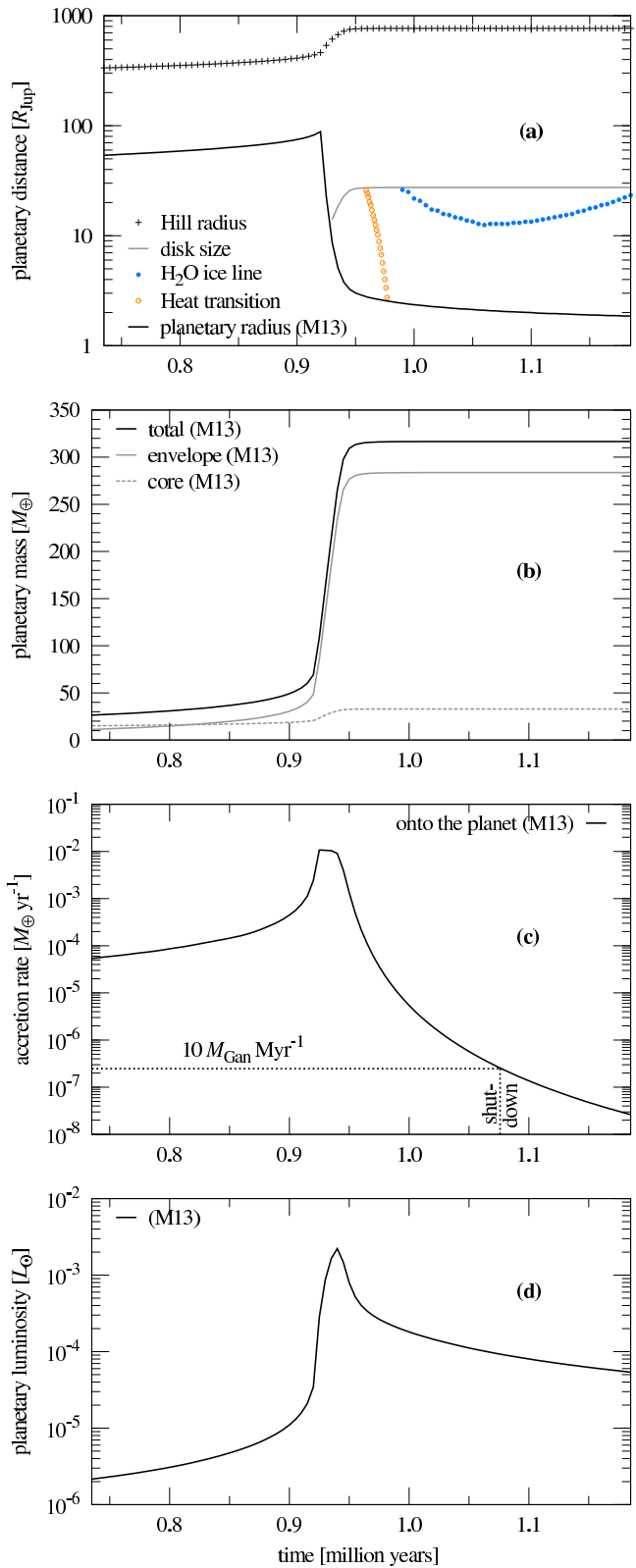

**Figure 1.** Evolution of a Jupiter-like model planet and its circumplanetary disk. Values taken from Mordasini (2013) are labeled "M13". **(a)** Circumplanetary disk properties. **(b)** Growth of the solid core, gaseous envelope, and total mass. **(c)** Total mass accretion rate. The dashed horizontal line indicates our fiducial shutdown rate for moon formation of $10\,M_{\rm Gan}\,{\rm Myr}^{-1}$. The dashed vertical line marks the corresponding shutdown for moon formation at about $1.08 \times 10^6$ yr. **(d)** Planetary luminosity evolution.

The total, instantaneous mass of solids in the disk at the time of moon formation shutdown is given as

$$M_{\rm s} = 2\pi X \left( \int_{r_{\rm in}}^{r_{\rm ice}} {\rm d}r\ r\Sigma_{\rm g}(r) + 3 \int_{r_{\rm ice}}^{r_{\rm cf}} {\rm d}r\ r\Sigma_{\rm g}(r) \right) \ , \quad (17)$$

where $r_{\rm in}$ is the inner truncation radius of the disk (assumed at Jupiter's corotation radius of $2.25\ R_{\rm Jup}$, Canup & Ward 2002), $r_{\rm ice}$ is the distance of the $H_2O$ ice line, and $r_{\rm cf}$ is the outer, centrifugal radius of the disk (Machida et al. 2008). Depending on the density of the disk gas and the size of the solid grains, the water sublimation temperature can vary by several degrees Kelvin (Lewis 1972; Lecar et al. 2006), but we adopt $170\,{\rm K}$ as our fiducial value (Hasegawa & Pudritz 2013).

We simulate the evolution of the $H_2O$ ice lines in the disks around young super-Jovian planets at 5.2 astronomical units (AU, the distance between the Sun and the Earth) from a Sun-like star, facilitating comparison of our results to the Jovian moon system. These planets belong to the observed population of super-Jovian planets at $\approx 1\,{\rm AU}$ around Sun-like stars (Hasegawa & Pudritz 2011, 2012; Howard 2013), and it has been shown that their satellite systems may remain intact during planet migration (Namouni 2010).

## 3. RESULTS AND PREDICTIONS

Figure 1(a) shows, on the largest radial scales, the Hill radius (black crosses) of the accreting giant planet. The planetary radius (black solid line) is well within the Hill sphere, but it is quite extensive for $0.9\,{\rm Myr}$, so much that the CPD (gray solid line) has not yet formed by that time. It only appears after $0.9\,{\rm Myr}$ of evolution of the system. Within that disk, we follow the time evolution of two features – the heat transition (orange open circles) and the $H_2O$ ice line (blue dots). The heat transition denotes the transition from the viscous to the irradiation heating regime in the disk (see Hasegawa & Pudritz 2011), and it appearsat the outer disk edge about $0.95 \times 10^6$ yr after the onset of accretion. It moves rapidly inwards and within $\approx 2 \times 10^4$ yr it reaches the inner disk edge, which sits roughly at the radius of the planet. At the same time ($\approx 0.99 \times 10^6$ yr after the onset of accretion), the $H_2O$ ice line appears at the outer disk radius and then moves slowly inward as the planet cools. The ice line reverses its direction of movement at $\approx 1.1 \times 10^6$ yr due to the decreasing gas surface densities, while the opacities are assumed to be constant through-out the disk, see Equation (13).

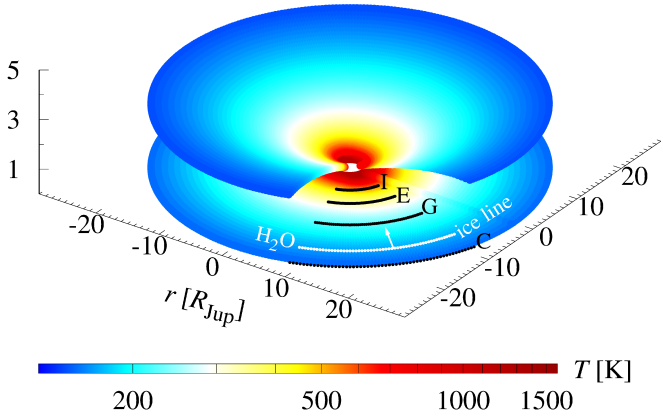

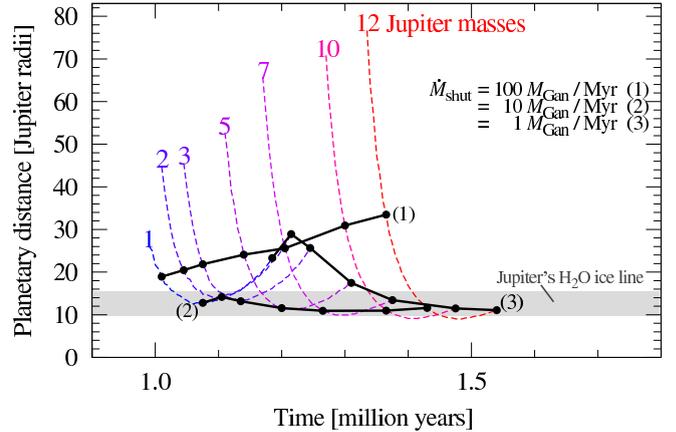

**Figure 2.** Disk temperatures around a forming Jupiter-like planet $10^6$ yr after the onset of accretion. The two disk levels represent the midplane and the photosphere (at a hight $z_s$ above the midplane). At a given radial distance ($r$) to the planet, measured in Jupiter radii, the midplane is usually warmer than the surface (see color bar). The orbits of Io, Europa, Ganymede, and Callisto (labelled I, E, G, and C, respectively) and the location of the instantaneous $H_2O$ ice line at $\approx 22.5\,R_{Jup}$ are indicated in the disk midplane. The arrow attached to the ice line indicates that it is still moving inward before it reaches its final location at roughly the orbit of Ganymede. This simulation assumes fiducial disk values ($k_s = 0.2, \kappa_P = 10^{-2}\,\mathrm{m}^2\,\mathrm{kg}^{-1}$), some $10^5$ yr before the shutdown of moon formation.

Figure 1(b) displays the mass evolution of the planetary core (gray dashed line) and atmosphere (gray solid line). Note that the rapid accumulation of the envelope and the total mass at around $0.93 \times 10^6$ yr corresponds to the runaway accretion phase. Panel (c) presents the total mass accretion rate onto the planet (black solid line). The dashed horizontal line shows an example for $\dot{M}_{shut}$ (here $10\,M_{Gan}\,\mathrm{Myr}^{-1}$), which corresponds to a time 1.08 Myr after the onset of accretion in that particular model. Note that shutdown accretion rates within one order of magnitude around this fiducial value occur 0.1 - 0.3 Myr after the runaway accretion phase, that is, after the planet has opened up a gap in the circumstellar disk. In panel (d), the planetary luminosity peaks during the runaway accretion phase and then dies off as the planet opens up a gap in the circumstellar disk, which starves the CPD.

Figure 2 shows a snapshot of the temperature structure of the disk surface and midplane around a Jupiter-mass planet, $10^6$ yr after the onset of mass accretion. The location of the instantaneous water ice line is indicated with a white dotted line, and the current positions of the Galilean satellites are shown with black dotted lines. Over the next hundred thousand years, the heating rates drop and the ice line moves inward to Ganymede's present orbit as the planet's accretion rate decreases due to its opening of a gap in the circumsolar disk. We argue that the growing Ganymede moved with the ice line trap and was parked in its present orbit when the circumjovian disk dissipated. Due to the rapid decrease of mass accretion onto the planet after gap opening (up to about an order of magnitude per $10^4$ yr), this process can be reasonably approximated as an instant shutdown on the time scales of planet formation (several $10^6$ yr), although it is truly a gradual process.

Figure 3 shows the radial positions of the ice lines around super-Jovian planets as a function of time and for a given disk surface absorptivity ($k_s$) and disk Planck opacity ($\kappa_P$). More massive planets have larger disks and are also hotter at a given time after the onset of accretion, which explains the larger distance and later occurrence of water ice around the more massive giants. Solid black lines connect epochs of equal accretion rates (1, 10, and 100 $M_{Gan}$ per Myr). Along any given ice line track, higher accretion rates correspond to earlier phases. The gray shaded region embraces the orbital radii of Europa and Ganymede, between which we expect the $H_2O$ ice line to settle. The $H_2O$ ice line around the 1 Jupiter mass model occurs after $\approx 0.99 \times 10^6$ yr at the outer edge of the disk, passes through the current orbit of Ganymede, and then begins to move outwards around $1.1 \times 10^6$ yr due to the decreasing gas surface densities (note, the opacities are assumed constant). In this graph, $\dot{M}_{shut} \approx 10\,M_{Gan}\,\mathrm{Myr}^{-1}$ can well explain the mentioned properties in the Galilean system.

**Figure 3.** Evolution of the $H_2O$ ice lines in the disks around super-Jovian gas planets. Black solid lines, labeled (1) - (3), indicate the locations of the $H_2O$ ice lines assuming different mass accretion rates for the shutdown of moon formation ($\dot{M}_{shut} \in \{100, 10, 1\} \times M_{Gan}\,\mathrm{Myr}^{-1}$). The shaded area embraces the orbits of Europa and Ganymede around Jupiter, where Jupiter's $H_2O$ ice line must have been at the time when the Galilean satellites completed formation. The most plausible shutdown rate for the Jovian system (black line with label 2) predicts ice lines between roughly 10 and 15 $R_{Jup}$ over the whole range of super-Jovian planetary masses. Simulations assume $k_s = 0.2$ and $\kappa_P = 10^{-2}\,\mathrm{m}^2\,\mathrm{kg}^{-2}$.

In Figure 4, we present the locations of the ice lines in a more global picture, obtained by performing 120 randomized disk simulations for each planet, where $\dot{M}_{shut}$ and $\kappa_P$ were drawn from a lognormal probability distribution. We also simulated several plausible surface absorptivities of the disk (D'Alessio et al. 2001; Makalkin & Dorofeeva 2014) ($0.1 \leq k_s \leq 0.5$), which resulted in ice line locations similar to those shown in Figure 4, where $k_s = 0.2$. The mean orbital radius of the ice line at the time of shutdown around the 1 $M_{Jup}$ planet is almost precisely at Ganymede's orbit around Jupiter, which we claim is no mere coincidence. Most importantly, despite a variation of $\dot{M}_{shut}$ by two orders of magnitude and considering more than one order of magnitude in planetary masses, the final distances of the $H_2O$ ice lines only vary between about 15 and 30 $R_{Jup}$. Hence, regardless of the actual value of $\dot{M}_{shut}$, the transition from rocky to icy moons around giant planets at several AU from Sun-like stars should occur at planetary distances similar to the one observed in the Galilean system.

We ascribe this result to the fact that the planetary lu-minosity is the dominant heat source at the time of moon formation shutdown. Planetary luminosity, in turn, is determined by accretion (and gravitational shrinking), hence a given $\dot{M}_{shut}$ translates into similar luminosities and similar ice line radii for all super-Jovian planets. Planets above $1\,M_{Jup}$ have substantially larger parts of their disks beyond their water ice lines (note the logarithmic scale in Figure 4) and thus have much more material available for the formation of giant, water-rich analogs of Ganymede and Callisto.

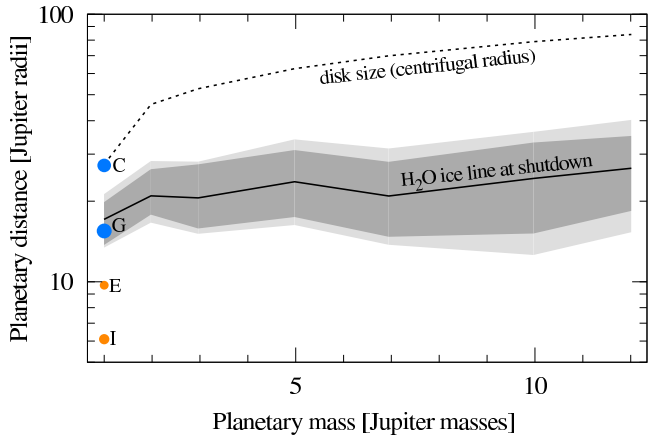

Figure 4. Distance of the $H_2O$ ice lines at the shutdown of moon formation around super-Jovian planets. The solid line indicates the mean, while shaded areas denote the statistical scatter (dark gray $1\sigma$, light gray $2\sigma$) in our simulations, based on the posterior distribution of the disk Planck mean opacity ($\kappa_P$) and the shutdown accretion rate for moon formation ($\dot{M}_{shut}$). The dashed line represents the size of the optically thick part of the circumplanetary disk, or its centrifugal radius. All planets are assumed to orbit a Sun-like star at a distance of 5.2 AU and $k_s$ is set to 0.2. Labeled circles at $1\,M_{Jup}$ denote the orbits of the Galilean satellite Io, Europa, Ganymede, and Callisto. Orange indicates rocky composition, blue represents $H_2O$-rich composition. Circle sizes scale with moon radii. Note that Ganymede sits almost exactly on the circumjovian ice line.

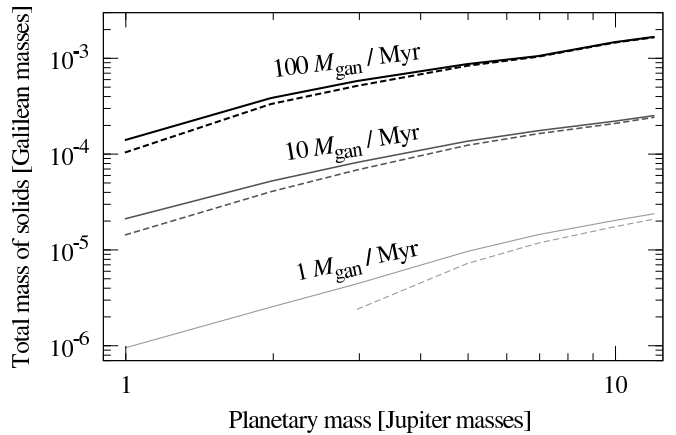

Figure 5. Instantaneous mass of solids in the disks around super-Jovian planets at moon formation shutdown. Solid and dashed lines refer to disk absorptivities of $k_s = 0.2$ and $0.4$, respectively. The ordinate scales in units of the total mass contained in the Galilean moons ($\approx 2.65\,M_{Gan}$). For any given shutdown rate, we find a linear increase in the mass of solids at the time of moon formation shutdown as a function of planetary mass, in agreement with previous simulations (Canup & Ward 2006; Sasaki et al. 2010). Super-Jovian planets of $10\,M_{Jup}$ should thus have moon systems with total masses of $\approx 10^{-3} \times M_{Jup}$, or 3 times the mass of Mars.

Figure 5 shows the total mass of solids at the time of moon formation shutdown around super-Jovian planets. Intriguingly, for any given shutdown accretion rate the total mass of solids scales proportionally to the planetary mass. This result is not trivial, as the mass of solids depends on the location of the $H_2O$ ice line at shutdown. Assuming that $\dot{M}_{shut}$ is similar among all super-Jovian planets, we confirm that the $M_T \propto 10^{-4}\,M_p$ scaling law observed in the solar system also applies for extrasolar super-Jupiters (Canup & Ward 2006; Sasaki et al. 2010).

In addition to the evolution of the $H_2O$ ice lines, we also tracked the movements of the heat transitions, a specific location within the disk, where the heating from planetary irradiation is superseded by viscous heating. Heat transitions cross the disk within only about $10^4$ yr (see Figure 1a), several $10^5$ yr before the shutdown of moon formation, and thereby cannot possibly act as moon traps. Their rapid movement is owed to the abrupt starving of the planetary disk due to the gap opening of the circumstellar disk, whereas the much slower photoevaporation of the latter yields a much slower motion of the circumstellar heat trap. The ineffectiveness of heat traps to reflect a key distinction between the processes of moon formation and terrestrial planet formation.

## 4. DISCUSSION

### 4.1. Accretion and Migration of both Planets and Moons

While Canup & Ward (2002) stated that accretion rates of $2\times10^{-7}\,M_{Jup}\,yr^{-1}$ (about $2.6\times10^4\,M_{Gan}\,Myr^{-1}$) best reproduced the disk conditions in which the Galilean satellites formed, our calculations predict a shutdown accretion rate that is considerably lower, closer to $10\,M_{Gan}\,Myr^{-1}$. The difference in these results is mainly owed to two facts. First, Canup & Ward (2002) only considered viscous heating.[6] Our additional heating terms (illumination from the planet, accretion onto the disk and stellar illumination) contribute additional heat, which imply smaller accretion rates to let the $H_2O$ ice lines move close enough to the Jupiter-like planet. Second, the parameterization of planetary illumination in the Canup & Ward (2002) model is different from ours. While they assume an $r^{-3/4}$ dependence of the midplane temperature from the planet ($r$ being the planetary radial distance), we do not apply any pre-described $r$-dependence. In particular, $T_m(r)$ cannot be described properly by a simple polynomial due to the different slopes of the various heat sources as a function of planetary distance.

Previous models assume that type I migration of the forming moons leads to a continuous rapid loss of proto-satellites into the planet (Canup & Ward 2002; Mosqueira & Estrada 2003a,b; Alibert et al. 2005; Sasaki et al. 2010). (Alibert et al. 2005) considered Jupiter's accretion disk as a closed system after the circumstellar accretion disk had been photo-evaporated, whereas Sasaki et al. (2010) described accretion onto Jupiter with an analytical model. In the Mosqueira & Estrada (2003a,b)(ME) model, satellites migrate via type I but perturb the gas as they migrate and eventually stall and open a gap, ensuring their survival. In opposition to the Canup & Ward (CW) theory, their model does not postulate "generations" of satellites, which are subsequently lost into the planet, because satellite formation doesn't start until the accretion inflow onto the planet wanes.

---

[6] Canup & Ward (2002) discuss the contribution of planetary luminosity to the disk's energy budget, but for their computations of the gas surface densities they ignore it.

There are two difficulties with the CW picture. First, type I migration can be drastically slowed down as growing giant moons get trapped by the ice lines or at the inner truncation radius of the disk[7]. Thus it is not obvious that a conveyor belt of moons into their host planets is ever established. Second, our Figure 5 also contradicts this scenario, because the instantaneous mass of solids in the disk during the end stages of moon formation (or planetary accretion) is not sufficient to form the last generation of moons. In other words, whenever the instantaneous mass of solids contained in the circumjovian disk was similar to the total mass of the Galilean moons, the correspondingly high accretion rates caused the $H_2O$ ice line to be far beyond the orbits of Europa and Ganymede.

We infer, therefore, that the final moon population around Jupiter and other Jovian or super-Jovian exoplanets must, at least to a large extent, have built during the ongoing, final accretion process of the planet, when it was still fed from the circumstellar disk.[8] In order to counteract the inwards flow due to type I migration, we suggest a new picture in which the circumplanetary $H_2O$ ice line and the inner cavity of Jupiter's accretion disk have acted as migration traps. This important hypothesis needs to be tested in future studies. The effect of an inner cavity will also need to be addressed, as it might have been essential to prevent Io and Europa from plunging into Jupiter.

In our picture, Io should have formed dry and its migration might have been stopped at the inner truncation radius of Jupiter's accretion disk, at a few Jupiter radii (Takata & Stevenson 1996). It did not form wet and then lose its water through tidal heating. Ganymede may have formed at the water ice line in the circumjovian disk, where it has forced Io and Europa in the 1:2:4 orbital mean motion resonance (Laplace et al. 1829). From a formation point of view, we suggest that Io and Europa be regarded as moon analogs of the terrestrial planets, whereas Ganymede and Callisto resemble the precursors of giant planets.

Our combination of planet formation tracks and a CPD model enables new constraints on planet formation from moon observations. As just one example, the "Grand Tack" (GT) model suggests that Jupiter migrated as close as about 1.5 AU to the Sun before it reversed its migration due to a mutual orbital resonance with Saturn (Walsh et al. 2011). In the proximity of the Sun, however, solar illumination should have depleted the circumjovian accretion disk from water ices during the end stages of Jupiter's accretion (Heller et al. 2015, in prep.). Thus, Ganymede and Callisto would have formed in a dry environment *during* the GT, which is at odds with their high $H_2O$ ice contents. They can also hardly have formed over millions of years (Mosqueira & Estrada 2003a) thereafter, because Jupiter's CPD (now truncated from its environment by a gap) still would have been dry. Alternatively, one might suggest that Callisto and Ganymede formed *after* the GT from newly accreted planetesimals into a still active, gaseous disk around Jupiter. But then Io and Europa might have been substantially enriched in water, too. Tanigawa et al. (2014, see their Figure 8) found that planetesimal accretion via gas drag is most efficient between 0.005 and 0.001 Hill radii ($R_H$) or about 4 to 8 $R_{Jup}$ where gas densities are relatively high.

To come straight to the point, our preliminary studies suggest that in the GT paradigm, the icy Galilean satellites must have formed *prior to* Jupiter's excursion to the inner solar system (Heller et al. 2015, in prep.). This illustrates the great potential of moons to constrain planet formation, which is particularly interesting for the GT scenario where the timing of migration and planetary accretion is yet hardly constrained otherwise (Raymond & Morbidelli 2014).

### 4.2. *Parameterization of the Disk*

Finally, we must address a technical issue, namely, our choice of the $\alpha$ parameter ($10^{-3}$). While this is consistent with many previous studies, how would a variation of $\alpha$ change our results? Magnetorotational instabilities might be restricted to the upper layers of CPDs, where they become sufficiently ionized (mostly by cosmic high-energy radiation and stellar X-rays). Magnetic turbulence and viscous heating in the disk midplane might thus be substantially lower than in our model (Fujii et al. 2014). On the other hand, Gressel et al. (2013) modeled the magnetic stresses in CPDs with a 3D magnetohydrodynamic model and inferred $\alpha$ values of 0.01 and larger, which would strongly enhance viscous heating. Obviously, sophisticated numerical simulations of giant planet accretion do not yet consistently describe the magnetic properties of the disks and the associated $\alpha$ values.

Given that circumstellar disks are almost certainly magnetized, CPD can be expected to have inherited magnetic fields from this source. This makes it likely that magnetized disk winds can be driven off the CPD (Fendt 2003; Pudritz et al. 2007) which can carry significant amounts of angular momentum. Even in the limit of very low ionization, Bai & Stone (2013) demonstrated that magnetized disk winds will transport disk angular momentum at the rates needed to allow accretion onto the central object. However, independent of these uncertainties, the final positions of the $H_2O$ ice line produced in our simulations turn out to depend mostly on planetary illumination, because viscous heating becomes negligible almost immediately following gap opening. Hence, even substantial variations of $\alpha$ by a factor of ten would hardly change our results for the ice line locations at moon formation shutdown since these must develop in radiatively dominated disk structure (but it would alter them substantially in the viscous-dominated regime before and during runaway accretion).

Our assumption of a constant Planck opacity throughout the disk is simplistic and ignores the effects of grain growth, grain distribution within the disk, as well as the evolution of the disk properties. In a more consistent model, $\kappa_P$ depends on both the planetary distance and distance from the midplane, which might entail significant modifications in the temperature distribution that we predict.

---

[7] An inner cavity can be caused by magnetic coupling between the rotating planet and the disk (Takata & Stevenson 1996), and it can be an important aspect to explain the formation of the Galilean satellites (Sasaki et al. 2010)

[8] This conclusion is similar to that proposed by the ME model, but for reasons that are very different.

## 5. CONCLUSIONS

We have demonstrated that ice lines imprint important structural features on systems of icy moons around massive planets. Given that observations show a strong concentration of super-Jovian planets at $\approx 1\,\mathrm{AU}$, we focused our analysis on the formation of massive moons in this planetary population.

After a forming giant planet opens up a gap in the circumstellar disk, its accretion rates and the associated viscous heating in the CPD drop substantially. We find that a heat transition crosses the CPD within $10^4\,\mathrm{yr}$, which is too fast for it to act as a moon migration trap. Alternatively, we propose that moon migration can be stalled at the $H_2O$ ice line, which moves radially on a $10^5\,\mathrm{yr}$ timescale. For Jupiter's final accretion phase, when the Galilean moons are supposed to form in the disk, our calculations show that the $H_2O$ line is at about the contemporary radial distance of Ganymede, suggesting that the most massive moon in the solar system formed at a circumplanetary migration trap. Moreover, dead zones might be present in the inner CPD regions (Gressel et al. 2013) where they act as additional moon migration traps, but this treatment is beyond the scope of this paper.

Our model confirms the mass scaling law for the most massive planets, which suggests that satellite systems with total masses several times the mass of Mars await discovery. Their most massive members will be rich in water and possibly parked in orbits at their host planet's $H_2O$ ice lines at the time of moon formation shutdown, that is, between 15 and 30 $R_{\mathrm{Jup}}$ from the planet. A Mars-mass moon composed of $50\,\%$ of water would have a radius of $\approx 0.7$ Earth radii (Fortney et al. 2007). Although we considered giant planet accretion beyond $1\,\mathrm{AU}$, super-Jovian planets are most abundant around $1\,\mathrm{AU}$ (Howard 2013) and their moon systems have been shown to remain intact during planet migration (Namouni 2010). Giant water-rich moons might therefore form an abundant population of extrasolar habitable worlds (Williams et al. 1997; Heller et al. 2014) and their sizes could make them detectable around photometrically quiet dwarf stars with the transit method (Kipping et al. 2012; Heller 2014). In a few cases, the transits of such giant moons in front of hot, young giant planets might be detectable with the *European Extremely Large Telescope*, with potential for follow-up observations of the planetary Rossiter-McLaughlin effect (Heller & Albrecht 2014).

More detailed predictions can be obtained by including the migration process of the accreting planet, which we will present in an upcoming paper. Ultimately, we expect that there will be a competition between the formation of water-rich, initially icy moons beyond the circumplanetary $H_2O$ ice line and the gradual heating of the disk (and loss of ices) during the planetary migration towards the star. Such simulations have the potential to generate a moon population synthesis with predictions for the abundance and detectability of large, water-rich moons around super-Jovian planets.

The report of an anonymous referee helped us to clarify several aspects of the manuscript. We thank C. Mordasini for sharing with us his planet evolution tracks, G. D'Angelo for discussions related to disk ionization, A. Makalkin for advice on the disk properties, and Y. Hasegawa for discussions of ice lines. R. Heller is supported by the Origins Institute at McMaster University and by the Canadian Astrobiology Program, a Collaborative Research and Training Experience Program funded by the Natural Sciences and Engineering Research Council of Canada (NSERC). R. E. Pudritz is supported by a Discovery grant from NSERC.

## 5.2 Conditions for Water Ice Lines and Mars-mass Exomoons Around Accreting Super-Jovian Planets at 1-20 AU From Sun-Like Stars (Heller & Pudritz 2015a)

Contribution:

RH did the literature research, worked out the mathematical framework, translated the math into computer code, performed the simulations, created all figures, led the writing of the manuscript, and served as a corresponding author for the journal editor and the referees.

# Conditions for water ice lines and Mars-mass exomoons around accreting super-Jovian planets at 1 - 20 AU from Sun-like stars

R. Heller[1],[*] and R. Pudritz[1]

Origins Institute, McMaster University, 1280 Main Street West, Hamilton, ON L8S 4M1, Canada
e-mail: rheller@physics.mcmaster.ca , e-mail: pudritz@physics.mcmaster.ca

## ABSTRACT

*Context.* The first detection of a moon around an extrasolar planet (an "exomoon") might be feasible with NASA's *Kepler* or ESA's upcoming *PLATO* space telescopes or with the future ground-based *European Extremely Large Telescope*. To guide observers and to use observational resources most efficiently, we need to know where the largest, most easily detected moons can form.
*Aims.* We explore the possibility of large exomoons by following the movement of water ($H_2O$) ice lines in the accretion disks around young super-Jovian planets. We want to know how the different heating sources in those disks affect the location of the $H_2O$ ice lines as a function of stellar and planetary distance.
*Methods.* We simulate 2D rotationally symmetric accretion disks in hydrostatic equilibrium around super-Jovian exoplanets. The energy terms in our semi-analytical framework – (1) viscous heating, (2) planetary illumination, (3) accretional heating of the disk, and (4) stellar illumination – are fed by precomputed planet evolution models. We consider accreting planets with final masses between 1 and 12 Jupiter masses at distances between 1 and 20 AU to a solar type star.
*Results.* Accretion disks around Jupiter-mass planets closer than about 4.5 AU to Sun-like stars do not feature $H_2O$ ice lines, whereas the most massive super-Jovians can form icy satellites as close as 3 AU to Sun-like stars. We derive an empirical formula for the total moon mass as a function of planetary mass and stellar distance and predict that super-Jovian planets forming beyond about 5 AU can host Mars-mass moons. Planetary illumination is the major heat source in the final stages of accretion around Jupiter-mass planets, whereas disks around the most massive super-Jovians are similarly heated by planetary illumination and viscous heating. This indicates a transition towards circumstellar accretion disks, where viscous heating dominates in the stellar vicinity. We also study a broad range of circumplanetary disk parameters for planets at 5.2 AU and find that the $H_2O$ ice lines are universally between about 15 and 30 Jupiter radii in the final stages of accretion when the last generation of moons is supposed to form.
*Conclusions.* If the abundant population of super-Jovian planets around 1 AU formed in situ, then these planets should lack the previously predicted population of giant icy moons, because those planets' disks did not host $H_2O$ ice in the final stages of accretion. But in the more likely case that these planets migrated to their current locations from beyond about 3 to 4.5 AU they might be orbited by large, water-rich moons. In this case, Mars-mass moons might be common in the stellar habitable zones. Future exomoon detections and non-detections can provide powerful constraints on the formation and migration history of giant exoplanets.

*Key words.* Accretion, accretion disks – Planets and satellites: formation – Planets and satellites: gaseous planets – Planets and satellites: physical evolution – Astrobiology

## 1. Introduction

Now that the detection of sub-Earth-sized objects has become possible with space-based photometry (Muirhead et al. 2012; Barclay et al. 2013) and almost 2000 extrasolar planets have been confirmed (Batalha et al. 2013; Rowe et al. 2014), technological and theoretical advances seem mature enough to find moons orbiting exoplanets. Natural satellites similar in size to Mars (0.53 Earth radii, $R_\oplus$) or Ganymede (0.41 $R_\oplus$) could be detectable in the available *Kepler* data (Kipping et al. 2012; Heller 2014), with the upcoming *PLATO* space mission (Simon et al. 2012; Rauer et al. 2014) or with the *European Extremely Large Telescope* (*E-ELT*) (Heller & Albrecht 2014). Exomoons at about 1 AU or closer to their star might be detected during stellar transits (Sartoretti & Schneider 1999; Szabó et al. 2006; Kipping et al. 2009). Young, self-luminous planets beyond 10 AU might reveal their satellites in the infrared through planetary transits of their moons, if the unresolved planet-moon binary can be directly imaged (Peters & Turner 2013; Heller & Albrecht 2014). Moon formation theories can guide observers and data analysts in their searches for exomoons. In turn, the first non-detections of exomoons (Brown et al. 2001; Pont et al. 2007; Kipping et al. 2013b,a, 2014) and possible future findings provide the first extrasolar observational constraints on satellite formation.

In a recent study, we developed a circumplanetary disk accretion model and simulated the radial motion of the water ($H_2O$) ice line around super-Jovian[1] exoplanets at 5.2 AU around Sun-like stars (Heller & Pudritz 2014). The $H_2O$ ice line is critical for the formation of large, possibly detectable moons, because here the surface density of solids increases sharply by a factor of 3 to 4 (Hayashi 1981). The composition and the masses of the Galilean moons are usually considered as records of the location of the $H_2O$ ice line and, more generally, of the temperature distri-bution in Jupiter's accretion disk (Pollack & Reynolds 1974; Lunine & Stevenson 1982). The inner moons Io and Europa turned out mostly rocky and comparatively light, whereas Ganymede and Callisto formed extremely rich in water ices (about 50 %) but became substantially more massive (Showman & Malhotra 1999).

---

[1] Throughout the paper, our reference to "super-Jovian" planets includes planets with masses between that of Jupiter ($M_{Jup}$) and 12 $M_{Jup}$, where the latter demarcates the transition into the brown dwarf regime.

Our main findings were that (1) super-Jovian planets at 5.2 AU around Sun-like stars have their $H_2O$ ices between about 15 and 30 Jupiter radii ($R_{Jup}$) during the late stages of accretion, when the final generation of moons form. This range is almost independent of the planetary mass ($M_p$). (2) With the most massive planets having the most widely extended disks, these disks host the largest reservoirs of $H_2O$ ices. In particular, the total instantaneous mass of solids ($M_{sld}$) in these disks scales roughly proportional with $M_p$. (3) The current orbital position of Ganymede is very close the mean radial location of the circumjovian $H_2O$ ice line in our simulations, suggesting a novel picture in which Ganymede formed at a circumplanetary ice line trap. (4) Heat transitions, however, transverse the accretion disks on a very short timescale ($\approx 10^4$ yr) once the planet opens up a gap in the circumstellar accretion disk and thereby drastically reduces the supply of material. This timescale is short compared to the satellite migration time scale ($10^5$ - $10^7$ yr) (Tanaka et al. 2002; Canup & Ward 2002, 2006; Sasaki et al. 2010; Ogihara & Ida 2012) and the time that is left for the final accretion until shutdown ($10^5$ yr). Hence, heat transitions around super-Jovian planets cannot act as moon traps, which indicates a different behavior of planet and moon formation (Menou & Goodman 2004; Hasegawa & Pudritz 2011; Kretke & Lin 2012).

We here extend the parameter space of our previous paper and consider planetary accretion between 1 and 20 AU from the star. This range is motivated by two facts. First, super-Jovians are the most abundant type of confirmed planets, that is, objects with accurate mass estimates at these distances (see Fig. 1)[2]. And second, this range contains the stellar habitable zone around Sun-like stars, which is of particular interest given the fact that giant, water-rich moons might be frequent habitable environments (Heller & Armstrong 2014; Heller et al. 2014). Our main goal is to locate the $H_2O$ ice lines in the circumplanetary accretion disks at the time of moon formation shutdown. Their radial separations from the planet will correspond to the orbital radii at or beyond which we can suspect the most massive, water-rich moons to reside. In particular, stellar illumination at only a few AU from the star will prevent the accretion disks from having $H_2O$ ice lines and therefore from hosting large moons. We determine these critical stellar distances below.

## 2. Methods

We use the framework developed in Heller & Pudritz (2014), which models a 2D axisymmetric accretion disk in hydrostatic equilibrium around the planet. It considers four heating terms, or energy fluxes, of the disk, namely (1) viscous heating ($F_{vis}$), (2) planetary illumination ($F_p$), (3) accretional heating of the disk ($F_{acc}$), and (4) stellar illumination, all of which determine the midplane and surface temperature of the disk as a function of planetary distance ($r$). This model (based on earlier work by Makalkin & Dorofeeva 1995, 2014) considers $F_{vis}$ as a distributed energy source within the disk, while $F_p$, $F_{acc}$, and stellar illumination are considered external heat sources. The model also includes an analytical treatment for the vertical radiative energy transfer, which depends on the Planck mean opacity ($\kappa_P$). In real disks, $\kappa_P$ will depend on the disk temperature and the composition of the solids. In particular, it will be a function of the radial distance to the planet and it will evolve in time as small particles stick together and coagulate. To reduce computational demands, we here assume a constant $\kappa_P$ throughout the disk, but we will test values over two orders of magnitude to explore the effects of changing opacities. The vertical gas density in the disk is approximated with an isothermal profile, which is appropriate because we are mostly interested in the very final stages of accretion when the disk midplane and the disk surface have similar temperatures. Furthermore, deviations between an isothermal and an adiabatic vertical treatment are significant only in the very dense and hot parts of the disk inside about 10 $R_{Jup}$. As the $H_2O$ ice line will always be beyond these distances, inaccuracies in our results arising from an isothermal vertical model are negligible.

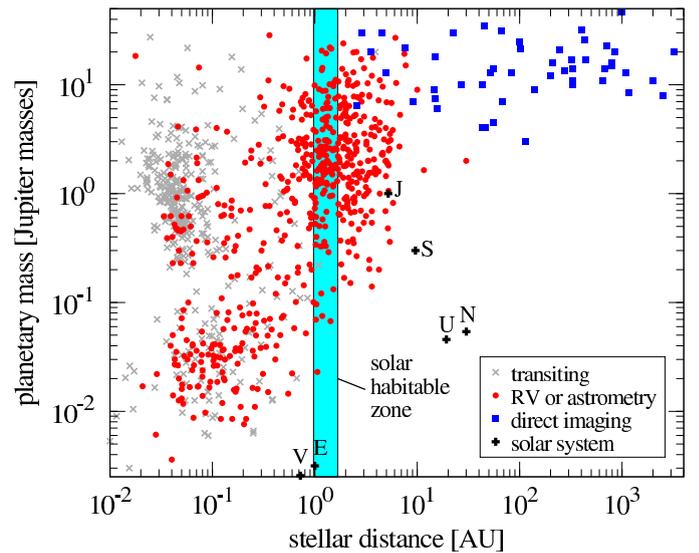

**Fig. 1.** Stellar distances and planetary masses of extrasolar planets listed on www.exoplanet.eu as of 5 April 2015. Symbols indicate the discovery method of each planet, six solar system planets are shown for comparison. Note the cluster of red dots around 1 AU along the abscissa and between 1 and 10 $M_{Jup}$ along the ordinate. These super-Jovian planets might be hosts of Mars-mass moons. The shaded region denotes the solar habitable zone defined by the runaway and maximum greenhouse as per Kopparapu et al. (2013).

The heating terms (1)-(3) are derived based on precomputed planet evolution models (provided by courtesy of C. Mordasini, Mordasini 2013), which give us the planetary mass, planetary mass accretion rate ($\dot{M}$), and planetary luminosity ($L_p$) as a function of time ($t$) (see Fig. 1 in Heller & Pudritz 2014). $L_p(t)$ is a key input parameter to our model as it determines $F_p(t)$, and it is sensitive to the planet's core mass. Yet, in the final stages of planetary accretion, $L_p$ differs by less than a factor of two for planetary cores between 22 to 130 Earth masses. With $F_p \propto L_p$ and temperature scaling roughly with $(F_p)^{1/4}$, uncertainties in the midplane or surface temperatures of the disk are lower than about 20 %, and uncertainties in the radial distance of the $H_2O$ ice line are as large as a few Jovian radii at most. For all our simulations the precomputed planetary models assume a final core mass of 33 Earth masses, which is about a factor of three larger than the mass of Jupiter's core (Guillot et al. 1997). With higher final core masses meaning higher values for $L_p$ at any given accretion rate,

**Table 1.** Parameterization of the circumplanetary disk as described in Heller & Pudritz (2014).

| Symbol | Meaning | Fiducial Value |
|--------|---------|---------------|
| | Constant or parameterized planetary disk parameters | |
| $r$ | distance to the planet | |
| $R_H$ | planetary Hill radius | |
| $r_{cf}$ | centrifugal radius | |
| $r_c$ | transition from optically thick to optically thin (viscously spread) outer part | $27.22/22 \times r_{cf}$ [a] |
| $r_d$ | outer disk radius | $R_H/5$ [b,c] |
| $\Lambda(r)$ | radial scaling of $\Sigma(r)$ | |
| $T_m(r)$ | midplane temperature | |
| $T_s(r)$ | surface temperature | |
| $\Sigma(r)$ | gas surface density | |
| $h(r)$ | effective half-thickness | |
| $\rho_0(r)$ | gas density in the midplane | |
| $\rho_s(r)$ | gas density at the radiative surface level | |
| $z_s(r)$ | radiative surface level, or photospheric height | |
| $q_s(r)$ | vertical mass coordinate at the radiative surface | |
| $c_s(T_m(r))$ | speed of sound | |
| $\alpha$ | Viscosity parameter | $0.001$ [d] |
| $\nu(r, T_m)$ | gas viscosity | |
| $\Gamma$ | adiabat exponent, or ratio of the heat capacities | $1.45$ |
| $\chi$ | dust enrichment relative to the protostellar, cosmic value | $10$ [c] |
| $X_d$ | dust-to-mass fraction | $0.006$ [e,f] |
| $\mu$ | mean molecular weight of the H/He gas | $2.34$ kg/mol [g] |
| | Variable planetary disk parameters | |
| $\kappa_P$ | Planck mean opacity | $10^{-3} - 10^{-1}$ m$^2$ kg$^{-1}$ [h] |
| $k_s$ | fraction of the solar radiation flux contributing to disk heating at $z \leq z_s$ | $0.1 - 0.5$ [c] |
| $\dot{M}_{shut}$ | shutdown accretion rate for moon formation | $100, 10, 1$ $M_{Gan}$/Myr |

**Notes.** See Heller & Pudritz (2014) and additional references for details about the respective parameter values or ranges: [a] Machida et al. (2008) [b] Sasaki et al. (2010) [c] Makalkin & Dorofeeva (2014) [d] Keith & Wardle (2014) [e] Lunine & Stevenson (1982) [f] Hasegawa & Pudritz (2013) [g] D'Angelo & Bodenheimer (2013) [h] Bell et al. (1997) .

we actually derive upper, or outer, limits for the special case of the H$_2$O ice line around a Jupiter-like planet 5.2 AU around a Sun-like star.

We interpolate the Mordasini (2013) tracks on a linear scale with a time step of 1 000 yr and evaluate the four heating terms as a function of $r$, which extends from Jupiter's co-rotation radius (2.25 $R_{Jup}$ in our simulations, Sasaki et al. 2010) out to the disk's centrifugal radius ($r_{cf}$). At that distance, the centrifugal force acting on a gas parcel equals the gravitational pull of the planet. We compute $r_{cf}$ using the analytical expression of Machida et al. (2008), which they derived by fitting a power law expression to their 3D hydrodynamical simulations of circumplanetary accretion disks. In this model, $r_{cf} \propto M_{Jup}^{1/3}$ for super-Jovian planets at a given stellar distance. Viscous heating is governed by the $\alpha$ viscosity parameter (Shakura & Sunyaev 1973), which we fix to a value of $10^{-3}$ in our simulations (for a discussion see Sect. 4 in Heller & Pudritz 2014). Note that even variations of $\alpha$ by an order of magnitude would only change our results for the circumplanetary H$_2$O ice line location during the final stages of planetary accretion by a few planetary radii at most, because then the disk is mostly heated by planetary illumination.

The sound velocity ($c_s$) in the disk midplane is evaluated as $c_s = 1.9 \, \text{km s}^{-1} \sqrt{T_m(r)/1\,000\,\text{K}}$ (Keith & Wardle 2014), which is an adequate approximation since ionization can be neglected in the late stages of moon formation when disk temperatures are usually below 1 000 K. At each time step, we assume a steady-state gas surface density ($\Sigma_g$) that is derived analytically by solving the continuity equation of the mass inflow onto a centrifugally supported, viscous disk with a uniform flux per area (Canup & Ward 2002). The equations of energy transport (Makalkin & Dorofeeva 2014) then allow us to derive the temperature profile both at the disk surface, where the optical depth $\tau = 2/3$, and in the disk midplane. This model invokes absorption of planetary illumination in the disk photosphere at a height $z_s(r)$ above the midplane, modeled by an absorption coefficient ($k_s$), as well as the transport of energy through an optically thick disk to the surface, modeled by $\kappa_P$. For the dust-to-gas ratio ($X_d$) we take a fiducial value of 0.006 (Lunine & Stevenson 1982; Hasegawa & Pudritz 2013) in the inner dry regions of the disk, and we assume that it jumps by a factor of three at the $H_2O$ ice line. The resulting mathematical framework contains several implicit functions, which we solve in an iterative, numerical scheme. Table 1 gives a complete overview of the parameters involved in the model. Our model neglects the effects of planetary migration, which is an adequate approximation because the planets don't migrate substantially within the $\approx 10^5$ yr required for moon formation. We also do not actually simulate the accretion and buildup of moons (Ogihara & Ida 2012).

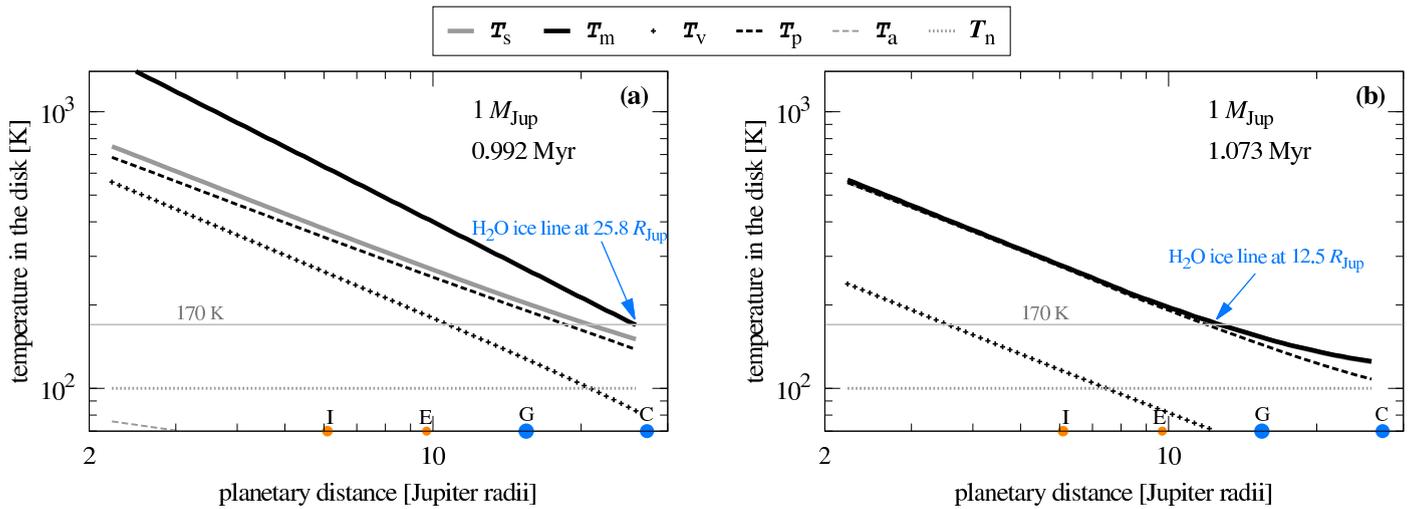

**Fig. 2.** Temperature structure in the disk around a Jupiter-mass planet 5.2 AU from a Sun-like star. Solid black lines indicate disk midplane temperatures, solid gray lines disk surface temperatures. The other lines indicate a hypothetical disk surface temperature assuming only one heat source (see legend). **(a)**: At 0.992 Myr in this particular simulation, the disk has sufficiently cooled to allow the appearance of an $H_2O$ ice line in the disk midplane at the outer disk edge at 25.8 $R_{Jup}$. The planetary accretion rate is about $4 \times 10^2 \, M_{Gan} \, Myr^{-1}$. **(b)**: At 1.073 Myr, when the planetary accretion rate has dropped to 10 $M_{Gan} \, Myr^{-1}$, the $H_2O$ ice line has settles between the orbits of Europa and Ganymede (see colored symbols). In these final stages of accretion, disk temperatures are governed by planetary illumination (see dashed line for $T_p$).

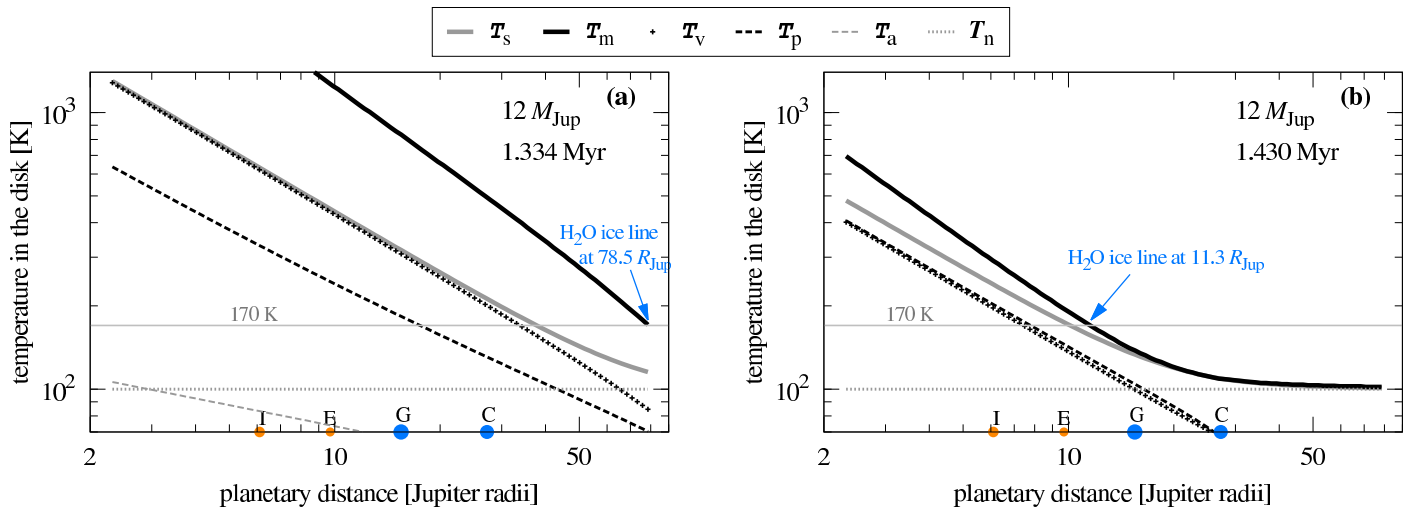

**Fig. 3.** Same as Fig. 2, but now for a 12 $M_{Jup}$ planet. Note that the accretion disk is substantially larger than in the 1 $M_{Jup}$ case. **(a)**: The $H_2O$ ice line appears much later than in the Jovian scenario, here at 1.334 Myr. The planetary accretion rate is about $10^3 \, M_{Gan} \, Myr^{-1}$ and viscous heating is dominant in this phase. **(b)**: Once the planetary accretion rate has dropped to 10 $M_{Gan} \, Myr^{-1}$ at 1.43 Myr, the $H_2O$ ice line is at a similar location as in the Jupiter-like scenario, that is, between the orbits of Europa and Ganymede.

As an extension of our previous study, where all planets were considered at 5.2 AU around a Sun-like star, we here place hypothetical super-Jovian planets at stellar distances between 1 and 20 AU to a solar type host. Although the precomputed planet formation tracks were calculated at 5.2 AU around a Sun-like star, we may still consider different stellar distances because the accretion rates through any annulus in the circumstellar disk are roughly constant at any time (and thereby similar to the instantaneous stellar accretion rate). The surface densities of gas and solids are naturally lower at larger stellar separations. Hence, the accretion rates provided by the tracks will actually overestimate $\dot{M}$ and $L_p$ in wider orbits at any given time. However, this will not affect our procedure as we are not primarily interested in the evolution as a function of absolute times but rather as a function of accretion rates. In particular, we introduced the shutdown accretion ($\dot{M}_{shut}$) in Heller & Pudritz (2014), which we use as a measure to consider as the final stages of moon formation around accreting giant planets. $\dot{M}_{shut}$ is a more convenient quantity (or independent variable) than time to refer to, because, first, the initial conditions of planet formation as well as the age of the system are often poorly constrained. And, second, accretion rates can be inferred more easily from observations than age. Hence, planetary accretion rates allow us to consider planet (and moon) formation at comparable stages, independent of the initial conditions.

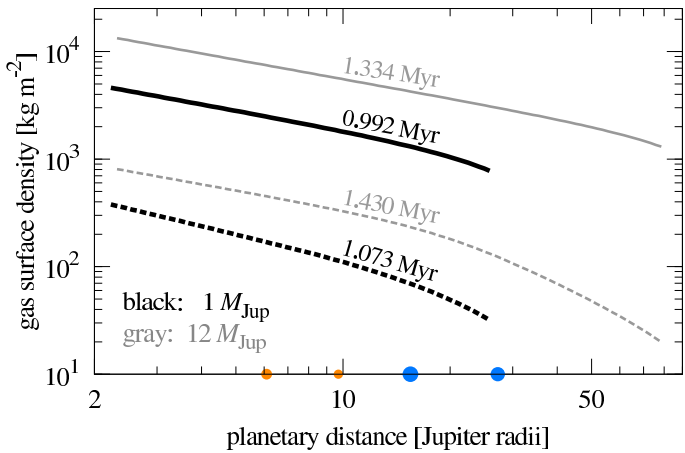

**Fig. 4.** Surface densities around the four test planets shown in Figs. 2 and 3. Black lines refer to the Jupiter-mass planets, gray lines to the 12 $M_{Jup}$-mass super-Jovian. Note that the disk of the Jupiter twin is much smaller than that of the super-Jovian and that both surface density distributions decrease with time. The radial positions of the Galilean moons are indicated at the bottom of the figure.

Variations of the stellar distance will affect both the size of the circumplanetary accretion disk (Machida et al. 2008) and the stellar heating term in our model (Makalkin & Dorofeeva 2014). Disks around close-in planets are usually smaller (though not necessarily less massive) than disks around planets in wide stellar orbits owing to their smaller Hill spheres and their lower average specific angular momentum. Moreover, substantial stellar illumination will prevent these disks from having $H_2O$ ice lines in the vicinity of the star. The temperature of the circumstellar accretion disk, in which the planet is embedded, is calculated under the assumption that the disk is transparent to stellar irradiation (Hayashi 1981) and that the stellar luminosity equals that of the contemporary Sun. In this model, the circumstellar $H_2O$ ice line is located at 2.7 AU from the star. We will revisit the faint young Sun (Sagan & Mullen 1972) as well as other parameterizations of the circumstellar disk in an upcoming paper (Heller et al. 2015, in prep.).

In Figs. 2 and 3 we show an application of our disk model to a Jupiter-mass and a 12 $M_{Jup}$ planet at 5.2 AU around a Sun-like star. All panels present the disk surface temperatures ($T_s$, solid gray lines) and disk midplane temperatures ($T_m$, solid black lines) as well as the contributions to $T_s$ from viscous heating ($T_v$, black crosses), planetary illumination ($T_p$, solid black lines), accretion onto the disk ($T_a$, gray dashed lines), and heating from the circumstellar accretion disk, or "nebula" ($T_n$, gray dotted lines). Any of these contributions to the disk surface temperature is computed assuming that all other contributions are zero. In other words, $T_v$, $T_p$, $T_a$, and $T_n$ depict the temperature of the disk photosphere in the hypothetical case that viscous heating, planetary illumination, accretional heating onto the disk, or the stellar illumination were the single energy source, respectively.[3] The different slopes of these curves, in particular of $T_v$ and $T_p$, lead to the appearance of heat transitions (not shown), which traverse the disk on relatively short time scales (Heller & Pudritz 2014). Also note that in all the simulations shown, the disk midplane is warmer than the disk surface. Only when accretion drops to about 10 $M_{Gan}$ Myr$^{-1}$ in panels (b) do $T_s$ and $T_m$ become comparable throughout the disk, both around the Jupiter- and the super-Jupiter-mass planet.

---

[3] As an example, $T_v$ is calculated setting $F_{acc} = F_p = T_{neb} = 0$ in Eq. (13) of Heller & Pudritz (2014).

Panels (a) in Figs. 2 and 3 are chosen at the time at which the $H_2O$ ice line first appears at the outer edge of the disk, while panels (b) illustrate the temperature distribution at the time when $\dot{M} = 10 M_{Gan}$, with $M_{Gan}$ as Ganymede's mass. During this epoch, the $H_2O$ ice line around the Jupiter-mass planet is safely between the orbital radii of Europa and Ganymede (see labeled arrow in Fig. 2b), which suggests that this corresponds to the shutdown phase of moon formation. The colored symbols in each panel depict the rocky (orange) or icy (blue) composition of the Galilean moons. Symbol sizes scale with the actual moon radii. In these simulations, the Planck mean opacity ($\kappa_P$) has been fixed at a fiducial value of $10^{-2}$ m$^2$ kg$^{-1}$ throughout the disk and the disk absorptivity is assumed to be $k_s = 0.2$.

A comparison of Figs. 2(a) and 3(a) shows that the centrifugal radius of the accretion disk around the Jovian planet ($r_c \approx 27 R_{Jup}$) is substantially smaller than the disk radius around the super-Jovian test planet ($r_c \approx 80 R_{Jup}$) – note the different distance ranges shown in the figures! Moreover, the midplanes and surfaces in the inner regions of the super-Jovian accretion disks are substantially hotter at any given distance around the super-Jupiter. Note also that planetary illumination is the main energy source around the Jupiter-mass planet in Figs. 2(a) and (b), while viscous heating plays an important role during the appearance of the $H_2O$ ice line around the super-Jovian planet in Fig. 3(a). In the final stages of accretion onto the 12 $M_{Jup}$ planet, shown in Fig. 3(b), viscous heating and planetary illumination are comparable throughout the disk.

In Fig. 4 we present the radial distributions of the gas surface densities around our Jovian and super-Jovian test planets. While solid lines refer to panels (a) in Figs. 2 and 3, dashed lines refer to panels (b), respectively. In particular, solid lines refer to that instant in time when the $H_2O$ ice lines first appear at the outer edges of the accretion disks around those two planets, whereas the dashed lines depict the accretion phase when $\dot{M} = 10 M_{Gan}$ Myr$^{-1}$, which is the phase when the $H_2O$ ice line around our Jupiter-mass test planet is in good agreement with the compositional gradient observed in the Galilean system (Heller & Pudritz 2014). At any given planetary distance in these particular states, the super-Jovian planet has a gas surface density that is higher by about a factor of five compared to the Jupiter-mass planet. Moreover, we find that the $H_2O$ ice lines in both cases need about $10^5$ yr to move radially from the outer disk edges to their final positions (see labels). Our values for $\Sigma_g$ are similar to those presented by Makalkin & Dorofeeva (2014) in their Fig. 4.

Similar to the procedure applied in Heller & Pudritz (2014), we performed a suite of simulations for planets with masses between 1 and 12 $M_{Jup}$ at 5.2 AU from a Sun-like star, where $\dot{M}_{shut}$ and $\kappa_P$ were randomly drawn from a Gaussian probability distribution. For $\log_{10}(\kappa_P/[m^2 kg^{-1}])$ we took a mean value of $-2$ with a standard variation of 1 (see Bell et al. 1997, and references therein), and for $\log_{10}(\dot{M}_{shut}/[M_{Gan} Myr^{-1}])$ we assumed a mean value of 1 with a standard variation of 1, which nicely reproduced the compositional $H_2O$ gradient in the Galilean moons (Heller & Pudritz 2014). As an extension of our previous simulations, we here consider various disk absorptivities ($k_s = 0.2$ and 0.4) and examine the total mass of solids in the accretion disks, both as a function of time and as a function of stellar distance. Above all, we study the disappearance of the circumplanetary $H_2O$ ice line in the vicinity of the star.

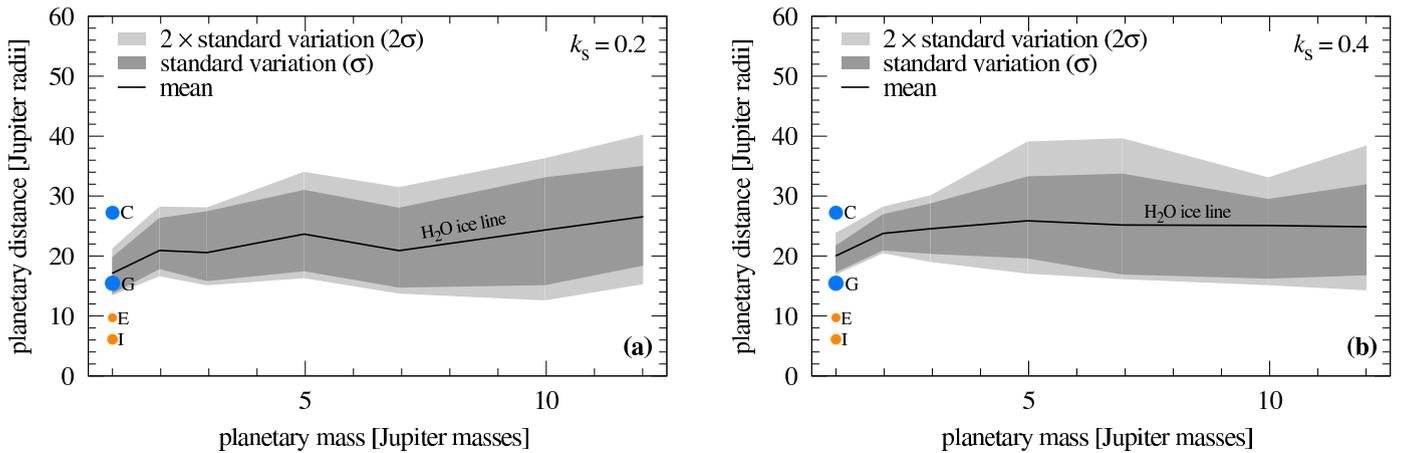

**Fig. 5.** Distance of the circumplanetary $H_2O$ ice lines as a function of planetary mass. A distance of 5.2 AU to a Sun-like star is assumed. The shaded area indicates the $1\sigma$ scatter of our simulations based on the posterior distribution of the disk Planck mean opacity ($\kappa_P$) and the shutdown accretion rate for moon formation ($\dot{M}_{shut}$). The labeled circles at $1\,M_{Jup}$ denote the orbital positions of Jupiter's moons Io, Europa, Ganymede, and Callisto. Orange indicates rocky composition, blue represents $H_2O$-rich composition. Circle sizes scale with the moons' radii. **(a)**: Disk reflectivity ($k_s$) is set to 0.2. **(b)**: Same model parameterization but now with disk reflectivity $k_s = 0.4$.

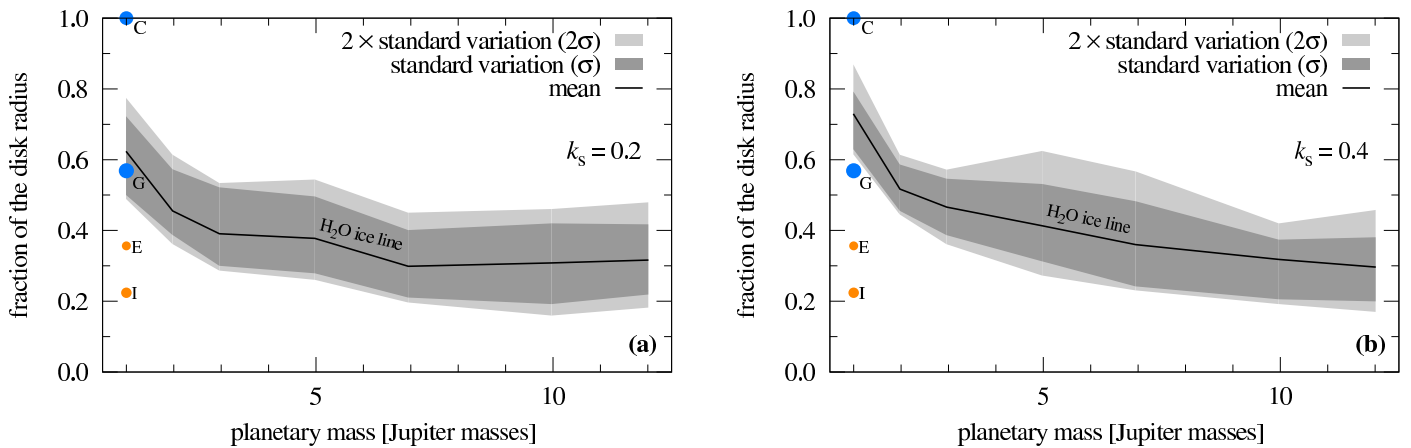

**Fig. 6.** Similar to Fig. 5, but now in units of fractional disk radius (see ordinate). **(a)**: Disk reflectivity ($k_s$) is set to 0.2. **(b)**: Same model parameterization but now with disk reflectivity $k_s = 0.4$. The negative slope in both panels indicates that the most massive giant planets have larger fractions of their disks fed with water ice and, consequently, relatively more space and material to form big moons.

## 3. $H_2O$ ice lines around planets of different masses

Figures 2(a) and (b) show that the temperature distribution in the late accretion disks around Jupiter-mass planets is determined mostly by the planetary illumination rather than viscous heating. This indicates a key difference between moon formation in circumplanetary accretion disks and planet formation in circumstellar disks, where the positions of ice lines have been shown to depend mostly on viscous heating rather than stellar illumination (Min et al. 2011; Hasegawa & Pudritz 2011). In this context, Figs. 3(a) and (b) depict an interesting intermediate case in terms of the mass of the central object, where both viscous heating and illumination around a $12\,M_{Jup}$ planet show comparable contributions towards the final stages of accretion in panel (b). Also note that in Fig. 3(b) the temperature distribution in the disk outskirts is determined by the background temperature provided by stellar illumination. This suggests that the extended disks around super-Jovian exoplanets in wide circumstellar orbits (beyond about 10 AU) might also feature CO and other ice lines due to the even weaker stellar illumination.

Figures 5(a) and (b) display the radial positions of the circumplanetary $H_2O$ ice lines around a range of planets with masses between 1 and $12\,M_{Jup}$ in the final stages of accretion, where we randomized $\kappa_P$ and $\dot{M}_{shut}$ as described above, and the stellar distance is 5.2 AU for all planets considered. Panel (a) shows the same data as Fig. 4 in Heller & Pudritz (2014), while panel (b) assumes $k_s = 0.4$. The larger absorptivity in panel (b) pushes the $H_2O$ ice lines slightly away from the accreting planets compared to panel (a) for all planets except for the $12\,M_{Jup}$ planet, which we ascribe to an insignificant statistical fluctuation. As a key result of our new simulations for $k_s = 0.4$, and in agreement with our previous study, the $H_2O$ ice line is between about 15 and $30\,R_{Jup}$ for all super-Jovian planets and almost independent of $M_p$. The effect of changing $k_s$ by a factor of two is moderate, pushing the $H_2O$ ice line outward by only a few $R_{Jup}$ on average.

Figure 6 presents a different visualization of these results, again with panel (a) assuming $k_s = 0.2$ and panel (b) assuming $k_s = 0.4$. Now the ordinate gives the circumplanetary distance in units of the fractional disk radius, and so the $H_2O$ ice lines are located closer to the planet with increasing $M_p$. This is a consequence of the larger disk sizes of the more massive planets. In other words, larger giant planets have larger fractions of the accretion disks beyond the $H_2O$ ice lines. Naturally, this means that these super-Jovians should form the most massive, icy moons.

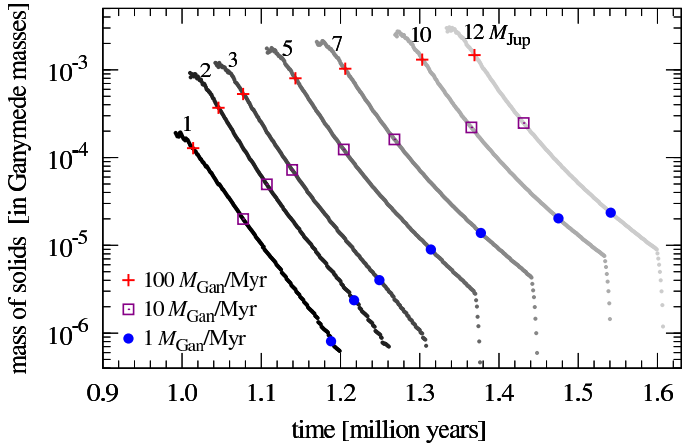

**Fig. 7.** Evolution of the total mass of solids ($M_{sld}$) in circumplanetary accretion disks of several test planets at 5.2 AU from a Sun-like star. Colored dots along the mass evolution tracks indicate accretion rates ($\dot{M}$). If moon formation around all of these planets shuts down at comparable disk accretion rates ($\dot{M}_{shut}$), then $M_{sld}(\dot{M}_{shut}) \propto M_p$ (see also Fig. 5 in Heller & Pudritz 2014). Compared to a Jupiter-mass planet, a 12 $M_{Jup}$ planet would then have 12 times the amount of solids available for moon formation.

In Fig. 7, we therefore analyze the evolution of $M_{sld}$ (in units of $M_{Gan}$) around these planets as a function of time. Indicated values for $\dot{M}$ along these tracks help to visualize $M_{sld}$ as a function of accretion rates (see crosses, squares, and circles), because time on the abscissa delivers an incomplete picture of the mass evolution due to the different timescales of these disks. We find that disks around lower-mass super-Jovians contain less solid mass at any given accretion rate than disks around higher-mass super-Jovians. Solids also occur earlier in time as the appearance of the disk itself is regulated by the shrinking of the initially very large planet in our models: lower-mass giant planets contract earlier (Mordasini 2013). Assuming that moon formation shuts down at similar accretion rates around any of the simulated planets, the example accretion rates (100, 10, 1 $M_{Gan}$ Myr$^{-1}$) indicate an increase of the mass of solids as a function of $M_p$. For a given $\dot{M}_{shut}$, this scaling is $M_{sld} \propto M_p$ (see also Fig. 5 in Heller & Pudritz 2014), which is in agreement with the scaling relation for the total moon mass around giant planets found by Canup & Ward (2006).

## 4. $H_2O$ ice lines and total mass of solids around planets at various stellar distances

Having examined the sensitivity of our results to $M_p$ and to the disk's radiative properties, we now turn to the question of what exomoon systems are like around super-Jovian exoplanets at very different locations in their disks than our own Jupiter at 5.2 AU. Here, we shall encounter some significant surprises.

Figure 8(a) shows the final circumplanetary distance of the $H_2O$ ice line around a Jupiter-mass planet between 2 and 20 AU from the star,[4] assuming a shutdown accretion rate of 100 $M_{Gan}$ Myr$^{-1}$. Different styles of the blue lines correspond to different disk opacities (see legend), while the solid black line indicates the radius of the circumplanetary accretion disk, following Machida et al. (2008). Circles indicate the radial distances of the Galilean moons around Jupiter, at 5.2 AU from the star. In this set of simulations, the $H_2O$ ice line always ends up between the orbits of Ganymede and Callisto, which is not in agreement with the observed $H_2O$ compositional gradient in the Galilean system. As stellar illumination decreases at larger distances while all other heating terms are constant for the given accretion rate, the ice lines move towards the planet at greater distances.

---

[4] We also simulated planets as close as 0.2 AU to the star, but their accretion disks are nominally smaller than the planetary radius, indicating a departure of our model from reality. Anyways, since the hypothetical disks around these planets do not harbor $H_2O$ ice lines and moon formation in the stellar vicinity is hard in the first place (Barnes & O'Brien 2002; Namouni 2010), we limit Fig. 10 to 2 AU.

Most importantly, however, we find that Jovian planets closer than about 4.8 AU do not have an $H_2O$ ice line in the first place. Hence, if the large population of Jupiter-mass planets around 1 AU (see Fig. 1) formed in situ and without substantial inward migration from the outer regions, then these giant planets should not have had the capacity to form giant, icy moons, that is, scaled-up versions of Ganymede or Callisto. These giant moons with masses up to that of Mars (suggested by Canup & Ward 2006; Heller et al. 2014; Heller & Pudritz 2014), may only be present if they have completed their own water-rich formation beyond about 4.8 AU from their star, before they migrated to their current circumstellar orbits together with their host planets.

In Fig. 8(b), the planetary accretion rate has dropped by a factor of ten, and the ice lines have shifted. For $\kappa_P = 10^{-2}$ m$^2$ kg$^{-1}$ and $\kappa_P = 10^{-1}$ m$^2$ kg$^{-1}$ they moved towards the planet. But for $\kappa_P = 10^{-3}$ m$^2$ kg$^{-1}$ they moved outward. The former two rates actually place the $H_2O$ ice line around Jupiter at almost exactly the orbit of Ganymede, which is in better agreement with observations. In these simulations, Jovian planets closer than about 4.5 AU do not have a circumplanetary $H_2O$ ice line.

In Fig. 9 we vary the stellar distance of a 12 $M_{Jup}$ planet. First, note that the disk (black solid line) is larger at any given stellar separation than in Fig. 8. Second, note that the $H_2O$ ice line for a given $\dot{M}_{shut}$ in Figs. 9(a) and (b), respectively, is further out than in the former case of a Jupiter-mass planet. This is due to both increased viscous heating and illumination from the planet for the super-Jovian object. Nevertheless, the much larger disk radius overcompensates for this effect and, hence, all these simulations suggests that accretion disks around the most massive planets can have $H_2O$ ice lines if the planet is not closer than about 3.8 AU (panel a) to 3.1 AU (panel b) to a Sun-like star. The latter value refers to a shutdown accretion rate of 10 $M_{Gan}$ Myr$^{-1}$, which is in good agreement with the water ice distribution in the Galilean moon system. Assuming that moon formation stops at comparable accretion rates around super-Jovian planets, we consider a critical stellar distance of about 3 AU a plausible estimate for the critical stellar distance of a 12 $M_{Jup}$ accreting planet to show a circumplanetary $H_2O$ ice line.

Figure 10 presents the total instantaneous mass of solids in circumplanetary accretion disk as a function of stellar distance for two different shutdown accretion rates and three different disk Planck opacities. The calculation of $M_{sld}$ follows Eq. (17) in Heller & Pudritz (2014), that is, we integrate the surface density of solids between the inner disk truncation radius and the outer centrifugal radius. Panel (a) for a Jupiter-mass planet demonstrates that indeed the amount of solids in its late accretion disk is negligible within about 4.5 AU from a Sun-like star. This gives us crucial insights into Jupiter's migration history within the Grand Tack framework (Walsh et al. 2011), which we will present in a forthcoming paper (Heller et al. 2015, in prep.). Moreover, both panels show that the effect of different disk opacities on $M_{sld}$ is small for a given $M_p$ and $\dot{M}_{shut}$.

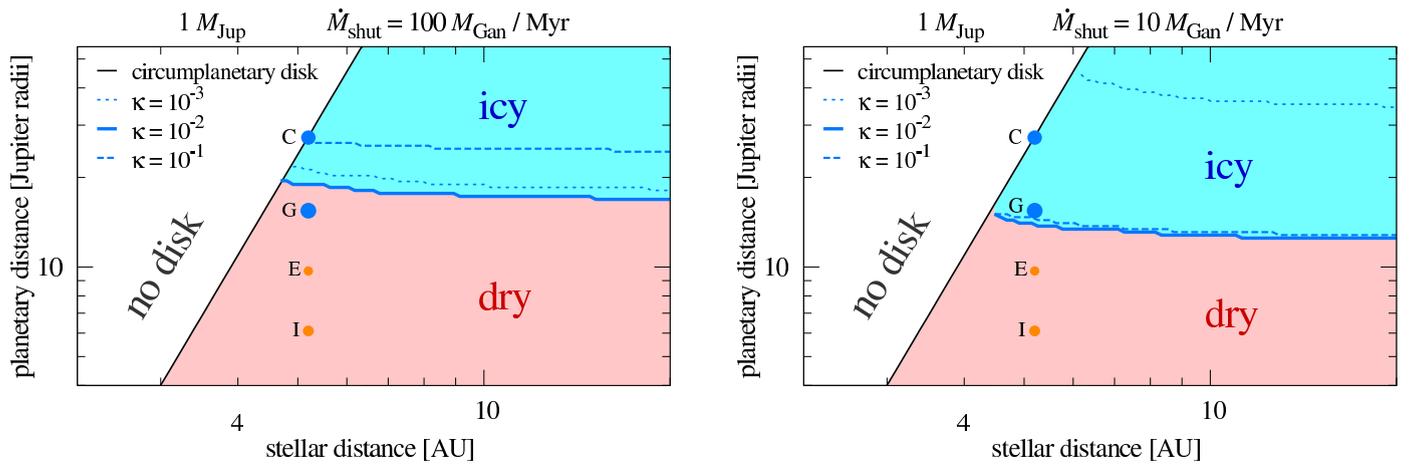

**Fig. 8.** Distances of the circumplanetary H$_2$O ice lines around a Jupiter-like planet (ordinate) as a function of distance to a Sun-like star (abscissa) at the time of moon formation shutdown. Each panel assumes a different shutdown formation rate (see panel titles). Three Planck opacities through the circumplanetary disk are tested in each panel (in units of m$^2$ kg$^{-1}$, see panel legends). The circumplanetary orbits of the Galilean satellites are represented by symbols as in Fig. 2. The black solid line shows the disk's centrifugal radius (Machida et al. 2008). **(a)** At $\dot{M}_{\rm shut} = 100\,M_{\rm Gan}$/Myr the H$_2$O ice lines at $\approx 5$ AU are about $5\,R_{\rm Jup}$ beyond Ganymede's current orbit. **(b)** At $\dot{M}_{\rm shut} = 10\,M_{\rm Gan}$/Myr values of $10^{-2}$ m$^2$ kg$^{-1}$ $\leq \kappa \leq 10^{-1}$ m$^2$ kg$^{-1}$ place the H$_2$O ice line slightly inside the current orbit of Ganymede and thereby seem most plausible.

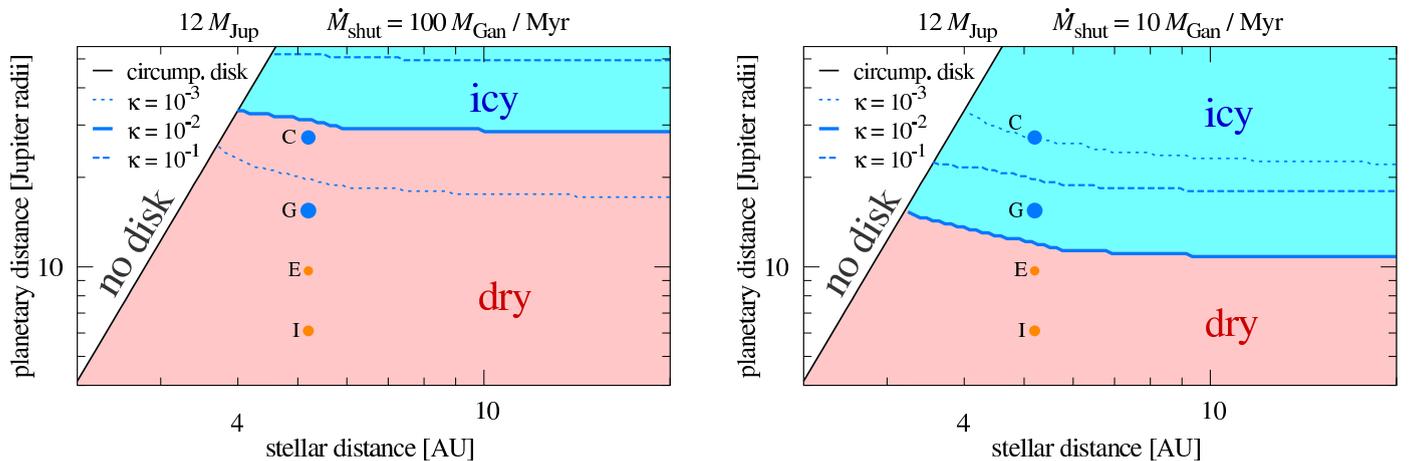

**Fig. 9.** Same as Fig. 8, but now for a 12 $M_{\rm Jup}$ mass planet.

Intriguingly, for any given disk parameterization (see legend) a circumjovian disk at 5.2 AU from a Sun-like star seems to harbor a relatively small amount of solids compared to disks around planets at larger stellar distances. The red dashed line indicates a best fit exponential function to the simulations of our fiducial disk with $\dot{M}_{\rm shut} = 10\,M_{\rm Gan}$ Myr$^{-1}$ and $\kappa_P = 10^{-2}$ m$^2$ kg$^{-1}$ beyond 5.2 AU, as an example. We choose this disk parameterization as it yields the best agreement with the radial location of the icy Galilean satellites (Heller & Pudritz 2014). It scales as

$$M_{\rm sld} = 10^{-5.3}\,M_{\rm Gan} \times \left(\frac{r_\star}{\rm AU}\right)^{1.16}\,,\qquad(1)$$

where $r_\star$ is the stellar distance. Assuming that moon formation stops at similar mass accretion rates around any super-Jovian planet and taking into account the previously known scaling of the total moon masses ($M_{\rm T}$) with $M_{\rm p}$, we deduce a more general estimate for the total moon mass:

$$M_{\rm T} = M_{\rm GM} \times \left(\frac{M_{\rm p}}{M_{\rm Jup}}\right) \times \left(\frac{r_\star}{5.2\ {\rm AU}}\right)^{1.16}\,,\ (r_\star \geq 5.2\ {\rm AU})\,,\qquad(2)$$

where $M_{\rm GM} = 2.65\,M_{\rm Gan}$ is the total mass of the Galilean moons. As an example, a 10 $M_{\rm Jup}$ planet forming at 5.2 AU around a Sunlike star should have a moon system with a total mass of about 10 $M_{\rm GM}$ or 6 times the mass of Mars. At a Saturn-like stellar distance of 9.6 AU, $M_{\rm T}$ would be doubled. Simulations by Heller et al. (2014) show that this mass will be distributed over three to six moons in about 90 % of the cases. Hence, if the most massive super-Jovian planets formed moon systems before they migrated to about 1 AU, where we observe them today (see Fig. 1), then Mars-mass moons in the stellar habitable zones might be very abundant.

## 5. Shutdown accretion rates and loss of moons

Canup & Ward (2002) argued that accretion rates of $2 \times 10^{-7}$ $M_{\rm Jup}$ yr$^{-1}$ (or about $2.6 \times 10^4$ $M_{\rm Gan}$ yr$^{-1}$) best reproduce the disk conditions in which the Galilean system formed. Based on the condition that the H$_2$O ice line needs to be between the orbits of Europa and Ganymede at the final stages of accretion, our calculations predict a shutdown accretion rate that is considerably lower, closer to 10 $M_{\rm Gan}$ Myr$^{-1}$ (see Fig. 8). The difference in these results is mainly owed to two facts.

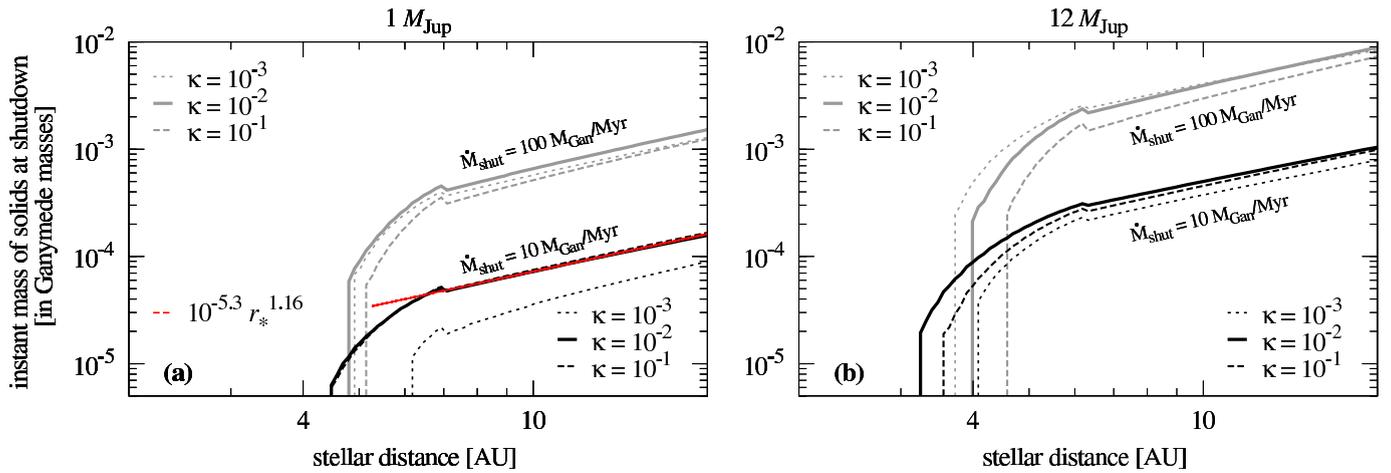

**Fig. 10.** Total instantaneous mass of solids in circumplanetary accretion disks in units of Ganymede masses and as a function of stellar distance. Labels indicate two different values for $\dot{M}_{shut}$, and different line types refer to different disk opacities (see legend). Black dashed lines in both panels indicate our fiducial reference disk. **(a):** The 1 $M_{Jup}$ test planet starts to have increasingly large amounts of solids only beyond 4.5 AU, depending on the accretion rate and disk opacity. The red dashed line indicates our fit as per Equation (1). **(b):** The 12 $M_{Jup}$ test planet starts to contain substantial amounts of solids as close as about 3.1 AU to the star, owed to its larger disk.

First, Canup & Ward (2002) only considered viscous heating and planetary illumination. Our additional heating terms (accretion onto the disk and stellar illumination) contribute additional heat, which imply smaller accretion rates to let the $H_2O$ ice lines move close enough to the Jupiter-like planet. Second, the parameterization of planetary illumination in the Canup & Ward (2002) model is different from ours. While Canup & Ward (2002) assume an $r^{-3/4}$ dependence of the midplane temperature from the planet ($r$ being the planetary radial distance), we do not apply any predescribed $r$-dependence. In particular, $T_m(r)$ cannot be described properly by a simple polynomial due to the different slopes of the various heat sources as a function of planetary distance (see the black solid lines in Figs. 2 and 3).

While our estimates of $\dot{M}_{shut}$ are about three orders of magnitude lower than the values proposed by Canup & Ward (2002), they are also one to two orders of magnitude lower than the values suggested by Alibert et al. (2005). Their moon formation model is similar to the so-called "gas-starved" disk model applied by Canup & Ward (2002). Accretion rates in their model were not derived from planet formation simulations (as in our case) but calculated using an analytical fit to previous simulations. Again, the additional energy terms in our model are one reason for the lower final accretion rates that we require in order to have the $H_2O$ ice line interior to the orbit of Ganymede.

Makalkin & Dorofeeva (2014) also studied Jupiter's accretion rates and their compatibility with the Galilean moon system. They found that values between $10^{-9}\,M_{Jup}\,yr^{-1}$ and $10^{-6}\,M_{Jup}\,yr^{-1}$ (about $10\,M_{Gan}\,Myr^{-1}$ to $10^4\,M_{Gan}\,Myr^{-1}$) satisfy these constraints. Obviously, our results are at the lower end of this range, while accretion rates up to three orders of magnitude higher should be reasonable according to Makalkin & Dorofeeva (2014). Yet, these authors also claimed that planetary illumination is negligible in the final states of accretion, which is not supported by our findings (see Fig. 2). We ascribe this discrepancy to the fact that they estimated the planetary luminosity ($L_p$) analytically, while in our model $L_p$, $R_p$, $M_p$, and $\dot{M}$ are coupled in a sophisticated planet evolution model (Mordasini 2013). In the Makalkin & Dorofeeva (2014) simulations, $10^{-7}\,L_\odot \lesssim L_p \lesssim 10^{-4}\,L_\odot$ (depending on assumed values for $R_p$, $M_p$, and $\dot{M}$), whereas in our case $L_p$ remains close to $10^{-4}\,L_\odot$ once a circumjovian $H_2O$ ice line forms, while $\dot{M}$ and $R_p$ evolve rapidly due to the planet's gap opening in the circumsolar disk (see Fig. 1 in Heller & Pudritz 2014).

We also calculate the type I migration time scales ($\tau_I$) of potential moons that form in our circumplanetary disks. Using Eq. (1) from Canup & Ward (2006) and assuming a Ganymede-mass moon, we find that $0.1\,Myr \lesssim \tau_I \lesssim 100\,Myr$ in the final stages of accretion, with the shortest time scales referring to close-in moons and early stages of accretion when the gas surface density is still high. Hence, since the remaining disk lifetime ($\approx 10^6\,yr$) is comparable or smaller than the type I migration time scale, migration traps might be needed to stop the moons from falling into the planet. As the gas surface density is decreasing by about an order of magnitude per $10^5\,yr$ (see Fig. 1(c) in Heller & Pudritz 2014) and since $\tau_I$ is inversely proportional to $\Sigma_g$, type I migration slows down substantially even if the protosatellites still grow. On the other hand, if a moon grows large enough to open up a gap in the circumplanetary accretion disk, then type II (inwards) migration might kick in, reinforcing the need for moon migration traps. All these issues call for a detailed study of moon migration under the effects of ice lines and other traps.

Another issue that could cause moon loss is given by their possible tidal migration. If a planet rotates very slowly, its corotation radius will be very wide and any moons would be forced to tidally migrate inwards (ignoring mean motion resonances for the time being). But if the planet rotates quickly and is not subject to tides raised by the star, such as Jupiter, then moons usually migrate outwards due to the planetary tides and at some point they might become gravitationally unbound. Barnes & O'Brien (2002, see their Fig. 2) showed that Mars-mass moons around giant planets do not fall into the planet and also remain bound for at least 4.6 Gyr if the planet is at least 0.17 AU away from a Sun-like star. Of course, details depend on the exact initial orbit of the moon and on the tidal parameterization of the system, but as we consider planets at several AU from the star, we conclude that loss of moons due to tidal inward or outward migration is not an issue. Yet, it might have an effect on the orbital distances where we can expect those giant moons to be found, so additional tidal studies will be helpful.

## 6. Discussion

If the abundant population of super-Jovian planets at about 1 AU and closer to Sun-like stars formed in situ, then our results suggest that these planets could not form massive, super-Ganymede-style moons in their accretion disks. These disks would have been too small to feature $H_2O$ ice lines and therefore the growth of icy satellites. The accretion disks around these planets might still have formed massive, rocky moons, similar in composition to Io or Europa, which would likely be in close orbits (thereby raising the issue of tidal evolution, Barnes & O'Brien 2002; Cassidy et al. 2009; Porter & Grundy 2011; Heller & Barnes 2013; Heller et al. 2014), since circumplanetary accretion disks at 1 AU are relatively small. If these large, close-orbit rocky moons exist, they also might be subject to substantial tidal heating. Alternatively, super-Jovians might have captured moons, e.g. via tidal disruption of binary systems during close encounters (Agnor & Hamilton 2006; Williams 2013), so there might exist independent formation channels for giant, possibly water-rich moons at 1 AU.

If, however, these super-Jovian planets formed beyond 3 to 4.5 AU and then migrated to their current locations, then they could be commonly orbited by Mars-mass moons with up to 50 % of water, similar in composition to Ganymede and Callisto. Hence, the future detection or non-detection of such moons will help to constrain rather strongly the migration history of their host planets. What is more, Mars-mass ocean moons at about 1 AU from Sun-like stars might be abundant extrasolar habitats (Williams et al. 1997; Heller & Barnes 2013; Heller et al. 2014).

Our results raise interesting questions about the formation of giant planets with satellite systems in the solar system and beyond. The field of moon formation around super-Jovian exoplanets is a new research area, and so many basic questions still need to be answered.

**(1) The "Grand Tack".** In the "Grand Tack" scenario (Walsh et al. 2011), the fully accreted Jupiter migrated as close as 1.5 AU to the Sun during the first few million years of the solar system, then got caught in a mean-motion orbital resonance with Saturn and then moved outward to about 5 AU. Our results suggest that the icy moons Ganymede and Callisto can hardly have formed during the several $10^5$ yr Jupiter spent inside 4.5 AU to the Sun. If they formed before Jupiter's tack, could their motion through the inner solar system be recorded in these moons today? Alternatively, if Callisto required $10^5$ - $10^6$ yr to form (Canup & Ward 2002; Mosqueira & Estrada 2003) and assuming that Jupiter's accretion disk was intact until after the tack, one might suppose that Ganymede and Callisto might have formed thereafter. But then how did Jupiter's accretion disk reacquire the large amounts of $H_2O$ that would then be incorporated into Callisto after all water had been sublimated during the tack? We will address these issues in a companion paper (Heller et al. 2015).

**(2) Migrating planets.** Future work will need to include the actual migration of the host planets, which we neglected in this paper. Namouni (2010) studied the orbital stability of hypothetical moons about migrating giant planets, but the formation of these moons was not considered. Yet, the timing of the accretion evolution, the gap opening, the movement within the circumstellar disk (and thereby the varying effect of stellar heating), and the shutdown of moon formation will be crucial to fully assess the possibility of large exomoons at about 1 AU. These simulations should be feasible within the framework of our model, but the precomputed planet evolution tracks would need to consider planet migration. Ultimately, magneto-hydrodynamical simulations of the circumplanetary accretion disks around migrating super-Jovian planets might draw a full picture.

**(3) Directly imaged planets.** Upcoming ground-based extremely large telescopes such as the *E-ELT* and the *Thirty Meter Telescope*, as well as the *James Webb Space Telescope* have the potential to discover large moons transiting directly imaged planets in the infrared (Peters & Turner 2013; Heller & Albrecht 2014). Exomoon hunters aiming at these young giant planets beyond typically 10 AU from the star will need to know how moon formation takes place in these possibly very extended circumplanetary accretion disks, under negligible stellar heating, and in the low-density regions of the circumstellar accretion disk.

**(4) Ice line traps.** If circumplanetary $H_2O$ ice lines can act as moon migration traps and if solids make up a substantial part of the final masses accreted by the planet, then the accretion rates onto giant planets computed under the neglect of moons might be incorrect in the final stages of accretion. The potential of the $H_2O$ ice line to act as a moon migration trap is new (Heller & Pudritz 2014) and needs to be tested. It will therefore be necessary to compute the torques acting on the accreting moons within the circumplanetary accretion disk as well as the possible gap opening in the circumplanetary disk by large moons, which might trigger type II migration (Canup & Ward 2002, 2006).

## 7. Conclusions

Planetary illumination is the dominant energy source in the late-stage accretion disks around Jupiter-mass planets at 5.2 AU from their Sun-like host stars (Fig. 2), while viscous heating can be comparable in the final stages of accretion around the most massive planets (Fig. 3).

At the time of moon formation shutdown, the $H_2O$ ice line in accretion disks around super-Jovian planets at 5.2 AU from Sun-like host stars is between roughly 15 and 30 $R_{Jup}$. This distance range is almost independent of the final planetary mass and weakly dependent on the disk's absorption properties (Fig. 5). With more massive planets having more extended accretion disks, this means that more massive planets have larger fractions (up to 70 %) of their disks beyond the circumplanetary $H_2O$ ice line (compared to about 25 % around Jupiter, see Fig. 6).

Jupiter-mass planets forming closer than about 4.5 AU to a Sun-like star do not have a circumplanetary $H_2O$ ice line (Fig. 8), depending on the opacity details of the circumstellar disk. A detailed application of this aspect to the formation of the Galilean satellites might help constraining the initial conditions of the Grand Tack paradigm. Due to their larger disks, the most massive super-Jovian planets can host an $H_2O$ ice line as close as about 3 AU to Sun-like stars (Fig. 9). With the circumstellar $H_2O$ ice line at about 2.7 AU in our model of an optically thin circumstellar disk (Hayashi 1981), the relatively small accretion disks around Jupiter-mass planets at several AU from Sun-like stars thus substantially constrain the formation of icy moons. The extended disks around super-Jovian planets, on the other hand, might still form icy moons even if the planet is close to the circumstellar $H_2O$ ice line.

We find an approximation for the total mass available for moon formation ($M_T$), which is a function of both $M_p$ and $r_\star$ (see Eq. 2). The linear dependence of $M_T \propto M_p$ has been known before, but the dependence on stellar distance ($M_T \propto r_\star^{1.16}$, for $r_\star \geq 5.2$ AU) is new. It is based on our finding that an accretion rate $\dot{M}_{shut} \approx 10 \, M_{Gan}/10^6$ yr (Fig. 10) yields the best results for the position of the circumjovian $H_2O$ ice line at the shutdown of the formation of the Galilean moons (Fig. 2b); and it assumes that this shutdown accretion rate is similar around all super-Jovian planets.

Our results suggest that the observed large population of super-Jovian planets at about 1 AU to Sun-like stars should not be orbited by water-rich moons if the planets formed in-situ. However, in the more plausible case that these planets migrated to their current orbits from beyond about 3 to 4.5 AU, they should be orbited by large, Mars-sized moons with astrobiological potential. As a result, future detections or non-detections of exomoons around giant planets can help to distinguish between the two scenarios because they are tracers of their host planets' migration histories.

*Acknowledgements.* We thank Sébastien Charnoz for his referee report which helped us to clarify several passages of this manuscript. We thank Christoph Mordasini for providing us with the precomputed planetary evolution tracks. René Heller is supported by the Origins Institute at McMaster University and by the Canadian Astrobiology Training Program, a Collaborative Research and Training Experience Program funded by the Natural Sciences and Engineering Research Council of Canada (NSERC). Ralph E. Pudritz is supported by a Discovery grant from NSERC. This work made use of NASA's ADS Bibliographic Services and of The Extrasolar Planet Encyclopaedia (www.exoplanet.eu). Computations have been performed with ipython 0.13.2 on python 2.7.2 (Pérez & Granger 2007), and all figures were prepared with gnuplot 4.6 (www.gnuplot.info).

## 5.3 The Formation of the Galilean Moons and Titan in the Grand Tack Scenario (Heller et al. 2015)

Contribution:

RH did the literature research, contributed to the mathematical framework, translated the math into computer code, created all figures, led the writing of the manuscript, and served as a corresponding author for the journal editor and the referees.

Letter to the Editor

# The formation of the Galilean moons and Titan
# in the Grand Tack scenario

R. Heller[1,2,*], G.-D. Marleau[3,**], and R. E. Pudritz[1,2]

[1] Origins Institute, McMaster University, 1280 Main Street West, Hamilton, ON L8S 4M1, Canada
[2] Department of Physics and Astronomy, McMaster University, rheller@physics.mcmaster.ca | pudritz@physics.mcmaster.ca
[3] Max-Planck-Institut für Astronomie, Königstuhl 17, 69117 Heidelberg, Germany, marleau@mpia.de

**ABSTRACT**

*Context.* In the Grand Tack (GT) scenario for the young solar system, Jupiter formed beyond 3.5 AU from the Sun and migrated as close as 1.5 AU until it encountered an orbital resonance with Saturn. Both planets then supposedly migrated outward for several $10^5$ yr, with Jupiter ending up at $\approx 5$ AU. The initial conditions of the GT and the timing between Jupiter's migration and the formation of the Galilean satellites remain unexplored.
*Aims.* We study the formation of Ganymede and Callisto, both of which consist of $\approx 50\%$ $H_2O$ and rock, in the GT scenario. We examine why they lack dense atmospheres, while Titan is surrounded by a thick $N_2$ envelope.
*Methods.* We model an axially symmetric circumplanetary disk (CPD) in hydrostatic equilibrium around Jupiter. The CPD is warmed by viscous heating, Jupiter's luminosity, accretional heating, and the Sun. The position of the $H_2O$ ice line in the CPD, which is crucial for the formation of massive moons, is computed at various solar distances. We assess the loss of Galilean atmospheres due to high-energy radiation from the young Sun.
*Results.* Ganymede and Callisto cannot have accreted their $H_2O$ during Jupiter's supposed GT, because its CPD (if still active) was too warm to host ices and much smaller than Ganymede's contemporary orbit. From a thermal perspective, the Galilean moons might have had significant atmospheres, but these would probably have been eroded during the GT in $< 10^5$ yr by solar XUV radiation.
*Conclusions.* Jupiter and the Galilean moons formed beyond $4.5 \pm 0.5$ AU and prior to the proposed GT. Thereafter, Jupiter's CPD would have been dry, and delayed accretion of planetesimals should have created water-rich Io and Europa. While Galilean atmospheres would have been lost during the GT, Titan would have formed after Saturn's own tack, because Saturn still accreted substantially for $\approx 10^6$ yr after its closest solar approach, ending up at about 7 AU.

**Key words.** Accretion, accretion disks – Planets and satellites: atmospheres – Planets and satellites: formation – Planets and satellites: physical evolution – Sun: UV radiation

## 1. Introduction

Recent simulations of the early solar system suggest that the four giant planets underwent at least two epochs of rapid orbital evolution. In the Grand Tack (GT) model (Walsh et al. 2011), they formed beyond the solar water ($H_2O$) ice line (Ciesla & Cuzzi 2006) between 3.5 AU and 8 AU. During the first few $10^6$ yr in the protostellar disk, when Jupiter had fully formed and Saturn was still growing, Jupiter migrated as close as 1.5 AU to the Sun within some $10^5$ yr, until Saturn grew enough mass to migrate even more rapidly, catching Jupiter in a 3:2 or a 2:1 mean motion resonance (Pierens et al. 2014). Both planets then reversed their migration, with Jupiter ending up at approximately 5 AU. The other gas planets also underwent rapid orbital evolution due to gravitational interaction, pushing Uranus and Neptune to and beyond about 10 AU, respectively, while Saturn settled at $\approx 7$ AU. A second period of rapid orbital evolution occurred $\approx 7 \times 10^8$ yr later, according to the Nice model (Tsiganis et al. 2005; Morbidelli et al. 2005; Gomes et al. 2005), when Jupiter and Saturn crossed a 2:1 mean motion resonance, thereby rearranging the architecture of the gas giants and of the minor bodies.

While the GT delivers adequate initial conditions for the Nice model, the initial conditions of the GT itself are not well constrained (Raymond & Morbidelli 2014). Details of the migrations of Jupiter and Saturn depend on details of the solar accretion disk and planetary accretion, which are also poorly constrained (Jacobson & Morbidelli 2014). We note, however, that the moons of the giant planets provide additional constraints on the GT that have scarcely been explored. As an example, using *N*-body simulations, Deienno et al. (2011) studied the orbital stability of the Uranian satellites during the Nice instability. Hypothetical moons beyond the outermost regular satellite Oberon typically got ejected, while the inner moons including Oberon remained bound to Uranus. Their results thus support the validity of the Nice model. Later, Deienno et al. (2014) found that one of three Jovian migration paths proposed earlier (Nesvorný & Morbidelli 2012) is incompatible with the orbital stability and alignment of the Galilean moons, yielding new constraints on the Nice model.

To our knowledge, no study has used the Galilean moons or Saturn's major moon Titan, which is surrounded by a thick nitrogen ($N_2$) atmosphere, to test the plausibility of the GT model. In this Letter, we focus on the early history of the icy Galilean moons, Ganymede and Callisto, and on Titan in the GT scenario; assuming that the GT scenario actually took place in one form or another, we identify new constraints on the timing of their formation. The novel aspect of our study is the evolution of the H$_2$O ice line in Jupiter's circumplanetary disk (CPD) under the effect of changing solar illumination during the GT. As the position and evolution of the ice line depends on irradiation from both the forming Jupiter and the Sun (Heller & Pudritz 2015a), we want to understand how the observed dichotomy of two mostly rocky and two very icy Galilean moons can be produced in a GT setting.

## 2. The icy Galilean moons in the Grand Tack model

### 2.1. Formation in the circumjovian accretion disk

Both Ganymede and Callisto consist of about 50 % rock and 50 % H$_2$O, while the inner Galilean satellites Io and Europa are mostly rocky. This has been considered a record of the temperature distribution in Jupiter's CPD at the time these moons formed (Pollack & Reynolds 1974; Sasaki et al. 2010). In particular, the H$_2$O ice line, which is the radial distance at which the disk is cool enough for the transition of H$_2$O vapor into solid ice, should have been between the orbits of rocky Europa at about 9.7 Jupiter radii ($R_{Jup}$) and icy Ganymede at about 15.5 $R_{Jup}$ from Jupiter.

We simulate a 2D axisymmetric CPD in hydrostatic equilibrium around Jupiter using a "gas-starved" standard disk model (Canup & Ward 2002, 2006) that has been modified to include various heat sources (Makalkin & Dorofeeva 2014), namely, (i) viscous heating, (ii) planetary illumination in the "cold-start scenario", (iii) direct accretion onto the CPD, and (iv) stellar illumination. The model is coupled to pre-computed planet evolution tracks (Mordasini 2013), and we here assume a solar luminosity 0.7 times its current value to take into account the faint young Sun (Sagan & Mullen 1972). Details of our semi-analytical model are described in Heller & Pudritz (2015a,b).

We evaluate the radial position of the circumjovian H$_2$O ice line at different solar distances of Jupiter and study several disk opacities ($\kappa_P$) as well as different shutdown accretion rates for moon formation ($\dot{M}_{shut}$), all of which determine the radial distributions of the gas surface density and midplane temperature in the CPD. We follow the CPD evolution after the planet opens up a gap in the circumstellar disk (CSD) until the planetary accretion rate ($\dot{M}_p$) in the pre-computed track drops to a particular value of $\dot{M}_{shut}$. In Heller & Pudritz (2015a,b) we found that $\dot{M}_{shut} = 10\,M_{Gan}\,\mathrm{Myr}^{-1}$ positions the H$_2$O ice line between Europa and Ganymede for a broad range of $\kappa_P$ values.

To assess the effect of heating on the position of the H$_2$O ice line in the CPD as a function of solar distance, we compare two CSD models. First, we use the model of Hayashi (1981, H81, his Eq. 2.3), which assumes that the CSD is mostly transparent in the optical. Second, we use the model for an optically thick disk of Bitsch et al. (2015, B15, their Eqs. A.3 and A.7), which takes into account viscous and stellar heating as well as radiative cooling and opacity transitions. We consider both a low-metallicity ($Z = 0.001$) and a solar-metallicity star ($Z = 0.02$) with a stellar accretion rate $\dot{M}_\star = 3.5 \times 10^{-8}\,M_\odot\,\mathrm{yr}^{-1}$ corresponding to Jupiter's runaway accretion phase (Mordasini 2013), in which the planet opens up a gap and moon formation shuts down.

Figure 1 shows the results of our calculations, assuming fiducial values $\kappa_P = 10^{-2}\,\mathrm{m}^2\,\mathrm{kg}^{-1}$ and $\dot{M}_{shut} = 10\,M_{Gan}\,\mathrm{Myr}^{-1}$ as well as a centrifugal CPD radius as per Machida et al. (2008). The abscissa denotes distance from the young Sun, the ordinate indicates distance from Jupiter. The vertical dotted line highlights the critical solar distance ($a_{crit}$) at which the CPD loses its H$_2$O ice line in the $Z = 2$ % H15 model. At 5.2 AU, Callisto's position (circle labeled "C") at the outer CPD edge is due to our specific CPD scaling (Heller & Pudritz 2015b), whereas Ganymede's position (circle labeled "G") near the H$_2$O ice line is not a fit but a result.[1] The CPD radius shrinks substantially towards the Sun owing to the increasing solar gravitational force, while the ice line recedes from the planet as a result of enhanced stellar heating. A comparison of the different CSD models (see legend) indicates that variations of stellar metallicity or solar illumination have significant effects on the circumjovian ice line, but the critical effect is the small CPD radius in the solar vicinity.

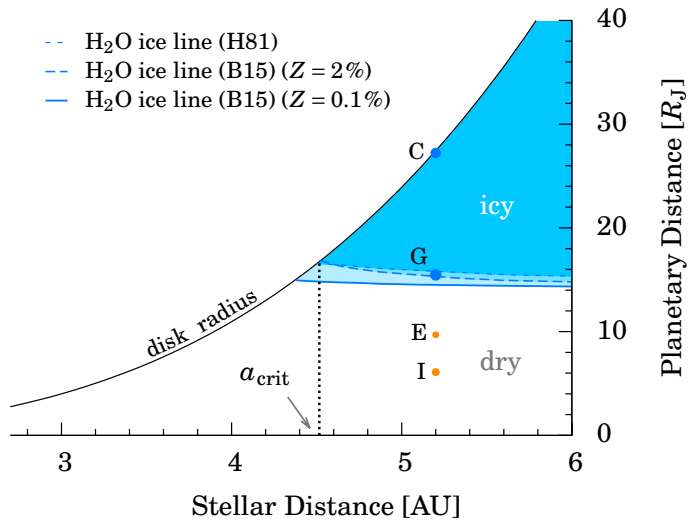

**Fig. 1.** Radial distance of Jupiter's H$_2$O ice line (roughly horizontal lines) and centrifugal disk radius (curved solid line) during its final accretion phase. Inside about 4.5 AU from the Sun, Jupiter's disk does not contain an H$_2$O ice line in any of the CSD models (Hayashi 1981; Bitsch et al. 2015, see legend). Locations of the Galilean moons are indicated by symbols at 5.2 AU. Symbol sizes scale with the physical radii of the moons (orange: rocky; blue: icy composition).

Most importantly, Fig. 1 suggests that Ganymede and Callisto cannot have accreted their icy components as long as Jupiter was closer than about $a_{crit} = 4.5$ AU to the faint Sun, where the H$_2$O ice line around Jupiter vanishes for both the H81 and the B15 model. We varied $\kappa_P$ and $\dot{M}_{shut}$ by an order of magnitude (not shown), which resulted in changes of this critical solar distance of $\lesssim 0.5$ AU. Moreover, closer than 5.2 AU from the Sun, Callisto's contemporary orbit around Jupiter would have been beyond the CPD radius, though still within Jupiter's Hill sphere.

Hence, large parts of both Ganymede and Callisto cannot have formed during the GT, which supposedly brought Jupiter much closer to the Sun than $4.5 \pm 0.5$ AU. Thereafter, Jupiter's accretion disk (if still present at that time) would have been void of H$_2$O because water first would have vaporized and then been photodissociated into hydrogen and oxygen. If Ganymede or Callisto had acquired their H$_2$O from newly accreted planetesimals after the GT (e.g. through gas drag within the CDP, Mosqueira & Estrada 2003), then Io (at 0.008 Hill radii, $R_H$) and Europa (at 0.01 $R_H$) would be water-rich, too, because planetesimal capture would have been efficient between 0.005 and 0.01 $R_H$ (Tanigawa et al. 2014). Hence, Ganymede and Callisto must have formed prior to the GT, at least to a large extent.

### 2.2. Atmospheric escape from the Galilean moons

Not only must a successful version of the GT model explain the observed H$_2$O ice contents in the Galilean moons, it also must be compatible with both the observed absence of thick atmospheres on the Galilean moons and the presence of a massive $N_2$ atmosphere around Saturn's moon Titan.

---
[1] Heller & Pudritz (2015b) argue that this suggests that Jupiter's H$_2$O ice line acted as a moon migration trap for Ganymede.

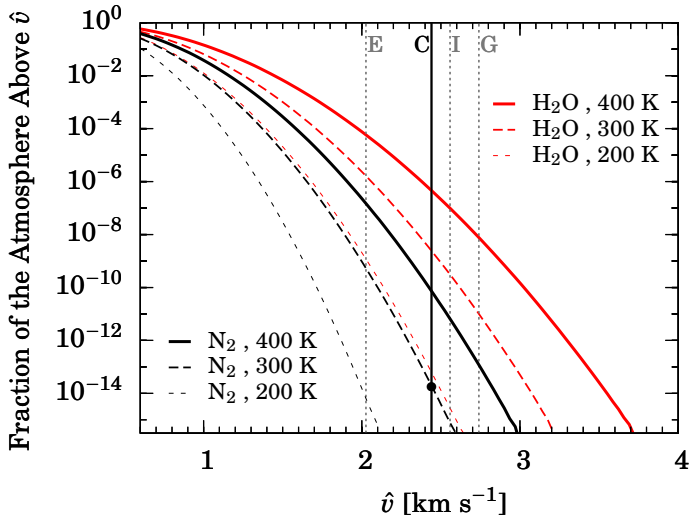

**Fig. 2.** Integrated Maxwell–Boltzmann velocity distribution of $N_2$ and $H_2O$ gas molecules with temperatures similar to Callisto's surface temperature during accretion. The black vertical line denotes Callisto's escape velocity, values for the other Galilean moons are indicated by gray vertical lines. The black circle refers to an example discussed in the text.

Several effects can drive atmospheric escape: (1) thermal (or "Jeans" escape, (2) direct absorption of high-energy (X-ray and ultraviolet, XUV) photons in the upper atmosphere, (3) direct absorption of high-energy particles from the solar wind, (4) impacts of large objects, and (5) drag of heavier gaseous components (such as carbon, oxygen, or nitrogen) by escaping lighter constituents (such as hydrogen) (Hunten et al. 1987; Pierrehumbert 2010).[2] Although additional chemical and weather-related effects can erode certain molecular species (Atreya et al. 2006), we identify the XUV-driven non-thermal escape during the GT as a novel picture that explains both the absence and presence of Galilean and Titanian atmospheres, respectively.

### 2.2.1. Thermal escape

As an example, we assess Callisto's potential to lose an initial $N_2$ or $H_2O$ vapor atmosphere via thermal escape. Our choice is motivated by Callisto's relatively low surface temperature during accretion, which was about 300 K at most, taking into account heating from Jupiter's accretion disk and from the accretion of planetesimals (Lunine & Stevenson 1982). The other moons might have been subject to substantial illumination by the young Jupiter, so the effect of thermal escape might be harder to assess.

We compute $F_{MB}(\hat{v}) = \int_{\hat{v}}^{\infty} dv f_{MB}(v)$, with $f_{MB}(v)$ as the Maxwell–Boltzmann velocity distribution, for various gas temperatures $v$. Then $F_{MB}^{norm}(\hat{v}) \equiv F_{MB}(\hat{v})/F_{MB}(0)$ is an approximation for the fraction of the atmosphere that is above a velocity $\hat{v}$.

In Figure 2 we compare $F_{MB}^{norm}(\hat{v})$ for $N_2$ (black lines) and $H_2O$ (red lines) molecules at temperatures between 200 K and 400 K (see legend) with Callisto's gravitational escape velocity (2.44 km s$^{-1}$, black vertical line). As an example, $F_{MB}^{norm}(\hat{v} > 2.44$ km s$^{-1}) \leq 10^{-14}$ for an $N_2$ troposphere at 300 K (see black circle), suggesting that a negligible fraction of the atmosphere would be beyond escape velocity.

---

[2] Atmospheric drag (5) can only occur if any of the other effects (1)–(4) is efficient for a lighter gas component.

Taking into account collisions of gas particles, their finite free path lengths, and the extent and temperature of the exosphere, Pierrehumbert (2010) estimates the loss time of $N_2$ from Titan as $\approx 10^{24}$ yr. For Callisto, which has about 80 % of Titan's mass, this timescale would only be reduced by a factor of a few (Pierrehumbert 2010, Eq. 8.37) even if it had been as close as 1.5 AU from the Sun. At that distance, Callisto would have received about 40 times higher irradiation than Titan receives today (assuming an optically thin CSD), and its exosphere temperatures might have been about $40^{1/4} \approx 2.5$ times higher than Titan's.

Hence, thermal escape cannot be the reason for the lack of an $N_2$ atmosphere on Callisto, even during the GT as close as 1.5 AU from the Sun.

### 2.2.2. XUV-driven non-thermal escape and drag

The X-ray and UV luminosities of the young Sun were as high as $10^2$ (Ribas et al. 2005) and $10^4$ (Zahnle & Walker 1982) times their current values, respectively, raising the question whether direct absorption of high-energy photons or atmospheric drag would have acted as efficient removal processes of an early Callistonian atmosphere. Such a non-thermal escape would have eroded a hypothetical initial $N_2$ atmosphere from Earth in only a few $10^6$ yr (Lichtenegger et al. 2010), owing to the high exobase temperatures (7000-8000 K) and the significant expansion of the thermosphere above the magnetopause (Tian et al. 2008). However, the Earth's primitive atmosphere was likely $CO_2$-rich, which cooled the thermosphere and limited the $H_2$ escape.

If Callisto initially had a substantial $CO_2$ or $H_2O$ steam atmosphere, perhaps provided by outgassing, both gases would have been photo-dissociated in the upper atmosphere, which then would have been dominated by escaping H atoms; $N_2$ and other gases would have been dragged beyond the outer atmosphere and lost from the moon forever. Lammer et al. (2014) simulated this gas drag for exomoons at 1 AU from young Sun-like stars and found that the H, O, and C inventories in the initially thick $CO_2$ and $H_2O$ atmosphere around moons 10 % the mass of the Earth (four times Ganymede's mass) would be lost within a few $10^5$ yr depending on the initial conditions and details of the XUV irradiation. Increasing the moon mass by a factor of five in their computations, non-thermal escape times increased by a factor of several tens. Given that strong a dependence of XUV-driven escape on a moon's mass, Ganymede and Callisto would most certainly have lost initial $N_2$ atmospheres during the GT (perhaps within $10^4$ yr), even if they approached the Sun as close as 1.5 AU rather than 1 AU as in Lammer et al. (2014).

### 2.3. Titan's atmosphere in the Grand Tack model

If XUV-driven atmospheric loss from the Galilean moons indeed occurred during the GT, this raises the question why Titan is still surrounded by a thick $N_2$ envelope, since Saturn supposedly migrated as close as 2 AU to the young Sun (Walsh et al. 2011). We propose that the key lies in the different formation timescales of the Galilean moons and Titan.

In the GT simulations of Walsh et al. (2011), Saturn accretes about 10 % of its final mass over several $10^5$ yr after its tack, when Jupiter is already fully formed. Hence, while the Galilean moons must have formed before Jupiter's GT (Sect. 2.1) and migrated towards the Sun, thereby losing any primordial atmospheres, Titan formed after Saturn's tack and on a longer timescale. Titan actually must have formed several $10^5$ yr after Saturn's tack, or it would have plunged into Saturn via its own type I migration within the massive CPD (Canup & Ward 2006; Sasaki et al. 2010) because there would be no ice line to trap it around Saturn (Heller & Pudritz 2015b).

The absence of $N_2$ atmospheres around the Galilean satellites supports the GT scenario and is compatible with the presence of a thick $N_2$ atmosphere around Titan. The Galilean satellites likely lost their primordial envelopes while approaching the Sun as close as 1.5 AU, whereas Titan formed after Saturn's tack at about 7 AU under less energetic XUV conditions. Its $N_2$ atmosphere then built up through outgassing of $NH_3$ accreted from the protosolar nebula (Mandt et al. 2014).

## 3. Discussion

Although the GT paradigm is still controversial, our results provide additional support for it. We show that the outcome of atmosphere-free icy Galilean satellites and of a thick $N_2$ atmosphere around Titan is possible in the GT scenario. This demonstrates how moon formation can be used to constrain the migration and accretion history of planets – in the solar system and beyond. The detection of massive exomoons around the observed exo-Jupiters at 1 AU from Sun-like stars would indicate that those gas giants migrated from beyond about 4.5 AU.

We tested two CSD models (H81 and B15) to study the effect of changing solar heating on the radial position of the $H_2O$ ice line in Jupiter's CPD. Both models yield similar constraints on the critical solar distance ($a_{crit}$) beyond which the icy Galilean satellites must have formed. We varied opacities, shutdown accretion rates, and stellar metallicities to estimate the dependence of our results on the uncertain properties of the protoplanetary and protosatellite disks and found that variations of $a_{crit}$ are $\lesssim 0.5$ AU. Residual heat from the moons' accretion, radiogenic decay, and tidal heating might have provided additional heat sources, which we neglected. Hence, $a_{crit} = 4.5 \pm 0.5$ AU must be considered a lower limit.

If the GT actually took place and our conclusions about the pre-GT formation of the Galilean moons are correct, their early thermal evolution needs to be readdressed. Radiogenic decay in the rocky components and residual heat from accretion have been considered the main heat sources that determined post-accretion internal differentiation (Kirk & Stevenson 1987). The total heat flux of several TW in both Ganymede and Callisto (Mueller & McKinnon 1988) translates into a few tens of $mW\,m^{-2}$ on the surface. Our results, however, suggest that sunlight might have been a significant external energy source. At 1.5 AU from the Sun, illumination might have reached several $W\,m^{-2}$, if the CSD was at least partly transparent in the optical. Near-surface temperatures might have been approximately 10 K higher for $\gtrsim 10^5$ yr (e.g. in Fig. 3a of Nagel et al. 2004). The GT might have significantly retarded the cooling of the Galilean satellites.

Alternatively, Ganymede and Callisto might have become $H_2O$-rich after the GT, maybe through ablation of newly accreted planetesimals (Mosqueira et al. 2010), but then a mechanism is required that either prevented Io and Europa from accreting significant amounts of icy planetesimals or that triggered the loss of accreted ice. Tidal heating and the release of large amounts of kinetic energy from giant impacts might account for that.

As H is an effective UV absorber (Glassgold et al. 2004) CSD gas might have shielded early Galilean atmospheres quite effectively. The net UV blocking effect depends on the disk scale height, the gas density profile in Jupiter's gap, and the residual gas flow (Fung et al. 2014). Ultraviolet photons might have been scattered deep into the gap, maybe even via a back-heating ef-fect from the dusty wall behind the gap (Hasegawa & Pudritz 2010). This aspect of our theory needs deeper investigation and dedicated CSD simulations with Jupiter migrating to 1.5 AU.

## 4. Conclusions

In this Letter we show that the Grand Tack model for the migration of Jupiter and Saturn, if valid, imposes important constraints on the formation of their massive icy moons, Ganymede, Callisto, and Titan: **(1)** Ganymede and Callisto (probably also Io and Europa) formed prior to the GT (Sect. 2.1), **(2)** their formation took place beyond $4.5 \pm 0.5$ AU from the Sun (Sect. 2.1), **(3)** the Galilean moons would have lost any primordial atmospheres during the GT via non-thermal XUV-driven escape due to the active young Sun (Sect. 2.2), and **(4)** Titan's thick $N_2$ atmosphere and constraints from moon migration in CPDs suggest that Titan formed after Saturn's tack (Sect. 2.3).

Detailed observations of the Galilean moons by ESA's upcoming JUpiter ICy moons Explorer (JUICE), scheduled for launch in 2022 and arrival at Jupiter in 2030 (Grasset et al. 2013), could deliver fundamentally new insights into the migration history of the giant planets. If Ganymede and Callisto formed prior to Jupiter's Grand Tack, then JUICE might have the capabilities to detect features imprinted during the moons' journey through the inner solar system.

*Acknowledgements.* We thank C. Mordasini, B. Bitsch, J. Blum, D. N. C Lin, W. Brandner, J. Leconte, C. P. Dullemond, E. Gaidos, and an anonymous referee for their helpful comments. RH is supported by the Origins Institute at McMaster U. and by the Canadian Astrobiology Program, a Collaborative Research and Training Experience Program by the Natural Sciences and Engineering Research Council of Canada (NSERC). REP is supported by an NSERC Discovery Grant.

## 5.4 The Nature of the Giant Exomoon Candidate Kepler-1625 b-i (Heller 2018c)

# The nature of the giant exomoon candidate Kepler-1625 b-i

René Heller[1]

Max Planck Institute for Solar System Research, Justus-von-Liebig-Weg 3, 37077 Göttingen, Germany; heller@mps.mpg.de

## ABSTRACT

The recent announcement of a Neptune-sized exomoon candidate around the transiting Jupiter-sized object Kepler-1625 b could indicate the presence of a hitherto unknown kind of gas giant moon, if confirmed. Three transits of Kepler-1625 b have been observed, allowing estimates of the radii of both objects. Mass estimates, however, have not been backed up by radial velocity measurements of the host star. Here we investigate possible mass regimes of the transiting system that could produce the observed signatures and study them in the context of moon formation in the solar system, i.e. via impacts, capture, or in-situ accretion. The radius of Kepler-1625 b suggests it could be anything from a gas giant planet somewhat more massive than Saturn ($0.4\,M_{\rm Jup}$) to a brown dwarf (BD) (up to $75\,M_{\rm Jup}$) or even a very-low-mass star (VLMS) ($112\,M_{\rm Jup} \approx 0.11\,M_\odot$). The proposed companion would certainly have a planetary mass. Possible extreme scenarios range from a highly inflated Earth-mass gas satellite to an atmosphere-free water-rock companion of about $180\,M_\oplus$. Furthermore, the planet-moon dynamics during the transits suggest a total system mass of $17.6^{+19.2}_{-12.6}\,M_{\rm Jup}$. A Neptune-mass exomoon around a giant planet or low-mass BD would not be compatible with the common mass scaling relation of the solar system moons about gas giants. The case of a mini-Neptune around a high-mass BD or a VLMS, however, would be located in a similar region of the satellite-to-host mass ratio diagram as Proxima b, the TRAPPIST-1 system, and LHS 1140 b. The capture of a Neptune-mass object around a $10\,M_{\rm Jup}$ planet during a close binary encounter is possible in principle. The ejected object, however, would have had to be a super-Earth object, raising further questions of how such a system could have formed. In summary, this exomoon candidate is barely compatible with established moon formation theories. If it can be validated as orbiting a super-Jovian planet, then it would pose an exquisite riddle for formation theorists to solve.

**Key words.** accretion, accretion disks – eclipses – planetary systems – planets and satellites: composition – planets and satellites: formation – planets and satellites: individual (Kepler-1625 b)

## 1. Introduction

The moons in the solar system serve as tracers of their host planets' formation and evolution. For example, the Earth's spin state, a key factor to our planet's habitability, is likely a result of a giant impact from a Mars-sized object into the proto-Earth (Cameron & Ward 1976), followed by the tidal interaction of the Earth-Moon binary (Touma & Wisdom 1994). The water contents and internal structures of the Galilean moons around Jupiter have been used to reconstruct the conditions in the accretion disk around Jupiter, in which they supposedly formed (Makalkin et al. 1999; Canup & Ward 2002; Heller et al. 2015). The moons around Uranus suggest a collisional tilting scenario for this icy gas giant (Morbidelli et al. 2012). It can thus be expected that the discovery of moons around extrasolar planets could give fundamentally new insights into the formation and evolution of exoplanets that cannot be obtained by exoplanet observations alone.

Another fascinating aspect of large moons beyond the solar system is their potential habitability (Reynolds et al. 1987; Williams et al. 1997; Scharf 2006; Heller et al. 2014). In fact, habitable moons could outnumber habitable planets by far, given their suspected abundance around gas giant planets in the stellar habitable zones (Heller & Pudritz 2015a).

Thousands of exoplanets have been found (Mayor & Queloz 1995; Morton et al. 2016), but no exomoon candidate has unequivocally been confirmed. Candidates have been presented based on a microlensing event (Bennett et al. 2014), asymmetries detected in the transit light curves of an exoplanet (Ben-Jaffel & Ballester 2014), and based on a single remarkable exoplanet transit in data from *CoRoT* (Lewis et al. 2015). Moreover, hints at an exomoon population have been found in the stacked lightcurves of the *Kepler* space telescope (Hippke 2015). The most recent and perhaps the most plausible and testable candidate has been announced by Teachey et al. (2018). In their search for the moon-induced orbital sampling effect (Heller 2014; Heller et al. 2016a) in exoplanet transit lightcurves from *Kepler*, they found the exomoon candidate Kepler-1625 b-i.

As the masses of Kepler-1625 b and its proposed companion remain unknown and the radius of the evolved host star is poorly constrained, the transiting object could indeed be a giant planet. It might also, however, be much larger than Jupiter and, thus, much more massive. Several Jupiter-sized transiting objects from *Kepler* that have been statistically validated but not confirmed through independent methods, such as stellar radial velocity (RV) measurements, later turned out to be very-low-mass stars (VLSMs) rather than planets (Shporer et al. 2017). Here, we report on the plausible masses of Kepler-1625 b and its satellite candidate and we show to what extent the possible scenarios would be compatible with planet and moon formation scenarios in the solar system.

## 2. Methods

### 2.1. System parameters and mass estimates

#### 2.1.1. Transit depths and structure models

Kepler-1625 (KIC 4760478, KOI-5084), at a distance of $2181^{+332}_{-581}$ pc, has been classified as an evolved G-type star with a mass of $M_\star = 1.079^{+0.100}_{-0.138} M_\odot$, a radius of $R_\star = 1.793^{+0.263}_{-0.488} R_\odot$, an effective temperature of $T_{\text{eff},\star} = 5548^{+83}_{-72}$ K (Mathur et al. 2017), and a Kepler magnitude of $K = 13.916$.[1] The transiting object Kepler-1625 b has an orbital period of $287.377314 \pm 0.002487$ d (Morton et al. 2016), which translates into an orbital semimajor axis of about 0.87 AU around the primary star. The next transit can be predicted to occur on 29 October 2017 at 02 : 34 : 51($\pm 00$ : 46 : 18) UT [2], and Teachey et al. (2018) have secured observations of this transit with the *Hubble Space Telescope*.

The mass of Kepler-1625 b is unknown. Nevertheless, a range of physically plausible masses can be derived from the observed transit depth and the corresponding radius ratio with respect to the star. The best model fits to the three transit lightcurves from Teachey et al. (2018) suggest transit depths for Kepler-1625 b (the primary) and its potential satellite (the secondary) of about $d_p = 4.3$ parts per thousand (ppt) and $d_s = 0.38$ ppt, which translate into a primary radius of $R_p = 1.18^{+0.18}_{-0.32} R_{\text{Jup}}$ and a secondary radius of $R_s = 0.35^{+0.05}_{-0.10} R_{\text{Jup}} = 0.99^{+0.15}_{-0.27} R_{\text{Nep}}$. The error bars are dominated by the uncertainties in $R_\star$.

Figure 1 shows a mass-radius curve for non-irradiated substellar objects at an age of 5 Gyr based on "COND" evolution tracks of Baraffe et al. (2003). Also indicated on the figure are the $1\sigma$ confidence range of possible radii of the primary. The surjective nature of the mass-radius relationship prevents a direct radius-to-mass conversion. As a consequence, as long as the mass of Kepler-1625 b is unknown, for example from longterm RV observations, the radius estimate is compatible with two mass regimes (see the blue bars at the bottom of Fig. 1). The low-mass regime extends from approximately $0.4 M_{\text{Jup}}$, (roughly 1.3 Saturn masses) to about $40 M_{\text{Jup}}$. The high-mass regime spans from $76 M_{\text{Jup}}$ to about $112 M_{\text{Jup}} \approx 0.11 M_\odot$. The entire range covers more than two orders of magnitude. Beyond that, the metallicity, age, and rotation state of Kepler-1625 b might have significant effects on its radius and on the curve shown in Fig. 1, allowing for an even wider range of plausible masses for a given radius.

The transition between gas giant planets and brown dwarfs (BDs), that is, between objects forming via core accretion that are unable to burn deuterium on the one hand and deuterium-burning objects that form through gravitational collapse on the other hand, is somewhere in the range between $10 M_{\text{Jup}}$ and $25 M_{\text{Jup}}$ (Baraffe et al. 2008). Objects more massive than about $85 M_{\text{Jup}}$ can ignite hydrogen burning to become VLMSs (Kumar 1963; Stevenson 1991). Based on the available radius estimates alone then, Kepler-1625 b could be anything from a gas giant to a VLMS.

Similarly, we may derive mass estimates for the proposed companion. The minimum plausible mass for an atmosphere-free object can be estimated by assuming a 50/50 water-rock composition, giving a mass of about $8 M_\oplus$ (Fortney et al. 2007). Most objects the size of this candidate do have gas envelopes (at least for orbital periods $\lesssim 50$ d; Rogers 2015), however, and a substantial atmosphere around the secondary would be likely. In fact, it could be as light as $1 M_\oplus$, considering the discovery of extremely low-density planets such as those around Kepler-51 (Masuda 2014). On the other end of the plausible mass range, if we consider the upper radius limit and an object composed of a massive core with a low-mass gas envelope akin to Neptune, we estimate a maximum mass of about $20 M_\oplus$ (Baraffe et al. 2008).

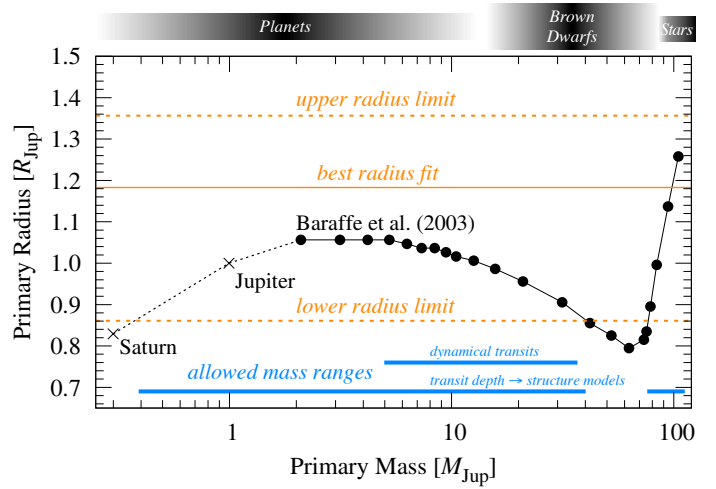

**Fig. 1.** Mass-radius isochrone for substellar objects and VLMSs at an age of 5 Gyr (Baraffe et al. 2003, black dots). Jupiter's and Saturn's positions are indicated with crosses. Horizontal lines indicate the range of possible radii for Kepler-1625 b estimated from the Teachey et al. (2018) transit lightcurves and from the uncertainties in the stellar radius. The blue bars at the bottom refer to the mass estimates that are compatible with the isochrone (lower intervals) and as derived from the dynamical signatures in the transits (upper interval).

### 2.1.2. Transit dynamics

Given the 287 d orbit of Kepler-1625 b, *Kepler* could have observed five transits during its four-year primary mission. Yet, only transits number two (T2), four (T4), and five (T5) of the transit chain were observed. The lightcurves by Teachey et al. (2018) suggest that T2 starts with the ingress of the proposed satellite, meaning that the satellite would have touched the stellar disk prior to its host. The transit T4 shows the opposite configuration, starting with the primary and ending with the unconfirmed secondary. Then, T5 indicates that the ingress of the moon candidate precedes the ingress of the planet. In contrast to T2, however, the planet leaves the stellar disk first whereas the proposed satellite would still be in transit for an additional 10 hr. This suggests a transit geometry in which the moon performs about half an orbit around the planet during the transit, between the two maximum angular deflections as seen from Earth. In fact, the 10 hr moon-only part of the transit suggest a sky-projected separation of $17.3 R_{\text{Jup}}$, close to the best fit for the orbital semimajor axis of $a_{\text{ps}} = 19.1^{+2.1}_{-1.9} R_p$ (Teachey et al. 2018). Thus, T5 carries important information about the possible planet-moon orbital period.

Assuming that a full orbit would take about twice the time required by the proposed satellite to complete T5 (i.e., its orbital period would be roughly $P_{\text{ps}} = 72$ hr), and assuming further an orbital semimajor axis of $a_{\text{ps}} = 19.1^{+2.1}_{-1.9} R_p$ (Teachey et al. 2018) together with the uncertainties in the planetary mass, Kepler's third law of motion predicts a barycentric mass of $17.6^{+19.2}_{-12.6} M_{\text{Jup}}$. This value would be compatible with the ten-Jupiter mass object described by Teachey et al. (2018), although it remains unclear how the total mass is shared between the primary and secondary. Despite our neglect of uncertainties in the orbital period, our estimate suggests that proper modeling of the dynamical transit

---

signature can deliver much tighter constraints on the masses of the two bodies than radius estimates alone (see Fig. 1).

### 2.2. Formation models

We have investigated the plausible masses of Kepler-1625 b and its proposed companion in the context of three moon formation scenarios that have been proposed for the solar system moons.

*Impacts*. Moon accretion from the gas and debris disk that forms after giant impacts between planet-sized rocky bodies is the most plausible scenario for the origin of the Earth-Moon system (Canup 2012) and for the Pluto-Charon binary (Canup 2005; Walsh & Levison 2015). A peculiar characteristic of these two systems is in their high satellite-to-host mass ratios of about $1.2 \times 10^{-2}$ for the Earth and $1.2 \times 10^{-1}$ for Pluto.

*In-situ accretion*. In comparison, the masses of the moon systems around the giant planets in the solar system are between $1.0 \times 10^{-4}$ and $2.4 \times 10^{-4}$ times the masses of their host planets. This scaling relation is a natural outcome of satellite formation in a "gas-starved" circumplanetary disk model (Canup & Ward 2006) and theory predicts that this relation should extend into the super-Jovian regime, where moons the mass of Mars would form around planets as massive as 10 $M_{\rm Jup}$ (Heller & Pudritz 2015b).

*Capture*. The retrograde orbit and the relative mass of Neptune's principal moon Triton cannot plausibly be explained by either of the above scenarios. Instead, Agnor & Hamilton (2006) proposed that Triton might be the captured remnant of a former binary system that was tidally disrupted during a close encounter with Neptune. The Martian moons Phobos and Deimos have also long been thought to have formed via capture from the asteroid belt. But recent simulations show that they also could have formed in a post-impact accretion disk very much like the protolunar disk (Rosenblatt et al. 2016).

To address the question of the origin of Kepler-1625 b and its potential satellite, we start by calculating the satellite-to-host mass ratios for a range of nominal scenarios for Kepler-1625 b and its proposed companion. These values shall serve as a first-order estimate of the formation regime which the system could have emerged from.

We then performed numerical calculations of accretion disks around giant planets as per Heller & Pudritz (2015b) to predict the satellite-to-host mass ratios in super-Jovian and BD regime, where satellites have not yet been observed. Our disk model is based on the gas-starved disk theory (Canup & Ward 2006; Makalkin & Dorofeeva 2014). While the conventional scenario assumes that the planetary luminosity is negligible for the temperature structure in the disk, we have included viscous heating, accretion heating, and the illumination of the disk by the young giant planet. The evolution of the planet is simulated using pre-computed planet evolution tracks provided by C. Mordasini (2013). We considered seven different giant planets with masses of 1, 2, 3, 5, 7, 10, and 12 Jupiter masses. Our model could also include the relatively weak stellar illumination, but we neglected it here to avoid unnecessary complexity.

We tested a range of reasonable disk surface reflectivities, $0.1 \leq k_{\rm s} \leq 0.3$, and assume a constant Planck opacity of $10^{-2}$ m$^2$ kg$^{-1}$ throughout the disk. Models in this range of the parameter space have been shown to reproduce the composition and orbital radii of the Galilean moons (Heller & Pudritz 2015b). The total amount of solids in the disk is determined by two contributions: (1) the initial solids-to-gas ratio of 1/100, which is compatible with the composition of the interstellar material and in rough agreement with the first direct measurement in a circumstellar disk (Zhang et al. 2017); (2) by the evolution of the circumplanetary water ice line (Heller & Pudritz 2015b), where the surface density of solids (rock and ice) exhibits a jump by a factor of about five as water vapor freezes out (Hayashi 1981). We measured the amount of solids during the final stages of planetary accretion, when the planet is supposed to open up a gap in the circumstellar disk, at which point the circumplanetary disk is essentially cut from further supply of dust and gas. We assumed that moon formation then halts and compare the final amount of solids to our model of the Jupiter-mass planet. We have not investigated the actual formation of moons from the dust and ice.

To test the plausibility of a capture scenario, we applied the framework of Williams (2013) and calculate the maximum satellite mass that can be captured by Kepler-1625 b during a close encounter with a binary,

$$M_{\rm cap} < 3 M_{\rm p} \left( \frac{G M_{\rm esc} \pi}{2 b v_{\rm enc} \Delta v} \right)^{3/2} - M_{\rm esc} , \tag{1}$$

where $G$ is the gravitational constant, $M_{\rm esc}$ is the escaping mass, $b$ is the encounter distance, $v_{\rm enc} = \sqrt{v_{\rm esc}^2 + v_{\infty}^2}$ is the encounter velocity, $v_{\infty}$ is the relative velocity of the planet and the incoming binary at infinity, $v_{\rm esc} = \sqrt{2 G M_{\rm p}/b}$ is the escape speed from the planet at the encounter distance $b$,

$$\Delta v > \sqrt{v_{\rm esc}^2 + v_{\infty}^2} - \sqrt{G M_{\rm p} \left( \frac{2}{b} - \frac{1}{a} \right)} , \tag{2}$$

is the velocity change experienced by $M_{\rm cap}$ during capture,

$$a < \frac{1}{2} \left( 0.5 a_{\star \rm p} \left( \frac{M_{\rm p}}{3 M_{\star}} \right)^{1/3} + b \right) \tag{3}$$

is the orbital semimajor axis of the captured mass around the planet, and $a_{\star \rm p}$ is the semimajor axis of the planet around the star. This set of equations is valid under the assumption that the captured mass would be on a stable orbit as long as its apoapsis were smaller than half the planetary Hill radius.

## 3. Results

Table 1 summarizes our nominal scenarios for Kepler-1625 b. Scenarios (1aa) to (1bb) refer to the lower stellar radius estimate, scenarios (2aa) and (2ab) to the nominal stellar radius, and scenarios (3aa) and (3ab) to the maximum stellar radius. In each case, possible values of $R_{\rm p}$ and $R_{\rm s}$ are derived consistently from the respective value of $R_{\star}$. Scenario (TKS) assumes the preliminary characterization of the transiting system as provided by Teachey et al. (2018), except that we explicitly adapted a companion mass equal to that of Neptune although the authors only describe "a moon roughly the size of Neptune". The last column in the table shows a list of the corresponding secondary-to-primary mass ratios. Figure 2 illustrates the locations of the said scenarios in a mass ratio diagram. The positions of the solar system planets are indicated as are the various formation scenarios of their satellites (in-situ accretion, impact, or capture). The locations of three examples of planetary systems are included for comparison, namely Proxima b (Anglada-Escudé et al. 2016), the seven planets around TRAPPIST-1 (Gillon et al. 2017), and LHS 1140 b (Dittmann et al. 2017).

We find that a Saturn-mass gas planet with either an Earth-mass gas moon (1aa) or a Neptune-mass water-rock moon (1ab)

**Table 1.** Possible scenarios of the nature of Kepler-1625 b and its proposed companion.

| Scenario | $R_\star [R_\odot]$ | $R_p [R_{Jup}]$ | $R_s [R_{Jup}]$ | $M_p [M_{Jup}]$ | $M_s [M_\oplus]$ | $M_s/M_p$ |
|---|---|---|---|---|---|---|
| (1aa) Saturn-mass gas planet, Earth-mass gas moon | 1.305[a] | 0.86[b] | 0.26[b] | 0.4[3] | 1[1] | 7.9×10⁻³ |
| (1ab) Saturn-mass gas planet, Neptune-mass water-rock moon | | | | | 17[2] | 1.3×10⁻¹ |
| (1ba) brown dwarf, Earth-mass gas moon | | | | 75[3] | 1[1] | 4.2×10⁻⁵ |
| (1bb) brown dwarf, Neptune-mass water-rock moon | | | | | 17[2] | 7.1×10⁻⁴ |
| (2aa) very-low-mass star, mini-Neptune planet | 1.793[a] | 1.18[b] | 0.35[b] | 91[3] | 10[4] | 3.5×10⁻⁴ |
| (2ab) very-low-mass star, super-Earth water-rock planet | | | | | 70[2] | 2.4×10⁻³ |
| (3aa) very-low-mass star, Neptune-like planet | 2.056[a] | 1.36[b] | 0.40[b] | 112[4] | 20[4] | 5.6×10⁻⁴ |
| (3ab) very-low-mass star, super-Saturn water-rock planet | | | | | 180[2] | 5.1×10⁻³ |
| (TKS) super-Jovian planet, Neptune-like moon | – | 1[5] | 0.35[5] | 10[5] | 17[5] | 5.4×10⁻³ |

**Notes.** [a] The stellar radius estimates are based on Mathur et al. (2017). [b] The radii of the transiting primary and proposed secondary were estimated from the lightcurves of Teachey et al. (2018). The corresponding masses of the objects were estimated using structure models and evolution tracks from the following references.

**References.** (1) Masuda (2014); (2) Fortney et al. (2007); (3) Baraffe et al. (2003); (4) Baraffe et al. (2008); (5) Teachey et al. (2018).

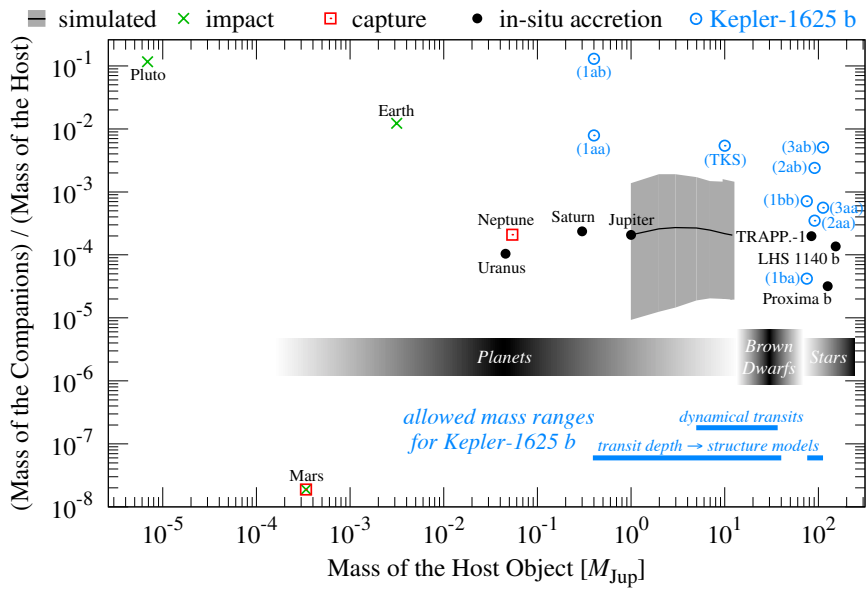

**Fig. 2.** Mass ratios of companions and hosts, i.e., moons around planets and planets around VLMSs. Host masses are shown along the abscissa, mass ratios on the ordinate. Solar system planets with moons are shown with different symbols to indicate the respective formation scenarios of their satellites (see legend above the panel). Three VLMSs with roughly Earth-mass planets (TRAPPIST-1, LHS 1140, Proxima Centauri) are plotted as examples of formation via accretion in the stellar regime. The solid black line expanding from Jupiter's position signifies simulations of moon formation in the Jupiter-mass regime, with the gray shaded region referring to uncertainties in the parameterization of the accretion disk. Possible scenarios for the planetary, BD, and VLMS nature of Kepler-1625 b are indicated with blue open circles (see Table 1 for details). The plausible mass range for Kepler-1625 b is shown with a blue line in the lower right corner and is the same as shown in Fig. 1.

would hardly be compatible with the common scaling law of the satellite masses derived from the gas-starved disk model. The mass ratio would rather be consistent with an impact scenario – just that the mass of the host object would be much higher than that of any host to this formation scenario in the solar system.

The remaining scenarios, which include a BD orbited by either an Earth-mass gas moon (1ba) or a Neptune-mass water-rock moon (1bb), as well as a VLMS orbited by either an inflated mini-Neptune (2aa), or a water-rock super-Earth (2ab), or a Neptune-like binary planet (3aa), or a water-rock super-Saturn-mass object (3ab) could all be compatible with in-situ formation akin to the formation of super-Earths or mini-Neptunes around VLMSs.

In Fig. 3, we plot the maximum captured mass, that is, the mass of what would now orbit Kepler-1625 b, over the mass that would have been ejected during a binary encounter with Kepler-1625 b. We adopted a parameterization that represents the (TKS) scenario, where $M_\star = 1.079 \, M_\odot$, $a_{\star p} = 0.87$ AU, $M_p = 10 \, M_{jup}$ and $b = 19.1 \, R_{Jup}$. We tested various plausible values for $v_\infty$, ranging from a rather small velocity difference of 1 km/s to a maximum value of $v_{orb} = 30$ km/s. The latter value reflects the Keplerian orbital speed of Kepler-1625 b, and crossing orbits would have relative speeds of roughly $e \times v_{orb}$, where $e$ is the heliocentric orbital eccentricity (Agnor & Hamilton 2006).

We find that the capture of a Neptune-mass object by Kepler-1625 b is possible at its current orbital location. The escaping object would have needed to be as massive as 5 $M_\oplus$ to 50 $M_\oplus$ for relative velocities at infinity between 1 km/s and 30 km/s. For nearly circular heliocentric orbits of Kepler-1625 b and the former binary, small values of $v_\infty$ would be more likely and prefer the ejection of a super-Earth or mini-Neptune companion from an orbit around the proposed companion to Kepler-1625 b.

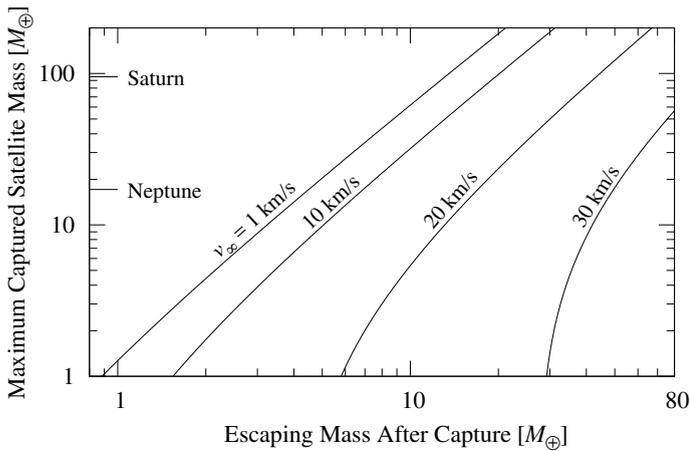

**Fig. 3.** Possible masses for the candidate around Kepler-1625 b in a capture scenario. The Keplerian speed at the orbit of Kepler-1625 b is about 30 km/s and provides a reasonable upper limit on the relative collision speed between it and a possible binary approaching on a close-encounter trajectory.

## 4. Discussion

The mass estimates derived from the analysis of the lightcurves presented by Teachey et al. (2018) permit an approximation of the total mass of Kepler-1625 b and its proposed companion. Photodynamical modeling (Huber et al. 2013) accounting for the orbital motion of the transiting system (Kipping 2011) as well as the upcoming *Hubble* observations will deliver much more accurate estimates, provided that the proposed companion around Kepler-1625 b can be validated. If this exomoon candidate is rejected, then the dynamical mass estimate of the planet-moon system (see Sect. 2.1.2) will naturally become meaningless and a mass-radius relation akin to Fig. 1 will be the only way to estimate the mass of Kepler-1625 b from photometry alone.

Consistent modeling of the transiting system based on T2, T4, and T5 could allow predictions of the relative orbital geometry of the transiting system during the upcoming transit (Heller et al. 2016b), depending on the as yet unpublished uncertainties in $P_{ps}/P_{\star p}$ by Teachey et al. (2018). If these predictions could be derived and made publicly available prior to the *Hubble* observations of T11 on 29 October 2017, then a confirmation of the proposed planet-moon system in the predicted orbital configuration would lend further credibility to a discovery claim.

The post-capture orbital stability at about 1 AU around Sun-like stars has been demonstrated for Earth-sized moons around giant planets (Porter & Grundy 2011). Most intriguingly, about half of the resulting binaries in these simulations were in a retrograde orbit, akin to Triton around Neptune. Hence, if future observations of Kepler-1625 b confirm the presence of a companion and if it is possible to determine the sense of orbital motion, then this could be a strong argument for a formation through capture. Measurements of the sense of orbital motion is impossible given even the best available photometric space-based resources nowadays available (Lewis & Fujii 2014; Heller & Albrecht 2014). Nevertheless, if Kepler-1625 b turns out to be a BD or VLMS, then its infrared spectrum could be used to determine its RV variation during the transit. Combined with the variation of the tangential motion of Kepler-1625 and its proposed companion that is available from the lightcurve, this would determine the sense of orbital motion (Oshagh et al. 2017).

The (TKS) scenario, suggesting a Neptune-sized moon in orbit around a 10 $M_{Jup}$ planet, would imply a companion-to-host mass ratio of about $5.4 \times 10^{-3}$, assuming a moon mass equal to that of Neptune. On the one hand, this is just a factor of a few smaller than the relative masses of the Earth-Moon system. On the other hand, this value is more than an order of magnitude larger than the $10^{-4}$ scaling relation established by the solar system giant planets. As a consequence, the satellite-to-host mass ratio does not indicate a preference for either the post-impact formation or in-situ accretion scenarios. In fact, Kepler-1625 b and its possible Neptune-sized companion seem to be incompatible with both.

If the companion around Kepler-1625 b can be confirmed and both objects can be validated as gas giant objects, then it would be hard to understand how these two gas planets could possibly have formed through either a giant impact or in-situ accretion at their current orbits around the star. Instead, they might have formed simultaneously from a primordial binary of roughly Earth-mass cores that reached the runaway accretion regime beyond the circumstellar iceline at about 3 AU (Goldreich & Tremaine 1980). These cores then may have started migrating to their contemporary orbits at about 0.87 AU as they pulled down their gaseous envelopes from the protoplanetary gas disk (Lin et al. 1996; Mordasini et al. 2015).

The case of a VLMS or massive BD with a super-Earth companion would imply a companion mass on the order of $10^{-4}$ times the host object, which reminds us of the proposed universal scaling relation for satellites around gas giant planets. Hence, this scenario might be compatible with a formation akin to the gas-starved disk model for the formation of the Galilean satellites around Jupiter, though in an extremely high mass regime.

## 5. Conclusions

We derived approximate lower and upper limits on the mass of Kepler-1625 b and its proposed exomoon companion by two independent methods. Firstly, we combined information from the Kepler transit lightcurves and evolution tracks of substellar objects. The radius of Kepler-1625 b is compatible with objects as light as a 0.4 $M_{Jup}$ planet (similar to Saturn) or as massive as a 0.11 $M_\odot$ star. The satellite candidate would be securely in the planetary mass regime with possible masses ranging between about 1 $M_\oplus$ for an extremely low-density planet akin to Kepler-51 b or c and about 20 $M_\oplus$ if it was metal-rich and similar in composition to Neptune. These uncertainties are dominated by the uncertainties in the stellar radius. Secondly, we inspected the dynamics of Kepler-1625 b and its potential companion during the three transits published by Teachey et al. (2018) and estimate a total mass of $17.6^{+19.2}_{-12.6} M_{Jup}$ for the binary. The error bars neglect uncertainties in our knowledge of the planet-moon orbital period, which we estimate from the third published transit to be about 72 hr.

We conclude that if the proposed companion around Kepler-1625 b is real, then the host is most likely a super-Jovian planet. In fact, a BD would also be compatible with both the mass-radius relationship for substellar objects and with the dynamical transit signatures shown in the lightcurves by Teachey et al. (2018). If the satellite candidate can be confirmed, then dynamical modeling of the transits can deliver even better mass estimates of this transiting planet-moon system irrespective of stellar RVs.

Our comparison of the characteristics of proposed exomoon candidate around Kepler-1625 b with those of the moon systems in the solar system reveal that a super-Jovian planet with a Neptune-sized moon would hardly be compatible with conventional moon formation models, for example, after a giant impact or via in-situ formation in the accretion disk around a gas giant primary. We also investigated formation through an encounter between Kepler-1625 b and a planetary binary system, which would have resulted in the capture of what has provisionally been dubbed Kepler-1625 b-i and the ejection of its former companion. Although such a capture is indeed possible at Kepler-1625 b's orbital distance of 0.87 AU from the star, the ejected object would have had a mass of a mini-Neptune itself. And so this raises the question how this ejected mini-Neptune would have formed around the Neptune-sized object that is now in orbit around Kepler-1625 b in the first place. If the proposed exomoon can be validated, then measurements of the binary's sense of orbital motion might give further evidence against or in favor of the capture scenario.

The upcoming *Hubble* transit observations could potentially allow a validation or rejection of the proposed exomoon candidate. If the moon is real, then dynamical transit modeling will allow precise mass measurements of the planet-moon system. If the moon signature turns out to be a ghost in the detrended *Kepler* data of Teachey et al. (2018), however, then the transit photometry will not enable a distinction between a single transiting giant planet and a VLMS. Stellar spectroscopy will then be required to better constrain the stellar radius, and thus the radius of the transiting object, and to determine the mass of Kepler-1625 b or Kepler-1625 B, as the case may be.

*Acknowledgements.* The author thanks Laurent Gizon, Matthias Ammler-von Eiff, Michael Hippke, and Vera Dobos for helpful discussions and feedback on this manuscript. This work was supported in part by the German space agency (Deutsches Zentrum für Luft- und Raumfahrt) under PLATO Data Center grant 50OO1501. This work made use of NASA's ADS Bibliographic Services. This research has made use of the NASA Exoplanet Archive, which is operated by the California Institute of Technology, under contract with the National Aeronautics and Space Administration under the Exoplanet Exploration Program.

**Chapter 6**

# New Detection and Validation Methods for Moons Around Extrasolar Planets

## 6.1 Detecting Extrasolar Moons Akin to Solar System Satellites with an Orbital Sampling Effect (Heller 2014)

# DETECTING EXTRASOLAR MOONS AKIN TO SOLAR SYSTEM SATELLITES WITH AN ORBITAL SAMPLING EFFECT

René Heller[1,2]

Origins Institute, McMaster University, Hamilton, ON L8S 4M1, Canada
rheller@physics.mcmaster.ca

## ABSTRACT

Despite years of high accuracy observations, none of the available theoretical techniques has yet allowed the confirmation of a moon beyond the solar system. Methods are currently limited to masses about an order of magnitude higher than the mass of any moon in the solar system. I here present a new method sensitive to exomoons similar to the known moons. Due to the projection of transiting exomoon orbits onto the celestial plane, satellites appear more often at larger separations from their planet. After about a dozen randomly sampled observations, a photometric orbital sampling effect (OSE) starts to appear in the phase-folded transit light curve, indicative of the moons' radii and planetary distances. Two additional outcomes of the OSE emerge in the planet's transit timing variations (TTV-OSE) and transit duration variations (TDV-OSE), both of which permit measurements of a moon's mass. The OSE is the first effect that permits characterization of multi-satellite systems. I derive and apply analytical OSE descriptions to simulated transit observations of the *Kepler* space telescope assuming white noise only. Moons as small as Ganymede may be detectable in the available data, with M stars being their most promising hosts. Exomoons with the 10-fold mass of Ganymede and a similar composition (about 0.86 Earth radii in radius) can most likely be found in the available *Kepler* data of K stars, including moons in the stellar habitable zone. A future survey with *Kepler*-class photometry, such as *Plato 2.0*, and a permanent monitoring of a single field of view over 5 years or more will very likely discover extrasolar moons via their OSEs.

*Keywords:* instrumentation: photometers – methods: analytical – methods: data analysis – methods: observational – methods: statistical – planets and satellites: detection

## 1. CONTEXT AND MOTIVATION

Although more than 1000 extrasolar planets have been found, no extrasolar moon has been confirmed. Various methods have been proposed to search for exomoons, such as analyses of the host planet's transit timing variation (TTV; Sartoretti & Schneider 1999; Simon et al. 2007), its transit duration variation (TDV; Kipping 2009a,b), direct photometric observations of exomoon transits (Tusnski & Valio 2011), scatter analyses of averaged light curves (Simon et al. 2012), a wobble of the planet-moon photocenter (Cabrera & Schneider 2007), mutual eclipses of the planet and its moon or moons (Cabrera & Schneider 2007; Sato & Asada 2009; Pál 2012), excess emission of transiting giant exoplanets in the spectral region between 1 and 4 μm (Williams & Knacke 2004), infrared emission from airless moons around terrestrial planets (Moskovitz et al. 2009; Robinson 2011), the Rossiter-McLaughlin effect (Simon et al. 2010; Zhuang et al. 2012), microlensing (Han & Han 2002), pulsar timing variations (Lewis et al. 2008), direct imaging of extremely tidally heated exomoons (Peters & Turner 2013), modulations of radio emission from giant planets (Noyola et al. 2013), and the generation of plasma tori around giant planets by volcanically active moons (Ben-Jaffel & Ballester 2014). Recently, Kipping et al. (2012) started the *Hunt for Exomoons with Kepler*

(HEK),[3] the first survey targeting moons around extrasolar planets. Their analysis combines TTV and TDV measurements of transiting planets with searches for direct photometric transit signatures of exomoons.

Exomoon discoveries are supposed to grant fundamentally new insights into exoplanet formation. The satellite systems around Jupiter and Saturn, for example, show different architectures with Jupiter hosting four massive moons and Saturn hosting only one. Intriguingly, the total mass of these major satellites is about $10^{-4}$ times their planet's mass, which can be explained by their common formation in the circumplanetary gas and debris disk (Canup & Ward 2006), and by Jupiter opening up a gap in the heliocentric disk during its own formation (Sasaki et al. 2010). The formation of Earth is inextricably linked with the formation of the Moon (Cameron & Ward 1976), and Uranus' natural satellites indicate a successive "collisional tilting scenario", thereby explaining the planet's unusual spin-orbit misalignment (Morbidelli et al. 2012). Further interest in the detection of extrasolar moons is triggered by their possibility to have environments benign for the formation and evolution of extrasolar life (Reynolds et al. 1987; Williams et al. 1997; Heller & Barnes 2013). After all, astronomers have found a great number of super-Jovian planets in the habitable zones (HZs) of Sun-like stars (Heller & Barnes 2014).

In this paper, I present a new theoretical method that allows the detection of extrasolar moons. It can be applied to discover and characterize multi-satellite systems and to measure the satellites' radii and orbital semi-major axes around their host planet, assuming roughly circular orbits. This assumption is justified because eccentric moon orbits typically circularize on a million year time scale due to tidal effects (Porter & Grundy 2011; Heller & Barnes 2013). The method does not depend on a satellite's direction of orbital motion (retrograde or prograde), and it relies on high-accuracy averaged photometric transit light curves. I refer to the physical phenomenon that generates the observable effect as the Orbital Sampling Effect (OSE). It causes three different effects in the phase-folded light curve, namely, (1) the photometric OSE, (2) TTV-OSE, and (3) TDV-OSE. Similarly to the photometric OSE, the scatter peak method developed by Simon et al. (2012) makes use of orbit-averaged light curves. But I will not analyze the scatter. While the scatter peak method was described to be more promising for moons in wide orbits, the OSE works best for close-in moons. Also, with an orbital semi-major axis spanning 82 % of the planet's Hill sphere, the example satellite system studied by Simon et al. (2012) would only be stable if it had a retrograde orbital motion (Domingos et al. 2006).

[1] Department of Physics and Astronomy, McMaster University
[2] Postdoctoral fellow of the Canadian Astrobiology Training Program

[3] www.cfa.harvard.edu/HEK

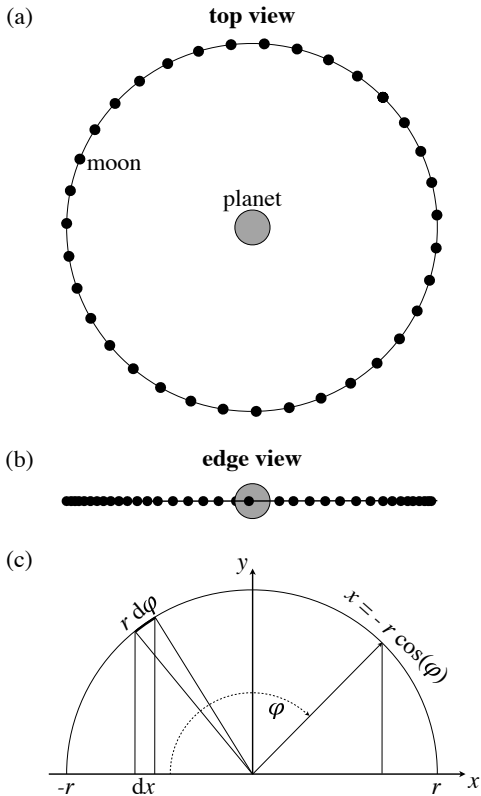

**Figure 1.** Geometry of a moon's OSE. Assuming a constant sampling frequency over one moon orbit (panel (a)), an observer in the moon's orbital plane would recognize a non-uniform projected density distribution (panel (b)). All snapshots combined in one sequence frame, the moon is more likely to occur at larger separations $x$ from the planet. The probability distribution $P_s(x)$ along the projected orbit can be constructed as $P_s(x) = r \, \mathrm{d}\varphi/\mathrm{d}x$ (panel (c)).

## 2. THE ORBITAL SAMPLING EFFECT

### 2.1. *Probability of Apparent Planet-moon Separation*

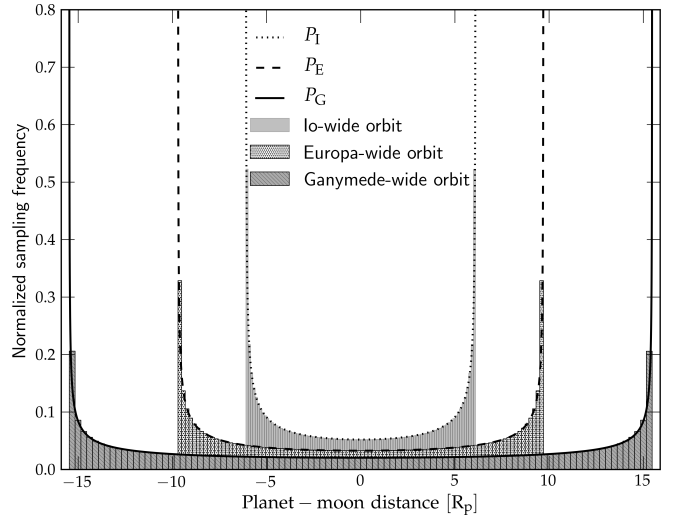

**Figure 2.** Normalized sampling frequency (or probability density $P_s(x)$) for three moons in an Io-wide, Europa-wide, and Ganymede-wide orbit in units of planetary radii. Bars show the results from a randomized numerical simulation, curves show the distribution according to Equation (4). The integral, that is, the area under each curve equals 1. This explains why moons on tighter orbits have a higher sampling frequency at a given planet-moon distance.

Imagine a moon orbiting a planet on a circular orbit. In a satellite system with $n$ moons, this particular moon shall be satellite number $s$, and its orbital semi-major axis around the planet be $a_{\mathrm{ps}}$.[4] Imagine further looking at the system from a top view position, such that the satellite's apparent path around the planet forms a circle. With a given, but arbitrary sampling frequency you take snapshots of the orbiting moon, and once the moon has completed one revolution, you stack up the frames to obtain a frame sequence. This sequence is depicted in panel (a) of Figure 1. An observer in the orbital plane of the satellite, taking snapshots with the same sampling frequency and stacking up a sequence frame from its edge view post, would see the moon's positions distributed along a line. The planet sits in the center of this line, which extends as far as the projected semi-major axis to either side (panel (b) of Figure 1). Due to the projection effect, the moon snapshots pile up toward the edges of the projected orbit. In other words, if the snapshots would be taken randomly from this edge-on perspective, then the moon would most likely be at an apparently wide separation from the planet. We can assume to observe most transiting exomoon systems in this edge view because the orbital plane of the planet-moon system should be roughly in the same plane as the orbital plane of the planet-moon barycenter around the star (Heller et al. 2011b).

The likelihood of a satellite to appear at an apparent separation $x$ from the planet can be described by a probability density $P_s(x)$, which is proportional to the "amount" of orbital path $r \times \mathrm{d}\varphi$, with $\varphi$ as the angular coordinate and $r$ as the orbital radius, divided by the projected part of this interval along the $x$-axis, $\mathrm{d}x$ (see

---

[4] Speaking about the orbital geometry of a planet-moon binary as depicted in Figure 1, I will refer to the moon's orbital radius around the planet as $r$. In case of a multi-satellite system, I designate the planet-satellite orbital semi-major axis of satellite number $s$ as $a_{\mathrm{ps}}$.

panel (c) of Figure 1). With $x = -r \cos(\varphi)$, we thus have

$$P_s(x) \propto \frac{r \, \mathrm{d}\varphi}{\mathrm{d}x} = r \frac{\mathrm{d}}{\mathrm{d}x} \arccos\left(\frac{-x}{r}\right)$$
$$= \frac{1}{r^2 \sqrt{1 - \left(\frac{x}{r}\right)^2}} \quad . \tag{1}$$

$P_s(x)$ must fulfill the condition

$$\int_{-r}^{+r} \mathrm{d}x \; P_s(x) = 1 \quad , \tag{2}$$

because it is a probability density, and the moon must be somewhere. With

$$\int_{-r}^{+r} \mathrm{d}x \; \frac{1}{\sqrt{1 - \left(\frac{x}{r}\right)^2}} = \int_{-r}^{+r} \mathrm{d}x \; r \frac{\mathrm{d}}{\mathrm{d}x} \arccos\left(\frac{-x}{r}\right) = \pi r \tag{3}$$

we thus have the normalized sampling frequency

$$P_s(x) = \frac{1}{\pi r \sqrt{1 - \left(\frac{x}{r}\right)^2}} \quad . \tag{4}$$

In Figure 2, I plot Equation (4) for a three-satellite system. The innermost moon with probability density $P_I$ is in an orbit as wide as Io's orbit around Jupiter, that is, it has a semi-major axis of 6.1 planetary radii ($R_p$). The central moon with normalized sampling frequency $P_E$ follows a circular orbit 9.7 $R_p$ from the planet, similar to Europa, and the outermost moon corresponds to Ganymede at 15.5 $R_p$ and with probability density $P_G$. The shaded areas illustrate a normalized suite of randomized measurements of the projected orbital separation $x$ that I have simulated with a computer. The lines follow Equation (4), where $r$ is replaced by the respective orbital semi-major axis $a_{ps}$, and they nicely match the randomized sampling. As the area under each curve equals 1, the innermost moon has a higher probability to appear at a given position within the interval $[-a_{pI}, +a_{pI}]$, with $a_{pI}$ the sky-projected orbital semi-major axis of the satellite in an Io-wide orbit, than any of the other moons. The normalized sampling frequency or probability density of apparent separation $P_s(x)$ is independent of the satellite radius.

### 2.2. The Photometric OSE

#### 2.2.1. The Photometric OSE in Averaged Transit Light Curves

From the perspective of a data analyst, it is appealing that the OSE method does not require modeling of the orbital evolution of the moon or moons during the transit or between transits, that is, during the circumstellar orbit. Assuming that the satellite is not in an orbital resonance with the circumstellar orbital motion, the OSE will smear out over many light curves and always yield a probability distribution as per Equation (4). What is more, this formula allows an analytic description of the actual effect in the light curve. In other words, once the phase-folded light curve is available after potential TTVs or TDVs – induced by the moons or by other planets – as well as red noise (Lewis 2013) have been removed, the OSE can be measured with a simple fit to the binned data points. I refer to this effect as the photometric OSE.

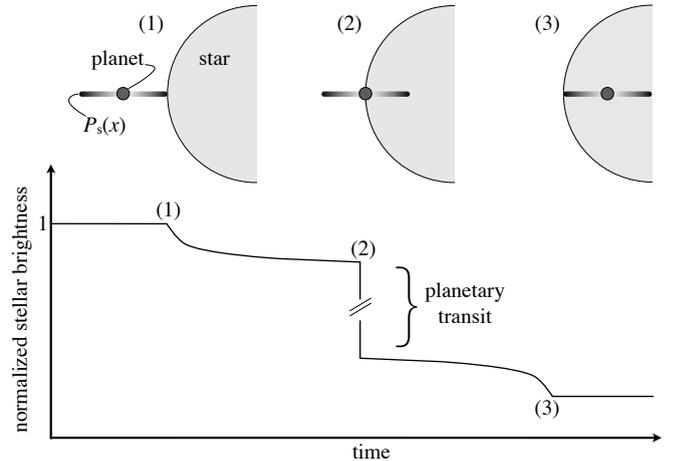

**Figure 3.** Photometric OSE during ingress. At epoch (1), a satellite's probability distribution $P_s(x)$ along its circumplanetary orbit touches the stellar disk, from then on causing a steep (but small) decline in stellar brightness (see lower panel). As circumplanetary orbits with lower values of $P_s(x)$ (visualized by lighter colors) enter the stellar disk, the brightness decrease weakens. At epochs (2), the planet enters the stellar disk and induces a dramatic decrease in stellar brightness, depicted by two slanted lines in the lower panel. From that point on, larger values of $P_s(x)$ enter the stellar disk, so the slope of the stellar flux decrease as the OSE increases until the moon orbits have completely entered the stellar disk at epochs (3).

Yet, what does the effect actually look like? Figure 3 visualizes the ingress of the planet-moon binary in front of the stellar disk from a statistical point of view. The satellite orbit is assumed to be coplanar with the circumstellar orbit, and the impact parameter (b) of the planet, corresponding to its minimal distance from the stellar center during the transit, in units of stellar radii, is zero. The thick shaded line drawn through the planet sketches the probability density $P_s(x)$ and describes a smoothed-out version of the frame shown in panel (b) of Figure 1. If one were able to repeatedly take snapshots of a certain moment during subsequent transits (for example at epochs (1), (2), or (3) in this sketch), then the averaged positions of the moon would scatter according to this shaded distribution. I call the part right of the planet the right wing of $P_s(x)$ (or $P^{rw}$) and that part of $P_s(x)$, which is left of the planet, the left wing of $P_s(x)$ (or $P^{lw}$).

As $P^{rw}$ touches the stellar disk with its highest values, visualized by dark shadings in Figure 3, it causes a steep (though weak) decrease in the averaged transit light curve, as depicted in the bottom panel. The decreasing probability distribution then proceeds over the stellar disk and causes a declining decrease in stellar brightness. At epochs (2), the planet enters the stellar disk and triggers a major brightness decrease, commonly known from a planetary transit light curve. Moving on to point (3), that part of the probability function that actually moves into the stellar disk has increasingly higher values toward $P^{lw}$, thereby inducing an increasing decline in stellar brightness. Note that the lower panel of Figure 3 visualizes the averaged light curve. During each particular transit, the moon can be anywhere along its projected orbit around the planet, and the decrease in stellar brightness follows a completely different curve.

### 2.2.2. Analytic Description of the Photometric OSE

Equation (4) allows for an analytic description of the photometric OSE. In Figure 3, the right wing of the probability distribution enters the stellar disk first, which corresponds to the right side of $P_s(x)$ shown in Figure 2. Stellar light is blocked between $a_{ps}$ to the right and $x'(t)$ to the left, where the latter variable describes the time dependence of the moving left edge of $P_s(x)$. To calculate the amount of blocked stellar light due to the photometric OSE during the ingress of a one-satellite system ($F_{OSE,in}^{(1)}$), I integrate $P_s(x)$ from $x'(t)$ to $a_{ps}$ and subtract this area from the normalized, apparent stellar brightness ($B_{OSE}$), which equals 1 out of transit. The amplitude of the blocked light is given by $(R_s/R_\star)^2$, where $R_\star$ is the stellar radius, and hence

$$
\begin{aligned}
F_{OSE,in}^{(1)} &= \left(\frac{R_s}{R_\star}\right)^2 \int_{x'(t)}^{a_{ps}} dx \, P_s(x) \\
&= \frac{1}{\pi}\left(\frac{R_s}{R_\star}\right)^2 \left[\pi - \arccos\left(\frac{-x'(t)}{a_{ps}}\right)\right] \\
&\qquad \left\{\text{for } -1 \le \frac{-x'(t)}{a_{ps}} \le 1\right\}
\end{aligned}
\tag{5}
$$

and $B_{OSE,in}^{(1)}(t) \equiv 1 - F_{OSE,in}^{(1)}(t)$ is the brightness during ingress. For $-x'(t)/a_{ps} < -1$, that is, as long as the probability distribution of the moon has not yet touched the stellar disk, the arccos() term is not defined, so I set $F_{OSE,in}^{(1)} = 0$ in that case. To parameterize $x'(t)$, I choose a coordinate system whose origin is in the center of the stellar disk. Time is 0 at the center of the transit, so that $x'(t) = -R_\star - v_{orb}t$, with $v_{orb} = 2\pi a_{\star b}/P_{\star b}$ as the circumstellar orbital velocity of the planet-moon barycentric mass $M_b$, assumed to be equal to the circumstellar orbital velocity of both the planet and the moon, $P_{\star b} = 2\pi\sqrt{a_{\star b}^3/(G(M_\star + M_b))}$ as the circumstellar orbital period of $M_b$, $G$ as Newton's gravitational constant, and $a_{\star b}$ as the orbital semi-major axis between the star and $M_b$. To consistently parameterize the transit light curve and the OSE, the star-planet system must thus be well-characterized. Assuming that the planet is much more massive than the moon, taking $M_b \approx M_p$ is justified.

Once the whole probability density of the moon has entered the stellar disk, $-x'(t)/a_{ps} > +1$ and the arccos() term is again not defined, so I take $F_{OSE,in}^{(1)} = (R_s/R_\star)^2$ in that case. During egress of the probability function, the moon, on average, uncovers a fraction $F_{OSE,eg}^{(1)}$ of the stellar disk. Its mathematical description is similar to Equation (5), except for $x'(t) = +R_\star - v_{orb}t$. Again, $F_{OSE,eg}^{(1)} = 0$ before the egress of the probability function and 1 after it has left the disk, so that the normalized, apparent stellar brightness becomes $B_{OSE}^{(1)}(t) = 1 - F_p(t) - F_{OSE,in}^{(1)}(t) + F_{OSE,eg}^{(1)}(t)$, with $F_p(t)$ as the stellar flux masked by the transiting planet (see Appendix A).

While Equation (5) is valid for one-satellite systems, it can be generalized to a system of $n$ satellites by subtracting the stellar flux that is blocked subsequently by

the integrated density functions via

$$
B_{OSE}^{(n)}(t) = \begin{cases}
1 - F_p(t) - \displaystyle\sum_{s=1}^{n} F_{OSE,in}^{(s)}(t) \\
\qquad + \displaystyle\sum_{s=1}^{n} F_{OSE,eg}^{(s)}(t) \\
\qquad\qquad\qquad \text{for } |x_p(t)| > R_\star \\[8pt]
1 - F_p(t) - \displaystyle\sum_{s=1}^{n} F_{OSE,in}^{(s)}(t) \\
\qquad + \displaystyle\sum_{s=1}^{n} F_{OSE,eg}^{(s)}(t) + A_{mask} \\
\qquad\qquad\qquad \text{for } |x_p(t)| \le R_\star
\end{cases}
\tag{6}
$$

where $x_p(t)$ is the position of the planet, and

$$
A_{mask} = \frac{2}{\pi}\sum_{s=1}^{n}\left(\frac{R_s}{R_p}\right)^2 \left[\arccos\left(\frac{-R_p}{a_{ps}}\right) - \arccos\left(\frac{+R_s}{a_{ps}}\right)\right]
\tag{7}
$$

compensates for those parts of the probability functions that do not contribute to the OSE because of planet-moon eclipses. This masking can only occur during the planetary transit when $|x_p(t)| \le R_\star$. Note that partial planet-moon eclipses as well as moon-moon eclipses are ignored (but treated by Kipping 2011a).

In this model, the ingress and egress of the moons are neglected, which is appropriate because even the largest moons that can possibly form within the circumplanetary disk around a 10-Jupiter-mass planet are predicted to have masses around ten times that of Ganymede, or $\approx 0.25 \, M_\oplus$ (Canup & Ward 2006), and even if they are water-rich their radii will be $< R_\oplus$. The effect of a moon's radial extension on the duration of the OSE will thus be $< R_\oplus/(5\,R_J) \approx 1.8\,\%$ for a sub-Earth-sized moon at a planet-moon orbital distance of 5 Jupiter radii ($R_J$) and $< 0.5\,R_\oplus/(10\,R_J) \approx 0.46\,\%$ for a Mars-sized moon with a semi-major axis of $10\,R_J$.

### 2.2.3. Numerical Simulations of the Photometric OSE

To simulate a light curve that contains a photometric OSE, I construct a hypothetical three-satellite system that is similar in scale to the three innermost moons of the Galilean system. The planet is assumed to have the 10-fold mass of Jupiter and a radius $R_p = 1.05$ Jupiter radii ($R_J$). According to the Canup & Ward (2006) model, I scale the satellite masses as $M_1 = 10\,M_I$, $M_2 = 10\,M_E$, and $M_3 = 10\,M_G$, with the indices I, E, and G referring to Io, Europa, and Ganymede, respectively. I derive the moon radii of this scaled-up system as per the Fortney et al. (2007) structure models for icy/rocky planets by assuming ice-to-mass fractions (imf) similar to those observed in the Jovian system, that is, $imf_1 = 0.02$, $imf_2 = 0.08$, and $imf_3 = 0.45$ (Canup & Ward 2009). The model then yields $R_1 = 0.62\,R_\oplus$, $R_2 = 0.52\,R_\oplus$, and $R_3 = 0.86\,R_\oplus$. The star is assumed to be a K star 0.7 times the mass of the Sun ($M_\odot$) with a radius ($R_\star$) of 0.64 solar radii ($R_\odot$) (for Sun-like metallicity at an age of 1 Gyr, following Bressan et al. 2012), and the planet-satellite system is placed into the center of the stellar HZ at 0.56 AU (derived from the model of Kopparapu et al. 2013). The impact parameter is $b = 0$, the moons' orbits are all in the orbital plane of the planet-moon barycenter around the star, and the satellites' orbital semi-major axes around the planet are $a_{p1} = a_{JI}R_p/R_J$, $a_{p2} = a_{JE}R_p/R_J$, and $a_{p3} = a_{JG}R_p/R_J$ for the innermost, the central, and the outermost satellite, respectively. The values of $a_{JI}$, $a_{JE}$, and $a_{JG}$ correspond to the semi-major axes of Io, Europa, and Ganymede around Jupiter, respectively, and consequently this system is similar to the one shown in Figure 2.

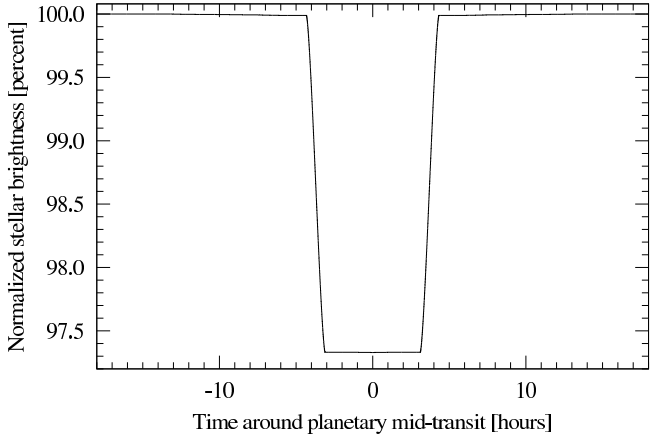

**Figure 4.** Simulated transit light curve of a hypothetical three-satellite system around a giant planet of 1.05 Jupiter radii and 10 Jupiter masses transiting a $0.64\,R_\odot$ K star in the HZ. On this scale, the photometric OSE is barely visible as a small decrease in stellar brightness just before planetary ingress and as a small delay in reaching 100 % of stellar brightness after planetary egress. Stellar limb darkening is neglected.

Figure 4 shows the transit light curve of this system, averaged over an arbitrary number of transits with infinitely small time resolution, excluding any sources of noise, and neglecting effects of stellar limb darkening. Individual transits are modeled numerically by assuming a random orbital positioning of the moons during each transit. In reality, a moon's relative position to the planet during a transit is determined by the initial conditions, say at the beginning of some initial transit, as well as by the orbital periods both around the star and around the planet-moon barycenter. I here only assume that the ratio of these periods is a low value integer. Then from a statistical point of view, positions during consecutive transits can be considered randomized.

In the simulations shown in Figure 4, the analytic description of Equation (6) is not yet applied – this pseudo phase-folded light curve is a randomized, purely numerical simulation. If one of the moons turns out to be in front of or behind the planet, as seen by the observer, then it does not cause a flux decrease in the light curve. On this scale, the combined OSE of the three satellites is hardly visible against the planetary transit because the depth of the satellites' features scale as ($\approx 0.64\,R_\oplus/(0.64\,R_\odot))^2 = 10^{-4}$, while the planet causes a depth of about 2.7 %, decreasing stellar brightness to roughly 97.3 %.

Figure 5 zooms into three parts of this light curve. The upper left panel highlights the respective ingresses of the three satellites. Each of the three impressions is caused by the right wing of the probability function of one of the moons, with the outermost moon (the third moon counting outwards from the planetary center) entering first, about 13 hr before the planetary mid-transit, the second moon following (ingress at about $-10$ hr), and the inner, or first, moon succeeding at roughly $-7.5$ hr along the abscissa. Each individual OSE ingress corresponds to the phase between moments (1) and (2) in Figure 3 and is tied to the photometric OSE of the preceding moon or moons. As the moons' probability functions enter the stellar disk, their OSE signals add up. At about $-4.5$ hr, the planetary ingress begins and causes a steep decrease in stellar brightness. The slight increase of the curve between $-5.5$ and $-4.5$ hr is caused by the right wing ($P_3^{\rm rw}$) of the outermost moon's probability function leaving the stellar disk even before the planet enters. This indicates that the projected orbital separation of the outermost satellite is larger than the radius of the stellar disk.

The upper right panel shows the egress of the planet and the three-satellite system from the stellar disk, which appears as a mirror-inverted version of the upper left panel. In this panel, the left wings of the satellites' probability functions that leave the stellar disk replace the right wings that enter the disk from the left panel.

The wide lower panel of Figure 5 illustrates the moons' photometric OSEs at the bottom of the planetary transit trough. Chronologically, this phase of the transit is between the ingress (upper left) and egress (upper right) phases of the planet-satellite system. The downtrend between about $-3$ and $-2.2$ hr visualizes the continued ingresses of $P_1^{\rm rw}$, $P_2^{\rm rw}$, and $P_3^{\rm rw}$ from the upper left panel. What is more, between about $-2.2$ hr and the center of the planetary transit curve, we witness an interplay of the satellites' $P_s(x)$ entering and leaving the stellar disk. In particular, the long curved feature between approximately $-2.2$ and 0 hr visualizes the egress of the right wing of the central moon ($P_2^{\rm rw}$) overlaid by the ingress of increasingly high sampling frequencies in the left wing of the central moon ($P_1^{\rm lw}$). The fact that the end of the ingress of $P_1^{\rm lw}$ and the beginning of the egress of $P_1^{\rm rw}$ almost coincide at planetary mid-transit means that the width of the projected semi-major axis of the innermost satellite equals almost exactly the stellar radius.

### 2.2.4. Emergence of the Photometric OSE in Transit Light Curves

Assuming the moons do not change their positions relative to the planet during individual transits, then their apparent separations are discrete in the sense that they are defined by one value. Only if numerous transits are observed and averaged will the moons' photometric OSEs appear, because the sampling frequencies, or the underlying density distributions $P_s(x)$, will take shape. Dropping the assumption of a fixed planet-satellite separation during transit, each individual transit will cause a dynamical imprint in the light curve. However, the photometric OSE will converge to the same analytical expression as given by Equation (6) after a large number of transits.

In Figure 6, I demonstrate the emergence of the photometric OSE in the hypothetical three-satellite system for an increasing number of transits $N$. Each panel shows the same time interval around planetary mid-transit as the upper left panel in Figure 5, that is, the ingress of the right wings of the probability distributions. The solid line shows the averaged transit light curve coming from my randomized transit simulations, and the dashed line shows the predicted OSE signal as per Equation (6). In order to draw the dashed line, I make use of the *known* satellite radii $R_s$ ($s = 1, 2, 3$), the planet-satellite semi-major axes $a_{\mathrm{p}s}$, the stellar radius, the planetary radius, and the orbital velocity of the transiting planet-moon system $v_{\mathrm{orb}}$. The dashed curve is thus no fit to the simulations but a prediction for this particular exomoon system.

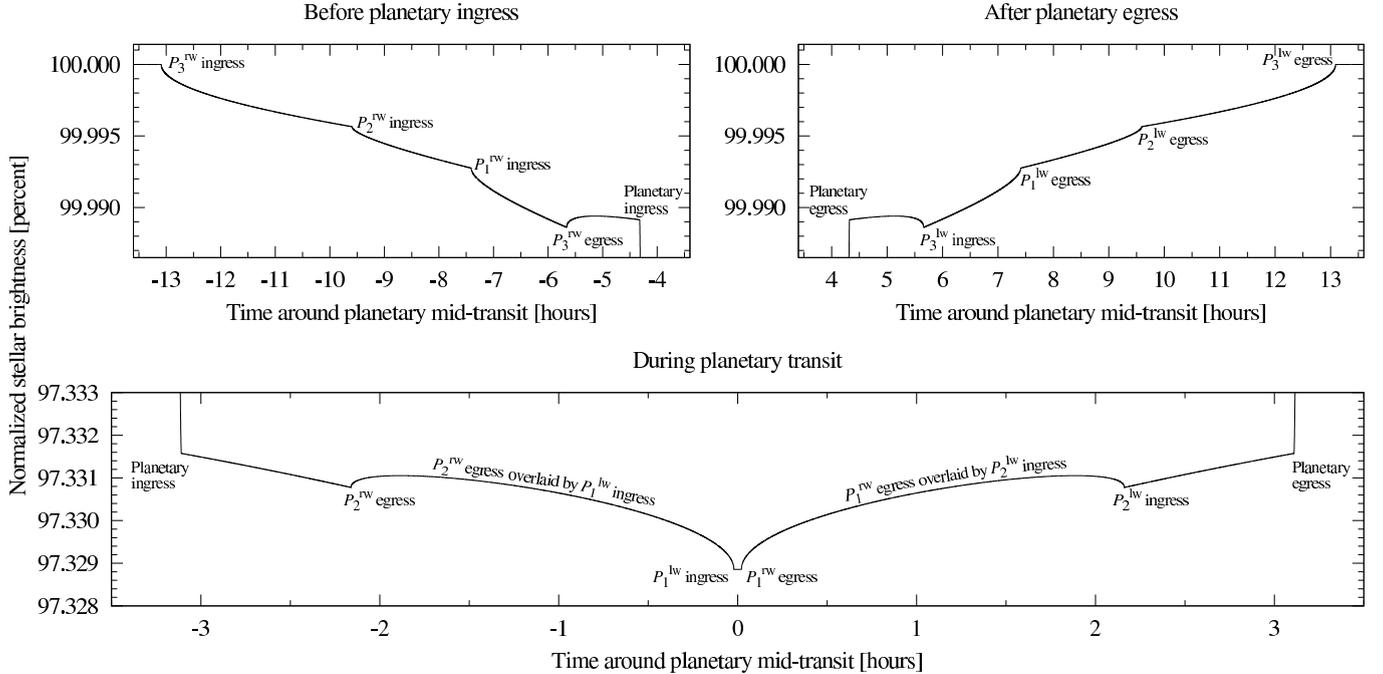

**Figure 5.** Photometric OSEs of a hypothetical three-satellite system around a super-Jovian planet transiting a $0.64 \, R_\odot$ K star in the HZ (zoom into Figure 4). The radii of the outermost (the 3rd), central (2nd), and innermost (1st) satellites are $R_3 = 0.86 \, R_\oplus$, $R_2 = 0.52 \, R_\oplus$, and $R_1 = 0.62 \, R_\oplus$, following planetary structure models (Fortney et al. 2007) and assuming ice-to-mass fractions of the Galilean moons. Due to the smaller range of orbits over which the probability function $P(a_{\mathrm{p}1})$ of the innermost satellite is spread, its dip in this averaged stellar light curve is deeper than the brightness decrease induced by the outermost satellite, although the innermost satellite is smaller ($R_3 > R_1$). An arbitrarily large number of transits has been averaged to obtain this curve, and no noise has been added.

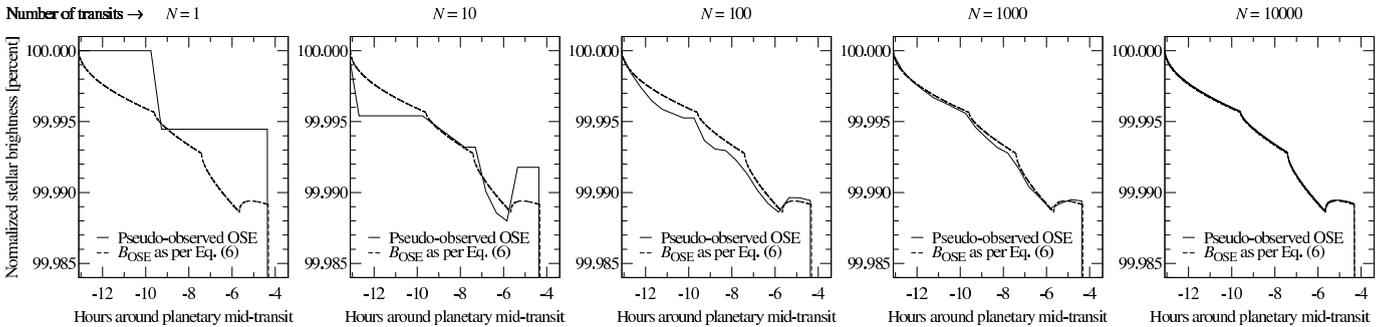

**Figure 6.** Emergence of the photometric OSE during the transit ingress of a hypothetical three-satellite system for an increasing number of averaged transits $N$ (zoom into upper left panel of Figure 5). Solid lines show the averaged noiseless light curves, while dashed curves illustrate the combined photometric OSEs of the three satellites as per Equation (6).

In the left-most panel, after the first transit of the planet-satellite system in front of the star ($N = 1$), only one of the three moons shows a transit, shortly after $-10$ hr. Referring to the picture given in Figure 3, this moon appears to the right of the planet and enters the stellar disk before the planet does. Examination of this moon's transit depth of $6 \times 10^{-5}$ reveals the second moon as the originator, because $(R_2/R_\star)^2 \approx 6 \times 10^{-5}$. As an increasing number of transits is collected in the second-to-left and the center panel, the photometric OSEs of the three-satellite system emerge. Obviously, between $N = 10$ and $N = 100$, a major improvement of the OSE signal strength occurs, suggesting that at least a few dozen transits are necessary to characterize this system. After $N = 1000$ transits, the noiseless averaged OSE curve becomes indistinguishable from the predicted function.

Whatever the precise value of the critical number of transits ($N_{\mathrm{OSE}}$) necessary to recover the satellite system from the solid curve in Figure 6, it is a principled threshold imposed by the very nature of the OSE. Noise added during real observations increases the number of tran-sits that is required to characterize the system to a value $N_{\rm obs}$, and the relation is deemed to be $N_{\rm obs} \geq N_{\rm OSE}$. To estimate realistic values for $N_{\rm obs}$, it is necessary to simulate noisy pseudo-observed data and try to recover the input systems. Section 3 is devoted to this task.

## 2.3. OSE of Exomoon-induced Transit Duration Variations (TDV-OSE)

Consider a single massive exomoon orbiting a planet. As a result of the two bodies' motion around their common barycenter, the planet's tangential velocity component with respect to the observer is different during each transit, and hence the duration of the planetary transit shows deviations from the mean duration during each transit (Kipping 2009a). While each individual planet-moon transit has its individual TDV offset, all TDV observations combined will reveal what I refer to as a TDV-OSE. In the reference system shown in Figure 1, the planet's velocity component is projected onto the $x$-axis and is given by $\dot{x}_{\rm p} = \omega r_{\rm p} \sin(\varphi)$, with $\omega = 2\pi/P_{\rm ps}$ as the planet-satellite orbital frequency and $P_{\rm ps}$ as the orbital period. The probability for a single planetary transit to show a certain projected velocity around the planet-satellite barycenter is then given by the sampling frequency

$$P_{\rm p}^{\rm TDV}(\dot{x}_{\rm p}) \propto \frac{{\rm d}\varphi}{{\rm d}\dot{x}_{\rm p}} = \frac{\rm d}{{\rm d}\dot{x}_{\rm p}} \arcsin\left(\frac{\dot{x}_{\rm p}}{\omega r_{\rm p}}\right)$$
$$= \frac{1}{\omega r_{\rm p}\sqrt{1 - \left(\frac{\dot{x}_{\rm p}}{\omega r_{\rm p}}\right)^2}} \ . \qquad (8)$$

This distribution describes the fraction of randomly sampled angles $\varphi$ that lies within an infinitesimal velocity interval $d\dot{x}_{\rm p}$ of the planet projected onto the $x$-axis, whereas $P_{\rm s}(x)$ in Equation (1) measures how much of an infinitely small orbital path element of a satellite lies in a projected planet-moon distance interval. In analogy to the normalization of the latter position probability in Eq. (2), the integral over $P_{\rm p}^{\rm TDV}(\dot{x}_{\rm p})$ must equal 1 between $-\omega r_{\rm p}$ and $+\omega r_{\rm p}$, which yields

$$P_{\rm p}^{\rm TDV}(\dot{x}_{\rm p}) = \frac{1}{\pi \omega r_{\rm p}\sqrt{1 - \left(\frac{\dot{x}_{\rm p}}{\omega r_{\rm p}}\right)^2}} \qquad (9)$$

and has a shape similar to the functions shown in Figure 2. Consequently, as the planet's velocity wobble induced by a single moon distributes within the interval $-\omega r_{\rm p} \leq \dot{x}_{\rm p} \leq \omega r_{\rm p}$ as per Equation (9), planetary TDVs will also distribute according to an OSE.

Assuming circular orbits, the spread of this TDV distribution is determined by the peak-to-peak TDV amplitude

$$\Delta_{\rm TDV} = 2\ t_{\rm T} \times \sqrt{\frac{a_{\star \rm p}}{a_{\rm ps}}}\ \sqrt{\frac{M_{\rm s}^2}{M_{\rm b}(M_{\rm b} + M_{\star})}}\quad , \qquad (10)$$

where $t_{\rm T}$ is the duration of the transit between first and fourth contact (Kipping 2009a). Then the TDV-OSE

distribution is given by

$$P_{\rm p}^{\rm TDV}(t) = \frac{1}{\pi \Delta_{\rm TDV}\sqrt{1 - \left(\frac{t}{\Delta_{\rm TDV}}\right)^2}}\quad , \qquad (11)$$

where $t$ is time. With $M_{\star} \gg M_{\rm p}$ and $M_{\rm p} \gg M_{\rm s}$ the right-most square root in Equation (10) simplifies to $\sqrt{M_{\rm s}^2/(M_{\rm p}M_{\star})}$. $M_{\star}$ can be determined via spectral classification, $M_{\rm p}$ may be accessible via radial velocity measurements, $t_{\rm T}$ is readily available from the light curve, $a_{\star \rm p}$ can be inferred by Kepler's third law, and $a_{\rm ps}$ can be measured by the photometric OSE. With $M_{\rm s}$ as the remaining free parameter, a fit of Equation (11) to the moon-induced planetary TDV distribution gives a direct measurement of the satellite mass. But note the limited applicability of the TTV and TDV methods for short-period moons (Section 6.3.8 in Kipping 2011b).

## 2.4. OSE of Exomoon-induced Transit Timing Variations (TTV-OSE)

There is, of course, also an OSE for the distribution of the TTVs. In circular one-satellite systems, the amplitude of this TTV-OSE distribution is given by

$$\Delta_{\rm TTV} = 2\ \times \frac{a_{\rm ps}M_{\rm s}}{M_{\rm b}\sqrt{a_{\star \rm p}G(M_{\star} + M_{\rm b})}} \qquad (12)$$

(Sartoretti & Schneider 1999; Kipping 2009a), and the distribution itself is given by

$$P_{\rm p}^{\rm TTV}(t) = \frac{1}{\pi \Delta_{\rm TTV}\sqrt{1 - \left(\frac{t}{\Delta_{\rm TTV}}\right)^2}}\quad . \qquad (13)$$

With all the parameters on the right-hand side of Eq. (13) known from radial velocity measurements, the photometric OSE and TDV-OSE, this third manifestation of the OSE can be used to further improve the confidence of any exomoon detection, as it needs to be consistent with all three OSE observations.

# 3. RECOVERING EXOMOON-INDUCED PHOTOMETRIC OSE WITH KEPLER

To assess the prospects of measuring the photometric OSE as described in Section 2.2.2, I generate a suite of pseudo-observed averaged light curves for a range of given star-planet-satellite systems and try to detect the injected exomoon signal. Following Kipping et al. (2009), I assume that the out-of-transit baseline is known to a sufficiently high degree, which is an adequate assumption for Kepler high-quality photometry with large amounts of out-of-transit data. My simulations also imply that red noise has been corrected for, which can be a time-consuming part of the data reduction. Stellar limb darkening is neglected, but this does not alter the photometric OSE substantially (Heller et al., in preparation).

## 3.1. Noise and Binning of Phase-Folded Light Curves

To simulate a Kepler-class photometry, I induce stellar brightness variations $\sigma_{\star}$, detector noise $\sigma_{\rm d}$, a quarter-to-quarter noise component $\sigma_{\rm q}$, and shot noise $\sigma_{\rm s}$ into the

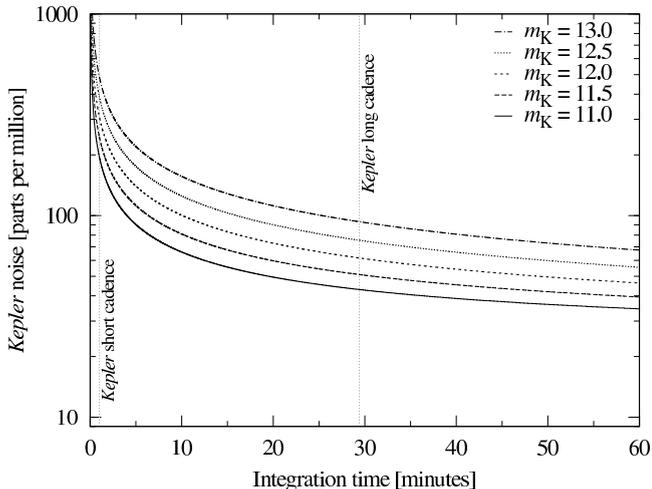

**Figure 7.** Simulated *Kepler* noise as per Equation (14), used for the synthesis of pseudo-observed transit light curves.

raw light curves that contain a moon-induced photometric OSE. The total noise is then given by

$$\sigma_{\mathrm{K}} = \sqrt{\sigma_\star^2 + \sigma_{\mathrm{d}}^2 + \sigma_{\mathrm{q}}^2 + \frac{1}{\Gamma_{\mathrm{ph}} t_{\mathrm{int}}}}, \qquad (14)$$

with $\sigma_\star = 19.5\,\mathrm{ppm}$, $\sigma_{\mathrm{D}} = 10.8\,\mathrm{ppm}$, $\sigma_{\mathrm{Q}} = 7.8\,\mathrm{ppm}$ (Gilliland et al. 2011),

$$\Gamma_{\mathrm{ph}} = 6.3 \times 10^8 \,\mathrm{hr}^{-1} \times 10^{-0.4(m_{\mathrm{K}}-12)} \qquad (15)$$

as *Kepler*'s photon count rate (Kipping et al. 2009), and $t_{\mathrm{int}}$ as the integration time. The stellar noise of 19.5 ppm is typical for a G-type star, while K and M dwarfs tend to show more intrinsic noise. Figure 7 visualizes the decrease of *Kepler* noise as a function of integration time and for five different stellar magnitudes. The *Kepler* short cadence and long cadence integration times at 1 and 29.4 min are indicated with vertical lines, respectively.

Figure 8 shows the pseudo phase-folded transit light curve of the hypothetical three-satellite system described in Section 2.2.3 after 100 transits. The K star is chosen to have a *Kepler* magnitude $m_{\mathrm{K}} = 12$. I simulate each individual transit and virtually "observe" the stellar brightness every 29.4 min, corresponding to the *Kepler* long-cadence mode, and add Gaussian noise following Equation (14). For each transit, I introduce a random timing offset to the onset of observations, so that each transit light curve samples different parts of the transit. Each individual light curve is normalized to 1 and added to the total pseudo phase-folded light curve. This pseudo phase-folded light curve is again normalized to 1. While dots in Figure 8 visualize my simulated *Kepler* measurements, the solid line depicts the analytic prediction of the OSE following Equation (6). Even in this noisy data, the OSE is readily visible with the unaided eye shortly before and after the planetary transit (upper two panels), suggesting that less than 100 transits are necessary to discover – yet maybe not to unambiguously characterize – extrasolar multiple moon systems with the photometric OSE. At the bottom of the pseudo phase-folded transit light curve, however, the OSE remains hardly noticeable (lower panel).

In Figure 9, I show the same simulated data, but now binned to intervals of 30 min (see Appendix B). Before and after planetary ingress, the OSE now becomes strongly apparent. Note that the standard deviations are substantially smaller than the scatter around the analytic model. This is not due to observational noise, but due to the discrete sampling of the transits. This scatter from the model decreases for an increasing number of transits, $N$ (see Figure 6). I also tried binning the short cadence *Kepler* data and used binning intervals of 10, 30, and 60 min, of which the 30 min binning of long cadence observations showed the most reliable results, at least for this particular star-planet-moon system. With a 60 min binning, the individual ingress and egress of the three moons are poorly sampled, while the 10 min sampling shows too much of a sampling scatter around the model.

### 3.2. *Recovery of Injected Exomoon OSE Signals*

Next, I evaluate the odds of characterizing extrasolar moons with the photometric OSE. Although multi-satellite systems can, in principle, be detected and characterized by the analytical OSE model (Equation (6)), I focus on a one-satellite system. I simulate a range of transits of a single exomoon orbiting a Jupiter- and 10-Jupiter-mass planet. In the former case, the moon is a Ganymede analog of $0.42\,R_\oplus$, in the latter case the moon's mass is scaled by a factor of ten and its radius of $0.86\,R_\oplus$ is derived from the Fortney et al. (2007) structure models using $\mathrm{imf}_1 = 0.45$. This moon corresponds to the outermost (or third) moon in the system simulated in Section 2.2.1. I study the detectability of these two moons around these two planets orbiting three different stars at 12th magnitude: a Sun-like star, a $0.7\,M_\odot$ K star as considered in the previous sections, and an $0.4\,M_\odot$-mass M dwarf with a radius of $0.36\,R_\odot$ (for solar metallicity at an age of 1 Gyr, derived from Bressan et al. 2012).

I start to simulate the pseudo phase-folded *Kepler* light curve of each of these hypothetical systems after $N = 5$ transits and fit the binned data with a $\chi^2$ minimization of the analytical model given by Equation (6), where the two free parameters are the satellite radius ($R_1$) and the orbital semi-major axis between the moon and the planet ($a_{\mathrm{p1}}$) (see Appendix C). This procedure is performed 100 times for a given $N$, and if both $R_1$ and $a_{\mathrm{p1}}$ are recovered with an error of less than 10 % in at least 68 of the 100 runs (corresponding to a $1\,\sigma$ confidence for the recovery rate), then the number of transits is stored as $N_{\mathrm{obs}}$. If the satellite cannot be recovered under these boundary conditions, then the number of transits $N$ is increased by an amount $\Delta N = \lfloor 10^{\log_{10}(N)+0.05} \rfloor$,[5] representing the series $N \in \{5, 6, 7, ..., 48, 54, 61, ..., 151, 169, 190, ...\}$, and I repeat my fit of the analytical model to 100 pseudo phase-folded light curves.

Figure 10 presents an example light curve of the Jupiter-mass planet and its Ganymede-like satellite in the K star toy system after $N = 100$ transits. The data points show the simulated *Kepler* observations, the solid line corresponds to the predicted light curve of the known system, and the dashed line indicates the best fit model. In this example, the best-fit radius of the moon of $0.4 \pm 0.0049\,R_\oplus$ is very close to the input value of $0.42\,R_\oplus$, and similarly the fit of the planet-moon orbital distance of $15.4 \pm 0.0487\,R_{\rm p}$ almost matches the input value of $15.47\,R_{\rm p}$. I consider the formal $1\sigma$ uncertainties of the $\chi^2$ fitting as physically unrealistic, which is why I repeat the fitting procedure 100 times to get more robust estimates of $N_{\rm obs}$.

---

[5] The notation $\lfloor x \rfloor$ denotes a rounding of the real number $x$ to the next integer.

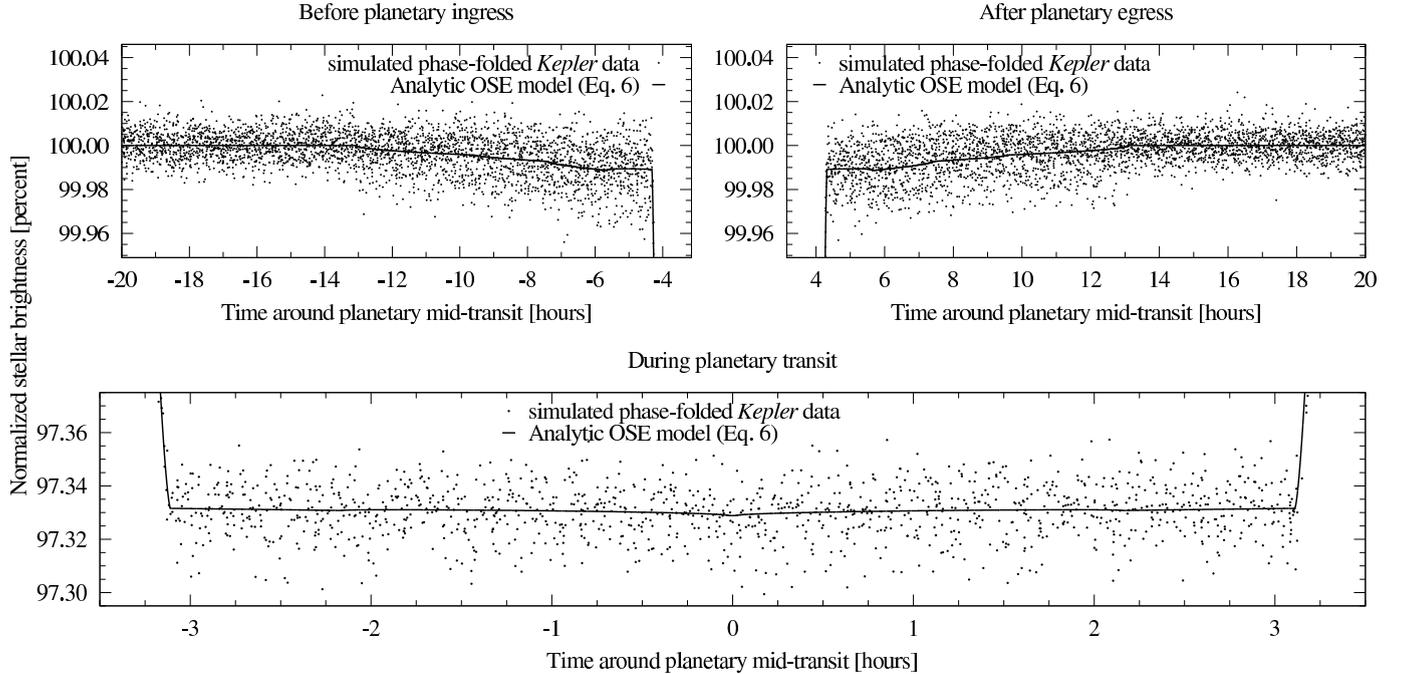

**Figure 8.** Photometric OSEs after 100 transits in simulated *Kepler* observations of a hypothetical three-satellite system around a Jupiter-sized planet 10 times the mass of Jupiter, transiting a $0.64\,R_\odot$ K star in the HZ (same as in Figure 5). Noise is simulated after Equation (14). The scale of the ordinate is much wider than in Figure 5. Although the OSE seems invisible at the bottom of the transit light curve (bottom panel) due to the noise, the three moons together mask a measurable amount of stellar light that triggers the fit.

## 4. RESULTS AND PREDICTIONS

Figure 11 shows the outcome of the data fitting. The upper table refers to transits of a Jupiter-sized host planet with a Ganymede-like moon, and the lower table lists the results for the super-Jovian planet with a super-Ganymede exomoon. Abscissae and ordinates of these charts indicate distance to the star and stellar mass, respectively, while the entries show $N_{\rm obs}$ as well as the equivalent observational time $t_{\rm obs}$, computed as $N_{\rm obs}$ times the orbital period around the star. Assuming circular orbits, I use the semi-analytic model of Domingos et al. (2006) to test all systems for orbital stability. While a Jupiter-mass planet 0.1 AU from a Sun-like star cannot hold a moon in a Ganymede-wide orbit, a few other configurations only allow the moon to be stable in retrograde orbital motion. The latter cases are labeled with an asterisk. Shaded regions indicate the locations of the respective stellar HZ following Kopparapu et al. (2013). Green cell borders demarcate the observation cycle of the *Kepler* space telescope that has been covered before the satellite's reaction wheel failure. Photometric OSEs of exomoons within these boundaries could be detectable in the available *Kepler* data of photometrically quiet stars.

Inspecting the upper panel for the Jupiter-Ganymede duet, three trends readily appear. First, large values of $N_{\rm obs}$ appear in the top row referring to a Sun-like host star, intermediate values for the K star in the center row, and small values for the M dwarf host star in the bottom line. This trend is explained by the ratio of the satellite radius to the stellar radius. The moon's transit is much deeper in the M dwarf light curve than in the light curve of the Sun-like star, hence, fewer transits are required to detect it. Second, going from short to wide stellar distances, $N_{\rm obs}$ decreases. This decline is caused by the decreasing orbital velocity of the planet-moon binary around the star. In wider stellar orbits, transits have a longer duration, and so the binary's passage of the stellar disk yields more data points. The binned data then has smaller error bars and allows for more reliable $\chi^2$ fitting. Third, $N_{\rm obs}$ converges to a minimum value in the widest orbits. For a Sun-like star, this minimum number of transits is about 120 at 0.9 AU and beyond, it is 34 for the $0.7\,M_\odot$ star at 0.5 AU and beyond, and roughly a dozen transits beyond 0.4 AU around a $0.4\,M_\odot$ star. In these regimes, white noise is negligible and the recovery of the injected moons depends mostly on the ratio $R_{\rm s}/R_\star$. $N_{\rm obs}$ is then comparable to the principled threshold $N_{\rm OSE}$ imposed by the OSE nature (see Section 2.2.4).

Translated into the required duty cycle of a telescope, timescales increase towards wider orbits, simply because the circumstellar orbital periods get longer. As an example, 107 transits were required in my simulations to discover a Ganymede-like satellite around a Jupiter-like planet orbiting a $0.7\,M_\odot$ star at 0.1 AU, corresponding to a monitoring over 4 yr. Only 34 transits were required for the planet-moon system at 0.5 AU around the same star, but this means an observation time of 14.4 yr. The odds of finding a Ganymede-sized moon transiting a G star in the available *Kepler* data are poor, with $t_{\rm obs} > 8.5\,{\rm yr}$ in any stellar orbit, and $t_{obs} > 100$ yr beyond about 0.8 AU. K stars are more promising candidates with $t_{obs}$ as small as 4 yr at a distance of 0.1 AU. M dwarfs, finally, show the best prospects for Ganymede-like exomoons, because $t_{obs} < 4$ yr at 0.2 AU. Around the $0.4 M_\odot$ star, this distance encases planet-moon binaries in the stellar HZ.

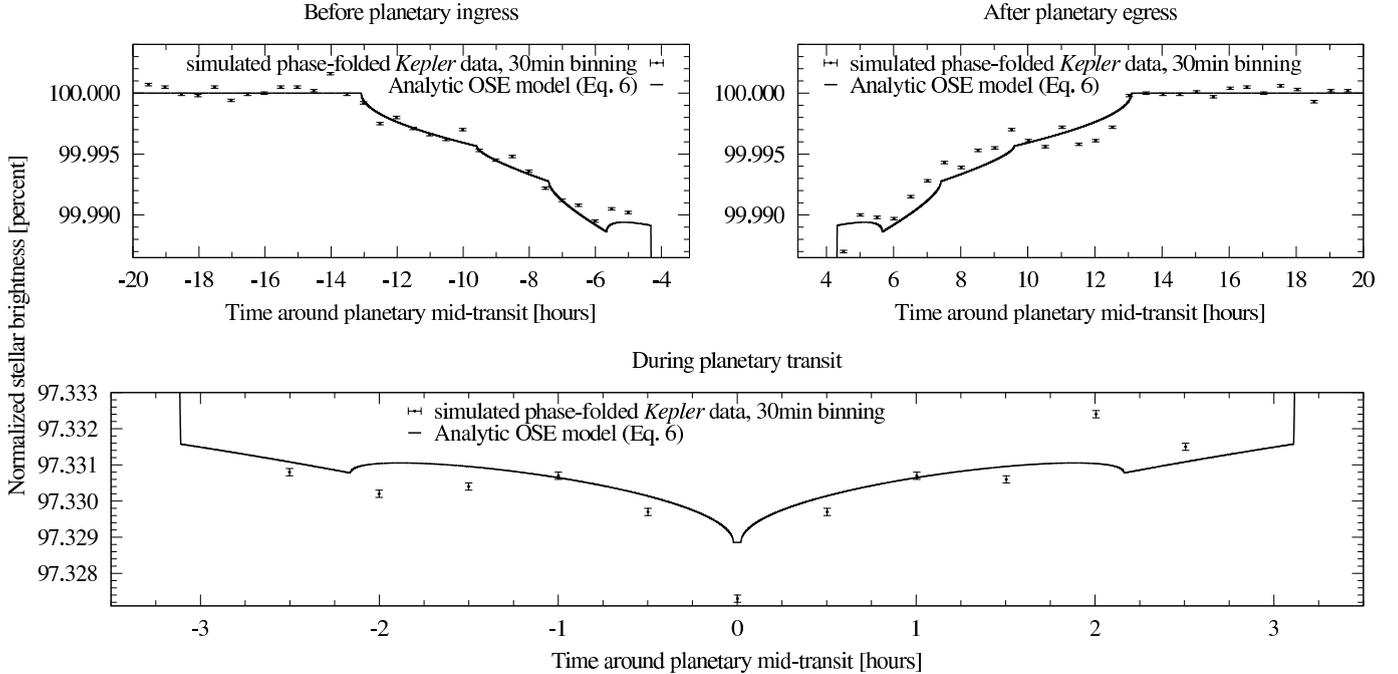

**Figure 9.** Same as Figure 8 but binned to intervals of 30 min. While the OSE of the three-satellite system clearly emerges in the ingress and egress parts of the probability functions $P_s(x)$ (upper two panels), it remains hardly visible at the bottom of the transit curve. Yet, the combined stellar transits of the moons still influence the depth of this light curve trough.

In the lower chart of Figure 11, a planet with the 10-fold mass of Jupiter and an exomoon of $0.86 R_\oplus$ is considered. In most cases, $N_{obs}$ for a given star and stellar distance is smaller than in the left panel, because the moon is larger and causes a transit signal that is better distinguishable from the noise. But different from the left chart, $N_{obs}$ does not strictly decrease towards low-mass stars. The $0.7 M_\odot$ K star shows the best prospects for this exomoon's photometric OSE detection in the available *Kepler* data. While a Sun-like host star could reveal the satellite after as few as 30 transits or 7.6 yr at 0.4 AU, the K star allows detection after 13 transits or 3.9 yr at the same orbital distance, but still 27 transits or 10.7 yr of observations would be required for the M star. The photometric OSE of such a hypothetical super-Ganymede moon could thus be measured in the available *Kepler* data for planets as far as 0.4 AU around K stars, thereby encompassing the stellar HZs.

Intriguingly, values of $N_{obs}$ are larger in the bottom line of the lower panel than in the bottom line of the upper panel, at a given stellar distance. This is counterintuitive, as the larger satellite radius (lower chart) should decrease the number of required transits, in analogy to the Sun-like and K star cases. The discrepancy in the M dwarf scenario, however, is both an artifact of my boundary conditions, which require the fitted values of $R_s$ and $a_{p1}$ to deviate less than 10 % from the genuine values of the injected test moons, and the very nature of the photometric OSE. For very small stellar radii, such as the M dwarf host star, and relatively large satellite radii, such as the one used in the right chart, the OSE scatter becomes comparatively larger than observational noise effects. Hence, the K star in the center line of the lower panel of Figure 11 actually yields the most promising odds for the detection of super-Ganymede-like exomoons.

My choice of a 10 % deviation in both $R_s$ and $a_{ps}$ between the genuine injected exomoon and the best fit is arbitrary. To test its credibility, I ran a suite of randomized planet-only transits and corresponding $\chi^2$ fits for a Jupiter-like planet orbiting the G, K, and M dwarf stars at 0.5 AU, respectively, which is roughly in the center of the top panel of Figure 11. I generated 100 white-noisy phase-folded light curves after 151 (for the G star), 34 (K dwarf), and 13 (M dwarf) transits, corresponding to the number of transits required to gather 68 % of the genuine satellite systems within 10 % of the injected moon parameters in my previous simulations. After these additional $3 \times 100$ "no moon" runs, the corresponding best-fit distributions turned out to be almost randomly distributed in the $R_s$-$a_{ps}$ plane with a slight clustering toward smaller satellite radii, and no 10 % × 10 % bin contained more than a few of the best fits. In contrast, if a moon were present, the best-fit systems would be distributed according to a Gaussian distribution around the genuine radius and planetary distance of the moon. I conclude that a by-chance clustering within 10 % of any given location in the $R_s$-$a_{ps}$ plane is $\lesssim 10^{-2}$. In turn, finding at least 68 % of the measurements within any 10 % × 10 % bin in the $R_s$-$a_{ps}$ plane makes a genuine moon system a highly probable explanation.

## 5. DISCUSSION

### 5.1. *Methodological Comparison with other Exomoon Detection Techniques*

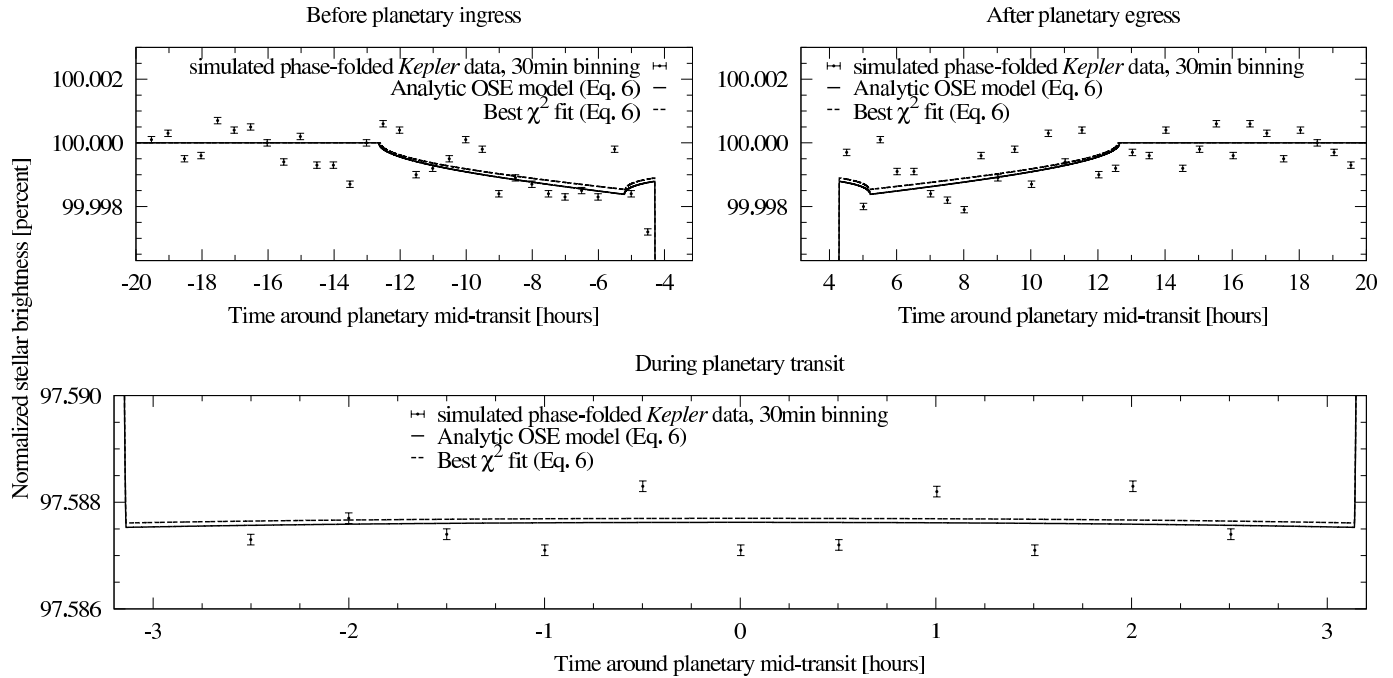

**Figure 10.** $\chi^2$ fit of the analytical OSE model via Equation (6) (dashed lines) to the binned, simulated *Kepler* data of a one-satellite system (data points) after $N = 100$ transits. The solid line shows the OSE light curve for the *known* input system. The moon is similar to Ganymede in terms of mass (0.025 $M_\oplus$), radius (0.42 $R_\oplus$), and planet-moon distance (15.47 $R_p$). In this simulation, the $\chi^2$ fit yields a moon radius $R_1 = 0.4 \pm 0.0049\ R_\oplus$ and planet-moon semi-major axis $a_{p1} = 15.4 \pm 0.0487\ R_p$.

### 5.1.1. *TTV/TDV-based Exomoon Searches*

While TTV and TDV refer to variations in the planet's transit light curve, the photometric OSE directly measures the decrease in stellar brightness caused by one or multiple moons. Combined TTV and TDV measurements allow computations of the planet-satellite orbital semi-major axis $a_{ps}$ and a moon's mass $M_s$ (Kipping 2009b,a). All descriptions of the TTV/TDV-based search for exomoons are restricted to one-satellite systems. The photometric OSE enables measurements of $a_{ps}$ and the satellite radii $R_s$ in multiple exomoon systems. In comparison to the TTV/TDV method, the OSE can be measured with analytical expressions (Equation (6) for the photometric OSE; Equation (11) for TDV-OSE; Equation (13) for TTV-OSE). Note that TTV and TDV still need to be removed prior to analyses of the photometric OSE.

A major distinction between TTV and TDV correction for the purpose of OSE analysis, compared to the actual detection of an exomoon via its TTV and TDV imposed on the planet, lies in the irrelevance of the TTV/TDV origin for the photometric OSE analysis. On the contrary, for TTV/TDV-based exomoon searches these corrections need to be accounted for in a consistent star-planet-satellite model of the system's orbital dynamics to exclude perturbing planets as the TTV/TDV source. If such a dynamical model is not applied, then still $a_{ps}$ and $M_s$ can be inferred if both TTV and TDV can be measured (Kipping 2009a,b) and a single moon is *assumed* as the originator of the signal. Exomoon detection via photometric OSE, on the other hand, can be achieved with much less computational power, practically within minutes, as the transit light curve is phase-folded without TTV/TDV corrections. Determination of $a_{ps}$ should be mostly unaffected, because TTV and TDV are supposed to be of the order of minutes (Kipping et al. 2009; Heller & Barnes 2013; Awiphan & Kerins 2013), whereas the photometric OSE extends as far as a few to tens of hours around the planetary transit (see Figures 9 and 10), depending on the mass of the host star and the semi-major axis of the planet-satellite barycenter around the star. The satellite radius is even more robust against uncorrected TTV/TDV, since it is derived from the shape and depth of the OSE signal, not from its duration (see Figure 5).

Ultimately, the photometric OSE allows for the detection and characterization of multi-satellite systems, whereas currently available models of the TTV/TDV strategy cannot unambiguously disentangle the underlying satellite architecture of multiple satellite systems. Since the number of satellites around a giant planet is supposed to vary with planetary mass and depending on the formation scenario (Sasaki et al. 2010), the photometric OSE is a promising alternative to TTV/TDV-based exomoon searches when it comes to understanding the formation history of extrasolar planets and moons.

In Sections 2.3 and 2.4, I examine the distribution of exomoon-induced TDV and TTV measurements. Combined with radial stellar velocity measurements and with the photometric OSE, they allow for a full parameterization of a star-planet-moon system. TTV-OSEs or TDV-OSEs alone may not unambiguously yield exomoon detections because they can be mimicked by planet-planet interactions (Mazeh et al. 2013). But if photometric OSEs indicate a satellite system, then TTV-OSE and TDV-OSE can be used to further strengthen the validity of the detection. In particular, TDV-OSE and TTV-OSE both offer the possibility of measuring a satellite's mass, which is inaccessible via the photometric OSE. In

## Exomoon detection after $N_{obs}$ transits

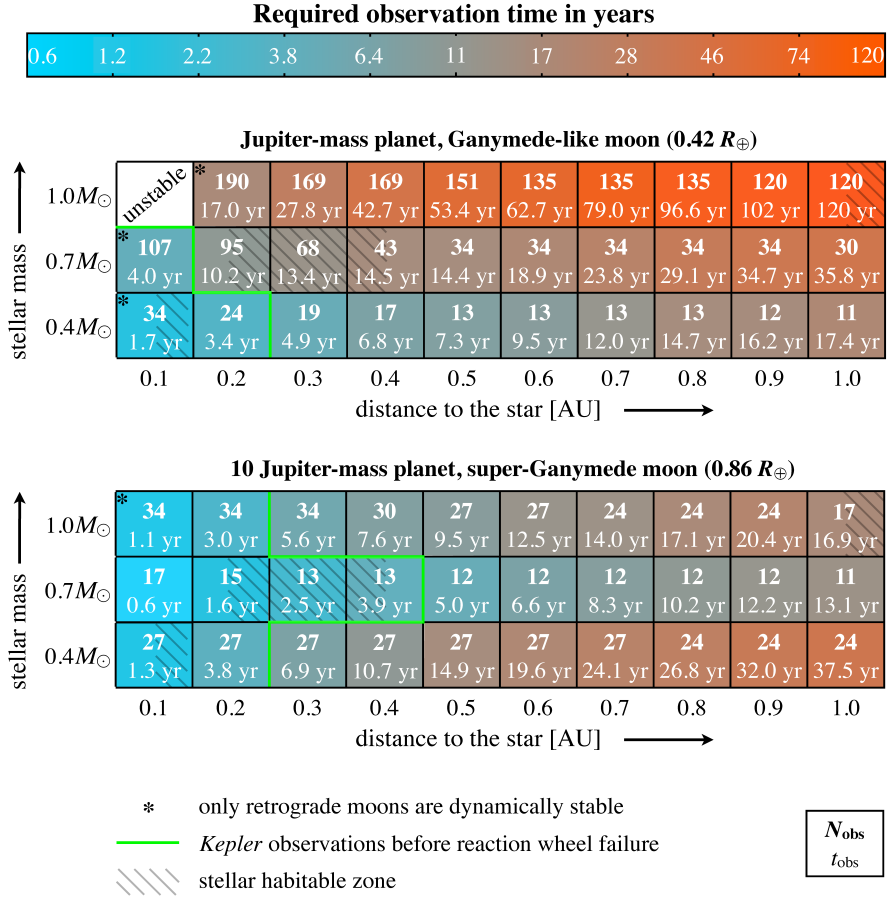

**Figure 11.** Number of transits required to detect the photometric OSE ($N_{obs}$, bold numbers) and equivalent observation time of a one-satellite system around a Jupiter-sized planet ( upper panel) and a 10 Jupiter-mass planet (lower panel). The three rows correspond to a 0.4, 0.7, and 1 $M_\odot$-mass host star, respectively, while the columns depict the semi-major axis of the planet-moon binary around the star. Striped areas indicate the stellar HZ, and green cell borders embrace the observation cycle that has been covered with *Kepler* before its reaction wheel failure. In all simulations, the moon is assumed to orbit in a Ganymede-wide orbit around the planet, that is, at about 15 $R_p$.

the spirit of Occam's razor, simultaneous observations of the photometric OSE, TTV-OSE, and TDV-OSE in longterm observations of a system would make an exomoon system the most plausible explanation, rather than a tilted transiting ring planet suffering planet-planet perturbations.

### 5.1.2. *Direct Photometric Exomoon Transits*

First, in comparison to direct observations of individual transits, the photometric OSE technique does not require dynamic modeling of the orbital movements of the star-planet-moon system, which drastically reduces the demands for computational power compared to photodynamic modeling (Kipping 2011a). Second, the amplitude of the photometric OSE signal in the phase-folded light curve is similar to the transit depth of a single exomoon transit, namely about $(R_s/R_\star)^2$. But in contrast to single-transit analyses (a method not applied by Kipping 2011a, by the way), OSE comes with a substantial increase in signal-to-noise by averaging over numerous transits (see Figures 8 and 9). Third, OSE has a more complex imprint in the phase-folded light curve – the ingress and egress patterns of the probability functions as well as a contribution to the total depth of the major transit trough (see Figure 5) – and thereby offers a larger "leverage" to tackle moon detections more securely. However, speed is only gained in exchange for loss of information, which is reasonable for the most likely cases considered in this paper (coplanarity, circularity, masses separated by orders of magnitudes, radii and distances differing by at least an order of magnitude).

### 5.1.3. *Other Techniques*

Besides TTV/TDV measurements and direct photometric transit observations, a range of other techniques to search for exomoons have been proposed (see Section 1). In comparison to direct imaging of a planet-moon binary's photocentric wobble, which requires an angular resolution of the order of microarcseconds for a planet-moon binary similar to Saturn and Titan (Cabrera & Schneider 2007)[6], the technical demands for photometric OSE measurements are much less restrictive. In other words, they are already available with the *Kepler* telescope and the upcoming *Plato 2.0* mission. Detections of planet-moon mutual eclipses require modeling of the system's orbital dynamics, which costs substantially more computing time than fitting Equation (6) to a phase-folded transit curve or Equations (11) and (13) to the distribution of TDV and TTV measurements. Mutual eclipses can also be mimicked or blurred by star spots, and they are hardly detectable for moon's as small as Ganymede orbiting a Jovian planet. And referring to direct imaging of tidally heated exomoons, a giant planet with a spot similar to Jupiter's Giant Red Spot could also mimic a satellite eclipse.

---

[6] Note that the authors use a mass ratio of 0.01 between Titan and Saturn to yield this threshold. However, the true mass ratio is actually about $2 \times 10^{-4}$, so the value for the required angular resolution might even be much smaller.

Detections of the Rossiter-McLaughlin effect of a Ganymede-sized moon around a Jupiter-like planet requires accuracies in radial velocity measurements of the order of a few centimeters per second (Simon et al. 2010) and will only be feasible for extremely quiet stars and with future technology (Zhuang et al. 2012).

In comparison to detections with microlensing, observations of exomoon-induced OSEs are reproducible. Announcements of possible exomoon detections around free-floating giant planets in the Galactic Bulge show the obstacles of this technique and should be treated with particular skepticism. Not only are microlensing observations irreproducible, but also from a formation point of view, a moon with about half the mass of Earth cannot possibly form in the protosatellite disk around a roughly Jupiter-sized planet (Canup & Ward 2006).

Exomoons orbiting exoplanets around pulsars constitute a bizarre family of hypothetical moons, but as the first confirmed exoplanets actually orbit a pulsar (Wolszczan & Frail 1992), they might exist. Analyses by Lewis et al. (2008) suggest that exomoons around pulsar planet PSR B1620−26 b, if they exist, need to be at least as massive as about 5 % the planetary mass, leaving satellites akin to solar system moons undetectable. More generally, their time-of-arrival technique can hardly access masses as small as Earth even in the most promising cases of planets and moons in wide orbits (Karen Lewis, private communication).

Direct imaging searches for extremely tidally heated exomoons also imply a yet unknown family of exomoons (Peters & Turner 2013), where the satellite is at least as large as Earth and orbits a giant planet in an extremely close, eccentric orbit. Exomoon-induced modulations of a giant planet's radio emission require the moon (not the planet) to be as large as Uranus (Noyola et al. 2013), that is, quite big. Such a moon does also not exist in the solar system.

#### 5.1.4. *Comparison of Detection Thresholds*

The detection limit of the combined TTV-TDV method using *Kepler* data is estimated to be as small as $0.2\,M_\oplus$ for moons around Saturn-like planets transiting relatively bright M stars with *Kepler* magnitudes $m_K < 11$ (Kipping et al. 2009). Note, however, that the TTV-TDV method is susceptible to the satellite-to-planet mass ratio $M_s/M_p$, not to the satellite mass in general. The HEK team achieves accuracies down to $M_s/M_p \approx 5\,\%$ (Kipping et al. 2013b,a, 2014; ?). Using the correlation of exomoon-induced TTV and TDV on planets transiting less bright M dwarfs ($m_K = 12.5$) in the stellar HZs, Awiphan & Kerins (2013) find less promising thresholds of 8 to $10\,M_\oplus$, and such an exomoon's host planet would need to be as light as $25\,M_\oplus$. Those systems would be considered planet binaries rather than planet-moon systems. Lewis (2011) simulated TTVs caused by the direct photometric transit signature of moons and found that these variations could indicate the presence of moons as small as $0.75\,R_\oplus$, with this limit increasing towards wide orbital separations. A similar threshold has been determined by Simon et al. (2012), based on their scatter-peak method applied to *Kepler* short cadence data.

Hence, depending on the actual analysis strategy of moon-induced TTV and TDV signals, and depending on the stellar apparent magnitude, planetary mass, and planet-moon orbital separation, a wide range of detection limits is possible. Most important, detection capabilities via TTV/TDV measurements as per Kipping et al. (2009) decrease for increasing planetary mass, which makes them most sensitive to very massive moons orbiting relatively light gas planets such as Saturn and Neptune. Yet, the most massive planets are predicted to host the most massive moons (Canup & Ward 2006; Williams 2013).

In comparison, the photometric OSE presented in this paper is not susceptible to planetary mass and can detect Ganymede- or Titan-sized moons around even the most massive planets. This technique is thus well-suited for the detection of extrasolar moons akin to solar system satellites. What is more, the photometric OSE is the first method to enable the detection and classification of multi-satellite systems.

Rings of giant planets could mimic the OSE of exomoons. However, planets at distances $\lesssim 1\,\mathrm{AU}$ will have small obliquities, or spin-orbit misalignments, due to the tidal interaction with the star. This "tilt erosion" (Heller et al. 2011b,a) will cause potential rings to be viewed edge-on during transits and so they will tend to be invisible. Also, a ring's transit signature will not generate an OSE, but its signal would look similar during every single transit. Analyses of the photometric scatter before the planetary ingress and after the planetary egress could be used to discriminate genuine exomoon systems from ring systems (Simon et al. 2012). As a ring system does not induce TTVs or TDVs, measurements of the TTV-OSE and/or TDV-OSE can be used as an independent method to confirm exomoon detections.

My predictions of the number $N_{\mathrm{obs}}$ of transits required to discover the hypothetical one-satellite system around a giant planet may be overestimations, because I used a fixed data binning of 30 min. This binning delivered the most reliable detections for the simulated Jupiter-satellite system in the HZ around a K star, but moon systems at different stellar orbital distances and around other host stars have different orbital velocities and their apparent trajectories differ in duration. Thus, more suitable data binning – for example a 10 min binning for short-period transiting planets – will yield more reliable fittings than derived in this report. My simulations are thought to cover a broad parameter space rather than a best fit for each individual hypothetical star-planet-exomoon system.

### 5.2. *Red Noise*

While my noise model assumes white noise only (Section 3.1), detrending real observations will have to deal with red noise (Lewis 2013). Instrumental effects such as CCD aging as well as red noise induced by stellar granulation and spots need to be removed or corrected for before the assumption of a light curve dominated by white noise becomes appropriate. In cases where red noise is comparable to white noise, $N_{obs}$ and $t_{obs}$ as presented in Figure 11 will increase substantially.

The results shown in Figure 11 do, however, still apply to a subset of photometrically quiet stars, such as the host star of transiting planet TrES-2b that has also been observed by *Kepler* (Kipping & Bakos 2011). Basri et al. (2010) have shown that about every second K dwarf and about 16 % of the M dwarfs in the *Kepler* sample are less active than the active Sun. Gilliland et al. (2011) found similar activity levels of K and M dwarfs but cautioned that the small sample of K and M dwarfs in the *Kepler* data as well as contamination by giant stars could spoil these rates.

An OSE-based exomoon survey focusing on quiet stars will automatically tend to avoid spotted stars. If nevertheless present, clearly visible signatures of big star spots can be removed from the individual transits before the light curve is detrended and phase-folded. As long as these removals are randomized during individual transits, no artificial statistical signal will be induced into the phase-folded curve. But if the circumstellar orbital plane of the planet-satellite system were substantially inclined against the stellar equator and if the star had spot belts, then spot crossings would occur at distinct phases during each single transit (see HAT-P-11 for an example, Sanchis-Ojeda & Winn 2011). Such a geometry would strongly hamper exomoon detections via their photometric OSE.

### 5.3. *OSE detections with Plato 2.0*

As the simulations in Section 4 show, *Kepler*'s photometry is sufficiently accurate to detect the photometric OSE of transiting exomoons around relatively quiet K and M dwarfs with intrinsic stellar noise below about 20 ppm. Hence, from a technological point of view, the *Plato 2.0* telescope with a detector noise similar to that of *Kepler* offers a near-future possibility to observe exoplanetary transits with similarly high accuracy.[7] However, even with arbitrarily precise photometry the number of observed transits determines the prospects of OSE detections. Given that *Plato 2.0* is planned to observe two star fields for two to three years (Rauer 2013), Figure 11 suggests that this mission could just deliver as many transits as are required to enable photometric OSE detections around M and K stars. Yet, the results presented in this paper strongly encourage longterm monitoring over at least five years of a given star sample to allow exomoons to imprint their OSEs into the transit light curves. If the survey strategy of *Plato 2.0* can be adjusted to observe one field for about five years or more, then the search for exomoons could become an additional science objective of this mission.

The *Transiting Exoplanet Survey Satellite* (TESS), however, is planned to have an observing duty cycle of only two years, and it will observe a given star for 72 days at most. Hence, TESS cannot possibly discover the photometric OSE of exomoons.

### 6. CONCLUSIONS

This paper describes a new method for the detection of extrasolar moons, which I refer to as the Orbital Sampling Effect (OSE). It is the first technique that allows for reproducible detections of extrasolar multiple satellite systems akin to those seen in the solar system. The OSE appears in three flavors: (1) the photometric OSE (Section 2.2), (2) the TDV-OSE (Section 2.3), and (3) the TTV-OSE (Section 2.4). The photometric OSE can reveal the satellite radii in units of stellar radii as well as the planet-moon orbital semi-major axes, but it cannot constrain the satellite masses. TDV-OSE and TTV-OSE can both constrain the satellite mass. Photometric OSE, TDV-OSE, and TTV-OSE offer important advantages over other established techniques for exomoon searches because (1) they do not require modeling of the moons' orbital movements around the planet-moon barycenter during the transit, (2) planet-moon semi-major axes, satellite radii, and satellite masses can be measured or fit with analytical expressions (Equations (6), (11), (13)), and (3) the photometric OSE is applicable to multi-satellite systems. TDV-OSE and TTV-OSE can also reveal the masses of moons in multi-satellite systems, but this parameterization is beyond the scope of this paper.

My simulations of photometric OSE detections with *Kepler*-class photometry show that Ganymede-sized exomoons orbiting Sun-like stars cannot possibly be discovered in the available *Kepler* data. However, they could be found around planets as far as 0.1 AU from a 0.7 solar mass K star or as far as 0.2 AU from a $0.4\,M_\odot$ M dwarf. The latter case includes planet-moon binaries in the stellar HZ. Exomoons with the 10-fold mass of Ganymede and Ganymede-like composition (implying radii around $0.86\,R_\oplus$) are detectable in the *Kepler* data around planets orbiting as far as 0.2 AU from a Sun-like host star, 0.4 AU from the K dwarf star, or about 0.2 AU from the M dwarf. The latter case comprises both the respective stellar HZ. What is more, such large moons are predicted to form locally around super-Jovian host planets (Canup & Ward 2006; Sasaki et al. 2010) and are therefore promising targets to search for.

To model realistic light curves or to fit real observations with a photometric OSE model, stellar limb darkening needs to be included into the simulations (Heller et al., in preparation). Effects on $N_{obs}$ are presumably small for planet-moon systems with low impact parameters, because the stellar brightness increases to roughly 60 % when the incoming moon has traversed only the first 5 % of the stellar radius during a transit (Claret 2004). What is more, effects of red noise have not been treated in this paper, and so the numbers presented in Figure 11 are restricted to systems where either (1) the host star is photometrically quiet on a $\approx 10$ hr timescale or (2) removal of red noise can be managed thoroughly. The prescriptions of the three OSE flavors delivered in this paper can be enhanced to yield the sky-projected angle between the orbital planes of the satellites and the diameter of the star. In principle, the photometric OSE allows measuring the inclinations of each satellite orbit separately. The effect of mutual moon eclipses will be small in most cases but offers further room for improvement. Ultimately, when proceeding to real observations, a Bayesian framework will be required for the statistical assessments of moon detections. As part of frequentists statistics, the $\chi^2$ method applied in this paper is only appropriate because I do not choose between different models since the injected moon architecture is known a priori.

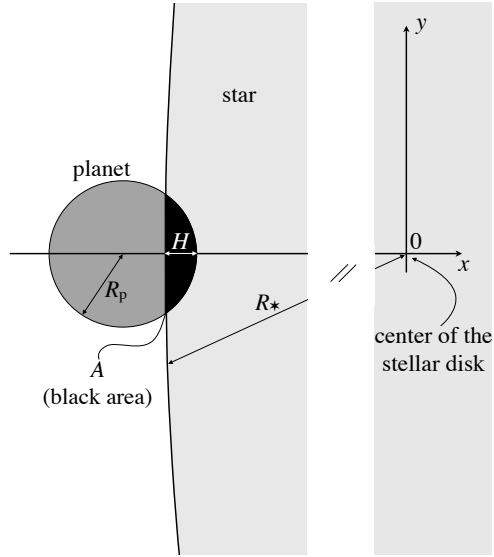

**Figure 12.** Parameterization of the planetary ingress.

Another application of the OSE technique, which is beyond the scope of this paper, lies in the parameterization of transiting binary systems. If not only the secondary constituent (in this paper the moon) shows an OSE but also the primary (in this paper the planet), then both orbital semi-major axes ($a_1$ and $a_2$) around the common center of mass can be determined. If the total binary mass $M_{\rm b} = M_1 + M_2$ were known from stellar radial velocity measurements, then it is principally possible to calculate the individual masses via $a_1/a_2 = M_2/M_1$ and substituting, for example, $M_1 = M_{\rm b} - M_2$. This proce­dure, however, would be more complicated than in the model presented in this paper, because the center of the primary transit could not be used as a reference any­more. Instead, as both the primary and the secondary orbit their common center of mass, this barycenter would need to be determined in each individual light curve and used as a reference for phase-folding.

To sum up, the photometric OSE, the TDV-OSE, and the TTV-OSE constitute the first techniques capable of detecting extrasolar multiple satellite systems akin to those around the solar system planets, in terms of masses, radii, and orbital distances from the planet, with cur­rently available technology. Their photometric OSE sig­nals should even be measurable in the available data, namely, that of the *Kepler* telescope. After the recent failure of the *Kepler* telescope, the upcoming *Plato 2.0* mission is a promising survey to yield further data for exomoon detections via OSE. To increase the likelihood of such detections, it will be useful to monitor a given field of view as long as possible, that is, for several years, rather than to visit multiple fields for shorter periods.

## APPENDIX

### A. PARAMETERIZATION OF THE PLANETARY INGRESS

During the ingress of the planet in front of the stellar disk, the planet blocks an increasing area $A$ of the stellar disk (black area in Fig. 12). With the star being substantially larger than the planet, the curvature of the stellar disk can be neglected and $A$ is determined by the height $H$ of the circular segment by

$$A = R_{\rm p}^2 \arccos\left(1 - \frac{H}{R_{\rm p}}\right) - \sqrt{2R_{\rm p}H - H^2}\,(R_{\rm p} - H) \quad . \tag{A1}$$

In my simulations, $H = H(t)$ is a function of time. The temporary increase (during ingress) and decrease (during egress) of $A$ is visualized in Figure 4 by the gradual decrease in stellar brightness between about $-4$ and $-3$ hr and its gradual increase between $+3$ and $+4$ hr, respectively.

### B. BINNING OF SIMULATED *KEPLER* DATA

The set of simulated, noisy brightness measurements shown in Figure 8 is given by data points $b_i$. I divide the simulated observations into time intervals with running index $j$. The mean value of brightness measurements in the

$j$th time bin is

$$\bar{b}_j = \frac{1}{K_j} \sum_{k=1}^{K_j} b_k \quad , \tag{B1}$$

with $K_j$ as the number of data points $b_k$ in bin $j$. The variance $s_j^2$ within each bin is given by

$$s_j^2 = \frac{1}{K_j - 1} \sum_{k=1}^{K_j} (b_k - \bar{b}_j)^2 \tag{B2}$$

and the standard deviation of the mean in that bin equals

$$\sigma_j = \frac{s}{\sqrt{K_j}} = \sqrt{\frac{1}{K_j(K_j - 1)} \sum_{k=1}^{K_j} (b_k - \bar{b}_j)^2} \quad . \tag{B3}$$

By increasing the bin width, say from 30 to 60 min, it is possible to collect more data points per interval and to increase $K_j$, which in turn decreases $\sigma_j$ and increases accuracy. However, this comes with a loss in time resolution. The best compromise I found, at least for a satellite system on orbits similar to those of the Galilean moons but transiting in the HZ around a K star, is a 30 min binning of long cadence *Kepler* data. Transits of systems on wider circumstellar orbits have a longer duration, and thus might yield best results with a binning longer than 30 min. Yet, their transits are less frequent, so this argument is only adequate for a comparable number of transits.

## C. $\chi^2$ MINIMIZATION

I fit my simulated *Kepler* observations of a one-satellite system with a brute force $\chi^2$ minimization technique, that is, I compute

$$\chi^2_{R_1, a_{p1}} = \frac{1}{K} \sum_{j=1}^{K} \frac{(b_j - m_j)^2}{\sigma_j^2} \tag{C1}$$

for the whole parameter space and search for the global minimum. The parameter grid I explore spans $0.1\,R_\oplus \leq R_1 \leq 2\,R_\oplus$ in increments of $0.02\,R_\oplus$ and $2\,R_p \leq a_{p1} \leq 30\,R_p$ in steps of $0.1\,R_p$. In Equation (C1), $K = \sum_j j$ is the number of binned data points to fit, $b_j$ denotes the binned simulated data points, and $m_j$ refers to the normalized brightness in bin $j$ predicted by the analytical model (Equation (6)) for the satellite's radius $R_1$ and semi-major axis $a_{p1}$ to be tested (see Figure 10).

I thank an anonymous reviewer for her or his valuable report. Karen Lewis' feedback also helped clarify several passages in this paper, and I thank Brian Jackson for his thoughtful comments. This work made use of NASA's ADS Bibliographic Services. Computations have been performed with `ipython 0.13.2` on `python 2.7.2` (Pérez & Granger 2007), and most figures were prepared with `gnuplot 4.4` (www.gnuplot.info).

## 6.2 Modeling the Orbital Sampling Effect of Extrasolar Moons (Heller et al. 2016a)

Contribution:

RH contributed to the literature research, worked out the mathematical framework, translated the math into computer code, performed the simulations shown in Fig. 6, created Figs. 1-4 and 6, led the writing of the manuscript, and served as a corresponding author for the journal editor and the referees.

# MODELING THE ORBITAL SAMPLING EFFECT OF EXTRASOLAR MOONS

René Heller
Max Planck Institute for Solar System Research, Justus-von-Liebig-Weg 3, 37077 Göttingen, Germany; heller@mps.mpg.de

Michael Hippke
Luiter Straße 21b, 47506 Neukirchen-Vluyn, Germany; hippke@ifda.eu

Brian Jackson
Carnegie Department of Terrestrial Magnetism, 5241 Broad Branch Road, NW, Washington, DC 20015, USA and
Department of Physics, Boise State University, Boise, ID 83725-1570, USA; bjackson@boisestate.edu
Published in The Astrophysical Journal 820:88 (11pp), 2016 April 1

## ABSTRACT

The orbital sampling effect (OSE) appears in phase-folded transit light curves of extrasolar planets with moons. Analytical OSE models have hitherto neglected stellar limb darkening and non-zero transit impact parameters and assumed that the moon is on a circular, co-planar orbit around the planet. Here, we present an analytical OSE model for eccentric moon orbits, which we implement in a numerical simulator with stellar limb darkening that allows for arbitrary transit impact parameters. We also describe and publicly release a fully numerical OSE simulator (PyOSE) that can model arbitrary inclinations of the transiting moon orbit. Both our analytical solution for the OSE and PyOSE can be used to search for exomoons in long-term stellar light curves such as those by *Kepler* and the upcoming *PLATO* mission. Our updated OSE model offers an independent method for the verification of possible future exomoon claims via transit timing variations and transit duration variations. Photometrically quiet K and M dwarf stars are particularly promising targets for an exomoon discovery using the OSE.

*Keywords:* instrumentation: photometers – methods: data analysis – methods: analytical – methods: observational – methods: statistical – planets and satellites: detection

## 1. CONTEXT AND MOTIVATION

The race toward the first detection of an extrasolar moon picks up pace. While the first attempts to detect exomoons were byproducts of planet-targeted observations (Brown et al. 2001; Pont et al. 2007; Maciejewski et al. 2010), high-accuracy space-based observations of thousands of transiting exoplanets and planet candidates by *Kepler* (Borucki et al. 1997; Batalha et al. 2013) now allow for dedicated exomoon searches (Kipping et al. 2012; Szabó et al. 2013; Hippke 2015). Upcoming data from the European Space Agency's *CHEOPS* and *PLATO* space missions will provide further promising avenues toward an exomoon detection (Hippke & Angerhausen 2015; Simon et al. 2015).

Exomoon detections will be highly valuable for our understanding of the origin and fate of planetary systems because they probe the substructures of planet formation that is not accessible through planet observations alone. As for the solar system, moons provide key insights into the formation of Earth (by a giant collision; Cameron & Ward 1976; Canup 2012), into the temperature distributions within the circumplanetary accretion disks of Jupiter and Saturn (Pollack & Reynolds 1974; Canup & Ward 2002; Sasaki et al. 2010; Heller & Pudritz 2015a), and into the cause of Uranus' tilted spin axis (by gradual collisional tilting; Morbidelli et al. 2012). As of today, no moon has been confirmed around a planet beyond the solar system. Hence, exoplanetary science suffers from a fundamental lack of knowledge about the fine structure of planetary systems.

Heller (2014) recently identified a new exomoon signature in planetary transit light curves, which occurs due to the additional darkening of the star by a transiting moon. This phenomenon, which we refer to as the photometric orbital sampling effect (OSE), occurs in phase-folded transit light curves, because an exomoon's sky-projected position with respect to its host planet is variable in subsequent transits but is statistically predictable for a large number of transits ($N \gtrsim 10$).[1] The photometric OSE is not sensitive to the satellite mass ($M_s$), but it is very sensitive to its radius ($R_s$). This makes the photometric OSE particularly sensitive to low-density, water-rich moons like the three most massive moons in the solar system, Ganymede and Callisto around Jupiter and Titan around Saturn. These moons would hardly be detectable in the available *Kepler* data by combined TTV and TDV measurements, which are sensitive to roughly Earth-mass moons (Szabó et al. 2013). The most advanced exomoon search as of today, the "Hunt for Exomoons with Kepler" (HEK; Kipping et al. 2012), using a photodynamical model, achieves much better detection limits down to a few Ganymede masses, depending on the mass of the host planet, amongst other parameters (Kipping et al. 2015). However, if giant planets were able to give their fully fledged, icy moon systems a piggyback ride to $\lesssim 1$ AU, where super-Jovian planets are abun-dant in radial velocity survey data (Dawson & Murray-Clay 2013; Heller & Pudritz 2015a), then the photometric OSE could be a promising method to find these exomoons. The mass-radius diagram in Figure 1 illustrates the detection threshold for moons transiting photometrically quiet M stars. Note that Mars-mass moons with a water-rock composition similar to Ganymede, Callisto, and Titan are significantly larger than Mars (see the open circle). Hence, they could be detectable around photometrically quiet stars. These moons are predicted to form frequently around the most massive super-Jovian planets (Heller & Pudritz 2015b).

---

[1] The OSE appears in three flavors (as per Heller 2014), one of which is the photometric OSE, which we focus on in this paper. The other two manifestations of the OSE appear in the planetary transit timing variations (TTV-OSE) and transit duration variations (TDV-OSE).

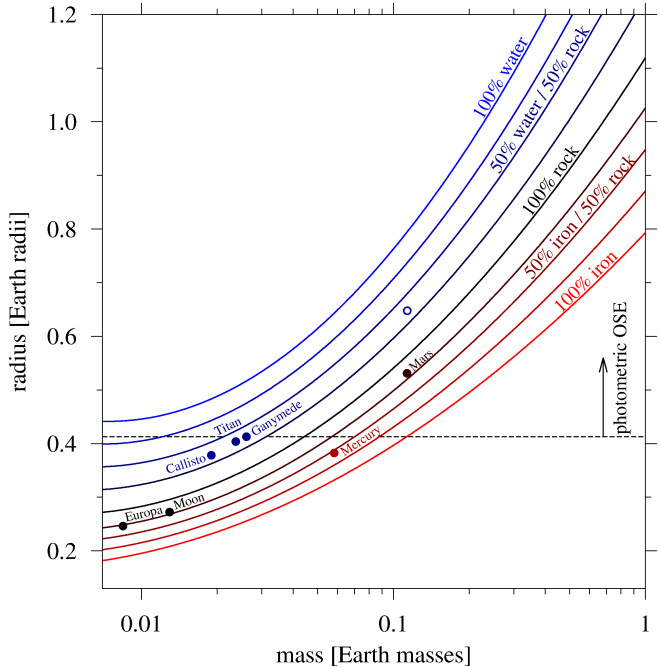

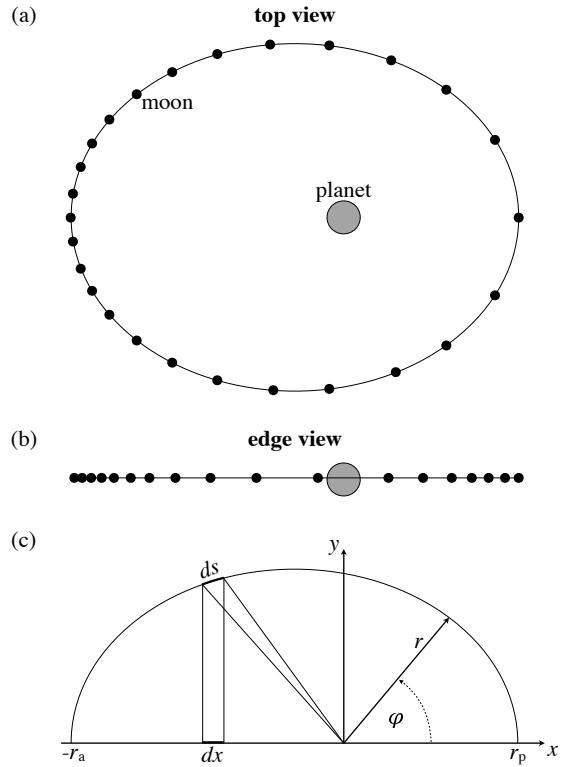

**Figure 1.** Prospective detection thresholds of the photometric OSE (horizontal dashed line) for moons around a photometrically quiet star after $\gtrsim 30$ transits (as per Heller 2014, Fig. 11 therein). Note that the photometric OSE is sensitive to an exomoon's radius but not to its mass. Hence, large (potentially low-density) moons are the most promising targets. The open circle denotes a Mars-mass moon with a Ganymede-like composition as predicted by Heller & Pudritz (2015b). The mass-radius relationship for moons of various compositions is according to Fortney et al. (2007).

**Figure 2.** Construction of an exomoon's orbital sampling frequency $\mathcal{P}_s(x)$ for elliptical orbits. **(a)** The moon's orbital position around the planet is measured with a constant sampling frequency, or frame rate. **(b)** The moon's variable orbital velocity yields an asymmetric probability density with respect to the planet. **(c)** The probability density can be derived as $P_s(x) = ds(x)/dx$. The special case of a circular orbit is given in Figure 1 of (Heller 2014).

Hippke (2015) searched for the photometric OSE in the archival data of the *Kepler* telescope and found indications for an OSE-like signal in the combined transit light curves of hundreds of planets and planet candidates with orbital periods > 35 d. The original description of the OSE by Heller (2014) was purely analytical and thereby fast, but it made several simplifications: stellar limb darkening was neglected; the planet–moon orbit was assumed to be non-eccentric ($e = 0$) and non-inclined ($i = 0°$) with respect to the circumstellar orbit; the planet–moon system was assumed to transit the star along the stellar diameter; that is, the transit impact parameter (b) was set to 0. Consequently, this model was not broadly applicable. Hippke (2015) performed a mostly numerical study that did include non-circular planet–moon orbits and arbitrary inclinations, but he did not explore a wide parameter range of the OSE.

We here first derive a novel equation for the orbital sampling frequency $\mathcal{P}_s(x)$ of exomoons on eccentric orbits ($e \geq 0$; Section 2.1). We then incorporate our formula into a numerical simulator that models the phase-folded transits of exoplanets with moons in front of stars with limb darkening (Section 2.2). Our simulations are compared to real *Kepler* data. We also perform purely numerical simulations for a wide range of inclined planet–moon orbits ($i \geq 0$), planetary impact parameters (b $\geq 0$), satellite radii ($R_s$), and semimajor axes of the satellite's orbit around the planet ($a$; Section 3).

## 2. A DYNAMICAL MODEL FOR THE PHOTOMETRIC OSE

### 2.1. Sampling Frequency for Eccentric Exomoon Orbits

In Heller (2014), the OSE was described for circular moon orbits only. Given that the eccentricities of the 10 largest moons in the solar system are all smaller than that of the Earth's Moon ($\approx 0.055$), this approximation seems appropriate. However, the architectures and physical properties of extrasolar planetary systems turned out to be very different from the solar system planets, as we recall the discoveries of terrestrial planets around pulsars (Wolszczan & Frail 1992), Jupiter-mass planets in extremely short-period orbits (Mayor & Queloz 1995), circumbinary planets (Doyle et al. 2011), and a large population of super-Earths with short orbital periods (Batalha et al. 2013). Hence, astronomers should not be surprised if they were to find that at least some exomoons follow significantly eccentric orbits. Eccentricities may, for example, be forced by gravitational interaction with the star (Heller 2012; Spalding et al. 2016), other planets (Ćuk 2007; Gong et al. 2013; Payne et al. 2013), or even the host planet (Goldreich 1963). Moreover, an analytical solution is always worthwhile as it can give a deeper understanding of the underlying physical processes that generate an observational feature.

We here derive the orbital sampling probability function for exomoons on eccentric circumplanetary orbits. Figure 2(a) shows the moon's circumplanetary orbit in a plane perpendicular to the reader's line of sight (solid line). If an observer were to take pictures of the moon's orbital position with a constant sampling rate, then the individual moon pictures would be distributed in a manner similar to the one depicted in Figure 2(a). Around periapsis, the moon's Keplerian velocity ($v$) is higher than around apoapsis. Thus, the probability of observing the moon, e.g. during a common stellar transit with its host planet, at a given orbital position is asymmetric with respect to the sky-projected distance to the planet. This fact is visualized in Figure 2(b), where the planet–moon system is sampled from a co-planar perspective. Due to the projection effect, the moon(s) pile(s) up toward the edges of the projected major axis, but the probability distribution to the left of the planet is different from what it looks to the right of the planet. Figure 2(c) shows how we construct the satellite's probability density $\mathcal{P}_\mathrm{s}(x)$ as a function of its projected distance ($x$) to the planet.

In its most general form, $\mathcal{P}_\mathrm{s}(x)$ describes what fraction of its orbital period ($P_\mathrm{ps}$) the satellite spends in an infinitely small interval ($dx$) on its sky-projected orbit along the $x$-axis (see Figure 2(c)). Hence, it is given as

$$\mathcal{P}_\mathrm{s}(x) \propto \frac{1}{P_\mathrm{ps}} \frac{dt}{dx} \ , \tag{1}$$

where $dt$ is an infinitesimal change in time. With $ds$ as an infinitely small interval on the moon's elliptical orbit, the Keplerian velocity is $v(x) = ds(x)/dt$, hence

$$\mathcal{P}_\mathrm{s}(x) \propto \frac{1}{P_\mathrm{ps} v(x)} \frac{ds(x)}{dx} \ , \tag{2}$$

and the challenge is then in finding $ds(x)/dx$. From Figure 2(a) we learn that $ds(x) = r(x) d\varphi(x)$. We use the parameterization of a Keplerian orbit

$$r(\varphi) = \frac{a(1-e^2)}{1 + e \cos(\varphi)} \tag{3}$$

and solve it for $\varphi(x)$ via

$$x(\varphi) = r(\varphi)\cos(\varphi) = \frac{a(1-e^2)}{\left(\dfrac{1}{\cos(\varphi)} + e\right)}$$

$$\Leftrightarrow \quad \varphi(x) = \arccos\left(\left[\frac{a}{x}(1-e^2) - e\right]^{-1}\right) \ . \tag{4}$$

Equation (2) then becomes

$$\mathcal{P}_\mathrm{s}(x) \propto \frac{r(x)}{P_\mathrm{ps} v(x)} \frac{d\varphi(x)}{dx} \tag{5}$$

$$= \frac{r(x)}{P_\mathrm{ps} v(x)} \underbrace{\frac{d}{dx} \arccos\left(\left[\frac{a}{x}(1-e^2) - e\right]^{-1}\right)}$$

$$= A\left(\sqrt{1 - \frac{1}{\left(\dfrac{A}{x} - e\right)^2}} \left(\frac{A}{x} - e\right)^2 x^2\right)^{-1} \ ,$$

where we introduced $A \equiv a(1-e^2)$. In Equation (5), $r(x)$ can be derived by inserting Equation (4) into (3), hence

$$r(x) = \frac{A}{1 - e\left(\dfrac{A}{x} - e\right)^{-1}} \ , \tag{6}$$

and $v(x)$ in Equation (5) is given by

$$v(r) = \sqrt{\mu\left(\frac{2}{r(x)} - \frac{1}{a}\right)} \ , \tag{7}$$

where $\mu = G(M_\mathrm{p} + M_\mathrm{s})$ and $G$ is Newton's gravitational constant.

So far, we assumed the line of sight to be parallel to the $y$-axis in Figure (2), that is, along the orbital semiminor axis ($b = a(1-e^2)^{1/2}$). Of course, the moon orbit can be rotated in the $x$-$y$ plane by an angle $\omega$. We can imagine that once $\omega = 90°$, the observer samples the moon orbit according to $\mathcal{P}_\mathrm{s}(y)$. In this case, the sky-projected orbit appears symmetric with an apparent radius ($\tilde{a}(e,\omega)$) equal to $b$. In general, we have

$$\tilde{a}(e, \omega) = \sqrt{\left(a\cos(\omega)\right)^2 + \left(b\sin(\omega)\right)^2}$$

$$= a\sqrt{\cos(\omega)^2 + (1-e)\sin(\omega)^2} \ . \tag{8}$$

We then replace $e$ with $\tilde{e} = e\cos(\omega)$ and $A$ with $\tilde{A} = \tilde{a}(1-\tilde{e}^2)$ in Equations (5)–(7) to obtain $\mathcal{P}_\mathrm{s}(x)$ for arbitrary orientations $\omega$.

At last, we can swap the proportionality sign in Equation (5) with an equality sign by normalizing

$$\int_{-r_\mathrm{a}}^{+r_\mathrm{p}} dx \ \mathcal{P}_\mathrm{s}(x) \equiv 1 \ , \tag{9}$$

where $r_\mathrm{a}$ and $r_\mathrm{p}$ are the moon's orbital radii at apoapsis and periapsis, respectively (see Figure 2(c)). We evaluate this integral numerically for arbitrary $a$, $e$, and $\omega$ and find an approximate solution

$$\mathcal{P}_{\rm s}(x) = 2(1 + e^4) \left(\frac{a}{\tilde{a}(e,\omega)}\right)^2 \frac{\tilde{r}(x)}{P_{\rm ps}\tilde{v}(x)}$$

$$\times \tilde{A} \left(\sqrt{1 - \frac{1}{\left(\frac{\tilde{A}}{x} - \tilde{e}\right)^2} \left(\frac{\tilde{A}}{x} - \tilde{e}\right)^2 x^2}\right)^{-1} , \qquad (10)$$

where $\tilde{r}(x)$ and $\tilde{v}(x)$ refer to Equations (6) and (7), respectively, but swapping $e$ for $\tilde{e}$ and $A$ for $\tilde{A}$.[2] The error in Equation 10 is $\ll 1\%$ for $e < 0.9$.

In Figure 3, we plot Equation (10) for a moon orbiting a planet on various eccentric orbits, all of which have a semimajor axis of 10 planetary radii ($R_{\rm p}$). The upper panel depicts the orbital geometries and the lower panel shows $\mathcal{P}_{\rm s}(x)$ in each case. Four cases with eccentricities between 0 and 0.4 assume $\omega = 0$ as presented in Figure 2, and the $e = 0.5$ case is rotated by $\omega = 90°$. Note that in the latter scenario, where the semiminor axis $b$ is substantially smaller than $a$ and the line of sight is parallel to $a$, $\mathcal{P}_{\rm s}(x)$ is significantly higher at a given planetary separation because the moon occupies a smaller circumplanetary region along the $x$-axis. This means that the OSE becomes particularly prominent in the phase-folded light curve of eccentric moon systems with $\omega$ close to 90° or 270°. On the other hand, it becomes relatively weak (or "smeared") for $\omega$ near 0° or 180°.

## 2.2. The Photometric OSE with Limb Darkening

### 2.2.1. Dynamical Simulations

We built a numerical model to simulate the stellar transit of an exoplanet for arbitrary probability functions $\mathcal{P}_{\rm s}(x)$ , including multiple functions in the case of multi-moon systems. As an update to Heller (2014), our simulator now considers stellar limb darkening and the transit impact parameter can be varied. In comparison to Heller (2014, Section 2.2.2 therein), where an analytical solution for the actual light curve (the stellar brightness $B_{\rm OSE}^{(n)}$ due to a transiting planet with $n$ exomoons) without stellar limb darkening has been derived, we do not derive an analytical solution for the light curve of the OSE with limb darkening. Instead, we simulate the OSE by generating a computer model of a limb-darkened stellar disk, approximated as a circle touching the edges of a square sized $1\,000 \times 1\,000$ pixels. Then we let the planet and the probability function(s) of the moon(s) transit. In the following, we refer to this approach as our "dynamical OSE model".

Once $\mathcal{P}_{\rm s}(x)$ enters the stellar disk, we multiply the stellar intensity ($I_\star^{\rm px}$) in any pixel that is covered with the probability density $\mathcal{P}_{\rm s}^{\rm px}(x)$ in this pixel, so that $dx \times I_\star^{\rm px} \times \mathcal{P}_{\rm s}^{\rm px}(x)$ is the relative amount of stellar brightness that is obscured in this pixel. Planet–moon eclipses are

---

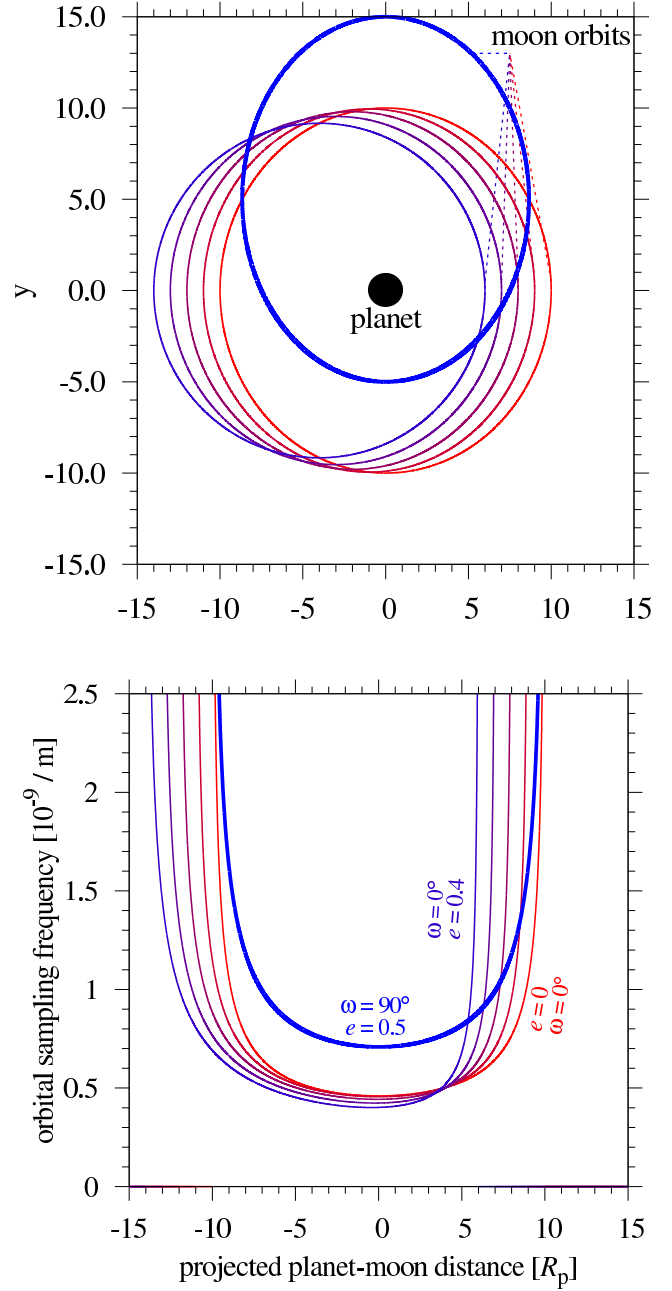

**Figure 3.** Different orbital eccentricities ($e$) in the planet–moon system (upper panel) cause different orbital sampling frequencies (lower panel). Six different values are shown: $e \in \{0, 0.1, 0.2, 0.3, 0.4, 0.5\}$.

automatically taken into account, because the planet is simulated as a black circle. Consequently, $I_\star^{\rm px} = 0$ inside the planetary radius and $\mathcal{P}_{\rm s}^{\rm px}(x)$ cannot contribute to the photometric OSE in the planetary shadow. At any given observation time $t$, the sum

$$F_{\rm OSE}^1(t) = \left(\frac{R_{\rm s}}{R_\star}\right)^2 \sum_{\rm px} dx \times I_\star^{\rm px} \times P_{\rm s}^{\rm px}(x) \qquad (11)$$

over all occulted pixels within the stellar radius ($R_\star$) gives us the relative loss in stellar brightness $B_{\rm OSE}^1(t) = 1 - F_{\rm OSE}^1(t)$ due to the photometric OSE of the first satel-lite. In the more general case of $n$ satellites,

$$F_{\mathrm{OSE}}^n(t) = \sum_{\mathrm{s}}^n \left[ \left( \frac{R_{\mathrm{s}}}{R_\star} \right)^2 \sum_{\mathrm{px}} dx \times I_\star^{\mathrm{px}} \times P_{\mathrm{s}}^{\mathrm{px}}(x) \right] \quad (12)$$

and $B_{\mathrm{OSE}}^n(t) = 1 - F_{\mathrm{OSE}}^n(t)$. We apply the nonlinear limb darkening law of Claret (2000, Equation (7) therein) and use stellar limb darkening coefficients (LDCs) listed in Claret & Bloemen (2011), which depend on stellar effective temperature ($T_{\star,\mathrm{eff}}$), metallicity ([Fe/H]), and surface gravity ($\log(g)$).

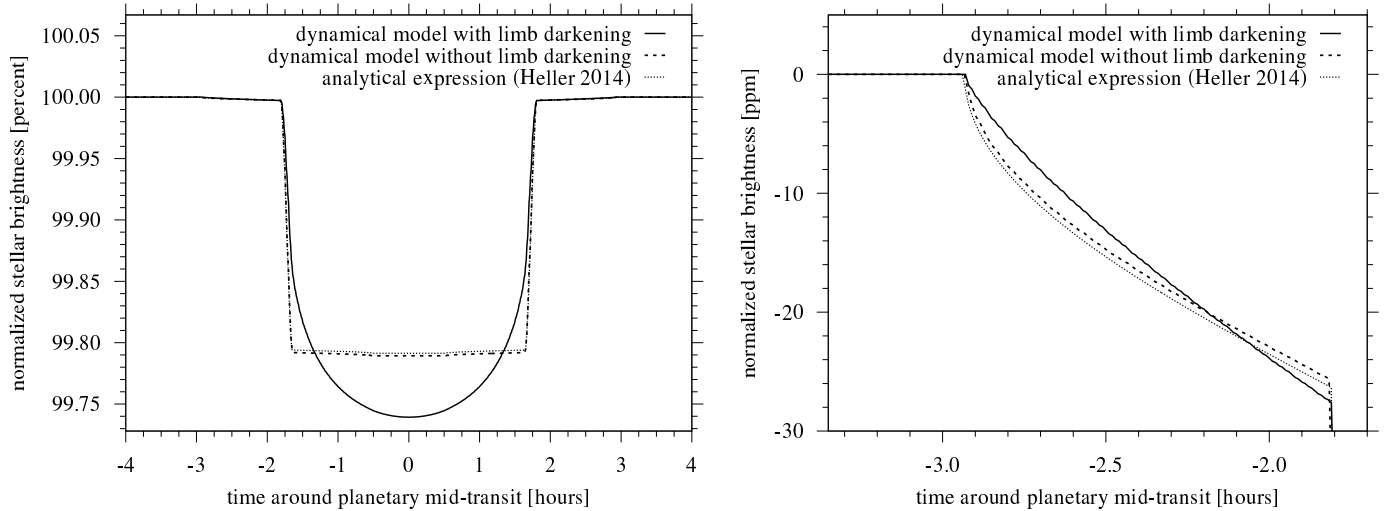

**Figure 4.** Comparison of the dynamical photometric OSE model with limb darkening (solid line), the dynamic photometric OSE model without limb darkening (dashed line), and the analytical expression for a non-limb-darkened star as per Eq. (6) in Heller (2014). *Left:* The OSE is hardly visible as a the slight brightness decrease in the wings of the planetary transit. *Right:* A zoom into the ingress of the moon. The dynamic model without limb darkening reproduces the analytical prediction very well. However, only the dynamic limb darkening model will be useful for fitting real observations. This hypothetical star–planet system is similar to the KOI 255.01 system, except that we here assume $\mathfrak{b} = 0$. The toy moon is as large as Ganymede (0.41 $R_\oplus$) and the planet–moon orbit is 15.47 $R_\mathrm{p}$ wide, equivalent to Ganymede's orbit around Jupiter.

As an example, Figure 4 shows the simulated transit of KOI 255.01 together with the OSE of a hypothetical Ganymede-sized moon at $a = 15.47\,R_\mathrm{p}$, equivalent to Ganymede around Jupiter. The 2.51 $R_\oplus$ super-Earth ($R_\oplus$ being the Earth's radius) is an interesting object as it transits a 0.53 $M_\odot$-mass M dwarf with a radius of $0.51 \pm 0.06\,R_\odot$ every $27.52197994 \pm 3.295 \times 10^{-5}$ d.[3] Hence, the photometric OSE of even a Ganymede-sized moon could be significant. In our simulations, the transit impact parameter is set to $\mathfrak{b} = 0.0$ to ease comparison with the analytic model, although it is really given as $\mathfrak{b} = 0.1244\,(+0.2132, -0.1243)$ in the Exoplanet Archive. The LDCs are $a_1 = 0.4354$, $a_2 = 0.2910$, and $a_3 = 0 = a_4$. In both panels of Figure 4, the solid line refers to our dynamical OSE model with limb darkening, the dashed line shows the dynamical model with the limb darkening option switched off, and the dotted line shows the analytical OSE model by Heller (2014), which also neglects limb darkening.

The left panel shows that the analytical solution is much less accurate than the dynamical OSE model inside the planetary transit trough because it neglects stellar limb darkening. In the wings of the transit curve, however, the analytical solution without stellar limb dark-ening and the dynamic OSE model with limb darkening differ by $< 3 \times 10^{-6}$ compared to a maximum transit depth of about $3 \times 10^{-5}$ just before the planetary ingress (right panel). The analytic model might thus offer sufficient accuracy for an initial OSE survey of a large data set. With such an approach, a first and preliminary search for OSE candidates within thousands of phase-folded light curves would be a matter of minutes. Even more encouraging, the dynamic OSE model with limb darkening almost resembles a straight line, at least in this configuration where the moon's semimajor axis is roughly as wide as the stellar diameter. A straight line fit would, of course, simplify an initial OSE search even further, as it would mostly depend on the moon's radius squared (in terms of maximum depth) and on the moon's semimajor axis (in terms of duration).

The right panel of Figure 4 also reveals that the dynamical OSE model causes a slightly smaller depth in the light curve for about two-thirds into the OSE ingress. This effect is caused by stellar limb darkening and the stellar brightness in the occulted regions being lower than the average brightness on the disk. In the final third of the OSE ingress, the dynamical OSE model then yields a deeper absorption because of the increasing stellar brightness towards the disk center. This division into two-thirds, in which the OSE model without stellar limb darkening is deeper than the one with limb darkening, and the one-third where things are reversed, is not a universal feature. Toward the stellar limb, e.g., the model without stellar limb darkening produces a deeper OSE signal during the entire transit.

### 2.2.2. Comparison with Kepler Data

We now apply our model to observations. This investigation is not an in-depth search for moons around a test planet, but it shall serve as an illustration of the OSE by reference to actual observed data. Our survey for confirmed, super-Neptune-sized *Kepler* planets with orbital periods $> 10$ d (to ensure Hill stability of an

---

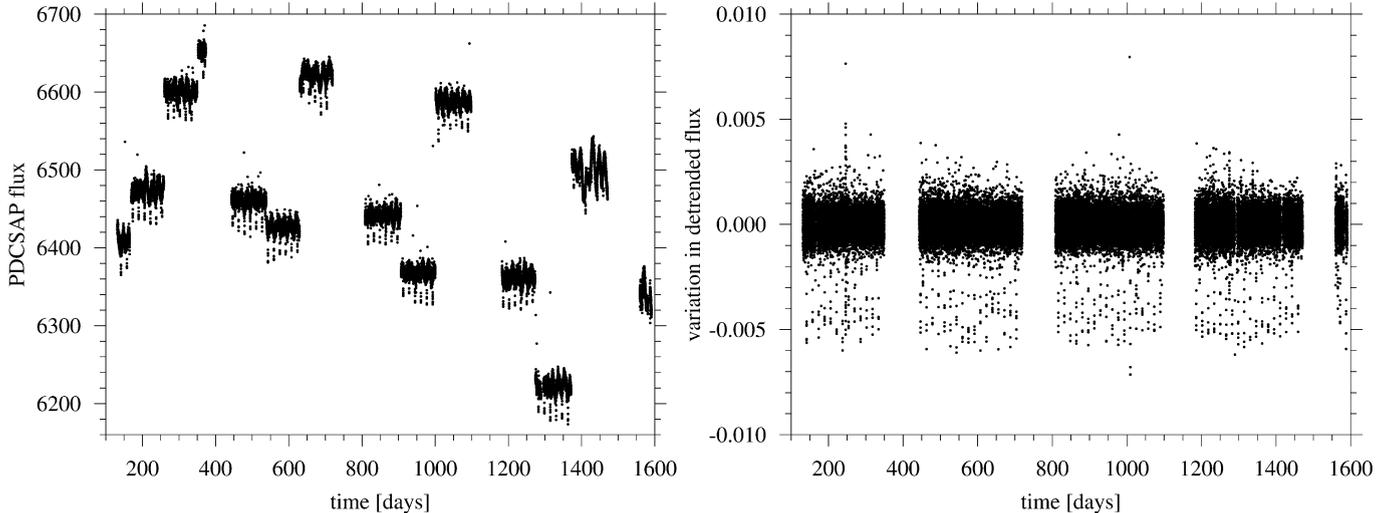

**Figure 5.** *Left:* Raw PDCSAP_FLUX for Kepler-229 c – individual quarters come offset from one another, and the approximately 90 transits are visible as large dips. *Right:* Detrended data.

moons) around a $\gtrsim 0.7\,R_\odot$ star (allowing detection of exomoons akin to the largest solar system moons) revealed Kepler-229 c (KOI 757.01) as a promising object. It is a $4.8\,R_\oplus$ planet transiting a $0.7\,R_\odot$ star every $16.96862$ d with an impact parameter $\mathfrak{b} = 0.25$ at a distance of about $0.117$ AU (Rowe et al. 2014). The stellar mass can then be estimated via Kepler's third law as $0.74\,M_\odot$, and the planetary Hill radius ($R_\mathrm{H}$) is about $78\,R_\oplus \approx 16\,R_\mathrm{p}$, assuming that Kepler-229 c's mass is similar to that of Neptune. Stellar LDCs are interpolated from Claret & Bloemen (2011) tables using the stellar effective temperature and surface gravity of Kepler-229 (Rowe et al. 2014), yielding $a_1 = 0.618423$, $a_2 = -0.522542$, $a_3 = 1.24769$, and $a_4 = -0.536854$.

We first retrieved as many quarters as were available among Q0-Q17 of the long-cadence[4] (30 minutes) publicly available *Kepler* data for Kepler-229 c.[5] We analyzed the PDCSAP_FLUX data, from which the *Kepler* mission has attempted to remove instrumental variability. Nevertheless, these data still exhibit significant variability unrelated to transits, as seen in Figure 5 (left panel). The creation of the PDCSAP fluxes by the *Kepler* mission involves the removal of common mode variability from the light curves attributable to instrumental effects, which can distort real astrophysical (such as stellar) variability but primarily at medium to long timescales. Since we consider relatively short-period planets, these possible distortions are unlikely to affect our analysis.

To condition each quarter's observations, we subtracted the quarter's mean value from all data points and then normalized by that mean. To these mean-subtracted, mean-normalized data, we applied a mean boxcar filter with a width equal to twice the transit duration plus 10 hr. This window is chosen to maximally remove non-transit variations while preserving the shape of the transit and OSE signals. The right panel of Figure 5 shows the resulting detrended data. Finally, we stitched together all quarters and masked out $10\sigma$ outliers.[6] Figure 6 (right panel) shows the final result for Kepler-229 c, after we folded the detrended data on a period of $16.968618$ d. Gray dots present the detrended, phase-folded data, and black dots with error bars show the binned data. The solid black line shows the direct output of our dynamical OSE simulator but for a planet without a moon, and the dashed red line shows the transit including the photometric OSE of an injected moon. A red cross on that curve at about 0.5 hr highlights the orbital configuration that is depicted in the left panel.

The left panel of Figure 6 illustrates our dynamical OSE model at work for Kepler-229 c. The planet along with the probability distribution of one hypothetical exomoon can be seen in transit. The injected moon has a radius of $0.7\,R_\oplus$, the vertical width of the moons $\mathcal{P}_\mathrm{s}(x)$ is to scale to both the planetary and the stellar radius. The moon orbit is set to $8\,R_\mathrm{p}$, which is $R_\mathrm{H}/2$ and therefore at the boundary of Hill stability for prograde moons (Domingos et al. 2006).

The inset in the right panel zooms into the transit light curve just about an hour prior to planetary ingress. In addition to the black solid (no moon) and red dashed lines ($0.7\,R_\oplus$-sized moon), we also show a blue dotted line indicating the OSE of a hypothetical Earth-sized moon. Note that the width of individual error bars of the binned data is $\approx 10^{-4}$, which corresponds to the depth of an Earth-sized moon's photometric OSE. With about 5 of such binned data points (or about 500 unbinned data points during the OSE ingress), a search for the exomoon-induced photometric OSE around this planet could yield statistical constraints on the presence of moons the size of Earth and smaller around this particular exoplanet. A more elaborate statistical analysis of exomoon effects in transit light curves is deferred to a future study (R. Heller et al. 2016, in preparation).

## 3. NUMERICAL SIMULATIONS OF THE PHOTOMETRIC OSE

---

[4] The OSE itself is an averaging effect, as it appears in the phase-folded light curve and only after several transits. Hence, for the OSE curve it does not make a difference if the data is taken in short cadence and then binned into 30 minute intervals or if it is taken in 30 minute intervals in the first place.

[5] http://archive.stsci.edu/kepler/data_search/search.php

[6] We define $\sigma$ to be the standard deviation estimated as $1.4826 \times$ the median absolute deviation (Bevington & Robinson 2003).

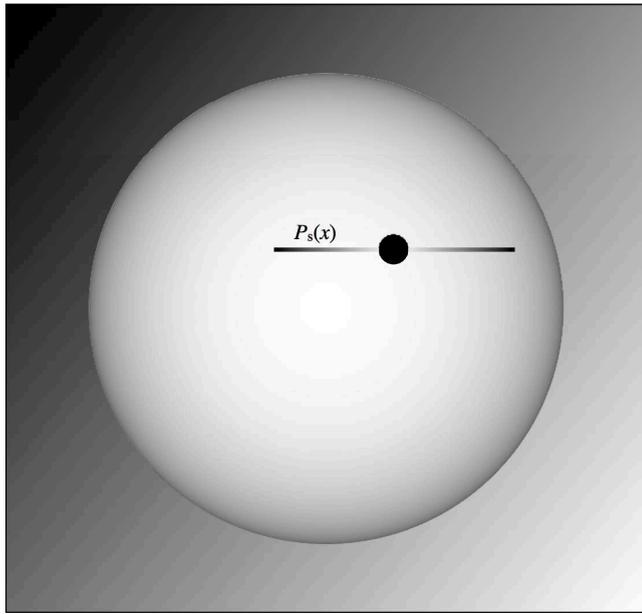

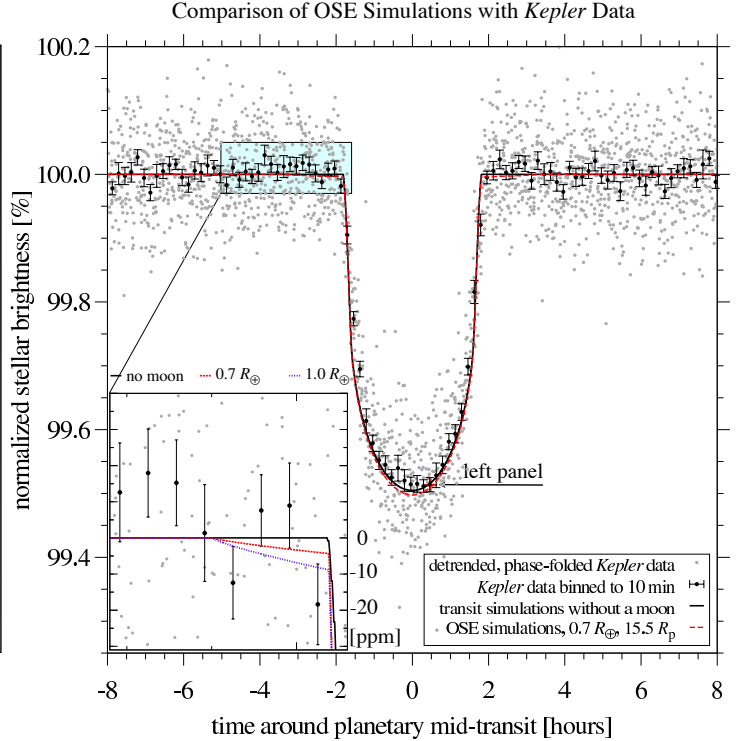

**Figure 6.** Visualization of our dynamical photometric OSE transit model. This example shows the transit of Kepler-229 c, a 4.8 $R_\oplus$ planet orbiting a 0.7 $R_\odot$ star every $\approx 17$ d with a transit impact parameter $b = 0.25$. *Left:* The planet (black circle) and the probability function (shaded horizontal strip) of a hypothetical 0.7 $R_\oplus$ moon with an orbital semimajor axis of 8 $R_p$ transit the limb-darkened star (large bright circle). *Right:* The red cross on the OSE curve at 0.5 hr refers to the moment shown in the left panel. The inset zooms into the wings of the transit ingress. Three models are shown: no moon (black solid), a 0.7 $R_\oplus$ moon (red dashed), and a 1 $R_\oplus$ moon (blue dotted), all moons with a semimajor axis of 8 $R_p$ around the planet.

To simulate the OSE in more complex configurations with inclined moon orbits, for which an analytical solution is not available, we wrote a numerical OSE simulator in `python`, which we call "PyOSE". The code and examples are publicly available[7] under the MIT license.[8] All of the following figures were generated with this code.

### 3.1. *Parameterization in* PyOSE

In PyOSE, the moon's orbital ellipse is defined by its eccentricity ($e$), circumplanetary semimajor axis ($a$), its orbital inclination with respect to the circumstellar orbit ($i_s$), the longitude of the ascending node ($\Omega$), and the argument of the periapsis ($\omega$). Figure 7 (top) shows a hypothetical 0.7 $R_\oplus$ exomoon around Kepler-229 c on a circular, inclined orbit ($i_s = 83°$, $\Omega = 30°$, $a = 8\,R_p$) at the time of the planetary mid-transit. In our numerical implementation, the planet–moon ensemble transits the star from left to right, which is an arbitrary choice. The motion of the planet and the moon around their common barycenter during transit can be simulated with PyOSE.

The center panel in Figure 7 shows a river plot (Carter et al. 2012) representation of 250 of these transit light curves, with one shown in each row. Each of these light curves corresponds to a different fixed position of the moon during the stellar transit. The orbital phase of the moon (along the ordinate in Figure 7) corresponds to the mean anomaly. Two horizontal gray regions at phases $\approx 0.1$ and $\approx 0.6$ show partial planet–moon eclipses. The black arrow at phase $\approx 0.2$ indicates the moon's position chosen in the top panel. For reference, the two vertical dashed lines illustrate the time of the planetary transit, which is omitted in our OSE simulations for clarity.

The bottom panel of Figure 7 shows a sum of the individual moon transits from the river plot, normalized by the number of transits. This procedure corresponds to a phase-folding of a hypothetical observed OSE.

Although PyOSE can simulate transits of planets with moons, we will focus on the moon's OSE signature in the following and study a range of moon transits for various orbital parameters. The transit model is the one presented by Mandel & Agol (2002). For each OSE curve, we chose to sample at least 100 moon transits equally spaced in time (not necessarily in space for $e \neq 0$; see Figure 2) to achieve convergence (Heller 2014). The limb-darkened stellar disk is represented by a numerical grid of $1\,000 \times 1\,000$ floating point values, and we calculate the total stellar brightness during transit for $\gtrsim 1\,000$ flux data points between first and last contact of the planetary silhouette with the stellar disk. With the moon having a radius of typically 1/100 the stellar radius in the shown simulations, or about 10 px, its transiting silhouette is represented by about 100 black pixels. The error of this approximation compared to a genuine black circle is $< 1\,\%$. PyOSE allows for arbitrarily large pixel grids at the cost of CPU time, which is proportional to the total number of pixels in the grid or to the square of $R_s$ (in units of pixels).

### 3.2. *Mutual Planet–Moon Eclipses*

We treat both the planet and the moon as black disks. Thus, it is irrelevant whether the satellite eclipses behind or in front of the planet.[9] For each of the simulated planet–moon transit configurations, PyOSE compares the distance between the planet and moon disk centers ($d_{\mathrm{ps}}$) to the sum of their radii. If $d_{\mathrm{ps}} < R_{\mathrm{p}} + R_{\mathrm{s}}$, a mutual eclipse occurs and PyOSE calculates the area ($A$) of the asymmetric lens defined by the intersection of the two circles as

$$
\begin{aligned}
A = {} & R_{\mathrm{s}}^2 \arccos\left(\frac{d_{\mathrm{ps}}^2 + R_{\mathrm{s}}^2 - R_{\mathrm{p}}^2}{2 d_{\mathrm{ps}} R_{\mathrm{s}}}\right) \\
& + R_{\mathrm{p}}^2 \arccos\left(\frac{d_{\mathrm{ps}}^2 + R_{\mathrm{p}}^2 - R_{\mathrm{s}}^2}{2 d_{\mathrm{ps}} R_{\mathrm{p}}}\right) \\
& - \left(\frac{1}{2}\sqrt{(-d_{\mathrm{ps}} + R_{\mathrm{s}} + R_{\mathrm{p}})(d_{\mathrm{ps}} + R_{\mathrm{s}} - R_{\mathrm{p}})}\right. \\
& \left. \times \sqrt{(d - R_{\mathrm{s}} + R_{\mathrm{p}})(d_{\mathrm{ps}} + R_{\mathrm{s}} + R_{\mathrm{p}})}\right)
\end{aligned}
\tag{13}
$$

This area does not contribute to the stellar blocking by the moon, as it is covered by the silhouette of the transiting planet.

### 3.3. Parameter Study

In the following, we study variations of the OSE signal due to variations in the parameterization of the star–planet–moon orbital and physical configuration. We use Kepler-229 c as a reference case and modify one parameter at a time, as specified below.

#### 3.3.1. The Moon's Semimajor Axis (a)

For circular moon orbits, changes in $a$ modify the shape of the OSE, while the area under the integral remains unaffected (see Equation 9). For $e \neq 0$ or $i_{\mathrm{s}} \neq 0$, however, both the shape and the integral will change. The left panel of Figure 8 visualizes this effect for two cases; one in which the moon is barely stable from a dynamical point of view ($a = 0.5\,R_{\mathrm{H}}$) and one for a close-in moon near the Roche limit ($a \approx 0.128\,R_{\mathrm{H}} \approx 2\,R_{\mathrm{p}}$). The duration of the OSE signal is longer for moons with larger semimajor axes. Planet–moon eclipses are visualized by bumps in the OSE curve of the moon at $a = 0.128\,R_{\mathrm{H}}$, near $\pm 0.07$ days. The moon in the wider orbit is not subject to planet–moon eclipses.

The right panel of Figure 8 shows the integral under the OSE curve as a function of $a$. This is an important quantity as it serves as a measure for the significance of the moon-induced OSE imprint on the light curve. The decline of the OSE integral up to $a \approx 0.6\,R_{\mathrm{H}}$ is mostly due to stellar limb darkening: the larger $a$, the more flux will be blocked closer to the stellar center for this specific orbital configuration. Beyond $0.6\,R_{\mathrm{H}}$, moon transits will occasionally be missed during planetary transits and the OSE signal decreases. The dashed part of the curve, corresponding to moons beyond 0.5 $R_H$, is physically implausible for prograde moons, and valid only for prograde moons (Domingos et al. 2006).

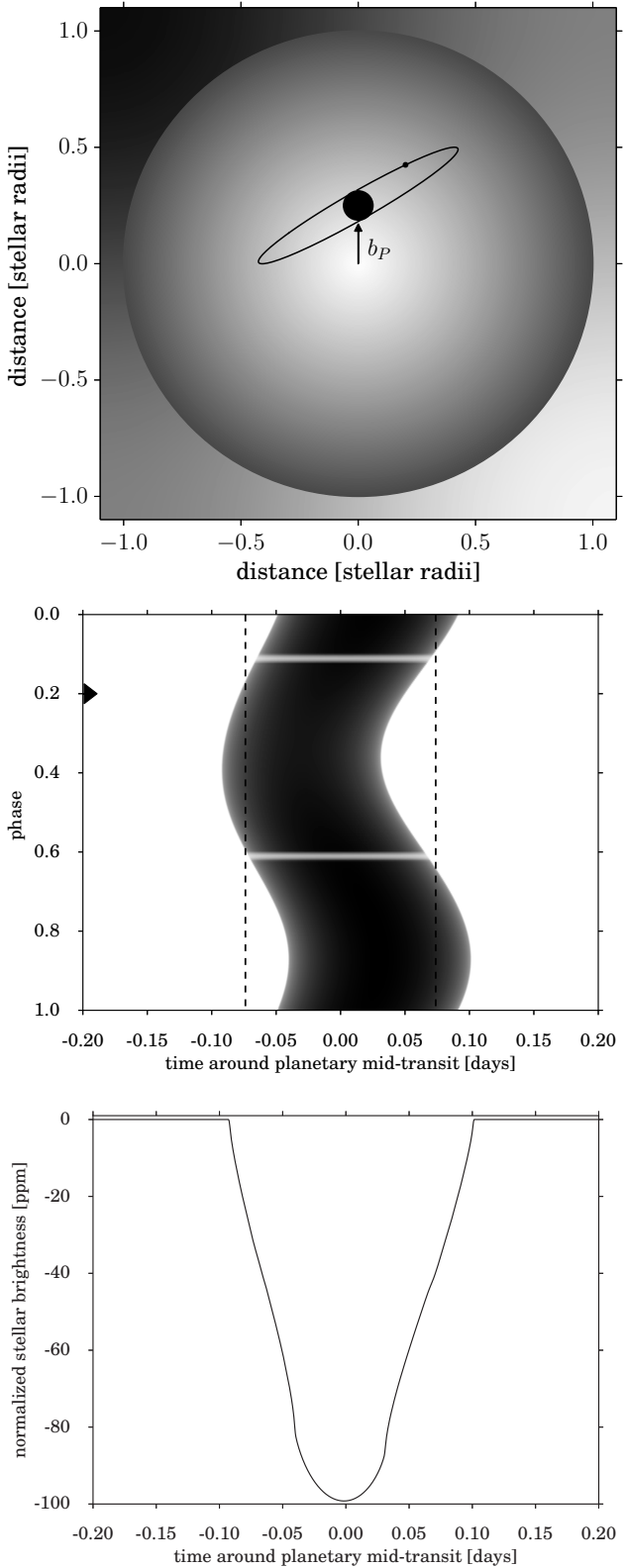

**Figure 7.** *Top:* Sky-projected view generated with PyOSE of both Kepler-229 c and a hypothetical exomoon in transit. The black arrow at phase 0.2 shows the moon's position chosen in the top panel. *Center:* River plot of the moon transit only. The black arrow at phase 0.2 shows the moon's position chosen in the top panel. *Bottom:* Average (phase-folded) transit light curve of the system after an arbitrarily large number of transits, where the moon orbit has been equally sampled in time.

---

[9] This fact makes the OSE insensitive to the satellite's sense of orbital motion around the planet (Heller & Albrecht 2014; Lewis & Fujii 2014).

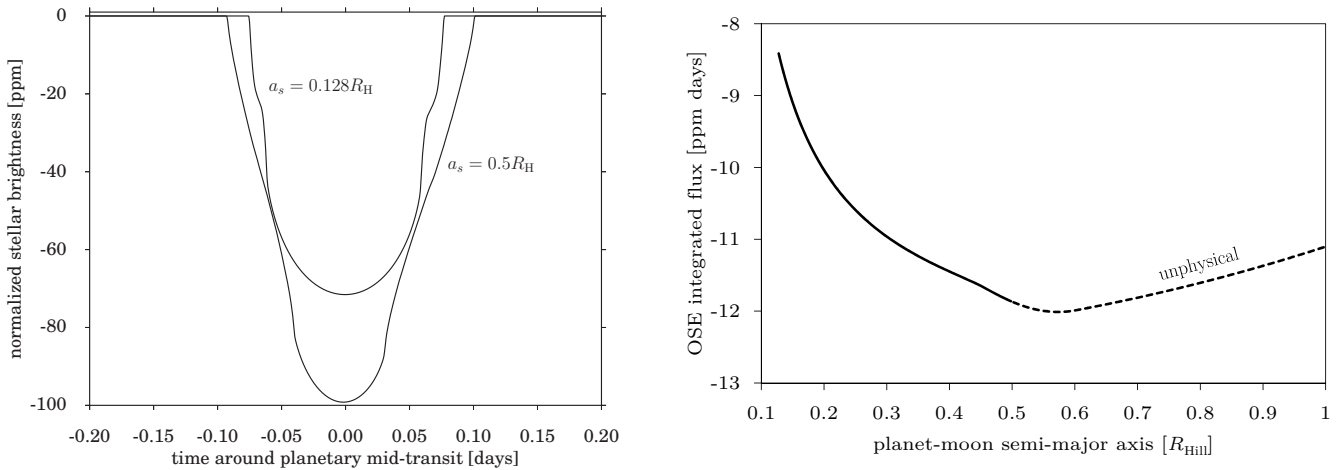

**Figure 8.** Variation of the OSE for different semimajor axes of a hypothetical exomoon around Kepler-229 c. *Left:* Photometric OSE for two cases where the satellite is at 0.128 and 0.5 $R_H$ around the planet. *Right:* Integral under the OSE curve as a function of the planet–moon orbital semimajor axis. The solid line shows the limiting cases at 0.128 and 0.5 $R_H$, corresponding to the two scenarios shown in the left panel. The dashed line represents moon orbits that are unphysically wide for prograde moons.

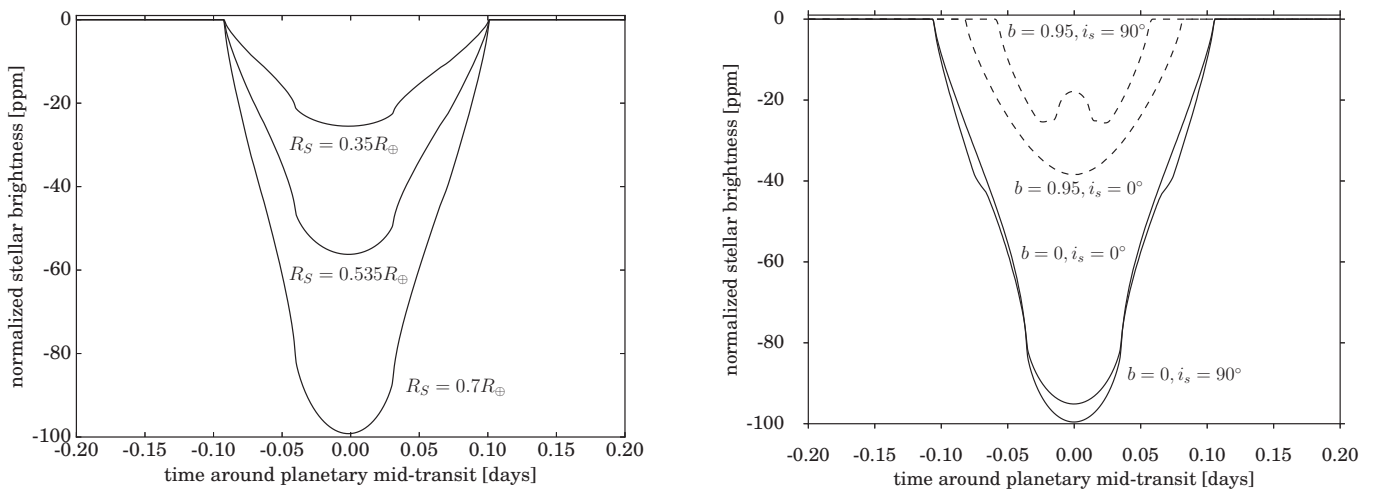

**Figure 9.** Variation of the OSE for different satellite radii. The inset zooms into the ingress of the phase-folded satellite transit, showing that the time of first contact between the satellite's silhouette and the stellar disk depends on the satellite radius.

**Figure 10.** Variation of the OSE for different planetary transit impact parameters and different inclinations of the satellite orbit. Solid lines refer to $b = 0$ and dashed lines relate to $b = 0.95$. For both cases, we show $i_s = 0°$ and 90°.

### 3.3.2. *The Moon's Radius ($R_s$)*

Larger moons naturally cause deeper transits. Keeping everything else fixed, variations in $R_s$ cause variations in the OSE amplitude roughly proportional to $R_s^2$, as illustrated in Figure 9 (see also the term $(R_s/R_\star)^2$ in Equation 11). Comparing the upper and lower curves, we see a signal increase by a factor of four (-25 ppm vs. -100 ppm) for a change in $R_s$ by a factor of two (from 0.35 to 0.7 $R_\oplus$). Changes in transit duration occur due to the different timings of the first and last contact of the planetary silhouette with the stellar disk (see inset in Figure 9).

### 3.3.3. *The Planetary Impact Parameter ($b$) and the Inclination of the Moon's Orbit ($i_s$)*

The planetary impact parameter and the inclination of the satellite orbit determine the fraction of planetary transits without moon transits, that is, planetary transits with the moon passing beyond the stellar disk. In Figure 10, we show four different scenarios of a hypothetical moon around Kepler-229 c. Solid lines refer to $b = 0$, dashed lines to $b = 0.95$, and we examine inclinations $i_s = 0°$ (face-on view) and $i_s = 90°$ (edge-on view). In the $b = 0.95$, $i_s = 90°$ case (upper dashed line), a bump around planetary mid-transit gives evidence of planet–moon eclipses. In the $b = 0.95$, $i_s = 0°$ and $b = 0$, $i_s = 0°$ cases, planet–moon eclipses do not occur. In the $b = 0$, $i_s = 90°$ case (lower solid line), the bump from planet–moon eclipses is very broad and deformed into two minor bumps at $\pm 0.07$ d in the moon's OSE.

### 3.4. *Multiple Exomoons*

Multiple moons are common in our solar system. Although photodynamical modeling (Kipping 2011) can tackle multi-satellite systems in principle (Kipping et al.

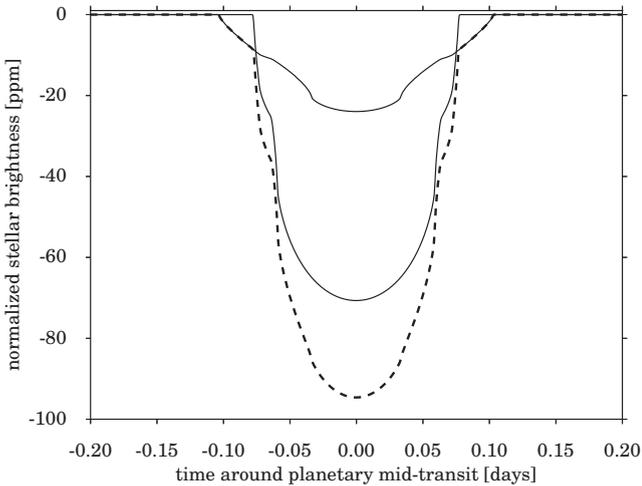

**Figure 11.** OSE of a multi-moon system. The two solid lines show the OSE of one $0.35\,R_\oplus$ moon at $0.5\,R_H$ and a second $0.7\,R_\oplus$ moon at $0.128\,R_H$. The dashed line shows the combined OSE.

2015), this is extremely time-consuming and a practicable approach still needs to be demonstrated. Following Kipping et al. (2014, Sect. 2.3 therein), the HEK team indeed restricts their exomoon search in the *Kepler* data to single-moon systems. Our Equation (12) is an analytical description of the additive OSE in multi-moon systems and describes how the dynamical OSE model handles multi-satellite systems. Our purely numerical simulator PyOSE can handle multi-satellite systems as well.

That being said, with real observational data, distinguishing between an OSE signal of a single large moon and a signal from several smaller moons may prove difficult, if not impossible. This is demonstrated in Figure 11, where we show the individual OSE signals from two moons (solid lines) and their combined OSE signature (dashed line). This dashed curve demonstrates that the photometric OSE is additive as long as moon–moon occultations can be neglected. [10] The dashed curve is also what would be observed in reality, noise effects aside. With noise taking into account, there would be substantial degeneracies among the model parameters and the number of moons.

## 4. DISCUSSION

In addition to the orbital and physical parameters treated above, there are several minor effects on the OSE, the most important being stellar limb darkening. We ran a series of simulations using different LDCs for FGK main-sequence stars. For a typical uncertainty in stellar temperature $\approx 100$K, the effect on the OSE signal is of the order of 1 ppm, that is, negligible for the vast majority of observable cases. Nevertheless, if LDCs cannot be constrained otherwise, there could be degenerate solutions between an exomoon-induced OSE and a sole planetary transit with a different LDC parameterization. This is, however, only an issue during planetary transit. Alternatively, one can search for OSE-like flux decreases before and after the planetary transit (Hippke 2015), which cannot possibly be made up for with different LDCs.

Some aspects of our OSE model might hardly be accessible with near-future technology. For moon orbits that are co-planar with the circumstellar orbit ($i_s = 90°$), moderate moon eccentricities will not cause an OSE signal much different from a circular moon orbit. Nevertheless, there are configurations in which these parameters make all the difference in determining the presence of an exomoon (see Fig. 10). Moreover, the discovery of the formerly unpredicted hot Jupiter population, the unsuspected dominance of super-Earth-sized exoplanets in orbits as short as that of Mercury, and the puzzling abundance of close-in planets with highly misaligned orbits suggests that the solar system does not present a reliable reference for extrasolar planetary systems. Hence, any model used to explore yet undiscovered exomoons will need to be able to search a large parameter space beyond the margins suggested by the solar system.

Both numerical simulations and observations will always be undersampled and only converge to the analytical solution. In terms of observations, this is because of the limited number of observed transits, usually < 100 for an observational campaign over a few years, and because of telescope downtime and observational windows. Beyond that, the orbital periods of both the planet–moon barycenter (around the star) and the moon(s) (around the planet–moon barycenter) *determine* a non-randomized sampling of the moon orbit in successive transits. If an exoplanet's orbital period were – by whatsoever reason – an integer multiple of its moon's period, then the moon were to appear at the same position relative to the planet in successive transits. The resulting phase-folded light curve would not display the OSE and therefore not converge to our solutions, but it would show two transits: one caused by the planet and one caused by the moon.

PyOSE can simulate large numbers of OSE curves for arbitrary star–planet–multi-moon configurations to test real observations. Beyond the functionalities demonstrated in this paper, PyOSE can add noise (parameterized or injected real noise).

Both our dynamical OSE model (Sect. 2.2.1) and our numerical OSE simulator (PyOSE, Sect. 3) are computationally inexpensive and easy to implement in computer code, which is crucial for the independent verification or rejection of possible exomoon signals. More advanced methods may suffer from a large parameter space to be explored, resulting in a huge number of simulations ($10^{11}$; Kipping et al. 2013). These could be difficult to verify. For an independent verification of an exomoon search, either the original code needs to be released for review, or an independent implementation is required. To exclude processing, runtime, and hardware errors (bit error rates are typically $10^{-14}$), a search based on TBs of data ultimately needs to be repeated on different hardware. The average computational burden for this is large, with an average of roughly 33,000 CPU hours required per candidate using photodynamical modeling. The monetary equivalent, e.g. using Amazon's EC2 on-demand facilities[11], is about $50,000 U.S. dollars (2015 December prices) for a single candidate check.

The dynamical OSE simulation in Figure 4 contains 48 data points and was computed within 10 to 14s on two modern computers.[12] Hence, this setup can generate a grid of about $10^4$ such OSE light curves per day. Our OSE model involves 11 independent parameters: $M_\star$, $R_\star$, the planet's orbital period around the star ($P_{\star p}$), $R_p$, $\mathfrak{b}$, $a_1$, $a_2$, $a_3$, $a_4$, $R_\star$, and $a$. For a well parameterized star–planet system, $M_\star$, $R_\star$, $P_{\star p}$, and $\mathfrak{b}$ can be observed and fit without considerations of any potential moon, assuming that moon-induced variations of the transit impact parameter (Kipping 2009) are negligible. If one were to carry out a search for the photometric OSE in the *Kepler* data, a limb darkening law with two LDCs should still do a good job in a first broad survey, leaving us with $R_p$, $R_\star$, and $a$ plus the two LDCs to be fit per phase-folded light curve. Testing 10 values per parameter would then imply a grid of $10^5$ OSE light curves per planet or planet candidate. With $10^4$ LCs simulated per day, one *Kepler* planet or planet candidate could be checked for an OSE signature within a week, given a standard desktop computer. If dedicated high-speed computational resources could be used, all *Kepler* planets and candidates (about 4000 as of today) could be checked for a photometric OSE within maybe a month. A simple straight line fit of the OSE, as suggested above, would dramatically reduce this time frame to much less than one day. A detailed statistical analysis, e.g. within a Bayesian framework (Kipping et al. 2012) and using an injection-retrieval method (Hippke 2015), could then be used to infer the significance of the best-fit model for each object.

---

[10] PyOSE currently neglects moon–moon occultations. They would only occur in < 1 % of the transits, depending on the exact geometry of the circumstellar and circumplanetary orbits.

## 5. CONCLUSION

We present a new formula to describe the orbital sampling frequency of a moon on an eccentric orbit around a planet, that is, the probability of a moon residing at a specific sky-projected distance from the planet (Equation 10). This formula assumes co-planar circumstellar and circumplanetary orbits. We implemented it in a dynamical OSE simulator with stellar limb darkening that can be applied to arbitrary transit impact parameters. In contrast to a previously derived framework, in which stellar limb darkening was neglected and $\mathfrak{b}$ was required to be zero (Heller 2014), our new dynamical OSE simulator can now be applied to observations (Figure 6).

Using an independent numerical OSE simulator dubbed PyOSE, we examined the moon's part of the OSE parameter space, spanned by its orbital semimajor axis, its physical radius, the inclination of its orbit with respect to the line of sight, the orbit's longitude of the ascending node, and the transit impact parameter of the planet (and therefore of the moon).

The OSE might give evidence of a multi-moon configuration, but the precise characterization of multi-satellite system will be extremely challenging using OSE only. Therefore, the OSE method could be used for preliminary analyses of a large number of systems, while more costly methods (Kipping et al. 2012; Heller & Albrecht 2014; Agol et al. 2015) could be used to focus on the most promising subset of targets. Beyond that, the OSE method can generally be used as an independent means to verify an exomoon claim via planetary TTV and TDV.

We thank the referee for diligent reports. René Heller has been supported by the Origins Institute at McMaster University, by the Canadian Astrobiology Program (a Collaborative Research and Training Experience Program funded by the Natural Sciences and Engineering Research Council of Canada), by the Institute for Astrophysics Göttingen, and by a Fellowship of the German Academic Exchange Service (DAAD). This work made use of NASA's ADS Bibliographic Services. Computations were performed with `ipython 0.13` on `python 2.7.2` (Pérez & Granger 2007), and figures were prepared with `gnuplot 4.6` (www.gnuplot.info).

## 6.3 How to Determine an Exomoon's Sense of Orbital Motion (Heller & Albrecht 2014)

Contribution:

RH did the literature research, worked out the mathematical framework, translated the math into computer code, created Figs. 1-4, led the writing of the manuscript, and served as a corresponding author for the journal editor and the referees.

# HOW TO DETERMINE AN EXOMOON'S SENSE OF ORBITAL MOTION

René Heller[1,2]

Origins Institute, McMaster University, Hamilton, ON L8S 4M1, Canada; rheller@physics.mcmaster.ca

AND

Simon Albrecht

Stellar Astrophysics Centre, Department of Physics and Astronomy, Aarhus University, Ny Munkegade 120, DK-8000 Aarhus C, Denmark; albrecht@phys.au.dk

Draft version September 26, 2014

## ABSTRACT

We present two methods to determine an exomoon's sense of orbital motion (SOM), one with respect to the planet's circumsolar orbit and one with respect to the planetary rotation. Our simulations show that the required measurements will be possible with the European Extremely Large Telescope (E-ELT). The first method relies on mutual planet-moon events during stellar transits. Eclipses with the moon passing behind (in front of) the planet will be late (early) with regard to the moon's mean orbital period due to the finite speed of light. This "transit timing dichotomy" (TTD) determines an exomoon's SOM with respect to the circumstellar motion. For the ten largest moons in the solar system, TTDs range between 2 and 12 s. The E-ELT will enable such measurements for Earth-sized moons around nearby stars. The second method measures distortions in the IR spectrum of the rotating giant planet when it is transited by its moon. This Rossiter-McLaughlin effect (RME) in the planetary spectrum reveals the angle between the planetary equator and the moon's circumplanetary orbital plane, and therefore unveils the moon's SOM with respect to the planet's rotation. A reasonably large moon transiting a directly imaged planet like $\beta$ Pic b causes an RME amplitude of almost $100\,\mathrm{m\,s^{-1}}$, about twice the stellar RME amplitude of the transiting exoplanet HD209458 b. Both new methods can be used to probe the origin of exomoons, that is, whether they are regular or irregular in nature.

Keywords: eclipses — methods: data analysis — methods: observational — planets and satellites: individual ($\beta$ Pic b) — techniques: photometric — techniques: radial velocities

## 1. CONTEXT

Although thousands of extrasolar planets and candidates have been found, some as small as the Earth's Moon (Barclay et al. 2013), no extrasolar moon has been detected. The first dedicated hunts for exomoons have now been initiated (Pont et al. 2007; Kipping et al. 2012; Szabó et al. 2013), and it has been shown that the Kepler or PLATO space telescopes may find large exomoons in the stellar light curves (Kipping et al. 2009; Heller 2014), if such moons exist.

The detection of exomoons would be precious from a planet formation perspective, as giant planet satellites carry information about the thermal and compositional properties in the early circumplanetary accretion disks (Canup & Ward 2006; Heller & Pudritz 2014). Moons can also constrain the system's collision history (see the Earth-Moon binary, Hartmann & Davis 1975) and bombardment record (see the misaligned Uranian system, Morbidelli et al. 2012), they can trace planet-planet encounters (see Triton's capture around Neptune, Agnor & Hamilton 2006), and even the migration history of close-in giant planets (Namouni 2010). Under suitable conditions, an exomoon observation could reveal the absolute masses and radii in a star-planet-moon system (Kipping 2010). What is more, moons may outnumber rocky planets in the stellar habitable zones (Heller & Barnes 2014) and therefore could be the most abundant species of habitable worlds (Williams et al. 1997; Heller et al. 2014).

A moon's sense of orbital motion (SOM) is crucial to determine its origin and orbital history. About a dozen techniques have been proposed to find an extrasolar moon (Heller 2014), but none of them can determine an exomoon's SOM with current technical equipment (Lewis & Fujii 2014). We here identify two means to determine an exomoon's SOM relative to the circumstellar orbit and with respect to the planet's direction of rotation. In our simulations, we use the European Extremely Large Telescope[3] (E-ELT) as an example for one of several ELTs now being built.

## 2. METHODS

### 2.1. An Exomoon's Transit Timing Dichotomy (TTD)

For our first new method to work, the moon needs to be large enough (and the star's photometric variability sufficiently low) to cause a direct transit signature in the stellar light curve (Sartoretti & Schneider 1999). Depending on the moon's orbital semi-major axis around the planet ($a_{\mathrm{ps}}$) and on the orbital alignment, some stellar transits of the planet-moon pair will then show mutual planet-moon eclipses. These events have been simulated (Cabrera & Schneider 2007; Sato & Asada 2009; Kipping 2011a; Pál 2012) and a planet-planet eclipse (Hirano et al. 2012) as well as mutual events in a stel-lar triple system (Carter et al. 2011) have already been found in the Kepler data.

[3] Construction of the E-ELT near the Paranal Observatory in Chile began in June 2014, with first light anticipated in 2024.

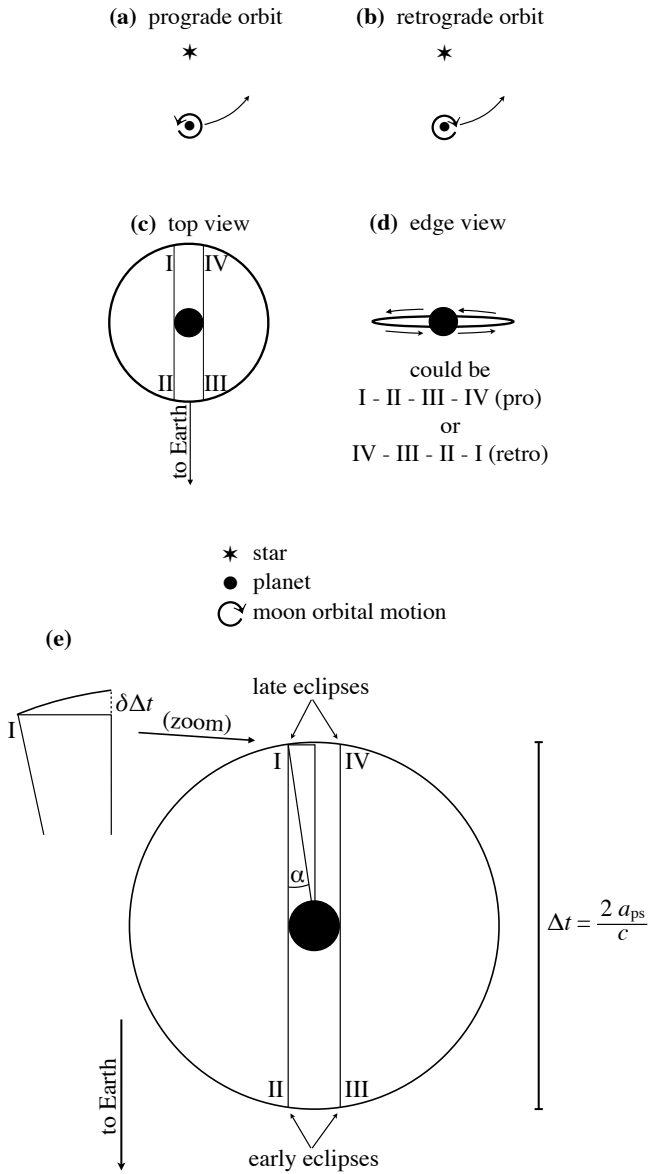

**Figure 1.** Orbital geometry of a star-planet-moon system in circular orbits. **(a)** Top view of the system's orbital motion. The moon's circumplanetary orbit is prograde with respect to the circumstellar orbit. **(b)** Similar to **(a)**, but now the moon is retrograde. **(c)** Top view of the moon's orbit around the planet. Roman numbers I to IV denote the ingress and egress of mutual planet-moon eclipses. **(d)** Edge view (as seen from Earth) of a circumplanetary moon orbit. **(e)** Transit timing dichotomy of mutual planet-moon events. Due to the finite speed of light, an Earth-bound observer witnesses events I and IV with a positive time delay $\Delta t$ compared to events II and III, respectively.

Figure 1 illustrates the difficulty in determining a moon's SOM. Panels (a) and (b) visualize the two possible scenarios of a prograde and a retrograde SOM with respect to the circumstellar motion. Panel (c) presents the four possible ingress and egress locations (arbitrarily labelled I, II, III, and IV) for a mutual planet-moon event during a stellar transit. Panel (d) shows the projection of the three-dimensional moon orbit on the two-dimensional celestial plane. If the moon transit is directly visible in the stellar light curve, then events I and II can be distin-guished from events III and IV, simply by determining whether the moon enters the stellar disk first and then performs a mutual event with the planet (III and IV) or the planet enters the stellar disk first before a mutual event (I and II). However, this inspection cannot discern event I from II or event III from IV. Therefore, prograde and retrograde orbits cannot be distinguished from each other.

Figure 1 (e) illustrates how the I/II and III/IV ambiguities can be solved. Due to the finite speed of light, there will be a time delay between events I and II, and between events III and IV. It will show up as a transit timing dichotomy (TTD) between mutual events where the moon moves in front of the planetary disk (early mutual events) or behind it (late mutual events). Events I and IV will be late by $\Delta t = 2a_{\mathrm{ps}}/c$ compared to events II and III, respectively. Imagine that two mutual events, either I and [II or III] or IV and [III or II], are observed during two different stellar transits and that the moon has completed $n$ circumplanetary orbits between the two mutual events. Then it is possible to determine the sequence of late and early eclipses, and therefore the SOM with respect to the circumstellar movement, if (1) the orbital period of the planetary satellite ($P_{\mathrm{ps}}$) can be determined independently with an accuracy $\delta P_{\mathrm{ps}} < P_{\mathrm{ps}}/n$, and if (2) the event mid-times can be measured with a precision $< \Delta t$. As a byproduct, measurements of $\Delta t$ yield an estimate for $a_{\mathrm{ps}} = \Delta t \times c$.

Concerning (1), the planet's transit timing variation (TTV) and transit duration variation (TDV) due to the moon combined may constrain the satellite mass ($M_{\mathrm{s}}$) and $P_{\mathrm{ps}}$ (Kipping 2011b). As an example, if two mutual events of a Jupiter-Ganymede system at 0.5 AU around a Sun-like star were observed after 10 stellar transits (or 3.5 yr), the moon ($P_{\mathrm{ps}} \approx 0.02$ yr) would have completed $n \approx 175$ circumplanetary orbits. Hence, $\delta P_{\mathrm{ps}} \lesssim 1$ hr would be required. A combination of TTV and TDV measurements might be able to deliver such an accuracy (Section 6.5.1 in Kipping 2011b).

For our estimates of $\Delta t$, we can safely approximate that a moon enters a mutual event at a radial distance $a_{\mathrm{ps}}$ to the planet, because

$$
\begin{aligned}
\delta \Delta t &= \frac{a_{\mathrm{ps}}}{c} \Big( 1 - \cos(\alpha) \Big) \\
&= \frac{a_{\mathrm{ps}}}{c} \left( 1 - \cos\left\{ \arcsin\left( \frac{R_{\mathrm{p}}}{a_{\mathrm{ps}}} \right) \right\} \right) \\
&\ll \Delta t \ ,
\end{aligned} \tag{1}
$$

with $c$ being the speed of light, $R_{\mathrm{p}}$ the planetary radius, and $\alpha$ defined by $\sin(\alpha) = R_{\mathrm{p}}/a_{\mathrm{ps}}$ as shown in Figure 1(e). For a moon at $15\,R_{\mathrm{p}}$ from its planet, such as Ganymede around Jupiter, $\alpha \approx 3.8°$ and $\delta \Delta t \approx 0.005$ s, which is completely negligible. Orbital eccentricities could also cause light travel times different from the one shown in Figure 1. But even for eccentricities comparable to Titan's value around Saturn, with 0.0288 the largest among the major moons in the solar system, TTDs would be affected by $< 5\,\%$, or fractions of a second. Significantly larger eccentricities are unlikely, as they will be tidally eroded within a million years (Porter & Grundy 2011; Heller & Barnes 2013).

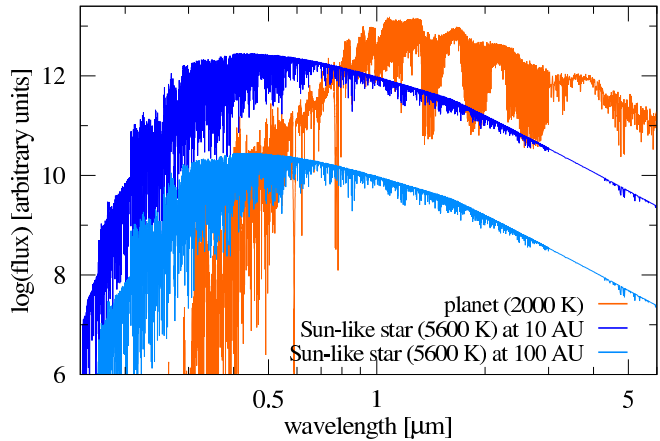

**Figure 2.** Theoretical emission spectra of a hot, Jupiter-sized planet (orange, upper) and a Sun-like star as reflected by the planet at 10 AU (blue, middle) and 100 AU (light blue, lower). For the planet, a Jupiter-like bond albedo of 0.3 is assumed.

### 2.2. *The Rossiter-McLaughlin effect (RME) in the Planetary Emission Spectrum*

In the solar system, all planets except Venus (Gold & Soter 1969) and Uranus rotate in the same direction as they orbit the Sun. One would expect that the orbital motion of a moon is aligned with the rotation of the planet. However, collisions, gravitational perturbations, capture scenarios etc. can substantially alter a satellite's orbital plane (Heller et al. 2014). Hence, knowledge about the spin-orbit misalignment, or obliquity, in a planet-exomoon system would be helpful in inferring its formation and evolution.

One such method to constrain obliquities is the Rossiter-McLaughlin effect (RME), a distortion in the rotationally broadened absorption lines caused by the partial occultation of the rotating sphere (typically a star) by a transiting body (usually another star or a planet). This distortion can either be measured directly (Albrecht et al. 2007; Collier Cameron et al. 2010) or be picked up as an RV shift during transit. The shape of this anomalous RV curve reveals the projection of the angle between the orbital normal of the occulting body (in our case the moon) and the rotation axis of the occulted body (here the planet). Originally observed in stellar binaries (Rossiter 1924; McLaughlin 1924), this technique experienced a renaissance in the age of extrasolar planets, with the first measurement taken by Queloz et al. (2000) for the transiting hot Jupiter HD209458 b. Numerous measurements[4] revealed planets in aligned, misaligned, and even retrograde orbits – strongly contrasting the architecture of the solar system (e.g. Albrecht et al. 2012). The *stellar* RME of exomoons has been studied before (Simon et al. 2010; Zhuang et al. 2012), but here we refer to the RME in the *planetary* infrared spectrum caused by the moon transiting a hot, young giant planet.

For this method to be effective, a giant planet's light needs to be measured directly. Starting with the planetary system around HR 8799 (Marois et al. 2008) and the planet candidate Fomalhaut b (Kalas et al. 2008), 18 giant exoplanets have now been directly imaged, most of which are hot ($> 1000$ K), and young ($< 100$ Myr). Upcoming instruments like SPHERE (Beuzit et al. 2006) and GPI (McBride et al. 2011) promise a rapid increase of this number. Stellar and planetary spectra can also be separated in velocity space without the need of spatial separation (Brogi et al. 2012; de Kok et al. 2013; Birkby et al. 2013), but for $\beta$ Pic-like systems, instruments like CRIRES can also separate stellar and planetary spectra spatially. Snellen et al. (2014) determined the rotation period of $\beta$ Pic b to be $8.1 \pm 1.0$ hours. The high rotation velocity ($\approx 25$ km s$^{-1}$) favors a large RME amplitude, but makes RV measurements more difficult.

Contamination of the planetary spectrum by the star via direct stellar light on the detector and stellar reflections from the planet might pose a challenge. Consequently, observations need to be carried out in the near-IR, where the planet is relatively bright and presents a rich forest of spectral absorption lines. Figure 2 shows that contamination of the planetary spectrum[5] by reflected star light becomes negligible beyond several 10 AU, with no need for additional cleaning.

## 3. RESULTS AND PREDICTIONS

### 3.1. *Transit Timing Dichotomies*

We computed the TTDs of the ten largest moons in the solar system, yielding values between about 2 and 12 s (Figure 3). Most intriguingly, the largest moons (Ganymede, Titan, and Callisto), which have the deepest solar transit signatures, also have the largest TTDs. This is owed to the location of the water ice line in the accretion disks around Jovian planets, which causes the most massive icy satellites to form beyond about 15 $R_{\rm Jup}$ (Heller & Pudritz 2014). Figure 3 indicates that timing precisions of 1 - 6 s need to be achieved for transit events with depths of only about $10^{-4}$, corresponding to the transit depth of an Earth-sized moon transiting a Sun-like star.

Precisions of 6 s in exoplanet transit mid-times have been achieved from the ground using the Baade 6.5 m telescope at Las Campanas Observatory in Chile (Winn

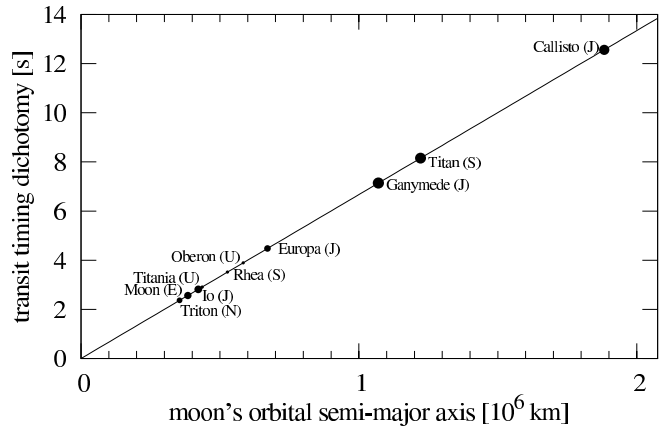

**Figure 3.** Transit timing dichotomies of the ten largest moons in the solar system. Moon radii are symbolized by circle sizes. The host planets Jupiter, Saturn, Neptune, Uranus, and Earth are indicated with their initials. Note that the largest moons, causing the deepest solar transits, induce the highest TTDs.

---

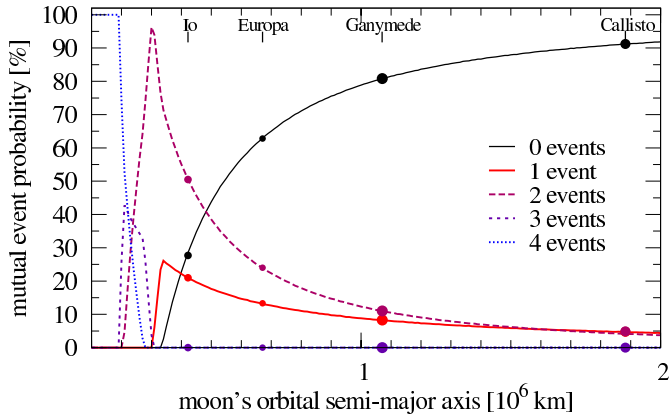

**Figure 4.** Probability of mutual planet-moon events as a function of the moon's planetary distance around a Jupiter-like planet, which is assumed to orbit a Sun-like star at 1 AU. The five curves indicate the frequency of 0, 1, 2, 3, or 4 mutual stellar transits as measured in our transit simulations. The orbits of Io, Europa, Ganymede, and Callisto are indicated with symbols along each curve.

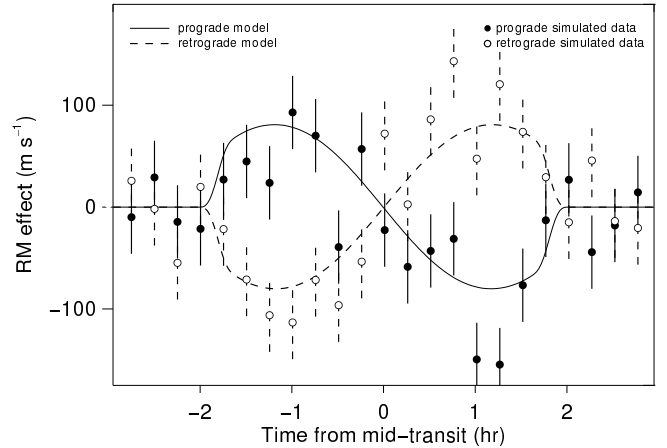

**Figure 5.** Simulated Rossiter-McLaughlin effect of a giant moon ($0.7\,R_\oplus$) transiting a hot, Jupiter-sized planet similar to $\beta$ Pic b. Solid and dashed lines correspond to a prograde and a retrograde coplanar orbit, respectively. Full and open circles indicate simulated E-ELT observations.

et al. 2009). On the one hand, the planet in these observations (WASP-4b) was comparatively large to its host star with a transit depth of about 2.4 %. On the other hand, the star was not particularly bright, with an apparent visual magnitude $m_V \approx 12.6$. The photon collecting power of the E-ELT will be $(39.3\,\mathrm{m}/6.5\,\mathrm{m})^2 \approx 37$ times that of the Baade telescope. And with improved data reduction methods, timing precisions of the order of seconds should be obtainable for transit depths of $10^{-4}$ with the E-ELT for very nearby transiting systems.

We calculate the probabilities of one to four mutual events ($\mathcal{P}_i$, $i \in \{1, 2, 3, 4\}$) during a stellar transit of a coplanar planet-moon system. The distance of the moon travelled on its circumplanetary orbit during the stellar transit is $s = 2R_\star P_{\ast b} a_{\mathrm{ps}}/(P_{\mathrm{ps}} a_{\star b})$, with $a_{\star b}$ denoting the circumstellar orbital semi-major axis of the planet-moon barycenter and $P_{\star b}$ being its circumstellar orbital period. We analyze a Jupiter-like planet at 1 AU from a Sun-like star and simulate $10^6$ transits for a range of possible moon semi-major axes, respectively, where the moon's initial orbital position during the stellar transit is randomized. As the stellar transit occurs, we follow the moon's circumplanetary orbit and measure the number of mutual events (of type I, II, III, or IV) during the transit, which can be 0, 1, 2, 3, or even 4.

For siblings of Io, Europa, Ganymede, and Callisto we find $\mathcal{P}_1 = 21$, 13, 8, and 5 % as well as $\mathcal{P}_2 = 50$, 24, 11, and 5 %, respectively (Figure 4). Notably, the probabilities for two mutual events (purple long-dashed line) are higher than the likelihoods for only one (red solid line) if $a_{\mathrm{ps}} \lesssim 1.6 \times 10^6$ km. For moons inside about halfway between Io and Europa, the probability of having no event (black solid line) is < 50 %, so it is more likely to have at least one mutual event during any transit than having none. Moons inside 200,000 km (half the semi-major axis of Io) can even have three or more events, allowing the TTD method to work with only one stellar transit. Deviations from well-aligned orbits due to high transit impact parameters or tilted moon orbits would naturally reduce the shown probabilities. Nevertheless, mutual events could obviously be common during transits.

A single stellar transit with ≥ 3 mutual events contains TTD information on its own. A two-event stellar transit only delivers TTD information if the events are either a combination of I and II or of III and IV. Moons in wide orbits cannot proceed from one conjunction to the other during one stellar transit. Hence, the contribution of TTD-containing events along $\mathcal{P}_2$ is zero beyond about $10^6$ km. In close orbits, the fraction of two-event cases with TTD information is $P_{2,\mathrm{TTD}} \approx R_{\mathrm{p}}(a_{\mathrm{ps}}\pi)^{-1}$. For Io, as an example, $P_{2,\mathrm{TTD}} = (6.1 \times \pi)^{-1} \approx 5$ %. So even for very close-in moons, two-event cases with TTD information are rare.

### 3.2. The Planetary Rossiter-McLaughlin Effect due to an Exomoon

We simulated the RME in the near-IR spectrum of a planet similar to $\beta$ Pic b assuming a planetary rotation speed of $25\,\mathrm{km\,s}^{-1}$ (Snellen et al. 2014) and a Jovian planetary radius ($R_{\mathrm{Jup}}$). The moon was placed in a Ganymede-like orbit ($a_{\mathrm{ps}} = 15\,R_{\mathrm{Jup}}$) and assumed to belong to the population of giant moons that form at the water ice lines around super-Jovian planets, with a radius of up to about 0.7 Earth radii ($R_\oplus$) (Heller & Pudritz 2014). Using the code of Albrecht et al. (2007, 2013), we simulated absorption lines of the rotating planet as distorted during the moon's transit with a cadence of 15 min. Focusing on the same spectral window (2.304 − 2.332 μm) as Snellen et al. (2014), we then convolved the planetary spectrum (Figure 2) with these distorted absorption lines. Employing the CRIRES Exposure Time Calculator and incorporating the increase of $(39.3\,\mathrm{m}/8.2\,\mathrm{m})^2 \approx 23$ in collecting area for the E-ELT, we obtain a S/N of 75 per pixel for a 15 min exposure – the same values as obtained by Snellen et al. (2014) for a similar calculation. The resulting pseudo-observed spectra are finally cross-correlated with the template spectrum, and a Gaussian is fitted to the cross-correlation functions to obtain RVs.

Pro- and retrograde coplanar orbits are clearly distinguishable in the resulting RME curve (Figure 5). In particular, the RME amplitude of ≈ $100\,\mathrm{m\,s}^{-1}$ is quite substantial. In comparison, the RME amplitude of HD209458 b is ≈ $40\,\mathrm{m\,s}^{-1}$ (Queloz et al. 2000), and re-cent RME detections probe down to $1\,\mathrm{m\,s^{-1}}$ (Winn et al. 2010).

## 4. DISCUSSION AND CONCLUSION

We present two new methods to determine an exomoon's sense of orbital motion. One method, which we refer to as the transit timing dichotomy, is based on a light travelling effect that occurs in subsequent mutual planet-moon eclipses during stellar transits. For the ten largest moons in the solar system, TTDs range between 2 and 12 s. If the planet-moon orbital period can be determined independently (e.g. via TTV and TDV measurements) and with an accuracy of $\lesssim 1\,\mathrm{hr}$, or if a very close-in moon shows at least three mutual events during one stellar transit, then TTD can uniquely determine the sequence of planet-moon eclipses. To resolve the TTD effect, photometric accuracies of $10^{-4}$ need to be obtained along with mid-event precisions $\lesssim 6\,\mathrm{s}$, which should be possible with the E-ELT. If an exomoon is self-luminous, e.g. due to extreme tidal heating (Peters & Turner 2013), and shows regular planet-moon transits and eclipses every few days, its TTD might be detectable without the need of stellar transits. Eccentricities as small as 0.001 can trigger tidal surface heating rates on large moons that could be detectable with the James Webb Space Telescope (Peters & Turner 2013; Heller et al. 2014). But TTD measurements do not require such an extreme scenario, which does not support the conclusion of Lewis & Fujii (2014) that mutual events can only be used to determine an exomoon's SOM if the moon's own brightness can be measured. In general, TTDs can be used to verify the prograde or retrograde motion of a moon with respect to the circumstellar motion.

The second method is based on measurements of the IR spectrum emitted by a young, luminous giant planet. We present simulations of the Rossiter-McLaughlin effect imposed by a transiting moon on a planetary spectrum. The largest moons that can possibly form in the circumplanetary accretion disks, halfway between Ganymede and Earth in terms of radii, can cause an RME that will be comfortably detectable with an IR spectrograph like CRIRES mounted to an ELT. A larger throughput or larger spectral coverage than the current non-cross-dispersed CRIRES will make it possible to determine the SOM of smaller moons, maybe similar to the ones we have in the solar system. A moon-induced planetary RME can determine the moon's orbital motion with respect to the planetary rotation.

Combined observations of an exomoon's TTD, its planetary RME as well as the moon's stellar RME (Simon et al. 2010; Zhuang et al. 2012) and the inclination between the moon's circumplanetary orbit and the planet's circumstellar orbit (Kipping 2011a) can potentially characterize an exomoon orbit in full detail.

We conclude by emphasizing that a moon's transit probability in front of a giant planet is about an order of magnitude higher than that of a planet around a star. A typical terrestrial planet at 0.5 AU from a Sun-like star has a transit probability of $R_{\odot}/0.5\,\mathrm{AU} \approx 1\%$. For comparison, the Galilean moons orbit Jupiter at distances of about 6.1, 9.7, 15.5, and $27.2\,R_{\mathrm{Jup}}$, implying transit probabilities of up to about 16%. With orbital periods of a few days, moon transits occur also much more frequently than for a common Kepler planet. Permanent, highly-accurate IR photometric monitoring of a few dozen directly imaged giant exoplanets thus has a high probability of finding an extrasolar moon.

The report of an anonymous referee was very helpful in clarifying several passages in this letter. We thank Tim-Oliver Husser for providing us with the PHOENIX models. René Heller is supported by the Origins Institute at McMaster University and by the Canadian Astrobiology Program, a Collaborative Research and Training Experience Program funded by the Natural Sciences and Engineering Research Council of Canada (NSERC). Funding for the Stellar Astrophysics Centre is provided by The Danish National Research Foundation (Grant agreement no.: DNRF106).

## 6.4 Predictable Patters in Planetary Transit Timing Variations and Transit Duration Variations due to Exomoons (Heller et al. 2016b)

Contribution:

RH contributed to the literature research, contributed to the mathematical framework, created Fig. 1, led the writing of the manuscript, and served as a corresponding author for the journal editor and the referees.

# Predictable patterns in planetary transit timing variations and transit duration variations due to exomoons

René Heller[1], Michael Hippke[2], Ben Placek[3], Daniel Angerhausen[4, 5], and Eric Agol[6, 7]

[1] Max Planck Institute for Solar System Research, Justus-von-Liebig-Weg 3, 37077 Göttingen, Germany; heller@mps.mpg.de
[2] Luiter Straße 21b, 47506 Neukirchen-Vluyn, Germany; hippke@ifda.eu
[3] Center for Science and Technology, Schenectady County Community College, Schenectady, NY 12305, USA; placekbh@sunysccc.edu
[4] NASA Goddard Space Flight Center, Greenbelt, MD 20771, USA; daniel.angerhausen@nasa.gov
[5] USRA NASA Postdoctoral Program Fellow, NASA Goddard Space Flight Center, 8800 Greenbelt Road, Greenbelt, MD 20771, USA
[6] Astronomy Department, University of Washington, Seattle, WA 98195, USA; agol@uw.edu
[7] NASA Astrobiology Institute's Virtual Planetary Laboratory, Seattle, WA 98195, USA

## ABSTRACT

We present new ways to identify single and multiple moons around extrasolar planets using planetary transit timing variations (TTVs) and transit duration variations (TDVs). For planets with one moon, measurements from successive transits exhibit a hitherto undescribed pattern in the TTV-TDV diagram, originating from the stroboscopic sampling of the planet's orbit around the planet–moon barycenter. This pattern is fully determined and analytically predictable after three consecutive transits. The more measurements become available, the more the TTV-TDV diagram approaches an ellipse. For planets with multi-moons in orbital mean motion resonance (MMR), like the Galilean moon system, the pattern is much more complex and addressed numerically in this report. Exomoons in MMR can also form closed, predictable TTV-TDV figures, as long as the drift of the moons' pericenters is sufficiently slow. We find that MMR exomoons produce loops in the TTV-TDV diagram and that the number of these loops is equal to the order of the MMR, or the largest integer in the MMR ratio. We use a Bayesian model and Monte Carlo simulations to test the discoverability of exomoons using TTV-TDV diagrams with current and near-future technology. In a blind test, two of us (BP, DA) successfully retrieved a large moon from simulated TTV-TDV by co-authors MH and RH, which resembled data from a known *Kepler* planet candidate. Single exomoons with a 10 % moon-to-planet mass ratio, like to Pluto-Charon binary, can be detectable in the archival data of the *Kepler* primary mission. Multi-exomoon systems, however, require either larger telescopes or brighter target stars. Complementary detection methods invoking a moon's own photometric transit or its orbital sampling effect can be used for validation or falsification. A combination of *TESS*, *CHEOPS*, and *PLATO* data would offer a compelling opportunity for an exomoon discovery around a bright star.

**Key words.** eclipses – methods: numerical – planets and satellites: detection – planets and satellites: dynamical evolution and stability – planets and satellites: terrestrial planets – techniques: photometric

## 1. Introduction

The search for moons around planets beyond the solar system is entering a critical phase. The first dedicated exomoon surveys have now been implemented using space-based highly accurate *Kepler* photometry (Kipping et al. 2012; Szabó et al. 2013; Hippke 2015) and more will follow in the near future using *CHEOPS* (Simon et al. 2015) and *PLATO* (Hippke & Angerhausen 2015). An important outcome of the first exomoon searches is that moons at least twice as massive as Ganymede, the most massive local moon, are rare around super-Earths (Kipping et al. 2015).

More than a dozen techniques have been proposed to search for exomoons: (i.) transit timing variations (TTVs; Sartoretti & Schneider 1999; Simon et al. 2007) and transit duration variations (TDVs; Kipping 2009a,b) of exoplanets; (ii.) the direct photometric transit signature of exomoons (Sartoretti & Schneider 1999; Brown et al. 2001; Szabó et al. 2006; Charbonneau et al. 2006; Pont et al. 2007; Tusnski & Valio 2011; Kipping 2011a); (iii.) microlensing (Han & Han 2002; Liebig & Wambsganss 2010; Bennett et al. 2014); (iv.) mutual eclipses of directly imaged, unresolved planet–moon binaries (Cabrera & Schneider 2007); (v.) the wobble of the photometric center of unresolved, directly imaged planet–moon systems (Cabrera & Schneider 2007; Agol et al. 2015) (vi.) time-arrival analyses of planet–moon systems around pulsars (Lewis et al. 2008); (vii.) planet–moon mutual eclipses during stellar transits (Sato & Asada 2009; Pál 2012); (viii.) the Rossiter-McLaughlin effect (Simon et al. 2010; Zhuang et al. 2012); (ix.) scatter peak analyses (Simon et al. 2012) and the orbital sampling effect of phase-folded light curves (Heller 2014; Heller et al. 2016); (x.) modulated radio emission from giant planets with moons (Noyola et al. 2014, 2016); (xi.) the photometric detection of moon-induced plasma torii around exoplanets (Ben-Jaffel & Ballester 2014); (xii.) and several other spectral (Williams & Knacke 2004; Robinson 2011; Heller & Albrecht 2014) and photometric

(Moskovitz et al. 2009; Peters & Turner 2013) analyses of the infrared light in exoplanet-exomoon systems.[1]

None of these techniques has delivered a secure exomoon detection as of today, which is partly because most of these methods are reserved for future observational technologies. Some methods are applicable to the available data, for example, from the *Kepler* primary mission, but they are extremely computer intense (Kipping et al. 2012). We present a novel method to find and characterize exomoons that can be used with current technologies and even the publicly available *Kepler* data. We identify a new pattern in the TTV-TDV diagram of exoplanet-exomoon systems that allows us to distinguish between single and multiple exomoons. Detection of multiple moons is naturally more challenging than the detection of single exomoons owing to the higher complexity of models that involve multiple moons (Heller 2014; Hippke & Angerhausen 2015; Kipping et al. 2015; Heller et al. 2016; Noyola et al. 2016).

## 2. Patterns in the TTV-TDV diagram

An exoplanet transiting a star can show TTVs (Sartoretti & Schneider 1999) and TDVs (Kipping 2009a,b) if accompanied by a moon. Various ways exist to measure TTVs, for example, with respect to the planet-moon barycenter or photocenter (Szabó et al. 2006; Simon et al. 2007, 2015). We utilize the barycentric TTV. On their own, TTVs and TDVs yield degenerate solutions for the satellite mass ($M_s$) and the orbital semimajor axis of the satellite around the planet ($a_s$). If both TTV and TDV can be measured repeatedly, however, and sources other than moons can be excluded, then the $M_s$ and $a_s$ root mean square (RMS) values can be estimated and the degeneracy be solved (Kipping 2009a,b).

This method, however, works only for planets (of mass $M_p$) with a single moon. Moreover, observations will always undersample the orbit of the moon and $P_s$ cannot be directly measured (Kipping 2009a). This is because the orbital period of the planet around the star ($P_p$) is $\gtrsim 9$ times the orbital period of the satellite around the planet ($P_s$) to ensure Hill stability. Strictly speaking, it is the circumstellar orbital period of the planet-moon barycenter ($P_B$) that needs to be $\gtrsim 9 P_s$, but for $M_s/M_p \to 0$ we have $P_p \to P_B$. Finally, a predictable pattern in TTV-TDV measurements has not been published to date.

We present new means to determine (1) the remainder of the division of the orbital periods of the moon and the planet for one-moon systems; (2) the TTV and TDV of the planet during the next transit for one-moon systems; and (3) the number of moons in multiple moon systems.

Our approach makes use of the fact that TTVs and TDVs are phase-shifted by $\pi/2$ radians, as first pointed out by Kipping (2009a). In comparison to the work of Kipping, however, we fold the TTV-TDV information into one diagram rather than treating this information as separate functions of time.

### 2.1. Exoplanets with a single exomoon

Imagine a transiting exoplanet with a single moon. We let $P_B$ be $f = n + r$ times $P_s$, where $n$ is an integer and $0 \le r \le 1$ is the remainder. In the Sun-Earth-Moon system, for example, we have $P_B \approx 365.25$ d, $P_s \approx 27.3$ d, and thus $P_B \approx 13.39 P_s$. Hence, $n = 13$ and $r \approx 0.39$.

We measure the transit timing and transit duration of the first transit in a series of transit observations. These data do not de-liver any TTV or TDV, because there is no reference value yet to compare our measurements to yet. After the second transit, we may obtain a TDV because the duration of the transit might be different from the first transit. We are still not able to obtain a TTV because determining the transit period requires two consecutive transits to be observed. Only after the third transit are we able to measure a variation of the transit period, i.e., our first TTV (and our second TDV). While we observe more and more transits, our average transit period and transit duration values change and converge to a value that is unknown a priori but that could be calculated analytically for one-moon systems if the system properties were known.

Figure 1 qualitatively shows such an evolution of a TTV-TDV diagram. Different from the above-mentioned procedure, measurements in each panel are arranged in a way to gradually form the same figure as shown in panel (e), as if the average TTV and TDV for $N_{obs} = \infty$ were known a priori. In each panel, $N_{obs}$ denotes the number of consecutive transit observations required to present those hypothetical data points. In panels (b)-(d), the first data point of the data series is indicated by a solid line. Generally, TTV and TDV amplitudes are very different from each other, so the final figure of the TTV-TDV diagram in panel (e) is an ellipse rather than a circle. However, if the figure is normalized to those amplitudes and if the orbits are circular, then the final figure is a circle as shown. The angle $\rho$ corresponds to the remainder of the planet-to-moon orbital period ratio, which can be deduced via $r = \rho/(2\pi)$. Knowledge of $\rho$ or $r$ makes it possible to predict the planet-moon orbital geometry during stellar transits, which enables dedicated observations of the maximum planet-moon apparent separation (at TTV maximum values) or possible planet-moon mutual eclipses (at TDV maximum values). Yet, $n$ remains unknown and it is not possible to determine how many orbits the moon has completed around the planet during one circumstellar orbit of their common barycenter. As an aside, $\rho$ cannot be used to determine the sense of orbital motion of the planet or moon (Lewis & Fujii 2014; Heller & Albrecht 2014).

If the moon's orbit is in an $n : 1$ orbital mean motion resonance (MMR) with the circumstellar orbit, then the moon always appears at the same position relative to the planet during subsequent transits. Hence, if $\rho = 0 = r$, there is effectively no TTV or TDV and the moon remains undiscovered. In an $(n + \frac{1}{2}) : 1$ MMR, we obtain $r = \frac{1}{2}$ and subsequent measurements in the TTV-TDV diagram jump between two points. Again, the full TTV-TDV figure is not sampled and the moon cannot be characterized. In general, in an $(n + \frac{1}{x}) : 1$ MMR, the diagram jumps between $x$ points and $r = \frac{1}{x}$.

Orbital eccentricities of the satellite ($e_s$) complicate this picture. Figure 2 shows the TTV-TDV diagram for a planet–moon system akin to the Earth-Moon binary around a Sun-like star. The only difference that we introduced is an $e_s$ value of 0.25. The resulting shape of the TTV-TDV can be described as an egg-shaped ellipsoid with only one symmetry axis that cannot be transformed into an ellipse by linear scaling of the axis. The orientation of that figure depends on the orientation of the periastron of the planet–moon system.

### 2.2. Exoplanets with multiple exomoons

For single moons on circular orbits, the TTV-TDV diagram can be calculated analytically if the TTV and TDV amplitudes are known (Sartoretti & Schneider 1999; Kipping 2009a,b). For systems with more than $N = 2$ bodies, however, analytical solutions do not exist. Hence, for systems with more than one moon (and one planet), we resort to numerical simulations.

---

[1] Naturally, these methods are not fully independent, and our enumeration is somewhat arbitrary.

### 2.2.1. From *N*-body simulations to TTVs and TDVs

In the following, we generate TTV-TDV diagrams of exoplanet-exomoons systems using a self-made, standard *N*-body integrator that calculates the Newtonian gravitational accelerations acting on *N* point masses.

The average orbital speeds of the major solar system moons are well known.[2] The initial planetary velocities in our simulations, however, are unknown and need to be calculated. We take $v_B = 0$ as the velocity of the planet–moon barycenter, assuming that the system is unperturbed by other planets or by the star. This is a reasonable assumption for planets at about 1 au from a Sun-like star and without a nearby giant planet. In the Earth-Moon system, for reference, perturbations from mostly Venus and Jupiter impose a forced eccentricity of just 0.0549 (Čuk 2007), which is undetectable for exomoon systems in the foreseeable future. We let the total mass of the planet–moon system, or the mass of its barycenter, be $m_B$. With index p referring to the planet, and indices s1, s2, etc. referring to the satellites, we have

$$v_B = \frac{v_p m_p + v_{s1} m_{s1} + v_{s2} m_{s2} + \dots}{m_B} = 0$$

$$\Leftrightarrow v_p = \frac{-v_{s1} m_{s1} - v_{s2} m_{s2} - \dots}{m_p} \qquad (1)$$

Our simulations run in fixed time steps. We find that $10^3$ steps per orbit are sufficient to keep errors in TTV and TDV below 0.1 s in almost all cases. In cases with very small moon-to-planet mass ratios, we need $10^4$ steps to obtain $< 0.1$ s errors. A TTV/TDV error of 0.1 s would generally be considered to be too large for studies dedicated to the long-term orbital evolution and stability of multiplanet systems. For our purpose of simulating merely a single orbit to generate the corresponding TTV-TDV diagram, however, this error is sufficient because the resulting errors in TTV and TDV are smaller than the linewidth in our plots. The processing runtime for $10^4$ steps in a five-moon system is $< 1$ s on a standard, commercial laptop computer. Our computer code, used to generate all of the following TTV-TDV figures, is publicly available with examples under an open source license.[3]

Under the assumption that the orbits of the planet–moon barycenter around the star and of the planet with moons around their local barycenter are coplanar, TTVs depend solely on the variation of the sky-projected position of the planet relative to the barycenter. On the other hand, TDVs depend solely on the variation of the orbital velocity component of the planet that is tangential to the line of sight.[4] For a given semimajor axis of the planet–moon barycenter ($a_B$) and $P_B$ around the star, we first compute the orbital velocity of the planet on a circular circumstellar orbit. We then take the stellar radius ($R_\star$) and convert variations in the relative barycentric position of the planet into TTVs and variations in its orbital velocity into TDVs. Each of these $\approx 10^3$ orbital configurations corresponds to one stellar transit of the planet–moon system. It is assumed that the planet–moon system transits across the stellar diameter, that is, with a transit impact parameter of zero, but our general conclusions would not be affected if this condition were lifted. In the Sun–Earth–Moon system, for example, Earth's maximum tangential displacement of 4 763 km from the Earth–Moon barycenter corresponds to a TTV of 159 s compared to a transit duration of about 13 hr. As we are interested in individual, consecutive TTV and TDV measurements, we use amplitudes rather than RMS values.

Our assumption of coplanar orbits is mainly for visualization purposes, but more complex configurations with inclined orbits should be revisited in future studies to investigate the effects of variations in the planetary transit impact parameter (Kipping 2009b) and the like. For this study, coplanar cases can be justified by the rareness of high moon inclinations ($i_s$) in the solar system; among the 16 largest solar system moons, only four have inclinations $> 1°$; Moon (5.5°), Iapetus (17°), Charon (120°), and Triton (130°).

### 2.2.2. TTV-TDV diagrams of exoplanets with multiple exomoons in MMR

In Fig. 3, we show the dynamics of a two-moon system around a Jupiter-like planet, where the moons are analogs of Io and Europa. The left panel shows the actual orbital setup in our *N*-body code, the right panel shows the outcome of the planetary TTV-TDV curve over one orbit of the outermost moon in a Europa-wide orbit. Numbers along this track refer to the orbital phase of the outer moon in units of percent, with 100 corresponding to a full orbit. In the example shown, both moons start in a conjunction that is perpendicular and "to the left" with respect to the line of sight of the observer (left panel), thereby causing a maximum barycentric displacement of the planet "to the right" from the perspective of the observer. Hence, if the planet were to transit the star in this initial configuration of our setup, its transit would occur too early with respect to the average transit period and with a most negative TTV (phase 0 in Fig. 3).

As an analytical check, we calculate the orbital phases of the outer moon at which the planetary motion reverses. The planetary deflection due to the Io-like moon is of functional form $f_I(t) \propto -k_I \cos(n_I t)$, where $n_I$ is the orbital mean motion of Io, $k_I$ is the amplitude of the planetary reflex motion due to the moon, and $t$ is time. The planetary displacement due to the Europa-like moon is $f_E(t) \propto -k_E \cos(n_E t)$, where $n_E$ is the orbital mean motion of Europa and $k_E$ is the amplitude. The total planetary displacement is then $f_{tot}(t) \propto f_I(t) + f_E(t)$ and the extrema can be found, where

$$0 \overset{!}{=} \frac{d}{dt} f_{tot}(t) \quad , \qquad (2)$$

that is, where the planetary tangential motion reverses and the planet "swings back". We replace $n_I$ with $2n_E$ and $k_I$ with $k_E$ (since $k_I = 1.17 \, k_E$) and find that Eq. (2) is roughly equivalent to

$$0 \overset{!}{=} 2 \sin(2n_E t) + \sin(n_E t) \quad , \qquad (3)$$

which is true at $n_E t = 0$, 1.824, $\pi$, 4.459, $2\pi$, 8.1 etc., corresponding to orbital phases $n_E t/(2\pi) = 0\%$, 29%, 50%, 71%, 100%, 129%. These values agree with the numerically derived values at the maximum displacements (Fig. 3, right panel) and they are independent of the spatial dimensions of the system.

Next, we explore different orbital period ratios in two-moon MMR systems. In Fig. 4, we show the TTV-TDV diagrams for a Jupiter-like planet (5.2 au from a Sun-like star) with an Io-plus a Europa-like moon, but in a 3:1 MMR (top panel) and a 4:1 MMR (bottom panel). Intriguingly, we find that the order of the MMR, or the largest integer in the MMR ratio, determines the number of loops in the diagram. Physically speaking, a loop describes the reverse motion of the planet due to the reversal of one of its moons. This behavior is observed in all our numerical simulations of MMRs of up to five moons (see Appendix A).

---

[4] In case of nonaligned orbits, the planet shows an additional TDV component (Kipping 2009b).

The sizes of any of these loops depends on the moon-to-planet mass ratios and on the semimajor axes of the moons. This dependency is illustrated in Fig. 5, where we varied the masses of the moons. In the upper (lower) panels, a 2:1 (4:1) MMR is assumed. In the left panels, the mass of the outer moon ($M_{s2}$) is fixed at the mass of Ganymede ($M_{Gan}$), while the mass of the inner moon ($M_{s1}$) is successively increased from 1 $M_{Gan}$ (black solid line) over 2 $M_{Gan}$ (blue dotted line) to 3 $M_{Gan}$ (red dashed line). In the right panels, the mass of the inner moon is fixed at 1 $M_{Gan}$, while the mass of the outer moon is varied accordingly. As an important observation of these simulations, we find that massive outer moons can make the loops extremely small and essentially undetectable.

### 2.2.3. Evolution of TTV-TDV diagrams in multiple moon systems

Orbital MMRs can involve librations of the point of conjunction as well as drifts of the pericenters of the moons. In fact, the perijoves (closest approaches to Jupiter) in the Io-Europa 2:1 MMR shows a drift of about $0°.7\,\mathrm{day}^{-1}$. Hence, the MMR is only valid in a coordinate system that rotates with a rate equal to the drift of the pericenters of the respective system. The libration amplitude of the pericenters on top of this drift has been determined observationally to be $0°.0247 \pm 0°.0075$ (de Sitter 1928). For multiple exomoon systems, these effects would smear the TTV-TDV figures obtained with our simulations, if the drift is significant on the timescale on which the measurements are taken.

If the moons are not in a MMR in the first place, then there is an additional smearing effect of the TTV-TDV figures because of the different loci and velocities of all bodies after one revolution of the outermost moon. We explore this effect using an arbitrary example, in which we add a second moon to the Earth-Moon system. We choose an arbitrary semimajor axis (50 % lunar) and mass (0.475% lunar), and reduced the Moon's mass to 5% lunar. The upper panel in Fig. 6 shows the TTV-TDV diagram after a single orbit of the outermost moon. As this system does not include a low-integer MMR, the TTV-TDV figure is very complex, involving an hourglass shape main figure with two minute loops around the origin (Fig. 6, center panel). As expected, the resulting TTV-TDV diagram after multiple moon orbits exhibits a smearing effect (Fig. 6, bottom panel). In this particular example system, the overall shape of the TTV-TDV figure actually remains intact, but much more substantial smearing may occur in other systems.

## 3. Blind retrieval of single and multiple moon systems

Next, we want to know whether the above-mentioned TTV-TDV patterns can actually be detected in a realistic dataset. Above all, observations only deliver a limited amount of TTV-TDV measurements per candidate system and white noise and read noise introduce uncertainties. To which extent does real TTV-TDV data enable the detection of exomoons, and permit us to discern single from multiple exomoon systems?

A $\chi^2$ test can determine the best-fitting model if the model parameters are known. In a realistic dataset, however, the number of parameters is generally unknown since the number of moons is unknown. Hence, we performed a Bayesian test, in which two of us (MH, RH) prepared datasets that were then passed to the other coauthors (BP, DA) for analyses. The preparation team kept the number of moons in the data secret but constrained it to be either 0, 1, or 2. It was agreed that any moons would be in circular, prograde, and stable orbits, that is, beyond the Roche radius but within $0.5\,R_H$ (Domingos et al. 2006). Our code allow us to simulate and retrieve eccentric moons as well, but for demonstration purpose we restrict ourselves to circular orbits in this study. Yet, eccentric and inclined orbits would need to be considered in a dedicated exomoon survey.

### 3.1. Parameterization of the planetary system

We chose to use an example loosely based on KOI-868, a system searched for exomoons by Kipping et al. (2015). We kept $P_p = 236\,\mathrm{d}$, $M_\star = 0.55 \pm 0.07\,M_\odot$, and $R_\star = 0.53 \pm 0.07\,R_\odot$, and the measured timing errors of ∼ 1.5 min per data point. A preliminary analysis suggested that a Moon-like moon could not possibly be detected about KOI-868 b with our approach as the TTV-TDV signals of the 0.32 $M_{Jup}$ planet would be too small. Hence, we assumed an Earth-mass planet ($M_p = 1\,M_\oplus \pm 0.34$). The mass uncertainty of the planet is based on realistic uncertainties from Earth-sized planets, for example, Kepler-20 f (Gautier et al. 2012) and requires radial velocity measurements of better than $1\,\mathrm{m\,s^{-1}}$ (Fressin et al. 2012). Our hypothesized planet would have a much smaller radius than KOI-868 b, potentially resulting in less accurate transit timing measurements than we assumed. This neglect of an additional source of noise is still reasonable, since there is a range of *Kepler* planets with very precise timing measurements, for example, Kepler-80 d with a radius of $R_p = 1.7 \pm 0.2\,R_\oplus$ and timing uncertainties ∼ 1 min (Rowe et al. 2014). To produce consistent TTV-TDV simulations, we adjust the transit duration to 0.3 d.

### 3.2. Parameterization of the one-moon system

In our first example, we assumed a heavy moon (0.1 $M_\oplus$) in a stable, circular Moon-wide orbit ($a_s = 3.84 \times 10^5$ km or 34 % the Hill radius of this planet, $R_H$). This high a mass yields a moon-to-planet mass ratio that is still slightly smaller than that of the Pluto-Charon system. Its sidereal period is 26.2 d, slightly shorter than the 27.3 d period of the Moon. The resulting TTV and TDV amplitudes are 23.1 min and ±1.6 min, respectively. A total of seven data points were simulated, which is consistent with the number of measurements accessible in the four years of *Kepler* primary observations. The data were simulated using our $N$-body integrator, then stroboscopically spread over the TTV-TDV diagram (see Sect. 2.1) and randomly moved in the TTV-TDV plane assuming Gaussian noise. The retrieval team treated $M_\star$, $R_\star$, $M_p$, and $a_B$ as fixed and only propagated the errors of $R_\star$ and $M_p$, which is reasonable for a system that has been characterized spectroscopically.

### 3.3. Parameterization of the two-moon system

Our preliminary simulations showed that multiple exomoon retrieval based on *Kepler*-style data quality only works in extreme cases with moons much larger than those known from the solar system or predicted by moon formation theories. Hence, we assumed an up-scaled space telescope with a theoretical instrument achieving ten times the photon count rate of *Kepler*, corresponding to a mirror diameter of $\approx 3.3$ m (other things being equal); this is larger than the *Hubble Space Telescope* (2.5 m), but smaller than the *James Webb Space telescope* (6.5 m). Neglecting other noise sources such as stellar jitter, our hypothetical telescope reaches ten times higher cadence than *Kepler* at the same noise level. We also found that seven data points do not sample the TTV-TDV figure of the planet sufficiently to reveal the second moon. Rather more than thrice this amount is necessary. We thus simulated 25 data points, corresponding to 15 years of observations. This setup is beyond the technological capacities that will be available within the next decade or so, and our investigations of two-moon systems are meant to yield insights into the principal methodology of multiple moon retrieval using TTV-TDV diagrams.

Our *N*-body simulations suggested that the second moon cannot occupy a stable, inner orbit if the more massive, outer moon has a mass $\gtrsim 0.01\,M_{\mathrm{p}}$, partly owing to the fact that both moons must orbit within $0.5\,R_{\mathrm{H}}$. We neglect exotic stable configurations such as the Klemperer rosette (Klemperer 1962) as they are very sensitive to perturbations. Instead, we chose masses of $0.02\,M_{\oplus}$ for the inner moon and $0.01\,M_{\oplus}$ for the outer moon, which is about the mass of the Moon. We set the semimajor axis of the outer moon ($a_{\mathrm{s2}}$) to $1.92 \times 10^5$ km, half the value of the Earth's moon. The inner moon was placed in a 1:2 MMR with a semimajor axis ($a_{\mathrm{s1}}$) of $1.2 \times 10^5$ km. The masses of the inner and outer satellites are referred to as $M_{\mathrm{s1}}$ and $M_{\mathrm{s2}}$, respectively.

### 3.4. Bayesian model selection and likelihood

In order to robustly select the model that best describes the given data, we employ Bayesian model selection (Sivia & Skilling 2006; Knuth et al. 2015), which relies on the ability to compute the Bayesian evidence

$$Z = \int_{\theta} \pi(\theta) \times L(\theta) d\theta. \qquad (4)$$

Here, $\pi(\theta)$ represents the prior probabilities for model parameters $\theta$ and quantifies any knowledge of a system prior to analyzing data. $L(\theta)$ represents the likelihood function, which depends on the sum of the square differences between the recorded data and the forward model.

The evidence is an ideal measure of comparison between competing models as it is a marginalization over all model parameters. As such, it naturally weighs the favorability of a model to describe the data against the volume of the parameter space of that model and thus aims to avoid the overfitting of data. It can be shown that the ratio of the posterior probabilities for two competing models with equal prior probabilities is equal to the ratio of the Bayesian evidence for each model. Therefore the model with the largest evidence value is considered to be more favorable to explain the data (Knuth et al. 2015).

We utilize the MultiNest algorithm (Feroz & Hobson 2008; Feroz et al. 2009, 2011, 2013) to compute log-evidences used in the model selection process, and posterior samples used for obtaining summary statistics for all model parameters. MultiNest is a variant on the Nested Sampling algorithm (Skilling 2006) and is efficient for sampling within many dimensional spaces that may or may not contain degeneracies. Nested sampling algorithms are becoming increasingly useful in exoplanet science, e.g. for the analysis of transit photometry (Kipping et al. 2012; Placek et al. 2014, 2015) or for the retrieval of exoplanetary atmospheres from transit spectroscopy (Benneke & Seager 2013; Waldmann et al. 2015).

**Table 1.** Normalized Bayesian evidence for our simulated datasets as per Eq. (6).

| no. of moons in model | 1-moon dataset | 2-moon dataset |
|---|---|---|
| 0 | 0 % | 0 % |
| 1 | 57.2 % | 0.00004 % |
| 2 | 42.8 % | 99.99996 % |

**Table 2.** Specification and blind-fitting results for the one-moon system.

| Parameter | True | Fitted |
|---|---|---|
| $M_{\mathrm{s1}}$ | $0.1\,M_{\oplus}$ | $0.12 \pm 0.03\,M_{\oplus}$ |
| $a_{\mathrm{s1}}$ | $384,399$ km | $413,600 \pm 88,450$ km |

**Table 3.** Specification and blind-fitting results for the two-moon system.

| Parameter | True | Fitted |
|---|---|---|
| $M_{\mathrm{s1}}$ | $0.02\,M_{\oplus}$ | $0.021\,M_{\oplus} \pm 0.002\,M_{\oplus}$ |
| $M_{\mathrm{s2}}$ | $0.01\,M_{\oplus}$ | $0.0097\,M_{\oplus} \pm 0.0012\,M_{\oplus}$ |
| $a_{\mathrm{s1}}$ | $120,000$ km | $131,000 \pm 16,560$ km |
| $a_{\mathrm{s2}}$ | $192,000$ km | $183,000 \pm 11,000$ km |

As inputs, MultiNest requires prior probabilities for all model parameters as well as the log-likelihood function. Priors for all model parameters were chosen to be uniform between reasonable ranges. For the single moon case, we explored a range of possible moon masses and semimajor axes with $0 \geq M_{\mathrm{s}} \geq 0.2\,M_{\oplus}$ and $10^4$ km $\geq a_{\mathrm{s}} \geq 5.5 \times 10^5$ km. The lower and upper limits for $a_{\mathrm{s}}$ correspond to the Roche lobe and to $0.5\,R_{\mathrm{H}}$, respectively. For the two-moon scenario, the prior probabilities for the orbital distances were kept the same but the moon masses were taken to range within $0 \geq M_{\mathrm{s}} \geq 0.03\,M_{\oplus}$. The upper mass limit has been determined by an *N*-body stability analysis.

Since both TDV and TTV signals must be fit simultaneously, a nearest neighbor approach was adopted for the log-likelihood function. For each data point, the nearest neighbor model point was selected, and the log-likelihood computed for that pair. This approach neglects the temporal information contained in the TTV-TDV measurements or, in our case, simulations. A fit of the data in the TTV-TDV plane is similar to a phase-folding technique as frequently used in radial velocity or transit searches for exoplanets. A fully comprehensive data fit would test all the possible numbers of moon orbits during each circumstellar orbit (*n*, see Sect. 2.1), which would dramatically increase the CPU demands. Assuming Gaussian noise for both TTV and TDV signals, the form of the log-likelihood was taken to be

$$\log L = -\frac{1}{2\sigma_{\mathrm{TTV}}^2} \sum_{i=1}^{N} \left( \mathcal{M}_{\mathrm{TTV},i} - \mathcal{D}_{\mathrm{TTV},i} \right)^2$$
$$-\frac{1}{2\sigma_{\mathrm{TDV}}^2} \sum_{i=1}^{N} \left( \mathcal{M}_{\mathrm{TDV},i} - \mathcal{D}_{\mathrm{TDV},i} \right)^2 \qquad (5)$$

where $\mathcal{M}_{\mathrm{TTV},i}$ and $\mathcal{M}_{\mathrm{TDV},i}$ are the TTV and TDV coordinates of the nearest model points to the $i^{\mathrm{th}}$ data points, $\mathcal{D}_{\mathrm{TTV},i}$ and $\mathcal{D}_{\mathrm{TDV},i}$, $N$ is the number of data points, and $\sigma^2$ is the signal variance. Ultimately, we normalize the evidence for the $i$th model as per

$$(\text{normalized evidence})_i = \frac{Z_i}{Z_1 + Z_2 + Z_3} \qquad (6)$$

to estimate the probability that a model correctly describes the data. The left-hand side of Eq. (6) would change if we were to investigate models with more than two moons.

## 4. Results

### 4.1. Blind retrieval of multiple exomoons

The results of our log-evidence calculations for the blind exomoon retrieval are shwon in Table 1. In the case where a one-moon system had been prepared for retrieval, the Bayesian log-evidences are $\log Z_0 = -463.44 \pm 0.02$, $\log Z_1 = -2.98 \pm 0.10$, and $\log Z_2 = -3.27 \pm 0.15$, indicating a slight preference of the one-moon model over both the two-moon and the zero-moon cases. The best fits to the data are shown in Fig. 7. While the difference in log-evidence between the one- and two-moon models is small, our retrieval shows that an interpretation with moon, be it a one- or a multiple system, is strongly favored over the planet-only hypothesis. A moon with zero mass is excluded at high confidence.

In the two-moon case, the log-evidences for the zero-, one-, and two-moon models are $\log Z_0 = -1894.49 \pm 0.02$, $\log Z_1 = -43.96 \pm 0.37$, and $\log Z_2 = -29.29 \pm 0.47$, respectively. The fits to the data are shwn in Fig. 8. In this case, the two-moon model is highly favored over both the zero- and one-moon models.

Once the most likely number of moons in the system has been determined, we were interested in the parameter estimates for the moons. For the one-moon case, our estimates are listed in Table 2. The one-moon model, which has the highest log-evidence, predicts $M_{s1} = 0.12 \pm 0.03\,M_\oplus$ and $a_{s1} = 413,600 \pm 88,450\,\text{km}$. The relatively large uncertainties indicate an $M_{1s}$–$a_{s1}$ degeneracy. Figure 9 shows the log-likelihood contours of the one-moon model applied to this dataset. Indeed, the curved probability plateau in the lower right corner of the plot suggests a degeneracy between the moon mass and orbital distance, which may be exacerbated by the small amplitude of the TDV signal.

The estimates for the two-moon case are shown in Table 3. The favored model in terms of log-evidence is indeed the two-moon model, predicting $M_{s1} = 0.021 \pm 0.002\,M_\oplus$ and $a_{s1} = 131,000 \pm 16,560\,\text{km}$. The outer moon is predicted to have $M_{s2} = 0.0097 \pm 0.0012\,M_\oplus$ and $a_{s2} = 183,000 \pm 11,000\,\text{km}$. Hence, both pairs of parameters are in good agreement with the true values. The log-likelihood landscape is plotted in Fig. 10. The landscape referring to the inner moon (left panel) shows a peak around the true value $(M_{s1}, a_{s1}) = (0.02\,M_\oplus, 120,000\,\text{km})$ rather than the above-mentioned plateau in the one-moon case, implying that the parameters of the inner moon are well constrained. The mass of the outer moon is tightly constrained as well, but $a_{s2}$ has large uncertainties visualized by the high-log-likelihood ridge in the right panel.

## 5. Discussion

Our exomoon search algorithm uses an $N$-body simulator to generate TTV-TDV diagrams based on $a_s$, $R_s$, and optionally $e_s$ and $i_s$). In search of one-moon systems, however, analytic solutions exist. In particular, the TTV-TDV ellipse of planets hosting one exomoon in a circular orbit can be analytically calculated using the TTV and TDV amplitudes via Eqs. (3) in Kipping (2009a) and (C7) in Kipping (2009b) and multiplying those RMS values by a factor $\sqrt{2}$. This would dramatically decrease the computation times, but computing times are short for our limited parameter range in this example study anyway.

TTVs can also be caused by additional planets (Agol et al. 2005), but the additional presence or absence of TDVs puts strong constraints on the planet versus moon hypothesis (Nesvorný et al. 2012). Planet-induced TDVs have only been seen as a result of an apsidal precession of eccentric planets caused by perturbations from another planet and in circumbinary planets (Doyle et al. 2011). Moreover, an outer planetary perturber causes TTV on a timescale that is longer than the orbital period of the planet under consideration (thereby causing sine-like TTV signals), whereas perturbations from a moon act on a timescale much shorter than the orbital period of the planet (thereby causing noise-like signals).

An important limitation of our method is in the knowledge of $M_p$, which is often poorly constrained for transiting planets. It can be measured using TTVs caused by other planets, but a TTV-TDV search aiming at exomoons would try to avoid TTV signals from other planets. Alternatively, stellar radial velocity measurements can reveal the total mass of the planet–moon system. If the moon(s) were sufficiently lightweight it would be possible to approximate the mass of the planet with the mass of the system. Joint mass-radius measurements have now been obtained for several hundred exoplanets. If the mass of the planet remains unknown, then TTV-TDV diagrams can only give an estimate of $M_s/M_p$ in one-moon systems. As an additional caveat, exomoon searches might need to consider the drift of the pericenters of the moons (Sect. 2.2.3). This would involve two more parameters beyond those used in our procedure: the initial orientation of the arguments of periapses and the drift rate. This might even result in longterm transit shape variations for sufficiently large moons on eccentric and/or inclined orbits. Finally, another important constraint on the applicability of our method is the implicit assumption that the transit duration is substantially shorter than the orbital period of the moon. For an analytical description see constraints $\alpha$3) (regarding TTV) and $\alpha$7) (regarding TDV) in Kipping (2011b).

TTV and TDV effects in the planetary transits of a star–planet–moon(s) system are caused by Newtonian dynamics. This dynamical origin enables measurements of the satellite masses and semimajor axes but not of their physical radii. The latter can be obtained from the direct photometric transit signature of the moons. In these cases, the satellite densities can be derived. Direct moon transits could potentially be observed in individual planet-moon transits or in phase-folded transit light curves. A combination of TTV-TDV diagrams with any of these techniques thus offers the possibility of deriving density estimates for exomoons. TTV-TDV diagrams and the orbital sampling effect are sensitive to multiple moon systems, so they would be a natural combination for multiple moon candidate systems.

We also point out an exciting, though extremely challenging, opportunity of studying exomoons on timescales shorter than the orbital period of a moon around a planet. From Earth, it is only possible to measure the angle $\rho$ (or the numerical remainder $r$ of $P_B/P_s$) in the TTV-TDV diagram (see Fig. 1). However, Katja Poppenhäger (private communication 2015) pointed out that a second telescope at a sufficiently different angle could observe transits of a given planet-moon system at different orbital phases of the moon. Hence, the orbit of the moon could be sampled on timescales smaller than $P_s$. Then $n$ could be determined, enabling a complete measurement of $\rho$, $P_s$, and $n$.

Beyond the star–planet–moon systems investigated in this paper, our concept is applicable to eclipsing binary stars with planets in satellite-type (S-type) orbits (Rabl & Dvorak 1988). Here, the TTV-TDV diagram of the exoplanet host star would provide evidence of the planetary system around it.

## 6. Conclusions

We identify new predictable features in the TTVs and TDVs of exoplanets with moons. First, in exoplanet systems with a single moon on a circular orbit, the remainder of the planet–moon orbital period ($0 \leq r \leq 1$) appears as a constant angle ($\rho$) in the TTV-TDV diagram between consecutive transits. The predictability of $\rho$ determines the relative position of the moon to the planet during transits. This is helpful for targeted transit observations of the system to measure the largest possible planet-moon deflections (at maximum TTV values) or observe planet-moon eclipses (at maximum TDV values).

Second, exoplanets with multiple moons in an orbital MMR exhibit loops in their TTV-TDV diagrams. These loops correspond the reversal of the tangential motion of the planet during its orbital motion in the planet–moon system. We find that the largest integer in the MMR of an exomoon system determines the number of loops in the corresponding TTV-TDV diagram, for example, five loops in a 5:3 MMR. The lowest number is equal to the number of orbits that need to be completed by the outermost moon to produce a closed TTV-TDV figure, for example, three in a 5:3 MMR.

Planetary TTV-TDV figures caused by exomoons are created by dynamical effects. As such, they are methodologically independent from purely photometric methods such as the direct transit signature of moons in individual transits or in phase-folded light curves. Our novel approach can thus be used to independently confirm exomoon candidates detected by their own direct transits, by planet–moon mutual eclipses during stellar transits, the scatter peak method, or the orbital sampling effect.

We performed blind retrievals of two hypothetical exomoon systems from simulated planetary TTV-TDV. We find that an Earth-sized planet with a large moon (10 % of the planetary mass, akin to the Pluto-Charon system) around an M dwarf star exhibits TTVs/TDVs that could be detectable in the four years of archival data from the *Kepler* primary mission. The odds of detecting exomoon-induced planetary TTVs and TDVs with *PLATO* are comparable. *PLATO* will observe about ten times as many stars as *Kepler* in total and its targets will be significantly brighter ($4 \leq m_V \leq 11$). The resulting gain in signal-to-noise over *Kepler* might, however, be compensated by *PLATO*'s shorter observations of its two long-monitoring fields (two to three years compared to four years of the *Kepler* primary mission; Rauer et al. 2014). Our blind retrieval of a multiple moon test system shows that the TTV-TDV diagram method works in principle, from a technical perspective. In reality, however, multiple moons are much harder to detect, requiring transit observations over several years by a space-based photometer with a collecting area slightly larger than that of the *Hubble Space Telescope*.

As modern space-based exoplanet missions have duty cycles of a few years at most, exomoon detections via planetary TTVs and TDVs will be most promising if data from different facilities can be combined into long-term datasets. Follow-up observations of planets detected with the *Kepler* primary mission might be attractive for the short term with four years of data being readily available. However, *Kepler* stars are usually faint. Long-term datasets should be obtained for exoplanets transiting bright stars to maximize the odds of an exomoon detec-tion. A compelling opportunity will be a combination of *TESS*, *CHEOPS*, and *PLATO* data. *TESS* (mid-2017 to mid-2019) will be an all-sky survey focusing on exoplanets transiting bright stars. *CHEOPS* (late 2017 to mid-2021) will observe stars known to host planets or planet candidates. Twenty percent of its science observation time will be available for open-time science programs, thereby offering a unique bridge between *TESS* and *PLATO* (2024 to 2030). The key challenge will be in the precise synchronization of those datasets over decades while the timing effects occur on a timescale of minutes.

*Note added in proof.* After acceptance of this paper, the authors learned that Montalto et al. (2012) used a TTV-TDV diagram to search for exomoons around WASP-3b. Awiphan & Kerins (2013) studied the correlation between the squares of the TTV and the TDV amplitudes of exoplanets with one moon.

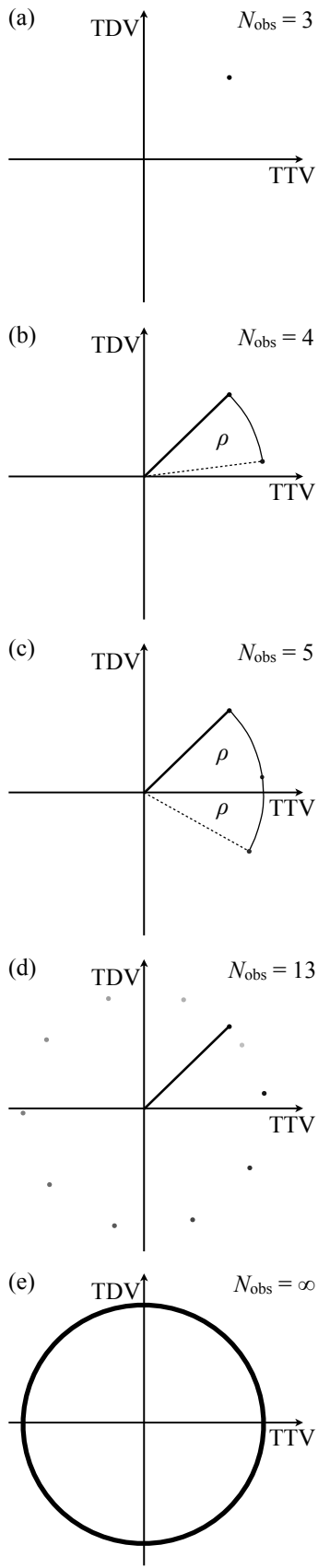

**Fig. 1.** Evolution of a TTV-TDV diagram of an exoplanet with 1 moon. TTV and TDV amplitudes are normalized to yield a circular figure.

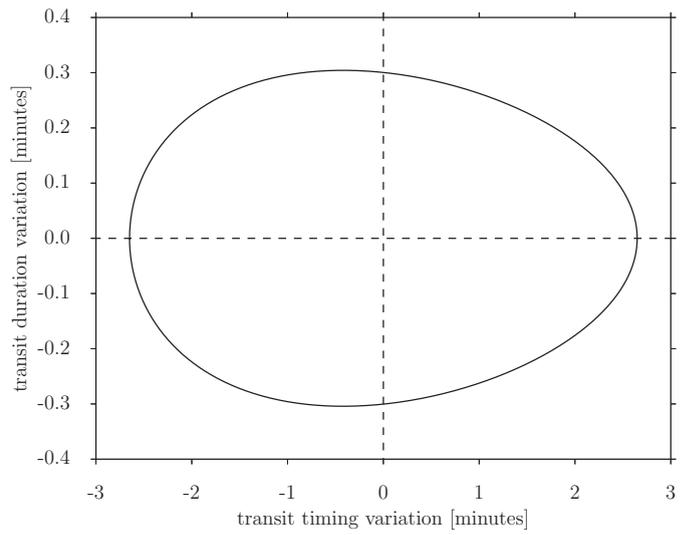

**Fig. 2.** TTV-TDV diagram for the Earth-Moon system transiting the Sun, but with the orbital eccentricity of the Moon increased to 0.25.

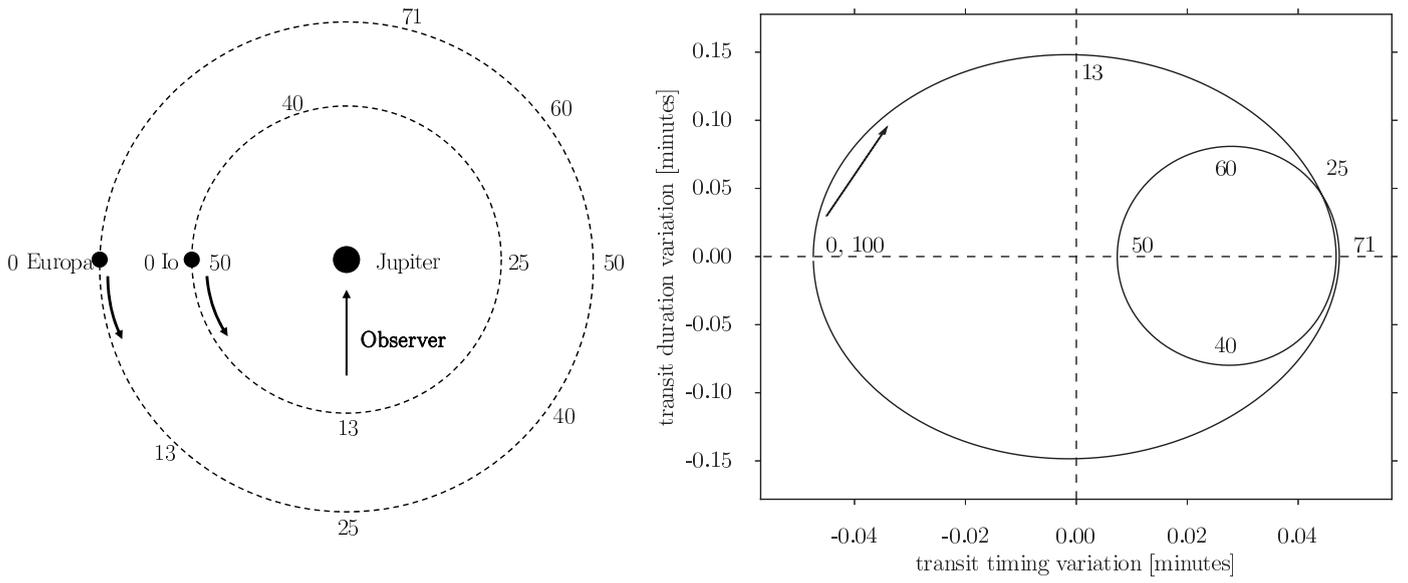

**Fig. 3.** Jupiter-Io-Europa system in a 2:1 MMR. Numbers denote the percental orbital phase of the outer moon, Europa. *Left*: Top-down perspective on our two-dimensional *N*-body simulation. The senses of orbital motion of the moons are indicated with curved arrows along their orbits. *Right*: TDV-TTV diagram for the same system. The progression of the numerical TTV-TDV measurements is clockwise.

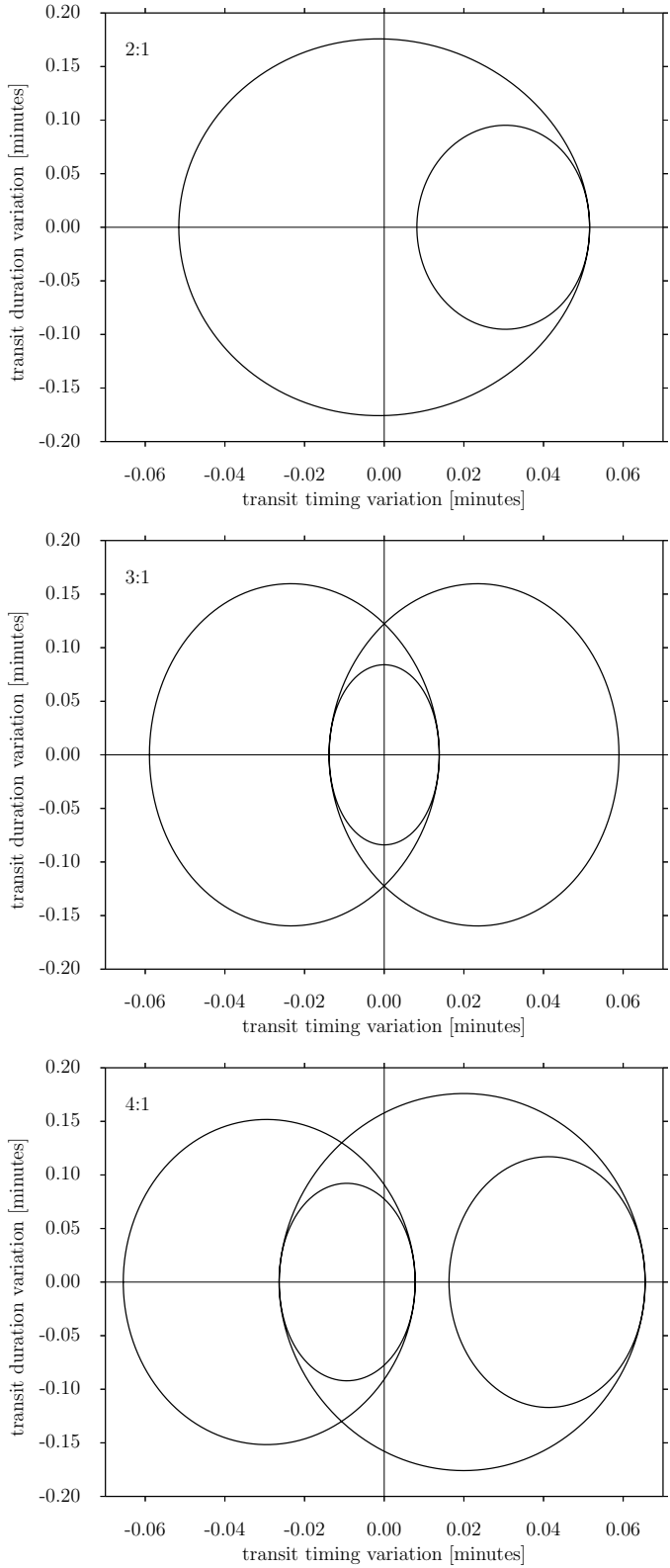

**Fig. 4.** TTV-TDV diagrams for a Jupiter-like planet with 2 moons akin to Io and Europa, but in different MMRs. A 2:1 MMR (top) produces 2 ellipses, a 3:1 MMR (center) 3, and a 4:1 MMR (bottom) 4.

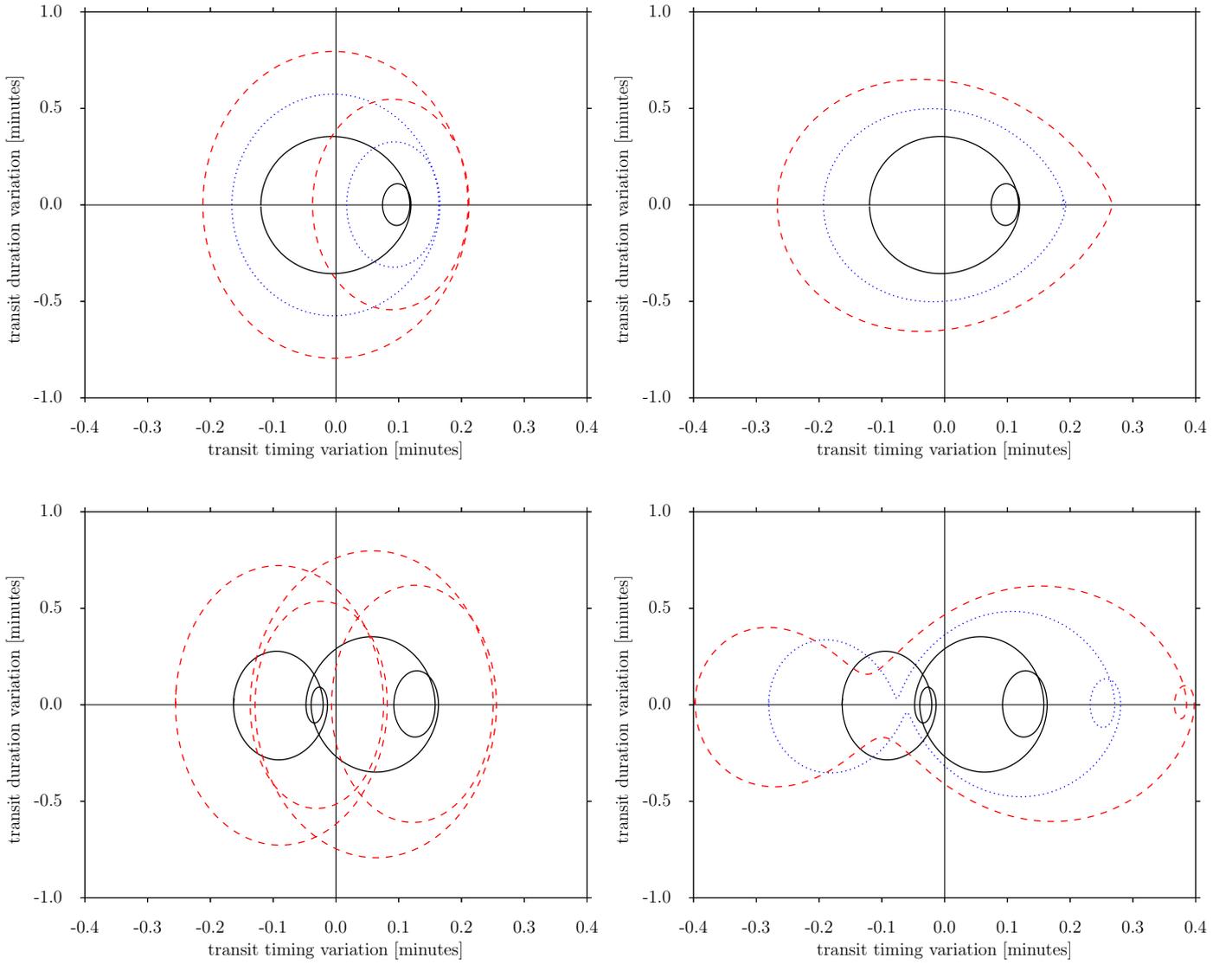

**Fig. 5.** TTV-TDV diagrams of a Jupiter-like planet with 2 moons. Top panels assume a 2:1 MMR akin to the Io-Europa resonance; bottom panels assume a 4:1 MMR akin to the Io-Ganymede resonance. Models represented by black solid lines assume that both moons are as massive as Ganymede. In the left panels, blue dotted lines assume the inner moon is twice as massive ($M_{s1} = 2\,M_{s2} = 2\,M_{\rm Gan}$), and red dashed lines assume the inner moon is thrice as massive. In the right panels, blue assumes the outer moon is twice as massive ($M_{s2} = 2\,M_{s1}$) and red assumes the outer moon is thrice as massive. The inner loop becomes invisibly small for low $M_{s1}/M_{s2}$ ratios in the top right panel.

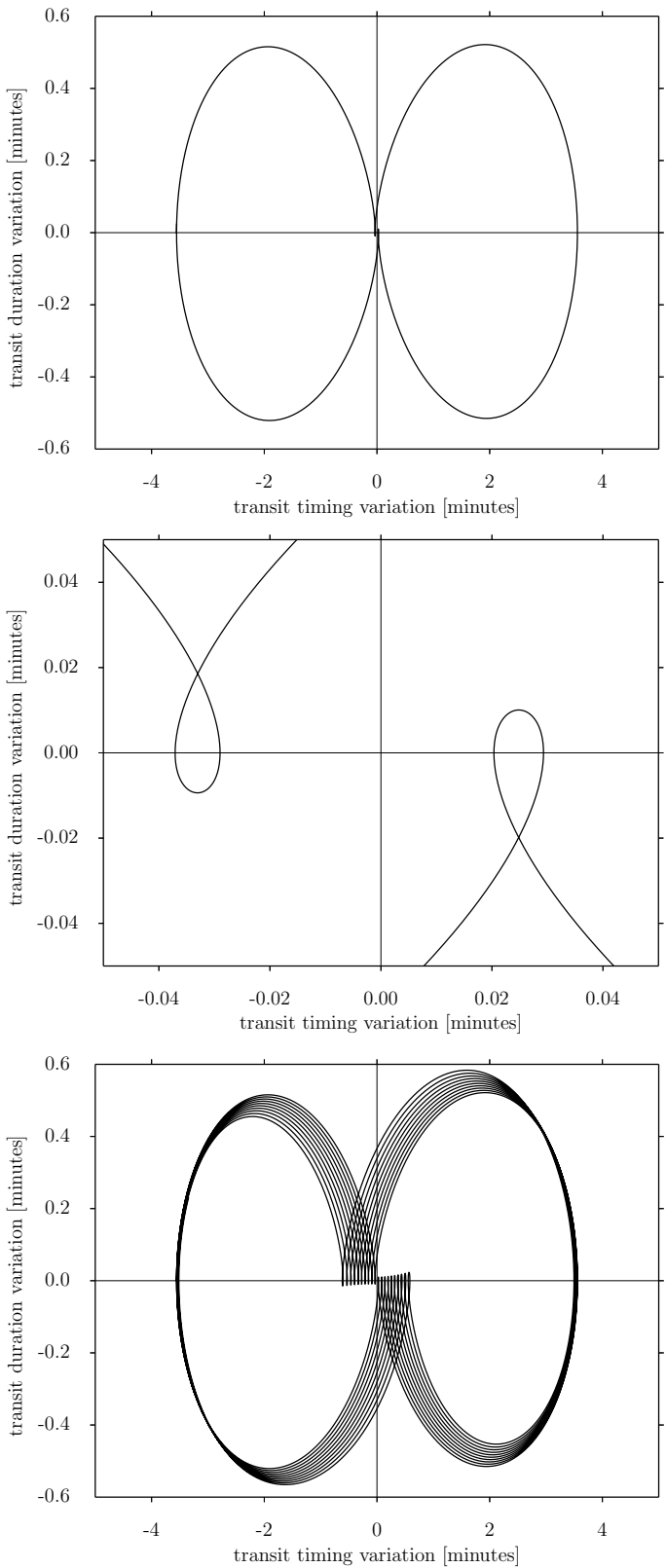

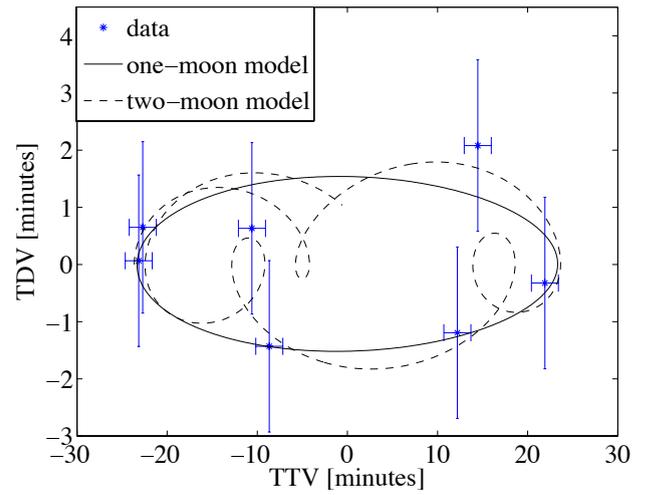

**Fig. 7.** One- and two-moon fits to the model generated data of one moon.

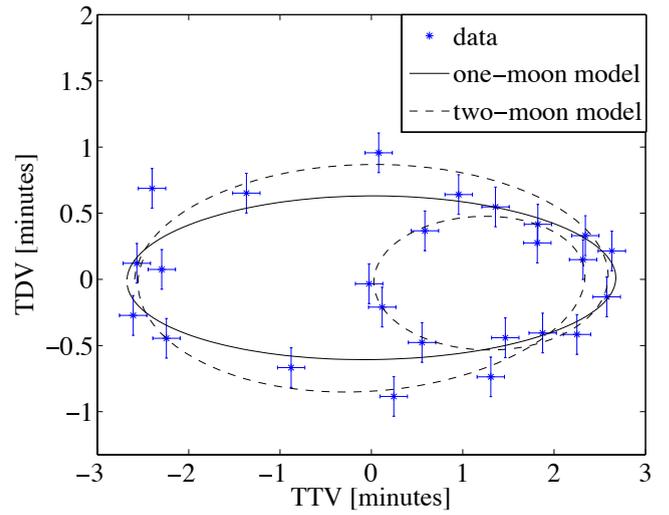

**Fig. 8.** One- and two-moon fits to the model generated data of two moons.

**Fig. 6.** *Top*: TTV-TDV diagram the Earth-Moon-like system involving a second satellite at half the distance of the Moon to Earth with 70 % of the lunar mass. *Center*: Zoom into the center region showing the turning points. *Bottom*: After sampling 10 orbits, a smearing effect becomes visible as the 2 moons are not in a MMR. Exomoon retrieval from such a more complex TTV-TDV figure needs to take into account the evolution of the system.

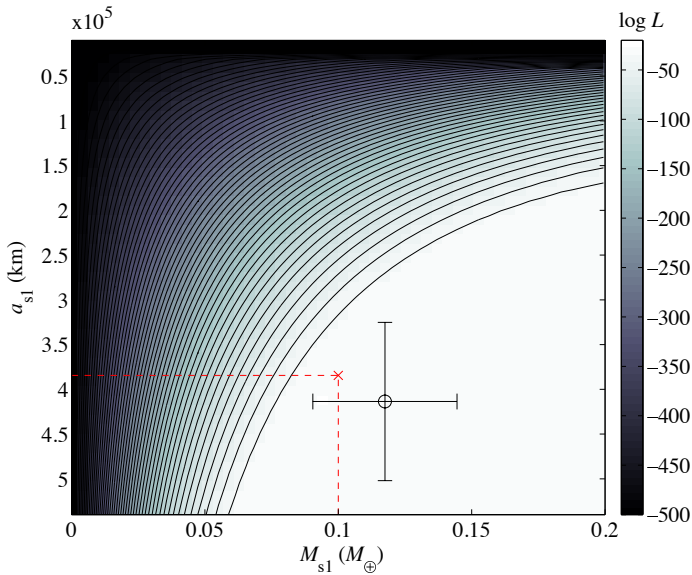

**Fig. 9.** Log-likelihood landscape of the one-moon model applied to the simulated one-moon data. Lighter areas indicate regions of high probability and the red cross indicates the true parameter values. One can clearly see a degeneracy between $a_{s1}$ and $M_{s1}$ in the form of a curved plateau in the bottom right portion of the plot.

## Appendix A: Higher-order MMRs

In the following, we present a gallery of TTV-TDV diagrams for planets with up to five moons in a chain of MMRs. This collection is complete in terms of the possible MMRs. In each case, a Jupiter-mass planet around a Sun-like star is assumed. All moon masses are equal to that of Ganymede. The innermost satellite is placed in a circular orbit at an Io-like semimajor axis ($a_{s1}$) with an orbital mean motion $n_{s1}$ around the planet. The outer moons, with orbital mean motions $n_{si}$ (where $2 \leq i \leq 5$), were subsequently placed in orbits $a_{si} = a_{s1}(n_{s1}/n_{si})^{2/3}$ corresponding to the respective MMR.

*Acknowledgements.* We thank Katja Poppenhäger for inspiring discussions and the referee for a swift and thorough report. E. A. acknowledges support from NASA grants NNX13AF20G, NNX13AF62G, and NASA Astrobiology Institute's Virtual Planetary Laboratory, supported by NASA under cooperative agreement NNH05ZDA001C. This work made use of NASA's ADS Bibliographic Services, of the Exoplanet Orbit Database, and of the Exoplanet Data Explorer at www.exoplanets.org.

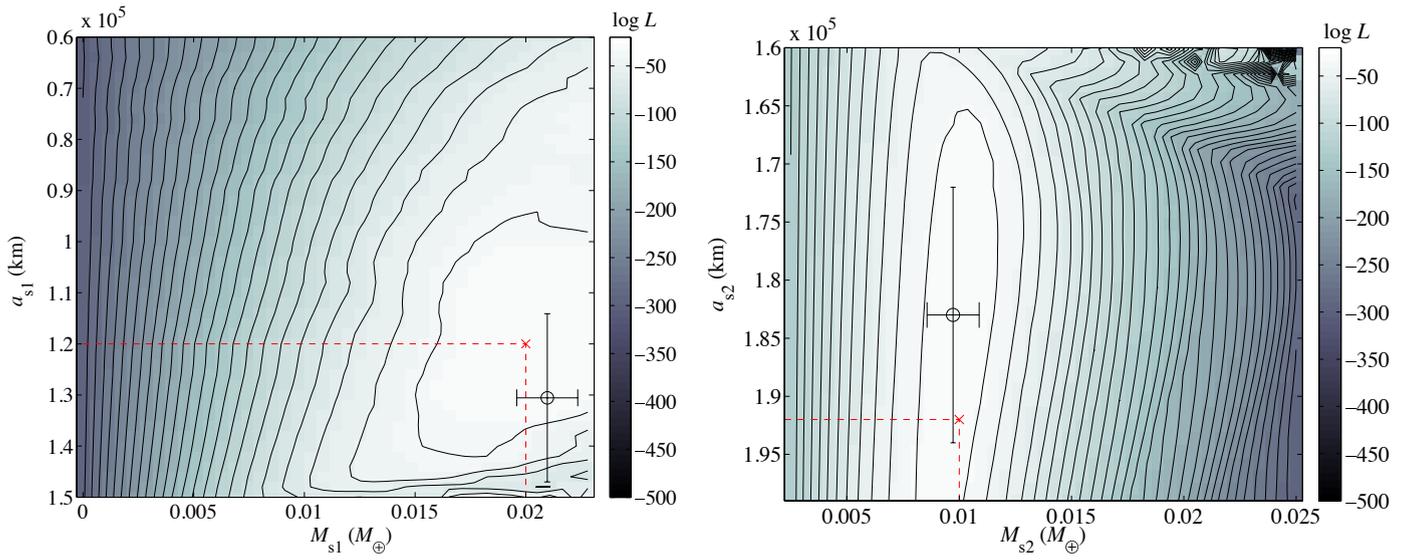

**Fig. 10.** Log-likelihood landscape of the two-moon model applied to simulated two-moon data. Lighter areas indicate regions of better fits. The red cross indicates the location of the simulated system to be retrieved. The parameters corresponding to the inner moon are shown in the left panel and those of the outer moon are shown in the right panel. The vertical ridge of high log-likelihood in the right panel indicates that the orbital distance of the outer moon ($a_{s2}$) is not as well constrained as that of the inner moon.

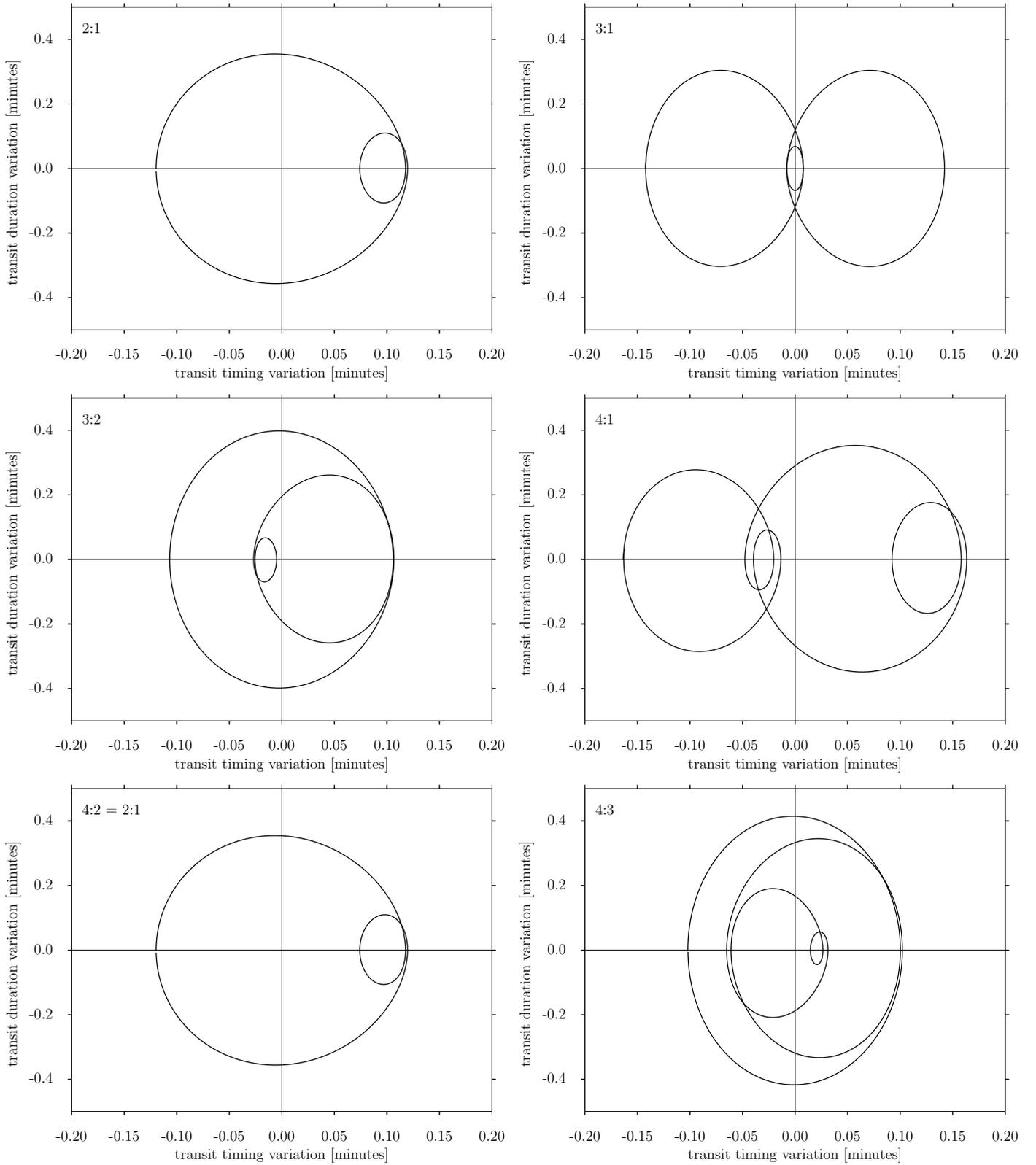

**Fig. A.1.** TTV-TDV diagrams of planets with two moons in MMRs.

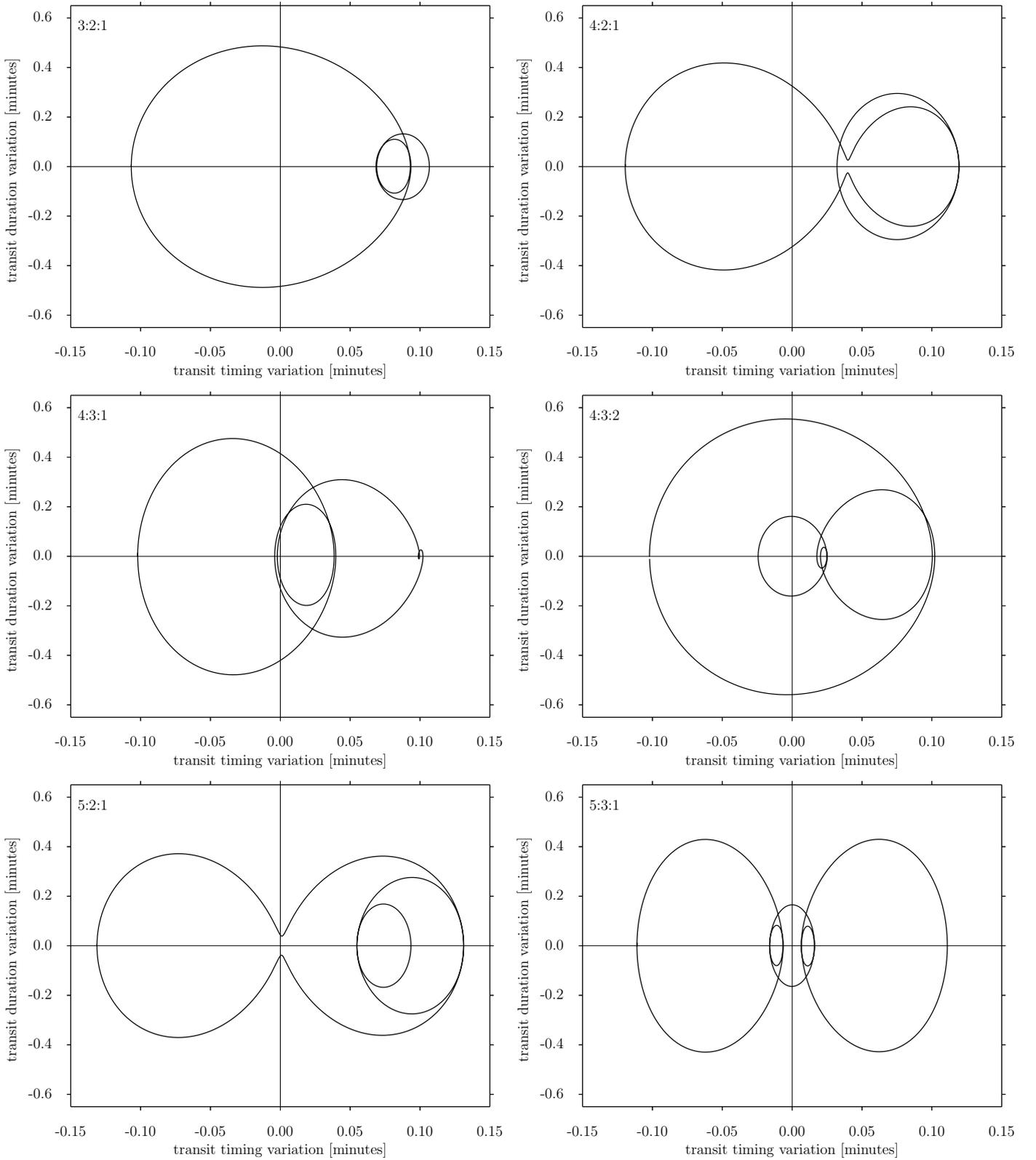

**Fig. A.2.** TTV-TDV diagrams of planets with three moons in MMRs.

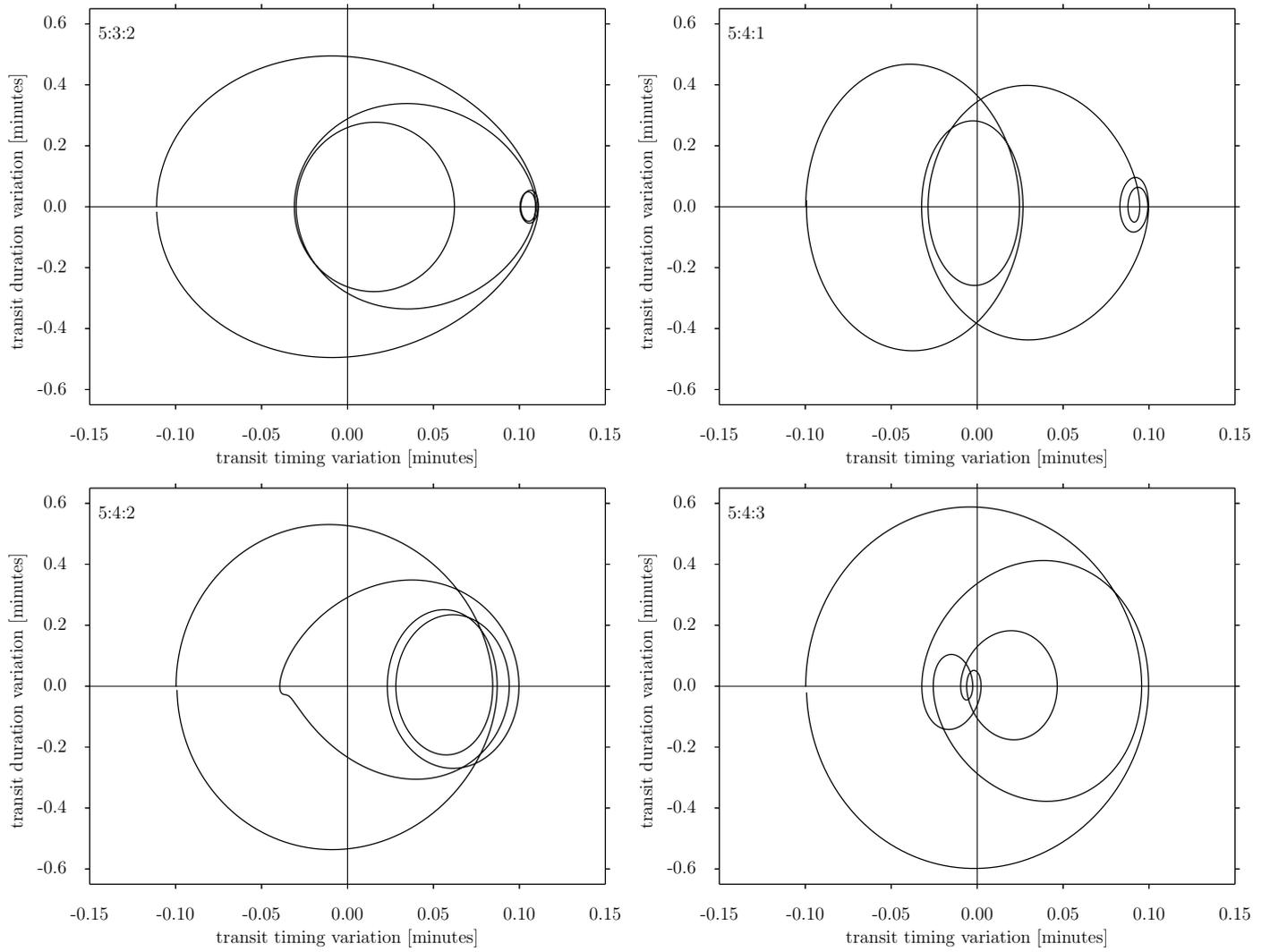

**Fig. A.2.** (continued)

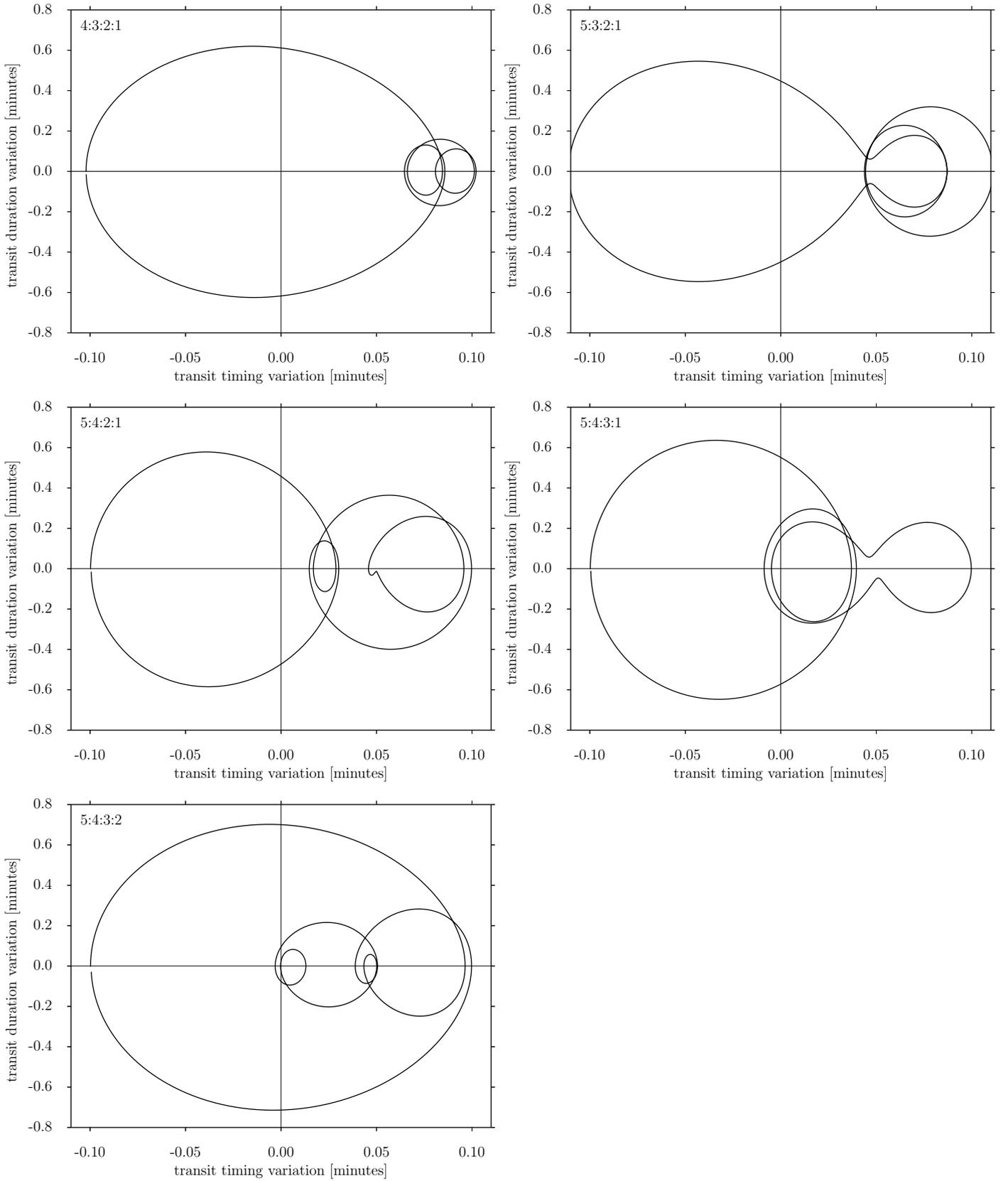

**Fig. A.3.** TTV-TDV diagrams of planets with four moons in MMRs.

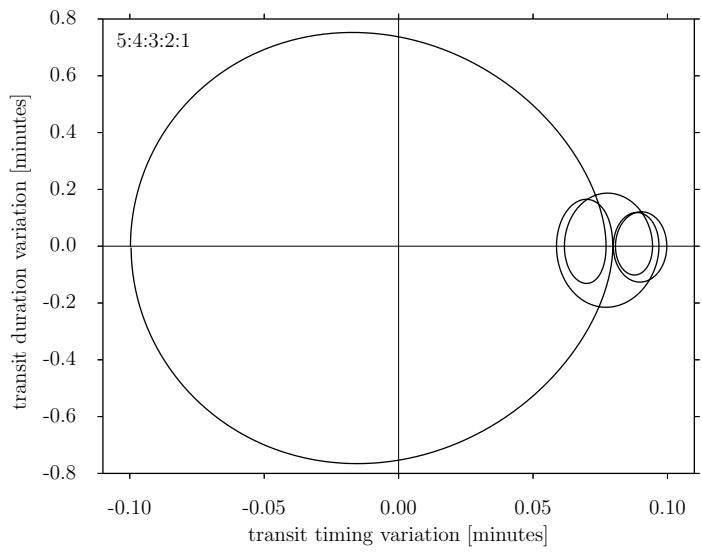

**Fig. A.4.** TTV-TDV diagram of a planets with five moons in MMRs.

## 6.5 Transits of Extrasolar Moons Around Luminous Giant Planets (Heller 2016)

# Transits of extrasolar moons around luminous giant planets
## (Research Note)

R. Heller[1]

Max Planck Institute for Solar System Research, Justus-von-Liebig-Weg 3, 37077 Göttingen, Germany; heller@mps.mpg.de

**ABSTRACT**

Beyond Earth-like planets, moons can be habitable, too. No exomoons have been securely detected, but they could be extremely abundant. Young Jovian planets can be as hot as late M stars, with effective temperatures of up to 2000 K. Transits of their moons might be detectable in their infrared photometric light curves if the planets are sufficiently separated ($\gtrsim$ 10 AU) from the stars to be directly imaged. The moons will be heated by radiation from their young planets and potentially by tidal friction. Although stellar illumination will be weak beyond 5 AU, these alternative energy sources could liquify surface water on exomoons for hundreds of Myr. A Mars-mass $H_2O$-rich moon around $\beta$ Pic b would have a transit depth of $1.5 \times 10^{-3}$, in reach of near-future technology.

**Key words.** Astrobiology – Methods: observational – Techniques: photometric – Eclipses – Planets and satellites: detection – Infrared: planetary systems

## 1. Introduction

Since the discovery of a planet transiting its host star (Charbonneau et al. 2000), thousands of exoplanets and candidates have been detected, mostly by NASA's *Kepler* space telescope (Rowe et al. 2014). Planets are natural places to look for extrasolar life, but moons around exoplanets are now coming more and more into focus as potential habitats (Reynolds et al. 1987; Williams et al. 1997; Scharf 2006; Heller et al. 2014; Lammer et al. 2014). Key challenges in determining whether an exomoon is habitable or even inhabited are in the extreme observational accuracies required for both a detection, e.g., via planetary transit timing variations plus transit duration variations (Sartoretti & Schneider 1999; Kipping 2009), and follow-up characterization, e.g., by transit spectroscopy (Kaltenegger 2010) or infrared (IR) spectral analyses of resolved planet-moon systems (Heller & Albrecht 2014; Agol et al. 2015). New extremely large ground-based telescopes with unprecedented IR capacities (*GTM*, 1st light 2021; *TMT*, 1st light 2022; *E-ELT*, 1st light 2024) could achieve data qualities required for exomoon detections (Quanz et al. 2015).

Here I investigate the possibility of detecting exomoons transiting their young, luminous host planets. These planets need to be sufficiently far away from their stars ($\gtrsim$ 10 AU) to be directly imaged. About two dozen of them have been discovered around any stellar spectral type from A to M stars, most of them at tens or hundreds of AU from their star. The young super-Jovian planet $\beta$ Pic b (11 ± 5 times as massive as Jupiter, Snellen et al. 2014) serves as a benchmark for these considerations. Its effective temperature is about 1700 K (Baudino et al. 2014), and contamination from its host star is sufficiently low to allow for direct IR spectroscopy with *CRIRES* at the *VLT* (Snellen et al. 2014). Exomoons as heavy as a few Mars masses have been predicted to form around such super-Jovian planets (Canup & Ward 2006; Heller et al. 2014), and they might be as large as 0.7 Earth radii for wet/icy composition (Heller & Pudritz 2015b). The transit depth of such a moon ($10^{-3}$) would be more than an order of magnitude larger than that of an Earth-sized planet around a Sun-like star. Photometric accuracies down to 1 % have now been achieved in the IR using *HST* (Zhou et al. 2015). Hence, transits of large exomoons around young giant planets are a compelling new possibility for detecting extrasolar moons.

## 2. Star-planet versus planet-moon transits

### 2.1. Effects of planet and moon formation on transits

Planets and moons form on different spatiotemporal scales. We thus expect that geometric transit probabilities, transit frequencies, and transit depths differ between planets (transiting stars) and moons (transiting planets). The $H_2O$ ice line, beyond which runaway accretion triggers formation of giant planets in the protoplanetary disk (Lissauer 1987; Kretke & Lin 2007), was at about 2.7 AU from the Sun during the formation of the local giant planets (Hayashi 1981). In comparison, the circum-Jovian $H_2O$ ice line, beyond which some of the most massive moons in the solar system formed, was anywhere between the orbits of rocky Europa and icy Ganymede (Pollack & Reynolds 1974) at 10 and 15 Jupiter radii ($R_J$), respectively. The ice line radius ($r_{ice}$), which is normalized to the physical radius of the host object ($R$), was 2.7 AU/12.5 $R_J \approx 480$ times larger in the solar accretion disk than in the Jovian disk. With $r_{ice}$ depending on $R$ and the effective temperature ($T_{eff}$) of the host object as per

$$r_{ice} \propto \sqrt{R^2 \, T_{eff}^4} \ , \qquad (1)$$

we understand this relation by comparing the properties of a young Sun-like star ($R = R_\odot$ [solar radius], $T_{eff} = 5000$ K) with those of a young, Jupiter-like planet ($R = R_J$, $T_{eff} = 1000$ K). This suggests a $\sqrt{10^2 \times 5^4} = 250$ times wider $H_2O$ ice line for the star, neglecting the complex opacity variations in both circumstellar and circumplanetary disks (Bitsch et al. 2015).

The mean geometric transit probabilities ($\bar{P}$) of the most massive moons should thus be larger than $\bar{P}$ of the most massive planets. Because of their shorter orbital periods, planet-moon transits of big moons should also occur more often than stellar transits of giant planets. In other words, big moons should exhibit higher transit frequencies ($\bar{f}$) around giant planets than planets around stars, on average. However, Eq. (1) does not give us a direct clue as to the mean relative transit depths ($\bar{D}$).

## 2.2. Geometric transit probabilities, transit frequencies, and transit depths

I exclude moons around the local terrestrial planets (most notably the Earth's moon) and focus on large moons around the solar system giant planets to construct an empirical moon sample representative of moons forming in the accretion disks around giant planets. This family of natural satellites has been suggested to follow a universal formation law (Canup & Ward 2006). We need to keep in mind, though, that these planets orbit the outer regions of the solar system, where stellar illumination is negligible for moon formation (Heller & Pudritz 2015a). However, many giant exoplanets are found in extremely short-period orbits (the "hot Jupiters"); and stellar radial velocity (RV) measurements suggest that giant planets around Sun-like stars can migrate to 1 AU, where we observe them today. Moreover, at least one large moon, Triton, has probably been formed through a capture rather than in-situ accretion (Agnor & Hamilton 2006).

For the planet sample, I first use all RV planets confirmed as of the day of writing. I exclude transiting exoplanets from my analysis as these objects are subject to detection biases. RV observations are also heavily biased (Cumming 2004), but we know that they are most sensitive to close-in planets because of the decreasing RV amplitude and longer orbital periods in wider orbits. Thus, planets in wide orbits are statistically underrepresented in the RV planet sample, but these planets are not equally underrepresented as planets in transit surveys. Transit surveys also prefer close-in planets, as the photometric signal-to-noise ratio scales as $\propto n_{\rm tr}$ ($n_{\rm tr}$ the number of transits, Howard et al. 2012).

Mean values of $\bar{P}$ (similarly of $\bar{f}$ and $\bar{D}$) are calculated as $\bar{P} = (\sum_i^{N_{\rm p}} P_i)/N_{\rm p}$, where $P_i$ is the geometric transit probability of each individual RV planet and $N_{\rm p}$ is the total number of RV planets. Standard deviations are measured by identifying two bins around $\bar{P}$: one in negative and one in positive direction, each of which contains $\frac{1}{2} 68\,\% = 34\,\%$ of all the planets in the distribution. The widths of these two bins are equivalent to asymmetric $1\,\sigma$ intervals of a skewed normal distribution.

The geometric transit probability $P = R/a$, with $a$ as the orbital semimajor axis of the planet. For the RV planets, $\bar{P} = 0.028\,(+0.026, -0.019)$.[1] The $\bar{P}$ value of an unbiased exoplanet population would be smaller as it would contain more long-period planets. On the contrary, no additional detection of a large solar system moon is expected, and $\bar{P} = 0.114\,(+0.080, -0.045)$, for the 20 largest moons of the solar system giant planets, likely reflects their formation scenarios.

The transit frequency $f \approx 1/(2\pi)\sqrt{(GM)/a^3}$, where $G$ is Newton's gravitational constant and $M$ the mass of the central object, assuming that $M$ is much larger than the mass of the companion. For the RV planets, $\bar{f} = 0.040\,(+0.052, -0.037)$/day.[2] The value of $\bar{f}$, which would be corrected for detection biases, would be smaller with long-period planets having lower frequen-cies. For comparison, $\bar{f}$ of the 20 largest moons of the solar system giant planets is $0.170\,(+0.111, -0.030)$/day.

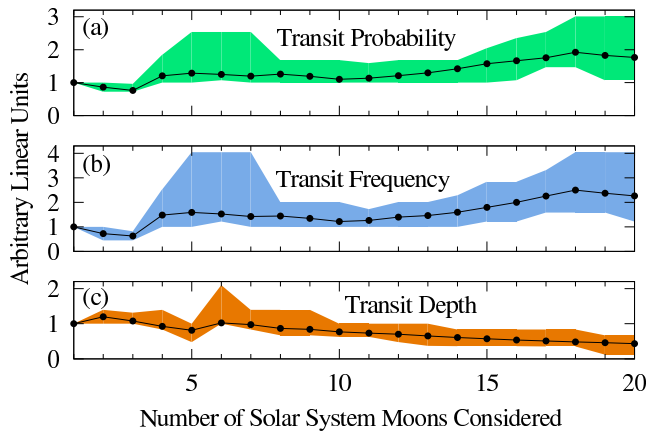

**Fig. 1.** Dependence of (a) the mean geometric transit probability, (b) mean transit frequency, and (c) mean transit depth of the largest solar system moons on the number of moons considered. The abscissa refers to the ranking ($N_{\rm s}$) of the moon among the largest solar system moons.

Stellar limb darkening, star spots, partial transits, etc. aside, the maximum transit depth depends only on $R$ and on the radius of the transiting object ($r$) as per $D = (r/R)^2$. In my calculations of $D$, I resort to the transiting exoplanet data because $r$ is not known for most RV planets. For the transiting planets, $\bar{D} = 3.00\,(+4.69, -2.40) \times 10^{-3}$.[3] It is not clear whether a debiased $\bar{D}$ value would be larger or smaller than that. This depends on whether long-period planets usually have larger or smaller radii than those used for this analysis. For comparison, $\bar{D}$ of the 20 largest moons of the solar system giant planets is $6.319\,(+3.416, -4.543) \times 10^{-4}$.

The choice of the 20 largest local moons for comparison with the exoplanet data is arbitrary and motivated by the number of digiti manus. Figure 1(a)-(c) shows $\bar{P}$, $\bar{f}$, and $\bar{D}$ as functions of the number of natural satellites ($N_{\rm s}$) taken into account. Solid lines denote mean values, shaded fields standard deviations. The first moon is the largest moon, Ganymede, the second moon Titan, etc., and the 20th moon Hyperion. The trends toward higher transit probabilities (a), higher transit frequencies (b) and lower transit depths (c) are due to the increasing amount of smaller moons in short-period orbits. The negative slope at moons 19 and 20 in (a) and (b) is due to Nereid and Hyperion, which are in wide orbits around Neptune and Saturn, respectively.

The key message of this plot is that $\bar{P}(N_{\rm s} = 20)$, $\bar{f}(N_{\rm s} = 20)$, and $\bar{D}(N_{\rm s} = 20)$ used above for comparison with the exoplanet data serve as adequate approximations for any sample of solar system moons that would have been smaller because variations in those quantities were limited to a factor of a few.

## 2.3. Evolution of exomoon habitability

An exomoon hosting planet needs to be sufficiently far from its star to enable direct imaging and to reduce contamination from IR stellar reflection. This planet needs to orbit the star beyond several AU, where alternative energy sources are required on its putative moons to keep their surfaces habitable, i.e., to prevent freezing of $H_2O$. This heat could be generated by (1) tides (Reynolds et al. 1987; Scharf 2006; Cassidy et al. 2009; Heller & Barnes 2013), (2) planetary illumination (Heller & Barnes 2015), (3) release of primordial heat from the moon's accretion (Kirk & Stevenson 1987), or (4) radiogenic decay in its mantle and/or core (Mueller & McKinnon 1988). All of these sources tend to subside on a Myr timescale, but (1) and (2) can compete with stellar illumination over hundreds of Myr in extreme, yet plausible, cases. (3) and (4) usually contribute $\ll 1\,\mathrm{W\,m^{-2}}$ at the surface even in very early stages. Water cannot be liquid under these conditions and the object is uninhabitable by definition. For a Mars-sized moon, which I consider as a reference case, the runaway greenhouse limit is at $266\,\mathrm{W\,m^{-2}}$, following a semianalytic model (Eq. 4.94 in Pierrehumbert 2010). If the combined illumination plus tidal heating (or any alternative energy source) exceed this limit, then the moon is uninhabitable.

---

[1] Information for both $R$ and $a$ was given for 398 RV planets listed on http://www.exoplanet.eu as of 1 October 2015.
[2] The orbital period was given for 612 RV planets
[3] Information for both $r$ and $R$ was given for 1202 transiting planets.

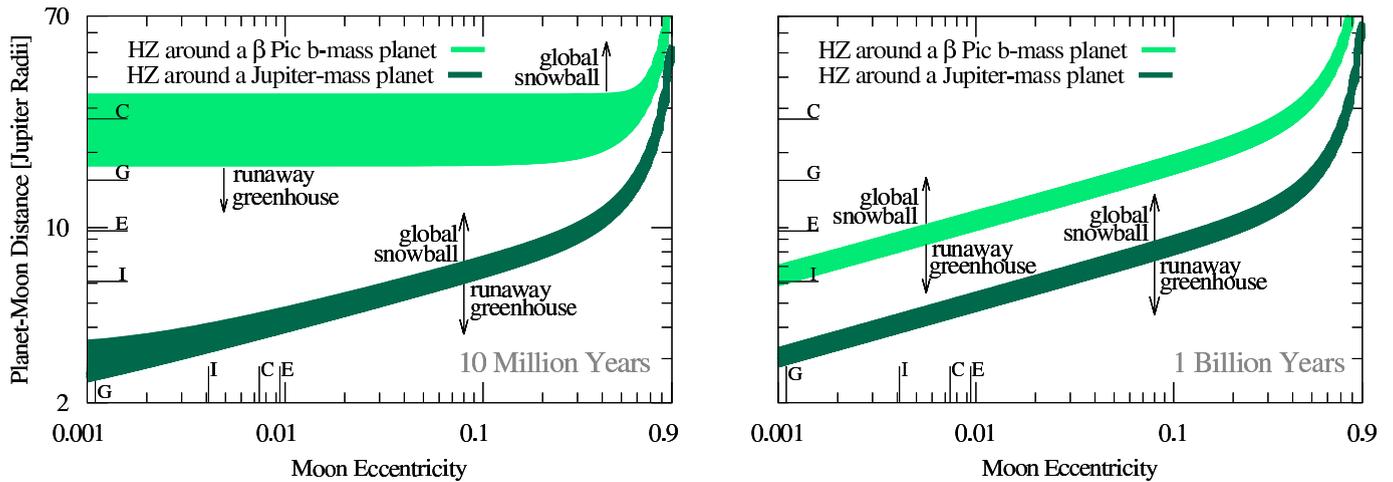

**Fig. 2.** Habitable zone (HZ) of a Mars-sized moon around a Jupiter-mass (light green) and a $\beta$ Pic b-like 11 $M_J$-mass (dark green) planet at $\geq 5$ AU from a Sun-like star. Inside the green areas, stellar illumination plus planetary illumination can liquify water on the moon surface. Values of the Galilean moons are denoted by their initials. *Left:* Both HZs refer to a planet at an age of 10 Myr, where planetary illumination is significant. *Right:* Both HZs refer to a planet at an age of 1 Gyr, where planetary illumination is weak and tidal heating is the dominant energy source on the moon.

On the other hand, there is a minim energy flux for a moon to prevent a global snowball state that is estimated to be 0.35 times the solar illumination absorbed by Earth ($83\,\mathrm{W\,m^{-2}}$, Kopparapu et al. 2013), which is only weakly dependent on the object's mass. I use the terms global snowball and runaway greenhouse limits to identify orbits in which a Mars-sized moon would be habitable, which is similar to an approach presented by Heller & Armstrong (2014). A snowball moon might still have subsurface oceans, but because of the challenges of even detecting life on the surface of an exomoon, I here neglect subsurface habitability. I calculate the total energy flux by adding the stellar visual illumination and the planetary thermal IR radiation absorbed by the moon to the tidal heating within the moon. Illumination is computed as by Heller & Barnes (2015), and I neglect both stellar reflected light from the planet and the release of primordial and radiogenic heat. Assuming an albedo of 0.3, similar to the Martian and terrestrial values, the absorbed stellar illumination by the moon at 5.2 AU from a Sun-like star is $9\,\mathrm{W\,m^{-2}}$. I consider a Jupiter-mass planet and a $\beta$ Pic b-like 11 $M_J$-mass planet, both in two states of planetary evolution; one, in which the system is 10 Myr old (similar to $\beta$ Pic b) and the planet is still very hot and inflated; and one, in which the system has evolved to an age of 1 Gyr and the planet is hardly releasing its thermal heat.

Planetary luminosities are taken from evolution models (Mordasini 2013), which suggest that the Jupiter-mass planet evolves from $R = 1.28\,R_J$ and $T_{\mathrm{eff}} = 536\,\mathrm{K}$ to $R = 1.03\,R_J$ and $T_{\mathrm{eff}} = 162\,\mathrm{K}$ over the said period. For the young 11 $M_J$-mass planet, I take $R = 1.65\,R_J$ (Snellen et al. 2014) and $T_{\mathrm{eff}} = 1\,700\,\mathrm{K}$ (Baudino et al. 2014). For the 1 Gyr version, I resort to the Mordasini (2013) tracks, predicting $R = 1.09\,R_J$ and $T_{\mathrm{eff}} = 480\,\mathrm{K}$. Tidal heating is computed by Heller & Barnes (2013), following an earlier work on tidal theory by Hut (1981).

Figure 2 shows the HZ for exomoons around a Jupiter-like (dark green areas) and a $\beta$ Pic b-like (light green areas) planet at ages of 10 Myr (left) and 1 Gyr (right). Planet-moon orbital eccentricities ($e$, abscissa) are plotted versus planet-moon distances (ordinate). In both panels, the HZ around the more massive planet is farther out for any given $e$, which is mostly due to the strong dependence of tidal heating on $M_p$. Most notably, the planetary luminosity evolution is almost negligible in the HZ around the 1 $M_J$ planet. In both panels, the corresponding dark green strip has a width of just 1 $R_J$, details depending on $e$.

The HZ around the young $\beta$ Pic b, on the other hand, spans from 18 $R_J$ to 33 $R_J$ for $e \lesssim 0.1$ (left panel). This wide range is due to the large amount of absorbed planetary thermal illumination, which is the main heat source on the moon in these early stages. Planetary radiation scales as $\propto a^{-2}$, causing a much smoother transition from a runaway greenhouse (at 18 $R_J$) to a snowball state (at 33 $R_J$) than on a moon that is fed by tidal heating alone, which scales as $\propto a^{-9}$. In the right panel, planetary illumination has vanished and tides have become the principal energy source around an evolved $\beta$ Pic b. However, the HZ around the evolved $\beta$ Pic b is still a few times wider (for any given $e$) than that around a Jupiter-like planet; note the logarithmic scale.

## 3. Discussion

A major challenge for exomoon transit observations around luminous giant planets is in the required photometric accuracy. Although *E-ELT* will have a collecting area 1 600 times the size of *Kepler*'s, it will have to deal with scintillation. The IR flux of an extrasolar Jupiter-sized planet is intrinsically low and comes with substantial white and red noise components. Photometric exomoon detections will thus depend on whether *E-ELT* can achieve photometric accuracies of $10^{-3}$ with exposures of a few minutes.[4] Adaptive optics and the availability of a close-by reference object, e.g., a low-mass star with an apparent IR brightness similar to the target, will be essential. Alternatively, systems with multiple directly imaged giant planets would provide particularly advantageous opportunities, both in terms of reliable flux calibrations and increased transit detection probabilities. Systems akin to the HR 8799 four-planet system (Marois et al. 2008, 2010) will be ideal targets.

Transiting moons could also impose RV anomalies on the planetary IR spectrum, known as the Rossiter-McLaughlin (RM) effect (McLaughlin 1924; Rossiter 1924). The RM reveals the sky-projected angle between the orbital plane of transiting object and the rotational axis of its host, which has now been measured for 87 extrasolar transiting planets[5]. *E-ELT* might be capable of RM measurements for large exomoons transiting giant exoplanets that can be directly imaged (Heller & Albrecht 2014).

Even nontransiting exomoons might be detectable in the IR RV data of young giant planets. The estimated RV $1\sigma$ confidence achievable on a $\beta$ Pic b-like giant exoplanet with a high-resolution ($\lambda/\Delta\lambda \approx 100\,000$, $\lambda$ being the wavelength of the observed light) near-IR spectrograph mounted to the *E-ELT* could be as low as $\approx 70\,\mathrm{m\,s^{-1}}$ in reasonable cases (Heller & Albrecht 2014). The RV amplitude of an Earth-mass exomoon in a Europa-wide ($a \approx 10 R_J$) orbit around a Jupiter-mass planet would be $43\,\mathrm{m\,s^{-1}}$ with a period of 3.7 d. RV detections of supermassive moons, if they exist, would thus barely be possible even with *E-ELT* IR spectroscopy. We should nevertheless recall that "hot Jupiters" had not been predicted prior to their detection (Mayor & Queloz 1995). In analogy, observational constraints could still allow for detections of a so-far unpredicted class of hot super-Ganymedes; they might indeed be hot due to enhanced tidal heating (Peters & Turner 2013) and/or illumination from the young planet (Heller & Barnes 2015).

Moon eccentricities tend to be tidally eroded to zero in a few Myr. Perturbations from other moons or from the star can maintain $e > 0$ for hundreds of Myr. With $e$ varying in time, exomoon habitability could be episodical. Uncertainties in the tidal quality factor ($Q$) and the 2nd order Love number ($k_2$), which scale the tidal heating as per $\propto k_2/Q$, can be up to an order of magnitude. However, because of the strong dependence on $a$, the HZ limits in Fig. 2 would be affected by $< 1 R_J$.

## 4. Conclusions

Large moons around the local giant planets transit their planets much more likely from a randomly chosen geometrical perspective and significantly more often than the RV exoplanets transit their stars. This is likely a fingerprint of planet and moon formation acting on different spatiotemporal scales. If the occurrence rate of planets per star is similar to the occurrence rate of large moons around giant planets (1-10 per system), the probability of observing a moon transiting a randomly chosen giant planet is at least four times higher than the probability of observing a planet transiting a randomly chosen star; planet-moon transits are at least four times more frequent than star-planets transits; the average transit depth of the transiting exoplanets is five times larger than the average transit depth of the twenty largest moons around the local giant planets. However, each solar system giant planet has at least one moon with a transit depth of $10^{-3}$. Following the gas-starved accretion disk model for moon formation (Canup & Ward 2006; Heller & Puditz 2015b), a super-Jovian exoplanet around the young super-Jovian exoplanet $\beta$ Pic b could have a transit depth of $\approx 1.5 \times 10^{-3}$, which is 18 times as deep as the transit of the Earth around the Sun.

Beyond the stellar HZ, more massive planets have wider circumplanetary HZs. The HZ for a Mars-sized moon around $\beta$ Pic b is between $18 R_J$ and $33 R_J$ for moon eccentricities $e \lesssim 0.1$. As the planet ages, the HZ narrows and moves in. After a few hundred Myr, tidal heating may become the moon's major energy source. Owing to the strong dependence of tidal heating on the planet-moon distance, the HZ around $\beta$ Pic b narrows to a few $R_J$ within 1 Gyr from now, details depending on $e$. An exomoon would have to reside in a very specific part of the $e$-$a$ space to be continuously habitable over a Gyr.

The high geometric transit probabilities of moons around giant planets, their higher transit frequencies, and the possibility of transit signals that are one order of magnitude deeper than that of the Earth around the Sun, make transit observations of moons around young giant planets a compelling science case for the upcoming *GMT*, *TMT*, and *E-ELT* extremely large telescopes. Most intriguingly, these exomoons could orbit their young planets in the circumplanetary HZ and therefore as cradles of life.

*Acknowledgements.* This study has been inspired by discussions with Rory Barnes, to whom I express my sincere gratitude. The reports of an anonymous referee helped to clarify several passages of this paper. I have been supported by the German Academic Exchange Service, the Institute for Astrophysics Göttingen, and the Origins Institute at McMaster University, Canada. This work made use of NASA's ADS Bibliographic Services.

---

[4] Transits of a moon that formed at the $H_2O$ ice line around a $\beta$ Pic b-like planet ($20 R_J$, Heller & Puditz 2015a) take about 1 hr:13 min.

[5] Holt-Rossiter-McLaughlin Encyclopaedia at http://www2.mps.mpg.de/homes/heller

## 6.6 How Eclipse Time Variations, Eclipse Duration Variations, and Radial Velocities Can Reveal S-type Planets in Close Eclipsing Binaries (Oshagh et al. 2017)

Contribution:

RH contributed to the literature research, computed the expected eclipse timing variations (ETVs) and eclipse duration variations (EDVs), wrote the equations into computer code, created Figs. 1, 3, 5, 8, and A.1, and contributed to the writing of the manuscript.

# How eclipse time variations, eclipse duration variations, and radial velocities can reveal S-type planets in close eclipsing binaries

M. Oshagh[1]⋆, R. Heller[2]†, and S. Dreizler[1]

[1]*Institut für Astrophysik, Georg-August-Universität, Friedrich-Hund-Platz 1, 37077 Göttingen, Germany*
[2]*Max Planck Institute for Solar System Research, Justus-von-Liebig-Weg 3, 37077 Göttingen, Germany*

## ABSTRACT

While about a dozen transiting planets have been found in wide orbits around an inner, close stellar binary (so-called "P-type planets"), no planet has yet been detected orbiting only one star (a so-called "S-type planet") in an eclipsing binary. This is despite a large number of eclipsing binary systems discovered with the *Kepler* telescope. Here we propose a new detection method for these S-type planets, which uses a correlation between the stellar radial velocities (RVs), eclipse timing variations (ETVs), and eclipse duration variations (EDVs). We test the capability of this technique by simulating a realistic benchmark system and demonstrate its detectability with existing high-accuracy RV and photometry instruments. We illustrate that, with a small number of RV observations, the RV-ETV diagrams allows us to distinguish between prograde and retrograde planetary orbits and also the planetary mass can be estimated if the stellar cross-correlation functions can be disentangled. We also identify a new (though minimal) contribution of S-type planets to the Rossiter-McLaughlin effect in eclipsing stellar binaries. We finally explore possible detection of exomoons around transiting luminous giant planets and find that the precision required to detect moons in the RV curves of their host planets is of the order of $cm\,s^{-1}$ and therefore not accessible with current instruments.

**Key words:** methods: numerical, observational– planetary system–stars: binaries–techniques: photometric, radial velocities, timing

## 1 INTRODUCTION

Among the detection of thousands of extrasolar planets and candidates around single stars, the *Kepler* telescope has also delivered the first transit observations of planets in stellar multiple systems, eleven in total as of today (Doyle et al. 2011; Welsh et al. 2012; Orosz et al. 2012a,b; Schwamb et al. 2013; Kostov et al. 2014; Welsh et al. 2015; Kostov et al. 2016). Most of these binaries show mutual eclipses, therefore allowing precise radius estimates of both stars and planets in a given system. All of these planets are classical "circumbinary planets", or P-type planets (Dvorak 1986), orbiting on the circumbinary orbit around an inner stellar binary (or their common center of mass) on a wide orbit. In other systems, the planetary transit cannot be observed directly, but stellar eclipse timing variations (ETVs) still signify the presence of one or more P-type planets (Beuermann et al. 2010;

Qian et al. 2010; Beuermann et al. 2011; Schwarz et al. 2011; Qian et al. 2012; Beuermann et al. 2012, 2013; Marsh et al. 2014; Schwarz et al. 2016), some of which are still being disputed (Horner et al. 2012, 2013; Bours et al. 2016).

Planets also exist in S-type configurations ("S" for satellite), where the planet orbits a star in a close orbit, while both the planet and its host star are gravitationally bound to an additional star.[1] Many of the known exoplanets are indeed S-type planets, mostly in wide-orbit binaries. A handful of S-type planets are orbiting stellar binaries with separations ≲ 20 AU (the tightest known binary which hosts a s-type planet is KOI-1257 system with the semi-major axis of 5.3 AU (Santerne et al. 2014)), but none of these systems are eclipsing. In other words, there has been no detection of an S-type planet in a stellar eclipsing binary.

---

⋆ E-mail: moshagh@astro.physik.uni-goettingen.de
† E-mail: heller@mps.mpg.de

---

[1] A third configuration is referred to as 'L-type' (or sometimes "T-type", "T" for Trojan), where the planet is located at either the $L_4$ or $L_5$ Lagrangian point (Dvorak 1986). Such planets have been suggested, and could possibly be detectable via eclipse time variations (Schneider & Doyle 1995; Schwarz et al. 2015), but remain undiscovered as of today.

Formation of S-type planets, particularly of giant planets, in close binaries may be hampered by gravitational and dynamical perturbations from the stars (Whitmire et al. 1998; Nelson 2000; Thébault et al. 2004; Haghighipour 2009; Thebault & Haghighipour 2014). However, to explain the discovered S-type planets in close (non-eclipsing) binaries, several solutions have been proposed. For instance, a binary orbit might be initially wide and tighten only after planet formation and migration have completed (Malmberg et al. 2007; Thébault et al. 2009; Thebault & Haghighipour 2014). Alternatively, S-type planets, giant or terrestrial, in close eclipsing binaries could build up through other formation scenarios, such as gravitational instabilities (Boss 1997) or in-situ core-accretion (Boss 2006; Mayer et al. 2007; Duchêne 2010).

Therefore, S-type planets in close eclipsing binaries might be expected to be rare from both a formation and an orbital stability point of view (Dvorak 1986; Holman & Wiegert 1999; Horner et al. 2012). Yet, the discoveries of planets around pulsars (Wolszczan & Frail 1992), of the previously unpredicted families of hot Jupiters (Mayor & Queloz 1995), and of circumbinary planets (Doyle et al. 2011) remind us that formation theories do not necessarily deliver ultimate constraints on the actual presence of planet populations. As a consequence, the discovery of an S-type planet in close eclipsing binaries would have major impact on our understanding of planet formation and evolution. The detection of almost 3,000 close eclipsing binaries with *Kepler* (Kirk et al. 2016) suggests that a dedicated search could indeed be successful. In Figure 1, we show the orbital period (P) distribution of these systems.[2] Note that binaries with $P < 10$ d have Hill spheres of about ten solar radii or smaller, so any planets might be subject to removal over billions of years. In systems with periods beyond 10 d, however, S-type planets might be long-lived and present today.

Here we report on a new method to detect S-type planets in close eclipsing binaries using radial velocities (RVs), eclipse timing variations (ETVs), and eclipse duration variation (EDVs). It is an extension of recently developed methods to find extrasolar moons around transiting exoplanets via the planet's transit timing variations (TTVs; Sartoretti & Schneider 1999) and transit duration variations (TDVs; Kipping 2009a) independent of the moon's direct photometric signature. In comparison to S-type exoplanet searches based on ETVs only, the combination of ETVs and EDVs provides a non-degenerate solution for the planet's mass ($M_p$) and its semi-major axis ($a$) around the secondary star (Kipping 2009a). The RV method, using proper treatment of the stellar cross-correlation functions (CCFs) as described in this paper, offers additional measurements of $M_p$, which must be consistent with the value derived from the ETV-EDV approach. What is more, the RV-ETV diagram offers a means to determine the sense of orbital motion of the planet (prograde or retrograde).

This paper is organised as follows: In Section 2 we describe the basic idea behind our method and we describe the simulations which we perform to evaluate the feasibility of our method. In Section 3. we present the results of our tests. We summarise our conclusions in Section 4.

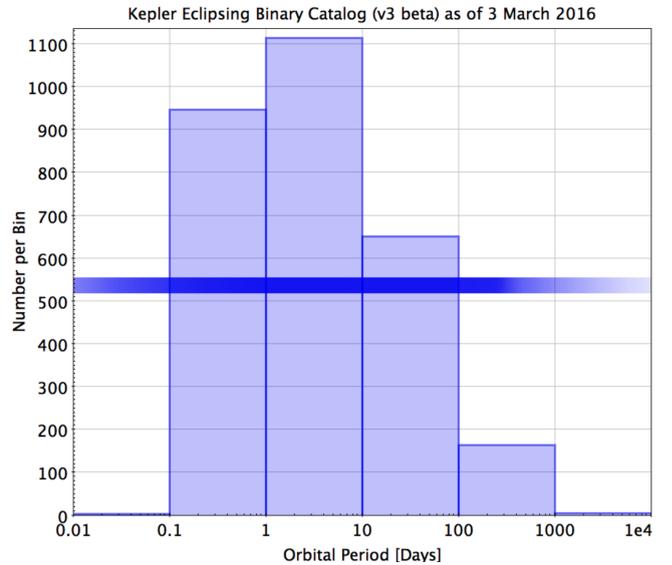

**Figure 1.** Period distribution of almost 3,000 stellar eclipsing binaries observed by *Kepler* space telescope from http://keplerebs.villanova.edu (Kirk et al. 2016). The horizontal bar represents a density distribution of the histogram with dark blue corresponding to the highest densities and white corresponding to zero.

## 2 METHODS

### 2.1 Principles of RV, ETV, and EDV correlations

Consider an eclipsing binary system consisting of a primary star, which shall be the more massive one, and a secondary star. Without loss of generality, the secondary star shall be orbited by a close-in planet. Then the planet will cause the secondary star to perform a curly, wobbly orbit around the center of mass of the stellar binary. Hence, the eclipses will not be exactly periodic: sometimes they will be too early, at other times they will be too late compared to the average eclipse period. The resulting ETVs are essentially equivalent to the previously predicted TTVs of transiting planets due to the gravitational perturbations of an extrasolar moon (Sartoretti & Schneider 1999). Additionally, the tangential orbital velocity component of the secondary star in the star-planet system causes EDVs, which are analogous to the transit duration variations of a transiting planet with a moon (Kipping 2009a). Both ETVs and EDVs (or TTVs and TDVs) are phase-shifted by $\pi/2$, that is, they form an ellipse in the ETV-EDV (or TTV-TDV) diagram (Heller et al. 2016). By using high-precision photometric time series, e.g. with the *Kepler* space telescope, very high-precision ETV-EDV measurements can be obtained. *Kepler's* median timing precision turned out to be $\lesssim 30$ s (Welsh et al. 2013) and ETV accuracies obtained by Borkovits et al. (2016) are typically below this level.

Beyond ETVs and EDVs, the planet imposes a periodic RV variation onto its host star. RV observations of the secondary star, however, would be challenging because, first, first, the spectrum of the secondary star would be diluted by the primary star; and second, the RV signal of the secondary star will be on top of a much more pronounced RV signal of the stellar binary. To tackle the latter aspect, we propose to obtain the RV measurements exactly just before or after eclipse. Thus, as the orbital geometry of the stellar binary will be consistent during each such RV measurement, the RV component of the stellar binary will be equal for each measurement as well. In particular, the binary's RV component will be zero (see Figure 2a) and the observed RV will be due to the secondary's motion around the secondary-planet center of mass only (see Figure 2d). By limiting ourselves to measurements just before or after eclipse, we also avoid contributions of light travel time effects to our ETV measurements (Woltjer 1922; Irwin 1952).

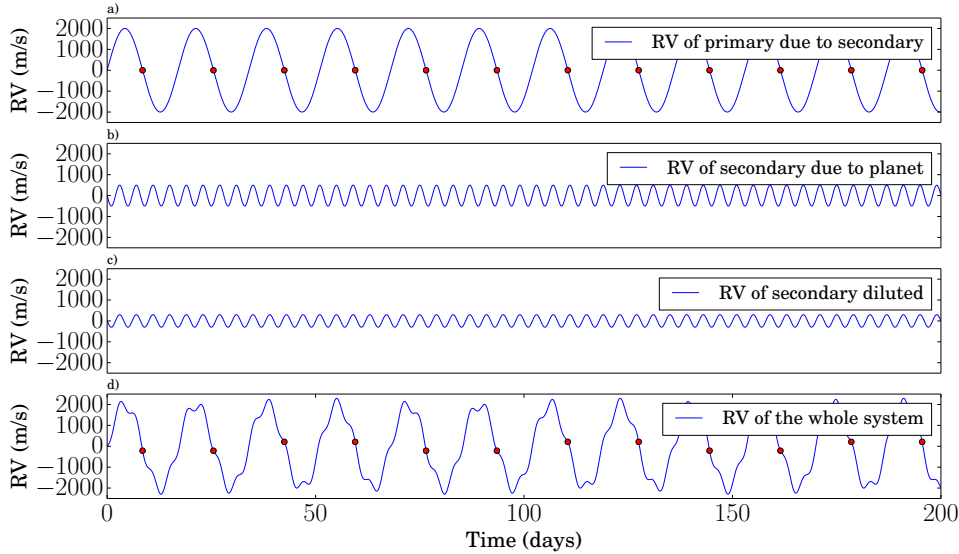

**Figure 2.** Radial velocity components of a binary system with an S-type planet around the secondary star. Panel **a)** shows the RV of the primary due to the gravitational pull of the secondary. Red dots denote times of eclipses, when the secondary star is in front of the primary (as seen from Earth) and the RV signal is zero. Panel **b)** illustrates the true RV variation of the secondary star due to its S-type planet and panel **c)** depicts the secondary's RV, which is diluted by the brightness contrast between the primary and the secondary. Panel **d)** shows the total RV of the system, i.e. the sum of **a)** and **c)**.

In Figure 3, we illustrate the correlation of ETVs with RVs and of EDVs with RVs acquired around eclipse. The upper panel refers to a prograde planetary orbit, where the planet's sense of orbital motion is the same as the stellar binary's sense of orbital motion; the lower panel depicts a retrograde planetary orbit. The key feature of the RV-ETV correlation is in the slope of the observed curve. A negative slope (upper panel) indicates a prograde planetary orbit, whereas a positive slope (lower panel) is suggestive of a retrograde orbit. We arbitrarily normalized both the RV and the EDV amplitudes to obtain a circle. In general, however, the RV-EDV figure will be an ellipse with its semi-major and semi-minor axis given by the RV and EDV amplitudes (or vice versa).

## 2.2 Feasibility tests for S-type planets and exomoons

To assess the feasibility of our method, we simulate observations of two toy systems.

### 2.2.1 A Sun-like primary and a K-dwarf secondary with an Jupiter-sized S-type planet

Our first test case consists of a Sun-like G-dwarf primary star and K-dwarf secondary star hosting an S-type Jupiter-style planet. The orbital period of the binary is chosen to be $50 \, \mathrm{d}$ and the planet is put in a $4 \, \mathrm{d}$ orbit. This choice is dominantly motivated by constraints on the long-term orbital stability of the planet. Numerical simulations by Domingos et al. (2006) showed that the outermost stable orbit of a prograde satellite is at about half the Hill radius around the secondary, which translates into a maximum orbital period for the planet of about $1/9$ the orbital period of the binary (Kipping 2009a).

We consider the CCFs of the publicly available HARPS spectra of 51 Peg (our primary) and HD 40307 (our secondary) as reference stars.[3] In Table 1 we list their physical parameters. As mentioned above, our hypothetical RV observations are supposed to be taken near eclipse when the primary's CCF will have zero RV. We therefore correct the CCF of 51 Peg to be centered around zero RV throughout our computations. The RV amplitude $(K)$ imposed by a $4 \, \mathrm{d}$ Jupiter-sized planet on an HD 40307 is about $150 \, \mathrm{m \, s^{-1}}$. Hence, we periodically shift the secondary's CCF according to the planet's orbital phase with an amplitude of $K = 150 \, \mathrm{m \, s^{-1}}$. We normalize the secondary CCF to the

---

[3] http://archive.eso.org/wdb/wdb/adp/phase3_spectral/form?collection_name=HARPS

Prograde Orbit of the Secondary with Companion

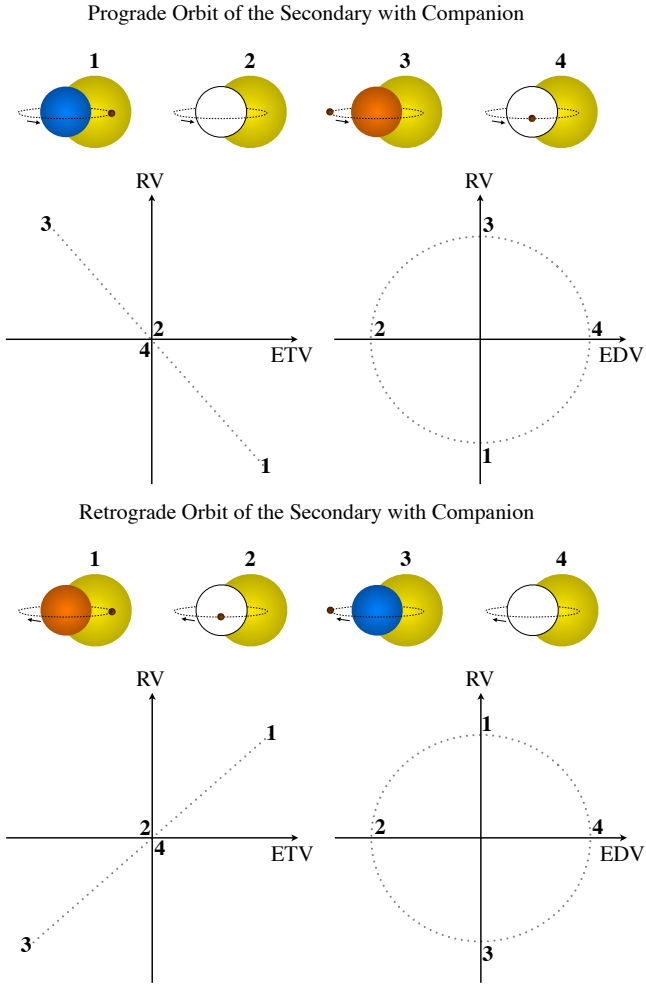

**Figure 3.** **Top**: Illustration of the RV-ETV and RV-EDV correlations for a prograde S-type planet (small black circle) around a secondary star (intermediate circle) eclipsing a primary star (big yellow circle). The colors of the secondary indicate its instantaneous RV. Blue means a negative RV-shift (toward the observer), white means zero RV, red means positive contribution (away from the observer). Arrows indicate the planet's direction of motion around the secondary. Images 1-4 depict four specific scenarios for the secondary-planet orbital geometry during eclipse, which can be found in the respective diagrams. **Bottom**: Same as top but for a retrograde planet.

total flux of the system to properly model the brightness contrast between the two stars (see Figure 4). We then add the shifted CCF of the secondary and the zero-RV CCF of the primary to obtain the combined CCF ($CCF_{tot}$). For different orbital phases of the planet around the secondary, we then fit a Gaussian function to $CCF_{tot}$ to approximate an observed RV measurement of the unresolved system. A similar approach was taken by Santos et al. (2002) to study the RV signal of HD 41004 AB.

To estimate the ETV and EDV amplitudes of the secondary star due its planet, we utilize the analytical framework of Kipping (2009a) and Kipping (2009b), which has originally been developed to estimate the TTVs and TDVs of transiting planets with moons. Each individual ETV and EDV measurement in our simulations is calculated by weigh-ing the respective amplitude with a sine function of the planetary orbital phase, ETV and EDV being shifted by $\pi/2$.

**Table 1.** Physical parameters of 51 Peg and HD 40307.

| Star | Mass $M_\odot$ | Radius $R_\odot$ | $T_{eff}$ K | $v\sin(i)$ km s$^{-1}$ | $m_V$ |
|------|------|------|------|------|------|
| 51 Peg | 1.11 | 1.266 | 5793 | 2.8 | 5.49 |
| HD 40307 | 0.77 | 0.68 | 4977 | < 1 | 7.17 |

#### 2.2.2 An M-dwarf and a primordially transiting hot Jupiter with a Neptune-sized exomoon

Our second test case consists of an M-dwarf and a luminous Jupiter-like transiting planet with a Neptune-mass invisible moon. The orbital periods are the same as we used for the first toy system and Hill stability is ensured. The Jupiter-like transiting planet cannot be considered as a canonical "hot Jupiter" because the illumination from the M-dwarf would be too weak to heat it up significantly. Hence, in order to make the planet visible in thermal emission, its intrinsic luminosity would need to be fed by primordial heat, which could be sufficient for detection during the first $\sim 100$ Myr after the formation of the system.

We model the CCF of the M-dwarf using a HARPS observed CCF of GJ 846, an M0.5 star (Henry et al. 2002) and with a mass of approximately 0.63 solar masses ($M_\odot$), assuming solar age and metallicity (Chabrier & Baraffe 2000). Since there is no direct observation of any hot Jupiter emission spectrum, there also is no observed CCF available for our simulations. Instead, we use the CCF of 51 Peg b observed by Martins et al. (2015) in reflected light. We consider a Gaussian function with same depth and width as the CCF of 51 Peg b reported by Martins et al. (2015) for the transiting planet.[4]

Our preliminary tests showed that no moon, which can reasonably form through in-situ accretion within the gas and dust accretion disk around a young giant planet (a few Mars masses at most; Canup & Ward 2006; Heller & Pudritz 2015), could induce a detectable RV or ETV-EDV signature. Hence, to explore the detection limit, we assumed an extreme version of a Neptune-sized satellite around the Jovian-style transiting planet, which essentially results in a planetary binary orbiting the M-dwarf host star. The $K$ amplitude of this planetary binary, which we use to shift the planetary CCF as a function of the planet's orbital phase around the common center of mass, is around 500 m s$^{-1}$.

### 2.3 In-eclipse RVs and the Rossiter-McLaughlin effect

As the primary star rotates, one hemisphere is moving toward the observer (and therefore subject to an RV blueshift), while the other hemisphere is receding (and therefore red-shifted). During eclipse of a prograde binary, the secondary star first occults part of the blue-shifted hemisphere and, thus, induces an RV red-shift. During the second half of an eclipse, assuming zero transit impact parameter, the RV shift is blue. The resulting RV signal, known as the Rossiter-McLaughlin effect (RM) (Rossiter 1924; McLaughlin 1924), has been used to determine the projected stellar rotation velocity ($v \sin i$) and the angle between the sky-projections of the stellar spin axis and the orbital plane of eclipsing binaries, and recently has been used for the transiting exoplanets in almost 100 cases as of today.[5]

---

[4] Note that 51 Peg b is orbiting on a shorter period orbit, therefore it is much brighter in the reflected light than our test planet in a 50 d orbit.

We suppose that the presence of an S-type planet around the secondary star could create an additional RV component in the RM signal because the light coming from the secondary star will show different RV shifts during subsequent eclipses (see Figure 3). To test our hypothesis, we use the publicly available SOAP-T software (Oshagh et al. 2013a)[6], which can simulate a rotating star being occulted by a transiting foreground object. SOAP-T generates both the photometric lightcurve and the RV signal and it can also deliver the time-resolved CCFs of the system.

We assume that the primary star rotates with the rotation period of 24 days, which is close to the Sun's value. We simulate the eclipse of the secondary and generate the CCFs of the combined three orbital phases of the secondary star around its common center of mass with its S-type planet. One orbital phase simply assumes a zero RV shift of the secondary, corresponding to an orbital geometry where the secondary and its planet are both on the observer's line of sight. The second scenario assumes the secondary is moving toward the observer with its $K$ amplitude (see Section 2.2) during eclipse and so the secondary's CCF is maximally blue-shifted in this eclipse simulation. We then estimate the observed RV by fitting a Gaussian to all those CCFs, and obtain the RM signal that is contaminated by a redshifted secondary. Finally, the third scenario assumes a maximum redshift of the secondary during eclipse.

## 3 RESULTS

In the following, we focus on our first test case of a Sun-like host with a K-dwarf companion and a Jovian S-type planet (Section 2.2.1). Our second test case of an M-dwarf with a planetary binary is considered in a separate Section 3.4.

### 3.1 RV correlations with ETVs and EDVs

We first explore the correlation between the secondary's RVs, as measured directly from the combined dG-dK CCF ($\text{CCF}_{\text{tot}}$), with their corresponding ETV and EDV observations. Figure 5 shows the RV-ETV and RV-EDV patterns for both a prograde (upper panels) and a retrograde (lower panels) S-type planet. Nominal error bars of 30 s (*Kepler*'s median timing precision; Welsh et al. 2013) for ETVs and EDVs as well as $1\,\mathrm{m\,s^{-1}}$ for the RV simulations (obtainable with HARPS) are shown in each panel. The sequence of the first five measurements is indicated with red symbols and numbers. As a reading example, consider measurement "1" in the upper row. A maximum ETV observation of $\sim +3$ min in panel (a)[7] means that the passage of the secondary star in front of the primary is maximally delayed. With regards to Figure 3, the secondary has its maximum deflection to the left (see configuration labelled "1") so that the prograde planet is maximally deflected to the right. The prograde motion of the planet then implies that the planet is receding from the observer while the star is approaching and, hence, the latter is subject to a maximal RV blue shift of about $-40\,\mathrm{m\,s^{-1}}$ in Figure 5a. Regarding the EDV component of "1" in Figure 5b, it must be zero as the secondary's motion around the local secondary-planet center of mass has no tangential velocity component with respect to the observer's line of sight.

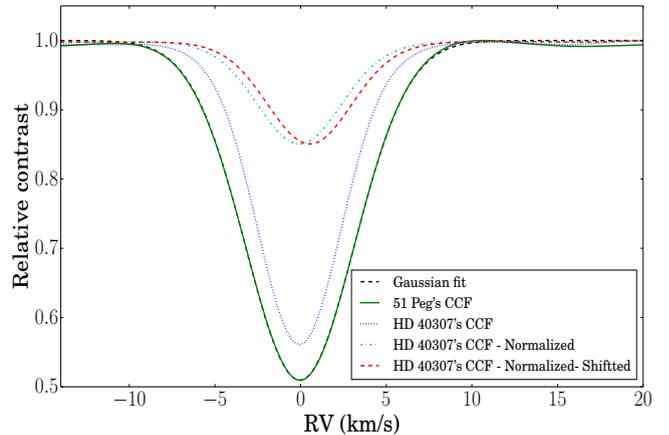

**Figure 4.** The solid green line represents the CCF of 51 Peg as observed with HARPS and normalized to its continuum. The black dashed line (used in Section 3.3) shows the best Gaussian fit to the CCF of 51 Peg. The dotted blue line shows the CCF of HD 40307 obtained with HARPS, and normalized to its continuum. The dashed dotted cyan line shows the CCF of HD 40307 at zero RV and normalized to the total flux of system. The dashed red line illustrates the same CCF of HD 40307 but shifted by the maximum RV induced by our S-type test planet ($K = 150\,\mathrm{m\,s^{-1}}$).

One intriguing result of these simulations is that the slopes of the RV-ETV figures for the prograde (panel a) and the retrograde planet (panel c) indeed exhibit different algebraic signs. The prograde scenario shows a negative slope, the retrograde case a positive slope, as qualitatively predicted in Section 2.1. Also note that the sequence of simultaneous RV-EDV measurements in the prograde case (panel b) is a horizontally mirrored version of the retrograde case (panel d). The main advantage of our method is that by obtaining a small number of RV observations we will be able to detect the S-type planet in the system, distinguish between the prograde and retrograde orbit.

---

[7] Sequences of both ETVs and EDVs always relate to an average eclipse period (for ETVs) or average eclipse duration (for EDVs). Hence, the first eclipse observation would, by definition, have no ETV or EDV. Only once multiple eclipses have been observed will the ETV-EDV measurements settle along the figures shown in Figure 5 and measurement "1" will turn out to have an ETV of about +3 min as in this example.

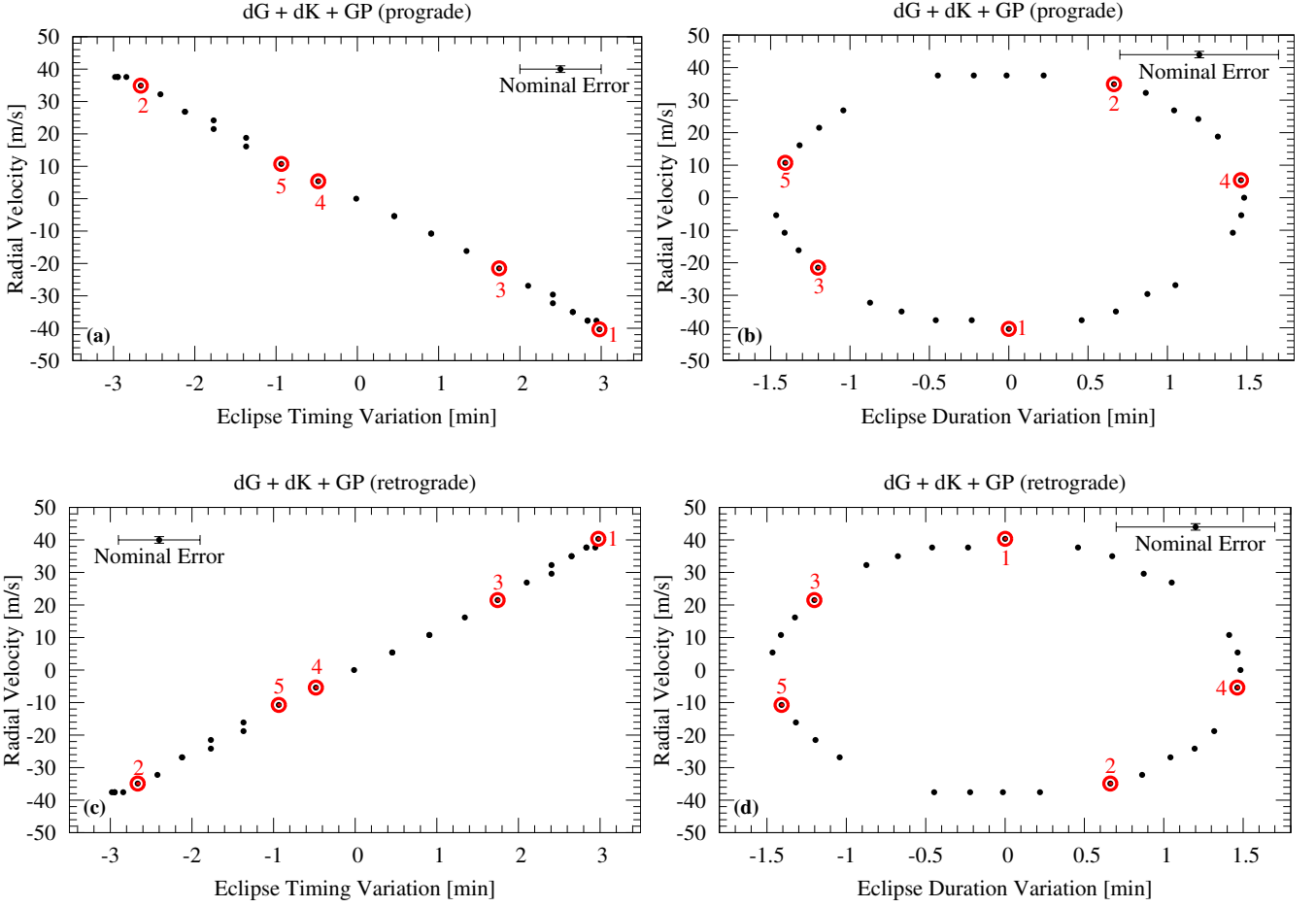

**Figure 5.** Correlation of the RVs, ETVs, and EDVs of the secondary K-dwarf star due to a giant planet (GP) companion, both orbiting a dG primary star. In each panel, the sequence of the first five measurements is labeled with red numbers. Nominal error bars of $\pm 1\,\mathrm{m\,s^{-1}}$ in RV and 30 s in both ETV and EDV are shown in each panel. Panels **(a)**-**(b)** assume a prograde sense of orbital motion of the dK-GP system. Panels **(c)**-**(d)** assume a retrograde sense of orbital motion.

In Appendix A, we also present the correlation between the Bisector Inverse Slope (BIS) (Queloz et al. 2001) and the FWHM of the CCFs and the ETV.

### 3.2   Planetary mass estimation

By fitting the Gaussian profile to CCF$_{\mathrm{tot}}$, we find that the RV variation of the secondary due to its Jupiter-sized S-type planet is strongly diluted by the primary star. We detect an RV amplitude of about $40\,\mathrm{m\,s^{-1}}$. Hence, the corresponding mass of the planet would be strongly underestimated. Our simulated RV observations are shown Figure 6.

To derive a more accurate measurement of planet's mass, we remove the contribution of primary light from CCF$_{\mathrm{tot}}$. We first fit a Gaussian function to the CCF of the primary's star and then use this fit as a template representing the CCF of primary (as shown in Figure 4 in dashed black line). We then remove this template Gaussian from all simulated CCF$_{\mathrm{tot}}$ measurements, which in reality would represent the observed CCF of the unresolved binary system.

We then fit a Gaussian function to all measurements in our simulated CCF time series, from which the primary's contribution has been removed, to estimate the RVs of the secondary due its S-type planet. We retrieve the RV variation amplitude induced by the S-type planet on the secondary star that is $85\,\% \pm 1\,\%$ of the actual value (see Figure 6), which translates into an underestimation of the minimum planetary mass by $15\,\% \pm 1\,\%$.

### 3.3   Distortions of the Rossiter-McLaughlin effect

While RV-ETV and RV-EDV correlations require the RV measurements to be taken either slightly before or after eclipse in order to minimize effects of the eclipsing star on the spectrum of the primary in the background (Section 2.1), we now consider the actual in-eclipse RVs of the system as described in Section 2.3.

In Figure 7 we show the impact of the S-type planet on the observed RM as derived from the total CCF of the dG-dK binary. Although the additional effect is only a few percent ($\sim 1.5\,\mathrm{m\,s^{-1}}$) of the total RM amplitude ($\sim 75\,\mathrm{m\,s^{-1}}$), it could be detectable with current high-precision spectrographs. If not taken into account properly, the asymmetry in the RM contributions could lead to inaccurate estimations of the spin-orbit angle of the eclipsing binary. But if the variation in the RM amplitude during subsequent eclipses can be correlated with the presence of an S-type planet, then it could serve as an additional means of verifying the S-type planet. That being said, note that there are several other phenomena that can affect the RM signal, such as gravitational microlensing (Oshagh et al. 2013c), the convective blueshift (Cegla et al. 2016), and occultations active region on the background star (Oshagh et al. 2016).

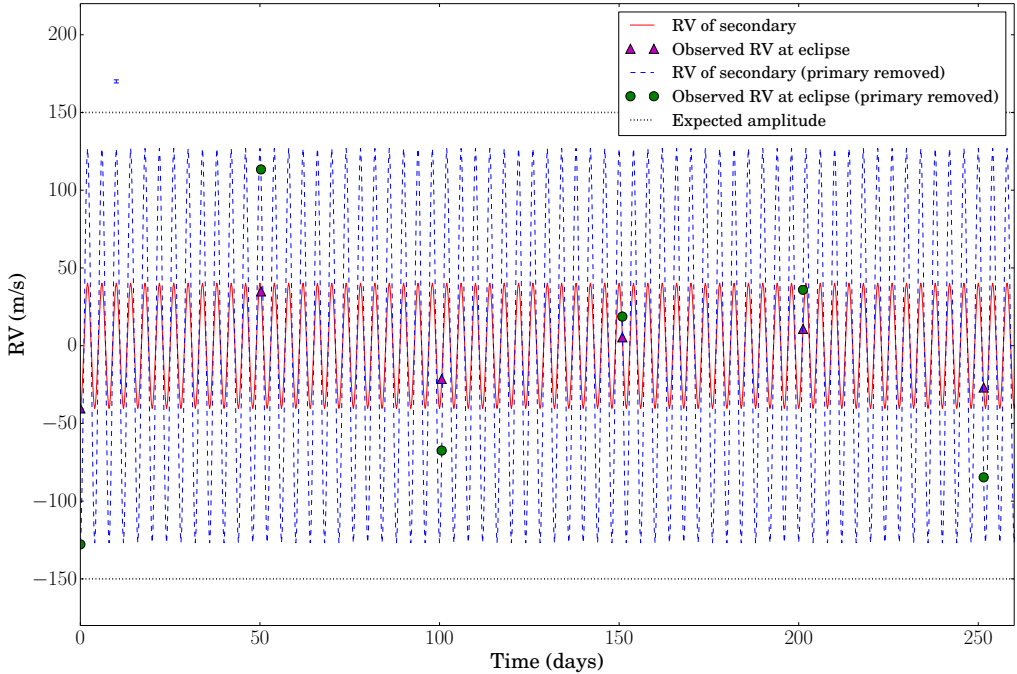

**Figure 6.** The red solid line represents the RV signal of the unresolved system derived by fitting a Gaussian profile to CCF$_{tot}$. Observed RVs near eclipses are represented with the magenta triangles. The dashed blue line displays the RV of the system after removing the contribution of the primary. Green circles show observed RVs near eclipses. The black dotted line at $K = 150\,\mathrm{m\,s^{-1}}$ presents the actual amplitude of RV signal due to S-type planet. A nominal error bar of $\pm 1\,\mathrm{m\,s^{-1}}$ is indicated in the upper left corner.

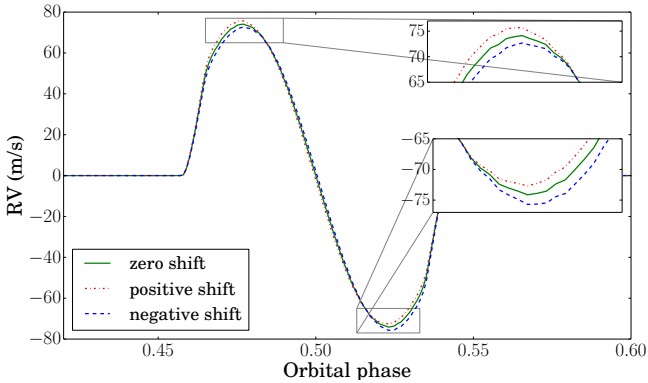

**Figure 7.** The green solid line shows the RM effect in the combined spectrum of a dG-dK binary with the K-dwarf star passing in front of the dG star. The dashed blue line refers to the same system, but now the K-dwarf star is subject to a maximum blue shift due to a Jupiter-sized S-type planet in a 4 d orbit. The dashed-dotted red line refers to the reverse motion, where the secondary is moving away from the observer.

### 3.4 Detectability of extremely massive moons around Jovian planets

Our simulations of an extremely massive moon around a transiting, luminous giant planet in orbit about an M-dwarf star are shown in Figure 8. The predicted RV amplitude of the observations is of order of cm s$^{-1}$. Note that the amplitude of the planet's actual RV motion of $500\,\mathrm{m\,s^{-1}}$ is much larger than the $150\,\mathrm{m\,s^{-1}}$ that we estimated for the K-dwarf due its S-type planet in our first test case. However, the flux ratio between a Jovian planet and its M-dwarf host star is much lower.

Hence, we find that the precision required to detect even a ridiculously massive exomoon, or essentially a binary planet, is unachievable with current RV facilities. Near future facilities such as ESPRESSO at the Very Large Telescope might be able to go down to the cm s$^{-1}$ accuracy level in favorable cases (Pepe et al. 2010), so future detections of highly massive exomoons or binary planets might be possible in principle, but will certainly be challenging.

### 4 CONCLUSION

We present a novel theoretical method to detect and verify S-type planets in stellar eclipsing binaries by correlating the RVs of the secondary star with its ETVs and EDVs. We test the applicability of our method by performing realistic simulations and find that it can be used to detect a short-period S-type planet (e.g. a hot Jupiter) around a K-dwarf or lower-mass secondary star in a moderately wide orbit ($\sim 50$ d) around a Sun-like primary star, e.g. using *Kepler* photometry and HARPS RV measurements. We also find that the RV-ETV diagram can be used to distinguish between prograde and retrograde S-type orbits. The sense of orbital motion is a key tracer of planet formation and migration, so the RV-ETV correlation identified in this paper could be very useful in studying the origin of close-in planets in binaries.

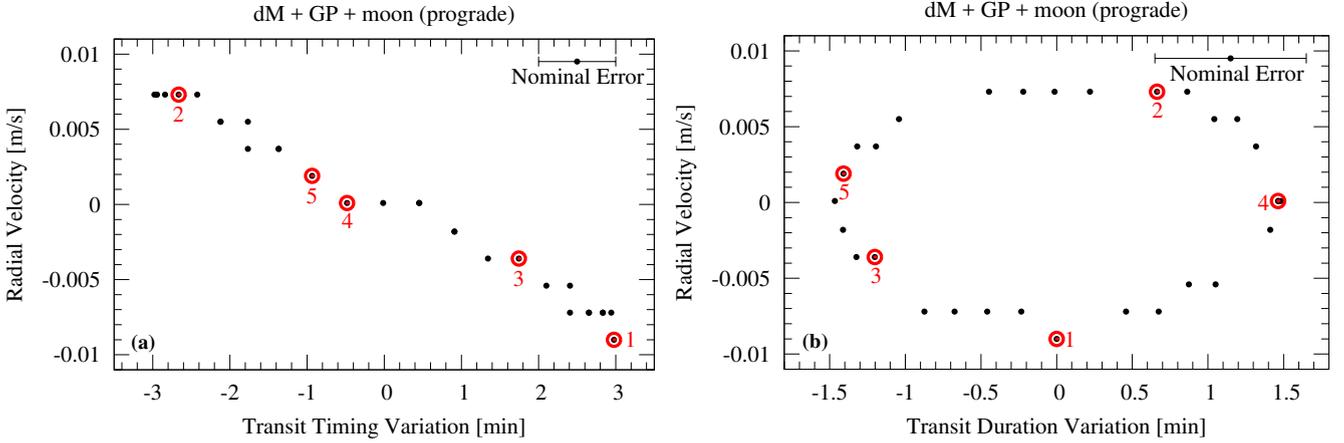

**Figure 8.** RVs, TTVs, TDVs of the transing planet due to a Neptune-sized moon. The host star is of spectral type M-dwarf. We assume a prograde sense of orbital motion of the exomoon system. Nominal error bars of 30 s in both TTV and TDV are shown in each panel. Nominal error bars of RV measurements are assumed to be in the order of $10 \, \mathrm{cm \, s^{-1}}$.

We show that the removal of the primary's CCF from the combined CCF of the unresolved stellar binary can yield realistic estimates of the planetary mass through RV measurements. With the ETV-EDV relation offering a methodologically independent measurement, we find that combined RV-ETV-EDV observations offer a means to both detect and confirm/validate S-type planets at the same time. ETV-EDV measurements also deliver the planet's orbital semi-major axis around the secondary star. In this paper, we propose that RV observations be taken near eclipse in order to correlate them with ETVs and EDVs. After about a dozen eclipses, or if additional RV measurements could be taken far from eclipse, it could be possible to securely identify the planet's orbital period around the secondary star. And if the secondary's mass can be estimated from its spectrum and using stellar classification schemes (e.g. stellar evolution models), then RVs could also yield an independent measurement of the planet's semi-major axis, which needs to be in agreement with the value derived from the ETV-EDV data.

Various physical phenomena can mimic RV-ETV and RV-EDV correlations, such as the stellar activity. Transiting planets crossing stellar active regions, for instance, can cause TTVs and TDVs (Oshagh et al. 2013b). It is also well-known that active regions on a rotating star affect the CCF, and thus produce RV variations, even if the planet does not transit (Queloz et al. 2001). We therefore presume that it is plausible that stellar activity produces some kind of correlation between RVs and ETVs or EDVs, although it might be very different from the patterns we predict for S-type planets.

Eclipsing binaries from *Kepler* are usually faint with typical *Kepler* magnitudes $11 \lesssim m_\mathrm{K} \lesssim 15$ (Borkovits et al. 2016). RV accuracy of $\sim 1 \, \mathrm{m \, s^{-1}}$ will be hard to achieve for many of these systems. The PLATO mission, scheduled for launch in 2025, will observe ten thousands of bright stars with $m_\mathrm{V} < 11$ (Rauer et al. 2014), many of which will turn out to be eclipsing binaries. PLATO will therefore discover targets that allow both more accurate ETV-EDV measurements and high-accuracy ground-based RV follow-up.

# APPENDIX A: VARIATIONS OF THE BISECTOR SPAN AND THE FULL WIDTH AT HALF MAXIMUM

As an extension of the data shown in Figure 5, we append figures of the bisector inverse slope (BIS) (Queloz et al. 2001) and of the full width at half maximum (FWHM) of the CCFs as a function of the stellar ETVs. The data refers to the prograde scenario of a Jupiter-sized S-type planet in a 4 d orbit around a K-dwarf star, both of which orbit a Sun-like primary star every 50 d (see Section 2.2.1). Figure A1 clearly reveals additional BIS-ETV and FWHM-ETV correlations. However, our follow-up simulations did not indicate a unique correlation between the planet's sense of orbital motion and either the BIS or the FWHM.

## ACKNOWLEDGEMENTS

MO acknowledges research funding from the Deutsche Forschungsgemeinschaft (DFG , German Research Foundation) - OS 508/1-1. This work made use of NASA's ADS Bibliographic Service. We would like to thank the anonymous referee for insightful suggestions.

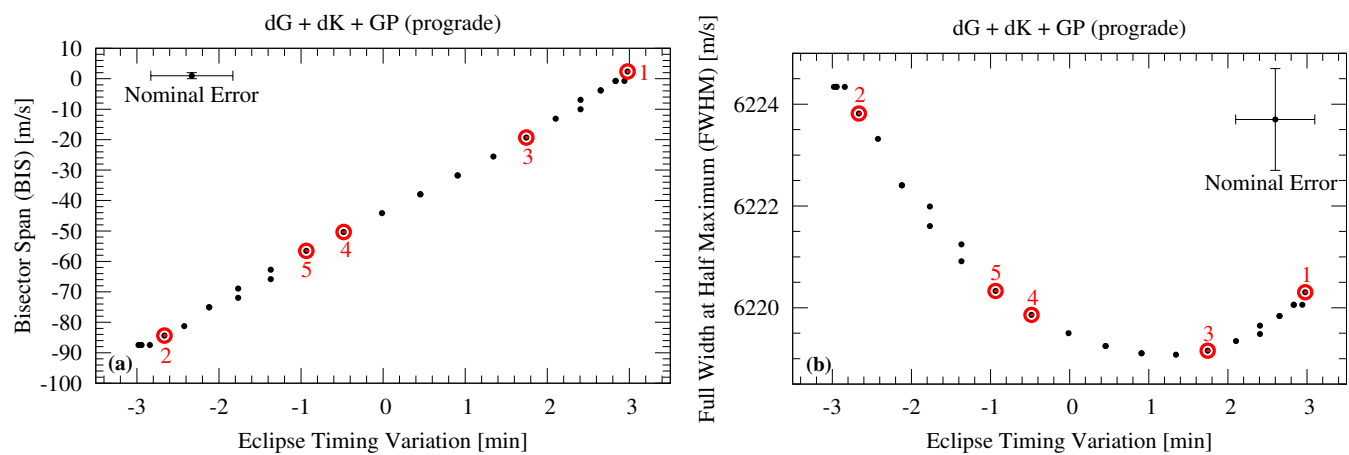

**Figure A1.** Supplementary information to Figure 5. Panel **(a)** shows variations of the bisector span and panel **(b)** the full width at half maximum of the CCF of the M-dwarf, both for a prograde scenario.

## 6.7 Revisiting the Exomoon Candidate around Kepler-1625 b (Rodenbeck et al. 2018)

Contribution:

RH contributed to the writing of the text, guided the direction of research, and contributed to the interpretation of the results.

# Revisiting the exomoon candidate signal around Kepler-1625 b

Kai Rodenbeck[1, 2], René Heller[2], Michael Hippke[3], and Laurent Gizon[2, 1]

[1] Institute for Astrophysics, Georg August University Göttingen, Friedrich-Hund-Platz 1, 37077 Göttingen, Germany
[2] Max Planck Institute for Solar System Research, Justus-von-Liebig-Weg 3, 37077 Göttingen, Germany
 e-mail: ⟨rodenbeck/heller/gizon⟩@mps.mpg.de
[3] Sonneberg Observatory, Sternwartestr. 32, 96515 Sonneberg, Germany e-mail: michael@hippke.org

**ABSTRACT**

*Context.* Transit photometry of the Jupiter-sized exoplanet candidate Kepler-1625 b has recently been interpreted to show hints of a moon. This exomoon, the first of its kind, would be as large as Neptune and unlike any moon we know from the solar system.
*Aims.* We aim to clarify whether the exomoon-like signal is indeed caused by a large object in orbit around Kepler-1625 b, or whether it is caused by stellar or instrumental noise or by the data detrending procedure.
*Methods.* To prepare the transit data for model fitting, we explore several detrending procedures using second-, third-, and fourth-order polynomials and an implementation of the Cosine Filtering with Autocorrelation Minimization (CoFiAM). We then supply a light curve simulator with the co-planar orbital dynamics of the system and fit the resulting planet-moon transit light curves to the Kepler data. We employ the Bayesian Information Criterion (BIC) to assess whether a single planet or a planet-moon system is a more likely interpretation of the light curve variations. We carry out a blind hare-and-hounds exercise using many noise realizations by injecting simulated transits into different out-of-transit parts of the original Kepler-1625 light curve: (1) 100 sequences with three synthetic transits of a Kepler-1625 b-like Jupiter-size planet and (2) 100 sequences with three synthetic transits of a Kepler-1625 b-like planet with a Neptune-sized moon.
*Results.* The statistical significance and characteristics of the exomoon-like signal strongly depend on the detrending method (polynomials versus cosines), the data chosen for detrending, and on the treatment of gaps in the light curve. Our injection-retrieval experiment shows evidence of moons in about 10 % of those light curves that do not contain an injected moon. Strikingly, many of these false-positive moons resemble the exomoon candidate, i.e. a Neptune-sized moon at about 20 Jupiter radii from the planet. We recover between about a third and half of the injected moons, depending on the detrending method, with radii and orbital distances broadly corresponding to the injected values.
*Conclusions.* A ΔBIC of −4.9 for the CoFiAM-based detrending is indicative of an exomoon in the three transits of Kepler-1625 b. This solution, however, is only one out of many and we find very different solutions depending on the details of the detrending method. We find it concerning that the detrending is so clearly key to the exomoon interpretation of the available data of Kepler-1625 b. Further high-accuracy transit observations may overcome the effects of red noise but the required amount of additional data might be large.

**Key words.** Planets and satellites: detection – Eclipses – Techniques: photometric – Methods: data analysis

## 1. Introduction

Where are they? – With about 180 moons discovered around the eight solar system planets and over 3,500 planets confirmed beyond the solar system, an exomoon detection could be imminent. While many methods have indeed been proposed to search for moons around extrasolar planets (Sartoretti & Schneider 1999; Han & Han 2002; Cabrera, J. & Schneider, J. 2007; Moskovitz et al. 2009; Kipping 2009; Simon et al. 2010; Peters & Turner 2013; Heller 2014; Ben-Jaffel & Ballester 2014; Agol et al. 2015; Forgan 2017; Vanderburg et al. 2018)[1], only a few dedicated surveys have actually been carried out (Szabó et al. 2013; Kipping et al. 2013b,a, 2014; Hippke 2015; Kipping et al. 2015; Lecavelier des Etangs et al. 2017; Teachey et al. 2018), one of which is the "Hunt for Exomoons with Kepler" (HEK for short; Kipping et al. 2012).

In the latest report of the HEK team, Teachey et al. (2018) find evidence for an exomoon candidate around the roughly Jupiter-sized exoplanet candidate Kepler-1625 b, which they provisionally refer to as Kepler-1625 b-i. Kepler-

1625 is a slightly evolved G-type star with a mass of $M_\star = 1.079^{+0.100}_{-0.138} M_\odot$ ($M_\odot$ being the solar mass), a radius of $R_\star = 1.793^{+0.263}_{-0.488} R_\odot$ (with $R_\odot$ as the solar radius), and an effective temperature of $T_{\mathrm{eff},\star} = 5548^{+83}_{-72}$ K (Mathur et al. 2017). Its Kepler magnitude of 15.756 makes it a relatively dim Kepler target.[2] The challenge of this tentative detection is in the noise properties of the data, which are affected by the systematic noise of the Kepler space telescope and by the astrophysical variability of the star. Although the exomoon signal did show up both around the ingress/egress regions of the phase-folded transits (known as the orbital sampling effect; Heller 2014; Heller et al. 2016a) generated by Teachey et al. (2018) and in the sequence of the three individual transits, it could easily have been produced by systematics or stellar variability, as pointed out in the discovery paper.

The noise properties also dictate a minimum size for an exomoon to be detected around a given star and with a given instrument. In the case of Kepler-1625 we calculate the root-mean-square of the noise level to be roughly 700 ppm. As a consequence, any moon would have to be at least about

---

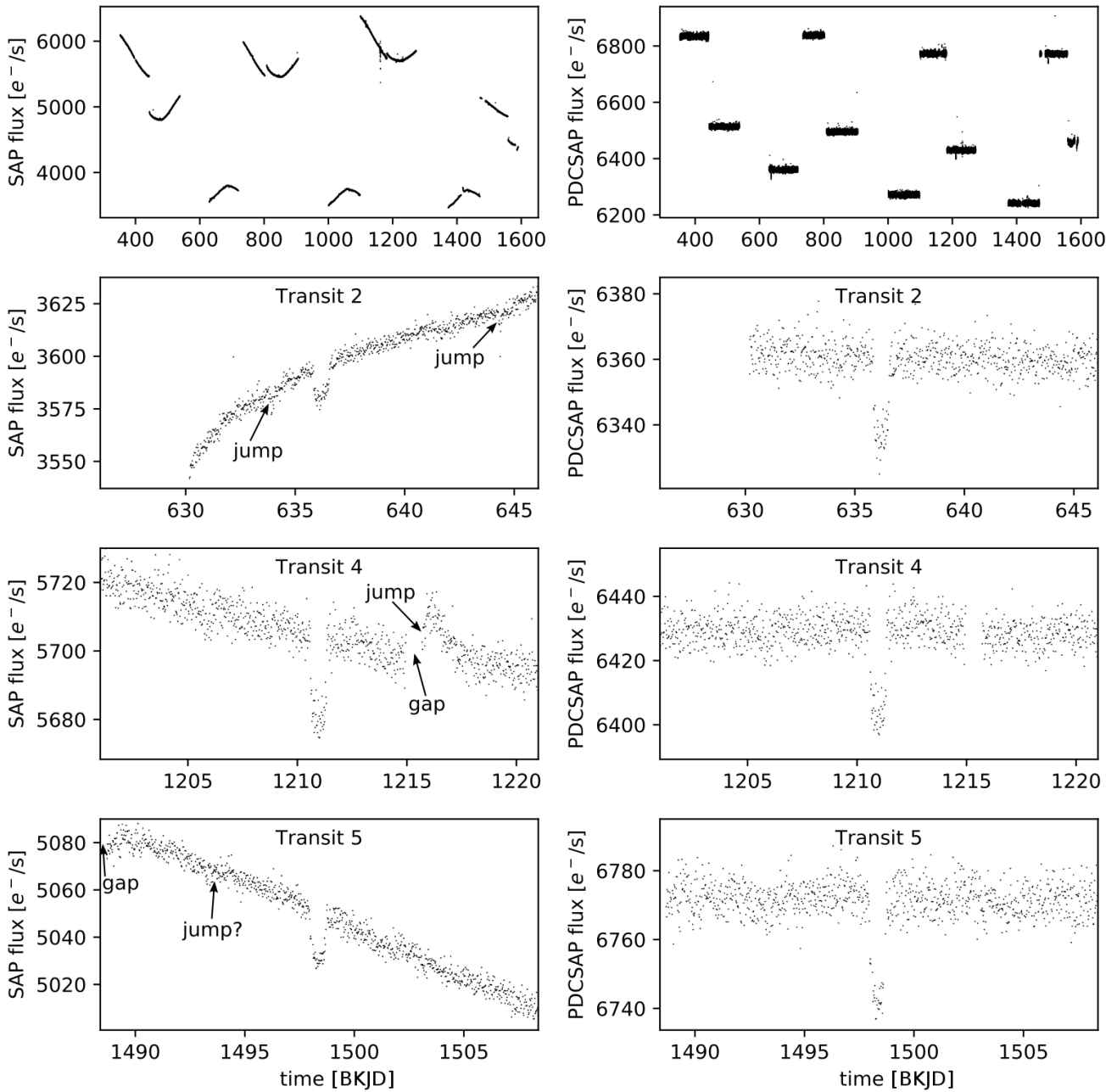

**Fig. 1.** Kepler light curve of Kepler-1625. *Left:* Simple Aperture Photometry (SAP) flux. *Right:* Pre-search Data Conditioning Simple Aperture Photometry (PDCSAP) flux. The top panels show the entire light curves, respectively. The second to fourth rows illustrate zooms into transits 2, 4, and 5 of Kepler-1625 b, respectively. These transits were shifted to the panel center and ±10 d of data are shown around the transit mid-points. Some examples of jumps and gaps in the light curve are shown. Time is in Barycentric Kepler Julian Date.

$\sqrt{700\,\mathrm{ppm}} \times 1.793\,R_\odot \approx 5.2\,R_\oplus$ ($R_\oplus$ being the Earth's radius) in size, about 30% larger than Neptune, in order to significantly overcome the noise floor in a single transit. The three observed transits lower this threshold by a factor of $\sqrt{3}$, suggesting a minimum moon radius of $\approx 3\,R_\oplus$. In fact, the proposed moon candidate is as large as Neptune, making this system distinct from any planet-moon system known in the solar system (Heller 2018).

Here we present a detailed study of the three publicly available transits of Kepler-1625 b. Our aim is to test whether the planet-with-moon hypothesis is favored over the planet-only hypothesis. In brief, we

1. develop a model to simulate photometric transits of a planet with a moon (see Sect. 2.2.2).

2. implement a detrending method following Teachey et al. (2018) and explore alternative detrending functions.

3. detrend the original Kepler-1625 light curve, determine the most likely moon parameters, and assess if the planet-with-moon hypothesis is favored over the planet-only hypothesis.

4. perform a blind injection-retrieval test. To preserve the noise properties of the Kepler-1625 light curve, we inject planet-with-moon and planet-only transits into out-of-transit parts of the Kepler-1625 light curve.

## 2. Methods

The main challenge in fitting a parameterized, noise-less model to observed data is in removing noise on time scales similar or larger than the time scales of the effect to be searched; at the same time, the structure of the effect shall be untouched, an approach sometimes referred to as "pre-whitening" of the data (Aigrain & Irwin 2004). The aim of this approach is to remove unwanted variations in the data, e.g. from stellar activity, systematics, or instrumental effects. This approach bears the risk of both removing actual signal from the data and of introducing new systematic variability. The discovery and refutal of the exoplanet interpretation of variability in the stellar radial velocities of α Centauri B serves as a warning example (Dumusque et al. 2012; Rajpaul et al. 2016). Recently developed Gaussian process frameworks, in which the systematics are modeled simultaneously with stellar variability, would be an alternative method (Gibson et al. 2012). This has become particularly important for the extended Kepler mission (K2) that is now working with degraded pointing accuracy (Aigrain et al. 2015).

That being said, Teachey et al. (2018) applied a pre-whitening technique to both the Simple Aperture Photometry (SAP) flux and the Pre-search Data Conditioning (PDCSAP) flux of Kepler-1625 to determine whether a planet-only or a planet-moon model is more likely to have caused the observed Kepler data. In the following, we develop a detrending and model fitting procedure that is based on the method applied by Teachey et al. (2018), and then we test alternative detrending methods.

During Kepler's primary mission, the star Kepler-1625 has been monitored for 3.5 years in total, and five transits could have been observed. This sequence of transits can be labeled as transits 1, 2, 3, 4, and 5. Due to gaps in the data, however, only three transits have been covered, which correspond to transits 2, 4, and 5 in this sequence. Figure 1 shows the actual data set that we discuss. The entire SAP (left) and PDCSAP (right) light curves are shown in the top panels, and close-up inspections of the observed transit 2, 4, and 5 are shown in the remaining panels. The time system used throughout the article is the Barycentric Kepler Julian Date (BKJD), unless marked as relative to a transit midpoint.

### 2.1. Detrending

A key pitfall of any pre-whitening or detrending method is the unwanted removal of signal or injection of systematic noise, the latter of which could mimic signal. In our case of an exomoon search, we know that the putative signal would be restricted to a time-window around the planetary mid-transit, which is compatible with the orbital Hill stability of the moon. This criterion defines a possible window length that we should exclude from our detrending procedures. For a nominal 10 Jupiter-mass planet in a 287 d orbit around a 1.1 $M_\odot$ star (as per Teachey et al. 2018), this window is about 3.25 days to both sides of the transit midpoint (see Appendix A).

Although this window length is astrophysically plausible to protect possible exomoon signals, many other choices are similarly plausible but they result in significantly different detrendings. Figure 2 illustrates the effect on the detrended light curve if two different windows around the midpoint of the planetary transit (here transit 5) are excluded from the fitting. We chose a fourth-order polynomial detrending function and a 7.5 d (blue symbols) or a 4 d (orange symbols) region around the midpoint to be excluded from the detrending, mainly for illustrative pur-poses. In particular, with the latter choice we produce a moon-like signal around the planetary transit similar to the moon signal that appears in transit 5 in Teachey et al. (2018). For the former choice, however, this signal does not appear in the detrended light curve.

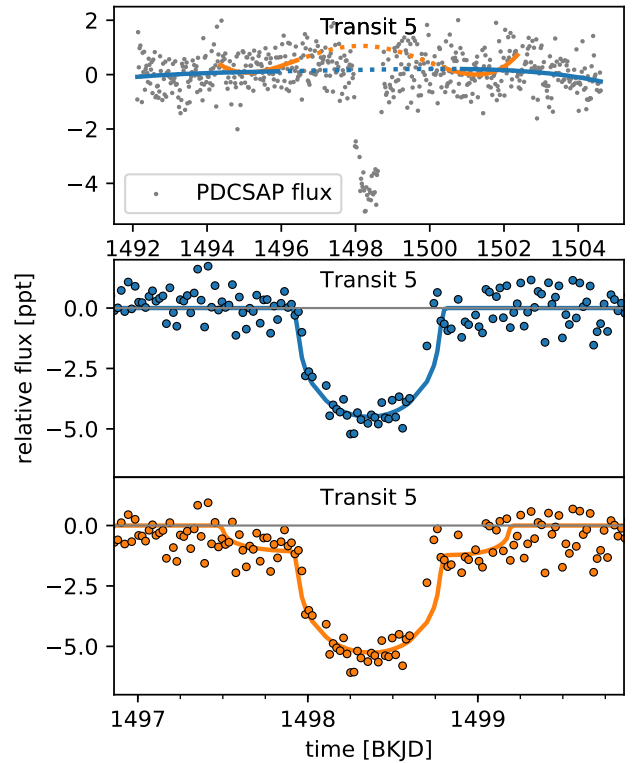

**Fig. 2.** Example of how the detrending procedure alone can produce an exomoon-like transit signal around a planetary transit. We use transit 5 of Kepler-1625 b as an example. *Top:* Gray dots indicate the Kepler PD-CSAP flux. The lines show a 4th-order polynomial fit for which we exclude 7.5 d (blue) or 4 d (orange) of data around the mid-point (dashed parts), respectively. *Center:* Dots show the detrended light curve derived from the blue polynomial fit in the top panel. The blue line illustrates a planet-only transit model. *Bottom:* Dots visualize the detrended light curve using the orange polynomial fit from the top panel. Note the additional moon-like transit feature caused by the overshooting of the orange polynomial in the top panel. The orange line shows a planet-moon transit model with moon parameters as in Table 1 (see Fig. 4 for transit dynamics). As an alternative interpretation, the blue detrending function filters out an actually existing moon signature while the orange detrending fit preserves the moon signal.

Teachey et al. (2018) use the Cosine Filtering with Autocorrelation Minimization (CoFiAM) detrending algorithm to detrend both the SAP and PDCSAP flux around the three transits of Kepler-1625 b. CoFiAM fits a series of cosines to the light curve, excluding a specific region around the transit. CoFiAM preserves the signal of interest by using only cosines with a period longer than a given threshold and therefore avoids the injection of artificial signals with periods shorter than this threshold. Teachey et al. (2018) also test polynomial detrending functions but report that this removes the possible exomoon signal. We choose to reimplement the CoFiAM algorithm as our primary detrending algorithm to remain as close as possible in our analysis to the work in Teachey et al. (2018). In our injection-retrieval test we also use polynomials of second, third, and fourth order for detrending. While low-order polynomials cannot generally fit the light curve as well as the series of cosines, the risk of injecting artificial signals may be reduced.

### 2.1.1. Trigonometric detrending

We implement the CoFiAM detrending algorithm as per the descriptions given by Kipping et al. (2013b) and Teachey et al. (2018). In the following, we refer to this reimplementation as trigonometric detrending as opposed to polynomial functions that we test as well (see Sect. 2.4.4).

The light curves around each transit are detrended independently. For each transit, we start by using the entire SAP flux of the corresponding quarter. We use the SAP flux instead of the PDCSAP flux to reproduce the methodology of Teachey et al. (2018) as closely as possible. The authors argue that the use of SAP flux avoids the injection of additional signals into the light curve that might have the shape of a moon signal. First, we remove outliers using a running median with a window length of 20 h and a threshold of 3 times the local standard deviation with the same window length. In order to achieve a fast convergence of our detrending and transit fitting procedures, we initially estimate the transit midpoints and durations by eye and identify data anomalies, e.g. gaps and jumps (e.g. the jump 2 d prior to transit 2 and the gap 4 d after transit 4, see Fig. 1).

Jumps in the light curve can have multiple reasons. The jumps highlighted around transit 2 in Fig. 1 are caused by a reaction wheel zero crossing event; the jump 5 d after transit 4 is caused by a change in temperature after a break in the data collection. Following Teachey et al. (2018), who ignore data points beyond gaps and other anomalous events for detrending, we cut the light curve around any of the transits as soon as it encounters the first anomaly, leaving us with a light curve of a total duration $D$ around each transit (see top left panel in Fig. 3). In Sect. 2.4.4, we investigate the effect of including data beyond gaps. The detrending is applied in two passes, using the first pass to get accurate transit parameters. In particular, we determine the duration ($t_T$) between the start of the planetary transit ingress and the end of the transit egress (Seager & Mallén-Ornelas 2003) and the second pass to generate the detrended light curve.

*First pass:* Using the estimated transit midpoints and durations, we calculate the time window ($t_c$, see top left panel in Fig. 3) around a given transit midpoint to be cut from the detrending fit as $t_c = f_{t_c} t_T$, where the factor $f_{t_c}$, relating the time cut around the transit to the transit duration, is an input parameter for the detrending algorithm. Specifically, $t_c$ denotes the total length of time around the transit excluded from the detrending. We fit the detrending function

$$\mathcal{G}^k(t, \vec{a}, \vec{b}) = a_0 + \sum_{l=1}^{k} a_l \cos\left(l\frac{2\pi}{2D}t\right) + b_l \sin\left(l\frac{2\pi}{2D}t\right) \quad (1)$$

to the light curve (excluding the region $t_c$ around the transit) by minimizing the $\chi^2$ between the light curve and $\mathcal{G}^k(t, \vec{a}, \vec{b})$, where $\vec{a} = (a_0, a_1, ..., a_k)$ and $\vec{b} = (b_1, b_2, ..., b_k)$ are the free model parameters to be fitted. The parameter $k$ is a number between 1 and $k_{max}$ = round($2D/t_p$), where $t_p = f_{t_p} t_T$ is the time scale below which we want to preserve possible signals. $f_{t_p}$ is an input parameter to the detrending algorithm. For each $k$ we divide the light curve by $\mathcal{G}^k$, giving us the detrended light curves $F^k$. We calculate the first-order autocorrelation according to the Durbin & Watson (1950) test statistic for each $F^k$ (excluding again the region around the transit). For each transit we select the $F^k$ with the lowest autocorrelation $F^k_{min}$ and combine these $F^k_{min}$ around each transit into our detrended light curve $F$. We fit the planet-only transit model to the detrended light curve $F$ and compute the updated transit midpoints and duration $t_T$.

*Second pass:* The second pass repeats the steps of the first pass, but using the updated transit midpoints and durations as input. The resulting detrended light curve $F$ is then used for our model fits with the ultimate goal of assessing whether an exomoon is a likely interpretation of the light curve signatures or not. We estimate the noise around each transit by taking the variance of $F$, excluding the transit region.

Figure 3 shows the detrending function as well as the detrended light curve for $f_{t_c} = 2.2$ and $f_{t_p} = 4.4$, corresponding to $t_c = 1.6$ d and $t_p = 3.1$ d.

### 2.2. Transit model

We construct two transit models, one of which contains a planet only and one of which contains a planet with one moon. We denote the planet-only model as $\mathcal{M}_0$ (the index referring to the number of moons) and the planet-moon model as $\mathcal{M}_1$. We do not consider models with more than one moon.

#### 2.2.1. Planet-only model

$\mathcal{M}_0$ assumes a circular orbit of the planet around its star. Given the period of that orbit ($P$) and the ratio between stellar radius and the orbital semimajor axis ($R_\star/a$), the sky-projected apparent distance to the star center relative to the stellar radius can be calculated as

$$z = \sqrt{\left[\frac{a}{R_\star}\sin\left(\frac{2\pi(t-t_0)}{P}\right)\right]^2 + \left[b\cos\left(\frac{2\pi(t-t_0)}{P}\right)\right]^2}, \quad (2)$$

where $b$ is the transit impact parameter and $t_0$ is the time of the transit midpoint. We use the python implementation of the Mandel & Agol (2002) analytic transit model by Ian Crossfield[3] to calculate the transit light curve based on the planet-to-star radius ratio ($r_p = R_p/R_\star$) and based on a quadratic limb-darkening law parametrized by the limb-darkening parameters $q_1$ and $q_2$ as given in Kipping (2013). We call this model light curve with zero moons $\mathcal{F}_0$.

#### 2.2.2. Planet-moon model

In our planet-moon model, we assume a circular orbit of the local planet-moon barycenter around the star with an orbital period $P_B$, a semimajor axis $a_B$, and a barycentric transit midpoint time $t_{0,B}$. The projected distance of the barycenter to the star center relative to the stellar radius is calculated the same way as in eq. 2. The planet and moon are assumed to be on circular orbits around their common center of mass with their relative distances to the barycenter determined the ratio of their masses $M_p$ and $M_s$ to the total mass $M_p + M_s$. The individual orbits of both the planet and the moon are defined by the total distance between the two objects $a_{ps}$, the planet mass $M_p$, the moon mass $M_s$ and by the time of the planet-moon conjunction $t_{0,s}$, that is, the time at which the moon is directly in front of the planet as seen from an observer on Earth.

This model is degenerate in terms of the sense of orbital motion of the moon. In other words, a given planet-moon transit light curve can be produced by both a prograde and a retrograde moon (Lewis & Fujii 2014; Heller & Albrecht 2014). We restrict ourselves to prograde moons. The planet mass is set to a nominal 10 Jupiter masses, as suggested by Teachey et al. (2018) and in agreement with the estimates of Heller (2018). This constraint simplifies the interpretation of the results substantially since the moon parameters are then unaffected by the planetary parameters. The moon mass is assumed to be much smaller than that of the planet. In fact, for a roughly Neptune-mass moon around a 10 Jupiter-mass planet, we expect a TTV amplitude of 3 to 4 minutes and a TDV amplitude of 6 to 7 minutes, roughly speaking. Hence, we simplify our model and set $M_s = 0$, which means that $a_{ps}$ is equal to the distance between the moon and the planet-moon barycenter, $a_s$. The moon is assumed to have a coplanar orbit around the planet and, thus, to have the same transit impact parameter as the planet.

---

[3] Available at http://www.astro.ucla.edu/~ianc/files as python.py.

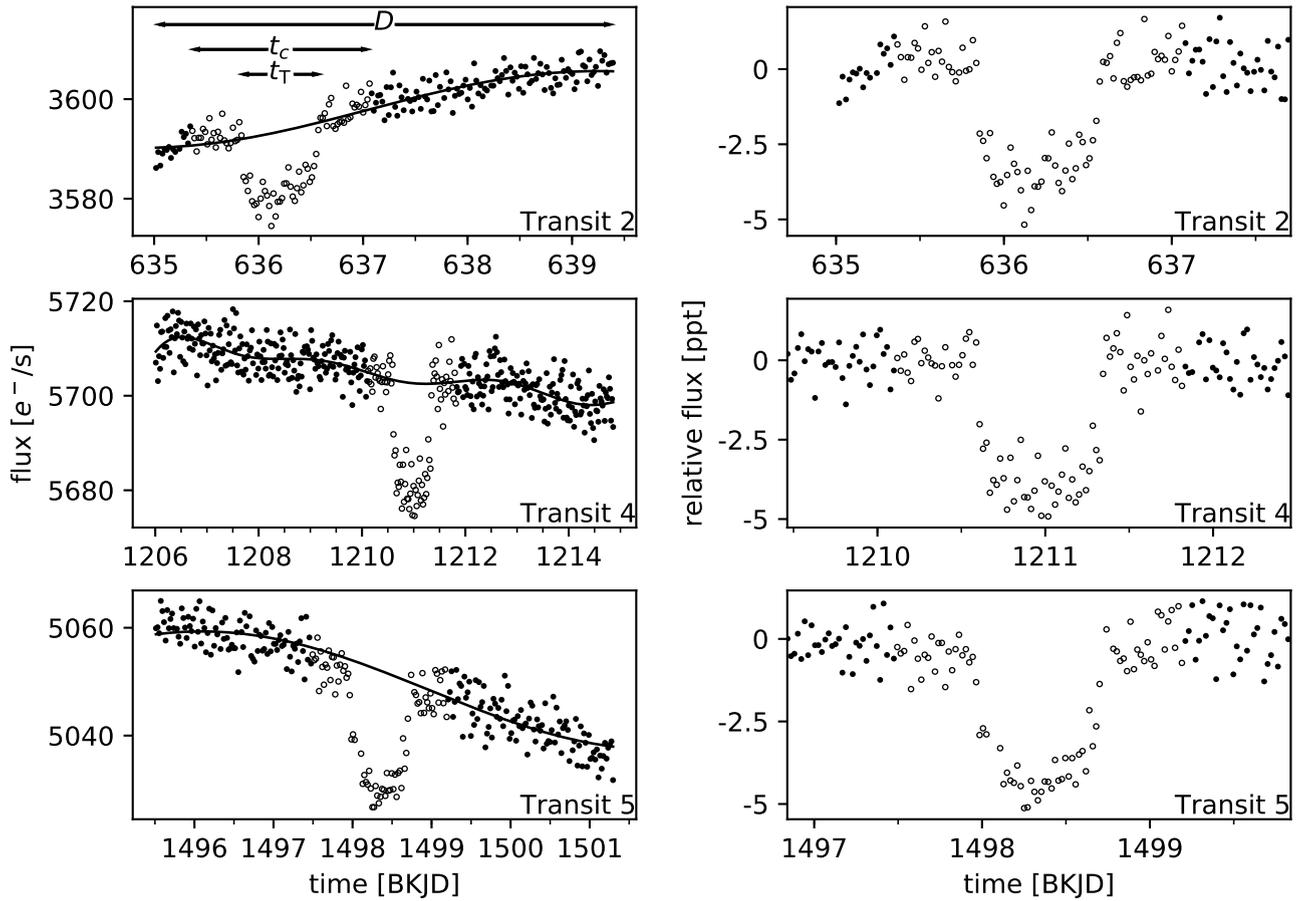

**Fig. 3.** *Left:* Kepler SAP flux around the transits used for the trigonometric detrending, our reimplementation of the CoFiAM algorithm. The data points denoted by open circles around the transits are excluded from the detrending fit. The black line shows the resulting light curve trend without the transit. *Right:* Detrended transit light curves as calculated by the trigonometric detrending.

With these assumptions the projected distance of the planet center to the star center relative to the stellar radius $z_p$ is equal to that of the barycenter $z_B$. The projected distance of the moon center to the star center relative to the stellar radius $z_s$ is given by

$$z_s^2 = \left[ \frac{a_B}{R_\star} \sin\left(\frac{2\pi(t - t_{0,B})}{P_B}\right) + \frac{a_{ps}}{R_\star} \sin\left(\frac{2\pi(t - t_{0,s})}{P_s}\right) \right]^2$$
$$+ \left[ b \cos\left(\frac{2\pi(t - t_{0,B})}{P_B}\right) \right]^2 , \tag{3}$$

where $P_s$ is the orbital period of the moon calculated from the fixed masses and $a_{ps}$.

We calculate the transit light curves of both bodies and combine them into the total model light curve, which we call $\mathcal{F}_1$. We use the limb-darkening parameter transformation from Kipping (2013). For computational efficiency, we do not consider planet-moon occultations. For the planet-moon system of interest, occultations would only occur only about half of the transits (assuming a random moon phase) even if the moon orbital plane would be perfectly parallel to the line of sight. Such an occultation would take about 1.5 h and would only affect 5-10 % of the total moon signal duration.

In Table 1 we give an overview of our nominal parameterization of the planet-moon model. In Fig. 4 we show the orbital dynamics of the planet and moon during transits 2, 4, and 5 using the nominal parameters in Table 1. This nominal parameterization was chosen to generate a model light curve that is reasonably close to the preferred model light curve found in Teachey et al. (2018), but it does not represent our most likely model fit to the data.

### 2.2.3. Finding the posterior probability distribution

We use the Markov Chain Monte Carlo (MCMC) implementation Emcee (Foreman-Mackey et al. 2013) to estimate the posterior probability distribution of the parameters for model $\mathcal{M}_{(i)}$ ($\mathcal{M}_0$ or $\mathcal{M}_1$). For this purpose, we need to formulate the proba-bility density of a light curve as well as the prior of the parameters.

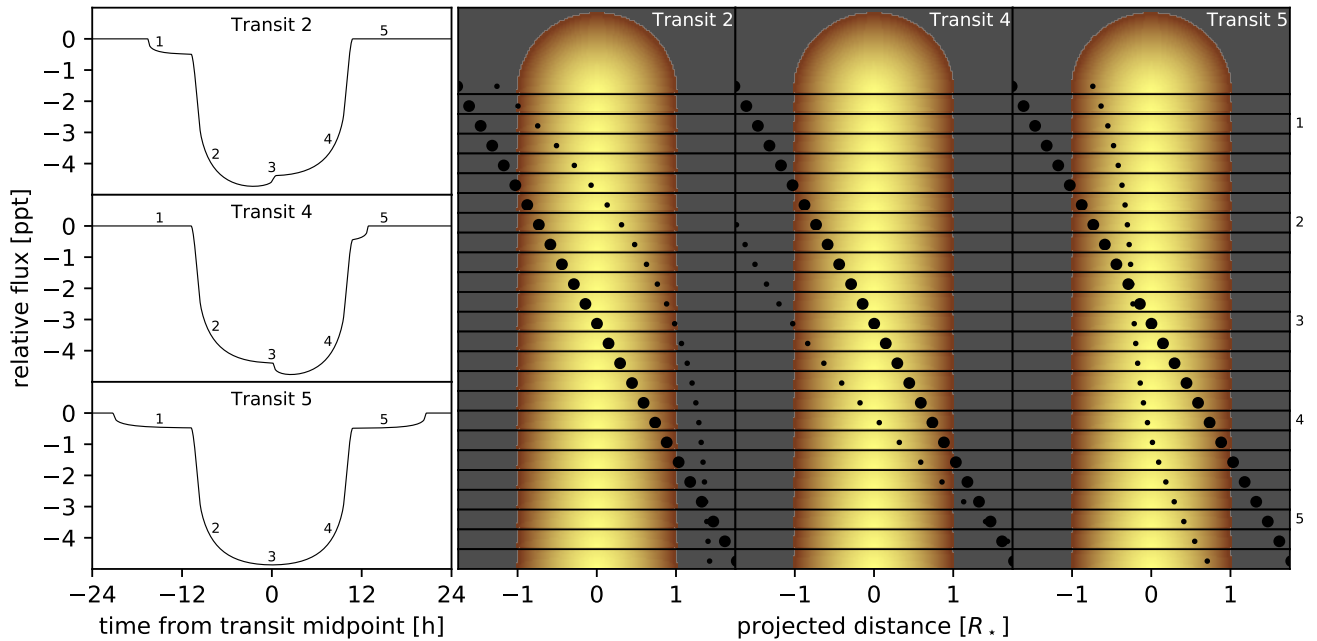

**Fig. 4.** *Left*: Example of a simulated planet-moon transit light curve for transits 2, 4, and 5 using the nominal parameterization given in Table 1. The relative flux is the difference to the out-of-transit model flux and is given in parts per thousand (ppt). *Right*: Visualization of the orbital configurations during transits 2 (left column), 4 (center column), and 5 (right column). Labels 1-5 in the light curves refer to configurations 1-5 (see labels along the vertical axis). An animation of this figure is available online.

**Table 1.** Nominal parameterization of the planet-moon model to reproduce the transit shape suggested by Teachey et al. (2018). The no-moon model uses the same parameter set (excluding the moon parameters), except that $R_\star$ and $a_B$ are combined into a single parameter $R_\star/a_B$.

| Parameter | Nominal Value | Description |
|---|---|---|
| $r_p$ | 0.06 | planet-to-star radius ratio |
| $a_B$ | 0.9 au | circumstellar semimajor axis of the planet-moon barycenter |
| $b$ | 0.1 | planetary transit impact parameter |
| $t_{0,B}$ | 61.51 d | transit midpoint of the planet-moon barycenter |
| $P_B$ | 287.35 d | circumstellar orbital period of the planet-moon barycenter |
| $R_\star$ | 1.8 $R_\odot$ | stellar radius |
| $q_1$ | 0.6 | 1st limb darkening coefficient |
| $q_2$ | 0.2 | 2nd limb darkening coefficient |
| $r_s$ | 0.02 | moon-to-star radius ratio |
| $a_s$ | 1871 $R_J$ | orbital semimajor axis of the planet-moon binary |
| $t_{0,s}$ | 1.86 d | time of planet-moon conjunction |

All three transits taken together, we have a total of $N$ detrended flux measurements (see Sect. 2.1.1). Given a set of pa-rameters $\vec{\theta}$, model $\mathcal{M}_i$ produces a model light curve $\mathcal{F}_i(t, \vec{\theta})$. We assume that the noise is uncorrelated (see Appendix B) and Gaussian with a standard deviation $\sigma_j$ at time $t_j$. This simplifies the joint probability density to a product of the individual probabilities. The joint probability density function of the detrended flux $F(t)$ is given by

$$p(F|\vec{\theta}, \mathcal{M}_i) = \prod_{j=1}^{N} \frac{1}{\sqrt{2\pi\sigma_j^2}} \exp\left(-\frac{\left(F(t_j) - \mathcal{F}_i(t_j, \vec{\theta})\right)^2}{2\sigma_j^2}\right). \quad (4)$$

The noise dispersion $\sigma_j$ has a fixed value for each transit.

Table 2 shows the parameter ranges that we explore. A prior is placed on the stellar mass according to the mass of $1.079^{+0.100}_{-0.138} M_\odot$ determined by Mathur et al. (2017). The stellar mass for a given parameter set is determined from the system's total mass using $P_B$ and $a_B$ and subtracting the fixed planet mass of 10 Jupiter masses.

A total of 100 walkers are initiated with randomly chosen parameters close to the estimated transit parameters. For the sake of fast computational performance, the walkers are initially separated into groups of 16 for the planet-only model and 24 for the planet-moon model (twice the number of parameters plus 2, respectively), temporarily adding walkers to fill the last group. To transform the initially flat distribution of walkers into a distribution according to the likelihood function, the walkers have to go through a so-called burn-in phase, the resulting model fits of which are discarded. We chose a burn-in phase for the walkers of 500 steps in both groups. Afterwards, we discard the temporarily added walkers, merge the walkers back together, and perform a second burn-in phase of 2 200 steps with a length determined by visual inspection. Finally, we initiate the main MCMC run with a total of 8 000 steps.

We run the MCMC code on the detrended light curve using both the planet-only and the planet-moon model.

**Table 2.** Parameter ranges explored with our planet-moon model. The ranges of the no-moon model parameters are the same for the shared parameter and is propagated to the derived parameter $R_\star/a$.

| Min. Value | | Parameter | | Max. Value |
|---|---|---|---|---|
| 0 | $\leq$ | $r_p$ | $\leq$ | 0.1 |
| 0.2 au | $\leq$ | $a_B$ | $\leq$ | 2 au |
| 0 | $\leq$ | $b$ | $\leq$ | 1 |
| $-P_B/2$ | $\leq$ | $t_{0,B}$ | $\leq$ | $P_B/2$ |
| 270 d | $\leq$ | $P_B$ | $\leq$ | 300 d |
| 0 | $\leq$ | $R_\star$ | $\leq$ | $4.3\,R_\odot$ |
| 0 | $\leq$ | $q_1$ | $\leq$ | 1 |
| 0 | $\leq$ | $q_1$ | $\leq$ | 1 |
| 0 | $\leq$ | $r_s$ | $\leq$ | $r_p$ |
| 0 | $\leq$ | $a_s$ | $\leq$ | $R_{Hill}/2$ |
| $-P_s/2$ | $\leq$ | $t_{0,s}$ | $\leq$ | $P_s/2$ |

## 2.3. Model selection

We use the Bayesian Information Criterion (BIC) to evaluate how well a model describes the observations in relation to the number of model parameters and data points. The BIC of a given model $\mathcal{M}_i$ with $m_i$ parameters is defined by Schwarz (1978) as

$$\text{BIC}(\mathcal{M}_i|F) = m_i \ln N - 2 \ln p(F|\vec{\theta}_{max}, \mathcal{M}_i), \quad (5)$$

where $\vec{\theta}_{max}$ is the set of parameters that maximizes the probability density function $p(F|\vec{\theta}, \mathcal{M}_i)$ for a given light curve $F$ and model $\mathcal{M}_i$.

The difference of the BICs between two models gives an indication as to which model is more likely. In particular, $\Delta\text{BIC}(\mathcal{M}_1, \mathcal{M}_0) \equiv \text{BIC}(\mathcal{M}_1) - \text{BIC}(\mathcal{M}_0) < 0$ if model $\mathcal{M}_1$ is more likely. We consider $\Delta\text{BIC} < 6$ (or $\Delta\text{BIC} > 6$) as strong evidence in favor of (or against) model $\mathcal{M}_1$ (see, e.g., Kass & Raftery 1995).

The best-fitting set of parameters derived from our MCMC runs ($\vec{\theta}_{max}$) is then used to calculate $\Delta\text{BIC}(\mathcal{M}_1, \mathcal{M}_0)$. For our calculations, we only use those parts of the light curve around the transits that could potentially be affected by a moon (3.25 d on each side of the transits, determined by the Hill radius $R_{Hill}$ and the orbital velocity of the planet-moon barycenter, see Appendix A).

## 2.4. Injection-retrieval test

In order to estimate the likelihood of an exomoon feature to be due to either a real moon or due to noise, we perform several injection-retrieval experiments. One of us (MH) injected two cases of transits into the out-of-transit parts of the original PDCSAP Kepler flux. In one case, a sequence of three planet-only transits (similar to the sequence of real transits 2, 4, and 5) was injected, where the planet was chosen to have a radius of 11 Earth radii. In another case, a sequence of three transits of a planet with moon with properties similar to the proposed Jupiter-Neptune system was injected. Author KR then applied the Baysian framework described above in order to evaluate the planet-only vs. the planet-with-moon hypotheses and in order to characterize the planet and (if present) its moon.

As an important trait of our experiment, KR did not know which of the light curves contained only a planet and which contained also a moon.

## 2.4.1. Transit injections into light curves

For the injection part, we use PyOSE (Heller et al. 2016a,b) to create synthetic planet and moon ensemble transits. This code numerically integrates the non-occulted areas of the stellar disk to calculate the instantaneous flux of the star, which makes it a computationally slow procedure. Hence we use the analytical model described Sect. 2.2 for the retrieval part. In our model, the moon's orbit is defined by its eccentricity ($e_s$, fixed at 0), $a_s$, its orbital inclination with respect to the circumstellar orbit ($i_s$, fixed at 90°), the longitude of the ascending node, the argument of the periapsis, and the planetary impact parameter ($b$, fixed at zero). Due to the small TTV and TDV amplitudes compared to the 29.4 min exposure of the Kepler long cadence data, we neglect the planet's motion around the planet/moon barycenter, although PyOSE can model this dynamical effect as well, and assume that the moon orbits the center of the planet.

Our numerical code creates a spherical limb-darkened star on a 2-dimensional grid of floating-point values. The sky-projected shapes of both the planet and the moon are modeled as black circles. The spatial resolution of the simulation is chosen to be a few million pixels so that the resulting light curve has a numerical error < 1 ppm that is negligible compared to the $\approx 700$ ppm noise level of the Kepler light curve. The initial temporal resolution of our model is equivalent to 1 000 steps per planetary transit duration, which we then downsample to the observed 29.4 min cadence. The creation of one such light curve of a planet with a moon takes about one minute on a modern desktop computer.

We create a set of 100 such transit simulations of the planet-moon ensemble, where the two bodies move consistently during and between transits. All orbits are modeled to be co-planar and mutual planet-moon occultations are also included. For each transit sequence, the initial orbital phase of the planet-moon system is chosen randomly.

With $P_B = 287.378949$ d and $P_s = 2.20833$ d, the moon advances by $\approx 0.13$ in phase between each subsequent planetary transit ($P_B/P_s \approx 130.13$). During a planetary transit, the moon advances by $\approx 0.36$ rad in phase (the planetary transit duration is $0.7869 \pm 0.0084$ d).

We also create a set of 100 such transits that only have a transiting planet without a moon. In these cases, the planetary radius was increased slightly to match the average transit depth of planet and moon.

## 2.4.2. Testing the model-selection algorithm on synthetic light curves with white noise only

As a first validation of our injection-retrieval experiment and our implementation of the Bayesian statistical framework, we generate a new set of white noise light curves to test only the model comparison part of our pipeline without any effects that could possibly arise from imperfect detrending. Any effects that we would see in our experiments with the real Kepler-1625 light curve but not in the synthetic light curves with noise only could then be attributed to the imperfect detrending of the time-correlated (red) noise.

MH generated 200 synthetic light curves with ten different levels of white noise, respectively, ranging from root mean squares of 250 ppm to 700 ppm in steps of 50 ppm. This results in a total of 2 000 synthetic light curves. MH used the method described in Sect. 2.4.1 to inject three transits of a planet only into 100 light curves per noise level and three transits of a planet with a moon into the remaining 100 light curves per noise level. The initial orbital phases were randomly chosen and are differ-planet-only system injected          planet+moon system injected ent from the ones used to generate the light curves in Sect. 2.4.3. MH delivered these light curves to KR without revealing their specific contents. KR then ran our model selection algorithm to find the $\Delta$BIC for each of the 2 000 systems. After the $\Delta$BICs were found, MH revealed the planet-only or the planet-moon nature of each light curve.

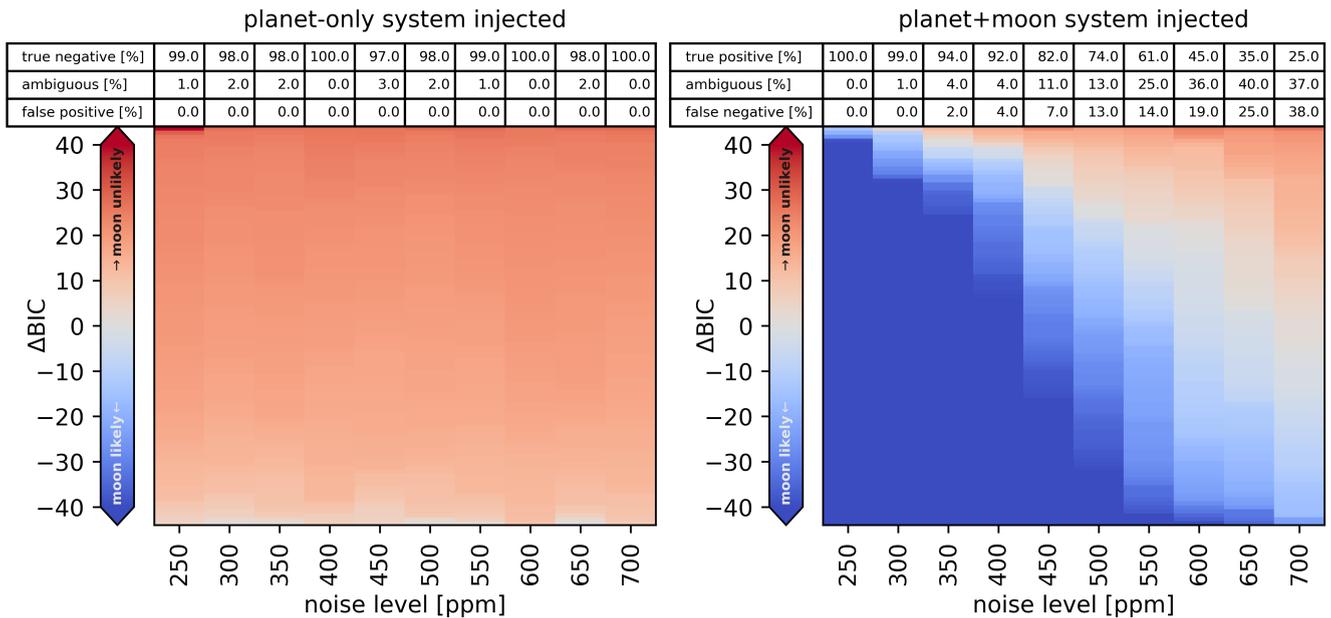

| true negative [%] | 99.0 | 98.0 | 98.0 | 100.0 | 97.0 | 98.0 | 99.0 | 100.0 | 98.0 | 100.0 |
|---|---|---|---|---|---|---|---|---|---|---|
| ambiguous [%] | 1.0 | 2.0 | 2.0 | 0.0 | 3.0 | 2.0 | 1.0 | 0.0 | 2.0 | 0.0 |
| false positive [%] | 0.0 | 0.0 | 0.0 | 0.0 | 0.0 | 0.0 | 0.0 | 0.0 | 0.0 | 0.0 |

| true positive [%] | 100.0 | 99.0 | 94.0 | 92.0 | 82.0 | 74.0 | 61.0 | 45.0 | 35.0 | 25.0 |
|---|---|---|---|---|---|---|---|---|---|---|
| ambiguous [%] | 0.0 | 1.0 | 4.0 | 4.0 | 11.0 | 13.0 | 25.0 | 36.0 | 40.0 | 37.0 |
| false negative [%] | 0.0 | 0.0 | 2.0 | 4.0 | 7.0 | 13.0 | 14.0 | 19.0 | 25.0 | 38.0 |

**Fig. 5.** Difference between the BIC of the planet-moon model and the no-moon model for 2 000 artificial white noise light curves at different noise levels, injected with simulated transits. On the left ($100 \times 10$ light curves) a planet and moon transit was injected, on the right ($100 \times 10$ light curves) only the planet. Each light curve consists of three consecutive transits. Each column is sorted by the $\Delta$BIC. The $\Delta$BIC threshold, over which a planet-moon or planet-only system is clearly preferred is $\pm 6$ with the state of systems with a $\Delta$BIC between those values considered to be ambiguous.

Fig. 5 shows the resulting $\Delta$BICs for each of the 2 000 light curves, separated into the planet-only (left panel) and planet-moon injected systems (right panel) and sorted by the respective white noise level (along the abscissa). Each vertical column contains 100 light curves, respectively. For a noise level of 250 ppm, as an example, our algorithm finds no false positive moons in the planet-only data, that is, no system with a $\Delta$BIC $< -6$, while 1 case remains ambiguous ($-6 < \Delta$BIC $< 6$) and the other 99 cases are correctly identified as containing no moons. In the case of an injected planet-moon system instead, the algorithm correctly retrieves the moon in 100 % of the synthetic light curves, that is, $\Delta$BIC $< -6$ for all systems.

More generally, for the simulated planet-only systems, the false positive rate is 0 % throughout all noise levels. Occasionally a system is flagged as ambiguous, but overall the algorithm consistently classifies planet-only systems correctly as having no moon. Referring to the injected planet-moon system (right panel), our false negative rate rises steadily with increasing noise level. In fact, it reaches parity with the true positive rate between about 650 ppm and 700 ppm.

In Fig. 6 we present $a_s$ and $R_s$ for each of the maximum-likelihood fits shown in Fig. 5. Each panel in Fig. 6 refers to one white noise level, that is, to one column in Fig. 5 of either the planet-only or the planet-moon injected system. In the case of an injected planet only (upper panels), the most likely values of $a_s$ are distributed almost randomly over the range of values that we explored. On the other hand, $R_s$ is constrained to a small range from about 1.5 $R_\oplus$ at 250 ppm to roughly 3 $R_\oplus$ at 700 ppm with the standard variation naturally increasing with the noise level.

The lower part of Fig. 6 shows the outcome of our planet-moon injection-retrievals from the synthetic light curves with white noise only. The correct parameters are generally recovered at all noise levels. In fact, we either recover the moon with a similar radius and orbital separation as the injection values (symbolized by blue points) or we find the moon to have very different radius and orbit while also rejection the hypothesis of its presence in the first place (symbolized by red points). The distribution of these false negatives in the $a_s$-$R_s$ plane resembles the distribution of the true negatives in the corresponding no-moon cases. The ambiguous runs with a $\Delta$BIC around 0 still mostly recover the injected moon parameters. This is especially clear for the 700 ppm level, with 50 % more ambiguous runs than true positives, where most of the runs still recover the injected parameters.

### 2.4.3. Transit injection into real out-of-transit data

We inject synthetic transits into the Kepler-1625 PDCSAP data prior to our own detrending (see Sect. 2.4.1). We use the PDCSAP flux instead of SAP flux because (1) it was easier for us to automate the anomaly detection and (2) PDCSAP flux has been cleaned from common systematics. Since the PDC pipeline removes many of the jumps in the data, we can focus on a single type of anomaly, that is gaps. Gaps are relatively easy to detect in an automated way, removing the requirement of visual inspection of each light curve. For the injection, we select out-of-transit parts of the Kepler-1625 light curve that have at least 50 d of mostly uninterrupted data (25 d to both sides of the designated time of transit injection), but accept the presence of occasional gaps with durations of up to several days during the injection process.

The set of 200 synthetic light curves was provided by MH to KR for blind retrieval without any disclosure as to which of the sequences have a moon. The time of mid-transit was communi-cated with a precision of 0.1 days to avoid the requirement of a pre-stage transit search. This is justified because (i.) the original transits of Kepler-1625 b have already been detected and (ii.) the transit are visible by-eye and do not necessarily need computer-based searches. We provide the 200 datasets to the community for reproducibility[4] and encourage further blind retrievals.

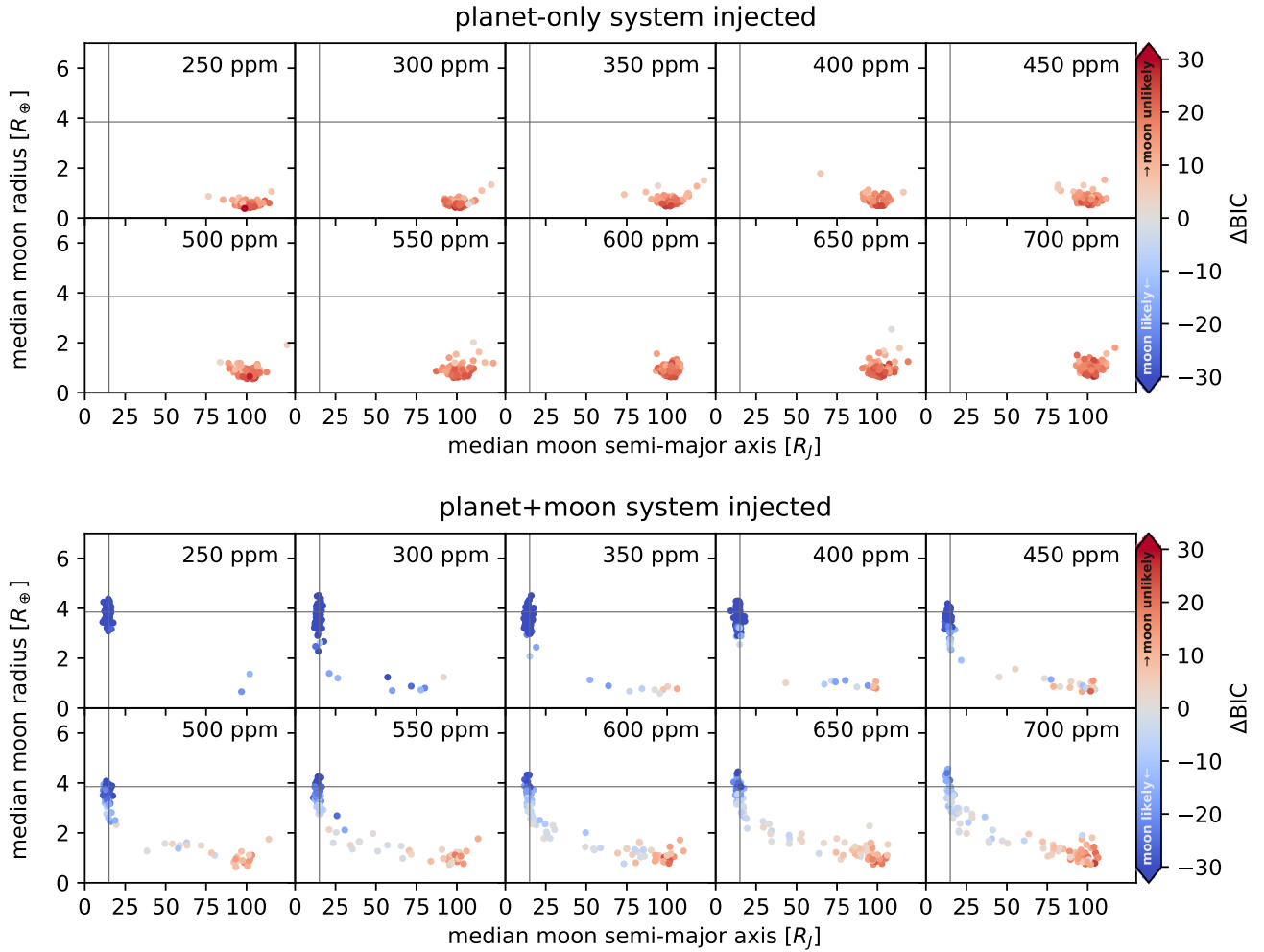

**Fig. 6.** Distribution of the median likelihood $R_s$ and $a_s$ for all the runs for the different noise levels, with the runs injecting planet and moon on the top and runs injecting only a planet in the bottom. The ΔBIC of the planet-moon model compared to the no-moon model for all runs is indicated by the color. Generally runs with a low ΔBIC (indicating the presence of a moon) also are in the vicinity of the injected parameter.

### 2.4.4. Detrending of the transit-injected light curves

The detrending procedure for our injection-retrieval experiment differs from the one used to detrend the original light curve around the Kepler-1625 b transits (see Sect. 2.1.1) in two regards.

First, we test the effect of the detrending function. In addition to the trigonometric function, we detrend the light curve by polynomials of second, third and fourth order.

In addition, we test if the inclusion or neglect of data beyond any gaps in the light curve affects the detrending. In one variation of our detrending procedure, we use the entire ± 25 d of data (excluding any data within $t_c$) around a transit midpoint. In another variation, we restrict the detrending to the data up to the nearest gap (if present) on both sides of the transit.

To avoid the requirement of time-consuming visual inspections of each light curve, we construct an automatic rule to determine the presence of gaps, which are the most disruptive kind of artifact to our detrending procedure. We define a gap as an interruption of the data of more than half a day. Whenever we do detect a gap, we cut another 12 h at both the beginning and the end of the gap, since our visual inspection of the data showed that many gaps are preceded or followed by anomalous trends (see e.g. the gap 4 d after transit 4 in Fig. 1).

We ignore any data points within $t_c$ around the transit midpoint (see Sect. 2.1.1). If a gap starts within an interval $[t_c/2, t_c/2 + 12\,\mathrm{h}]$ around the transit midpoint, then we lift our constraint of dismissing a 12 h interval around gaps and use all the data within $[t_c/2, t_c/2 + 12\,\mathrm{h}]$ plus any data up to 12 h around the next gap.

If all these cuts result in no data points for the detrending procedure to one side of one of the three transits in a sequence, then we ignore the entire sequence for our injection-retrieval experiment. This is the case for 40 out of the 200 artificially injected light curves. This high loss rate of our experimental data is a natural outcome of the gap distribution in the original Kepler-1625 light curve. We exclude these 40 light curves for all variations of the detrending procedure that we investigate. All things combined, these constraints produce synthetic light curves with gap characteristics similar to the original Kepler-1625 b transits (see Fig. 3), that is, we allow the simulation of light curves with gaps close to but not ranging into the transits. The four detrending functions and our two ways of treating gaps yield a total of eight different detrending methods that we investigate (see Table 3).

**Table 3.** Definition of the detrending identifiers in relation to the respective detrending functions that we explored in our transit injection-retrieval experiment of the Kepler-1625 data. We define a gap as any empty parts in the light curve that show more than 12 h between consecutive data points. The trigonometric function refers to our reimplementation of the CoFiAM algorithm. P2 to P4 refer to polynomials of second to fourth order. T refers to our trigonometric detrending. G stands for the inclusion of data beyond gaps, N stands for the exclusion of data beyond gaps.

| Identifier | Detrending Function | Reject Data Beyond Gap? |
|---|---|---|
| P2/G | 2nd order polynomial | yes |
| P2/N | 2nd order polynomial | no |
| P3/G | 3rd order polynomial | yes |
| P3/N | 3rd order polynomial | no |
| P4/G | 4th order polynomial | yes |
| P4/N | 4th order polynomial | no |
| T/G | trigonometric | yes |
| T/N | trigonometric | no |

## 3. Results

Our first result is a reproduction of a detrended transit light curve of Kepler-1625 b that has the same morphology and moon characterization as the one proposed by Teachey et al. (2018) and that has a negative $\Delta$BIC. We explore the variation of the free parameters of our trigonometric detrending procedure, $f_{t_p}$ and $f_{t_c}$, and identify such a detrended light curve for $f_{t_p} = 4.4$ and $f_{t_c} = 2.2$. Figure 7 shows the resulting light curve.

In Fig. 8 we show the results of our MCMC analysis of this particular light curve, which yields a moon with $a_s = 16.3^{+5.0}_{-1.9} \, R_J$ and $R_s = 2.87^{+0.61}_{-0.94} \, R_\oplus$. While both the moon radius and semimajor axis are well constrained, the distribution of the initial planet-moon orbital conjunction ($t_{0,s}$) fills out almost the entire allowed parameter range from $-1/2 \, P_s$ to $+1/2 \, P_s$. The planetary radius is $0.863^{+0.072}_{-0.051} \, R_J$, the stellar radius is $R_\star = 1.57^{+0.11}_{-0.09} \, R_\odot$, and the density is $\rho_\star = 0.26^{+0.04}_{-0.05} \, \rho_\odot$.

The point of maximum likelihood in the resulting MCMC distribution is at $a_s = 14.7 \, R_J$, $R_s = 3.4 \, R_\oplus$, $R_\star = 1.57 \, R_\odot$, $\rho_\star = 0.23 \, \rho_\odot$ and $R_p = 8.63 \, R_J$. The $\Delta$BIC($\mathcal{M}_1, \mathcal{M}_0$) we found is -4.954, indicating moderate evidence in favor of an exomoon being in the light curve.

### 3.1. Injection-retrieval experiment

In Fig. 9 we show the $\Delta$BIC for the 160 simulated Kepler light curves that were not rejected by our detrending method due to gaps very close to a transit. The left panel shows our results for the analysis of planet-only injections and the right panel refers to planet-moon injections. The tables in the panel headers list the true negative, false positive, true positive, and false negative rates as well as the rates of ambiguous cases. With "positive" ("negative"), we here refer to the detection (non-detection) of a moon.

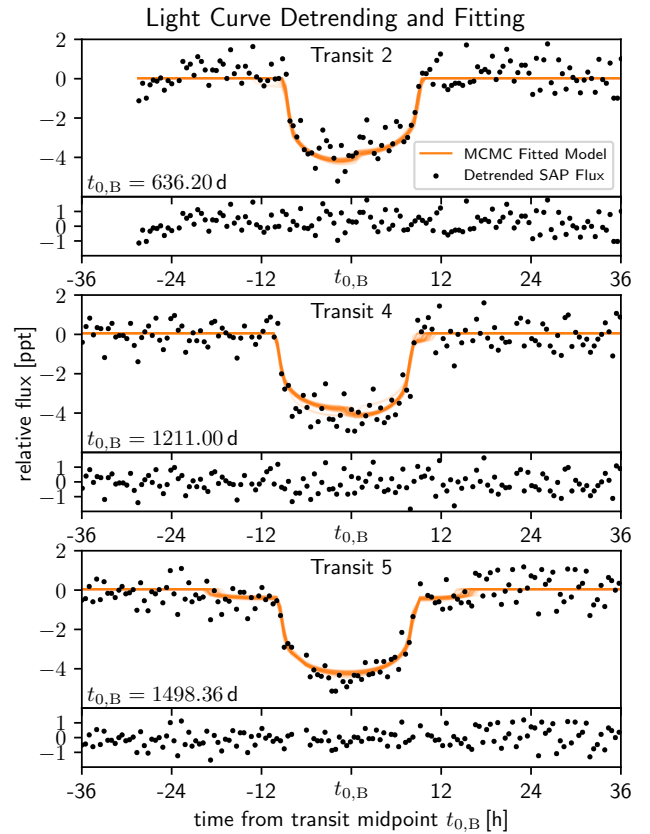

**Fig. 7.** The observed 2nd, 4th, and 5th transits of Kepler 1625 b. Black dots refer to our detrended light curve from the trigonometric detrending procedure, and orange curves are the model light curves generated using the 100 best fitting parameter sets of the MCMC run. The $\Delta$BIC, calculated from the most likely parameters, is −4.954.

In particular, we find the true negative rate (left panel, $\Delta$BIC $\geq 6$) to be between 65 % and 87.5 % and the true positive rate (right panel, $\Delta$BIC $\leq -6$) to be between 31.25 % and 46.25 % depending on the detrending method, respectively.

The rates of false classifications is between 8.75 % and 17.5 % for the injected planet-only systems with a falsely detected moon (false positives) and between 30 % and 41.25 % for the injected planet-moon systems with a failed moon recovery (false negatives).

The rates of classification as a planet-moon system depends significantly on the treatment of gaps during the detrending procedure. Whenever the light curve is cut at a gap, the detection rates for a moon increase – both for the false positives and for the true positives. Among all the detrending methods, this effect is especially strong for the trigonometric detrending. The false positive rate increases by almost a factor of two from 8.75 % (T/N) to 16.25 % (T/G) and the true positive rate increases by 15 % to 46.25 %. The effect on the true negative rate is strongest for the trigonometric detrending, decreasing from 87.5 % when the light curve is not cut at gaps (T/N) to 72.5 % if the light curve is cut (T/G). The false negative rate for the second order polynomial detrending decreases from 41.25 % (P2/N) to 30 % (P2/G) when gaps are cut, while the false negative rates of the other detrending methods remain almost unaffected.

Of all the light curves with an injected planet only, 21.25 % have an ambiguous classification with at least one of the detrending methods showing a negative and a different method showing a positive $\Delta\mathrm{BIC}$ above the threshold. For the light curves with an injected planet-moon system, there are 18.75 % with ambiguous classification and another 18.75 % of the injected planet-moon systems are classified unanimously as true positives by all detrending methods.

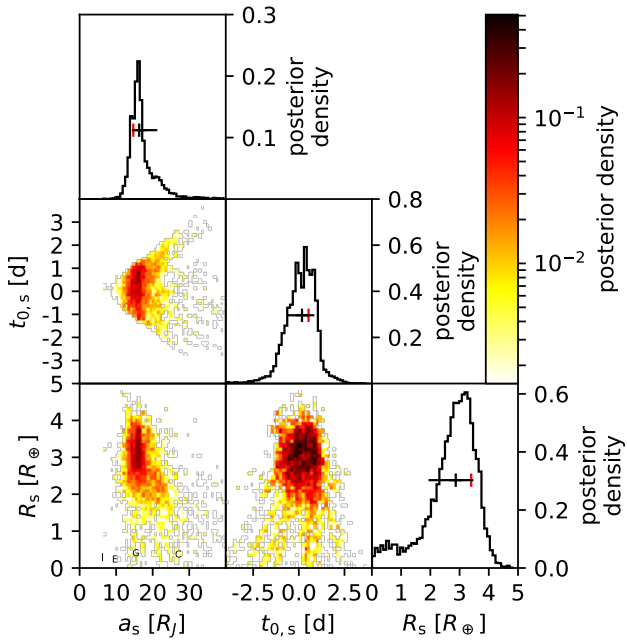

**Fig. 8.** Posterior probability distribution of the moon parameters generated by the MCMC algorithm for the light curve detrended by the trigonometric detrending. The black vertical lines show the median of the posterior distribution, the black horizontal lines indicate the $1\sigma$ range around the median. The red vertical lines show the point of maximum likelihood. The locations of the Galilean moons are included in the lower left panel for comparison.

Fig. 10 shows the distribution of the retrieved moon parameters $a_s$ and $R_s$ as well as the corresponding $\Delta\mathrm{BIC}$ (see color scale) for each of the detrending methods.

For the light curves with an injected planet-moon system (lower set of panels), the maximum likelihood values of $a_s$ and $R_s$ of the true moons (blue) generally cluster around the injected parameters. In particular, we find that the moon turns out to be more likely (deeper-blue dots) when it is fitted to have a larger radius. The parameters of the false positives (blue dots in the upper set of panels) are more widely spread out, with moon radii ranging between 2 and 5 $R_\oplus$ and the moon semimajor axes spread out through essentially the entire parameter range that we explored. The clustering of median $a_s$ at around 100 $R_J$ is an artifact of taking the median over a very unlocalized distribution along $a_s$. For the polynomial detrending methods there are a certain number of what one could refer to as mischaracterized true positives. In these cases the $\Delta\mathrm{BIC}$-based planet-only vs. planet-moon classification is correct but the maximum likelihood values are very different from the injected ones.

The correctly identified planet-only systems show a similar distribution of $a_s$ and $R_s$ as in our experiment with white noise only and a 700 ppm amplitude (Fig. 5).

Most surprisingly, and potentially most worryingly, the false positives (blue dots in the upper set of panels in Fig. 10) cluster around the values of the moon parameters found by Teachey et al. (2018), in particular if the light curve is cut at the first gap.

## 4. Discussion

In this article we compare several detrending methods of the light curve of Kepler-1625, some of which were used by Teachey et al. (2018) in their characterization of the exomoon candidate around Kepler-1625 b. However, we do not perform an exhaustive survey of all available detrending methods, such as Gaussian processes (Aigrain et al. 2016).

We show that the sequential detrending and fitting procedure of transit light curves is prone to introducing features that can be misinterpreted as signal, in our case as an exomoon. This "prewhitening" method of the data has thus to be used with caution. Our investigations of a polynomial-based fitting and of a trigonometric detrending procedure show that the resulting best-fit model depends strongly on the specific detrending function, e.g. on the order of the polynomial or on the minimum time scale (or wavelength) of a cosine. This is crucial for any search of secondary effects in the transit light curves – moons, rings, evaporating atmospheres etc. – and is in stark contrast to a claim by Aizawa et al. (2017), who stated that neither the choice of the detrending function nor the choice of the detrending window of the light curve would have a significant effect on the result. We find that this might be true on a by-eye level but not on a level of 100 ppm or below. Part of the difference between our findings and those of Aizawa et al. (2017) could be in the different time scales we investigate. While they considered the effect of stellar flairs on time scales of less than a day, much less than the ~2 day transit duration of their specific target, our procedure operates on various time scales of up to several weeks. Moreover, we develop a dynamical moon model to fit multiple transits, whereas Aizawa et al. (2017) study only a single transit.

Since the actual presence and the putative orbital position of a hypothetical exomoon around Kepler-1625 b is unknown a priori, it is unclear how much of the light curve would need to be protected from (or neglected for) the pre-fit detrending process in order to avoid a detrending of a possible moon signal itself. In turn, we show that in the case of Kepler-1625 different choices for this protected time scale around the transit yield different confidences and different solutions for a planet-moon system. We find that the previously announced solution by Teachey et al. (2018) is only one of many possibilities with similar likelihoods (specifically: Bayesian Information Criteria). This suggests, but by no means proves, that all of these solutions could, in fact, be due to red noise artifacts (e.g. stellar or instrumental) rather than indicative of a moon signal.

Our finding of higher true positive rate compared to a false positive rate from injection-retrieval experiments could be interpreted as slight evidence in favor of a genuine exomoon. This interpretation, however, depends on the number of transiting planets and planet candidates around stars with similar noise characteristics that were included in the Teachey et al. (2018) search. Broadly speaking, if more than a handful of similar targets were studied, the probability of at least one false positive detection becomes quite likely.

## 5. Conclusions

We investigated the detrending of the transit light curve of Kepler-1625 b with a method very similar to the one used by Teachey et al. (2018) and then applied a Bayesian framework with MCMC modeling to search for a moon. Our finding of a $\Delta\mathrm{BIC}$ of $-4.954$ favors the planet-moon over the planet-only hypothesis. Although significant, this tentative detection fails to cross the threshold of $-6$, which we would consider strong evi-dence of a moon. Our ΔBIC value would certainly change if we could include the additional data from the high-precision transit observations executed in October 2017 with the Hubble Space Telescope (Teachey et al. 2018) in our analysis. Moreover, by varying the free parameters of our detrending procedure, we also find completely different solutions for a planet-moon system, i.e. different planet-moon orbital configurations during transits and different moon radii or planet-moon orbital semimajor axes.

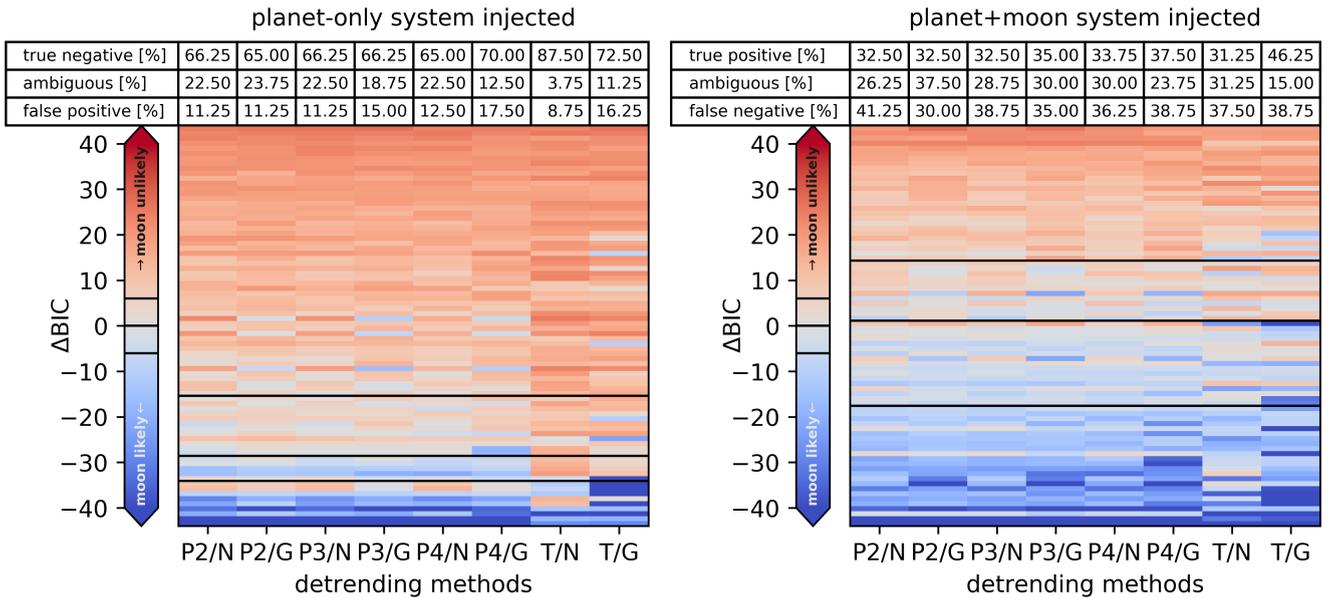

**Fig. 9.** Difference between the BIC of the planet-moon model and the no-moon model using different detrending methods for 160 light curves, generated using the PDCSAP flux of Kepler 1625, injected with three simulated transits. On the left (80 light curves) a planet and moon transit was injected, on the right (80 light curves) only the planet. Each light curve consists of three consecutive transits. Each row of 8 detrending methods uses the same light curve. The rows are sorted by their mean ΔBIC, with black lines indicating the ΔBIC = {−6, 0, 6} positions for the mean ΔBIC per row.

As an extension to this validation of the previously published work, we performed 200 injection-retrieval experiments into the original out-of-transit parts of the Kepler light curve. We also extended the previous work by exploring different detrending methods, such as second-, third-, and fourth-order polynomials as well as trigonometric methods and find false-positive rates between 8.75 % and 16.25 %, depending on the method. Surprisingly, we find that the moon radius and planet-moon distances of these false positives are very similar to the ones measured by Teachey et al. (2018). In other words, in 8.75 % to 16.25 % of the light curves that contained an artificially injected planet only, we find a moon that is about as large as Neptune and orbits Kepler-1625 b at about 20 $R_J$.

To sum up, we find tentative statistical evidence for a moon in this particular Kepler light curve of Kepler-1625, but we also show that the significant fraction of similar light curves, which contained a planet only, would nevertheless indicate a moon with properties similar to the candidate Kepler-1625 b-i. Clearly, stellar and systematic red noise components are the ultimate barrier to an unambiguous exomoon detection around Kepler-1625 b and follow-up observations have the potential of solving this riddle based on the framework that we present.

Of all the detrending methods we investigated, the trigonometric method, which is very similar to the CoFiAM method of Teachey et al. (2018), can produce the highest true positive rate. At the same time, however, this method also ranks among the ones producing the highest false positive rates as well. To conclude, we recommend that any future exomoon candidate be detrended with as many different detrending methods as possible to evaluate the robustness of the classification.

*Acknowledgements.* We thank James Kuszlewicz and Jesper Schou for useful discussions. This work was supported in part by the German Aerospace Center (DLR) under PLATO Data Center grant 50OL1701. This paper includes data collected by the Kepler mission. Funding for the Kepler mission is provided by the NASA Science Mission directorate. This work has made use of data provided by NASA and the Space Telescope Science Institute. K.R. is a member of the International Max Planck Research School for Solar System Science at the University of Göttingen. K.R. contributed to the analysis of the simulated light curves, to the interpretation of the results, and to the writing of the article.

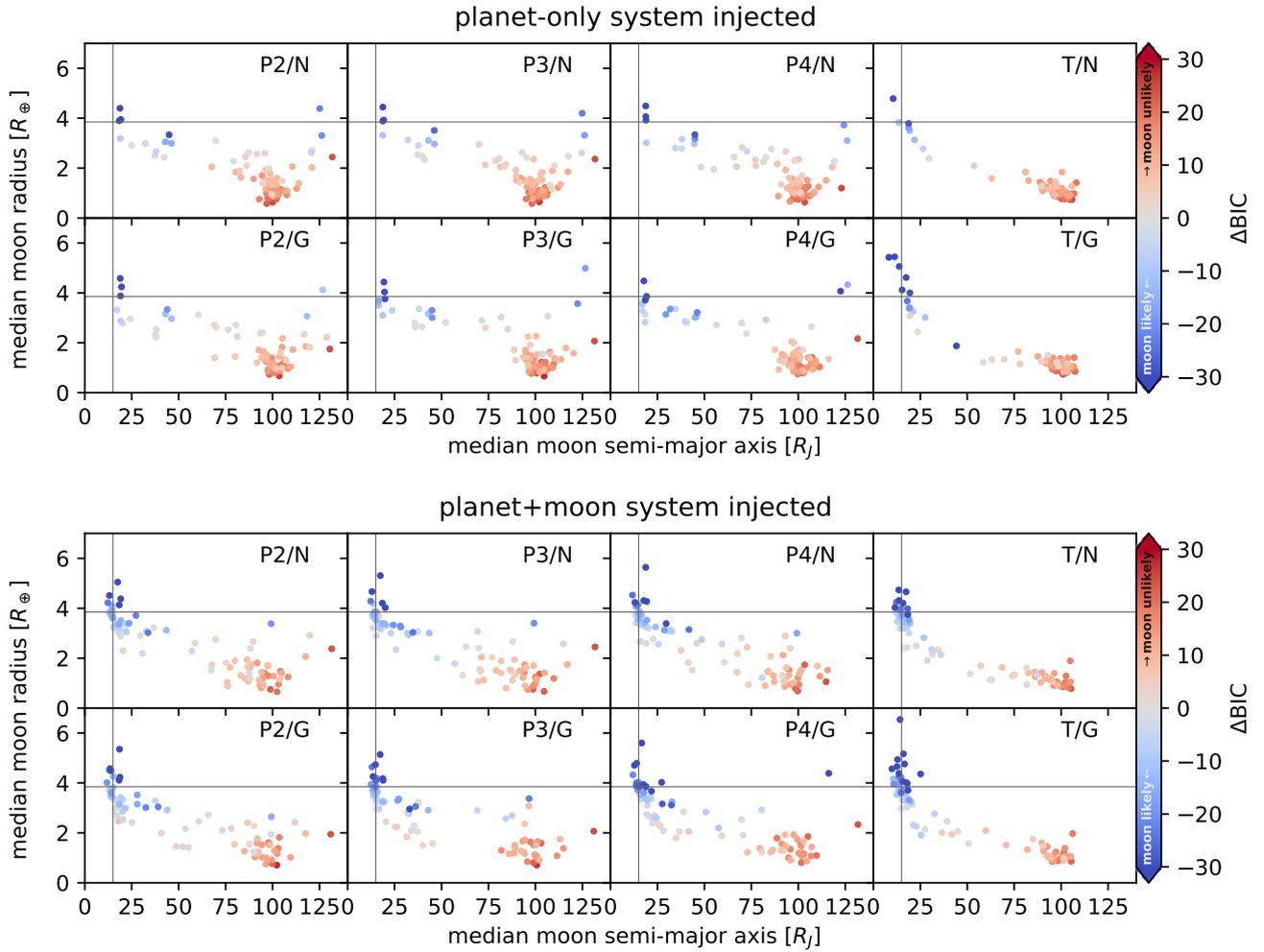

**Fig. 10.** Distribution of the median likelihood $R_s$ and $a_s$ for the transits injected into different parts of the Kepler-1625 light curve, using different detrending methods. The $\Delta$BIC of the planet-only model compared to the planet-moon model is indicated by the symbol color. The values of the moon semimajor axis (abscissa) and radius (ordinate) suggested by Teachey et al. (2018) are indicated with thin, gray lines in each sub-panel.

## Appendix A: Effect of the window length on the Bayesian Information Criterion

Given the constraint of orbital stability, a moon can only possibly orbit its planet within the planet's Hill sphere. Hence, transits may only occur within a certain time interval around the midpoint of the planetary transit. This time $t_{\mathrm{Hill}}$ can be calculated as

$$t_{\mathrm{Hill}} = \frac{\eta\, R_{\mathrm{Hill}}}{v_{\mathrm{orbit}}} = \eta \frac{P}{2\pi} \sqrt{\frac{M_{\mathrm{p}}}{3 M_{\star}}} \ , \qquad (A.1)$$

where $v_{\mathrm{orbit}}$ is the orbital velocity of the planet-moon system around the star, $M_{\mathrm{p}}$ and $M_{\star}$ the planet and star mass, $P$ the orbital period of the planet-moon system, and $R_{\mathrm{Hill}}$ is the Hill radius of the planet. $\eta$ is a factor between 0 and 1, which has been numerically determined for prograde moons ($\eta \approx 0.5$) and for retrograde moons ($\eta \approx 1$), details depending on the orbital eccentricities (Domingos et al. 2006). We focus on prograde moons and choose $\eta = 0.5$. For a 10 $M_{\mathrm{J}}$ planet in a 287 d orbit around a 1.1 $M_{\odot}$ star the Hill time is $t_{\mathrm{Hill}} = 3.25$ d.

As shown in Fig. 2, the length of the light curve, which is neglected for the polynomial fit has a strong effect on the resulting detrended light curve. Figure A.1 shows the effect that different cutout times $t_{\mathrm{c}}$ and detrending base lines $D$ can have on whether a moon is detected or not.

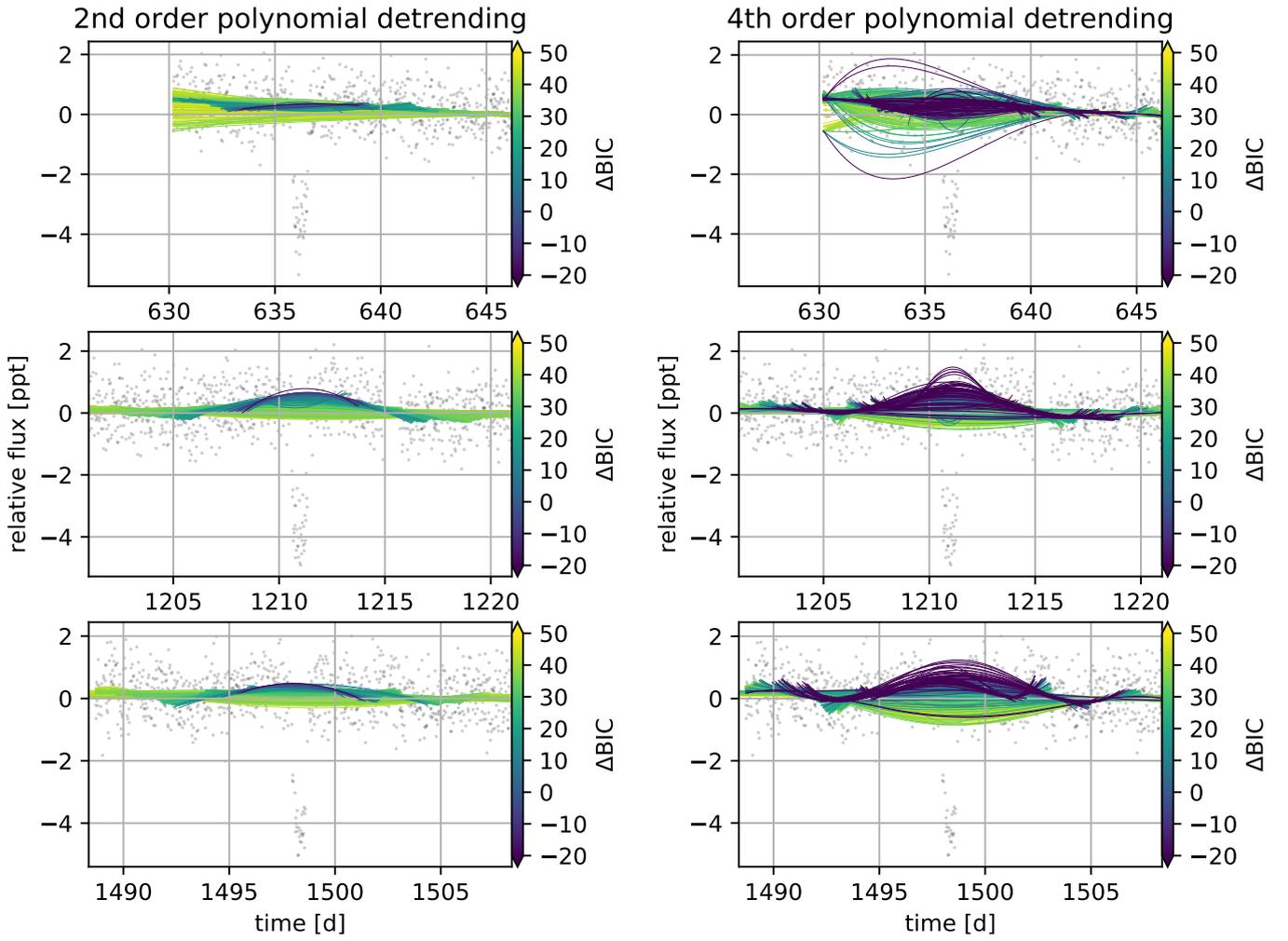

**Fig. A.1.** Detrending for different cutout times $t_c$ and base length $D$, color coded by the resulting $\Delta$BIC using a 2nd- and 4th-order polynomial function. While some of the detrending models corresponding to a large negative $\Delta$BIC are clearly results of wrong detrending, it is much less clear for many other detrending models.

## Appendix B: Autocorrelation of Detrended Light Curves

The autocorrelations of the detrended light curves are shown in Fig. B.1. For all three transits, the autocorrelation is close to zero, except for the zero-lag component. This suggests that it is reasonable to model the noise covariance matrix as a diagonal matrix (see Sect. 2.2.3).

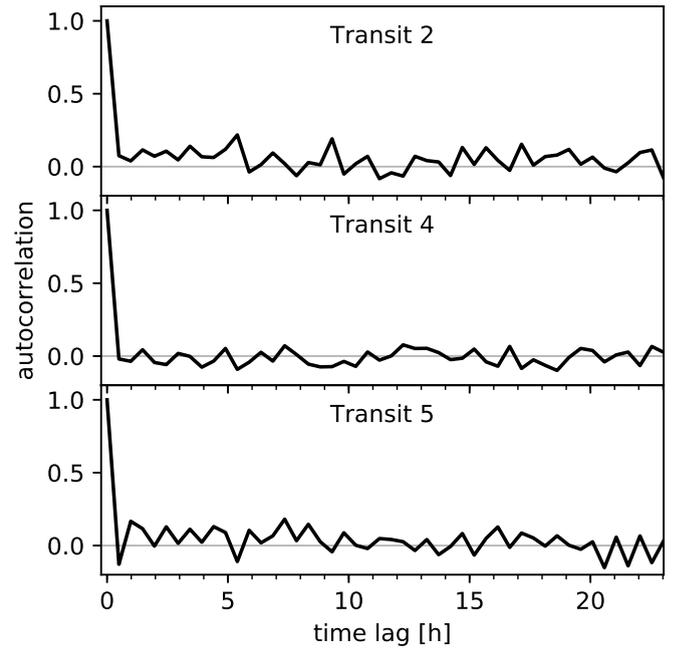

**Fig. B.1.** The autocorrelation of the difference between the detrended light curve and the best fitting model.

## 6.8 An Alternative Interpretation of the Exomoon Signal in the Combined Kepler and Hubble Data of Kepler-1625 (Heller et al. 2019)

Contribution:

RH guided the work, did the literature research, created Fig. 5, led the writing of the manuscript, and served as a corresponding author for the journal editor and the referees.

# An alternative interpretation of the exomoon candidate signal in the combined *Kepler* and *Hubble* data of Kepler-1625

René Heller[1], Kai Rodenbeck[2, 1], and Giovanni Bruno[3]

[1] Max Planck Institute for Solar System Research, Justus-von-Liebig-Weg 3, 37077 Göttingen, Germany
    heller@mps.mpg.de, rodenbeck@mps.mpg.de
[2] Institute for Astrophysics, Georg August University Göttingen, Friedrich-Hund-Platz 1, 37077 Göttingen, Germany
[3] INAF, Astrophysical Observatory of Catania, Via S. Sofia 78, 95123 Catania, Italy, giovanni.bruno@inaf.it

## ABSTRACT

*Context.* *Kepler* and *Hubble* photometry of a total of four transits by the Jupiter-sized exoplanet Kepler-1625 b have recently been interpreted to show evidence of a Neptune-sized exomoon. The key arguments were an apparent drop in stellar brightness after the planet's October 2017 transit seen with *Hubble* and its 77.8 min early arrival compared to a strictly periodic orbit.
*Aims.* The profound implications of this first possible exomoon detection and the physical oddity of the proposed moon, i.e., its giant radius prompt us to examine the planet-only hypothesis for the data and to investigate the reliability of the Bayesian information criterion (BIC) used for detection.
*Methods.* We combined *Kepler*'s Pre-search Data Conditioning Simple Aperture Photometry (PDCSAP) with the previously published *Hubble* light curve. In an alternative approach, we performed a synchronous polynomial detrending and fitting of the *Kepler* data combined with our own extraction of the *Hubble* photometry. We generated five million parallel-tempering Markov chain Monte Carlo (PTMCMC) realizations of the data with both a planet-only model and a planet-moon model, and compute the BIC difference ($\Delta$BIC) between the most likely models, respectively.
*Results.* The $\Delta$BIC values of $-44.5$ (using previously published *Hubble* data) and $-31.0$ (using our own detrending) yield strong statistical evidence in favor of an exomoon. Most of our orbital realizations, however, are very different from the best-fit solutions, suggesting that the likelihood function that best describes the data is non-Gaussian. We measure a 73.7 min early arrival of Kepler-1625 b for its *Hubble* transit at the $3\,\sigma$ level. This deviation could be caused by a 1 d data gap near the first *Kepler* transit, stellar activity, or unknown systematics, all of which affect the detrending. The radial velocity amplitude of a possible unseen hot Jupiter causing the Kepler-1625 b transit timing variation could be approximately $100\,\mathrm{m\,s^{-1}}$.
*Conclusions.* Although we find a similar solution to the planet-moon model to that previously proposed, careful consideration of its statistical evidence leads us to believe that this is not a secure exomoon detection. Unknown systematic errors in the *Kepler*/*Hubble* data make the $\Delta$BIC an unreliable metric for an exomoon search around Kepler-1625 b, allowing for alternative interpretations of the signal.

**Key words.** eclipses – methods: data analysis – planets and satellites: detection – planets and satellites: dynamical evolution and stability – planets and satellites: individual: Kepler-1625 b – techniques: photometric

## 1. Introduction

The recent discovery of an exomoon candidate around the transiting Jupiter-sized object Kepler-1625 b orbiting a slightly evolved solar mass star (Teachey et al. 2018) came as a surprise to the exoplanet community. This Neptune-sized exomoon, if confirmed, would be unlike any moon in the solar system, it would have an estimated mass that exceeds the total mass of all moons and rocky planets of the solar system combined. It is currently unclear how such a giant moon could have formed (Heller 2018).

Rodenbeck et al. (2018) revisited the three transits obtained with the *Kepler* space telescope between 2009 and 2013 and found marginal statistical evidence for the proposed exomoon. Their transit injection-retrieval tests into the out-of-transit *Kepler* data of the host star also suggested that the exomoon could well be a false positive. A solution to the exomoon question was supposed to arrive with the new *Hubble* data of an October 2017 transit of Kepler-1625 b (Teachey & Kipping 2018).

The new evidence for the large exomoon by Teachey & Kipping (2018), however, remains controversial. On the one hand, the *Hubble* transit light curve indeed shows a significant decrease in stellar brightness that can be attributed to the previously suggested moon. Perhaps more importantly, the transit of Kepler-1625 b occurred 77.8 min earlier than expected from a sequence of strictly periodic transits, which is in very good agreement with the proposed transit of the exomoon candidate, which occurred before the planetary transit. On the other hand, an upgrade of *Kepler*'s Science Operations Center pipeline from version 9.0 to version 9.3 caused the exomoon signal that was presented in the Simple Aperture Photometry (SAP) measurements in the discovery paper (Teachey et al. 2018) to essentially vanish in the SAP flux used in the new study of Teachey & Kipping (2018). This inconsistency, combined with the findings of Rodenbeck et al. (2018) that demonstrate that the characterization and statistical evidence for this exomoon candidate depend strongly on the methods used for data detrending, led us to revisit the exomoon interpretation in light of the new *Hubble* data.

Here we address two questions. How unique is the proposed orbital solution of the planet-moon system derived with the Bayesian information criterion (BIC)? What could be the reason for the observed 77.8 min difference in the planetary transit timing other than an exomoon?

## 2. Methods

Our first goal was to fit the combined *Kepler* and *Hubble* data with our planet-moon transit model (Rodenbeck et al. 2018) and to derive the statistical likelihood for the data to represent the model. In brief, we first model the orbital dynamics of the star-planet-moon system using a nested approach, in which the planet-moon orbit is Keplerian and unperturbed by the stellar gravity. The transit model consists of two black circles, one for the planet and one for the moon, that pass in front of the limb-darkened stellar disk. The variations in the stellar brightness are computed using Ian Crossfield's `python` code of the Mandel & Agol (2002) analytic transit model.[1] The entire model contains 16 free parameters and it features three major updates compared to Rodenbeck et al. (2018): (1) Planet-moon occultations are now correctly simulated, (2) the planet's motion around the local planet-moon barycenter is taken into account, and (3) inclinations between the circumstellar orbit of the planet-moon barycenter and the planet-moon orbit are now included.

We used the `emcee` code[2] of Foreman-Mackey et al. (2013) to generate Markov chain Monte Carlo (MCMC) realizations of our planet-only model ($\mathcal{M}_0$) and planet-moon model ($\mathcal{M}_1$) and to derive posterior probability distributions of the set of model parameters ($\vec{\theta}$). We tested both a standard MCMC sampling with 100 walkers and a parallel-tempering ensemble MCMC (PTMCMC) with five temperatures, each of which has 100 walkers. As we find a better convergence rate for the PTMCMC sampling, we use it in the following. Moreover, PTMCMC can sample both the parameter space at large and in regions with tight peaks of the likelihood function. The PTMCMC sampling is allowed to walk five million steps.

The resulting model light curves are referred to as $\mathcal{F}_i(t, \vec{\theta})$, where $t$ are the time stamps of the data points from *Kepler* and *Hubble* ($N$ measurements in total), for which time-uncorrelated standard deviations $\sigma_j$ at times $t_j$ are assumed, following the suggestion of Teachey & Kipping (2018). This simplifies the joint probability density of the observed (and detrended) flux measurements ($F(t)$) to the product of the individual probabilities for each data point,

$$p(F|\vec{\theta}, \mathcal{M}_i) = \prod_{j=1}^{N} \frac{1}{\sqrt{2\pi\sigma_j^2}} \exp\left(-\frac{\left(F(t_j) - \mathcal{F}_i(t_j, \vec{\theta})\right)^2}{2\sigma_j^2}\right). \quad (1)$$

We then determined the set of parameters ($\vec{\theta}_{\mathrm{max}}$) that maximizes the joint probability density function ($p(F|\vec{\theta}_{\mathrm{max}}, \mathcal{M}_i)$) for a given light curve $F(t_j)$ and model $\mathcal{M}_i$ and calculated the BIC (Schwarz 1978)

$$\mathrm{BIC}(\mathcal{M}_i|F) = m_i \ln N - 2 \ln p(F|\vec{\theta}_{\mathrm{max}}, \mathcal{M}_i). \quad (2)$$

The advantage of the BIC in comparison to $\chi^2$ minimization, for example, is in its relation to the number of model parameters

($m_i$) and data points. The more free parameters in the model, the stronger the weight of the first penalty term in Eq. (2), thereby mitigating the effects of overfitting. Details of the actual computer code implementation or transit simulations aside, this Bayesian framework is essentially what the *Hunt for Exomoons with Kepler* survey used to identify and rank exomoon candidates (Kipping et al. 2012), which ultimately led to the detection of the exomoon candidate around Kepler-1625 b after its first detection via the orbital sampling effect (Heller 2014; Heller et al. 2016a).

### 2.1. Data preparation

In a first step, we used *Kepler*'s Pre-search Data Conditioning Simple Aperture Photometry (PDCSAP) and the *Hubble* Wide Field Camera 3 (WFC3) light curve as published by Teachey & Kipping (2018) based on their quadratic detrending. Then we executed our PTMCMC fitting and derived the $\Delta$BIC values and the posterior parameter distributions.

In a second step, we did our own extraction of the *Hubble* light curve including an exponential ramp correction for each *Hubble* orbit. Then we performed the systematic trend correction together with the transit fit of a planet-moon model. Our own detrending of the light curves is not a separate step, but it is integral to the fitting procedure. For each calculation of the likelihood, we find the best fitting detrending curve by dividing the observed light curve by the transit model and by fitting a third-order polynomial to the resulting light curve. Then we remove the trend from the original light curve by dividing it through the best-fit detrending polynomial and evaluate the likelihood. We also performed a test in which the detrending parameters were free PTMCMC model parameters and found similar results for the parameter distributions but at a much higher computational cost. We note that the resulting maximum likelihood is (and must be) the same by definition if the PTMCMC sampling converges.

Kepler-1625 was observed by *Hubble* under the GO program 15149 (PI Teachey). The observations were secured from October 28 to 29, 2017, to cover the $\sim 20\,\mathrm{hr}$ transit plus several hours of out-of-transit stellar flux (Teachey & Kipping 2018). The F130N filter of WFC3 was used to obtain a single direct image of the target, while 232 spectra were acquired with the G141 grism spanning a wavelength range from 1.1 to 1.7 μm. Due to the faintness of the target, it was observed in staring mode (e.g. Berta et al. 2012; Wilkins et al. 2014) unlike the most recent observations of brighter exoplanet host stars, which were monitored in spatial scanning mode (McCullough & MacKenty 2012). Hence, instead of using the IMA files as an intermediate product, we analyzed the FLT files, which are the final output of the `calwfc3` pipeline of *Hubble* and allow a finer manipulation of the exposures during consecutive nondestructive reads. Each FLT file contains measurements between about 100 and 300 electrons per second, with exposure times of about 291 seconds.

We used the centroid of the stellar image to calculate the wavelength calibration, adopting the relations of Pirzkal et al. (2016). For each spectroscopic frame, we first rejected the pixels flagged by `calwfc3` as "bad detector pixels", pixels with unstable response, and those with uncertain flux value (Data Quality condition 4, 32, or 512). Then we corrected each frame with the flat field file available on the Space Telescope Science Institute (STScI) website[3] by following the prescription of the WFC3 online manual. We performed the background subtraction on a column-by-column basis. Due to a number of contaminant stars in the observation field (Fig. 1, top panel), we carefully selected a region on the detector that was as close as possible to the spectrum of Kepler-1625, close to row 150 in spatial dimension, and far from any contaminant. For each column on the detector, we applied a $5\sigma$ clipping to reject the outliers and then calculated the median background flux value in that column. Following STScI prescriptions, we also removed pixels with an electron-per-second count larger than 5. An example for the background behavior is shown in the bottom panel of Fig. 1.

---

[1] Available at www.astro.ucla.edu/~ianc/files
[2] Available at http://dfm.io/emcee/current/user/pt

[3] www.stsci.edu/hst/wfc3/analysis/grism_obs/calibrations/wfc3_g141.html

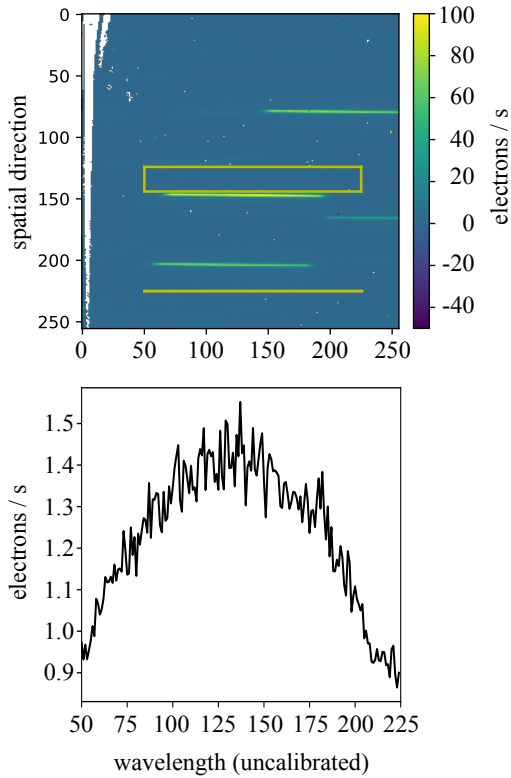

**Fig. 1.** *Top*: Example of a WFC3 exposure of Kepler-1625. The abscissa shows the column pixel prior to wavelength calibration. The yellow box indicates the region used for background estimation. The spectrum of Kepler-1625 is at the center of the frame, around row 150 in the spatial direction, while several contaminant sources are evident in other regions of the detector. The color bar illustrates the measured charge values. *Bottom*: Background value measured across the rows of the same frame.

We inspected each frame with the `image registration` package (Baker & Matthews 2001) to search drifts in both axes of the detector with respect to the very last frame, and then extracted the spectrum of Kepler-1625 by performing optimal extraction (Horne 1986) on the detector rows containing the stellar flux. This procedure automatically removes bad pixels and cosmic rays from the frames by correcting them with a smoothing function. We started the extraction with an aperture of a few pixels centered on the peak of the stellar trace and gradually increased its extension by one pixel per side on the spatial direction until the flux dispersion reached a minimum.

We performed another outlier rejection by stacking all the one-dimensional spectra along the time axis. We computed a median-filtered version of the stellar flux at each wavelength bin and performed a $3\sigma$ clipping between the computed flux and the median filter. Finally, we summed the stellar flux across all wavelength bins from 1.115 to 1.645 $\mu$m to obtain the band-integrated stellar flux corresponding to each exposure.

Before performing the PTMCMC optimization, we removed the first *Hubble* orbit from the data set and the first data point of each *Hubble* orbit, as they are affected by stronger instrumental effects than the other observations (Deming et al. 2013) and cannot be corrected with the same systematics model. We also removed the last point of the 12th, 13th, and 14th *Hubble* orbit since they were affected by the passage of the South Atlantic Anomaly (as highlighted in the proposal file, available on the STScI website).

### 2.2. Proposed unseen planet

#### 2.2.1. Mass-orbit constraints for a close-in planet

According to Teachey et al. (2018), the 2017 *Hubble* transit of Kepler-1625 b occurred about 77.8 min earlier than predicted, an effect that could be astrophysical in nature and is referred to as a transit timing variation (TTV). As proposed by Teachey et al. (2018), this TTV could either be interpreted as evidence for an exomoon or it could indicate the presence of a hitherto unseen additional planet. Various planetary configurations can cause the observed TTV effect such as an inner planet or an outer planet. At this point, no stellar radial velocity measurements of Kepler-1625 exist that could be used to search for additional nontransiting planets in this system.

In the following, we focus on the possibility of an inner planet with a much smaller orbital period than Kepler-1625 b simply because it would have interesting observational consequences. We use the approximation of Agol et al. (2005) for the TTV amplitude ($\delta t$) due to a close inner planet, which would impose a periodic variation on the position of the star, and solve their expression for the mass of the inner planet ($M_{p,in}$) as a function of its orbital semimajor axis ($a_{p,in}$),

$$M_{p,in} = \delta t \, M_\star \frac{a_{p,out}}{a_{p,in}} \frac{1}{P_{p,out}} \,, \tag{3}$$

where $a_{out} = 0.87$ AU is the semimajor axis of Kepler-1625 b. The validity of this expression is restricted to coplanar systems without significant planet-planet interaction and with $a_{out} \gg a_{in}$, so that TTVs are only caused by the reflex motion of the star around its barycenter with the inner planet.

As we show in Sect. 3.2, the proposed inner planet could be a hot Jupiter. The transits of a Jupiter-sized planet, however, would be visible in the *Kepler* data. As a consequence, we can estimate the minimum orbital inclination ($i$) between Kepler-1625 b and the suspected planet to prevent the latter from showing transits. This angle is given as per $i = \arctan(R_\star/a_{p,in})$ and we use $R_\star = 1.793^{+0.263}_{-0.488} R_\odot$ (Mathur et al. 2017).

#### 2.2.2. Orbital stability

We can exclude certain masses and orbital semimajor axes for an unseen inner planet based on the criterion of mutual Hill stability. This instability region depends to some extent on the unknown mass of Kepler-1625 b. Mass estimates can be derived from a star-planet-moon model, but these estimates are irrelevant if the observed TTVs are due to an unseen planetary perturber. Hence, we assume a nominal Jupiter mass ($M_{Jup}$) for Kepler-1625 b.

The Hill sphere of a planet with an orbital semimajor axis $a_p$ around a star with mass $M_\star$ can be estimated as $R_H = a_p(M_p/[3M_\star])^{1/3}$, which suggests $R_H = 125 R_{Jup}$ for Kepler-1625 b. We calculate the Hill radius of the proposed inner planet accordingly, and identify the region in the mass-semimajor axis diagram of the inner planet that would lead to an overlap of the Hill spheres and therefore to orbital instability.

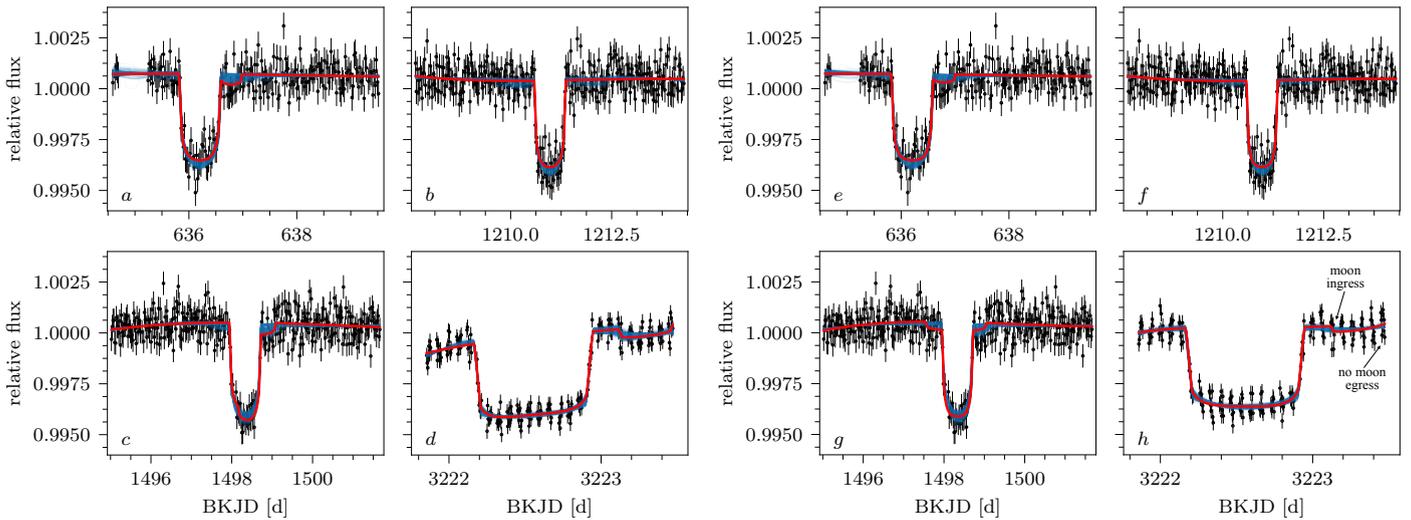

**Fig. 2.** Orbital solutions for Kepler-1625 b and its suspected exomoon based on the combined *Kepler* and *Hubble* data. (*a,b,c*) *Kepler* PDCSAP flux and (*d*) the quadratic detrending of the *Hubble* data from Teachey & Kipping (2018). The blue curves show 1 000 realizations of our PTMCMC fitting of a planet-moon model. Our most likely solution (red line) is very similar to the one found by Teachey & Kipping (2018), but differs significantly from the one initially found by Teachey et al. (2018). (*e,f,g*) *Kepler* PDCSAP flux and (*h*) our own detrending of the *Hubble* light curve (in parallel to the fitting). The ingress and egress of the model moon are denoted with arrows and labels in panel *h* as an example.

## 3. Results

### 3.1. PTMCMC sampling and ΔBIC

Regarding the combined data set of the *Kepler* and *Hubble* data as detrended by Teachey & Kipping (2018), we find a ΔBIC of −44.5 between the most likely planet-only and the most likely planet-moon solution. A combination of the *Kepler* and *Hubble* light curves based on our own extraction of the WFC3 data yields a ΔBIC of −31.0. Formally speaking, both of these two values can be interpreted as strong statistical evidence for an exomoon interpretation. The two values are very different, however, which suggests that the detrending of the *Hubble* data has a significant effect on the exomoon interpretation. In other words, this illustrates that the systematics are not well-modeled and poorly understood.

In Fig. 2a-d, we show our results for the PTMCMC fitting of our planet-moon model to the four transits of Kepler-1625 b including the *Hubble* data as extracted and detrended by Teachey & Kipping (2018) using a quadratic fit. Although our most likely solution shows some resemblance to the one proposed by Teachey & Kipping (2018), we find that several aspects are different. As an example, the second *Kepler* transit (Fig. 2b) is fitted best without a significant photometric moon signature, that is to say, the moon does not pass in front of the stellar disk[4], whereas the corresponding best-fit model of Teachey & Kipping (2018) shows a clear dip prior to the planetary transit (see their Fig. 4). What is more, most of our orbital solutions (blue lines) differ substantially from the most likely solution (red line). In other words, the orbital solutions do not converge and various planet-moon orbital configuration are compatible with the data, though with lower likelihood.

In Fig. 2e-h, we illustrate our results for the PTMCMC fitting of our planet-moon model to the four transits of Kepler-1625 b including our own extraction and detrending of the *Hubble* transit. Again, the orbital solutions (blue lines) do not converge. A comparison of panels d and h shows that the different extraction and detrending methods do have a significant effect on the individual flux measurements, in line with the findings of Rodenbeck et al. (2018). Although the time of the proposed exomoon transit is roughly the same in both panels, we find that the best-fit solution for the data detrended with our own reduction procedure does not contain the moon egress (panel h), whereas the best-fit solution of the data detrended by Teachey & Kipping (2018) does contain the moon egress (panel d). A similar fragility of this particular moon egress has been noted by Teachey & Kipping (2018) as they explored different detrending functions (see their Fig. 3).

Our Fig. 3 illustrates the distribution of the differential likelihood for the planet-moon model between the most likely model parameter set ($\vec{\theta}_{max}$) and the parameter sets ($\vec{\theta}'$) found after five million steps of our PTMCMC fitting procedure, $p(\vec{\theta}'|F, \mathcal{M}_1) - p(\vec{\theta}_{max}|F, \mathcal{M}_1)$. For the combined *Kepler* and *Hubble* data detrended by Teachey & Kipping (2018) (left panel) and for our own *Hubble* data extraction and detrending (right panel), we find that most model solutions cluster around a differential likelihood that is very different from the most likely solution, suggesting that the most likely model is, in some sense, a statistical outlier. We initially detected this feature after approximately the first one hundred thousand PTMCMC fits. Hence, we increased the number of PTMCMC samplings to half a million and finally to five million to make sure that we sample any potentially narrow peaks of the likelihood function near the best-fit model at $p(\vec{\theta}'|F, \mathcal{M}_1) - p(\vec{\theta}_{max}|F, \mathcal{M}_1) = 0$ with sufficient accuracy. We find, however, that this behavior of the differential likelihood distribution clustering far from the best-fit solution persists, irrespective of the available computing power devoted to the sampling.

---

[4] Martin et al. (2019) estimate that failed exomoon transits should actually be quite common for misaligned planet-moon systems, such as the one proposed by Teachey & Kipping (2018).

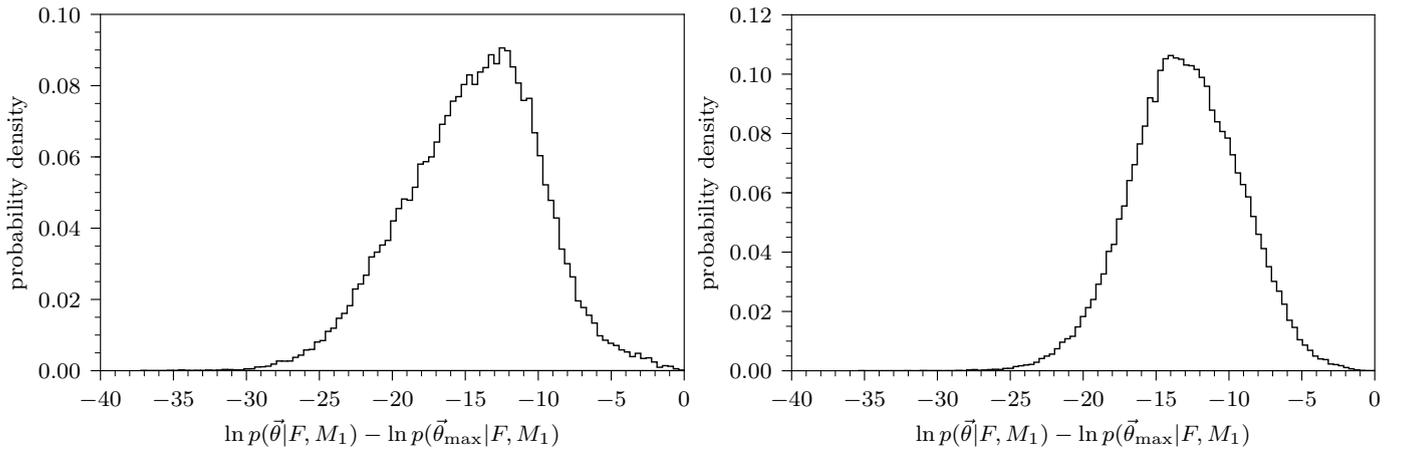

**Fig. 3.** Differential likelihood distribution between the most likely planet-moon model and the other solutions using $10^6$ steps of our PTMCMC fitting procedure. *Left*: Results from fitting our planet-moon transit model to the original data from Teachey & Kipping (2018). *Right*: Results from fitting our planet-moon transit model to our own detrending of the *Kepler* and WFC3 data. In both panels the most likely model is located at 0 along the abscissa by definition. In both cases the models do not converge to the best-fit solution, suggesting that the best-fit solution could in fact be an outlier.

**Table 1.** Results of our PTMCMC fitting procedure to the combined *Kepler* and *Hubble* data. The *Hubble* data was either based on the photometry extracted by Teachey & Kipping (2018, TK18b, central column) or based on our own extraction (right column).

|  | TK18b *HST* photometry | Our *HST* photometry |
|---|---|---|
| $r_s$ [%] | $1.9^{+0.5}_{-0.5}$ | $1.6^{+0.5}_{-0.5}$ |
| $a_{ps}$ [$R_\star$] | $2.9^{+1.3}_{-0.6}$ | $2.9^{+1.5}_{-0.6}$ |
| $P_s$ [d] | $27^{+15}_{-13}$ | $29^{+17}_{-15}$ |
| $\varphi$ [rad] | $3.3^{+2.0}_{-2.3}$ | $3.2^{+2.2}_{-2.2}$ |
| $f_M$ [%] | $1.9^{+1.6}_{-0.8}$ | $2.0^{+1.7}_{-1.0}$ |
| $i_s$ [rad] | $-0.2^{+2.4}_{-2.3}$ | $-0.1^{+2.3}_{-2.2}$ |
| $\Omega_s$ [rad] | $-0.1^{+2.1}_{-2.1}$ | $-0.2^{+2.0}_{-1.8}$ |

**Notes.** Figure 4 illustrates quite clearly that the posterior distributions are not normally distributed and often not even representative of skewed normal distributions. The confidence intervals stated in this table have thus to be taken with care.

Figure 4 shows the posterior distributions of the moon parameters of our planet-moon model. The top panel refers to our PTMCMC fitting of the combined *Kepler* and *Hubble* data (*Hubble* data as detrended and published by Teachey & Kipping 2018), and the bottom panel shows our PTMCMC fitting of the *Kepler* data combined with our own extraction and detrending of the *Hubble* light curve. The respective median values and standard deviations are noted in the upper right corners of each sub-panel and summarized in Table 1.

A comparison between the upper and lower corner plots in Fig. 4 reveals that the different detrending and fitting techniques have a significant effect on the resulting posterior distributions, in particular for $i_s$ and $\Omega_s$, the two angles that parameterize the orientation of the moon orbit. At the same time, however, the most likely values (red dots above the plot diagonal) and median values (blue crosses below the plot diagonal) of the seven parameters shown are well within the $1\sigma$ tolerance.

The following features can be observed in both panels of Fig. 4. The moon-to-star radius ratio (Col. 1) shows an approximately normal distribution, whereas the scaled planet-moon orbital semimajor axis (Col. 2) shows a more complicated, skewed distribution. The solutions for the orbital period of the exomoon candidate (Col. 3) show a comb-like structure owing to the discrete number of completed moon orbits that would fit a given value of the moon's initial orbital phase (Col. 4), which is essentially unconstrained. The moon-to-planet mass ratio (Col. 5) then shows a skewed normal distribution with a tail of large moon masses. Our results for the inclination $i_s$ between the satellite orbit (around the planet) and the line of sight, and for the longitude of the ascending node of the moon orbit are shown in Cols. 6 and 7. The preference of $i_s$ being either near 0 or near $\pm\pi$ (the latter is equivalent to a near-coplanar retrograde moon orbit) illustrates the well-known degeneracy of the prograde/retrograde solutions available from light curve analyses (Lewis & Fujii 2014; Heller & Albrecht 2014).

## 3.2. Transit timing variations

Next we consider the possibility of the transits being caused by a planet only. Neglecting the *Hubble* transit, our PTMCMC sampling of the three *Kepler* transits with our planet-only transit model gives an orbital period of $P = 287.3776 \pm 0.0024$ d and an initial transit midpoint at $t_0 = 61.4528 \pm 0.0092$ d in units of the Barycentric Kepler Julian Day (BKJD), which is equal to BJD $- 2,454,833.0$ d. The resulting transit time of the 2017 *Hubble* transit is $3222.6059 \pm 0.0182$ d.

Our planet-only model for the 2017 *Hubble* transit gives a transit midpoint at $3222.5547 \pm 0.0014$ d, that is $73.728(\pm2.016)$ min earlier than the predicted transit midpoint. This is in agreement with the measurements of Teachey & Kipping (2018), who found that the *Hubble* transit occurred 77.8 min earlier than predicted. This observed early transit of Kepler-1625 b has a formal $\sim 3\sigma$ significance. We note, however, that this $3\sigma$ deviation is mostly dictated by the first transit observed with *Kepler* (see Fig. S12 in Teachey et al. 2018). We also note that this transit was preceded by a $\sim 1$ d observational gap in the light curve, about 0.5 d prior to the transit, which might affect the local detrending of the data and the determination of the transit mid-point of a planet-only model. Moreover, with most of the TTV effect being due to the large deviation from the linear ephemeris of the first transit, stellar (or any other systematic) variability could have a large (but unknown) effect on the error bars that go into the calculations.

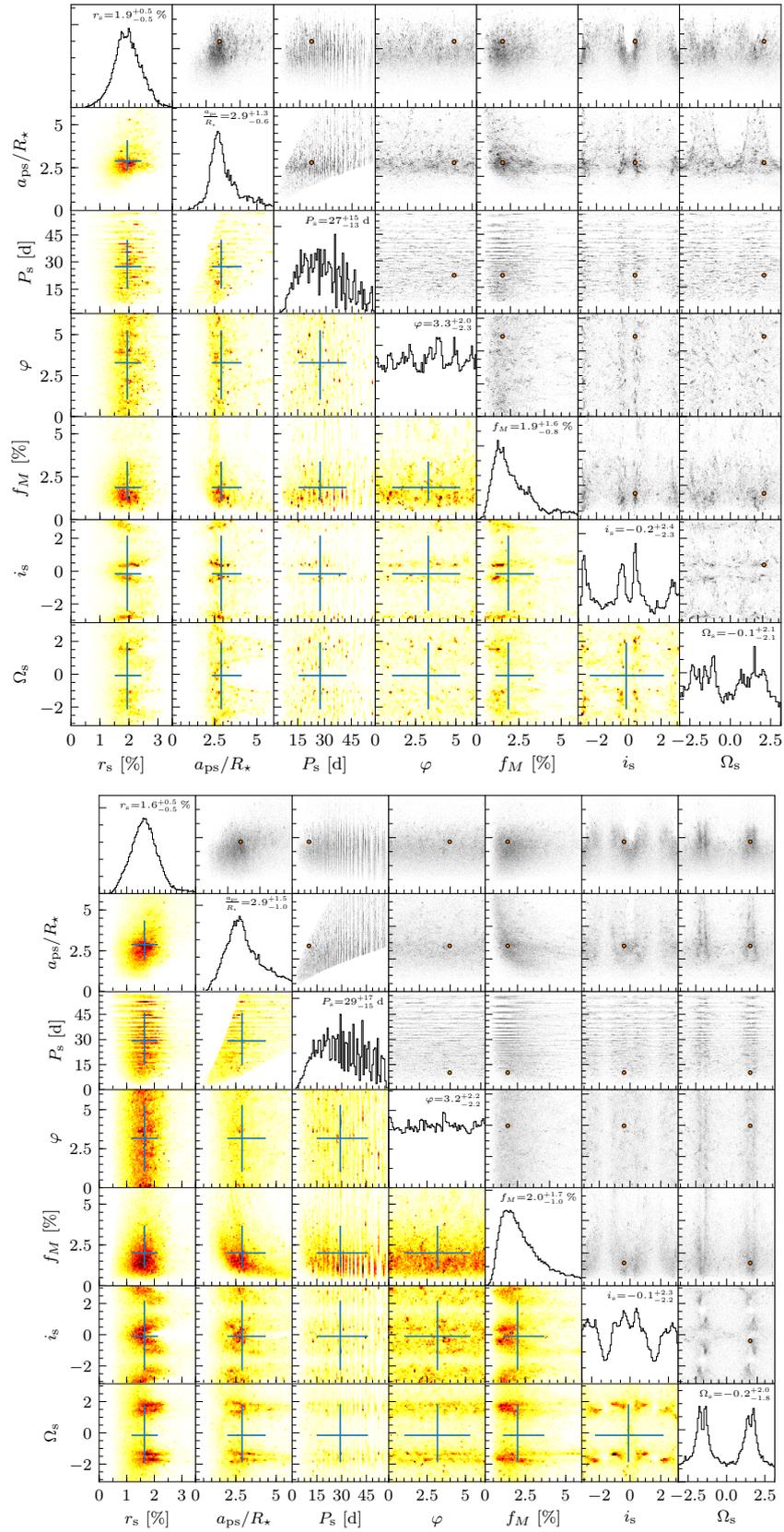

**Fig. 4.** Posterior distributions of a parallel tempering ensemble MCMC sampling of the combined *Kepler* and WFC3 data with our planet-moon model. *Top*: Results for the original data from Teachey & Kipping (2018). *Bottom*: Results for our own detrending of the *Kepler* and WFC3 data. In both figures, scatter plots are shown with black dots above the diagonal, and projected histograms are shown as colored pixels below the diagonal. The most likely parameters are denoted with an orange point in the scatter plots. Histograms of the moon-to-star radius ratio $r_s$, scaled semimajor axis of the planet-moon system ($a_{ps}/R_\star$), satellite orbital period ($P_s$), satellite orbital phase ($\varphi$), moon-to-planet mass ratio ($f_M$), orbital inclination of the satellite with respect to our line of sight ($i_s$), and the orientation of the ascending node of the satellite orbit ($\Omega_s$) are shown on the diagonal. Median values and standard deviations are indicated with error bars in the histograms.

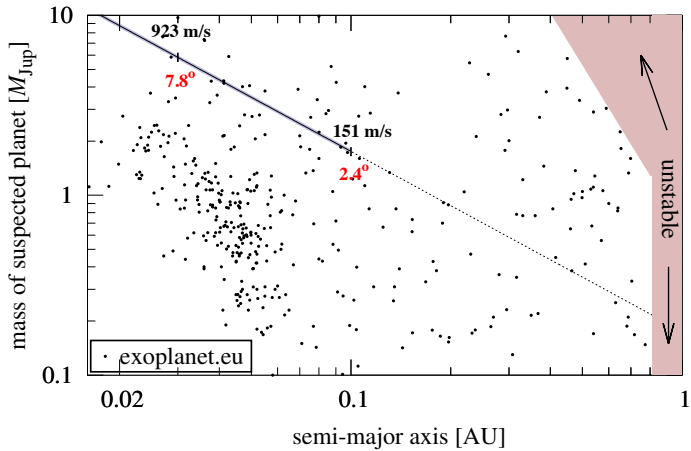

**Fig. 5.** Mass estimate for the potential inner planet around Kepler-1625 based on the observed TTV of 73.728 min. The thin pale blue fan around the solid curve shows the 1 $\sigma$ tolerance interval of ±2.016 min. Values for semimajor axes > 0.1 AU are poor approximations and thus shown with a dashed line. Black points show masses and semimajor axes of all planets from exoplanet.eu (as of 26 October 2018) around stars with masses between 0.75 $M_\odot$ and 1.25 $M_\odot$. A conservative estimate of a dynamically unstable region for the suspected inner planet, where its Hill sphere would touch the Hill sphere of Kepler-1625 b with an assumed mass of 1 $M_{\rm Jup}$, is shaded in pale red. RV amplitudes and minimum orbital inclination with respect to Kepler-1625 b are noted along the curve for the planetary mass estimate.

In Fig. 5 we show the mass of an unseen inner planet that is required to cause the observed 73.7 min TTV amplitude from our PTMCMC fit as a function of its unknown orbital semimajor axis. The mass drops from 5.8 $M_{\rm Jup}$ at 0.03 AU to 1.8 $M_{\rm Jup}$ at 0.1 AU. Values beyond 0.1 AU cannot be assumed to fulfill the approximations made for Eq. (3) and are therefore shown with a dashed line. The actual TTV amplitude of Kepler-1625 b could even be higher than the ~ 73 min that we determined for the *Hubble* transit, and thus the mass estimates shown for a possible unknown inner planet serve as lower boundaries.

The resulting radial velocity amplitudes of the star of 923 m s$^{-1}$ (at 0.03 AU) and 151 m s$^{-1}$ (at 0.1 AU), respectively, are indicated along the curve. Even if the approximations for a coplanar, close-in planet were not entirely fulfilled, our results suggest that RV observations of Kepler-1625 with a high-resolution spectrograph attached to a very large (8 m class) ground-based telescope could potentially reveal an unknown planet causing the observed TTV of Kepler-1625 b. Also shown along the curve in Fig. 5 are the respective minimum orbital inclinations (rounded mean values shown) between Kepler-1625 b and the suspected close-in planet required to prevent Kepler-1625 b from transiting the star. The exact values are $i = 7.8^{+1.1}_{-2.0}$ degrees at 0.03 AU and $i = 2.4^{+0.3}_{-0.6}$ degrees at 0.1 AU.

The pale red shaded region is excluded from a dynamical point of view since this is where the planetary Hill spheres would overlap. The extent of this region is a conservative estimate because it assumes a mass of 1 $M_{\rm Jup}$ for Kepler-1625 b and neglects any chaotic effects induced by additional planets in the system or planet-planet cross tides etc. The true range of unstable orbits is probably larger. The black dots show all available exoplanet masses and semimajor axes from the Exoplanet Encyclopaedia, continuous in- and near-transit monitoring of the target is re-which illustrates that the suspected planet could be more massive than most of the known hot Jupiters.

## 4. Conclusions

With a $\Delta$BIC of −44.5 (using published *Hubble* data of Teachey & Kipping 2018) or −31.0 (using our own *Hubble* extraction and detrending) between the most likely planet-only model and the most likely planet-moon model, we find strong statistical evidence for a roughly Neptune-sized exomoon. In both cases of the data detrending, the most likely orbital solution of the planet-moon system, however, is very different from most of the other orbital realizations of our PTMCMC modeling and the most likely solutions do not seem to converge. In other words, the most likely solution appears to be an outlier in the distribution of possible solutions and small changes to the data can have great effects on the most likely orbital solution found for the planet-moon system. As an example, we find that the two different detrending methods that we explored produce different interpretations of the transit observed with *Hubble*: in one case our PTMCMC sampling finds the egress of the moon in the light curve, in the other case it does not (Fig. 2).

Moreover, the likelihood of this best-fit orbital solution is very different from the likelihoods of most other solutions from our PTMCMC modeling. We tested both a standard MCMC sampling and a parallel-tempering MCMC (Foreman-Mackey et al. 2013); the latter is supposed to explore both the parameter space at large and the tight peaks of the likelihood function in detail. Our finding of the nonconvergence could imply that the likelihood function that best describes the data is non-Gaussian. Alternatively, with the BIC being an asymptotic criterion that requires a large sample size by definition (Stevenson et al. 2012), our findings suggest that the available data volume is simply too small for the BIC to be formally applicable. We conclude that the $\Delta$BIC is an unreliable metric for an exomoon detection for this data set of only four transits and possibly for other data sets of *Kepler* as well.

One solution to evaluating whether the BIC or an alternative information criterion such as the Akaike information criterion (AIC; Akaike 1974) or the deviance information criterion (DIC; Spiegelhalter et al. 2002) is more suitable for assessing the likelihoods of a planet-only model and of a planet-moon model could be injection-retrieval experiments of synthetic transits (Heller et al. 2016b; Rodenbeck et al. 2018). Such an analysis, however, goes beyond the scope of this paper.

We also observe the TTV effect discovered by Teachey & Kipping (2018). If the early arrival of Kepler-1625 b for its late-2017 transit was caused by an inner planet rather than by an exomoon, then the planet would be a super-Jovian mass hot Jupiter, the exact mass limit depending on the assumed orbital semimajor axis. For example, the resulting stellar radial velocity amplitude would be about 900 m s$^{-1}$ for a 5.8 $M_{\rm Jup}$ planet at 0.03 AU and about 150 m s$^{-1}$ for a 1.8 $M_{\rm Jup}$ planet at 0.1 AU. From the absence of a transit signature of this hypothetical planet in the four years of *Kepler* data, we conclude that it would need to have an orbital inclination of at least $i = 7.8^{+1.1}_{-2.0}$ (if it were at 0.03 AU) or $i = 2.4^{+0.3}_{-0.6}$ degrees (if it were at 0.1 AU). If its inclination is not close to 90°, at which point its effect on the stellar RV amplitude would vanish, then the hypothesis of an unseen inner planet causing the Kepler-1625 TTV could be observationally testable.

Ground-based photometric observations are hardly practicable to answer the question of this exomoon candidate because quired over at least two days. Current and near-future space-based exoplanet missions, on the other hand, will likely not be able to deliver the signal-to-noise ratios required to validate or reject the exomoon hypothesis. With a *Gaia* G-band magnitude of $m_G = 15.76$ (Gaia Collaboration et al. 2016, 2018) the star is rather faint in the visible regime of the electromagnetic spectrum and the possible moon transits are therefore beyond the sensitivity limits of the *TESS*, *CHEOPS*, and *PLATO* missions. 2MASS observations suggest that Kepler-1625 is somewhat brighter in the near-infrared (Cutri et al. 2003), such that the *James Webb Space Telescope* (launch currently scheduled for early 2021) should be able to detect the transit of the proposed Neptune-sized moon, for example via photometric time series obtained with the NIRCam imaging instrument.

All things combined, the fragility of the proposed photometric exomoon signature with respect to the detrending methods, the unknown systematics in both the *Kepler* and the *Hubble* data, the absence of a proper assessment of the stellar variability of Kepler-1625, the faintness of the star (and the resulting photometric noise floor), the previously stated coincidence of the proposed moon's properties with those of false positives (Rodenbeck et al. 2018), the existence of at least one plausible alternative explanation for the observed TTV effect of Kepler-1625 b, and the serious doubts that we have about the ΔBIC as a reliable metric at least for this particular data set lead us to conclude that the proposed moon around Kepler-1625 b might not be real. We find that the exomoon hypothesis heavily relies on a chain of delicate assumptions, all of which need to be further investigated.

A similar point was raised by Teachey & Kipping (2018), and our analysis is an independent attempt to shed some light on the "unknown unknowns" referred to by the authors. For the time being, we take the position that the first exomoon has yet to be detected as the likelihood of an exomoon around Kepler-1625 b cannot be assessed with the methods used and data currently available.

*Acknowledgements.* The authors thank Kevin Stevenson, Hannah Wakeford and Megan Sosey for their help with the data analysis, and Nikole Lewis for the feedback on the manuscript. The authors would also like to thank the referee for a challenging and constructive report. This work was supported in part by the German space agency (Deutsches Zentrum für Luft- und Raumfahrt) under PLATO Data Center grant 50OO1501. This work made use of NASA's ADS Bibliographic Services and of the Exoplanet Encyclopaedia (http://exoplanet.eu). RH wrote the manuscript, proposed Figs. 2 - 4, generated Fig. 5, and guided the work. KR derived the star-planet-moon orbital simulations and the respective statistics and generated Figs. 2 - 4. GB performed the light curve extraction from the WFC3 *Hubble* data and generated Fig. 1. All authors contributed equally to the interpretation of the data.

# Chapter 7

# Astrophysical Constraints on the Habitability of Extrasolar Moons

## 7.1 Exomoon Habitability Constrained by Illumination and Tidal Heating (Heller & Barnes 2013a)

Contribution:

RH did the literature research, worked out the mathematical framework, translated the math into computer code, performed all computations and simulations, created Figs. 1-2, 4-13, A.1, B.1, and C.1, led the writing of the manuscript, and served as a corresponding author for the journal editor and the referees. RH also contributed to the writing and to the illustration of a press release of the Leibniz Institute for Astrophysics in Potsdam to support the communication of this research to the public (www.aip.de/en/news/press/exomoons).

# Exomoon habitability constrained by illumination and tidal heating

René Heller[I] , Rory Barnes[II,III]

[I] Leibniz-Institute for Astrophysics Potsdam (AIP), An der Sternwarte 16, 14482 Potsdam, Germany, rheller@aip.de
[II] Astronomy Department, University of Washington, Box 951580, Seattle, WA 98195, rory@astro.washington.edu
[III] NASA Astrobiology Institute – Virtual Planetary Laboratory Lead Team, USA

## Abstract

The detection of moons orbiting extrasolar planets ("exomoons") has now become feasible. Once they are discovered in the circumstellar habitable zone, questions about their habitability will emerge. Exomoons are likely to be tidally locked to their planet and hence experience days much shorter than their orbital period around the star and have seasons, all of which works in favor of habitability. These satellites can receive more illumination per area than their host planets, as the planet reflects stellar light and emits thermal photons. On the contrary, eclipses can significantly alter local climates on exomoons by reducing stellar illumination. In addition to radiative heating, tidal heating can be very large on exomoons, possibly even large enough for sterilization. We identify combinations of physical and orbital parameters for which radiative and tidal heating are strong enough to trigger a runaway greenhouse. By analogy with the circumstellar habitable zone, these constraints define a circumplanetary "habitable edge". We apply our model to hypothetical moons around the recently discovered exoplanet Kepler-22b and the giant planet candidate KOI211.01 and describe, for the first time, the orbits of habitable exomoons. If either planet hosted a satellite at a distance greater than 10 planetary radii, then this could indicate the presence of a habitable moon.

Key Words: Astrobiology – Extrasolar Planets – Habitability – Habitable Zone – Tides

## 1. Introduction

The question whether life has evolved outside Earth has prompted scientists to consider habitability of the terrestrial planets in the Solar System, their moons, and planets outside the Solar System, that is, extrasolar planets. Since the discovery of the first exoplanet almost two decades ago (Mayor & Queloz 1995), roughly 800 more have been found, and research on exoplanet habitability has culminated in the targeted space mission *Kepler*, specifically designed to detect Earth-sized planets in the circumstellar irradiation habitable zones (IHZs, Huang 1959; Kasting et al. 1993; Selsis et al. 2007; Barnes et al. 2009)[1] around Sun-like stars. No such Earth analog has been confirmed so far. Among the 2312 exoplanet candidates detected with *Kepler* (Batalha et al. 2012), more than 50 are indeed in the IHZ (Borucki et al. 2011; Kaltenegger & Sasselov 2011; Batalha et al. 2012), yet most of them are significantly larger and likely more massive than Earth. Habitability of the moons around these planets has received little attention. We argue here that it will be possible to constrain their habitability on the data available at the time they will be discovered.

Various astrophysical effects distinguish investigations of exomoon habitability from studies on exoplanet habitability. On a moon, there will be eclipses of the star by the planet (Dole 1964); the planet's stellar reflected light, as well the planet's thermal emission, might affect the moon's climate; and tidal heating can provide an additional energy source, which must be considered for evaluations of exomoon habitability (Reynolds et al. 1987; Scharf 2006; Debes & Sigurdsson 2007; Cassidy et al. 2009; Henning et al. 2009). Moreover, tidal processes between the moon and its parent planet will determine the orbit and spin evolution of the moon. Earth-sized planets in the IHZ around low-mass stars tend to become tidally locked, that is, one hemisphere permanently faces the star (Dole 1964; Goldreich 1966; Kasting et al. 1993), and they will not have seasons because their obliquities are eroded (Heller et al. 2011a,b). On moons, however, tides from the star are mostly negligible compared to the tidal drag from the planet. Thus, in most cases exomoons will be tidally locked to their host planet rather than to the star (Dole 1964; Gonzalez 2005; Henning et al. 2009; Kaltenegger 2010; Kipping et al. 2010) so that (*i*.) a satellite's rotation period will equal its orbital period about the planet, (*ii*.) a moon will orbit the planet in its equatorial plane (due to the Kozai mechanism and tidal evolution, Porter & Grundy 2011), and (*iii.*) a moon's rotation axis will be perpendicular to its orbit about the planet. A combination of (*ii.*) and (*iii.*) will cause the satellite to have the same obliquity with respect to the circumstellar orbit as the planet.

---

[1] A related but more anthropocentric circumstellar zone, termed "ecosphere", has been defined by Dole (1964, p. 64 therein). Whewell (1853, Chapter X, Section 4 therein) presented a more qualitative discussion of the so-called "Temperate Zone of the Solar System".

More massive planets are more resistive against the tidal erosion of their obliquities (Heller et al. 2011b); thus massive host planets of exomoons can maintain significant obliquities on timescales much larger than those for single terrestrial planets. Consequently, satellites of massive exoplanets could be located in the IHZ of low-mass stars while, firstly, their rotation would not be locked to their orbit around the star (but to the planet), and secondly, they could experience seasons if the equator of their host planet is tilted against the orbital plane. Both aspects tend to reduce seasonal amplitudes of stellar irradiation (Cowan et al. 2012) and thereby stabilize exomoon climates.

An example is given by a potentially habitable moon in the Solar System, Titan. It is the only moon known to have a substantial atmosphere. Tides exerted by the Sun on Titan's host planet, Saturn, are relatively weak, which is why the planet could maintain its spin-orbit misalignment, or obliquity, $\psi_p$ of 26.7° (Norman 2011). Titan orbits Saturn in the planet's equatorial plane with no significant tilt of its rotation axis with respect to its circumplanetary orbit. Thus, the satellite shows a similar obliquity with respect to the Sun as Saturn and experiences strong seasonal modulations of insolation as a function of latitude, which leads to an alternation in the extents and localizations of its lakes and potential habitats (Wall et al. 2010). While tides from the Sun are negligible, Titan is tidally synchronized with Saturn (Lemmon et al. 1993) and has a rotation and an orbital period of ≈16d. Uranus, where $\psi_p \approx 97.9°$ (Harris & Ward 1982), illustrates that even more extreme scenarios are possible.

No exomoon has been detected so far, but it has been shown that exomoons with masses down to 20% the mass of Earth ($M_\oplus$) are detectable with the space-based *Kepler* telescope (Kipping et al. 2009). Combined measurements of a planet's transit timing variation (TTV) and transit duration variation (TDV) can provide information about the satellite's mass ($M_s$), its semi-major axis around the planet ($a_{ps}$) (Sartoretti & Schneider 1999; Simon et al. 2007; Kipping 2009a), and possibly about the inclination (*i*) of the satellite's orbit with respect to the orbit around the star (Kipping 2009b). Photometric transits of a moon in front of the star (Szabó et al. 2006; Lewis 2011; Kipping 2011a; Tusnski & Valio 2011), as well as mutual eclipses of a planet and its moon (Cabrera & Schneider 2007; Pál 2012), can provide information about its radius ($R_s$), and the photometric scatter peak analysis (Simon et al. 2012) can offer further evidence for the exomoon nature of candidate objects. Finally, spectroscopic investigations of the Rossiter-McLaughlin effect can yield information about the satellite's orbital geometry (Simon et al. 2010; Zhuang et al. 2012), although relevant effects require accuracies in stellar radial velocity of the order of a few centimeters per second (see also Kipping 2011a). Beyond, Peters & Turner (2013) suggest that direct imaging of extremely tidally heated exomoons will be possible with next-generation space telescopes. It was only recently that Kipping et al. (2012) initiated the first dedicated hunt for exomoons, based on *Kepler* observations. While we are waiting for their first discoveries, hints to exomoon-forming regions around planets have already been found (Mamajek et al. 2012).

In Section 2 of this paper, we consider general aspects of exomoon habitability to provide a basis for our work, while Section 3 is devoted to the description of the exomoon illumination plus tidal heating model. Section 4 presents a derivation of the critical orbit-averaged global flux and the description of habitable exomoon orbits, ultimately leading to the concept of the "habitable edge". In Section 5, we apply our model to putative exomoons around the first Neptune-sized[2] planet in the IHZ of a Sun-like star, Kepler-22b, and a much more massive, still to be confirmed planet candidate, the "Kepler Object of Interest" (KOI) 211.01[3], also supposed to orbit in the IHZ. We summarize our results with a discussion in Section 6. Detailed illustrations on how we derive our model are placed into the appendices.

## 2. Habitability of exomoons

So far, there have been only a few published investigations on exomoon habitability (Reynolds et al. 1987; Williams et al. 1997; Kaltenegger 2000; Scharf 2006; Porter & Grundy 2011). Other studies were mainly concerned with the observational aspects of exomoons (for a review see Kipping et al. 2012), their orbital stability (Barnes & O'Brien 2002; Domingos et al. 2006; Donnison 2010; Weidner & Horne 2010; Quarles et al. 2012; Sasaki et al. 2012), and eventually with the detection of biosignatures (Kaltenegger 2010). Thus, we provide here a brief overview of some important physical and biological aspects that must be accounted for when considering exomoon habitability.

---

[2] Planets with radii between $2R_\oplus$ and $6R_\oplus$ are designated Neptune-sized planets by the *Kepler* team (Batalha et al. 2012).

[3] Although KOI211.01 is merely a planet candidate we talk of it as a planet, for simplicity, but keep in mind its unconfirmed status.

Williams et al. (1997) were some of the first who proposed that habitable exomoons could be orbiting giant planets. At the time of their writing, only nine extrasolar planets, all of which are giant gaseous objects, were known. Although these bodies are not supposed to be habitable, Williams et al. argued that possible satellites of the jovian planets 16 Cygni B and 47 Ursae Majoris could offer habitats, because they orbit their respective host star at the outer edge of the habitable zone (Kasting et al. 1993). The main counter arguments against habitable exomoons were (*i*.) tidal locking of the moon with respect to the planet, (*ii*.) a volatile endowment of those moons, which would have formed in a circumplanetary disk, that is different from the abundances available for planets forming in a circumstellar disk, and (*iii*.) bombardment of high-energy ions and electrons within the magnetic fields of the jovian host planet and subsequent loss of the satellite's atmosphere. Moreover, (*iv*.) stellar forcing of a moon's upper atmosphere will constrain its habitability.

Point (*i*.), in fact, turns out as an advantage for Earth-sized satellites of giant planets over terrestrial planets in terms of habitability, by the following reasoning: Application of tidal theories shows that the rotation of extrasolar planets in the IHZ around low-mass stars will be synchronized on timescales ≪1Gyr (Dole 1964; Goldreich 1966; Kasting et al. 1993). This means one hemisphere of the planet will permanently face the star, while the other hemisphere will freeze in eternal darkness. Such planets might still be habitable (Joshi et al. 1997), but extreme weather conditions would strongly constrain the extent of habitable regions on the planetary surface (Heath & Doyle 2004; Spiegel et al. 2008; Heng & Vogt 2011; Edson et al. 2011; Wordsworth et al. 2011). However, considering an Earth-mass exomoon around a Jupiter-like host planet, within a few million years at most the satellite should be tidally locked to the planet – rather than to the star (Porter & Grundy 2011). This configuration would not only prevent a primordial atmosphere from evaporating on the illuminated side or freezing out on the dark side (*i*.) but might also sustain its internal dynamo (*iii*.). The synchronized rotation periods of putative Earth-mass exomoons around giant planets could be in the same range as the orbital periods of the Galilean moons around Jupiter (1.7d–16.7d) and as Titan's orbital period around Saturn (≈16d) (NASA/JPL planetary satellite ephemerides)[4]. The longest possible length of a satellite's day compatible with Hill stability has been shown to be about $P_{*p}/9$, $P_{*p}$ being the planet's orbital period about the star (Kipping 2009a). Since the satellite's rotation period also depends on its orbital eccentricity around the planet and since the gravitational drag of further moons or a close host star could pump the satellite's eccentricity (Cassidy et al. 2009; Porter & Grundy 2011), exomoons might rotate even faster than their orbital period.

Finally, from what we know about the moons of the giant planets in the Solar System, the satellite's enrichment with volatiles (*ii*.) should not be a problem. Cometary bombardment has been proposed as a source for the dense atmosphere of Saturn's moon Titan, and it has been shown that even the currently atmosphere-free jovian moons Ganymede and Callisto should initially have been supplied with enough volatiles for an atmosphere (Griffith & Zahnle 1995). Moreover, as giant planets in the IHZ likely formed farther away from their star, that is, outside the snow line (Kennedy & Kenyon 2008), their satellites will be rich in liquid water and eventually be surrounded by substantial atmospheres.

The stability of a satellite's atmosphere (*iv*.) will critically depend on its composition, the intensity of stellar extreme ultraviolet radiation (EUV), and the moon's surface gravity. Nitrogen-dominated atmospheres may be stripped away by ionizing EUV radiation, which is a critical issue to consider for young (Lichtenberger et al. 2010) and late-type (Lammer et al. 2009) stars. Intense EUV flux could heat and expand a moon's upper atmosphere so that it can thermally escape due to highly energetic radiation (*iii*.), and if the atmosphere is thermally expanded beyond the satellite's magnetosphere, then the surrounding plasma may strip away the atmosphere nonthermally. If Titan were to be moved from its roughly 10AU orbit around the Sun to a distance of 1AU (AU being an astronomical unit, i.e., the average distance between the Sun and Earth), then it would receive about 100 times more EUV radiation, leading to a rapid loss of its atmosphere due to the moon's smaller mass, compared to Earth. For an Earth-mass moon at 1AU from the Sun, EUV radiation would need to be less than 7 times the Sun's present-day EUV emission to allow for a long-term stability of a nitrogen atmosphere. $CO_2$ provides substantial cooling of an atmosphere by infrared radiation, thereby counteracting thermal expansion and protecting an atmosphere's nitrogen inventory (Tian 2009).

A minimum mass of an exomoon is required to drive a magnetic shield on a billion-year timescale ($M_s \gtrsim 0.1 M_\oplus$, Tachinami et al. 2011); to sustain a substantial, long-lived atmosphere ($M_s \gtrsim 0.12 M_\oplus$, Williams et al. 1997; Kaltenegger 2000); and to drive tectonic activity ($M_s \gtrsim 0.23 M_\oplus$, Williams et al. 1997), which is necessary to maintain plate tectonics and to support the carbon-silicate cycle. Weak internal dynamos have been detected in Mercury and Ganymede (Kivelson et al. 1996; Gurnett et al. 1996), suggesting that satellite masses $> 0.25 M_\oplus$ will be adequate for considerations of exomoon habitability. This lower limit, however, is not a fixed number. Further sources of energy – such as radiogenic and tidal heating, and the effect of a moon's composition and structure – can alter our limit in either direction. An upper mass limit is given by the fact that increasing mass leads to high pressures in the moon's interior, which will increase the mantle viscosity and depress heat transfer throughout the mantle as well as in the core. Above a critical mass, the dynamo is strongly suppressed and becomes too weak to generate a magnetic field or sustain plate tectonics. This maximum mass can be placed around $2M_\oplus$ (Gaidos et al. 2010; Noack & Breuer 2011; Stamenković et al. 2011). Summing up these conditions, we expect approximately Earth-mass moons to be habitable, and these objects could be detectable with the newly started *Hunt for Exomoons with Kepler* (HEK) project (Kipping et al. 2012).

---

[4] Maintained by Robert Jacobson, http://ssd.jpl.nasa.gov.

### 2.1 Formation of massive satellites

The largest and most massive moon in the Solar System, Ganymede, has a radius of only ≈$0.4R_\oplus$ ($R_\oplus$ being the radius of Earth) and a mass of ≈$0.025M_\oplus$. The question as to whether much more massive moons could have formed around extrasolar planets is an active area of research. Canup & Ward (2006) have shown that moons formed in the circumplanetary disk of giant planets have masses ≲$10^{-4}$ times that of the planet's mass. Assuming satellites formed around Kepler-22b, their masses will thus be $2.5\times10^{-3} M_\oplus$ at most, and around KOI211.01 they will still weigh less than Earth's Moon. Mass-constrained in situ formation becomes critical for exomoons around planets in the IHZ of low-mass stars because of the observational lack of such giant planets. An excellent study on the formation of the Jupiter and the Saturn satellite systems is given by Sasaki et al. (2010), who showed that moons of sizes similar to Io, Europa, Ganymede, Callisto, and Titan should build up around most gas giants. What is more, according to their Fig. 5 and private communication with Takanori Sasaki, formation of Mars- or even Earth-mass moons around giant planets is possible. Depending on whether or not a planet accretes enough mass to open up a gap in the protostellar disk, these satellite systems will likely be multiple and resonant (as in the case of Jupiter), or contain only one major moon (see Saturn). Ogihara & Ida (2012) extended these studies to explain the compositional gradient of the jovian satellites. Their results explain why moons rich in water are farther away from their giant host planet and imply that capture in 2:1 orbital resonances should be common.

Ways to circumvent the impasse of insufficient satellite mass are the gravitational capture of massive moons (Debes & Sigurdsson 2007; Porter & Grundy 2011; Quarles et al. 2012), which seems to have worked for Triton around Neptune (Goldreich et al. 1989; Agnor & Hamilton 2006); the capture of Trojans (Eberle et al. 2011); gas drag in primordial circumplanetary envelopes (Pollack et al. 1979); pull-down capture trapping temporary satellites or bodies near the Lagrangian points into stable orbits (Heppenheimer & Porco 1977; Jewitt & Haghighipour 2007); the coalescence of moons (Mosqueira & Estrada 2003); and impacts on terrestrial planets (Canup 2004; Withers & Barnes 2010; Elser et al. 2011). Such moons would correspond to the irregular satellites in the Solar System, as opposed to regular satellites that form in situ. Irregular satellites often follow distant, inclined, and often eccentric or even retrograde orbits about their planet (Carruba et al. 2002). For now, we assume that Earth-mass extrasolar moons – be they regular or irregular – exist.

### 2.2 Deflection of harmful radiation

A prominent argument against the habitability of moons involves high-energy particles, which a satellite meets in the planet's radiation belt. Firstly, this ionizing radiation could strip away a moon's atmosphere, and secondly it could avoid the buildup of complex molecules on its surface. In general, the process in which incident particles lose part of their energy to a planetary atmosphere or surface to excite the target atoms and molecules is called sputtering. The main sources for sputtering on Jupiter's satellites are the energetic, heavy ions $O^+$ and $S^+$, as well as $H^+$, which give rise to a steady flux of $H_2O$, $OH$, $O_2$, $H_2$, $O$, and $H$ from Ganymede's surface (Marconi 2007). A moon therefore requires a substantial magnetic field that is strong enough to embed the satellite in a protective bubble inside the planet's powerful magnetosphere. The only satellite in the Solar System with a substantial magnetic shield of roughly 750nT is Ganymede (Kivelson et al. 1996). The origin of this field is still subject to debate because it can only be explained by a very specific set of initial and compositional configurations (Bland et al. 2008), assuming that it is generated in the moon's core.

For terrestrial planets, various models for the strength of global dipolar magnetic fields $B_{\rm dip}$ as a function of planetary mass and rotation rate exist, but none has proven exclusively valid. Simulations of planetary thermal evolution have shown that $B_{\rm dip}$ increases with mass (Tachinami et al. 2011; Zuluaga & Cuartas 2012) and rotation frequency (Lopez-Morales et al. 2012). The spin of exomoons will be determined by tides from the planet, and rotation of an Earth-sized exomoon in the IHZ can be much faster than rotation of an Earth-sized planet orbiting a star. Thus, an exomoon could be prevented from tidal synchronization with the host star – in support of an internal dynamo and thus magnetic shielding against energetic irradiation from the planet and the star. Some studies suggest that even extremely slow rotation would allow for substantial magnetic shielding, provided convection in the planet's or moon's mantle is strong enough (Olson & Christensen 2006). In this case, tidal locking would not be an issue for magnetic shielding.

The picture of magnetic shielding gets even more complicated when tidal heating is considered, which again depends on the orbital parameters. In the Moon, tidal heating, mostly induced by the Moon's obliquity of 6.68° against its orbit around Earth, occurs dominantly in the core (Kaula 1964; Peale & Cassen 1978). On Io, however, where tidal heating stems from Jupiter's effect on the satellite's eccentricity, dissipation occurs mostly in the mantle (Segatz et al. 1988). In the former case, tidal heating might enhance the temperature gradient between the core and the mantle and thereby also enhance convection and finally the strength of the magnetic shielding; in the latter case, tidal heating might decrease convection. Of course, the magnetic properties of terrestrial worlds will evolve and, when combined with the evolution of EUV radiation and stellar wind from the host star, define a time-dependent magnetically restricted habitable zone (Khodachenko et al. 2007; Zuluaga et al. 2012).

We conclude that radiation of highly energetic particles does not ultimately preclude exomoon habitability. In view of possible deflection due to magnetic fields on a massive satellites, it is still reasonable to consider the habitability of exomoons.

## 2.3 Runaway greenhouse

On Earth, the thermal equilibrium temperature of incoming and outgoing radiation is 255K. However, the mean surface temperature is 289K. The additional heating is driven by the greenhouse effect (Kasting 1988), which is a crucial phenomenon to the habitability of terrestrial bodies. The strength of the greenhouse effect depends on numerous variables – most importantly on the inventory of greenhouse gases, the albedo effect of clouds, the amount of liquid surface water, and the spectral energy distribution of the host star.

Simulations have shown that, as the globally absorbed irradiation on a water-rich planetary body increases, the atmosphere gets enriched in water vapor until it gets opaque. For an Earth-like body, this imposes a limit of about 300W/m² to the thermal radiation that can be emitted to space. If the global flux exceeds this limit, the body is said to be a runaway greenhouse. Water vapor can then leave the troposphere through the tropopause and reach the stratosphere, where photodissociation by stellar UV radiation allows the hydrogen to escape to space, thereby desiccating the planetary body. While boiling oceans, high surface temperatures, or high pressures *can* make a satellite uninhabitable, water loss *does* by definition. Hence, we will use the criterion of a runaway greenhouse to define an exomoon's habitability.

Surface temperatures strongly depend on the inventory of greenhouse gases, for example, $CO_2$. The critical energy flux $F_{RG}$ for a runaway greenhouse, however, does not (Kasting 1988; Goldblatt & Watson 2012). As in Barnes et al. (2013), who discussed how the interplay of stellar irradiation and tidal heating can trigger a runaway greenhouse on exoplanets, we will use the semi-analytical approach of Pierrehumbert (2010) for the computation of $F_{RG}$:

$$F_{RG} = o\ \sigma_{SB} \left( \frac{l}{R \ln \left( P' / \sqrt{\frac{2 P_0 g_s (M_s, R_s)}{k_0}} \right)} \right)^4 \qquad (1)$$

with

$$P' = P_{ref} \exp \left\{ \frac{l}{R\ T_{ref}} \right\}\ , \qquad (2)$$

$P_{ref} = 610.616$Pa, $l$ is the latent heat capacity of water, $R$ is the universal gas constant, $T_{ref} = 273.13$K, $o = 0.7344$ is a constant designed to match radiative transfer simulations, $\sigma_{SB}$ is the Stefan-Boltzmann constant, $P_0 = 104$Pa is the pressure at which the absorption line strengths of water vapor are evaluated, $g_s = GM_s/R_s^2$ is the gravitational acceleration at the satellite's surface, and $k_0 = 0.055$ is the gray absorption coefficient at standard temperature and pressure. Recall that the runaway greenhouse does not depend on the composition of the atmosphere, other than it contains water. As habitability requires water and Eq. (1) defines a limit above which the satellite will lose it, the formula provides a conservative limit to habitability.

In addition to the maximum flux $F_{RG}$ to allow for a moon to be habitable, one may think of a minimum flux required to prevent the surface water from freezing. On terrestrial exoplanets, this freezing defines the outer limit of the stellar IHZ. On exomoons, the extra light from the planetary reflection and thermal emission as well as tidal heating in the moon will move the circumstellar habitable zone away from the star, whereas eclipses will somewhat counterbalance this effect. While it is clear that a moon under strong tidal heating will not be habitable, it is not clear to what extent it might actually support habitability (Jackson et al. 2008). Even a relatively small tidal heating flux of a few watts per square meter could render an exomoon inhospitable; see Io's global volcanism, where tidal heating is a mere 2W/m² (Spencer et al. 2000). Without applying sophisticated models for the moon's tidal heating, we must stick to the irradiation aspect to define an exomoon's circumstellar habitable zone. At the outer edge of the stellar IHZ, the host planet will be cool and reflected little stellar flux. Neglecting tidal heating as well thermal emission and reflection from the planet, the minimum flux for an Earth-like moon to be habitable will thus be similar to that of an Earth-like planet at the same orbital distance to the star. Below, we will only use the upper flux limit from Eq. (1) to constrain the orbits of habitable exomoons. This will lead us to the concept of the circumplanetary "habitable edge".

## 3. Energy reservoirs on exomoons

Life needs liquid water and energy, but an oversupply of energy can push a planet or an exomoon into a runaway greenhouse and thereby make it uninhabitable. The critical, orbit-averaged energy flux for an exomoon to turn into a runaway greenhouse is around 300W/m², depending on the moon's surface gravity and atmospheric composition (Kasting 1988; Kasting et al. 1993; Selsis et al. 2007; Pierrehumbert 2010; Goldblatt & Watson 2012). An exomoon will thus only be habitable in a certain range of summed irradiation and tidal heat flux (Barnes et al. 2013).

We consider four energy reservoirs and set them into context with the IHZ: (*i.*) stellar illumination, (*ii.*) stellar reflected light from the planet, (*iii.*) thermal radiation from the planet, and (*iv.*) tidal heating on the moon. Here, primordial heat from the moon's formation and radiogenic decay is neglected, and it is assumed that the moon's rotation is tidally locked to its host planet, as is the case for almost all the moons in the Solar System. Our irradiation model includes arbitrary orbital eccentricities $e_{*p}$ of the planet around the star[5]. While we compute tidal heating on the satellite as a function of its orbital eccentricity $e_{ps}$ around the planet, we assume $e_{ps} = 0$ in the parametrization of the moon's irradiation. This is appropriate because typically $e_{ps} \ll 0.1$. Bolmont et al. (2011) studied the tidal evolution of Earth-mass objects around brown dwarfs, a problem which is similar to an Earth-mass moon orbiting a jovian planet, and found that tidal equilibrium occurs on very short timescales compared to the lifetime of the system. For non-zero eccentricities, ($e_{ps} \neq 0$), the moon will not be tidally locked. But since $e_{ps} \ll 0.1$, the moon's rotation will librate around an equilibrium orientation toward the planet, and the orbital mean motion will still be almost equal to the rotational mean motion (for a review on Titan's libration, see Sohl et al. 1995). By reasons specified by Heller et al. (2011b), we also assume that the obliquity of the satellite with respect to its orbit around the planet has been eroded to 0°, but we allow for arbitrary inclinations $i$ of the moon's orbit with respect to the orbit of the planet-moon barycenter around the star. If one assumed that the moon always orbits above the planet's equator, that would imply that $i$ is equal to the planetary obliquity $\psi_p$, which is measured with respect to the planet's orbit around the star. We do not need this assumption for the derivation of our equations, but since $\psi_p \approx i$ for all the large moons in the Solar System, except Triton, observations or numerical predictions of $\psi_p$ (Heller et al. 2011b) can provide reasonable assumptions for $i$.

In our simulations, we consider two prototype moons: one rocky Earth-mass satellite with a rock-to-mass fraction of 68% (similar to Earth) and one water-rich satellite with the tenfold mass of Ganymede and an ice-to-mass fraction of 25% (Fortney et al. 2007). The remaining constituents are assumed to be iron for the Earth-mass moon and silicates for the Super-Ganymede. The more massive and relatively dry moon represents what we guess a captured, Earth-like exomoon could be like, while the latter one corresponds to a satellite that has formed in situ. Note that a mass of $10M_G$ ($M_G$ being the mass of Ganymede) corresponds to roughly $0.25M_\oplus$, which is slightly above the detection limit for combined TTV and TDV with *Kepler* (Kipping et al. 2009). Our assumptions for the Super-Ganymede composition are backed up by observations of the Jupiter and Saturn satellite systems (Showman & Malhotra 1999; Grasset et al. 2000) as well as terrestrial planet and satellite formation studies (Kuchner 2003; Ogihara & Ida 2012). These papers show that in situ formation naturally generates water-rich moons and that such objects can retain their water reservoir for billions of years against steady hydrodynamic escape. Concerning the habitability of the water-rich Super-Ganymede, we do not rely on any assumptions concerning possible life forms in such water worlds. Except for the possible strong heating in a water-rich atmosphere (Matsui & Abe 1986; Kuchner 2003), we see no reason why ocean moons should not be hospitable, in particular against the background that life on Earth arose in (possibly hot) oceans or freshwater seas.

---

[5] In the following, a parameter index "∗" will refer to the star, "p" to the planet, and "s" to the satellite. The combinations "∗p" and "ps", e.g., for the orbital eccentricities $e_{*p}$ and $e_{ps}$, refer to systems of a star plus a planet and a planet plus a satellite, respectively. For a vector, e.g. $\vec{r}_{ps}$, the first letter indicates the starting point (in this case the planet) and the second index locates the endpoint (here the star).

For the sake of consistency, we derive the satellites' radii $R_s$ from planetary structure models (Fortney et al. 2007). In the case of the Earth-mass moon, we obtain $R_s = 1 R_\oplus$, and for the much lighter but water-dominated Super-Ganymede $R_s = 0.807 R_\oplus$. Equation (1) yields a critical flux of 295W/m² for the Earth-mass moon, and 266W/m² for the Super-Ganymede satellite. The bond albedo of both moons is assumed to be 0.3, similar to the mean albedo of Earth and of the Galilean satellites (Clark 1980). In the following, we call our Earth-like and Super-Ganymede satellites our "prototype moons". Based on the summary of observations and the model for giant planet atmospheres provided by Madhusudhan & Burrows (2012), we also use a bond albedo of 0.3 for the host planet, although higher values might be reasonable due to the formation of water clouds at distances 1AU from the host star (Burrows et al. 2006a). Mass and radius of the planet are not fixed in our model, but we will mostly refer to Jupiter-sized host planets.

### 3.1. Illumination

The total bolometric illumination on a moon is given by the stellar flux ($f_*$), the reflection of the stellar light from the planet ($f_r$), and the planetary thermal emission ($f_t$). Their variation will be a function of the satellite's orbital phase $0 \leq \varphi_{ps}(t) = (t-t_0)/P_{ps} \leq 1$ around the planet (with $t$ being time, $t_0$ as the starting time [0 in our simulations], and $P_{ps}$ as the period of the planet-moon orbit), the orbital phase of the planet-moon duet around the star ($\varphi_{*p}$, which is equivalent to the mean anomaly $\mathfrak{M}_{*p}$ divided by $2\pi$), and will depend on the eccentricity of the planet around the star ($e_{*p}$), on the inclination ($i$) of the two orbits, on the orientation of the periapses ($\eta$), as well as on longitude and latitude on the moon's surface ($\phi$ and $\theta$).

In Fig. 1, we show the variation of the satellite's illumination as a function of the satellite's orbital phase $\varphi_{ps}$. For this plot, the orbital phase of the planet-moon pair around the star $\varphi_{*p} = 0$ and $i = 0$. Projection effects due to latitudinal variation have been neglected, starlight is assumed to be plane-parallel, and radii and distances are not to scale.

In our irradiation model of a tidally locked satellite, we neglect clouds, radiative transfer, atmospheric circulation, geothermal flux[6], thermal inertia, and so on, and we make use of four simplifications:

- (*i*.) We assume the planet casts no penumbra on the moon. There is either total illumination from the star or none. This assumption is appropriate since we are primarily interested in the key contributions to the moon's climate.
- (*ii*.) The planet is assumed to be much more massive than the moon, and the barycenter of the planet-moon binary is placed at the center of the planet. Even if the planet and the moon had equal masses, corrections would be small since the range between the planet-moon barycenter and the star $\gg a_{ps}$.
- (*iii*.) For the computation of the irradiation, we treat the moon's orbit around the planet as a circle. The small eccentricities which we will consider later for tidal heating will not modify our results significantly.
- (*iv*.) The distance between the planet-satellite binary and the star does not change significantly over one satellite orbit, which is granted when either $e_{*p}$ is small or $P_{ps} \ll P_{*p}$.

In the following, we present the general results of our mathematical derivation. For a more thorough description and discussions of some simple cases, see Appendices A and B.

### 3.1.1 Illumination from the star

The stellar flux on the substellar point on the moon's surface will have a magnitude $L_*/(4\pi r_{s*}(t)^2)$, where $L_*$ is stellar luminosity and $\vec{r}_{s*}$ is the vector from the satellite to the star. We multiply this quantity with the surface normal $\vec{n}_{\phi,\theta}/n_{\phi,\theta}$ on the moon and $\vec{r}_{s*}/r_{s*}$ to include projection effects on a location ($\phi,\theta$). This yields

$$f_*(t) = \frac{L_*}{4\pi \, \vec{r}_{s*}(t)^2} \frac{\vec{r}_{s*}(t)}{r_{s*}(t)} \frac{\vec{n}_{\phi,\theta}(t)}{n_{\phi,\theta}(t)} \; . \tag{3}$$

---

[6] Tidal heating will be included below, but we will neglect geothermal feedback between tidal heating and irradiation.

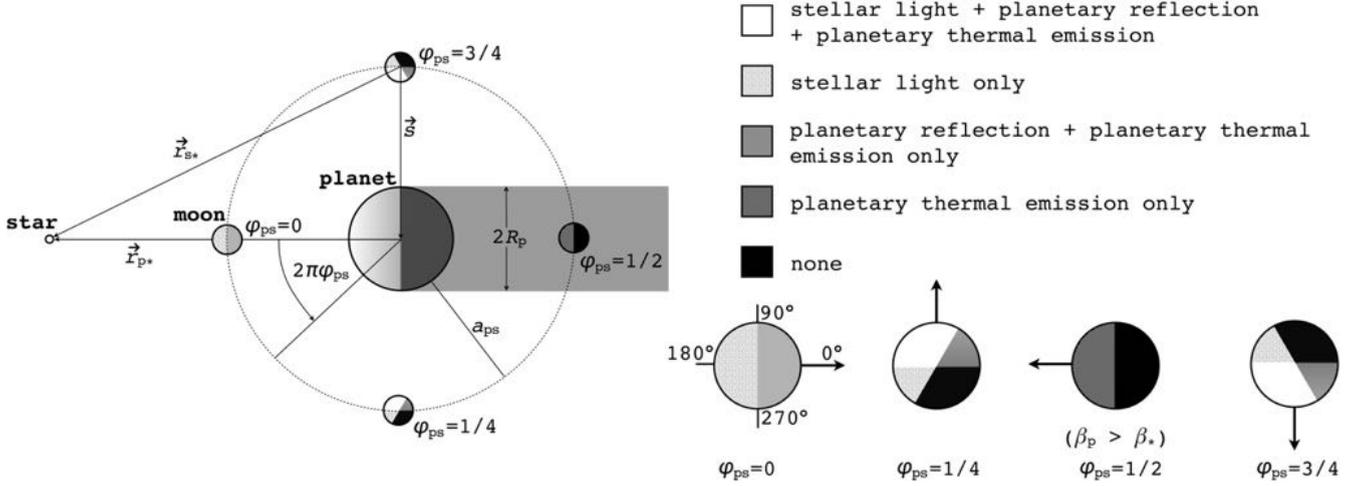

**FIG. 1.** Geometry of the triple system of a star, a planet, and a moon with illuminations indicated by different shadings (pole view). For ease of visualization, the moon's orbit is coplanar with the planet's orbit about the star and the planet's orbital position with respect to the star is fixed. Combined stellar and planetary irradiation on the moon is shown for four orbital phases. Projection effects as a function of longitude $\phi$ and latitude $\theta$ are ignored, and we neglect effects of a penumbra. Radii and distances are not to scale, and starlight is assumed to be plane-parallel. In the right panel, the surface normal on the subplanetary point is indicated by an arrow. For a tidally locked moon this spot is a fixed point on the moon's surface. For $\varphi_{\mathrm{ps}} = 0$ four longitudes are indicated.

If $\vec{r}_{\mathrm{s*}}$ and $\vec{n}$ have an antiparallel part, then $f_* < 0$, which is meaningless in our context, and we set $f_*$ to zero. The task is now to find $\vec{r}_{\mathrm{s*}}(t)$ and $\vec{n}_{\phi,\theta}(t)$. Therefore, we introduce the surface vector from the subplanetary point on the satellite to the planet, $\vec{s} \equiv \vec{n}_{0,0}$, and the vector from the planet to the star, $\vec{r}_{\mathrm{p*}}(t)$, which gives $\vec{r}_{\mathrm{s*}}(t) = \vec{r}_{\mathrm{p*}}(t) + \vec{s}(t)$ (see Fig. 1). Applying Kepler's equations of motion, we deduce $\vec{r}_{\mathrm{p*}}(t)$; and with a few geometric operations (see Appendix A) we obtain $\vec{n}_{\phi,\theta}(t)$:

$$\vec{r}_{\mathrm{p*}}(t) \;=\; -a_{*\mathrm{p}} \begin{pmatrix} \tilde{c} - e_{*\mathrm{p}} \\[4pt] \sqrt{1 - e_{*\mathrm{p}}^2}\,\tilde{s} \\[4pt] 0 \end{pmatrix} \qquad (4) \qquad\qquad \vec{n}_{\phi,\theta}(t) \;=\; a_{\mathrm{ps}} \begin{pmatrix} -\bar{s}S\tilde{C} + \bar{c}(\tilde{C}cC - \tilde{S}s) \\[4pt] -\bar{s}S\tilde{S} + \bar{c}(\tilde{S}cC - \tilde{C}s) \\[4pt] \bar{s}C + \bar{c}cS \end{pmatrix} \qquad (5)$$

with

$$
\begin{aligned}
c &= \cos\left(2\pi(\varphi_{\mathrm{ps}}(t) + \frac{\phi}{360^\circ})\right) & s &= \sin\left(2\pi(\varphi_{\mathrm{ps}}(t) + \frac{\phi}{360^\circ})\right) \\[4pt]
\bar{c} &= \cos\left(\theta\frac{\pi}{180^\circ})\right) & \bar{s} &= \sin\left(\theta\frac{\pi}{180^\circ})\right) \\[4pt]
\tilde{c} &= \cos\left(E_{*\mathrm{p}}(t)\right) & \tilde{s} &= \sin\left(E_{*\mathrm{p}}(t)\right) \\[4pt]
C &= \cos\left(i\frac{\pi}{180^\circ}\right) & S &= \sin\left(i\frac{\pi}{180^\circ}\right) \\[4pt]
\tilde{C} &= \cos\left(\eta\frac{\pi}{180^\circ}\right) & \tilde{S} &= \sin\left(\eta\frac{\pi}{180^\circ}\right) \qquad (6)
\end{aligned}
$$

where $i$, $\eta$, $0 \le \phi \le 360^\circ$, and $0 \le \theta \le 90^\circ$ are provided in degrees, and $\phi$ and $\theta$ are measured from the subplanetary point (see Fig. 1).

$$E_{*\mathrm{p}}(t) - e_{*\mathrm{p}} \sin\left(E_{*\mathrm{p}}(t)\right) = \mathfrak{M}_{*\mathrm{p}}(t) \qquad\qquad (7)$$

defines the eccentric anomaly $E_{*\mathrm{p}}$ and

$$\mathfrak{M}_{*\mathrm{p}}(t) = 2\pi\frac{(t - t_0)}{P_{*\mathrm{p}}} \qquad\qquad (8)$$

is the mean anomaly. The angle $\eta$ is the orientation of the lowest point of the moon's inclined orbit with respect to the star at periastron (see Appendix A). *Kepler's equation* (Eq. 7) is a transcendental function which we solve numerically.

To compute the stellar flux over one revolution of the moon around the planet, we put the planet-moon duet at numerous orbital phases around the star (using a fixed time step $dt$), thus $\tilde{c}$ and $\tilde{s}$ will be given. At each of these positions, we then evolve $\varphi_{ps}$ from 0 to 1. With this parametrization, the moon's orbit around the planet will always start at the left, corresponding to Fig. 1, and it will be more facile to interpret the phase functions. If we were to evolve the moon's orbit consistently, $2\pi\mathfrak{M}_{*p}$ would have to be added to the arguments of $c$ and $s$. Our simplification is appropriate as long as $\vec{r}_{*p}$ does not change significantly over one satellite orbit. Depending on the orientation of an eventual inclination between the two orbits and depending on the orbital position of the planet-moon system around the star, the can be eclipsed by the planet for a certain fraction of $\varphi_{ps}$ as seen from the moon. This phenomenon might have significant impacts on exomoon climates. Eclipses occur if the perpendicular part

$$r_\perp = \sin\left(\arccos\left(\frac{\vec{r}_{s*}\,\vec{r}_{p*}}{|\vec{r}_{s*}||\vec{r}_{p*}|}\right)\right)|\vec{r}_{s*}| \qquad (9)$$

of $\vec{r}_{s*}$ with respect to $\vec{r}_{p*}$ is smaller than the radius of the planet and if $|\vec{r}_{s*}| > |\vec{r}_{p*}|$, that is, if the moon is behind the planet as seen from the star and not in front of it. The angular diameters of the star and the planet, $\beta_*$ and $\beta_p$, respectively, are given by

$$\beta_* = 2\arctan\left(\frac{R_*}{a_{*p} + a_{ps}}\right)$$
$$\beta_p = 2\arctan\left(\frac{R_p}{a_{ps}}\right) . \qquad (10)$$

If $\beta_p > \beta_*$ then the eclipse will be total. Otherwise the stellar flux will be diminished by a factor $[1 - (\beta_p/\beta_*)^2]$.

### 3.1.2 Illumination from the planet

We now consider two contributions to exomoon illumination from the planet, namely, reflection of stellar light ($f_r$) and thermal radiation ($f_t$). If the planet's rotation period is $\lesssim 1$d, then the stellar irradiation will be distributed somewhat smoothly over longitude. However, for a planet which is tidally locked to the star, the illuminated hemisphere will be significantly warmer than the back side. In our model, the bright side of the planet has a temperature $T_{eff,p}^b$, and the dark back side has a temperature $T_{eff,p}^d$ (see Appendix B). With $dT = T_{eff,p}^b - T_{eff,p}^d$ as the temperature difference between the hemispheres and $\alpha_p$ as the planet's bond albedo, that is, the fraction of power at all wavelengths scattered back into space, thermal equilibrium yields

$$p(T_{eff,p}^b) \equiv (T_{eff,p}^b)^4 + (T_{eff,p}^b - dT)^4 - T_{eff,*}^4 \frac{(1-\alpha_p)R_*^2}{2\,\vec{r}_{*p}^2} = 0 . \qquad (11)$$

For a given $dT$, we search for the zero points of the polynomial $p(T_{eff,p}^b)$ numerically. In our prototype system at 1AU from a Sun-like star and choosing $dT = 100$K, Eq. (11) yields $T_{eff,p}^b = 291$ K and $T_{eff,p}^d = 191$ K. Finally, the thermal flux received by the moon from the planet turns out as

$$f_t(t) = \frac{R_p^2\sigma_{SB}}{a_{ps}^2}\cos\left(\frac{\phi\pi}{180°}\right)\cos\left(\frac{\theta\pi}{180°}\right)\times\left[(T_{eff,p}^b)^4\xi(t) + (T_{eff,p}^d)^4(1-\xi(t))\right] , \quad (12)$$

where

$$\xi(t) = \frac{1}{2}\left\{1 + \cos\left(\vartheta(t)\right)\cos\left(\Phi(t) - \nu_{*p}(t)\right)\right\} \qquad (13)$$

weighs the contributions from the two hemispheres,

$$\nu_{*\mathrm{p}}(t) = \arccos\left(\frac{\cos(E_{*\mathrm{p}}(t)) - e_{*\mathrm{p}}}{1 - e_{*\mathrm{p}}\cos(E_{*\mathrm{p}}(t))}\right) \tag{14}$$

is the true anomaly,

$$\Phi(t) = 2\arctan\left(\frac{s_{\mathrm{y}}(t)}{\sqrt{s_{\mathrm{x}}^2(t) + s_{\mathrm{y}}^2(t) + s_{\mathrm{x}}(t)}}\right)$$

$$\vartheta(t) = \frac{\pi}{2} - \arccos\left(\frac{s_{\mathrm{y}}(t)}{\sqrt{s_{\mathrm{x}}^2(t) + s_{\mathrm{y}}^2(t) + s_{\mathrm{z}}^2(t)}}\right) , \tag{15}$$

and $s_{\mathrm{x,y,z}}$ are the components of $\vec{s} = (s_{\mathrm{x}}, s_{\mathrm{y}}, s_{\mathrm{z}})$ (see Appendix B).

Additionally, the planet reflects a portion $\pi R_{\mathrm{p}}^2 \alpha_{\mathrm{p}}$ of the incoming stellar light. Neglecting that the moon blocks a small fraction of the starlight when it passes between the planet and the star ($< 1\%$ for an Earth-sized satellite around a Jupiter-sized planet), we find that the moon receives a stellar flux

$$f_{\mathrm{r}}(t) = \frac{R_*^2 \sigma_{\mathrm{SB}} T_{\mathrm{eff},*}^4}{r_{\mathrm{p}*}^2} \frac{\pi R_{\mathrm{p}}^2 \alpha_{\mathrm{p}}}{a_{\mathrm{ps}}^2} \cos\left(\frac{\phi\pi}{180°}\right) \cos\left(\frac{\theta\pi}{180°}\right) \xi(t) \tag{16}$$

from the planet.

In Fig. 2, we show how the amplitudes of $f_{\mathrm{l}}(t)$ and $f_{\mathrm{r}}(t)$ compare. Therefore, we neglect the time dependence and compute simply the maximum possible irradiation on the moon's subplanetary point as a function of the moon's orbit around the planet, which occurs in our model when the moon is over the substellar point of the planet. Then it receives maximum reflection and thermal flux at the same time. For our prototype system, it turns out that $f_{\mathrm{l}} > f_{\mathrm{r}}$ at a given planet-moon distance only if the planet has an albedo $\leq 0.1$, which means that it needs to be almost black. The exact value, $\alpha_{\mathrm{p}} = 0.093$ in this case, can be obtained by comparing Eqs. (12) and (16) (see Appendix B). For increasing $\alpha_{\mathrm{p}}$, stellar reflected flux dominates more and more; and for $\alpha_{\mathrm{p}} \gtrsim 0.6$, $f_*$ is over a magnitude stronger than $f_{\mathrm{t}}$.

The shapes of the curves can be understood intuitively, if one imagines that at a fixed semi-major axis (abscissa) the reflected flux received on the moon increases with increasing albedo (ordinate), whereas the planet's thermal flux increases when it absorbs more stellar light, which happens for decreasing albedo.

The shaded area in the upper left corner of the figure indicates where the sum of maximum $f_{\mathrm{t}}$ and $f_{\mathrm{r}}$ exceeds the limit of $295\,\mathrm{W/m}^2$ for a runaway greenhouse on an Earth-sized moon. Yet a satellite in this part of the parameter space would not necessarily be uninhabitable, because firstly it would only be subject to intense planetary radiation for less than about half its orbit, and secondly eclipses could cool the satellite half an orbit later. Moons at $a_{\mathrm{ps}} \leq 4R_{\mathrm{p}}$ are very likely to experience eclipses. Note that a moon's orbital eccentricity $e_{\mathrm{ps}}$ will have to be almost perfectly zero to avoid intense tidal heating in such close orbits (see Section 3.2).

Since we use $a_{\mathrm{ps}}$ in units of planetary radii, $f_{\mathrm{t}}$ and $f_{\mathrm{r}}$ are independent of $R_{\mathrm{p}}$. We also show a few examples of Solar System moons, where we adopted 0.343 for Jupiter's bond albedo (Hanel et al. 1981), 0.342 for Saturn (Hanel et al. 1983), 0.32 for Uranus (Neff et al. 1985; Pollack et al. 1986; Pearl et al. 1990), 0.29 for Neptune (Neff et al. 1985; Pollack et al. 1986; Pearl & Conrath 1991), and 0.3 for Earth. Flux contours are not directly applicable to the indicated moons because the host planets Jupiter, Saturn, Uranus, and Neptune do not orbit the Sun at 1AU, as assumed for our prototype exomoon system[7]. Only the position of Earth's moon, which receives a maximum of $0.35\,\mathrm{W/m}^2$ reflected light from Earth, reproduces the true Solar System values. The Roche radius for a fluid-like body (Weidner & Horne 2010, and references therein) is indicated with a gray line at $2.07R_{\mathrm{p}}$.

---

[7] At a distance of 5.2AU from the Sun, Europa receives roughly $0.5\,\mathrm{W/m}^2$ reflected light from Jupiter, when it passes the planet's subsolar point. Jupiter's thermal flux on Europa is negligible.

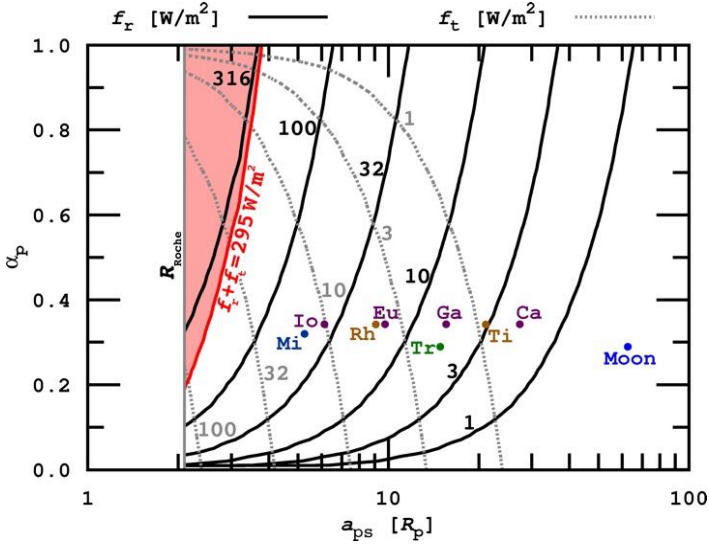

**FIG. 2.** Contours of constant planetary flux on an exomoon as a function of the planet-satellite semi-major axis $a_\mathrm{ps}$ and the planet's bond albedo $\alpha_\mathrm{p}$. The planet-moon binary orbits at 1AU from a Sun-like host star. Values depict the maximum possible irradiation in terms of orbital alignment, i.e., on the subplanetary point on the moon, and when the moon is over the substellar point of the planet. For $\alpha_\mathrm{p} \approx 0.1$ contours of equal $f_\mathrm{r}$ and $f_\mathrm{t}$ intersect, i.e., both contributions are equal. An additional contour is added at 295W/m², where the sum of $f_\mathrm{r}$ and $f_\mathrm{t}$ induces a runaway greenhouse on an Earth-sized moon. Some examples from the Solar System are given: Miranda (Mi), Io, Rhea (Rh), Europa (Eu), Triton (Tr), Ganymede (Ga), Titan (Ti), Callisto (Ca), and Earth's moon (Moon).

### 3.1.3 The circumstellar habitable zone of exomoons

We next transform the combined stellar and planetary flux into a correction for the IHZ, for which the boundaries are proportional to $L_*^{1/2}$ (Selsis et al. 2007). This correction is easily derived if we restrict the problem to just the direct and the reflected starlight. Then, we can define an "effective luminosity" $L_\mathrm{eff}$ that is the sum of the direct starlight plus the orbit-averaged reflected light. We ignore the thermal contribution as its spectral energy distribution will be much different from the star and, as shown below, the thermal component is the smallest for most cases. Our IHZ corrections are therefore only lower limits. From Eqs. (3), (13), and (16) one can show that

$$L_\mathrm{eff} = L_* \left(1 + \frac{\alpha_\mathrm{p} R_\mathrm{p}^2}{8 a_\mathrm{ps}^2}\right), \tag{17}$$

where we have averaged over the moon's orbital period. For realistic moon orbits, this correction amounts to 1% at most for high $\alpha_\mathrm{p}$ and small $a_\mathrm{ps}$. For planets orbiting F dwarfs near the outer edge of the IHZ, a moon could be habitable about 0.05AU farther out due to the reflected planetary light. In Fig. 3, we show the correction factor for the inner and outer boundaries of the IHZ due to reflected light as a function of $\alpha_\mathrm{p}$ and $a_\mathrm{ps}$.

### 3.1.4 Combined stellar and planetary illumination

With Eqs. (3), (12), and (16), we have derived the stellar and planetary contributions to the irradiation of a tidally locked moon in an inclined, circular orbit around the planet, where the orbit of the planet-moon duet around the star is eccentric. Now, we consider a satellite's total illumination

$$f_\mathrm{s}(t) = f_*(t) + f_\mathrm{t}(t) + f_\mathrm{r}(t) . \tag{18}$$

For an illustration of Eq. (18), we choose a moon that orbits its Jupiter-sized host planet at the same distance as Europa orbits Jupiter. The planet-moon duet is in a 1AU orbit around a Sun-like star, and we arbitrarily choose a temperature difference of d$T = 100$K between the two planetary hemispheres. Equation (18) does not depend on $M_\mathrm{s}$ or $R_\mathrm{s}$, so our irradiation model is not restricted to either the Earth-sized or the Super-Ganymede prototype moon.

In Fig. 4 we show $f_\mathrm{s}(t)$ as well as the stellar and planetary contributions for four different locations on the moon's surface. For all panels, the planet-moon duet is at the beginning of its revolution around the star, i.e. $\mathfrak{M}_{*\mathrm{p}} = 0$, and we set $i = 0$. Although $\mathfrak{M}_{*\mathrm{p}}$ would slightly increase during one orbit of the moon around the planet, we fix it to zero, so the moon starts and finishes over the illuminated hemisphere of the planet (similar to Fig. 1). The upper left panel depicts the subplanetary point, with a pronounced eclipse around $\varphi_\mathrm{ps} = 0.5$. At a position 45° counterclockwise along the equator (upper right panel), the stellar contribution is shifted in phase, and $f_\mathrm{t}$ as well as $f_\mathrm{r}$ are diminished in magnitude (note the logarithmic scale!) by a factor cos(45°). In the lower row, where $\phi = 90°$, $\theta = 0°$ (lower left panel) and $\phi = 180°$, $\theta = 80°$ (lower right panel), there are no planetary contributions. The eclipse trough has also disappeared because the star's occultation by the planet cannot be seen from the antiplanetary hemisphere.

For Fig. 5, we assume a similar system, but now the planet-moon binary is at an orbital phase $\varphi_{*\mathrm{p}} = 0.5$, corresponding to $\mathfrak{M}_{*\mathrm{p}} = \pi$, around the star. We introduce an eccentricity $e_{*\mathrm{p}} = 0.3$ as well as an inclination of 45° between the two orbital planes. The first aspect shifts the stellar and planetary contributions by half an orbital phase with respect to Fig. 4. Considering the top view of the system in Fig. 1, this means eclipses should now occur when the moon is to the left of the planet because the star is to the right, "left" here meaning $\varphi_{ps} = 0$. However, the non-zero inclination lifts the moon out of the planet's shadow (at least for this particular orbital phase around the star), which is why the eclipse trough disappears. Due to the eccentricity, stellar irradiation is now lower because the planet-moon binary is at apastron. An illustration of the corresponding star-planet orbital configuration is shown in the pole view (left panel) of Fig. 6.

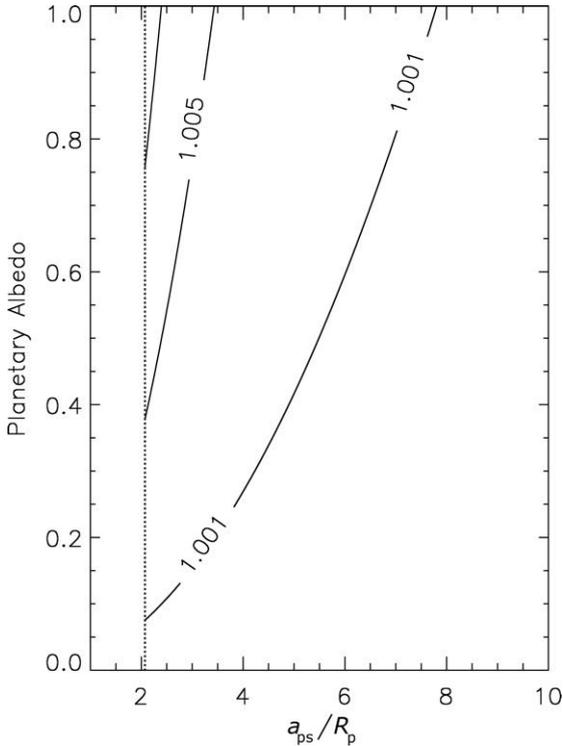

**FIG. 3.** Contours of the correction factor for the limits of the IHZ for exomoons, induced by the star's reflected light from the planet. Since we neglect the thermal component, values are lower limits. The left-most contour signifies 1.01. The dotted vertical line denotes the Roche lobe.

The eccentricity-driven cooling of the moon is enhanced on its northern hemisphere, where the inclination induces a winter. Besides, our assumption that the moon is in the planet's equatorial plane is equivalent to $i = \psi_p$; thus the planet also experiences northern winter. The lower right panel of Fig. 5, where $\phi = 180°$ and $\theta = 80°$, demonstrates a novel phenomenon, which we call an "antiplanetary winter on the moon". On the satellite's antiplanetary side there is no illumination from the planet (as in the lower two panels of Fig. 4); and being close enough to the pole, at $\theta > 90° - i$ for this occasion of northern winter, there will be no irradiation from the star either, during the whole orbit of the moon around the planet. In Fig. 6, we depict this constellation in the edge view (right panel). Note that antiplanetary locations close to the moon's northern pole receive no irradiation at all, as indicated by an example at $\phi = 180°$, $\theta = 80°$ (see arrow). Of course there is also a "proplanetary winter" on the moon, which takes place just at the same epoch but on the proplanetary hemisphere on the moon. The opposite effects are the "proplanetary summer", which occurs on the proplanetary side of the moon at $\mathfrak{M}_{*p} = 0$, at least for this specific configuration in Fig. 5, and the "antiplanetary summer".

Finally, we compute the average surface flux on the moon during one stellar orbit. Therefore, we first integrate $d\varphi_{ps} f_s(\varphi_{ps})$ over $0 \le \varphi_{ps} \le 1$ at an initial phase in the planet's orbit about the star ($\varphi_{*p} = 0$), which yields the area under the solid lines in Fig. 4. We then step through $\approx 50$ values for $\varphi_{*p}$ and again integrate the total flux. Finally, we average the flux over one orbit of the planet around the star, which gives the orbit-averaged flux $F_s(\phi,\theta)$ on the moon. In Fig. 7, we plot these values as surface maps of a moon in four scenarios. The two narrow panels to the right of each of the four major panels show the averaged flux for $-1/4 \le \varphi_{*p} \le +1/4$ and $+1/4 \le \varphi_{*p} \le +3/4$, corresponding to northern summer (ns) and southern summer (ss) on the moon, respectively.

In the upper left panel, the two orbits are coplanar. Interestingly, the subplanetary point at $\phi = 0 = \theta$ is the "coldest" spot along the equator (if we convert the flux into a temperature) because the moon passes into the shadow of the planet when the star would be at zenith over the subplanetary point. Thus, the stellar irradiation maximum is reduced (see the upper left panel in Fig. 4). The contrast between polar and equatorial irradiation, reaching from 0 to $\approx 440\mathrm{W/m}^2$, is strongest in this panel. In the upper right panel, the subplanetary point has turned into the "warmest" location along the equator. On the one hand, this is due to the inclination of 22.5°, which is why the moon does not transit behind the planet for most of the orbital phase around the star. On the other hand, this location gets slightly more irradiation from the planet than any other place on the moon. In the lower left panel, the average flux contrast between equatorial and polar illumination has decreased further. Again, the subplanetary point is slightly warmer than the rest of the surface. In the lower right panel, finally, where the moon's orbital inclination is set to 90°, the equator has become the coldest region of the moon, with the subplanetary point still being the warmest location along 0° latitude.

While the major panels show that the orbit-averaged flux contrast decreases with increasing inclination, the side panels indicate an increasing irradiation contrast between seasons. Exomoons around host planets with obliquities similar to that of Jupiter with respect to the Sun ($\psi_p \approx 0$) are subject to an irradiation pattern corresponding to the upper left panel of Fig. 7. The upper right panel depicts an irradiation pattern of exomoons around planets with obliquities similar to Saturn (26.7°) and is qualitatively in good agreement with the yearly illumination pattern of Titan as simulated by Mitchell (2012, see his Fig. 1c). Exomoons around a planet with a Uranus-like obliquity (97.9°) will have an irradiation similar to the lower right panel.

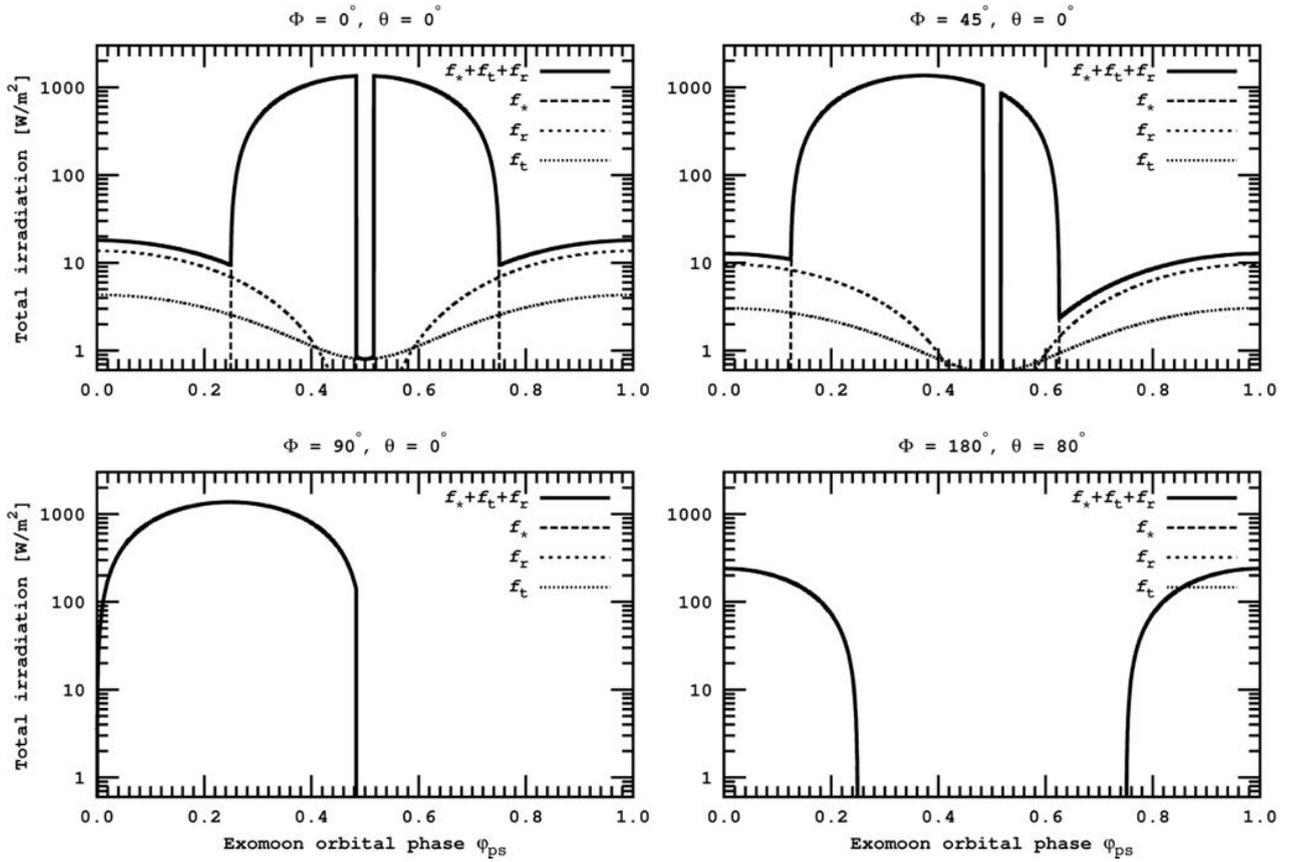

**FIG. 4.** Stellar and planetary contributions to the illumination of our prototype moon as a function of orbital phase $\varphi_{ps}$. Tiny dots label the thermal flux from the planet ($f_t$), normal dots the reflected stellar light from the planet ($f_r$), dashes the stellar light ($f_*$), and the solid line is their sum. The panels depict different longitudes and latitudes on the moon's surface. The upper left panel is for the subplanetary point, the upper right 45° counterclockwise along the equator, the lower left panel shows a position 90° counterclockwise from the subplanetary point, and the lower right is the antiplanetary point.

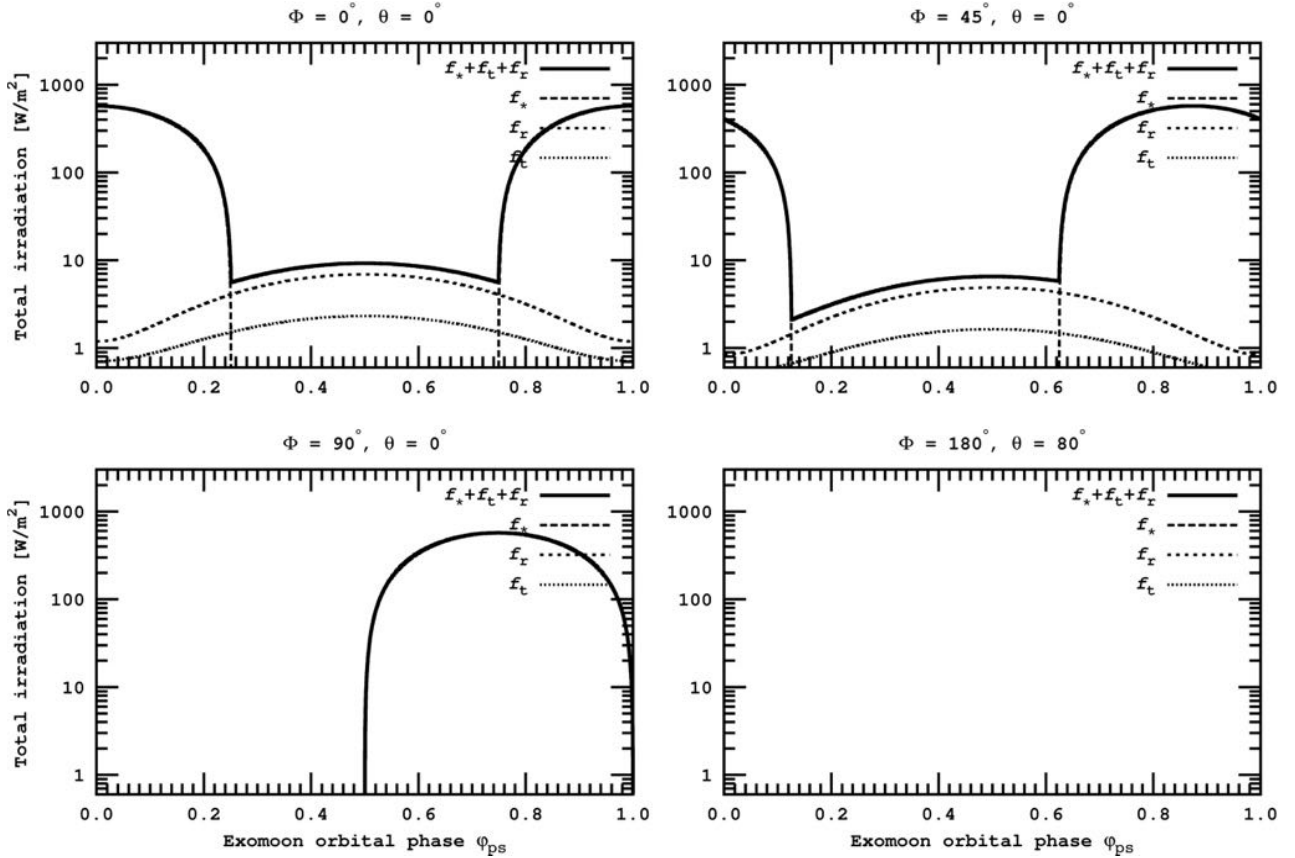

**FIG. 5.** Stellar and planetary contributions to the illumination of our prototype moon as in Fig. 4 but at a stellar orbital phase $\varphi_{*p} = 0.5$ in an eccentric orbit ($e_{*p} = 0.3$) and with an inclination $i = \pi/4 \triangleq 45°$ of the moon's orbit against the circumstellar orbit.

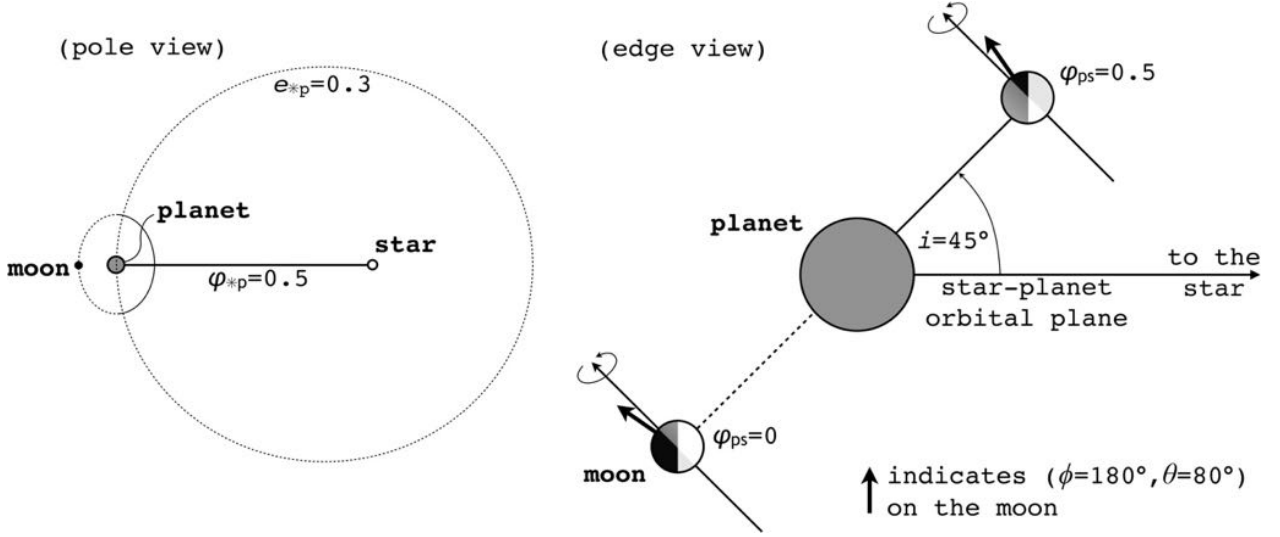

**FIG. 6.** Illustration of the antiplanetary winter on the moon with the same orbital elements as in Fig. 5. The arrow in the edge view panel indicates the surface normal at $\phi=180°$, $\theta=80°$, i.e., close to pole and on the antiplanetary side of the moon. For all orbital constellations of the moon around the planet ($\varphi_{ps}$ going from 0 to 1), this location on the moon receives neither irradiation from the star nor from the planet (see lower right panel in Fig. 5). Shadings correspond to the same irradiation patterns as in Fig. 1.

The typical orbit-averaged flux between 300 and 400W/m² in Fig. 7 is about a quarter of the solar constant. This is equivalent to an energy redistribution factor of 4 over the moon's surface (Selsis et al. 2007), indicating that climates on exomoons with orbital periods of a few days (in this case 3.55d, corresponding to Europa's orbit about Jupiter) may be more similar to those of freely rotating planets rather than to those of planets that are tidally locked to their host star.

### 3.2 Tidal heating

Tidal heating is an additional source of energy on moons. Various approaches for the description of tidal processes have been established. Two of the most prominent tidal theories are the "constant-time-lag" (CTL) and the "constant-phase-lag" (CPL) models. Their merits and perils have been treated extensively in the literature (Ferraz-Mello et al. 2008; Greenberg 2009; Efroimsky & Williams 2009; Hansen 2010; Heller et al. 2010,2011b; Lai 2012) and it turns out that they agree for low eccentricities. To begin with, we arbitrarily choose the CTL model developed by Hut (1981) and Leconte et al. (2010) for the computation of the moon's instantaneous tidal heating, but we will compare predictions of both CPL and CTL theory below. We consider a tidal time lag $\tau_s=638$s, similar to that of Earth (Lambeck 1977; Neron de Surgy & Laskar 1997), and an appropriate second-order potential Love number of $k_{2,s}=0.3$ (Henning et al. 2009).

In our two-body system of the planet and the moon, tidal heating on the satellite, which is assumed to be in equilibrium rotation and to have zero obliquity against the orbit around the planet, is given by

$$\dot{E}_{\mathrm{tid},s}^{\mathrm{eq}} = \frac{Z_s}{\beta^{15}(e_{ps})}\left[f_1(e_{ps}) - \frac{f_2^2(e_{ps})}{f_5(e_{ps})}\right] \ , \tag{19}$$

where

$$Z_s \equiv 3G^2 k_{2,s} M_p^2 (M_p+M_s)\frac{R_s^5}{a_{ps}^9}\ \tau_s \tag{20} \qquad \text{and} \qquad \beta(e_{ps}) = \sqrt{1-e_{ps}^2} \ ,$$

$$f_1(e_{ps}) = 1+\frac{31}{2}e_{ps}^2+\frac{255}{8}e_{ps}^4+\frac{185}{16}e_{ps}^6+\frac{25}{64}e_{ps}^8 \ ,$$

$$f_2(e_{ps}) = 1+\frac{15}{2}e_{ps}^2+\frac{45}{8}e_{ps}^4\ +\frac{5}{16}e_{ps}^6 \ ,$$

$$f_5(e_{ps}) = 1+3e_{ps}^2\ +\frac{3}{8}e_{ps}^4 \tag{21}$$

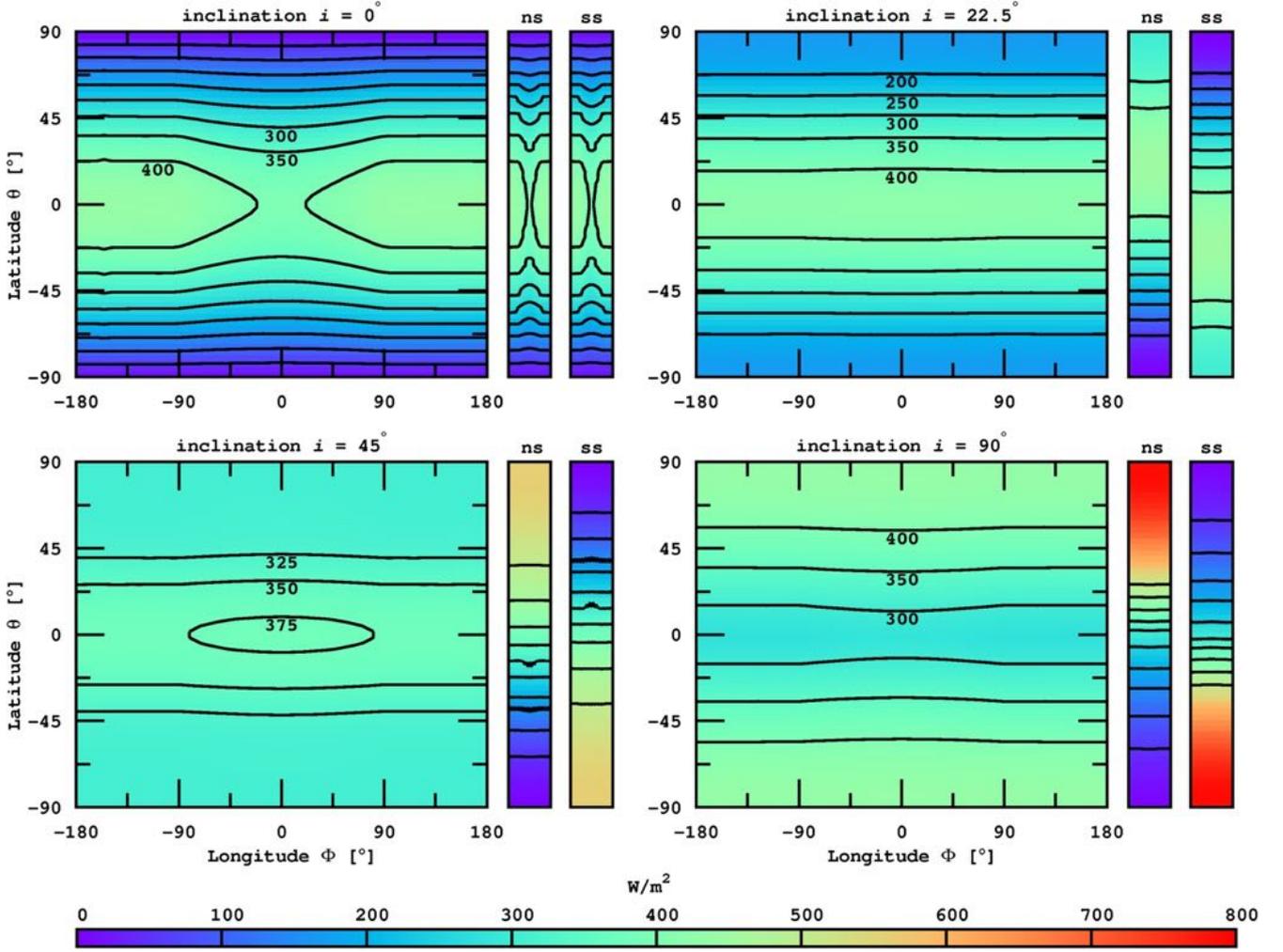

**FIG. 7.** Illumination of our prototype exomoon (in W/m²) averaged over the orbit of the planet-moon duet around their host star. Major panels present four different orbital inclinations: $i = 0°$ (upper left), $i = 22.5°$ (upper right), $i = 45°$ (lower left), and $i = 90°$ (lower right). The two bars beside each major panel indicate averaged flux for the northern summer (ns) and southern summer (ss) on the moon. Contours of constant irradiation are symmetric about the equator; some values are given.

Here, $G$ is Newton's gravitational constant, and $M_p$ is the planet's mass. The contribution of tidal heating to the moon's energy flux can be compared to the incoming irradiation when we divide $\dot{E}_{\mathrm{tid,s}}^{\mathrm{eq}}$ by the surface of the moon and define its surface tidal heating $h_s \equiv \dot{E}_{\mathrm{tid,s}}^{\mathrm{eq}}/(4\pi R_s^2)$. We assume that $h_s$ is emitted uniformly through the satellite's surface.

To stress the importance of tidal heating on exomoons, we show the sum of $h_s$ and the absorbed stellar flux for four star-planet-moon constellations as a function of $e_{\mathrm{ps}}$ and $a_{\mathrm{ps}}$ in Fig. 8. In the upper row, we consider our Earth-like prototype moon, in the lower row the Super-Ganymede. The left column corresponds to a Jupiter-like host planet, the right column to a Neptune-like planet, both at 1AU from a Sun-like star. Contours indicate regions of constant energy flux as a function of $a_{\mathrm{ps}}$ and $e_{\mathrm{ps}}$. Far from the planet, illustrated by a white area at the right in each plot, tidal heating is negligible, and the total heat flux on the moon corresponds to an absorbed stellar flux of 239W/m², which is equal to Earth's absorbed flux. The right-most contour in each panel depicts a contribution by tidal heating of 2W/m², which corresponds to Io's tidal heat flux (Spencer et al. 2000). Satellites left of this line will not necessarily experience enhanced volcanic activity, since most of the dissipated energy would go into water oceans of our prototype moons, rather than into the crust as on Io. The blue contour corresponds to a tidal heating of 10W/m², and the red line demarcates the transition into a runaway greenhouse, which occurs at 295W/m² for the Earth-mass satellite and at 266W/m² for the Super-Ganymede. For comparison, we show the positions of some prominent moons in the Solar System, where $a_{\mathrm{ps}}$ is measured in radii of the host planet. Intriguingly, both the Earth-like exomoon and the Super-Ganymede, orbiting either a Jupiter- or a Neptune-mass planet, would be habitable in a Europa- or Miranda-like orbit (in terms of fractional planetary radius and eccentricity), while they would enter a runaway greenhouse state in an Io-like orbit.

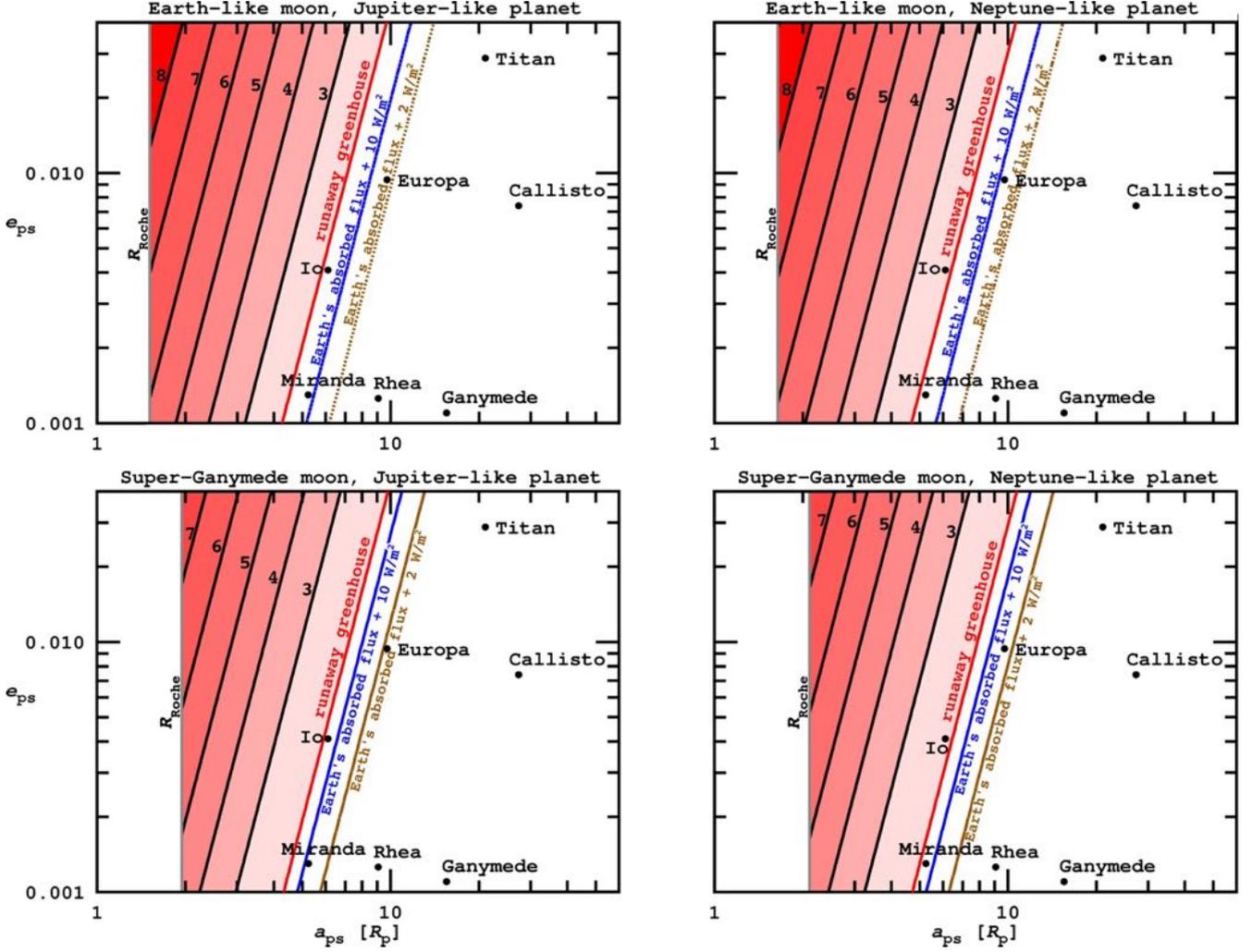

**FIG. 8.** Contours of summed absorbed stellar irradiation and tidal heating (in logarithmic units of W/m²) as a function of semi-major axis $a_{ps}$ and eccentricity $e_{ps}$ on an Earth-like (upper row) and a Super-Ganymede (lower row) exomoon. In the left panels, the satellite orbits a Jupiter-like planet, in the right panels a Neptune-mass planet, in both cases at 1AU from a Sun-like host star. In the white area at the right, tidal heating is negligible and absorbed stellar flux is 239W/m². The right-most contours in each panel indicate Io's tidal heat flux of 2W/m², a tidal heating of 10W/m², and the critical flux for the runaway greenhouse (295W/m² for the Earth-like moon and 266W/m² for the Super-Ganymede). Positions of some massive satellites in the Solar System are shown for comparison.

The Roche radii are $\approx 1.5R_p$ for an Earth-type moon about a Jupiter-like planet (upper left panel in Fig. 8), $\approx 1.6R_p$ for an Earth-type moon about a Neptune-mass planet (upper right), $\approx 1.9R_p$ for the Super-Ganymede about a Jupiter-class host (lower left), and $\approx 2.1R_p$ for the Super-Ganymede orbiting a Neptune-like planet (lower right). The extreme tidal heating rates in the red areas may not be realistic because we assume a constant time lag $\tau_s$ of the satellite's tidal bulge and ignore its dependence on the driving frequency as well as its variation due to the geological processes that should appear at such enormous heat fluxes.

In Fig. 8, irradiation from the planet is neglected; thus decreasing distance between planet and moon goes along with increasing tidal heating only. An Earth-like exomoon (upper panels) could orbit as close as $\approx 5R_p$, and tidal heating would not induce a runaway greenhouse if $e_{ps} \lesssim 0.001$. If its orbit has an eccentricity similar to Titan ($e_{ps} \approx 0.03$), then the orbital separation needed to be $\gtrsim 10R_p$ to prevent a runaway greenhouse. Comparison of the left and right panels shows that for a more massive host planet (left) satellites can be slightly closer and still be habitable. This is because we draw contours over the fractional orbital separation $a_{ps}/R_p$ on the abscissa. In the left panels $10R_p \approx 7 \times 10^5$km, whereas in the right panel $10R_p \approx 2.5 \times 10^5$km. When plotting over $a_{ps}/R_p$, this discrepancy is somewhat balanced by the much higher mass of the host planet in the left panel and the strong dependence of tidal heating on $M_p$ (see Eq. 20).

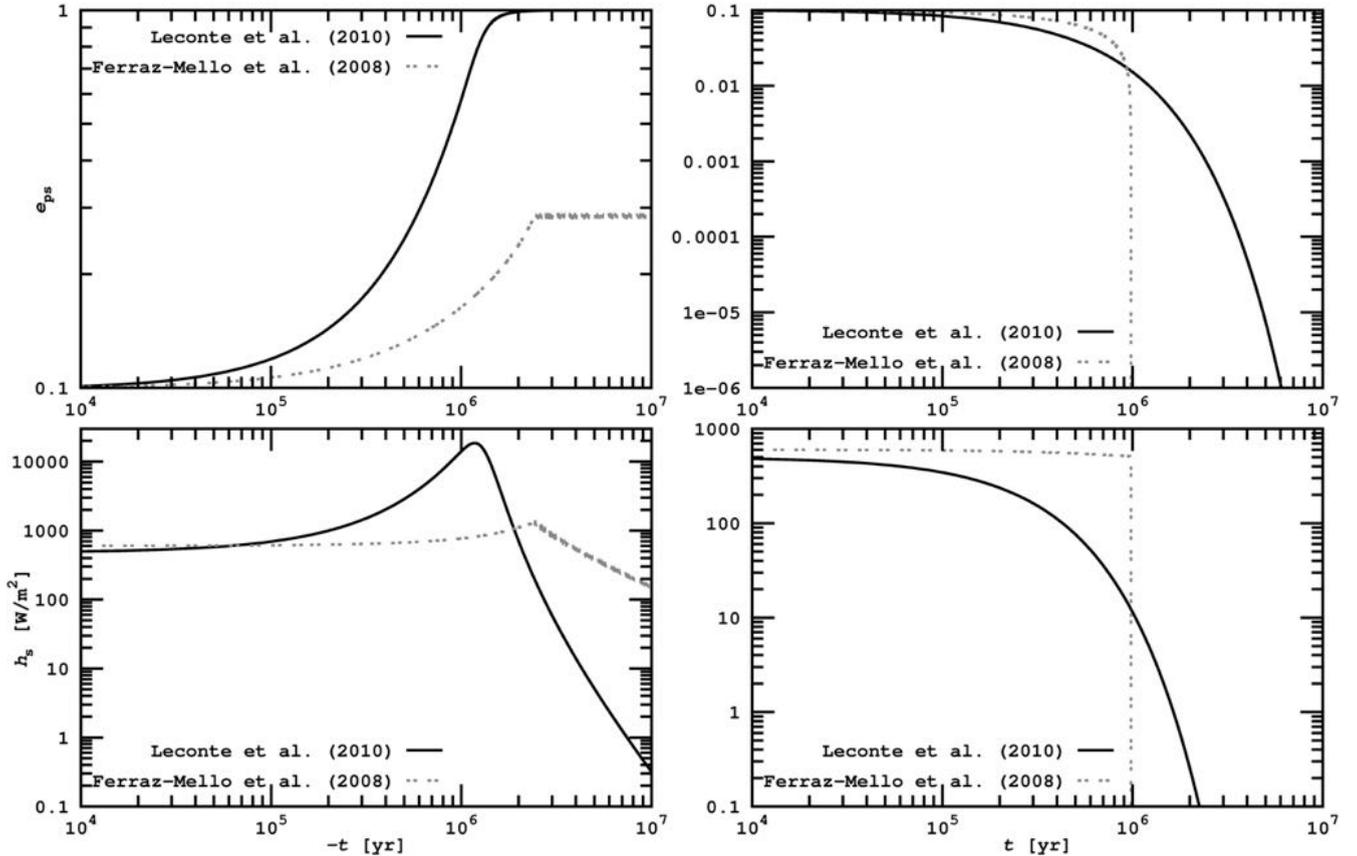

**FIG. 9.** Evolution of the orbital eccentricity (upper row) and the moon's tidal heating (lower row) following the two-body tidal models of Leconte et al. (2010) (a "constant-time-lag" model, solid line) and Ferraz-Mello et al. (2008) (a "constant-phase-lag" model, dashed line). Initially, an Earth-sized moon is set in an eccentric orbit ($e_{ps} = 0.1$) around a Jupiter-mass planet at the distance in which Europa orbits Jupiter. In the left panels evolution is backward, in the right panels into the future. Both tidal models predict that free eccentricities are eroded and tidal heating ceases after <10Myr.

### 3.2.1 Planet-moon orbital eccentricity

Tidal heating on the moon will only be significant, if $e_{ps} \neq 0$. If the eccentricity is *free*, that is, the moon is not significantly perturbed by other bodies, then $e_{ps}$ will change with time due to tidal damping, and orbit-averaged equations can be applied to simulate the tidal evolution. For those cases considered here, $de_{ps}/dt < 0$. Similar to the approach used by Heller et al. (2011b), we apply a CTL model (Leconte et al. 2010) and a CPL model (Ferraz-Mello et al. 2008) to compute the tidal evolution, albeit here with a focus on exomoons rather than on exoplanets.

In Fig. 9, we show the evolution of our Earth-sized prototype moon in orbit around a Jupiter-like planet. The upper two panels show the change of eccentricity $e_{ps}(t)$, the lower two panels of tidal surface heating $h_s(t)$. In the left panels, evolution is backward until $-10^7$yr; in the right panels, evolution is forwards until $+10^7$yr. The initial eccentricity is set to 0.1, and for the distance between the planet and the moon we choose the semi-major axis of Europa around Jupiter. Going backward, eccentricity increases[8] as does tidal heating. When $e_{ps} \to 1$ at $\approx -1$Myr, the moon is almost moving on a line around the planet, which would lead to a collision (or rather to an ejection because we go backward in time). At this time, tidal heating has increased from initially 600W/m² to 1300W/m². The right panels show that free eccentricities will be damped to zero within <10Myr and that tidal heating becomes negligible after $\approx$2Myr.

Although $e_{ps}$ is eroded in <10Myr in these two-body simulations, eccentricities can persists much longer, as the cases of Io around Jupiter and Titan around Saturn show. Their eccentricities are not free, but they are *forced*, because they are excited by interaction with other bodies. The origin of Titan's eccentricity $e_{ps} = 0.0288$ is still subject to debate. As shown by Sohl et al. (1995), tidal dissipation would damp it on timescales shorter than the age of the Solar System. It could only be primordial if the moon had a methane or hydrocarbon ocean that is deeper than a few kilometers. However, surface observations by the *Cassini Huygens* lander negated this assumption. Various other possibilities have been discussed, such as collisions or close encounters with massive bodies (Smith et al. 1982; Farinella et al. 1990) and a capture of Titan by Saturn ≪ 4.5 Gyr ago. The origin of Io's eccentricity $e_{ps} = 0.0041$ lies in the moon's resonances with Ganymede and Callisto (Yoder 1979). Such resonances may also appear among exomoons.

---

[8] For the CPL model, $e_{ps}$ converges to 0.285 for $-t \gtrsim 2$Myr. This result is not physical but owed to the discontinuities induced by the phase lags, in particular by $\varepsilon_{1,s}$ and $\varepsilon_{1,p}$ (for details see Ferraz-Mello et al. 2008; Heller et al. 2011b). We have tried other initial eccentricities, which all led to this convergence.

Using the publicly available *N*-body code *Mercury* (Chambers 1999)[9], we performed *N*-body experiments of a hypothetical satellite system to find out whether forced eccentricities can drive a tidal greenhouse in a manner analogous to the volcanic activity of Io. We chose a Jupiter-mass planet with an Earth-mass exomoon at the same distance as Europa orbits Jupiter, placed a second exomoon at the 2:1 external resonance, and integrated the resulting orbital evolution. In one case, the second moon had a mass equal to that of Earth, in the other case a mass equal to that of Mars. For the former case, we find the inner satellite could have its eccentricity pumped to 0.09 with a typical value of 0.05. For the latter, the maximum eccentricity is 0.05 with a typical value near 0.03. Although these studies are preliminary, they suggest that massive exomoons in multiple configurations could trigger a runaway greenhouse, especially if the moons are of Earth-mass, and that their circumstellar IHZ could lie further away from the star. A comprehensive study of configurations that should also include Cassini states and damping to the fixed-point solution is beyond the scope of this study but could provide insight into the likelihood that exomoons are susceptible to a tidal greenhouse.

## 4. Orbits of habitable exomoons

By analogy with the circumstellar habitable zone for planets, we can imagine a minimum orbital separation between a planet and a moon to let the satellite be habitable. The range of orbits for habitable moons has no outer edge, except that Hill stability must be ensured. Consequently, habitability of moons is only constrained by the inner edge of a circumplanetary habitable zone, which we call the "habitable edge". Moons inside the habitable edge are in danger of running into a greenhouse by stellar and planetary illumination and/or tidal heating. Satellites outside the habitable edge with their host planet in the circumstellar IHZ are habitable by definition.

Combining the limit for the runaway greenhouse from Section 2.2 with our model for the energy flux budget of extrasolar moons from Section 3, we compute the orbit-averaged global flux $\bar{F}_s^{glob}$ received by a satellite, which is the sum of the averaged stellar ($\bar{f}_*$), reflected ($\bar{f}_r$), thermal ($\bar{f}_t$), and tidal heat flux ($h_s$). Thus, in order for the moon to be habitable

$$F_{RG} > \bar{F}_s^{glob} = \bar{f}_* + \bar{f}_r + \bar{f}_t + h_s$$
$$= \frac{L_*(1 - \alpha_s)}{16\pi a_{*p}^2 \sqrt{1 - e_{*p}^2}} \left(1 + \frac{\pi R_p^2 \alpha_p}{2a_{ps}^2}\right) + \frac{R_p^2 \sigma_{SB}(T_p^{eq})^4}{a_{ps}^2} \frac{(1 - \alpha_s)}{4} + h_s \quad , \qquad (22)$$

where the critical flux for a runaway greenhouse $F_{RG}$ is given by Eq. (1), the tidal heating rate $h_s \equiv \dot{E}_{tid,s}^{eq}/(4\pi R_s^2)$ by Eq. (19), and the planet's thermal equilibrium temperature $T_p^{eq}$ can be determined with Eq. (11) and using $dT = 0$. Note that the addend "1" in brackets implies that we do not consider reduction of stellar illumination due to eclipses. This effect is treated in a companion paper (Heller 2012).

In Fig. 10, we show the $F_{RG} = \bar{F}_s^{glob}$ orbits of both the Earth-like (blue lines) and the Super-Ganymede (black lines) prototype moon as a function of the planet-moon semi-major axis and the mass of the giant host planet, which orbits the Sun-like star at a distance of 1 AU.[10] Contours are plotted for various values of the orbital eccentricity, which means that orbits to the left of a line induce a runaway greenhouse for the respective eccentricity of the actual moon. These innermost, limiting orbits constitute the circumplanetary habitable edges.

When the moon is virtually shifted toward the planet, then illumination from the planet and tidal heating increase, reaching $F_{RG}$ at some point. With increasing eccentricity, tidal heating also increases; thus $F_{RG}$ will be reached farther away from the planet. Blue lines appear closer to the planet than black contours for the same eccentricity, showing that more massive moons can orbit more closely to the planet and be prevented from becoming a runaway greenhouse. This is a purely atmospheric effect, determined only by the moon's surface gravity $g_s(M_s, R_s)$ in Eq. (1). We also see that, for a fixed eccentricity, moons can orbit closer to the planet – both in terms of fractional and absolute units – if the planet's mass is smaller.

---

[9] Download via www.arm.ac.uk/~jec .
[10] For consistency, we computed the planetary radius (used to scale the abscissa) as a function of the planet's mass (ordinate) by fitting a high-order polynomial to the Fortney et al. (2007) models for a giant planet at 1 AU from a Sun-like star (see line 17 in their Table 4).

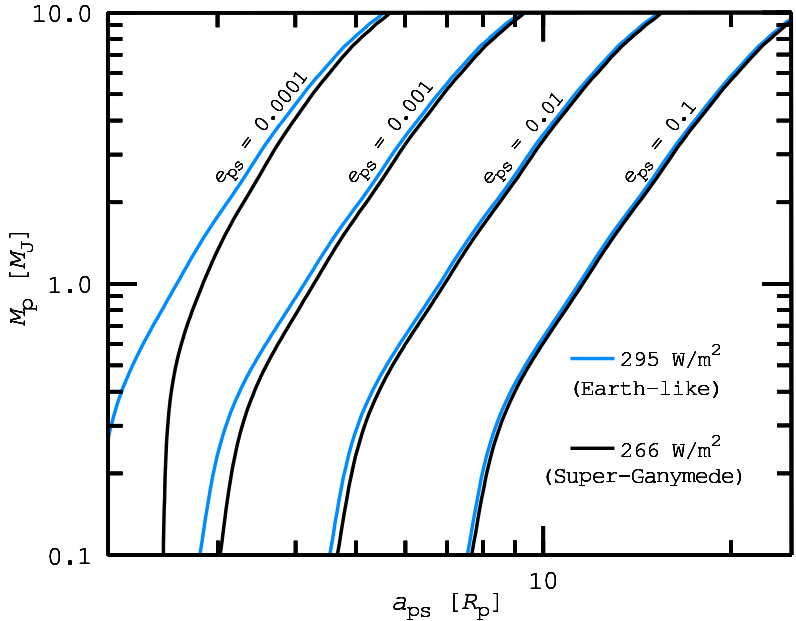

**FIG. 10.** Innermost orbits to prevent a runaway greenhouse, i.e., the "habitable edges" of an Earth-like (blue lines) and a Super-Ganymede (black lines) exomoon. Their host planet is at 1 AU from a Sun-like star. Flux contours for four eccentricities of the moons' orbits from $e_{ps} = 10^{-4}$ to $e_{ps} = 10^{-1}$ are indicated. The larger $e_{ps}$, the stronger tidal heating and the more distant from the planet will the critical flux be reached.

Estimating an exomoon's habitability with this model requires a well-parametrized system. With the current state of technology, stellar luminosity $L_*$ and mass $M_*$ can be estimated by spectral analysis and by using stellar evolution models. In combination with the planet's orbital period $P_{*p}$, this yields $a_{*p}$ by means of Kepler's third law. The planetary mass $M_p$ could be measured with the radial-velocity (RV) method and assuming it is sufficiently larger than the moon's mass. Alternatively, photometry could determine the mass ratios $M_p/M_*$ and $M_s/M_p$. Thus, if $M_s$ were known from spectroscopy and stellar evolution models, then all masses were accessible (Kipping 2010). Combining RV and photometry, the star-planet orbital eccentricity $e_{*p}$ can be deduced (Mislis et al. 2012); and by combination of TTV and TDV, it is possible to determine $M_s$ as well as $a_{ps}$ (Kipping 2009a). Just like the planetary radius, the moon's radius $R_s$ can be determined from photometric transit observations if its transit can directly be observed. The satellite's second-order tidal Love number $k_{2,s}$ and its tidal time lag $\tau_s$, however, would have to be assumed. For this purpose, Earth or Solar System moons could serve as reference bodies. If the age of the system (i.e., of the star) was known or if the evolution of the satellite's orbit due to tides could be observed, then tidal theory could give a constraint on the product of $k_{2,s}$ and $\tau_s$ (or $k_{2,s}/Q_s$ in CPL theory) (Lainey et al. 2009; Ferraz-Mello 2012). Then the remaining free parameters would be the satellite's albedo $\alpha_s$ and the orbital eccentricity of the planet-moon orbit $e_{ps}$. $N$-body simulations of the system would allow for an average value of $e_{ps}$.

We conclude that combination of all currently available observational and theoretical techniques can, in principle, yield an estimation of an exomoon's habitability. To that end, the satellite's global average energy flux $\bar{F}_s^{glob}$ (Eq. 22) needs to be compared to the critical flux for a runaway greenhouse $F_{RG}$ (Eq. 1).

## 5. Application to Kepler-22b and KOI211.01

We now apply our stellar-planetary irradiation plus tidal heating model to putative exomoons around Kepler-22b and KOI211.01, both in the habitable zone around their host stars. The former is a confirmed transiting Neptune-sized planet (Borucki et al. 2012), while the latter is a much more massive planet candidate (Borucki et al. 2011). We choose these two planets that likely have very different masses to study the dependence of exomoon habitability on $M_p$.

For the moon, we take our prototype Earth-sized moon and place it in various orbits around the two test planets to investigate a parameter space as broad as possible. We consider two planet-moon semi-major axes: $a_{ps} = 5R_p$, which is similar to Miranda's orbit around Uranus, and $a_{ps} = 20R_p$, which is similar to Titan's orbit around Saturn[11]. We also choose two eccentricities, namely, $e_{ps} = 0.001$ (similar to Miranda) and $e_{ps} = 0.05$ (somewhat larger than Titan's value). With this parametrization, we cover a parameter space of which the diagonal is spanned by Miranda's close, low-eccentricity orbit around Uranus and Titan's far but significantly eccentric orbit around Saturn (see Fig. 8). We must keep in mind, however, that strong additional forces, such as the interaction with further moons, are required to maintain eccentricities of 0.05 in the assumed close orbits for a long time, because tidal dissipation will damp $e_{ps}$. Finally, we consider two orbital inclinations for the moon, $i = 0°$ and $i = 45°$.

---

[11] Note that in planet-moon binaries close to the star Hill stability requires that the satellite's orbit is a few planetary radii at most. Hence, if they exist, moons about planets in the IHZ around M dwarfs will be close to the planet (Barnes & O'Brien 2002; Domingos et al. 2006; Donnison 2010; Weidner & Horne 2010).

### 5.1 Surface irradiation of putative exomoons about Kepler-22b

Orbiting its solar-mass, solar-luminosity ($T_{eff,*} = 5518$K, $R_* = 0.979R_\odot$, $R_\odot$ being the radius of the Sun) host star at a distance of ≈0.85AU, Kepler-22b is in the IHZ of a Sun-like star and has a radius $2.38 \pm 0.13$ times the radius of Earth (Borucki et al. 2012). For the computation of tidal heating on Kepler-22b's moons, we take the parametrization presented in Borucki et al. (2012), and we assume a planetary mass of $25M_\oplus$, consistent with their photometric and radial-velocity follow-up measurements but not yet well constrained by observations. This ambiguity leaves open the question whether Kepler-22b is a terrestrial, gaseous, or transitional object.

In Fig. 11, we show the photon flux, coming both from the absorbed and re-emitted illumination as well as from tidal heating, at the upper atmosphere for an Earth-sized exomoon in various orbital configurations around Kepler-22b. Illumination is averaged over one orbit of the planet-moon binary around the star. In the upper four panels $a_{ps} = 5R_p$ (similar to the Uranus-Miranda semi-major axis), while in the lower four panels $a_{ps} = 20R_p$ (similar to the Saturn-Titan semi-major axis). The left column shows co-orbital simulations ($i = 0°$); in the right column $i = 45°$. In the first and the third line $e_{ps} = 0.001$; in the second and fourth line $e_{ps} = 0.05$.

In orbits closer than ≈5$R_p$ even very small eccentricities induce strong tidal heating of exomoons around Kepler-22b. For $e_{ps} = 0.001$ (upper row), a surface heating flux of roughly 6250W/m² should induce surface temperatures well above the surface temperatures of Venus, and for $e_{ps} = 0.05$ (second row), tidal heating is beyond $10^7$W/m², probably melting the whole hypothetical moon (Léger et al. 2011). For orbital distances of 20$R_p$, tidal heating is 0.017W/m² for the low-eccentricity scenario; thus the total flux is determined by stellar irradiation (third line). However, tidal heating is significant for the $e_{ps} = 0.05$ case (lower line), namely, roughly 42W/m².

Non-inclined orbits induce strong variations of irradiation over latitude (left column), while for high inclinations, seasons smooth the distribution (right column). As explained in Section 3.1.4, the subplanetary point for co-planar orbits is slightly cooler than the maximum temperature due the eclipses behind the planet once per orbit. But for tilted orbits, the subplanetary point becomes the warmest spot.

We apply the tidal model presented in Heller et al. (2011b) to compute the planet's tilt erosion time $t_{ero}$ and assess whether its primordial obliquity $\psi_p$ could still persist today. Due to its weakly constrained mass, the value of the planet's tidal quality factor $Q_p$ is subject to huge uncertainties. Using a stellar mass of $1M_\odot$ and trying three values $Q_p = 10^2$, $10^3$ and $10^4$, we find $t_{ero} = 0.5$Gyr, 5Gyr, and 50Gyr, respectively. The lowest $Q_p$ value is similar to that of Earth, while the highest value corresponds approximately to that of Neptune.[12] Thus, if Kepler-22b turns out mostly gaseous and provided that it had a significant primordial obliquity, the planet and its satellites can experience seasons today. But if Kepler-22b is terrestrial and planet-planet perturbations in this system can be neglected, it will have no seasons; and if its moons orbit above the planet's equator, they would share this tilt erosion.

### 5.2 Surface irradiation of putative exomoons about KOI211.01

KOI211.01 is a Saturn- to Jupiter-class planet candidate (Borucki et al. 2011). In the following, we consider it as a planet. Its radius corresponds to $0.88R_J$, and it has an orbital period of 372.11d around a 6072K main-sequence host star, which yields an estimate for the stellar mass and then a semi-major axis of 1.05AU. The stellar radius is $1.09R_\odot$, and using models for planet evolution (Fortney et al. 2007), we estimate the planet's mass to be $0.3M_J$. This value is subject to various uncertainties because little is known about the planet's composition, the mass of a putative core, the planet's atmospheric opacity and structure, and the age of the system. Thus, our investigations will serve as a case study rather than a detailed prediction of exomoon scenarios around this particular planet. Besides KOI211.01, some ten gas giants have been confirmed in the IHZ of their host stars, all of which are not transiting. Thus, detection of their putative moons will not be feasible in the foreseeable future.

In Fig. 12, we present the flux distribution on our prototype Earth-like exomoon around KOI211.01 in the same orbital configurations as in the previous subsection. Tidal heating in orbits with $a_{ps} \lesssim 5R_p$ can be strong, depending on eccentricity, and make the moon uninhabitable. For $e_{ps} = 0.001$, tidal heating is of the order of $1 \text{W/m}^2$ and thereby almost negligible for the moderate flux distribution (first line). However, it increases to over $3200 \text{W/m}^2$ for $e_{ps} = 0.05$ and makes the moon uninhabitable (second line). Tidal heating is negligible at semi-major axes $\geq 20 R_p$ even for a significant eccentricity of 0.05.

---

[12] See Gavrilov & Zharkov (1977) and Heller et al. (2010) for discussions of $Q$ values and Love numbers for gaseous substellar objects.

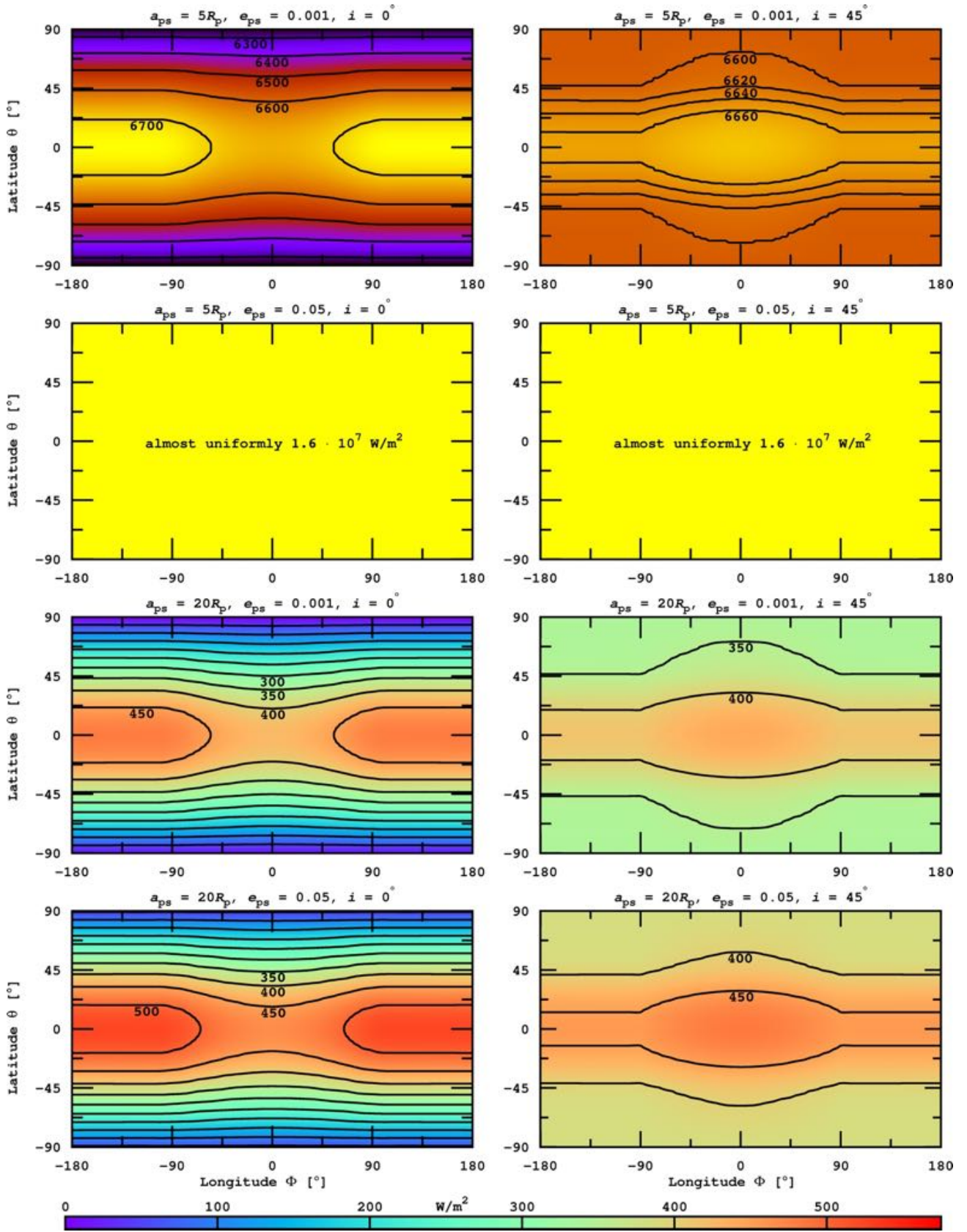

**FIG. 11.** Orbit-averaged flux (in units of W/m²) at the top of an Earth-sized exomoon's atmosphere around Kepler-22b for eight different orbital configurations. Computations include irradiation from the star and the planet as well as tidal heating. The color bar refers only to the lower two rows with moderate flux.

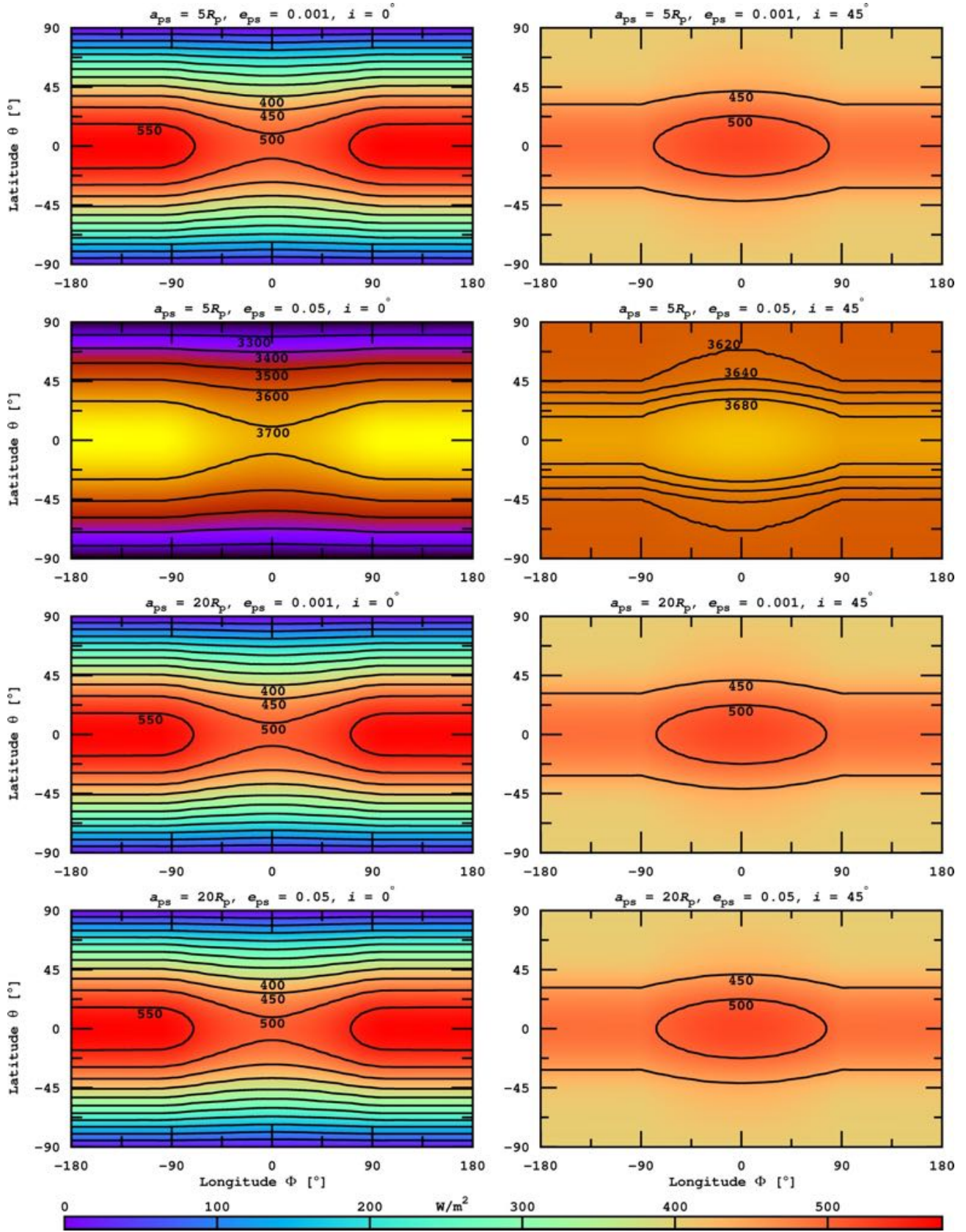

**FIG. 12.** Orbit-averaged flux (in units of W/m²) at the top of an Earth-sized exomoon's atmosphere around KOI211.01 for eight different orbital configurations. Computations include irradiation from the star and the planet as well as tidal heating. The color bar refers only to the first and the lower two rows with moderate flux.

By comparison of Fig. 11 with Fig. 12, we see that in those cases where tidal heating can be neglected and stellar irradiation dominates the energy flux, namely, in the third line of both figures, irradiation is higher on comparable moons around KOI211.01. But in the lower panel line, the total flux on our Kepler-22b prototype satellite becomes comparable to the one around KOI211.01, which is because of the extra heat from tidal dissipation. Surface flux distributions in the low-eccentricity and the moderate-eccentricity state on our KOI211.01 exomoons are virtually the same at $a_{ps} = 20 R_p$.

Since KOI211.01 is a gaseous object, we apply a tidal quality factor of $Q_p = 10^5$, which is similar to, but still lower than, the tidal response of Jupiter. We find that $t_{ero}$ of KOI211.01 is much higher than the age of the Universe. Thus, satellites of KOI211.01 will experience seasons, provided the planet had a primordial obliquity and planet-planet perturbations can be neglected.

The radius of KOI211.01 is about 4 times greater than the radius of Kepler-22b; thus moons at a certain multitude of planetary radii distance from the planet, say $5 R_p$ or $20 R_p$ as we considered, will be effectively much farther away from KOI211.01 than from Kepler-22b. As tidal heating strongly depends on $a_{ps}$ and on the planet's mass, a quick comparison between tidal heating on exomoons around Kepler-22b and KOI211.01 can be helpful. For equal eccentricities, the fraction of tidal heating in two moons about Kepler-22b and KOI211.01 will be equal to a fraction of $Z_s$ from Eq. (20). By taking the planet masses assumed above, thus $M_{KOI} \approx 3.81 M_{Ke}$, and assuming that the satellite mass is much smaller than the planets' masses, respectively, we deduce

$$\frac{Z_s^{\text{Ke}}}{Z_s^{\text{KOI}}} = \frac{M_{Ke}^2}{M_{KOI}^2} \frac{(M_{Ke} + M_s)}{(M_{KIO} + M_s)} \left(\frac{a_{ps}^{\text{KOI}}}{a_{ps}^{\text{Ke}}}\right)^9 \overset{M_s \text{ negligible}}{\approx} \frac{4^9}{3.81^3} \approx 5000 \ , \quad (23)$$

where indices 'Ke' and 'KOI' refer to Kepler-22b and KOI211.01, respectively. The translation of tidal heating from any panel in Fig. 11 to the corresponding panel in Fig. 12 can be done with a division by this factor.

### 5.3 Orbits of habitable exomoons

We now set our results in context with observables to obtain a first estimate for the magnitude of the TTV amplitude induced by habitable moons around Kepler-22b and KOI211.01. For the time being, we will neglect the actual detectability of such signals with *Kepler* but discuss it in Section 6 (see also Kipping et al. 2009).

To begin with, we apply our method from Section 4 and add the orbit-averaged stellar irradiation on the moon to the averaged stellar-reflected light, the averaged thermal irradiation from the planet, and the tidal heating (in the CTL theory). We then compare their sum $\bar{F}_s^{\text{glob}}$ (see Eq. 22) to the critical flux for a runaway greenhouse $F_{RG}$ on the respective moon. In Fig. 13, we show the limiting orbits ($\bar{F}_s^{\text{glob}} = F_{RG}$) for a runaway greenhouse of an Earth-like (upper row) and a Super-Ganymede (lower row) exomoon around Kepler-22b (left column) and KOI211.01 (right column). For both moons, we consider two albedos ($\alpha_s = 0.3$, gray solid lines; and $\alpha_s = 0.4$, black solid lines) and three orbital eccentricities ($e_{ps} = 0.001$, 0.01, 0.1). Solid lines define the habitable edge as described in Section 4. Moons in orbits to the left of a habitable edge are uninhabitable, respectively, because the sum of stellar and planetary irradiation plus tidal heating exceeds the runaway greenhouse limit. Dashed lines correspond to TTV amplitudes and will be discussed below.

#### 5.3.1 Orbit-averaged global flux

Let us first consider the three black solid lines in the upper left panel, corresponding to an Earth-like satellite with $\alpha_s = 0.4$ about Kepler-22b. Each contour represents a habitable edge for a certain orbital eccentricity of the moon; that is, in these orbits the average energy flux on the moon is equal to $F_{RG} = 295 \text{W/m}^2$. Assuming a circular orbit around the star ($e_{*p} = 0$), the orbit-averaged stellar flux absorbed by a moon with $\alpha_s = 0.4$ is

$$\frac{L_*(1 - \alpha_s)}{16 \pi a_{*p}^2} \approx 227 \text{ W/m}^2 \ , \qquad (29)$$

using the parametrization presented in Section 5.1. Thus, black contours indicate an additional heating of $\approx 68 \text{W/m}^2$. The indicated eccentricities increase from left to right. This is mainly due to tidal heating, which increases for larger eccentrici-ties but decreases at larger separations $a_{ps}$. Thus, the larger $e_{ps}$, the farther away from the planet the moon needs to be to avoid becoming a greenhouse. If a moon is close to the planet and shall be habitable, then its eccentricity must be small enough.

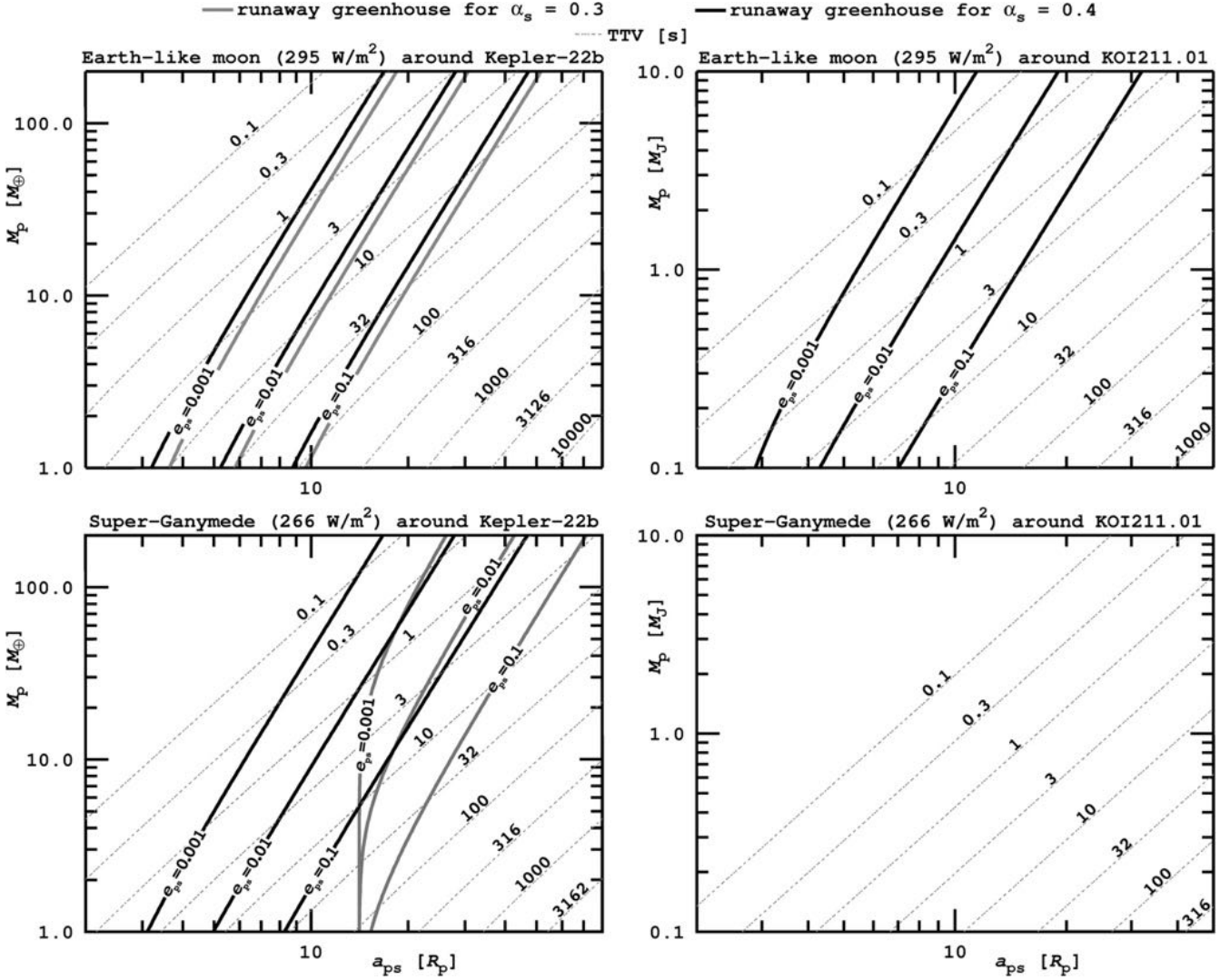

**FIG. 13.** Habitable edges for an Earth-like moon (upper row) and a Super-Ganymede (lower row) exomoon in orbit around Kepler-22b (left column) and KOI211.01 (right column). Masses of both host planets are not well constrained; thus abscissae run over several decades (in units of $M_\oplus$ for Kepler-22b and $M_J$ for KOI211.01). We consider two albedos $\alpha_s = 0.3$ (gray solid lines) and $\alpha_s = 0.4$ (black solid lines) and three eccentricities ($e_{ps} = 0.001, 0.01, 0.1$) for both moons. No gray solid lines are in the right column because both prototype moons would not be habitable with $\alpha_s = 0.3$ around KOI211.01. TTV amplitudes (in units of seconds) for coplanar orbits are plotted with dashed lines.

Next, we compare the black lines to the gray lines, which assume an albedo of 0.3 for an Earth-like satellite. With this albedo, it absorbs ≈265W/m² of stellar irradiation. Consequently, a smaller amount of additional heat is required to push it into a runaway greenhouse, so the corresponding habitable edges are farther away from the planet, where both tidal heating and irradiation from the planet are lower.

In the lower left panel, we consider a Super-Ganymede around Kepler-22b. Its critical flux of 266W/m² is smaller than for an Earth-like moon, so for an albedo of 0.4 only 39W/m² of additional heating is required for the moon to turn into a runaway greenhouse. Yet, compared to the Earth-like satellite in the upper left panel, the habitable edges (black lines) are still slightly closer to the planet, because tidal heating in the smaller Super-Ganymede is much weaker at a given orbital-semi major axis. More important, gray lines are at much larger distances in this panel because with an albedo of 0.3 the satellite's absorbed stellar flux (265W/m²) is almost as high as the critical flux (266W/m²). This is irrespective of whether our Ganymede-like object is orbiting a planet.

Next, we consider moons in orbit about KOI211.01, shown in the right column of Fig. 13. First note the absence of gray solid lines. For moons with an albedo of $\alpha_s = 0.3$, the orbit-averaged stellar irradiation will be 315 W/m², with the parametrization presented in Sect 5.2. This means that stellar irradiation alone is larger than the critical flux of both our Earth-like (295 W/m², upper panel) and Super-Ganymede (266 W/m², lower panel) prototype moons. Thus, both moons with $\alpha_s = 0.3$ around KOI211.01 are not habitable, irrespective of their distance to the planet.

Moons with $\alpha_s = 0.4$ around KOI211.01 absorb 270 W/m², which is less than the critical flux for the Earth-like moon (upper right panel) but more than the Super-Ganymede moon could bear (lower right panel). Thus, the black lines are absent from the latter plot but show up in the upper right. Contours for given eccentricities are closer to the planet than their counterparts in the case of Kepler-22b. This is because we plot the semi-major axis in units of planetary radii – while KOI211.01 has a much larger radius than Kepler-22b – and tidal heating, which strongly depends on the *absolute* distance between the planet and the moon.

### 5.3.2 Transit timing variations of habitable exomoons

Now we compare the habitable edges as defined by Eq. (22) to the amplitude of the TTV induced by the moon on the planet. TTV amplitudes $\Delta$ in units of seconds are plotted with dashed lines in Fig. 13. Recall, however, that $\Delta$ would not directly be measured by transit observations. Rather the root-mean-square of the TTV wave would be observed (Kipping 2009a). We apply the Kipping et al. (2012) equations, assuming that the moon's orbit is circular and that both the circumstellar and the circumplanetary orbit are seen edge-on from Earth.

Each panel of Fig. 13 shows how the TTV amplitude decreases with decreasing semi-major axis of the satellite (from right to left) and with increasing planetary mass (from bottom to top). Thus, for a given planetary mass, a moon could be habitable if its TTV signal is sufficiently large. Comparison of the upper and the lower panels in each column shows how much larger the TTV amplitude of an Earth-like moon is with respect to our Super-Ganymede moon.

Corrections due to an inclination of the moon's orbit and an accidental alignment of the moon's longitude of the periapses are not included but would be small in most cases. For non-zero inclinations, the TTV amplitude $\Delta$ will be decreased. This decrease is proportional to $\sqrt{1 - \cos(i_s)^2 \sin(\varpi_s)^2}$, where $i_s$ is measured from the circumstellar orbital plane normal to the orbital plane of the planet-moon orbit, and $\varpi_s$ denotes the orientation of the longitude of the periapses (see Eq. (6.46) in Kipping 2011b). Thus, corrections to our picture will only be relevant if both the moon's orbit is significantly tilted and $\varpi_s \approx 90°$. In the worst case of $\varpi_s \approx 90°$, an inclination of 25° reduces $\Delta$ by <10%, while for $\varpi_s \approx 45°$, $i_s$ could be as large as 40° to produce a similar correction. Simulations of the planet's and the satellite's tilt erosion can help assess whether substantial misalignments are likely (Heller et al. 2011b). While such geometric blurring should be small for most systems, prediction of the individual TTV signal for a satellite in an exoplanet system – as we suggest in Fig. 13 – is hard because perturbations of other planets, moons, or Trojans could affect the TTV.

## 6. Summary and discussion

Our work yields the first translation from observables to exomoon habitability. Using a scaling relation for the onset of the runaway greenhouse effect, we have deduced constraints on exomoon habitability from stellar and planetary irradiation as well as from tidal heating. We determined the orbit-averaged global energy budget $\bar{F}_s^{\text{glob}}$ for exomoons to avoid a runaway greenhouse and found that for a well parametrized system of a star, a host planet, and a moon, Eq. (22) can be used to evaluate the habitability of a moon. By analogy with the circumstellar habitable zone (Kasting et al. 1993), these rules define a circumplanetary "habitable edge". To be habitable, moons must orbit their planets outside the habitable edge.

Application of our illumination plus tidal heating model shows that an Earth-sized exomoon about Kepler-22b with a bond albedo of 0.3 or higher would be habitable if (*i.*) the planet's mass is $\approx 10 M_\oplus$, (*ii.*) the satellite would orbit Kepler-22b with a semi-major axis $\gtrsim 10 R_p$, and (*iii.*) the moon's orbital eccentricity $e_{ps}$ would be <0.01 (see Fig. 13). If Kepler-22b turns out more massive, then such a putative moon would need to be farther away to be habitable. Super-Ganymede satellites of Kepler-22b meet similar requirements, but beyond that their bond albedo needs to be $\gtrsim 0.4$.

Super-Ganymede or smaller moons around KOI211.01 will not be habitable, since their critical flux for a runaway greenhouse effect ($\lesssim 266$ W/m²) is less than the orbit-averaged irradiation received by the star (270 W/m²). But Earth-like or more massive moons can be habitable if (*i.*) the planet's mass $\approx M_J$, (*ii.*) the satellite's bond albedo $\gtrsim 0.4$, (*iii.*) the satellite orbits KOI211.01 with a semi-major axis $\gtrsim 10 R_p$, and (*iv.*) the moon's orbital eccentricity $e_{ps} < 0.01$. If the planet turns out less massive, then its moons could be closer and have higher eccentricities and still be habitable.

Kepler-22 and KOI211 both are mid-G-type stars, with *Kepler* magnitudes 11.664 (Borucki et al. 2012) and 14.99 (Borucki et al. 2011), respectively. As shown by Kipping et al. (2009), the maximum *Kepler* magnitude to allow for the detection of an Earth-like moon is about 12.5. We conclude that such moons around Kepler-22b are detectable if they exist, and their habitability could be evaluated with the methods provided in this communication. Moons around KOI211.01 will not be detectable within the 7 year duty cycle of *Kepler*. Nevertheless, our investigation of this giant planet's putative moons serves as a case study for comparably massive planets in the IHZ of their parent stars.

Stellar flux on potentially habitable exomoons is much stronger than the contribution from the planet. Nevertheless, the sum of thermal emission and stellar reflected light from the planet can have a significant impact on exomoon climates. Stellar reflection dominates over thermal emission as long as the planet's bond albedo $\alpha_p \gtrsim 0.1$ and can reach some 10 or even 100W/m$^2$ once the moon is in a close orbit ($a_{ps} \lesssim 10R_p$) and the planet has a high albedo ($\alpha_p \gtrsim 0.3$) (Section 3.1.2). Precise values depend on stellar luminosity and the distance of the planet-moon duet from the star. Our calculations for Kepler-22b and KOI211.01 show that the limiting orbits for exomoons to be habitable are very sensitive to the satellite's albedo.

Due to the weak tides from its host star, KOI211.01 can still have a significant obliquity, and if its moons orbit the planet in the equatorial plane they could have seasons and are more likely to be discovered by transit duration variations of the transit impact parameter (TDV-TIP, Kipping 2009b). For Kepler-22b, the issue of tilt erosion cannot be answered unambiguously until more about the planet's mass and composition is known.

If a moon's orbital inclination is small enough, then it will be in the shadow of the planet for a certain time once per planet-moon orbit (see also Heller 2012). For low inclinations, eclipses can occur about once every revolution of the moon around the planet, preferentially when the subplanetary hemisphere on the moon would experience stellar irradiation maximum (depending on $e_{*p}$). Eclipses have a profound impact on the surface distribution of the moon's irradiation. For low inclinations, the subplanetary point on the moon will be the "coldest" location along the equator, whereas for moderate inclinations it will be the "warmest" spot on the moon due to the additional irradiation from the planet. Future investigations will clarify whether this may result in enhanced sub*planetary* weathering instabilities, that is, runaway CO$_2$ drawdown rates eventually leading to very strong greenhouse forcing, or sub*planetary* dissolution feedbacks of volatiles in sub*planetary* oceans, as has been proposed for exoplanets that are tidally locked to their host stars and thus experience such effects at the fixed sub*stellar* point (Kite et al. 2011).

We predict seasonal illumination phenomena on the moon, which emerge from the circumstellar season and planetary illumination. They depend on the location on the satellite and appear in four versions, which we call the "proplanetary summer", "proplanetary winter", "antiplanetary summer", and "antiplanetary winter". The former two describe seasons due to the moon's obliquity with respect to the star with an additional illumination from the planet; the latter two depict the permanent absence of planetary illumination during the seasons.

For massive exomoons with $a_{ps} \lesssim 10R_p$ around Kepler-22b and around KOI211.01, tidal heating can be immense, presumably making them uninhabitable if the orbits are substantially non-circular. On the one hand, tidal heating can be a threat to life on exomoons, in particular when they are in close orbits with significant eccentricities around their planets. If the planet-moon duet is at the inner edge of the circumstellar IHZ, small contributions of tidal heat can render an exomoon uninhabitable. Tidal heating can also induce a thermal runaway, producing intense magmatism and rapid resurfacing on the moon (Běhounková et al. 2011). On the other hand, we can imagine scenarios where a moon becomes habitable only because of tidal heating. If the host planet has an obliquity similar to Uranus, then one polar region will not be illuminated for half the orbit around the star. Moderate tidal heating of some tens of watts per square meter might be just adequate to prevent the atmosphere from freezing out. Or if the planet and its moon orbit their host star somewhat beyond the outer edge of the IHZ, then tidal heating might be necessary to make the moon habitable in the first place. Tidal heating could also drive long-lived plate tectonics, thereby enhancing the moon's habitability (Jackson et al. 2008). An example is given by Jupiter's moon Europa, where insolation is weak but tides provide enough heat to sustain a subsurface ocean of liquid water (Greenberg et al. 1998; Schmidt et al. 2011). On the downside, too much tidal heating can render the body uninhabitable due to enhanced volcanic activity, as it is observed on Io.

Tidal heating has a strong dependence on the moon's eccentricity. Eccentricities of exomoons will hardly be measurable even with telescopes available in the next decade, but it will be possible to constrain $e_{ps}$ by simulations. Therefore, once exomoon systems are discovered, it will be necessary to search for further moons around the same planet to consistently simulate the $N$-body ($N > 2$) evolution with multiple-moon interaction, gravitational perturbations from other planets, and the gravitational effects of the star. Such simulations will also be necessary to simulate the long-term evolution of the orientation $\eta$ of the moon's inclination $i$ with respect to the periastron of the star-related orbit, because for signifiant eccentricities $e_{*p}$ it will make a big difference whether the summer of either the northern or southern hemisphere coincides with minimum or maximum distance to the star. Technically, these variations refer to the apsidal precession (the orientation of the star-related eccentricity, i.e., of $a_{*p}$) and the precession of the planet's rotation axis, both of which determine $\eta$. For Saturn, and thus Titan, the corresponding time scale is of order 1Myr (French et al. 1993), mainly induced by solar torques on both Saturn's oblate figure and the equatorial satellites. Habitable exomoons might preferably be irregular satellites (see Section 2) for which Carruba et al. (2002) showed that their orbital parameters are subject to particularly rapid changes, driven by stellar perturbations.

We find that more massive moons can orbit more closely to the planet and be prevented from becoming a runaway greenhouse (Section 2.2). This purely atmospheric effect is shared by all terrestrial bodies. Similar to the circumstellar habitable zone of extrasolar planets (Kasting et al. 1993), we conclude that more massive exomoons may have somewhat wider habitable zones around their host planets – of which the inner boundary is defined by the habitable edge and the outer boundary by Hill stability – than do less massive satellites. In future investigations, it will be necessary to include simulations of the moons' putative atmospheres and their responses to irradiation and tidal heating. Thus, our irradiation plus tidal heating model should be coupled to an energy balance or global climate model to allow for more realistic descriptions of exomoon habitability. As indicated by our basal considerations, the impact of eclipses and planetary irradiation on exomoon climates can be substantial. In addition to the orbital parameters which we have simulated here, the moons' climates will depend on a myriad bodily characteristics

Spectroscopic signatures of life, so-called "biosignatures", in the atmospheres of inhabited exomoons will only be detectable with next-generation, several-meter-class space telescopes (Kaltenegger 2010; Kipping et al. 2010). Until then, we may primarily use our knowledge about the orbital configurations and composition of those worlds when assessing their habitability. Our method allows for an evaluation of exomoon habitability based on the data available at the time they will be discovered. The recent detection of an Earth-sized and a sub-Earth-sized planet around a G-type star (Fressin et al. 2012) suggests that not only the moons' masses and semi-major axes around their planets can be measured (e.g., by combined TTV and TDV, Kipping 2009a) but also their radii by direct photometry. A combination of these techniques might finally pin down the moon's inclination (Kipping 2009b) and thus allow for precise modeling of its habitability based on the model presented here.

Results of ESA's *Jupiter Icy Moons Explorer* ("*JUICE*") will be of great value for characterization of exomoons. With launch in 2022 and arrival at Jupiter in 2030, one of the mission's two key goals will be to explore Ganymede, Europa, and Callisto as possible habitats. Therefore, the probe will acquire precise measurements of their topographic distortions due to tides on a centimeter level; determine their dynamical rotation states (i.e., forced libration and nutation); characterize their surface chemistry; and study their cores, rocky mantles, and icy shells. The search for water reservoirs on Europa, exploration of Ganymede's magnetic field, and monitoring of Io's volcanic activity will deliver fundamentally new insights into the planetology of massive moons.

Although our assumptions about the moons' orbital characteristics are moderate, that is, they are taken from the parameter space mainly occupied by the most massive satellites in the Solar System, our results imply that exomoons might exist in various habitable or extremely tidally heated configurations. We conclude that the advent of exomoon observations and characterization will permit new insights into planetary physics and reveal so far unknown phenomena, analogous to the staggering impact of the first exoplanet observations 17 years ago. If observers feel animated to use the available *Kepler* data, the *Hubble Space Telescope,* or meter-sized ground-based instruments to search for evidence of exomoons, then one aim of this communication has been achieved.

## Appendix

### Appendix A: Stellar irradiation

We include here a thorough explanation for the stellar flux $f_*(t)$ presented in Section 3.1.1. To begin with, we assume that the irradiation on the moon at a longitude $\phi$ and latitude $\theta$ will be

$$f_*(t) = \frac{L_*}{4\pi \, \overrightarrow{r}_{\mathrm{s*}}(t)^2} \, \frac{\overrightarrow{r}_{\mathrm{s*}}(t)}{r_{\mathrm{s*}}(t)} \, \frac{\overrightarrow{n}_{\phi,\theta}(t)}{n_{\phi,\theta}(t)} \ , \qquad\qquad\qquad \text{(A.1)}$$

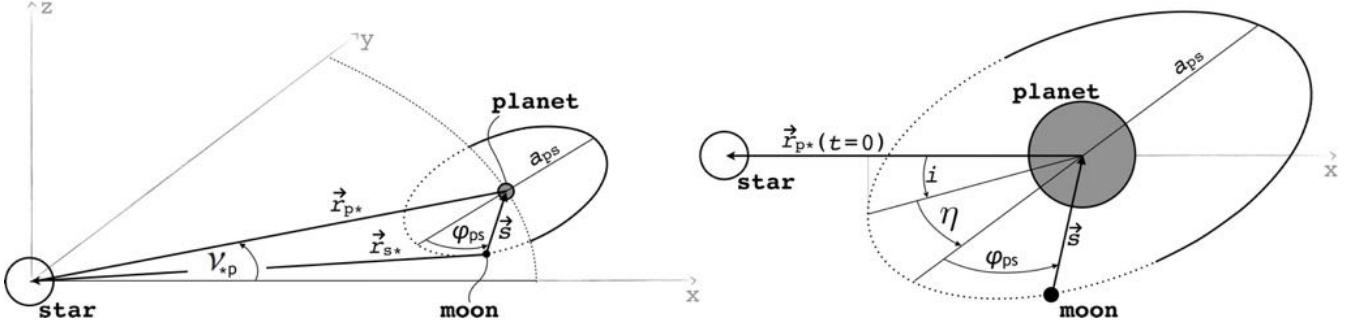

**FIG. A.1.** Geometry of the triple system of a star, a planet, and a moon. In the left panel the planet-moon duet has advanced by an angle $\nu_{*p}$ around the star, and the moon has progressed by an angle $2\pi\varphi_{ps}$. The right panel shows a zoom-in to the planet-moon binary. As in the left panel, time has proceeded, and a projection of $\vec{r}_{p*}(t)$ at time $t = 0$ has been included to explain the orientation of the moon's orbit, which is inclined by an angle $i$ and rotated against the star-moon periapses the by an angle $\eta$.

with $L_*$ as the stellar luminosity and $\vec{r}_{s*}(t)$ as the vector from the moon to the star (see Fig. A.1). The product of the surface normal $\vec{n}_{\phi,\theta}/n_{\phi,\theta}$ on the moon and $\vec{r}_{s*}(t)/r_{s*}(t)$ accounts for projection effects on the location $(\phi,\theta)$, and we set $f_*$ to zero in those cases where the star shines from the back. Figure A.1 shows that $\vec{r}_{s*}(t) = \vec{r}_{p*}(t) + \vec{s}(t)$. While the Keplerian motion, encapsulated in $\vec{r}_{p*}(t)$, is deduced in Section 3.1.1, we focus here on the moon's surface normal $\vec{n}_{\phi,\theta}$.

In the right panel of Fig. A.1, we show a close-up of the star-planet-moon geometry at $t = 0$, which corresponds to the initial configuration of the orbit evolution. The planet is at periastron ($\varphi_{*p} = 0$) and for the case where $\eta = 0$ the vector $\vec{s} = \vec{n}_{0,0}(t)$ from the subplanetary point on the moon to the planet would have the form

$$
a_{\text{ps}}
\begin{pmatrix}
\cos\left(2\pi\dfrac{(t-\tau)}{P_{\text{ps}}}\right)\cos\left(i\dfrac{\pi}{180°}\right) \\[2mm]
\sin\left(2\pi\dfrac{(t-\tau)}{P_{\text{ps}}}\right) \\[2mm]
\cos\left(2\pi\dfrac{(t-\tau)}{P_{\text{ps}}}\right)\sin\left(i\dfrac{\pi}{180°}\right)
\end{pmatrix},
\tag{A.2}
$$

where all angles are provided in degrees. The angle $\eta$ depicts the orientation of the lowest point of the moon's orbit with respect to the projection of $\vec{r}_{p*}$ on the moon's orbital plane at $t = 0$. At summer solstice on the moon's nothern hemisphere, the true anomaly $\nu_{*p}$ equals $\eta$ (see Eq. B.8). This makes $\eta$ a critical parameter for the seasonal variation of stellar irradiation on the moon because it determines how seasons, induced by orbital inclination $i$, relate to the changing distance to the star, induced by star-planet eccentricity $e_{*p}$. In particular, if $\eta = 0$ then northern summer coincides with the periastron passage about the star and nothern winter occurs at apastron, inducing distinctly hot summers and cold winters. It relates to the conventional orientation of the ascending node $\Omega$ as $\eta = \Omega + 270°$ and can be considered as the climate-precession parameter.

For $\eta \neq 0$, we have to apply a rotation $M(\eta)\colon \mathbb{R}^3 \to \mathbb{R}^3$ of $\vec{s}$ around the $z$-axis $(0,0,1)$, which is performed by the rotation matrix

$$
M(\eta) =
\begin{pmatrix}
\cos(\eta) & -\sin(\eta) & 0 \\[2mm]
\sin(\eta) & \cos(\eta) & 0 \\[2mm]
0 & 0 & 1
\end{pmatrix}.
\tag{A.3}
$$

With the abbreviations introduced in Eq. (6) (with $\phi = 0$ for the time being), we obtain

$$
\vec{s}(t) \equiv \vec{n}_{0,0}(t) = a_{\text{ps}}
\begin{pmatrix}
\tilde{C}cC - \tilde{S}s \\[2mm]
\tilde{S}cC - \tilde{C}s \\[2mm]
cS
\end{pmatrix}.
\tag{A.4}
$$

This allows us to parametrize the surface normal $\vec{s}(t)/s(t)$ of the subplanetary point on the moon for arbitrary $i$ and $\eta$.

Finally, we want to find the surface normal $\vec{n}_{\phi,\theta}(t)/n_{\phi,\theta}(t)$ for *any* location $(\phi,\theta)$ on the moon's surface. Therefore, we have to go two steps on the moon's surface, each one parametrized by one angle: one in longitudinal direction $\phi$ and one in latitudinal direction $\theta$. The former one can be walked easily, just by adding $2\pi\phi/360°$ in the sine and cosine arguments in Eq. (A.2) (see first line in Eq. 6), thus $\vec{n}_{\phi,0} = \vec{s}(\varphi_{ps} + 2\pi\phi/360°)$. This is equivalent to a shift along the moon's equator. For the second step, we know that if $\theta = 90°$, then the surface normal will be along the rotation axis

$$\vec{N} = a_{ps} \begin{pmatrix} -S\tilde{C} \\ -S\tilde{S} \\ C \end{pmatrix} \qquad (A.5)$$

of the satellite; that is, we are standing on the north pole. The vector $\vec{n}_{\phi,\theta}$ can then be obtained by tilting $\vec{s}(\varphi_{ps} + 2\pi\phi/360°)$ by an angle $\theta$ toward $\vec{N}$. With $N = a_{ps} = s$, we then derive

$$\vec{n}_{\phi,\theta}(t) = a_{ps}\sin(\theta)\frac{\vec{N}}{N} + a_{ps}\cos(\theta)\frac{\vec{s}(\varphi_{ps} + 2\pi\phi/360°)}{s(\varphi_{ps} + 2\pi\phi/360°)} = \sin(\theta)\vec{N} + \cos(\theta)\vec{s}(\varphi_{ps} + 2\pi\phi/360°)$$

$$= a_{ps} \begin{pmatrix} -\bar{s}S\tilde{C} + \bar{c}(\tilde{C}cC - \tilde{S}s) \\ -\bar{s}S\tilde{S} + \bar{c}(\tilde{S}cC - \tilde{C}s) \\ \bar{s}C + \bar{c}cS \end{pmatrix} , \qquad (A.6)$$

which solves Eq. (A.1).

## Appendix B: Planetary irradiation

In addition to stellar light, the moon will receive thermal and stellar-reflected irradiation from the planet, $f_t(t)$ and $f_r(t)$, respectively. One hemisphere on the planet will be illuminated by the star and the other one will be dark. On the bright side of the planet, there will be both thermal emission from the planet as well as starlight reflection, while on the dark side the planet will only emit thermal radiation, though with a lower intensity than on the bright side due to the lower temperature.

We begin with the thermal part. The planet's total thermal luminosity $L_{th,p}$ will be the sum of the radiation from the bright side and from the dark side. On the bright side, the planet shall have a uniform temperature $T_{eff,p}^b$, and on the dark side its temperature shall be $T_{eff,p}^d$. Then thermal equilibrium between outgoing thermal radiation and incoming stellar radiation yields

$$L_{th,p} = 2\pi R_p^2 \sigma_{SB}\left((T_{eff,p}^b)^4 + (T_{eff,p}^d)^4\right)$$
$$= \pi R_p^2(1 - \alpha_p)\frac{4\pi R_*^2 \sigma_{SB} T_{eff,*}^4}{4\pi \bar{r}_{p*}^2} + W_p , \qquad (B.1)$$

where the first term in the second line describes the absorbed radiation from the star and the second term ($W_p$) can be any additional heat source, for example, the energy released by the gravitation-induced shrinking of the gaseous planet. For Jupiter, Saturn, and Neptune, which orbit the Sun at distances >5AU, $W_p$ is greater than the incoming radiation. However, for gaseous planets in the IHZ it will be negligible once the planet has reached an age $\gtrsim 100$Myr (Baraffe et al. 2003).[13] Owed to the negligibility for our purpose and for simplicity we set $W_p = 0$.

In our model, we want to parametrize the two hemispheres by a temperature difference $dT \equiv T_{eff,p}^b - T_{eff,p}^d$. We define

$$p(T_{eff,p}^b) \equiv (T_{eff,p}^b)^4 + (T_{eff,p}^b - dT)^4 - T_{eff,*}^4 \frac{(1 - \alpha_p)R_*^2}{2\bar{r}_{*p}^2} = 0 \qquad (B.2)$$

---

[13] Note that Barnes et al. (2013) identified 100Myr as the time required for a runaway greenhouse to sterilize a planet.

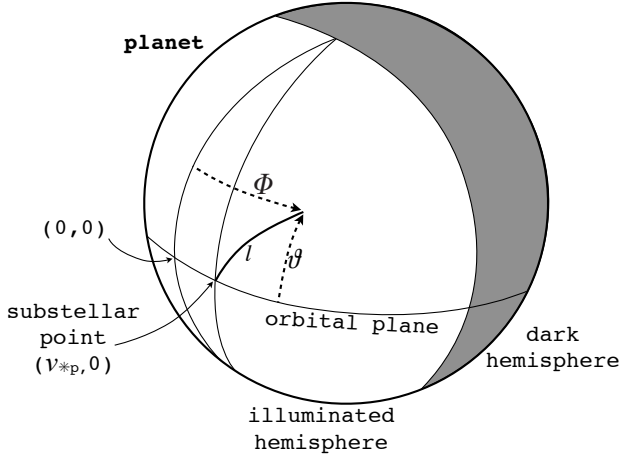

**FIG. B.1.** Geometry of the planetary illumination. The sub-satellite point on the planetary surface is at $(\Phi, \vartheta)$. The angular distance $l$ of the moon from the substellar point $(\nu_{*p}, 0)$ determines the amount of light received by the moon from the two different hemispheres.

The time dependence of the irradiation is now packed into $\xi(t)$, which serves as a weighting function for the two contributions from the bright and the dark side. It is given by

$$\xi(t) = \frac{1}{2}\left(1 + \cos\left(l(t)\right)\right) , \qquad (B.5)$$

where

$$l(t) = \arccos\left\{\cos\left(\vartheta(t)\right)\cos\left(\Phi(t) - \nu_{*p}(t)\right)\right\} \qquad (B.6)$$

is the angular distance between the moon's projection on the planetary surface and the substellar point on the planet. In other words, $l(t)$ is an orthodrome on the planet's surface, determined by $\Phi(t)$ and $\vartheta(t)$ (see Fig. B.1). This yields

$$\xi(t) = \frac{1}{2}\left\{1 + \cos\left(\vartheta(t)\right)\cos\left(\Phi(t) - \nu_{*p}(t)\right)\right\} . \quad (B.7)$$

Since the subplanetary point lies in the orbital plane of the planet, it will be at a position $(\nu_{*p}(t), 0)$ on the planetary surface, where

$$\nu_{*p}(t) = \arccos\left(\frac{\cos(E_{*p}(t)) - e_{*p}}{1 - e_{*p}\cos(E_{*p}(t))}\right) \qquad (B.8)$$

is the true anomaly. Moreover, with $s_x$, $s_y$, and $s_z$ as the components of $\vec{s} = (s_x, s_y, s_z)$ we have

$$\Phi(t) = 2\arctan\left(\frac{s_y(t)}{\sqrt{s_x^2(t) + s_y^2(t) + s_x(t)}}\right)$$

$$\vartheta(t) = \frac{\pi}{2} - \arccos\left(\frac{s_y(t)}{\sqrt{s_x^2(t) + s_y^2(t) + s_z^2(t)}}\right) \qquad (B.9)$$

and, once d$T$ is given, search for the first zero point of $p(T_{\text{eff,p}}^{\text{b}})$ above

$$T_{\text{eff,p}}^{\text{eq}} = T_{\text{eff},*}\left(\frac{(1-\alpha_{\text{p}})R_*^2}{4\,\vec{r}_{*p}^2}\right)^{1/4} , \qquad (B.3)$$

which yields $T_{\text{eff,p}}^{\text{b}}$ and $T_{\text{eff,p}}^{\text{d}}$. In our prototype exomoon system, we assume d$T = 100$K, which is equivalent to fixing the efficiency of heat redistribution from the bright to the dark hemisphere on the planet (Burrows et al. 2006b; Budaj et al. 2012).

From Eq. (B.1), we can deduce the thermal flux on the subplanetary point on the moon. With increasing angular distance from the subplanetary point, $f_t$ will decrease. We can parametrize this distance on the moon's surface by longitude $\phi$ and latitude $\theta$, which gives

$$f_{\text{t}} = \frac{R_{\text{p}}^2\sigma_{\text{SB}}}{a_{\text{ps}}^2}\cos\left(\frac{\phi\pi}{180°}\right)\cos\left(\frac{\theta\pi}{180°}\right)$$

$$\times\left[(T_{\text{eff,p}}^{\text{b}})^4\xi(t) + (T_{\text{eff,p}}^{\text{d}})^4\left(1 - \xi(t)\right)\right] . \ (B.4)$$

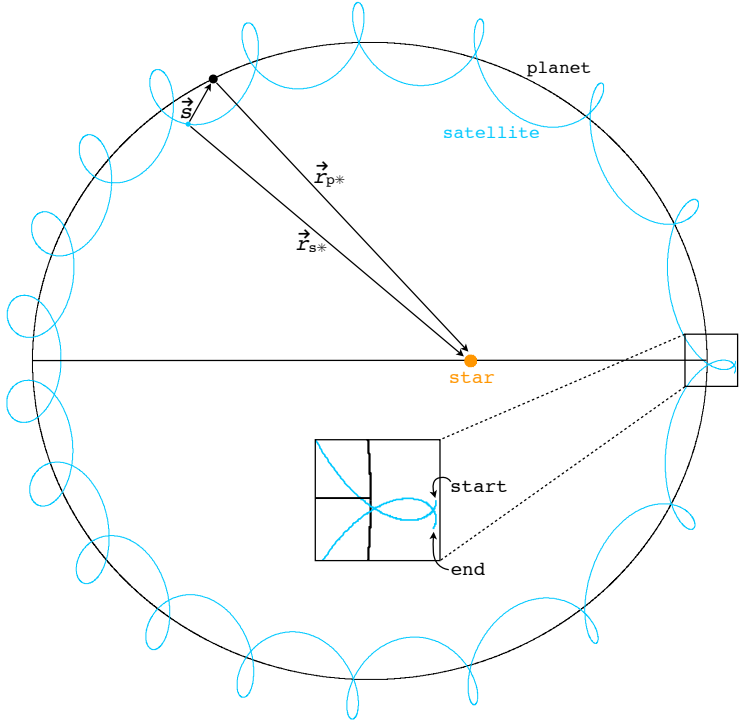

**FIG. C.1.** Orbits of an exomoon and an exoplanet around their common host star as computed with our `exomoon.py` software.

and thus determined $f_t(t)$.

We now consider the reflected stellar light from the planet. For the derivation of $f_r(t)$, we assumed that the planet is divided in two hemispheres, one of which is the bright side and one the dark side. Now the bright part coincides with the hemisphere from which the planet receives stellar reflected light, while there is no contribution to the reflectance from the dark side. Thus, the deduction of the geometrical part of $f_r(t)$, represented by $\xi(t)$, goes analogously to $f_t(t)$. We only have to multiply the stellar flux received by the planet $R_\ast^2 \sigma_{BS} T_{eff,\ast}^4 / r_{p\ast}^2$ with the amount $\pi R_p^2 \alpha_p$ that is reflected from the planet and weigh it with the squared distance decrease $a_{ps}^{-2}$ between the planet and its satellite. Consideration of projectional effects of longitude and latitude yields

$$f_r(t) = \frac{R_\ast^2 \sigma_{SB} T_{eff,\ast}^4}{r_{p\ast}^2} \frac{\pi R_p^2 \alpha_p}{a_{ps}^2} \quad \text{(B.10)}$$
$$\times \cos\left(\frac{\phi\pi}{180°}\right) \cos\left(\frac{\theta\pi}{180°}\right) \xi(t) .$$

In Section 3.1.2, we have compared thermal radiation and stellar reflection from the planet as a function of the planet-moon distance $a_{ps}$ and the planet's albedo $\alpha_p$ (see Fig. 2). To compute the amplitudes of $f_t$ and $f_r$ over the moon's orbit around the planet, we assume that the moon is over the substellar point on the planet where the satellite receives both maximum reflected and thermal radiation from the planet. Then $\xi(t) = 1$ and equating Eq. (B.4) with (B.10) allows us to compute that value of $\alpha_p$ for which the two contributions will be similar. We neglect projectional effects of longitude and latitude and derive

$$\alpha_p = \frac{1}{\pi} \left(\frac{r_{p\ast}}{R_\ast}\right)^2 \left(\frac{T_{eff,p}^b}{T_{eff,\ast}}\right)^4 \bigg|_{f_t = f_r} . \quad \text{(B.11)}$$

For our prototype system this yields $\alpha_p = 0.093$. For higher planetary albedo, stellar reflected light will dominate irradiation on the moon.

## Appendix C: Computer code of our model: `exomoon.py`

Finally, we make the computer code `exomoon.py`, which we set up to calculate the phase curves (Figs. 4 and 5) and surface maps (Figs. 7, 11, and 12), publicly available. It can be downloaded from www.physics.mcmaster.ca/~rheller or requested via email. The code is written in the programming language `python` and optimized for use with `ipython` (Pérez & Granger 2007). Care has been taken to make it easily human-readable, and a downloadable manual is available, so it can be modified by non-expert users. The output format are ascii tables, which can be accessed with `gnuplot` and other plotting software.

In brief, the program has four operation modes, which allow the user to compute (*i.*) phase curves of $f_\ast(\varphi_{ps})$, $f_r(\varphi_{ps})$, and $f(\varphi_{ps})$, (*ii.*) orbit-averaged flux maps of exomoon surfaces, (*iii.*) the orbit of the planet-moon duet around their common host star, and (*iv.*) the runaway greenhouse flux $F_{RG}$. In Fig. C.1, we show an example for such an orbit calculation. The moon's orbit is inclined by 45° against the plane spanned by the planet and the star, and the stellar orbit has an eccentricity $e_{\ast p} = 0.3$. Near periastron (at the right of the plot), where $\mathfrak{M}_{\ast p} = 0$, the orbital velocity of the planet-moon duet is greater than at apastron. This is why the moon's path is more curly at the left, where $\mathfrak{M}_{\ast p} = \pi$. Note that the starting point and the final point of the moon's orbit at the very right of the figure do not coincide! This effect induces TTVs of the planet.

## Acknowledgements

René Heller received funding from the Deutsche Forschungsgemeinschaft (reference number schw536/33-1). Rory Barnes acknowledges support from the NASA Astrobiology Institute's Cooperative agreement No. NNH05ZDA001C that supports the Virtual Planetary Lab, as well as NSF grant AST-1108882. This work has made use of NASA's Astrophysics Data System Bibliographic Services and of Jean Schneider's exoplanet database (www.exoplanet.eu). Computations have been performed with `ipython 0.13` on `python 2.7.2` and figures have been prepared with `gnuplot 4.4` (www.gnuplot.info) as well as with `gimp 2.6` (www.gimp.org). We thank the two anonymous referees for their expert counsel.

## 7.2 Exomoon Habitability Constrained by Energy Flux and Orbital Stability (Heller 2012)

# Exomoon habitability constrained by energy flux and orbital stability

## R. Heller[1]

Leibniz-Institut für Astrophysik Potsdam (AIP), An der Sternwarte 16, 14482 Potsdam, Germany, e-mail: rheller@aip.de

### ABSTRACT

*Context*. Detecting massive satellites that orbit extrasolar planets has now become feasible, which led naturally to questions about the habitability of exomoons. In a previous study we presented constraints on the habitability of moons from stellar and planetary illumination as well as from tidal heating.

*Aims*. Here I refine our model by including the effect of eclipses on the orbit-averaged illumination. I then apply an analytic approximation for the Hill stability of a satellite to identify the range of stellar and planetary masses in which moons can be habitable. Moons in low-mass stellar systems must orbit their planet very closely to remain bounded, which puts them at risk of strong tidal heating.

*Methods*. I first describe the effect of eclipses on the stellar illumination of satellites. Then I calculate the orbit-averaged energy flux, which includes illumination from the planet and tidal heating to parametrize exomoon habitability as a function of stellar mass, planetary mass, and planet-moon orbital eccentricity. The habitability limit is defined by a scaling relation at which a moon loses its water by the runaway greenhouse process. As a working hypothesis, orbital stability is assumed if the moon's orbital period is less than 1/9 of the planet's orbital period.

*Results*. Due to eclipses, a satellite in a close orbit can experience a reduction in orbit-averaged stellar flux by up to about 6 %. The smaller the semi-major axis and the lower the inclination of the moon's orbit, the stronger the reduction. I find a lower mass limit of ≈ 0.2 $M_\odot$ for exomoon host stars that allows a moon to receive an orbit-averaged stellar flux comparable to the Earth's, with which it can also avoid the runaway greenhouse effect. Precise estimates depend on the satellite's orbital eccentricity. Deleterious effects on exomoon habitability may occur up to ≈ 0.5 $M_\odot$ if the satellite's eccentricity is ≳ 0.05.

*Conclusions*. Although the traditional habitable zone lies close to low-mass stars, which allows for many transits of planet-moon binaries within a given observation cycle, resources should not be spent to trace habitable satellites around them. Gravitational perturbations by the close star, another planet, or another satellite induce eccentricities that likely make any moon uninhabitable. Estimates for individual systems require dynamical simulations that include perturbations among all bodies and tidal heating in the satellite.

**Key words.** Astrobiology – Planets and satellites: general – Eclipses – Celestial mechanics – Stars: low-mass

## 1. Introduction

The detection of dozens of Super-Earths and Jupiter-mass planets in the stellar habitable zone naturally makes us wonder about the habitability of their moons (Williams et al. 1997; Kaltenegger 2010). So far, no extrasolar moon has been confirmed, but dedicated surveys are underway (Kipping et al. 2012). Several studies have addressed orbital stability of extrasolar satellite systems (Donnison 2010; Weidner & Horne 2010) and tidal heating has been shown to be an important energy source in satellites (Reynolds et al. 1987; Scharf 2006; Cassidy et al. 2009). In a recent study (Heller & Barnes 2012, HB12 in the following), we have extended these concepts to the illumination from the planet, i.e. stellar reflected light and thermal emission, and presented a model that invokes the runaway greenhouse effect to constrain exomoon habitability.

Earth-mass moons about Jupiter-mass planets have been shown to be dynamically stable for the lifetime of the solar system in systems where the stellar mass is greater than > 0.15 $M_\odot$ (Barnes & O'Brien 2002). I consider here whether such moons could actually be habitable. Therefore, I provide the first study of exomoon habitability that combines effects of stellar illumination, reflected stellar light from the planet, the planet's thermal emission, eclipses, and tidal heating with constraints by orbital stability and from the runaway greenhouse effect.

## 2. Methods

### 2.1. Orbit-averaged energy flux

To assess the habitability of a satellite, I estimated the global average

$$\bar{F}_{\rm s}^{\rm glob} = \frac{L_* \, (1 - \alpha_{\rm s})}{16\pi a_{*\rm p}^2 \sqrt{1 - e_{*\rm p}^2}} \left( x_{\rm s} + \frac{\pi R_{\rm p}^2 \alpha_{\rm p}}{2a_{\rm ps}^2} \right)$$
$$+ \frac{R_{\rm p}^2 \sigma_{\rm SB}(T_{\rm p}^{\rm eq})^4}{a_{\rm ps}^2} \frac{(1 - \alpha_{\rm s})}{4} + h_{\rm s} \qquad (1)$$

of its energy flux over one stellar orbit, where $L_*$ is stellar luminosity, $a_{*\rm p}$ the semi-major axis of the planet's orbit about the star, $a_{\rm ps}$ the semi-major axis of the satellite's orbit about the planet, $e_{*\rm p}$ the circumstellar orbital eccentricity, $R_{\rm p}$ the planetary radius, $\alpha_{\rm p}$ and $\alpha_{\rm s}$ are the albedos of the planet and the satellite, respectively, $T_{\rm p}^{\rm eq}$ is the planet's thermal equilibrium temperature, $\sigma_{\rm SB}$ the Stefan-Boltzmann constant, and $x_{\rm s}$ is the fraction of the satellite's orbit that is *not* spent in the shadow of the planet. Tidal heating $h_{\rm s}$ depends on the satellite's eccentricity $e_{\rm ps}$, on its radius $R_{\rm s}$, and strongly on $a_{\rm ps}$. I avoid repeating the various approaches available for a parametrization of $h_{\rm s}$ and refer the reader to HB12 and Heller et al. (2011), where we discussed the constant-time-lag model by Leconte et al. (2010) and the constant-phase-lag model from Ferraz-Mello et al. (2008). For this study, I arbitrarily chose the constant-time-lag model with a tidal time lag of the model satellites similar to that of the Earth, i.e. $\tau_s = 638\,\mathrm{s}$ (Neron de Surgy & Laskar 1997).

In HB12 we derived Eq. (1) for $x_s = 1$, thereby neglecting the effect of eclipses on the average flux. Here, I explore the decrease of the average stellar flux on the satellite due to eclipses. Eclipses occur most frequently – and will thus have the strongest effect on the moon's climate – when the satellite's circum-planetary orbit is coplanar with the circumstellar orbit. If the two orbits are also circular and eclipses are total

$$x_s = 1 - R_p/(\pi a_{ps}) \,.$$ (2)

Applying the Roche criterion for a fluid-like body (Weidner & Horne 2010), I derive

$$1 - \frac{1}{2.44\pi} \frac{R_p}{R_s} \left(\frac{M_s}{M_p}\right)^{1/3} < x_s \leq 1 \,,$$ (3)

where $x_s = 1$ can occur when a moon's line of nodes never crosses the planet's disk, i.e. in wide orbits. For an Earth-sized satellite about a Jupiter-sized planet Eq. (3) yields $x_s > 79\,\%$. Due to tidal effects, the moon's semi-major axis will typically be $> 5\,R_p$, then $x_s > 1 - 1/(5\pi) \approx 93.6\,\%$. In other words, eclipses will cause an orbit-averaged decrease in direct stellar illumination of $\approx 6.4\,\%$ at most in realistic scenarios. Using a constraint similar to Eq. (3), Scharf (2006) derived eclipsing effects of the same order of magnitude.

Eclipses will be most relevant for exomoons in low-mass stellar systems for two reasons. Firstly, tides raised by the star on the planet will cause the planet's obliquity to be eroded in $\ll 1$ Gyr (Heller et al. 2011), and since moons will orbit their planets in the equatorial plane (Porter & Grundy 2011), eclipses will then be most likely to occur. Secondly, exomoons in low-mass star systems must be close to their planet to ensure stability (see Sect. 2.3), hence, eclipses will always be total. Below, I compute the effect of an eclipse-induced decrease of stellar irradiation for some examples.

### 2.2. The circumstellar habitable zone

The irradiation habitable zone (IHZ) is defined as the circumstellar distance range in which liquid surface water can persist on a terrestrial planet (Dole 1964). On the one hand, if a planet is too close to the star, its atmosphere will become saturated with $H_2O$. Photodissociation then drives an escape of hydrogen into space, which desiccates the planet by turning it into a runaway greenhouse (Kasting et al. 1993). On the other hand, if the planet is too far away from the star, condensation of the greenhouse gas $CO_2$ will let any liquid surface water freeze (Kasting et al. 1993). This picture can be applied to exomoons as well. However, instead of direct stellar illumination only, the star's reflected light from the planet, the planet's thermal emission, and tidal heating in the moon need to be taken into account for the global energy flux (HB12).

Analytic expressions exist that parametrize the width of the IHZ as a function of stellar luminosity. I used the set of equations from Selsis et al. (2007) to compute the Sun-Earth equivalent distance $l_\oplus$ between the host star and the planet – or in this case: the moon. To a certain extent, the constraints on exomoon habitability will therefore still allow for moons with massive carbon dioxide atmospheres and strong greenhouse effect to exist in the outermost regions of the IHZ but I will not consider these special cases.

Exomoon habitability can be constrained by an upper limit for the orbit-averaged global flux. Above a certain value, typically around $300\,\mathrm{W/m^2}$ for Earth-like bodies, a satellite will be subject to a runaway greenhouse effect. This value does not depend on the atmospheric composition, except that it contains water (Kasting 1988). As in HB12, I used the semi-analytic expression from Pierrehumbert (2010) to compute this limit, $F_{RG}$. For computations of the satellite's radius, I used the model of Fortney et al. (2007) with a 68 % rock-to-mass fraction, similar to the Earth.

### 2.3. Orbital stability

A body's Hill radius $R_H$ is the distance range out to which the body's gravity dominates the effect on a test particle. In realistic scenarios the critical semi-major axis for a satellite to remain bound to its host planet is merely a fraction of the Hill radius, i.e. $a_{ps} < f\,R_H$ (Holman & Wiegert 1999), with a conservative choice for prograde satellites being $f = 1/3$ (Barnes & O'Brien 2002). If the combined mass of a planet-moon binary measured by radial-velocity is much smaller than the stellar mass, but still much larger than the satellite's mass, then $P_{ps}/P_{*p} \lesssim 1/9$, where $P_{*p}$ is the circumstellar period of the planet-moon pair and $P_{ps}$ is the satellite's orbital period about the planet (Kipping 2009). With $M_* \gg M_p \gg M_s$ and the planet-moon barycenter orbiting at an orbital distance $l_\oplus$ to the star, Kepler's third law gives

$$a_{ps} < l_\oplus \left[\frac{1}{81} \frac{(M_p + M_*)}{(M_* + M_p)}\right]^{1/3} \,.$$ (4)

With this dynamical restriction in mind, we can think of scenarios in which the mass of the star is very small, so that $l_\oplus$ lies close to it, and a moon about a planet in the IHZ must be very close to the planet to remain gravitationally bound. Then there will exist a limit (in terms of minimum stellar mass) at which the satellite needs to be so close to the planet that its tidal heating becomes so strong that it will initiate a greenhouse effect and will not be habitable. Below, I determine this minimum mass for stars to host potentially habitable exomoons.

The $P_{ps}/P_{*p} \lesssim 1/9$ constraint is used as a working hypothesis, backed up by numerical studies (Barnes & O'Brien 2002; Domingos et al. 2006; Cassidy et al. 2009) and observations of solar system satellites. But more extreme scenarios such as retrograde, Super-Ganymede exomoons in a wide orbit about planets in the IHZ of low-mass stars are theoretically possible (Namouni 2010). I did not consider the evolution of the planet's radius, its co-rotation radius, or the satellite's orbit about the planet. Indeed, those aspects would further reduce the likelihood of habitable satellites for given star-planet masses, because they would remove moons from the planet's orbit (Barnes & O'Brien 2002; Domingos et al. 2006; Sasaki et al. 2012).

## 3. Results

### 3.1. Eclipse-induced decrease of stellar illumination

In Fig. 1 I show two examples for $x_s < 1$.[1] The ordinate gives the perpendicular displacement $r_\perp$ of the satellite from the planet in units of planetary radii and as a function of circumstellar orbital phase $\varphi_{*p}$. Thus, if $r_\perp < 1$ and if the moon is behind the planet (and not in front of it), an eclipse occurs. Effects of the planet's penumbra are neglected.

---

[1] Computations were performed with my python code exomoon.py. Download via www.aip.de/People/RHeller.

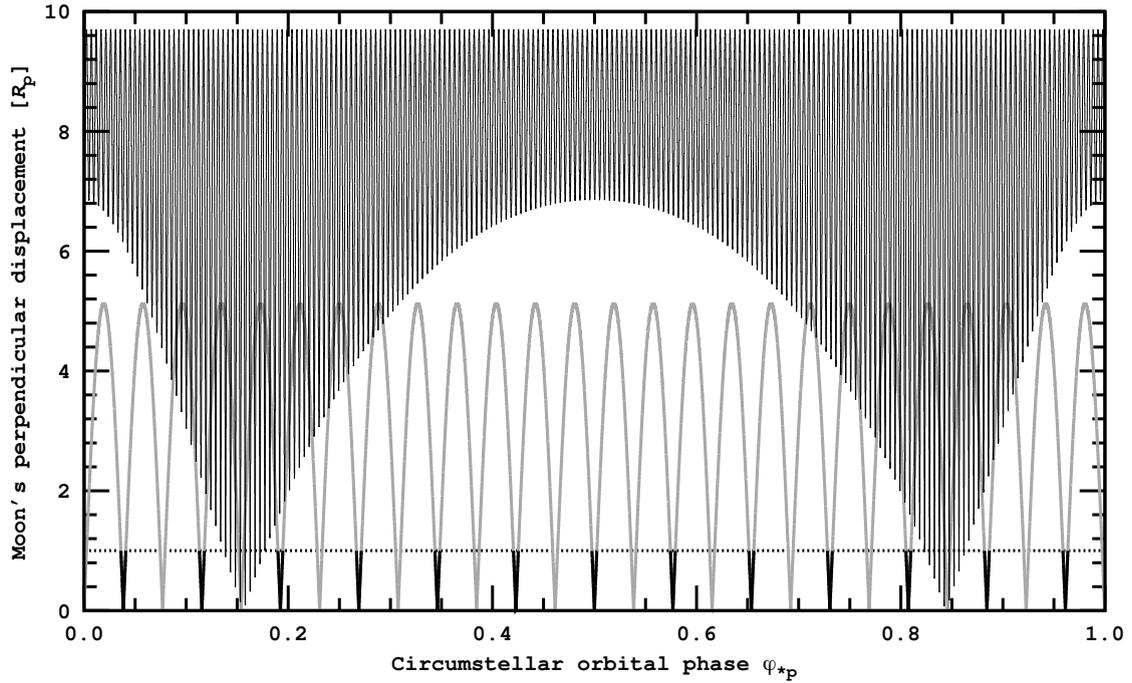

**Fig. 1.** Perpendicular displacement of a moon as seen from the star for two scenarios. The thin and highly oscillating curve corresponds to the orbit of a satellite in a Europa-wide orbit (in units of $R_p$) about a Jupiter-sized planet at 1 AU from a Sun-like star. The satellite's orbit about the Jovian planet is inclined by $i = 45°$. The thicker gray line shows the Miranda-wide orbit of a satellite orbiting a Neptune-mass object at $\approx 0.16$ AU from a $0.4\,M_\odot$ star, i.e. in the center of the IHZ. Here, the moon orbits in the same plane as the planet orbits the low-mass host star, i.e. $i = 0°$. Time spent in eclipse is emphasized with a thick, black line. Both orbits are normalized to the circumstellar orbital phase $\varphi_{*p}$.

In one scenario (thin, highly oscillating line) I placed a satellite in a Europa-wide orbit, i.e. at $9.7\,R_p$, about a Jupiter-sized planet in 1 AU distance from a Sun-like star. I applied a stellar orbital eccentricity of 0.3 and the moon's orbit is tilted by 45° against the circumstellar orbit. In this specific scenario, the eccentricity causes the summer ($0 \leq \varphi_{*p} \leq 0.18$ and $0.82 \leq \varphi_{*p} < 1$) to be shorter than the winter ($0.18 \leq \varphi_{*p} \leq 0.82$), and the inclination causes eclipses to occur only during a small part of the 365 d-period circumstellar orbit, namely around $\varphi_{*p} \approx 0.18$ and $\varphi_{*p} \approx 0.82$. Owing to the relatively wide satellite orbit, occultations are short compared to the moon's orbital period. Thus, effects of eclipses on the satellite's climate are small, in this case I find $x_s \approx 99.8\%$.

In the second scenario (gray line), I put a satellite in a close orbit at $5.1\,R_p$, comparable to Miranda's orbit about Uranus, about a Neptune-mass planet, which orbits a $0.4\,M_\odot$ star in the center of the IHZ at $\approx 0.16$ AU. The planet's circumstellar orbit with a period of roughly 36 days is circular and the moon's three-day orbit is coplanar with the stellar orbit. In this case, transits (thick black line) occur periodically once per satellite orbit and effects on its climate will be significant with $x_s \approx 93.8\%$.

### 3.2. Minimum stellar masses for habitable exomoons

In Fig. 2 contours present limiting stellar masses (abscissae) and host planetary masses (ordinate) for satellites to be habitable. I considered two moons: one with a mass ten times that of Ganymede ($10\,M_{Ga}$, left panel) and one with the mass of the Earth ($M_\oplus$, right panel), respectively. Variation of the satellite's mass changes the critical orbit-averaged flux $F_{RG}$ (see title of each panel). Contours of $\bar{F}_s^{glob} = F_{RG}$ are drawn for five different eccentricities of the satellite, four of which are taken from the solar system: $e_{Io} = 0.0041$, $e_{Ca} = 0.0074$, $e_{Eu} = 0.0094$, and $e_{Ti} = 0.0288$ for Io, Callisto, Europa, and Titan, respectively. I also plot a threshold for a hypothetical eccentricity of 0.05. Moons in star-planet systems with masses $M_*$ and $M_p$ left to a contour for a given satellite eccentricity are uninhabitable. As a confidence estimate of the tidal model, dashed contours provide the limiting mass combinations for strongly reduced ($\tau_s = 63.8$ s) and strongly enhanced ($\tau_s = 6380$ s) tidal dissipation in the moon for the $e_{ps} = 0.05$ example.

Satellites with the lowest eccentricities can be habitable even in the lowest-mass stellar systems, i.e. down to $0.1\,M_\odot$ and below for extremely circular planet-moon orbits. With increasing eccentricity, however, the satellite needs to be farther away from the planet to avoid a greenhouse effect, consequently it needs to be in a wider orbit about its planet, which in turn means that the planet needs to be farther away from the star for the moon to satisfy Eq. (4). Thus, the star needs to be more massive to have a higher luminosity and to reassure that the planet-moon binary is within the IHZ. This trend explains the increasing minimum stellar mass for increasing eccentricities in Fig. 2. For $e_{ps} = 0.05$ this minimum stellar mass is between $0.3\,M_\odot$ for a $10\,M_{Gan}$-mass moon (left panel) and $0.36\,M_\odot$ for an Earth-mass satellite (right panel). Uncertainties in the parametrization of tidal dissipation increase this limit to almost $0.5\,M_\odot$.

## 4. Conclusions

Eclipses of moons that are in tight orbits about their planet can substantially decrease the satellite's orbit-averaged stellar illumination. Equation (2) can be used to compute the reduction by total eclipses for circular and coplanar orbits. In orbits similar to the closest found in the solar system, this formula yields a reduction of about 6.4 %. In orbits wider than 32 planetary radii, the reduction is < 1 %. In tight orbits, illumination from the planet partly compensates for this reduction (HB12).

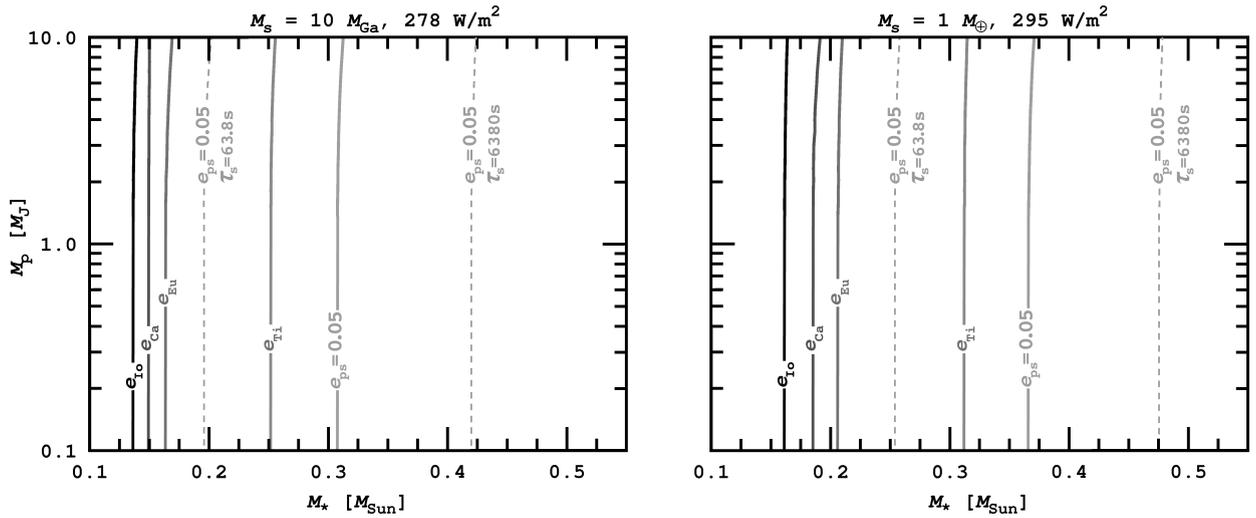

**Fig. 2.** Contours of maximum stellar masses (abscissa) for two possibly habitable moons with given planetary host masses (ordinate). Solid lines correspond to the satellite's time lag $\tau_s$ of 638 s and one out of five eccentricities, namely those of Io, Callisto, Europa, Titan, and 0.05, from left to right. Dashed lines refer to $e_{ps} = 0.05$ and variation of $\tau_s$ by a factor of ten. A satellite with host star and host planet masses located left to its respective eccentricity contour is uninhabitable.

M dwarfs with masses ≲ 0.2 $M_\odot$ can hardly host habitable exomoons. These moons' orbits about their planet would need to be almost perfectly circularized, which means that they would need to be on the very massive moon in that satellite system. Moreover, the nearby star will excite substantial eccentricities in the moon's orbit. This increases the minimum mass of host stars for habitable moons. To further constrain exomoon habitability, it will be necessary to simulate the eccentricity evolution of satellites with a model that considers both N-body interaction and tidal evolution. Such simulations are beyond the scope of this communication and will be conducted in a later study.

The full Kozai cycle and tidal friction model of [Porter & Grundy](2011) demonstrates very fast orbital evolution of satellites that orbit M stars in their IHZ. The star forces the satellite orbit to be eccentric, thus the moon's semi-major axis will shrink due to the tides on timescales much shorter than one Myr. Eventually, its orbit will be circular and very tight. Indeed, a moon's circum-planetary period in the IHZ about an M0 star will be even shorter than the $P_{*p}/9$ I used as a working hypothesis (see Fig. 3 in [Porter & Grundy](2011)).

For low-mass stars whose IHZs are close in and whose planets' obliquities are eroded in ≪ 1 Gyr, eclipses will reduce the orbit-averaged stellar illumination of satellites by about a few percent. To compensate for this cooling, the planet-moon duet would need to be closer to the star, but then the moon would become uninhabitable by initiating a greenhouse effect. This contradiction makes my estimates for the minimum stellar mass for a habitable moon star more robust.

Additionally, moons about low-mass stars will most likely not experience seasons because they would orbit their planet in its equatorial plane ([Porter & Grundy](2011)). The paucity of Jovian planets in the IHZ of low-mass stars ([Borucki et al.](2011)) further decreases the chances for habitable Earth-sized exomoons in low-mass stellar systems.

I conclude that stars with masses ≲ 0.2 $M_\odot$ cannot host habitable moons, and stars with masses up to 0.5 $M_\odot$ can be affected by the energy flux and orbital stability criteria combined in this paper. A model that couples gravitational scattering with tidal evolution is required to further constrain exomoon habitability about low-mass stars.

*Acknowledgements.* I thank the anonymous referee and the editor Tristan Guillot for their valuable comments. I am deeply grateful to Rory Barnes for our various discussions on the subject of this paper. This work has made use of NASA's Astrophysics Data System Bibliographic Services. Computations were performed with `ipython 0.13` ([Pérez & Granger 2007](#)) on `python 2.7.2` and figures were prepared with `gnuplot 4.4` ([www.gnuplot.info](http://www.gnuplot.info)) as well as with `gimp 2.6` ([www.gimp.org](http://www.gimp.org)).

## 7.3 A Dynamically-packed Planetary System around GJ 667 C with Three Super-Earths in its Habitable Zone (Anglada-Escudé et al. 2013)

Contribution:

RH contributed to the literature research, performed the computations of the tidal evolution of planet GJ 667 C f, created Fig. 14, and contributed to the writing of the manuscript (in particular to the exomoon part in Sect. 9.6). RH also contributed to the writing and to the illustration of a press release of the University of Göttingen to support the communication of this research to the public (http://uni-goettingen.de/en/891.html?archive=true&archive_id=4510&archive_source=presse).

# A dynamically-packed planetary system around GJ 667C with three super-Earths in its habitable zone ⋆ ⋆⋆

Guillem Anglada-Escudé[1], Mikko Tuomi[2,3], Enrico Gerlach[4], Rory Barnes[5], René Heller[6], James S. Jenkins[7], Sebastian Wende[1], Steven S. Vogt[8], R. Paul Butler[9], Ansgar Reiners[1], and Hugh R. A. Jones[2]

[1]  Universität Göttingen, Institut für Astrophysik, Friedrich-Hund-Platz 1, 37077 Göttingen, Germany
[2]  Centre for Astrophysics, University of Hertfordshire, College Lane, Hatfield, Hertfordshire AL10 9AB, UK
[3]  University of Turku, Tuorla Observatory, Department of Physics and Astronomy, Väisäläntie 20, FI-21500, Piikkiö, Finland
[4]  Technical University of Dresden, Institute for Planetary Geodesy, Lohrmann-Observatory, 01062 Dresden, Germany
[5]  Astronomy Department, University of Washington, Box 951580, WA 98195, Seattle, USA
[6]  Leibniz Institute for Astrophysics Potsdam (AIP), An der Sternwarte 16, 14482 Potsdam, Germany
[7]  Departamento de Astronomía, Universidad de Chile, Camino El Observatorio 1515, Las Condes, Casilla 36-D Santiago, Chile
[8]  UCO/Lick Observatory, University of California, Santa Cruz, CA 95064, USA
[9]  Carnegie Institution of Washington, Department of Terrestrial Magnetism, 5241 Broad Branch Rd. NW, 20015 Washington D.C., USA

submitted Feb 2013

## ABSTRACT

*Context.* Since low-mass stars have low luminosities, orbits at which liquid water can exist on Earth-sized planets are relatively close-in, which produces Doppler signals that are detectable using state-of-the-art Doppler spectroscopy.
*Aims.* GJ 667C is already known to be orbited by two super-Earth candidates. We have applied recently developed data analysis methods to investigate whether the data supports the presence of additional companions.
*Methods.* We obtain new Doppler measurements from HARPS extracted spectra and combined them with those obtained from the PFS and HIRES spectrographs. We used Bayesian and periodogram-based methods to re-assess the number of candidates and evaluated the confidence of each detection. Among other tests, we validated the planet candidates by analyzing correlations of each Doppler signal with measurements of several activity indices and investigated the possible quasi-periodic nature of signals.
*Results.* Doppler measurements of GJ 667C are described better by six (even seven) Keplerian-like signals: the two known candidates (b and c); three additional few-Earth mass candidates with periods of 92, 62 and 39 days (d, e and f); a cold super-Earth in a 260-day orbit (g) and tantalizing evidence of a ∼ 1 $M_⊕$ object in a close-in orbit of 17 days (h). We explore whether long-term stable orbits are compatible with the data by integrating $8{\times}10^4$ solutions derived from the Bayesian samplings. We assess their stability using secular frequency analysis.
*Conclusions.* The system consisting of six planets is compatible with dynamically stable configurations. As for the solar system, the most stable solutions do not contain mean-motion resonances and are described well by analytic Laplace-Lagrange solutions. Preliminary analysis also indicates that masses of the planets cannot be higher than twice the minimum masses obtained from Doppler measurements. The presence of a seventh planet (h) is supported by the fact that it appears squarely centered on the only island of stability left in the six-planet solution. Habitability assessments accounting for the stellar flux, as well as tidal dissipation effects, indicate that three (maybe four) planets are potentially habitable. Doppler and space-based transit surveys indicate that 1) dynamically packed systems of super-Earths are relatively abundant and 2) M-dwarfs have more small planets than earlier-type stars. These two trends together suggest that GJ 667C is one of the first members of an emerging population of M-stars with multiple low-mass planets in their habitable zones.

**Key words.** Techniques : radial velocities – Methods : data analysis – Planets and satellites : dynamical evolution and stability – Astrobiology – Stars: individual : GJ 667C

## 1. Introduction

Since the discovery of the first planets around other stars, Doppler precision has been steadily increasing to the point where objects as small as a few Earth masses can currently be detected around nearby stars. Of special importance to the exoplanet searches are low-mass stars (or M-dwarfs) nearest to the Sun. Since low-mass stars are intrinsically faint, the orbits at which a rocky planet could sustain liquid water on its surface (the so-called habitable zone, Kasting et al. 1993) are typically closer to the star, increasing their Doppler signatures even more. For this reason, the first super-Earth mass candidates in the habitable zones of nearby stars have been detected around M-dwarfs (e.g. GJ 581, Mayor et al. 2009; Vogt et al. 2010)).

Concerning the exoplanet population detected to date, it is becoming clear that objects between 2 $M_\oplus$ and the mass of Neptune (also called super-Earths) are very common around all G, K, and M dwarfs. Moreover, such planets tend to appear in close in/packed systems around G and K dwarfs (historically preferred targets for Doppler and transit surveys) with orbits closer in than the orbit of Mercury around our Sun. These features combined with a habitable zone closer to the star, point to the existence of a vast population of habitable worlds in multi-planet systems around M-dwarfs, especially around old/metal-depleted stars (Jenkins et al. 2013).

GJ 667C has been reported to host two (possibly three) super-Earths. GJ 667Cb is a hot super-Earth mass object in an orbit of 7.2 days and was first announced by Bonfils (2009). The second companion has an orbital period of 28 days, a minimum mass of about 4.5 $M_\oplus$ and, in principle, orbits well within the liquid water habitable zone of the star (Anglada-Escudé et al. 2012; Delfosse et al. 2012). The third candidate was considered tentative at the time owing to a strong periodic signal identified in two activity indices. This third candidate (GJ 667Cd) would have an orbital period between 74 and 105 days and a minimum mass of about 7 $M_\oplus$. Although there was tentative evidence for more periodic signals in the data, the data analysis methods used by both Anglada-Escudé et al. (2012) and Delfosse et al. (2012) studies were not adequate to properly deal with such high multiplicity planet detection. Recently, Gregory (2012) presented a Bayesian analysis of the data in Delfosse et al. (2012) and concluded that several additional periodic signals were likely present. The proposed solution, however, contained candidates with overlapping orbits and no check against activity or dynamics was done, casting serious doubts on the interpretation of the signals as planet candidates.

Efficient/confident detection of small amplitude signals requires more specialized techniques than those necessary to detect single ones. This was made explicitly obvious in, for example, the re-analysis of public HARPS data on the M0V star GJ 676A. In addition to the two signals of gas giant planets reported by Forveille et al. (2011), Anglada-Escudé & Tuomi (2012) (AT12 hereafter) identified the presence of two very significant low-amplitude signals in closer-in orbits. One of the main conclusions of AT12 was that correlations of signals already included in the model prevent detection of additional low-amplitude using techniques based on the analysis of the residuals only. In AT12, it was also pointed out that the two additional candidates (GJ 676A d and e) could be confidently detected with 30% less measurements using Bayesian based methods.

In this work, we assess the number of Keplerian-like signals around GJ 667C using the same analysis methods as in Anglada-Escudé & Tuomi (2012). The basic data consists of 1) new Doppler measurements obtained with the HARPS-TERRA software on public HARPS spectra of GJ 667C (see Delfosse et al. 2012, for a more detailed description of the dataset), and 2) Doppler measurements from PFS/Magellan and HIRES/Keck spectrometers (available in Anglada-Escudé & Butler 2012). We give an overview of GJ 667C as a star and provide updated parameters in Section 2. The observations and data-products used in later analyses are described in Section 3. Section 4 describes our statistical tools, models and the criteria used to quantify the significance of each detection (Bayesian evidence ratios and log–likelihood periodograms). The sequence and confidences of the signals in the Doppler data are given in section 5 where up to seven planet-like signals are spotted in the data. To promote Doppler signals to planets, such signals must be validated against possible correlations with stellar activity. In section 6, we discuss the impact of stellar activity on the significance of the signals (especially on the GJ 667Cd candidate) and we conclude that none of the seven candidates is likely to be spurious. In section 7, we investigate if all signals were detectable in subsets of the HARPS dataset to rule out spurious detections from quasi-periodic variability caused by stellar activity cycles. We find that all signals except the least significant one are robustly present in both the first and second-halves of the HARPS observing campaign independently. A dynamical analysis of the Bayesian posterior samples finds that a subset of the allowed solutions leads to long-term stable orbits. This verification steps allows us promoting the first six signals to the status of planet candidates. In Section 8 we also investigate possible mean-motion resonances (MMR) and mechanisms that guarantee long-term stability of the system. Given that the proposed system seems physically viable, we discusses potential habitability of each candidate in the context of up-to-date climatic models, possible formation history, and the effect of tides in Section 9. Concluding remarks and a summary are given in Section 10. The appendices describe additional tests performed on the data to double-check the significance of the planet candidates.

## 2. Properties of GJ 667C

GJ 667C (HR 6426 C), has been classified as an M1.5V star (Geballe et al. 2002) and is a member of a triple system, since it is a common proper motion companion to the K3V+K5V binary pair, GJ 667AB. Assuming the HIPPARCOS distance to the GJ 667AB binary ($\sim$ 6.8 pc van Leeuwen 2007), the projected separation between GJ 667C and GJ 667AB is $\sim$ 230 AU. Spectroscopic measurements of the binary have revealed a metallicity significantly lower than the Sun (Fe/H =-0.59±0.10 Cayrel de Strobel 1981). The galactic kinematics of GJ 667 are compatible with both thin and thick disk populations and there is no clear match to any known moving group or stream (Anglada-Escudé et al. 2012). Spectrocopic studies of the GJ 667AB pair (Cayrel de Strobel 1981) show that they are on the main sequence, indicating an age between 2 and 10 Gyr. Following the simple models in Reiners & Mohanty (2012), the low activity and the estimate of the rotation period of GJ 667C (P> 100 days, see Section 6) also support an age of > 2 Gyr. In conclusion, while the age of the GJ 667 system is uncertain, all analyses indicate that the system is old.

We performed a spectroscopic analysis of GJ 667C using high resolution spectra obtained with the UVES/VLT spectrograph (program 87.D-0069). Both the HARPS and the UVES spectra show no $H_\alpha$ emission. The value of the mean S-index measurement (based on the intensity of the CaII H+K emission lines) is 0.48 ± 0.02, which puts the star among the most inactive objects in the HARPS M-dwarf sample (Bonfils et al. 2013). By comparison, GJ 581(S=0.45) and GJ 876 (S=0.82) are RV-stable enough to detect multiple low-mass planets around them, while slightly more active stars like GJ 176 (S=1.4), are stable enough to detect at least one low-mass companion. Very active and rapidly rotating M-dwarfs, such GJ 388 (AD Leo) or GJ 803 (AU Mic), have S-indices as high as 3.7 and 7.8, respectively. A low activity level allows one to use a large number of atomic and molecular lines for the spectral fitting without accounting for magnetic and/or rotational broadening. UVES observations of GJ 667C were taken in service mode in three exposures during the night on August 4th 2011. The high resolution mode with a slit width of 0.3″ was used to achieve a resolving power of $R \sim 100\,000$. The observations cover a wavelength range from 640 nm to 1020 nm on the two red CCDs of UVES.

**Table 1.** Parameter space covered by the grid of synthetic models.

|           | Range          | Step size |
|-----------|----------------|-----------|
| $T_{eff}$ [K] | 2,300 – 5,900  | 100       |
| $\log(g)$ | 0.0 – +6.0     | 0.5       |
| $[Fe/H]$  | -4.0 – -2.0    | 1.0       |
|           | -2.0 – +1.0    | 0.5       |

**Table 2.** Stellar parameters of GJ 667C

| Parameters | Value | Ref. |
|------------|-------|------|
| R.A. | 17 18 57.16 | 1 |
| Dec | -34 59 23.14 | 1 |
| $\mu_{R.A.}$ [mas yr$^{-1}$] | 1129.7(9.7) | 1 |
| $\mu_{Dec.}$ [mas yr$^{-1}$] | -77.0(4.6) | 1 |
| Parallax [mas] | 146.3(9.0) | 1 |
| Hel. RV [km s$^{-1}$] | 6.5(1.0) | 2 |
| V [mag] | 10.22(10) | 3 |
| J [mag] | 6.848(21) | 4 |
| H [mag] | 6.322(44) | 4 |
| K [mag] | 6.036(20) | 4 |
| $T_{eff}$[K] | 3350(50) | 5 |
| [Fe/H] | -0.55(10) | 5 |
| $\log g$ [g in cm s$^{-1}$] | 4.69(10) | 5 |
| Derived quantities | | |
| UVW$_{LSR}$ [km s$^{-1}$] | (19.5, 29.4,-27.2) | 2 |
| Age estimate | > 2 Gyr | 5 |
| Mass [M$_\odot$] | 0.33(2) | 5 |
| $L_\ast/L_\odot$ | 0.0137(9) | 2 |

The spectral extraction and reduction were done using the ESOREX pipeline for UVES. The wavelength solution is based, to first order, on the Th-Ar calibration provided by ESO. All orders were corrected for the blaze function and also normalized to unity continuum level. Afterwards, all orders were merged together. For overlapping orders the redder ends were used due to their better quality. In a last step, an interactive removal of bad pixels and cosmic ray hits was performed.

The adjustment consists of matching the observed spectrum to a grid of synthetic spectra from the newest PHOENIX/ACES grid (see Husser et al. 2013). The updated codes use a new equation of state that accounts for the molecular formation rates at low temperatures. Hence, it is optimally suited for simulation of spectra of cool stars. The 1D models are computed in plane parallel geometry and consist of 64 layers. Convection is treated in mixing-length geometry and from the convective velocity a micro-turbulence velocity (Edmunds 1978) is deduced via $v_{mic} = 0.5 \cdot v_{conv}$. The latter is used in the generation of the synthetic high resolution spectra. An overview of the model grid parameters is shown in Table 1. Local thermal equilibrium is assumed in all models.

First comparisons of these models with observations show that the quality of computed spectra is greatly improved in comparison to older versions. The problem in previous versions of the PHOENIX models was that observed spectra in the $\epsilon$- and $\gamma$-TiO bands could not be reproduced assuming the same effective temperature parameter (Reiners 2005). The introduction of the new equation of state apparently resolved this problem. The new models can consistently reproduce both TiO absorption bands together with large parts of the visual spectrum at very high fidelity (see Fig. 1).

As for the observed spectra, the models in our grid are also normalized to the local continuum. The regions selected for the fit were chosen as unaffected by telluric contamination and are marked in green in Fig. 1. The molecular TiO bands in the region between 705 nm to 718 nm ($\epsilon$-TiO) and 840 nm to 848 nm ($\gamma$-TiO) are very sensitive to $T_{eff}$ but almost insensitive to $\log g$. The alkali lines in the regions between 764 nm to 772 nm and 816 nm to 822 nm (K- and Na-atomic lines, respectively) are sensitive to $\log g$ and $T_{eff}$. All regions are sensitive to metallicity. The simultaneous fit of all the regions to all three parameters breaks the degeneracies, leading to a unique solution of effective temperature, surface gravity and metallicity.

As the first step, a three dimensional $\chi^2$-map is produced to determine start values for the fitting algorithm. Since the model grid for the $\chi^2$-map is discrete, the real global minimum is likely to lie between grid points. We use the parameter sets of the three smallest $\chi^2$-values as starting points for the adjustment procedure. We use the IDL *curvefit*-function as the fitting algorithm. Since this function requires continuous parameters, we use three dimensional interpolation in the model spectra. As a fourth free parameter, we allow the resolution of the spectra to vary in order to account for possible additional broadening (astrophysical or instrumental). For this star, the relative broadening is always found to be < 3% of the assumed resolution of UVES, and is statistically indistinguishable from 0. More information on the method and first results on a more representative sample of stars will be given in a forthcoming publication.

As already mentioned, the distance to the GJ 667 system comes from the HIPPARCOS parallax of the GJ 667AB pair and is rather uncertain (see Table 2). This, combined with the luminosity-mass calibrations in Delfosse et al. (2000), propagates into a rather uncertain mass (0.33±0.02 $M_\odot$) and luminosity estimates (0.0137 ± 0.0009 $L_\odot$) for GJ 667C (Anglada-Escudé et al. 2012). A good trigonometric parallax measurement and the direct measurement of the size of GJ 667C using interferometry (e.g. von Braun et al. 2011) are mostly needed to refine its fundamental parameters. The updated values of the spectroscopic parameters are slightly changed from previous estimates. For example, the effective temperature used in Anglada-Escudé et al. (2012) was based on evolutionary models using the stellar mass as the input which, in turn, is derived from the rather uncertain parallax measurement of the GJ 667 system. If the spectral type were to be understood as a temperature scale, the star should be classified as an M3V-M4V instead of the M1.5V type assigned in previous works (e.g. Geballe et al. 2002). This mismatch is a well known effect on low metallicity M dwarfs (less absorption in the optical makes them appear of earlier type than solar metallicity stars with the same effective temperature). The spectroscopically derived parameters and other basic properties collected from the literature are listed in Table 2.

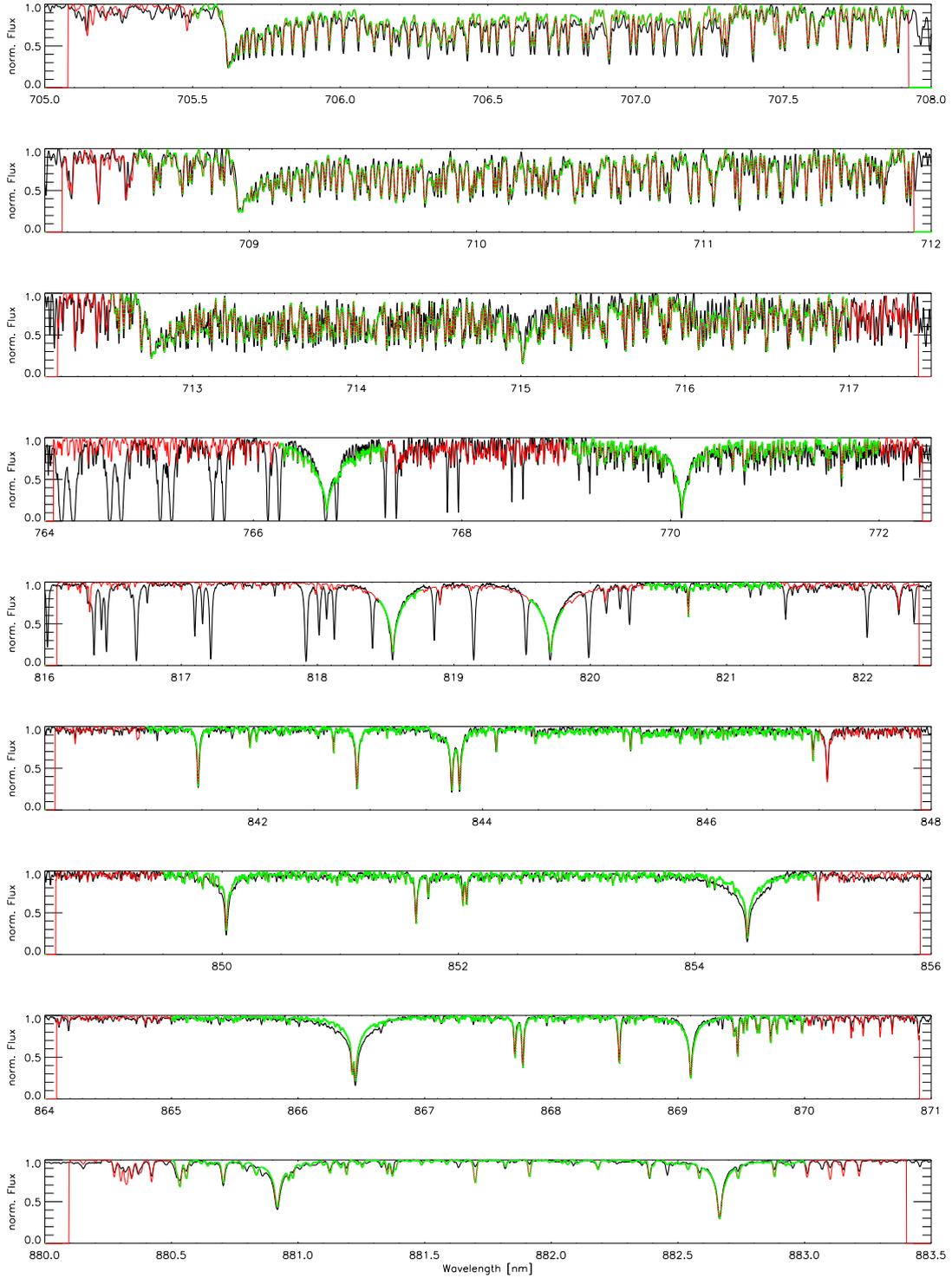

**Fig. 1.** Snapshots of the wavelength regions used in the spectral fit to the UVES spectrum of GJ 667C. The observed spectrum is represented in black, the green curves are the parts of the synthetic spectrum used in the fit. The red lines are also from the synthetic spectrum that were not used to avoid contamination by telluric features or because they did not contain relevant spectroscopic information. Unfitted deep sharp lines- especially on panels four and five from the top of the page- are non-removed telluric features.

## 3. Observations and Doppler measurements

A total of 173 spectra obtained using the HARPS spectrograph (Pepe et al. 2002) have been re-analyzed using the HARPS-TERRA software (Anglada-Escudé & Butler 2012). HARPS-TERRA implements a least-squares template matching algo-rithm to obtain the final Doppler measurement. This method and is especially well suited to deal with the highly blended spectra of low mass stars. It only replaces the last step of a complex spectral reduction procedure as implemented by the HARPS Data Reduction Software (DRS). Such extraction is automatically done by the HARPS-ESO services and includes all the necessary steps from 2D extraction of the echelle orders, flat and dark corrections, and accurate wavelength calibration using several hundreds of Th Ar lines across the HARPS wavelength range (Lovis & Pepe 2007). Most of the spectra (171) are extracted from the ESO archives and have been obtained by other groups over the years (e.g., Bonfils et al. 2013; Delfosse et al. 2012) covering from June 2004 to June 2010. To increase the time baseline and constrain long period trends, two additional HARPS observations were obtained between March 5th and 8th of 2012. In addition to this, three activity indices were also extracted from the HARPS spectra. These are: the S-index (proportional to the chromospheric emission of the star), the full-width-at-half-maximum of the mean line profile (or FWHM, a measure of the width of the mean stellar line) and the line bisector (or BIS, a measure of asymmetry of the mean stellar line). Both the FWHM and BIS are measured by the HARPS-DRS and were taken from the headers of the corresponding files. All these quantities might correlate with spurious Doppler offsets caused by stellar activity. In this sense, any Doppler signal with a periodicity compatible with any of these signals will be considered suspicious and will require a more detailed analysis. The choice of these indices is not arbitrary. Each of them is thought to be related to an underlying physical process that can cause spurious Doppler offsets. For example, S-index variability typically maps the presence of active regions on the stellar surface and variability of the stellar magnetic field (e.g., solar-like cycles). The line bisector and FWHM should have the same period as spurious Doppler signals induced by spots corotating with the star (contrast effects combined with stellar rotation, suppression of convection due to magnetic fields and/or Zeeman splitting in magnetic spots). Some physical processes induce spurious signals at some particular spectral regions (e.g., spots should cause stronger offsets at blue wavelengths). The Doppler signature of a planet candidate is constant over all wavelengths and, therefore, a signal that is only present at some wavelengths cannot be considered a credible candidate. This feature will be explored below to validate the reality of some of the proposed signals. A more comprehensive description of each index and their general behavior in response to stellar activity can be found elsewhere (Baliunas et al. 1995; Lovis et al. 2011). In addition to the data products derived from HARPS observations, we also include 23 Doppler measurements obtained using the PFS/Magellan spectrograph between June 2011 and October 2011 using the Iodine cell technique, and 22 HIRES/Keck Doppler measurements (both RV sets are provided in Anglada-Escudé & Butler 2012) that have lower precision but allow extending the time baseline of the observations. The HARPS-DRS also produces Doppler measurements using the so-called cross correlation method (or CCF). In the Appendices, we show that the CCF-Doppler measurements actually contain the same seven signals providing indirect confirmation and lending further confidence to the detections.

## 4. Statistical and physical models

The basic model of a radial velocity data set from a single telescope-instrument combination is a sum of $k$ Keplerian signals, with $k = 0, 1, ...,$ a random variable describing the instrument noise, and another describing all the excess noise in the data. The latter noise term, sometimes referred to as stellar RV jitter (Ford 2005), includes the noise originating from the stellar surface due to inhomogeneities, activity-related phenomena, and can also include instrumental systematic effects. Following Tuomi (2011), we model these noise components as Gaussian random variables with zero mean and variances of $\sigma_i^2$ and $\sigma_l^2$, where the former is the formal uncertainty in each measurement and the latter is the jitter that is treated as a free parameter of the model (one for each instrument $l$).

Since radial velocity variations have to be calculated with respect to some reference velocity, we use a parameter $\gamma_l$ that describes this reference velocity with respect to the data mean of a given instrument. For several telescope/instrument combinations, the Keplerian signals must necessarily be the same but the parameters $\gamma_l$ (reference velocity) and $\sigma_l^2$ (jitter) cannot be expected to have the same values. Finally, the model also includes a linear trend $\dot{\gamma}$ to account for very long period companions (e.g., the acceleration caused by the nearby GJ 667AB binary). This model does not include mutual interactions between planets, which are known to be significant in some cases (e.g. GJ 876, Laughlin & Chambers 2001). In this case, the relatively low masses of the companions combined with the relatively short time-span of the observations makes these effects too small to have noticeable impact on the measured orbits. Long-term dynamical stability information is incorporated and discussed later (see Section 8). Explicitly, the baseline model for the RV observations is

$$v_l(t_i) = \gamma_l + \dot{\gamma}(t_i - t_0) + \sum_{j=1}^{k} f(t_i, \boldsymbol{\beta}_j) + g_l\left[\boldsymbol{\psi}; t_i, z_i, t_{i-1}, r_{i-1}\right], \quad (1)$$

where $t_0$ is some reference epoch (which we arbitrarily choose as $t_0 = 2450000$ JD), $g$ is a function describing the specific noise properties (instrumental and stellar) of the $l$-th instrument on top of the estimated Gaussian uncertainties. We model this function using first order moving average (MA) terms Tuomi et al. (2013, 2012) that and on the residual $r_{i-1}$ to the previous measurement at $t_{i-1}$, and using linear correlation terms with activity indices (denoted as $z_i$). This component of the model is typically parameterized using one or more "nuisance parameters" $\boldsymbol{\psi}$ that are also treated as free parameters of the model. Function $f$ represents the Doppler model of a planet candidate with parameters $\boldsymbol{\beta}_j$ (Period $P_j$, Doppler semi-amplitude $K_j$, mean anomaly at reference epoch $M_{0,j}$, eccentricity $e_j$, and argument of the periastron $\omega_j$).

The Gaussian white noise component of each measurement and the Gaussian jitter component associated to each instrument enter the model in the definition of the likelihood function $L$ as

$$L(m|\boldsymbol{\theta}) = \prod_{i=1}^{N} \frac{1}{\sqrt{2\pi(\sigma_i^2 + \sigma_l^2)}} \exp\left\{\frac{-\left[m_i - v_l(t_i)\right]^2}{2(\sigma_i^2 + \sigma_l^2)}\right\}, \quad (2)$$

where $m$ stands for *data* and $N$ is the number of individual measurements. With these definitions, the posterior probability density $\pi(\boldsymbol{\theta}|m)$ of parameters $\boldsymbol{\theta}$ given the data $m$ ($\boldsymbol{\theta}$ includes the orbital elements $\boldsymbol{\beta}_j$, the slope term $\dot{\gamma}$, the instrument dependent constant offsets $\gamma_l$, the instrument dependent jitter terms $\sigma_l$, and a number of nuisance parameters $\boldsymbol{\psi}$), is derived from the Bayes' theorem as

$$\pi(\boldsymbol{\theta}|m) = \frac{L(m|\boldsymbol{\theta})\pi(\boldsymbol{\theta})}{\int L(m|\boldsymbol{\theta})\pi(\boldsymbol{\theta})d\boldsymbol{\theta}}. \quad (3)$$

This equation is where the prior information enters the model through the choice of the prior density functions $\pi(\boldsymbol{\theta})$. This way, the posterior density $\pi(\boldsymbol{\theta}|m)$ combines the new information provided by the new data $m$ with our prior assumptions for the parameters. In a Bayesian sense, finding the most favored model and allowed confidence intervals consists of the identification and exploration of the higher probability regions of the posterior density. Unless the model of the observations is very simple (e.g., linear models), a closed form of $\pi(\theta|m)$ cannot be derived analytically and numerical methods are needed to explore its properties. The description of the adopted numerical methods are the topic of the next subsection.

### 4.1. Posterior samplings and Bayesian detection criteria

Given a model with $k$ Keplerian signals, we draw statistically representative samples from the posterior density of the model parameters (Eq. 3) using the adaptive Metropolis algorithm Haario et al. (2001). This algorithm has been used successfully in e.g. Tuomi (2011), Tuomi et al. (2011) and Anglada-Escudé & Tuomi (2012). The algorithm appears to be a well suited to the fitting of Doppler data in terms of its relatively fast convergence – even when the posterior is not unimodal (Tuomi 2012) – and it provides samples that represent well the posterior densities. We use these samples to locate the regions of maximum *a posteriori* probability in the parameter space and to estimate each parameter confidence interval allowed by the data. We describe the parameter densities briefly by using the maximum *a posteriori* probability (MAP) estimates as the most probable values, i.e. our preferred solution, and by calculating the 99% Bayesian credibility sets (BCSs) surrounding these estimates. Because of the caveats of point estimates (e.g., inability to describe the shapes of posterior densities in cases of multimodality and/or non-negligible skewness), we also plot marginalized distributions of the parameters that are more important from a detection and characterization point of view, namely, velocity semi-amplitudes $K_j$, and eccentricities $e_j$.

The availability of samples from the posterior densities of our statistical models also enables us to take advantage of the signal detection criteria given in Tuomi (2012). To claim that any signal is significant, we require that 1) its period is well-constrained from above and below, 2) its RV amplitude has a density that differs from zero significantly (excluded from the 99% credibility intervals), and 3) the posterior probability of the model containing $k+1$ signals must be (at least) 150 times greater than that of the model containing only $k$ signals.

The threshold of 150 on condition (3) might seem arbitrary, and although posterior probabilities also have associated uncertainties (Jenkins & Peacock 2011), we consider that such a threshold is a conservative one. As made explicit in the definition of the posterior density function $\pi(\theta|m)$, the likelihood function is not the only source of information. We take into account the fact that all parameter values are not equally probable prior to making the measurements via prior probability densities. Essentially, our priors are chosen as in Tuomi (2012). Of special relevance in the detection process is the prior choice for the eccentricities. Our functional choice for it (Gaussian with zero mean and $\sigma_e = 0.3$) is based on statistical, dynamical and population considerations and it is discussed further in the appendices (Appendix A). For more details on different prior choices, see the dedicated discussion in Tuomi & Anglada-Escudé (2013).

### 4.2. Log–Likelihood periodograms

Because the orbital period (or frequency) is an extremely non-linear parameter, the orbital solution of a model with $k + 1$ signals typically contains many hundreds or thousands of local likelihood maxima (also called independent frequencies). In any method based on stochastic processes, there is always a chance that the global maxima of the target function is missed. Our log–likelihood periodogram (or log–L periodogram) is a tool to systematically identify periods for new candidate planets of higher probability and ensure that these areas have been well explored by the Bayesian samplings (e.g., we always attempt to start the chains close to the five most significant periodicities left in the data). A log–L periodogram consists of computing the improvement of the logarithm of the likelihood (new model with $k + 1$ planets) compared to the logarithm of the likelihood of the null hypothesis (only $k$ planets) at each test period. Log–L periodograms are represented as period versus $\Delta \log L$ plots, where log is always the natural logarithm. The definition of the likelihood function we use is shown in Eq. 2 and typically assumes Gaussian noise sources only (that is, different jitter parameters are included for each instrument and g=0 in Eq. 1).

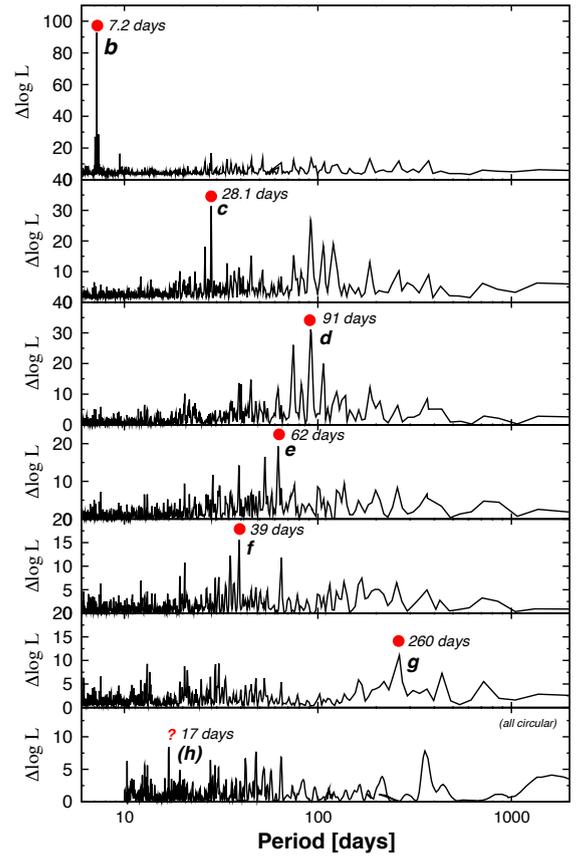

**Fig. 2.** Log–likelihood periodograms for the seven candidate signals sorted by significance. While the first six signals are easily spotted, the seventh is only detected with log–L periodograms if all orbits are assumed to be circular.

$\Delta \log L$ can also be used for estimating the *frequentist* false alarm probability (FAP) of a solution using the likelihood-ratio test for nested models. This FAP estimates what fraction of times one would recover such a significant solution by an unfortunate arrangement of Gaussian noise. To compute this FAP from $\Delta \log L$ we used the up-to-date recipes provided by Baluev (2009). We note that that maximization of the likelihood involves solving for many parameters simultaneously: orbital parameters of the new candidate, all orbital parameters of the already detected signals, a secular acceleration term $\dot{\gamma}$, a zero-point $\gamma_l$ for each instrument, and jitter terms $\sigma_l$ of each instrument (see Eq. 1). It is, therefore, a computationally intensive task, espe-cially when several planets are included and several thousand of test periods for the new candidate must be explored.

As discussed in the appendices (see Section A.1), allowing for full Keplerian solutions at the period search level makes the method very prone to false positives. Therefore while a full Keplerian solution is typically assumed for all the previously detected $k$-candidates, the orbital model for the $k + 1$-candidate is always assumed to be circular in our standard setup. This way, our log–L periodograms represent a natural generalization of more classic hierarchical periodogram methods. This method was designed to account for parameter correlations at the detection level. If such correlations are not accounted for, the significance of new signals can be strongly biased causing both false positives and missed detections. In the study of the planet hosting M-dwarf GJ 676A (Anglada-Escudé & Tuomi 2012) and in the more recent manuscript on GJ 581 (Tuomi & Jenkins 2012), we have shown that -while log–L periodograms represent an improvement with respect to previous periodogram schemes- the aforementioned Bayesian approach has a higher sensitivity to lower amplitude signals and is less prone to false positive detections. Because of this, the use of log–L periodograms is not to quantify the significance of a new signal but to provide visual assessment of possible aliases or alternative high-likelihood solutions.

Log–L periodograms implicitly assume flat priors for all the free parameters. As a result, this method provides a quick way of assessing the sensitivity of a detection against a choice of prior functions that are different from uniform. As discussed later, the sixth candidate is only confidently spotted using log–L periodograms (our detection criteria is FAP< 1%) when the orbits of all the candidates are assumed to be circular. This is the *red line* beyond which our detection criteria becomes strongly dependent on our choice of prior on the eccentricity. The same applies to the seventh tentative candidate signal.

## 5. Signal detection and confidences

As opposed to other systems analyzed with the same techniques (e.g. Tau Ceti or HD 40307, Tuomi et al. 2012, 2013), we found that for GJ 667C the simplest model ($g = 0$ in equation 1) already provides a sufficient description of the data. For brevity, we omit here all the tests done with more sophisticated parameterizations of the noise (see Appendix C) that essentially lead to unconstrained models for the correlated noise terms and the same final results. In parallel with the Bayesian detection sequence, we also illustrate the search using log–L periodograms. In all that follows we use the three datasets available at this time : HARPS-TERRA, HIRES and PFS. We use the HARPS-TERRA Doppler measurements instead of CCF ones because TERRA velocities have been proven to be more precise on stable M-dwarfs (Anglada-Escudé & Butler 2012).

The first three periodicities (7.2 days, 28.1 days and 91 days) were trivially spotted using Bayesian posterior samplings and the corresponding log–L periodograms. These three signals were already reported by Anglada-Escudé et al. (2012) and Delfosse et al. (2012), although the last one (signal d, at 91 days) remained uncertain due to the proximity of a characteristic timescale of the star's activity. This signal is discussed in the context of stellar activity in Section 6. Signal d has a MAP period of 91 days and would correspond to a candidate planet with a minimum mass of $\sim 5$ $M_\oplus$.

After that, the log–L periodogram search for a fourth signal indicates a double-peaked likelihood maximum at 53 and

**Table 3.** Relative posterior probabilities and log-Bayes factors of models $\mathcal{M}_k$ with $k$ Keplerian signals given the combined HARPS-TERRA, HIRES, and PFS RV data of GJ 667C. Factor $\Delta$ indicates how much the probability increases with respect to the best model with one less Keplerian and $P$ denotes the MAP period estimate of the signal added to the solution when increasing $k$. Only the highest probability sequence is shown here (reference solution). A complete table with alternative solutions corresponding to local probability maxima is given in Appendix B.2

| $k$ | $P(\mathcal{M}_k|d)$ | $\Delta$ | $\log P(d|\mathcal{M}_k)$ | $P$ [days] | ID |
|---|---|---|---|---|---|
| 0 | $2.7\times10^{-85}$ | – | -602.1 | – | |
| 1 | $3.4\times10^{-48}$ | $1.3\times10^{37}$ | -516.0 | 7.2 | |
| 2 | $1.3\times10^{-35}$ | $3.9\times10^{12}$ | -486.3 | 91 | |
| 3 | $8.9\times10^{-18}$ | $6.7\times10^{17}$ | -444.5 | 28 | |
| 4 | $1.9\times10^{-14}$ | $2.1\times10^{3}$ | -436.3 | 53 | |
| 4 | $1.2\times10^{-14}$ | $1.3\times10^{3}$ | -436.7 | 62 | |
| 5 | $1.0\times10^{-7}$ | $5.5\times10^{6}$ | -420.0 | 39, 53 | |
| 5 | $1.0\times10^{-8}$ | $5.3\times10^{5}$ | -422.3 | 39, 62 | |
| 6 | $4.1\times10^{-3}$ | $4.0\times10^{4}$ | -408.7 | 39, 53, 256 | |
| 6 | $4.1\times10^{-4}$ | $4.0\times10^{3}$ | -411.0 | 39, 62, 256 | |
| 7 | 0.057 | 14 | -405.4 | 17, 39, 53, 256 | |
| 7 | 0.939 | 230 | -402.6 | 17, 39, 62, 256 | |

62 days -both candidate periods receiving extremely low false-alarm probability estimates (see Figure 2). Using the recipes in Dawson & Fabrycky (2010), it is easy to show that the two peaks are the yearly aliases of each other. Accordingly, our Bayesian samplings converged to either period equally well giving slightly higher probability to the 53-day orbit (×6). In both cases, we found that including a fourth signal improved the model probability by a factor >$10^3$. In appendix B.2 we provide a detailed analysis and derived orbital properties of both solutions and show that the precise choice of this fourth period does not substantially affect the confidence of the rest of the signals. As will be shown at the end of the detection sequence, the most likely solution for this candidate corresponds to a minimum mass of 2.7 $M_\oplus$ and a period of 62 days.

After including the fourth signal, a fifth signal at 39.0 days shows up conspicuously in the log–L periodograms. In this case, the posterior samplings always converged to the same period of 39.0 days without difficulty (signal f). Such a planet would have a minimum mass of $\sim 2.7$ $M_\oplus$. Given that the model probability improved by a factor of $5.3\times10^5$ and that the FAP derived from the log-L periodogram is 0.45%, the presence of this periodicity is also supported by the data without requiring further assumptions.

The Bayesian sampling search for a sixth signal always converged to a period of 260 days that also satisfied our detection criteria and increased the probability of the model by a factor of $4 \times 10^3$. The log–L periodograms did spot the same signal as the most significant one but assigned a FAP of $\sim$20% to it. This apparent contradiction is due to the prior on the eccentricity. That is, the maximum likelihood solution favors a very eccentric orbit for the Keplerian orbit at 62 days ($e_e \sim 0.9$), which is unphysical and absorbs variability at long timescales through aliases. To investigate this, we performed a log–L periodogram search assuming circular orbits for all the candidates. In this case, the 260-day period received a FAP of 0.5% which would then qualify as a significant detection. Given that the Bayesian detection criteria are well satisfied and that the log–L periodograms also provide substantial support for the signal, we also include it in

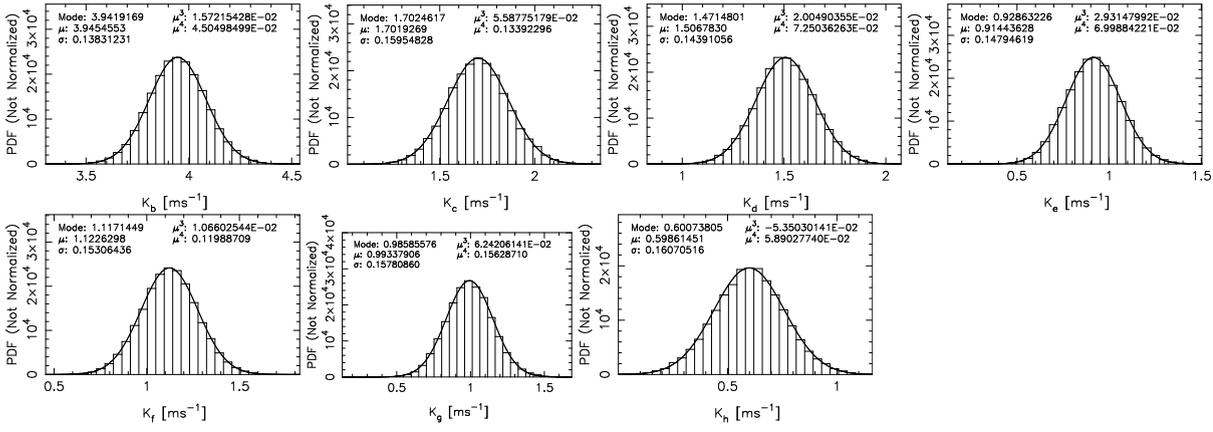

**Fig. 3.** Marginalized posterior densities for the Doppler semi-amplitudes of the seven reported signals.

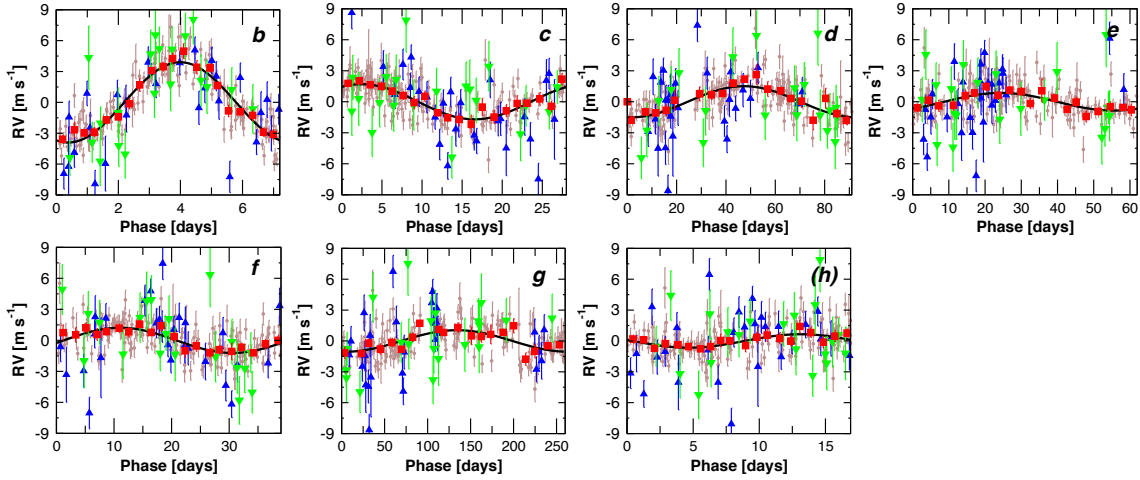

**Fig. 4.** RV measurements phase-folded to the best period for each planet. Brown circles are HARPS-TERRA velocities, PFS velocities are depicted as blue triangles, and HIRES velocities are green triangles. Red squares are averages on 20 phase bins of the HARPS-TERRA velocities. The black line corresponds to the best circular orbital fit (visualization purposes only).

**Table 4.** Reference orbital parameters and their corresponding 99% credibility intervals. While the angles $\omega$ and $M_0$ are unconstrained due to strong degeneracies at small eccentricities, their sum $\lambda = M_0 + \omega$ is better behaved and is also provided here for reference.

| | b | (h) | c | f | e* |
|---|---|---|---|---|---|
| P [days] | 7.2004 [7.1987, 7.2021] | 16.946 [16.872, 16.997] | 28.140 [28.075, 28.193] | 39.026 [38.815, 39.220] | 62.24 [61.69, 62.79] |
| e | 0.13 [0.02, 0.23] | 0.06 [0, 0.38] | 0.02 [0, 0.17] | 0.03 [0, 0.19] | 0.02 [0, 0.24] |
| K [m s⁻¹] | 3.93 [3.55, 4.35] | 0.61 [0.12, 1.05] | 1.71 [1.24, 2.18] | 1.08 [0.62, 1.55] | 0.92 [0.50, 1.40] |
| $\omega$ [rad] | 0.10 [5.63, 0.85] | 2.0 [0, 2$\pi$] | 5.1 [0, 2$\pi$] | 1.8 [0, 2$\pi$] | 0.5 [0, 2$\pi$] |
| $M_0$ [rad] | 3.42 [2.32, 4.60] | 5.1 [0, 2$\pi$] | 0.3 [0, 2$\pi$] | 5.1 [0, 2$\pi$] | 4.1 [0, 2$\pi$] |
| $\lambda$ [deg] | 201[168, 250] | 45(180)† | 308(99)† | 34 (170)† | 262(150)† |
| M sin $i$ [M$_\oplus$] | 5.6 [4.3, 7.0] | 1.1 [0.2, 2.1] | 3.8 [2.6, 5.3] | 2.7 [1.5, 4.1] | 2.7 [1.3, 4.3] |
| a [AU] | 0.0505 [0.0452, 0.0549] | 0.0893 [0.0800, 0.0977] | 0.125 [0.112, 0.137] | 0.156 [0.139, 0.170] | 0.213 [0.191, 0.232] |

| | d | g | Other model parameters | |
|---|---|---|---|---|
| P [days] | 91.61 [90.72, 92.42] | 256.2 [248.3, 270.0] | $\dot{\gamma}$ [m s⁻¹ yr⁻¹] | 2.07 [1.79, 2.33] |
| e | 0.03 [0, 0.23] | 0.08 [0, 0.49] | $\gamma_{HARPS}$ [m s⁻¹] | -30.6 [-34.8, -26.8] |
| K [m s⁻¹] | 1.52 [1.09, 1.95] | 0.95 [0.51, 1.43] | $\gamma_{HIRES}$ [m s⁻¹] | -31.9 [-37.0,, -26.9] |
| $\omega$ [rad] | 0.7 [0, 2$\pi$] | 0.9 [0, 2$\pi$] | $\gamma_{PFS}$ [m s⁻¹] | -25.8 [-28.9, -22.5] |
| $M_0$ [rad] | 3.7 [0, 2$\pi$] | 4.1 [0, 2$\pi$] | $\sigma_{HARPS}$ [m s⁻¹] | 0.92 [0.63, 1.22] |
| | | | $\sigma_{HIRES}$ [m s⁻¹] | 2.56 [0.93, 5.15] |
| $\lambda$ [deg] | 251(126)† | 285(170)† | $\sigma_{PFS}$ [m s⁻¹] | 1.31 [0.00, 3.85] |
| M sin $i$ [M$_\oplus$] | 5.1 [3.4, 6.9] | 4.6 [2.3, 7.2] | | |
| a [AU] | 0.276 [0.246, 0.300] | 0.549 [0.491, 0.601] | | |

**Notes.** † Values allowed in the full range of $\lambda$. Full-width-at-half-maximum of the marginalized posterior is provided to illustrate the most likely range (see Figure 10). * Due to the presence of a strong alias, the orbital period of this candidate could be 53 days instead. Such an alternative orbital solution for planet e is given in Table B.2.

the model (signal g). Its amplitude would correspond to a planet with a minimum mass of 4.6 $M_\oplus$.

When performing a search for a seventh signal, the posterior samplings converged consistently to a global probability maximum at 17 days (M sin $i \sim 1.1$ $M_\oplus$) which improves the model probability by a factor of 230. The global probability maximum containing seven signals corresponds to a solution with a period of 62 days for planet e. This solution has a total probability $\sim 16$ times larger than the one with $P_e = 53$ days. Although such a difference is not large enough to make a final decision on which period is preferred, from now on we will assume that our reference solution is the one with $P_e = 62.2$ days. The log–L periodogram also spotted the same seventh period as the next favored one but only when all seven candidates were assumed to have circular orbits. Given that this seventh signal is very close to the Bayesian detection limit, and based on our experience on the analysis of similar datasets (e.g., GJ 581 Tuomi & Jenkins 2012), we concede that this candidate requires more measurements to be securely confirmed. With a minimum mass of only $\sim 1.1$ $M_\oplus$, it would be among the least massive exoplanets discovered to date.

As a final comment we note that, as in Anglada-Escudé et al. (2012) and Delfosse et al. (2012), a linear trend was always included in the model. The most likely origin of such a trend is gravitational acceleration caused by the central GJ 667AB binary. Assuming a minimum separation of 230 AU, the acceleration in the line-of-sight of the observer can be as large as 3.7 m s$^{-1}$, which is of the same order of magnitude as the observed trend of $\sim 2.2$ m s$^{-1}$ yr$^{-1}$. We remark that the trend (or part of it) could also be caused by the presence of a very long period planet or brown dwarf. Further Doppler follow-up, astrometric measurements, or direct imaging attempts of faint companions might help addressing this question.

In summary, the first five signals are easily spotted using Bayesian criteria and log–L periodograms. The global solution containing seven-Keplerian signals prefers a period of 62.2 days for signal e, which we adopt as our reference solution. Still, a period of 53 days for the same signal cannot be ruled out at the moment. The statistical significance of a 6th periodicity depends on the prior choice for the eccentricity, but the Bayesian odds ratio is high enough to propose it as a genuine Keplerian signal. The statistical significance of the seventh candidate (h) is close to our detection limit and more observations are needed to fully confirm it.

## 6. Activity

In addition to random noise (white or correlated), stellar activity can also generate spurious Doppler periodicities that can mimic planetary signals (e.g., Lovis et al. 2011; Reiners et al. 2013). In this section we investigate whether there are periodic variations in the three activity indices of GJ 667C (S-index, BIS and FWHM are described in Section 3). Our general strategy is the following : if a significant periodicity is detected in any of the indices and such periodicity can be related to any of the candidate signals (same period or aliases), we add a linear correlation term to the model and compute log–L periodograms and new samplings of the parameter space. If the data were better described by the correlation term rather than a genuine Doppler signal, the overall model probability would increase and the planet signal in question should decrease its significance (even disappear).

Log–L periodogram analysis of two activity indices (S-index but specially the FWHM) show a strong periodic variability at 105 days. As discussed in Anglada-Escudé et al. (2012) and

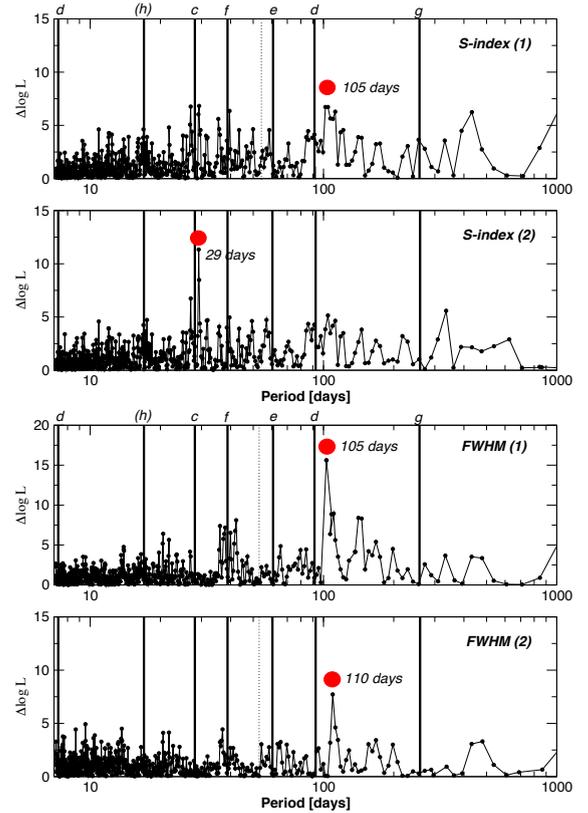

**Fig. 5. Top two panels** Log–L periodograms for up to 2 signals in the S-index. The most likely periods of the proposed planet candidates are marked as vertical lines. **Bottom two panels.** Log–L periodograms for up to 2 signals in the FWHM. Given the proximity of these two signals, it is possible that both of them originate from the same feature (active region corotating with the star) that is slowly evolving with time.

Delfosse et al. (2012), this cast some doubt on the candidate at 91 days (d). Despite the fact that the 91-day and 105-day periods are not connected by first order aliases, the phase sampling is sparse in this period domain (see phase–folded diagrams of the RV data for the planet d candidate in Fig. 4) and the log–L periodogram for this candidate also shows substantial power at 105 days. From the log–L periodograms in Figure 2, one can directly obtain the probability ratio of a putative solution at 91 days versus one with a period of 105 days when no correlation terms are included. This ratio is $6.8 \times 10^4$, meaning that the Doppler period at 91 days is largely favoured over the 105-day one. All Bayesian samplings starting close to the 105-day peak ended-up converging on the signal at 91 day. We then applied our validation procedure by inserting linear correlation terms in the model (g=$C_F \times$ FWHM$_i$ or g=$C_S \times S_i$), in eq. 1) and computed both log–L periodograms and Bayesian samplings with $C_F$ and $C_S$ as free parameters. In all cases the $\Delta \log L$ between 91 and 105 days slightly increased, thus supporting the conclusion that the signal at 91 days is of planetary origin. For example, the ratio of likelihoods between the 91 and 105 day signals increased from $6.8 \times 10^4$ to $3.7 \times 10^6$ when the correlation term with the FWHM was included (see Figure 6). The Bayesian samplings including the correlation term did not improve the model probability (see Appendix C) and still preferred the 91-day signal without any doubt. We conclude that this signal is not directly related to the stellar activity and therefore is a valid planet candidate.

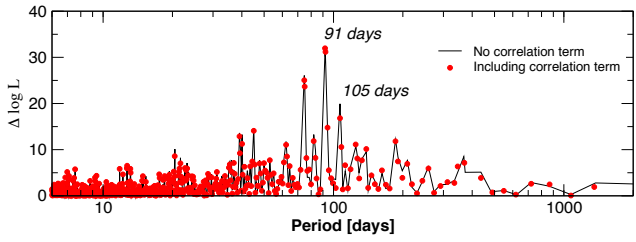

**Fig. 6.** Log–likelihood periodograms for planet d (91 days) including the correlation term (red dots) compared to the original periodogram without this term (black line). The inclusion of the correlation term increases the contrast between the peaks at 91 and 105 days, favoring the interpretation of the 91 days signals as a genuine planet candidate.

Given that activity might induce higher order harmonics in the time-series, all seven candidates have been analyzed and double-checked using the same approach. Some more details on the results from the samplings are given in the Appendix C.2. All candidates -including the tentative planet candidate h- passed all these validation tests without difficulties.

## 7. Tests for quasi-periodic signals

Activity induced signals and superposition of several independent signals can be a source of confusion and result in detections of "apparent" false positives. In an attempt to quantify how much data is necessary to support our preferred global solution (with seven planets) we applied the log-L periodogram analysis method to find the solution as a function of the number of data points. For each dataset, we stopped searching when no peak above FAP 1% was found. The process was fully automated so no human-biased intervention could alter the detection sequence. The resulting detection sequences are show in Table 5. In addition to observing how the complete seven-planet solution slowly emerges from the data one can notice that for $N_{obs} <100$ the $k = 2$ and $k = 3$ solutions converge to a strong signal at $\sim 100$ days. This period is dangerously close to the activity one detected in the FWHM and S-index time-series. To explore what could be the cause of this feature (perhaps the signature of a quasi-periodic signal), we examined the $N_{obs}=75$ case in more detail and made a supervised/visual analysis of that subset.

The first 7.2 days candidate could be easily extracted. We then computed a periodogram of the residuals to figure out if there were additional signals left in the data. In agreement with the automatic search, the periodogram of the residuals (bottom of Figure 7) show a very strong peak at ∼100 days. The peak was so strong that we went ahead and assessed its significance. It had a very low FAP (< 0.01%) and also satisfied our Bayesian detectability criteria. We could have searched for additional companions, but let us assume we stopped our analysis here. In this case, we would have concluded that two signals were strongly present (7.2 days and 100 days). Because of the proximity with a periodicity in the FWHM-index ( 105 days), we would have expressed our doubts about the reality of the second signal so only one planet candidate would have been proposed (GJ 667Cb).

With 228 RV measurements in hand (173 HARPS-TERRA, 23 PFS and 22 from HIRES) we know that such a conclusion is no longer compatible with the data. For example, the second and third planets are very consistently detected at 28 and 91 days. We investigated the nature of that 100 day signal using synthetic subsets of observations as follows. We took our preferred seven-planet solution and generated the exact signal we would expect if we only had planet c (28 days) in the first 75 HARPS-TERRA measurements (without noise). The periodogram of such a noiseless time-series (top panel in Fig. 7) was very different from the real one. Then, we sequentially added the rest of the signals. As more planets were added, the periodogram looked closer to the one from the real data. This illustrates that we would have reached the same wrong conclusion even with data that had negligible noise. How well the general structure of the periodogram was recovered after adding all of the signals is rather remarkable (comparing the bottom two panels in Fig. 7). While this is not a statistical proof of significance, it shows that the *periodogram* of the residuals from the 1-planet fit (even with only 75 RVs measurements) is indeed consistent with the proposed seven-planet solution without invoking the presence of quasi-periodic signals. This experiment also shows that, until all stronger signals could be well-decoupled (more detailed investigation showed this happened at about $N_{obs} \sim 140$), proper and robust identification of the correct periods was very difficult. We repeated the same exercise with $N_{obs}=100$, 120 and 173 (all HARPS measurements) and obtained identical behavior without exception (see panels in Figure 8). Such an effect is not new and the literature contains several examples that cannot be easily explained by simplistic aliasing arguments- e.g., see GJ 581d (Udry et al. 2007; Mayor et al. 2009) and HD 125612b/c (Anglada-Escudé et al. 2010; Lo Curto et al. 2010). The fact that all signals detected in the velocity data of GJ 667C have similar amplitudes – except perhaps candidate b which has a considerably higher amplitude – made this problem especially severe. In this sense, the currently available set of observations are a sub-sample of the many more that might be obtained in the future, so it might happen that one of the signals we report "bifurcates" into other periodicities. This experiment also suggests that spectral information beyond the most trivial aliases can be used to verify and assess the significance of future detections (under investigation).

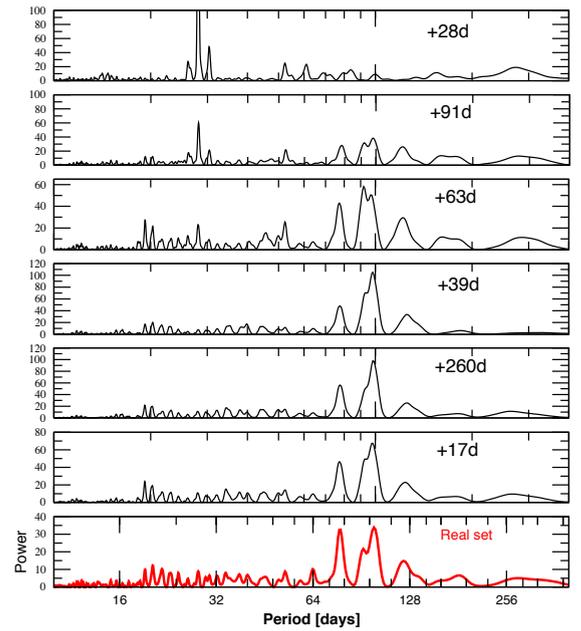

**Fig. 7.** Sequence of periodograms obtained from synthetic noiseless data generated on the first 75 epochs. The signals in Table 4 were sequentially injected from top to bottom. The bottom panel is the periodogram to the real dataset after removing the first 7.2 days planet candidate.

**Table 5.** Most significant periods as extracted using log–L periodograms on subsamples of the first $N_obs$ measurements. Boldfaced values indicate coincidence with a signal of our seven-planet solution (or a first order yearly alias thereof). A parenthesis in the last period indicates a preferred period that did not satisfy the frequentist 1% FAP threshold but did satisfy the Bayesian detectability criteria.

| $N_{obs}$ | 1 | 2 | 3 | 4 | 5 | 6 | 7 |
|---|---|---|---|---|---|---|---|
| 50 | **7.2** | 101.5 | – | – | – | – | – |
| 75 | **7.2** | 103.0 | – | – | – | – | – |
| 90 | **7.2** | **28.0** | 104.1 | – | – | – | – |
| 100 | **7.2** | **91.2** | **28.0** | **54.4**[a] | – | – | – |
| 120 | **7.2** | **91.6** | **28.0** | – | – | – | – |
| 143 | **7.2** | **91.6** | **28.0** | **53.6**[a] | **35.3**[a] | **(260)** | – |
| 160 | **7.2** | **28.1** | **91.0** | **38.9** | **53.4**[a] | **275** | **(16.9)** |
| 173 | **7.2** | **28.1** | **91.9** | **61.9** | **38.9** | **260** | **(16.9)** |

**Notes.** [a] 1 year$^{-1}$ alias of the preferred period in Table 4.

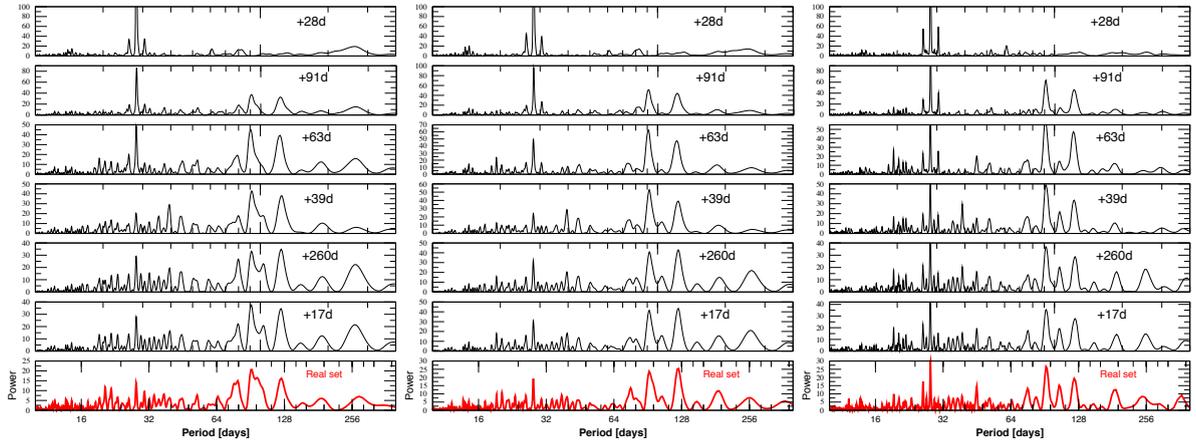

**Fig. 8.** Same as 7 but using the first 100 epochs (left), first 120 (center) and all of them (right).

## 7.1. Presence of individual signals in one half of the data

As an additional verification against quasi-periodicity, we investigated if the signals were present in the data when it was divided into two halves. The first half corresponds to the first 86 HARPS observations and the second half covers the remaining data. The data from PFS and HIRES were not used for this test. The experiment consists of removing all signals except for one, and then computing the periodogram on those residuals (first and second halfs separately). If a signal is strongly present in both halfs, it should, at least, emerge as substantially significant. All signals except for the seventh one passed this test nicely. That is, in all cases except for h, the periodograms prominently display the non-removed signal unambiguously. Besides demonstrating that all signals are strongly present in the two halves, it also illustrates that any of the candidates would have been trivially spotted using periodograms if it had been orbiting alone around the star. The fact that each signal was not spotted before (Anglada-Escudé et al. 2012; Delfosse et al. 2012) is a consequence of an inadequate treatment of signal correlations when dealing with periodograms of the residuals only. Both the described Bayesian method and the log-likelihood periodogram technique are able to deal with such correlations by identifying the combined global solution at the period search level. As for other multiplanet systems detected using similar techniques (Tuomi et al. 2013; Anglada-Escudé & Tuomi 2012), optimal exploration of the global probability maxima at the signal search level is essential to properly detect and assess the significance of low mass multiplanet solutions, especially when several signals have similar amplitudes close to the noise level of the measurements.

Summarizing these investigations and validation of the signals against activity, we conclude that

- Up to seven periodic signals are detected in the Doppler measurements of GJ 667C data, with the last (seventh) signal very close to our detection threshold.
- The significance of the signals are not affected by correlations with activity indices and we could not identify any strong wavelength dependence with any of them.
- The first six signals are strongly present in subsamples of the data. Only the seventh signal is unconfirmed using half of the data alone. Our analysis indicates that any of the six stronger signals would have been robustly spotted with half the available data if each had been orbiting alone around the host star.
- Signal correlations in unevenly sampled data are the reason why Anglada-Escudé et al. (2012) and Delfosse et al. (2012) only reported three of them. This is a known problem when assessing the significance of signals using periodograms of residuals only (see Anglada-Escudé & Tuomi 2012, as another example).

Given the results of these validation tests, we promote six of the signals (b, c, d, e, f, g) to planet candidates. For economy of language, we will refer to them as *planets* but the reader must remember that, unless complementary and independent verification of a Doppler signal is obtained (e.g., transits), they should be called *planet candidates*. Verifying the proposed planetary system against dynamical stability is the purpose of the next section.

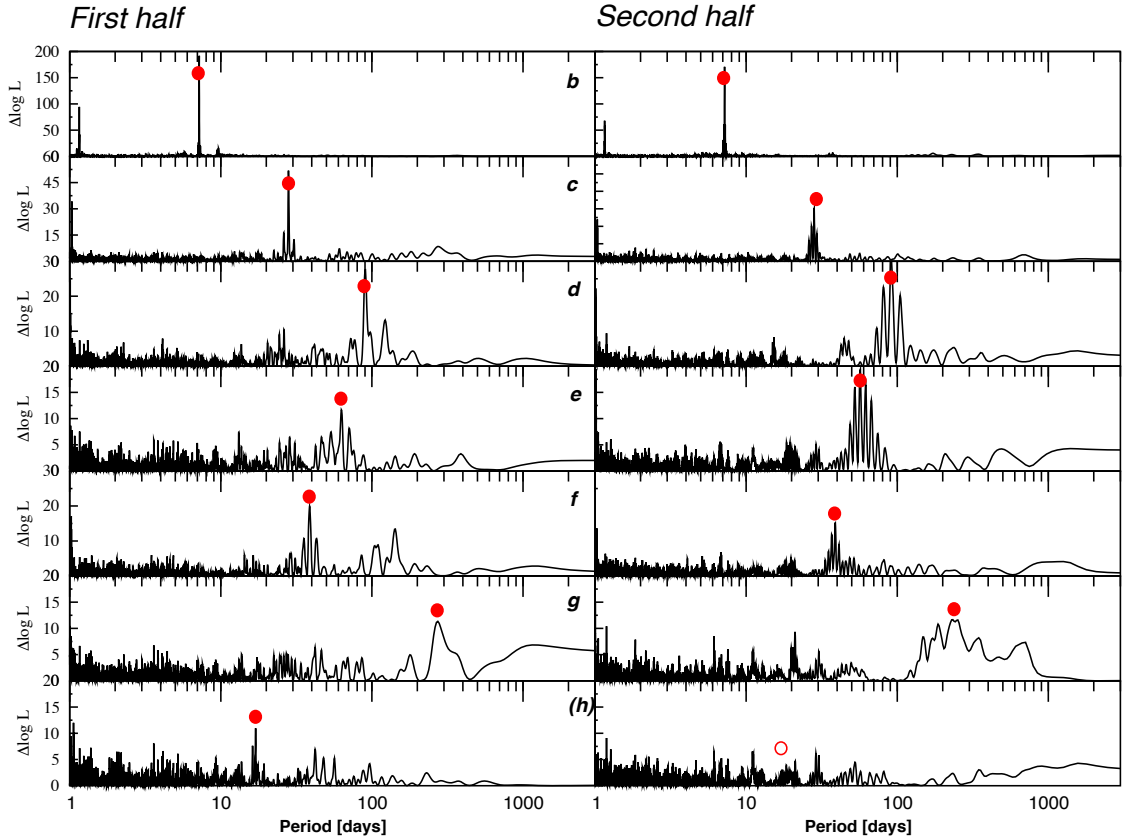

**Fig. 9.** Periodograms on first and second half of the time series as obtained when all signals except one were removed from the data. Except for signal h, all signals are significantly present in both halves and could have been recovered using either half if they had been in single planet systems.

## 8. Dynamical analysis

One of the by-products of the Bayesian analysis described in the previous sections, are numerical samples of statistically allowed parameter combinations. The most likely orbital elements and corresponding confidence levels can be deduced from these samples. In Table 4 we give the orbital configuration for a planetary system with seven planets around GJ 667C, which is preferred from a statistical point of view. To be sure that the proposed planetary system is physically realistic, it is necessary to confirm that these parameters not only correspond to the solution favored by the data, but also constitute a dynamically stable configuration. Due to the compactness of the orbits, abundant resonances and therefore complex interplanetary interactions are expected within the credibility intervals. To slightly reduce this complexity and since evidence for planet h is weak, we split our analysis and present- in the first part of this section- the results for the six-planet solution with planets b to g. The dynamical feasibility of the seven-planet solution is then assessed by investigating the semi-major axis that would allow introducing a seventh planet with the characteristics of planet h.

### 8.1. Finding stable solutions for six planets

A first thing to do is to extract from the Bayesian samplings those orbital configurations that allow stable planetary motion over long time scales, if any. Therefore we tested the stability of each configuration by a separate numerical integration using the symplectic integrator SABA$_2$ of Laskar & Robutel (2001) with a step size $\tau = 0.0625$ days. In the integration, we included a correction term to account for general relativistic precession. Tidal effects were neglected for these runs. Possible effects of tides are discussed separately in Section 9.4. The integration was stopped if any of the planets went beyond 5 AU or planets approached each other closer than $10^{-4}$ AU.

The stability of those configurations that survived the integration time span of $10^4$ orbital periods of planet g (i. e. $\approx 7000$ years), was then determined using frequency analysis (Laskar 1993). For this we computed the stability index $D_k$ for each planet $k$ as the relative change in its mean motion $n_k$ over two consecutive time intervals as was done in Tuomi et al. (2013). For stable orbits the computed mean motion in both intervals will be almost the same and therefore $D_k$ will be very small. We note that this also holds true for planets captured inside a mean-motion resonance (MMR), as long as this resonance helps to stabilize the system. As an index for the total stability of a configuration we used $D = \max(|D_k|)$. The results are summarized in Figure 10. To generate Figure 10, we extracted a sub-sample of 80,000 initial conditions from the Bayesian samplings. Those configurations that did not reach the final integration time are represented as gray dots. By direct numerical integration of the remaining initial conditions, we found that almost all configurations with $D < 10^{-5}$ survive a time span of 1 Myr. This corresponds to ~0.3 percent of the total sample. The most stable orbits we found ($D < 10^{-6}$) are depicted as black crosses.

In Figure 10 one can see that the initial conditions taken from the integrated 80,000 solutions are already confined to a very narrow range in the parameter space of all possible orbits. This means that the allowed combinations of initial $a$ and $\lambda$ are already quite restricted by the statistics. By examining Figure 10 one can also notice that those initial conditions that turned out to be long-term stable are quite spread out along the areas where the density of Bayesian states is higher. Also, for some of the candidates (d, f and g), there are regions were no orbit was found with D<$10^{-5}$. The paucity of stable orbits at certain regions indicate areas of strong chaos within the statistically allowed ranges (likely disruptive mean-motion resonances) and illustrate that the dynamics of the system are far from trivial.

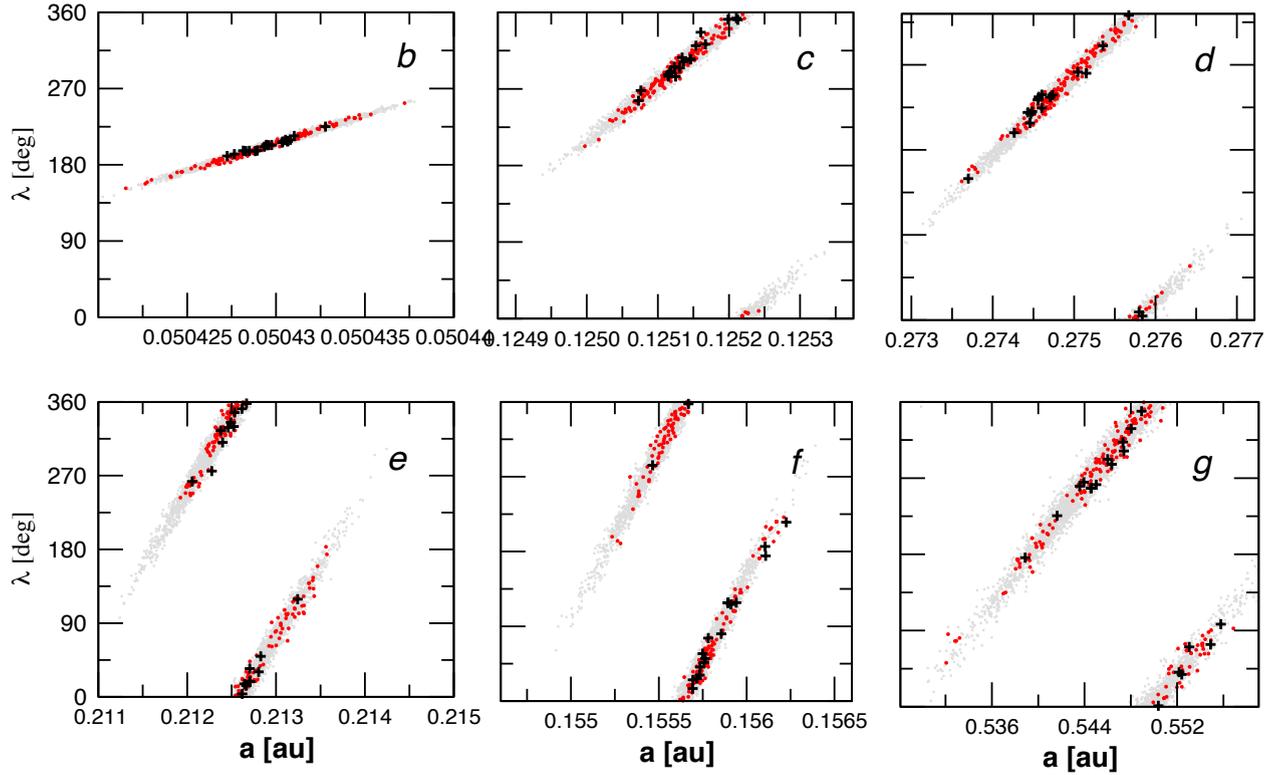

**Fig. 10.** Result of the stability analysis of 80,000 six-planet solutions in the plane of initial semi-major axis $a$ vs. initial mean longitude $\lambda$ obtained from a numerical integration over $T \approx 7000$ years. Each initial condition is represented as a point. Initial conditions leading to an immediate disintegration of the system are shown as gray dots. Initial conditions that lead to stable motion for at least 1 Myr are shown as red points (D< $10^{-5}$). Black crosses represent the most stable solutions (D< $10^{-6}$), and can last over many Myr.

**Table 6.** Astrocentric orbital elements of solution $S_6$.

| Planet | $P$ (d) | $a$ (AU) | $e$ | $\omega$ (°) | $M_0$ (°) | $M \sin i$ ($M_\oplus$) |
|--------|---------|----------|-----|--------------|-----------|-------------------------|
| b | 7.2006 | 0.05043 | 0.112 | 4.97 | 209.18 | 5.94 |
| c | 28.1231 | 0.12507 | 0.001 | 101.38 | 154.86 | 3.86 |
| f | 39.0819 | 0.15575 | 0.001 | 77.73 | 339.39 | 1.94 |
| e | 62.2657 | 0.21246 | 0.001 | 317.43 | 11.32 | 2.68 |
| d | 92.0926 | 0.27580 | 0.019 | 126.05 | 243.43 | 5.21 |
| g | 251.519 | 0.53888 | 0.107 | 339.48 | 196.53 | 4.41 |

The distributions of eccentricities are also strongly affected by the condition of dynamical stability. In Figure 11 we show the marginalized distributions of eccentricities for the sample of all the integrated orbits (gray histograms) and the distribution restricted to relatively stable orbits (with D< $10^{-5}$, red histograms). We see that, as expected, stable motion is only possible with eccentricities smaller than the mean values allowed by the statistical samples. The only exceptions are planets b and g. These two planet candidates are well separated from the other candidates. As a consequence, their probability densities are rather unaffected by the condition of long-term stability. We note here that the information about the dynamical stability has been used only *a posteriori*. If we had used long-term dynamics as a prior (e.g., assign 0 probability to orbits with D>$10^{-5}$), moderately eccentric orbits would have been much more strongly suppressed than with our choice of prior function (Gaussian distribution of zero mean and $\sigma = 0.3$, see Appendix A.1). In this sense, our prior density choice provides a much softer and uninformative constraint than the dynamical viability of the system.

In the following we will use the set of initial conditions that gave the smallest $D$ for a detailed analysis and will refer to it as $S_6$. In Table 6, we present the masses and orbital parameters of $S_6$, and propose it as the favored configuration. To double check our dynamical stability results, we also integrated $S_6$ for $10^8$ years using the HNBody package (Rauch & Hamilton 2002) including general relativistic corrections and a time step of $\tau = 10^{-3}$ years.[1]

---

[1] Publicly available at http://janus.astro.umd.edu/HNBody/

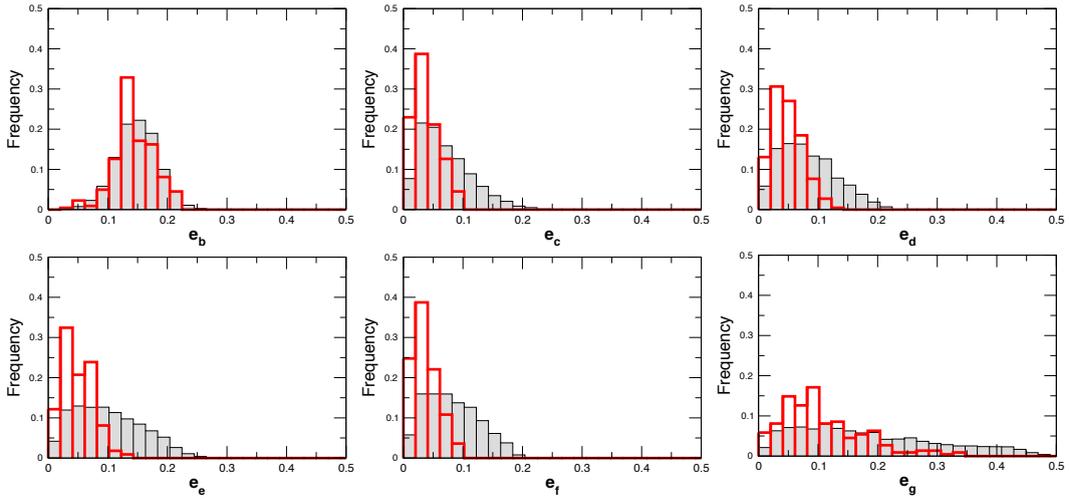

**Fig. 11.** Marginalized posterior densities for the orbital eccentricities of the six planet solution (b, c, d, in the first row; e, f, g in the second) before (gray histogram) and after (red histogram) ruling out dynamically unstable configurations.

### 8.2. Secular evolution

Although the dynamical analysis of such a complex system with different, interacting resonances could be treated in a separate paper, we present here a basic analysis of the dynamical architecture of the system. From studies of the Solar System, we know that, in the absence of mean motion resonances, the variations in the orbital elements of the planets are governed by the so-called secular equations. These equations are obtained after averaging over the mean longitudes of the planets. Since the involved eccentricities for GJ 667C are small, the secular system can be limited here to its linear version, which is usually called a Laplace-Lagrange solution (for details see Laskar (1990)). Basically, the solution is obtained from a transformation of the complex variables $z_k = e_k e^{i\varpi_k}$ into the proper modes $u_k$. Here, $e_k$ are the eccentricities and $\varpi_k$ the longitudes of the periastron of planet $k = b, c, \ldots, g$. The proper modes $u_k$ determine the secular variation of the eccentricities and are given by $u_k \approx e^{i(g_k t + \phi_k)}$.

Since the transformation into the proper modes depends only on the masses and semi-major axes of the planets, the secular frequencies will not change much for the different stable configurations found in Figure 10. Here we use solution $S_6$ to obtain numerically the parameters of the linear transformation by a frequency analysis of the numerically integrated orbit. The secular frequencies $g_k$ and the phases $\phi_k$ are given in Table 7. How well the secular solution describes the long-term evolution of the eccentricities can be readily seen in Figure 12.

**Table 7.** Fundamental secular frequencies $g_k$, phases $\phi_k$ and corresponding periods of the six-planet solution.

| $k$ | $g_k$ [deg/yr] | $\phi_k$ [deg] | Period [yr] |
|---|---|---|---|
| 1 | 0.071683 | 356.41 | 5022.09 |
| 2 | 0.184816 | 224.04 | 1947.88 |
| 3 | 0.991167 | 116.46 | 363.21 |
| 4 | 0.050200 | 33.63 | 7171.37 |
| 5 | 0.656733 | 135.52 | 548.17 |
| 6 | 0.012230 | 340.44 | 29435.80 |

From Figure 12, it is easy to see that there exists a strong secular coupling between all the inner planets. From the Laplace-

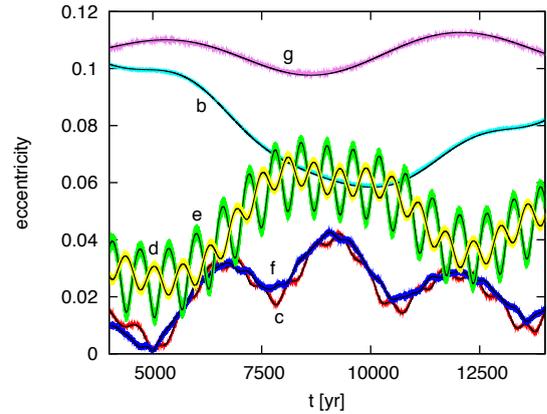

**Fig. 12.** Evolution of the eccentricities of solution $S_6$. Colored lines give the eccentricity as obtained from a numerical integration. The thin black lines show the eccentricity of the respective planet as given by the linear, secular approximation. Close to each line we give the name of the corresponding planet.

Lagrange solution, we find that the long-term variation of the eccentricities of these planets is determined by the secular frequency $g_1 - g_4$ with a period of $\approx 17000$ years. Here, the variation in eccentricity of planet b is in anti-phase to that of planets c to f due to the exchange of orbital angular momentum. On shorter time scales, we easily spot in Figure 12 the coupling between planets d and e with a period of $\approx 600$ years ($g_1 - g_5$), while the eccentricities of planets c and f vary with a period of almost 3000 years ($g_1 + g_4$). Such couplings are already known to prevent close approaches between the planets (Ferraz-Mello et al. 2006). As a result, the periastron of the planets are locked and the difference $\Delta\varpi$ between any of their $\varpi$ librates around zero.

Although the eccentricities show strong variations, these changes are very regular and their maximum values remain bounded. From the facts that 1) the secular solution agrees so well with numerically integrated orbits, and 2) at the same time the semi-major axes remain nearly constant (Table 8), we can conclude that $S_6$ is not affected by strong MMRs.

Nevertheless, MMRs that can destabilize the whole system are within the credibility intervals allowed by the samplings and not far away from the most stable orbits. Integrating some of the initial conditions marked as chaotic in Figure 10 one finds that, for example, planets d and g are in some of these cases temporarily trapped inside a 3:1 MMR, causing subsequent disintegration of the system.

**Table 8.** Minimum and maximum values of the semi-major axes and eccentricities during a run of $S_6$ over 10 Myr.

| $k$ | $a_{min}$ | $a_{max}$ | $e_{min}$ | $e_{max}$ |
|---|---|---|---|---|
| b | 0.050431 | 0.050433 | 0.035 | 0.114 |
| c | 0.125012 | 0.125135 | 0.000 | 0.060 |
| f | 0.155582 | 0.155922 | 0.000 | 0.061 |
| e | 0.212219 | 0.212927 | 0.000 | 0.106 |
| d | 0.275494 | 0.276104 | 0.000 | 0.087 |
| g | 0.538401 | 0.539456 | 0.098 | 0.116 |

### 8.3. Including planet h

After finding a non-negligible set of stable six-planet solutions, it is tempting to look for more planets in the system. From the data analysis, one even knows the preferred location of such a planet. We first considered doing an analysis similar to the one for the six-planet case using the Bayesian samples for the seven-planet solution. As shown in previous sections, the subset of stable solutions found by this approach is already small compared to the statistical samples in the six-planet case (~ 0.3%). Adding one more planet (five extra dimensions) can only shrink the relative volume of stable solutions further. Given the large uncertainties on the orbital elements of h, we considered this approach too computationally expensive and inefficient.

As a first approximation to the problem, we checked whether the distances between neighboring planets are still large enough to allow stable motion. In Chambers et al. (1996) the mean lifetime for coplanar systems with small eccentricities is estimated as a function of the mutual distance between the planets, their masses and the number of planets in the system. From their results, we can estimate the expected lifetime for the seven-planet solution to be at least $10^8$ years.

Motivated by this result, we explored the phase space around the proposed orbit for the seventh planet. To do this, we use solution $S_6$ and placed a fictitious planet with 1.1 $M_\oplus$ (the estimated mass of planet h as given in Table 4) in the semi-major axis range between 0.035 and 0.2 AU (step size of 0.001 AU) varying the eccentricity between 0 and 0.2 (step size of 0.01). The orbital angles $\omega$ and $M_0$ were set to the values of the statistically preferred solution for h (see Table 4). For each of these initial configurations, we integrated the system for $10^4$ orbits of planet g and analyzed stability of the orbits using the same secular frequency analysis. As a result, we obtained a value of the chaos index $D$ at each grid point. Figure 13 shows that the putative orbit of h appears right in the middle of the only island of stability left in the inner region of the system. By direct numerical integration of solution $S_6$ together with planet h at its nominal position, we found that such a solution is also stable on Myr timescales. With this we conclude that the seventh signal detected by the Bayesian analysis also belongs to a physically viable planet that might be confirmed with a few more observations.

### 8.4. An upper limit for the masses

Due to the lack of reported transit, only the minimum masses are known for the planet candidates. The true masses depend on the unknown inclination $i$ of the system to the line-of-sight. In all the analysis presented above, we implicitly assume that the GJ 667C system is observed edge-on ($i = 90°$) and that all true masses are equal to the minimum mass M sin $i$. As shown in the discussion on the dynamics, the stability of the system is fragile in the sense that dynamically unstable configurations can be found close in the parameter space to the stable ones. Therefore, it is likely that a more complete analysis could set strong limitations on the maximum masses allowed to each companion. An exploration of the total phase space including mutual inclinations would require too much computational resources and is beyond the scope of this paper. To obtain a basic understanding of the situation, we only search for a constraint on the maximum masses of the $S_6$ solution assuming co-planarity. Decreasing the inclination of the orbital plane in steps of 10°, we applied the frequency analysis to the resulting system. By making the planets more massive, the interactions between them become stronger, with a corresponding shrinking of the areas of stability. In this experiment, we found that the system remained stable for at least one Myr for an inclination down to $i = 30°$. If this result can be validated by a much more extensive exploration of the dynamics and longer integration times (in prep.), it would confirm that the masses of all the candidates are within a factor of 2 of the minimum masses derived from Doppler data. Accordingly, c,f and e would be the first dynamically confirmed super-Earths (true masses below 10 $M_\oplus$) in the habitable zone of a nearby star.

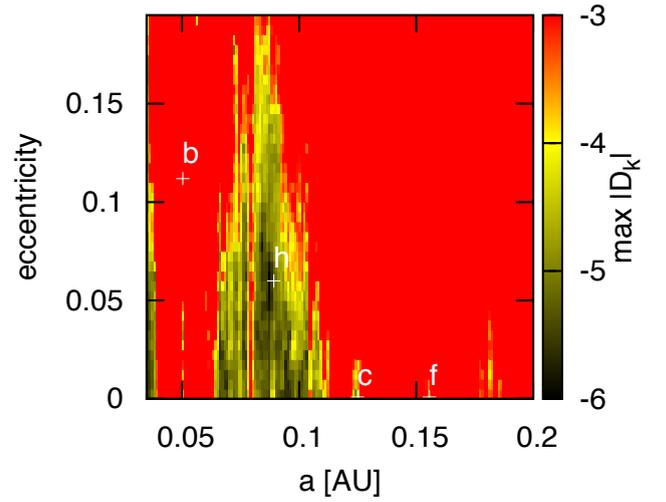

**Fig. 13.** Stability plot of the possible location of a 7th planet in the stable $S_6$ solution (Table 5). We investigate the stability of an additional planet with 1.1 Earth masses around the location found by the Bayesian analysis. For these integrations, we varied the semi-major axis and eccentricity of the putative planet on a regular grid. The orbital angles $\omega$ and $M_0$ were set to the values of the statistically preferred solution, while the inclination was fixed to zero. The nominal positions of the planets as given in Table 6 are marked with white crosses.

## 9. Habitability

Planets h–d receive 20–200% of the Earth's current insolation, and hence should be evaluated in terms of potential habitability. Traditionally, analyses of planetary habitability begin with determining if a planet is in the habitable zone (Dole 1964; Hart 1979; Kasting et al. 1993; Selsis et al. 2007; Kopparapu et al. 2013), but many factors are relevant. Unfortunately, many aspects can-not presently be determined due to the limited characterization derivable from RV observations. However, we can explore the issues quantitatively and identify those situations in which habitability is precluded, and hence determine which of these planets *could* support life. In this section we provide a preliminary analysis of each potentially habitable planet in the context of previous results, bearing in mind that theoretical predictions of the most relevant processes cannot be constrained by existing data.

### 9.1. The Habitable Zone

The HZ is defined at the inner edge by the onset of a "moist greenhouse," and at the outer edge by the "maximum greenhouse" (Kasting et al. 1993). Both of these definitions assume that liquid surface water is maintained under an Earth-like atmosphere. At the inner edge, the temperature near the surface becomes large enough that water cannot be confined to the surface and water vapor saturates the stratosphere. From there, stellar radiation can dissociate the water and hydrogen can escape. Moreover, as water vapor is a greenhouse gas, large quantities in the atmosphere can heat the surface to temperatures that forbid the liquid phase, rendering the planet uninhabitable. At the outer edge, the danger is global ice coverage. While greenhouse gases like $CO_2$ can warm the surface and mitigate the risk of global glaciation, $CO_2$ also scatters starlight via Rayleigh scattering. There is therefore a limit to the amount of $CO_2$ that can warm a planet as more $CO_2$ actually cools the planet by increasing its albedo, assuming a moist or runaway greenhouse was never triggered.

We use the most recent calculations of the HZ (Kopparapu et al. 2013) and find, for a 1 Earth-mass planet, that the inner and outer boundaries of the habitable zone for GJ 667C lie between 0.095–0.126 AU and 0.241–0.251 AU respectively. We will adopt the average of these limits as a working definition of the HZ: $0.111 - 0.246$ AU. At the inner edge, larger mass planets resist the moist greenhouse and the HZ edge is closer in, but the outer edge is almost independent of mass. Kopparapu et al. (2013) find that a 10 $M_\oplus$ planet can be habitable 5% closer to the star than a 1 $M_\oplus$ planet. However, we must bear in mind that the HZ calculations are based on 1-dimensional photochemical models that may not apply to slowly rotating planets, a situation likely for planets c, d, e, f and h (see below).

From these definitions, we find that planet candidate h ($a = 0.0893$ AU) is too hot to be habitable, but we note its semi-major axis is consistent with the most optimistic version of the HZ. Planet c ($a = 0.125$ AU) is close to the inner edge but is likely to be in the HZ, especially since it has a large mass. Planets f and e are firmly in the HZ. Planet d is likely beyond the outer edge of the HZ, but the uncertainty in its orbit prevents a definitive assessment. Thus, we conclude that planets c, f, and e are in the HZ, and planet d might be, *i.e.* there up to four potentially habitable planets orbiting GJ 667C.

Recently, Abe et al. (2011) pointed out that planets with small, but non-negligible, amounts of water have a larger HZ than Earth-like planets. From their definition, both h and d are firmly in the HZ. However, as we discuss below, these planets are likely water-rich, and hence we do not use the Abe et al. (2011) HZ here.

### 9.2. Composition

Planet formation is a messy process characterized by scattering, migration, and giant impacts. Hence precise calculations of planetary composition are presently impossible, but see Bond et al. (2010); Carter-Bond et al. (2012) for some general trends. For habitability, our first concern is discerning if a planet is rocky (and potentially habitable) or gaseous (and uninhabitable). Unfortunately, we cannot even make this rudimentary evaluation based on available data and theory. Without radii measurements, we cannot determine bulk density, which could discriminate between the two. The least massive planet known to be gaseous is GJ 1214 b at 6.55 $M_\oplus$ (Charbonneau et al. 2009), and the largest planet known to be rocky is Kepler-10 b at 4.5 $M_\oplus$ (Batalha et al. 2011). Modeling of gas accretion has found that planets smaller than 1 $M_\oplus$ can accrete hydrogen in certain circumstances (Ikoma et al. 2001), but the critical mass is likely larger (Lissauer et al. 2009). The planets in this system lie near these masses, and hence we cannot definitively say if any of these planets are gaseous.

Models of rocky planet formation around M dwarfs have found that those that accrete from primordial material are likely to be sub-Earth mass (Raymond et al. 2007) and volatile-poor (Lissauer 2007). In contrast, the planets orbiting GJ 667C are super-Earths in a very packed configuration summing up to $> 25$ $M_\oplus$ inside 0.5 AU. Therefore, the planets either formed at larger orbital distances and migrated in (*e.g.* Lin et al. 1996), or additional dust and ice flowed inward during the protoplanetary disk phase and accumulated into the planets Hansen & Murray (2012, 2013). The large masses disfavor the first scenario, and we therefore assume that the planets formed from material that condensed beyond the snow-line and are volatile rich. If not gaseous, these planets contain substantial water content, which is a primary requirement for life (and negates the dry-world HZ discussed above). In conclusion, these planets could be terrestrial-like with significant water content and hence are potentially habitable.

### 9.3. Stellar Activity and habitability

Stellar activity can be detrimental to life as the planets can be bathed in high energy photons and protons that could strip the atmosphere or destroy ozone layers. In quiescence, M dwarfs emit very little UV light, so the latter is only dangerous if flares occur frequently enough that ozone does not have time to be replenished (Segura et al. 2010). As already discussed in Section 2, GJ 667C appears to be relatively inactive (indeed, we would not have been able to detect planetary signals otherwise), and so the threat to life is small today. If the star was very active in its youth- with mega-flares like those on the equal mass star AD Leo (Hawley & Pettersen 1991)- any life on the surface of planets might have been difficult during those days (Segura et al. 2010). While M dwarfs are likely to be active for much longer time than the Sun (West et al. 2008; Reiners & Mohanty 2012), GJ 667C is not active today and there is no reason to assume that life could not form after an early phase of strong stellar activity.

### 9.4. Tidal Effects

Planets in the HZ of low mass stars may be subject to strong tidal distortion, which can lead to long-term changes in orbital and spin properties (Dole 1964; Kasting et al. 1993; Barnes et al. 2008; Heller et al. 2011), and tidal heating (Jackson et al. 2008; Barnes et al. 2009, 2013). Both of these processes can affect hab-itability, so we now consider tidal effects on planets c, d, e, f and h.

| | CPL | | | | CTL | | | |
|---|---|---|---|---|---|---|---|---|
| | base | | max | | base | | max | |
| | $t_{lock}$ | $t_{ero}$ | $t_{lock}$ | $t_{ero}$ | $t_{lock}$ | $t_{ero}$ | $t_{lock}$ | $t_{ero}$ |
| h | 0.07 | 0.08 | 18.2 | 20.4 | 0.55 | 0.77 | 66.9 | 103 |
| c | 0.62 | 0.69 | 177 | 190 | 4.7 | 8.1 | 704 | 1062 |
| f | 2.2 | 2.3 | 626 | 660 | 18.5 | 30.1 | 2670 | 3902 |
| e | 14.2 | 15.0 | 4082 | 4226 | 129 | 210 | $> 10^4$ | $> 10^4$ |
| d | 70.4 | 73 | $> 10^4$ | $> 10^4$ | 692 | 1094 | $> 10^4$ | $> 10^4$ |

**Table 9.** Timescales for the planets' tidal despinning in units of Myr. "CPL" denotes the constant-phase-lag model of Ferraz-Mello et al. (2008), "CTL" labels the constant-time-lag model of Leconte et al. (2010).

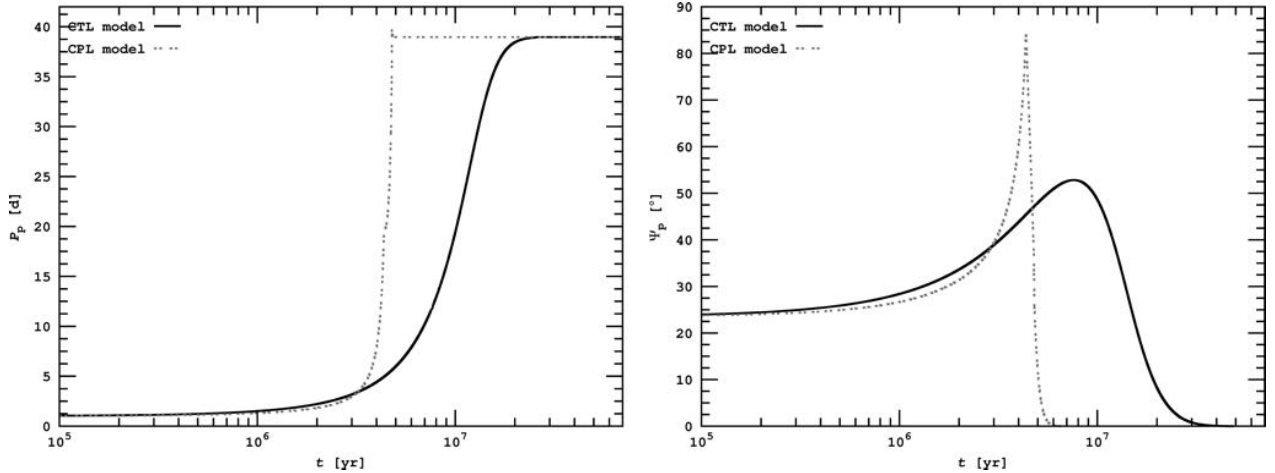

**Fig. 14.** Tidal evolution of the spin properties of planet GJ 667C f. Solid lines depict predictions from constant-time-lag theory ("CTL"), while dashed lines illustrate those from a constant-phase-lag model ("CPL"). All tracks assume a scenario similar to the "base" configuration (see text and Table 9). *Left*: Despinning for an assumed initial rotation period of one day. The CPL model yields tidal locking in less than 5 Myr, and CTL theory predicts about 20 Myr for tidal locking. *Right*: Tilt erosion of an assumed initial Earth-like obliquity of 23.5°. Time scales for both CPL and CTL models are similar to the locking time scales.

Tides will first spin-lock the planetary spin and drive the obliquity to either 0 or $\pi$. The timescale for these processes is a complex function of orbits, masses, radii and spins, (see *e.g.* Darwin 1880; Hut 1981; Ferraz-Mello et al. 2008; Leconte et al. 2010) but for super-Earths in the HZ of a $\sim 0.3\ M_\oplus$ star, Heller et al. (2011) found that tidal locking should occur in $10^6$–$10^9$ years. We have used both the constant-time-lag and constant-phase-lag tidal models described in Heller et al. (2011) and Barnes et al. (2013) (see also Ferraz-Mello et al. 2008; Leconte et al. 2010), to calculate how long tidal locking could take for these planets. We consider two possibilities. Our baseline case is very similar to that of Heller et al. (2011) in which the planets initially have Earth-like properties: a 1-day rotation period, an obliquity of 23.5° and the current tidal dissipation of the Earth (a tidal $Q$ of 12-Yoder (1995) or time lag of 638 s-Lambeck (1977); Neron de Surgy & Laskar (1997)). We also consider an extreme, but plausible, case that maximizes the timescale for tidal locking: 8-hour rotation period, obliquity of 89.9° and a tidal $Q$ of 1000 or time lag of 6.5 s. In Table 9 we show the time for the obliquity to erode to 1°, $t_{ero}$, and the time to reach the pseudo-synchronous rotation period, $t_{lock}$.

In Figure 14, we depict the tidal evolution of the rotation period (left panel) and obliquity (right panel) for planet f as an example. The assumed initial configuration is similar to the "base"

scenario. Time scales for rotational locking and tilt erosion are similar to those shown in Table 9.[2]

As these planets are on nearly circular orbits, we expect tidally-locked planets to be synchronously rotating, although perhaps when the eccentricity is relatively large pseudo-synchronous rotation could occur (Goldreich 1966; Murray & Dermott 1999; Barnes et al. 2008; Correia et al. 2008; Ferraz-Mello et al. 2008; Makarov & Efroimsky 2013). From Table 9 we see that all the planets h–f are very likely synchronous rotators, planet e is likely to be, but planet d is uncertain. Should these planets have tenuous atmospheres (< 0.3 bar), then they may not support a habitable surface (Joshi et al. 1997). Considerable work has shown that thicker atmospheres are able to redistribute dayside heat to produce clement conditions (Joshi et al. 1997; Joshi 2003; Edson et al. 2011; Pierrehumbert 2011; Wordsworth et al. 2011). As we have no atmospheric data, we assert that tidal locking does not preclude habitability for any of the HZ planets.

During the tidal despinning, tidal heat can be generated as dissipation transforms rotational energy into frictional heat. In some cases, the heating rates can be large enough to trigger a runaway greenhouse and render a planet uninhabitable (Barnes et al. 2013). Tidal heating is maximized for large radius planets that rotate quickly and at high obliquity. Using the

---

[2] Note that evolution for the CPL model is faster with our parameterization. In the case of GJ 581 d, shown in Heller et al. (2011), the planet was assumed to be less dissipative in the CPL model ($Q_p = 100$) and evolution in the CPL model was slower.

Leconte et al. (2010) model and the Earth's dissipation, we find that tidal heating of the HZ planets will be negligible for most cases. Consider an extreme version of planet h, which is just interior to the HZ. Suppose it has the same tidal dissipation as the Earth (which is the most dissipative body known), a rotation period of 10 hr, an eccentricity of 0.1, and an obliquity of 80°. The Leconte et al. (2010) model predicts such a planet would have a tidal heat flux of nearly 4000 W m$^{-2}$. However, that model also predicts the flux would drop to only 0.16 W m$^{-2}$ in just $10^6$ years. The timescale for a runaway greenhouse to sterilize a planet is on the order of $10^8$ years (Watson et al. 1981; Barnes et al. 2013), so this burst of tidal heating does not forbid habitability.

After tidal locking, the planet would still have about 0.14 W m$^{-2}$ of tidal heating due to the eccentricity (which, as for the other candidates, can oscillate between 0 and 0.1 due to dynamical interactions). If we assume an Earth-like planet, then about 90% of that heat is generated in the oceans, and 10% in the rocky interior. Such a planet would have churning oceans, and about 0.01 W m$^{-2}$ of tidal heat flux from the rocky interior. This number should be compared to 0.08 W m$^{-2}$, the heat flux on the Earth due entirely to other sources. As $e = 0.1$ is near the maximum of the secular cycle, see § 8, the actual heat flux is probably much lower. We conclude that tidal heating on planet h is likely to be negligible, with the possibility it could be a minor contributor to the internal energy budget. As the other planets are more distant, the tidal heating of those planets is negligible. The CPL predicts higher heating rates and planet c could receive $\sim 0.01$ W m$^{-2}$ of internal heating, but otherwise tidal heating does not affect the HZ planets.

### 9.5. The Weather Forecast

Assuming planets c, f and e have habitable surfaces (see Figure 15), what might their climates be like? To first order we expect a planet's surface temperature to be cooler as semi-major axis increases because the incident flux falls off with distance squared. However, albedo variations can supersede this trend, *e.g.* a closer, high-albedo planet could absorb less energy than a more distant low-albedo planet. Furthermore, molecules in the atmosphere can trap photons near the surface via the greenhouse effect, or scatter stellar light via Rayleigh scattering, an antigreenhouse effect. For example, the equilibrium temperature of Venus is actually lower than the Earth's due to the former's large albedo, yet the surface temperature of Venus is much larger than the Earth's due to the greenhouse effect. Here, we speculate on the climates of each HZ planet based on the our current understanding of the range of possible climates that HZ planets might have.

Certain aspects of each planet will be determined by the redder spectral energy distribution of the host star. For example, the "stratosphere" is expected to be isothermal as there is negligible UV radiation (Segura et al. 2003). On Earth, the UV light absorbed by ozone creates a temperature inversion that delineates the stratosphere. HZ calculations also assume the albedo of planets orbiting cooler stars are lower than the Earth's because Rayleigh scattering is less effective for longer wavelengths, and because the peak emission in the stellar spectrum is close to several $H_2O$ and $CO_2$ absorption bands in the near infrared. Therefore, relative to the Earth's insolation, the HZ is farther from the star. In other words, if we placed the Earth in orbit around an M dwarf such that it received the same incident radiation as the modern Earth, the M dwarf planet would be hotter as it would have a lower albedo. The different character of the light can also impact plant life, and we might expect less productive photosynthesis (Kiang et al. 2007), perhaps relying on pigments such as chlorophyll d (Mielke et al. 2013) or chlorophyll f (Chen et al. 2010).

Planet c is slightly closer to the inner edge of the HZ than the Earth, and so we expect it to be warmer than the Earth, too. It receives 1230 W m$^{-2}$ of stellar radiation, which is actually less than the Earth's solar constant of 1360 W m$^{-2}$. Assuming synchronous rotation and no obliquity, then the global climate depends strongly on the properties of the atmosphere. If the atmosphere is thin, then the heat absorbed at the sub-stellar point cannot be easily transported to the dark side or the poles. The surface temperature would be a strong function of the zenith angle of the host star GJ 667C. For thicker atmospheres, heat redistribution becomes more significant. With a rotation period of $\sim 28$ days, the planet is likely to have Hadley cells that extend to the poles (at least Titan, with a similar rotation period, is a guide), and hence jet streams and deserts would be unlikely. The location of land masses is also important. Should land be concentrated near the sub-stellar point, then silicate weathering is more effective, and cools the planet by drawing down $CO_2$ (Edson et al. 2012).

Planet f is a prime candidate for habitability and receives 788 W m$^{-2}$ of radiation. It likely absorbs less energy than the Earth, and hence habitability requires more greenhouse gases, like $CO_2$ or $CH_4$. Therefore a habitable version of this planet has to have a thicker atmosphere than the Earth, and we can assume a relatively uniform surface temperature. Another possibility is an "eyeball" world in which the planet is synchronously rotating and ice-covered except for open ocean at the sub-stellar point (Pierrehumbert 2011). On the other hand, the lower albedo of ice in the IR may make near-global ice coverage difficult (Joshi & Haberle 2012; Shields et al. 2013).

Planet e receives only a third the radiation the Earth does, and lies close to the maximum greenhouse limit. We therefore expect a habitable version of this planet to have $> 2$ bars of $CO_2$. The planet might not be tidally locked, and may have an albedo that evolves significantly due to perturbations from other planets. From this perspective planet e might be the most Earth-like, experiencing a day-night cycle and seasons.

Finally planet d is unlikely to be in the habitable zone, but it could potentially support sub-surface life. Internal energy generated by, *e.g.*, radiogenic heat could support liquid water below an ice layer, similar to Europa. Presumably the biologically generated gases could find their way through the ice and become detectable bio-signatures, but they might be a very small constituent of a large atmosphere, hampering remote detection. While its transit probability is rather low ($\sim 0.5\%$), its apparent angular separation from the star is $\sim 40$ milliarcseconds. This value is the baseline inner working angle for the Darwin/ESA high-contrast mission being considered by ESA (Cockell et al. 2009) so planet d could be a primary target for such a mission.

### 9.6. Moons in the habitable zone of GJ 667C

In addition to planets, extrasolar moons have been suggested as hosts for life (Reynolds et al. 1987; Williams et al. 1997; Tinney et al. 2011; Heller & Barnes 2013). In order to sustain a substantial, long-lived atmosphere under an Earth-like amount of stellar irradiation (Williams et al. 1997; Kaltenegger 2000), to drive a magnetic field over billions of years (Tachinami et al. 2011), and to drive tectonic activity, Heller & Barnes (2013) concluded that a satellite in the stellar HZ needs a mass $\gtrsim 0.25\,M_\oplus$ in order to maintain liquid surface water.

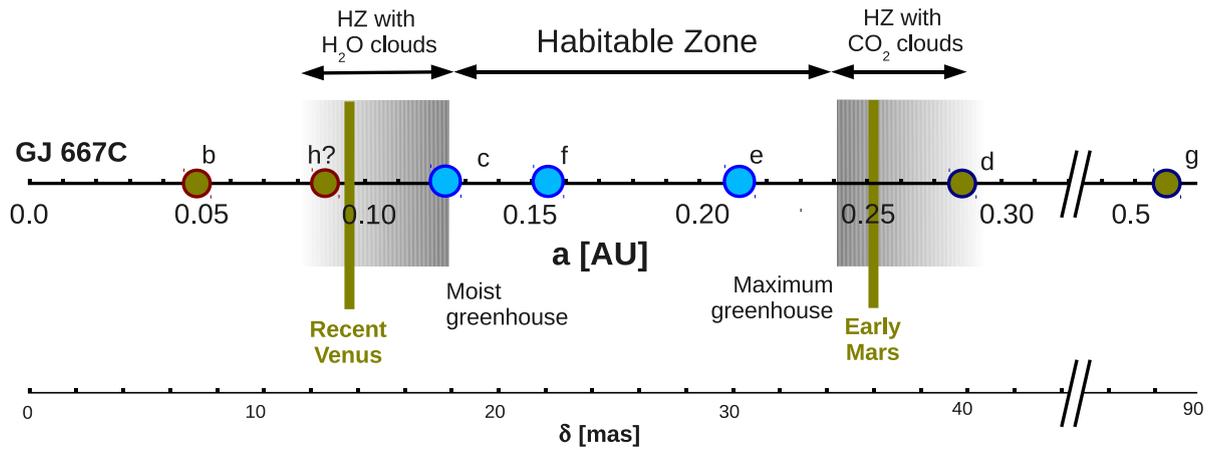

**Fig. 15.** Liquid water habitable zone of GJ 667C with the proposed seven candidates as estimated using the updated relations in Kopparapu et al. (2013). Three of the reported planets lie within the HZ. The newly reported planets f and e are the most comfortably located within it. The inner edge is set by the moist greenhouse runaway limit and the outer edge is set by the snow ball runaway limit. The empirical limits set by a recent uninhabitable Venus and an early habitable Mars are marked in brown solid lines. The presence of clouds of water (inner edge) or $CO_2$ (outer edge) might allow habitable conditions on planets slightly outside the nominal definition of the habitable zone (Selsis et al. 2007).

If potential moons around planets GJ 667C c, f, or e formed in the circumplanetary disk, then they will be much less massive than the most massive satellites in the Solar System (Canup & Ward 2006) and thus not be habitable. However, if one of those planets is indeed terrestrial then impacts could have created a massive moon as happened on Earth (Cameron & Ward 1976). Further possibilities for the formation of massive satellites are summarized in Heller & Barnes (2013, Sect. 2.1).

As the stellar HZ of GJ 667C is very close to this M dwarf star, moons of planets in the habitable zone would have slightly eccentric orbits due to stellar perturbations. These perturbations induce tidal heating and they could be strong enough to prevent any moon from being habitable (Heller 2012). Moons around planet d, which orbits slightly outside the stellar HZ, could offer a more benign environment to life than the planet itself, if they experience weak tidal heating of, say, a few watts per square meter (see Jupiter's moon Io, Reynolds et al. 1987; Spencer et al. 2000).

Unless some of these planets are found to transit, there is no currently available technique to identify satellites (Kipping 2009; Kipping et al. 2012). The RV technique is only sensitive to the combined mass of a planet plus its satellites so it might be possible that some of the planets could be somewhat lighter– but host a massive moon.

## 10. Conclusions

We describe and report the statistical methods and tests used to detect up to seven planet candidates around GJ 667C using Doppler spectroscopy. The detection of the first five planets is very robust and independent of any prior choice. In addition to the first two already reported ones (b and c Anglada-Escudé & Butler 2012; Delfosse et al. 2012) we show that the third planet also proposed in those papers (planet d) is much better explained by a Keplerian orbit rather than an activity-induced periodicity. The next two confidently detected signals (e and f) both correspond to small super-Earth mass objects with most likely periods of 62 and 39 days. The detection of the 6th planet is weakly dependent on the prior choice of the orbital eccentricity. The statistical evidence for the 7th candidate (planet h) is tentative and requires further Doppler follow-up for confirmation. Gregory (2012) proposed a solution for the system with similar characteristics to the one we present here but had fundamental differences. In particular, he also identified the first five stronger signals but his six-planet solution also included a candidate periodicity at 30 days- which would be dynamically unstable- and activity was associated to the signal at 53 days without further discussion or verification. The difference in our conclusions are due to a slightly different choice of priors (especially on the eccentricity), more data was used in our analysis -only HARPS-CCF data was used by Gregory (2012)-, and we performed a more thorough investigation of possible activity-related periodicities.

Numerical integration of orbits compatible with the posterior density distributions show that there is a subset of configurations that allow long-term stable configurations. Except for planets b and g, the condition of dynamical stability dramatically affects the distribution of allowed eccentricities indicating that the lower mass planet candidates (c, e, f) must have quasi-circular orbits. A system of six planets is rather complex in terms of stabilizing mechanisms and possible mean-motion resonances. Nonetheless, we identified that the physically allowed configurations are those that avoid transient 3:1 MMR between planets d and g. We also found that the most stable orbital solutions are well described by the theory of secular frequencies (Laplace-Lagrange solution). We investigated if the inclusion of a seventh planet system was dynamically feasible in the region disclosed by the Bayesian samplings. It is notable that this preliminary candidate appears around the center of stability. Additional data should be able to confirm this signal and provide detectability for longer period signals.

The closely packed dynamics keeps the eccentricities small but non-negligible for the lifetime of the system. As a result, potential habitability of the candidates must account for tidal dissipation effects among others. Dynamics essentially affect 1) the total energy budget at the surface of the planet (tidal heating), 2) synchronization of the rotation with the orbit (tidal locking), and 3) the timescales for the erosion of their obliquities. These dynamical constraints, as well as predictions for potentially habitable super-Earths around M dwarf stars, suggest that at least three planet candidates (planets c, e and f) could have remained habitable for the current life-span of the star. Assuming a rocky composition, planet d lies slightly outside the cold edge of the stellar HZ. Still, given the uncertainties in the planet parameters and in the assumptions in the climatic models, its potential habitability cannot be ruled out (e.g., ocean of liquid water under a thick ice crust, or presence of some strong green-house effect gas).

One of the main results of the Kepler mission is that high-multiplicity systems of dynamically-packed super-Earths are quite common around G and K dwarfs (Fabrycky et al. 2012). The putative existence of these kinds of compact systems around M-dwarfs, combined with a closer-in habitable zone, suggests the existence of a numerous population of planetary systems with several potentially-habitable worlds each. GJ 667C is likely to be among first of many of such systems that may be discovered in the coming years.

*Acknowledgements.* We acknowledge the constructive discussions with the referees of this manuscript. The robustness and confidence of the result greatly improved thanks to such discussions. G. Anglada-Escudé is supported by the German Federal Ministry of Education and Research under 05A11MG3. M. Tuomi acknowledges D. Pinfield and RoPACS (Rocky Planets Around Cool Stars), a Marie Curie Initial Training Network funded by the European Commission's Seventh Framework Programme. E. Gerlach would like to acknowledge the financial support from the DFG research unit FOR584. R. Barnes is supported by NASA's Virtual Planetary Laboratory under Cooperative Agreement Number NNH05ZDA001C and NSF grant AST-1108882. R. Heller receives funding from the Deutsche Forschungsgemeinschaft (reference number scho394/29-1). J.S. Jenkins also acknowledges funding by Fondecyt through grant 3110004 and partial support from CATA (PB06, Conicyt), the GEMINI-CONICYT FUND and from the Comité Mixto ESO-GOBIERNO DE CHILE. S. Weende acknowledges DFG funding by SFB-963 and the GrK-1351 A. Reiners acknowledges research funding from DFG grant RE1664/9-1. S.S. Vogt gratefully acknowledges support from NSF grant AST-0307493. This study contains data obtained from the ESO Science Archive Facility under request number ANGLADA36104. We also acknowledge the efforts of the PFS/Magellan team in obtaining Doppler measurements. We thank Sandy Keiser for her efficient setup of the computer network at Carnegie/DTM. We thank Dan Fabrycky, Aviv Ofir, Mathias Zechmeister and Denis Shulyak for useful and constructive discussions. This research made use of the Magny Cours Cluster hosted by the GWDG, which is managed by Georg August University Göttingen and the Max Planck Society. This research has made extensive use of the SIMBAD database, operated at CDS, Strasbourg, France; and NASA's Astrophysics Data System. The authors acknowledge the significant efforts of the HARPS-ESO team in improving the instrument and its data reduction software that made this work possible. We also acknowledge the efforts of the teams and individual observers that have been involved in observing the target star with HARPS/ESO, HIRES/Keck, PFS/Magellan and UVES/ESO.

**Table A.1.** Reference prior probability densities and ranges of the model parameters.

| Parameter | $\pi(\theta)$ | Interval | Hyper-parameter values |
|---|---|---|---|
| $K$ | Uniform | $[0, K_{max}]$ | $K_{max} = 5$ m s$^{-1}$ |
| $\omega$ | Uniform | $[0, 2\pi]$ | – |
| $e$ | $\propto \mathcal{N}(0, \sigma_e^2)$ | $[0,1]$ | $\sigma_e = 0.3$ |
| $M_0$ | Uniform | $[0, 2\pi]$ | – |
| $\sigma_J$ | Uniform | $[0, K_{max}]$ | (*) |
| $\gamma$ | Uniform | $[-K_{max}, K_{max}]$ | (*) |
| $\phi$ | Uniform | $[-1, 1]$ | |
| $\log P$ | Uniform | $[\log P_0, \log P_{max}]$ | $P_0 = 1.0$ days $P_{max} = 3000$ days |

**Notes.** * Same $K_{max}$ as for the $K$ parameter in first row.

# Appendix A: Priors

The choice of uninformative and adequate priors plays a central role in Bayesian statistics. More classic methods, such as weighted least-squares solvers, can be derived from Bayes theorem by assuming uniform prior distributions for all free parameters. Using the definition of Eq. 3, one can note that, under coordinate transformations in the parameter space (e.g., change from $P$ to $P^{-1}$ as the free parameter) the posterior probability distribution will change its shape through the Jacobian determinant of this transformation. This means that the posterior distributions are substantially different under changes of parameterizations and, even in the case of least-square statistics, one must be very aware of the prior choices made (explicit, or implicit through the choice of parameterization). This discussion is addressed in more detail in Tuomi & Anglada-Escude (2013). For the Doppler data of GJ 667C, our reference prior choices are summarized in Table A.1. The basic rationale on each prior choice can also be found in Tuomi (2012), Anglada-Escudé & Tuomi (2012) and Tuomi & Anglada-Escude (2013). The prior choice for the eccentricity can be decisive in detection of weak signals. Our choice for this prior ($\mathcal{N}(0, \sigma_e^2)$) is justified using statistical, dynamical and population arguments.

### A.1. Eccentricity prior : statistical argument

Our first argument is based on statistical considerations to minimize the risk of false positives. That is, since $e$ is a strongly non-linear parameter in the Keplerian model of Doppler signals (especially if $e > 0.5$), the likelihood function has many local maxima with high eccentricities. Although such solutions might appear very significant, it can be shown that, when the detected amplitudes approach the uncertainty in the measurements, these high-eccentricity solutions are mostly spurious.

To illustrate this, we generate simulated sets of observations (same observing epochs, no signal, Gaussian white noise injected, 1m s$^{-1}$ jitter level), and search for the maximum likelihood solution using the log–L periodograms approach (assuming a fully Keplerian solution at the period search level, see Section 4.2). Although no signal is injected, solutions with a formal false-alarm probability (FAP) smaller than 1% are found in 20% of the sample. On the contrary, our log–L periodogram search for circular orbits found 1.2% of false positives, matching the expectations given the imposed 1% threshold. We performed an additional test to assess the impact of the eccentricity prior on the detection completeness. That is, we injected one Keplerian signal ($e = 0.8$) at the same observing epochs with amplitudes of 1.0 m s$^{-1}$ and white Gaussian noise of 1 m s$^{-1}$ . We then performed the log–L periodogram search on a large number of these datasets ($10^3$). When the search model was allowed to be a fully Keplerian orbit, the correct solution was only recovered 2.5% of the time, and no signals at the right period were spotted assuming a circular orbit. With a $K = 2.0$ m s$^{-1}$ , the situation improved and 60% of the orbits were identified in the full Keplerian case, against 40% of them in the purely circular one. More tests are illustrated in the left panel of Fig. A.1. When an eccentricity of 0.4 and K=1 m s$^{-1}$ signal was injected, the completeness of the fully Keplerian search increased to 91% and the completeness of the circular orbit search increased to 80%. When a $K > 2$ m s$^{-1}$ signal was injected, the orbits were identified in > 99% of the cases. We also obtained a histogram of the semi-amplitudes of the false positive detections obtained when no signal was injected. The histogram shows that these amplitudes were smaller than 1.0 m s$^{-1}$ with a 99% confidence level (see right panel of Fig. A.1). This illustrates that statistical tests based on point estimates below the noise level do not produce reliable assessments on the significance of a fully Keplerian signal. For a given dataset, information based on simulations (e.g., Fig. A.1) or a physically motivated prior is necessary to correct such detection bias (Zakamska et al. 2011).

In summary, while a uniform prior on eccentricity only looses a few very eccentric solutions in the low amplitude regime, the rate of false positive detections ($\sim 20\%$) is unacceptable. On the other hand, only a small fraction of highly eccentric orbits are *missed* if the search is done assuming strictly circular orbits. This implies that, for log–likelihood periodogram searches, circular orbits should always be assumed when searching for a new low-amplitude signals and that, when approaching amplitudes comparable to the measurement uncertainties, assuming circular orbits is a reasonable strategy to avoid false positives. In a Bayesian sense, this means that we have to be very skeptic of highly eccentric orbits when searching for signals close to the noise level. The natural way to address this self-consistently is by imposing a prior on the eccentricity that suppresses the likelihood of highly eccentric orbits. The log–Likelihood periodograms indicate that the strongest possible prior (force $e = 0$), already does a very good job so, in general, any function less informative than a delta function ($\pi(e) = \delta(0)$) and a bit more constraining than a uniform prior ($\pi(e) = 1$) can provide a more optimal compromise between sensitivity and robustness. Our particular functional choice of the prior (Gaussian distribution with zero-mean and $\sigma = 0.3$) is based on dynamical and population analysis considerations.

### A.2. Eccentricity prior : dynamical argument

From a physical point of view, we expect *a priori* that eccentricities closer to zero are more probable than those close to unity when multiple planets are involved. That is, when one or two planets are clearly present (e.g. GJ 667Cb and GJ 667Cc are solidly detected even with a flat prior in $e$), high eccentricities in the remaining lower amplitude candidates would correspond to unstable and therefore physically impossible systems.

Our prior for $e$ takes this feature into account (reducing the likelihood of highly eccentric solutions) but still allows high eccentricities if the data insists so (Tuomi 2012). At the sampling level, the only orbital configurations that we explicitly forbid is that we do not allow for orbital crossings. While a rather weak limitation, this requirement essentially removes all extremely eccentric multiplanet solutions ($e > 0.8$) from our MCMC samples. This requirement does not, for example, re-move orbital configurations with close encounters between planets, and the solutions we receive still have to be analyzed by numerical integration to make sure that they correspond to stable systems. As shown in Section 8, posterior numerical integration of the samplings illustrate that our prior function was, after all, rather conservative.

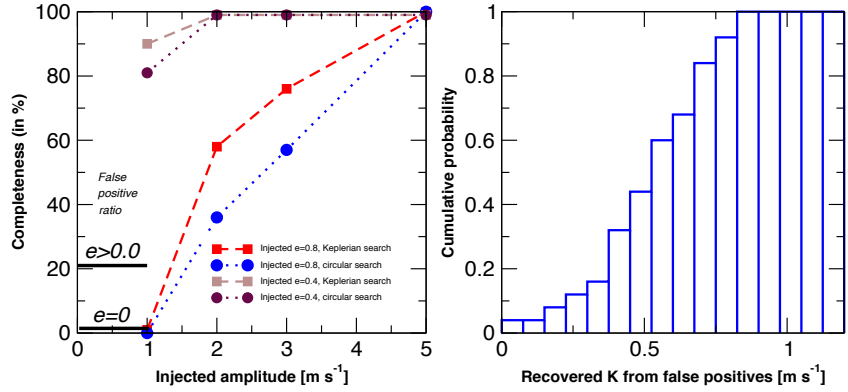

**Fig. A.1.** Left panel. Detection completeness as a function of the injected signal using a fully Keplerian search versus a circular orbit search. Red and blue lines are for an injected eccentricity of 0.8, and the brown and purple ones are for injected signals with eccentricity of 0.4. Black horizontal lines on the left show the fraction of false-positive detections satisfying the FAP threshold of 1% using both methods (Keplerian versus circular). While the completeness is slightly enhanced in the low $K$ regime, the fraction of false positives is unacceptable and, therefore, the implicit assumptions of the method (e.g., uniform prior for $e$) are not correct. **Right panel**. Distribution of semi-amplitudes $K$ for these false positive detections. Given that the injected noise level is 1.4 m s$^{-1}$ (1 m s$^{-1}$ nominal uncertainty, 1 m s$^{-1}$ jitter), it is clear that signals detected with fully Keplerian log–L periodograms with $K$ below the noise level cannot be trusted.

### A.3. Eccentricity prior : population argument

To investigate how realistic our prior choice is compared to the statistical properties of the known exoplanet populations, we obtained the parameters of all planet candidates with $M \sin i$ smaller than 0.1 M$_{jup}$ as listed in The Extrasolar Planets Encyclopaedia [3] as at 2012 December 1. We then produced a histogram in eccentricity bins of 0.1. The obtained distribution follows very nicely a Gaussian function with zero mean and $\sigma_e = 0.2$, meaning that our prior choice is more uninformative (and therefore, more conservative) than the current distribution of detections. This issue is the central topic of Tuomi & Anglada-Escude (2013), and a more detailed discussion (plus some illustrative plots) can be found in there.

## Appendix B: Detailed Bayesian detection sequences

In this section, we provide a more detailed description of detections of seven signals in the combined HARPS-TERRA, PFS, and HIRES data. We also show that the same seven signals (with some small differences due to aliases) are also detected independently when using HARPS-CCF velocities instead of HARPS TERRA ones. The PFS and HIRES measurements used are again those provided in Anglada-Escudé & Butler (2012).

### B.1. HARPS-CCF, PFS and HIRES

First, we perform a detailed analysis with the CCF values provided by Delfosse et al. (2012) in combination with the PFS and HIRES velocities. When increasing the number of planets, parameter $k$, in our statistical model, we were able to determine the relative probabilities of models with $k = 0, 1, 2, ...$ rather easily. The parameter spaces of all models could be sampled with our Markov chains relatively rapidly and the parameters of the signals converged to the same periodicities and RV amplitudes regardless of the exact initial states of the parameter vectors.

Unlike in Anglada-Escudé et al. (2012) and Delfosse et al. (2012), we observed immediately that $k = 2$ was not the number of signals favored by the data. While the obvious signals at 7.2 and 28.1 days were easy to detect using samplings of the parameter space, we also quickly discovered a third signal at a period of 91 days. These signals correspond to the planets GJ 667C b, c, and d of Anglada-Escudé et al. (2012) and Delfosse et al. (2012), respectively, though the orbital period of companion d was found to be 75 days by Anglada-Escudé et al. (2012) and 106 days by Delfosse et al. (2012). The periodograms also show considerable power at 106 days corresponding to the solution of Delfosse et al. (2012). Again, it did not provide a solution as probable as the signal at 91 days and the inclusion of linear correlation terms with activity did not affect its significance (see also Sec. 6).

We could identify three more signals in the data at 39, 53, and 260 days with low RV amplitudes of 1.31, 0.96, and 0.97 ms$^{-1}$, respectively. The 53-day signal had a strong alias at 62 days and so we treated these as alternative models and calculated their probabilities as well. The inclusion of $k = 6$ and $k = 7$ signals at 260 and 17 days improved the models irrespective of the preferred choice of the period of planet e (see Table B.1). To assess the robustness of the detection of the 7-th signal, we started alternative chains at random periods. All the chains that show good signs of convergence (bound period) unequivocally suggested 17 days for the last candidate. Since all these signals satisfied our Bayesian detection criteria, we denoted all of them to planet candidates.

We performed samplings of the parameter space of the model with $k = 8$ but could not spot a clear 8-th periodicity. The period of such 8th signal converged to a broad probability maximum between 1200 and 2000 days but the corresponding RV amplitude received a posterior density that did not differ significantly from zero. The probability of the model with $k = 8$ only exceeded the probability of $k = 7$ by a factor of 60.

**Table B.1.** Relative posterior probabilities and log-Bayes factors of models $\mathcal{M}_k$ with $k$ Keplerian signals derived from the combined HARPS-CCF, HIRES, and PFS RV data on the left and HARPS-TERRA HIRES, PFS on the right. Factor $\Delta$ indicates how much the probability increases with respect to the best model with one less Keplerian signal. $P_s$ denotes the MAP period estimate of the signal added to the solution when increasing the number $k$. For $k = 4, 5, 6$, and 7, we denote all the signals on top of the three most significant ones at 7.2, 28, and 91 days because the 53 and 62 day periods are each other's yearly aliases and the relative significance of these two and the signal with a period of 39 days are rather similar.

| | HARPS-CCF, PFS, HIRES | | | | HARPS-TERRA, PFS, HIRES | | | |
|---|---|---|---|---|---|---|---|---|
| $k$ | $P(\mathcal{M}_k\|d)$ | $\Delta$ | $\log P(d\|\mathcal{M}_k)$ | $P_s$ [days] | $P(\mathcal{M}_k\|d)$ | $\Delta$ | $\log P(d\|\mathcal{M}_k)$ | $P_s$ [days] |
| 0 | $2.2\times10^{-74}$ | – | -629.1 | – | $2.7\times10^{-85}$ | – | -602.1 | – |
| 1 | $2.4\times10^{-40}$ | $1.1\times10^{34}$ | -550.0 | 7.2 | $3.4\times10^{-48}$ | $1.3\times10^{37}$ | -516.0 | 7.2 |
| 2 | $1.3\times10^{-30}$ | $5.6\times10^{9}$ | -526.9 | 28 | $1.3\times10^{-35}$ | $3.9\times10^{12}$ | -486.3 | 91 |
| 3 | $8.7\times10^{-21}$ | $6.5\times10^{9}$ | -503.6 | 91 | $8.9\times10^{-18}$ | $6.7\times10^{17}$ | -444.5 | 28 |
| 4 | $5.1\times10^{-17}$ | $5.9\times10^{3}$ | -494.2 | 39 | $1.5\times10^{-14}$ | $1.7\times10^{3}$ | -436.4 | 39 |
| 4 | $1.0\times10^{-14}$ | $1.2\times10^{6}$ | -488.9 | 53 | $1.9\times10^{-14}$ | $2.1\times10^{3}$ | -436.2 | 53 |
| 4 | $2.0\times10^{-17}$ | $2.3\times10^{3}$ | -495.2 | 62 | $1.2\times10^{-14}$ | $1.3\times10^{3}$ | -436.7 | 62 |
| 5 | $8.0\times10^{-9}$ | $7.6\times10^{5}$ | -474.7 | 39, 53 | $1.0\times10^{-7}$ | $5.5\times10^{6}$ | -420.0 | 39, 53 |
| 5 | $5.4\times10^{-12}$ | $5.2\times10^{2}$ | -482.0 | 39, 62 | $1.0\times10^{-8}$ | $5.3\times10^{5}$ | -422.3 | 39, 62 |
| 6 | $3.4\times10^{-4}$ | $4.3\times10^{4}$ | -463.3 | 39, 53, 256 | $4.1\times10^{-3}$ | $4.0\times10^{4}$ | -408.7 | 39, 53, 256 |
| 6 | $1.3\times10^{-7}$ | 16 | -471.2 | 39, 62, 256 | $4.1\times10^{-4}$ | $4.0\times10^{3}$ | -411.0 | 39, 62, 256 |
| 7 | 0.998 | $2.9\times10^{3}$ | -454.6 | 17, 39, 53, 256 | 0.057 | 14 | -405.4 | 17, 39, 53, 256 |
| 7 | $1.5\times10^{-3}$ | 4.3 | -461.2 | 17, 39, 62, 256 | 0.939 | 230 | -402.6 | 17, 39, 62, 256 |

**Table B.2.** Seven-Keplerian solution of the combined RVs of GJ 667C with HARPS-CCF data. MAP estimates of the parameters and their 99% BCSs. The corresponding solution derived from HARPS-TERRA data is given in Table 4. Note that each dataset prefers a different alias for planet f (53 versus 62 days).

| Parameter | b | h | c | f |
|---|---|---|---|---|
| $P$ [days] | 7.1998 [7.1977, 7.2015] | 16.955 [16.903, 17.011] | 28.147 [28.084, 28.204] | 39.083 [38.892, 39.293] |
| $e$ | 0.10 [0, 0.25] | 0.16 [0, 0.39] | 0.02 [0, 0.20] | 0.03 [0, 0.18] |
| $K$ [ms$^{-1}$] | 3.90 [3.39, 4.37] | 0.80 [0.20, 1.34] | 1.60 [1.09, 2.17] | 1.31 [0.78, 1.85] |
| $\omega$ [rad] | 0.2 [0, 2$\pi$] | 2.3 [0, 2$\pi$] | 2.3 [0, 2$\pi$] | 3.6 [0, 2$\pi$] |
| $M_0$ [rad] | 3.2 [0, 2$\pi$] | 6.0 [0, 2$\pi$] | 2.9 [0, 2$\pi$] | 2.8 [0, 2$\pi$] |
| | e | d | g | |
| $P$ [days] | 53.19 [52.73, 53.64] | 91.45 [90.81, 92.23] | 256.4 [248.6, 265.8] | |
| $e$ | 0.13 [0, 0.19] | 0.12 [0, 0.29] | 0.18 [0, 0.49] | |
| $K$ [ms$^{-1}$] | 0.96 [0.48, 1.49] | 1.56 [1.11, 2.06] | 0.97 [0.41, 1.53] | |
| $\omega$ [rad] | 0.8 [0, 2$\pi$] | 3.0 [0, 2$\pi$] | 6.2 [0, 2$\pi$] | |
| $M_0$ [rad] | 5.9 [0, 2$\pi$] | 5.4 [0, 2$\pi$] | 1.0 [0, 2$\pi$] | |
| $\gamma_1$ [ms$^{-1}$] (HARPS) | -32.6 [-37.3, -28.2] | | | |
| $\gamma_2$ [ms$^{-1}$] (HIRES) | -33.3 [-38.9., -28.2] | | | |
| $\gamma_3$ [ms$^{-1}$] (PFS) | -27.7 [-31.0, -24.0] | | | |
| $\dot{\gamma}$ [ms$^{-1}$a$^{-1}$] | 2.19 [1.90, 2.48] | | | |
| $\sigma_{J,1}$ [ms$^{-1}$] (HARPS) | 0.80 [0.20, 1.29] | | | |
| $\sigma_{J,2}$ [ms$^{-1}$] (HIRES) | 2.08 [0.54, 4.15] | | | |
| $\sigma_{J,3}$ [ms$^{-1}$] (PFS) | 1.96 [0.00, 4.96] | | | |

We therefore conclude that the combined data set with HARPS-CCF RVs was in favor of $k = 7$. The corresponding orbital parameters of these seven Keplerian signals are shown in Table B.2. Whether there is a weak signal at roughly 1200-2000 days or not remains to be seen when future data become available. The naming of the seven candidate planets (b to h) follows the significance of detection.

### B.2. HARPS-TERRA, PFS and HIRES (reference solution)

The HARPS-TERRA velocities combined with PFS and HIRES velocities contained the signals of GJ 667C b, c, and d very clearly. We could extract these signals from the data with ease and obtained estimates for orbital periods that were consistent with the estimates obtained using the CCF data in the previous subsection. Unlike for the HARPS-CCF data, however, the 91 day signal was more significantly present in the HARPS-TERRA data and it corresponded to the second most significant periodicity instead of the 28 day one. Also, increasing $k$ improved the statistical model significantly and we could again detect all the additional signals in the RVs.

As for the CCF data, we branched the best fit solution into the two preferred periods for the planet e (53/62 days). The solution and model probabilities are listed on the right-side of Table B.1. The only significant difference compared to the HARPS-CCF one is that the 62-day period for planet e is now preferred. Still, the model with 53 days is only seventeen times less probable than the one with a 62 days signal, so we cannot confidently rule out that the 53 days one is the real one. For simplicity and to avoid confusion, we use the 62-day signal in our reference solution and is the one used to analyze dynamical stability and habitability for the system. As an additional check, preliminary dynamical analysis of solutions with a period of 53 days for planet e showed very similar behaviour to the reference solution in terms of long-term stability (similar fraction of dynamically stable orbits and similar distribution for the $D$ chaos indices). Finally, we made several efforts at sampling the eight-Keplerian model with different initial states. As for the CCF data, there are hints of a signal in the ∼ 2000 day period domain that could not be constrained.

## Appendix C: Further activity-related tests

### C.1. Correlated noise

The possible effect of correlated noise must be investigated together with the significance of the reported detections (e.g. GJ 581; Baluev 2012; Tuomi & Jenkins 2012). We studied the impact of correlated noise by adding a first order Moving Average term (MA(1), see Tuomi et al. 2012) to the model in Eq. 1 and repeated the Bayesian search for all the signals. The MA(1) model proposed in (Tuomi et al. 2012) contains two free parameters: a characteristic time-scale $\tau$ and a correlation coefficient $c$. Even for the HARPS data set with the greatest number of measurements, the characteristic time-scale could not be constrained. Similarly, the correlation coefficient (see e.g. Tuomi et al. 2013, 2012) had a posteriori density that was not different from its prior, i.e. it was found to be roughly uniform in the interval [-1,1], which is a natural range for this parameter because it makes the corresponding MA model stationary. While the $k = 7$ searches lead to the same seven planet solution, models containing such noise terms produced lower integrated probabilities, which suggests over-parameterization. When this happens, one should use the simplest model (principle of parsimony) and accept that the noise is better explained by the white noise components only. Finally, very low levels of correlated noise are also consistent with the good match between synthetic and real periodograms of subsamples in Section 7.

### C.2. Including activity correlation in the model

Because the HARPS activity-indices (S-index, FWHM, and BIS) were available, we analyzed the data by taking into account possible correlations with any of these indices. We added an extra component into the statistical model taking into account linear correlation with each of these parameters as $F(t_i, C) = C \, z_i$ (see Eq. 1), where $z_i$ is the any of the three indices at epoch $t_i$.

In Section 6, we found that the log–L of the solution for planet d slightly improved when adding the correlation term. The slight improvement of the likelihood is compatible with a consistently positive estimate of $C$ for both FWHM and the S-index obtained with the MC samplings (see Fig. C.1, for an example distribution of $C_{S-index}$ as obtained with a $k = 3$ model). While the correlation terms were observed to be mostly positive in all cases, the 0 value for the coefficient was always within the 95% credibility interval. Moreover, we found that the integrated model probabilities decreased when compared to the model with the same number of planets but no correlation term included. This means that such models are over-parameterized and, therefore, they are penalized by the principle of parsimony.

### C.3. Wavelength dependence of the signals

Stellar activity might cause spurious Doppler variability that is wavelength dependent (e.g., cool spots). Using the HARPS-TERRA software on the HARPS spectra only, we extracted one time-series for each echelle aperture (72 of them). This process results is $N_{obs} = 72 \times 173 = 12456$ almost independent RV measurements with corresponding uncertainties. Except for calibration related systematic effects (unknown at this level of precision), each echelle aperture can be treated as an independent instrument. Therefore, the radial velocities $\gamma_l$ and jitter terms $\sigma_l$ of each aperture were considered as free parameters. To assess the wavelength dependence of each signal, the Keplerian model of the i-th planet (one planet at a time) also contained one semi-amplitude $K_{i,l}$ per echelle aperture and all the other parameters ($\omega_i$, $M_{0,i}$ and $P_i$) were common to all apertures. The resulting statistical model has $72 \times 3 + 5 \times k - 1$ parameters when $k$ Keplerian signals are included and one is investigated (250 free parameters for $k = 7$). We started the posterior samplings in the vicinity of the solutions of interest because searching the period space would have been computationally too demanding. Neglecting the searches for periodicities, we could obtain relatively well converged samples from the posterior densities in a reasonable time-scale (few days of computer time).

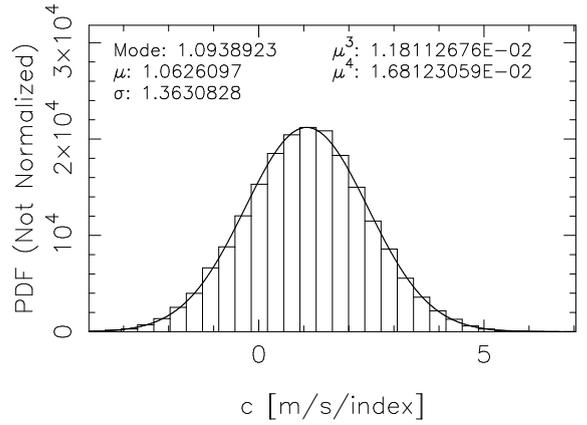

**Fig. C.1.** Value of the linear correlation parameter of the S-index ($C_S$) with the radial velocity data for a model with $k = 3$ Keplerians (detection of planet d).

The result is the semi-amplitude $K$ of each signal measured as a function of wavelength. Plotting this $K$ against central wavelength of each echelle order enabled us to visually assess if signals had strong wavelength dependencies (i.e. whether there were signals only present in a few echelle orders) and check if the obtained solution was consistent with the one derived from the combined Doppler measurements. By visual inspection and applying basic statistical tests, we observed that the scatter in the amplitudes for all the candidates was consistent within the error bars and no significant systematic trend (e.g. increasing $K$ towards the blue or the red) was found in any case. Also, the weighted means of the derived amplitudes were fully consistent with the values in Table 4. We are developing a more quantitative version of these tests by studying reported activity-induced signals on a larger sample of stars. As examples of low–amplitude wavelength-dependent signals ruled out using similar tests in the HARPS wavelength range see : Tuomi et al. (2013) on HD 40307 (K3V), Anglada-Escudé & Butler (2012) on HD 69830 (G8V) and Reiners et al. (2013) on the very active M dwarf AD Leo (M3V).

**Table C.2.** HARPS-TERRA Doppler measurements of GJ 667C. Measurements are in the barycenter of the Solar System and corrected for perspective acceleration. The median has been velocity has been subtracted for cosmetic purposes. Nominal uncertainties in FWHM and BIS are 2.0 and 2.35 times the corresponding $\sigma_{cef}$ (see Section 4.5 in Zechmeister et al. 2013). No consistent CCF measurement could be obtained for the last two spectra because of conflicting HARPS-DRS software versions (different binary masks) used to produce them. Except for those two spectra and according to the ESO archive documentation, all CCF measurements were generated using the default M2 binary mask.

| BJD (days) | RV$_{TERRA}$ (m s$^{-1}$) | $\sigma_{TERRA}$ (m s$^{-1}$) | RV$_{ccf}$ (m s$^{-1}$) | $\sigma_{ccf}$ (m s$^{-1}$) | FWHM (km s$^{-1}$) | BIS (m s$^{-1}$) | S-index (−) | $\sigma_S$ (−) |
|---|---|---|---|---|---|---|---|---|
| 2453158.764366 | -3.10 | 0.95 | -3.11 | 1.05 | 3.0514 | -7.93 | 0.4667 | 0.0095 |
| 2453201.586794 | -11.88 | 1.25 | -11.8 | 1.09 | 3.0666 | -9.61 | 0.4119 | 0.0074 |
| 2453511.798846 | -7.61 | 0.89 | -9.22 | 1.07 | 3.0742 | -7.42 | 0.5915 | 0.0088 |
| 2453520.781048 | -3.92 | 1.17 | -0.37 | 1.23 | 3.0701 | -11.99 | 0.4547 | 0.0082 |
| 2453783.863348 | 0.25 | 0.61 | 0.34 | 0.65 | 3.0743 | -13.31 | 0.4245 | 0.0053 |
| 2453810.852282 | -3.48 | 0.55 | -3.00 | 0.54 | 3.0689 | -10.62 | 0.4233 | 0.0044 |
| 2453811.891816 | 2.20 | 1.08 | 0.24 | 1.02 | 3.0700 | -9.37 | 0.4221 | 0.0066 |
| 2453812.865858 | -0.34 | 0.71 | -0.56 | 0.72 | 3.0716 | -9.78 | 0.4125 | 0.0054 |
| 2453814.849082 | -10.16 | 0.49 | -10.06 | 0.47 | 3.0697 | -10.63 | 0.4848 | 0.0042 |
| 2453816.857459 | -9.15 | 0.52 | -9.89 | 0.65 | 3.0698 | -12.20 | 0.4205 | 0.0051 |
| 2453830.860468 | -6.96 | 0.56 | -7.29 | 0.59 | 3.0694 | -11.59 | 0.4729 | 0.0052 |
| 2453832.903068 | -0.49 | 0.64 | -0.35 | 0.68 | 3.0706 | -13.33 | 0.4930 | 0.0058 |
| 2453834.884977 | -1.50 | 0.72 | -1.68 | 0.57 | 3.0734 | -8.20 | 0.4456 | 0.0049 |
| 2453836.887788 | -6.99 | 0.48 | -6.24 | 0.48 | 3.0723 | -8.27 | 0.4864 | 0.0044 |
| 2453861.796371 | 6.38 | 0.59 | 7.84 | 0.59 | 3.0780 | -11.47 | 0.6347 | 0.0060 |
| 2453862.772051 | 6.69 | 0.76 | 8.00 | 0.74 | 3.0768 | -12.54 | 0.5534 | 0.0065 |
| 2453863.797178 | 4.57 | 0.59 | 4.58 | 0.56 | 3.0759 | -10.71 | 0.4891 | 0.0051 |
| 2453864.753954 | 1.21 | 0.68 | 2.52 | 0.65 | 3.0783 | -9.21 | 0.4854 | 0.0055 |
| 2453865.785606 | -1.85 | 0.61 | -2.55 | 0.55 | 3.0752 | -7.73 | 0.4815 | 0.0050 |
| 2453866.743120 | -1.36 | 0.58 | -2.32 | 0.49 | 3.0770 | -7.49 | 0.5277 | 0.0045 |
| 2453867.835652 | -0.48 | 0.66 | -0.05 | 0.64 | 3.0816 | -10.55 | 0.4708 | 0.0055 |
| 2453868.813512 | 2.34 | 0.56 | 0.62 | 0.61 | 3.0754 | -10.01 | 0.4641 | 0.0053 |
| 2453869.789495 | 3.85 | 0.63 | 4.73 | 0.65 | 3.0795 | -12.71 | 0.4837 | 0.0055 |
| 2453870.810097 | 2.37 | 0.88 | 2.82 | 0.81 | 3.0813 | -10.48 | 0.4567 | 0.0062 |
| 2453871.815952 | -1.11 | 0.61 | -3.03 | 0.81 | 3.0790 | -9.16 | 0.5244 | 0.0068 |
| 2453882.732970 | -2.96 | 0.52 | -4.17 | 0.51 | 3.0795 | -8.09 | 0.5121 | 0.0047 |
| 2453886.703550 | -4.54 | 0.58 | -3.78 | 0.48 | 3.0757 | -10.11 | 0.4607 | 0.0042 |
| 2453887.773514 | -5.97 | 0.48 | -3.98 | 0.44 | 3.0700 | -10.94 | 0.4490 | 0.0041 |
| 2453917.737524 | -4.12 | 0.88 | -2.44 | 1.14 | 3.0666 | -10.91 | 0.5176 | 0.0084 |
| 2453919.712544 | 0.98 | 0.99 | 0.69 | 1.17 | 3.0774 | -8.01 | 0.4324 | 0.0073 |
| 2453921.615825 | -1.67 | 0.49 | -1.24 | 0.51 | 3.0671 | -9.87 | 0.4305 | 0.0043 |
| 2453944.566259 | -2.02 | 0.98 | -2.16 | 1.00 | 3.0776 | -9.25 | 0.6143 | 0.0079 |
| 2453947.578821 | 3.89 | 1.68 | 5.83 | 2.43 | 3.0806 | -8.54 | 0.7079 | 0.0134 |
| 2453950.601834 | -1.01 | 0.89 | 1.65 | 0.92 | 3.0780 | -11.80 | 0.5612 | 0.0071 |
| 2453976.497106 | 2.40 | 0.61 | 3.52 | 0.60 | 3.0791 | -12.74 | 0.5365 | 0.0054 |
| 2453979.594316 | -2.67 | 0.95 | -0.48 | 1.19 | 3.0776 | -9.20 | 0.5517 | 0.0091 |
| 2453981.555311 | -4.77 | 0.64 | -4.29 | 0.57 | 3.0749 | -13.12 | 0.5339 | 0.0055 |
| 2453982.526504 | -4.36 | 0.81 | -2.88 | 0.69 | 3.0717 | -11.84 | 0.4953 | 0.0061 |
| 2454167.866839 | -1.87 | 0.62 | -2.51 | 0.61 | 3.0798 | -10.14 | 0.5141 | 0.0053 |
| 2454169.864835 | -0.10 | 0.59 | -0.04 | 0.63 | 3.0793 | -11.94 | 0.4729 | 0.0052 |
| 2454171.876906 | 5.17 | 0.71 | 6.08 | 0.58 | 3.0744 | -7.24 | 0.4893 | 0.0050 |
| 2454173.856452 | -1.18 | 0.83 | -1.44 | 0.61 | 3.0746 | -10.33 | 0.4809 | 0.0052 |
| 2454194.847290 | 1.27 | 0.59 | 0.85 | 0.69 | 3.0756 | -8.43 | 0.4586 | 0.0054 |
| 2454196.819157 | -3.57 | 0.79 | -3.06 | 0.79 | 3.0759 | -12.33 | 0.4809 | 0.0061 |
| 2454197.797125 | -3.83 | 0.86 | -4.71 | 0.97 | 3.0726 | -9.12 | 0.4584 | 0.0069 |
| 2454198.803823 | -4.06 | 0.76 | -4.99 | 0.79 | 3.0708 | -9.33 | 0.5685 | 0.0068 |
| 2454199.854238 | 0.18 | 0.55 | 0.97 | 0.51 | 3.0714 | -10.66 | 0.4652 | 0.0044 |
| 2454200.815699 | 1.30 | 0.60 | 2.55 | 0.57 | 3.0708 | -10.26 | 0.4468 | 0.0047 |
| 2454201.918397 | 0.54 | 0.79 | 2.31 | 0.63 | 3.0681 | -11.27 | 0.4690 | 0.0056 |
| 2454202.802697 | -2.96 | 0.69 | -3.23 | 0.66 | 3.0696 | -8.49 | 0.4954 | 0.0056 |
| 2454227.831743 | -1.26 | 0.84 | 0.47 | 0.95 | 3.0619 | -9.96 | 0.4819 | 0.0071 |
| 2454228.805860 | 3.35 | 0.68 | 5.19 | 0.65 | 3.0651 | -15.03 | 0.4603 | 0.0055 |
| 2454229.773888 | 7.44 | 1.29 | 7.23 | 1.28 | 3.0708 | -6.34 | 0.5213 | 0.0082 |
| 2454230.845843 | 1.51 | 0.58 | 1.97 | 0.62 | 3.0631 | -8.92 | 0.4409 | 0.0053 |
| 2454231.801726 | -0.57 | 0.62 | -1.15 | 0.55 | 3.0704 | -8.86 | 0.5993 | 0.0055 |
| 2454232.721251 | -0.63 | 1.15 | -2.17 | 1.41 | 3.0719 | -9.70 | 0.3737 | 0.0079 |
| 2454233.910349 | -1.27 | 1.29 | -2.10 | 1.68 | 3.0687 | -12.12 | 0.5629 | 0.0112 |
| 2454254.790981 | -1.89 | 0.74 | -1.48 | 0.66 | 3.0672 | -8.39 | 1.2169 | 0.0093 |
| 2454253.728334 | 0.99 | 0.79 | 1.65 | 0.84 | 3.0773 | -10.30 | 0.4509 | 0.0062 |
| 2454254.755898 | -2.64 | 0.54 | -3.25 | 0.52 | 3.0779 | -7.99 | 0.4426 | 0.0046 |
| 2454255.709350 | -2.92 | 0.74 | -2.83 | 0.72 | 3.0775 | -7.36 | 0.4829 | 0.0059 |

**Table C.2.** continued.

| BJD (days) | RV$_{TERRA}$ (m s$^{-1}$) | $\sigma_{TERRA}$ (m s$^{-1}$) | RV$_{ccf}$ (m s$^{-1}$) | $\sigma_{ccf}$ (m s$^{-1}$) | FWHM (km s$^{-1}$) | BIS (m s$^{-1}$) | S-index (–) | $\sigma_S$ (–) |
|---|---|---|---|---|---|---|---|---|
| 2454256.697674 | -0.21 | 0.97 | -0.45 | 0.84 | 3.0775 | -9.19 | 0.4608 | 0.0063 |
| 2454257.704446 | 2.93 | 0.66 | 2.39 | 0.70 | 3.0766 | -11.09 | 0.4549 | 0.0055 |
| 2454258.698322 | 4.19 | 0.83 | 5.19 | 0.63 | 3.0799 | -9.57 | 0.4760 | 0.0052 |
| 2454291.675565 | -5.58 | 1.16 | -4.45 | 1.35 | 3.0802 | -9.95 | 0.4298 | 0.0086 |
| 2454292.655662 | -4.37 | 0.75 | -1.25 | 0.76 | 3.0820 | -11.83 | 0.4487 | 0.0056 |
| 2454293.708786 | 0.89 | 0.63 | 2.84 | 0.59 | 3.0732 | -11.52 | 0.5344 | 0.0056 |
| 2454295.628628 | 3.05 | 0.92 | 3.67 | 1.03 | 3.0786 | -6.85 | 0.4975 | 0.0072 |
| 2454296.670395 | -4.68 | 0.75 | -3.99 | 0.74 | 3.0703 | -7.79 | 0.5453 | 0.0067 |
| 2454297.631678 | -5.53 | 0.63 | -4.81 | 0.55 | 3.0725 | -10.38 | 0.5212 | 0.0053 |
| 2454298.654206 | -5.39 | 0.67 | -6.73 | 0.71 | 3.0743 | -5.18 | 0.5718 | 0.0066 |
| 2454299.678909 | -1.46 | 0.85 | -2.26 | 0.92 | 3.0785 | -6.46 | 0.5299 | 0.0070 |
| 2454300.764649 | 0.14 | 0.74 | -0.07 | 0.63 | 3.0693 | -12.07 | 0.4803 | 0.0057 |
| 2454314.691809 | -0.53 | 1.88 | -2.89 | 2.22 | 3.0756 | -12.48 | 0.3823 | 0.0102 |
| 2454315.637551 | 3.41 | 1.12 | 2.31 | 1.47 | 3.0701 | -10.42 | 0.4835 | 0.0091 |
| 2454316.554926 | 5.78 | 0.96 | 6.61 | 1.12 | 3.0746 | -6.02 | 0.4402 | 0.0069 |
| 2454319.604048 | -6.64 | 0.79 | -7.01 | 0.59 | 3.0694 | -7.54 | 0.4643 | 0.0052 |
| 2454320.616852 | -5.58 | 0.65 | -6.49 | 0.69 | 3.0698 | -3.94 | 0.4611 | 0.0057 |
| 2454340.596942 | -1.52 | 0.60 | -0.55 | 0.55 | 3.0691 | -10.11 | 0.4480 | 0.0048 |
| 2454342.531820 | -2.39 | 0.66 | -1.74 | 0.54 | 3.0667 | -9.95 | 0.4573 | 0.0048 |
| 2454343.530662 | 0.55 | 0.64 | 1.39 | 0.61 | 3.0669 | -7.25 | 0.4900 | 0.0055 |
| 2454346.551084 | -0.17 | 1.01 | -0.82 | 1.14 | 3.0677 | -5.48 | 0.5628 | 0.0086 |
| 2454349.569500 | -5.24 | 0.65 | -4.02 | 0.77 | 3.0658 | -11.12 | 0.3809 | 0.0058 |
| 2454522.886464 | -1.68 | 0.70 | -1.11 | 0.61 | 3.0688 | -9.85 | 0.5582 | 0.0056 |
| 2454524.883089 | 4.38 | 0.69 | 3.05 | 0.69 | 3.0668 | -8.66 | 0.4779 | 0.0057 |
| 2454525.892144 | 1.96 | 0.72 | 0.69 | 0.58 | 3.0692 | -8.77 | 0.4202 | 0.0047 |
| 2454526.871196 | -1.08 | 0.54 | 0.30 | 0.52 | 3.0717 | -9.58 | 0.4898 | 0.0046 |
| 2454527.897962 | -2.69 | 0.64 | -3.31 | 0.65 | 3.0689 | -8.46 | 0.4406 | 0.0052 |
| 2454528.903672 | -2.80 | 0.71 | -5.04 | 0.74 | 3.0679 | -8.40 | 0.4666 | 0.0058 |
| 2454529.869217 | 0.48 | 0.63 | -0.10 | 0.62 | 3.0664 | -8.23 | 0.4255 | 0.0050 |
| 2454530.878876 | 1.40 | 0.68 | 1.38 | 0.53 | 3.0667 | -7.31 | 0.4331 | 0.0044 |
| 2454550.901932 | -6.95 | 0.70 | -6.46 | 0.58 | 3.0680 | -5.44 | 0.4330 | 0.0047 |
| 2454551.868783 | -3.73 | 0.65 | -3.44 | 0.53 | 3.0654 | -7.36 | 0.4287 | 0.0045 |
| 2454552.880221 | 0.24 | 0.59 | -0.25 | 0.50 | 3.0665 | -9.12 | 0.4342 | 0.0042 |
| 2454554.846366 | 2.14 | 0.57 | 1.73 | 0.68 | 3.0699 | -5.49 | 0.4116 | 0.0052 |
| 2454555.870790 | -2.84 | 0.58 | -2.26 | 0.58 | 3.0663 | -7.66 | 0.4704 | 0.0050 |
| 2454556.838936 | -4.14 | 0.59 | -3.47 | 0.51 | 3.0686 | -11.20 | 0.4261 | 0.0043 |
| 2454557.804592 | -4.56 | 0.66 | -4.37 | 0.60 | 3.0650 | -8.87 | 0.4306 | 0.0049 |
| 2454562.905075 | 0.67 | 0.70 | 0.80 | 0.57 | 3.0668 | -7.11 | 0.4709 | 0.0051 |
| 2454563.898808 | -1.37 | 0.71 | -1.39 | 0.53 | 3.0656 | -11.93 | 0.4127 | 0.0046 |
| 2454564.895759 | -2.63 | 0.85 | -2.82 | 0.71 | 3.0680 | -8.67 | 0.5068 | 0.0061 |
| 2454568.891702 | 3.27 | 0.87 | 4.85 | 1.02 | 3.0735 | -11.20 | 0.4682 | 0.0069 |
| 2454569.881078 | -0.46 | 0.83 | 0.46 | 0.78 | 3.0720 | -16.02 | 0.4939 | 0.0061 |
| 2454570.870766 | -1.70 | 0.72 | -1.21 | 0.88 | 3.0715 | -10.68 | 0.4606 | 0.0063 |
| 2454583.933324 | 0.44 | 1.00 | 1.56 | 1.11 | 3.0711 | -17.51 | 0.5177 | 0.0087 |
| 2454587.919825 | -0.50 | 0.90 | -1.42 | 1.10 | 3.0824 | -7.86 | 0.4602 | 0.0078 |
| 2454588.909632 | 4.05 | 0.98 | 3.18 | 1.05 | 3.0828 | -6.95 | 0.5501 | 0.0080 |
| 2454590.901964 | 4.22 | 0.93 | 4.19 | 0.93 | 3.0758 | -9.05 | 0.4707 | 0.0073 |
| 2454591.900611 | 1.69 | 0.91 | -1.27 | 0.96 | 3.0753 | -7.39 | 0.5139 | 0.0075 |
| 2454592.897751 | -2.50 | 0.68 | -2.50 | 0.63 | 3.0757 | -8.84 | 0.4741 | 0.0057 |
| 2454593.919961 | -2.30 | 0.74 | -2.58 | 0.65 | 3.0680 | -12.41 | 0.5039 | 0.0063 |
| 2454610.878230 | 9.08 | 0.88 | 10.36 | 0.95 | 3.0671 | -9.46 | 0.4037 | 0.0069 |
| 2454611.856581 | 5.49 | 0.56 | 6.40 | 0.54 | 3.0650 | -8.37 | 0.4296 | 0.0050 |
| 2454616.841719 | 4.81 | 0.91 | 5.15 | 0.88 | 3.0713 | -8.09 | 0.3999 | 0.0065 |
| 2454617.806576 | 8.12 | 0.93 | 7.30 | 1.33 | 3.0753 | -14.38 | 0.4948 | 0.0086 |
| 2454618.664475 | 10.67 | 1.76 | 7.01 | 2.51 | 3.0854 | -7.21 | 0.6755 | 0.0135 |
| 2454639.867730 | 3.14 | 1.06 | 4.26 | 1.10 | 3.0588 | -8.27 | 0.4083 | 0.0083 |
| 2454640.723804 | 5.06 | 0.64 | 7.07 | 0.66 | 3.0705 | -13.61 | 0.4387 | 0.0055 |
| 2454642.676950 | -0.81 | 0.47 | 1.56 | 0.61 | 3.0704 | -10.27 | 0.4720 | 0.0053 |
| 2454643.686130 | -2.06 | 0.72 | -4.52 | 0.76 | 3.0709 | -9.26 | 0.4809 | 0.0064 |
| 2454644.732044 | -1.19 | 0.46 | -1.85 | 0.56 | 3.0680 | -8.64 | 0.5097 | 0.0054 |
| 2454646.639658 | 5.74 | 1.11 | 5.01 | 0.95 | 3.0737 | -10.14 | 0.4316 | 0.0066 |
| 2454647.630210 | 5.37 | 0.68 | 3.28 | 0.72 | 3.0693 | -6.35 | 0.4938 | 0.0062 |
| 2454648.657090 | 2.58 | 0.92 | 0.96 | 0.94 | 3.0720 | -8.85 | 0.4597 | 0.0068 |
| 2454658.650838 | -4.20 | 0.97 | -3.30 | 0.88 | 3.0714 | -13.06 | 0.4193 | 0.0065 |
| 2454660.650214 | -0.82 | 1.13 | -0.40 | 1.06 | 3.0728 | -10.20 | 0.4224 | 0.0074 |
| 2454661.760056 | 1.72 | 0.73 | 1.76 | 0.84 | 3.0737 | -11.56 | 0.4238 | 0.0065 |

**Table C.2.** continued.

| BJD (days) | RV$_{TERRA}$ (m s$^{-1}$) | $\sigma_{TERRA}$ (m s$^{-1}$) | RV$_{ecf}$ (m s$^{-1}$) | $\sigma_{ecf}$ (m s$^{-1}$) | FWHM (km s$^{-1}$) | BIS (m s$^{-1}$) | S-index (–) | $\sigma_S$ (–) |
|---|---|---|---|---|---|---|---|---|
| 2454662.664144 | 3.30 | 0.72 | 2.58 | 0.97 | 3.0713 | -8.43 | 0.4675 | 0.0070 |
| 2454663.784376 | -1.92 | 0.93 | -1.14 | 0.78 | 3.0643 | -11.52 | 0.3811 | 0.0061 |
| 2454664.766558 | -1.00 | 1.51 | 0.0 | 1.85 | 3.0765 | -9.85 | 0.4702 | 0.0106 |
| 2454665.774513 | -1.88 | 0.87 | -2.51 | 0.85 | 3.0695 | -9.89 | 0.4183 | 0.0065 |
| 2454666.683607 | -0.37 | 0.87 | 0.36 | 0.79 | 3.0717 | -9.64 | 0.4098 | 0.0060 |
| 2454674.576462 | 4.82 | 1.01 | 6.41 | 1.39 | 3.0901 | -6.38 | 0.4226 | 0.0083 |
| 2454677.663487 | 7.37 | 1.78 | 8.63 | 3.11 | 3.1226 | -4.66 | 0.4452 | 0.0117 |
| 2454679.572671 | 2.94 | 1.26 | -1.23 | 1.48 | 3.0822 | -5.41 | 0.5622 | 0.0103 |
| 2454681.573996 | 2.51 | 0.89 | 2.86 | 1.10 | 3.0780 | -6.26 | 0.4443 | 0.0075 |
| 2454701.523392 | -0.50 | 0.68 | -0.28 | 0.67 | 3.0719 | -4.48 | 0.5141 | 0.0058 |
| 2454708.564794 | -0.12 | 0.86 | -0.67 | 0.79 | 3.0803 | -12.80 | 0.5160 | 0.0062 |
| 2454733.487290 | 8.06 | 3.51 | 10.75 | 3.89 | 3.0734 | -3.79 | 0.5017 | 0.0146 |
| 2454735.499425 | 0.00 | 1.04 | -2.22 | 1.19 | 3.0720 | -11.44 | 0.4337 | 0.0072 |
| 2454736.550865 | -3.28 | 0.91 | -4.99 | 1.05 | 3.0671 | -8.70 | 0.4647 | 0.0075 |
| 2454746.485935 | -4.49 | 0.58 | -5.00 | 0.53 | 3.0611 | -13.01 | 0.4259 | 0.0045 |
| 2454992.721062 | 6.84 | 0.79 | 7.80 | 0.65 | 3.0748 | -10.71 | 0.4826 | 0.0053 |
| 2455053.694541 | -3.20 | 0.84 | -3.32 | 1.09 | 3.0741 | -11.73 | 0.4427 | 0.0078 |
| 2455276.882590 | 0.27 | 0.74 | 1.86 | 0.77 | 3.0732 | -11.03 | 0.4699 | 0.0061 |
| 2455278.827303 | 1.84 | 0.92 | 1.02 | 0.85 | 3.0760 | -8.41 | 0.5883 | 0.0074 |
| 2455280.854800 | 5.26 | 0.76 | 4.41 | 0.87 | 3.0793 | -12.50 | 0.4817 | 0.0065 |
| 2455283.868014 | -0.69 | 0.68 | -0.09 | 0.61 | 3.0793 | -8.23 | 0.5411 | 0.0054 |
| 2455287.860052 | 4.99 | 0.72 | 5.42 | 0.64 | 3.0779 | -11.76 | 0.5366 | 0.0056 |
| 2455294.882720 | 8.56 | 0.69 | 6.81 | 0.58 | 3.0775 | -8.71 | 0.5201 | 0.0051 |
| 2455295.754277 | 10.15 | 1.06 | 8.21 | 0.98 | 3.0743 | -8.94 | 0.5805 | 0.0076 |
| 2455297.805750 | 4.95 | 0.64 | 3.95 | 0.70 | 3.0779 | -10.19 | 0.4614 | 0.0057 |
| 2455298.813775 | 2.52 | 0.75 | 2.95 | 0.67 | 3.0807 | -7.72 | 0.5828 | 0.0061 |
| 2455299.785905 | 3.74 | 1.60 | 4.62 | 2.19 | 3.0793 | -10.36 | 0.4187 | 0.0106 |
| 2455300.876852 | 5.07 | 0.60 | 5.78 | 0.75 | 3.0792 | -9.53 | 0.5104 | 0.0060 |
| 2455301.896438 | 9.54 | 0.99 | 8.40 | 1.32 | 3.0774 | -12.07 | 0.4395 | 0.0085 |
| 2455323.705436 | 8.56 | 0.86 | 8.82 | 0.78 | 3.0702 | -7.78 | 0.4349 | 0.0067 |
| 2455326.717047 | 2.17 | 1.05 | 0.67 | 1.27 | 3.0649 | -9.48 | 0.5955 | 0.0103 |
| 2455328.702599 | 1.56 | 1.02 | 1.83 | 1.01 | 3.0658 | -8.48 | 0.5077 | 0.0089 |
| 2455335.651717 | 1.01 | 0.92 | 3.02 | 1.22 | 3.0593 | -10.79 | 0.4685 | 0.0092 |
| 2455337.704618 | 7.58 | 1.03 | 6.13 | 1.24 | 3.0725 | -11.55 | 0.4859 | 0.0090 |
| 2455338.649293 | 13.01 | 1.97 | 12.32 | 2.54 | 3.0687 | -5.74 | 0.4969 | 0.0134 |
| 2455339.713716 | 6.70 | 1.03 | 5.13 | 1.57 | 3.0700 | -11.07 | 0.4760 | 0.0097 |
| 2455341.789626 | -0.40 | 0.63 | 0.01 | 0.71 | 3.0812 | -11.27 | 0.4916 | 0.0061 |
| 2455342.720036 | 4.80 | 0.91 | 5.68 | 1.13 | 3.0718 | -7.25 | 0.4674 | 0.0076 |
| 2455349.682257 | 6.55 | 0.78 | 4.00 | 0.97 | 3.0685 | -4.609 | 0.4787 | 0.0073 |
| 2455352.601155 | 12.92 | 1.11 | 14.24 | 1.35 | 3.0700 | -7.48 | 0.4530 | 0.0093 |
| 2455354.642822 | 7.52 | 0.60 | 8.38 | 0.65 | 3.0663 | -10.63 | 0.4347 | 0.0057 |
| 2455355.576777 | 6.41 | 0.97 | 4.76 | 0.96 | 3.0681 | -9.41 | 0.4278 | 0.0074 |
| 2455358.754723 | 8.54 | 1.17 | 10.00 | 1.30 | 3.0619 | -11.04 | 0.3527 | 0.0095 |
| 2455359.599377 | 6.89 | 0.90 | 6.90 | 0.97 | 3.0724 | -11.39 | 0.3649 | 0.0070 |
| 2455993.879754 | 12.28 | 1.04 | – | – | – | – | 0.5179 | 0.0086 |
| 2455994.848576 | 15.43 | 1.30 | – | – | – | – | 0.6712 | 0.0120 |

## 7.4 Runaway Greenhouse Effect on Exomoons due to Irradiation from Hot, Young Giant Planets (Heller & Barnes 2015)

Contribution:

RH did the literature research, contributed to the mathematical framework, created Figs. 1, 4, and 5, led the writing of the manuscript, and served as a corresponding author for the journal editor and the referees.

# Runaway greenhouse effect on exomoons due to irradiation from hot, young giant planets

R. Heller[1] and R. Barnes[2,3]

[1] McMaster University, Department of Physics and Astronomy, 1280 Main Street West, Hamilton (ON) L8S 4M1, Canada
rheller@physics.mcmaster.ca
[2] University of Washington, Department of Astronomy, Seattle, WA 98195, USA
rory@astro.washington.edu
[3] Virtual Planetary Laboratory, USA

**ABSTRACT**

*Context.* The *Kepler* space telescope has proven capable of detecting transits of objects almost as small as the Earth's Moon. Some studies suggest that moons as small as 0.2 Earth masses can be detected in the *Kepler* data by transit timing variations and transit duration variations of their host planets. If such massive moons exist around giant planets in the stellar habitable zone (HZ), then they could serve as habitats for extraterrestrial life.
*Aims.* While earlier studies on exomoon habitability assumed the host planet to be in thermal equilibrium with the absorbed stellar flux, we here extend this concept by including the planetary luminosity from evolutionary shrinking. Our aim is to assess the danger of exomoons to be in a runaway greenhouse state due to extensive heating from the planet.
*Methods.* We apply pre-computed evolution tracks for giant planets to calculate the incident planetary radiation on the moon as a function of time. Added to the stellar flux, the total illumination yields constraints on a moon's habitability. Ultimately, we include tidal heating to evaluate a moon's energy budget. We use a semi-analytical formula to parametrize the critical flux for the moon to experience a runaway greenhouse effect.
*Results.* Planetary illumination from a 13-Jupiter-mass planet onto an Earth-sized moon at a distance of ten Jupiter radii can drive a runaway greenhouse state on the moon for about 200 Myr. When stellar illumination equivalent to that received by Earth from the Sun is added, then the runaway greenhouse holds for about 500 Myr. After 1000 Myr, the planet's habitable edge has moved inward to about 6 Jupiter radii. Exomoons in orbits with eccentricities of 0.1 experience strong tidal heating; they must orbit a 13-Jupiter-mass host beyond 29 or 18 Jupiter radii after 100 Myr (at the inner and outer boundaries of the stellar HZ, respectively), and beyond 13 Jupiter radii (in both cases) after 1000 Myr to be habitable.
*Conclusions.* If a roughly Earth-sized, Earth-mass moon be detected in orbit around a giant planet, and if the planet-moon duet would orbit in the stellar HZ, then it will be crucial to recover the orbital history of the moon. If, for example, such a moon around a 13-Jupiter-mass planet has been closer than 20 Jupiter radii to its host during the first few hundred million years at least, then it might have lost substantial amounts of its initial water reservoir and be uninhabitable today.

**Key words.** Astrobiology – Celestial mechanics – Planets and satellites: general – Radiation mechanisms: general

## 1. Introduction

The advent of exoplanet science in the last two decades has led to the compelling idea that it could be possible to detect a moon orbiting a planet outside the solar system. By observational selection effects, such a finding would reveal a massive moon, because its signature in the data would be most apparent. While the most massive moon in the solar system – Jupiter's satellite Ganymede – has a mass roughly 1/40 the mass of Earth ($M_\oplus$), a detectable exomoon would have at least twice the mass of Mars, that is, $1/5\,M_\oplus$ (Kipping et al. 2009). Should these relatively massive moons exist, then they could be habitats for extrasolar life.

One possible detection method relies on measurements of the transit timing variations (TTV) of the host planet as it periodically crosses the stellar disk (Sartoretti & Schneider 1999; Simon et al. 2007; Kipping 2009a; Lewis 2013). To ultimately pin down a satellite's mass and its orbital semi-major axis around its host planet ($a_{ps}$), it would also be necessary to measure the transit duration variation (TDV, Kipping 2009a,b). As shown by Awiphan & Kerins (2013), *Kepler*'s ability to find an exomoon is crucially determined by its ability to discern the TDV signal, as it is typically weaker than the TTV signature. Using TTV and TDV observations together, it should be possible to detect moons as small as $0.2\,M_\oplus$ (Kipping et al. 2009).

Alternatively, it could even be possible to observe the direct transits of large moons (Szabó et al. 2006; Tusnski & Valio 2011; Kipping 2011), as the discovery of the sub-Mercury-sized planet Kepler-37b by Barclay et al. (2013) recently demonstrated. Now that targeted searches for extrasolar moons are underway (Kipping et al. 2012; Kipping et al. 2013; Kipping et al. 2013) and the detection of a roughly Earth-mass moon in the stellar habitable zone (Dole 1964; Kasting et al. 1993, HZ in the following) has become possible, we naturally wonder about the conditions that determine their habitability. Indeed, the search for spectroscopic biosignatures in the atmospheres of exomoons will hardly be possible in the near future, because the moon's transmission spectrum would need to be separated from that of the planet (Kaltenegger 2010). But the possible detection of radio emission from intelligent species on exomoons still allows the hypothesis of life on exomoons to be tested.

The idea of habitable moons has been put forward by Reynolds et al. (1987) and Williams et al. (1997). Both studies concluded that tidal heating can be a key energy source if a moon orbits its planet in a close, eccentric orbit (for tidal heating in exomoons see also Scharf 2006; Debes & Sigurdsson 2007; Cassidy et al. 2009; Henning et al. 2009; Heller 2012; Heller & Barnes 2013). Reflected stellar light from the planet and the planet's own thermal emission can play an additional role in a moon's energy flux budget (Heller & Barnes 2013; Hinkel & Kane 2013). Having said that, eclipses occur frequently in close satellite orbits which are coplanar to the circumstellar orbit. These occultations can significantly reduce the average stellar flux on a moon (Heller 2012), thereby affecting its climate (Forgan & Kipping 2013). Beyond that, the magnetic environment of exomoons will affect their habitability (Heller & Zuluaga 2013).

Here, we investigate another effect on a moon's global energy flux. So far, no study focused on the impact of radial shrinking of a gaseous giant planet and the accompanying thermal illumination of its potentially habitable moons.[1] As a giant planet converts gravitational energy into heat (Baraffe et al. 2003; Leconte & Chabrier 2013), it may irradiate a putative Earth-like moon to an extent that makes the satellite subject to a runaway greenhouse effect. Atmospheric mass loss models suggest that desiccation of an Earth-sized planet (or, in our case, of a moon) in a runaway greenhouse state occurs as fast as within 100 million years (Myr). Depending on the planet's surface gravitation, initial water content, and stellar XUV irradiation in the high atmosphere (Barnes et al. 2013, see Sect. 2 and Appendix B therein), this duration can vary substantially. But as water loss is a complex process – not to forget the possible storage of substantial amounts of water in the silicate mantle, the history of volcanic outgassing, and possible redelivery of water by late impacts (Lammer 2013) – we here consider moons in a runaway greenhouse state to be temporarily uninhabitable, rather than desiccated forever.

## 2. Methods

In the following, we consider a range of hypothetical planet-moon binaries orbiting a Sun-like star during different epochs of the system's life time. As models for in-situ formation of exomoons predict that more massive host planets will develop more massive satellites (Canup & Ward 2006; Sasaki et al. 2010; Ogihara & Ida 2012), we focus on the most massive host planets that can possibly exist, that is, Jovian planets of roughly 13 Jupiter masses.[2] Following Canup & Ward (2006) and Heller et al. (2013), such a massive planet can grow Mars- to Earth-sized moons in its circumplanetary disk.

In Fig. 1, we show the detections of stellar companions with masses between that of Uranus and 13 $M_J$, where $M_J$ denotes Jupiter's mass.[3] While the ordinate measures planetary mass ($M_p$), the abscissa depicts orbit-averaged stellar illumination received by the planet, which we compute via

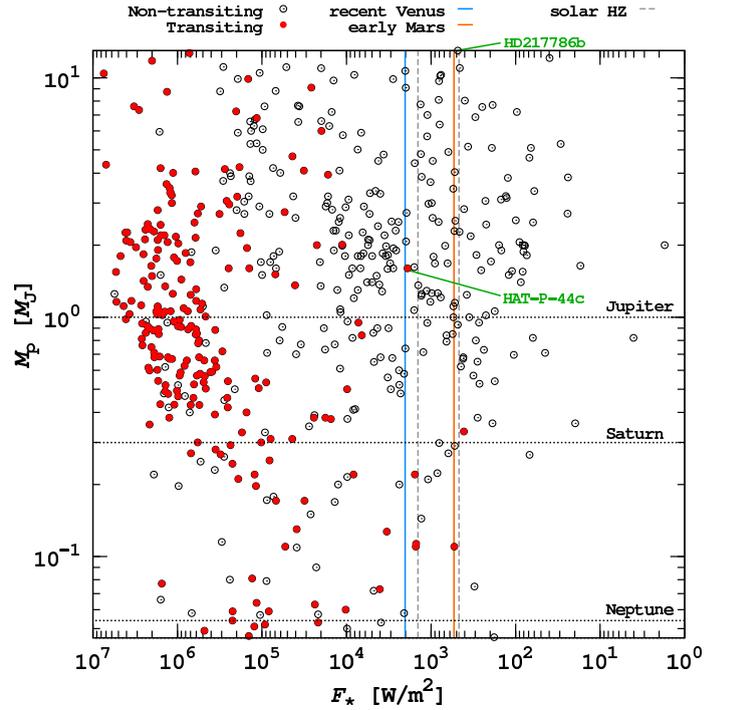

**Fig. 1.** Planetary masses as a function of orbit-averaged stellar illumination for objects with masses between that of Uranus and 13 times that of Jupiter. Non-transiting objects are shown as open circles; red circles correspond to transiting objects, which could allow for the detection of exomoons via TTV or TDV techniques. HD 217786 b is labeled because it could host Mars- to Earth-sized moons in the stellar HZ. HAT-P-44c is the most massive transiting planet in the recent Venus/early Mars HZ, whose width independent of the host star is denoted by the solid vertical lines. The dashed vertical lines denote the inner and outer HZ borders of the solar system (Kopparapu et al. 2013). Obviously, some ten super-Jovian planets reside in or close to the stellar HZs of their stars, but most of which are not known to transit their stars.

$$F_\star = \frac{\sigma_{SB} T_\star^4}{\sqrt{1 - e_{\star p}^2}} \left( \frac{R_\star}{a_{\star p}} \right)^2 . \qquad (1)$$

Here, $\sigma_{SB}$ is the Stefan-Boltzmann constant, $T_\star$ the stellar effective temperature, $R_\star$ the stellar radius, $e_{\star p}$ the orbital eccentricity of the star-planet system, and $a_{\star p}$ the planet's orbital semi-major axis.

Dashed vertical lines illustrate the inner and outer boundaries of the solar HZ, which Kopparapu et al. (2013) locate at $1.0512 \times S_{eff,\odot}$ for the runaway greenhouse effect and at $0.3438 \times S_{eff,\odot}$ for the maximum possible greenhouse effect ($S_{eff,\odot} = 1360\,W\,m^{-2}$ being the solar constant). For stars other than the Sun or planetary atmospheres other than that of Earth, the HZ limits can be located at different flux levels. Planets shown in this plot orbit a range of stars, most of which are on the main-sequence. The two solid vertical lines denote a flux interval which is between the recent Venus and early Mars HZ boundaries independent of stellar type, namely between $1.487 \times S_{eff,\odot}$ (recent Venus) and $0.393 \times S_{eff,\odot}$ (early Mars) (Kopparapu et al. 2013). The purpose of Fig. 1 is to confirm that super-Jovian planets (and possibly their massive moons) exist in or near the HZ of main-sequence stars in general, rather than to explic-itly identify giant planets in the HZ of their respective host stars. Yet, the position of HD 217786 b is indicated in Fig. 1, because it roughly corresponds to our hypothetical test planet in terms of orbit-averaged stellar illumination at the outer HZ boundary (483 W m$^{-2}$ = 0.355 × $S_{\text{eff},\odot}$) and planetary mass ($M_{\text{p}}$ = 13 ± 0.8 $M_{\text{J}}$, Moutou et al. 2011). We also highlight HAT-P-44 c because it is the most massive transiting planet in the recent Venus/early Mars HZ (Hartman et al. 2013) and could thus allow for TTV and TDV detections of its massive moons if they exist.

---

[1] We include a term $W_p$ for an additional source of illumination from the planet onto the moon in our orbit-averaged Eq. (B1) in Heller & Barnes (2013). This term can be attributed to heating form the planet. In Heller & Zuluaga (2013), we use methods developed here, but illumination from Neptune-, Saturn-, and Jupiter-like planets is weak.

[2] Given only the mass, such an object might well be a brown dwarf and not a planet. While we *assume* that this hypothetical object is a giant planet and that it can be described by the Baraffe et al. (2003) evolution models (see below).

[3] Data from www.exoplanet.eu as of Sep. 25, 2013.

### 2.1. Exomoon illumination from a star and a cooling planet

The electromagnetic spectrum of a main-sequence star is very different from that of a giant planet. M, K, and G dwarf stars have effective temperatures between ≈ 3000 K and ≈ 6000 K, whereas planets typically are cooler than 1000 K. Earth-like planets near the inner edge of the HZ around Sun-like stars are thought to have bolometric albedos as low as 0.18, increasing up to about 0.45 towards the outer HZ edge (Kopparapu et al. 2013). The inner edge value would likely be significantly higher if clouds were included in the calculation (Yang et al. 2013). As a proxy, we here assume an optical albedo $A_{\text{s,opt}}$ = 0.3, while an Earth-like planet orbiting a cool star with an effective temperature of 2500 K has an infrared albedo $A_{\text{s,IR}}$ = 0.05 (Kasting et al. 1993; Selsis et al. 2007; Kopparapu et al. 2013)[4]. We therefore use a dualpass band to calculate the total illumination absorbed by the satellite

$$F_{\text{i}} = \frac{L_*(t)(1 - A_{\text{s,opt}})}{16\pi a_{\text{*p}}^2} + \frac{L_{\text{p}}(t)(1 - A_{\text{s,IR}})}{16\pi a_{\text{ps}}^2} \ , \qquad (2)$$

where the first term on the right side of the equation describes the stellar flux absorbed by the moon, and the second term denotes the absorbed illumination from the planet. $L_*(t)$ and $L_{\text{p}}(t)$ are the stellar and planetary bolometric luminosities, respectively, while $a_{\text{*p}}$ is the orbital semi-major axes of the planet around the star.[5] The factor 16 in the denominators indicates that we assume effective redistribution of both stellar and planetary irradiation over the moon's surface (see Sect. 2.1 in Selsis et al. 2007). Stellar reflected light from the planet is neglected (for a description see Heller & Barnes 2013).

To parametrize the luminosities of the star and the cooling planet, we use the cooling tracks from Baraffe et al. (1998) and Baraffe et al. (2003), respectively. Figure 2 shows the evolution of the luminosity for a Sun-like star and the luminosity of three giant planets with 13, 5, and 1 $M_{\text{J}}$. Note the logarithmic scale: at an age of 0.2 billion years (Gyr), a Jupiter-mass planet emits about 0.3 % the amount of radiation of a 13 $M_{\text{J}}$ planet. These giant planets models assume arbitrarily large initial temperatures and radii. Yet, to actually assess the luminosity evolution of a giant exoplanet during the first ≈ 500 Myr, its age, mass, and luminosity would need to be known (Mordasini 2013). Our study is by necessity illustrative. Once actual exomoons are discovered more realistic giant planet models may be more appropriate.

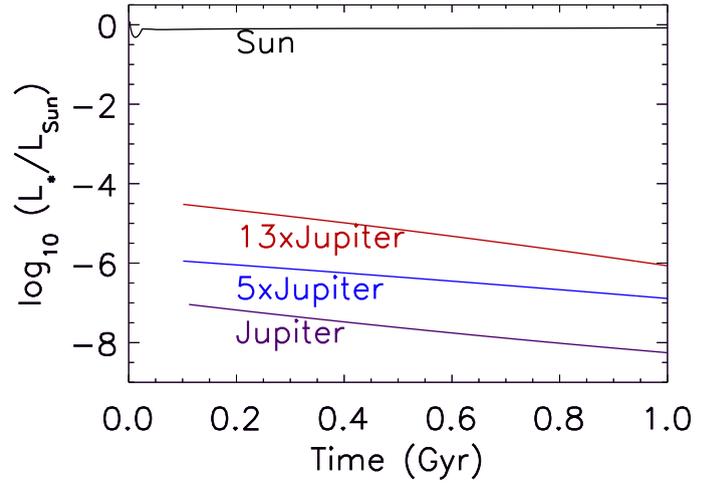

### 2.2. Tidal heating in exomoons

Tidal heating has been identified as a possible key source for a moon's energy flux budget (see Sect. 1). Hence, we will also explore the combined effects of stellar and planetary illumination on extrasolar satellites plus the contribution of tidal heating.

While tidal theories were initially developed to describe the tidal flexing of rigid bodies in the solar system (such as the Moon and Jupiter's satellite Io, see Darwin 1879, 1880; Peale et al. 1979; Segatz et al. 1988), the detection of bloated giant planets in close circumstellar orbits has triggered new efforts on realistic tidal models for gaseous objects. Nowadays, various approaches exist to parametrize tidal heating, and two main realms for bodies with an equilibrium tide (Zahn 1977) have emerged: constant-phase-lag (CPL) models (typically applied to rigid bodies, see Ferraz-Mello et al. 2008; Greenberg 2009), and the constant-time-lag (CTL) models (typically applied to gaseous bodies, see Hut 1981; Leconte et al. 2010).

We here make use of the CPL model of Ferraz-Mello et al. (2008) to estimate the tidal surface heating $F_{\text{t}}$ on our hypothetical exomoons. This model includes tidal heating from both circularization (up to second order in eccentricity) and tilt erosion (that is, tidally-induced changes in obliquity $\psi_{\text{s}}$, Heller et al. 2011). For simplicity, we assume that $F_{\text{t}}$ distributes evenly over the satellite's surface, although observations of Io, Titan, and Enceladus suggest that tidal heat can be episodic and heat would leave a planetary body through hot spots (Ojakangas & Stevenson 1986; Spencer et al. 2000; Sotin et al. 2005; Spencer et al. 2006; Porco et al. 2006; Tobie et al. 2008; Běhounková et al. 2012), that is, volcanoes or cryovolcanoes. The obliquity of a satellite in an orbit similar to Jupiter's Galilean moons and Saturn's moon Titan, with orbital periods ≲ 16 d, is eroded in much less than 1 Gyr. In Heller & Barnes (2013), we thus assume $\psi_{\text{s}}$ = 0. We treat the moon's orbit to have an instantaneous eccentricity $e_{\text{ps}}$. Tidal heating from circularization implies that $e_{\text{ps}}$ approaches zero, but it can be forced by stellar, planetary, or even satellite perturbations to remain non-zero. Alternatively, it can be the remainder of an extremely large initial eccentricity, or be caused by a recent impact. Note also that Titan's orbital eccentricity of roughly 0.0288 is still not understood by any of these processes (Sohl et al. 1995).

---

[4] Strong infrared absorption of gaseous $H_2O$ and $CO_2$ in the atmospheres of terrestrial worlds and the absence of Rayleigh scattering could further decrease $A_{\text{s,IR}}$. These effects, however, would tend to warm the stratosphere rather than the surface. Stratospheric warming could then be re-radiated to space without contributing much to warming the surface.

[5] Equation (2) implies that the planet is much more massive than the moon, so that the planet-moon barycenter coincides with the planetary center of mass. It is also assumed that $a_{\text{ps}} \ll a_{\text{*p}}$ and that both orbits are circular.

### 2.3. The runaway greenhouse effect

To assess whether our hypothetical satellites would be habitable, we compare the sum of total illumination $F_i$ (Eq. 2) and the globally averaged tidal heat $F_t$ to the critical flux for a runaway greenhouse, $F_{crit}$. The latter value is given by the semi-analytical approach of Pierrehumbert (2010, Eq. 4.94 therein) (see also Eq. 1 in Heller & Barnes 2013), which predicts the initiation of the runaway greenhouse effect based on the maximum possible outgoing longwave radiation from a planetary body with an atmosphere saturated in water vapor.

We study two hypothetical satellites. In the first case of a rocky Earth-mass, Earth-sized moon, we obtain a critical flux of $F_{crit} = 295$ W/m$^2$ for the onset of the runaway greenhouse effect. In the second case, we consider an icy moon of 0.25 $M_\oplus$[6] and an ice-to-mass ratio of 25 %. Using the structure models of Fortney et al. (2007), we deduce the satellite radius (0.805 times the radius of Earth) and surface gravity (3.75 m/s$^2$) and finally obtain $F_{crit} = 266$ W/m$^2$ for this Super-Ganymede. If one of our test moons undergoes a total flux that is beyond its critical limit, then it can be considered temporarily uninhabitable.

### 2.4. An exomoon menagerie

With decreasing distance from the planet, irradiation from the planet and tidal forces on the moon will increase. While an Earth-sized satellite in a wide circumplanetary orbit may essentially behave like a freely rotating planet with only the star as a relevant light source, moons in close orbits will receive substantial irradiation from the planet – at the same time be subject to eclipses – and eventually undergo enormous tidal heating.

To illustrate combined effects of illumination from the planet and tidal heat as a function of distance to the planet, we introduce an exomoon menagerie (Barnes & Heller 2013). It consists of the following specimen (colors in brackets refer to Figs. 4 and 5):

- Tidal Venus (red):
  $F_t \geq F_{crit}$ (Barnes et al. 2013)

- Tidal-Illumination Venus (orange):
  $F_i < F_{crit} \; \wedge \; F_t < F_{crit} \; \wedge \; F_i + F_t \geq F_{crit}$

- Super-Io (hypothesized by Jackson et al. 2008, yellow):
  $F_t > 2$ W/m$^2$ $\wedge \; F_i + F_t < F_{crit}$

- Tidal Earth (blue):
  $0.04$ W/m$^2 < F_t < 2$ W/m$^2$ $\wedge \; F_i + F_t < F_{crit}$

- Earth-like (green):
  $F_t < 0.04$ W/m$^2$ and within the stellar HZ

Among these states, a Tidal Venus and a Tidal-Insolation Venus are uninhabitable, while a Super-Io, a Tidal Earth, and an Earth-like moon could be habitable. The 2 and 0.04 W/m$^2$ limits are taken from examples in the solar system, where Io's extensive volcanism coincides with an endogenic surface flux of roughly 2 W/m$^2$ (Spencer et al. 2000). Williams et al. (1997) estimated that tectonic activity on Mars came to an end when its outgoing energy flux through the surface fell below 0.04 W/m$^2$.

For our menageries, we consider the rocky Earth-type moon orbiting a giant planet with a mass 13 times that of Jupiter. We investigate planet-moon binaries at two different distances to a Sun-like star. In one configuration, we will assume that the planet-moon duet orbits a Sun-like host at 1 AU. In this scenario, the planet-moon system is close to the inner edge of the stellar HZ (Kopparapu et al. 2013). In a second setup, the binary is assumed to orbit the star at a distance equivalent to 0.331 $S_{eff,\odot}$, which is the average of the maximum greenhouse and the early Mars limits computed by Kopparapu et al. (2013). In this second configuration, the binary is assumed at a distance of 1.738 AU from a Sun-like host star, that is, at the outer edge of the stellar HZ.

To explore the effect of tidal heating, we consider four different orbital eccentricities of the planet-moon orbit: $e_{ps} \in \{10^{-4}, 10^{-3}, 10^{-2}, 10^{-1}\}$. For the Tidal Venus and the Tidal-Illumination Venus satellites, we assume a tidal quality factor $Q_s = 100$, while we use $Q_s = 10$ for the others. Our choice of larger $Q_s$, corresponding to lower dissipation rates, for the Tidal Venus and the Tidal-Illumination Venus moons is motivated by the effect we are interested in, namely volcanism, which we assume independent of tidal dissipation in a possible ocean. Estimates for the tidal dissipation in dry solar system objects yields $Q_s \approx 100$ (Goldreich & Soter 1966). On the other moons, tidal heating is relatively weak, and we are mostly concerned with the runaway greenhouse effect, which depends on the surface energy flux. Hence, dissipation in the ocean is crucial, and because Earth's dissipation constant is near 10, we choose the same value for moons with moderate and weak tidal heating. Ultimately, we consider all these constellations at three different epochs, namely, at ages of 100, 500, and 1000 Myr.

In total, two stellar distances of the planet-moon duet, four orbital eccentricities of the planet-moon system, and three epochs yield 24 combinations, that is, our circumplanetary exomoon menageries.

## 3. Results

### 3.1. Evolution of stellar and planetary illumination

In Fig. 3, we show how absorbed stellar flux (dashed black line), illumination from the planet (dashed red line), and total illumination (solid black line) evolve for a moon at $a_{ps} = 10 R_J$ from a 13 $M_J$ planet.[7] At that distance, irradiation from the planet alone can drive a runaway greenhouse for about 200 Myr on both the Earth-type moon and the Super-Ganymede. What is more, when we include stellar irradiation from a Sun-like star at a distance of 1 AU, then the total illumination at 10 $R_J$ from the planet is above the runaway greenhouse limit of an Earth-like satellite for about 500 Myr. Our hypothetical Super-Ganymede would be in a runaway greenhouse state for roughly 600 Myr.

At a distance of 15 $R_J$, flux from the planet would be $(10/15)^2 = 0.\overline{4}$ times the red dashed line shown in Fig. 3. After 200 Myr, illumination from the planet would still be $0.\overline{4} \times 310$ W/m$^2$, and with the additional 190 W/m$^2$ absorbed from the star, the total irradiation would still sum up to about 328 W/m$^2$ at an age of 200 Myr. Consequently, even at a distance similar to that of Ganymede from Jupiter, an Earth-sized moon could undergo a runaway greenhouse effect around a 13 $M_J$ planet over several hundred million years. At a distance of 20 $R_J$, total illumination would be 268 W/m$^2$ after 200 Myr, and our Super-Ganymede test moon would still be uninhabitable. Clearly, thermal irradiation from a super-Jupiter host planet can have a major effect on the habitability of its moons.

---

[6] This mass corresponds to about ten times the mass of Ganymede and constitutes roughly the detection limit of *Kepler* (Kipping et al. 2009).

[7] For comparison, Io, Europa, Ganymede, and Callisto orbit Jupiter at approximately 6.1, 9.7, 15.5, and 27.2 Jupiter radii. In-situ formation of moons occurs mostly between roughly 5 and 30 $R_p$, from planets the mass of Saturn up to planets with the 10-fold mass of Jupiter (Sasaki et al. 2010; Heller et al. 2013).

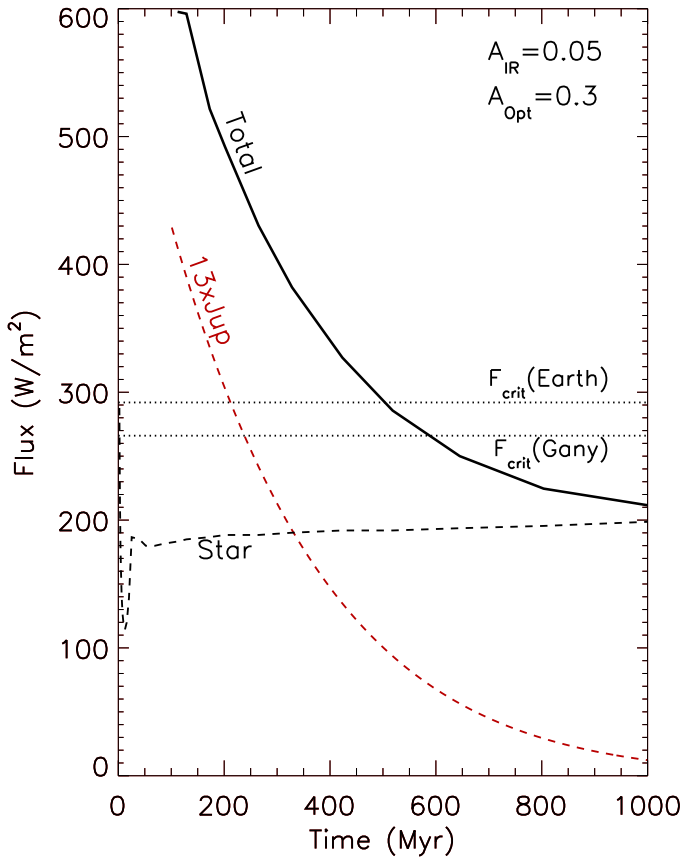

**Fig. 3.** The total illumination $F_i$ absorbed by a moon (thick black line) is composed of the absorbed flux from the star (black dashed line) and from its host planet (red dashed line). The critical values for an Earth-type and a Super-Ganymede moon to enter the runaway greenhouse effect ($F_{crit}$) are indicated by dotted lines at 295 and 266 W/m$^2$, respectively. Both moons orbit a 13 $M_J$ planet at a distance of 10 $R_J$, at 1 AU from a Sun-like host star. Illumination from the planet alone can trigger a runaway greenhouse effect for the first $\approx 200$ Myr.

### 3.2. Evolution of illumination and tidal heating

In Fig. 4, we show four examples for our circumplanetary exomoon menageries. Abscissae and ordinates denote the distance to the planet, which is chosen to be located at the center at (0,0) [note the logarithmic scale!]. Colors illustrate a Tidal Venus (red), Tidal-Illumination Venus (orange), Super-Io (yellow), Tidal Earth (blue), and an Earth-like (green) state of an Earth-sized exomoon (for details see Sect. 2.4). Light green depicts the Hill radius for retrograde moons, dark green for a prograde moons. In all panels, the planet-moon binary has an age of 500 Myr. The two panels in the left column show planet-moon systems at the inner edge of the stellar HZ; the two panels to the right show the same systems at the outer edge of the HZ. In the top panels, $e_{ps} = 10^{-1}$ and tidal heating is strong; in the bottom panels $e_{ps} = 10^{-4}$ and tidal heating is weak.

The outermost stable satellite orbit is only a fraction of the planet's Hill radius and depends, amongst others, on the distance of the planet to the star (Domingos et al. 2006): the closer the star, the smaller the planet's Hill radius. This is why the green circles are smaller at the inner edge of the HZ (left panels). In this particular constellation, the boundary between the dark and the light green circles, that is, the Hill radii for moons in prograde orbits, is at 70 $R_J$, while the outermost stable orbit for retrograde moons is at 132 $R_J$. At the outer edge of the stellar HZ (right panels), these boundaries are 121 at and 231 $R_J$, respectively.

A comparison between the top and bottom panels shows that tidal heating, triggered by the substantial eccentricity in the upper plots, can have a dramatic effect on the circumplanetary, astrophysical conditions. The Tidal Earth state (blue) in a highly eccentric orbit (upper panels) can be maintained between roughly 21 and 36 $R_J$ both at the inner and the outer edge of the stellar HZ.[8] But this state does not exist in the extremely low eccentricity configuration at the inner HZ edge (lower left panel). In the latter constellation, the region of moderate tidal heating is inside the circumplanetary sphere in which stellar illumination plus illumination from the planet are strong enough to trigger a runaway greenhouse effect on the moon (red and orange circles for a Tidal Venus and Tidal-Illumination Venus state, respectively). In the lower right panel, a very thin rim of a Tidal Earth state exists at roughly 8 $R_J$.

Each of these four exomoon menageries has its own circumplanetary death zone, that is, a range of orbits in which an Earth-like moon would be in a Tidal Venus or Tidal-Illumination Venus state. In Heller & Barnes (2013), we termed the outermost orbit, which would just result in an uninhabitable satellite, the "habitable edge". Inspection of Fig. 4 yields that the habitable edges for $e_{ps} = 10^{-1}$ (top) and $e_{ps} = 10^{-4}$ (bottom) are located at around 20 and 12 $R_J$ with the planet-moon binary at the inner edge of the stellar HZ (left column), and at 15 and 8 $R_J$ with the planet-moon pair orbiting at the outer edge of the HZ (right column, top and bottom panels), respectively.

Note that for moons with surface gravities lower than that of Earth, the critical energy flux for a runaway greenhouse effect would be smaller. Although tidal surface heating in the satellite would also be smaller, as it is proportional to the satellite's radius cubed, the habitable edge would be even farther away from the planet (see Fig. 10 in Heller & Barnes 2013). For Mars- to Earth-sized moons, and in particular for our Super-Ganymede hypothetical satellite, the habitable edges would be even farther away from the planet than depicted in Fig. 4.

Figure 5, finally, shows our whole model grid of 24 exomoon menageries. The four examples from Fig. 4 can be found at the very top and the very bottom of the diagram in the center – though now on a linear rather than on a logarithmic scale. The left, center, and right graphics show a range of exomoon menageries at ages of 100, 500, and 1000 Myr, respectively. In all three epochs, highly eccentric exomoon orbits are shown at the top ($e_{ps} = 10^{-1}$), almost circular orbits at the bottom ($e_{ps} = 10^{-4}$). The four circumplanetary circles at the left illustrate menageries with the planet-moon system assumed at the inner edge of the HZ, while the four menageries at the right visualize the planet-moon duet at the outer edge of the HZ.

In the highly eccentric cases at the top, tidal heating dominates the circumplanetary orbital conditions for exomoon hab-itability. In the system's youth at 100 Myr, the circumplanetary habitable edge is as far as 29 $R_{\rm J}$ (upper left panel, at the inner edge of the HZ) or 18 $R_{\rm J}$ (right panel in the leftmost diagram, outer edge of the HZ) around the planet. The region for Tidal Earth moons spans from roughly 29 to 36 $R_{\rm J}$ at the inner edge of the HZ and from 21 to 36 $R_{\rm J}$ at the outer edge of the stellar HZ. At an age of 1000 Myr (diagram at the right), when illumination from the planet has decreased by more than one order of magnitude (see red line in Fig. 2), the habitable edge has moved inward to roughly 19 $R_{\rm J}$ at the inner edge and 16 $R_{\rm J}$ at the outer edge of the HZ, while the Tidal Earth state is between 21 to 36 $R_{\rm J}$ in both the inner and outer HZ edge cases.

---

[8] Tidal heating in the moon does not depend on the stellar distance.

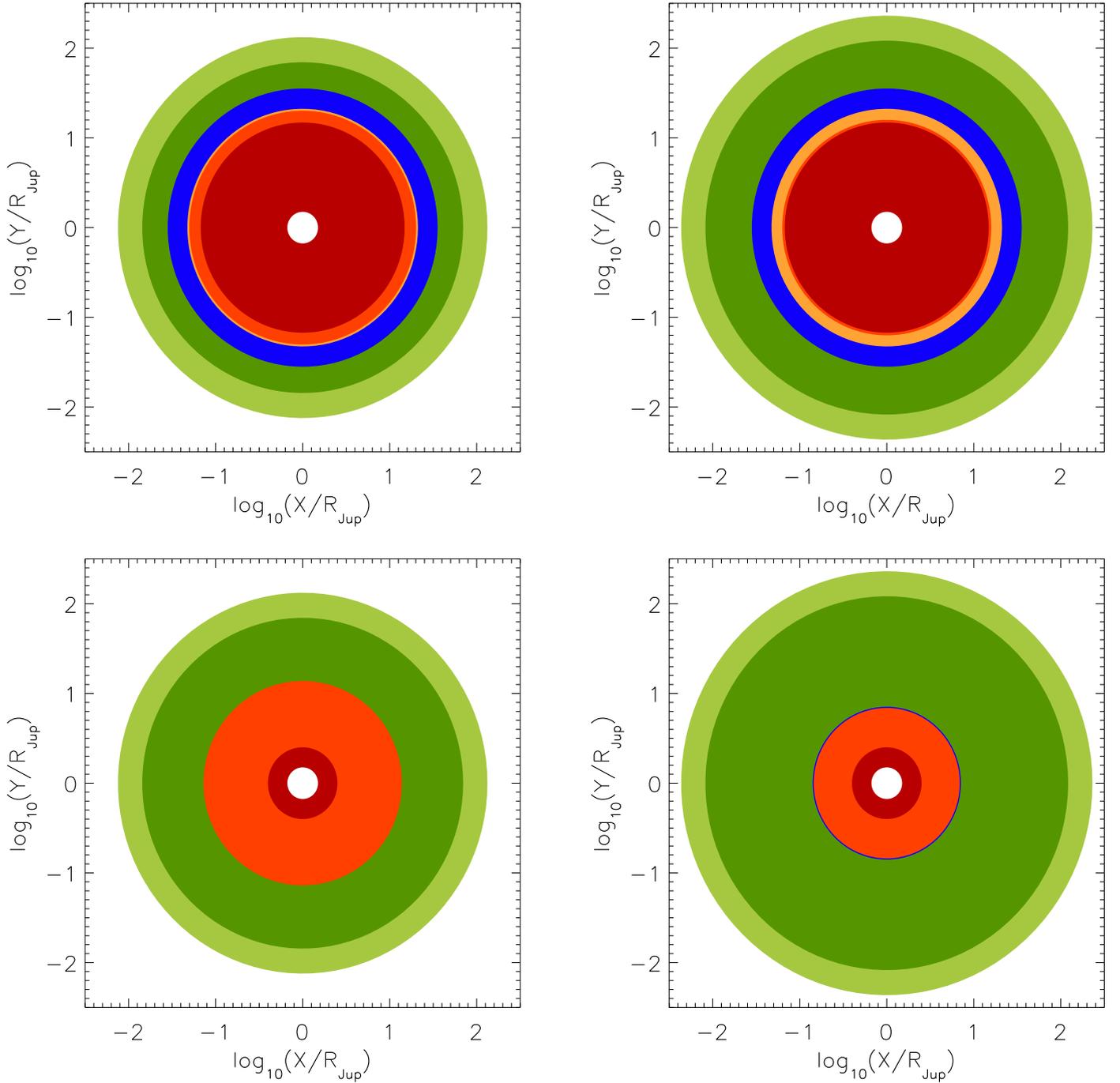

**Fig. 4.** Circumplanetary exomoon menageries for Earth-sized satellites around a 13 Jupiter-mass host planet at an age of 500 Myr. In each panel, the planet's position is at (0,0), and distances are shown on a logarithmic scale. In the left panels, the planet-moon binary orbits at a distance of 1 AU from a Sun-like star; in the right panels, the binary is at the outer edge of the stellar HZ at 1.738 AU. In the upper two panels, $e_{\rm ps} = 10^{-1}$; in the lower two panels, $e_{\rm ps} = 10^{-4}$. Starting from the planet in the center, the white circle visualizes the Roche radius, and the exomoon types correspond to Tidal Venus (red), Tidal-Illumination Venus (orange), Super-Io (yellow), Tidal Earth (blue), and Earth-like (green) states (see Sect. 2.4 for details). Dark green depicts the Hill sphere of prograde Earth-like moons, light green for retrograde Earth-like moons. Note the larger Hill radii at the outer edge of the HZ (right panels)!

In the low-eccentricity scenarios at the bottom, tidal heating has a negligible effect, and thermal flux from the planet dominates the evolution of the circumplanetary conditions. At 100 Myr, the Tidal-Illumination Venus state, which is just inside the planetary habitable edge, ranges out to 28 $R_J$ at the inner edge of the HZ (lower left) and to 14 $R_J$ at the outer edge of the HZ (lower right menagerie in the left-most diagram). At an age of 1000 Myr, those values have decreased to 5 and 2.5 $R_J$ for the planet-moon duet at the inner and outer HZ boundaries, respectively (bottom panels in the rightmost sketch). Moreover, at 1000 Myr illumination from the planet has become weak enough that moons with substantial tidal heating in these low-eccentricity scenarios could exist between 5 and 6 $R_J$ (Tidal Earth) at the inner HZ edge or between 2.5 and 6 $R_J$ (Super-Io and Tidal Earth) at the outer HZ boundary.

## 4. Conclusions

Young and hot giant planets can illuminate their potentially habitable, Earth-sized moons strong enough to make them uninhabitable for several hundred million years. Based on the planetary evolution models of Baraffe et al. (2003), thermal irradiation from a 13 $M_J$ planet on an Earth-sized moon at a distance of 10 $R_J$ can trigger a runaway greenhouse effect for about 200 Myr. The total flux of Sun-like irradiation at 1 AU plus thermal flux from a 13 $M_J$ planet will force Earth-sized moons at 10 or 15 $R_J$ into a runaway greenhouse for 500 or more than 200 Myr, respectively. A Super-Ganymede moon 0.25 times the mass of Earth and with an ice-to-mass fraction of 25 % would undergo a runaway greenhouse effect for longer periods at the same orbital distances or, equivalently, for the same periods at larger separations from the planet. Even at a distance of 20 $R_J$ from a 13 $M_J$ host giant, it would be subject to a runaway greenhouse effect for about 200 Myr, if it receives an early-Earth-like illumination from a young Sun-like star. In all these cases, the moons could lose substantial amounts of hydrogen and, consequently, of water. Such exomoons would be temporarily uninhabitable and, perhaps, uninhabitable forever.

If tidal heating is included, the danger for an exomoon to undergo a runaway greenhouse effect increases. The habitable edge around young giant planets, at an age of roughly 100 Myr and with a mass 13 times that of Jupiter, can extend out to about 30 $R_J$ for moons in highly eccentric orbits ($e_{ps} \approx 0.1$). That distance encompasses the orbits of Io, Europa, Ganymede, and Callisto around Jupiter. Beyond the effects we considered here, that is, stellar irradiation, thermal irradiation from the planet, and tidal heating, other heat sources in moons may exist. Consideration of primordial thermal energy (or "sensible heat"), radioactive decay, and latent heat from solidification inside an exomoon would increase the radii of the circumplanetary menageries and push the habitable edge even further away from the planet.

Our estimates for the instantaneous habitability of exomoons should be regarded as conservative because (*i.*) the minimum separation for an Earth-like moon from its giant host planet to be habitable could be even larger than we predict. This is because a giant planet's luminosity at 1 AU from a Sun-like star can decrease more slowly than in the Baraffe et al. (2003) models used in our study (Fortney et al. 2007). (*ii.*) The runaway greenhouse limit, which we used here to assess instantaneous habitability, is a conservative approach itself. A moon with a total energy flux well below the runaway greenhouse limit may be uninhabitable as its surface is simply too hot. It could be caught in a moist greenhouse state with surface temperatures up to the critical point of water, that is, 647 K if an Earth-like inventory of water and surface $H_2O$ pressure are assumed (Kasting 1988). (*iii.*) Moons with surface gravities smaller than that of Earth will have a critical energy flux for the runaway greenhouse effect that is also smaller than Earth's critical flux. Hence, their corresponding Tidal Venus and Tidal-Illumination Venus states would reach out to wider orbits than those depicted in Figs. 4 and 5. (*iv.*) Tidal heating could be stronger than the values we derived with the Ferraz-Mello et al. (2008) CPL tidal model. As shown in Heller et al. (2011), the CTL theory of Leconte et al. (2010) includes terms of higher orders in eccentricity and yields stronger tidal heating than the CPL of Ferraz-Mello et al. (2008). Yet, such mathematical extensions may not be physically valid (Greenberg 2009), and parametrization of a planet's or satellite's tidal response with a constant tidal quality factor $Q$ or a fixed tidal time lag $\tau$ remains uncertain (Leconte et al. 2010; Heller et al. 2011; Efroimsky & Makarov 2013).

Massive moons can bypass an early runaway greenhouse state if they form after the planet has cooled sufficiently. Possible scenarios for such a delayed formation include gravitational capture of one component of a binary planet system, the capture of Trojans, gas drag or pull-down mechanisms, moon mergers, and impacts on terrestrial planets (for a review, see Sect. 2.1 in Heller & Barnes 2013). Alternatively, a desiccated moon in the stellar HZ could be re-supplied by cometary bombardment later. Reconstruction of any such event would naturally be difficult.

Once extrasolar moons will be discovered, assessments of their habitability will depend not only on their current orbital configuration and irradiation, but also on the history of stellar and planetary luminosities. Even if a Mars- to Earth-sized moon would be found about a Jupiter- or super-Jupiter-like planet at, say, 1 AU from a Sun-like star, the moon could have lost substantial amounts of its initial water reservoir and be uninhabitable today. In the most extreme cases, strong thermal irradiation from the young, hot giant host planet could have desiccated the moon long ago by the runaway greenhouse effect.

*Acknowledgements.* The referee reports of Jim Kasting, Nader Haghighipour, and an anonymous reviewer significantly improved the quality of this study. We thank Jorge I. Zuluaga for additional comments on the manuscript and Jean Schneider for technical support. René Heller is funded by the Canadian Astrobiology Training Program and a member of the Origins Institute at McMaster University. Rory Barnes acknowledges support from NSF grant AST-1108882 and the NASA Astrobiology Institute's Virtual Planetary Laboratory lead team under cooperative agreement No. NNH05ZDA001C. This work has made use of NASA's Astrophysics Data System Bibliographic Services and of Jean Schneiders exoplanet database www.exoplanet.eu.

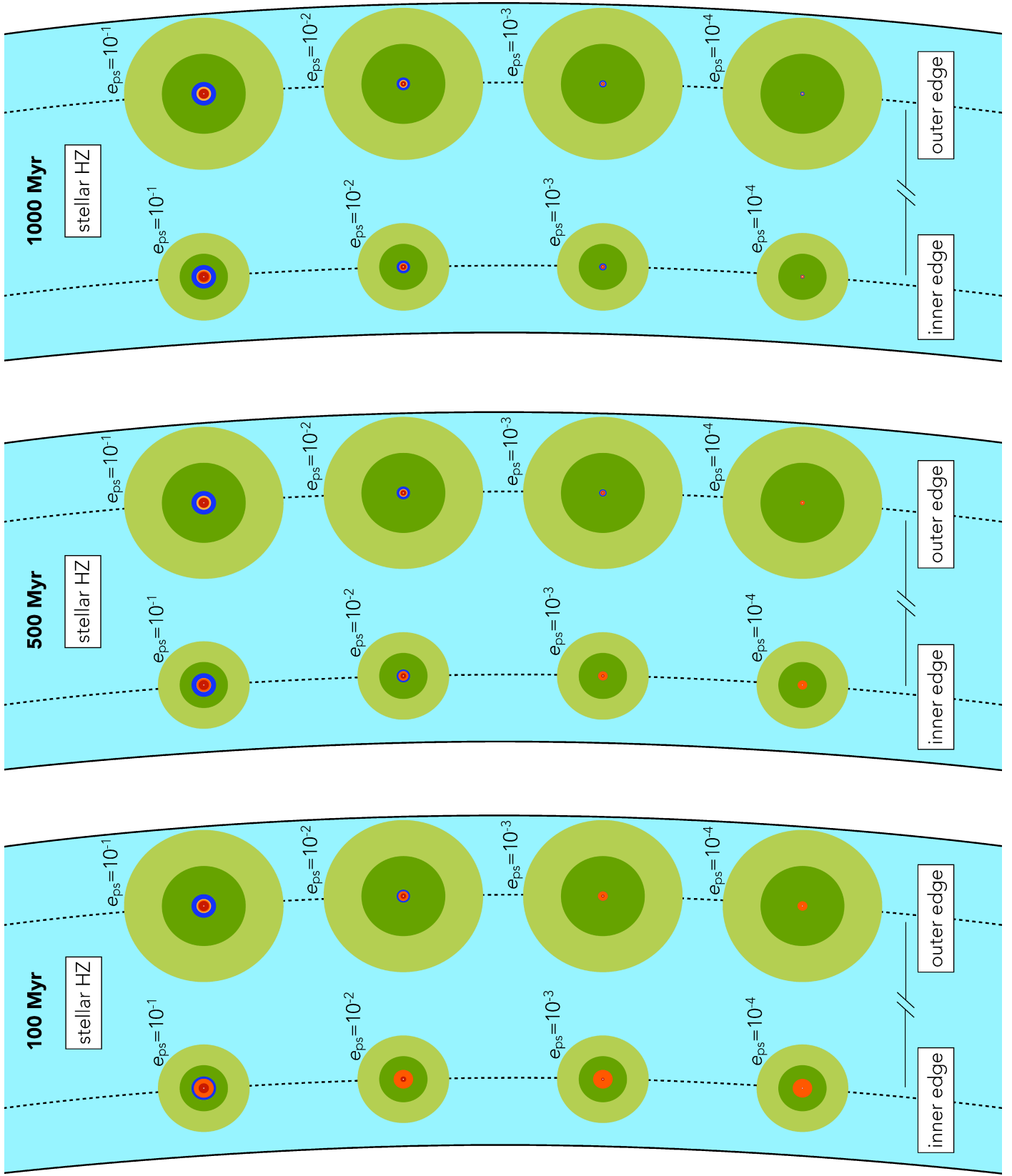

**Fig. 5.** Exomoon menageries at 100 (left), 500 (center), and 1000 Myr (right) for a hypothetical Earth-like satellite around a 13 $M_J$ planet. For each epoch, one suite of menageries is at the inner edge of the HZ (left) and another one at the outer edge of the HZ around a Sun-like star. Four different orbital eccentricities $e_{ps}$ of the satellite are indicated. Distances from the central planet in each menagerie are on a linear scale, and absolute values can be estimated by comparison with the four examples shown on a logarithmic scale in Fig. 4. The Hill sphere for prograde moons (boundary between dark and light green) is at 70 $R_J$ at the inner HZ boundary and at 121 $R_J$ at the outer HZ edge.

## 7.5 Magnetic Shielding of Exomoons Beyond the Circumplanetary Habitable Edge (Heller & Zuluaga 2013)

Contribution:

RH contributed to the literature research, contributed to the mathematical investigations, contributed to the writing of the computer code and to the generation of the figures, led the writing of the manuscript, and served as a corresponding author for the journal editor and the referees.

# MAGNETIC SHIELDING OF EXOMOONS BEYOND THE CIRCUMPLANETARY HABITABLE EDGE

René Heller[1]

McMaster University, Department of Physics and Astronomy, Hamilton, ON L8S 4M1, Canada; rheller@physics.mcmaster.ca

AND

Jorge I. Zuluaga[2]

FACom - Instituto de Física - FCEN, Universidad de Antioquia, Calle 70 No. 52-21, Medellín, Colombia; jzuluaga@fisica.udea.edu.co
Draft version September 5, 2013

## ABSTRACT

With most planets and planetary candidates detected in the stellar habitable zone (HZ) being super-Earths and gas giants, rather than Earth-like planets, we naturally wonder if their moons could be habitable. The first detection of such an exomoon has now become feasible, and due to observational biases it will be at least twice as massive as Mars. But formation models predict moons can hardly be as massive as Earth. Hence, a giant planet's magnetosphere could be the only possibility for such a moon to be shielded from cosmic and stellar high-energy radiation. Yet, the planetary radiation belt could also have detrimental effects on exomoon habitability. We here synthesize models for the evolution of the magnetic environment of giant planets with thresholds from the runaway greenhouse (RG) effect to assess the habitability of exomoons. For modest eccentricities, we find that satellites around Neptune-sized planets in the center of the HZ around K dwarf stars will either be in an RG state and not be habitable, or they will be in wide orbits where they will not be affected by the planetary magnetosphere. Saturn-like planets have stronger fields, and Jupiter-like planets could coat close-in habitable moons soon after formation. Moons at distances between about 5 and 20 planetary radii from a giant planet can be habitable from an illumination and tidal heating point of view, but still the planetary magnetosphere would critically influence their habitability.

*Keywords:* astrobiology – celestial mechanics – methods: analytical – planets and satellites: magnetic fields – planets and satellites: physical evolution – planet-star interactions

## 1. INTRODUCTION

The search for life on worlds outside the solar system has experienced a substantial boost with the launch of the *Kepler* space telescope in 2009 March (Borucki et al. 2010). Since then, thousands of planet candidates have been detected (Batalha et al. 2013), several tens of which could be terrestrial and have orbits that would allow for liquid surface water (Kaltenegger & Sasselov 2011). The concept used to describe this potential for liquid water, which is tied to the search for life, is called the "habitable zone" (HZ) (Kasting et al. 1993). Yet, most of the *Kepler* candidates, as well as most of the planets in the HZ detected by radial velocity measurements, are giant planets and not Earth-like. This leads us to the question whether these giants can host terrestrial exomoons, which may serve as habitats (Reynolds et al. 1987; Williams et al. 1997).

Recent searches for exomoons in the *Kepler* data (Kipping et al. 2013a,b) have fueled the debate about the existence and habitability of exomoons and incentivized others to develop models for the surface conditions on these worlds. While these studies considered illumination effects from the star and the planet, as well as eclipses, tidal heating (Heller 2012; Heller & Barnes 2013b,a), and the transport of energy in the moon's atmosphere (Forgan & Kipping 2013), the magnetic environment of exomoons has hitherto been unexplored.

Moons around giant planets are subject to high-energy radiation from (1) cosmic particles, (2) the stellar wind, and (3) particles trapped in the planet's magnetosphere (Baumstark-Khan & Facius 2002). Contributions (1) and (2) are much weaker inside the planet's magnetosphere than outside, but effect (3) can still have detrimental consequences. The net effect (beneficial or detrimental) on a moon's habitability depends on the actual orbit, the extent of the magnetosphere, the intrinsic magnetosphere of the moon, the stellar wind, etc.

Stellar mass-loss and X-ray and extreme UV (XUV) radiation can cause the atmosphere of a terrestrial world to be stripped off. Light bodies are in particular danger, as their surface gravity is weak and volatiles can escape easily (Lammer et al. 2013). Mars, for example, is supposed to have lost vast amounts of $CO_2$, N, O, and H (the latter two formerly bound as water) (McElroy 1972; Pepin 1994; Valeille et al. 2010). Intrinsic and extrinsic magnetic fields can help moons sustain their atmospheres and they are mandatory to shield life on the surface against galactic cosmic rays (Grießmeier et al. 2009). High-energy radiation can also affect the atmospheric chemistry, thereby spoiling signatures of spectral biomarkers, especially of ozone (Segura et al. 2010).

Understanding the evolution of the magnetic environments of exomoons is thus crucial to asses their habitability. We here extend models recently applied to the evolution of magnetospheres around terrestrial planets under an evolving stellar wind (Zuluaga et al. 2013). Employed on giant planets, they allow us a first approach toward parameterizing the potential of a giant planet's magnetosphere to affect potentially habitable moons.

## 2. METHODS

In the following, we model a range of hypothetical systems to explore the potential of a giant planet's mag-netosphere to embrace moons that are habitable from a tidal and energy budget point of view.

### 2.1. *Bodily Characteristics of the Star*

M dwarf stars are known to show strong magnetic bursts, eventually coupled with the emission of XUV radiation as well as other high-energy particles (Gurzadian 1970; Silvestri et al. 2005). Moreover, stars with masses below 0.2–0.5 solar masses ($M_\odot$) cannot possibly host habitable moons, since stellar perturbations excite hazardous tidal heating in the satellites (Heller 2012). We thus concentrate on stars more similar to the Sun. G dwarfs, however, are likely too bright and too massive to allow for exomoon detections in the near future. As a compromise, we choose a $0.7\,M_\odot$ K dwarf star with solar metallicity $Z = 0.0152$ and derive its radius ($R_\star$) and effective temperature ($T_{\rm eff,\star}$) at an age of 100 Myr (Bressan et al. 2012): $R_\star = 0.597\,R_\odot$ ($R_\odot$ being the solar radius), $T_{\rm eff,\star} = 4270$ K. These values are nearly constant over the next couple of Gyr.

### 2.2. *Bodily Characteristics of the Planet*

We use planetary evolution models of Fortney et al. (2007) to explore two extreme scenarios, between which we expect most giant planets: (1) mostly gaseous with a core mass $M_c = 10$ Earth masses ($M_\oplus$) and (2) planets with comparatively massive cores. Class (1) corresponds to larger planets for given planetary mass ($M_p$). Models for suite (2) are constructed in the following way. For $M_p < 0.3\,M_{\rm Jup}$, we interpolate between radii of planets with core masses $M_c = 10, 25, 50$ and $100\,M_\oplus$ to construct Neptune-like worlds with a total amount of 10 % hydrogen (H) and helium (He) by mass. For $M_p > 0.3\,M_{\rm Jup}$, we take the precomputed $M_c = 100\,M_\oplus$ grid of models. These massive-core planets (2) yield an estimate of the *minimum* radius for given $M_p$. To account for irradiation effects on planetary evolution, we apply the Fortney et al. (2007) models for planets at 1 AU from the Sun.

As it is desirable to compare our scaling laws for the magnetic properties of giant exoplanets with known magnetic dipole moments of solar system worlds, we start out by considering a Neptune-, a Saturn-, and a Jupiter-class host planet. For the sake of consistency, we attribute total masses of 0.05, 0.3, and 1 $M_{\rm Jup}$, as well as $M_c = 10$, 25, and again $10\,M_\oplus$, respectively.

### 2.3. *Bodily Characteristics of the Moon*

*Kepler* has been shown capable of detecting moons as small as $0.2\,M_\oplus$ combining measurements of the planet's transit timing variation and transit duration variation (Kipping et al. 2009). The detection of a planet as small as 0.3 Earth radii ($R_\oplus$), almost half the radius of Mars (Barclay et al. 2013), around a K star suggests that direct transit measurements of Mars-sized moons may be possible with current or near-future technology (Sartoretti & Schneider 1999; Szabó et al. 2006; Kipping 2011). Yet, in-situ formation of satellites is restricted to a few times $10^{-4}\,M_p$ at most (Canup & Ward 2006; Sasaki et al. 2010; Ogihara & Ida 2012). For a Jupiter-mass planet, this estimate yields a satellite of about $0.03\,M_\oplus \approx 0.3$ times the mass of Mars. Alternatively, moons can form via a range of other mechanisms (for a review, see Section 2.1 in Heller & Barnes 2013b). We therefore suspect moons of roughly the mass and size of Mars to exist and to be detectable in the near future.

Following Fortney et al. (2007) and assuming an Earth-like rock-to-mass fraction of 68 %, we derive a radius of 0.94 Mars radii or $0.5\,R_\oplus$ for a Mars-mass exomoon. Tidal heating in the moon is calculated using the model of Leconte et al. (2010) and assuming an Earth-like time lag of the moon's tidal bulge $\tau_s = 638$ s as well as a second degree tidal Love number $k_{2,s} = 0.3$ (Heller et al. 2011).

### 2.4. *The Stellar Habitable Zone*

We investigate a range of planet-moon binaries located in the center of the stellar HZ. Therefore, we compute the arithmetic mean of the inner HZ edge (given by the moist greenhouse effect) and the outer HZ edge (given by the maximum greenhouse) around a K dwarf star (Section 2.1) using the model of Kopparapu et al. (2013). For this particular star, we localize the center of the HZ at 0.56 AU.

### 2.5. *The Runaway Greenhouse and Io Limits*

The tighter a moon's orbit around its planet, the more intense the illumination it receives from the planet and the stronger tidal heating. Ultimately, there exists a minimum circumplanetary orbital distance, at which the moon becomes uninhabitable, called the "habitable edge" (Heller & Barnes 2013b). As tidal heating depends strongly on the orbital eccentricity $e_{ps}$, amongst others, the radius of the HE also depends on $e_{ps}$.

We consider two thresholds for a transition into an uninhabitable state: (1) When the moon's tidal heating reaches a surface flux similar to that observed on Jupiter's moon Io, that is $2\,{\rm W\,m^{-2}}$, then enhanced tectonic activity as well as hazardous volcanism may occur. Such a scenario could still allow for a substantial area of the moon to be habitable, as tidal heat leaves the surface through hot spots, (see Io and Enceladus, Ojakangas & Stevenson 1986; Spencer et al. 2006; Tobie et al. 2008), and there may still exist habitable regions on the surface. Hence, we consider this "Io-limit HE" (Io HE) as a pessimistic approach. (2) When the moon's global energy flux exceeds the critical flux to become a runaway greenhouse (RG), then any liquid surface water reservoirs can be lost due to photodissociation into hydrogen and oxygen in the high atmosphere (Kasting 1988). Eventually, hydrogen escapes into space and the moon will be desiccated forever. We apply the semi-analytic RG model of Pierrehumbert (2010) (see Equation 1 in Heller & Barnes 2013b) to constrain the innermost circumplanetary orbit at which a moon with given eccentricity would just be habitable. For our prototype moon, this model predicts a limit of $269\,{\rm W\,m^{-2}}$ above which the moon would transition to an RG state. We call the corresponding critical semi-major axis the "runaway greenhouse HE" (RG HE) and consider it as an optimistic approach.

We calculate the Io and RG HEs for $e_{ps} \in \{0.1, 0.01, 0.001\}$ using Equation (22) from Heller & Barnes 2013b) and introducing two modifications. First, the planetary surface temperature ($T_p$) depends on the equilibrium temperature ($T_p^{\rm eq}$) due to absorbed stellar light and on an additional component ($T_{\rm int}$) from internal heating: $T_p = ([T_p^{\rm eq}]^4 + T_{\rm int}^4)^{1/4}$. Second, we use a

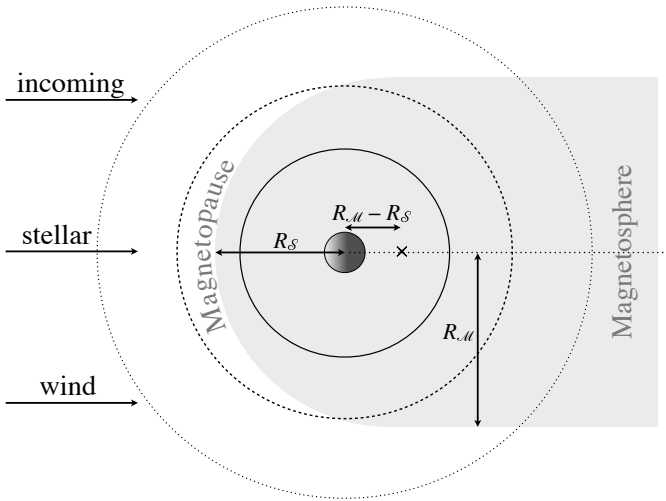

**Figure 1.** Sketch of the planetary magnetosphere. $R_\mathcal{S}$ denotes the standoff distance, $R_\mathcal{M}$ labels the radius of the magnetopause. A range of satellite orbits illustrates how a moon can periodically dive into and out of the planetary magnetosphere. Conceptually, the dashed orbit resembles that of Titan around Saturn, with occasional shielding and exposition to the solar wind (Bertucci et al. 2008). At orbital distances $\gtrsim 2R_\mathcal{S}$, the fraction of the moon's orbit spent outside the planet's magnetospheric cavity reaches $\approx 80\%$.

Bond albedo $\alpha_{\rm opt} = 0.3$ for the stellar illumination absorbed by the moon and $\alpha_{\rm IR} = 0.05$ for the light absorbed from the relatively cool planet (Heller & Barnes 2013a).

### 2.6. Planetary Dynamos

We apply scaling laws for the magnetic field strength (Olson & Christensen 2006) that consider convection in a spherical conducting shell inside a giant planet and a convective power $Q_{\rm conv}$ (Equation (28) in Zuluaga et al. 2013), provided by the Fortney et al. (2007) models. The ratio between inertial forces to Coriolis forces is crucial in determining the field regime — be it dipolar- or multipolar-dominated — for the dipole field strength on the planetary surface. We scale the ratio between dipolar and total field strengths following Zuluaga & Cuartas (2012).

The core density ($\rho_{\rm c}$) is estimated by solving a polytropic model of index 1 (Hubbard 1984). Radius and extent of the convective region are estimated by applying a semi-empirical scaling relationship for the dynamo region (Grießmeier 2006). Thermal diffusivity $\kappa$ is assumed equal to $10^{-6}$ for all planets (Guillot 2005). Electrical conductivity $\sigma$ is assumed to be $6 \times 10^4$ for planets rich in H and He, and $\sigma = 1.8 \times 10^4$ for the ice-rich giants (Olson & Christensen 2006).

We have verified that our model predicts dynamo regions that are similar to results obtained by more sophisticated analyses. For Neptune, our model predicts a dynamo radius $R_{\rm c} = 0.77$ planetary radii ($R_{\rm p}$) and $\rho_{\rm c} = 3000\,{\rm kg\,m^{-3}}$, in good agreement with Kaspi et al. (2013). We also ascertained the dynamo scaling laws to reasonably reproduce the planetary dipole moments of Ganymede, Earth, Uranus, Neptune, Saturn, and Jupiter (J. I. Zuluaga et al., in preparation). For the mass range considered here, the predicted dipole moments agree within a factor of two to six. Discrepancies of this magnitude are sufficient for our estimation of magnetospheric properties, because they scale with $M_{\rm dip}^{1/3}$. Significant underestimations arise for Neptune and Uranus, which have strongly non-dipolar surface fields.

### 2.7. Evolution of the Magnetic Standoff Distance

The shape of the planet's magnetosphere can be approximated as a combination of a semi-sphere with radius $R_\mathcal{M}$ and a cylinder representing the tail region (Figure 1). The planet is at a distance $R_\mathcal{M} - R_\mathcal{S}$ off the sphere's center, with

$$R_\mathcal{S} = \left(\frac{\mu_0 f_0^2}{8\pi^2}\right)^{1/6} \mathcal{M}^{1/3} P_{\rm sw}^{-1/6} \qquad (1)$$

being the standoff distance, $\mu_0 = 4\pi \times 10^{-7}\,{\rm N\,A^{-2}}$ the vacuum magnetic permeability, $f_0 = 1.3$ a geometric factor, $\mathcal{M}$ planetary magnetic dipole moment, and $P_{\rm sw} \propto n_{\rm sw} v_{\rm sw}$ the dynamical pressure of the stellar wind. Number densities $n_{\rm sw}$ and velocities $v_{\rm sw}$ of the stellar wind are calculated using a hydrodynamical model (Parker 1958). Both quantities evolve as the star ages. Thus, we use empirical formulae (Grießmeier et al. 2007) to parameterize their time dependence.

Observations of stellar winds from young stars are challenging, and thus the models only cover stars older than 700 Myr. We extrapolate these models back to 100 Myr, although there are indications that stellar winds "saturate" when going back in time (J. Linsky 2012, private communication).

## 3. RESULTS

### 3.1. Evolution of Magnetic Standoff Distance versus Runaway Greenhouse and Io Habitable Edges

Figure 2 visualizes the evolution of $R_\mathcal{S}$ (thick blue line) as a snail curling around the planet, indicated by a dark circle in the center. Stellar age ($t_\star$) is denoted in units of Gyr along the snail, starting at 0.1 Gyr at "noon" and ending after 4.6 Gyr at "midnight". $R_\mathcal{S}$, as well as the RG and Io HEs, are given in units of $R_{\rm p}$. At $t_\star = 0.1$ Gyr ($t_\star = 4.6$ Gyr), $R_{\rm p} = 0.385$, 1.023, and 1.195 $R_{\rm Jup}$ (0.329, 0.862, and 1.056 $R_{\rm Jup}$) for the Neptune-, Saturn-, and Jupiter-like planets, respectively.

In panel (a) for the Neptune-like host, $R_\mathcal{S}$ starts very close to the planet and even inside the RG HE for the $e_{\rm ps} = 0.001$ case (thin black snail), at roughly $2\,R_{\rm p}$ from the planetary center. This means, at an age of 0.1 Gyr any Mars-like satellite with $e_{\rm ps} = 0.001$ would need to orbit beyond $2\,R_{\rm p}$ to avoid transition into a RG state, where it would not be affected by the planet's magnetosphere. As the system ages, wider orbits are enshrouded by the planetary magnetic field, until after 4.6 Gyr, $R_\mathcal{S}$ reaches as far as the Io HE for $e_{\rm ps} = 0.1$ (thick dashed line).[1] In summary, moons in low-eccentricity orbits around Neptune-like planets can be close the planet and be habitable from an illumination and tidal heating point of view, but it will take at least 300 Myr in our specific case until they get coated by the planet's magnetosphere. Exomoons on more eccentric orbits will either be uninhabitable or affected by the planet's magnetosphere after more than 300 Myr.

---

[1] Note that as the planet shrinks, the RG and Io HEs move outward, too. This is because of our visualization in units of planetary radii, while the values of the HEs remain constant in secular units.

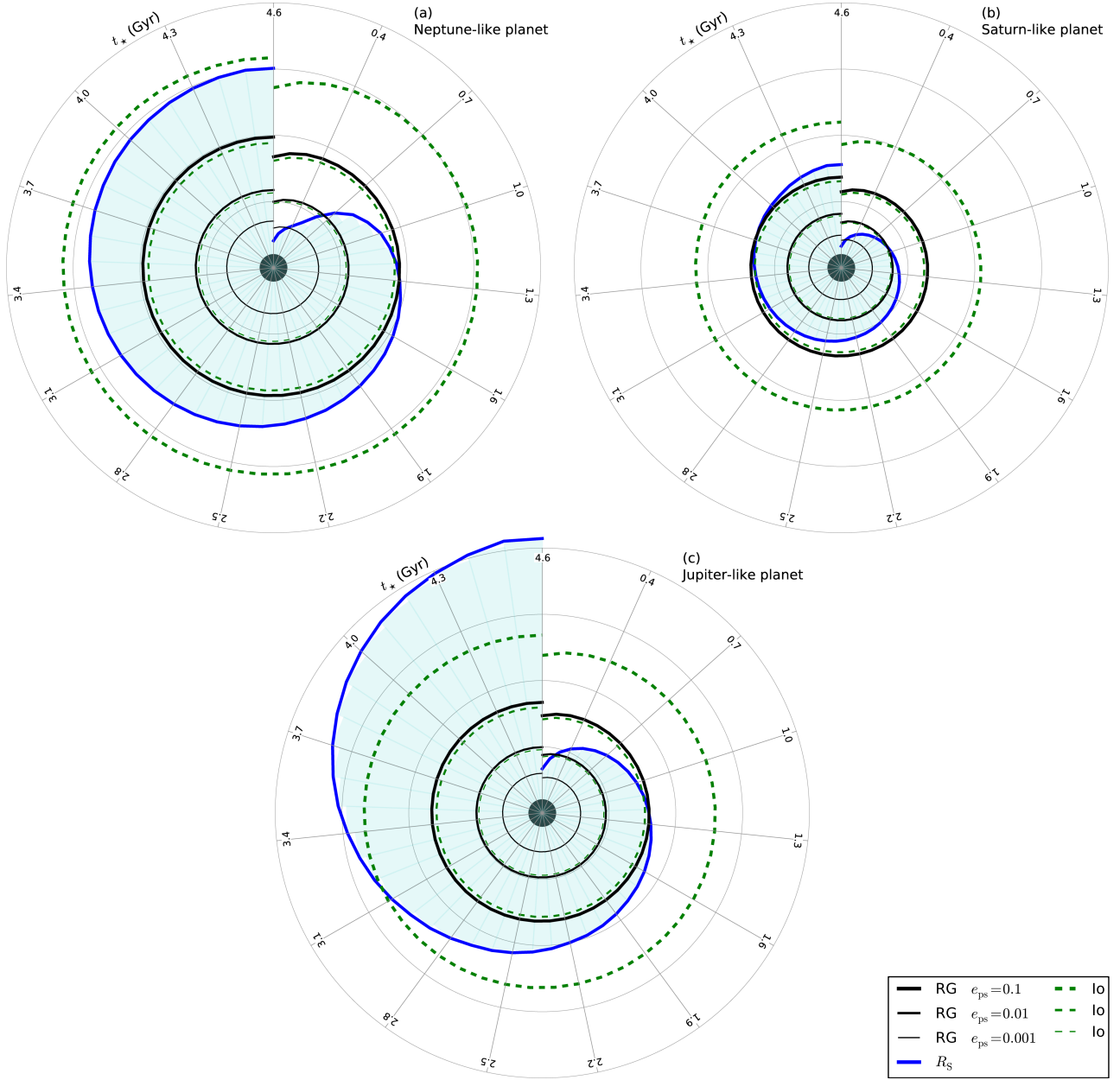

**Figure 2.** Evolution of the magnetic shielding (blue curves) compared to the RG HEs (solid black lines) and Io-like HEs (dashed green lines). Thick lines correspond to HEs for $e_{\rm ps} = 0.1$, intermediate thickness to $e_{\rm ps} = 0.01$, and thin circles to $e_{\rm ps} = 0.001$. Thin gray lines denote distances in intervals of 5 planetary radii. The filled circle in the center symbolizes the planetary radius. Panel (a) shows a Neptune-like host, (b) a Saturn-like planet, and (c) a Jupiter-like planet. In all computations, a Mars-sized moon is assumed, and the planet-moon binary orbits a K dwarf in the center of the stellar HZ.

Moving on to Figure 2(b) and the Saturn-like host, we find that $R_{\mathcal{S}}$ reaches the $e_{\rm ps} = 0.001$ RG HE after roughly 200 Myr, but it transitions all the other HEs substantially later than in the case of a Neptune-like host. After $\approx 1$ Gyr, the $e_{\rm ps} = 0.01$ RG HE and the Io HE for $e_{\rm ps} = 0.01$ are transversed. After 4.6 Gyr, all orbits except for the Io HE at an eccentricity of 0.1 are covered by the planet's magnetic field. In conclusion, close-in Mars-sized moons around Saturn-like planets can be magnetically affected early on and be habitable from a RG or tidal heating point of view, if their orbital eccentricities are small.

Finally, Figure 2(c) shows that exomoons at the RG

HE of Jovian planets for $e_{\rm ps} = 0.001$ will be bathed in the planetary magnetosphere at stellar ages as young as 100 Myr. After roughly 1.3 Gyr, $R_{\mathcal{S}}$ transitions the RG HE for $e_{\rm ps} = 0.1$ and the Io HE for $e_{\rm ps} = 0.01$. Even the Io HE with an eccentricity of $e_{\rm ps} = 0.1$ is covered at around 3.1 Gyr. After 4.6 Gyr, the planet's magnetic shield reaches as far as 21 $R_{\rm p}$. Clearly, exomoons about Jupiter-like planets face the greatest prospects of interference with the planetary magnetosphere, even in orbits that are sufficiently wide to ensure negligible tidal heating.

### 3.2. *Minimum Magnetic Dipole Moment*

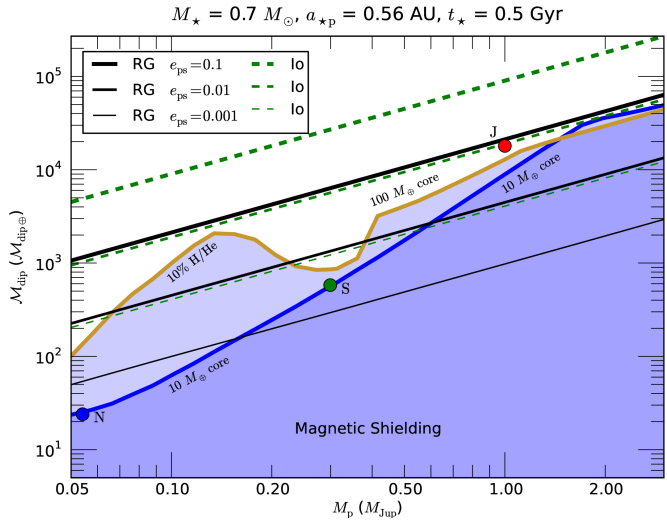

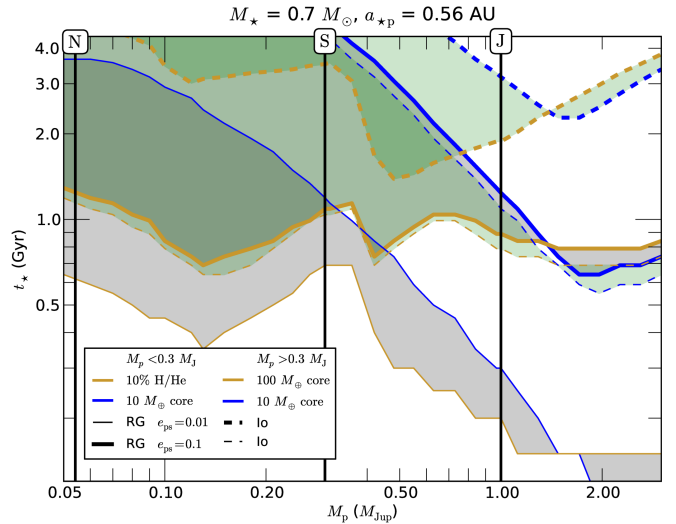

**Figure 3.** Magnetic dipole moments (ordinate) for a range of planetary masses (abscissa). The straight lines depict $\mathcal{M}_{dip}$ as it would be required at an age of 0.5 Gyr in order to shield a moon at a given HE (solid: runaway greenhouse; dashed: Io-like heating). Low-mass giants obviously fail to protect their moons beyond most HEs, while Jupiter-like planets offer a range of shielded orbits beyond the RG and Io HEs. Neptune, Saturn, and Jupiter are indicated with filled circles.

**Figure 4.** Time required by the planet's magnetic standoff radius $R_S$ to envelop the Io (dashed) and RG (solid) HEs. Shaded regions illustrate uncertainties coming from planetary structure models, with the boundaries corresponding to a high- and a low-core mass planetary model, respectively (brown and blue lines). In general, the magnetic standoff distance around more massive planets requires less time to enshroud moons at a given HE.

Looking at the standoff radii for the three cases in Figure 2, we wonder how strong the magnetic dipole moment $\mathcal{M}_{dip}$ would need to be after 0.5 Gyr, when the atmospheric buildup should have mostly ceased, in order to magnetically enwrap the moon at a given HE. This question is answered in Figure 3.

While planets with relatively massive cores (brown solid line) shield a wider range of orbits for given $M_p$ below roughly $1 M_{Jup}$, the low-mass core model (blue solid line) catches up for more massive giants. What is more, our model tracks for the predicted $M_{dip}$, which we expect to be located between the thick brown and thick blue line, "overtake" the RG and Io-like HEs for a range of eccentricities. HE contours that fall within the shaded area, are magnetically protected, while moons above a certain HE are habitable from a tidal and illumination point of view.

We finally examine temporal aspects of magnetic shielding in Figure 4. The question answered in this plot is: "How long would it take a planet to magnetically coat its moon beyond the habitable edges?". Again, planetary mass is along the abscissa, but now stellar life time $t_\star$ is along the ordinate. Clearly, the magnetic standoff radius of lower-mass planets requires more time to expand out to the respective HEs. While low-mass giants with low-mass cores (blue lines) require up to 3.5 Gyr to reach the RG HE for $e_{ps} = 0.01$, equally mass planets but with a high-mass core (thin brown line) could require as few as 650 Myr.

Planets more massive than roughly $0.3 M_{Jup}$, will coat their moons at the RG HE for $e_{ps} = 0.01$ as early as 1 Gyr after formation. For lower eccentricities, time scales decrease. As the RG HE is more inward to the planet than the Io HE, it is coated earlier than the Io limit for given $e_{ps}$. RG and Io HEs for $e_{ps} = 0.001$ are not shown as they are covered earlier than $\approx 300$ Myr in all cases.

## 4. CONCLUSION

Mars-sized exomoons of Neptune-sized exoplanets in the stellar HZ of K stars will hardly be affected by planetary magnetospheres if these moons are habitable from an illumination and tidal heating point of view. While the magnetic standoff distance expands for higher-mass planets, ultimately Jovian hosts can enshroud their massive moons beyond the HE, depending on orbital eccentricity. In any case, exomoons beyond about $20 R_p$ will be habitable in terms of illumination and tidal heating, and they will not be coated by the planetary magnetosphere within about 4.5 Gyr. Moons between 5 and $20 R_p$ can be habitable, depending on orbital eccentricity, and be affected by the planetary magnetosphere at the same time.

Uncertainties in the parameterization of tidal heating cause uncertainties in the extent of both the RG and Io HEs. Once a potentially habitable exomoon would be discovered, detailed interior models for the satellite's behavior under tidal stresses would need to be explored.

In a forthcoming study, we will examine the evolution of planetary dipole fields, and we will apply our methods to planets and candidates from the *Kepler* sample. Obviously, a range of giant planets resides in their stellar HZs, and these planets need to be prioritized for follow-up search on the potential of their moons to be habitable.

The referee report of Jonathan Fortney substantially improved the quality of this study. We have made use of NASA's ADS Bibliographic Services. Computations have been performed with `ipython 0.13` on `python 2.7.2` (Pérez & Granger 2007). RH receives funding from the Canadian Astrobiology Training Program. JIZ is supported by CODI-UdeA and Colciencias.

## 7.6 Superhabitable Worlds (Heller & Armstrong 2014)

Contribution:

RH did the literature research, worked out the mathematical framework for the concept of the exomoon menagerie, translated the math into computer code, performed all computations, created all figures, led the writing of the manuscript, and served as a corresponding author for the journal editor and the referees.

# Superhabitable Worlds

René Heller[1], John Armstrong[2]

[1] McMaster University, Department of Physics and Astronomy, Hamilton, ON L8S 4M1, Canada (rheller@physics.mcmaster.ca)
[2] Department of Physics, Weber State University, 2508 University Circle, Ogden, UT 84408-2508 (jcarmstrong@weber.edu)

To be habitable, a world (planet or moon) does not need to be located in the stellar habitable zone (HZ), and worlds in the HZ are not necessarily habitable. Here, we illustrate how tidal heating can render terrestrial or icy worlds habitable beyond the stellar HZ. Scientists have developed a language that neglects the possible existence of worlds that offer more benign environments to life than Earth does. We call these objects "superhabitable" and discuss in which contexts this term could be used, that is to say, which worlds tend to be more habitable than Earth. In an appendix, we show why the principle of mediocracy cannot be used to logically explain why Earth should be a particularly habitable planet or why other inhabited worlds should be Earth-like.

Superhabitable worlds must be considered for future follow-up observations of signs of extraterrestrial life. Considering a range of physical effects, we conclude that they will tend to be slightly older and more massive than Earth and that their host stars will likely be K dwarfs. This makes Alpha Centauri B, member of the closest stellar system to the Sun that is supposed to host an Earth-mass planet, an ideal target for searches of a superhabitable world.

## 1. Introduction

A substantial amount of research is conducted and resources are spent to search for planets that could be habitats for extrasolar life. Engineers and astronomers have developed expensive instruments and large ground-based telescopes, such as the *High Accuracy Radial Velocity Spectrograph* (HARPS) at the 3.6m ESO telescope and the *Ultraviolet-Visual Echelle Spectrograph* (UVES) at the *Very Large Telescope* (VLT), and launched the *CoRoT* and *Kepler* space telescopes with the explicit aim to detect and characterize Earth-sized planets. Even larger facilities are being planned or constructed, such as the *European Extremely Large Telescope* (E-ELT) and the *James Webb Space Telescope* (JWST), and an ever growing community of scientists is working to solve not only the observational but also the theoretical and laboratory challenges.

At the theoretical front, the concept of the stellar "habitable zone" (HZ) has been widely used to identify potentially habitable planets (Huang, 1959; Dole, 1964; Kasting et al., 1993). To the confusion of some, planets that reside within a star's HZ are often called "habitable planets." However, a planet in the HZ need not be habitable in the sense that it has at least some niches that allow for the existence of liquid surface water. Naturally, as Earth is the only inhabited world we know, this object usually serves as a reference for studies on habitability. Instruments are being designed in a way to detect and characterize Earth-like planets and spectroscopic signatures of life in Earth-like atmospheres (Des Marais et al., 2002; Kaltenegger and Traub, 2009; Kaltenegger et al., 2010; Rauer et al., 2011; Kawahara et al., 2012). However, other worlds can offer conditions that are even more suitable for life to emerge and to evolve. Besides planets, moons could be habitable, too (Reynolds et al., 1987; Williams et al., 1997; Kaltenegger, 2010; Porter and Grundy, 2011; Heller and Barnes, 2013a). To find a habitable and ultimately an inhabited world, a characterization concept is required that is biocentric rather than geo- or anthropocentric.

In Section 2 of this paper, we illustrate how tidal heating can make planets inside the stellar HZ uninhabitable, and how it can render exoplanets and exomoons beyond the HZ habitable. Section 3 is devoted to conditions that could make a world more hospitable for life than Earth is. We call these objects "superhabitable worlds." Though our considerations are anticipatory, they still rely on the assumption that life needs liquid water. Our conclusions on the nature and prospects for finding superhabitable worlds are presented in Section 4. In Appendix A, we disentangle confusions between planets in the HZ and habitable planets, and we address related disorder that emerges from language issues. Appendix B is dedicated to the principle of mediocracy – in particular why it cannot explain that Earth is a typical, inhabited world.

## 2. Habitability in and Beyond the Stellar Habitable Zone

### 2.1 The stellar habitable zone

A natural starting point towards the characterization of a world's habitability is computing its absorbed stellar energy flux. This approach has led to what is called the "stellar habitable zone." The oldest record of a description of a circumstellar zone suitable for life traces back to Whewell (1853, Chap. X, Section 4), who, referring to the local stellar system in a qualitative way, called this distance range the "Temperate Zone of the Solar System." More than a century later, Huang (1959) presented a more general discussion of the "Habitable Zone of a Star," which considers time scales of stellar evolution, dynamical constraints in stellar multiple systems, and the stellar galactic orbit. A much broader, less anthropocentric elaboration on habitability has then been given by Dole (1964), who termed the circumstellar habitable zone "ecosphere."[1] The most widely used concept as of today is the one presented by Kasting et al. (1993), who applied a one-dimensional climate model and identified the $CO_2$ feedback to ensure the inner and the outer edges of the stellar HZs. The inner edge is defined by the activation of the moist or runaway greenhouse process, which desiccates the planet by evaporation of atmospheric hydrogen; the outer edge is defined by $CO_2$ freeze out, which breaks down the greenhouse effect whereupon the planet transitions into a permanent snowball state. Extensions of this concept to include orbital eccentricities have been given by Selsis et al. (2007) and Barnes et al. (2008), and a recent revision of the input model used by Kasting et al. (1993) has been presented by Kopparapu et al. (2013).

Considering the aging of the star, which involves a steady increase of stellar luminosity as long as the star is on the main sequence, the distance range within the HZ that is habitable for a certain period (say over the last 4.6 Gyr in the case of the Solar System) has been termed the continuous habitable zone (CHZ) (Kasting et al., 1993; Rushby et al., 2013). From an observational point of view, the CHZ provides a more useful tool because life needs time to evolve to a certain level such that it modifies its atmosphere on a global scale. Life seems to have appeared relatively early after the formation of Earth. Chemical and fossil indicators for early life can be found in sediments that date back to about $3.5 - 3.8$ Gyr ago (Schopf, 1993; Mojzsis et al., 1996; Schopf, 2006; Basier et al., 2006), that is, less than about 1 Gyr after Earth had formed. If correct, then life would have recovered within 100 Myr or so after the Late Heavy Bombardment (LHB) on Earth (Gomes et al., 2005). However, life required billions of years before it modified Earth's atmosphere substantially and imprinted substantial amounts of bio-relevant signatures in the atmospheric transmission spectrum. Stars more massive than the Sun have shorter lifetimes. Thus, although the lifetime of a 1.4 solar-mass ($M_\odot$) star is still about 4.5 Gyr, superhabitable planets will tend to orbit stars that are as massive as the Sun at most.

Further modifications of the circumstellar HZ include effects of tidal heating (Jackson et al., 2008a; Barnes et al., 2009), orbital evolution due to tides (Barnes et al., 2008), tidal locking of planetary rotation (Dole, 1964; Kasting et al., 1993), planetary obliquity (Spiegel et al., 2009), loss of seasons due to tilt erosion (Heller et al., 2011), land-to-ocean fractional coverage on planets (Spiegel et al., 2008), stellar irradiation in eccentric orbits (Dressing et al., 2010; Spiegel et al., 2010), the formation of water clouds on tidally locked planets (Yang et al., 2013), and the dependence of the ice-albedo feedback on the stellar spectrum and the planetary atmosphere (Joshi and Haberle, 2012; von Paris et al., 2013). These studies show that planets in the HZ of stars with masses $M_\star \lesssim 0.5\ M_\odot$ can be subject to enormous tidal heating, substantial variations in their semi-major axis, loss of seasons, and tidal locking. Above all, they demonstrate that the circumstellar HZ, though a helpful working concept, does not define a planet's habitability. For one and the same star, two different planets can have a different HZ, depending on a myriad of bodily and orbital characteristics.

*2.2 A terrestrial menagerie*

Accounting for some of these effects, we can imagine a menagerie of terrestrial worlds. In Fig. 1, we show these planets, all of which are assumed to have a mass 1.5 times that of Earth ($M_P = 1.5\ M_\oplus$), a radius 1.12 that of Earth[2], and a host star similar to Gl581 (Mayor et al., 2009). Rather than discussing the exact orbital limits for any of these hypothetical worlds, we shall illustrate here the range of possible scenarios. Irradiation from the star is given by $F_i$, tidal heat flux by $F_t$ (computed with the Leconte et al., 2010 tidal equilibrium model), and the critical flux for the planet to initiate a runaway greenhouse effect by $F_{RG}$ (Goldblatt and Watson, 2012). Using the analytical expression given in Pierrehumbert (2010), we estimate $F_{RG} = 301$ W/m² for our test planet, and we compute the HZ boundaries with the model of Kopparapu et al. (2013). Following the approach of Barnes et al. (2013) and Heller and Barnes (2013b), we identify the following members of the menagerie:

- **Tidal Venus:** $F_t \geq F_{RG}$ (Barnes et al., 2013)
- **Insolation Venus:** $F_i \geq F_{RG}$
- **Tidal-Insolation Venus:** $F_t < F_{RG}$, $F_i < F_{RG}$, $F_t + F_i \geq F_{RG}$
- **Super-Io:** $F_t > 2$ W/m², $F_t + F_i < F_{RG}$ (hypothesized by Jackson et al., 2008b)
- **Tidal Earth:** 0.04 W/m² $< F_t < 2$ W/m², $F_t + F_i < F_{RG}$ and within the HZ
- **Super-Europa:** 0.04 W/m² $< F_t < 2$ W/m² and beyond the HZ
- **Earth twin:** $F_t < 0.04$ W/m² and within the HZ
- **Snowball Earth:** $F_t < 0.04$ W/m² and beyond the HZ

Among these worlds, a Tidal Venus, an Insolation Venus, and a Tidal-Insolation Venus are uninhabitable by definition, while a Super-Io, Tidal Earth, Super-Europa, and an Earth Twin could be habitable. The surface of a Snowball Earth is also uninhabitable because it is so cold that even atmospheric $CO_2$ would condense, and the warming greenhouse effect could not operate to maintain liquid surface water. Note that the 2 and 0.04 W/m² thresholds are taken from the Solar System, where it has been observed that Io's global volcanism coincides with a surface flux of 2 W/m² (Spencer et al., 2000). Moreover, Williams et al. (1997) suggested that tectonic activity on Mars ceased when its endogenic surface flux fell below 0.04 W/m². Concerning the Super-Europa class, note that O'Brien et al. (2002) estimated Europa's tidal heat flux to about 0.8 W/m².

---

[1] Dole notes that the term "ecosphere" goes back to Strughold (1955).

[2] The radius is derived with an assumed Earth-like rock-to-mass fraction of 0.68 and using the analytical expression provided by Fortney et al. (2007).

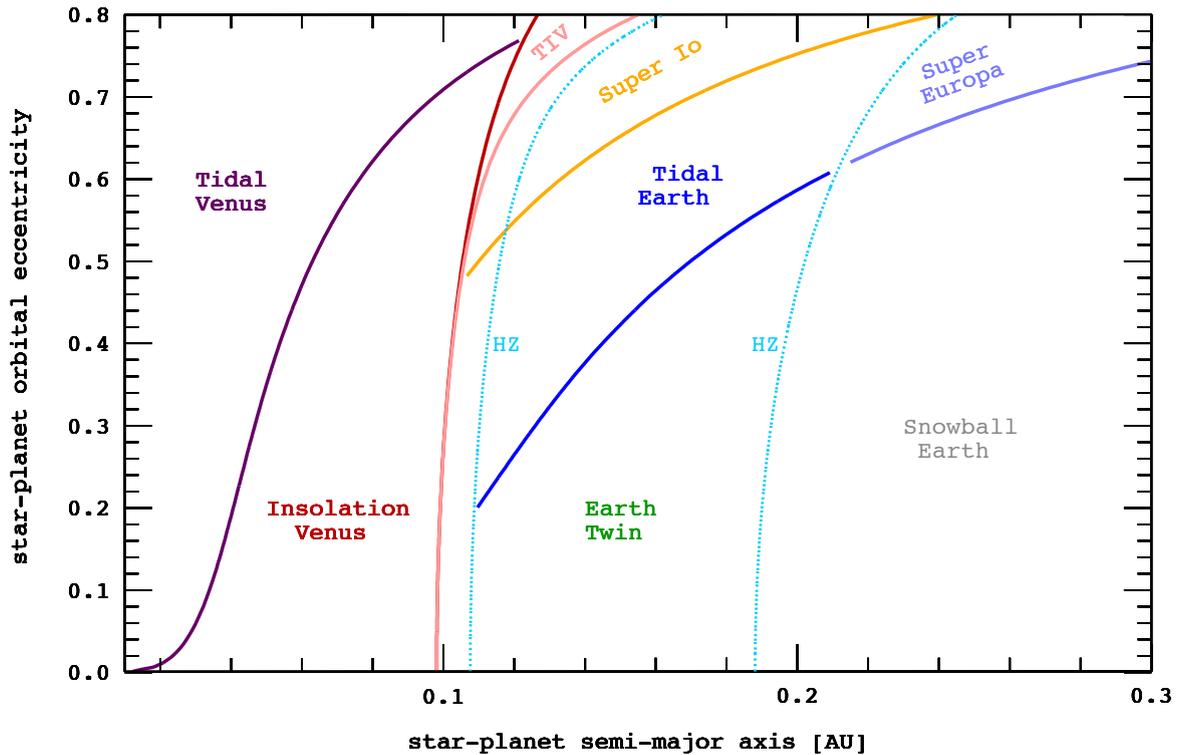

**Fig. 1**: Menagerie of terrestrial planets based on stellar irradiation and tidal heating. A planet with a mass of 1.5 $M_\oplus$ in orbit around a star similar to Gl581 is assumed. Dashed lines indicate the borders of the HZ following Kopparapu et al. (2013). The inner edge is constituted by the runaway greenhouse effect, the outer limit by the maximum greenhouse effect. Note that tidal heating can potentially heat planets beyond the HZ and open a class of Super Europas.

This menagerie illustrates that terrestrial planets can be located in the HZ and yet be uninhabitable. Tidal heating during the planet's orbital circularization can be an additional heat source that causes a planet to enter a runaway greenhouse state. What is more, tidal heating could make a world habitable *beyond* the HZ, possibly the Super-Europa planets in our menagerie. Elevated orbital eccentricities would induce tidal friction in these planets, which would transform orbital energy into heat. Such highly eccentric orbits would tend to be circularized, and hence perturbations from other planets or stars in the system would be required to maintain substantial eccentricities. Then tidal heating could partly compensate for the reduced stellar illumination beyond the stellar HZ and potentially maintain liquid water reservoirs.

### 2.3 Habitable exomoons beyond the stellar habitable zone

In exomoons beyond the stellar HZ, tidal heat could even become the major source of energy to allow for liquid water – be it on the surface or below (Reynolds et al., 1987; Scharf, 2006; Debes and Sigurdsson, 2007; Cassidy et al., 2009; Henning et al., 2009; Heller and Barnes, 2013a,b). Imagine a moon the size and mass of Earth in orbit around a planet the size and mass of Jupiter, and assume that this binary orbits a star of solar luminosity at a distance of 1 AU. If the moon is in a wide orbit, say beyond 20 planetary radii from its host, then it will hardly receive stellar reflected light or thermal emission from the planet (Heller and Barnes, 2013a,b), its orbit-averaged stellar illumination will not be substantially reduced by eclipses behind the planet (Heller, 2012), tidal heating will be insignificant[3], and the moon will essentially be heated by illumination absorbed from the star. But as the moon is virtually shifted into a closer orbit around the planet, illumination from the planet and tidal heating increase, and the total energy flux can become large enough to render the moon uninhabitable. The critical orbit, in which the total energy flux equals the critical flux for the moon to enter the runaway greenhouse effect, has been termed the circumplanetary "habitable edge" (Heller and Barnes, 2013a). Moons inside the habitable edge are uninhabitable. Imagine further that the planet-moon binary is virtually shifted away from the star. Due to the reduced stellar illumination, the habitable edge moves inward towards the planet because tidal heat and illumination from the planet can outbalance the loss of stellar illumination. When the planet-moon system is shifted even beyond the stellar HZ, then the moon will *need* to be close enough to the planet such that it will prevent transition into a snowball state. In this sense, giant planets beyond the stellar HZ have their own circumplanetary HZ, defined by illumination from the planet and tidal heating in the moon.

---

[3] Only if the moon's rotation is fast after formation, then it can experience tidal heating in a wide orbit due to the deceleration towards synchronous rotation. In this particular constellation of an Earth-like moon around a Jupiter-like planet at 1 AU from a Sun-like star, this tidal locking takes less than 4.5 Gyr, even in the widest possible orbits (Hinkel and Kane, 2013). Moreover, tidal heating can be substantial even beyond 20 planetary radii from the host planet if the moon's orbital eccentricity is large. However, circularization will damp it within a few Myr (Porter and Grundy, 2011; Heller and Barnes, 2013a).

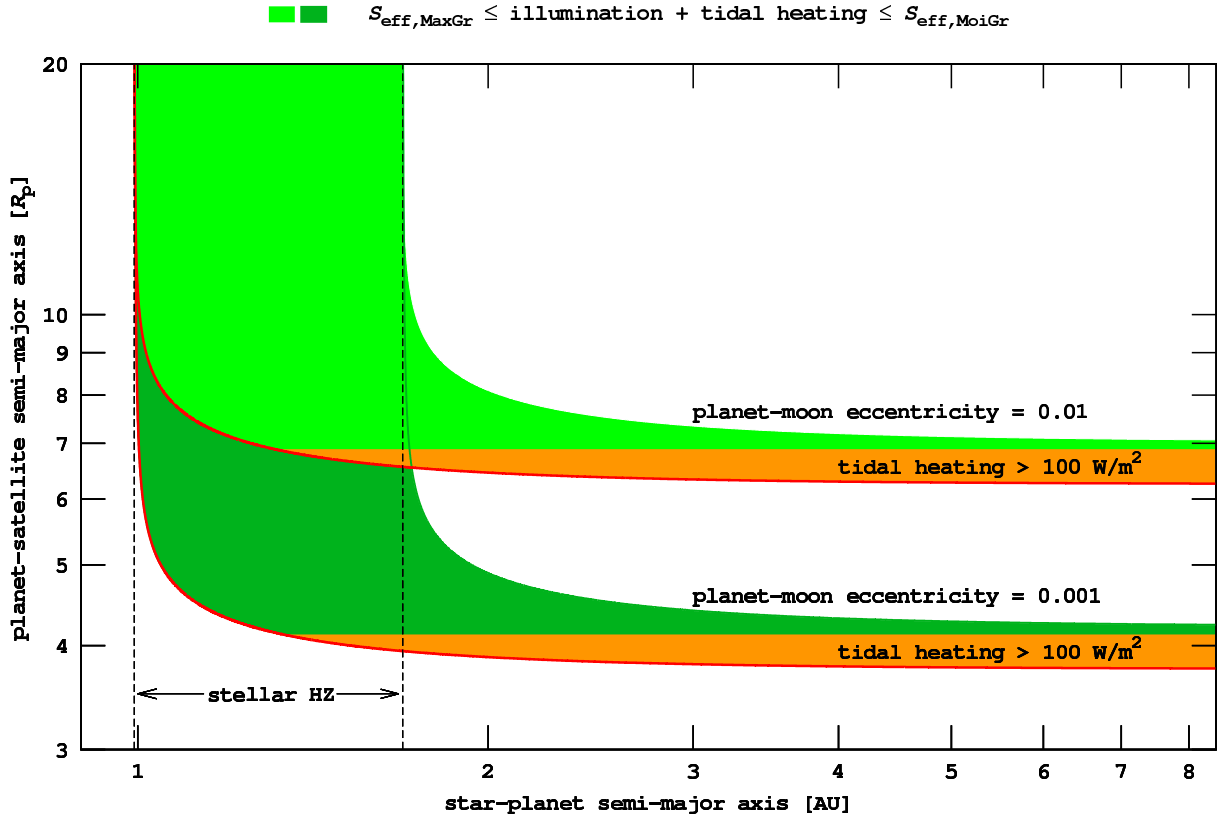

**Fig. 2:** Habitable orbits for an Earth-like exomoon around a Jupiter-like planet around a solar luminosity star. Green areas ilustrate orbits, in which the total energy flux of absorbed illumination and tidal heating is above the maximum greenhouse limit $S_{\text{eff,MaxGr}}$ and below the moist greenhouse threshold $S_{\text{eff,MoiGr}}$. Within these stripes, orbits with tidal heating rates above 100 W/m² are highlighted in orange. The circumplanetary habitable edge, here defined by the moist greenhouse, is indicated with a red line.

In Fig. 2, we illustrate this scenario for two different orbital eccentricities of the planet-moon binary, 0.01 and 0.001. The abscissa denotes stellar distance of the planet-moon system, and the ordinate shows the distance between the planet and its satellite. Green areas denote orbits, in which the total flux – composed of stellar plus planetary illumination and tidal heating – varies between the minimum and maximum energy flux ($S_{\text{eff,MaxGr}}$ and $S_{\text{eff,MoiGr}}$, respectively) identified by Kopparapu et al. (2013) to define the solar HZ. To compute the total energy flux, we chose the same model as in Heller and Barnes (2013a)[4]. As the planet-moon system is assumed at increasing stellar distances, the habitable edge (red line) moves closer to the planet. Ultimately, beyond the stellar HZ, the satellite must be closer to its planet than a certain maximum distance such that it receives enough tidal heating. Moving to the outer regions of the star system, stellar irradiation vanishes and tidal heat becomes the dominant source of energy. Comparison of the two stripes in Fig. 2 indicates that moons in orbits with only small orbital eccentricities would need to be closer to the planet to experience substantial tidal heating.

Note that the orbital eccentricities of the Galilean satellites around Jupiter are all larger than 0.001, and that Titan's eccentricity around Saturn is 0.0288. While the reason for Titan's enhanced eccentricity remains unclear (Sohl et al., 1995), the eccentricities of the major Jovian moons are not free but forced, that is, they are excited by the satellites' gravitational interaction (Yoder, 1979). Thus, as rocky and icy exomoons are predicted to exist around extrasolar Jovian planets (Sasaki et al., 2010; Ogihara and Ida, 2012), and if these moons encounter substantial orbital perturbations by other moons, then possibly many habitable exomoons in and beyond the stellar HZ await their discovery.

---

[4] This model, which includes a tidal theory presented by Leconte et al. (2010), neglects the feedback between tidal heating and the rheology of the moon. Yet, it has been shown that increasing tidal heat can melt a terrestrial body, thereby shutting down tidal heating itself (Zahnle et al., 2007; Henning et al., 2009; Remus et al., 2012).

### 3. Physical Characteristics of Superhabitable Worlds

All exoplanets detected so far are either subject to stellar irradiation that is very different from the amount or spectral distribution currently received by Earth, or they have masses larger than a few Earth masses. This has led astrobiologists to speculate about extremophile life forms that could cope with more bizarre conditions and maybe survive on a planet that is more hostile than Earth (for a brief review see Dartnell, 2011). The word "bizarre" is here to be understood from an anthropocentric point of view. From a potpourri of habitable worlds that may exist, Earth might well turn out as one that is marginally habitable[5], eventually bizarre from a biocentric standpoint. In other words, it is not clear why Earth should offer the most suitable regions in the physicochemical parameter space that can be tolerated by living organisms. Such an anthropocentric assumption could mislead research for extrasolar habitable planets because planets could be non-Earth-like but yet offer more suitable conditions for the emergence and evolution of life than Earth did or does, that is, they could be superhabitable.

   As to how superhabitable planets could look like or under which conditions a world could occupy a more benign zone within the physicochemical volume, we now discuss planetary characteristics that are relevant to planetary habitability (for a broader review see Gaidos et al., 2005; Lammer et al., 2009). These considerations will allow us to deduce quantitative estimates for superhabitable worlds. Instead of elaborating on extremophile or even completely different forms of life, we will still stick to liquid water as a pre-requisite for life and explore more comfortable environments as those found on Earth. Thus, our extensions of habitability towards superhabitability are incremental and still carry a geocentric flavor.

   What could we understand under a superhabitable world? So far, the term has not been in use, and thus its meaning remains obscure. We propose a context family in which it might be used with reason.[6]

- Habitable surface area: An Earth-sized planet on which the surface area that permits liquid water is larger than that of Earth (Spiegel et al., 2009; Pierrehumbert, 2010, § 1.9.1) could be regarded as superhabitable.

- Total surface area: A more uneven surface, or simply a larger planet with more space for living forms, could make a planet superhabitable. Due to the higher surface gravity of a more massive planet, however, both characteristics tend to exclude one another. To increase a planet's habitability, the body cannot be arbitrarily large. As mass typically increases with increasing radius for terrestrial planets, plate tectonics will cease to operate at a certain mass (see below). Moreover, a terrestrial planet that is much heavier than Earth might not get rid of its primordial hydrogen atmosphere, which could hamper the emergence of life (Huang, 1960). A planet slightly larger than Earth can, however, still be regarded superhabitable. Note that Earth, the only inhabited planet known so far, is the largest terrestrial planet in the Solar System.

- Land-to-ocean-fraction and distribution: The amount of surface water compared to the amount of land is not only crucial for planetary climate but also for the emergence and diversification of life. Giant continents, as Earth's Gondwana about 500 Myr ago, may have vast deserts in their interiors, as they are not subject to the moderating effect of oceans. In contrast, planets with more fractionate continents and archipelagos should favor superhabitable environments due to their enhanced richness in habitats. Earth's shallow waters have a higher biodiversity than the deep oceans (Gray, 1997). Hence, we expect that planets with shallow waters rather than those with deep extended oceans tend to be superhabitable.

   What is more, Abe et al. (2011) found that planets dryer than Earth should have wider stellar HZs. At the inner HZ boundary, these "Dune" planets are more tolerant against transition into the runaway greenhouse effect, because their low-humidity equatorial regions can emit above the critical flux, assumed for an atmosphere saturated in water. But still the atmosphere is somewhat opaque in the infrared and thus exerts a global greenhouse effect, which prevents water at poles from freezing. On dry planets at the outer HZ boundary, the low humidity in the tropics hampers formation of clouds and thus snowfall. Dry planets will thus tend to have lower albedos than frozen aqua planets (such as Earth), and they will effectively absorb more stellar illumination and be less susceptible to transitioning into a snowball state. In addition, daytime temperatures will be higher on dryer planets at the outer HZ regions due to their smaller thermal inertia.

   Combined with the shallow-waters argument, considerations of dry planets thus suggest that planets with a lower fractional surface coverage of water, and with bodies of liquid water that are distributed over many reservoirs rather than combined in one big ocean, can be considered superhabitable.

---

[5] Note that Earth is located at the very inner margin of the solar habitable zone (Kopparapu et al., 2013).

[6] We here understand superhabitability as a state in which a terrestrial world is generally more habitable than Earth. Conventionally, habitability is considered a binary condition, an "on/off" or "1/0" state, just as a sow is in pig or not. In this sense, we discuss the prospects of a sow being pregnant with several farrows, a state more fertile than only "on" or "1".

- **Plate tectonics**: On Earth, plate tectonics drive the carbon-silicate cycle. In this planet-wide geochemical reaction, near-surface weathering of calcium silicate ($CaSiO_3$) rocks leads to the formation of quartz-like minerals, that is, silicon dioxide ($SiO_2$). At the same time, carbon dioxide ($CO_2$, for example from the atmosphere) combines with the residual carbon atoms to form calcium carbonate ($CaCO_3$). When subducted to deeper sediments, elevated pressures and temperatures reverse this reaction, ultimately leading to volcanic outgassing of $CO_2$. If this cycle stopped or if it never started on a hypothetical terrestrial, water-rich planet, then silicate weathering would draw down atmospheric $CO_2$, which could lead to a global snowball state. On a planet that receives more stellar illumination or has other internal heat sources (for example tidal or radiogenic heating), this collapse could be avoided. The period over which radiogenic heating is strong enough to maintain plate tectonics increases with increasing planetary mass (Walker et al., 1981). To a certain degree, more massive terrestrial planets should thus tend to be superhabitable.

  However, planets with masses several times that of Earth develop high pressures in their mantle, and the resulting enhanced viscosities make plate tectonics less likely (Noack and Breuer, 2011). Moreover, a stagnant lid forms at the core-mantle-boundary that allows only a reduced heat flow from the core and thereby also frustrates tectonics (Stamenković et al., 2011). Too high a mass thus impedes plate tectonics and therefore also subduction that is required for the carbon-silicate. Yet, "propensity of plate tectonics seems to have a peak between 1 and 5 Earth masses" (Noack and Breuer, 2011), which, of course, depends on composition and primordial heat reservoir. We conclude that planets with masses up to about 2 $M_\oplus$ tend to be superhabitable from the tectonic point of view.

- **Magnetic shielding**: To allow for surface life, a world must be shielded against high-energy radiation from interstellar space (termed "cosmic radiation") and from the host star (Baumstark-Khan and Facius, 2002). Too strong an irradiation could destroy molecules relevant for life, or it could strip off the world's atmosphere, an effect to which low-mass terrestrial worlds are particularly prone (Luhmann et al., 1992). Protection can be achieved by a global magnetic field, whether it is intrinsic as on Earth or extrinsic as may be the case on moons (Heller and Zuluaga, 2013), and by the atmosphere. While a giant planet's magnetosphere can shield a moon against cosmic rays and stellar radiation, it may itself induce a bombardment of the moon with ionized particles that are trapped in the planet's radiation belt (see Jupiter; Fischer et al., 1996).

  To sustain an intrinsic magnetic field strong enough for protection over billions of years, a terrestrial world needs to have a liquid, rotating, and convecting core. Within Earth, this liquid is composed of molten iron alloys in the outer core, that is, between 800 and 3000 km from its center. Less massive planets or moons will have weaker, short-lived magnetic shields. Williams et al. (1997) estimated a minimum mass of 0.07 $M_\oplus$ for a world under solar irradiation to retain atmospheric oxygen and nitrogen over 4.5 Gyr. Beyond that, the dipole component of the magnetic moment $\mathcal{M}$ depends on the core radius $r_o$, rotation frequency $\Omega$, and the thickness $D$ of the core rotating shell where convection occurs via $\mathcal{M} \approx r_o^3\, D^{5/9}\, \Omega^{7/6}$ (Olson and Christensen, 2006; López-Morales et al., 2011), which implies that tidally locked planets and moons in wide orbits may have weak magnetic shielding.

- **Climatic thermostat**: A more reliable global thermostat that impedes ice ages and snowball states would prevent an existing ecosystem from experiencing mass extinctions, which would decelerate or even frustrate evolution. There should exist atmospheric and geological processes whose interplay constitutes a thermostat that makes a planet superhabitable.

  Triggered by the recent discoveries of super-Earth planets in or near the stellar HZ, recycling mechanisms of atmospheric $CO_2$ and $CH_4$ have been proposed for potentially water-rich planets (Kaltenegger et al., 2013).[7] These planets are predicted to be completely covered by a deep liquid water ocean on top of high-pressure ices and without direct contact to the rocky interior. On such worlds, an Earth-like carbon-silicate cycle cannot possibly operate as there would be no $CO_2$ weathering. Alternatively, lattices of high-pressure water molecules could trap $CO_2$ as guest molecules, a chemical substance known as carbon clathrate, and provide an effective climatic thermostat by moderating the $H_2O$ and $CO_2$ levels in water-rich super-Earths. A similar clathrate mediation has been shown possible for $CH_4$ instead of $CO_2$ (Levi et al., 2013), that is, methane clathrate. Clathrate convection could be an effective mechanism to transport $CH_4$ and/or $CO_2$ from a water-rich planet's silicate-iron core through a high-pressure ice-mantle into the ocean and, ultimately, into the atmosphere (Fu et al., 2010).

- **Surface temperature**: On worlds with substantial atmospheres, in other words with surface pressures $P$ at least as high as those on Mars (where 1 mb $\lesssim P \lesssim$ 10 mb), surface temperatures will generally be different from the thermal equilibrium temperature given by stellar irradiation and planetary albedo alone (Selsis et al., 2007; Leconte et al., 2013). The biodiversity, or the richness of families and genera, seems to have multiplied during warmer epochs on Earth (Mayhew et al., 2012), indicating that worlds warmer than Earth could be more habitable. A slightly warmer version of Earth might have extended tropical zones that would allow for more biological variance. This is suggested by both the "cradle model" and the "museum model" used in evolutionary biology. The former approach suggests that rapid diversification occurred recently and rapidly in the tropics, while the latter theory claims that the tropics provide particularly favorable circumstances for slow accumulation and preservation of diversity over time (McKenna and Farrell, 2006; Moreau and Bell, 2013).

---

[7] As candidates for such water-rich planets, Levi et al. (2013) propose Kepler-11b, Kepler-18, and Kepler-20b. Kaltenegger et al. (2013) suggest Kepler-62e and f.

However, warming Earth does not necessarily yield increased biodiversity. Warming on short timescales causes mass extinction, which can currently be witnessed on Earth. Only a planet that is warm compared to Earth on a Gyr timescale or a world that warms gently over millions and billions of years could have more extended surface regions suitable for liquid water and biodiversity.

On the downside, with fewer temperate zones and no arctic regions, an enormous range of life forms known from Earth could not exist. Above all, a world that is substantially warmer than Earth might have anoxic oceans. On Earth, Oceanic Anoxic Events occurred in periods of warm climate, with average surface temperatures above 25°C compared to pre-industrial 14°C (IPCC, 1995), and resulted in extensive extinctions like the Permian/Triassic around 250 Myr ago (Wignall and Twitchett, 1996). While the concatenation of circumstances that led to extinctions during hot periods is complicated and may reflect problems of Earth's ecosystem, it cannot be excluded that a world moderately warmer than Earth could be superhabitable. A colder planet, however, can be assumed to be less habitable as less energy input would slow down chemical reactions and metabolism on a global scale.

- **Biological diversification**: An inhabited planet whose flora and fauna are more diverse than they are on Earth could reasonably be termed superhabitable as it empirically shows that its environment is more benign to life. An evolutionary explosion, such as the Cambrian one on Earth, could occur earlier in a planet's history than it did on Earth – or simply long enough ago to make the respective planet more diversely inhabited than Earth is today. Alternatively, evolution could have progressed faster on other planets. Jumps in diversification or accelerated evolution can be triggered by nearby supernovae and by enhanced radiogenic or ultraviolet radiation.

- **Multihabitability and panspermia**: Stellar systems could be more habitable than the Solar System if there were more than one terrestrial planet or moon in the HZ (Borucki et al., 2013; Anglada-Escudé et al., 2013[8]). If, for example, the Moon-forming impact had distributed the mass more evenly between Earth and the Moon, then both objects might have been habitable. Alternatively, in a hypothetical Solar System analog in which only the orbits of Mars and Venus would be exchanged, there could exist three habitable planets. With the possibility of massive moons about giant planets, there might also exist satellite systems with several habitable exomoons. Such stellar systems could be called "multihabitable." Impacts of comets, asteroids, or other interplanetary debris might trigger exchange of material between those worlds. This exchange could then induce mutual fertilization among multiple habitable worlds, a process known as panspermia (Hoyle and Wickramasinghe, 1981; Weber and Greenberg, 1985). Worlds in multihabitable systems, whether they are planets or moons, could thus be regarded as superhabitable because they have a higher probability to be inhabited.

- **Localization in the stellar habitable zone**: Recent work emphasized that Earth is scraping at the very inner edge of the Sun's HZ (Kopparapu et al., 2013; Worsworth and Pierrehumbert, 2013). Terrestrial worlds that are located more towards the center of the stellar HZ could be considered superhabitable. These objects would be more resistant against transitioning into a moist or runaway greenhouse state (at the inner edge of the HZ) than Earth is.

- **Age**: From a biological point of view, older worlds can be assumed to be more habitable because Earth experienced a steady increase in biodiversity as it aged (Mayhew et al., 2012). This diversification indicates that non-intelligent life itself is able to modify an environment so as to make it more habitable for its ancestors.[9] A stronger claim has been put forward by what is now known as the Gaia hypothesis, which suggests that the global biosphere as a whole can be regarded as a creature controlling "the global environment to suit its needs" (Lovelock, 1972). Whether considered as a global entity or not, Earth's ecosystem obviously influences global geochemical processes, which has perpetually led to an increase in biodiversity over billions of years. As an example, note that after the Great Oxygen Event about 2.5 Gyr ago (Anbar et al., 2007)[10], which was likely induced by oceanic algae, Earth's surface became more habitable, allowing life to conquer the continents about 360 to

---

[8] In the case of GL 667C, it is entertaining to imagine how the structure of that system might influence the development of an intelligent species' astronomy and human space flight activities, with three potentially habitable worlds and three complete stellar systems to study up-close.

[9] Intriguingly, now that the first form of life on Earth able to call itself intelligent it causes a drastic decrease in biodiversity. But even in case evolution typically leads to intelligent life, then if an intelligence destroyed itself, it can be assumed that the respective ecosystem would be able to recover on a Myr or Gyr timescale, of course depending on the magnitude of the caused extinction and the environmental effects left behind by the intelligence.

[10] Analyses of chromium isotopes and redox-sensitive metals of drill cores from South Africa by Crowe et al. (2013) indicate a first slight increase in atmospheric oxygen about 3 Gyr ago, which could be related to the emergence of oxygenic photosynthesis.

480 Myr ago (Kenrick and Crane, 1997). Therefore, older planets should tend to be more habitable, or superhabitable if inhabited.

- **Stellar mass**: The mass of a star on the main sequence determines its luminosity, its spectral energy distribution, and its lifetime. The Sun emits most of its light between 400 nm and 700 nm, which is the part of the spectrum visible to the human eye. This is also the spectral range in which plants and other organisms perform oxygenic photosynthesis. On worlds orbiting stars with masses ≤ 0.6 $M_\odot$ (known as M dwarfs, Baraffe and Chabrier, 1996), these forms of life might not have the capacity to properly harvest energy for their survival because their stars have their radiation maxima in the infrared. However, Miller et al. (2005) found a free-living cyanobacterium that is able to use near-infrared photons at wavelengths > 700 nm. This discovery, as well as the ability of the oxygenic photosynthetic cyanobacterium Acaryochloris marina to use chlorophyll d for harvesting photons at 750 nm (Chen and Blankenship, 2011), suggests that – of course provided that many other conditions are met – oxygenic photosynthesis on planets orbiting cool stars is possible. Discussing the results of Kiang et al. (2007a,b) and Stomp et al. (2007), Raven (2007) also concluded that photosynthesis can occur on exoplanets in the HZ of M dwarfs. Ultimately, the transmissivity of the planet's atmosphere needs to be appropriate to allow an adequate amount of spectral energy to arrive at the planet's surface.

  We will not go deeper in possible extremophile life – extremophile from the standpoint of an Earthling – and, for the time being, consider M stars as less likely hosts for superhabitable planets. However, these reflections show that stars slightly less massive than the Sun could still provide the appropriate spectral energy distribution for photosynthesis.

- **Stellar UV irradiation**: Stellar UV radiation can damage deoxyribonucleic acid (DNA) and thus impede the emergence of life. Today, Earth has a substantial stratospheric ozone column that absorbs solar irradiation almost completely between 200 and 285 nm (UVC) and most of the radiation between 280 and 315 nm (UVB). During the Archean (3.8 to 2.5 Gyr ago), this ozone shield did not exist, but yet life managed to form. We can assume that terrestrial planets with anoxic primordial atmospheres would be more habitable than early Earth if they received less hazardous UV irradiation.

  M stars remain very active and emit a lot of X-ray and UV radiation during about the first Gyr of their lifetime (Scalo et al., 2007). The activity-driven XUV flux of G stars, such as the Sun, falls off much more rapidly, but their quiescent UV flux is enhanced with respect to K and M dwarfs. What is more, while the UV flux of young M stars is generally much stronger than that of young Sun-like stars, quiescent UV radiation from evolved M dwarfs may be too weak for some essential biochemical compounds to be synthesized (Guo et al., 2010). Thus, they do not seem to offer superhabitable primordial environments. K stars offer a convenient compromise between moderate initial and long-term high-energy radiation. This is supported by considerations of the weighted irradiance spectrum of complex carbon-based molecules, indicating that planets in the HZs of K main sequence stars experience particularly favorable UV environments (Cockell, 1999). This indicates that K dwarf stars are favorable host stars for superhabitable planets.

- **Stellar lifetime**: With a planet's tendency to be superhabitable increasing with age, the star must burn long enough for existing life forms to evolve. Stars less massive than the Sun have longer lifetimes, and planets or moons can spend more time within the HZ before they transition inside the expanding inner edge (Rushby et al., 2013). Against the background from the two previous items and accounting for the relatively stable spectral radiance once they have settled on the main-sequence, we propose that K dwarfs are more likely to host superhabitable planets than the Sun or M dwarfs.

- **Early planetary bombardment**: The nature of Earth is closely coupled to its bombardment history. From the lunar forming impact (Cameron and Ward, 1976) to the Late Heavy Bombardment (Gomes et al., 2005), the impact history influenced the surface environment, delivery of organic molecules and volatiles (Chyba and Sagan, 1992; Raymond et al., 2009), and spin/orbital evolution of Earth. This means that the history of Earth's evolution is closely coupled to the orbital dynamics of the planetary system. It is possible that the LHB itself is responsible for Earth's habitability, since it helped deliver water and other volatiles to Earth's surface from farther out in the Solar System.

  While the exact cause of the LHB is uncertain, the debate has focused on effects of a continuous, though gradually tapering, history of impacts versus a spiked delivery of material caused by changes in orbital dynamics (Ryder, 2002). Either way, the system architecture played a large role in determining the extent of these impacts (Raymond et al., 2004). Is it possible that a system with more dynamical instability early in a planet's history would result in a longer, more extensive LHB, or – in the case of a stochastic LHB – a sequence of LHB-type events? Such a history could have little effect on the ongoing evolution of marine or subterranean microbes, yet result in a richer volatile inventory for the host planet or moon, and even encourage multihabitability by enhancing transfer of material between planets in the same system.

- **Planetary spin**: The initial spin-orbit misalignment, or obliquity, and rotation rate of a planet are largely due to the random events that lead to a planet's formation (Miguel and Brunini, 2010), but the subsequent evolution is tightly coupled to orbital dynamics. Conventional wisdom suggests that Earth is an "ideal" habitable world, since it has a large – and presumably rare

– moon to stabilize its tilt relative against the orbital forcing from the Sun and other planets (Laskar et al., 1993). However, there are a couple of assumptions in this: 1) that a stable spin is required or even desired for a habitable planet, and 2) that this effect is not mitigated by the crucial role the Moon has had on the evolution of Earth's spin rate. For example, studies have indicated that Earth's rotation axis can be stable without the presence of a massive satellite (Lissauer et al., 2012), and that such stability is perhaps not desirable (Spiegel et al., 2009; Armstrong et al., 2013). In the latter case, planets with a large tilt can break the ice-albedo feedback at locations further from the star, keeping the planet from entering the "snowball Earth" stage (Williams and Kasting, 1997), and systems with varying tilts could provide slow but steady changes in ecosystems that encourage evolution of life.

It is uncertain whether any given spin rate is desirable for life, as long as it helps keep the surface uniformly habitable, while radical changes in such a spin rate might be detrimental. Did the existence of the Moon encourage life to evolve by changing the diurnal and tidal cycles, or was this an impediment to evolution? Could moderate changes of a world's obliquity or rotation rate even force life to adapt to a broader range of environmental conditions, thereby triggering more diverse evolution? Ultimately, is it possible that a terrestrial planet without a massive moon, or a planet more subject to changes in spin, could be superhabitable?

- **Orbital dynamics**: It is occasionally claimed that Earth is habitable largely owing to its stable, circular orbit. However, climate studies indicate a range of dramatic shifts in climate due to subtle changes in Earth's orbit. These oscillations of obliquity, precession, orbital eccentricity, and rotation period – mainly driven by gravitational interaction with the Sun, the Moon, Jupiter, and Saturn – are known as Milankovitch cycles (Hays et al., 1976; Berger, 1976). The very stability of our orbit, in these cases, makes such events treacherous, as Earth may experience only subtle changes again to help rectify the problem. In fact, it is entirely possible that such stability might put the brakes on biological evolution. Planets with eccentric orbits would still provide a range of seasonally viable habitats while perhaps acting as a "vaccine" against life-threatening snowball events. Tidal heating in planets or moons on eccentric orbits may even act as a buffer against transition into a global snowball state (Reynolds et al., 1987; Scharf, 2006; Barnes et al., 2009). Planets with large swings in eccentricity can also influence the planetary tilt, which has its own, perhaps positive, impacts on the habitability of a planet. We thus claim that moderate variations in the orbital elements of a terrestrial world need not necessarily hamper the evolution or inhibit the formation of life. Consequently, we see no terminating argument that Earth's configuration in an almost circular orbit with mild changes in its orbital elements should be considered the most benign situation. Planets that undergo soft variations in their orbital configurations may still be superhabitable.

- **Atmosphere**: The atmosphere of an exoplanet or exomoon is essential to its surface life, as it serves as a mediator of transport for water and, to a lesser degree, nutrients. Atmospheric composition and the gases' partial pressures will determine surface temperatures and, hence, have a key role in shaping the environment and providing the preconditions for formation and evolution of life.

Just as an example of how an atmosphere different from that of Earth could make an otherwise similar world superhabitable, note that (i.) enhanced atmospheric oxygen concentration allows a larger range of metabolic networks (Berner et al., 2007); (ii.) variations in the atmospheric oxygen concentration seem to constrain the maximum possible body size of living forms (Harrison et al., 2010; Payne et al., 2011); and (iii.) there are no known multicellular organisms that are strictly anaerobic. Today, Earth's atmosphere contains about 21% oxygen by volume or partial pressure ($pO_2$). Limited by runaway wildfires for $pO_2 > 35\%$ and lack of fire at $pO_2 < 15\%$ (Belcher and McElwain, 2008), a range of oxygen partial pressures is compatible with an ecosystem broadly similar to Earth's. Obviously, atmospheric oxygen contents can be much greater than on Earth today, and worlds with oxygen-rich atmospheres could be entitled superhabitable, because of items (i.) – (iii.).

While atmospheres less massive than that of Earth would offer weaker shielding against high-energy irradiation from space, weaker balancing of day-night temperature contrasts, retarded global distribution of water, etc., somewhat more massive atmospheres could induce positive effects for habitability. Again, this indicates that planets slightly more massive than Earth should tend to be superhabitable because, first, they acquire thicker atmospheres and, second, their initially extended hydrogen atmospheres can envelope gaseous nitrogen and thereby prevent its loss due to non-thermal ion puck up under an initially strong stellar UV irradiation (Lammer, 2013).

Some of the conditions listed in this Section are already, or will soon be, accessible remotely (namely, orbital and bodily characteristics of extrasolar planets or moons), some will be modeled and thereby constrained (such as orbital evolution and composition), and others will remain hidden and induce random effects on habitability (climate history, radiogenic heating, ocean salinity, former presence of meanwhile ejected planets or satellites, etc.) from the viewpoint of an observer. This list is by far not complete, and it is not our goal to provide such a complete list. However, it is supposed to illustrate that a range of physical characteristics and processes can make a world exhibit more benign environments than Earth does. Given the amount of planets that exist in the Galaxy, it is therefore reasonable to predicate that worlds with more comfortable settings for life than Earth exist.

Earth might still be rare, but this does not make the emergence and existence of extraterrestrial life impossible or even very unlikely because superhabitable worlds exist.

## 4. Conclusions

Utilization of any flavor of the HZ concept implies that a planet is either in the HZ and habitable or outside it and uninhabitable. Resuming our considerations from Section 2, our results are threefold: (i.) Extensions of the HZ concept which include tidal heating, show that planets ("Super-Europas" in our terminology) can exist beyond the HZ and still be habitable. (ii.) Fed by tidal heating, moons of planets beyond the HZ can be habitable. (iii.) Intriguingly, none of all the discussed concepts for the HZ describes a circumstellar distance range that would make a planet a more suitable place for life than Earth currently is.

Terrestrial planets that are slightly more massive than Earth, that is, up to 2 or 3 $M_\oplus$, are preferably superhabitable due to the longer tectonic activity, a carbon-silicate cycle that is active on a longer timescale, enhanced magnetic shielding against cosmic and stellar high-energy radiation, their larger surface area, a smoother surface allowing for more shallow seas, their potential to retain atmospheres thicker than that of Earth, and the positive effects of non-intelligent life on a planet's habitability, which can be observed on Earth. Higher biodiversity made Earth more habitable in the long term. If this is a general feature of inhabited planets, that is to say, that planets tend to become more habitable once they are inhabited, a host star slightly less massive than the Sun should be favorable for superhabitability. These so-called K dwarf stars have lifetimes that are longer than the age of the Universe. Consequently, if they are much older than the Sun, then life has had more time to emerge on their potentially habitable planets and moons, and – once occurred – it would have had more time to "tune" its ecosystem to make it even more habitable.

The K1V star Alpha Centauri B (αCenB), member of the closest stellar system to the Sun and supposedly hosting an Earth-mass planet in a 3.235-day orbit (Dumusque et al., 2012), provides an ideal target for searches of planets in the HZ and, ultimately, for superhabitable worlds. Age estimates for αCenB, derived via asteroseismology, chromospheric activity, and gyrochronology (Thévenin et al., 2002; Eggenberger et al., 2004; Thoul et al., 2008; Miglio and Montalbán, 2005; Bazot et al., 2012), show the star to be slightly evolved compared to the Sun, with estimates being 4.85 ±0.5 Gyr, 6.52 ±0.3 Gyr, 6.41 Gyr, 5.2 – 8.9 Gyr, and 5.0 ±0.5 Gyr, respectively. Radiation effects of the stellar primary Alpha Centauri A have been shown to be small and should not induce significant climatic variations on planets about αCenB (Forgan, 2012). If life on a planet or moon in the HZ of αCenB evolved similarly as it did on Earth and if this planet had the chance to collect water from comets and planetesimals beyond the snowline (Wiegert and Holman, 1997; Haghighipour and Raymond, 2007), then primitive forms of life could already have flourished in its waters or on its surface when the proto-Earth collided with a Mars-sized object, thereby forming the Moon.

Eventually, just as the Solar System turned out to be everything but typical for planetary systems, Earth could turn out everything but typical for a habitable or, ultimately, an inhabited world. Our argumentation can be understood as a refutation of the Rare Earth hypothesis. Ward and Brownlee (2000) claimed that the emergence of life required an extremely unlikely interplay of conditions on Earth, and they concluded that complex life would be a very unlikely phenomenon in the Universe. While we agree that the occurrence of another truly Earth-like planet is trivially impossible, we hold that this argument does not constrain the emergence of other inhabited planets. We argue here in the opposite direction and claim that Earth could turn out to be a marginally habitable world. In our view, a variety of processes exists that can make environmental conditions on a planet or moon more benign to life than is the case on Earth.

## Appendix A: Usage and Meaning of Terms Related to Habitability

Discussions about habitability suffer from diverging understanding of the terms "habitability," "habitable," etc. Recall that a planet in the stellar illumination HZ, as it is defined by physicists and astronomers (see Section 2), need not necessarily be habitable. It is thus precipitate, if not simply false, to state that the planet Gl581d is "habitable, but not much like home" (Schilling, 2007). Analogously, a world such as a tidally heated moon outside the HZ need not necessarily be uninhabitable. Claiming that "Being inside the habitable zone is a necessary but not sufficient condition for habitability" (Selsis et al., 2007) can be wrong, depending on the meaning of the word "habitable." If that statement means that habitable planets are in the HZ by definition, then the sentence is tautological. If, however, it means that a planet needs to be in the HZ to provide liquid surface water, then it can be proven wrong.

Confusions from blurred pictures are not restricted to the qualitative. As an example, quantitative problems occur in discussions about the occurrence rate of planets similar to Earth that orbit Sun-like stars. The parameter $\eta_\oplus$ has been introduced to quantify their abundance. Unfortunately, different understandings of $\eta_\oplus$ occur in the literature. It has been used as "fraction of stars with Earth-mass planets in the habitable zone" (Howard et al., 2009), "the fraction of Sun-like stars that have planets like Earth" (Catanzarite and Shao, 2011), "the fraction of Sun-like stars with Earth-like planets in their habitable zones" (O'Malley-James et al., 2012), "the fraction of habitable planets for all Sun-like stars" (Catanzarite and Shao, 2011), "the fraction of Sun-like stars that have at least one planet in the habitable zone" (Lunine et al., 2008), the "frequency of Earth-mass planets in the habitable zone" (Wittenmyer et al., 2011), the fraction of "Earth-like planets with $M \sin i$ = 0.5-2$M_{Earth}$ and $P < 50$ days"[11] (Howard et al., 2010), "the frequency of habitable planets orbiting M dwarfs" (Bonfils et al., 2013b), "the frequency of 1 $< m \sin i < 10$ M$_\oplus$ planets in the habitable zone of M dwarfs"[12] (Bonfils et al., 2013a), "the frequency of terrestrial planets in the habitable zone [...] of solar-like stars in our galaxy" (Jenkins, 2012), and "the number of planets with 0.1$M_\oplus < M_p < 10M_\oplus$ in the 3 Gyr CHZ (a < 0.02AU)"[13] (Agol, 2011). The latter two definitions stand out because Jenkins (2012) restricts $\eta_\oplus$ to the Milky Way, and Agol (2011) introduces $\eta_\oplus$ as a total count, and yet, he uses it as a frequency.

Intriguingly, (i.) as it is not clear whether a planet must be similar to Earth to be habitable, (ii.) as the definitions diverge in their reference to the stellar type, and (iii.) as it sometimes remains obscure what "Earth-like planets" are in the respective context, none of these understandings is equivalent to at least one of the others, except for the Howard et al. (2009) and Wittenmyer et al. (2011) explanations. As a consequence, different estimates for $\eta_\oplus$ *must* occur. Although physical, observational, and systematic effects play a role, a quantitative divergence of estimates for $\eta_\oplus$ will remain as long as there is no consensus about the meaning, that is, the usage, of this word or variable. This problem is not physical, but it is a logical consequence of the diverging understanding of $\eta_\oplus$. Imagine a situation in which all the authors of the mentioned studies sit around a desk to discuss their values for $\eta_\oplus$ and the implications! If they were not aware of the meaning/usage drift of "their" respective $\eta_\oplus$, then their dialogue would founder on a language problem.

The crux of the matter lies in the meaning of any of these terms, which again depends on the context in which any term is used. Following the Austrian philosopher Ludwig Wittgenstein and his *Philosophische Untersuchungen* (Wittgenstein, 1953), many logical problems occur when terms are alienated from their ancestral use and then unreasonably applied in other contexts. Ultimately, as astrobiology is an interdisciplinary science, it is exposed to those dangers of confusion and contradiction to a special degree. In this communication, we shall not infringe the use of language and terminology but unravel possible perils. In other words, we ought to be descriptive rather than normative (Wittgenstein, 1953, §124). To answer the question of whether a planet is habitable, it must be clear what we understand a habitable planet to be. And following semantic holism, a doctrine in the philosophy of language, the term "habitable" then is defined by its usage in the language.[14]

## Appendix B: An Algebraic Approach to Superhabitable Planets

Astronomers have developed an inclination to evaluate habitability in terms of geocentric conditions. Expressions such as "Earth-like," "Earth analog," "Earth twin," "Earth-sized," and "Earth-mass" are often used to evaluate a planet's habitability. Although being a natural body of reference, if other inhabited worlds exist – and obviously some scientists assume that and look for them – then it would be presumptuous to claim that they need to be Earth-like or that Earth offers the most favorable conditions. We can use set algebra to discern and display planet families. This somewhat unconventional approach would allow us to identify Earth as one sort of an habitable and inhabited world and to become acquainted with superhabitable worlds.

### *Appendix B.1: Set theory*

Consider a set T of terrestrial planets. We assume that any solar or extrasolar planet will either be an element of T or not. Planets have been detected with masses of about 5 to 10 Earth masses, and they likely constitute a transitional regime between terrestrial and, as the case may be, icy or gaseous. They may still have their bulk mass in solid form but also have a substantial atmosphere. Nevertheless, we use a sharp classification here for simplicity. We concentrate here on the genuine terrestrial planets. As an example, Earth ($e_\oplus$) is a terrestrial planet ($e_\oplus \in$ T), whereas Jupiter is not.

The elements of T are the terrestrial planets: T = {t $\in$ T | t terrestrial} (see Fig. A). Some of these planets will be habitable and thus be an element of the set of habitable, terrestrial planets H = {h $\in$ T | h habitable} (dotted area). The complement of this set is the set of uninhabitable, terrestrial planets U = $\bar{H}$ = {u $\in$ T | u uninhabitable} (blank area). There are no planets that are both habitable and uninhabitable. Hence, the union of H and U is equal to the terrestrial planets: H $\cap$ U = T. Beyond, there will be a set of Earth-like planets E = {e $\in$ T | e Earth-like} (vertically striped area). Our intuition, trained by the usage of the term "Earth-like" in literature, by talks, and conversations, suggests that Earth-like planets are habitable. For the time being, we prefer to take a more general point of view and allow Earth-like planets also to be uninhabitable. E thus overlaps with U in Fig.

---

[11] In this context, M is planetary mass, i is the inclination of the planet's orbital plane with respect to an Earth-based observer's line of sight, M$_{Earth}$ is an Earth-mass, and $P$ is the planet's orbital period about the star.

[12] Here, m is planetary mass and $i$ the inclination of the planet's orbital plane with respect to an Earth-based observer's line of sight.

[13] In this context, $M_p$ is planetary mass, "CHZ" is an abbreviation for the "continuous habitable zone", and a is the planet's orbital semi-major axis.

[14] (Wittgenstein, 1953, §43): "Die Bedeutung eines Wortes ist sein Gebrauch in der Sprache."

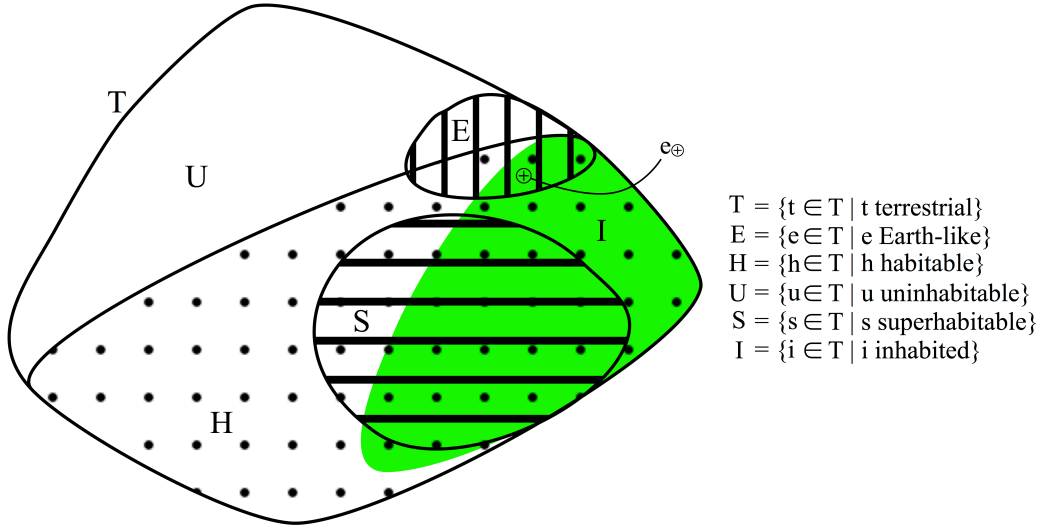

$T = \{t \in T \mid t \text{ terrestrial}\}$
$E = \{e \in T \mid e \text{ Earth-like}\}$
$H = \{h \in T \mid h \text{ habitable}\}$
$U = \{u \in T \mid u \text{ uninhabitable}\}$
$S = \{s \in T \mid s \text{ superhabitable}\}$
$I = \{i \in T \mid i \text{ inhabited}\}$

**Fig. A:** Set of terrestrial worlds T and subsets. The set membership of Earth $e_\oplus \in (E \cap I)$ is indicated with a symbol. This graphic visualizes our claim that habitable planets (H) need not be Earth-like (E), and that there may well exist a set of superhabitable worlds (S). The cardinality of S may be greater than that of E, and the fraction of planets inside S that are actually inhabited (I, green) may be greater than the fraction of Earth-like, inhabited planets. For this purpose, S is depicted to be larger than E, and $(E \cap I)$ is chosen to be smaller with respect to E than $(S \cap I)$ with respect to S.

A. Yet, to be inhabited, a terrestrial planet must also be habitable. Thus, the set $I = \{i \in T \mid i \text{ inhabited}\}$ (green area) of inhabited planets is a subset of H: $I \subseteq H$. Note that the equality is only valid if all the habitable planets were indeed inhabited. It is reasonable to assume that there exists at least one terrestrial planet that is habitable but yet uninhabited. Thus, we can securely state $I \subset H \Leftrightarrow (t \in I \Rightarrow t \in H)$.[15] With Earth being Earth-like, habitable, and inhabited, we have $e_\oplus \in (H \cap I \cap E)$. Finally, we propose that there exists a set $S = \{s \in T \mid s \text{ superhabitable}\}$ (horizontally striped area) of terrestrial planets, whose elements (that is, superhabitable planets) offer more comfortable environments to life than Earth does. From a statistical perspective, this statement reads:

A randomly chosen element $s \in S$ is more likely to be inhabited than a randomly chosen element $e \in E$.     (1)

Alternatively, with $p$ being the probability of a planet to be inhabited:

$$p(s) > p(e) \qquad (s \in S, e \in E) \qquad (2)$$

In Fig. A, we insinuate sentences (1) and (2) by plotting the relative area of $(S \cap I)$ to S larger than the relation of $(E \cap I)$ to E. An equivalent sentence to (2) is

$$|S \cap I| / |S| = p(s) > |E \cap I| / |E| = p(e) \qquad (s \in S, e \in E), \qquad (3)$$

where $|X|$ is the number of elements, or "cardinality," of X.

Sentences (1) – (3) say nothing about the absolute number of inhabited worlds from sets E and S, which corresponds to the size of the areas of E and S in Fig. A. Perhaps there are only two superhabitable planets in our galactic neighborhood, both of which are inhabited, and it may be that there are one hundred Earth-like planets in a similar volume, of which, say, ten are inhabited. Then still (2) is true because $p(s) = 2/2 = 1 > p(e) = 10/100 = 0.1$. But there would be five times as many Earth-like planets with life than there are superhabitable planets.

In debates about habitable planets, it is subliminally assumed that there are more Earth-like inhabited planets than there are non-Earth-like inhabited planets: $|E \cap I| > |\bar{E} \cap I|$. However, the numbers $|E \cap I|$, $|E \cap \bar{I}|$, $|(\bar{E} \cap T) \cap I|$, and $|(\bar{E} \cap T) \cap \bar{I}|$ are truly not known, say for a local volume of 100 pc about the Sun. There are only the following constraints: $|E \cap I| \geq 1$ and $|(\bar{E} \cap T) \cap I| \geq 30 = |\{CoRoT-7b, Kepler-10b, 55Cnc e, Kepler-18b, Kepler-20e, Kepler-20f, Kepler-36b, Kepler-42b, Kepler-42c, Kepler-42d,

---

[15] This question of equality is related to the question how long it took life to occur on Earth after the planet became habitable. In fact, planets may generally become inhabited very shortly after becoming habitable. This would allow one to advocate the $I \subseteq H$ relation or even $I = H$.

Kepler-62c, and others}][16] (Léger et al., 2009; Batalha et al., 2011; Winn et al., 2011; Cochran et al., 2011; Fressin et al., 2012; Carter et al., 2012; Muirhead et al., 2012; Borucki et al., 2013). More terrestrial planet candidates are known, but they lack either radius or mass determinations (for example Gl581d, GJ667Cb to h, GJ1214b, HD 88512, and Alpha Centauri Bb).

The possible existence of S has fundamental observational implications. Were it possible to describe S and predict the characteristics of its elements s, as we attempt in this communication, then the search for extraterrestrial life could be made more efficient. Assume two planets were found; one (ê) being Earth-like and another one (ŝ) being member of S. Then it would be more reasonable to spend research resources on ŝ rather than on ê in order to find extrasolar life. And intriguingly, ŝ could be less Earth-like than ê. Ultimately, a superhabitable world may already have been detected but not yet noticed as such.

*Appendix B.2: The principle of mediocracy*

The principle of mediocracy claims that, if an item is drawn at random from one of several categories, it is likelier to come from the most numerous category than from any of the other less numerous categories (Section 1 in Kukla, 2010). As an example, consider the cardinalities of two sets A and B were known; $|A| < |B|$ and $A \cap B = \varnothing$, where $\varnothing$ is the empty set. Further, $A \cup B = M = \{m \mid m \in A \vee m \in B\}$ is the set of all elements. Then if an arbitrary element $\dot{m} \in M$ were drawn, it would be more likely to come from B than from A. This is all the principle of mediocracy states. In this reading, it comes as a truism. Note that the proportion of A and B, that is, the prior $|A| < |B|$, is known and it is the probability for the drawing that is inferred: $p(\dot{m} \in B) > p(\dot{m} \in A)$.

In a second reading of the principle of mediocracy, and this is the one subliminally applied in modern searches for inhabited planets, the functions of the prior and the drawing are reversed. Here, $\dot{m}$ (which in our example from Section B.1 is Earth, $e_\oplus$) has already been drawn. It is recognized as an element of a certain set (E $\cap$ I), and it is claimed that this set is more abundant than the other one. In the terrestrial worlds scenario (Section B.1), this ventured conclusion reads "$e_\oplus \in (E \cap I) \Rightarrow |E \cap I| > |\bar{E} \cap I|$". We paraphrase it because it is not justified. Given that we have almost no antecedent knowledge of E, $\bar{E}$ , and I, this claim is not logic.[17]

What is more, it is not logical to state that the choice of $e_\oplus$ has been random; humans have not chosen the Earth by random from a set T (Mash, 1993). To make things worse, even if we could have chosen $e_\oplus$ randomly from T and if our assumption were correct, then what could we conclude from only one drawing? Numerous drawings, in other words observations and knowledge about inhabitance of many Earth-like and non-Earth-like planets, would be required to reconstruct the prior with statistical significance. Hence, Earth cannot be justified as a reference for astrobiological investigations with the principle of mediocracy. The claim "$e_\oplus \in (E \cap I) \Rightarrow |E \cap I| > |\bar{E} \cap I|$" remains arbitrary and current searches for life might not be designed optimally.

To conclude, the principle of mediocracy cannot explain why Earth should be considered a particularly benign, inhabited world. When applied to our set of terrestrial worlds, the principle simply states that a randomly chosen world most likely comes from the most numerous subset of worlds. In this understanding, the cardinality of the subsets of terrestrial worlds is the prior – it is know before the drawing – and the probability of affiliation with any subset can be predicted. Yet, concluding that inhabited worlds are most likely Earth-like is not logical, because, first, the roles of the prior (here: the inhabited worlds) and the drawing (here: Earth) are reversed and, second, Earth has not been drawn (by whom?) at random.

## Acknowledgments

René Heller thanks Morten Mosgaard for board and lodge on the Danish island Langeland where this study has been initiated. René Heller is funded by the Canadian Astrobiology Training Program and a member of the Origins Institute at McMaster University. Discussions with Rory Barnes have been a valuable stimulation to this study and we appreciate Alyssa Cobb's helpful comments on the manuscript. Computations have been performed with ipython 0.13 (Pérez and Granger, 2007) on python 2.7.2 and figures have been prepared with gnuplot 4.6 (www.gnuplot.info). This work has made use of NASA's Astrophysics Data System Bibliographic Services. Our collaboration has been inspired by a question of John Armstrong asked online during an AbGradCon talk 2012.

---

[16] www.exoplanet.eu/catalog as of July 15, 2013

[17] Yet, it could be true.

## 7.7 Formation, Habitability, and Detection of Extrasolar Moons (Heller et al. 2014)

Contribution:

RH invited all co-authors to this review, structured the content of this manuscript, and moderated the correspondence among the co-authors. RH guided the literature research, contributed to the calculations illustrated in Figs. 7, 8, and 13, created Figs. 1, 7, 8, and 13, led the writing of the manuscript, and served as a corresponding author for the journal editor and the referees. RH also created an illustration of Europa, Enceladus, Ganymede, and Titan (similar to Fig. 1 in the paper) that was chosen as a journal cover for the September 2014 issue of *Astrobiology* (www.liebertpub.com/toc/ast/14/9). An alternative representation made by RH with a to-scale version of these four moons was selected as an Astronomy Picture of the Day on 19 September 2014 (https://apod.nasa.gov/apod/ap140919.html).

# Formation, Habitability, and Detection of Extrasolar Moons

René Heller[1], Darren Williams[2], David Kipping[3], Mary Anne Limbach[4,5], Edwin Turner[4,6], Richard Greenberg[7], Takanori Sasaki[8], Émeline Bolmont[9,10], Olivier Grasset[11], Karen Lewis[12], Rory Barnes[13,14], Jorge I. Zuluaga[15]

## Abstract

The diversity and quantity of moons in the Solar System suggest a manifold population of natural satellites to exist around extrasolar planets. Of peculiar interest from an astrobiological perspective, the number of sizable moons in the stellar habitable zones may outnumber planets in these circumstellar regions. With technological and theoretical methods now allowing for the detection of sub-Earth-sized extrasolar planets, the first detection of an extrasolar moon appears feasible. In this review, we summarize formation channels of massive exomoons that are potentially detectable with current or near-future instruments. We discuss the orbital effects that govern exomoon evolution, we present a framework to characterize an exomoon's stellar plus planetary illumination as well as its tidal heating, and we address the techniques that have been proposed to search for exomoons. Most notably, we show that natural satellites in the range of 0.1 to 0.5 Earth mass (i.) are potentially habitable, (ii.) can form within the circumplanetary debris and gas disk or via capture from a binary, and (iii.) are detectable with current technology.

**Key Words**: Astrobiology – Extrasolar Planets – Habitability – Planetary Science – Tides

## 1. Introduction

Driven by the first detection of an extrasolar planet orbiting a Sun-like star almost 20 years ago (Mayor and Queloz, 1995), the search for these so-called exoplanets has nowadays achieved the detection of over one thousand such objects and several thousand additional exoplanet candidates (Batalha et al., 2013). Beyond revolutionizing mankind's understanding of the formation and evolution of planetary systems, these discoveries allowed scientists a first approach towards detecting habitats outside the Solar System. Ever smaller and lighter exoplanets were found around Sun-like stars, the record holder now smaller than Mercury (Barclay et al., 2013). Furthermore, ever longer orbital periods can be traced, now encompassing Earth-sized planets in the stellar habitable zones (Quintana et al. 2014) and beyond. While the realm of extrasolar planets is being explored in ever more detail, a new class of objects may soon become accessible to observations: extrasolar moons. These are the natural satellites of exoplanets, and based on our knowledge from the Solar System planets, they may be even more abundant.

Moons are tracers of planet formation, and as such their discovery around extrasolar planets would fundamentally reshape our understanding of the formation of planetary systems. As an example, the most massive planet in the Solar System, Jupiter, has four massive moons – named Io, Europa, Ganymede, and Callisto – whereas the second-most massive planet around the Sun, Saturn, hosts only one major moon, Titan. These different architectures were likely caused by different termination time scales of gas infall onto the circumplanetary disks, and they show evidence that Jupiter was massive enough to open up a gap in the circumsolar primordial gas and debris disk, while Saturn was not (Sasaki et al., 2010).

Besides clues to planet formation, exomoons excite the imagination of scientists and the public related to their possibility of being habitats for extrasolar life (Reynolds et al., 1987; Williams et al., 1997; Heller and Barnes, 2013). This idea has its roots in certain Solar System moons, which may – at least temporarily and locally – provide environments benign for certain organisms found on Earth. Could those niches on the icy moons in the Solar System be inhabited? And in particular, shouldn't there be many more moons outside the Solar System, some of which are not only habitable beyond a frozen surface but have had globally habitable surfaces for billions of years?

[1] Origins Institute, McMaster University, Department of Physics and Astronomy, Hamilton, ON L8S 4M1, Canada; rheller@physics.mcmaster.ca
[2] Penn State Erie, The Behrend College School of Science, 4205 College Drive, Erie, PA 16563-0203; dmw145@psu.edu
[3] Harvard-Smithsonian Center for Astrophysics, 60 Garden Street, Cambridge, MA 02138, USA; dkipping@cfa.harvard.edu
[4] Department of Astrophysical Sciences, Princeton University, Princeton, NJ 08544, USA; mapeters@princeton.edu
[5] Department of Mechanical and Aerospace Engineering, Princeton University, Princeton, NJ 08544, USA
[6] The Kavli Institute for the Physics and Mathematics of the Universe, The University of Tokyo, Kashiwa 227-8568, Japan; elt@astro.princeton.edu
[7] Lunar and Planetary Laboratory, University of Arizona, 1629 East University Blvd, Tucson, AZ 85721-0092, USA; greenberg@lpl.arizona.edu
[8] Department of Astronomy, Kyoto University, Kitashirakawa-Oiwake-cho, Sakyo-ku, Kyoto 606-8502, Japan; takanori@kusastro.kyoto-u.ac.jp
[9] Université de Bordeaux, LAB, UMR 5804, 33270 Floirac, France; emeline.bolmont@obs.u-bordeaux1.fr
[10] CNRS, LAB, UMR 5804, F-33270, Floirac, France
[11] Planetology ad Geodynamics, University of Nantes, France; olivier.grasset@univ-nantes.fr
[12] Earth and Planetary Sciences, Tokyo Institute of Technology, Japan; karen.michelle.lewis@gmail.com
[13] Astronomy Department, University of Washington, Box 351580, Seattle, WA 98195, USA; rory@astro.washington.edu
[14] NASA Astrobiology Institute – Virtual Planetary Laboratory Lead Team, USA
[15] FACom - Instituto de Física - FCEN, Universidad de Antioquia, Calle 70 No. 52-21, Medellín, Colombia; jzuluaga@fisica.udea.edu.co

While science on extrasolar moons remains theoretical as long as no such world has been found, predictions can be made about their abundance, orbital evolution, habitability, and ultimately their detectability. With the first detection of a moon outside the Solar System on the horizon (Kipping et al., 2012; Heller 2014), this paper summarizes the state of research on this fascinating, upcoming frontier of astronomy and its related fields.

To begin with, we dedicate Section 2 of this paper to the potentially habitable icy moons in the Solar System. This section shall provide the reader with a more haptic understanding of the possibility of moons being habitats. In Section 3, we tackle the formation of natural satellites, thereby focussing on the origin of comparatively massive moons roughly the size of Mars. These moons are suspected to be a bridge between worlds that can be habitable in terms of atmospheric stability and magnetic activity on the one hand, and that can be detectable in the not-too-far future on the other hand. Section 4 is devoted to the orbital evolution of moons, with a focus on the basics of tidal and secular evolution in one or two-satellite systems. This will automatically lead us to tidal heating and its effects on exomoon habitability, which we examine in Section 5, together with aspects of planetary evolution, irradiation effects on moons, and magnetic protection. In Section 6, we outline those techniques that are currently available to search for and characterize exomoons. Section 7 presents a summary and Section 8 an outlook.

## 2. Habitable Niches on Moons in the Solar System

Examination of life on Earth suggests that ecosystems require liquid water, a stable source of energy, and a supply of nutrients. Remarkably, no other planet in the Solar System beyond Earth presently shows niches that combine all of these basics. On Mars, permanent reservoirs of surface water may have existed billions of years ago, but today they are episodic and rare. Yet, we know of at least three moons that contain liquids, heat, and nutrients. These are the Jovian companion Europa, and the Saturnian satellites Enceladus and Titan. Ganymede's intrinsic and induced magnetic dipole fields, suggestive of an internal heat reservoir and a liquid water ocean, make this moon a fourth candidate satellite to host a subsurface habitat. What is more, only four worlds in the Solar System other than Earth are known to show present tectonic or volcanic activity. These four objects are not planets but moons: Jupiter's Io and Europa, Saturn's Enceladus, and Neptune's Triton.

Naturally, when discussing the potential of yet unknown exomoons to host life, we shall begin with an inspection of the Solar System moons and their prospects of being habitats. While the exomoon part of this review is dedicated to the surface habitability of relatively large natural satellites, the following Solar Systems moons are icy worlds that could only be habitable below their frozen surfaces, where liquid water may exist.

### 2.1 Europa

Europa is completely surrounded by a global ocean that contains over twice the liquid water of Earth on a body about the size of Earth's Moon. Its alternative heat source to the weak solar irradiation is tidal friction. Tidal heating tends to turn itself off by circularizing orbits and synchronizing spins. However, Europa's orbit is coupled to the satellites Io and Ganymede through the Laplace resonance (Peale et al., 1979; Greenberg, 1982). The orbital periods are locked in a ratio of 1:2:4, so their mutual interactions maintain eccentricities. As a result, enough heat is generated within Europa to maintain a liquid subsurface ocean about 10 to 100 km deep, which appears to be linked to the surface.

Its surface is marked by dark lines and splotches, and the low rate of impact craters suggests that the surface is younger than 50 Myr (Zahnle et al., 2003). Tectonics produce linear features (cracks, ridges, and bands), and thermal effects produce splotches (chaotic terrain) (Fig. 1). These global scale lineaments roughly correlate with the patterns of expected tidal stress on the ice shell (Helfenstein and Parmentier, 1983; Greenberg et al., 1998) and may record past deviations from uniform synchronous rotation. Double ridges likely form on opposite sides of cracks due to the periodic tidal working over each 80 hr orbit, thereby squeezing up material from the crack and out over the surface. This process would be especially effective if the crack extended from the ocean (Greenberg et al., 1998). Although tidal tension is adequate to crack ice, at depth the overburden pressure counteracts this stress. Therefore, it is unlikely that the ice is thicker than 10 km if ridges form in this way.

The most distinctive crack patterns associated with tidal stress are the cycloids, chains of arcs, each about 100 km long, connected at cusps, and extending often over 1000 km or more. Cycloids are ubiquitous on Europa, and they provided the first observational evidence for a liquid water ocean, because they fit the tidal stress model so well and because adequate tidal distortion of Europa would be impossible if all the water were frozen (Hoppa et al., 1999; Groenleer and Kattenhorn, 2008). Given that Europa's surface age is less than 1% of the age of the Solar System, and that many cycloids are among the freshest features, it seemed evident that the ocean must still exist today. Confirmation came with measurements of Jupiter's magnetic field near the satellite, which showed distortions consistent with the effect of a near-surface electrical conducting salty ocean (Kivelson et al., 2000). Active plumes observed at the moon's south pole deliver further evidence of a liquid water reservoir, but it remains unclear whether these waters that feed these geysers are connected to the global subsurface ocean or they are local (Roth et al., 2014).

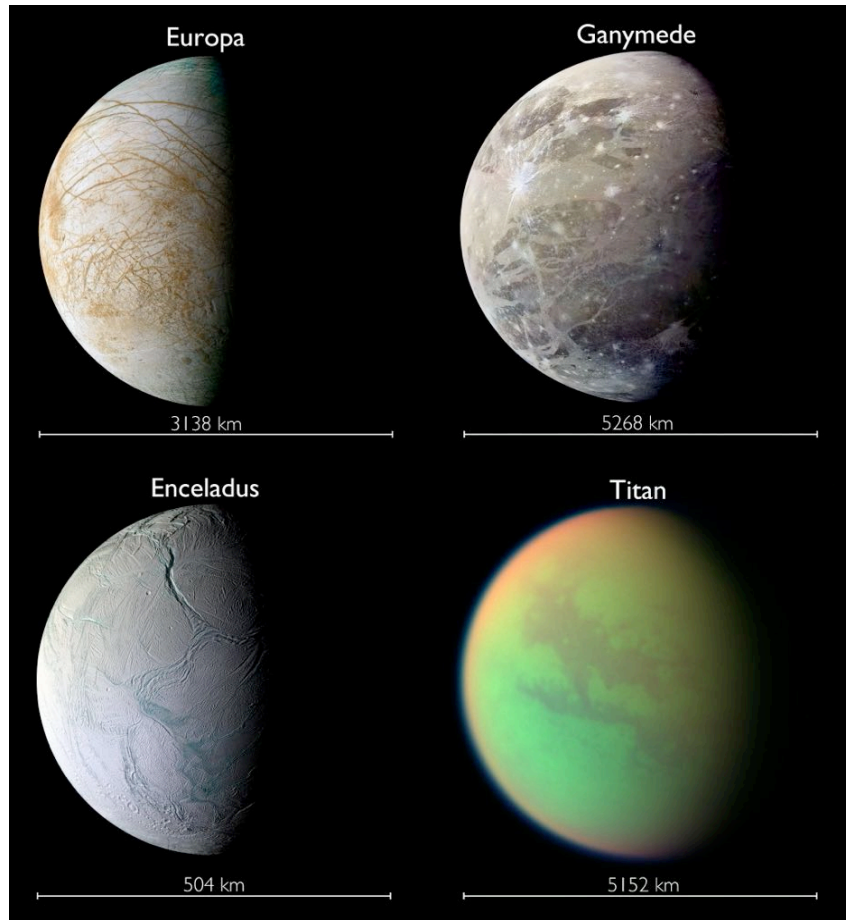

**Figure 1**: Europa, Enceladus, Ganymede, and Titan are regarded as potentially habitable moons. Global lineaments on Europa's surface and ridges on Enceladus indicate liquid water as close as a few kilometers below their frozen surfaces. Ganymede's surface is much older with two predominant terrains, bright, grooved areas and older, heavily cratered, dark regions. Titan has a dense nitrogen atmosphere and liquid methane/ethane seas on its surface. While the atmosphere is intransparent to the human eye, the lower right image contains information taken in the infrared. Note the different scales! Moon diameters are indicated below each satellite. [Image credits: NASA/JPL/Space Science Institute/Ted Stryk]

Large plates of Europa's surface ice have moved relative to one another. Along some cracks, often hundreds of kilometers long, rifts (called dilation bands) have opened up and filled with new striated ice (Tufts et al., 2000). The dilation can be demonstrated by reconstructing the surface, matching opposite sides, like pieces of a picture puzzle. The apparent mobility of large plates of surface ice shows that the cracks must penetrate to a fluid or low-viscosity layer, again indicating that the ice is less than 10 km thick.

Chaotic terrain covers nearly half of Europa's surface (Riley et al., 2000) and appears to have been thermally disrupted, leaving a lumpy matrix between displaced rafts, on whose surfaces fragments of the previous surface are clearly visible. The crust appears to have melted, allowing blocks of surface ice to float, move, and tilt before refreezing back into place (Carr et al., 1998). Only modest, temporary, local or regional concentrations of tidal heat are required for substantial melt-through. If 1% of the total internal heat flux were concentrated over an area about 100 km across, the center of such a region would melt through in only a few thousand years and cause broad exposure over tens of km wide in $10^4$ yr (O'Brien et al., 2000).

If the ice were thick, the best prospects for life would be near the ocean floor, where volcanic vents similar to those on the Earth's ocean floor could support life by the endogenic substances and tidal heat. Without oxygen, organisms would require alternative metabolisms whose abundance would be limited (Gaidos et al., 1999). For a human exploration mission, Europan life would only be accessible after landing on the surface, penetrating about 10 km of viscous ice, and diving 100 km or more to the ocean floor, where the search would begin for a hypothetical volcanic vent. If the ice is thin enough to be permeable, the odds that life is present and detectable increase. At the surface, oxidants – especially oxygen and hydrogen peroxide – are produced by impacts of energetic charged particles trapped in the Jovian magnetic field (Hand et al., 2007). Although this radiation must also sterilize the upper 10 cm of ice, organisms might be safe below that level. Photosynthesis would be possible down to a depth of a few meters, and the oxidants, along with organic compounds from cometary debris, are mixed to that depth by micrometeoroid impacts. Any active crack, periodically opening and closing with the tide, might allow the vertical flow of liquid water from the ocean and back down. Organisms living in the ice or the crack might take advantage of the access to near-surface oxidants and organics as well as oceanic substances and the flow of warm (0°C) water (Greenberg et al., 2002). Based on the size of the double ridges, cracks probably remain active for tens of thousands of years, and so organisms would have time to thrive. But once a crack freezes, they would need to hibernate in the ice or migrate into the ocean or to another active site. Exposure of water at the surface would allow some oxygen to enter the ocean directly. The gradual build-up of frozen ocean water over the surface exposes fresh ice to the production of oxidants and also buries ever deeper the previously oxygenated ice. Based on resurfacing rates of the various geological processes, oxygen may enter the ocean at a rate of about $3 \times 10^{11}$ moles/yr (Greenberg, 2010), equivalent to the respiration requirements of 3 million tons of terrestrial fish. Hence, this delivery rate could allow a concentration of oxygen adequate to support complex life, which is intrinsically less efficient than microorganisms, due to their extra operational overhead. Moreover, with this oxygen source, an ecosystem could be independent of photosynthesis. In the ocean, interaction of the oxygen with the rocky or clay seafloor could be significant and eventually result in a drastic decrease in the pH of the water (Pasek and Greenberg, 2012). How much this process could affect life depends on the efficiency of the contact between the water and the rock. Moreover, if organisms consume the oxygen fast enough, they could ameliorate the acidification.

## *2.2 Ganymede*

With its 2634 km in radius, the most massive moon in the Solar System, Ganymede, features old, densely cratered terrain and widespread regions that may have been subject to tectonic resurfacing. The great variety in geologic and geomorphic units has been dated over a range of several billions of years, and shows evidence of past internal heat release. Besides Mercury and Earth, Ganymede is one of only three solid bodies in the Solar System that generate a magnetic dipole field. It also possesses a small moment of inertia factor of 0.3115 (Schubert et al., 2004)[16], which is indicative of a highly differentiated body. This moon is thought to have (i) an iron-rich core, in which a liquid part must be present to generate the magnetic field, (ii) a silicate shell, (iii) a hydrosphere at least 500 km thick, and (iv) a tenuous atmosphere (Anderson et al., 1996; Kivelson et al., 2002; Sohl et al., 2002; Spohn and Schubert, 2003).

There is no evidence of any present geologic activity on Ganymede. Locally restricted depressions, called "paterae", may have formed through cryovolcanic processes, and at least one of them is interpreted as an icy flow that produced smooth bright lanes within the grooved terrain (Giese et al., 1998; Schenk et al., 2004). This suggests that cryovolcanism and tectonic processes played a role in the formation of bright terrain. There is no geologic evidence on Ganymede that supports the existence of shallow liquid reservoirs. And analyses of magnetic field measurements collected by the Galileo space probe are still inconclusive regarding an interpretation of a global subsurface ocean (Kivelson et al., 1996). The challenge is in the complex interaction among four field components, namely, a possibly induced field, the moon's intrinsic field, Jupiter's magnetosphere, and the plasma environment. But still, the presence of an ocean is in agreement with geophysical models, which predict that tidal dissipation and radiogenic energy keep the water liquid (Spohn and Schubert, 2003; Hussmann et al., 2006; Schubert et al. 2010).

While Europa with its relatively thin upper ice crust and a global ocean, which is likely in contact with the rocky ocean floor, has been referred to as a class III habitat (Lammer et al., 2009)[17], Ganymede may be a class IV habitat. In other words, its hydrosphere is split into (i) a high-pressure ice layer at the bottom that consists of various water-rich ices denser than liquid water, (ii) a subsurface water ocean that is likely not in direct contact with the underlying silicate floor, and (iii) an ice-I layer forming the outer crust of the satellite. In this model, Ganymede serves as an archetype for the recently suggested class of water world extrasolar planets (Kuchner, 2003; Léger et al., 2004; Kaltenegger et al., 2013; Levi et al., 2013). Ganymede's liquid layer could be up to 100 km thick (Sohl et al., 2010), and it is not clear whether these deep liquid oceans can be habitable. Chemical and energy exchanges between the rocky layer and the ocean, which are crucial for habitability, cannot be ruled out, but they require efficient transport processes through the thick high-pressure icy layer. Such processes are indeed possible (Sohl et al., 2010) but not as clear-cut as the exchanges envisaged for Europa, where they probably prevailed

---

[16] This dimensionless measure of the moment of inertia equals $I_{Cal}/(M_{Cal} \, R_{Cal}^2)$, where $I_{Cal}$ is Callisto's moment of inertia, $M_{Cal}$ its mass, and $R_{Cal}$ its radius. This factor is 0.4 for a homogeneous spherical body but less if density increases with depth.
[17] See cases 3 and 4 in their Fig. 11.

until recent times.

### 2.3 Enceladus

Another potential moon habitat is Enceladus. With an average radius of merely 250 km, Saturn's sixth-largest moon should be a cooled, dead body, if it were not subject to intense endogenic heating due to tidal friction. Contrary to the case of Europa, Enceladus' orbital eccentricity of roughly 0.0047 cannot be fully explained by gravitational interactions with its companion moons. While perturbations from Dione may play a role, they cannot explain the current thermal flux observed on Enceladus (Meyer and Wisdom, 2007). The current heat flux may actually be a remainder from enhanced heating in the past, and it has been shown that variations in the strength of this heat source likely lead to episodic melting and resurfacing events (Běhounková et al., 2012).

Given the low surface temperatures on Enceladus of roughly 70 K, accompanied by its low surface gravity of 0.114 m/s, viscous relaxation of its craters is strongly inhibited. The existence of a large number of shallow craters, however, suggests that subsurface temperatures are around 120 K or higher and that there should have been short periods of intense heating with rates up to 0.15 W/m$^2$ (Bland et al., 2012). These conditions agree well with those derived from studies of its tectonically active regions, which yield temporary heating of 0.11 to 0.27 W/m$^2$ (Bland et al., 2007; Giese et al., 2008).

Enceladus' internal heat source results in active cryovolcanic geysers on the moon's southern hemisphere (Porco et al., 2006; Hansen et al., 2006). The precise location coincides with a region warm enough to make Enceladus the third body, after Earth and Io, whose geological energy flow is accessible by remote sensing. Observations of the *Cassini* space probe during close flybys of Enceladus revealed temperatures in excess of 180 K along geological formations that have now become famous as "tiger stripes" (Porco et al, 2006; Spencer et al., 2008) (Fig. 1). Its geological heat source is likely strong enough to sustain a permanent subsurface ocean of liquid saltwater (Postberg et al., 2011). In addition to water (H$_2$O), traces of carbon dioxide (CO$_2$), methane (CH$_4$), ammonia (NH$_3$), salt (NaCl), and $^{40}$Ar have been detected in material ejected from Enceladus (Waite et al., 2006; Waite et al., 2009; Hansen et al., 2011). The solid ejecta, about 90% of which fall back onto the moon (Hedman et al, 2009; Postberg et al., 2011), cover its surface with a highly reflective blanket of μm-sized water ice grains. With a reflectivity of about 0.9, this gives Enceladus the highest bond albedo of any body in the Solar System (Howett et al., 2010).

While Europa's subsurface ocean is likely global and may well be in contact with the moon's silicate floor, Enceladus' liquid water reservoir is likely restricted to the thermally active south polar region. It is not clear if these waters are in contact with the rocky core (Tobie et al., 2008; Zolotov, 2007) or if they only form pockets in the satellite's icy shell (Lammer et al., 2009).

All of these geological activities, and in particular their interference with liquid water, naturally open up the question of whether Enceladus may be habitable. Any ecosystem below the moon's frozen ice shield would have to be independent from photosynthesis. It could also not be based on the oxygen (O$_2$) or the organic compounds produced by surface photosynthesis (McKay et al., 2008). On Earth, indeed, such subsurface ecosystems exist. Two of them rely on methane-producing microorganisms (methanogens), which themselves feed on molecular hydrogen (H$_2$) released by chemical reactions between water and olivine rock (Stevens and McKinley, 1995; Chapelle et al., 2002). A third such anaerobic ecosystem is based on sulfur-reducing bacteria (Lin et al, 2006). The H$_2$ required by these communities is ultimately produced by the radioactive decay of long-lived uranium (U), thorium (Th), and potassium (K). It is uncertain whether such ecosystems could thrive on Enceladus, in particular due to the possible lack of redox pairs under a sealed ocean (Gaidos et al., 1999). For further discussion of possible habitats on Enceladus, see the work of McKay et al. (2008).

What makes Enceladus a particularly interesting object for the in-situ search for life is the possibility of a sample return mission that would not have to land on the moon's surface (Reh et al., 2007; Razzaghi et al., 2007; Tsou et al., 2012). Instead, a spacecraft could repeatedly dive through the Enceladian plumes and collect material that has been ejected from the subsurface liquid water reservoir. This icy moon may thus offer a much more convenient and cheap option than Mars from which to obtain biorelevant, extraterrestrial material. What is more, once arrived in the Saturnian satellite system, a spacecraft could even take samples of the upper atmosphere of Titan (Tsou et al., 2012).

### 2.4 Titan

With its nitrogen-dominated atmosphere and surface pressures of roughly 1.5 bar, Titan is the only world in the Solar System to maintain a gaseous envelope at least roughly similar to that of Earth in terms of composition and pressure. It is also the only moon beyond Earth's Moon and the only object farther away from the Sun than Mars, from which a spacecraft has returned in-situ surface images. Footage sent back from the surface by the Huygens lander in January 2005 show pebble-sized, rock-like objects some ten centimeters across (mostly made of water and hydrocarbon ices) on a frozen ground with compression properties similar to wet clay or dry sand (Zarnecki et al., 2005). Surface temperatures around 94 K imply that water is frozen and cannot possibly play a key role in the weather cycle as it does on Earth. Titan's major atmospheric constituents are molecular nitrogen (N$_2$, 98.4 %), CH$_4$ (1.4 %), molecular hydrogen (0.1 %), and smaller traces of acetylene ($C_2H_2$) as well as ethane ($C_2H_6$) (Coustenis et al., 2007). The surface is hidden to the human eye under an optically thick photochemical haze (Fig. 1), which is composed of various hydrocarbons. While the surface illumination is extremely low, atmospheric $C_2H_2$ could act as a mediator and transport the energy of solar ultra-violet radiation and high-energy particles to the surface, where it could undergo exothermic reactions (Lunine, 2010). Intriguingly, methane should be irreversibly destroyed by photochemical processes on a timescale of 10 to 100 million years (Yung et al., 1984; Atreya et al., 2006). Hence, its abundance suggests that it is continuously resupplied. Although possibilities of methanogenic life on Titan have been hypothesized (McKay and Smith, 2005; Schulze-Makuch and Grinspoon, 2005), a biological origin of methane seems unlikely because its $^{12}C/^{13}C$ ratio is not enhanced with respect to the Pee Dee Belemnite inorganic standard value (Niemann et al., 2005). Instead, episodic cryovolcanic activity could release substantial amounts of methane from Titan's crust (LeCorre et al., 2008), possibly driven by outgassing from internal reservoirs of clathrate hydrates (Tobie et al., 2006).

In addition to its thick atmosphere, Titan's substantial reservoirs of liquids on its surface make it attractive from an astrobiological perspective (Stofan et al., 2007; Hayes et al., 2011). These ponds are mostly made of liquid ethane (Brown et al., 2008) and they feed a weather cycle with evaporation of surface liquids, condensation into clouds, and precipitation (Griffith et al., 2000). Ultimately, the moon's non-synchronous rotation period with respect to Saturn as well as its substantial orbital eccentricity of 0.0288 (Sohl et al., 1995; Tobie et al., 2005) point towards the presence of an internal ocean, the composition and depth of which is unknown (Lorenz et al., 2008; Norman, 2011). If surface life on Titan were to use non-aqueous solvents, where chemical reactions typically occur with much higher rates than in solid or gas phases, it would have to rely mostly on ethane. However, laboratory tests revealed a low solubility of organic material in ethane and other non-polar solvents (McKay, 1996), suggesting also a low solubility in liquid methane. While others have argued that life in the extremely cold hydrocarbon seas on Titan might still be possible (Benner et al., 2004), the characterization of such ecosystems on extrasolar moons will not be possible for the foreseeable future. In what follows, we thus exclusively refer to habitats based on liquid water.

## 3. Formation of Moons

Ganymede, the most massive moon in the Solar System, has a mass roughly 1/40 that of Earth or 1/4 that of Mars. Supposing Jupiter and its satellite system would orbit the Sun at a distance of one astronomical unit (AU), Ganymede's ices would melt but it would not be habitable since its gravity is too small to sustain a substantial atmosphere. Terrestrial sized objects must be larger than 1 to 2 Mars masses, or 0.1 to 0.2 Earth masses ($M_\oplus$), to have a long-lived atmosphere and moist surface conditions in the stellar habitable zones (Williams et al., 1997). Hence, to assess the potential of extrasolar moons to have habitable surfaces, we shall consult satellite formation theories and test their predictions for the formation of Mars- to Earth-sized moons.

Ganymede as well as Titan presumably formed in accretion disks surrounding a young Jupiter and Saturn (Canup and Ward, 2002), and apparently there wasn't enough material or accretion efficiency in the disks to form anything larger. Disks around similar giant planets can form much heavier moons if the disk has a lower gas-to-solid ratio or a high viscosity parameter α (Canup and Ward, 2006). Assuming disks similar to those that existed around Jupiter and Saturn, giant exoplanets with masses five to ten times that of Jupiter might accrete moons as heavy as Mars.

The orbit of Triton, the principal moon of Neptune, is strongly suggestive of a formation via capture rather than agglomeration of solids and ices in the early circumplanetary debris disk. The satellite's orbital motion is tilted by about 156° against the equator of its host planet, and its almost perfectly circular orbit is embraced by various smaller moons, some of which orbit Neptune in a prograde sense (that is, in the same direction as the planet rotates) and others which have retrograde orbits. With Triton being the seventh-largest moon in the Solar System and with Earth- to Neptune-sized planets making up for the bulk part of extrasolar planet discoveries (Batalha et al. 2013; Rowe et al. 2014), the capture of substantial objects into stable satellites provides another reasonable formation channel for habitable exomoons.

Moving out from the Solar System towards moons orbiting extrasolar planets, and eventually those orbiting planets in the stellar habitable zones, we ask ourselves: "How common are massive exomoon systems around extrasolar planets?" Ultimately, we want to know the frequency of massive moons the size of Mars or even Earth around those planets. In the following, we discuss recent progress towards addressing these questions from the formation theory point of view.

### 3.1 In-situ formation

#### 3.1.1 Planetary disk models

Various models for satellite formation in circumplanetary protosatellite disks have been proposed recently. One is called the solids enhanced minimum mass model (Mosqueira and Estrada, 2003a, 2003b; Estrada et al., 2009), another one is an actively supplied gaseous accretion disk model (Canup and Ward, 2002, 2006, 2009), and a third one relies on the viscous spreading of a massive disk inside the planet's Roche limit (Crida and Charnoz, 2012). All these models have been applied to study the formation of the Jovian and Saturnian satellite systems. In the Mosqueira and Estrada model, satellite formation occurs once sufficient gas has been removed from an initially massive subnebula and turbulence in the circumplanetary disk subsides. Satellites form from solid materials supplied by ablation and capture of planetesimal fragments passing through the massive disk. By contrast, in the Canup and Ward picture satellites form in the accretion disk at the very end of the host planet's own accretion, which should reflect the final stages of growth of the host planets. Finally, the Crida and Charnoz theory assumes that satellites originate at the planet's Roche radius and then move outward due to gravitational torques experienced within the tidal disk. While the latter model may explain some aspects of satellite formation around a wide range of central bodies, that is, from bodies as light as Pluto to giant planets as heavy as Saturn, it has problems reproducing the Jovian moon system.

The actively supplied gaseous accretion disk model postulates a low-mass, viscously evolving protosatellite disk with a peak surface density near 100 g/cm² that is continuously supplied by mass infall from the circumstellar protoplanetary disk. The temperature profile of the circumplanetary disk is dominated by viscous heating in the disk, while the luminosity of the giant planet plays a minor role. "Satellitesimals", that is, proto-moons are assumed to form immediately from dust grains that are supplied by gas infall. Once a satellite embryo has grown massive enough, it may be pushed into the planet through type I migration driven by to satellite-disk interaction (Tanaka et al., 2002). The average total mass of all satellites resulting from a balance between disposal by type I migration and repeated satellitesimal accretion is universally of the order of $10^{-4}$ $M_p$, where $M_p$ is the host planet mass (Canup and Ward, 2006). Beyond a pure mass scaling, Sasaki et al. (2010) were able to explain the different architectures of the Jovian and the Saturnian satellite systems by including an inner cavity in the circum-Jovian disk, while Saturn's disk was assumed to open no cavity, and by considering Jupiter's gap opening within the solar disk. Ogihara and Ida (2012) improved this model by tracing the orbital evolution of moonlets embedded in the circumplanetary disk with an $N$-body method, thereby gaining deeper insights into the compositional evolution of moons.

### 3.1.2 Differences in the Jovian and Saturnian satellite systems

As a gas giant grows, its gravitational perturbations oppose viscous diffusion and pressure gradients in the gas disk and thereby open up a gap in the circumstellar protoplanetary disk (Lin and Papaloizou, 1985). Since the critical planetary mass for the gap opening at 5 - 10 AU is comparable to Jupiter's mass, it is proposed that Jupiter's final mass is actually determined by the gap opening (Ida and Lin, 2004). While the theoretically predicted critical mass for gap opening generally increases with orbital radius, Saturn has a mass less than one-third of Jupiter's mass in spite of its larger orbital radius. It is conventionally thought that Saturn did not open up a clear gap and the infall rate of mass onto Saturn and its protosatellite disk decayed according to the global depletion of the protoplanetary disk.[18] Since core accretion is generally slower at larger orbital radii in the protoplanetary disk (Lin and Papaloizou, 1985), it is reasonable to assume that gas accretion onto Saturn proceeded in the dissipating protoplanetary disk, while Jupiter's growth was truncated by the formation of a clear gap before the disk lost most of its gas.

Although gap opening may have failed to truncate the incident gas completely, it would certainly reduce the infall rate by orders of magnitude (D'Angelo et al., 2003). After this truncation, the protosatellite disk would be quickly depleted on its own viscous diffusion timescale of about $(10^{-3}/\alpha)10^3$ years, where $\alpha$ represents the strength of turbulence (Shakura and Sunyaev, 1973) and has typical values of $10^{-3}$ to $10^{-2}$ for turbulence induced by magneto-rotational instability (Sano et al., 2004). On the other hand, if Saturn did not open up a gap, the Saturnian protosatellite disk was gradually depleted on the much longer viscous diffusion timescales of the protoplanetary disk, which are observationally inferred to be $10^6$ to $10^7$ years (Meyer et al., 2007). The different timescales come from the fact that the protosatellite disk is many orders of magnitude smaller than the protoplanetary disk. The difference would significantly affect the final configuration of satellite systems, because type I migration timescales of protosatellites (approximately $10^5$ years) are in between the two viscous diffusion timescales (Tanaka et al., 2002). Jovian satellites may retain their orbital configuration frozen in a phase of the protosatellite disk with relatively high mass at the time of abrupt disk depletion, while Saturnian satellites must be survivors against type I migration in the final less massive disk.

The different disk masses reflected by the two satellite systems indicate that their disks also had different inner edges. By analogy with the observationally inferred evolution of the inner edges of disks around young stars (Herbst and Mundt, 2005), Sasaki et al. (2010) assumed (i.) a truncated boundary with inner cavity for the Jovian system and (ii.) a non-truncated boundary without cavity for the Saturnian system. Condition (i.) may be appropriate in early stages of disk evolution with high disk mass and accretion rate and strong magnetic field, while in late stages with a reduced disk mass, condition (ii.) may be more adequate. The Jovian satellite system may have been frozen in a phase with condition (i.), since the depletion timescale of its protosatellite disk is much shorter than typical type I migration timescales of protosatellites. In this system, type I migration of satellites is terminated by a local pressure maximum near the disk edge. On the other hand, the Saturnian system that formed in a gradually dissipating protosatellite disk may reflect a later evolution phase with condition II.

---

[18] The recently proposed Grand Tack model (Walsh et al., 2011), however, suggests that Jupiter and Saturn actually opened up a common gap (Pierens and Raymond, 2011).

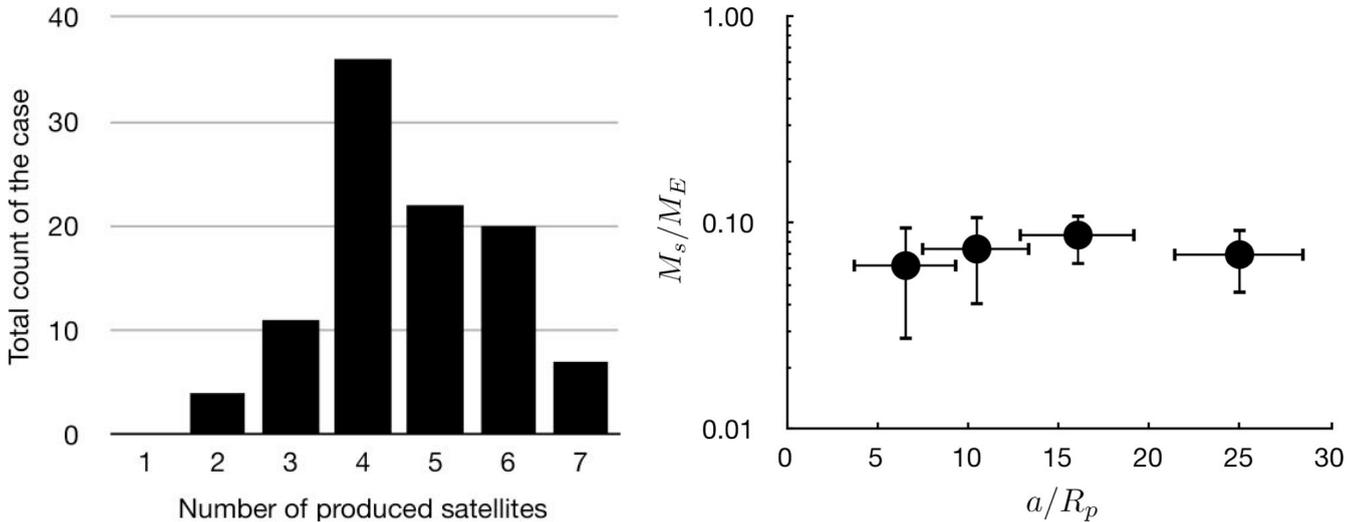

**Figure 2**: Results after 100 simulations of moon formation around a 10 $M_{Jup}$ planet. Left panel: Multiplicity distribution of the produced satellite systems for satellite masses $M_s > 10^{-2}$ $M_\oplus$. Right panel: Averaged $M_s$ and semi-major axis ($a$) are shown as filled circles, their standard deviations are indicated by error bars for the 36 four-satellite-systems.

Sasaki et al. (2010) explained how these different final formation phases between Jupiter and Saturn produce the significantly different architectures of their satellite systems, based on the actively supplied gaseous accretion protosatellite disk model (Canup and Ward, 2002). They applied the population synthesis planet formation model (Ida and Lin, 2004, 2008) to simulate growth of protosatellites through accretion of satellitesimals and their inward orbital migration caused by tidal interactions with the gas disk. The evolution of the gas surface density of the protosatellite disk is analytically given by a balance between the infall and the disk accretion. The surface density of satellitesimals is consistently calculated with accretion by protosatellites and supply from solid components in the incident gas. They assumed that the solid component was distributed over many very small dust grains so that the solids can be delivered to the protosatellite disk according to the gas accretion to the central planet. The model includes type I migration and regeneration of protosatellites in the regions out of which preceding runaway bodies have migrated leaving many satellitesimals. Resonant trapping of migrating protosatellites is also taken into account. If the inner disk edge is set, the migration is halted there and the migrated protosatellites are lined up in resonances from the inner edge to the outer regions. When the total mass of the trapped satellites exceeds the disk mass, the halting mechanism is not effective, such that they release the innermost satellite into the planet.

Sasaki et al. (2010) showed that in the case of the Jovian system, a few satellites of similar masses were likely to remain in mean motion resonances. These configurations form by type I migration, temporal stopping of the migration near the disk inner edge, and quick truncation of gas infall by gap opening in the solar nebula. On the other hand, the Saturnian system tended to end up with one dominant body in its outer regions caused by the slower decay of gas infall associated with global depletion of the solar nebula. Beyond that, the compositional variations among the satellites was consistent with observations with more rocky moons close to the planet and more water-rich moons in wider orbits.

### 3.1.3 Formation of massive exomoons in planetary disks

Ultimately, we want to know whether much more massive satellites, for example the size of Mars, can form around extrasolar giant planets. If the Canup and Ward (2006) mass scaling law is universal, these massive satellites could exist around super-Jovian gas giants. To test this hypothesis, we applied the population synthesis satellite formation model of Sasaki et al. (2010) to a range of hypothetical "super-Jupiters" with tenfold the mass of Jupiter. The results from 100 runs are depicted in Fig. 2, showing that in about 80% of all cases, four to six large bodies are formed. In the right panel, we show the averaged semi-major axes (abscissa) and masses (ordinate), along with the 1σ standard deviations, of those 36 systems that contain four large satellites. In these four-satellite-systems, the objects reach roughly the mass of Mars, and they are composed of rocky materials as they form in massive protosatellite disks with high viscous heating. We conclude that massive satellites around extrasolar gas giants can form in the circumplanetary disk and that they can be habitable if they orbit a giant planet in the stellar habitable zone.

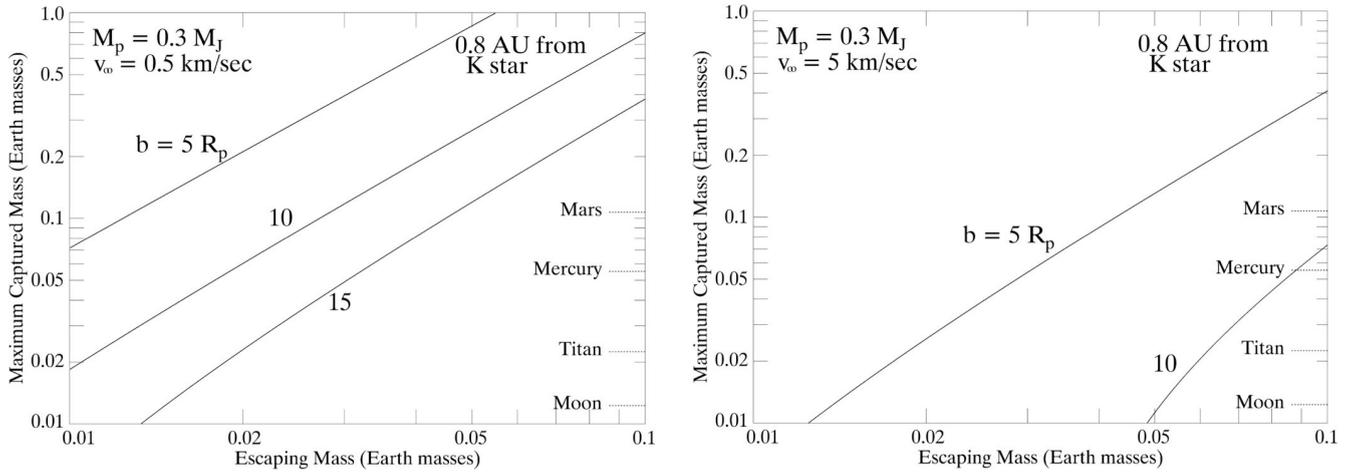

**Figure 3**: Maximum captured mass (ordinate) as a function of escaping mass (abscissa) and encounter distance $b = 5\ R_p$, 10 $R_p$, and 15 $R_p$ (contours). The curves are calculated from Eq. (3) with the planetary mass set to 0.3 $M_{\mathrm{Jup}}$ and the distance from the K star $a_\star = 0.8$ AU in both panels. The encounter speed at infinity $v_\infty = 0.5$ km/s in the left panel and $v_\infty = 5$ km/s in the right panel.

### 3.2 Formation by capture

Other ways of forming giant moons include collision and capture, both of which require strong dynamical mixing of giant planets and terrestrial objects in a protoplanetary disk. A gas giant planet can migrate both inward and outward through a protoplanetary disk once it grows large enough to clear a gap in the disk. The migrating gas giant is then able to sweep through terrestrial material, as is thought to have occurred in the Solar System (Gomes et. al, 2005), possibly enabling close encounters that might result in the capture of an over-sized moon. In all capture scenarios, the approaching mass must be decelerated below the planet's escape velocity and inserted into a bound, highly elliptical orbit that can later be circularized by tides (Porter and Grundy, 2011). The deceleration may result from the impactor striking the planet, a circumplanetary disk (McKinnon and Leith, 1995), or a fully formed satellite, but in each of these cases, the intercepting mass – gas or solid – must be comparable to the mass of the impactor to make a significant change in its encounter trajectory. For an impact with the planet to work, the impactor must tunnel deep enough through a planet's outer atmosphere to release enough energy but still shallow enough to avoid disintegration. An impact with a massive satellite already in orbit around the planet might work if the collision is head-on and the impactor-to-satellite mass ratio is smaller than a few, which makes this formation scenario unlikely.

Irrespective of the actual physical reason or history of a possible capture, Porter and Grundy (2011) followed the orbital evolution of captured Earth-mass moons around a range of giant planets orbiting in the HZ of M, G, and F stars. Stellar perturbations on the planet-satellite orbit were treated with a Kozai Cycle and Tidal Friction model, and they assumed their satellites to start the planetary entourage in highly elliptical orbits with eccentricities $e_{ps} > 0.85$ and apoapses beyond 0.8 times the planetary Hill radius

$$R_{\mathrm{Hill,p}} = \left(\frac{M_p}{3M_\star}\right)^{1/3} a_{\star p} \qquad , \tag{1}$$

where $M_\star$ is the stellar mass, and $a_{\star p}$ is the orbital semi-major axis of the star-planet system. Initial orbital inclinations and moon spin states were randomized, which allowed investigations of a potential preference for prograde or retrograde orbital stability. Most importantly, their results showed that captured exomoons in stellar HZs tend to be more stable the higher the stellar mass. While about 23% of their captured satellites around Neptune- and Jupiter-sized planets in the HZs of an M0 star remained stable, roughly 45% of such planet-moon binaries in the HZ of a Sun-like star survived and about 65% of similar scenarios in the HZ of an F0 star stabilized over one billion years. This effect is related to the extent of the planet's Hill sphere, which scales inversely proportional to stellar distance while stellar illuminations goes with one over distance squared. No preference for either pro- or retrograde orbits was found. Typical orbital periods of the surviving moons were 0.9, 2.1, and 3.6 days for planet-moon binaries in the HZs of M, G, and F stars, respectively.

A plausible scenario for the origin of these captures was recently discussed by Williams (2013, W13 hereafter), who showed that a massive moon could be captured if it originally belonged to a binary-terrestrial object (BTO) that was tidally disrupted from a close-encounter with a gas-giant planet. During the binary-exchange interaction, one of the BTO members is ejected while the other is captured as a moon. In fact, it is this mechanism that gives the most reasonable formation scenario for Neptune's odd principal moon Triton (Agnor and Hamilton, 2006). The first requirement for a successful capture is for the BTO to actually form in the first place. The second requirement is to pass near enough to the planet (inside five to ten planet radii) to be tidally disrupted. This critical distance is where the planet's gravity exceeds the self-gravity of the binary, and is expressed in W13 as

$$\left(\frac{b}{R_{\rm p}}\right) \lesssim \left(\frac{a_{\rm B}}{R_1}\right)\left(\frac{3\rho_{\rm p}}{\rho_1}\right)^{1/3} \quad , \tag{2}$$

where the subscript "1" refers to the more massive component of the binary, "p" refers to the planet, and $a_{\rm B}$ is the binary separation. Setting $a_{\rm B}/R_1 = 10$, and using densities $\rho_{\rm p}$ and $\rho_1$ appropriate for gaseous and solid planets, respectively, yields $b \approx 9.3\ R_{\rm p}$, which shows that the encounter must occur deeper than the orbits of many of the major satellites in the Solar System.

A second requirement for a successful binary-exchange is for the BTO to rotate approximately in the encounter plane as it passes the planet. The binary spin thereby opposes the encounter velocity of the retrograde-moving binary mass so that its final velocity may drop below the escape velocity from the planet if the encounter velocity of the BTO at infinity is small, say ≲ 5 km/s. Finally, the distance and mass of the host star are important because they determine how large of a satellite orbit is stable given the incessant stellar pull. Implementing these physical and dynamical requirements yields an analytic upper limit on the captured mass $m_1$ as a function of the escaping mass $m_2$, in addition to other parameters such as planet mass $M_{\rm p}$, impact parameter $b$, encounter velocity $v_{\rm enc}$, and the periapsis velocity $v_{\rm peri}$ of the newly captured mass (W13). Expressed in compact form, this relation is

$$m_1 < 3M_{\rm p}\left(\frac{G\,m_2\,\pi}{2\,b\,v_{\rm enc}(v_{\rm enc}-v_{\rm peri})}\right)^{3/2} - m_2 \quad , \tag{3}$$

with $G$ as Newton's gravitational constant and secondary expressions for $v_{\rm enc}$ and $v_{\rm peri}$ given in Eqs. (4) and (5) of W13.

It is apparent from this expression that the heaviest moons (large $m_1$) will form from the closest encounters (small $b$) occurring at low velocity (small $v_{\rm enc}$). The dependence on planet mass is not as straightforward. It appears from Eq. (3) that larger planets should capture heavier masses, but the encounter velocity and periapsis velocity in the denominator both increase with $M_{\rm p}$, making the dependence on planet mass inverted. This is borne out in Fig. 7 of W13, which shows the size of the captured mass to decrease as planet mass increases with the impact parameter $b$ held constant. Therefore, in general, it is easier to capture a moon around a Saturn- or Neptune-class planet than around a Jupiter or a super-Jupiter because the encounter speeds tend to be smaller.

As detailed in W13, (see his Figs. 3 to 7), moons the size of Mars or even Earth are possibly formed if the ratio of captured mass to escaping mass is not too large. The limiting ratio depends on encounter details such as the size and proximity of both the planet and the star, as well as the encounter velocity. Yet, a ratio > 10:1 could yield a moon as big as Earth around a Jupiter at 2 AU from the Sun (W13, right side of Fig. 5 therein). Mars-sized moons can possibly form by ejecting a companion the size of Mercury or smaller.

Expanding the cases considered in W13, we assume here a Saturn-mass planet in the habitable zone (0.8 AU) around a K-star and vary the encounter distance $b$ as well as the encounter velocity to examine whether velocities as large as 5 km/s make a binary-exchange capture impossible. According to the left panel of Fig. 3, losing an object the size of Mercury (≈ 0.05 $M_{\oplus}$) in a 0.5 km/s encounter would result in the capture of a Mars-sized moon, with tidal disruption occurring as far as 15 $R_{\rm p}$ from the planet. The right panel of Fig. 3 shows that such an exchange is also possible at ten times this encounter velocity, provided the approach distance is reduced by a factor of three. This is a promising result given the broad range of encounter speeds and approach distances expected between dynamically interacting planets in developing planetary systems. To sum up, close encounters of binary-terrestrial objects can reasonably provide a second formation channel for moons roughly the mass of Mars.

## 4. Orbital Dynamics

Once the supply of the circumplanetary disk with incident gas and dust has ceased several million years after the formation of the planetary system or once a satellite has been captured by a giant planet, orbital evolution will be determined by the tidal interaction between the planet and the moons, gravitational perturbations among multiple satellites, the gravitational pull from the star, and perturbations from other planets. These effects give rise to a range of phenomena, such as spin-orbit resonances, mean motion resonances among multiple satellites, chaos, ejections, and planet-satellite mergers.

As an example, a giant planet orbiting its host star closer than about 0.5 AU will have its rotation frequency $\Omega_p$ braked and ultimately synchronized with its orbital motion $n_{\star p}$ around the star. This rotational evolution, $\Omega_p = \Omega_p(t)$ ($t$ being time), implies an outward migration of the planet's corotation radius, at which $\Omega_p = n_{ps}$, with $n_{ps}$ as the orbital mean motion of a satellite around the planet. Due to the exchange of orbital and rotational momentum via the planet's tidal bulge raised by a moon, satellites inside the corotation radius will spiral towards the planet and perhaps end in a collision (see Phobos' fall to Mars; Efroimsky and Lainey, 2007), while satellites beyond the corotation radius will recede from their planets and eventually be ejected (Sasaki et al., 2012). Hence, the evolution of a planet's corotation radius due to stellar-induced tidal friction in the planet affects the stability of moon systems. Barnes and O'Brien (2002) considered Earth-mass moons subject to an incoming stellar irradiation similar to the solar flux received by Earth and found that these satellites can follow stable orbits around Jupiter-like planets if the host star's mass is greater than 0.15 $M_\odot$, with $M_\odot$ as one solar mass. From another perspective, satellite systems around giant planets that orbit their stars beyond 0.6 AU should still be intact after 5 Gyr. Cassidy et al. (2009), however, claimed that an Earth-sized moon could even follow a stable orbit around a hot Jupiter if the planet is rotationally synchronized to its orbit around the star. Then the tidal forcing frequencies raised by the moon on the planet would be in a weakly dissipative regime, allowing for a slow orbital evolution of the moon. Combining the results of Barnes and O'Brien (2002) with those of Domingos et al. (2006), Weidner and Horne (2010) concluded that 92% of the transiting exoplanets known at that time (almost all of which have masses between that of Saturn and a few times that of Jupiter) could not have prograde satellites akin to Earth's Moon. Further limitations on the orbital stability of exomoons are imposed by the migration-ejection instability, causing giant planets in roughly 1-day orbits to lose their moons during migration within the circumstellar disk (Namouni, 2010).

Investigations on the dynamical stability of exomoon systems have recently been extended to the effects of planet-planet scattering. Gong et al. (2013) found that planetary systems whose architectures are the result of planet-planet scattering and mergers would have lost their initial satellite systems. Destruction of moon systems would be particularly effective for scattered hot Jupiters and giant planets on eccentric orbits. Most intriguingly, in their simulations, the most massive giant planets were not hosts of satellite systems, if these planets were the product of former planet-planet mergers. In a complimentary study, Payne et al. (2013) explored moons in tightly packed giant planet systems, albeit during less destructive planet-planet encounters. For initially stable planet configurations, that is, planet architectures that avoid planet-planet mergers or ejections, they found that giant exoplanets in closely packed systems can very well harbor exomoon systems.

Orbital effects constrain the habitability of moons. Heller (2012) concluded that stellar perturbations would force exomoons into elliptical orbits, thereby generating substantial amounts of tidal heat in the moons. Giant planets in the HZs around low-mass stars have small Hill radii, and so their moons would necessarily have small semi-major axes. As a consequence of both the requirement for a close orbit and the stellar-induced orbital eccentricities, Mars- to Earth-sized exomoons in the HZs of stars with masses below 0.2 to 0.5 $M_\odot$ would inevitably be in a runaway greenhouse state due to their intense tidal heating.

To illustrate the evolution in satellite systems in the following, we simulate the orbital evolution of hypothetical exomoon systems and discuss secular and tidal processes in more detail. Stellar perturbations are neglected. At first, we consider single satellite systems and then address multiple satellite systems.

### 4.1 Orbital evolution in single satellite systems

Imagine a hypothetical Earth-mass satellite orbiting a Jupiter-mass planet. Both the tides raised on the planet by the satellite (the planetary tide) and the tides raised on the satellite by the planet (the satellite tide) can dissipate energy from this two-body system. For our computations, we use the constant-time-lag model, which is a standard equilibrium tidal model (Hut, 1981). As in the work of Bolmont et al. (2011,2012), we solve the secular equations for this system assuming tidal dissipation in the moon to be similar to that on Earth, where $\Delta\tau_\oplus = 638$s is the time lag between Earth's tidal bulge and the line connecting the two centers of mass (Neron de Surgy and Laskar, 1997), and $k_{2,\oplus} \times \Delta\tau_\oplus = 213$ s, with $k_{2,\oplus}$ as the Earth's second-degree tidal Love number. For Jupiter, we use the value given by Leconte et al. (2010) ($k_{2,Jup} \times \Delta\tau_{Jup} = 2.5 \times 10^{-2}$ s) and include the evolution of the planetary radius $R_{Jup}$, of $k_{2,J}$, and of the radius of gyration squared ($r_{g,Jup}$)$^2 = I_{Jup}/(M_{Jup} R_{Jup}^2)$, where $I_{Jup}$ is Jupiter's moment of inertia, following the evolutionary model of Leconte and Chabrier (2012, 2013). This model was scaled so that the radius of Jupiter at the age 4.5 Gyr is equal to its present value.

The left panel of Fig. 4 shows the evolution of the semi-major axes (top panel) and spin-orbit misalignments, or "obliquities" (bottom panel), of various hypothetical Earth-mass satellites around a Jupiter-mass planet for different initial semi-major axes. The orbital eccentricity $e_{ps}$ is zero and the moon is in synchronous rotation, so once the obliquity is zero only the planetary tide influences the evolution of the system. Not shown in these diagrams is the rotation of the Jupiter, which spins up due to contraction and which is modified by the angular momentum transfer from the satellite's orbit (Bolmont et al., 2011). Figure 4 shows that the corotation radius (ensemble of dashed blueish lines) – defined as the distance at which the satellite's orbital frequency equals the planetary rotation frequency – shrinks due to the spin-up of the planet. A satellite initially interior to the corotation radius migrates inward, because the bulge of the Jupiter is lagging behind the position of the satellite, and eventually falls onto the planet. A satellite initially exterior to the corotation radius migrates outward. Most notably, an Earth-mass satellite can undergo substantial tidal evolution even over timescales that span the age of present Jupiter. The bottom left panel of Fig. 4 shows the evolution of the obliquities of the satellites, whose initial value is set to 11.5° for all tracks. In all cases, the obliquities rapidly eroded to zero, in less than a year for a very close-in satellite and in a few $10^4$ yr for a satellite as far as $7 \times 10^{-3}$ AU. Thus, in single satellite systems, satellite obliquities are likely to be zero. The timescale of evolution of the satellite's rotation period is similar to the obliquity evolution timescale, so a satellite gets synchronized very quickly (for $e_{ps} \approx 0$) or pseudo-synchronized (if $e_{ps}$ is substantially non-zero).

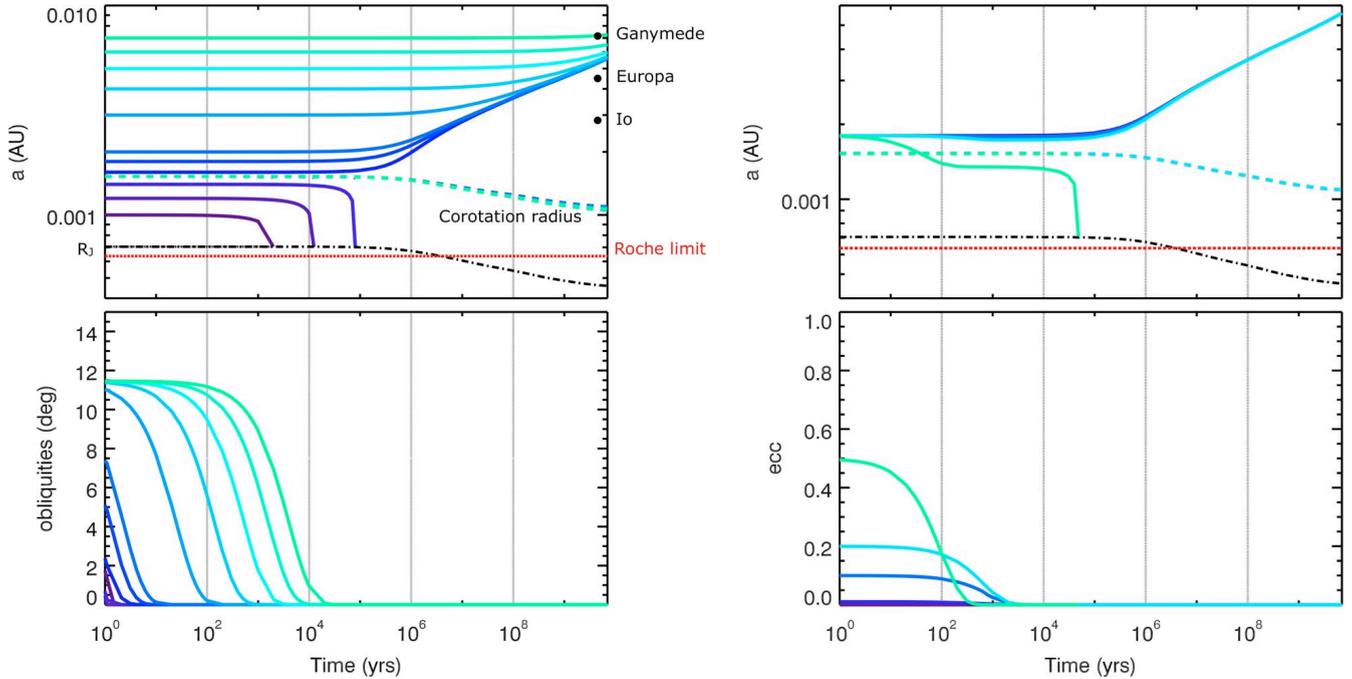

**Figure 4**: Tidal evolution of an Earth-sized moon orbiting a Jupiter-sized planet. Left panels: Semi-major axis (top) and obliquity (bottom) evolution for different initial semi-major axes, while all other initial parameters are equal. The black dashed dotted line in the top panel represents the planetary radius; four overlapping dashed lines indicate the corotation radii. A red dashed line represents the Roche limit. Right panels: Semi-major axis (top) and eccentricity (bottom) evolution for the same system but for different initial eccentricities.

The right panel of Fig. 4 shows the evolution of semi-major axes (top panel) and eccentricities (bottom panel) of the same Earth-Jupiter binaries as in the left panel, but now for different initial eccentricities. In the beginning of the evolution, the eccentricity is damped by both the planetary and the satellite tides in all simulations. What is more, all solid tracks start beyond their respective dashed corotation radius, so the satellites should move outward. However, the satellite corresponding to the light blue curve undergoes slight inward migration during the first $10^5$ years, which is due to the satellite tide that causes eccentricity damping. There is a competition between both the planetary and the satellite tides, and when the eccentricity is > 0.08 then the satellite tide determines the evolution causing inward migration. But if $e_{ps} < 0.08$, then the planetary tide dominates and drives outward migration (Bolmont et al., 2011). The planet with initial $e_{ps} = 0.5$ undergoes rapid inward migration such that it is interior to the corotation radius when its eccentricity approaches zero. Hence, it then transfers orbital angular momentum to the planet until it merges with its host.

### 4.2 Orbital evolution in multiple-satellite systems

We consider now a system of two satellites orbiting the same planet. One moon is a Mars-mass satellite in a close-in orbit at 5 $R_{Jup}$, the other one is an Earth-mass satellite orbiting farther out, between 10 and 50 $R_{Jup}$. All simulations include the structural evolution of the Jupiter-mass planet as in the previous subsection. We use a newly developed code, based on the *Mercury* code (Chambers, 1999), which takes into account tidal forces between both satellites and the planet as well as secular interaction. We also consider tides raised by each of the satellites on the planet, assuming these planetary bulges are independent, and the tide raised by the planet in each of the two satellites. The code consistently computes the orbital evolution as well as the evolution of the rotation period of the three bodies and their obliquities (Bolmont et al., 2014). The inner satellite is assumed to have an initial eccentricity of 0.05, an orbital inclination of 2° with respect to the planet's equatorial plane, and an obliquity of 11.5°. The outer satellite has an initial eccentricity of 0.1, an inclination of 5° to the planet's equator, and an obliquity of 7°. Tidal dissipation parameters for the satellite and the planet are assumed as above.

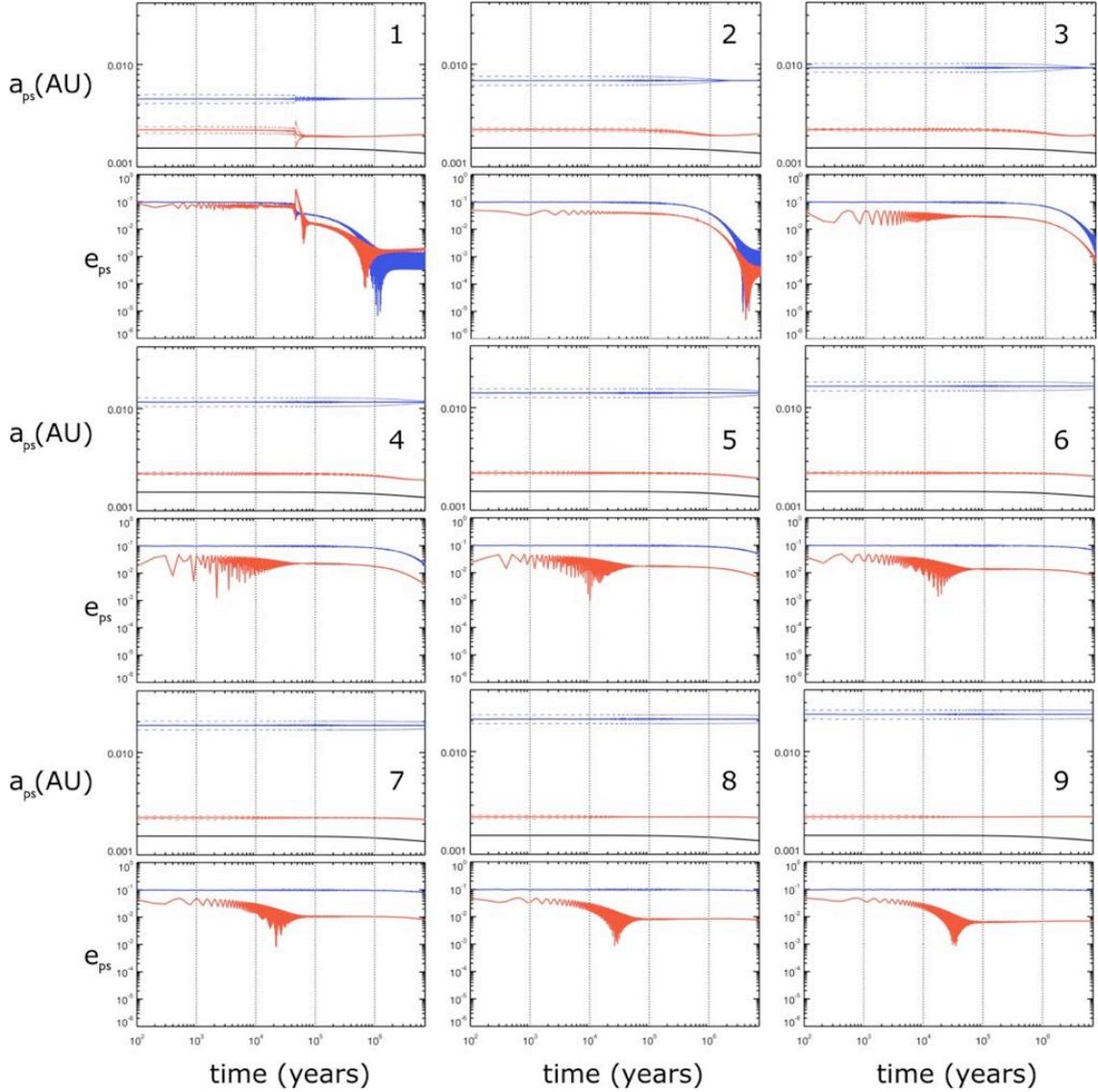

**Figure 5**: Evolution of the semi-major axes ($a_{ps}$) and eccentricities ($e_{ps}$) of the two satellites orbiting a Jupiter-mass planet. Red lines correspond to the inner Mars-mass satellite, blue lines to the outer Earth-mass satellite, and black lines represent the corotation distance. Each panel depicts a different initial distance for the outer satellite, increasing from panels 1 to 9.

  Figure 5 shows the results of these simulations, where each panel depicts a different initial semi-major axis of the outer moon. In all runs, the outer massive satellite excites the eccentricity of the inner moon. Excitations result in higher eccentricities if the outer satellite is closer in. Both eccentricities decrease at a similar pace, which is dictated by the distance of the outer and more massive satellite: the farther away the outer satellite, the more slowly the circularization. At the same time, the inner smaller satellite undergoes inward migration due to the satellite tide. When its eccentricity reaches values below a few $10^{-3}$, the planetary tide causes outward migration because the moon is beyond the corotation radius (panels 1 to 3). In panels 4 to 9, $e_{ps}$ of the outer moon decreases on timescales > 10 Myr, so the system does not reach a state in which the planetary tide determines the evolution to push the inner satellite outward.

  Panel 1 of Fig. 5 shows an instability between $10^4$ and $10^5$ yr. The inner satellite eccentricity is excited to values close to the outer satellite eccentricity. At a few $10^4$ yr, $e_{ps}$ of the inner satellite is excited to even higher values, but the satellite tide quickly damps it. This rapid circularization is accompanied by a substantial decrease of the inner moon's semi-major axis. Once the eccentricity reaches values of $10^{-2}$, evolution becomes smoother. Both eccentricities then decrease and dance around a kind of "equilibrium" at $e_{ps} \approx 10^{-3}$. On even longer timescales, both eccentricities are expected to decrease further towards zero (Mardling, 2007).

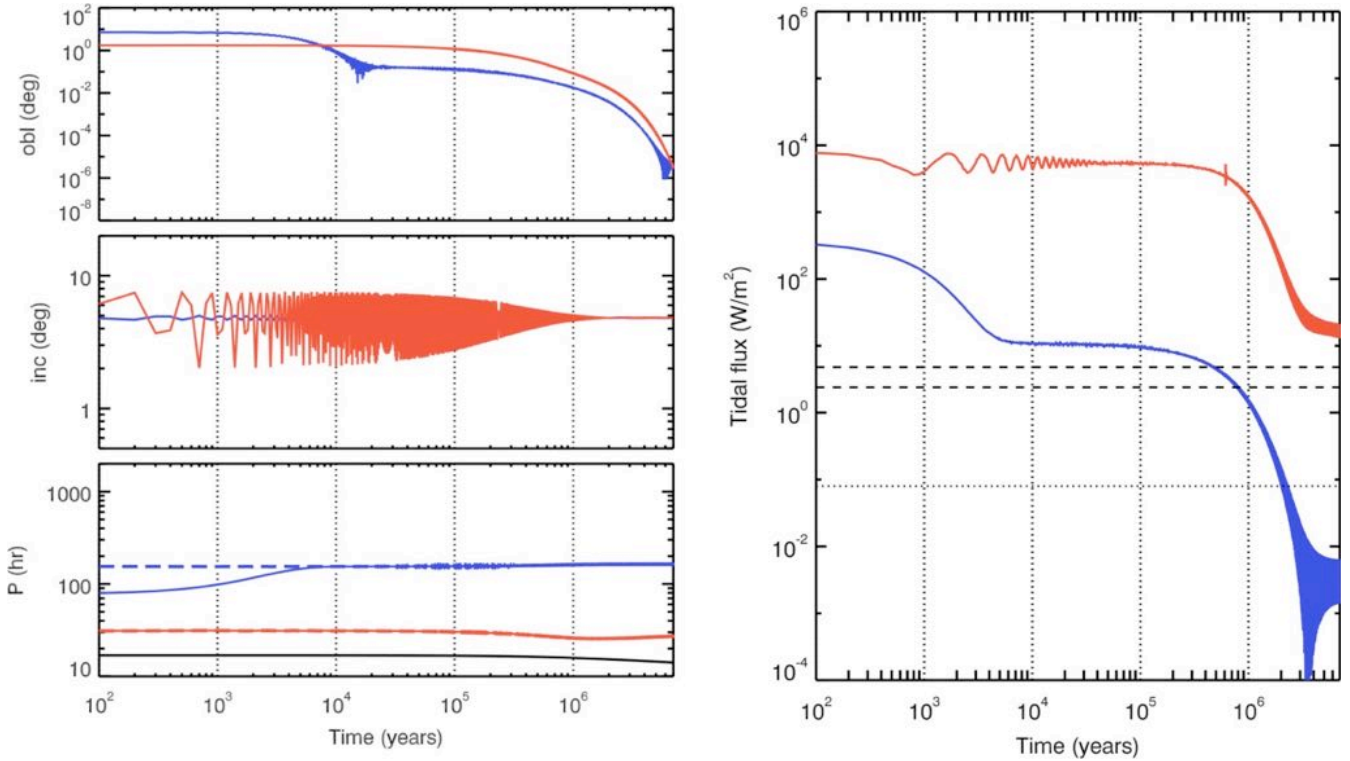

**Figure 6**: Evolution of the obliquity (top left), inclination (center left), rotation period (bottom left), and internal heat flux (right) of a Mars- (red lines) and an Earth-like (blue lines) satellite orbiting a Jupiter-mass planet (configuration of panel 2 in Fig. 9). In the bottom left panel, the black line represents the rotation period of the planet, and the dashed lines correspond to the pseudo-synchronization period of the satellites. In the right panel, the dotted black horizontal line corresponds to the internal heat flux of Earth (0.08 W/m$^2$), the dashed black lines correspond to tidal surface heating on Io (2.4 - 4.8 W/m$^2$).

Figure 6 shows the evolution of the obliquities, inclinations, rotation periods, and tidal surface heat flux for both satellites in the scenario of panel 2 of Fig. 5. The inner satellite reaches pseudo-synchronization in < 100 yr, during that time its obliquity decreases from initially 11.5° to about 2°. The outer satellite reaches pseudo-synchronization in a few 1000 yr, while its obliquity decreases to a few $10^{-1}$ degrees within some $10^4$ yr. Once pseudo-synchronization is reached on both satellites, their obliquities slowly decrease to very low values between $10^4$ and $10^7$ yr in Fig. 6. The inclination of the small inner satellite oscillates around the value of the more massive outer one, whose inclination does not vary significantly during the first $10^7$ yr because the planetary tides are not efficient enough to adjust the orbital planes to the planet's equatorial plane.

Moving on to the tidal heat flux in the right panel of Fig. 6, note that the internal heat flux through the Earth's surface is around 0.08 W/m$^2$ (Pollack et al., 1993) and Io's value is between 2.4 and 4.8 W/m$^2$ (Spencer et al., 2000). Initially, the tidal heat flux is extremely large in both satellites. While being pseudo-synchronized during the first $\approx 10^4$ yr, the tidal heat flux of the outer satellite decreases from some $10^2$ to about 10 W/m$^2$. After $10^4$ yr, the decay of tidal heating in both satellites is due to the circularization and tilt erosion (Heller et al., 2011b) on a 10 Myr timescale. The tidal surface heat flux of the outer satellite is about an order of magnitude lower than Earth's internal heat flux through its surface, but the tidal heat going through a surface unit on the inner moon is several times the tidal surface heat flux on Io, probably causing our Mars-mass test moon to show strong tectonic activity for as long as its first 100 Myr after formation.

The long-term evolution of this system might lead to interesting configurations. For example, the outward migration of the inner satellite decreases the distance between the two satellites, eventually resulting in an orbital resonance (Bolmont et al., 2014). In such a resonance, the eccentricities of both satellites would be excited and thereby reignite substantial tidal heating. Not only would such a resurgence affect the satellites' potentials to be, remain, or become habitable, but their thermal emission could even prevail against the host planet's thermal emission and make the moon system detectable by near-future technology (Section 6.4).

## 5. Effects of Orbital and Planetary Evolution on Exomoon Habitability

Predictions of exomoons the size of Mars orbiting giant planets (Section 3) and the possible detection of exomoons roughly that size with the *Kepler* space telescope or near-future devices (Section 6) naturally make us wonder about the habitability of these worlds. Tachinami et al. (2011) argued that a terrestrial world needs a mass ≥ 0.1 $M_\oplus$ to sustain a magnetic shield on a billion-year timescale, which is necessary to protect life on the surface from high-energy stellar and interstellar radiation. Further constraints come from the necessity to hold a substantial, long-lived atmosphere, which requires satellite masses $M_s$ ≥ 0.12 $M_\oplus$ (Williams et al., 1997; Kaltenegger, 2000). Tectonic activity over billions of years, which is mandatory to entertain plate tectonics and to promote the carbon-silicate cycle, requires $M_s$ ≥ 0.23 $M_\oplus$ (Williams et al., 1997). Hence, formation theory, current technology, and constraints from terrestrial world habitability point towards a preferred mass regime for habitable moons that can be detected with current or near-future technology, which is between 0.1 and 0.5 $M_\oplus$.

### 5.1 The global energy flux on moons

While these mass limits for a habitable world were all derived from assuming a cooling, terrestrial body such as Mars, exomoons have an alternative internal heat source that can retard their cooling and thus maintain the afore-mentioned processes over longer epochs than in planets. This energy source is tidal heating, and its effects on the habitability of exomoons have been addressed several times in the literature (Reynolds et al, 1987; Scharf, 2006; Henning et al, 2009; Heller, 2012; Heller and Barnes, 2013, 2014; Heller and Zuluaga, 2013; Heller and Armstrong, 2014). In their pioneering study, Reynolds et al. (1987) point out the remarkable possibility that water-rich extrasolar moons beyond the stellar HZ could be maintained habitable mainly due to tidal heating rather than stellar illumination. Their claim was supported by findings of plankton in Antarctica lakes, which require an amount of solar illumination that corresponds to the flux received at the orbit of Neptune. Heller and Armstrong (2014) advocated the idea that different tidal heating rates allow exomoons to be habitable in different circumplanetary orbits, depending on the actual distance of the planet-moon system from their common host star.

Assuming an exomoon were discovered around a giant planet in the stellar HZ, then a first step towards assessing its habitability would be to estimate its global energy flux $\bar{F}_s^{\rm glob}$. Were the moon to orbit its planet too closely, then it might be undergoing a moist or runaway greenhouse effect (Kasting, 1988; Kasting et al., 1993) and be temporarily uninhabitable or even desiccated and uninhabitable forever (Heller and Barnes, 2013, 2014). In addition to the orbit-averaged flux absorbed from the star ($\bar{f}_\star$), the moon absorbs the star's reflected light from the planet ($\bar{f}_r$), the planet's thermal energy flux ($\bar{f}_t$), and it can undergo substantial tidal heating ($h_s$, the amount of tidal heating going through a unit area on the satellite surface). The total global average flux at the top of a moon's surface is then given by

$$
\begin{aligned}
\bar{F}_s^{\rm glob} &= \bar{f}_\star \;+\; \bar{f}_r \;+\; \bar{f}_t \;+\; h_s \;+\; W_s \\
&= \frac{L_\star\,(1-\alpha_{\rm s,opt})}{4\pi a_{\rm *p}^2 \sqrt{1-e_{\rm *p}^2}}\left(\frac{x_s}{4}+\frac{\pi R_p^2 \alpha_p}{f_s\,2\,a_{\rm ps}^2}\right) \;+\; \frac{L_p(1-\alpha_{\rm s,IR})}{4\pi a_{\rm ps}^2 f_s \sqrt{1-e_{\rm ps}^2}} \;+\; h_s \;+\; W_s \\
&= \frac{R_\star \sigma_{\rm SB}^2 T_{\rm eff,\star}^4\,(1-\alpha_{\rm s,opt})}{a_{\rm *p}^2 \sqrt{1-e_{\rm *p}^2}}\left(\frac{x_s}{4}+\frac{\pi R_p^2 \alpha_p}{f_s\,2\,a_{\rm ps}^2}\right) \;+\; \frac{R_p^2 \sigma_{\rm SB} T_{\rm p,eff}^4 (1-\alpha_{\rm s,IR})}{a_{\rm ps}^2 f_s \sqrt{1-e_{\rm ps}^2}} \;+\; h_s \;+\; W_s
\end{aligned}
\tag{4}
$$

where $L_\star$ and $L_p$ are the stellar and planetary luminosities, respectively, $\alpha_{\rm s,opt}$ and $\alpha_{\rm s,IR}$ the satellite's optical and infrared albedo, respectively (Heller and Barnes, 2014), $\alpha_p$ is the planetary bond albedo, $a_{\rm *p}$ and $a_{\rm p,s}$ are the star-planet[19] and planet-satellite semi-major axes, respectively, $e_{\rm *,p}$ and $e_{\rm p,s}$ the star-planet and planet-satellite orbital eccentricities, respectively, $R_p$ is the planetary radius, $x_s$ the fraction of the satellite orbit that is *not* spent in the planetary shadow (Heller, 2012), $\sigma_{\rm SB}$ the Stefan-Boltzmann constant, $T_{\rm p,eff}$ is the planet's effective temperature, $h_s$ is the tidal surface heating of the satellite, and $W_s$ denotes contributions from other global heat sources such as primordial thermal energy (or "sensible heat"), radioactive decay, and latent heat from solidification. The planet's effective temperature $T_{\rm eff}$ is a function of long timescales as young, hot giant planets cool over millions of years; and it depends on the planetary surface temperature $T_{\rm p,eq}$, triggered by thermal equilibrium between absorbed and re-emitted stellar light, as well as on the internal heating contributing a surface component $T_{\rm p,int}$. Hence, $T_{\rm p,eff} = ([T_{\rm p,eq}]^4 + [T_{\rm p,int}]^4)^{1/4}$ (Heller and Zuluaga, 2013). The factor $f_s$ in the denominator of the terms describing the reflected and thermal irradiation from the planet accounts for the efficiency of the flux distribution over the satellite surface. If the planet rotates freely with respect to the planet, then $f = 4$, and if the satellite is tidally locked with one hemisphere pointing at the planet permanently, then $f = 2$ (Selsis et al, 2007). Another factor of 2 in the denominator of the term related to the stellar light reflected from the planet onto the satellite accounts for the fact that only half of the planet is actually starlit. The split-up of absorbed stellar and planetary illumination in Eq. (4), moderated by the different optical and infrared albedos, has been suggested by Heller and Barnes (2014) and is owed to the different spectral regimes, in which the energy is emitted from the sources and absorbed by the moon.

---

[19] It is assumed that the planet is much more massive than its moons and that the barycenter of the planet and its satellite system is close to the planetary center.

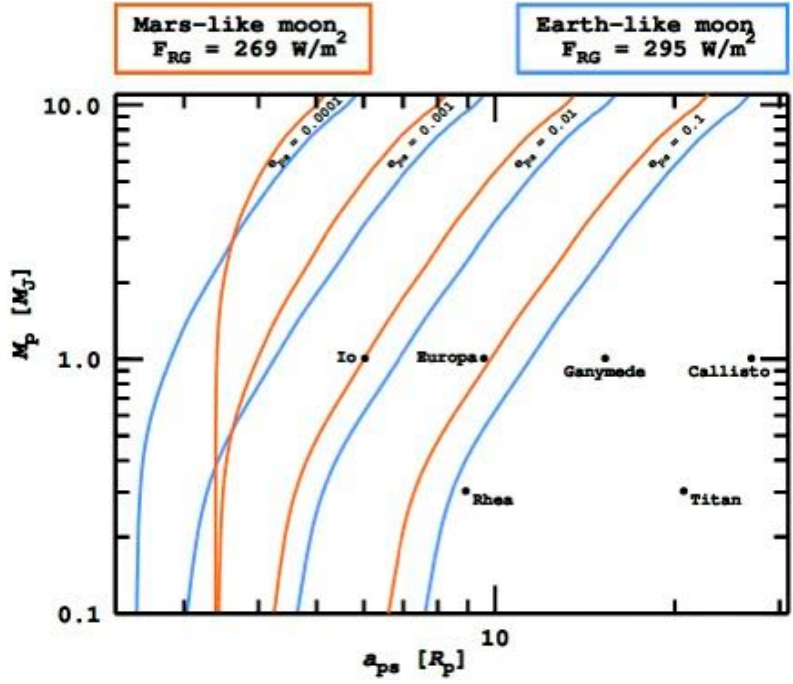

**Figure 7**: Circumplanetary habitable edges (HEs) for a Mars-mass (orange lines) and an Earth-mass (blue lines) exomoon orbiting a range of host planets (masses indicated along the ordinate). All system are assumed to be at 1 AU from a Sun-like star. HEs are indicated for four different orbital eccentricities: $e_{ps} \in \{0.0001, 0.001, 0.01, 0.1\}$. The larger $e_{ps}$, the further away the moons need to be around a given planet to avoid transition into a runaway greenhouse due to extensive tidal heating. Examples for the orbital distances found in the Jovian and Saturnian satellite systems are indicated with labeled dots.

A progenitor version of this model has been applied to examine the maximum variations in illumination that moons in wide circumplanetary orbits can undergo, indicating that these fluctuations can be up to several tens of W/m² (Hinkel and Kane, 2013). Taking into account stellar illumination and tidal heating, Forgan and Kipping (2013) applied a one-dimensional atmosphere model to assess the redistribution of heat in the Earth-like atmospheres of Earth-like exomoons. In accordance with the results of Hinkel and Kane, they found that global mean temperatures would vary by a few Kelvin at most during a circumplanetary orbit. While global flux or temperature variations might be small on moons with substantial atmospheres, the local distribution of stellar and planetary light on a moon can vary dramatically due to eclipses and the moon's tidal locking (Heller, 2012; Heller and Barnes, 2013; Forgan and Yotov, 2014).

Equation (4) can be used to assess an exomoon's potential to maintain liquid reservoirs of water, if $\bar{F}_s^{\mathrm{glob}}$ is compared to the critical flux $F_{\mathrm{RG}}$ for a water-rich world with an Earth-like atmosphere to enter a runaway greenhouse state. We use Eq. (8) of Fortney et al. (2007b) and an Earth-like rock-to-mass fraction of 68% to calculate the two satellites' radii $R_s = 0.5\ R_\oplus$ and $1\ R_\oplus$, respectively, where $R_\oplus$ symbolizes the radius of Earth. Then, using the semi-analytic atmosphere model of Pierrehumbert (2010, see also Eq. 1 in Heller and Barnes, 2013), we compute $F_{\mathrm{RG}} = 269$ W/m² and 295 W/m² for a Mars-mass and and Earth-mass exomoon, respectively. The closer an exomoon is to its host planet, the stronger tidal heating and the stronger illumination from the planet. Ultimately, there exists a minimum circumplanetary orbit at which the moon transitions into a runaway greenhouse state, thereby becoming uninhabitable. This critical orbit has been termed the circumplanetary "habitable edge", or HE for short (Heller and Barnes, 2013). Giant planets in the stellar HZ will not have an outer habitable edge for their moons, because moons in wide orbits will essentially behave like free, terrestrial planets. Beyond about 15 $R_{\mathrm{jup}}$, tidal heating will be weak or negligible; illumination from the planet will be very low; eclipses will be infrequent and short compared to the moon's circumplanetary orbital period; and magnetic effects from the planet will be weak. Thus, by definition of the stellar HZ, exomoons could be habitable even at the planetary Hill radius.

As an application of Eq. (4), we assume a planet-moon binary at 1 AU from a Sun-like star. Following Heller and Barnes

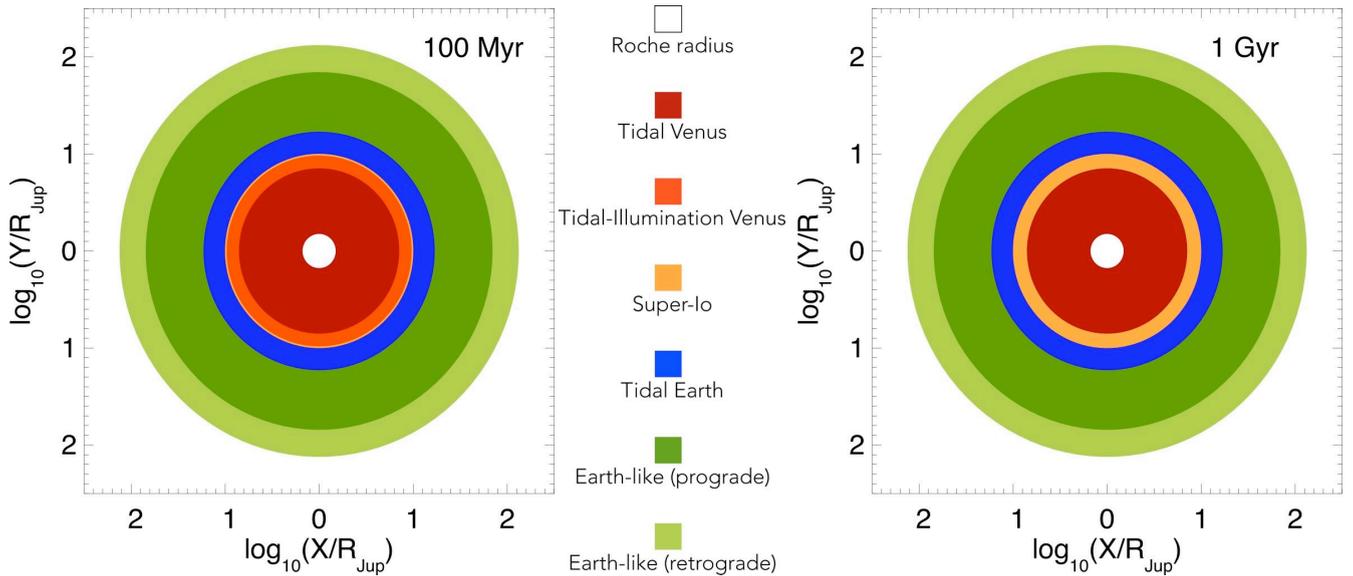

**Figure 8**: Circumplanetary exomoon menageries for Mars-sized satellites around a 10 $M_{Jup}$ host planet at ages of 100 Myr (left panel) and 1 Gyr (right panel). The planet is assumed to orbit in the middle of the HZ of a 0.7 $M_\odot$ star, and the moon orbits the planet with an eccentricity of $10^{-3}$. In each panel, the planet's position is at (0,0), and distances are shown on logarithmic scales. Note that exomoons in a Tidal Venus or a Tidal-Illumination Venus state are in a runaway greenhouse state and thereby uninhabitable.

(2014), we assume that the exomoon's albedo to starlight is $\alpha_{s,opt}$ = 0.3, while that due to the planet's thermal emission is $\alpha_{s,IR}$ = 0.05. This difference is due to the expectation that the planet will emit primarily in the infrared, where both Rayleigh scattering is less effective and molecular absorption bands are more prominent, decreasing the albedo (Kopparapu et al., 2013). Figure 7 shows the HEs for the Mars- and Earth-like satellites described above (orange and blue lines, respectively), assuming four different orbital eccentricities, $e_{ps} \in \{0.0001, 0.001, 0.01, 0.1\}$ from left to right. Along the abscissa, semi-major axis $a_{ps}$ is given in units of planetary radii $R_p$, which depends on planetary mass $M_p$, which is depicted along the ordinate. To intertwine abscissa and ordinate with one another, we fit a polynomial to values predicted for giant planets with a core mass of 10 $M_\oplus$ at 1 AU from a Sun-like star (Table 4 in Fortney et al., 2007a). Examples for Solar System moons in this mass-distance diagram are indicated with black dots.

For $e_{ps} \gtrsim 0.01$, we find the HEs of the Mars-like satellite are up to a few planetary radii closer to the planet than the HEs for the Earth-like moon. This is due to the strong dependence of tidal heating on the satellite radius, $h_s \propto R_s^3$. Thus, although the Earth-like moon is more resistant against a runaway greenhouse state than the Mars-like satellite, in the sense that 295 W/m$^2$ > 269 W/m$^2$, it is more susceptible to enhanced tidal heating in a given orbit. We also see a trend of both suites of HEs moving outward from the planet as $M_p$ increases. This is because, firstly, $h_s \propto M_p^3$ for $M_p \gg M_s$ and, secondly, we measure the abscissa in units of planetary radii and radii of super-Jovian planets remain approximately constant around $R_{Jup}$. For $e_{ps} \lesssim$ 0.01, we find that the relative positions of the HEs corresponding to the Mars- and Earth-like satellite strongly depend on $M_p$. Again, this is owed to the strong dependence of $h_s$ on $M_p$ and explains why the more resistive Earth-like moon, in terms of a transition to a runaway greenhouse state, can be closer to less massive host planets than the Mars-like moon: in low-eccentricity orbits around low-mass planets, tidal heating is negligible and the HEs depend mostly on additional illumination from the planet.

Orbital eccentricities of exomoons will typically be small due to tidal circularization of their orbits (Porter and Grundy, 2011; Heller and Barnes, 2013). And given that the highest eccentricity among the most massive satellites in the Solar System is 0.0288 for Titan, Fig. 7 suggests that exomoons the mass of Mars or Earth orbiting a giant planet at about 1 AU around a Sun-like star can be habitable if their orbital semi-major axis around the planet-moon barycenter ≳ 10 $R_p$, with closer orbits still being habitable for lower eccentricities and lower-mass host planets.

*5.2 Runaway greenhouse due to "inplanation"*

In this section, we examine the evolution of incident planetary radiation ("inplanation") on the habitability of exomoons, encapsulated in $L_p$ in Eq. (4). Following the prescriptions of Heller and Barnes (2014), we approximate Eq. (4) by considering stellar light, the evolution of inplanation due to the radial shrinking of young gaseous giant planets (Baraffe et al., 2003;

Leconte and Chabrier, 2013), and assuming constant tidal heating. While Heller and Barnes (2014) considered several configurations of star-planet-moon systems, we limit our example to that of a Mars-sized moon (see Section 5.1) around a 10 $M_{\mathrm{Jup}}$ planet that orbits a K dwarf in the middle of the HZ. Further, we arbitrarily consider a moon with $e_{\mathrm{ps}} = 10^{-3}$, a small value that is similar to those of the Galilean satellites. Heller and Barnes (2014) found that Earth-sized moons orbiting planets less massive than 5 $M_{\mathrm{Jup}}$ are safe from desiccation by inplanation, so our example exomoon may be at risk of losing its water from inplanation.

As in the previous subsection, our test exomoon may be considered habitable if the total flux is < 269 W/m². What is more, the duration of the runaway greenhouse is important. Barnes et al. (2013) argued that ≈ 100 Myr are sufficient to remove the surface water of a terrestrial body, based on the pioneering work of Watson et al. (1981). While it is possible to recover habitability after a runaway greenhouse, we are aware of no research that has examined how a planet evolves as energy sources drop below the limit. However, we conjecture that the flux must drop well below the runaway greenhouse limit to sufficiently weaken a $CO_2$ greenhouse to permit stable surface water.

In Fig. 8, we classify exomoons with different surface fluxes as a function of semi-major axis of the planet-moon orbit. We assume satellites of Earth-like properties in terms of overall tidal response, and that 90% of the energy is dissipated in a putative ocean, with the remainder in the solid interior. For this example, we use the "constant-phase-lag" tidal model, as described by Heller et al. (2011a), and assume a tidal quality factor $Q_s$ of 10, along with $k_{2,s} = 0.3$. We classify planets according to the scheme presented by Barnes and Heller (2013) and Heller and Barnes (2014), which the authors refer to as the "exomoon menagerie". If tidal heating alone is strong enough to reach the runaway greenhouse limit, the moon is a "Tidal Venus" (Barnes et al., 2013), and the orbit is colored red. If the tidal heating flux is less, but the total flux is still sufficient to trigger the runaway greenhouse, the moon is a "Tidal-Illumination Venus," and the orbit is orange. These two types of moons are uninhabitable. If the surface flux from the rocky interior is between the runaway limit and that observed on Io (≈ 2 W/m²; Veeder et al., 1994; Spencer et al., 2000), then we label the moon a "Super-Io" (see Jackson et al., 2008; Barnes et al., 2009a, 2009b), and the orbit is yellow. If the tidal heating of the interior is less than Io's but larger than a theoretical limit for tectonic activity on Mars (0.04 W/m²; Williams et al., 1997), then we label the moon a "Tidal Earth," and the orbit is blue. For lower tidal heat fluxes, the moon is considered Earth-like, and the orbit is green. The outer rim of this sphere containing Earth-like moons is determined by stability criteria found by Domingos et al. (2006). Satellites orbiting their planet in the same direction as the planet orbits the star, that is, in a prograde sense, are bound to the planet as far as about 0.49 $R_{\mathrm{Hill,p}}$. Moons orbiting their planet in the opposite direction, in a retrograde sense, can follow stable orbits out to 0.93 $R_{\mathrm{Hill,p}}$, depending on eccentricities.

The left panel of Fig. 8 shows these classes for a 100 Myr old planet, the right after 1 Gyr. As expected, all boundaries are the same except for the Tidal-Illumination Venus limit, which has moved considerably inward (note the logarithmic scale!). This shrinkage is due to the decrease in inplanation, as all other heat sources are constant[20]. In this case, moons at distances ≲ 10 $R_{\mathrm{Jup}}$ are in a runaway greenhouse for at least 100 Myr, assuming they formed contemporaneously with the planet, and thus are perilously close to permanent desiccation. Larger values of planet-moon eccentricity, moon mass, and/or moon radius increase the threat of sterilization. Thus, should our prototype moon be discovered with $a_{\mathrm{ps}} \lesssim 10$ $R_{\mathrm{Jup}}$, then it may be uninhabitable due to an initial epoch of high inplanation.

### 5.3 Magnetic environments of exomoons

Beyond irradiation and tidal heating, the magnetic environments of moons determine their surface conditions. It is now widely recognized that magnetic fields play a role in the habitability of exoplanetary environments (Lammer et al., 2010). Strong intrinsic magnetic fields can serve as an effective shield against harmful effects of cosmic rays and stellar high energy particles (Grießmeier et al., 2005). Most importantly, an intrinsic magnetic shield can help to protect the atmosphere of a terrestrial planet or moon against non-thermal atmospheric mass loss, which can obliterate or desiccate a planetary atmosphere, see Venus and Mars (Zuluaga et al., 2013). To evaluate an exomoon's habitability, assessing their magnetic and plasmatic environments is thus crucial (Heller and Zuluaga, 2013).

To determine the interaction of a moon with the magnetic and plasmatic environment of its host planet, we need to estimate the size and shape of the planetary magnetosphere. Magnetospheres are cavities within the stellar wind, created by the intrinsic magnetic field of the planet (Fig. 9). The scale of a magnetosphere is given by the distance between the planetary center and the planetary magnetopause. This so-called "standoff radius" $R_S$ has been measured for giant planets in the Solar System (see Table 1). For extrasolar planets, however, $R_S$ can only be estimated from theoretical and semi-empirical models, which predict that $R_S$ scales with the dynamic pressure of the stellar wind $P_{\mathrm{sw}} \propto n_{\mathrm{sw}} v_{\mathrm{sw}}^2$, with $n_{\mathrm{sw}}$ as the particle density and $v_{\mathrm{sw}}$ as the particle velocity, and with the planetary dipole moment $\mathcal{M}$ as

$$R_S \; \propto \; \mathcal{M}^{1/3} \; \times \; P_{\mathrm{sw}}^{-1/\alpha_{\mathrm{mag}}} \qquad . \tag{5}$$

---

[20] Without external perturbations, tidal heating should also dissipate as the eccentricity is damped, but here we do not consider orbital evolution.

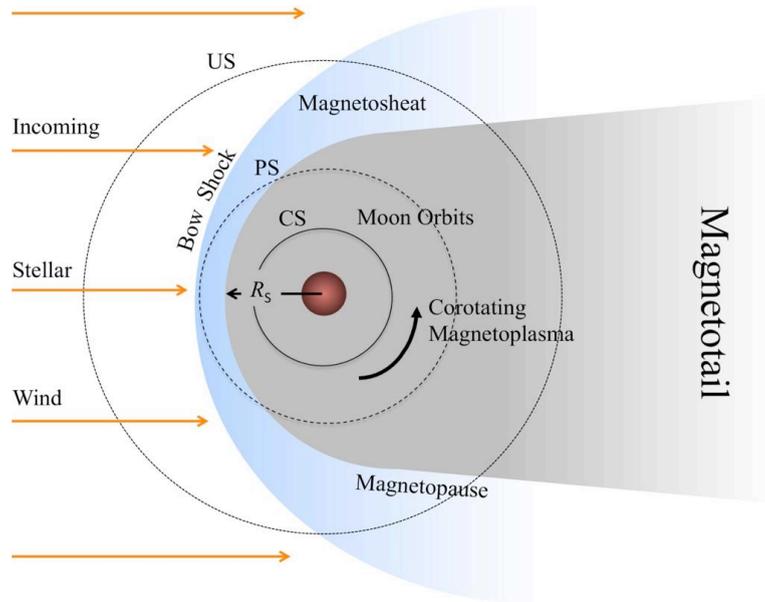

**Figure 9**: Schematic representation of a planetary magnetosphere. Its roughly spherical dayside region has a standoff radius $R_S$ to the planetary center. Between the bow shock and the magnetopause lies the magnetosheat, a region of shocked stellar wind plasma and piled up interplanetary magnetic fields. Inside the magnetosphere, plasma is dragged by the co-rotating magnetic field, thereby creating a particle stream called the "co-rotating magnetoplasma". Moons at orbital distances < $R_S$ (completely shielded, CS) can be subject to this magnetospheric wind and, hence, experience similar effects as unshielded (US) moons exposed to stellar wind. Partially shielded (PS) moons that spend most of their orbital path inside the magnetosphere will experience both effects of interplanetary magnetic fields and stellar wind periodically.

Here, $\alpha_{\mathrm{mag}}$ depends on the magnetospheric compressibility, that is, the contribution to internal stress provided by magnetospheric plasma. Values for $\alpha_{\mathrm{mag}}$ range from 6 in the case of relatively empty magnetospheres such as that of Earth, Uranus, and Neptune (Zuluaga et al., 2013; Heller and Zuluaga, 2013), to 4 if important magnetospheric plasma sources are present as is the case for Jupiter and Saturn (Huddleston et al., 1998; Arridge et al., 2006). Predictions of $R_S$ for the Solar System giant planets are listed in Table 1, together their actually observed dipole moments and present values of the solar wind properties.

As a giant planet ages, its internal heat source recedes and thus the planet's internal dynamo as well as its magnetic field weaken[21]. Yet, the stellar wind also weakens as the stellar activity decreases. The combination of both effects induces a net expansion of the magnetosphere over billions of years. Figure 10 shows the evolution of $R_S$ for a Jupiter-analog in the center of the Sun's HZ, following methods presented by Heller and Zuluaga (2013): $M_p = M_{\mathrm{Jup}}$, the planetary core mass $M_{\mathrm{core}} = 10$ $M_\oplus$, and the planet's rotation $P_{\mathrm{rot}} = 10$ h. Magnetospheres of HZ giant planets could be very compressed during the first billion years, thereby leaving their moons unshielded. As an example, compare the blue spiral in Figure 10, which denotes $R_S$, with the orbits of the Galilean moons. While an Io-analog satellite would be coated by the planetary magnetic field as early as 500 Myr after formation, a moon in a Europa-wide orbit would be protected after about 2 Gyr, and a satellite in a Ganymede-wide orbit would only be shielded after 3.5 Gyr. This evolutionary effect of an exomoon's magnetic environment can thus impose considerable constraints on its habitability. Heller and Zuluaga (2013), while focussing on moons around giant planets in the HZ of K stars, concluded that exomoons beyond 20 planetary radii from their host will hardly ever be shielded by their planets' magnetospheres. Moons in orbits between 5 and 20 $R_p$ could be shielded after several hundred Myr or some Gyr, and they could also be habitable from an illumination and tidal point of view. Moons closer to their planets than 5 $R_p$ could be shielded during the most hazardous phase of the stellar wind, that is, during the first ≈ 100 Myr, but they would likely not be habitable in the first place, due to enormous tidal heating and strong planetary illumination (Heller and Barnes, 2014).

---

[21] The evolution of the magnetic field strength could be much more complex if the interplay between the energy available energy for the dynamo and the Coriolis forces are taken into account (Zuluaga and Cuartas, 2012).

| Planet | $\mathcal{M}$ ($\mathcal{M}_\oplus$) | $R_S^{\text{obs}}$ ($R_p$) | $\bar{R}_S^{\text{th}}$ ($R_p$) | $\bar{R}_S^{\text{th,HZ}}$ ($R_p$) | Moons | $a_{\text{ps}}$ ($R_p$) | Shielding | |
|--------|------|-------|-------|-----------|-------|-------|--------|-----|
| | | | | | | | **Actual** | **HZ** |
| Jupiter | 1800 | $45-80$ | $38-48$ | $22-24$ | Io | 5.9 | CS | CS |
| | | | | | Europa | 9.4 | CS | CS |
| | | | | | Ganymede | 15.0 | CS | CS |
| | | | | | Callisto | 26.3 | CS | **PS** |
| Saturn | 580 | $17-30$ | $17-23$ | $8.5-9.0$ | Enceladus | 4.0 | CS | CS |
| | | | | | Titan | 20.3 | **PS** | **US** |
| Uranus | 50 | 18 | $23-33$ | $8.9-9.3$ | Titania | 17 | CS | **US** |
| | | | | | Oberon | 22.8 | **PS** | **US** |
| Neptune | 24 | $23-26$ | $21-32$ | $7.2-7.6$ | Triton | 14 | CS | **US** |

**Table 1**: Measured magnetic properties of giant planets in the Solar System. $\mathcal{M}$: Magnetic dipole moment (Bagenal, 1992; Guillot, 2005). $R_S^{\text{obs}}$: observed range of magnetospheric standoff radii (Arridge et al., 2006; Huddleston et al., 1998). Note that standoff distances vary with solar activity. $R_S^{\text{th}}$: range of predicted average standoff distances. Predicted standoff distances depend on magnetospheric compressibility (see text). $R_S^{\text{th,HZ}}$: extrapolation of $R_S^{\text{th}}$ assuming that the planet would be located at 1 AU from the Sun. Semi-major axes $a_{\text{ps}}$ of selected major moons are shown for comparison. The final two columns depict the shielding status (see Fig. 13) of the moons for their actual position in the Solar System and assuming a distance of 1 AU to the Sun.

### 5.3.1 Unshielded exomoons

The fraction of a moon's orbit spent inside the planet's magnetospheric cavity defines three different shielding conditions: (i.) unshielded (US), (ii.) partially shielded (PS), and (iii.) completely shielded (CS) (Fig. 9). Depending on its membership to any of these three classes, different phenomena will affect an exomoon, some of which will be conducive to habitability and others of which will threaten life.

Assuming that the dayside magnetosphere is approximately spherical and that the magnetotail is cylindrical, a moon orbiting its host planet at a distance of about 2 $R_S$ will spend more than 50% of its orbit outside the planetary magnetosphere. To develop habitable surface conditions, such a partially shielded exomoon would need to have an intrinsic magnetic field similar in strength to that required by a terrestrial planet at the same stellar distance (Zuluaga et al., 2013; Vidotto et al., 2013). For moons in the HZ of Sun-like stars, intrinsic magnetic dipole moments larger than about that of Earth would allow for habitable surface conditions. Moons near the HZs around less massive stars, however, which show enhanced magnetic activity, need intrinsic dipole moments that are several times larger to prevent an initial satellite atmosphere from being exposed to the strong stellar wind (Vidotto et al., 2013). As formation models predict that moons can hardly be as massive as Earth (see Section 3), the maximum magnetic field attainable by an hypothetical exomoon may be insufficient to prevent atmospheric erosion (Williams et al., 1997). Perhaps moons with alternative internal heat sources could drive a long-lived, sufficiently strong internal dynamo.

Ganymede may serve as an example to illustrate this peril. It is the only moon in the Solar System with a strong intrinsic magnetic field (Ness et al., 1979). With a dipole moment about $2 \times 10^{-3}$ times that of Earth, a Ganymede-like, unshielded moon (see "US" zone in Fig. 9) orbiting a giant planet in the HZ of a Sun-like star would only have a standoff radius of about 1.5 Ganymede radii at an age of 1 Gyr. This would expose a potentially thin and extended atmosphere of such a moon to the high-energy particles from the stellar wind, eventually stripping off the whole atmosphere.

### 5.3.2 Partially shielded exomoons

Inside the planetary magnetosphere, the main threat to a partially shielded exomoon is that of the so-called co-rotating magnetoplasma (Neubauer et al., 1984). Magnetospheric plasma, which is dragged by the planetary magnetic field to obtain the same rotational angular velocity as the planet, will produce a "magnetospheric wind". Around gas giants akin to those in the Solar System, the circumplanetary angular velocity of the co-rotating plasma is typically much higher than the Keplerian velocity of the moons. Hence, the magnetospheric wind will hit exomoon atmospheres and may erode them in orbits as wide as a significant fraction of the standoff distance. This effect could be as intense as effects of the direct stellar wind. For instance, it has been estimated that Titan, which is partially shielded by Saturn's magnetosphere, could have lost 10% of its present atmospheric content just because of charged particles from the planet's co-rotating plasma (Lammer and Bauer, 1993).

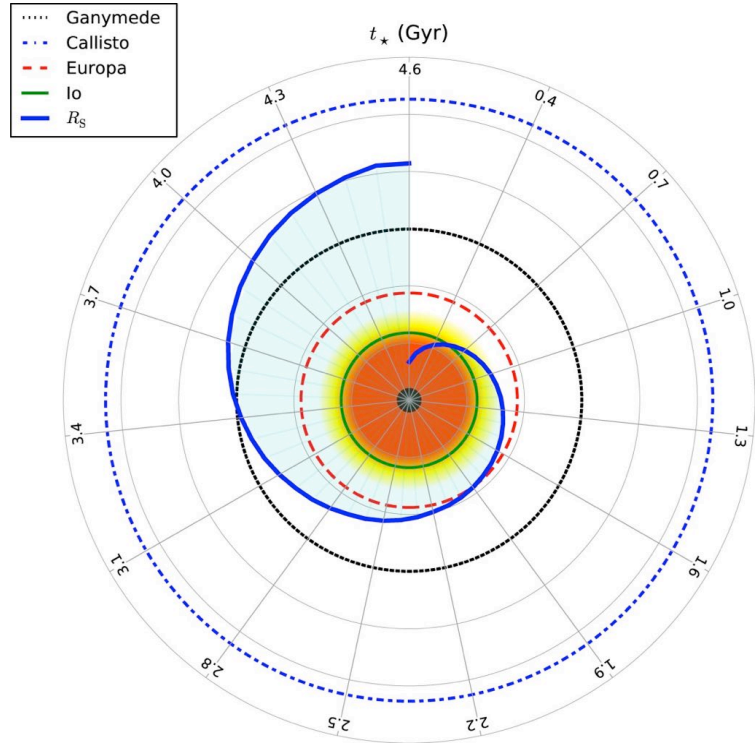

**Figure 10**: Evolution of the standoff radius $R_S$ (thick blue line) around a Jupiter-like planet in the center of the stellar HZ of a Sun-like star. For comparison, the orbits of the Galilean moons are shown (see legend). Angles with respect the vertical line encode time in Gyr. Time starts at the "12:00" position in 100 Myr and advances in steps of 0.3 Gyr up to the present age of the Solar System. Thin gray circles denote distances in intervals of 5 planetary radii. The filled circle in the center denotes the planetary radius.

Towards the inner edge of the HZ around low-mass stars, the stellar wind flux is extremely strong. What is more, the planetary magnetosphere will be small, thereby increasing the density of the magnetospheric wind. Other sources of plasma such as stellar wind particles, planetary ionosphere gasses, and ions stripped off from other moons could also be greatly enhanced.

*5.3.3 Completely shielded exomoons*

Moons that are completely shielded by the planetary magnetic shield are neither exposed to stellar wind nor to significant amounts of cosmic high-energy particles. What is more, being well inside the magnetosphere, the co-rotating plasma will flow at velocities comparable to the orbital velocity of the moon. Hence, no strong interactions with the satellite atmosphere are expected.

On the downside, planetary magnetic fields can trap stellar energetic particles as well as cosmic rays with extremely high energies. Around Saturn and Jupiter, electrons, protons, and heavy ions with energies up to hundreds of GeV populate the inner planetary magnetospheres within about ten planetary radii, thereby creating what is known as radiation belts. Around Jupiter, as an example, the flux of multi-MeV electrons and protons at a distance of about 15 $R_{Jup}$ to the planet is between $10^6$ and $10^8$ cm$^{-2}$ s$^{-1}$ (Divine and Garrett, 1983), that is four to six orders of magnitude larger than that of solar high-energy particles received by Earth. Exposed to these levels of ionizing radiation, moon surfaces could absorb about 1 to 10 J kg$^{-1}$

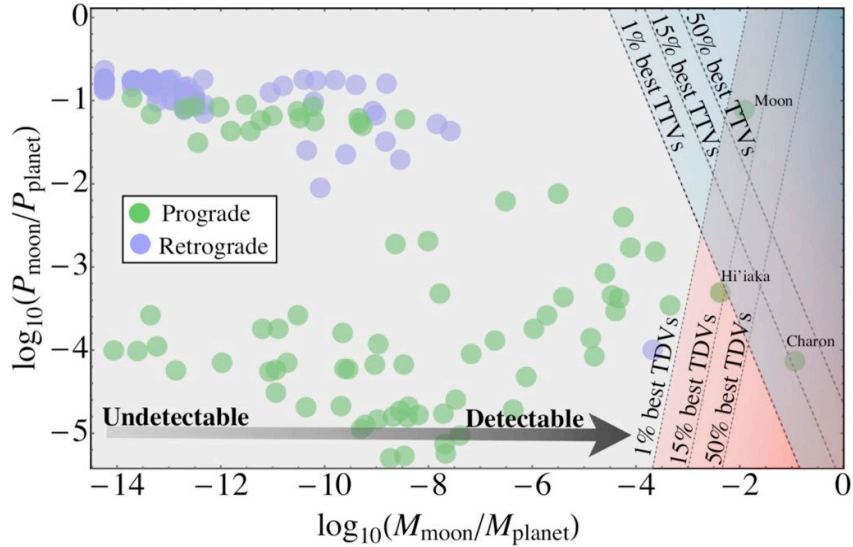

**Figure 11**: Period-ratio versus mass-ratio scatter plot of the Solar System moons. Transit timing and duration variations (TTV and TDV) exhibit complementary sensitivities with the period-ratio. Using the *Kepler* timing measurements from Ford et al. ([2012](#)), one can see that the tip of the observed distribution is detectable. The above assumes a planetary period of 100 days and a baseline of 4.35 years of *Kepler* data.

min⁻¹ (National Research Council, [2000](#))[22] , which is many orders of magnitude above the maximum tolerable radiation levels for most microorganisms found on Earth (Baumstark-Khan and Facius, [2002](#)).

   Magnetic field intensities inside compressed magnetospheres will be larger for a given dipole moment of a Solar System giant planet. Moreover, stellar high-energy particle fluxes, feeding radiation belts, will be also larger. As a result, the flux of high energy particles accelerated inside the magnetosphere of giant planets in the HZ could be strongly enhanced with respect to the already deadly levels expected on Jupiter's moon Europa.

<div align="center">

**6. Detection Methods for Extrasolar Moons**

</div>

Within the arena of transiting planets, there are two basic effects which may betray the presence of an exomoon: (i.) dynamical effects which reveal the mass ratio between the satellite and the planet ($M_s/M_p$), and (ii.) eclipse effects which reveal the radius ratio between the satellite and the star ($R_s/R_\star$). Detecting both effects allows for a measurement of the bulk density and thus allows one to distinguish between, say, icy moons versus rocky moons.

   Beyond the possibility to detect extrasolar satellites when using the stellar transits of a planet-moon system, several other methods have been proposed over the past decade**.** Cabrera and Schneider ([2007](#)) suggested that an exomoon might induce a wobble of a directly imaged planet's photocenter and that planet-moon eclipses might be detectable for directly imaged planets (see also Sato and Asada [2010](#); Pál [2012](#)). Excess emission of transiting giant exoplanets in the spectrum between 1 and 4 µm (Williams and Knacke [2004](#)) or enhanced infrared of terrestrial planets (Moskovitz et al. [2009](#); Robinson [2011](#)) might also indicate the presence of a satellite. Further approaches consider the Rossiter-McLaughlin effect (Simon et al. [2010](#); Zhuang et al. [2012](#)), pulsar timing variations (Lewis et al. [2008](#)), microlensing (Han and Han [2002](#)), modulations of a giant planet's radio emission (Noyola et al. [2014](#)), and the the generation of plasma tori around giant planets caused by the tidal activity of a moon (Ben-Jaffel and Ballester [2014](#)). In particular, the upcoming launch of the James Webb Space Telescope (*JWST*) inspired Peters and Turner ([2013](#)) to propose the possibility of detecting an exomoon's thermal emission. In the following, we discuss the dynamical effects that may reveal an exomoon as well as the prospects of direct photometric transit observations and constraints imposed by white and red noise. Possibilities of detecting extremely tidally heated exomoons via their thermal emission will also be illustrated.

*6.1 Dynamical effects on transiting exoplanets*

<hr />

[22] Report "Preventing the Forward Contamination of Europa" prepared by the Task Group on the Forward Contamination of Europa, Space Studies Board, Commission on Physical Sciences, Mathematics, and Applications, National Research Council. Available online at http://www.nap.edu/openbook.php?isbn=NI000231

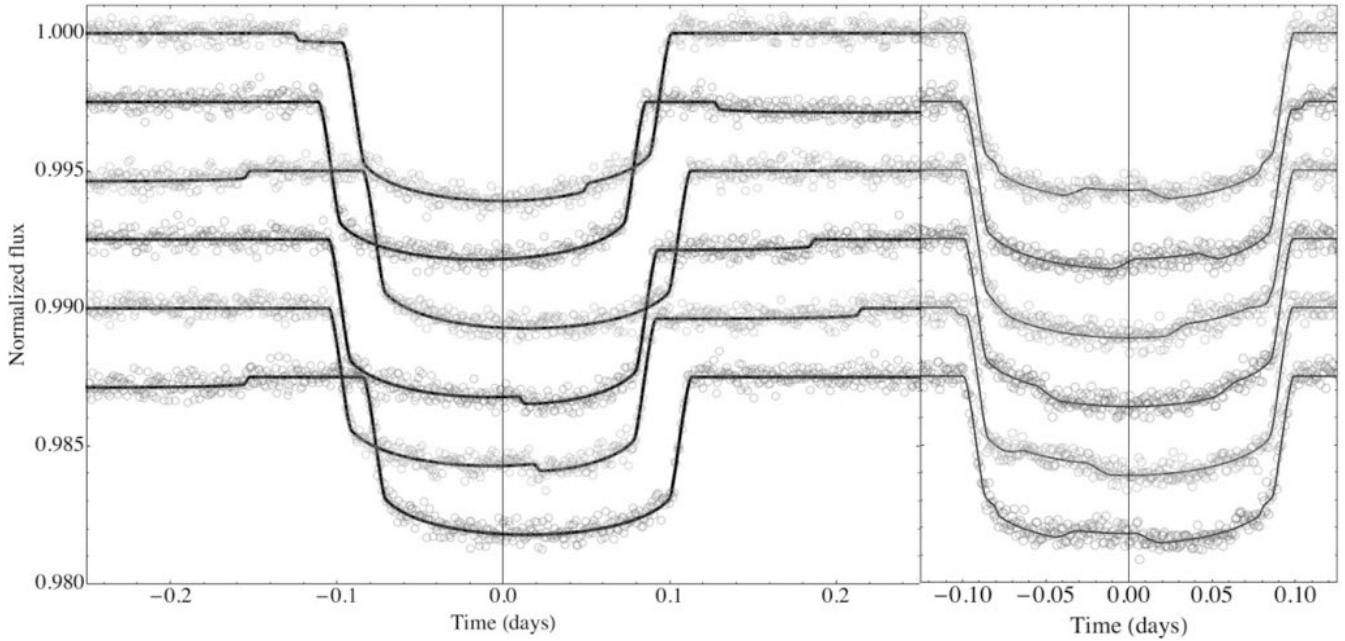

**Figure 12**: (*Left*) Six simulated transits using LUNA (Kipping, [2011b]) of a HZ Neptune around an M2 star with an Earth-like moon on a wide orbit (90% of the Hill radius). The moon can be seen to exhibit "auxiliary transits" and induce TTVs. (*Right*) Same as left, except the moon is now on a close-in orbit (5% of the Hill radius), causing "mutual events". Both plots show typical *Kepler* noise properties for a 12th-magnitude star observed in short-cadence.

The first technique ever proposed to detect an exomoon comes from Sartoretti and Schneider ([1999]), which falls into the dynamical category and is now often referred to as transit timing variations, or simply TTV, although it was not referred to as this in the original paper. For a single moon system, the host planet and the companion orbit a common barycenter, which itself orbits the host star on a Keplerian orbit. It therefore follows that the planet does not orbit the star on a Keplerian orbit and will sometimes transit slightly earlier or later depending upon the phase of the moon. These deviations scale proportional to $P_P \times (M_s/M_p) \times (a_s/a_p)$ and can range from a few seconds to up to a few hours for terrestrial moons. The original model of Sartoretti and Schneider ([1999]) deals with circular, coplanar moons but extensions to non-coplanar and eccentric satellites have been proposed since (Kipping, [2009a], [2011a]). Three difficulties with employing TTV in isolation are that, firstly, other effects can be responsible (notably perturbing planets); secondly, the TTV waveform is guaranteed to be below the Nyquist frequency and thus undersampled, making a unique determination of $M_s$ untenable; and, thirdly, moons in close orbits blur their planet's TTV signature as the planet receives a substantial tangential acceleration during each transit (Kipping [2011a]).

TTV can be thought of as being conceptually analogous to the astrometric method of finding planets, since it concerns changes in a primary's position due to the gravitational interaction of a secondary. Astrometry, however, is not the only dynamical method of detecting planets. Notably, Doppler spectroscopy of the host star to measure radial velocities has emerged as one of the work horses of exoplanet detection in the past two decades. Just as with astrometry, an exomoon analogy can be devised by measuring changes in a planet's transit duration, as a proxy for its velocity (Kipping, [2009a]). Technically though, one is observing tangential velocity variations rather than those in the radial direction. Transit duration variations due to velocity variations, dubbed TDV-V, scale as $(P_P/P_s) \times (M_s/M_p) \times (a_s/a_p)$ and vary from seconds to tens of minutes in amplitude for terrestrial satellites.

Just as radial velocity is complementary to astrometry, TDV-V and TTV are complementary since TTV is more sensitive to wide-orbit moons (sensitivity scales as $\propto a_s$) and TDV is more sensitive to close-orbit moons (sensitivity scales as $\propto a_s^{-1/2}$). Furthermore, should one detect both signals, the ratio of their root-mean-square amplitudes (that is, their statistical scatter) yields a direct measurement of $P_s$, which can be seen via inspection of the aforementioned scalings. This provides a powerful way of measuring $P_s$ despite the fact that the signals are undersampled. Finally, TDV-V leads TTV by a $\pi/2$ phase shift in amplitude (Kipping [2009a]) providing a unique exomoon signature, which even for undersampled data can be detected with cross-correlation techniques (Awiphan and Kerins, [2013]). Just as with TTV though, a TDV-V signal in isolation suffers from both model and parameter degeneracies.

An additional source of transit duration variations comes from non-coplanar satellite systems, where the planet's reflex motion causes position changes not only parallel to the transit chord (giving TTV) but also perpendicular to it. These variations cause the apparent impact parameter of the transit to vary leading to transit impact parameter induced transit duration variations, known as TDV-TIP (Kipping, 2009b). This is generally a small effect of order of seconds but can become much larger for grazing or near-grazing geometries. Should the effect be detected, it can be shown to induce a symmetry breaking that reveals whether a moon is prograde or retrograde in its orbit, thereby providing insights into the history and formation of the satellite.

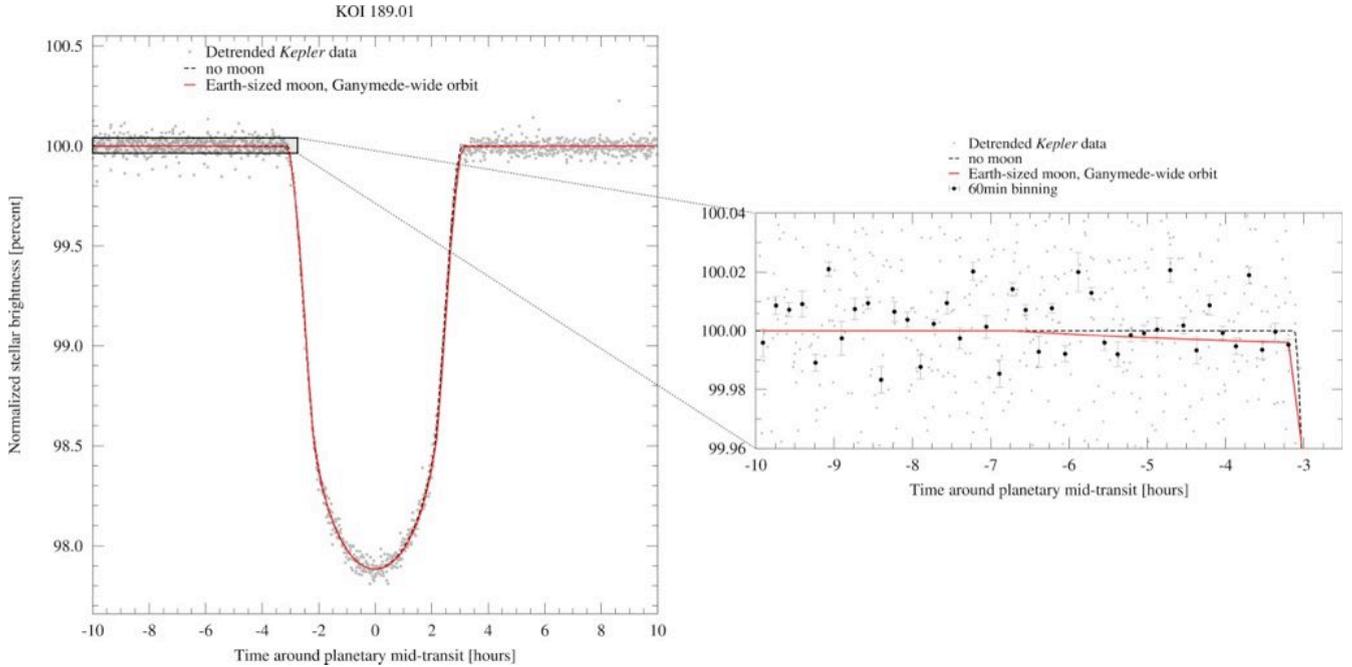

**Figure 13**: Transit of the almost Jupiter-sized planet candidate KOI189.01 around a star of about 0.7 solar radii. Gray dots indicate the original phase-folded *Kepler* light curve, the black dashed line indicates a model for the transit assuming a planet only, and the red line assumes an Earth-sized moon in an orbit that is 15 planetary radii wide. Black dots in the right panel indicate data that is binned to 60 minutes. The photometric orbital sampling effect appears in the right panel as a deviation between the red solid and the black dashed line about 6.5 hours before the planetary mid-transit.

For all of the aforementioned techniques, multiple moons act to cancel out the overall timing deviations, except for resonant cases. In any case, the limited number of observables (two or three) make the detection of multiple moons untenable with timing data alone. Despite this, there are reasons to be optimistic for the detection of a large quasi-binary type exomoon. Kipping et al. (2009) and Awiphan and Kerins (2013) estimated that the *Kepler* space telescope is sensitive to telluric moons. Recent statistics on the timing sensitivity of *Kepler* by Ford et al. (2012) reveal that the tip of the known population of moons in our own Solar System would be detectable with *Kepler* too, as shown in Fig. 11.

In terms of detectable habitable exomoons, Kipping et al. (2009) took into account orbital stability criteria from Barnes and O'Brien (2002) and found that moons more massive than about 0.2 $M_\oplus$ and orbiting a Saturn-mass planet in the center of the stellar HZ can be detectable with *Kepler*-class photometry if the star is more massive than about 0.3 $M_\odot$ and has a *Kepler* magnitude ≤ 12 (Fig. 6 of Kipping et al, 2009). More massive stars would need to be brighter because the planetary transit and its TTV and TDV signals become relatively dimmer. Simulations by Awiphan and Kerins (2013) suggest that moons in the habitable zone of a 12.5 *Kepler* magnitude M dwarf with 0.5 times the mass of the Sun would need to be as massive as 10 $M_\oplus$ and to orbit a planet less massive than about 1/4 the mass of Saturn to be detectable via their planet's TTV and TDV. Such a system would commonly be referred to as a planetary binary system rather than a planet-moon binary because the common center of gravity would be outside the primary's radius. From this result and similar results (Lewis, 2011b), the detection of a HZ moon in the ≈ 1400 d of *Kepler* data is highly challenging.

*6.2 Direct eclipse effects*

While dynamical effects of the planet can yield the system's orbital configuration, a transit of the moon itself not only provides a measurement of its radius but also has the potential to distort the transit profile shape leading to the erroneous derivation of TTVs and TDVs. Modeling the moon's own transit signal is therefore both a critical and inextricable component of hunting for exomoons.

Exomoons imprint their signal on the stellar light curve via two possible effects. The first is that the moon, with a wide sky-projected separation from the planet, transits the star and causes a familiar transit shape. These "auxiliary transits" can occur anywhere from approximately 93% of the Hill radius away from the planet (Domingos et al., 2006) to being ostensibly on-top of the planet's own transit (see Fig. 12 for examples). Since the phase of the moon will be unique for each transit epoch, the position and duration of these auxiliary transits will vary, which can be thought of as TTV and TDV of the moon itself, magnified by a factor of $M_p/M_s$. The position and duration variations of the moon will also be in perfect anti-phase with the TTVs and TDVs of the planet, providing a powerful confirmation tool.

The second eclipse effect caused by exomoons are so-called "mutual events". This is where the moon and planet appear separated at the start of a transit but then the moon eclipses (either in front or behind) the planet at some point during the full transit duration. This triple-overlap means that the amount of light being blocked from the star actually decreases during the mutual event, leading to what appears to be an inverted-transit signal (Fig. 12). Mutual events can occur between two planets too (Ragozzine and Holman, 2010), but the probability is considerably higher for a moon. The probability of a mutual event scales as $\propto R_p/a_s$ and thus is highest for close-in moons. One major source of false-positives here are starspots crossings (Rabus et al., 2009), which appear almost identical, but of course do not follow Keplerian motion.

Accounting for all of these eclipse effects plus all of the timing effects is arguably required to mount a cogent and expansive search for exomoons. Self-consistent modeling of these phenomena may be achieved with full "photodynamic" modeling, where disc-integrated fluxes are computed at each time stamp for positions computed from a three-body (or more) dynamical model. Varying approaches to moon-centric photodynamical modeling exist in the literature including a re-defined TTV for which the photocenter (Simon et al., 2007), circular/coplanar photodynamical modeling (Sato and Asada, 2009; Tusnski and Valio, 2011), and full three-dimensional photodynamical modeling (Kipping, 2011b) have been used.

Another way of detecting an exomoon's direct transit signature is by folding all available transits of a given system with the circumstellar orbital period into a single, phase-folded light curve. If more than a few dozen transits are available, then the satellites will show their individual transit imprints (Heller 2014). Each moon causes an additional transit dip before and after the planetary transit and thereby allows measurements of its radius and planetary distance. This so-called orbital sampling effect loses the temporal information content exploited in rigorous photodynamical modeling, meaning that the satellite's period cannot be determined directly. And without the satellite period, one cannot compute the planetary density via the trick described by Kipping (2010). Despite this, the computational efficiency of the OSE method makes it attractive as a quick method for identifying exomoon candidate systems, perhaps in the readily available *Kepler* data or the upcoming *Plato* space mission. But for such a detection, dozens of transits in front of a photometrically quiet M or K dwarf are required (Heller 2014). What is more, to ultimately confirm the presence of an exomoon, additional evidence would be required to exclude photometric variations due other phenomena, both astrophysical (e.g., stellar activity, rings) and instrumental (e.g., red noise). A confirmation could be achieved with photodynamical modeling as applied by the HEK team. As an example for the photometric OSE, Figure 13 presents the phase-folded Kepler light curve (gray dots) of the almost Jupiter-sized planetary candidate KOI189.01, transiting a ≈0.7 $R_\odot$ K star. Note how the "no moon" model (black dashed line) and the model for an Earth-sized moon (red solid line) diverge in the right panel! The latter is based on an improved model of Heller (2014), now including stellar limb darkening, amongst others, and is supposed to serve as a qualitative illustration but not as a statistical fit to the Kepler data for the purpose of this paper.

### 6.3 Photometric noise

One major factor constraining the detection of moons of transiting planets is the type of photometric noise contaminating the light curve.[23] A first detailed analysis of noise effects on the detection of exomoons with *Kepler* was performed by Kipping et al. (2009), showing that photon noise and instrument noise strongly increase for stellar apparent magnitudes ≳ 13. Combing shot noise, *Kepler*'s instrumental noise, and stellar variability with arguments from orbital stability, they obtained a lower limit of about 0.2 $M_\oplus$ for the detection of moons orbiting Saturn-like planets in the stellar HZ of bright M stars (with *Kepler* magnitudes < 11) by using TTV and TDV. This mass detection limit increases for Neptune- and Jupiter-like planets because of these planets' higher densities (see Fig. 3 in Kipping et al., 2009). What is more, absolute mass detection limits are hard to generalize because the TTV-TDV combined method constrains the planet-to-satellite mass ratio $M_s/M_p$. In their targeted search for exomoons, Kipping et al. (2013a) achieved accuracies as good as $M_s/M_p \approx 4\%$.

Lewis (2011a) investigated the effects of filtering on Sun-like stellar noise for a variant of TTV, photometric transit timing, and found that realistic photometric noise suppresses the detection of moons on wide circumplanetary orbits. In another work, Lewis (2013) confirmed that these noise sources also hamper the detection of moons around planets that follow distant circumstellar orbits. As a result, exomoons hidden in the noise of the *Kepler* data will need to have radii ≳ 0.75 $R_\oplus$ to be detectable by direct eclipse effects. This value is comparable to the results of Simon et al. (2012), who found that moons of close-in planets, that is, with circumstellar orbital periods ≤ 10 days, could be detected by using their scatter peak method on *Kepler* short cadence data if the moons have radii ≥ 0.7 $R_\oplus$.

---

[23] In the case of moon detection through perturbation of the Rossiter-McLaughlin effect (Simon et al., 2010; Zhuang et al., 2012) one needs to consider types of noise that contaminate radial velocity measurements.

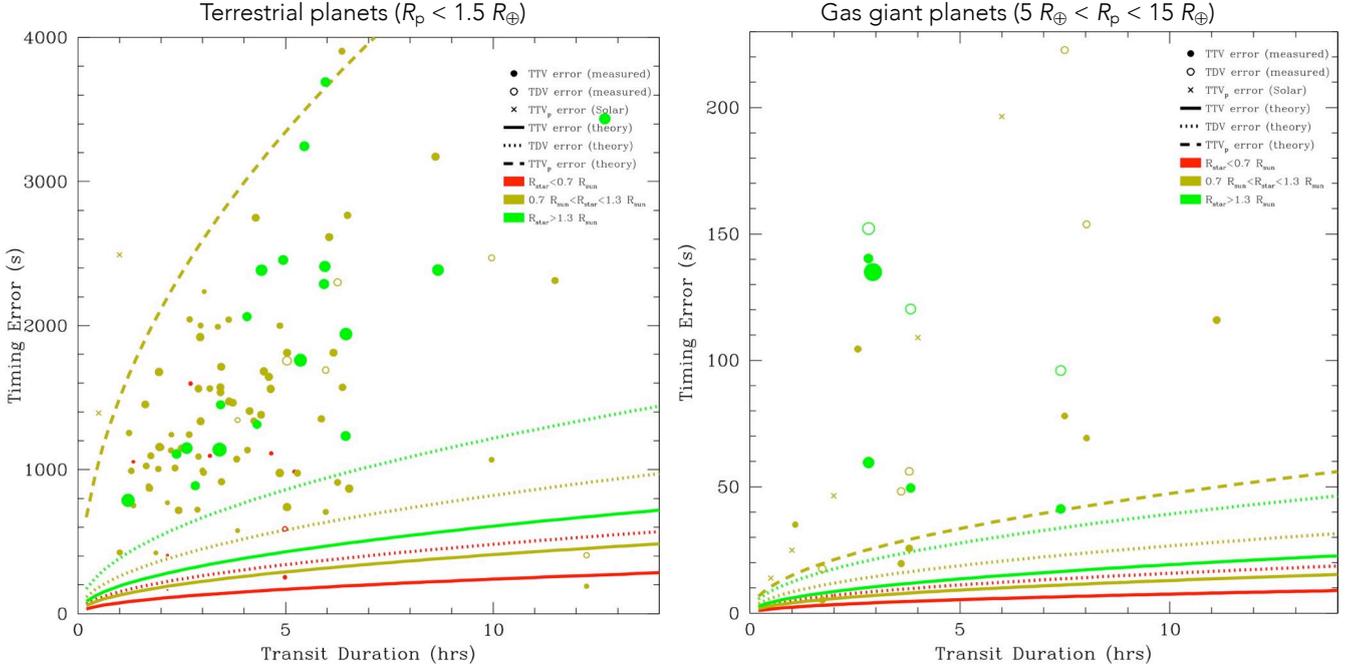

**Figure 14**: Calculated timing errors of the transit mid-time (filled circles) and duration (open circles) for *Kepler* planet candidates around stars with magnitudes < 13.5 (Mazeh et al., 2013). Planetary radii are estimated by assuming an Earth-like composition (left panel) or a mostly gaseous composition (right panel). Point size is proportional to the stellar radius, with color indicating if the radius is smaller than 0.7 solar radii ($R_\odot$) (red), between 0.7 $R_\odot$ and 1.3 $R_\odot$ (gold), or larger than 1.3 $R_\odot$ (green). Predictions of the errors in transit mid-time (solid line) and duration (dotted line) following Carter et al. (2008) assume a central transit and white noise for the case of a 1 $R_\oplus$ or 10 $R_\oplus$ radius planet orbiting a 0.7 $R_\odot$ (red), $R_\odot$ (gold), or 1.3 $R_\odot$ (green) star. Photometric transit timing errors, dominated by white noise (dashed line) and realistic solar-type noise (crosses) assume a 12th magnitude G dwarf and relative photometric precision of $2 \times 10^{-5}$ in a 6.5 hour exposure (Lewis, 2013).

Uncorrelated noise that follows a Gaussian probability distribution

$$f(x) = \frac{1}{\sigma\sqrt{2\pi}}e^{\frac{(x-\mu)^2}{2\sigma^2}}$$

(6)

where σ is the standard deviation, *x* is the perturbation to the light curve due to noise, and μ is the mean of the noise (usually zero), is called "white noise", and it makes moon detection relatively straightforward. At a given time, *f(t)* is assumed to be uncorrelated with the past. As a result, white noise contains equal power at all spectral frequencies. White noise or noise that is nearly white is produced by many instrumental or physical effects, such as shot noise.

Processes in stars, in Earth's atmosphere, and in telescopes can lead to long-term correlated trends in photometry that show an overabundance of long-wavelength components. In astrophysics, these effects are collectively referred to as red noise (Carter and Winn, 2009). Stars pulsate, convect, show spots and rotate, each of which leaves a distinctive red noise signature on the light curve (Aigrain et al., 2004). Solar oscillations have a typical amplitude of a few parts per million and period on the order of five minutes, while for other stars the specific details depend on the stellar size and structure (Christensen-Dalsgaard, 2004). In Sun-like stars, stellar convection leads to granulation, which produces red noise with characteristic timescales between minutes and hours. Finally, starspots can cause red noise in transit light curves, first through spot crossing events and second through spot evolution or rotational modulation. If a planet passes in front of a starspot group during transit, a temporary short-term increase in brightness results. This effect was predicted (Silva, 2003) and has finally been observed in many transiting systems (Pont et al., 2007; Sanchis-Ojeda and Winn, 2011; Kundurthy et al., 2011). Evolution and rotational modulation of starspots can cause long-term photometric variation over tens of days. This variation has been observed and modeled for a range of stars including CoRoT-2a (Lanza et al., 2009a) and CoRoT-4a (Lanza et al.,

2009b).

Ground-based observations can be substantially contaminated with red noise induced by Earth's atmosphere (Pont et al., 2006), for example, by the steadily changing value of the air mass over the course of a night. The operation of the telescope, as well as the telescope surroundings, can also introduce red noise, as has been noted by the ground-based WASP (Street et al., 2004) and HATNet surveys (Bakos et al., 2002, 2009). For the space-based *Kepler* and *CoRoT* missions, a number of different processes can lead to trends and jumps in the data, including effects from spacecraft motion, motion of the image on the CCD, and cosmic rays (Auvergne et al., 2009; *Kepler Characteristics Handbook*[24]). When a spacecraft rotates to reposition solar panels or to downlink data, or when it passes into Earth's shadow[25], the varying solar irradiation on various parts of the spacecraft's surface can cause transient photometric perturbations. In addition, the target's image can move on the CCD, and, finally, damage due to cosmic rays can lead to degradation of the CCDs. Cosmic rays can be isolated as well as certain specific events such as the passage of the south atlantic anomaly in the case of *CoRoT*, or the three large coronal mass ejections in *Kepler's* quarter 12. While processing pipelines work to reduce effects from cosmic rays (*Kepler Data Processing Handbook*[26]), they are never completely removed in all cases.

### 6.3.1 Effects of red noise on searches for exomoons

While uncertainties in the transit mid-time and transit duration are dominated by errors in the ingress and egress parts of the light curve, relatively short lengths of time, red noise acts over longer periods. Hence, these techniques should be quite robust to red noise. However, transit mid-time (filled circles in Fig. 14) and duration (open circles) errors (Mazeh et al., 2013) of *Kepler* Objects of Interest are up to many factors above the values predicted by assuming white noise only. While inclination, stellar luminosity, stellar radius, and planetary radius can explain some of this discrepancy, some of it is undoubtedly due to red noise.

Carter and Winn (2009) investigated the effect of red noise on both mid-time and duration errors with a power spectrum following a power law, showing that parameters can be effectively recovered through use of a wavelet transform. Alternatively, Kipping et al. (2009) and later Awiphan and Kerins (2013) modeled red noise as white noise plus a set of longer period sinusoids. While Kipping et al. (2009) ended up neglecting the difference between white and red noise, their simulations show that the distribution of transit mid-times became slightly non-gaussian, in particular, the tails of the distribution became fatter (Fig. 2 of Kipping et al., 2009). Awiphan and Kerins (2013) confirmed this effect and concluded that this variety of red noise decreases moon detectability. Other recent studies suggest that the presence of starspots, especially near the stellar limb, can alter transit durations and timings (Silva-Válio, 2010; Barros et al., 2013), reducing sensitivity for spotted stars. Mazeh et al. (2013) confirmed this trend in transit timing errors for numerous Kepler Objects of Interest. In particular, they found that TTVs correlate with stellar rotation in active stars, indicating again that noise components due to starspots could alter timing results.

In addition to the standard transit timing technique, photometric transit timing has been proposed (Szabó et al., 2006), a technique that measures timing perturbations with respect to the average transit time weighted by the dip depth. Using solar photometric noise from the Solar and Heliosphere Observatory as a proxy, Lewis (2013) found that realistic red stellar photometric noise dramatically degraded moon detectability for this technique (see crosses in Fig. 14), compared to white noise for the case where the transit duration was longer than three hours. What is more, this degradation was not completely reversed by filtering, independent of the filtering method. This result is unfortunate, as the photometric transit timing statistic is proportional to the moon radius squared (Simon et al., 2007) as opposed to the moon's mass (Sartoretti and Schneider, 1999), but unsurprising as photometric transit timing uses data from both within and outside the planetary transit, compared to transit timing variation, which only uses shorter sections of data.

While Tusnski and Valio (2011) assumed white photometric noise, other projects have started to consider the effect of red noise in their analyses. The "Hunt for Exomoons with Kepler" (HEK) (Kipping et al., 2012) addresses the effect of red noise in a number of ways. First, they reject candidates with high levels of correlated noise (Kipping et al., 2013a). Second, a high-pass cosine filter is used to remove long-term trends from the data. The possibility of modeling the effects of starspots has also been suggested (Kipping, 2012). To help distinguish between star spot crossings and mutual events, rules have been proposed, for example, it is required that a mutual event have a flat top (Kipping et al., 2012). Also, to allow for secure detection, it is required that a given moon is detected by several detection methods. In an attempt to quantify the true error on detected moon properties, fake moon transits were injected into real data for the case of Kepler-22 b (Kipping et al., 2013b), indicating that an Earth-like moon, if present, would have been detected.

---

[24] http://archive.stsci.edu/kepler/manuals/Data_Characteristics.pdf

[25] *Kepler* orbits the Sun in an Earth-trailing orbit, where it avoids transits in the shadow of Earth. At an altitude of 827 km, *CoRoT* is in a polar orbit around the Earth, and occasionally passes the Earth's shadow.

[26] http://archive.stsci.edu/kepler/manuals/KSCI-19081-001_Data_Processing_Handbook.pdf

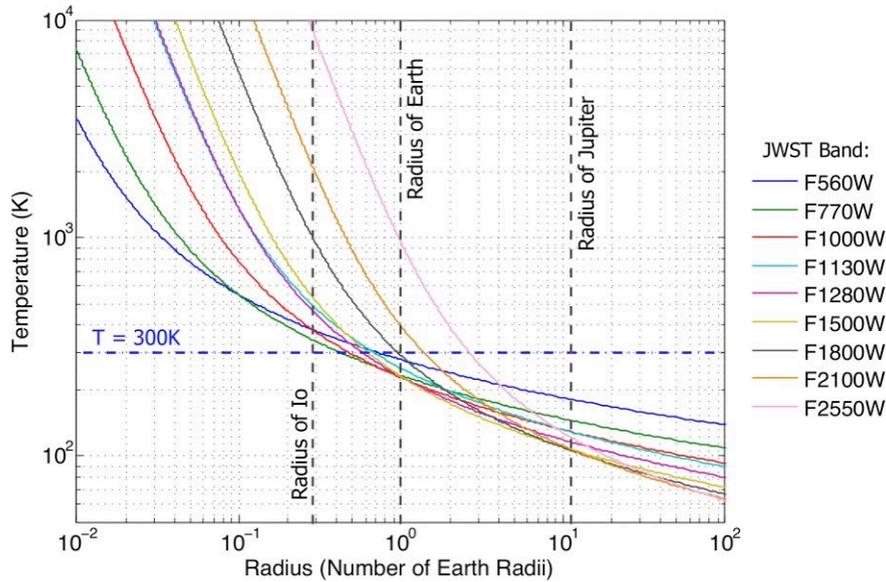

**Figure 15**: 5σ detection limits of JWST's MIRI for tidally heated exomoons with radii along the abscissa and effective temperatures along the ordinate (following Turner and Peters, 2013). 10,000s of integration are assumed for a star 3 pc from the Sun. MIRI's nine imaging bands are indicated with different line colors, their names encoding the wavelength in units of microns times 100. Dashed vertical black lines denote the radii of Io, Earth, and Jupiter, which is roughly equal to that of a typical brown dwarf.

### 6.4 Direct imaging of extrasolar moons

#### 6.4.1 Principles of direct imaging

Direct imaging of exoplanets, especially those in the stellar HZ, is extremely difficult because of the very small star-planet angular separation and the high contrast ratio between the star and planet. In fact, all exoplanets that have been directly imaged are well separated from their host star, and are young systems that are still hot from formation, rather than being heated by stellar irradiation, with effective temperatures around 1000 K. Examples of directly imaged exoplanets include the HR8799 planets, β Pic b, LkCa15b, κ And b, and GJ 504b (Marois et al., 2008; Lagrange et al., 2008; Kraus and Ireland, 2011; Carson et al., 2013; Kuzuhara et al., 2013). Intuitively, one would expect exomoons to be even more difficult to be imaged directly. For exomoons similar to those found in the Solar System, this is likely the case. However, satellites that are heated by sources other than stellar irradiation, such as tidal heating, can behave completely differently.

There has been considerable discussion in the literature of the existence of tidally heated exomoons (THEMs) that are extrasolar analogs to Solar System objects such as Io, Europa, and Enceladus (Peale et al., 1979; Yoder and Peale, 1981; Ross and Schubert, 1987; Ross and Schubert, 1989; Nimmo et al., 2007; Heller and Armstrong 2014). Yet, the possibility of imaging the thermal emission from unresolved THEMs was proposed only recently by Peters and Turner (2013)[27]. From an observational point of view, direct imaging of exomoons has several advantages over exoplanet direct imaging. THEMs may remain hot and luminous for timescales of order the stellar main sequence lifetime and thus could be visible around both young and old stars. Additionally, THEMs can be quite hot even if they receive negligible stellar irradiation, and therefore they may be luminous even at large separations from the system primary. This will reduce or even eliminate the inner-working angle requirement associated with exoplanet high contrast imaging. Furthermore, tidal heating depends so strongly on orbital and physical parameters of the THEM that plausible systems with properties not very different from those occurring in the Solar System will result in terrestrial planet sized objects with effective temperatures up to 1000 K or even higher in extreme but physically permissible cases. The total luminosity of a THEM due to tidal heating is given by

$$L_{\mathrm{tid}} = \left( \frac{6272\pi^7 G^5}{9747} \right)^{1/2} \left( \frac{R_s^7 \rho_s^{9/2}}{\mu_s Q_s} \right) \left( \frac{e_{\mathrm{ps}}^2}{\beta_{\mathrm{ps}}^{15/2}} \right) , \tag{7}$$

where $\mu_s$ is the moon's elastic rigidity, $Q_s$ is the moon's tidal dissipation function (or quality factor), $e_{\mathrm{ps}}$ is the eccentricity of the planet-satellite orbit, $\beta_{\mathrm{ps}}$ is the satellite's orbital semi-major axis in units of Roche radii, $\rho_s$ is the satellite density, and $R_s$ is the satellite radius (Scharf, 2006; Peters and Turner, 2013). Equation (7) is based on the tidal heating equations originally derived by Reynolds et al. (1987) and Segatz et al. (1988) and assumes zero obliquity. The terms that depend on the exomoon's physical properties and those that describe its orbit are grouped separately. Although $\beta_{\mathrm{ps}}$ is grouped with the orbital terms, in addition to its linear dependence on the moon's semi-major axis $a_{\mathrm{ps}}$, it also scales with the planetary mass and the satellite density as $(M_p/\rho_s)^{1/3}$ for a fixed $a_{\mathrm{ps}}$.

---

[27] For exoplanets, direct imaging refers to the imaging of a planet that is spatially resolved from its host star. In the model proposed by Peters and Turner (2013), the tidally heated moon is not spatially resolved from its host planet, but it rather adds to the thermal flux detected at the circumstellar orbital position of the planet. Despite this unresolved imaging, we here refer to this exomoon detection method as direct imaging.

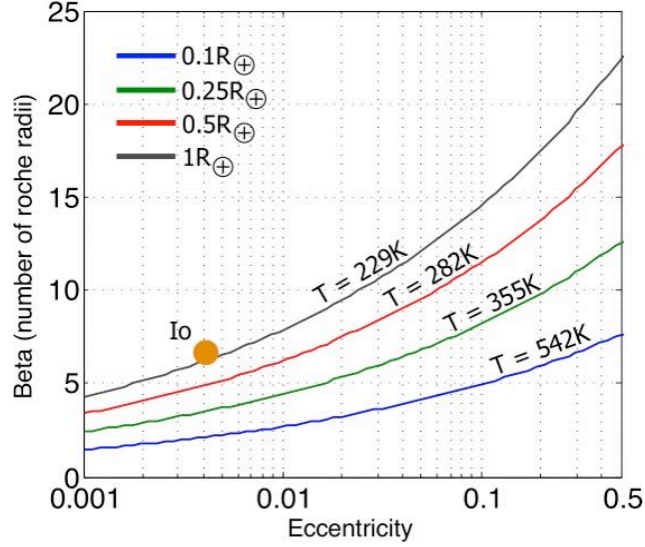

**Figure 16**: Minimum detectable eccentricity (abscissa) and semi-major axis in units of Roche radii ($\beta_{ps}$, ordinate) for tidally-heated exomoons between 0.1 and 1 $R_\oplus$ in size (Turner and Peters, 2013). The assumed bodily characteristics are described around Eqs. (5) - (6). The orange dot indicates Io's eccentricity and $\beta_{ps}$, but note that its size is 0.066 $R_\oplus$. Satellites below their respective curve will be detectable by MIRI as far as 3 pc from Earth, provided they are sufficiently separated from their star.

Following Peters and Turner (2013), the scaling relation for Eq. (7) relative to the luminosity of Earth ($L_\oplus = 1.75 \times 10^{24}$ ergs/s) is

$$L_s \approx L_\oplus \left[ \left( \frac{R_s}{R_\oplus} \right)^7 \left( \frac{\rho_s}{\rho_\oplus} \right)^{\frac{9}{2}} \left( \frac{36}{Q_s} \cdot \frac{10^{11} \frac{\mathrm{dynes}}{\mathrm{cm}^2}}{\mu_s} \right) \right] \times \left[ \left( \frac{e_{\mathrm{ps}}}{0.0028} \right)^2 \left( \frac{\beta_{\mathrm{ps}}}{8} \right)^{-\frac{15}{2}} \right] . \tag{8}$$

Note that Eq. (8) adopts $Q_s = 36$, $\mu_s = 10^{11}$ dynes/cm² as for Io (Peale et al., 1979; Segatz et al., 1988), and Earth's radius and density as reference values. The reference values of $\beta_{ps}$ and $e_{ps}$ were then chosen to give $L_\oplus$. If we assume that THEMs are blackbodies, we can use the exomoon's luminosity to approximate its effective temperature. The blackbody assumption is a reasonable approximation, but in general we would expect THEMs to emit excess light at bluer wavelengths due to hotspots on the surface. Additionally, exomoons with atmospheres are likely to have absorption lines in their spectrum. Assuming a blackbody, we can define a satellite's effective temperature ($T_s$) from the luminosity via the Stefan-Boltzmann law as a scaling relation relative to a 279 K exomoon, this temperature corresponding to the equilibrium temperature of Earth:

$$T_s \approx 279K \left[ \left( \frac{R_s}{R_\oplus} \right)^{\frac{5}{4}} \left( \frac{\rho_s}{\rho_\oplus} \right)^{\frac{9}{8}} \left( \frac{36}{Q_s} \cdot \frac{10^{11} \frac{\mathrm{dynes}}{\mathrm{cm}^2}}{\mu_s} \right)^{1/4} \right] \times \left[ \left( \frac{e_{\mathrm{ps}}}{0.0028} \right)^{\frac{1}{2}} \left( \frac{\beta_{\mathrm{ps}}}{8} \right)^{-\frac{15}{8}} \right] \tag{9}$$

(Peters and Turner, 2013). As an example, Eq. (9) yields $\approx$ 60 K for Io, corresponding to the effective temperature of Io if it were not irradiated by the Sun. Note that Eq. (9) adopts the same reference values as Eq. (8) and that it assumes only tidal heating with no additional energy sources, such as stellar irradiation or interior radiogenic heat.

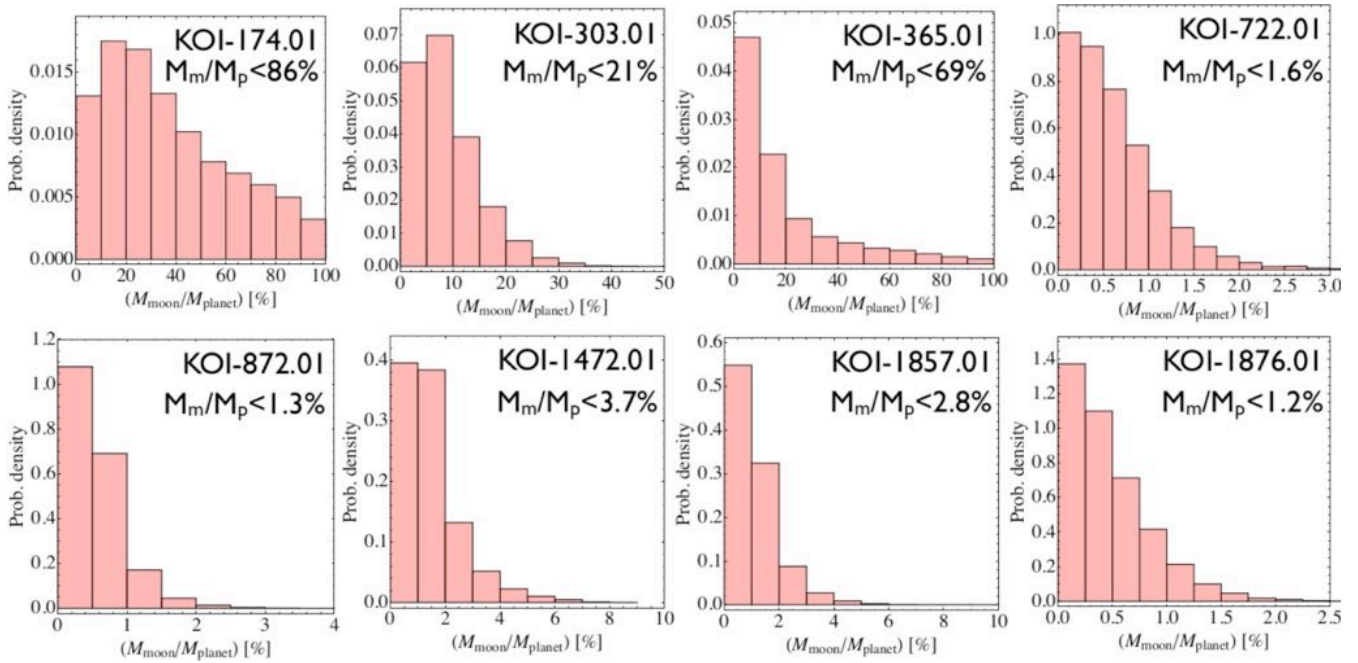

**Figure 17**: Moon-to-planet mass ratio constraints derived so far by the HEK project (Nesvorný et al., 2012; Kipping et al., 2013a). Kepler-22b has also been studied and yields $M_s < 0.5\ M_\oplus$ but is not shown here as the paper is still under review at the time of writing.

An exomoon's brightness is expected to be variable for multiple reasons, including eclipses of the exomoon behind its host planet (Heller, 2012), phase curve variations due to temperature variation across an object's surface (Heller and Barnes, 2013), and volcanism.

### 6.4.2 Direct imaging with future facilities

If THEMs exist and are common, are they detectable with current and future instrumentation? As shown by Peters and Turner (2013), Spitzer's IRAC could detect an exomoon the size of Earth with a surface temperature of 850 K and at a distance of five parsecs (pc) from Earth. Future instruments, such as JWST's Mid-Infrared Instrument (MIRI), have even more potential for direct imaging of exomoons. Figure 15 shows minimum temperatures and radii of exomoons detectable with MIRI at the 5σ confidence level in 10,000 seconds for a star 3 pc from the Sun (Glasse et al., 2010). Note that some exomoons shown here have temperatures and radii similar to Earth's! Thus, it is plausible that some of these hypothetical exomoons are not only directly visible with JWST, but they could potentially be habitable, in the sense of having surface temperatures that would allow liquid water to be present.

A 300 K, Earth-radius THEM is only $3 \times 10^5$ fainter at a wavelength of $\lambda \approx 14\ \mu m$ than a Sun-like star, but at the same time it can be at a large distance from it's host star. For example, at 30 AU projected separation, it would be 15″ from the star at a distance of 2 pc. At $\lambda \approx 14\ \mu m$, $\lambda/D = 0.44″$ for a D = 6.5-m telescope, which means this moon would be at 30 × λ/D. This far away from the star, the airy rings are about $3 \times 10^5$ times fainter than the core of the star and about the same intensity as the 300 K, Earth-radius moon, indicating that the detection should be possible. This example is not at the limit of JWST's sensitivity and inner working angle. The most challenging THEM detection JWST will be capable of making is a 300 K THEM as far as 4 pc from the Sun. If there is such an Earth-sized 300 K moon orbiting αCen, MIRI will be able to detect it in 8 of its 9 spectral bands with better than 15 σ signal-to-noise in a $10^4$ s integration. Thus, directly imaging a 300 K, Earth-radius moon that is tidally heated is potentially much easier than resolving an Earth-like exoplanet orbiting in the HZ of its primary!

We can ask the question of what orbital parameters would give rise to the temperatures of moons shown in Fig. 15. The minimum detectable temperatures by MIRI for four different radii satellite ($R_s = 0.1\ R_\oplus$, 0.25 $R_\oplus$, 0.5 $R_\oplus$, and $R_\oplus$) at 3 pc were calculated to be $T_s$ = 542 K, 355 K, 282 K, and 229 K, respectively (see Fig. 15). We can use Eq. (9) to calculate the expected orbital parameters $\beta_{ps}$ and $e_{ps}$ for these four objects. Figure 16 shows $\beta_{ps}$ and $e_{ps}$ corresponding to these four cases, indicating that a roughly Earth-sized exomoon (black line) at a distance of ten Roche radii, with an eccentricity similar to that of Titan ($e_{ps}$ = 0.0288), and an effective temperature equal to the reader's room temperature could be detectable with JWST.

### 6.5 Results of exomoon searches to date

Compared to their planetary cousins, relatively few searches for exomoons have been conducted to date. The great enabler in this field has been the *Kepler* mission offering continuous, precise photometric data for years of the same targets. The downside of *Kepler* data is that the field is deep (and thus faint) and dominated by Sun-like stars, whereas K- and M-dwarfs provide a significant boost to sensitivity thanks to their smaller sizes.

Within the current literature, one can find claims that various observations are consistent with an exomoon without specifically claiming an unambiguous exomoon detection. For example, Szabo et al. (2013) suggested that many of *Kepler*'s hot-Jupiters show high TTVs that could be due to exomoons or equally other planets in the system. Others used observations of transiting exoplanets to perform serendipitous searches for exomoons, such as Brown et al. (2001) around HD 209458b, Charbonneau et al. (2006) around HD 149026b, Maciejewski et al. (2010) around WASP-3b, and Montalto et al. (2012) around WASP-3b, none of which revealed evidence for an extrasolar satellite.

At the time of writing, the only known systematic survey is the "Hunt for Exomoons with Kepler" (HEK) (Kipping et al., 2012). Using photodynamic modeling, the HEK team has surveyed seventeen *Kepler* planetary candidates for evidence of an extrasolar moon so far. Due to the large number of free parameters, the very complex and multimodal parameter space, the demands of full photodynamic modeling, and the need for careful Bayesian model selection, the computational requirements reported by the HEK project have been staggering compared to the a typical planet-only analysis. For example, Kepler-22b required 50 years of modern CPU time (Kipping et al., 2013b).

So far, the HEK team reports no compelling evidence for an exomoon, but they have derived strong constraints of $M_s/M_p <$ 4% for five cases (Kipping et al., 2013a; Nesvorný et al., 2012), translating into satellite masses as small as 0.07 $M_\oplus$ (Fig. 17), and $M_s < 0.5$ $M_\oplus$ for the additional case of the habitable-zone planet Kepler-22b (Kipping et al., 2013b). In the latter example, an absolute mass constraint is possible thanks to the availability of radial velocities. In their latest release, Kipping et al. (2014) added another eight null detections of moons around planets transiting M dwarfs, with sensitivities to satellites as tiny as 0.36 $M_\oplus$ in the case of KOI3284.01. Despite the recent malfunction with the *Kepler* spacecraft, there is a vast volume of unanalyzed planetary candidates for moon hunting and still relatively few conducted so far. In the next one to two years then, we should see about 100 systems analyzed, which will provide a statistically meaningful constraint on the occurrence rate of large moons, $\eta_{\leftmoon}$.

## 7. Summary and Conclusions

This review highlights a remarkable new frontier of human exploration of space: the detection and characterization of moons orbiting extrasolar planets. Starting from the potentially habitable icy satellites of the Solar System (Section 2), we show that the formation of very massive satellites the mass of Mars is possible by in-situ formation in the circumplanetary disk and how a capture during binary exchange can result in a massive moon, too (Section 3). After discussing the complex orbital evolution of single and multiple moon systems, which are governed by secular perturbations and tidal evolution (Section 4), we explore illumination, tidal heating, and magnetic effects on the potential of extrasolar moons to host liquid surface water (Section 5). Finally, we explain the currently available techniques for searching for exomoons and summarize a recently initiated survey for exomoons in the data of the *Kepler* space telescope (Section 6).

From a formation point of view, habitable exomoons can exist. Mars-sized exomoons can form by either in-situ formation in the circumplanetary disks of super-Jovian planets (Section 3.1) or by gravitational capture from a former planet-moon or planet-planet binary (Section 3.2). This mass of about 0.1 $M_\oplus$ is required to let any terrestrial world be habitable by atmospheric and geological considerations (Section 5). Stellar perturbations on these moons' orbits, their tidal orbital evolution, tidal heating in the moons, and satellite-satellite interactions prevent moons in the HZs of low-mass stars from being habitable (Section 4). Moreover, while more massive giant planets should tend to produce more massive moons, strong inplanation from young, hot planets onto their moons can initiate a runaway greenhouse effect on the moons and make them at least temporarily uninhabitable (Section 5.2). With this effect becoming increasingly severe for the most massive giant planets, we identify here the formation and habitability of moons as competing processes. Although growing moons can accumulate more mass from the disks around more massive giant planets, the danger of a planet-induced runaway greenhouse effect on the moons increases, too. Finally, the host planet's intrinsic magnetic field and the stellar wind affect the moon's habitability by regulating the flux of high-energy particles (Section 5.3.2).

Exomoons in the *Kepler* data may be detectable if they belong to a class of natural satellites that does not exist in the Solar System. While the most massive known moon, Ganymede, has a mass roughly 1/40 $M_\oplus$, exomoons would need to have about ten times this mass to be traceable via their planet's TTV and TDV (Section 6.1). Another detection channel would be through a moon's direct photometric transit, which could be measurable in the *Kepler* data for moons as small as Mars or even Ganymede (Sections 6.2 and 6.3).

Combining the key predictions from formation, detection, and habitability sections, we find a favorable mass regime for the first extrasolar moons to be detected. Notably, there is a mass overlap of the most massive satellites that (i.) can possibly form in protosatellite disks around super-Jovian planets, (ii.) can be detected with current and near-future technology, and (iii.) can be habitable in terms of atmospheric stability. This regime is roughly between one and five times the mass of Mars or $0.1 - 0.5 \, M_\oplus$. Heavier moons would have better odds to be habitable (Heller and Armstrong, 2014), and they would be easier to detect, though it is uncertain whether they exist.

Although the observational challenges of exomoon detection and characterization are huge, our gain in understanding planet formation and evolution would be enormous. The moons of Earth, Jupiter, Saturn, and Neptune have proven to be fundamentally important tracers of the formation of the Solar System – most notably of the formation of Earth and life – and so the characterization of exomoon systems engenders the power of probing the origin of individual extrasolar planets (Withers and Barnes, 2010). Just as the different architectures of the Jovian and Saturnian satellite systems are records of Jupiter's gap opening in the early gas disk around the Sun, of an inner cavity in Jupiter's very own circumplanetary disk, and of the evolution of the $H_2O$ snow line in the disks, the architectures of exomoon systems could trace their planets' histories, too. We conclude that any near-future detection of an exomoon, be it in the *Kepler* data or by observations with similar accuracy, could fundamentally challenge formation and evolution theories of planets and satellites.

Another avail of exomoon detections could lie in a drastic increase of potentially habitable worlds. With most known planets in the stellar HZ being gas giants between the sizes of Neptune and Jupiter rather than terrestrial planets, the moons of giant planets could actually be the most numerous population of habitable worlds.

## 8. Outlook

On the theoretical front, improvements of our understanding of planet and satellite formation might help to focus exomoon searches on the most promising host planets. The formation and movement of water ice lines in the disks around young giant planets, as an example, has a major effect on the total mass of solids available for moon formation. Preliminary studies show that super-Jovian planets might host water-rich giant moons the size of Mars or even larger (Heller and Pudritz, in prep.), in agreement with the scaling relation suggested by Canup and Ward (2006). Hence, even future null detections of moons around the biggest planets could help to assess the conditions in circumplanetary disks. As an example, the absence of giant moons could indicate hotter planetary environments, in which the formation of ices is prevented, than are currently assumed. Earth-mass moons, however, are very hard to form with any of the viable formation theories, and consequently the suggested low abundance or absence (Kipping et al., 2014) of such moons may not help to constrain models immediately. However, in the most favorable conditions, where a somewhat super-Jovian-mass planet with an entourage of an oversized Galilean-style moon system transits a photometrically quiet M or K dwarf star, a handful of detections might be feasible with the available *Kepler* data (Kipping et al., 2012; Heller 2014) and therefore might confirm the propriety of moon formation theories for extrasolar planets.

Another largely unexplored aspect of habitable moon formation concerns the capture and orbital stability of moons during planet migration. A few dozen giant planets near, or in, their host stars' HZs have been detected by radial velocity techniques (see Fig. 1 in Heller and Barnes, 2014), most of which cannot possibly have formed at their current orbital locations, as giant planets' cores are assumed to form beyond the circumstellar snow line (Pollack et al., 1996; Ward, 1997). This prompts the compelling question of whether migrating giant planets, which end up in their host star's HZ, can capture terrestrial planets into stable satellite orbits during migration. A range of studies have addressed the post-capture evolution or stability of moons (Barnes and O'Brien, 2002; Domingo et al., 2006; Donnison et al., 2006; Porter and Grundy, 2011; Sasaki et al., 2012; Williams, 2013) and the loss of moons during planet migration towards the hot Jupiter regime (Namouni, 2010). But dynamical simulations of the capture scenarios around migrating giant planets near the HZ are still required in order to guide searches for habitable exomoons.

NASA's all-sky survey with the *Transiting Exoplanet Survey Satellite* (*TESS*), to be launched in 2017, will use the transit method to search for planets orbiting bright stars with orbital periods ≤ 72 days (Deming et al., 2009). As stellar perturbations play a key role in the orbital stability of moons orbiting giant planets (see Section 4), *TESS* will hardly find such systems. The proposed ESA space mission *PLAnetary Transits and Oscillations of stars* (*PLATO 2.0*) (Rauer et al., 2013), with launch envisioned for 2022-2024, however, has been shown capable of finding exomoons as small as $0.5 \, R_\oplus$, that is, the size of Mars (Simon et al., 2012; Heller 2014). Since the target stars will be mostly K and M dwarfs in the solar neighborhood, precise radius and mass constraints on potentially discovered exomoons will be possible (Kipping, 2010). What is more, *PLATO*'s long pointings of two to three years would allow for exomoon detections in the HZs of K and M dwarfs.

Future exomoon surveys may benefit by expanding their field of view while keeping a continuous staring mode and using the redder part of the spectrum to look at K and M dwarfs. Targeting these low-mass stars increases the relative dip in the light curves due to an exomoon as well as the transit frequency, and it decreases the minimum transit duration of a planet capable of hosting long-lived stable moons. Searches for exomoons around specific transiting planets will need to target relatively inactive stars or stars with noise spectra that are amenable to filtering (Carter and Winn, 2009) in order to reduce the effect of starspot modulation and errors due to spot crossing events. While recently developed procedures of white and red noise filtering in stellar light curves represent major steps towards secure moon detections, they do not directly address additional timing noise due to spots on the stellar limb (Silva-Válio, 2010; Barros et al., 2013) and the empirical correlation between the timing error and the slope of the out of transit light curve (Mazeh et al., 2013). Further progress, either through the modeling-based approach used by the HEK project, the filtering approach suggested by Carter and Winn (2009), or the observational approach used by Mazeh et al. (2013) is required to securely detect and constrain moons of *Kepler* planets. Finally, telescopes capable of multi-color measurements will help discriminate between starspot occupations on the one hand and planet/moon transits on the other hand.

*JWST* will be sensitive to a hypothetical family of tidally heated exomoons (THEMs) at distances as far as 4 pc from the Sun and as a cool as 300 K. These moons would need to be roughly Earth-sized and, together with the host planets, separated from their host star by at least 0.5". Future surveys that aim to detect smaller and colder THEMs would need to operate in the mid- to far-infrared with sub-microjansky sensitivity. Although THEMs need not be close to their host star to be detectable, instruments able to probe smaller inner working angles are likely to detect more THEMs simply because there will presumably be more planets closer into the star that can host THEMs. A preliminary imaging search for THEM around the nearest stars with the use of archival Spitzer IRAC data has recently been completed and will deliver the first constraints on the occurrence of THEMs (Limbach and Turner in prep).

Once an exomoon has been detected around a giant planet in the stellar HZ, scientists will try to discern whether it is actually inhabited. Kaltenegger (2010) showed that *JWST* would be able to spectroscopically characterize the atmospheres of transiting exomoons in nearby M dwarf systems, provided their circumplanetary orbits are wide enough to allow for separate transits. In her simulations, the spectroscopic signatures of the biologically relevant molecules $H_2O$, $CO_2$, and $O_3$ could be observable for HZ exomoons transiting M5 to M9 stars as far as 10 pc from the Sun. Although stellar perturbations on those moon's orbits could force the latter to be eccentric and thereby generate substantial tidal heating (Heller, 2012), the threat of a runaway greenhouse effect would be weak for moons in wide orbits because tidal heating scales inversely to a high power in planet-moon distance.

Exomoon science will benefit from results of the *Jupiter Icy Moons Explorer* (*JUICE*) (*JUICE* Assessment Study Report[28], 2011; Grasset et al., 2013), the first space mission designated to explore the emergence of habitable worlds around giant planets. Scheduled to launch in 2022 and to arrive at Jupiter in 2030, *JUICE* will constrain tidal processes and gravitational interaction in the Galilean system and thereby help to calibrate secular-tidal models for the orbital evolution of moons around extrasolar planets. As an example, the tidal response of Ganymede's icy shell will be measured by laser altimetry and radio science experiments. The amplitudes of periodic surface deformations are suspected to be about 7 to 8 m in case of a shallow subsurface ocean, but only some 10 cm if the ocean is deeper than roughly 100 km or not present at all. With Ganymede being one of the three solid bodies in the Solar System – besides Mercury and Earth – that currently generate a magnetic dipole field (Kivelson et al., 2002), *JUICE* could constrain models for the generation of intrinsic magnetic fields on extrasolar moons, which could be crucial for their habitability (Baumstark-Khan and Facius, 2002; Heller and Zuluaga, 2013). Measuring the bodily and structural properties of the Galilean moons, the mission will also constrain formation models for moon systems in general.

In view of the unanticipated discoveries of planets around pulsars (Wolszczan and Frail, 1992), Jupiter-mass planets in orbits extremely close to their stars (Mayor and Queloz, 1995), planets orbiting binary stars (Doyle et al., 2011), and small-scale planetary systems that resemble the satellite system of Jupiter (Muirhead et al., 2012), the discovery of the first exomoon beckons, and promises yet another revolution in our understanding of the universe.

## Acknowledgements

The helpful comments of two referees are very much appreciated. We thank Alexis Carlotti, Jill Knapp, Matt Mountain, George Rieke, Dave Spiegel and Scott Tremaine for useful conversations and Ted Stryk for granting permission to use a reprocessed image of Europa. René Heller is supported by the Origins Institute at McMaster University and by the Canadian Astrobiology Training Program, a Collaborative Research and Training Experience Program funded by the Natural Sciences and Engineering Research Council of Canada (NSERC). Darren Williams is a member of the Center for Exoplanets and Habitable Worlds, which is supported by the Pennsylvania State University, the Eberly College of Science, and the Pennsylvania Space Grant Consortium. Takanori Sasaki was supported by a grant for the Global COE Program, "From the Earth to 'Earths'", MEXT, Japan, and Grant-in-Aid for Young Scientists (B), JSPS KAKENHI Grant Number 24740120. Rory Barnes acknowledges support from NSF grant AST-1108882 and the NASA Astrobiology Institute's Virtual Planetary Laboratory lead team under cooperative agreement no. NNH05ZDA001C. Jorge I. Zuluaga is supported by CODI/UdeA. This research has been supported in part by World Premier International Research Center Initiative, MEXT, Japan. This work has made use of NASA's Astrophysics Data System Bibliographic Services.

---

[28] Yellow Book available at http://sci.esa.int/juice

No competing financial interests exist.

## 7.8 Exomoon Habitability and Tidal Evolution in Low-mass Star Systems (Zollinger et al. 2017)

Contribution:

RH contributed to the literature research and to the mathematical investigations, assisted in the generation of the figures, contributed to the writing of the manuscript, and contributed to the interpretation of the results.

# Exomoon Habitability and Tidal Evolution in Low-Mass Star Systems

Rhett R. Zollinger,[1]* John C. Armstrong,[2]† René Heller[3]‡

[1]*Southern Utah University, Cedar City, Utah 84720, USA*
[2]*Weber State University, Ogden, Utah 84408, USA*
[3]*Max Planck Institute for Solar System Research, Justus-von-Liebig-Weg 3, 37077 Göttingen, Germany*

## ABSTRACT

Discoveries of extrasolar planets in the habitable zone (HZ) of their parent star lead to questions about the habitability of massive moons orbiting planets in the HZ. Around low-mass stars, the HZ is much closer to the star than for Sun-like stars. For a planet-moon binary in such a HZ, the proximity of the star forces a close orbit for the moon to remain gravitationally bound to the planet. Under these conditions the effects of tidal heating, distortion torques, and stellar perturbations become important considerations for exomoon habitability.

Utilizing a model that considers both dynamical and tidal interactions simultaneously, we performed a computational investigation into exomoon evolution for systems in the HZ of low-mass stars ($\lesssim 0.6\ M_\odot$). We show that dwarf stars with masses $\lesssim 0.2\ M_\odot$ cannot host habitable exomoons within the stellar HZ due to extreme tidal heating in the moon. Perturbations from a central star may continue to have deleterious effects in the HZ up to $\approx 0.5\ M_\odot$, depending on the host planet's mass and its location in the HZ, amongst others. In addition to heating concerns, torques due to tidal and spin distortion can lead to the relatively rapid inward spiraling of a moon. Therefore, moons of giant planets in HZs around the most abundant type of star are unlikely to have habitable surfaces. In cases with lower intensity tidal heating the stellar perturbations may have a positive influence on exomoon habitability by promoting long-term heating and possibly extending the HZ for exomoons.

**Key words:** planets and satellites: dynamical evolution and stability – planets and satellites: physical evolution

## 1 INTRODUCTION

The exploration of the moons of Jupiter and Saturn has provided immense understanding of otherworldly environments. Some of these moons have reservoirs of liquids, nutrients, and internal heat (Squyres et al. 1983; Hansen et al. 2006; Brown et al. 2008; Saur et al. 2015) – three basic components that are essential for life on Earth. Recent technological and theoretical advances in astronomy and biology now raise the question of whether life might exist on any of the moons beyond the solar system ("exomoons"). Given the abundant population of moons in our system, exomoons may be even more numerous than exoplanets (Heller & Pudritz 2015).

No moon outside the Solar System has been detected, but the first detection of an extrasolar moon appears to be on the horizon (Kipping et al. 2009, 2012; Heller et al. 2014) now that modern techniques enable detections of sub-Earth sized extrasolar planets (Muirhead et al. 2012; Barclay et al. 2013). A recent review of current theories on the formation, detection, and habitability of exomoons suggests that natural satellites in the range of 0.1-0.5 Earth masses (i) are potentially habitable, (ii) can form within the circumplanetary debris and gas disk or via capture from a binary, and (iii) are detectable with current technology (Heller et al. 2014). Considering the potential for current observation, we explore the expected properties of such exomoons in this paper.

When investigating planet habitability, the primary heat source is typically the radiated energy from a central star. Considerations of stellar radiation and planet surface temperatures have led to the adoption of a concept referred to as the stellar "habitable zone" (HZ) that is the region around a star in which an Earth-like planet with an Earth-like atmosphere can sustain liquid surface water (Kasting et al. 1993). Tidal heating is typically less important as an energy source for planets in the HZ, though it might be relevant for eccentric planets in the HZs around low-mass stars of spectral type M (Barnes et al. 2008). For the moon of a large planet, tidal heating can work as an alternative internal heat source as well. Several studies have addressed the importance of tidal heating and its effects on the habitability of exomoons (Reynolds et al. 1987a; Scharf 2006; Henning et al. 2009; Porter & Grundy 2011a; Heller 2012; Forgan & Kipping 2013; Heller & Barnes 2013; Heller & Zuluaga 2013; Heller & Barnes 2015; Dobos & Turner 2015; Dobos et al. 2017). Tidal heat can potentially maintain internal heating over several Gyr (see Jupiter's moon Io; Spencer et al. 2000a), which contributes to surface heating and potentially drives important internal processes such as plate tectonics.

* E-mail: rhettzollinger@suu.edu
† E-mail: jcarmstrong@weber.edu
‡ E-mail: heller@mps.mpg.de

Tidal interactions between a massive moon and its host planet become particularly interesting for planets in the HZs of M dwarfs, which are smaller, cooler, fainter, and less massive than the Sun. They are the predominant stellar population of our Galaxy (Chabrier & Baraffe 2000), and as such, these low-mass stars may be the most abundant planet hosts in our Galaxy (Petigura et al. 2013; Dressing & Charbonneau 2013). Their lower core temperatures and decreased energy output result in a HZ that is much closer to the star in comparison to Sun-like stars. If a planet in the HZ around an M dwarf star has a massive moon, the close-in orbital distances could potentially influence the orbit of the moon around the planet (Heller 2012). The relatively close central star can serve as a continual source of gravitational perturbation to the moon's orbit, which has important implications to tidal heating and orbital evolution of moons in these systems.

With the possible abundance of planetary systems around low-mass stars and their unique characteristics for habitability, we are interested in exploring their potential for exomoon habitability. To accomplish this we employed a distinctive approach to simultaneously consider both orbital and tidal influences. In this paper, we review some theories on tidal interactions between two massive bodies and then conduct a computational investigation into the importance of these interactions on the long-term evolution of hypothetical exomoons. For our study we focus on massive moons around large planets in the HZ of low-mass stars ($\lesssim 0.6\ M_\odot$) which demonstrates the long-term effect of gravitational perturbations from a central star.

## 2  DYNAMICAL AND TIDAL EVOLUTION

Tidal bulges raised in both a planet and satellite will dissipate energy and apply torques between the two bodies. The rate of dissipation strongly depends on the distance between the two objects. As a result of the tidal drag from the planet, a massive satellite will be subject to essentially four effects on its spin-orbital configuration:

(i)  **Semi-major axis:** Tidal torques can cause a moon to either spiral in or out (Barnes & O'Brien 2002; Sasaki et al. 2012). The direction of the spiral depends on the alignment between the planet's tidal bulge (raised by the moon) and the line connecting the two centers of mass. If the planet's rotation period is shorter than the orbital period of the satellite, the bulge will lead (assuming prograde orbits) and the moon will slowly spiral outward. This action could eventually destabilize the moon's orbit, leading to its ejection. On the other hand, if the planet's rotation period is longer, the bulge will lag and the moon will slowly spiral inward. As this happens, the tidal forces on the moon become increasingly greater. If the inward migration continues past the Roche limit, the satellite can be disintegrated.

(ii)  **Eccentricity:** Non-circular orbits will be circularized over the longterm. The timescale for which the eccentricity is damped can be estimated as (Goldreich & Soter 1966)

$$\tau_e \approx \frac{4}{63} \frac{M_s}{M_p} \left( \frac{a}{R_s} \right)^5 \frac{Q'_s}{n}\ , \qquad (1)$$

where $n$ is the orbital mean motion frequency, $Q'_s$ is the tidal dissipation function, $a$ is the semi-major axis, $M_s$ and $R_s$ are the mass and radius of the satellite, and $M_p$ the mass of the host planet.

(iii)  **Rotation frequency:** The moon will have its rotation frequency braked and ultimately synchronized with its orbital motion around the planet (Dole 1964; Porter & Grundy 2011b; Sasaki et al. 2012), a state that is commonly known as tidal locking.

(iv)  **Rotation axis:** Any initial spin-orbit misalignment (or obliquity) will be eroded, causing the moon's rotation axis to be perpendicular to its orbit about the planet. In addition, a moon will inevitably orbit in the equatorial plane of the planet due to both the Kozai mechanism and tidal evolution (Porter & Grundy 2011b). The combination of all these effects will result in the satellite having the same obliquity as the planet with respect to the circumstellar orbit. As for the host planet, massive planets are more likely to maintain their primordial spin-orbit misalignment than small planets (Heller et al. 2011b). Therefore, satellites of giant planets are more likely to maintain an orbital tilt relative to the star than even a single terrestrial planet at the same distance from a star.

The issue of tidal equilibrium in three-body hierarchical star-planet-moon systems in the Keplerian limit has recently been treated by Adams & Bloch (2016), who demonstrated that a moon in tidal equilibrium will have its circumplanetary orbit widened until it ultimately gets ejected by stellar perturbations. *N*-body gravitational perturbations from the star set an upper limit on the maximum orbital radius of the moon around its planet at about half the planetary Hill radius for prograde moons (Domingos et al. 2006a). However, moons can be ejected from their planet even from close orbits around their planets, e.g. when the planet-moon system migrates towards the star and stellar perturbations increase the moon's eccentricity fatally (Spalding et al. 2016). Perturbations from other planets can also destruct exomoon systems during planet-planet encounters (Gong et al. 2013; Payne et al. 2013; Hong et al. 2015).

### 2.1  Tidal Heating

As a result of the tidal orbital evolution, orbital energy is transformed into heat in either the planet or the moon or both. Several quantitative models for tidal dissipation have been proposed, the most widely used family of which is commonly referred to as equilibrium tide models (Darwin 1879; Hut 1981; Efroimsky & Lainey 2007; Ferraz-Mello et al.

2008; Hansen 2010). However, the exact mechanisms of tidal dissipation invoke feedback processes between the tidal response of the body and the material it consists of (Henning et al. 2009; Henning & Hurford 2014), which are neglected in most equilibrium tide models. A conventional model that assumes a constant phase lag between the tidal bulge and the line between the centers of mass quantifies the tidal heating ($H$) of a satellite as

$$H = \frac{63}{4} \frac{(GM_p)^{3/2} M_p R_s^5}{Q_s'} a^{-15/2} e^2,$$ (2)

where $G$ is the gravitational constant and $e$ is the satellite's eccentricity around the planet (Peale & Cassen 1978; Jackson et al. 2008b). Equation (2) implies that tidal heating drops off quickly with increasing distance and altogether ceases for circular orbits ($e = 0$).[1]

In Eqs. (1) and (2), $Q'$ parameterizes the physical response of a body to tides (Peale et al. 1979). Its specific usefulness is that it encapsulates all the uncertainties about the tidal dissipation mechanisms. For solid bodies, this function can be related to the rigidity $\mu$ and the standard $Q$-value, $Q' = Q(1 + 19\mu/2g\rho R)$, where $g$ is the gravitational acceleration at the surface of the body and $\rho$ is its mean density. The dissipation function can also be defined in terms of the tidal Love number ($k_L$) as $Q' = 3Q/2k_L$.

The tidal heating in Jupiter's moons Io and Europa works to circularize their orbits relative to Jupiter. Interestingly, estimates for their eccentricity damping timescales are considerably less than the age of the Solar System (Murray & Dermott 1999, p. 173). Yet, their eccentricities are noticeably not zero (0.0043 and 0.0101, respectively). This inconsistency has been explained by the observed Laplace resonance between them and the satellite Ganymede (Peale et al. 1979; Yoder & Peale 1981). The orbital periods of the three bodies are locked in a ratio of 1:2:4, so their gravitational interactions continually excite the orbits and maintain their non-zero eccentricities. This example demonstrates the need to include external gravitational influences when considering the long-term tidal evolution of non-isolated planet-moon systems. For systems in the HZ of low-mass star systems, external perturbations from the relatively close star could serve to maintain non-zero satellite eccentricity in much the same way as the orbital resonances in the Galilean moon system (Heller 2012).

Equation (2) represents the energy being tidally dissipated by the whole of a satellite. However, to assess the surface effects of tidal heating on a potential biosphere it is necessary to consider the heat flux through the satellite's surface. Assuming the energy eventually makes its way to the moon's surface, the tidal surface heat flux ($h$) can be represented as

$$h_{conv} = H/4\pi R_s^2 .$$ (3)

Note in the definition that a subscript is used to identify Eq. (3) as the surface heat flux based on the conventional tidal model for $H$. Below, we will define another tidal heat flux based on a different tidal model.

In terms of the effects of tidal heating on a moon's surface habitability, Barnes et al. (2009) proposed a conservative limit based on observations of Io's surface heat flux of about $h = 2\,\mathrm{W\,m^{-2}}$ (Spencer et al. 2000b; McEwen et al. 2004), which results in intense global volcanism and a lithosphere recycling timescale on order of $10^5$ yr (Blaney et al. 1995; McEwen et al. 2004). Such rapid resurfacing could preclude the development of a biosphere, and thus $2\,\mathrm{W\,m^{-2}}$ can be considered as a pessimistic maximum rate of tidal heating to still allow habitable environments.

Tidal heating is not the only source of surface heat flow in a terrestrial body. Radiogenic heating, which comes from the radioactive decay of U, Th, and K, is an additional source for surface heat flow. Barnes et al. (2009) used this combination to also set a lower limit of $h_{min} \equiv 0.04\,\mathrm{W\,m^{-2}}$ for the total surface heat flux of a terrestrial body by considering that internal heating must be sufficient to drive the vitally important plate tectonics. This value was based on theoretical studies of Martian geophysics which suggest that tectonic activity ceased when the radiogenic[2] heat flux dropped below this value (Williams et al. 1997). Even though the processes which drive plate tectonics on Earth are not fully understood (Walker et al. 1981; Regenauer-Lieb et al. 2001), it is accepted that an adequate heat source is essential. The phenomenon of plate tectonic is considered important for habitability because it drives the carbon-silicate cycle thereby stabilizing atmospheric temperatures and $CO_2$ levels on timescales of ~$10^8$ yr. For reference, the Earth's combined outward heat flow (which includes both tidal and radiogenic heat) is $0.065\,\mathrm{W\,m^{-2}}$ through the continents and $0.1\,\mathrm{W\,m^{-2}}$ through the ocean crust (Zahnle et al. 2007), mostly driven by radiogenic heating in the Earth.

Radiogenic heating scales as the ratio of volume to area (Barnes et al. 2009). Consequently, for most cases involving closely orbiting bodies that are significantly smaller than the Earth, it is believed that tidal heating probably dominates (Jackson et al. 2008b,a). The theoretical moons considered in this study meet these conditions. Therefore, we assume that radiogenic heating is negligible and that the total surface heat flux is equal to the tidal flux. We also adopt the heating flux limits for habitability presented above, so that $h_{min} < h < h_{max}$ could allow exomoon surface habitability.

## 2.2 Global Energy Flux

Investigations of exomoon habitability can be distinguished from studies on exoplanet habitability in that a variety of astrophysical effects can be considered in addition to the illumination received by a parent star. For example, a moon's climate can be affected by the planet's stellar reflected light and its thermal emission. Moons also experience eclipses of the star by the planet, and tidal heating can provide an additional energy source that is typically less substantial for planets. Heller & Barnes (2013) considered these effects individually, and then combined them to compute the orbit-averaged global flux ($F_{glob}$) received by a satellite. More specifically, this computation summed the averaged stellar, reflected, thermal, and tidal heat flux for a satellite. In their study, they provided a convenient definition for the global flux as

$$F_{\rm glob} = \frac{L_\star(1-\alpha_s)}{16\pi a_{\star p}^2 \sqrt{1-e_{\star p}^2}} \left(1 + \frac{\pi R_{\rm p}^2 \alpha_{\rm p}}{2a_{\rm ps}^2}\right) + \frac{R_{\rm p}^2 \sigma_{\rm SB}(T_{\rm p}^{\rm eq})^4}{a_{\rm ps}^2} \frac{1-\alpha_s}{4} + h_s \tag{4}$$

where $L_\star$ is the luminosity of the star, $a_{\star p}$ is the semi-major axis of the planet about the star and $a_{\rm ps}$ is the satellite's semimajor axis about the planet, $\alpha$ is the Bond albedo, $e$ is eccentricity, $\sigma_{\rm SB}$ is the Stefan-Boltzmann constant, $h_s$ is the tidal heat flux in the satellite, and $T_{\rm p}^{\rm eq}$ is the planet's thermal equilibrium temperature.

---

[1] This model breaks down for large $e$ (Greenberg 2009) and it neglects effects of tidal heating from obliquity erosion (Heller et al. 2011b) and from rotational synchronization.

[2] At the orbital distance of Mars, any contribution from tidal heating would be very low. Therefore, the total internal heating is essential equal to the radiogenic heating in this case.

As an analogy with the circumstellar HZ for planets, there is a minimum orbital separation between a planet and moon that will allow the satellite to be habitable. Moons inside this minimum distance are in danger of runaway greenhouse effects by stellar and planetary illumination and/or tidal heating. There is not a corresponding maximum separation distance (other than stability limits) because satellites with host planets in the stellar HZ are habitable by definition. The benefit of Eq. (4) is that it can be used to explore the minimum distance. This is accomplished by comparing the global flux to estimates of the critical flux for a runaway greenhouse ($F_{\rm RG}$). Heller & Barnes (2013) discussed a useful definition for $F_{\rm RG}$ originally derived by Pierrehumbert (2010). Applying that definition to an Earth-mass exomoon gives a critical flux of 295 W m$^{-2}$ for a water-rich world with an Earth-like atmosphere to enter a runaway greenhouse state. In comparison to the conservative limit of 2 W m$^{-2}$ for the tidal heating discussed above, the runaway greenhouse limit defines the ultimate limit on habitability. Moons with a higher top-of-the-atmosphere energy flux cannot be habitable by definition, since all surface water will be vaporized.

### 2.3 Coupling tidal and gravitational effects

Heller (2012) suggested that low-mass stars cannot possibly host habitable moons in the stellar habitable zones because these moons must orbit their planets in close orbits to ensure Hill stability. In these close orbits they would be subject to devastating tidal heating which would trigger a runaway greenhouse effect and make any initially water-rich moon uninhabitable. This tidal heating was supposed to be excited, partly, by stellar perturbations. While tidal processes in the planet-moon system would work to circularize the satellite orbit, the stellar gravitational interaction would force the moon's orbital eccentricity around the planet to remain non-zero. However, Heller (2012) acknowledged that his model did not couple the tidal evolution with the gravitational scattering of a hypothetical satellite system so the extent of the gravitational influence of the star was surmised, but not tested. The need therefore remains to simulate the eccentricity evolution of satellites about low-mass stars with a model that considers both *N*-body gravitational acceleration and tidal interactions.

We can estimate the stellar mass below which no habitable moons can exist for a given system age and planet mass.

Neglecting all atmospheric effects on a water-rich, Earth-like object, we have

$$a_{\rm HZ} \approx \left(\frac{L_\star}{L_\odot}\right)^{1/2} {\rm AU} \approx \left(\frac{M_\star}{M_\odot}\right)^{7/4} {\rm AU} \ , \tag{5}$$

where $L_\odot$ and $M_\odot$ are the luminosity and mass of the Sun, respectively. The planetary Hill radius can be approximated as

$$R_{\rm Hill} \approx M_\star^{17/12} \left(\frac{M_{\rm p}}{3}\right)^{1/3} \frac{\rm AU}{M_\odot^{21/12}} \ . \tag{6}$$

Using the timescale for tidal orbital decay as used in Barnes & O'Brien (2002; Eq. (7) therein), we have

$$a_{\rm crit} \approx \left(\frac{13\tau_{\rm dec}}{2} \frac{3k_{2,\rm p}M_sR_{\rm p}^5}{Q_{\rm p}} \sqrt{\frac{G}{M_{\rm p}}} + R_{\rm p}^{13/2}\right)^{2/13} \overset{!}{<} R_{\rm Hill} \ , \tag{7}$$

for tidal love number $k_{2,\rm p}$ for the planet, where the latter relation must be met to allow for tidal survival over a time $\tau_{\rm dec}$. Hence

$$M_\star \overset{!}{>} \left(a_{\rm crit} \left(\frac{3}{M_{\rm p}}\right)^{1/3} \frac{M_\odot^{7/4}}{\rm AU}\right)^{12/17} \ . \tag{8}$$

For a Jupiter-like planet ($Q_{\rm p} = 10^5$, $k_{2,\rm p} = 0.3$) with an Earth-like moon, and assuming tidal survival for 4.5 Gyr, we estimate a minimum stellar mass of 0.18 $M_\odot$.

For a computational study, many popular and well tested computer codes are available to simulate the gravitational (dynamical) evolution of many bodied systems (Chambers & Migliorini 1997; Rauch & Hamilton 2002). Such codes are particularly useful for studying the long-term stability of planetary systems. These codes, however, do not include tidal interactions in their calculations. A modification of the *Mercury N*-body code (Chambers & Migliorini 1997) to include tidal effects (Bolmont et al. 2015) has recently been used to simulate, for the first time, the evolution of multiple moons around giant exoplanets (Heller et al. 2014). Yet, these orbital calculations neglected stellar effects in the planet-moons system.

Useful derivations for a tide model that provides the accelerations from tidal interactions at any point in a satellites orbit derivations were presented by Eggleton et al. (1998), whose work was based on the equilibrium tide model by Hut (1981). Their particular interest was to consider tidal interactions between binary stars. Eggleton et al. (1998) derived from first principles equations governing the quadrupole tensor of a star distorted by both rotation and the presence of a companion in a possibly eccentric orbit. The quadrupole distortion produces a non-dissipative acceleration $f_{\rm QD}$. They also found a functional form for the dissipative force of tidal friction which can then be expressed as the acceleration due to tidal fiction $f_{\rm TF}$. These acceleration terms are useful because they can be added directly to the orbital equation of motion for the binary:

$$\ddot{\boldsymbol{r}} = -\frac{GM\boldsymbol{r}}{r^3} + \boldsymbol{f}_{\rm QD} + \boldsymbol{f}_{\rm TF} \ , \tag{9}$$

where $\boldsymbol{r}$ is the distance vector between the two bodies and $M$ is the combined mass. This enables the evaluation of both the dynamical and tidal evolution of a binary star system.

## 2.4 A Self-Consistent Evolution Model for Planetary Systems

Mardling & Lin (2002) were the first to recognize that the formulations by Eggleton et al. (1998) provided a powerful method for calculating the complex evolution of not just binary stars, but planetary systems as well. Based on their formulations, Mardling & Lin (2002) presented an efficient method for calculating self-consistently the tidal plus *N*-body evolution of a many-bodied system. Their work had a particular focus on planets, yet they emphasized that the method did not assume any specific mass ratio and that their schemes were entirely general. As such, they could be applied to any system of bodies.

The Mardling & Lin (2002) method lends itself best to a hierarchical (Jacobi) coordinate system. Since our primary interest involves the evolution of moons around a large planet, we use the planet's position as the origin of a given system. The orbit of the next closest body will be a moon, whose position is referred to the planet. The orbit of a third body is then referred to the center of mass of the planet and innermost moon, while the orbit of a fourth body is referred to the center of mass of the other three bodies. This system has the advantage that the relative orbits are simply perturbed Keplerian orbits so the osculating orbital elements are easy to calculate (Murray & Dermott 1999).

Mardling & Lin (2002) parameterized the acceleration terms to create equations of motion for systems containing up to four bodies. Let the masses of the four objects be $m_1, m_2, m_3,$ and $m_4$. For our study, $m_1$ always represented a planet and $m_2$ represented a moon. The third mass, $m_3$, represented a central star for 3-body systems, which are created by simply setting $m_4$ equal to zero. The planet and moon (being the closest pair of bodies in the system) are endowed with structure that is specified by their radii $S_1$ and $S_2$, moments of inertia $I_1$ and $I_2$, spin vectors $\mathbf{\Omega}_1$ and $\mathbf{\Omega}_2$, their quadrupole apsidal motion constants (or half the appropriate Love numbers for planets with some rigidity) $k_1$ and $k_2$, and their *Q*-values $Q_1$ and $Q_2$. Body 3 is assumed to be structureless, meaning it is treated as a point mass. We should note one necessary correction to Eq. (7) in Mardling & Lin (2002), the $m_3\boldsymbol{\beta}_{34}$ term in the first set of brackets should be replaced with $m_4\boldsymbol{\beta}_{34}$.

The total angular momentum and total energy for each system were calculated as well as the evolution of the spin vectors for bodies 1 and 2 (Eqs. 9, 10, and 15 in Mardling & Lin 2002). We should also note one minor correction to Eq. (14) in Mardling & Lin (2002), the factor $S_1^5$ should be replaced by $S_2^5$.

The acceleration from quadrupole distortion $f_{QD}$ was derived from a potential and under prevailing circumstances would conserve total energy. On the other hand, the acceleration due to tidal friction $f_{TF}$ cannot be derived from a potential and represents the effects of a slow dissipation of orbital energy. The dissipation within the moon causes change in its orbital semi-major axis and spin vector. We defined the rate of energy loss from tidal heating $\dot{E}_{tide}$ using Eq. (71) in Mardling & Lin (2002). While total orbital energy is not conserved, we still expect conservation of total angular momentum. Since $\dot{E}_{tide}$ represents the tidal heating in a satellite according to this tidal model, the surface heat flux in the satellite can be defined as

$$h = \dot{E}_{tide}/4\pi S_2^2 \ . \tag{10}$$

### 2.4.1 The Simulation Code

Using the equations of motion defined in the previous subsection, we designed a computer program that has the ability to simultaneously consider both dynamical and tidal effects, and with it, we simulated exomoon tidal evolution in low-mass star systems. The program code was written in C++ and a Bulirsch-Stoer integrator with an adaptive timestep (Press et al. 2002) was used to integrate the equations of motion.

The robustness of the code was evaluated by tracking the relative error in total angular momentum, defined as $(L_{out} - L_{in})/L_{in}$, where $L_{in}$ is the system angular momentum at the start of a simulation and $L_{out}$ is the angular momentum at a later point. The size of the relative error could be controlled by adjusting an absolute tolerance parameter. However, as is often the case with direct orbit integrators, the problem of systematic errors in the semi-major axis existed. The simulations were monitored to ensure a maximum allowed error of $10^{-9}$ for the total angular momentum. The algorithm efficiency required about seven integration steps per orbit, for the smallest orbit. Relative errors in total energy $((E_{out} - E_{in})/E_{in})$ were also tracked, although, conservation of mechanical energy was not expected for our systems and, hence, not included as a performance constraint in our simulations.

A primary drawback to directly integrating orbital motion is the extensive computational processing times required to simulate long-term behavior. With our particular evolution model the situation is compounded by extra calculations for tidal interactions. Early tests for code efficiency indicated a processing time of roughly two days per Myr of simulated time for 3-body systems (two extended objects and a third point mass object). Integration time significantly increases when additional bodies are considered, so a maximum of 3 bodies was chosen for this initial study. Since the integration timestep is effectively controlled by the object with the shortest orbital period, the exact processing times varied significantly between wide orbit and short orbit satellites. For our study the size of the orbit was determined from the theoretical habitable zone around a low-mass star (see subsection 3.1.2). From these results we predicted an ability to simulate satellite systems with timescales on order of $10^7$ yr with our limited computational resources.

## 3 SIMULATING EXOMOON EVOLUTION: SETUP

### 3.1 System Architectures and Physical Properties

As part of our investigation into the evolution of exomoons around giant planets in the HZ of low-mass stars, we evaluated two different system architectures. The first involved a minimal 2-body system consisting of only a planet and a moon; the second was a 3-body system of one planet, one moon, and a central star. For each system, the planet and moon were given structure while a star was treated as a point mass (which significantly reduce the required number of calculations per timestep). For each 3-body simulation, a corresponding 2-body simulation was performed for the same planet-moon binary. Comparing the two simulations would provide a baseline for determining the stellar contribution to a moon's long-term evolution.

**Table 1.** Physical properties for a hypothetical Mars-like exomoon. The parameters $A$, $B$ and $C$ are the principal moments of inertia.

| Parameter | Value |
|---|---|
| Mass ($M$) | $0.107\ M_\oplus$ |
| Mean Radius ($R$) | $0.532\ R_\oplus$ |
| Love Number ($k_L$) | 0.16 |
| Bond Albedo ($\alpha$) | 0.250 |
| Dissipation Factor ($Q$) | 80 |
| $C/MR^2$ | 0.3662 |
| $C/A$ | 1.005741 |
| $C/B$ | 1.005044 |

Although thousands of extrasolar planets have been detected, very little information is known about their internal structure and composition. Notwithstanding, work is being conducted to offer a better understanding (for example, see Unterborn et al. (2016) and Dorn et al. (2017)). Recognizing that limits for tidal dissipation depend critically on these properties, without this information, objects in the solar system provide the best guide for hypothesizing the internal structure and dynamics of extrasolar bodies. For this reason, we used known examples from our Solar System to model the hypothetical extrasolar satellite systems.

### 3.1.1 The Exomoon Model

Formation models for massive exomoons show reasonable support for the formation of moons with roughly the mass of Mars around super-Jovian planets (Canup & Ward 2006; Heller & Pudritz 2015), which is near the current detection limit of ~0.1 $M_\oplus$ (Heller 2014; Kipping et al. 2015) and also lies in the preferred mass regime for habitable exomoons (see Sect. 1). For these reasons, we chose to model the physical structure of our hypothetical exomoons after planet Mars.

The specific physical properties used in our moon model are shown in Table 1. The Bond albedo represents current estimates for Mars, although one could argue that an Earth-like value of 0.3 might be equally appropriate since we consider habitable exomoons. Bond albedos are not actually used in the evolution simulations, they are utilized afterwards to estimate the global flux $F_{glob}$ received by the moon, see Eq. (4). Habitability considerations from the global flux are made in comparison to the critical flux for a runaway greenhouse ($F_{RG}$, see subsection 2.2). For a Mars-mass exomoon, the critical flux is $F_{RG} = 269\ \mathrm{W\,m^{-2}}$. We decided to use the lower Bond albedo of Mars, keeping in mind that it would produce slightly higher estimates of $F_{glob}$. If the simulation results showed a global flux > $269\ \mathrm{W\,m^{-2}}$ we would then also consider an Earth-like Bond albedo.

The Love number and dissipation factor are also based on recent estimates for Mars (Yoder et al. 2003; Bills et al. 2005; Lainey et al. 2007; Konopliv et al. 2011; Nimmo & Faul 2013). Their specific values represent the higher dissipation range of the estimates. This choice produces a slightly faster tidal evolution and also tests the extent of the gravitational influence of the star against a slightly higher rate of energy loss that continually works to circularize the orbit of the moon. The principal moments of inertia also reflect estimates for Mars (Bouquillon & Souchay 1999) and represent a tri-axial ellipsoid for the overall shape of the body. We realize that a Mars-like exomoon orbiting close to a giant planet would undoubtedly develop a different bodily shape than the current shape of Mars. If we assume a constant moment of inertia, then the principal moments only become important in calculating the evolution of the moon's spin vector (see Eq. 15 in Mardling & Lin 2002). To minimize this importance, we set the moon's initial obliquity to zero and started each simulation with the moon in synchronous rotation. Under these conditions there is minimal change to the moon's spin vector and the given triaxial shape is effective at keeping the moon tidally locked to the planet as its orbit slowly evolves. In that sense, another choice in shape could have equally served the same purpose.

We only considered prograde motion for the moon relative to the spin of the planet. In keeping with our use of local examples, we modeled the moon's orbital distance after known satellite orbits around giant planets in the Solar System. The large solar system moons Io, Europa, Ganymede, Titan, and Callisto happen to posses roughly evenly spaced intervals for orbital distance in terms of their host planet's radius $R_p$ (5.9, 9.6, 15.3, 21.0, and 26.9 $R_p$, respectively). In reference to this natural spacing, when discussing orbital distances of a moon we will often refer to them as Io-like or Europa-like orbits, etc.

Considering stability constraints, and assuming our satellite systems are not newly formed, we would not expect high eccentricities for stable exomoon orbits. For this reason we use an initial eccentricity of 0.1 for our 2-body and 3-body models. We then monitor the moon's evolution as the tidal dissipation works to circularize the orbit.

### 3.1.2 The Planet Models

Moon formation theories suggest that massive terrestrial moons will most likely be found around giant planets. With this consideration, we chose to model our hypothetical planet after the two most massive planets in our Solar System, specifically, planets Jupiter and Saturn. The physical properties used in our planet models are shown in Table 2. The Bond albedos, Love numbers, and dissipation factors are based on current estimates (Gavrilov & Zharkov 1977; Hanel et al. 1981, 1983; Meyer & Wisdom 2007; Lainey et al. 2009). We chose values that represent more substantial heating in the planets, similar to our choice for the Mars-like moons.

The normalized moments of inertia were also based on recent estimates (Helled 2011; Helled et al. 2011) and the three equal principal moments imply a spherical shape for the planet. This shape may be unrealistic in that a planet orbiting close to a star with a massive moon is unlikely to maintain a truly spherical shape. With spherical bodies we also do not match the exact shapes of the solar system planets, i.e., we ignore their oblateness that results from their short rotation periods (about 10 hrs for Jupiter, 11 hrs for Saturn). However, the planet's exact shape is not particularly import for the purposes of our simulations. We start each planet with zero obliquity and synchronous rotation relative to its orbit around the central star, which results in a rotation period that ranges from about 15 to 120 days. This setup is consistent with the prediction that planets in the HZ of low-mass stars will most likely be tidally locked to the star.

**Table 2.** Physical properties for hypothetical giant exoplanets. The planet shape is assumed spherical with principal moments of inertia $A = B = C$.

| Jupiter-like | |
|---|---|
| **Parameter** | **Value** |
| Mass ($M$) | 318 $M_\oplus$ |
| Mean Radius ($R$) | 11.0 $R_\oplus$ |
| Bond Albedo ($\alpha$) | 0.343 |
| Love Number ($k_L$) | 0.38 |
| Dissipation Factor ($Q$) | 35000 |
| $C/MR^2$ | 0.263 |

| Saturn-like | |
|---|---|
| **Parameter** | **Value** |
| Mass ($M$) | 95.2 $M_\oplus$ |
| Mean Radius ($R$) | 9.14 $R_\oplus$ |
| Bond Albedo ($\alpha$) | 0.342 |
| Love Number ($k_L$) | 0.341 |
| Dissipation Factor ($Q$) | 18000 |
| $C/MR^2$ | 0.21 |

To define the HZ for a given star we followed the work by Kopparapu et al. (2013), who estimated a variety of limits for the inner and outer edges around stars with effective temperatures ($T_{eff}$) in the range 2600 K $\leq$ $T_{eff}$ $\leq$ 7200 K. For this study we used their conservative limits corresponding to a moist greenhouse for the inner boundary and a maximum greenhouse for the outer boundary. In order to calculate these limits it is necessary to define the effective temperature and radius of the star. For our purposes, however, it was more convenient to scale the HZ based on stellar mass. The stellar radius can be estimated from the mass using an empirical relation derived from observations of eclipsing binaries (Gorda & Svechnikov 1999):

$$\log_{10} \frac{R_\star}{R_\odot} = 1.03 \log_{10} \frac{M_\star}{M_\odot} + 0.1. \tag{11}$$

Note that Eq. (11) is valid only when $M_\star \lesssim M_\odot$, which is valid for the dwarf stars we consider. The stellar luminosity can be estimated from the mass by

$$\lambda = 4.101\mu^3 + 8.162\mu^2 + 7.108\mu + 0.065, \tag{12}$$

where $\lambda = \log_{10}(L/L_\odot)$ and $\mu = \log_{10}(M_\star/M_\odot)$ (Scalo et al. 2007). With luminosity and radius, the stellar effective temperature can be calculated using the familiar relationship

$$L = 4\pi\sigma_{SB}R_\star^2\ T_{eff}^4. \tag{13}$$

An online database of confirmed exoplanet detections[3] lists almost 3,500 confirmed planets in total. Of those, only about 8 percent have host stars with mass $\lesssim 0.6\ M_\odot$. This relatively small sample is most likely due to selection bias as detection techniques have evolved. The majority of these detections occurred in recent years and the number is expected to grow. While several large planets can be found orbiting low-mass stars, only a small number orbit inside the HZ. The percentage of these systems is questionable since a significant number of confirmed planets in the database are missing key parameters such as planet mass or orbit distance. One known planet in particular corresponds nicely with our Saturn model. The planet HIP 57050 b has a mass of 0.995 ($\pm$0.083) $M_{Sat} \times \sin(i)$ (Haghighipour et al. 2010), where $M_{Sat}$ is the mass of Saturn and $i$ is the unknown inclination between the normal of the planetary orbital plane and our line of sight. Since no other information is available about the physical properties of these giant planets, only direct comparisons relating to their masses can be made. None of the low-mass HZ candidates matches directly with Jupiter, but two have masses about double that of Jupiter (GJ 876 b and HIP 79431 b). This at least supports the plausibility for the existence of Jupiter mass planets in the HZ of dwarf stars.

Each simulated system consisted of only one planet. Each planet started with a circular orbit relative to the star at a distance inside the stellar HZ. To conserve computational processing time, we limited exploration of the HZ to just two specific orbital distances per planetary system. The first location was in the center of the HZ for a given star mass, otherwise defined as

$$a_{center} = (\text{inner edge} + \text{outer edge})/2 . \tag{14}$$

This particular location was to serve as a reference point for the next round of simulations. If most of our hypothetical satellites already experienced intense tidal heating at the center, then the next location should be further out in the zone. On the other hand, if the surface heating rates were below the proposed maximum for habitability ($h_{max} = 2\,\text{W}\,\text{m}^{-2}$), then we would move inward for the next round. As we will show, after simulating satellite systems in the center of the HZ it became clear that the second round of simulations should involve the inner HZ.

While the innermost edge was a reasonable option to explore, an Earth-equivalent distance had obvious attraction. By 'Earth-equivalent' we refer to the Earth's relative position in the Sun's HZ as compared to the total width of the zone. Conservative estimates by Kopparapu et al. (2013) place the inner edge of the Sun's HZ at 0.99 AU and the outer edge at 1.67 AU. With these boundaries we define Earth's relative location within the solar HZ as

$$d_{rel} = 1 - (1.67\,\text{AU} - 1\,\text{AU})/(1.67\,\text{AU} - 0.99\,\text{AU}) = 0.0147 . \tag{15}$$

For a given star, we use this relative location and the width of its HZ to define a planet's Earth-equivalent orbital distance as

$$a_{eq} = \text{inner edge} + (\text{outer edge} - \text{inner edge}) \times d_{rel} . \tag{16}$$

### 3.1.3 The Star Models

Information relating to a star's mass, radius, and effective temperature is required to calculate its circumstellar HZ and to estimate the global flux received by a moon. In our simulations the stars are treated as point masses, and so the stellar mass is the only physical characteristic used in the actual simulation of a system. For our study we considered a star mass range from 0.075 $M_\odot$ to 0.6 $M_\odot$. These would consist of mostly M spectral type stars and possibly some late K type stars, with typical surface temperatures less than 4,000 K. This range of stars is sometimes referred to as red dwarf stars.

**Table 3.** Eccentricity damping timescale estimates for a Mars-like moon at different orbital distances.

| Jupiter-like Host Planet | |
|---|---|
| **Orbital Distance** | $\tau_e$ **(years)** |
| Io-like ($5.9R_{\mathrm{Jup}}$) | $4 \times 10^5$ |
| Europa-like ($9.6R_{\mathrm{Jup}}$) | $9 \times 10^6$ |
| Ganymede-like ($15.3R_{\mathrm{Jup}}$) | $2 \times 10^8$ |
| Titan-like ($21.0R_{\mathrm{Jup}}$) | $1 \times 10^9$ |
| Callisto-like ($26.9R_{\mathrm{Jup}}$) | $7 \times 10^9$ |

| Saturn-like Host Planet | |
|---|---|
| **Orbital Distance** | $\tau_e$ **(years)** |
| Io-like ($5.9R_{\mathrm{Sat}}$) | $6 \times 10^5$ |
| Europa-like ($9.6R_{\mathrm{Sat}}$) | $1 \times 10^7$ |
| Ganymede-like ($15.3R_{\mathrm{Sat}}$) | $3 \times 10^8$ |
| Titan-like ($21.0R_{\mathrm{Sat}}$) | $2 \times 10^9$ |
| Callisto-like ($26.9R_{\mathrm{Sat}}$) | $1 \times 10^{10}$ |

### 3.2 Estimates for Eccentricity Damping Timescales

We can estimate the eccentricity damping timescales ($\tau_e$) for our hypothetical satellite systems following Eq. (1). The timescales are summarized in Table 3 for the two planet models and for the various moon orbital distances considered. It was fortunate to see estimates around 1 Myr since integration times of order $10^7$ yr represent achievable computational processing times. Results showed that we were able to simulate complete tidal evolution for moons with Io-like and Europa-like orbital distances. However, it was unlikely that we could show any significant evolution for the wider orbits on these timescales. It should be noted that Eq. (1) was derived from a two body calculation. As such, it does not take into account any perturbing effects from additional bodies. Since the degree to which the central star would influence a moon's orbit was unknown, we decided to include the wider orbits in our considerations.

## 4    2-BODY ORBITAL EVOLUTION: TIDES VERSUS NO-TIDES

Figure 1 demonstrates some typical evolutions for Mars-like moons around a Jupiter-like planet. Solid curves are simulations that included tidal interactions. The lower, middle, and upper red curves represent Io-like, Europa-like, and Ganymede-like moon orbital distances, respectively. As expected, the rate of change depended strongly on the moon's orbital distance. When repeated without tidal interactions, all simulations had the same result, with no significant change to the orbits (see dashed lines). Simulations for

Titan-like and Callisto-like orbital distances showed very little change over 10 Myr whether tides were considered or not. Their results are nearly identical to the dashed curves in each plot so they were not included in Fig. 1.

Simulations for a Saturn-like host planet were also conducted. The satellite evolutions are similar to those in Figure 1 for a Jupiter-like host planet, with one exception. An Io-like moon orbit around Saturn experienced significantly greater change in semi-major axis than the Io-like moon orbit around Jupiter. In just 7 Myr the semi-major axis decayed to 75% of its initial value with a Saturn-like host planet, compared to about 96% with a Jupiter-like host. The explanation is related to our approach of normalizing planet-moon distances by the radius of the respective host planet. A moon's initial orbital distance is defined by $R_p$, the planet radius (an 'Io-like' orbit is 5.9 $R_p$). Because of Saturn's smaller radius a moon with an Io-like orbital distance is actually closer to the planet than an Io-like orbit around Jupiter, since 5.9 $R_{\mathrm{Sat}} < 5.9 R_{\mathrm{Jup}}$. Following this formulation for the orbital distances, identical moons around each planet will experience different heating rates.

While different tidal heating rates explain part of the discrepancy in semi-major axis evolution between Saturn-like and Jupiter-like host planets, it does not account for all of it. Figure 1(a) shows the Io-like orbit around the Jupiter-like host was circularized after $\approx 2.5$ Myr, at which point the tidal heating ceases in our synchronized moon with zero spin-orbit alignment. Therefore, any change in the semi-major axis after $\approx 2.5$ Myr is no longer attributed to tidal dissipation in the moon. Instead, tidal dissipation must occur in the planet, which itself is synchronized to the star, causing the planet's tidal bulge to lag behind the moon in our systems. This evolution makes the moon spiral towards the planet. Ultimately, the moon will be destroyed near the planetary Roche radius (Barnes & O'Brien 2002). Tidal evolution of moons in Europa-wide orbits are much slower, with a fractional decrease of the semi-major axis of $\approx 0.1\%$ per 10 Myr.

Using the corresponding slope in Fig. 1(b) for the Jupiter system, we roughly estimate a total lifetime of $\approx 200$ Myr for the moon with an Io-like orbit. As a comparison of this timescale to theoretical predictions, Goldreich & Soter (1966) derived a relation for the change in semi-major axis of a satellite as

$$\frac{da}{dt} = \frac{9}{2}\left(\frac{G}{M_p}\right)^{1/2}\frac{R_p^5}{Q_p'}\frac{M_s}{a^{11/2}} \ . \tag{17}$$

This model assumes that the planet's $Q$-value is frequency-independent, which is only a fair approximation over a very narrow range of frequencies and therefore implicitly limits the model to low eccentricities and inclinations (Greenberg 2009).

As an alternative, Mardling (2011) suggested to use the constant time lag ($\tau$) model to study the tidal evolution of hot Jupiters based on the knowledge of tidal dissipation in Jupiter. This is done by assuming that $\tau$ (rather than $Q$) is common to Jupiter-mass planets. Mardling (2011) propose to approximate the $Q$-value of a planet with orbital period $P$ via $Q/Q_J = P/P_{\mathrm{Io}}$, where $Q_J$ and $P_{\mathrm{Io}} = 1.77$ d are the tidal dissipation constant of Jupiter and the orbital period of Io, respectively. The shortest orbital period for our Jupiter-like planets was 13 d, which gives us an estimated $Q$-value 7.3

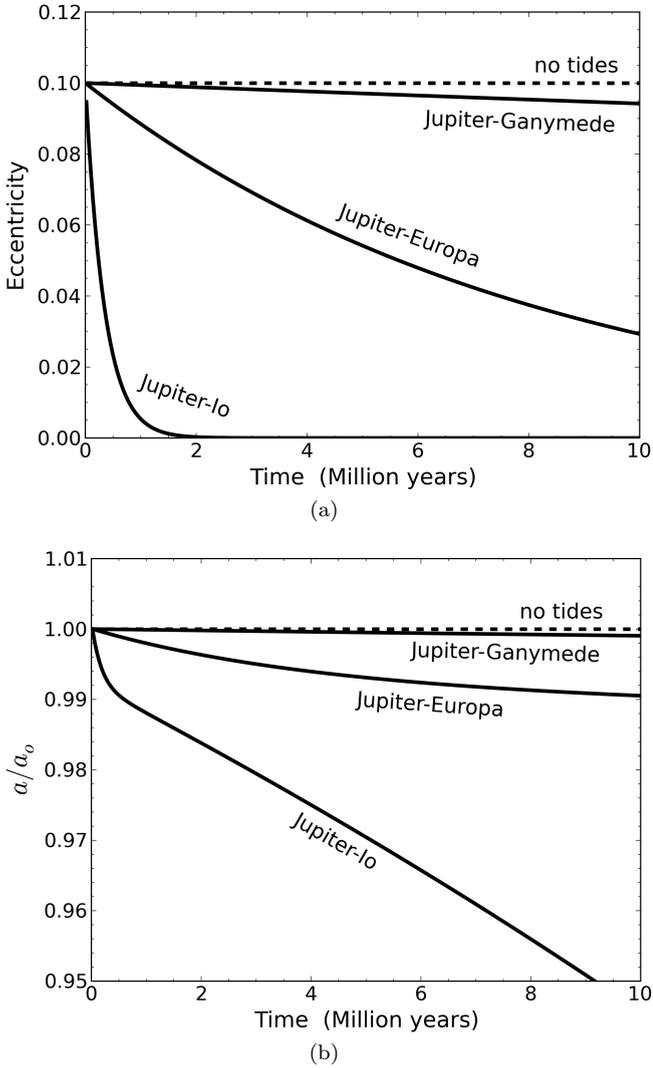

(a)

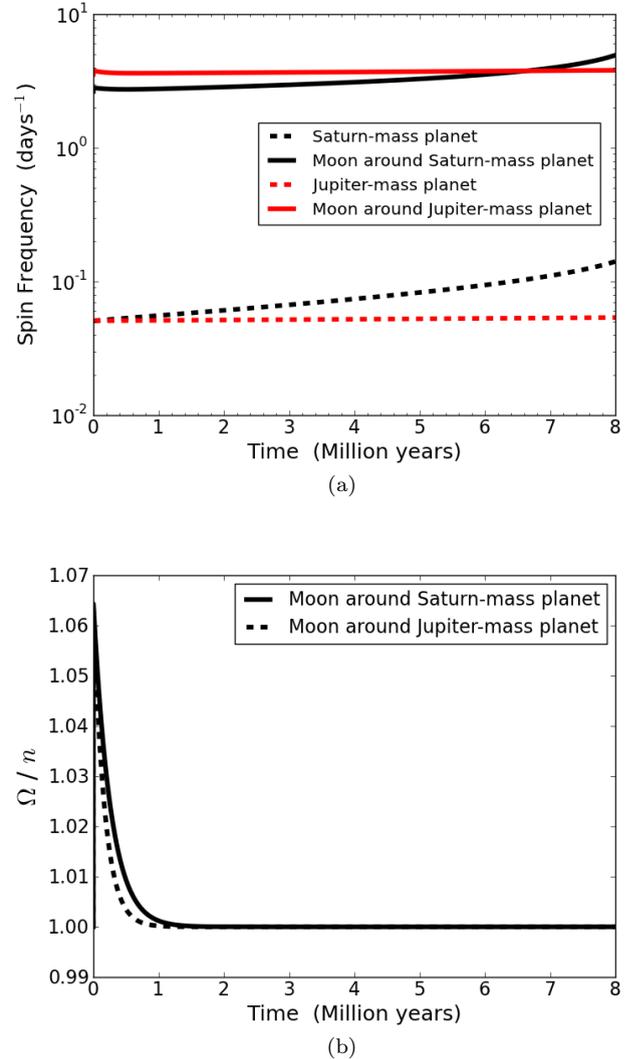

(a)

(b)

**Figure 1.** Orbital evolution of a single Mars-like moon around a Jupiter-like planet in four different cases. The dashed line assumes no tides. The solid lines all assume an initial eccentricity of 0.1 but with the moon starting at different semi-major axes from the planet, i.e. in an Io-wide, a Europa-wide, and in a Ganymede-wide orbit around the planet (see labels).

(b)

**Figure 2.** Spin evolution of a moon and its host planet, assuming an Io-like moon orbit. (**a**) Spin frequencies of the planets (dashed lines) and moons (solid lines) assuming host planets akin to Jupiter (red) and Saturn (black). (**b**) Ratio of spin magnitude ($\Omega$) and mean motion ($n$) for the moons represented in plot a.

times that for Jupiter. Applying this result and using the values listed in Table 2, Eq. (17) estimates a timescale of $\approx 560$ Myr for the orbital decay. While a few times longer than the rough estimate from our simulations, they agree on the order of magnitude.

A graphical representation of the spin evolution for the Io-like systems is provided in Fig. 2. Comparing the planet spin rates (dashed lines) in Fig. 2(a), the Saturn-like planet (black dashed line) had a noticeable increase relative the Jupiter-like host (red dashed line), note the log scaling for the ordinate axis. This difference is explained by the unequal distances for the otherwise identical Mars-like moons as well as the unequal moments of inertia for the planets. The short satellite orbit around Saturn combined with its lower moment of inertia leads to the small, but noticeable increase in spin. That small increase causes the moon to almost double its initial spin rate. On the other hand, the wider orbit and more massive Jupiter host planet caused very little change for both the moon and planet in this system.

The tidally locked state of the moons throughout the simulated time are demonstrated in Fig. 2(b). A perfectly synchronized spin rate would produce a value of 1 for the ratio between spin magnitude and mean orbital motion. Notice that the ratio is not exactly 1 for the first few Myr, although the difference is small. This discrepancy can be explained by the shapes of the orbit during that time. Referring to the eccentricity plotted in Fig. 1(a), the spin cannot completely synchronize with the orbital period until the eccentricity approaches zero.

As a final observation, one similarity between the Jupiter and Saturn-like systems is that there is little change in the orbital elements over a period of 10 Myr for Ganymede-like orbits and greater. This was not particularly surprising as the result was predicted from the damping timescales. At these wider orbits the dissipation rate would be much lower for a given eccentricity due to the $a^{-15/2}$ dependence of tidal heating. On the other hand, a central star's potential for exciting the eccentricity is still untested, so we included the wider orbits in our 3-body analysis.

**Table 4.** 3-body stability summary for a Jupiter-like host planet. The moon semi-major axes are presented as fractions of $R_{\mathrm{Hill}}$. Gray shaded cells represent unstable moon orbits.

**Table 5.** 3-body stability summary for a Saturn-like host planet. The moon semi-major axes are presented as fractions of $R_{\mathrm{Hill}}$. Gray shaded cells represent unstable moon orbits.

| **Earth-Equivalent Planet Orbit** | | | | | | **Earth-Equivalent Planet Orbit** | | | | | |
|---|---|---|---|---|---|---|---|---|---|---|---|
| Star Mass | Moon Semi-major Axis | | | | | Star Mass | Moon Semi-major Axis | | | | |
| | $5.9\,R_{\mathrm{Jup}}$ | $9.6\,R_{\mathrm{Jup}}$ | $15.3\,R_{\mathrm{Jup}}$ | $21.0\,R_{\mathrm{Jup}}$ | $26.9\,R_{\mathrm{Jup}}$ | | $5.9\,R_{\mathrm{Sat}}$ | $9.6\,R_{\mathrm{Sat}}$ | $15.3\,R_{\mathrm{Sat}}$ | $21.0\,R_{\mathrm{Sat}}$ | $26.9\,R_{\mathrm{Sat}}$ |
| $0.1\,M_{\odot}$ | 0.54 | 0.88 | 1.4 | 1.9 | 2.5 | $0.1\,M_{\odot}$ | 0.66 | 1.1 | 1.7 | 2.3 | 3.0 |
| $0.2\,M_{\odot}$ | 0.32 | 0.52 | 0.82 | 1.1 | 1.4 | $0.2\,M_{\odot}$ | 0.39 | 0.63 | 1.0 | 1.4 | 1.8 |
| $0.3\,M_{\odot}$ | 0.25 | 0.41 | 0.66 | 0.91 | 1.2 | $0.3\,M_{\odot}$ | 0.31 | 0.50 | 0.80 | 1.1 | 1.4 |
| $0.4\,M_{\odot}$ | 0.20 | 0.33 | 0.53 | 0.72 | 0.93 | $0.4\,M_{\odot}$ | 0.25 | 0.40 | 0.64 | 0.88 | 1.1 |
| $0.5\,M_{\odot}$ | 0.16 | 0.26 | 0.41 | 0.57 | 0.72 | $0.5\,M_{\odot}$ | 0.19 | 0.32 | 0.50 | 0.69 | 0.88 |
| $0.6\,M_{\odot}$ | 0.12 | 0.20 | 0.31 | 0.43 | 0.55 | $0.6\,M_{\odot}$ | 0.15 | 0.24 | 0.38 | 0.53 | 0.67 |

| **Planet Orbit in Center of HZ** | | | | | | **Planet Orbit in Center of HZ** | | | | | |
|---|---|---|---|---|---|---|---|---|---|---|---|
| Star Mass | Moon Semi-major Axis | | | | | Star Mass | Moon Semi-major Axis | | | | |
| | $5.9\,R_{\mathrm{Jup}}$ | $9.6\,R_{\mathrm{Jup}}$ | $15.3\,R_{\mathrm{Jup}}$ | $21.0\,R_{\mathrm{Jup}}$ | $26.9\,R_{\mathrm{Jup}}$ | | $5.9\,R_{\mathrm{Sat}}$ | $9.6\,R_{\mathrm{Sat}}$ | $15.3\,R_{\mathrm{Sat}}$ | $21.0\,R_{\mathrm{Sat}}$ | $26.9\,R_{\mathrm{Sat}}$ |
| $0.1\,M_{\odot}$ | 0.38 | 0.61 | 0.98 | 1.3 | 1.7 | $0.1\,M_{\odot}$ | 0.46 | 0.75 | 1.2 | 1.6 | 2.1 |
| $0.2\,M_{\odot}$ | 0.22 | 0.36 | 0.57 | 0.79 | 1.0 | $0.2\,M_{\odot}$ | 0.27 | 0.44 | 0.70 | 0.96 | 1.2 |
| $0.3\,M_{\odot}$ | 0.18 | 0.29 | 0.46 | 0.63 | 0.80 | $0.3\,M_{\odot}$ | 0.22 | 0.35 | 0.56 | 0.77 | 0.98 |
| $0.4\,M_{\odot}$ | 0.14 | 0.23 | 0.37 | 0.50 | 0.65 | $0.4\,M_{\odot}$ | 0.17 | 0.28 | 0.45 | 0.61 | 0.79 |
| $0.5\,M_{\odot}$ | 0.11 | 0.18 | 0.29 | 0.39 | 0.51 | $0.5\,M_{\odot}$ | 0.14 | 0.22 | 0.35 | 0.48 | 0.62 |
| $0.6\,M_{\odot}$ | 0.09 | 0.14 | 0.22 | 0.30 | 0.39 | $0.6\,M_{\odot}$ | 0.10 | 0.17 | 0.27 | 0.37 | 0.47 |

## 5  3-BODY SIMULATIONS

We continue with systems consisting of one star, one planet, and one moon. The Mars-like moon was given an initial eccentricity of 0.1, measured relative to the planet. We used this high value for eccentricity to cover a wide range of formation possibilities and allowed the eccentricity to decay as a result of tidal dissipation in the moon and planet. We monitored the simulations to see if the moon orbits would settle to steady, non-zero eccentricities long after the orbits should have circularized due to the tidal heating. For each satellite system we also ran a nearly identical simulation with the moon instead starting with a circular orbit. In this way we could test if the stellar perturbation raised the eccentricity to the same steady state value as was achieved following the decay from a higher initial value. For each 3-body simulation, we ran a corresponding 2-body simulation with just the planet and moon to show what the evolution would be without the influence of the star.

### 5.1  3-body Stability Considerations

With a central star ranging in mass from $0.075\,M_{\odot}$ to $0.6\,M_{\odot}$ and our confined planetary orbits within the relatively close stellar HZs, the exomoon Hill stability became very limited.

We defined the stability region by the Hill radius ($R_{\mathrm{Hill}}$), given by the relation

$$R_{\mathrm{Hill}} = a_{\mathrm{P}} \left( \frac{M_{\mathrm{p}}}{3M_{\star}} \right)^{1/3}, \qquad (18)$$

where $a_{\mathrm{P}}$ is the semimajor axis of the planet's orbit around the star, $M_{\mathrm{p}}$ and $M_{\star}$ are the masses of the planet and star, respectively. An extended study of moon orbital stability was conducted by Domingos et al. (2006b), who showed that the actual stability region depends upon the eccentricity and orientation of the moon's orbit. For prograde moons the stable region is:

$$a_{\mathrm{s,max}} = 0.4895 R_{\mathrm{Hill}}(1.0000 - 1.0305\,e_{\mathrm{p}} - 0.2738\,e_{\mathrm{s}}), \qquad (19)$$

where $e_{\mathrm{p}}$ and $e_{\mathrm{s}}$ are the orbital eccentricities of the planet and satellite, respectively. Following Eqs. (18) and (19), we expected a stability limit of $\sim 0.4\,R_{\mathrm{Hill}}$ for our chosen moon orbits, and so we only simulated systems for which the moon's semi-major axis about the planet was $< 0.5\,R_{\mathrm{Hill}}$. Our initial tests showed that for even the tightest exomoon orbit, which was an Io-like orbit at $a = 5.9\,R_{\mathrm{p}}$, no stable systems were found around stars less than $0.1\,M_{\odot}$. We therefore limited extended simulations to stellar masses $\geq 0.1\,M_{\odot}$, using intervals of $0.1\,M_{\odot}$ for our systems (i.e., the considered star masses were 0.1, 0.2, 0.3, 0.4, 0.5, and $0.6\,M_{\odot}$).

The stable region around a planet slowly increased with stellar mass (i.e. with increasingly larger HZ distances). As explained in Sect. 3.1.1, we considered a discrete range of moon orbital distances for each star-planet pair. Not surprising, many of the wider moon orbits were unstable for the lowest star masses. This instability resulted in the moon becoming unbound from the planet. We found that the 3-body simulations were only able to maintain long-term stability for a satellite orbit of $a_{\mathrm{s}} \lesssim 0.4\,R_{\mathrm{Hill}}$, which was the predicted value described above. A summary of the stable 3-body systems is provided in Tables 4 and 5. Note in the tables that the moon semi-major axes are modeled after those of the Solar System moons Io, Europa, Ganymede, Titan, and Callisto, respectively. The planet distances from the star represented in the tables range from about 0.05 AU for the lowest stellar mass to about 0.4 AU for the highest stellar mass.

Due to perturbations from the third body the eccentricity of the moon experienced short-term oscillations with periods relating to the moon's orbital period around the planet and the planet's orbital period around the star. Because of this, when reporting orbital parameters and surface heat flux for the moon the values were first averaged over the moon's orbit and then averaged again over multiple planet orbits.

### 5.2 Extended Simulation Results

After completing our investigations into system stability and short-term behavior, we were left with a set of 3-body systems for which we could expect stability and reasonably predict the long-term behavior. In other words, our systems were feasable from a strictly dynamical perspective. The integrated times for long-term simulations were extended until one of three results was achieved: (1) the average surface heat flux fell well below the proposed maximum limit for habitability, $h_{max} = 2\,W\,m^{-2}$; (2) the average surface heat flux settled to a reasonably constant value; or (3) the tidal evolution was slow enough that it was not practical to continue the simulation further.

Examples of tidal evolution plots for Io-like and Europa-like moons with a Jupiter-like host planet are shown in Figs. 3 and 4. Similar plots for a Saturn-like host planet are included in Figs. 5 and 6. Plots for Ganymede-like, Callisto-like, and Titan-like moon simulations are not included since there was little to no change in the surface heat flux of those moons during the integrated time periods. Displayed in these figures is the average surface heat flux ($h$) in the moon as a function of time, with $h$ defined by Eq. (10). The individual plots in each figure represent low and high limit mass values considered for the central star. Note that a greater star mass signifies a larger orbital distance for the planet. Each plot includes four curves with two solid ones referring to 3-body systems and two dashed ones referring to 2-body systems. Both pair of curves contains one case with zero initial eccentricity and another scenario with an initial eccentricity of 0.1. Comparing the 2-body and 3-body curves is useful in demonstrating the star's influence on the long-term tidal evolution of the moon.

### 5.3 Discussion

The focus of this study involves the long-term behavior of the solid red curves in Figs. 3 through 6. With them we test the extent to which a low-mass star has influence over a moon's evolution in the HZ, as a consequence of the short distance of the HZ. This is done by comparing the solid red curve to the dashed red curve, which represents an isolated planet-moon binary that began the simulation with the same initial conditions, minus the star. The dashed red curve shows what the solid red curve would look like without the influence of the central star.

Since tidal heating works to circularize the orbits, initially circular orbits (black curves) should continue with little to no heating, unless there are significant perturbations to the orbit. The purpose of these simulations was to test if the central star would excite the heating to the same steady state value that was achieved through the slow decay of an initially higher value. If that result was observed, then there is confidence that a non-zero, steady-state value for the surface heat flux was the product of perturbations from the star, and not simply based on our choice of initial conditions or an external artifact of our computer code. This anticipated result was indeed observed for Io-like and Europa-like orbital distances, but not for the wider orbits (Ganymede-like, Titan-like, and Callisto-like orbits). This difference between the eccentric and circular simulations for wide orbit systems remained constant throughout a test period of 10 Myr.

Our specific interest relates to the difference in the solid red and dashed red curves by the end of each simulation. When the slope of the solid red line deviates from the dashed red line and eventually levels off, it suggests a sustainable value for the surface heat flux. This indicates that the star has significant influence on the moon, and the continual perturbations are restricting the moon's eccentricity from reaching zero. This behavior is observed for all the Io-like and Europa-like orbital distances. For our systems the timescales necessary to achieve the constant state were more than double the estimates listed in Table 3. For convenience we will refer to Io and Europa like orbits as "short" orbits. At these distances the initially circular simulations eventually reached the same approximate steady state values as the non-circular systems. This result helps to confirm the degree of excitation to which the central star can influence the tidal evolution of a moon.

For the shorter orbits, our results show that perturbations from a low-mass star can prevent circularization and are able to maintain tidal heating in the moon for extended periods of time. While this effect was predicted (Heller 2012), this is the first instance in which it was actually been tested with an evolution model that simultaneously and self-consistently considered gravitational and tidal effects.

At orbital distances $a_s \gtrsim 15R_p$, little change to the surface heat flux occurred over 10 Myr. For convenience, reference to these distances as "wide" orbits will correspond to Ganymede, Titan, and Callisto-like moon orbits. In wide orbits, we found little deviation from the 2-body models, which was expected based on estimates of eccentricity damping timescales and early 2-body test runs. Tidal heat fluxes for these orbits are $\lesssim 0.1\,W\,m^{-2}$, even with the influence of the star. This is an order of magnitude less than the shorter orbit moons and well below $h_{max}$. Such low heating rates would not be expected to noticeably affect the orbit over the timescales considered.

An examination of Fig. 5 shows unique behavior that is not observed in the other figures. In these simulations the surface heat flux rapidly increases near the end of the simulated time. The cause for this behavior was already discussed in Sect. 4. Using 2-body models, we found that Mars-like moons with Io-like orbits and Saturn-like host planets experience significant evolution in their semi-major axis as a result of tidal torques and exchanges of angular momentum. Our 3-body results indicate that the star's influence does little to change this long-term behavior. In this case, the rapid rise in tidal heating is due to the equally rapid decay of the moon's semi-major axis as the moon spirals ever closer to the planet. The sudden end to the plots in Fig. 5 reflects the early termination of the simulations when the moon approached the Roche radius at $2\,R_p$, where it would experience tidal disruption. Note that the complete inward spiral occurred in $< 10\,\mathrm{Myr}$.

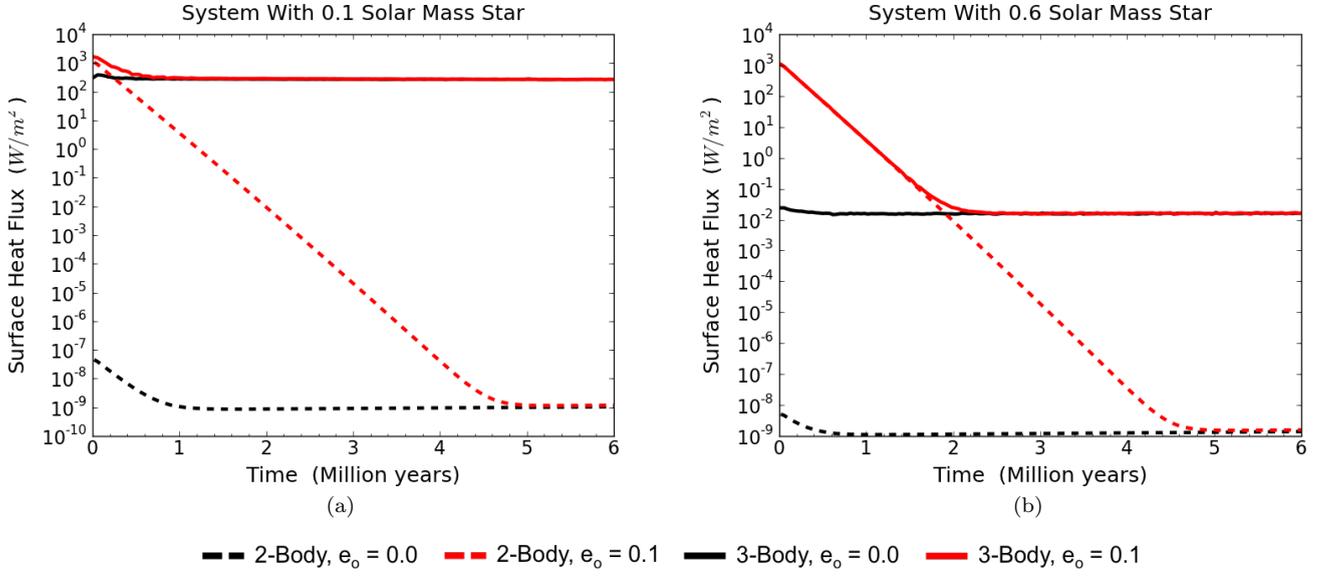

**Figure 3.** Examples of the orbital evolution of a Mars-mass moon in an Io-like orbit around a Jupiter-like planet.

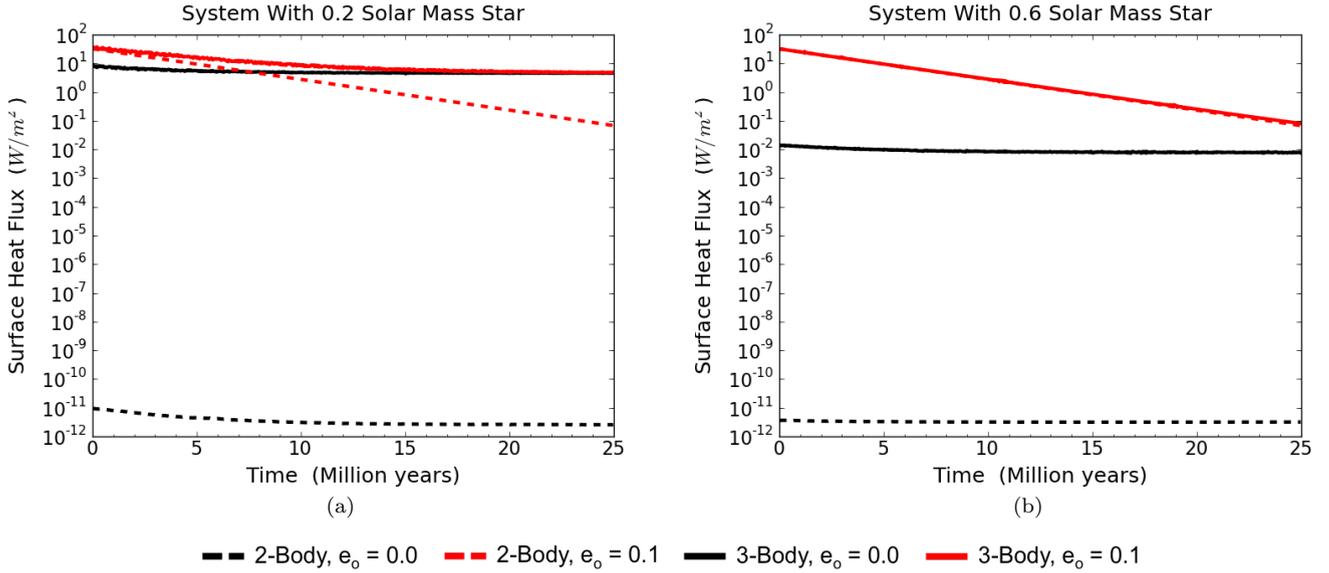

**Figure 4.** Examples of the orbital evolution of a Mars-mass moon in a Europa-like orbit around a Jupiter-like planet.

### 5.3.1  *A Complete Simulation Summary*

Our results for the 3-body simulations of Io-like and Europa-like satellite systems in the center of the HZ are summarized in Table 6, while the results for systems at Earth-equivalent distances are summarized in Table 7. Our interest is in heating values which are developed after noticeable tidal evolution and may be maintained for extended periods of time by stellar perturbations. The tables do not include results for wider orbits since, as discussed previously, there was no significant orbital evolution in these systems during the considered time periods. Therefore, surface heating values for wide orbit moons would simply reflect our choice for the initial orbital parameters of those systems and would not allow for any conclusive statements as to their ultimate tidal evolution. It is worth noting that very few wide orbit moons are even stable in our low-mass star range, and of those, all wide orbit systems had surface heat flux values below $2\,\mathrm{W\,m^{-2}}$ for moon eccentricities at or below 0.1.

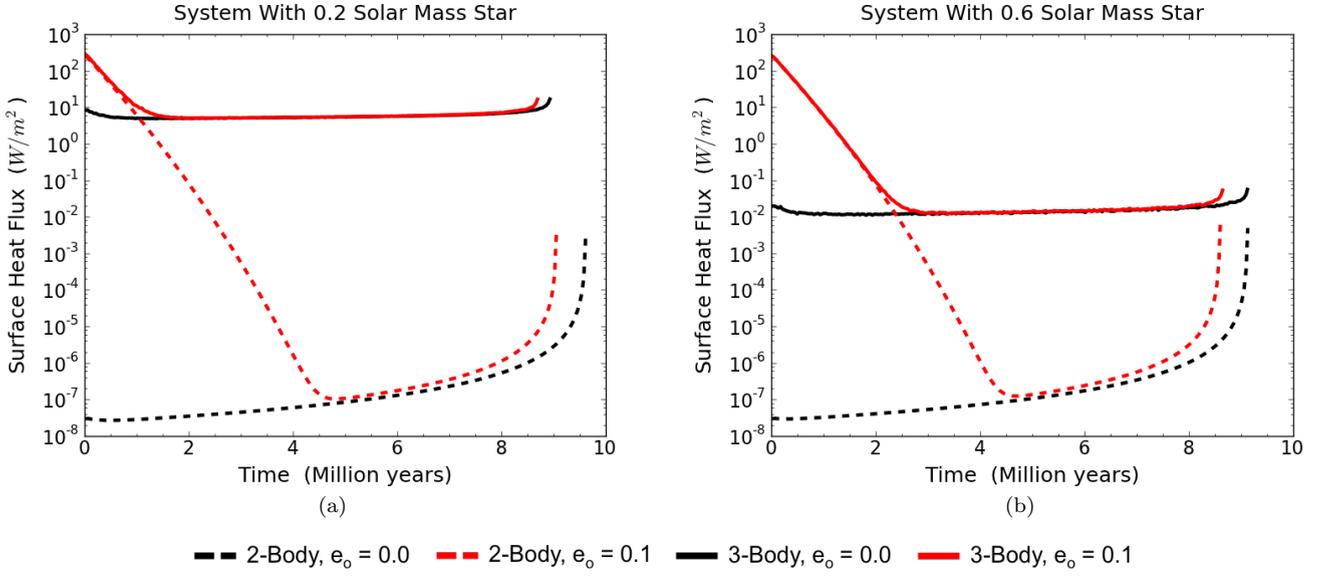

**Figure 5.** Examples of the orbital evolution of a Mars-mass moon in an Io-like orbit around a Saturn-like planet.

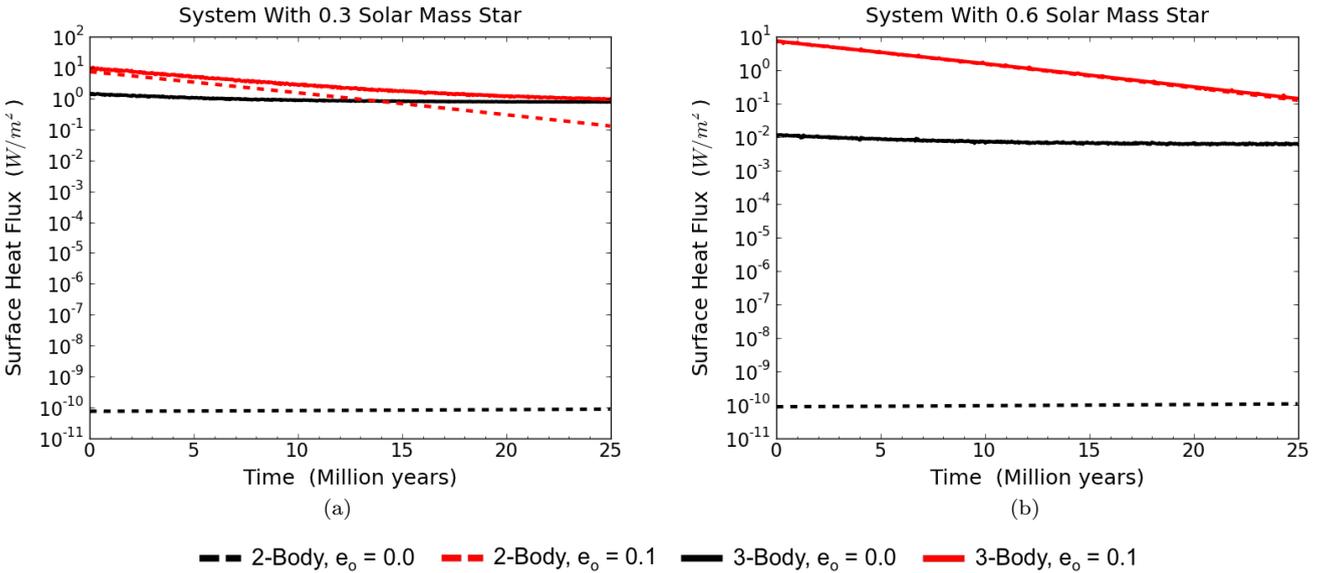

**Figure 6.** Examples of the orbital evolution of a Mars-mass moon in a Europa-like orbit around a Saturn-like planet.

Tables 6 and 7 show the average values at the end of each simulation and only include simulations that started with $e_0 = 0.1$ for the moon. Included is the moon's average surface heat flux, the average eccentricity, and the average semi-major axis (in comparison to its initial value). The initial value for the semi-major axis is listed as the moon distance. We also included the 'conventional' surface heat flux $h_{\mathrm{conv}}$. This value was calculated with Eq. (3) using the simulation results for the final eccentricity and semi-major axis and serves as a comparison between our chosen model and the conventional model for tidal heating. In regards to overall exomoon habitability, we included the orbit-averaged global flux ($F_{\mathrm{glob}}$) received by a satellite, as defined by Eq. (4). The total integration time considered for each simulation is also included.

Results from a 2-body simulation for each moon orbital distance were also included in Tables 6 and 7 to demonstrate the influence of the low-mass star in comparison to an iso-lated planet-moon system. The planet's spin rate in these simulations reflected a tidally locked rotation around a 0.6 solar mass star, which is the slowest of all considered planet spins.

**Table 6.** 3-body moon evolution summary for systems in the center of the stellar HZ. Values represent the average at the end of each simulation. Shaded rows are 2-body planet/moon systems. Red text indicates steady state values above $h_{max}$. Values of ** indicate the moon spiraled into the planet.

| Jupiter-like Host Planet | | | | | | | | |
|---|---|---|---|---|---|---|---|---|
| Moon Distance ($R_{Jup}$) | Star Mass ($M_\odot$) | Planet Distance (*AU*) | $h$ (W/m$^2$) | $h_{conv}$ (W/m$^2$) | $e$ | $a/a_0$ | $F_{glob}$ (W/m$^2$) | Sim. Time ($10^6$ Years) |
| | – | – | 0.00 | 0.00 | 0.000 | 0.95 | – | 10 |
| | 0.1 | 0.051 | 265 | 422 | 0.030 | 0.82 | 371 | 10 |
| Io-like | 0.2 | 0.11 | 6.46 | 11.1 | 0.008 | 0.94 | 113 | 10 |
| 5.9 | 0.3 | 0.16 | 1.51 | 2.59 | 0.004 | 0.95 | 107 | 10 |
| | 0.4 | 0.22 | 0.38 | 0.66 | 0.002 | 0.95 | 106 | 10 |
| | 0.5 | 0.30 | 0.08 | 0.14 | 0.001 | 0.95 | 107 | 10 |
| | 0.6 | 0.41 | 0.02 | 0.03 | 0.000 | 0.95 | 109 | 10 |
| | – | – | 0.07 | 0.07 | 0.005 | 0.99 | – | 25 |
| | 0.2 | 0.11 | 4.94 | 7.22 | 0.049 | 0.98 | 109 | 25 |
| Europa-like | 0.3 | 0.16 | 0.97 | 1.53 | 0.022 | 0.99 | 105 | 25 |
| 9.6 | 0.4 | 0.22 | 0.29 | 0.39 | 0.011 | 0.99 | 104 | 25 |
| | 0.5 | 0.30 | 0.12 | 0.12 | 0.006 | 0.99 | 105 | 25 |
| | 0.6 | 0.41 | 0.08 | 0.08 | 0.005 | 0.99 | 108 | 25 |

| Saturn-like Host Planet | | | | | | | | |
|---|---|---|---|---|---|---|---|---|
| Moon Distance ($R_{Sat}$) | Star Mass ($M_\odot$) | Planet Distance (*AU*) | $h$ (W/m$^2$) | $h_{conv}$ (W/m$^2$) | $e$ | $a/a_0$ | $F_{glob}$ (W/m$^2$) | Sim. Time ($10^6$ Years) |
| | – | – | ** | ** | ** | ** | ** | 10 |
| | 0.2 | 0.11 | ** | ** | ** | ** | ** | 10 |
| Io-like | 0.3 | 0.16 | ** | ** | ** | ** | ** | 10 |
| 5.9 | 0.4 | 0.22 | ** | ** | ** | ** | ** | 10 |
| | 0.5 | 0.30 | ** | ** | ** | ** | ** | 10 |
| | 0.6 | 0.41 | ** | ** | ** | ** | ** | 10 |
| | – | – | 0.13 | 0.13 | 0.012 | 0.97 | – | 25 |
| | 0.3 | 0.16 | 0.97 | 1.38 | 0.041 | 0.97 | 105 | 25 |
| Europa-like | 0.4 | 0.22 | 0.29 | 0.36 | 0.020 | 0.97 | 104 | 25 |
| 9.6 | 0.5 | 0.30 | 0.15 | 0.15 | 0.013 | 0.97 | 106 | 25 |
| | 0.6 | 0.41 | 0.14 | 0.14 | 0.013 | 0.97 | 108 | 25 |

As discussed in the previous subsection, moons with initial Io-like orbits and Saturn-like host planets spiraled into the planet after $\approx 10$ Myr. Over that same time period, distortion torques were much less effective at evolving the semi-major axis for Io-like orbits around Jupiter-like hosts. Note that for a given star mass, the two planets were at equal distances from the star and they both started with a tidally locked rotation relative to the star. This results in approximately equal spin rates for the two planets. The relative effectiveness of the torques is then due to the difference in orbital distances for the satellites. While moons with a Jupiter host evolved much slower in comparison to a Saturn host, their inward migration was still noticeable. If we assume the inward migration rate will only increase, we estimate a total lifetime < 200 Myr for a Mars-like moon with an initial Io-like orbit and Jupiter-like host planet in the HZ of a low-mass star.

Lifetime estimates for the inward spiral of all the Io-like and Europa-like moon orbit simulations are shown in Fig. 7, which shows results from simulations that started with circular moon orbits and thereby limits orbital effects of tidal heating and emphasizes evolution due to distortion torques. Simulations that started with eccentric orbits have shorter estimates for moon lifetimes. Estimated lifetimes for Europa-like orbits are significantly larger than those for Io-like orbits, around a maximum of 1 Gyr for Saturn-like host planets and several Gyr for a Jupiter-like host. Figure 7 indicates that the lifetime of moons at larger orbital distances are more sensitive to the planet's position in the HZ since they are more weakly bound to the planet and can experience greater influence from the central star. The sensitivity, however, becomes negligible for HZ distances around 0.6 solar mass stars.

One result that stands out in Tables 6 and 7 is the difference between the conventional model for surface heat flux and our chosen model. Slightly lower conventional values apply to systems involving higher eccentricities. As mentioned in Sect. 2.1, the conventional model can break down for high eccentricities, while our chosen model is still appropriate at large values. For systems involving lower eccentricities, $h_{conv}$ is consistently higher than $h$. The difference in the two values scales down for the lowest eccentricities until there is little difference between them. The greatest total difference was less than a factor of 2. Since the exact mechanisms of tidal dissipation are still poorly understood, some difference between two separate tidal models is not surprising.

A similarity between the two tidal models is a direct dependence on the tidal Love number $k_L$ and an inverse dependence on the dissipation factor $Q$. Higher values for $Q$ decrease the heating estimates and represent lower dissipation rates in the moon. Higher values for $k_L$ produce higher estimates for tidal heating and represent an increased susceptibility of the moon's shape to change in response to a tidal potential. This would also produce a larger acceleration due to the quadrupole moment of the moon, resulting in a shorter inward spiral towards the planet. Recent measurements of these two parameters for the solar system planet Mars contain significant uncertainty (e.g. $k_L = 0.148 \pm 0.017$; $Q = 88 \pm 16$; Nimmo & Faul 2013). While these uncertainties can be considered in the simulation results for a single orbit, the long term effects of varying $k_L$ and $Q$ would require each simulation to be repeated with the different values.

**Table 7.** 3-body moon evolution summary for systems at Earth-equivalent distances. Values represent the average at the end of each simulation. Shaded rows are 2-body planet/moon systems. Red text indicates steady state values above $h_{max}$. Values of ** indicates the moon spiraled into the planet.

| Moon Distance ($R_{Jup}$) | Star Mass ($M_\odot$) | Planet Distance ($AU$) | $h$ (W/m²) | $h_{conv}$ (W/m²) | $e$ | $a/a_0$ | $F_{glob}$ (W/m²) | Sim. Time ($10^6$ Years) |
|---|---|---|---|---|---|---|---|---|
| Jupiter-like Host Planet ||||||||
| Io-like 5.9 | – | – | 0.00 | 0.00 | 0.000 | 0.95 | – | 10 |
| | 0.2 | 0.076 | 72.7 | 121 | 0.024 | 0.92 | 292 | 10 |
| | 0.3 | 0.11 | 15.4 | 26.7 | 0.013 | 0.94 | 234 | 10 |
| | 0.4 | 0.15 | 3.68 | 6.45 | 0.006 | 0.94 | 223 | 10 |
| | 0.5 | 0.21 | 0.76 | 1.35 | 0.003 | 0.95 | 221 | 10 |
| | 0.6 | 0.29 | 0.14 | 0.25 | 0.001 | 0.95 | 222 | 10 |
| Europa-like 9.6 | – | – | 0.07 | 0.07 | 0.005 | 0.99 | – | 25 |
| | 0.3 | 0.11 | 13.3 | 19.9 | 0.056 | 0.91 | 229 | 25 |
| | 0.4 | 0.15 | 2.42 | 3.94 | 0.035 | 0.98 | 218 | 25 |
| | 0.5 | 0.21 | 0.48 | 0.77 | 0.015 | 0.99 | 217 | 25 |
| | 0.6 | 0.29 | 0.15 | 0.18 | 0.008 | 0.99 | 219 | 25 |

| Moon Distance ($R_{Sat}$) | Star Mass ($M_\odot$) | Planet Distance ($AU$) | $h$ (W/m²) | $h_{conv}$ (W/m²) | $e$ | $a/a_0$ | $F_{glob}$ (W/m²) | Sim. Time ($10^6$ Years) |
|---|---|---|---|---|---|---|---|---|
| Saturn-like Host Planet ||||||||
| Io-like 5.9 | – | – | ** | ** | ** | ** | ** | 10 |
| | 0.2 | 0.076 | ** | ** | ** | ** | ** | 10 |
| | 0.3 | 0.11 | ** | ** | ** | ** | ** | 10 |
| | 0.4 | 0.15 | ** | ** | ** | ** | ** | 10 |
| | 0.5 | 0.21 | ** | ** | ** | ** | ** | 10 |
| | 0.6 | 0.29 | ** | ** | ** | ** | ** | 10 |
| Europa-like 9.6 | – | – | 0.13 | 0.13 | 0.012 | 0.97 | – | 25 |
| | 0.4 | 0.15 | 2.64 | 3.71 | 0.068 | 0.96 | 218 | 25 |
| | 0.5 | 0.21 | 0.51 | 0.70 | 0.029 | 0.97 | 217 | 25 |
| | 0.6 | 0.29 | 0.18 | 0.19 | 0.015 | 0.97 | 219 | 25 |

It is important to compare the 3-body results with the provided 2-body result (the shaded rows) for each moon distance in Tables 6 and 7. In each case, the difference in tidal heating between the 3-body and 2-body models decreases with increasing star mass. In other words, the influence of the star can be seen to decrease as the HZ moves outward for higher mass stars. Other than these general similarities, we will discuss each table separately. A graphical representation of the surface heat flux in each table, as a function of stellar mass, is also provided in Fig. 8.

### Satellites Systems at the Center of the HZ (Table 6)

In the tables, surface heat flux values highlighted in red are above $h_{max} = 2\,\mathrm{W\,m^{-2}}$ and represent a tidally evolved,

steady state value from the 3-body simulations. Continuing with the assumption that exomoon habitability might be in jeopardy above $h_{max}$, the table indicates that most of the moons would have the potential for habitability.

The highest surface heat flux involves an Io-like orbit around a Jupiter-like planet with a 0.1 solar mass central star. This was the only stable orbit for a 0.1 $M_\odot$ star and represents the tightest 3-body simulation (smaller star systems were considered, but none were stable). The effects of such a tight system can be seen by the long-term excitation of the moon in comparison to an isolated planet-moon binary whose orbit was completely circularized during the same time period. It is interesting that the final eccentricity is not unreasonably large ($e = 0.035$), yet it is more than enough to generate extreme heating when combined with the short orbital distance. Heating rates for star masses of 0.2 $M_\odot$ are significantly lower in comparison. However, the surface fluxes are still about 3 times more than the most geologically active surface in our solar system, so we will exclude them from habitability considerations as well. Notice that all systems with a 0.2 $M_\odot$ star had heating rates well above our maximum for habitability.

There were mixed results for a star mass of 0.3 $M_\odot$. Both planet models had stable Europa-like moon orbits and the moons were comfortably below $h_{max}$. For Io-like moon orbits the surface heat flux was below $h_{max}$ for our model, but just above that for the conventional model. Given the inherent uncertainty in the tidal heating estimates, rather than discussing which model gives the better estimate, it is more relevant at this point to discuss the global flux ($F_{glob}$) observed. Excluding the previously mentioned case of extreme tidal heating, the global flux for all the satellites is ≈ 110 W m$^{-2}$. Although the global flux does not lead to a direct estimate for the exomoon's surface temperature, a useful comparison can be made with the critical flux ($F_{RG}$) estimate of 269 W m$^{-2}$ for a runaway greenhouse in a Mars-mass exomoon. Clearly, the habitability of exomoons at this location is not at risk based on runaway greenhouse conditions from the global energy flux. In this case, tidal heating may actually be beneficial in warming large bodies of surface water that would otherwise freeze, which might be helpful also beyond the stellar HZ (Reynolds et al. 1987b; Heller & Armstrong 2014).

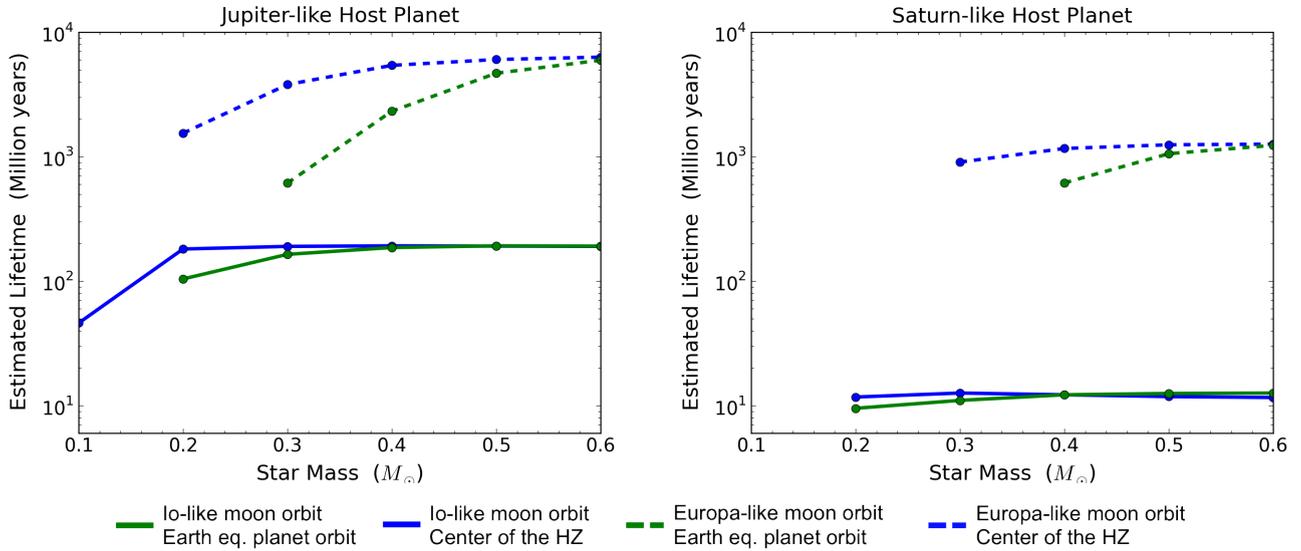

**Figure 7.** Estimated lifetimes from each 3-body simulation for the moon to completely spiral into the host planet.

Central star masses of 0.4 $M_\odot$ mark a cutoff for all exomoon habitability concerns based on intense tidal heating alone, since all the hypothetical exomoons are comfortably below $h_{max}$ at this point. The star does influence the short orbit moons, but the influence may be seen as a benefit in promoting surface activity in the exomoons rather than a restriction for habitability.

For the tidally evolved short orbits there is not a large difference in the final satellite heating rates between the two planet models and a given star mass. For example, with a star mass of 0.3 $M_\odot$ and an Io-like moon orbit, the moon around Jupiter had $h$ = 1.49 W m$^{-2}$. In comparison, the moon around Saturn had $h$ = 1.56 W m$^{-2}$. Some examples match even closer. The same cannot be said for the orbital elements, yet these parameters would have to vary in order to compensate for the unique physical characteristics of the planet. Given the physical differences and the corresponding difference in orbital distances for the moons, we did not expect the tidally evolved heating rates to be so similar.

As explained earlier, the 3-body simulation results involving the center of the HZ helped determine our second location for exploration inside the zone. Considering that the majority of the satellites had heating rates below $h_{max}$ and that the global flux was well below the critical flux for a runaway greenhouse, it was clear that the next round of simulations should involve the inner HZ.

### Satellites Systems at Earth-equivalent Distances (Table 7)

For the second round of 3-body simulations we essentially moved the planet-moon binary closer to the star in the prior 3-body systems. By decreasing the orbital distance from the star the Hill radius of the planet also decreased. This led to a reduction in stable satellite systems, especially for wider orbits.

All stable satellite systems involving a central star of mass 0.4 $M_\odot$ and below experience surface heating greater than $h_{max}$. Since the surface heat fluxes represent tidally evolved, steady state values, the heating has the potential to be maintained for extended periods of time. There is only small differences in final heating rates between specific moon orbits around Jupiter or Saturn-like host planets.

In regards to global flux, extreme tidal heating caused one satellite to be above the critical flux of 269 W m$^{-2}$. In that situation the extreme heating was already enough to rule it out for habitability. The rest of the satellites are comfortably below the critical flux, leaving the surface heat flux as our primary consideration for exomoon habitability in these systems. At least, that is the case for the chosen physical parameters. An interesting future study would be to vary parameters in regards to the global flux and then investigate changes to the tidal evolution of each system.

## 6  CONCLUSIONS

We performed a computational exploration into specific characteristics of putative exomoon systems. Our study focused on satellite systems in the habitable zone (HZ) of low-mass stars, which contain significantly reduced HZ distances

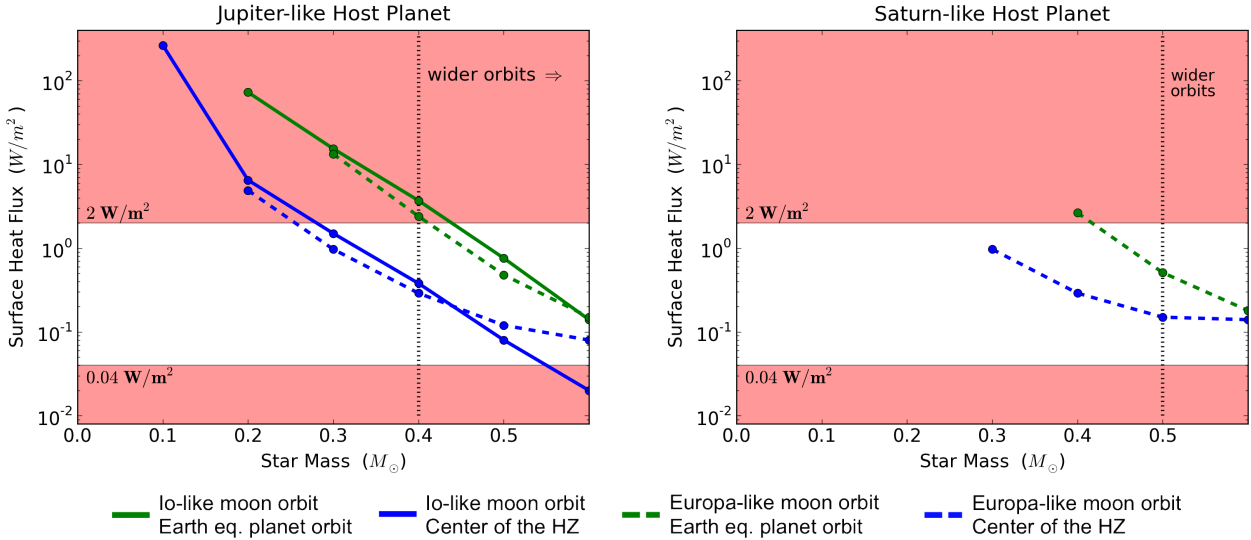

**Figure 8.** Graphical representations of the surface heat fluxes listed in Tables 6 and 7. The red shaded regions represent exomoon tidal heating rates above $h_{max} \equiv 2 \, \mathrm{W \, m^{-2}}$ or below $h_{min} \equiv 0.04 \, \mathrm{W \, m^{-2}}$. The dashed vertical lines at 0.4 and 0.5 $M_\odot$ are where Ganymede-like orbits (wider orbits) begin to be gravitationally stable towards higher stellar masses.

in comparison to Sun-like stars. For our planet-moon binaries the close proximity of the star presents dynamical restrictions to the stability of the moon, forcing it to orbit close to the planet to remain gravitationally bound. At short orbital distances, tidal heating and tidal torques between the planet and moon become more substantial. The relatively close star can also influence the long-term tidal evolution of the moon by continually exciting its orbit through gravitational perturbations. Key results from the computational simulations are highlighted as follows:

● **M dwarfs with masses $\lesssim 0.2 \, M_\odot$ cannot host habitable exomoons within the stellar habitable zone.**

Considering planets up to one Jupiter mass, stars with masses $\lesssim 0.2 \, M_\odot$ can support a Mars-like satellite out to distances of $\sim 10 \, R_p$ (Europa-like orbits). Though these systems are gravitationally stable, the close proximity to the star continually excites non-zero eccentricities in the moon's orbit. With limited orbital distances, tidal heating in the moon was significant, resulting in surface heating rates well above the proposed limit of $h_{max} = 2 \, \mathrm{W \, m^{-2}}$. Considering the excessively high heating estimates these systems are unlikely to host habitable exomoons. This conclusion is consistent with our theoretical estimate based on Eqs. (5) - (8).

Our results confirm that exomoon habitability is more complex than traditional definitions for planet habitability, which are based primarily on irradiation from a host star. Massive moons in the stellar HZ are not necessarily habitable by definition. Since the intense heating rates in the hypothetical exomoons are maintained by stellar perturbations, moving the host planet slightly outside the HZ should reduce the stellar influence. In this case, tidal heating in exomoons beyond the stellar HZ could make up for the reduced stellar illumination so that adequate surface temperatures for liquid surface water could be maintained.

● **In the mass range $0.2 \, M_\odot < M_\star < 0.5 \, M_\odot$, perturbations from the central star may continue to influence the long-term tidal evolution of exomoons in the stellar habitable zone.**

For stellar masses $\gtrsim 0.3 \, M_\odot$, the distinction for habitable exomoons became less defined when based primarily on tidal evolution. Our results suggest that the host planet's location in the HZ has to be taken into consideration. Results from simulations involving Earth-equivalent distances (the inner HZ) show that M dwarfs with masses $\lesssim 0.4 \, M_\odot$ promote surface heating beyond our accepted limit for habitability. In comparison, planetary orbits in the center of the HZ are within the established limits for habitability. Since a star's influence on a moon decreases with distance, so does its ability to excite higher heating rates. Therefore, this result for the center of HZs can be applied, by extension, to outer HZs.

Our adopted maximum limit for habitability ($h_{max} = 2 \, \mathrm{W \, m^{-2}}$) is based on a single example – Jupiter's volcanically active moon Io. Although there is little doubt that Io does not support a habitable environment, there is no evidence that a Mars-sized or even an Earth-sized exomoon could not remain habitable given the same internal heating rate. This is especially true considering the uncertainty in the exact mechanisms of tidal dissipation and the efficiency of plate tectonics in terrestrial bodies. For these reasons, $h_{max}$ should be considered more as a unit of comparison than a hard cutoff for habitable exomoons. The ultimate constraint on surface habitability will be given by the runaway greenhouse limit. With this consideration and treating the HZ as a whole, perturbations from a central star may continue to have deleterious effect on exomoon habitability up to $\approx 0.5 \, M_\odot$.

In contrast to the perils of intense tidal heating, perturbations from the star may actually have a positive influence on the habitability of exomoons. Figure 8 shows many surface heating rates below $h_{max}$, yet above the proposed minimum for habitability ($h_{min} = 0.04 \, \mathrm{W \, m^{-2}}$). These results sug-gest that satellite systems around stars in this mass range would need to be considered on a case by case basis depending on the planet mass and the specific location in the HZ. Moderate tidal heating might actually help to sustain tectonic activity (thus possibly a carbon-silicate cycle) and an internal magnetic dynamo that might help to shield the moon from high-energy radiation (Heller & Zuluaga 2013) on a Gyr timescale.

• **Considerations of global energy flux do not restrict habitability of exomoons in the HZs around stars with masses above 0.2 $M_\odot$.**

The conclusions thus far are in agreement with predictions made by Heller (2012) who followed considerations of energy flux and gravitational stability. This study, which focused on tidal evolution, has verified those predictions using a tidal model that considered both *N*-body interaction and tidal evolution. Similar considerations for energy flux were incorporated within this study with the global averaged flux ($F_{\mathrm{glob}}$) listed in Tables 6 and 7. Compared to the critical flux of 269 W m$^{-2}$ for a runaway greenhouse on a Mars-mass satellite, the 3-body global flux results support the conclusion that star masses $\lesssim 0.2\ M_\odot$ are unlikely to host habitable exomoons. Above that mass, exomoon habitability was not constrained by global energy flux.

• **Torques due to spin and tidal distortion between the planet and moon can result in rapid inward spiraling of a moon for orbital distances $\lesssim 6\ R_{\mathrm{p}}$.**

In specific simulations involving a Saturn-like host planet and Io-like ($a \sim 6\ R_{\mathrm{p}}$) moon orbits, distortion torques resulted in the complete inward spiral of a moon in < $10^7$ yr. The inward migration was connected to the assumption that the giant host planet was tidally locked to the star. While the orbital decay rate was slower for more massive Jupiter-like host planets, a conservative estimate for the maximum lifetime of Io-like moon orbits was only 200 Myr. Compared to the geological age of the Earth, this is a short lifespan. Assuming a Mars-like moon in an Io-like orbit was initially habitable, implications for the development of life may be considerable.

### 6.1 Future Work

For the sake of minimizing computational demands we assumed coplanar orbits and no spin-orbit misalignments of the moons. However, we expect new effects to arise in more complex configurations, such as spin-orbit resonances between the circumstellar and circumplanetary orbits as well as substantial effects on the longterm evolution of tidal heating. Important observational predictions can be obtained by studying the obliquity evolution of initially misaligned planet spins due to tidal interaction with the star. Will this "tilt erosion" (Heller et al. 2011a) tend to align the moons' orbits with the circumstellar orbit? These investigations could help predicting and interpreting possible variations of the planetary transit impact parameter due to the presence of an exomoon (Kipping 2009).

Of specific interest are retrograde satellites, which we neglected in this study. These satellites can form through direct capture or tidal disruption during planet-planet encounters (Agnor & Hamilton 2006; Williams 2013), and they can be orbitally stable as far as the Hill radius (Domingos et al. 2006b). Hence, we expect that retrograde moons could still be present or habitable in some cases that we identified as unstable or even uninhabitable for prograde moons. What is more, the detection of an exomoon in a very wide circumplanetary orbit near or beyond about 0.5 Hill radii might only be explained by a retrograde moon.

Owing to computational restrictions, the number of bodies considered was limited to three. A study with additional bodies would be of interest. The effects of orbital resonances for mutliple moons could be considered as well as the influence of additional planets in the system just outside the HZ. It was also necessary to limit many physical parameters such as the moments of inertia, tidal dissipation factors ($Q$), and tidal Love numbers ($k_L$). More fundamentally, these parameters are treated as constants in our applied tidal model. However, under the effect of tidal heating, the rheology and tidal response of a moon (or planet) can change substantially (Henning & Hurford 2014; Dobos & Turner 2015). Hence, more advanced tidal theory needs to be incorporated in our mathematical treatment of tidal plus *N*-body interaction to realistically model tidal effects in the regime of enhanced tidal heating. Modifying these values can coincide with variations in other physical parameters such as mass and radius for the extended bodies, which should noticeably affect the dynamical stability and tidal evolution of the systems.

One result from this study is the hypothetical existence of extremely tidally heated moons. Peters & Turner (2013) proposed the direct imaging of tidally heated exomoons. Closer examination is warranted to see if our computer model would be useful in providing orbital constraints on directly detectable exomoons. When considering extreme heating in a massive body, the issue of inflation may become important. Inflation is a physical response that was not incorporated into our model. Planetary inflation was considered by Mardling & Lin (2002) and future plans include the integration of this effect into our tidal evolution code.

### 7 ACKNOWLEDGEMENTS

Rhett Zollinger was supported by funding from the Utah Space Grant Consortium. René Heller has been supported by the Origins Institute at McMaster University, by the Canadian Astrobiology Program (a Collaborative Research and Training Experience Program funded by the Natural Sciences and Engineering Research Council of Canada), by the Institute for Astrophysics Göttingen, by a Fellowship of the German Academic Exchange Service (DAAD), and by the German space agency (Deutsches Zentrum für Luft- und Raumfahrt) under PLATO grant 50OO1501. This work made use of NASA's ADS Bibliographic Services. We would like to thank Benjamin C. Bromley, from the University of Utah, for his valuable insight and advice regarding this work. Finally, we would like to thank the reviewers for their helpful suggestions and useful insights.

This paper has been typeset from a TeX/LaTeX file prepared by the author.

## 7.9 The Effect of Multiple Heat Sources on Exomoon Habitable zones (Dobos et al. 2017)

Contribution:

RH assisted in the literature research, translated parts of the mathematical framework into computer code, generated Figs. 1-2, helped to structure Figs. 3-4, and contributed to the writing of the manuscript.

# The Effect of Multiple Heat Sources on Exomoon Habitable Zones

Vera Dobos[1, 2], René Heller[3], and Edwin L. Turner[4, 5]

1 Konkoly Thege Miklós Astronomical Institute, Research Centre for Astronomy and Earth Sciences, Hungarian Academy of Sciences, H–1121 Konkoly Thege Miklós út 15-17, Budapest, Hungary
2 Geodetic and Geophysical Institute, Research Centre for Astronomy and Earth Sciences, Hungarian Academy of Sciences, H–9400 Csatkai Endre u. 6-8., Sopron, Hungary
3 Max Planck Institute for Solar System Research, Justus-von-Liebig-Weg 3, 37077 Göttingen, Germany
4 Department of Astrophysical Sciences, Princeton University, 08544, 4 Ivy Lane, Peyton Hall, Princeton, NJ, USA
5 The Kavli Institute for the Physics and Mathematics of the Universe (WPI), University of Tokyo, 227-8583, 5-1-5 Kashiwanoha, Kashiwa, Japan

## ABSTRACT

With dozens of Jovian and super-Jovian exoplanets known to orbit their host stars in or near the stellar habitable zones, it has recently been suggested that moons the size of Mars could offer abundant surface habitats beyond the solar system. Several searches for such exomoons are now underway, and the exquisite astronomical data quality of upcoming space missions and ground-based extremely large telescopes could make the detection and characterization of exomoons possible in the near future. Here we explore the effects of tidal heating on the potential of Mars- to Earth-sized satellites to host liquid surface water, and we compare the tidal heating rates predicted by tidal equilibrium model and a viscoelastic model. In addition to tidal heating, we consider stellar radiation, planetary illumination and thermal heat from the planet. However, the effects of a possible moon atmosphere are neglected. We map the circumplanetary habitable zone for different stellar distances in specific star-planet-satellite configurations, and determine those regions where tidal heating dominates over stellar radiation. We find that the 'thermostat effect' of the viscoelastic model is significant not just at large distances from the star, but also in the stellar habitable zone, where stellar radiation is prevalent. We also find that tidal heating of Mars-sized moons with eccentricities between 0.001 and 0.01 is the dominant energy source beyond 3–5 AU from a Sun-like star and beyond 0.4–0.6 AU from an M3 dwarf star. The latter would be easier to detect (if they exist), but their orbital stability might be under jeopardy due to the gravitational perturbations from the star.

**Key words.** astrobiology – methods: numerical – planets and satellites: general – planets and satellites: interiors

## 1. Introduction

Although no exomoon has been discovered as of today, it has been shown that Mars-mass exomoons could exist around super-Jovian planets at Sun-like stars, from a formation point of view (Heller et al. 2014; Heller and Pudritz 2015a,b). If they exist, these moons could be detectable with current or near-future technology (Kipping et al. 2009; Heller 2014; Hippke and Angerhausen 2015) and they would be potentially habitable (Williams et al. 1997; Heller 2014; Lammer et al. 2014).

For both planets and moons, it is neither sufficient nor even necessary to orbit their star in the habitable zone (HZ) – the circumstellar region where the climate on an Earth-like planet would allow the presence of liquid surface water (Kasting 1993). Instead, tidal heating can result in a suitable surface temperature for liquid water at large stellar distances (Reynolds et al. 1987; Peters and Turner 2013; Heller and Armstrong 2014; Dobos and Turner 2015). Moreover, most of the water – and even most of the liquid water – in the solar system can be found beyond the location of the past snowline around the young Sun (Kereszturi 2010), which is supposed to have been at about 2.7 AU during the final stages of the solar nebula (Hayashi 1981). Hence, these large water reservoirs could be available for life on the surfaces of large exomoons at wide stellar separations.

However, observations of moons orbiting at several AU from their stars cannot be achieved with the conventional transit method, because transit surveys can only detect exomoons with significant statistical certainty after multiple transits in front of their star. This implies relatively small orbital separations from the star, typically < 1 AU (Szabó et al. 2006). It might, however, be possible to image exomoons with extreme tidal heating (Peters and Turner 2013) or to observe the transits of moons around luminous giant planets (Heller and Albrecht 2014; Heller 2016; Sengupta and Marley 2016) at several AU from their host stars.

At this wide distance from its host star, tidal heating in a large exomoon would be key to longterm surface habitability. Tidal heating of exomoons is usually calculated from equilibrium tide models, e.g. through parameterization with a tidal dissipation factor ($Q$) and with a uniform rigidity of the rocky material of the moon ($\mu$), both of which are considered constants. This family of the tidal equilibrium models is called the "constant-phase-lag" (CPL) models because they assume a constant lag of the tidal phases between the tidal bulge of the moon and the line connecting the moon and its planetary perturber (Efroimsky and Makarov 2013). Alternatively, equilibrium tides can be described by a "constant-time-lag" (CTL) model, where it is assumed that the tidal bulge of the moon lags the line between the two centers of mass by a constant time.

In reality, however, all these parameters depend on temperature. Another problem with equilibrium tide models is that the values of the parameters (e.g. of $Q$) are difficult to calculate (Remus et al. 2012) or constrain observationally even for solar system bodies: $Q$ can vary several orders of magnitude for different bodies: from $\approx 10$ for rocky planets to $> 1000$ for giant planets (see e.g. Goldreich and Soter 1966).

For these reasons, Dobos and Turner (2015) applied a viscoelastic model previously developed by Henning et al. (2009) to predict the tidal heating in hypothetical exomoons. Viscoelastic models are more realistic than tidal equilibrium models, as they consider the temperature feedback between the tidal heating and the viscosity of the material. Due to the phase transition at the melting of rocky material, viscoelastic models result in more moderate temperatures than do tidal equilibrium models, and so they predict a wider circumplanetary habitable zone (Forgan and Dobos 2016).

In this work, we compare the effects of two kinds of tidal heating models, a CTL model and a viscoelastic model, on the habitable edge for moons within the circumstellar HZ. We also calculate the contribution of viscoelastic model to the total energy flux of hypothetical exomoons with an emphasis on moons far beyond the stellar habitable zone.

## 2. Method

### 2.1. Energy flux at the moon's top of the atmosphere

We neglect any greenhouse or cloud feedbacks by a possible moon atmosphere and rather focus on the global energy budget. The following energy sources are included in our model: stellar irradiation, planetary reflectance, thermal radiation of the planet, and tidal heating. The globally averaged energy flux on the moon is estimated as per (Heller et al. 2014, Eq. 4)

$$\overline{F_{\rm s}}^{\rm glob} = \frac{L_* \left(1 - \alpha_{\rm s,opt}\right)}{4\pi a_{\rm p}^2 \sqrt{1 - e_{\rm p}^2}} \left(\frac{x_{\rm s}}{4} + \frac{\pi R_{\rm p}^2 \alpha_{\rm p}}{f_{\rm s} 2 a_{\rm s}^2}\right) + \frac{L_{\rm p} \left(1 - \alpha_{\rm s,IR}\right)}{4\pi a_{\rm s}^2 f_{\rm s} \sqrt{1 - e_{\rm s}^2}} + h_{\rm s} + W_{\rm s} \,, \tag{1}$$

where $L_*$ and $L_{\rm p}$ are the stellar and planetary luminosities, respectively, $\alpha_{\rm p}$ is the Bond albedo of the planet, $\alpha_{\rm s,opt}$ and $\alpha_{\rm s,IR}$ are the optical and infrared albedos of the moon, $a_{\rm p}$ and $a_{\rm s}$ are the semi-major axes of the planet's orbit around the star and of the moon's orbit around the planet[1], $e_{\rm p}$ and $e_{\rm s}$ are the eccentricities of the planet's and the satellite's orbit, $R_{\rm p}$ is the radius of the planet, $x_{\rm s}$ is the fraction of the satellite's orbit that is not spent in the shadow of the host planet, and $f_{\rm s}$ describes the efficiency of the flux distribution on the surface of the satellite. The first and second terms of this equation account for illumination effects from the star and the planet, whereas $h_{\rm s}$ indicates the tidal heat flux through the satellite's surface. $W_{\rm s}$ denotes arbitrary additional energy sources such as residual internal heat from the moon's accretion or heat from radiogenic decays, but we use $W_{\rm s} = 0$. We also neglect eclipses, thus $x_{\rm s} = 1$, and we assume that the satellite is not tidally locked to its host planet, hence $f_{\rm s} = 4$. The planetary radius is calculated from the mass, using a polynomial fit to the data given by Fortney et al. (2007, Table 4, line 17). The albedos of both the planet and the moon (in the optical and also in the infrared) were set to 0.3, that is, to Earth-like values.

[1] We assume that the moon's mass is negligible compared to the planetary mass and that the barycenter of the planet-moon system is in the center of the planet.

We estimate the runaway-greenhouse limit of the globally averaged energy flux ($F_{\rm RG}$), which defines the circumplanetary habitable edge interior to which a moon becomes uninhabitable, using a semi-analytic model of Pierrehumbert (2010) as described in Heller and Barnes (2013). In this model, the outgoing radiation on top of a water-rich atmosphere is calculated using an approximation for the wavelength-dependent absorption spectrum of water. The runaway greenhouse limit then depends exclusively on the moon's surface gravity (i.e. its mass and radius) and on the fact that there is enough water to saturate the atmosphere with steam. Our test moons are assumed to be rocky bodies with masses between the mass of Mars (0.1 Earth masses, $M_\oplus$) and 1 $M_\oplus$. Our test host planets are gas giants around either a sun-like star or an M dwarf star.

We apply the model of Kopparapu et al. (2014) to calculate the borders of the circumstellar HZ, which are different for the 0.1 and the 1 Earth-mass moons. The corresponding energy fluxes are then used to define and evaluate the habitability of our test moons, both inside and beyond the stellar HZ.

### 2.2. Tidal heating models

Tidal heat flux is calculated using two different models: a viscoelastic one with temperature-dependent tidal $Q$ and a CTL one, the latter of which converges to a CPL model for small eccentricities as the time-lag of the principal tide becomes $1/Q$ (Heller et al. 2011). For the CTL model we use the same framework and parameterization as in the work of Heller and Barnes (2013), which goes back to Leconte et al. (2010) and Hut (1981). In particular, we use the following parameters: $k_2 = 0.3$ (as in Henning et al. 2009; Heller and Barnes 2013), and tidal time lag, $\tau = 638$s, which was measured for the Earth (Lambeck 1977; Neron de Surgy and Laskar 1997).

In this model, the tidal heating function is linear in both $\tau \sim 1/(nQ)$ and $k_2$. Hence, changes in these parameters usually do not have dramatic effects on the tidal heating. In contrast, changes in the radius or mass of the tidally distorted object, or in the orbital eccentricity or semi-major axis result in significant changes of the tidal heating rates.

For the viscoelastic tidal heating calculations, we use the model described by Dobos and Turner (2015), which was originally developed by Henning et al. (2009). Tidal heat flux is calculated from

$$h_{\rm s,visc} = -\frac{21}{2} Im(k_2) \frac{R_{\rm p}^5 n_{\rm s}^5 e_{\rm s}^2}{G} \,, \tag{2}$$

where $Im(k_2)$ is the complex Love number, which describes the structure and rheology of the satellite (Segatz et al. 1988). In the Maxwell model, the value of $Im(k_2)$ is given by (Henning et al. 2009)

$$-Im(k_2) = \frac{57 \eta \omega}{4 \rho g R_{\rm m} \left[1 + \left(1 + \frac{19\mu}{2\rho g R_{\rm m}}\right)^2 \frac{\eta^2 \omega^2}{\mu^2}\right]} \,, \tag{3}$$

where $\eta$ is the viscosity, $\omega$ is the orbital frequency, and $\mu$ is the shear modulus of the satellite. The temperature dependency of the viscosity and the shear modulus is described by Fischer and Spohn (1990) and Moore (2003); and the adapted values and equations are listed by Dobos and Turner (2015). Since only rocky bodies like the Earth are considered as satellites in this work, the solidus and liquidus temperatures, at which the material of the rocky body starts melting and becomes completely liquid, were chosen to be 1600 K and 2000 K, respectively. We assume that disaggregation occurs at 50% melt fraction, which implies a breakdown temperature of 1800 K.

The viscoelastic tidal heating model also describes the convective cooling of the body. The iterative method described by Henning et al. (2009) was used for our calculations of the convective heat loss:

$$q_{\mathrm{BL}} = k_{\mathrm{therm}} \frac{T_{\mathrm{mantle}} - T_{\mathrm{surf}}}{\delta(T)}, \qquad (4)$$

where $k_{\mathrm{therm}}$ is the thermal conductivity ($\sim 2\,\mathrm{W/mK}$), $T_{\mathrm{mantle}}$ and $T_{\mathrm{surf}}$ are the temperatures in the mantle and on the surface, respectively, and $\delta(T)$ is the thickness of the conductive layer. We use $\delta(T) = 30$ km as a first approximation, and then for the iteration

$$\delta(T) = \frac{d}{2 a_2} \left( \frac{Ra}{Ra_{\mathrm{c}}} \right)^{-1/4} \qquad (5)$$

is used, where $d$ is the mantle thickness ($\sim 3000$ km), $a_2$ is the flow geometry constant ($\sim 1$), $Ra_{\mathrm{c}}$ is the critical Rayleigh number ($\sim 1100$) and $Ra$ is the Rayleigh number which can be expressed as

$$Ra = \frac{\alpha\, g\, \rho\, d^4\, q_{\mathrm{BL}}}{\eta(T)\, \kappa\, k_{\mathrm{therm}}}. \qquad (6)$$

with $\alpha$ ($\sim 10^{-4}$) as the thermal expansivity, $\kappa = k_{\mathrm{therm}}/(\rho\, C_p)$ as the thermal diffusivity, and $C_p = 1260\,\mathrm{J/(kg\,K)}$. The iteration of the convective heat flux ends when the difference of the last two values becomes smaller than $10^{-10}\mathrm{W/m}^2$. Once the stable equilibrium temperature is found, we compute the tidal heat flux.

This viscoelastic model was already used together with a climate model by Forgan and Dobos (2016) with the aim of determining the location and width of the circumplanetary habitable zone for exomoons. The 1D latitudinal climate model included eclipses, the carbonate-silicate cycle and the ice-albedo feedback of the moon (in addition to tidal heating, stellar and planetary radiation). The ice-albedo positive loop in the climate model along with eclipses result in a relatively close-in outer limit for circumplanetary habitability, if the orbit of the satellite is not inclined. The climate model, however, can only be used for bodies of similar sizes to the Earth. In this work we investigate the effect of the viscoelastic model on smaller exomoons, as well, and for this reason, instead of a climate model, we apply an orbit-averaged illumination model. Beside solar-like host stars, we made calculations also for M dwarfs.

Both the CPL and the viscoelastic models are valid only for small orbital eccentricities, that is, for $e \lesssim 0.1$. For larger eccentricities, the instantaneous tidal heating in the deformed body can differ strongly from the orbit-averaged tidal heating rate and the frequency spectrum of the decomposed tidal potential could involve a wide range of frequecies (Greenberg 2009). Both aspects go beyond the approximations involved in the models, and hence our results for $e \gtrsim 0.1$ need to be taken with a grain of salt.

## 2.3. Simulation setup

We consider various reference systems of a star, a planet, and a moon.

### 2.3.1. Comparison of equilibrium and viscoelastic tides

First, we examine the different effects of tidal heating in either a CTL or in a viscoelastic tidal model on the location of the circumplanetary habitable edge (see Section 3.1). We consider a star with a sun-like radius ($R_\star = R_\odot$) and effective temperature ($T_{\mathrm{eff}} = 5778$ K), a 5 Jupiter-mass gas giant at 1 AU orbital distance with an eccentricity of 0.1, and a 0.5 $M_\oplus$ moon. The orbital period and eccentricity of the moon are varied between 1 and 20 days and between 0.01 and 1, respectively.

### 2.3.2. Viscoelastic tides beyond the stellar habitable zone

Second, we map the circumplanetary habitable zone over a wide range of circumstellar orbits (see Section 3.2) using four different star-planet-moon configurations. The star is either sun-like (see Section 2.3.1) or an M3 class main sequence star ($M_\star = 0.36\,M_\odot$, $R_\star = 0.39\,R_\odot$, $T_{\mathrm{eff}} = 3250$ K and $L_\star = 0.0152\,L_\odot$, Kaltenegger and Traub 2009, Table 1). The planet is a Jupiter-mass gas giant, and the satellite's mass is either 0.1 $M_\oplus$ or 1 $M_\oplus$ (i.e. a Mars or Earth analogue). The density of the moon is that of the Earth. The stellar luminosity and the temperature values are used for our calculations of the HZ boundaries, and the stellar radius and temperature values are required for the incident stellar flux calculation.

## 3. Results

### 3.1. The habitable edge

We calculated the total energy flux (using Eq. 1) at the top of the moon's atmosphere as a function of distance to the planet in both the CTL (Fig. 1) and the viscoelastic (Fig. 2) frameworks. In both Figs. 1 and 2, colours show the amount of the total flux received by the moon, and the white contours at 288 W/m² indicate the habitable edge defined by the runaway-greenhouse limit (Heller & Barnes 2013). Interestingly, in the CTL model the habitable edge is located closer in to the planet than in the viscoelastic model.

Black contour curves show the tidal heat flux alone at 288, 100 and 2 W/m², the latter is the global mean heat flow from tides on Io as measured with the Galileo spacecraft (Spencer et al. 2000). This curve in Fig. 2 has a different slope than the one labeled 'tidal heat at 100 W/m²'. The different slope is caused by the changing equations in the viscoelastic model. The tidal flux (and also the convective cooling flux) is described by different formulae below or above certain temperatures (namely the solidus, the breakdown and the liquidus temperatures). In other words, if the tidal heating is low, different equations will be used, than in the case of higher tidal forces.

Figs. 1 and 2 show substantial differences. The viscoelastic model predicts moderate ($\leq 2$ W/m²) heating beyond 5.3 days of an orbital period for $e = 0.01$, while the CTL model needs the moon to be as close as 3.5 days to generate the same tidal heating with the same orbital eccentricity. In other words, the viscoelastic model predicts significant tidal heating in wider orbits. At close orbits, however, the CTL model products extremely high fluxes. With a 1.6 day orbital period (similar to Io, see arrows at the top), our test moon with e = 0.01 would generate a tidal heat flux that is sufficient to trigger a runaway greenhouse effect. In this case, stellar illumination is not even required to make such a moon uninhabitable. In contrast, with the viscoelastic model the moon would need to have a 1 day orbital period to trigger the same effect.

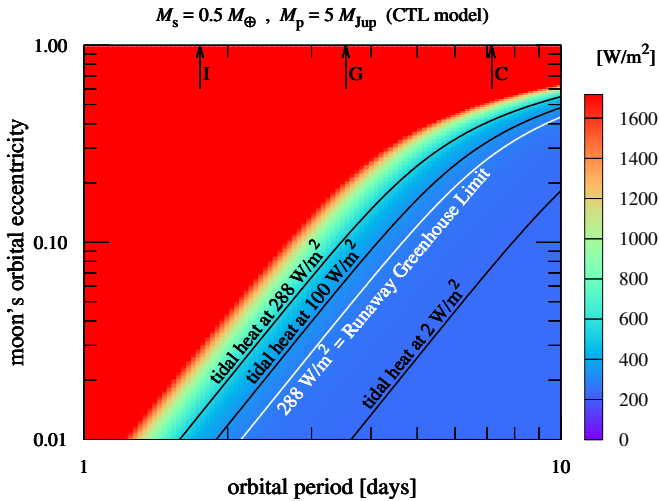

**Fig. 1.** Energy flux at the top of the atmosphere of a 0.5 Earth-mass moon orbiting a 5 Jupiter-mass planet at 1 AU distance from a Sun-like star; tidal heating is calculated with a CTL model. The planetary orbit has 0.1 eccentricity. The stellar radiation, the planetary reflectance, thermal radiation of the planet and tidal heating were considered as energy sources. White contour curve indicates the runaway greenhouse limit considering all energy sources, while the black curves show the tidal heating flux alone. The arrows at the top of the figure indicate the orbital periods of Io (I), Ganymede (G) and Callisto (C). Note that the CTL model is valid only for small orbital eccentricities. Heating contours in the upper half of the panel (above 0.1 along the ordinate) could be off by orders of magnitude in real cases.

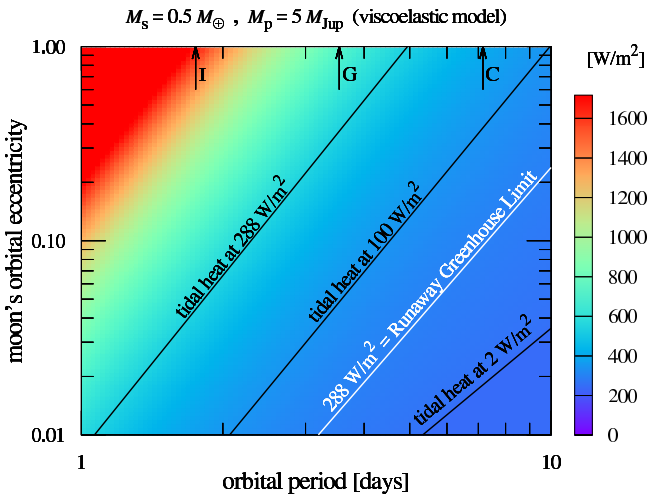

**Fig. 2.** Energy flux at the top of the atmosphere of a 0.5 Earth-mass moon orbiting a 5 Jupiter-mass planet at 1 AU distance from a Sun-like star; tidal heating is calculated with a viscoelastic model. The same colours, contours and signatures were applied as in Fig. 1. Note that the viscoelastic model is valid only for small orbital eccentricities. Heating contours in the upper half of the panel (above 0.1 along the ordinate) could be off by orders of magnitude in real cases.

A comparison of these two plots shows the 'thermostat effect' of the viscoelastic model described by Dobos and Turner (2015). The CTL model yields lower tidal heating rates than the viscoelastic model in the weak-to-moderate heating regime

(tidal heat ≤ 100 W/m²), while above 100 W/m² the viscoelastic model produces lower heating rates, where the CTL model runs away. As a consequence, in this specific example of a 0.5 Earth-mass moon around a 5 Jupiter-mass planet, the habitable edge (red contour curve in the figures) is located at a larger distance from the planet than for the CTL model. It is caused by the stronger tidal heating rate below 100 W/m².

Stellar and planetary illuminations can also have important effects. They can break the feedback loop between tidal heating and convective cooling: in an extreme case the system will not even find a stable convective heat transport rate. However, the most relevant cases of tidally heated exomoons will involve systems orbiting so far from the star that illumination heating is irrelevant and around planets so old that planetary illumination is also small. In any cases, such systems will be the ones that can be most easily understood, if they exist.

### 3.2. Circumplanetary tidal habitable zone

In addition to the inner habitable edge, we also want to locate the outer boundary of the circumplanetary habitable zone for different satellite sizes and stellar classes. Figs. 3 and 4 show the circumplanetary habitable zones for our Jupiter-like test planet with a moon around either a sun-like or an M dwarf star, respectively. The left and right panels of the figures show the cases of 0.1 and 1 Earth-mass satellites, respectively. The eccentricity of the moon's orbit is 0.001 in the top panels and 0.01 in the bottom panels.

Grey horizontal lines in each panel indicate zero tidal heat flux. Above these lines, only stellar radiation is relevant since the reflected and thermal radiation from the planet are also negligibly small. Green areas illustrate habitable surface conditions with tidal heating rates below 100 W m⁻², whereas diagonally striped orange areas visualize exomoon habitability with tidal heating rates above 100 W m⁻² (color coding adopted from Heller and Armstrong 2014).

As expected, for smaller satellite masses the HZ is located closer to the planet. At zero tidal flux, small plateaus are present which are consequences of the viscoelastic tidal heating model. At the horizontal line tidal forces are turned on, hence there is a change in the position of the green area. At the plateau the tidal heat flux elevates from zero to about 10 W/m² or more. The sudden change is caused by the fact that the tidal heating model gives result only if the tidal heat flux and the convective cooling flux has stable equilibrium. It means that if tidal forces are weak, then the convective cooling will be weak too, or not present in the body at all, hence there is no equilibrium. In other words, tidal heating is insufficient to drive convection in the body. There will still be some heating, but a different heat transport mechanism (probably conduction) would be in play. A more accurate model would consider all heat transport mechanisms at all temperatures and would probably not exhibit such discontinuities.

In the inspected cases the two dominating energy sources are the stellar radiation and the tidal heating. Since these two effects are physically independent, one would not expect them to be of comparable importance except in a small fraction of cases. In general, however, one effect will dominate the other. To explore this interplay between stellar illumination and tidal heating in more detail, we calculated the distance from the star at which tidal heating (for a given hypothetical moon and orbit) equals stellar heating. These critical distances are indicated with blue dashed curves in Figs. 3 and 4. We find that beyond 3-5 AU around a G2 star and beyond 0.4-0.6 AU around an M3 star, tidal heating for moons with eccentricities between 0.001 and 0.01 will be the dominant source of energy rather than stellar radiation (see the overlapping of the grey and dashed blue curves). This is the case for both satellite masses investigated. For smaller stellar distances, the dominant heat source depends on the circumplanetary distance of the moon.

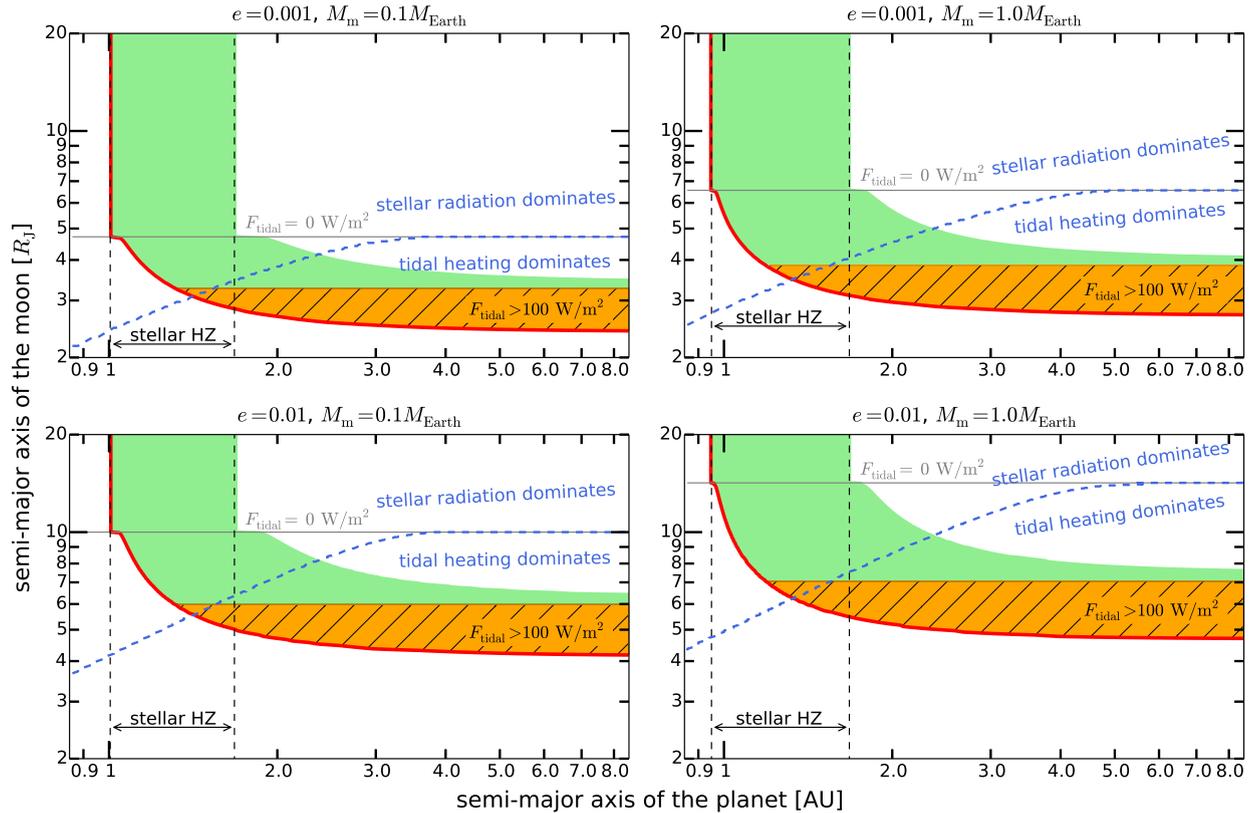

**Fig. 3.** Tidal heating habitable zone for 0.1 and 1 Earth-mass exomoons (left and right panels, respectively) around a Jupiter-mass planet hosted by a G2 star as functions of the semi-major axes of the planet and the moon. The red curve indicates the runaway greenhouse limit (habitable edge). The green and the diagonally striped orange areas cover the habitable region, where the striped orange colour indicates that the tidal heating flux is larger than $100\,\mathrm{W/m^2}$. The orbital eccentricity of the moon is 0.001 in the top panels, and 0.01 in the bottom panels. Vertical dashed lines indicate the inner and outer boundaries of the circumstellar habitable zone. At the dashed blue contour curve the tidal heating flux equals to the stellar radiation flux, hence it separates the tidal heating dominated and the stellar radiation dominated regime. At the grey horizontal line the tidal flux is zero.

Based on the fact that Io is a global volcano world, one might take the $2\,\mathrm{W/m^2}$ limit as a conservative limit for Earth-like surface habitability. But other worlds might still be habitable even in a state of extreme tidal heating near $100\,\mathrm{W/m^2}$. In fact, the surface habitability of rocky, water-rich planets or moons has not been studied in this regime of extreme tidal heating to our knowledge. Nevertheless, although $100\,\mathrm{W/m^2}$ seems a lot of internal heating compared to the internal heat flux on Earth, which is less than $0.1\,\mathrm{W/m^2}$ (Zahnle et al. 2007), it could still allow the presence of liquid surface water as long as the total energy flux is below the runaway greenhouse limit. The striped orange colour indicates that the body might become volcanic or tectonically active in this regime.

It is supposed that the gravitational pull of an M dwarf can force an exomoon whose orbit is within the stellar HZ into a very eccentric circumplanetary orbit (Heller 2012). The resulting tidal heating might ultimately prevent such moons from being habitable. Since the planetary Hill sphere in the stellar HZ of M dwarfs is not so large, any moon needs to be in a tight orbit. Prograde (regular) moons are only stable out to about 0.5 Hill radius (Domingos et al. 2006). In Fig. 4, the area beyond this line is shown as a shaded region in grey and labeled as 'Hill unstable'. As a consequence of the closeness of this Hill unstable limit, the circumplanetary volume for habitable satellites is much smaller in systems of M dwarfs than in systems with sun-like stars, in particular if the moons have substantial eccentricities. Note how the circumplanetary space between the red (runaway greenhouse) curve and the Hill unstable region becomes smaller as the moon's eccentricity or its mass increases.

## 4. Discussion

From the results shown in Section 3, the following general findings can be concluded.

The circumplanetary habitable edge calculated with the viscoelastic model (Fig. 2) is located at a larger distance than in the CTL model (Fig. 1). If the outer boundary of the circumplanetary habitable zone is defined by Hill stability, than it means that the viscoelastic model reduces the habitable environment. However, Forgan and Yotov (2014) showed that the moon can enter into a snowball phase, if the ice-albedo feedback and eclipses are also taken into account. It means that the outer boundary will be significantly closer to the planet, than in the case when it is defined by Hill stability. Since the viscoelastic model resulted in higher tidal fluxes below $100\,\mathrm{W/m^2}$, than the CTL model, we expect that the outer boundary defined by the snowball state will be farther from the planet. Altogether, the circumplanetary HZ is not likely to be thinner in the viscoelastic case, but it is located in a larger distance from the planet.

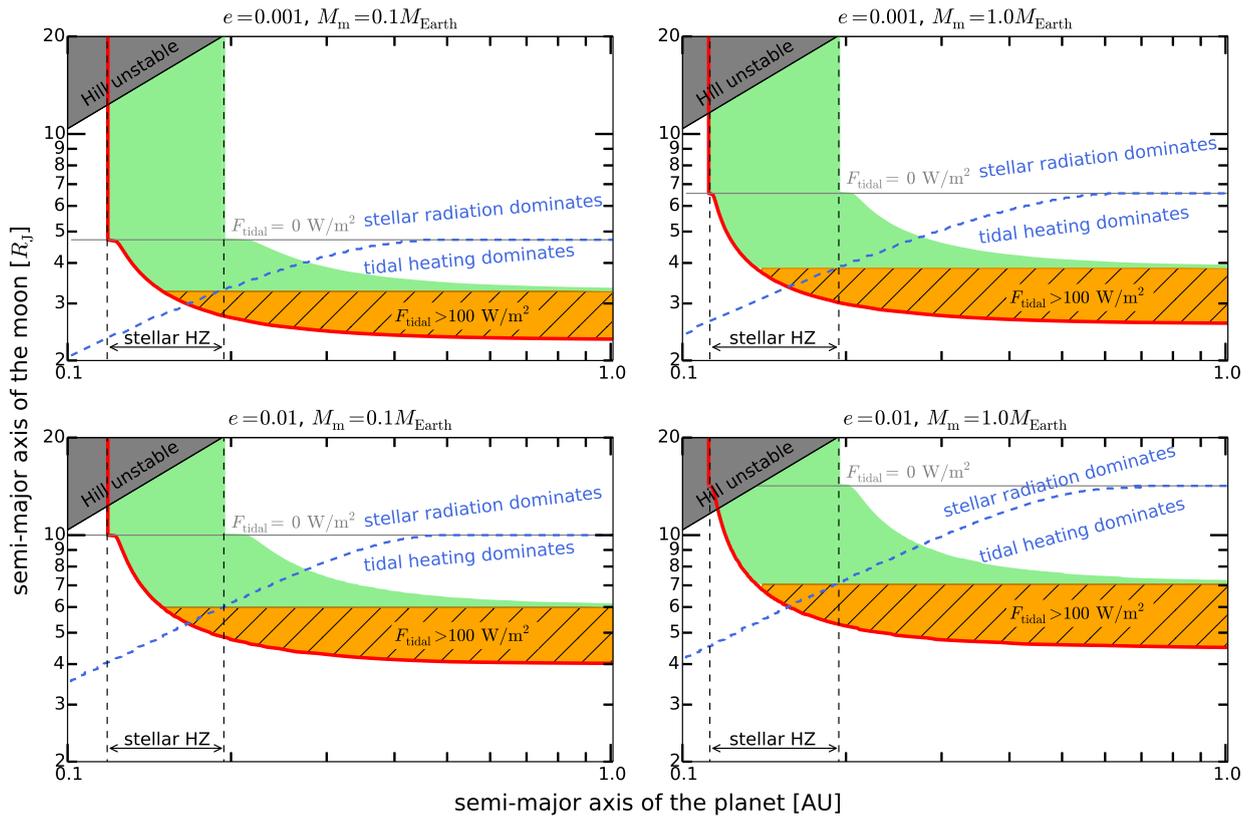

**Fig. 4.** Tidal heating habitable zone for 0.1 and 1 Earth-mass exomoons (left and right panels, respectively) around a Jupiter-mass planet hosted by an M3 dwarf as functions of the semi-major axes of the planet and the moon. The orbital eccentricity of the moon is 0.001 in the top panels, and 0.01 in the bottom panels. The same colours and contours were applied as in Fig. 3. The grey area in the upper left corner covers those cases where the moon's orbit is considered unstable.

Tidal heating in close-in orbits is more moderate with the viscoelastic model and does not generate extremely high rates as predicted by the CTL model.

Moons that are primarily heated by tides, that is, moons beyond the stellar habitable zone, have a wider circumplanetary range of orbits for habitability with the viscoelastic model than with the CTL model. This is in agreement with the findings of Forgan and Dobos (2016). In Heller and Armstrong (2014, Fig. 2), it was shown that the circumplanetary habitable zone calculated with the CTL model thins out dramatically as the planet-moon binary is virtually shifted away from their common host star. This is due to the very strong dependence of tidal heating in the CTL model on the moon's semi-major axis.

In the calculations we considered 0.1, 0.5 and 1 Earth-mass moons. Lammer et al. (2014) found that about 0.25 Earth masses (or 2.5 Mars masses) are required for a moon to hold an atmosphere during the first 100 Myrs of high stellar XUV activity, assuming a moon in the habitable zone around a Sun-like star. Considering this constraint, a 0.1 Earth-mass moon may not be habitable in the stellar HZ because of atmospheric loss, however, at larger stellar distances the stellar wind and strong stellar activity do not present such severe danger to habitability.

From a formation point of view, smaller moons are more probable to form around super-Jovian planets than Earth-mass satellites, according to Canup and Ward (2006), who gave an upper limit of $10^{-4}$ to the mass ratio of the satellite system and the planet. However, moons can also originate from collisions rather than the circumplanetary disk, as in the case of the Moon. In such cases, larger mass ratios are reasonable, too.

We have considered a large range of orbital eccentricities for the moon with values up to 1. Beyond the fact that our models are physically plausible only for moderate values ($\leq 0.1$), tidal circularization will act to decrease eccentricities to zero on time scales that are typically much shorter than 1 Gyr (Porter and Grundy 2011; Heller and Barnes 2013). Hence, even moderate eccentricities can only be expected in real exomoon systems if the star has a significant effect on the moon's orbit (Heller 2012), if other planets can act as orbital perturbers (Gong et al. 2013; Payne et al. 2013; Hong et al. 2015), if other massive moons are present around the same planet, or if the planet-moon system has migrated through orbital resonances with the circumstellar orbit (Namouni 2010; Spalding et al. 2016).

Hot spots (hot surface areas generated by geothermical heat, that could generate volcanoes) might be important on both observational and astrobiological grounds. Regarding observations, hot spots produce variability in the brightness and spectral energy distribution of thermal emission from the moon. As Io illustrates, hot spots can potentially be big and prominently hot (see Spencer et al. 2000, Fig. 4). This could let one determine the moon's spin period and may indicate a geologically 'active' body. Hot spots also allow liquid water to locally exist and persist over long periods of time even if the mean surface temperature is far below the freezing point. Enceladus may provide an example for such phenomenon in the Solar System, where tidal heating maintains a liquid (and probably global) ocean below the ice cover, and contributes to the eruption of plumes at the southern region of the satellite (Thomas et al. 2016).

On the other hand, hot spots can be effective in conducting the internal heat. The calculated fluxes in this paper are average surface fluxes, but in reality hot spots are probable to form. These areas are much higher in temperature, meaning that other areas on the surface must be somewhat colder than the average. As a result, the temperature of the surface (excluding the vicinity of the hot spots) can be lower than that is calculated.

Two end-member models exist for spatial distribution of tidal heat dissipation. In model A the dissipation mostly occurs in the deep-mantle of the body, and in model B it occurs in the astheno-sphere, which is a thin layer in the upper mantle (Segatz et al. 1988). In model A the surface heat flux is higher at the polar regions, while in model B the flux is mostly distributed between $-45\,°$ and $+45\,°$ latitudes. From volcano measurements of Io it seems that model B is more consistent with the location of hot spots (Hamilton et al. 2011, 2013; Rathbun and Spencer 2015). The global distribution of volcanoes on the surface is random (Poisson distribution), but closer to the equator they are more widely spaced (uniform distribution, Hamilton et al. 2013).

On Io the total power output of volcanoes and paterae is about $5 \cdot 10^{13}$ W (Veeder et al. 2012a,b). About 62% of the flux is going through volcanoes in Io (Veeder et al. 2012a), which means that there is a large temperature difference in the volcanic areas and lowlands. For moons of larger sizes than Io, or with much stronger tidal force, it is probable that even higher percentage of energy will leave through volcanoes. A geological model is needed to estimate the connection between the tidal heating flux and the energy output of volcanoes and hot spots, which is beyond the scope of this paper.

Detectibility of Mars-to-Earth size satellites is not investigated in this work, but detections could be possible in the *Kepler* data (Kipping et al. 2009; Heller 2014) or with the PLATO (Hippke and Angerhausen 2015) or CHEOPS (Simon et al. 2015) missions, or with the E-ELT (Heller and Albrecht 2014; Heller 2016; Sengupta and Marley 2016). Naturally, larger moons are easier to detect than smaller moons, and so we expect an observational selection bias for the first known exomoons to be large and potentially habitable.

## 5. Conclusion

In this work we improved the viscoelastic tidal heating model for exomoons (Dobos and Turner 2015) by adding the stellar radiation, the planetary reflectance and the planet's thermal radiation to the energy budget. We found that the 'thermostat effect' of the viscoelastic model is robust even with the inclusion of these additional energy sources. This temperature regulation in the viscoelastic tidal heating model is caused by the melting of the satellite's inner material, since the phase transition prevents the temperature from rising to extreme heights.

We investigated the circumplanetary tidal heating HZ for a few representative configurations. We showed that the extent of the tidal HZ is considerably wide even at large distances from the stellar HZ, predicting more habitable satellite orbits than the CTL models. In a previous study with a CTL model, Heller and Armstrong (2014) found that the tidal HZ thinned out for large stellar distances. We also showed that if tidal heating is present in the moon, then beyond 3–5 AU distance from solar-like stars this will be the dominating energy source, and for M3 main sequence dwarfs tidal heating dominates beyond 0.4–0.6 AU distance. At smaller stellar distances the semi-major axes of the moon defines whether stellar radiation or tidal heating dominates.

From an observational point of view, M dwarfs are better candidates to detect habitable exomoons, since both the stellar and the circumplanetary tidal HZs and closer to the star, meaning that the orbital period of the planet is shorter, hence more transits can be observed. However, in the stellar HZ, the possible habitable orbits for exomoons are constrained by Hill stability. At larger stellar distances the circumplanetary tidal HZ and the Hill unstable region do not overlap, so there is no such constraint.

## Acknowledgements

We thank Duncan Forgan for a very helpful referee report. VD thanks László L. Kiss for the helpful discussion. This work was supported in part by the German space agency (Deutsches Zentrum für Luft- und Raumfahrt) under PLATO grant 50OO1501. This research has been supported in part by the World Premier International Research Center Initiative, MEXT, Japan. VD has been supported by the Hungarian OTKA Grants K104607, K119993, and the Hungarian National Research, Development and Innovation Office (NKFIH) grant K-115709.

## 7.10 Exploring Exomoon Surface Habitability with an Idealized Three-dimensional Climate Model (Haqq-Misra & Heller 2018)

Contribution:

RH did the literature research to embed this study in the context of exomoon science, assisted in the conception of the figures and in their interpretation, and contributed to the writing of the manuscript.

# Exploring exomoon atmospheres with an idealized general circulation model

Jacob Haqq-Misra[1]★ and René Heller[2]

[1]*Blue Marble Space Institute of Science, 1001 4th Ave Suite 3201, Seattle, WA 98154, USA*
[2]*Max Planck Institute for Solar System Research, Justus-von-Liebig-Weg 3, 37077 Göttingen, Germany*

**ABSTRACT**

Recent studies have shown that large exomoons can form in the accretion disks around super-Jovian extrasolar planets. These planets are abundant at about 1 AU from Sun-like stars, which makes their putative moons interesting for studies of habitability. Technological advances could soon make an exomoon discovery with *Kepler* or the upcoming *CHEOPS* and *PLATO* space missions possible. Exomoon climates might be substantially different from exoplanet climates because the day-night cycles on moons are determined by the moon's synchronous rotation with its host planet. Moreover, planetary illumination at the top of the moon's atmosphere and tidal heating at the moon's surface can be substantial, which can affect the redistribution of energy on exomoons. Using an idealized general circulation model with simplified hydrologic, radiative, and convective processes, we calculate surface temperature, wind speed, mean meridional circulation, and energy transport on a 2.5 Mars-mass moon orbiting a 10-Jupiter-mass at 1 AU from a Sun-like star. The strong thermal irradiation from a young giant planet causes the satellite's polar regions to warm, which remains consistent with the dynamically-driven polar amplification seen in Earth models that lack ice-albedo feedback. Thermal irradiation from young, luminous giant planets onto water-rich exomoons can be strong enough to induce water loss on a planet, which could lead to a runaway greenhouse. Moons that are in synchronous rotation with their host planet and do not experience a runaway greenhouse could experience substantial polar melting induced by the polar amplification of planetary illumination and geothermal heating from tidal effects.

**Key words:** planets and satellites: terrestrial planets – planets and satellites: atmospheres – hydrodynamics – astrobiology

## 1 INTRODUCTION

Low-mass stars are conventionally thought to exhibit the most promising odds for the detection of terrestrial planets, at least from an observational point of view. Their low masses enable detections of low-mass companions like Earth-mass planets using radial velocity measurements, and their small radii allow findings of small transiting objects in photometric time series. The recent detections of sub-Earth-sized planets around the M dwarf stars TRAPPIST-1 (Gillon et al. 2016; Gillon et al. 2017), Proxima Centauri (Anglada-Escudé et al. 2016), and LHS 1140 (Dittmann et al. 2017) serve as impressive benchmark discoveries.

Even cooler dwarfs exist. Brown dwarfs (BDs), with masses between about 13 and 75 Jupiter masses ($M_{\rm J}$) cannot fuse hydrogen, but their eternal shrinking nevertheless con-verts significant amounts of gravitational energy into heat for billions of years. From the perspective of BD formation, one can expect that satellites of BDs should commonly form in the dusty, gaseous disks around accreting BDs (Payne & Lodato 2007). At the transitional mass regime to giant planets, models of moon formation have shown that Mars-sized moons should commonly form around the most massive super-Jovian gas giant planets (Heller & Pudritz 2015b,a). Photometric accuracies of $10^{-2}$ have now been achieved on BDs using the *Hubble Space Telescope* (Zhou et al. 2016), and an improvement of about one order of magnitude should allow the detection of moons transiting luminous giant planets that can be directly imaged around their host star (Cabrera & Schneider 2007; Heller & Albrecht 2014; Heller 2016).

The search for exomoons has recently become an active area of research that several groups are now competing in, mostly using space-based stellar photometry of exoplanet-exomoon transits (Kipping et al. 2012; Szabó et al. 2013;

Simon et al. 2015; Hippke 2015) but also using alternatives such as radio emission from giant planets with magnetic fields that are perturbed by moons (Lukić 2017) or space-based coronographic methods such as spectroastrometry (Agol et al. 2015). With the first tentative detection of exomoons or exorings recently claimed in the literature (Mamajek et al. 2012; Bennett et al. 2014; Udalski et al. 2015; Aizawa et al. 2017; Hill et al. 2017; Teachey et al. 2018; Rodenbeck et al. 2018), a first unambiguous discovery could thus be imminent. Observational biases as well as dynamical constraints will preferentially reveal large, massive moons beyond 0.1 AU around their star (Szabó et al. 2006; Heller 2014), similar to the moon system that we investigate in this study.

The first step in estimating the climate conditions on a moon is in the identification of the relevant energy sources. Different from planets, moons receive stellar reflected light from their planetary host (Heller & Barnes 2015) as well as the planet's own thermal emission (Heller & Barnes 2015). In particular, a super-Jovian planet's own luminosity can desiccate its initially water-rich moons over several 100 Myr and make it ultimately uninhabitable. But exomoons around giant planets can also be subject to significant tidal heating (Reynolds et al. 1987; Scharf 2006; Cassidy et al. 2009; Heller & Barnes 2013); see Io around Jupiter for a prominent example in the solar system (Peale et al. 1979).

For moons around giant planets in the stellar habitable zone (HZ), the reflected plus thermal planetary light onto the moon is a significant source of energy ($\gtrsim 10\,\mathrm{W\,m^{-2}}$) if the moon is closer than about 10 Jupiter radii ($R_\mathrm{jup}$) to its host planet. Global top-of-atmosphere flux maps showed that eclipses[1] can decrease the average energy flux on the moon's subplanetary point by tens of $\mathrm{W\,m^{-2}}$ relative to the moon's antiplanetary hemisphere (Heller & Barnes 2013). Eclipses can cause a maximum decrease of the globally averaged stellar flux of $\approx 6.4\,\%$ at most (Heller 2012).

In analogy to the stellar HZ, Heller & Barnes (2013) defined a circumplanetary "habitable edge" (HE) as a critical distance to the planet interior to which a moon with an Earth-like atmosphere (mostly $N_2$, small amounts $CO_2$, see Kasting 1993) and a substantial water reservoir experiences a runaway greenhouse effect and is therefore at least temporarily uninhabitable. In top-of-atmosphere energy flux calculations, moons orbiting planets in the stellar HZ encounter an inner HE but no outer HE. Only beyond the stellar HZ does an outer HE appear around the planet (Reynolds et al. 1987; Heller & Armstrong 2014). One-dimensional latitudinal energy balance models suggest that moons near the outer edge of the stellar HZ can face an outer HE owing to eclipses and an ice-albedo effect if they orbit their star near the outer edge of the HZ (Forgan & Yotov 2014; Forgan & Dobos 2016).

Hinkel & Kane (2013) used the Heller & Barnes (2013) model to study the effect of global energy flux variations for hypothetical exomoons orbiting four confirmed giant exoplanets in or near the stellar HZ ($\mu$ Ara b, HD 28185 b, BD+14 4559 b, and HD 73534). Their focus was on the orbital eccentricity of the planet-moon barycenter around the star with the result that fluctuations of tens of $\mathrm{W\,m^{-2}}$ can occur on moons orbiting on highly eccentric stellar orbits.

Significant improvements in exomoon climate simulations were presented by Forgan & Kipping (2013), who used a 1D latitudinal energy balance model to assess exomoon surface temperatures under the effect of tidal heating and eclipses. Forgan & Yotov (2014) advanced this model by also including planetary illumination, and Forgan & Dobos (2016) showed yet another update including a global carbonate-silicate cycle and a viscoelastic tidal heating model. In any of the previous studies that estimated surface temperatures on exomoons (Hinkel & Kane 2013; Forgan & Kipping 2013; Forgan & Yotov 2014; Forgan & Dobos 2016), maximum and minimum surface temperatures on exomoons varied by several degrees Kelvin (K) over the circumplanetary orbit at most, while variations due to a moon's changing stellar distance on its circumplanetary orbit or due to eclipses were a mere $\approx 0.1\,\mathrm{K}$.

General circulation models (GCMs) have been applied to model the effects of eclipses on the atmosphere and surface conditions on Titan, which shows up to 6 hr long eclipses during $\approx 20$ consecutive orbits around Saturn around equinox. Tokano (2016) showed that eclipse-induced cooling of Titan's surface, averaged over one orbit around Saturn, is usually $< 1\,\mathrm{K}$ on the pro-Saturnian hemisphere.

Here we present the first simulations of exomoon climates using an idealized GCM (Haqq-Misra & Kopparapu 2015). This GCM improves upon previous 1D studies by allowing us to calculate the energy transfer not only as a function of latitude but also of longitude and height in the exomoon atmosphere. Our main objective is to determine whether exomoon climates are principally different from exoplanet climates, that is: how do planetary illumination and tidal heating contribute to surface habitability in an exomoon atmosphere? And ultimately, could these climatic effects of the different heat sources possibly be observed with near-future technology?

## 2    METHODS

### 2.1    Choice of the simulated systems

Most of the major moons in the solar system are in synchronous rotation with their host planet; that is, one and the same hemisphere faces the planet permanently (except maybe for libration effects). The star, however, does not a have fixed position in the reference frame of such a moon, as the satellite rotates with its circumplanetary orbital period ($P_\mathrm{ps}$). Hence, while stellar illumination can be averaged over the day and night side of the moon (longitudinally but not latitudinally), the planet will always shine on the moon's subplanetary point. This is a principal difference between the illumination effects experienced by a planet and a moon.

Furthermore, there will be planet-moon eclipses. But they will only be relevant to the global climate if the moon is in a very close orbit that is nearly coplanar with the circumstellar orbit. Beyond Io's orbit around Jupiter, the orbit-averaged flux decrease will be a few percent at most, and beyond ten planetary radii around a Jovian planet eclipses will be completely negligible (Heller 2012). We therefore neglect planet-moon eclipses in our simulations.

---

[1] With an eclipse we here refer to a moon moving through the stellar shadow cast by the planet.

As an additional heat source, we consider geothermal energy, which can be produced via tidal heating, radiogenic decays in the rocky part of the moon, or through release of primordial heat from the moon's accretion. We do not simulate the production of these heat sources in our model, but use interesting and reasonable fiducial values in our GCM simulations.

The moon's mass is chosen as 0.25 Earth masses ($M_\oplus$), which we consider as an optimal value in terms of exomoon formation around accreting giant planets (Canup & Ward 2006; Heller et al. 2014; Heller & Pudritz 2015b), exomoon detectability (Kipping et al. 2009; Lewis 2011; Heller 2014; Heller & Albrecht 2014; Kipping et al. 2015), and exomoon habitability (Lammer et al. 2014).[2] As these moons should be water-rich, their radii should be about 0.7 Earth radii ($R_\oplus$).

We consider two moon orbits, one as wide as Europa's orbit around Jupiter ($10\,R_{\rm Jup}$) and one twice that value. The former choice is supported by simulations of moon formation around Jupiter-mass planets, which suggest that icy moons can migrate to ~$10\,R_{\rm Jup}$ within the circumplanetary disks (Sasaki et al. 2010; Ogihara & Ida 2012). The latter choice is motivated by the recent prediction that the most massive, water-rich moons around super-Jovian planets beyond 1 AU should form near the circumplanetary ice line at about 20 to $30\,R_{\rm Jup}$ (Heller & Pudritz 2015b,a). The orbital periods related to these two semi-major axes of the satellite orbit ($a_{\rm ps}$) are 1.175 and 3.324 d. We refer to these orbits as our "short-period" and "long-period" cases, respectively. In all cases, the moon's spin-orbit misalignment with respect to its circumplanetary orbit is assumed to be zero and variations of planetary illumination due to the moon being on an eccentric orbit are neglected.

As for the geothermal heat budget of the satellite ($F_{\rm g}$), we consider three cases of 0, 10, and $100\,{\rm W\,m^{-2}}$. In those cases where tidal heating is a major contribution, the highest values will only be reached in very close-in orbits within $\lesssim 10\,R_{\rm Jup}$ around our test planet. If the moon under consideration is the only major satellite, tidal processes will usually act to circularize its orbit (Goldreich 1963), to erode its obliquity (Heller et al. 2011), and to lock its rotation rate with the orbital mean motion (Makarov et al. 2016). These processes generate tidal heating in the moon, which will gradually decay over time. Tidal surface heating rates on moons could be > $100\,{\rm W\,m^{-2}}$ for about $10^6$ yr for a single moon on an initially eccentric orbit Heller & Barnes (2013). In this sense, the physical conditions that we model could preferably correspond to young systems rather than evolved systems. The system age, however, is not an input parameter in our models and our results are not restricted to young systems to begin with. In fact, large extrasolar moons have not been detected unambiguously so far, and so it remains unclear whether they are only subject to significant tidal heating when they are young. If the moon is member of a multi-satellite system, for example, then mutual interaction and resonances could maintain significant orbital eccentricities and extend this timescale by orders of magnitude (Heller et al. 2014; Zollinger et al. 2017).

As for the planetary illumination absorbed by the moon, planet evolution tracks show that the luminosities of young super-Jovian planets 10 times the mass of Jupiter can be as high as $10^{-5}$ to $10^{-3}$ solar luminosities ($L_\odot$), depending on the planet's core mass, amongst others (Mordasini 2013). Even at the lower end of this range, a moon in a Europa-wide orbit at about $10\,R_{\rm Jup}$ would absorb about $500\,{\rm W\,m^{-2}}$ (maybe for some ten Myr), thereby easily triggering a runaway greenhouse effect on the moon.[3] Hence, we consider four cases of planetary illumination onto the satellite's sub-planetary point, namely, 0, 10, 100, and $500\,{\rm W\,m^{-2}}$ (the latter one only in the short-period moon orbit). All these cases are summarized in Table 1.

## 2.2 Climate model

We use an atmospheric GCM to simulate the climate of an Earth-like moon in orbit around a super-Jovian planet. This GCM was developed by the Geophysical Fluid Dynamics Laboratory (GFDL), based upon their 'Flexible Modeling System' (FMS), with idealized physical components (Frierson et al. 2006, 2007a; Haqq-Misra et al. 2011; Haqq-Misra & Kopparapu 2015). The dynamical core uses a spectral decomposition method with T42 resolution to solve the Navier-Stokes (or 'primitive') equations of motion. We use a shallow penetrative adjustment scheme to perform convective adjustment in the model (Frierson et al. 2007a), which provides a computationally efficient method for restoring vertical balance in lieu of a more explicit representation of convective processes. The GCM surface is bounded with a diffusive boundary layer scheme and a 50 m thick thermodynamic ocean layer with a fixed heat capacity. This is analogous to assuming that the moons are fully covered with a static, uniform ocean[4]; hence, topography is neglected. Our assumption of aquamoon conditions with no topography or ice also means that we neglect ice-albedo feedback. These simplifications allow for computational efficiency, and they allow us to examine fundamental changes in climate structure without any positive feedback processes causing the model to become numerically unstable.

The GCM includes two-stream gray radiative transfer, which uses a gray-gas radiative absorber with a specified vertical profile to mimic a greenhouse effect (Frierson et al. 2006). The model atmosphere is transparent to incoming stellar (*i.e.*, shortwave) radiation, so that incoming starlight penetrates the atmosphere and warms the surface (with a fixed surface albedo of 0.31 for shortwave radiation). Stellar radiation is averaged across the surface (so there is no diurnal cycle). Infrared (*i.e.*, longwave) radiation is absorbed by

---

[2] However, Awiphan & Kerins (2013) showed that photometric red noise from stellar variability might make it difficult to detect even Earth-mass moons in the HZ around low-mass stars with *Kepler*.

[3] This case is particularly interesting in view of the possible detection of moons around young, self-luminous giant planets via direct imaging with the *European Extremely Large Telescope* (Heller & Albrecht 2014).

[4] This is motivated by our assumption that our test moons formed in the icy parts of the accretion disks around their giant host planets at several AU from their Sun-like star. The initial $H_2O$ ice content of the moons would then have liquefied as the planets and moons migrated to about 1 AU, where ~100 super-Jovian exoplanets are known today.

**Table 1.** Simulated exomoon systems: initialization parameters and global average quantities

| Case | Initialization parameters | | | | Average quantities | | | |
|------|------|------|------|------|------|------|------|------|
| | $a_{\rm ps}$ | $P_{\rm ps}$ | $F_{\rm g}$ | $F_{\rm t}$ | $T_{\rm surf}$ | $\Delta T_{\rm pole}$ | $F_{\rm OLR}$ | $q_{\rm strat}$ |
| $1^{a}$ | $20\,R_{\rm Jup}$ | 3.324 d | $0\,{\rm W\,m^{-2}}$ | $0\,{\rm W\,m^{-2}}$ | 289.5 K | 0.0 K | $236.4\,{\rm W\,m^{-2}}$ | $7.6\times10^{-6}$ |
| 2 | $20\,R_{\rm Jup}$ | 3.324 d | $0\,{\rm W\,m^{-2}}$ | $10\,{\rm W\,m^{-2}}$ | 290.0 K | 0.7 K | $239.4\,{\rm W\,m^{-2}}$ | $7.6\times10^{-6}$ |
| 3 | $20\,R_{\rm Jup}$ | 3.324 d | $10\,{\rm W\,m^{-2}}$ | $0\,{\rm W\,m^{-2}}$ | 291.7 K | 4.0 K | $246.0\,{\rm W\,m^{-2}}$ | $8.3\times10^{-6}$ |
| 4 | $20\,R_{\rm Jup}$ | 3.324 d | $10\,{\rm W\,m^{-2}}$ | $10\,{\rm W\,m^{-2}}$ | 292.2 K | 5.0 K | $249.0\,{\rm W\,m^{-2}}$ | $8.1\times10^{-6}$ |
| 5 | $20\,R_{\rm Jup}$ | 3.324 d | $10\,{\rm W\,m^{-2}}$ | $100\,{\rm W\,m^{-2}}$ | 295.9 K | 11.9 K | $276.3\,{\rm W\,m^{-2}}$ | $3.9\times10^{-5}$ |
| $6^{b}$ | $10\,R_{\rm Jup}$ | 1.175 d | $0\,{\rm W\,m^{-2}}$ | $0\,{\rm W\,m^{-2}}$ | 287.9 K | 0.0 K | $234.9\,{\rm W\,m^{-2}}$ | $1.4\times10^{-5}$ |
| 7 | $10\,R_{\rm Jup}$ | 1.175 d | $0\,{\rm W\,m^{-2}}$ | $10\,{\rm W\,m^{-2}}$ | 288.4 K | 0.6 K | $237.9\,{\rm W\,m^{-2}}$ | $1.7\times10^{-5}$ |
| 8 | $10\,R_{\rm Jup}$ | 1.175 d | $10\,{\rm W\,m^{-2}}$ | $0\,{\rm W\,m^{-2}}$ | 290.0 K | 3.8 K | $244.4\,{\rm W\,m^{-2}}$ | $2.0\times10^{-5}$ |
| 9 | $10\,R_{\rm Jup}$ | 1.175 d | $10\,{\rm W\,m^{-2}}$ | $10\,{\rm W\,m^{-2}}$ | 290.5 K | 4.5 K | $247.5\,{\rm W\,m^{-2}}$ | $2.3\times10^{-5}$ |
| 10 | $10\,R_{\rm Jup}$ | 1.175 d | $10\,{\rm W\,m^{-2}}$ | $100\,{\rm W\,m^{-2}}$ | 294.9 K | 13.8 K | $275.5\,{\rm W\,m^{-2}}$ | $6.8\times10^{-5}$ |
| 11 | $10\,R_{\rm Jup}$ | 1.175 d | $10\,{\rm W\,m^{-2}}$ | $500\,{\rm W\,m^{-2}}$ | 310.0 K | 45.5 K | $398.3\,{\rm W\,m^{-2}}$ | $3.3\times10^{-3}$ |
| 12 | $10\,R_{\rm Jup}$ | 1.175 d | $100\,{\rm W\,m^{-2}}$ | $500\,{\rm W\,m^{-2}}$ | 321.9 K | 60.8 K | $482.2\,{\rm W\,m^{-2}}$ | $1.4\times10^{-2}$ |

**Notes.** In all cases the planet-moon binary orbits at 1 AU from a Sun-like star, $M_{\rm p} = 10\,M_{\rm Jup}$, $M_{\rm s} = 0.25\,M_{\oplus}$, and $R_{\rm s} = 0.7\,M_{\oplus}$. The parameter $F_{\rm g}$ is uniform geothermal heating at all latitudes. The parameter $F_{\rm t}$ is absorbed planetary illumination, with a latitudinal distribution of $F_{\rm t}|\cos\lambda|$ when $90^{\circ} < \lambda < 270^{\circ}$ and zero otherwise. Mean values from the set of simulations are shown for global surface temperature $T_{\rm surf}$, change in polar surface temperature $\Delta T_{\rm pole}$, global outgoing longwave radiative flux at the top of the atmosphere $F_{\rm OLR}$, and global specific humidity at the 50 hPa level $q_{\rm strat}$.
$^{(a,b)}$ We also refer to 1 and 4 as our "slow rotator control" and "rapid rotator control" cases, respectively.

the gray-gas atmosphere in proportion to the optical depth at each model layer, with the surface values of optical depth tuned to reproduce an Earth-like value when the GCM is configured with Earth-like parameters. Furthermore, water vapor is decoupled from the gray radiative transfer scheme, so that water vapor feedback is neglected. The GCM is also cloud-free, although we remove excess moisture and energy from the atmosphere through large-scale condensation to the surface.

Even with these idealized assumptions, this GCM remains capable of representing surface and tropospheric temperature profiles, as well as the large-scale circulation features, of Earth today (Frierson et al. 2006, 2007a; Haqq-Misra et al. 2011). The model maintains symmetry about the equator due to the lack of a seasonal cycle, which can be interpreted as a mean annual climate state. The model was originally developed to explore the role of moisture on tropospheric static stability (Frierson et al. 2006) and the transport of static energy in moist climates (Frierson et al. 2007a). We also note that this model has been used to demonstrate that a realistic tropospheric profile can be maintained by eddy fluxes alone, even in the absence of stratospheric ozone warming (Haqq-Misra et al. 2011). Even so, our application of this model to exomoons implies that our results should be interpreted qualitatively, as a conservative estimate with regard to surface temperature values, runaway greenhouse thresholds, and large-scale dynamics.

The use of a gray-gas absorber allows us to avoid the problem of choosing a specific atmospheric composition, as any particular choice of greenhouse gases (such as carbon dioxide or methane) will yield a unique GCM solution. Although many GCM studies of exoplanet atmospheres use band-dependent (*e.g.*, non-gray) radiation with cloud parameterizations (Yang et al. 2014; Kopparapu et al. 2016; Wolf et al. 2017; Leconte et al. 2013b; Popp et al. 2016; Godolt et al. 2015; Haqq-Misra et al. 2018), such an approach introduces a new set of free parameters for deter-mining the appropriate mix of greenhouse and inert gases. We instead choose to use an idealized gray-gas GCM for our study of exomoon habitability, as others have done for qualitatively exploring the runaway greenhouse threshold (Ishiwatari et al. 2002). Likewise, the assumption that water vapor is radiatively neutral allows our model to remain stable at high stellar flux levels without initiating a runaway greenhouse state; thus, our assumption of a cloud-free atmosphere and a simplified convection scheme limits the use of our GCM for quantitatively determining the runaway greenhouse threshold. For example, clouds beneath the subplanetary point could help to delay the onset of a runaway greenhouse (Yang et al. 2014), although rapid rotation may weaken this effect (Kopparapu et al. 2016). We emphasize that our radiation limits, particular for identifying a runaway greenhouse, should be interpreted in a qualitative sense in order to guide more sophisticated investigations with less-idealized GCMs.

We fix the moon into synchronous rotation with its planet so that the subplanetary point is centered at the moon's equator, and we set the moon to have a circular orbit and an obliquity of zero. We include an additional source of infrared heating at the top of the atmosphere, centered on the subplanetary point, which we use to represent heating by the super-Jovian planet in our simulations. Specifically, we set the downward infrared flux at the top of the atmosphere equal to $F_{\rm t}|\cos\lambda|$ when $90^{\circ} < \lambda < 270^{\circ}$, and zero otherwise (where $\lambda$ is longitude). Because the atmosphere is absorbing to infared radiation (both in upward and downward directions), this planetary infrared flux is absorbed by the uppermost atmospheric layers, with none of this infrared radiation penetrating through to the surface in any of our simulations. This upper atmosphere absorption of infrared radiation from the host planet is the primary feature that distinguishes an exomoon climate from an Earth-like exoplanet climate.

We also add geothermal heating uniformly at the bot-tom of the atmosphere, as a representation of tidal heating. The infrared flux from the surface follows the Stefan-Boltzman law, with the geothermal heating term, $F_g$, added as a secondary source of surface warming. Geothermal heating due to tidal heating may also appear on synchronously rotating exoplanets (Haqq-Misra & Kopparapu 2015), particularly those with highly eccentric orbits. Geothermal heating can contribute to habitability by increasing surface temperature, while it can also alter atmospheric circulation patterns by driving stronger poleward transport. Geothermal heating provides a secondary mechanisms that may contribut to climatic features on exomoons.

Each case in Table 1 was initialized from a state of rest and run for a period of 3,000 d in total. The first 1,000 d were discarded, and the average of the subsequent 2,000 d of runtime were used to analyze our cases. The model reaches a statistically steady state within about 500 d of initialization, which takes approximately 4 h to complete using a Linux workstation (6 cores at 2.0 GHz). In our presentation of results, we refer to our cases with a 1.175 d rotation rate as 'short-period' and our cases with a 3.324 d rotation rate as 'long-period.' We also refer to our two cases with geothermal heating and absorbed planetary illumination set to zero as our 'control' cases. Our set of twelve cases provide an overview of the dependence of an exomoon's climate on the properties of its host planet.

## 3 RESULTS

For all of our control and experimental cases, we calculate global average values of surface temperature, outgoing longwave flux at the top of the atmosphere, and stratospheric specific humidity (Table 1). Our control cases (1 and 6) both show global average surface temperatures similar to that of Earth today. Our experimental cases (2-5 and 7-12) show an increase in temperature as geothermal and planetary fluxes increase, with a corresponding increase in stratospheric water vapor and outgoing infrared radiation.

### 3.1 Surface habitability

Surface temperature and winds are shown for the two control cases in Figure 1, with the slow rotator on the top row, the rapid rotator on the middle row, and the difference between the two on the bottom row. The change in rotation rate from the slow to rapid rotator results in an increase in the strength of the easterly equatorial jet, which also corresponds to an increase in the equator-to-pole temperature contrast. This expected behavior corroborates the classic results of Williams & Halloway (1982), who find that even slower rotation rates will result in a global meridional circulation cell that spans the entire hemisphere. Lacking geothermal or planetary heating, these control cases represent Earth-like climate states for a smaller planet at different rotation rates.

The addition of infrared planetary illumination to the top of the atmosphere and geothermal heating to the surface can lead to departures from an Earth-like climate state. Figure 2 shows surface temperature and winds for the rapid rotator experiments with increasing contributions of infrared heating. The top row shows cases with modest geothermal

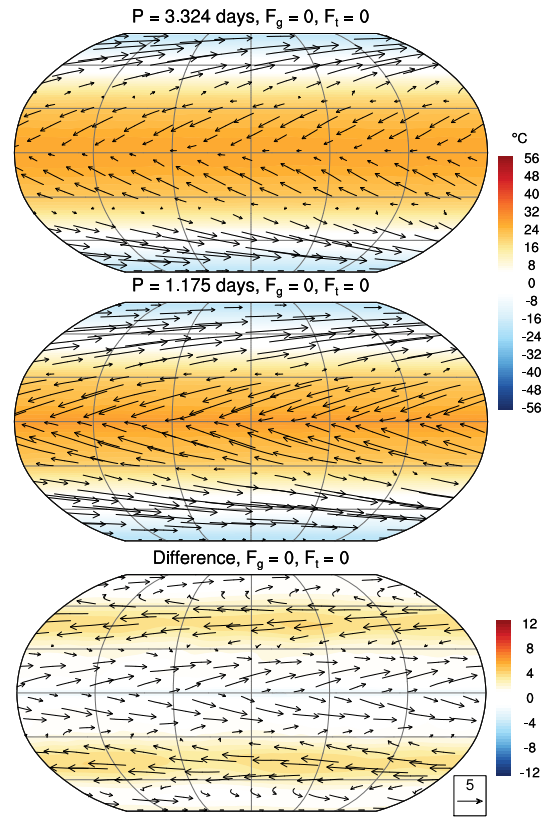

**Figure 1.** Time average of surface temperature deviation from the freezing point of water (shading) and horizontal wind (vectors) for the $P = 3.324$ d (top panel) and $P = 1.175$ d (middle panel) control cases. The bottom row shows the difference of the first row minus the second. The subplanetary point falls on the center of each panel. The lengths of the vectors are proportional to the local wind speeds, and a reference vector with a length of $5\,\mathrm{m\,s^{-1}}$ is shown on the last panel. (The color scale is chosen to match the maximum temperature range shown in Figures 2 and 3 to ease comparison.)

and planetary heating, which leads to both warming of the poles and destabilization of the easterly equatorial jet. As additional heating is applied, shown in the bottom rows of Figure 2, the predominantly zonal flow of winds becomes disturbed in favor of a pattern dominated by large-scale vortices. Figure 2 also includes dark contours showing the 'animal habitable zone' limits of 0 and 50 degrees Celsius, which represents the temperature range where complex animal life could survive (Ward & Brownlee 2000; Edson et al. 2012). The two cases in the top row of Figure 2 both show that the freezing line of 0 degrees Celsius sits near the boundary between midlatitude and polar regions, around 60 degrees latitude, so that the planet's habitable real estate is confined to the midlatitude and equatorial regions. The bottom left panel of Figure 2 shows a case where the entire surface is within the animal habitability limits, as a result from strong thermal heating from the host planet. The most extreme case, shown in the bottom right panel of Figure 2, has the 50 degree Celsius contour at about 30 degrees latitude,

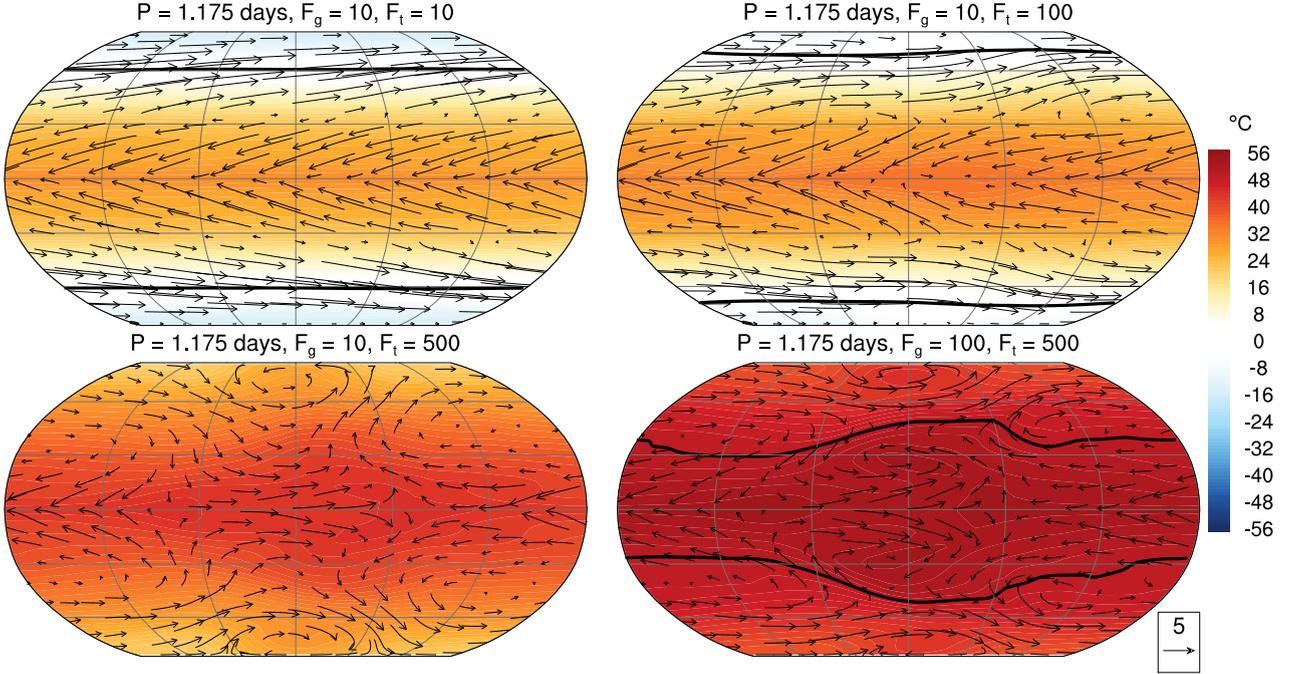

**Figure 2.** Time average of surface temperature deviation from the freezing point of water (shading) and horizontal surface wind (vectors) for the $P = 1.175$ d experiments. The geothermal ($F_g$) and top-of-the atmosphere heat flux from the planet ($F_t$) are shown on top of each panel. Dark coutours indicate the 'animal habitability' temperature bounds of 0 and 50 degrees Celsius where complex life could survive. The lengths of the vectors are proportional to the local wind speeds, and a reference vector with a length of $5\,\mathrm{m\,s^{-1}}$ is shown on the last panel.

with regions closer to the equator being too warm to support complex life.

Our set of GCM cases also illustrate the propensity for an exomoon atmosphere to enter a runaway greenhouse as a result of strong thermal heating aloft. As we discussed above, our idealized GCM neglects water vapor feedback and relies upon gray-gas radiative transfer, so our consideration of greenhouse states should be interpreted as a qualitative and conservative estimate. Further GCM development with band-dependent radiative transfer, cloud paramterization, and more realistic convective processes will be required to identify quantitative thresholds for when we should expect to observe a runaway greenhouse state on an exomoon. Additionally, a dry moon with a desert surface and little standing water will remain stable past these radiation limits (Abe et al. 2011; Leconte et al. 2013b), so our moist GCM calculations also serve as a conservative estimate of the runaway greenhouse threshold. Given these caveats, it still remains instructive to consider how our exomoon simulations compare with theoretical limits for expecting a runaway greenhouse.

The long-period cases (1-5) show a maximum outgoing infrared flux of $276.3\,\mathrm{W\,m^{-2}}$, which falls within the stable radiation limits calculated from one-dimensional (Kasting 1988; Goldblatt & Watson 2012; Ramirez et al. 2014) and three-dimensional (Leconte et al. 2013a; Wolf & Toon 2014) climate models. By contrast, the short-period cases (6-12) include experiments with an outgoing infrared flux of $400\,\mathrm{W\,m^2}$ or larger (cases 11 and 12), which exceeds stable radiation limits and indicates the climate should be in a runaway greenhouse state (Kasting 1988; Goldblatt & Watson

2012; Ramirez et al. 2014; Leconte et al. 2013a). The two cases (11 and 12) with $F_t = 500\,\mathrm{W\,m^{-2}}$ are also both within the runaway greenhouse regime, which suggests that their surface temperatures, as shown in the bottom row of Figure 2, would continue to increase in a GCM with raditively-coupled water vapor.

Water loss can also occur prior to the runaway greenhouse, due to the photodissociation of water vapor as the stratosphere becomes wet, in a process sometimes known as a 'moist greenhouse' (Kasting 1988; Kopparapu et al. 2017). The moist greenhouse state can be inferred by amount of water vapor that crosses the tropopause and reaches the stratosphere, which we indicate using specific humidity (the ratio of moist to total air), $q_{\mathrm{strat}}$, near the model top (see Table 1). Calculations with other models indicate that atmospheres enter a moist greenhouse and begin to rapidly lose water to space when stratospheric specific humidity exceeds a threshold of $q_{\mathrm{strat}} \approx 10^{-3}$ (Kasting 1988; Kopparapu et al. 2013; Wolf & Toon 2014; Kopparapu et al. 2016). As a qualitative approach to this problem, our results illustrate that thermal heating from the host planet, as well as geothermal heating from tides, are both plausible mechanisms for heating an exomoon atmosphere to the point of initiating water loss.

## 3.2   Polar amplification of warming

We compare our potentially habitable long-peroid and short-peroid cases with $F_g = 10\,\mathrm{W\,m^{-2}}$ and $F_t = 100\,\mathrm{W\,m^{-2}}$ with the corresponding control cases in Figure 3. These cases are within stable radiation limits and are not at risk of entering a runaway greenhouse or otherwise losing water to space due to a wet stratosphere. Both panels in Figure 3 show that warming is concentrated toward the poles, with a more modest degree of warming at the tropics and midlatitudes. This 'polar amplification' is a well-known process that occurs in climate models, particularly in simulations of global warming, and is an expected consequence of imposing additional heating on an atmosphere.

Polar amplification is commonly attributed to the warming that results from ice-albedo feedback in polar regions, which accelerates the loss of ice, thereby reducing albedo and continuing to accelerate the rate of warming (Polyakov et al. 2002; Holland & Bitz 2003). However, polar amplification is also present in idealized models that lack sea ice feedback entirely (Alexeev et al. 2005; Langen & Alexeev 2007; Alexeev & Jackson 2013), which suggests that atmospheric heat transport alone can provide an explanatory mechanism. Alexeev et al. (2005) demonstrated that polar amplification should occur when a GCM is forced with an additional source of uniform surface warming, which in principle could be casued by both infrared and visible sources. Alexeev & Jackson (2013) argued that both ice-albedo feedback and meridional energy transport contribute to polar amplification, with the effects of energy transport being masked when ice-albedo feedback is present.

Our exomoon simulations further illustrate the capability of GCMs to show polar amplification in the absence of ice-albedo feedback. Previous idealized GCMs have shown polar amplification from uniform thermal heating sources, and our results demonstrate that similar polar amplification can be obtained from the non-uniform thermal heating of planetary illumination. Polar amplification on ice-free planets occurs as a response to enhancement of the meridional circulation on a warmer planet (Lu et al. 2007), which leads to increased poleward transport of energy and moisture (Alexeev et al. 2005).

We calculate the polar amplification, $\Delta T_{pole}$ as the difference in the mean temperature at the north pole between each experiment and the corresponding control case. The values of $\Delta T_{pole}$ are shown in Table 1. A constant uniform geothermal heating of $F_g = 10\,W\,m^{-2}$ with no planetary illumination (cases 3 and 8) yields a polar amplification of about $4\,K$, with the outgoing longwave radiation also about $10\,W\,m^{-2}$ greater than the control case. This indicates that geothermal heating is entirely absorbed and re-radiated by the lower, thicker layers of the atmosphere. Conversely, a non-uniform planetary illumination of $F_t = 10\,W\,m^{-2}$ with no geothermal heating (cases 2 and 7) shows a smaller polar amplification of about $0.6\,K$. Cases 2 and 7 show an increase in outgoing longwave radiation of only about $3\,W\,m^{-2}$ compared to the control; however, this is expected because the non-uniform distribution of planetary heating ($F_t |\cos \lambda|$ when $90° < \lambda < 270°$) results in a net warming of $F_t/\pi \approx 3.2\,W\,m^{-2}$. Even so, the total polar amplification of $0.6\,K$ in cases 2 and 7 is nearly seven times less than the polar amplification of $4\,K$ in cases 3 and 8. This indicates that some planetary illumination is absorbed by the uppermost layers of the atmosphere, with the remainder of this energy contributing to surface warming. However, these cases (2 and 7) still demonstrate that polar amplification can occur from a non-uniform upper-atmospheric heating source. The experiments with equal geothermal and planetary heating of $F_g = 10\,W\,m^{-2}$ and $F_t = 10\,W\,m^{-2}$ (cases 4 and 9) show that the value of $\Delta T_{pole}$ is equal to the sum of the polar amplification when each heating source is considered in isolation. Likewise, the value of $F_{OLR}$ for case 4 (and 9) equals the sum of the outgoing longwave radiation terms from cases 2 and 3 (7 and 8). Polar amplification continues to increase when $F_t = 100\,W\,m^{-2}$ and greater (cases 5, 10-12), with corresponding increases in $F_{OLR}$ that indicate a further penetration depth for incoming planetary illumination. These results emphasize that polar amplification in GCMs that lack ice albedo feedback can still occur with both uniform surface and non-uniform stratospheric warming.

The expansion of the meridional overturning (i.e., Hadley) circulation is shown in Fig. 4 for the $P = 3.324\,d$ experiments. The top row of Fig. 4 shows the control case, while the bottom row shows the experiment with $F_g = 10\,W\,m^{-2}$ and $F_t = 100\,W\,m^{-2}$. The left column of Fig. 4 shows the global average, while the middle and right columns separate the atmosphere into the hemispheres east and west, respectively, of the subplanetary point. The purpose of this decomposition is to show that the atmosphere responds to a fixed heating source by altering both the direction and width of the Hadley circulation in each hemisphere relative. This prediction originates from the shallow water model of Geisler (1981), which demonstrated that the Hadley circulation should change directions on either side of a fixed heating source. Although the Hadley circulation appears to weaken and maintain its width when planetary and geothermal heating is applied (left column), the hemispheric decomposition shows that the Hadley circulation in eastern hemisphere reaches fully to the poles while the western hemisphere shows a Hadley circulation with the opposite direction. The zonal wind pattern in these atmospheres remains relatively consistent between the two cases, with prominent upper-level jets associated with the descending branch of the global Hadley cell that appear identical in the hemispheric decomposition.

We also demonstrate this Hadley cell expansion or the $P = 1.175\,d$ experiments in Fig. 5, which shows the control experiment along with three other cases of increasing planetary illumination and geothermal heating. Fig. 5 shows that the global average Hadley circulation tends to diminish as planetary illumination increases, but the hemispheric decomposition shows that the Hadley circulation is actually expanding poleward. The two hemispheres show circulations with opposite direction and approximately the same strength. Geothermal heating does not significantly alter the circulation strength, although it does change the morphology of the ascending branch of the Hadley circulation. Two upper-level midlatitude jets are present when $F_t \leq 100\,W\,m^{-2}$, while a third equatorial jet emerges when $F_t = 500\,W\,m^{-2}$. Strong geothermal heating of $F_g = 100\,W\,m^{-2}$ tends to sharpen the equatorial jet and raise the altitude of the midlatitude jets.

We further demonstrate this behavior in our results by examining the vertically integrated flux of moist static energy as a function of latitude, following Frierson et al. (2007b) and Kaspi & Showman (2015). Moist static energy, $m$, represents the combination of dry static energy and latent

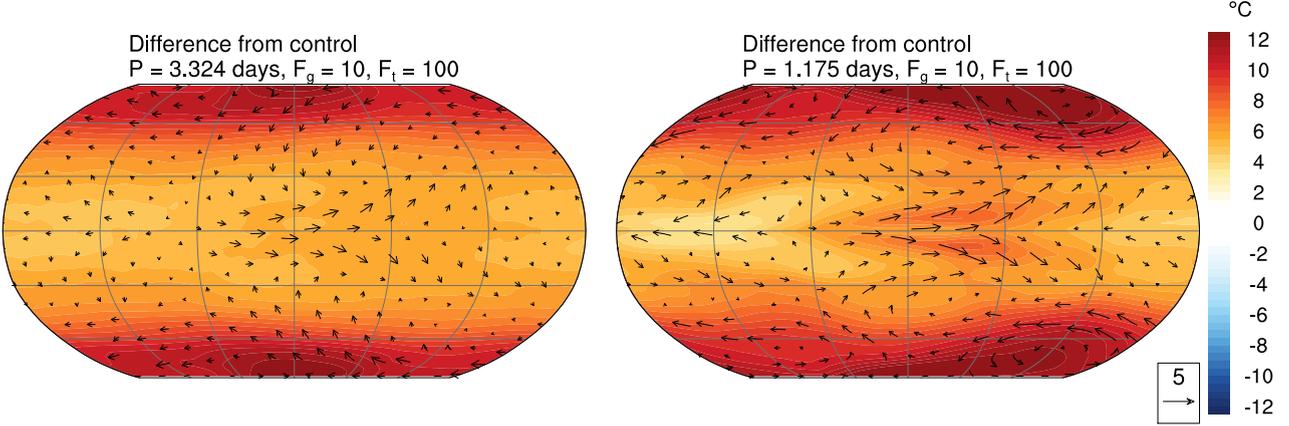

**Figure 3.** Difference from control cases of the time average of surface temperature (shading) and horizontal wind (vectors) for the $P = 3.324$ d (left panel) and $P = 1.175$ d (right panel) experiments with $F_g = 10\,\mathrm{W\,m^{-2}}$ and $F_t = 100\,\mathrm{W\,m^{-2}}$. The lengths of the vectors are proportional to the local wind speeds, and a reference vector with a length of $5\,\mathrm{m\,s^{-1}}$ is shown on the last panel.

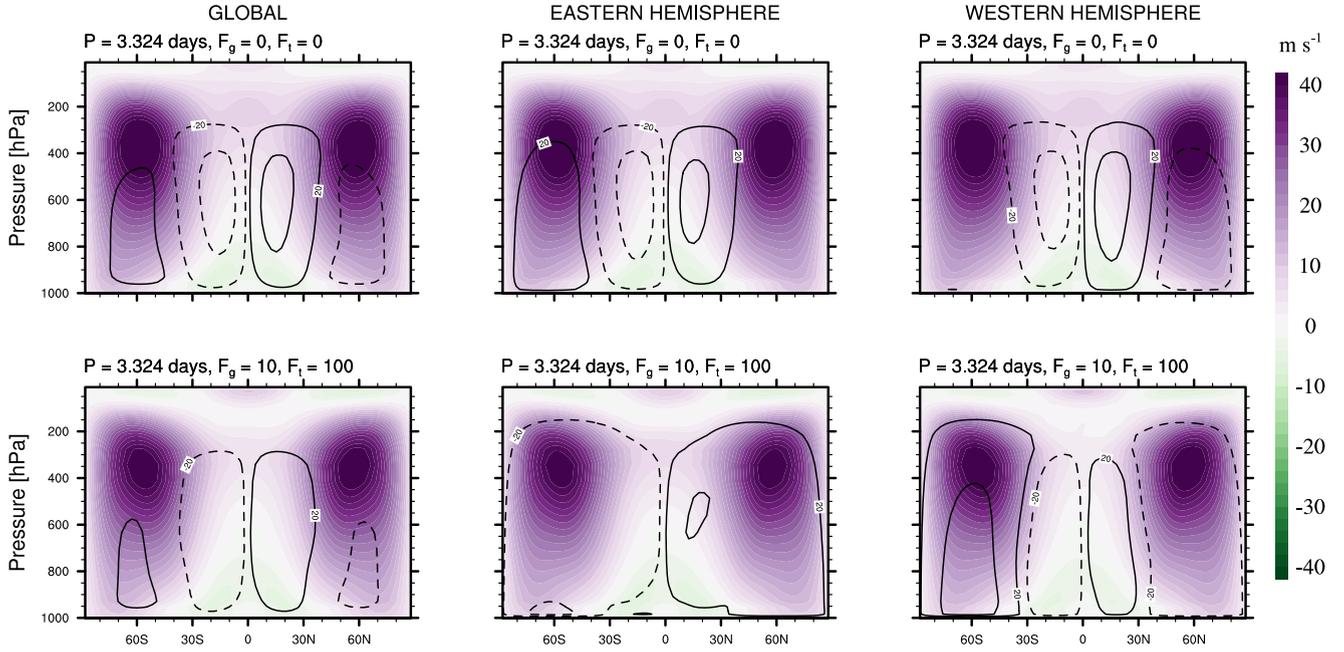

**Figure 4.** The mean meridional circulation (line contours) and zonal mean zonal wind (shading) are averaged across the entire planet (first column), eastern hemisphere (second column), and western hemisphere (third column) from the subplanetary point for the $P = 3.324$ d experiments with $F_g = F_t = 0\,\mathrm{W\,m^{-2}}$ (top row) and $F_g = 10\,\mathrm{W\,m^{-2}}$ and $F_t = 100\,\mathrm{W\,m^{-2}}$ (bottom row). Contours are drawn at an interval of $\pm\{20, 100, 300\} \times 10^9\,\mathrm{kg\,s^{-1}}$. Solid contours indicate positive (northward) circulation, and dashed contours indicate negative (southward) circulation.

energy as

$$m = c_p T + \Phi + L_v q, \tag{1}$$

where $c_p$ is the specific heat capacity of air, $\Phi$ is geopotential height, $L_v$ is the enthalpy of vaporization, and $q$ is specific humidity. We decompose moist static energy $m$ into a sum of time mean $\overline{m}$ and eddy $m'$ contributions as $m = \overline{m} + m'$. This allows us to write the meridional moist static energy flux as

$$\overline{vm} = \overline{v}\,\overline{m} + \overline{v'm'}, \tag{2}$$

where $\overline{vm}$ represents meridional mean energy transport and $\overline{v'm'}$ represents meridional eddy energy transport. The vertically integrated flux of $m$ is defined as

$$\overline{M} = 2\pi a \cos\phi \int_0^{p_s} \frac{\overline{vm}}{g}\,dp, \tag{3}$$

where $a$ is planetary radius, $\phi$ is latitude, and the overbar denotes a zonal and time mean. Equation (3) gives the total value of $\overline{M}$ from all dynamical contributions. We can likewise separate the mean and eddy contributions to $\overline{M}$ by replacing the meridional static energy flux $\overline{vm}$ in Eq. (3) with $\bar{v}\bar{m}$ or $\overline{v'm'}$, respectively.

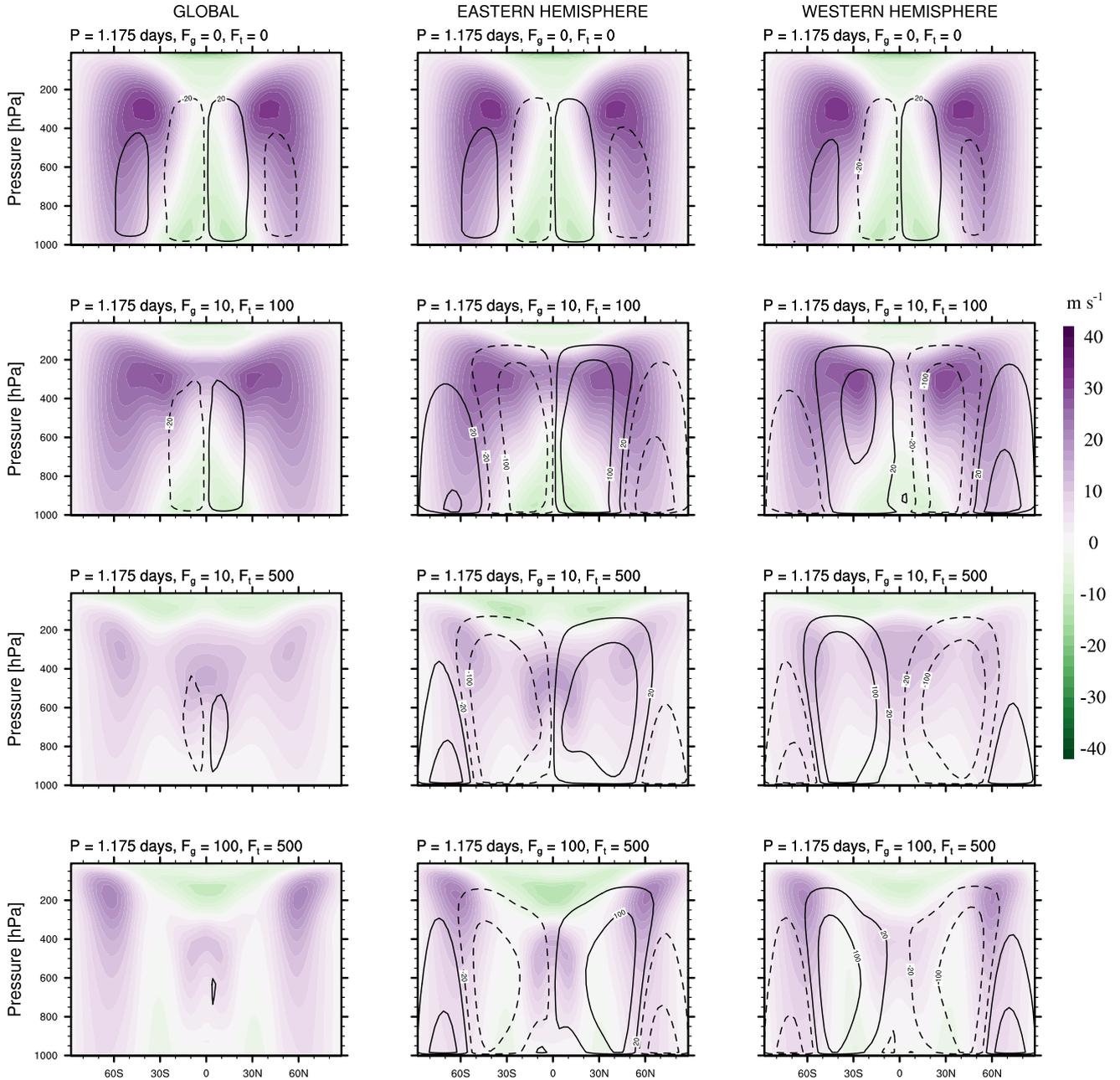

**Figure 5.** The mean meridional circulation (line contours) and zonal mean zonal wind (shading) are averaged across the entire planet (first column), eastern hemisphere (second column), and western hemisphere (third column) from the subplanetary point for the $P = 1.175$ d experiments with $F_g = F_t = 0$ W m$^{-2}$ (first row), $F_g = 10$ W m$^{-2}$ and $F_t = 100$ W m$^{-2}$ (second row), $F_g = 10$ W m$^{-2}$ and $F_t = 500$ W m$^{-2}$ (third row), and $F_g = 100$ W m$^{-2}$ and $F_t = 500$ W m$^{-2}$ (last row). Contours are drawn at an interval of $\pm\{20, 100, 3000\} \times 10^9$ kg s$^{-1}$. Solid contours indicate positive (northward) circulation, and dashed contours indicate negative (southward) circulation.

We present the total, mean, and eddy fluxes of $\overline{M}$ in Figure 6 to show the difference between our control cases and our experiments with $F_g = 10$ W m$^{-2}$ and $F_t = 100$ W m$^{-2}$. Both panels show a poleward increase in $\overline{M}$ when geothermal and planetary heating are added, with all of this contribution due to increases in the mean component of $\overline{M}$ at latitudes $\phi > 50°$. By contrast, the eddy component of $\overline{M}$ decreases in the range $30° < \phi < 50°$ when heating is induced. Polar amplification, and an associated increase in moisture, occurs as a result of intensification of the mean poleward transport of static and latent energy fluxes from both surface geothermal and top-of-atmosphere planetary heating.

The moist static energy flux also provides an explanation for the equatorial band of warming in our difference plots shown in Figure 3, particularly in the rapid rotating case. Figure 6 also shows a decrease in $\overline{M}$ at tropical latitudes in the range $0° < \phi < 15°$, which leads to warming and an accumulation of moisture beneath the subplanetary point. This point corresponds to a maximum in the rising motion of the mean meridional circulation (not shown), as a result of planetary illumination. Although our GCM does not include cloud processes, convective processes at the subplanetary point should lead to cloud formation, which could contribute to an expansion of the inner habitable region around the host planet (Yang et al. 2014). In general, the circulation patterns of climates with a fixed source of heating are not easily characterized by longitudinally-averaged mean meridional circulation functions, due to hemispheric reversals in the direction of these circulation patterns (Haqq-Misra & Kopparapu 2015). Nevertheless, we can still expect strong rising motion beneath the subplanetary point, with both zonal and meridional transport toward the opposing hemisphere.

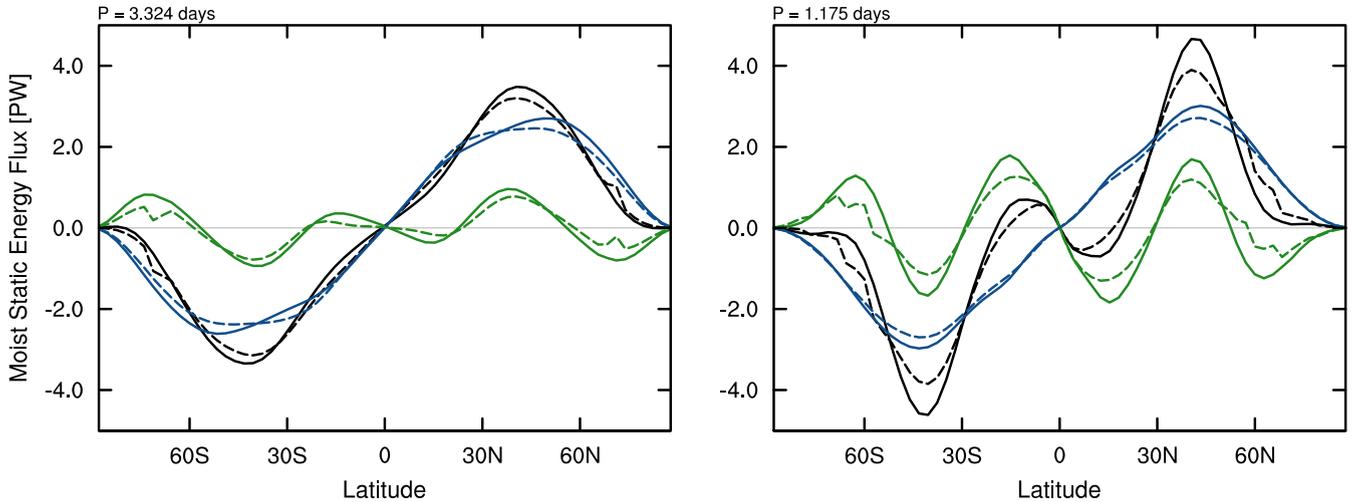

**Figure 6.** Vertically integrated moist static energy flux ($\overline{M}$) for the $P = 3.324$ d (left panel) and $P = 1.175$ d (right panel) experiments. The total $\overline{M}$ (black curves), mean contribution to $\overline{M}$ (green curves), and eddy contribution to $\overline{M}$ (blue curves) are shown for control cases with $F_g = F_t = 0$ (solid) and experiments with $F_g = 10\,\mathrm{W\,m^{-2}}$ and $F_t = 100\,\mathrm{W\,m^{-2}}$ (dashed).

### 3.3 Vertical structure of the atmosphere

The redistribution of energy from planetary illumination causes warming of both the surface as well as the polar stratosphere. Figure 7 shows the vertical structure of mean zonal temperature for the short-period cases with $F_g = F_t = 0\,\mathrm{W\,m^{-2}}$ (left panel) and $F_g = 10\,\mathrm{W\,m^{-2}}$ and $F_t = 100\,\mathrm{W\,m^{-2}}$ (right panel). The height of the tropopause is also shown as a dark curve in both panels of Figure 7, which follows the World Meteorological Organization definition of the tropopause as the altitude at which the lapse rate equals $-2\,\mathrm{K\,km^{-1}}$. For the control case (left panel), the tropical tropopause extends up to about $100\,\mathrm{hPa}$ due to convective heating by both moist and dry processes (Haqq-Misra et al. 2011). The height of the extratropical tropopause is determined by the balance between warming from latent and sensible heating in the troposphere below with warming in the stratosphere from the poleward transport of the Brewer-Dobson circulation (Haqq-Misra et al. 2011).

When planetary and geothermal heating are included (Figure 7, right panel), the boundaries of the tropical tropopause sharpen and act to widen the extratropical zone while narrowing the tropics. Increased warming beneath the subplanetary point drives stronger convection, which causes the tropopause to extend higher. This increase in convective heating also increases the poleward flux of moist static energy, which causes the extratropical tropopause to steepen. Warming in the stratosphere occurs primarily in the polar regions, which is driven by poleward energy transport processes such as the Brewer-Dobson circulation (not shown). This stratospheric warming competes with tropospheric warming in the tropics, which results in the polar height of the tropopause remaining relatively constant between the two cases shown. One interpretation of this behavior is that the expanded extratropics are analogous to an increase in efficiency of the moon's 'radiator fins,' which provide a means of transporting and dissipating energy from the warming tropics (Pierrehumbert 1995). Planetary illumination thus serves to sharpen the distinction between tropical and extratropical climate zones by redistributing energy poleward both along the surface and aloft in the stratosphere.

The effect of planetary illumination on the atmospheric structure is also evident from examining vertical temperature profiles along the equator (Fig. 8, left panel). For all profiles, the long-period cases show a colder stratosphere but a warmer troposphere than the short-period cases with the same planetary and geothermal heating. These differences correspond to the enhanced meridional transport of energy and moisture in the long-period cases, which also drives stronger and narrower jets in the short-period cases (Williams & Halloway 1982). The effect of planetary illumination causes a stratospheric temperature inversion, analogous to the ozone-driven stratospheric inversion on Earth today. The warmest cases with $F_t = 500 \, \mathrm{W\,m^{-2}}$ show a profile with increasing slope that begins approaching an isothermal atmosphere from the additional planetary and geothermal heating.

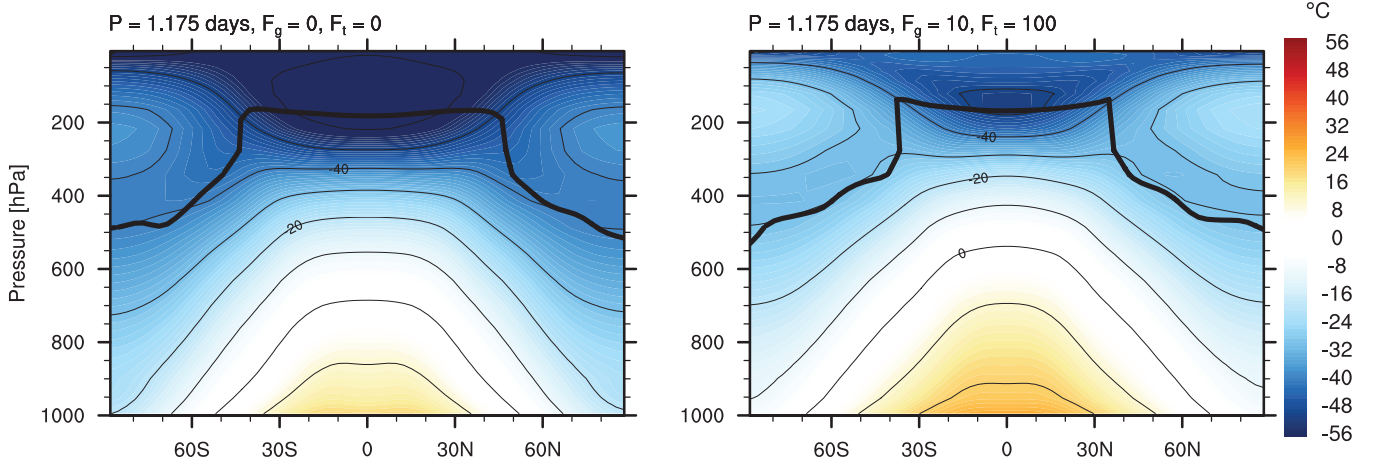

**Figure 7.** Time average of mean zonal temperature (shading and contours) for the $P = 1.175$ d experiments with $F_g = F_t = 0 \, \mathrm{W\,m^{-2}}$ (left panel) and $F_g = 10 \, \mathrm{W\,m^{-2}}$ and $F_t = 100 \, \mathrm{W\,m^{-2}}$ (right panel). The dark curve shows the height of the tropopause. (The color scale is chosen to match previous figures, and contours are drawn every 10 °C.)

Some planetary illumination is absorbed in the uppermost layers of the model atmosphere, while the rest contributes to surface warming. Fig. 8 (right panel) shows vertical profiles of the temperature tendency due to radiation (or radiative heating) along the equator. This reflects the direct change in temperature due to radiation alone, neglecting physical processes such as convection, boundary layer diffusion, and dynamical heating. The green curves with $F_t = 10 \, \mathrm{W\,m^{-2}}$ show modest warming in the upper atmosphere of about $0.1 \, \mathrm{K\,day^{-1}}$, with strong cooling in the middle troposphere and surface heating of about $2.8 \, \mathrm{K\,day^{-1}}$. The blue curves show increased upper atmosphere warming to $\sim 0.2 \, \mathrm{K\,day^{-1}}$ with stronger cooling in the middle troposphere. The blue curves also show a lower radiative heating rate at the surface, even though these cases show a higher surface temperature. The decrease in direct surface radiative heating is accounted for by increased diffusion of the boundary layer, which in turn leads to increased surface warming. The structure of the boundary layer is also evident from the left panel of Fig. 8 as a change in lapse rate near $\sim 950 \, \mathrm{hPa}$. The middle troposphere is characterized by strong convection, which acts to restore the radiative cooling.

The more extreme cases with $F_t = 500 \, \mathrm{W\,m^{-2}}$ also continue this trend, with radiative cooling of $\sim 1.2 \, \mathrm{K\,day^{-1}}$ indicating a transfer of energy to the vertical diffusion of the atmosphere's boundary layer—which thereby leads to surface warming. Note also that case 12 shows reduced upper-atmosphere absorption due to the large geothermal flux of $F_g = 100 \, \mathrm{W\,m^{-2}}$ along with stronger cooling (and thus convection) in the middle troposphere. The gray-gas radiative absorber in this GCM assumes a specified vertical profile as a function of optical depth, which represents a greenhouse effect in the troposphere and stratosphere. The upper atmosphere absorption of planetary illumination on an actual exomoon would depend upon the atmospheric composition and pressure, among other factors, which could be explored with other GCMs that use band-dependent radiative transfer. Comparison of several different GCMs in a similar exomoon configuration would provide more robust constraints on the expected heating profile from planetary illumination.

## 4 DISCUSSION

In general, these simulations illustrate that the potential habitability of an exomoon depends upon the thermal energy emitted by its host planet. We find stable climate states for both slow and rapid rotators with $F_g \leq 10 \, \mathrm{W\,m^{-2}}$ and $F_t \leq 100 \, \mathrm{W\,m^{-2}}$, which indicates that both geothermal heating and planetary illumination could provide an additional source of warming for an exomoon. However, strong thermal illumination by the host planet ($F_t = 500 \, \mathrm{W\,m^{-2}}$ in our experiments) would likely lead to an onset of a runaway greenhouse and the loss of all standing water.

On the one hand, the habitability of some exomoon systems may therefore be precluded based upon the presence of a luminous host planet, although planetary illumination itself does not necessarily limit an exomoon's habitability (Heller 2016). On the other hand, we expect polar amplification of warming in all cases, which may suggest that exomoons in orbit around a luminous host planet may be less likely to develop polar ice caps. This tendency for an exomoon to have warmer poles due to planetary illumination suggests that such bodies may have a greater fractional habitable area than Earth today (Spiegel et al. 2008), potentially improving the prospects of an origin of life (Heller & Armstrong 2014). This prediction of polar warming on exomoons could eventually translate into observables from the circumstellar phase curves of an exomoon, if the planet's contribution to the combined planet-moon phase curve can be filtered out (Cowan et al. 2012; Forgan 2017).

To a lesser extent, the rotational period of the exomoon also contributes to differences in surface habitability. The short-period cases tend to show a greater amount of warming along the equator, near the subplanetary point (Figure 3, right panel), which could serve as an additional source of en-ergy to maintain regional habitable conditions. Dynamical changes in atmospheric jet structure that result from differences in rotation rate also contribute to changes in surface wind patterns, as well as the general circulation, which will likely correspond to significant contrasts in resulting cloud patterns. The large-scale circulation also shows an opposite directional sense in the hemispheres east and west of the subplanetary illumination point, which could also impact the probable location of clouds.

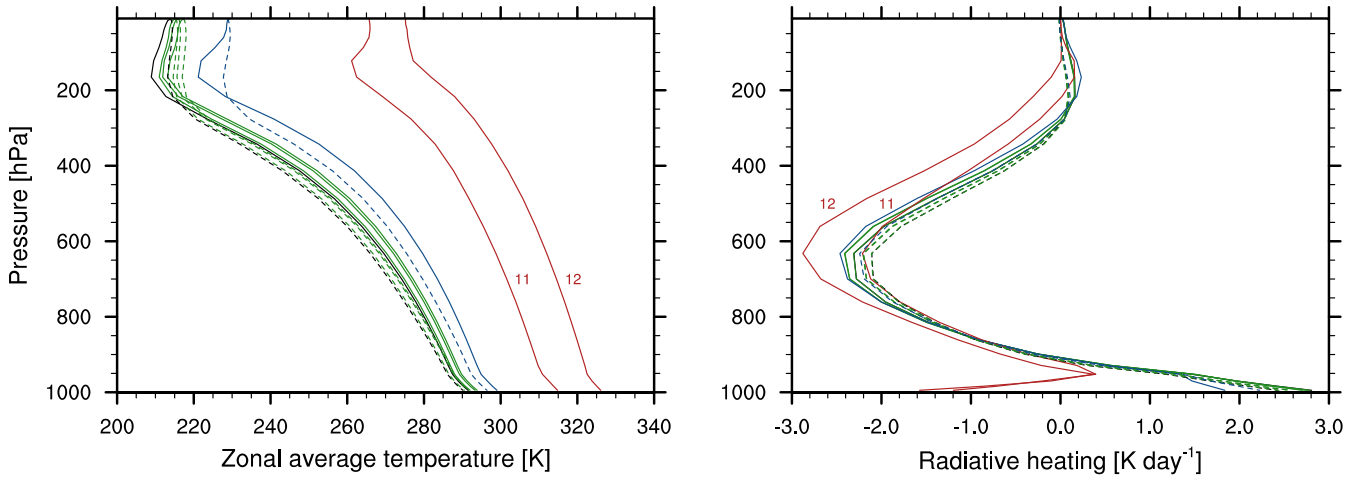

**Figure 8.** Vertical profiles of the zonal mean temperature (left) and radiative heating (right) at the equator for all simulations listed in Table 1. Solid curves indicate short-period cases (1.175 d) and dashed curves indicate long-period cases (3.324 d). Black curves show control experiments, while green indicates all simulations with $F_t \leq 10\,\mathrm{W\,m^{-2}}$. Blue curves show simulations with $F_g = 10\,\mathrm{W\,m^{-2}}$ and $F_t = 100\,\mathrm{W\,m^{-2}}$, while red curves show cases with $F_t = 500\,\mathrm{W\,m^{-2}}$. Case numbers as per Table 1 are given for some of the lines.

The hemispheric differences in the Hadley circulation (Figs. 4 and 5) show similarities to simulations of terrestrial planets in synchronous rotation around low mass stars, where the host star is fixed upon a substellar point on the planet. Haqq-Misra et al. (2011) used the idealized FMS GCM to demonstrate that the Hadley circulation shows direction in the opposite direction when comparing the hemispheres east and west of the substellar point for planets with 1 d and 230 d rotation periods. Haqq-Misra et al. (2018) also find the same hemispheric circulation patterns in an analysis the Community Earth System Model (CESM), which includes band-dependent radiation, cloud processes, and other physical processes. Synchronously rotating planets drive such a circulation when their stellar energy source is fixed to a single location; however, our exomoon calculations demonstrate that such a circulation can also be obtained when planetary illumination is fixed, even if the moon otherwise experiences variations in incoming starlight.

These idealized calculations provide a qualitative description of surface temperature and winds on an exomoon, but the application of more sophisticated GCMs will help to identify particular threshold where a runaway greenhouse and other water loss processes occur. Non-gray radiative transfer will allow for particular atmospheric compositions to be examined, such as a mixture of nitrogen, carbon dioxide, and water vapor that is characteristic of Earth-like atmospheres. Implementation of a cloud scheme into the GCM will also provide important insights into habitabil-ity, as clouds could help to delay the onset of a runaway greenhouse state (Yang et al. 2014).

In terms of the odds of an actual detection of the climatic effects described in this paper, this could in principle be possible if the moon's electromagnetic spectrum (either reflection or emission) could be separated from that of the planet. This might be possible in very fortunate cases where a large moon is transiting its luminous giant planet (Heller & Albrecht 2014; Heller 2016) or where both the planet and its moon transit their common low-mass host star (Kaltenegger 2010). Alternatively, if the moon is subject to extreme tidal heating, it could even outshine its host planet in the infrared and therefore directly present its emission spectrum (Peters & Turner 2013) while the planet would still dominate the visible part of the spectrum, where it reflects much more light than the moon. The technological requirements, however, will go beyond the ones offered by the James Webb Space Telescope (Kaltenegger 2010) and might not be accessible within the next decade.

## 5  CONCLUSIONS

We present the first GCM simulations of the atmospheres of moons with potentially Earth-like surface conditions. Our simulations illustrate the effects of tidal heating and of planetary illumination on the atmospheres of large exomoons that could be abundant around super-Jovian planets in the stellar habitable zones.

Most of the energy from planetary thermal heating and geothermal heating is transported toward the poles as a result of enhanced meridional transport of moisture and energy. This suggests that polar ice melt may be less prevalent on exomoons that are in synchronous rotation with their host planet. In general, these calculations further illustrate that the poleward expansion of the Hadley circulation enhances meridional energy transport and can lead to polar amplification of warming, even in the absence of ice albedo feedback.

This polar heat transport could increase the fraction of the surface that allows the presence of liquid surface water by compensating for the lower stellar flux per area received at the poles of a moon. In other words, illumination from the planet might be beneficial for the development of life on exomoons. Future observations that are able to distinguish exomoons from their host planet may be able to detect the absence of polar ice caps due to polar amplification of planetary illumination, such as analysis of photometric phase-curves.

## ACKNOWLEDGMENTS

The authors thank Ravi Kopparapu for helpful feedback on a previous version of the manuscript. J.H. acknowledges funding from the NASA Astrobiology Institute's Virtual Planetary Laboratory under awards NNX11AC95G and NNA13AA93A, as well as the NASA Habitable Worlds program under award NNX16AB61G. R.H. has been supported by the German space agency (Deutsches Zentrum für Luft- und Raumfahrt) under PLATO Data Center grant 50OO1501, by the Origins Institute at McMaster University, and by the Canadian Astrobiology Program, a Collaborative Research and Training Experience Program funded by the Natural Sciences and Engineering Research Council of Canada (NSERC). Any opinions, findings, and conclusions or recommendations expressed in this material are those of the authors and do not necessarily reflect the views of NASA or NSERC.

# Part III

# Outlook and Appendix

# Chapter 8

# Towards an Exomoon Detection

## 8.1 New Constraints on the Exomoon Candidate Host Planet Kepler-1625 b

Of all the exomoon cases put forward in the literature so far, the proposed Neptune-sized exomoon around the Jupiter-sized transiting planet candidate Kepler-1625 b is currently the most actively debated object (see Sect. 1.2). And it will most likely remain a source of debate for some time to come due to the star's faintness, which makes conclusive photometric follow-up observations extremely challenging.[1] So far, the scientific debate has focussed on the three transits observed in the four years of data from Kepler and one additional transit observed with *Hubble*. On 25 October 2017, we have started looking at Kepler-1625 and its proposed planet-moon system from a different perspective and this survey is still ongoing. Our preliminary results are the subject of this section.

### 8.1.1 Motivation for Radial Velocity Measurements of Kepler-1625

To give a little bit of a historical context, the discovery of an exomoon candidate around Kepler-1625 b was announced on 26 July 2017 by Teachey et al.[2] Soon thereafter, we (Principal Investigator: R. Heller) requested Director's Discretionary Time (DDT) at the 3.5 m telescope at the Calar Alto Observatory to take spectra of the star just before and just after the transit on 29 October 2017. At that time, Kepler-1625 b was known only as a transit planet candidate, that is to say, it was estimated to have a very low false positive probability (FPP) but it had not been confirmed as a planet by an independent method. The FPP of giant planets in wide orbits has been shown to be particularly high (Morton et al. 2016) and Kepler-1625 b is a giant planet on a wide orbit, with an orbital period of about 287 d. Hence, there was a non-negligible probability of Kepler-1625 b being a false positive caused by, e.g., an unrelated eclipsing stellar binary in the background. Alternatively, the object might in fact be transiting in front of Kepler-1625 but the relatively large error bars on the physical radius of the star propagate to the radius of the transiting object and, in this case, permit it to be as large as a very-low-mass star or a brown dwarf (Heller 2018c). Moreover, only a few transiting planets with similar orbital periods had been confirmed by stellar radial velocity (RV) measurements at the time. As a consequence, the announcement by Teachey et al. made it a highly conclusive case to determine the mass of Kepler-1625 b and its moon using RVs and to confirm the planetary nature of the transiting object irrespective of the moon candidate.

The purpose of our initial observations in late 2017 was to determine the star's proper radial velocity component with respect to the solar system since its RV variation due to the planet (and potentially its moon) is very close to zero at the time of the transit. Our first service mode observations were successfully taken on 25 and 31 October 2017. Further observations were proposed and granted through

---

[1] The characterization of the star by the Gaia Data Release 2 (Gaia Collaboration et al. 2016, 2018) derives a parallax of 0.40645 (±0.03580) mas, which implies a distance of $d = 2460^{+238}_{-199}$ pc, and the photometric $g$ mean magnitude is 15.76.

[2] The original pre-print is freely available at https://arxiv.org/abs/1707.08563v1.

Table 8.1: CARMENES observations of Kepler-1625.

| observation date | observation time | BJD | S/N |
|---|---|---|---|
| 25.10.17 | 18h 14min | 2458052.26775 | 5.3721 |
|  | 18h 37min | 2458052.28352 | 4.266 |
|  | 18h 59min | 2458052.29859 | 5.1383 |
| 31.10.17 | 18h 31min | 2458058.28172 | 6.9242 |
|  | 19h 04min | 2458058.30206 | 4.8779 |
|  | 19h 27min | 2458058.31763 | 5.3678 |
| 28.04.18 | 03h 41min | 2458236.66051 | 3.5338 |
|  | 04h 03min | 2458236.67569 | 3.4044 |
|  | 04h 24min | 2458236.69211 | 7.0565 |
| 04.06.18 | 02h 14min | 2458273.60239 | 3.4891 |
|  | 02h 36min | 2458273.61722 | 3.5922 |
|  | 02h 58min | 2458273.63306 | 3.5064 |
| 30.06.18 | 01h 10min | 2458299.54861 | 0.8296 |
|  | 01h 39min | 2458299.57736 | 3.8821 |
|  | 02h 01min | 2458299.59385 | 4.3231 |
|  | 02h 24min | 2458299.60959 | 3.9163 |
| 09.08.18 | 21h 56min | 2458340.42432 | 3.2388 |
|  | 22h 18min | 2458340.43902 | 4.2685 |
|  | 22h 38min | 2458340.45349 | 4.3744 |
| 23.10.18 | 18h 21min | 2458415.27194 | 2.9084 |
|  | 18h 49min | 2458415.29215 | 2.9188 |
|  | 19h 12min | 2458415.30818 | 2.8022 |

**Notes.**
The first spectrum on 30.06.18 was not analyzed due to poor data quality.
This table was prepared by courtesy of A. Timmermann.

an Open Time proposal (PI: R. Heller) and another DDT proposal (PI: R. Heller) and the data were acquired through one orbital cycle of Kepler-1625 b. We took three spectra with 20 min exposures per night, which we then combined into one spectrum per night that we used for our subsequent RV analysis. An overview of these observations is shown in Table 8.1 and their expected contribution to the RV curve over one cycle is illustrated in Fig. 8.1.

### 8.1.2  Data Analysis and System Characterization

The data reduction of the CARMENES spectra was performed using the SERVAL software (Zechmeister et al. 2018)[3] by courtesy of A. Timmermann and in cooperation with M. Zechmeister, the results of which will soon be submitted for peer review. Figure 8.2 shows our best fit of a one-planet Keplerian model with eccentricity ($e$), which takes into account the boundary condition of the known transit time. The orbital period and the stellar mass were fixed at 287.38 d and 1.1 solar masses, respectively. The resulting planetary mass is roughly 5 Jupiter masses and the orbital eccentricity is about 0.4. These values are preliminary and we are currently working to deduce robust error bars using Markov chain Monte Carlo simulations (Foreman-Mackey et al. 2013). That said, all things combined we are very certain that Kepler-1625 b is indeed a transiting planet, whether it has a moon or not. At the time of writing, our results suggest that among the more than 3100 confirmed transiting exoplanets, 2345 of which have been discovered with Kepler[4], there are only nine with orbital periods longer than that of Kepler-1625 b. And so while our RV measurements cannot be used to validate or reject the exomoon hypothesis directly, we find new constraints on the mass of Kepler-1625 b or the combined planet-moon mass, as the case may be. Any future characterization of this system that would take into account additional photometric or RV observations will need to be consistent with our new results.

---

[3] Available at https://github.com/mzechmeister/serval.
[4] NASA Exoplanet Archive (https://exoplanetarchive.ipac.caltech.edu/docs/counts_detail.html) as of 28 August 2019.

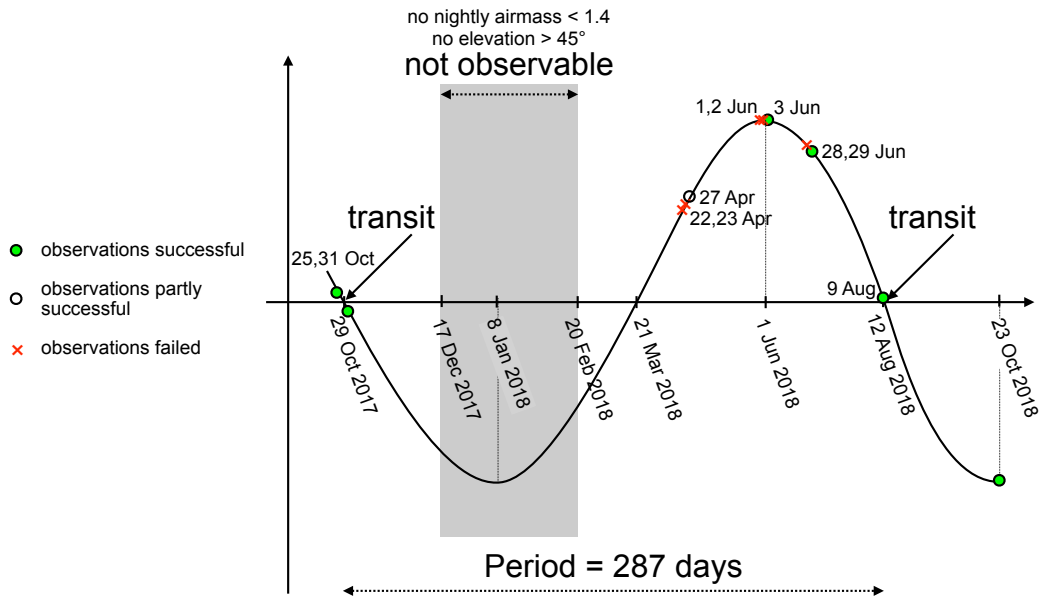

Figure 8.1: A sketch of the expected RV curve of Kepler-1625 due to the presence of its giant transiting planet (plus its hypothetical moon, as the case may be). Symbols on the curve indicate whether the proposed observations were successful or not (see legend at the left). The dates shown along the curve refer to the beginning of the night of the proposed observations.

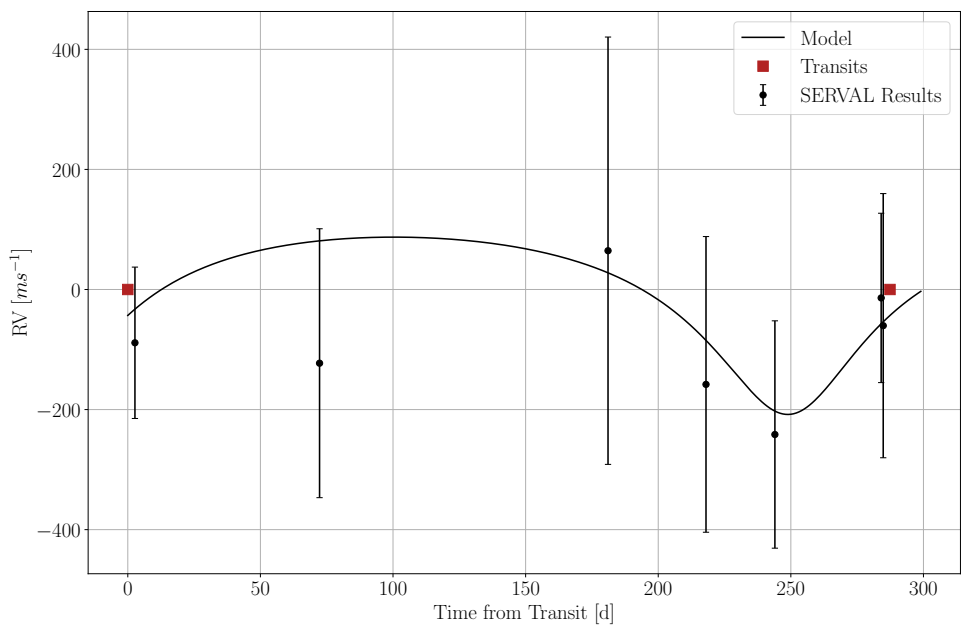

Figure 8.2: Best fit of a one-planet Keplerian orbit (solid line) to our RV measurements with CARMENES (points with error bars). This solution includes the known transit times (dark red squares) as boundary conditions and it suggest a giant planet with about 5 Jupiter masses on an eccentric orbit with $e \approx 0.4$. Image credit: A. Timmermann.

## 8.2 Exomoon Indicators in High-precision Transit Light Curves

### 8.2.1 Overview on Transit Timing, Transit Duration, and Planet Radius Variations

In this section we present our ongoing study of three indicators of an exomoon that emerge if well-established planet-only models are fitted to a light curve that contains a planet-moon system: transit timing variations (TTVs), transit duration variations (TDVs), and planetary radius variations (PRVs). We re-evaluate under realistic conditions the previously proposed exomoon signatures in the TTV and TDV series (Sartoretti & Schneider 1999; Kipping 2009a). We simulate light curves of a transiting exoplanet with a single moon, taking into account stellar limb darkening, orbital inclinations, planet-moon occultations, and noise from both stellar granulation and instrumental effects. These model light curves are then fitted with a planet-only transit model, pretending we wouldn't know that there is a moon, and we explore the resulting TTV, TDV, and PTV series for evidence of the moon. Our results are preliminary and will soon be submitted to a peer-reviewed journal. This work was done in collaboration with K. Rodenbeck in the context of this PhD thesis (Rodenbeck 2019).

As we have seen in Sect. 6, and in particular in Sects. 6.7 (Rodenbeck et al. 2018) and 6.8 (Heller et al. 2019), the dynamical modeling of planet-moon systems is computationally demanding. Considering the large computational load required for the statistical vetting of exomoon candidates in a star-planet-moon framework and taking into account the large numbers of transiting planets that need to be examined, tools for an uncomplicated identification of the most promising exomoon candidates could be beneficial to streamline more detailed follow-up studies. Many ways have been proposed to look for moons in transit light curves (for a review see Heller 2018a). Sartoretti & Schneider (1999) suggested that a massive moon can distract the transits of its host planet from strict periodicity and induce transit timing variations (TTVs). Kipping (2009a) identified an additional effect of the moon on the planet's sky-projected velocity component tangential to the line of sight that results in transit duration variations (TDVs). These effects relate to the planet's position with respect the planet-moon barycenter, which is modeled on a Keplerian orbit around the star, and they depend on the planet-satellite orbital semi-major axis ($a_{\rm ps}$) and on the satellite's mass ($M_{\rm s}$) but they do not depend on the satellite's radius ($R_{\rm s}$). Heller et al. (2016b) proposed that the combined planetary TTV and TDV effects can produce distinct ellipsoidal patterns in the TTV-TDV diagram if the moon is sufficiently small to avoid any effects on the photometric transit center.

Szabó et al. (2006), however, noted that the stellar dimming caused by a moon with a sufficiently large radius can affect the photometric center of the transit light curve, defined as $\tau = \sum_i t_i \Delta m_i / \sum_i \Delta m_i$ with $t_i$ as the times when the stellar differential magnitudes ($\Delta m_i$) are measured. For sufficiently small (but arbitrarily massive) moons, $\tau$ would coincide with the transit midpoint of the planet but for large moons $\tau$ can be substantially offset from the position of the midpoint of the planetary transit. Simon et al. (2007) derived an analytical estimate of the photometric TTV effect (dubbed "TTV$_{\rm p}$") and compared it to the magnitude of the barycentric TTV effect (dubbed "TTV$_{\rm b}$"). They define the TTV$_{\rm p}$ as the difference between TTV$_{\rm b}$ and $\tau$.

Here we simulate many different transit light curves of a planet with a moon of non-negligible radius and then fit the resulting data with a planet-only transit model to obtain the TTVs and TDVs of a hypothetical sequence of transits. Our aim is to provide the exoplanet community with a tool to inspect their TTV-TDV distributions for possible exomoon candidates without the need of developing a fully consistent, photodynamical transit modelling of a planet with a moon (Kipping 2011; Rodenbeck et al. 2018). Our most recent investigations, which we present below, predict that the previously described ellipse in the TTV-TDV diagram of an exoplanet with a moon (Heller et al. 2016b; see Sect. 6.4) emerges only for high-density moons. Low-density moons, however, distort the sinusoidal shapes of the TTV and the TDV series due to their photometric contribution to the combined planet-moon transit. Sufficiently large moons can distort the previously proposed ellipse in the TTV-TDV into very complicated patterns, which are much harder to discriminate from the TTV-TDV distribution of a planet without a moon. After all, we identify a new effect that appears in the sequence of planetary radius ($R_{\rm p}$) measurements, which could be used to identify exomoon candidates in large exoplanet surveys. Sufficiently large moons can nevertheless produce periodic apparent PRVs of their host planets that could be observable in the archival Kepler data or with the PLATO mission. We demonstrate how the periodogram of the sequence of planetary radius measurements can indicate the presence of a moon and propose that PRVs could be a more promising means to identify exomoons in large exoplanet surveys. An inspection of a limited amount of exoplanets from Kepler reveals substantial PRVs of the Saturn-sized planet Kepler-856 b although an exomoon could only ensure Hill stability in a very narrow orbital range.

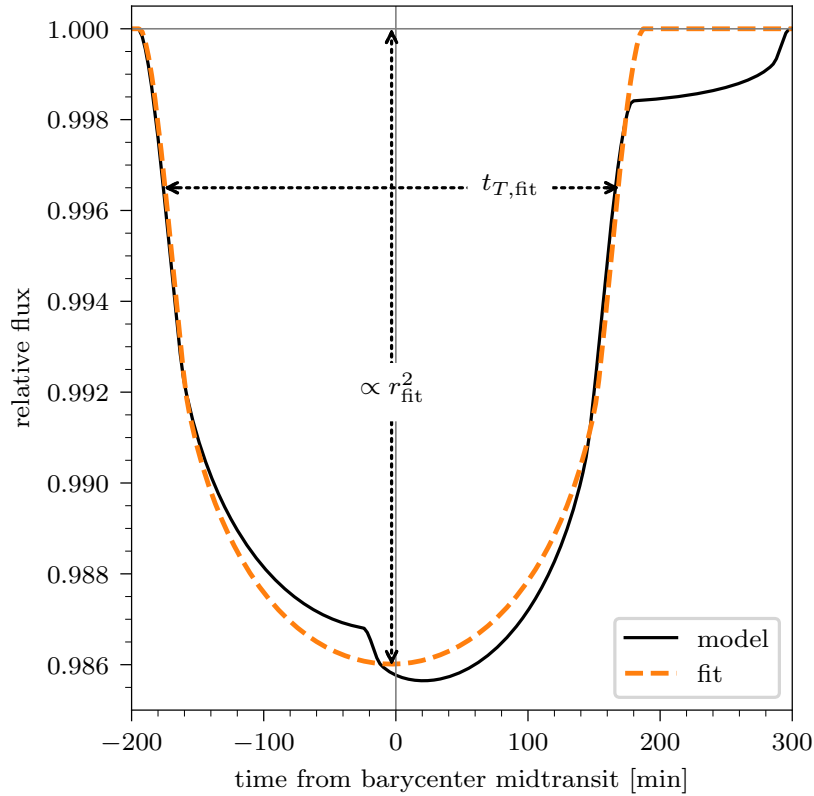

Figure 8.3: Example of a planet-only fit (dashed orange line) to the planet-moon model (solid black line). The star is sun-like, the planet is Jupiter-sized and in a 30 d orbit around the star, and the moon is Neptune-sized and in a 3.55 d orbit around the planet. In this example, the fitted planet-to-star radius ratio ($r_{\rm fit}$), which goes into the calculation of the transit depth with a power of 2, is slightly overestimated due to the moon's photometric signature.

### 8.2.2 Transit Model and Noise Properties of Simulated Light Curves

Figure 8.3 demonstrates how the transit light curve of an exoplanet with a moon (solid black line) differs ever so slightly from the light curve expected for a single planet (orange dashed line). For one thing, the midpoint of the planetary transit changes between transits with respect to the planet-moon barycenter, thereby causing a TTV$_{\rm b}$ effect (Sartoretti & Schneider 1999). Moreover, the additional dimming of the star by a sufficiently large moon could act to shift the photometric center of the combined transit, an effect referred to as TTV$_{\rm p}$ (Szabó et al. 2006; Simon et al. 2007). The same principle of the moon's photometric (rather than the barycentric) effect should be applicable to the TDV, though it has not been explored in the literature so far. And finally, Fig. 8.3 illustrates how the planet-only fit to the planet-moon model overestimates the radius of the planet. For single planets, the transit depth is roughly proportional to $R_{\rm p}^2$, although details depend on the stellar limb darkening (Heller 2019). But a moon can affect the radius estimate for the planet and therefore cause planetary

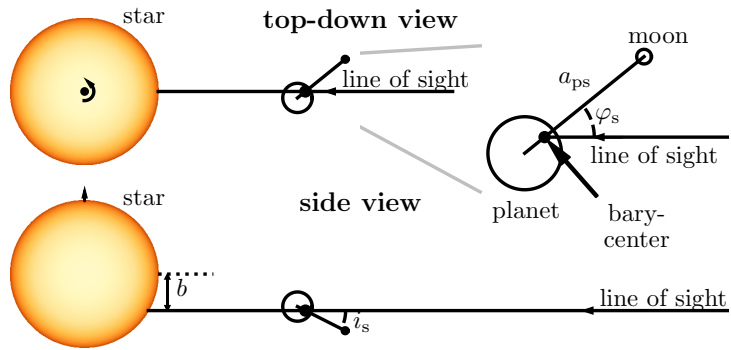

Figure 8.4: Some orbital parameters for the planet-moon system. $\varphi_s$ is the orbital phase relative to the line of sight at the time of mid-transit.

radius variations (PRVs) in a sequence of transit measurements.

We investigate both the barycentric and the photometric contributions to the measured TTVs and TDVs of a planet-with-moon transit light curve that is fitted with a planet-only model. To separate the $TTV_b$ and $TTV_p$ effects, we simulate transits of a hypothetical planet-moon system with a zero-radius or zero-mass moon, respectively, and then fit the simulated light curves with a planet-only transit model. Finally, we assume a realistic moon with both mass and radius and redo our fit to the resulting light curve.

Synthetic exoplanet-exomoon transits are generated using our implementation of the Mandel & Agol (2002) analytic transit model in a fashion similar to the one presented in Rodenbeck et al. (2018). As an improvement to Rodenbeck et al. (2018), however, our model now includes planet-moon occultations and the possibility of inclining the planet-moon orbit with respect to the line-of-sight. The planet-only transit model is fitted to the simulated light curve using the planet-to-star radius ratio ($r_p$), transit duration ($t_T$), and the orbital phase ($\varphi_b$), while the orbital period ($P_b$) and the limb darkening coefficients ($q_1$ and $q_2$ for the quadratic limb darkening model as per Kipping 2013) are kept constant at the respective values (for details see Table A.1 in Rodenbeck 2019). Figure 8.3 shows an example of a fitted planet-only light curve (orange dashed line) to the simulated noiseless transit of a planet-moon system. We also investigate the possibility of detecting PRVs. Variations of the planetary transit depth have been observed before (Holczer et al. 2016) but to our knowledge they have not been treated in the context of exomoons so far.

We compare the effect of photon noise between two space missions: Kepler (Borucki et al. 2010) and PLATO (Rauer et al. 2014), the latter of which is currently scheduled for launch in 2026. For both missions we test stars of two different apparent magnitudes, namely 8 and 11. The two stellar magnitudes provide us with two possible amplitudes of the white noise level per data point, which is composed of photon noise and telescope noise. For Kepler long cadence (29.4 min) observations, we obtain a white noise level of 9 parts per million (ppm) at magnitude 8 and 36 ppm at magnitude 11 (Koch et al. 2010). For PLATO we estimate 113 ppm at magnitude 8 and 448 ppm at magnitude 11, both at a cadence of 25 s. These estimates for PLATO assume an optimal target coverage by all 24 normal cameras.

We also simulate a time-correlated granulation noise component according to Gilliland et al. (2011), where the granulation power spectrum is modeled as a Lorentzian distribution in the frequency domain and then transformed into the time domain. The granulation amplitude and time scale depend on the star's surface gravity and temperature. We use two stellar types, a sun-like star and a red dwarf with similar properties as $\epsilon$ Eridani ($\epsilon$ Eri).

We investigate several hypothetical star-planet-moon systems to explore the observability of the resulting TTV, TDV, and PRV effects. We ensure orbital stability of the moons by arranging them sufficiently deep in the gravitational potential of their host planet, that is, at $< 0.5$ times the planetary

Hill radius. This distance has been shown to guarantee orbital stability of both pro- and retrograde moons based on numerical $N$-body simulations (Domingos et al. 2006). We choose stellar limb darkening coefficients ($q_1$, $q_2$) for both the sun and a K2V star akin to $\epsilon$ Eri from Claret & Bloemen (2011).

For case 1, and for all of its sub cases 1a - 1p, we choose the planet to be a Jupiter-sized and the moon to be Earth-sized. The orbit period of the moon can take two values, either the orbital period of Europa (3.55 d) or that of Io (1.77 d). The barycenter of the planet-moon system is in a $P_b = 30$ d orbit around its Sun-like star. In test cases 2a - 2d, we simulate Neptune-sized moons and vary both orbital periods involved to test the effect of the number and duration of the transits. We set $P_b$ to be either 30 d orbit or close to twice that value. In fact, however, we shorten the 60 d orbit by 1.69 d to 58.31 d for the Io-like orbit and by 1.59 d to 58.41 d for the Europa-wide orbit in order for the moon to show the same orbital advancement between transits, that is, for the remainder of $P_b/P_s$ to be the same (Heller et al. 2016b). Cases 3a - 3d refer to a Saturn-sized planet and a super-Earth moon, cases 4a - 4d to a Neptune-sized planet and an Earth-sized moon, cases 5a - 5d to an Earth-Moon analog (though at either a 30 d or a 58.31 d orbital period around their star). In cases 6 - 9 we essentially redo all these cases except for the star to resemble $\epsilon$ Eri-like.

In cases 1a - 1p, we study the effect of the moon's orbital phase on the resulting transit shape and generate transits of planet-moon systems for orbital moon phases ranging from 0 to 1. We allow for different orbital inclinations of the moon orbit ($i_s$) in these test cases (see Fig. 8.4), which may prevent occultations if the planet-moon orbit is sufficiently wide for a given inclination. As we show below, the inclination has an effect on the measured transit parameters if the line connecting the planet and the moon is parallel to the line of sight, that is, if the planet and the moon have a conjunction during the transit. For test cases 2a - 9d we choose the lowest inclination possible without causing a planet-moon occultation.

As a first application of our search for exomoon indicators, we explore the time series of the fitted transit depth from Holczer et al. (2016). We then do a by-eye vetting of the transit depth time series, of the autocorrelation functions (ACF), and of the periodograms for each of the 2598 Kepler Objects of Interest (KOIs) listed in Holczer et al. (2016) and identify KOI-1457.01 (Kepler-856 b) as an interesting candidate. In brief, this is a roughly Saturn-sized validated planet with a false positive probability of $2.9 \times 10^{-5}$ (Morton et al. 2016) and an orbital period of about 8 d around an $\epsilon$ Eri-like host star.

## 8.2.3  TTVs, TDVs, and PRVs from Planet-Moon Light Curves without Noise

Figure 8.5 shows the TDV (panel a), TTV (panel b), PRV (panel c), and combined TTV-TDV (panel d) effects. Each of these four panels is divided into three subplots showing the contributions of the photometric distortion of the light curve due to the moon (top subplots, moon mass set to zero), of the barycentric motion of the planet due to its moon (center subplots, moon radius set to zero), and the combined effect as derived from the planet-only fit to the simulated planet-moon transit light curve (bottom subplots). All panels assume a Jupiter-sized planet at an orbital period of $P_b = 30$ d, an Earth-like moon with an orbital period of $P_s = 3.55$ d ($= P_{Eu}$), and a sun-like host star. Blue lines refer to an orbital inclination of $i_s = 0$, that is to say, to co-planar orbital configurations. This results in occultations near moon phases of 0, when the moon is in front of the planet as seen from Earth (see Fig. 8.4), and 0.5, when the planet is behind the moon. Orange lines refer to $i_s = 0.2$ rad and do not produce planet-moon occultations.

The central panels of Fig. 8.5, which illustrate the barycentric effects, correspond to the frameworks of Sartoretti & Schneider (1999) (for the TTVs), Kipping (2009a) (for the TDVs), and Heller et al. (2016b) (for the TTV-TDV figure). The contribution of the photometric distortion of the light curves by the moon (upper subplots), however, results in a very different fit of the TDV, TTV, PRV, and TTV-TDV figures (bottom subplots) than proposed in these papers. In particular, the fitted TTV-TDV figure does not resemble at all the ellipse that one would expect based on the barycentric contribution only. As a consequence, the TTV-TDV figure might not be as straightforward an exomoon indicator as proposed by Heller et al. (2016b).

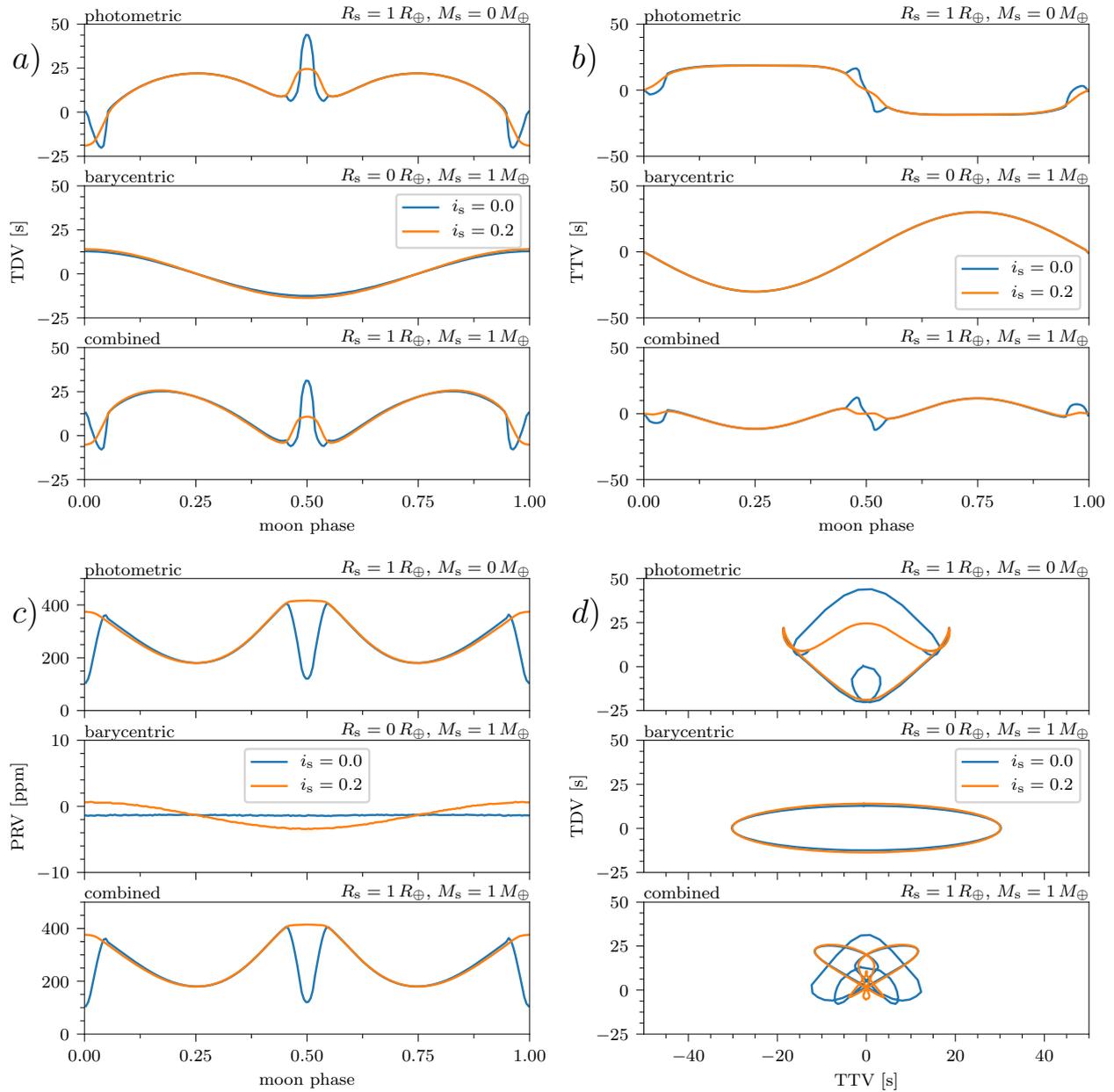

Figure 8.5: Exomoon indicators for a hypothetical Jupiter-Earth planet-moon system in a 30 d orbit around a sun-like star as measured from noiseless simulated light curves by fitting a planet-only model. **(a)** Transit duration variation (TDV). **(b)** Transit timing variation (TTV). **(c)** Planetary radius variation (PRV). **(d)** TTV vs. TDV diagram. Blue lines refer to a coplanar planet-moon system with $i_s = 0$ and orange lines depict the effects for an inclined planet-moon system with $i_s = 0.2\,\mathrm{rad}$. In each panel, the top subplot shows the photometric contribution, the center subplot the barycentric contribution, and the bottom subplot the combined and measured effect.

The variation of the photometric TDV signal as a function of the moon's orbital phase ($0 \leq \varphi_s \leq 1$) shows a relatively complicated behavior with minima when the planet and the moon align along the line of sight (moon phases of 0 or 1) and maxima at moon phases of 0.25, 0.5, and 0.75. For comparison, the variation of the barycentric TDV signal with changing moon phase is strongest when the planet and moon align at either a moon phase of 0 or 0.5. Both contributions combined, this leads to a complicated pattern for the TDVs as derived from the light curve fits.

The pattern of the photometric contribution to the measured TTV (Fig. 8.5b) is shift by half an orbital moon phase (or by $\varphi_s = \pi$) relative to the barycentric effect (panel a), though with roughly the same amplitude. The resulting TTV signal in the fitted model light curves then turns out to be extremely small (panel c).

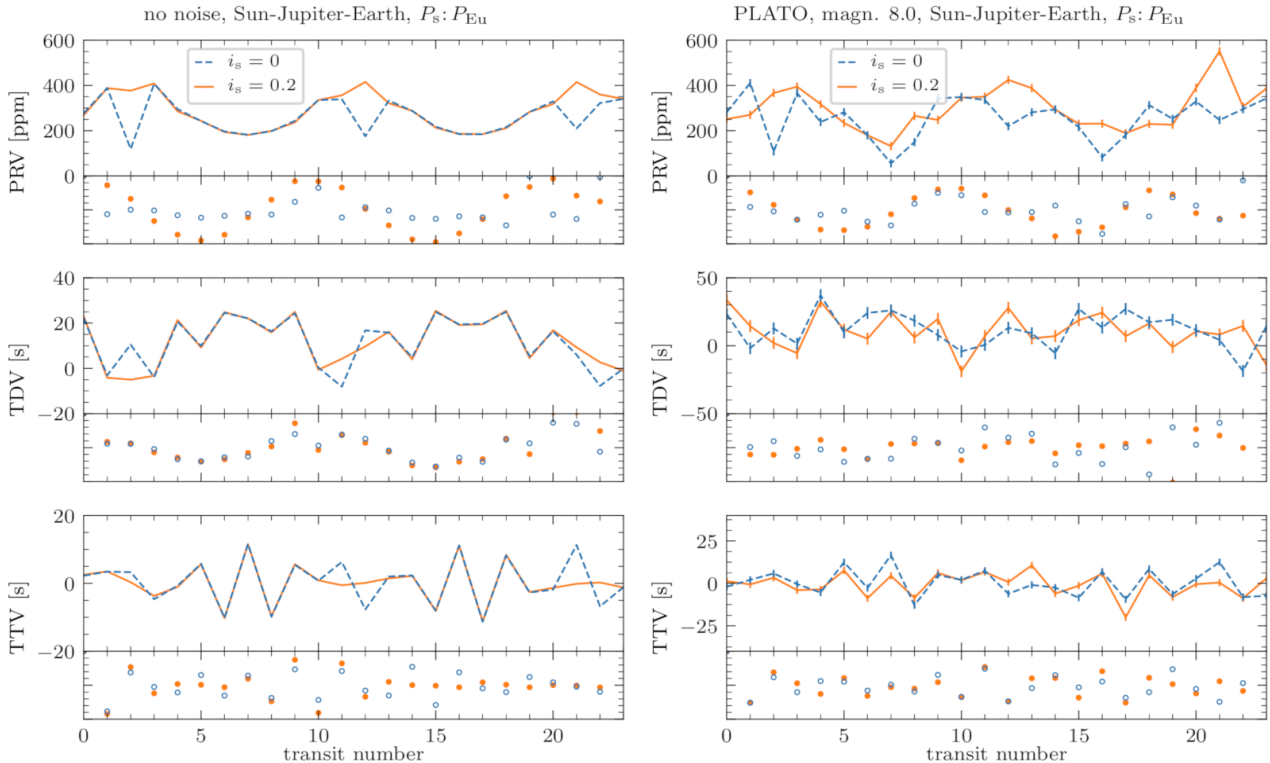

Figure 8.6: The PRV, TDV and TTV effects fitted for the transit sequence of an Earth-sized moon in a Europa-like orbit (3.55 d) around a Jupiter-sized planet. The planet-moon barycenter has a 30 d orbit around a sun-like star and the transits (not shown) were sampled with a PLATO-like cadence of 25 s. *Left:* Without noise. *Right:* With noise contributions from stellar granulation and PLATO-like instrumental noise for an $m_V = 8$ star. The subpanels show the autocorrelation. In all panels, blue symbols refer to a coplanar planet-moon system with $i_s = 0$ and orange symbols depict the effects for an inclined planet-moon system with $i_s = 0.2$ rad.

The photometric PRV depends strongly on the projected separation of planet and moon, with $\varphi_s = 0$ corresponding to the moon being precisely in front of the planet as seen from Earth, $\varphi_s = \pi/2 \equiv 90° \equiv$ "moon phase of 0.25" corresponding to the moon being separated as much as possible from the planet and as "late" as possible for the transit etc. (see Fig. 8.4) The larger the sky-projected apparent separation between the planet and the moon, that is, the better the moon's photometric dip in the light curve is separated from the planet's photometric signature, the smaller the PRV. On the other hand, if the two bodies are sufficiently close to cause a planet-moon occultation, then the PRV drops significantly (see the dip at moon phase 0.5 in Fig. 8.5c).

Interestingly, there is a difference in amplitude depending on whether the moon passes behind or in front the planet. This is due to the slightly different transit shapes of the combined planet-moon system. If the moon's effective (tangential) speed across the stellar disk is lower than that of the planet, the overall transit shape resembles that of a single planet except for an additional signature in the wings of the transit light curve. Note that this asymmetry of the PRV effect cannot help to discriminate between prograde and retrograde moon orbits, which is both a theoretical and an observational challenge for exomoon characterization (Lewis & Fujii 2014; Heller & Albrecht 2014). Moreover, in practice this effect will hardly be observable when noise is present. Note that the barycentric PRV shows no variation as a function of the moon phase in the case of $i_s = 0$ but some variation in the case of $i_s = 0.2$ rad. This can be explained by the planet crossing the star at a higher or lower impact parameter due to its movement around the planet-moon barycenter. At different transit impact parameters then, the resulting transit depth will be different due to the stellar limb darkening effect (Heller 2019).

### 8.2.4 TTVs, TDVs, and PRVs from Planet-Moon Light Curves with PLATO-like Noise

The panels at the left of Fig. 8.6 show the PRV (top), TDV (center), and TTV (bottom) effects, respectively, as measured from a sequence of simulated light curves without noise. Just like in Fig. 8.3, we modeled the planet moon transits as such but fitted a planet-only model. The number of the transit in this sequence is shown along the abscissa. Also shown are the respective autocorrelations below each of these three time series. The model systems used for these sequences are our test cases 1b ($i_s = 0$) and 1j ($i_s = 0.2$ rad), that is a Jupiter-sized, Jupiter-mass planet with an Earth-sized moon in a 3.55 d (Europa-like) orbit. The center of mass of this planet-moon binary is in a 30 d orbit around a sun-like star.

The combination of these particular orbital periods around the star ($P_b$) and around the planet-moon barycenter ($P_s$) happen to make the moon jump by a phase of 0.448 circumplanetary (or circumbarycentric) orbits between successive transits. This well-known undersampling results in an observed TTV signal period of under 2 $P_b$ (Kipping 2009a). The TDV series is also subject to this aliasing effect, but the autocorrelation suggests some periodicity. Finally, the PRV signal, whose period is just a half of $P_s$, is not affected by the undersampling. Hence, the periodicity can be observed even by eye in the top panel and, of course, also in the corresponding autocorrelation.

Moving on to the addition of noise in our simulated light curves, the panels at the right of Fig. 8.6 illustrate the same PRV (top), TDV (center), and TTV (bottom) series as in the panels at the left, but now as if the host star were an $m_V = 8$ solar type star observed with PLATO. In both the TDV and the TTV panels we see that the magnitude of the error caused by the noise in the light curves is comparable to the TTV and TDV amplitudes. As a consequence, both the two time series and their autocorrelations show only weak hints of periodic variations. For comparison, the PRV signal of the slightly inclined system ($i = 0.2$ rad, orange dashed line) is still clearly visible, whereas the signal for the $i = 0$ system is somewhat weaker.

To get a more quantitative handle on the possible periodicities in the PRV, TTV, and TDV sequences, we investigate the periodograms of the data. In Fig. 8.7 we present the periodograms of the time series and of the autocorrelations of the PRV (top), TTV (center), and TDV (bottom) sequences. Each panel is based on 20 randomized noise realizations of 24 transits of the same system as in Fig. 8.6. The PRVs of each realization show a clear peak at around 9.5 times the orbital period of the barycenter around the star. This peak occurs in the periodograms of both the time series and of the autocorrelation of the time series. The periodograms of the TDV series and TDV autocorrelation, however, do not show any significant signals. The TTVs show a peak in the periodogram close to the period corresponding to the Nyquist frequency (for details see Rodenbeck 2019).

### 8.2.5 The PRV Effect of Different Star-Planet-Moon Systems

The results presented in Figs. 8.6 and 8.7 refer to only two specific test cases, that is, 1b and 1f, respectively. Next, we extend our investigations to all other test cases, while Fig. 8.8 shows the results for a subsection of these cases as a cut through the parameter space.[5] All results shown in Fig. 8.8 assume synchronous observations of an $m_V = 8$ star with all 24 normal cameras of PLATO. We focus on the PRV because our investigations to this point have indicated that the PRV effect is more pronounced than the TTV and TDV effects.

The four systems in the upper row in Fig. 8.8 refer to cases 2b (solid orange), 2d (dashed orange), 6b (blue solid), and 6d (blue dashed). The second line of panels shows cases 3b (solid orange), 3d (dashed orange), 7b (solid blue) and 7d (dashed blue). The third line of panels shows cases 4b (solid orange), 4d (dashed orange), 8b (blue solid), and 8d (blue dashed). The bottom row of panels refers to cases 5b (orange solid), 5d (orange dashed), 9b (blue solid), and 9d (blue dashed).

---

[5]A complete description of all test cases is given in Rodenbeck (2019).

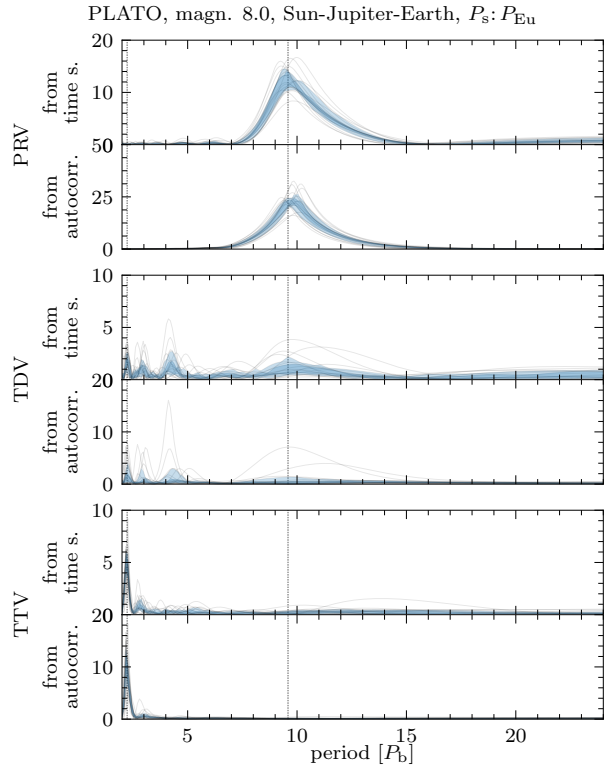

Figure 8.7: Periodograms of the PRV (top), TDV (center), and TTV (bottom) effects and their autocorrelations measured from a sequence of 20 simulated light curves for a Jupiter-Earth planet moon system similar to the one used in Fig. 8.6. The 20 randomized noise realizations contain both white noise and granulation equivalent to an $m_V = 8$ star observed with PLATO at a 25 s cadence.

In cases 2, 3, 6, and 7 involving the large Neptune and super-Earth-sized moons, the PRV signal is always visible in the time series (left column). The autocorrelation function of the PRV time series (center column) for systems with $P_b = 60$ d, however, is truncated at 6 times the orbital period due to the low number of transits available at that period. This truncation is caused by our our restriction of the autocorrelation function to a lag time of at least half the observed number of transits. For comparison, in those cases that assume $P_b = 30$ d and, thus, twice the number of transits available for the analysis, the signal in the corresponding periodogram of the autocorrelation is very clear.

For cases 4 and 8 involving an Earth-sized moon, the detectability of the PRV signal depends on the moon's orbital period and the size of its star. For the case of a Europa-like orbital period of the moon and the $\epsilon$ Eridani-like star, a marginally significant peak in the periodogram is present, but no significant peak can be seen if the star is sun-like or the moon's orbital period is that of Io. In none of the cases with a Moon-sized moon (cases 5 and 9) we were able to produce a significant PRV signal.

In the right column of panels in Fig. 8.8, we observe that our test moons with a Europa-like orbital period all cause a peak in the periodograms at around 9.5 $P$b. For all the other cases involving a moon with an Io-like period (all subcases a and c), the position of the peak in the periodgrams of the autocorrelation is near 12 $P_b$, which is at the edge of the number of orbital periods we deem reasonable for analysis.

The positions of the peak in the periodogram of the PRV autocorrelation function corresponds to the remainder of the fraction between the circumstellar period of the planet and the circumplanetary period of the moon. If the nominal value of this resulting peak position is smaller than two planetary orbital periods, then the actual position of the peak is shifted to a position longer than two times $P_b$.

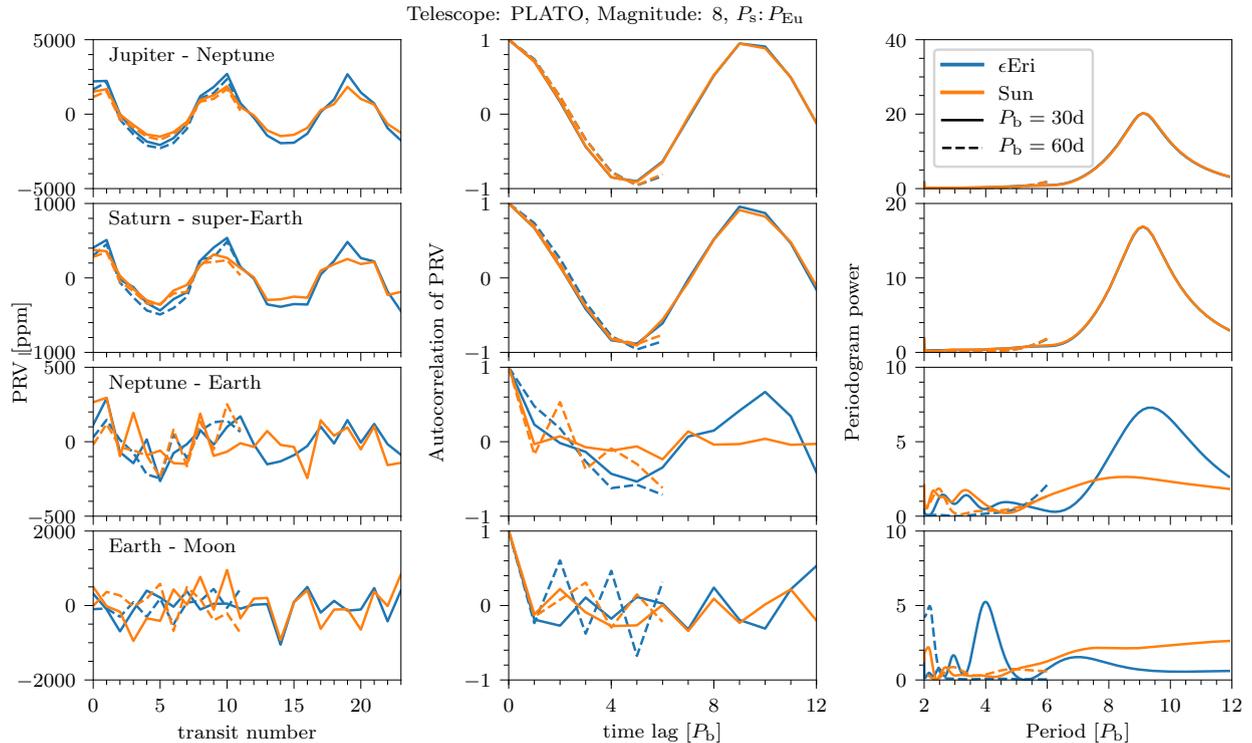

Figure 8.8: Transit sequence of the planet radius variation (PRV, left), autocorrelation function of the PRV transit sequence (center), and periodogram of the of autocorrelation (right) for various planet-moon configurations around an $m_V = 8$ star as observed with PLATO (see Sect. 8.2.5). Blue lines refer to an $\epsilon$ Eridani-like star, orange lines to a sun-like star. In the Jupiter-Neptune and Saturn/Super-Earth cases the PRV signal is clearly visible in the measured PRV time series. Consequently the signal is also visible in the autocorrelation. Due to only calculating the AC up to a time lag of have the observation length, a peak in the periodograms is only visible in the $P_B = 30$ d cases and not the $P_B = 60$ d cases. In the Neptune-Earth cases the signal in the PRV time series is barely visible, while the autocorrelation and periodogram for the $P_B = 30$ d cases show a clear signal. There is no visible signal in any of the Earth-Moon cases for in the corresponding time series, autocorrelation, and periodogram.

### 8.2.6 Kepler-856 b as an Example for PRVs

As a first application of our novel PRV method, we searched for PRVs in the published values of the transit depth sequences of 2598 Kepler Objects of Interest (KOIs), or transiting exoplanet candidates, from Holczer et al. (2016). As a general result of our by-eye inspection of those KOIs with the highest periodogram power, we noticed that stellar rotation is the major cause of false positives. In the case of Kepler-856 b (KOI-1457.01), however, the interpretation of a false positive is less evident based purely on the inspection of the light curve. Figure 8.9 shows our vetting sheet constructed for our PRV survey with Kepler. Kepler-856 b is a Saturn-sized planet with unknown mass in a 8 d orbit around a sun-like star. The transit sequence has been used to statistically validate Kepler-856 b as a planet with less than 1 % probability of being a false positive (Morton et al. 2016).

Figure 8.9(a) shows the Pre-Search Conditioning Simple Aperture Photometry (PDCSAP) Kepler flux (Jenkins et al. 2010), which has been corrected for the systematic effects of the telescope, normalized by the mean of each Kepler quarter. We note that the apparent periodic variation of the maximum transit depth is probably not the PRV effect that we describe. The apparent variation of the maximum transit depth visible in Fig. 8.9(a) is likely an aliasing effect due to the finite exposure time of the telescope.

Figure 8.9(b) shows the series of the transit depths of Saturn-sized object Kepler-856 b, panel (c)

the autocorrelation function, panel (d) the periodogram power of the transit depth series (blue) and of the autocorrelation function (orange), and panel (e) shows the periodograms of the 14 Kepler quarters of the light curve (gray) together with their mean values (black). We used panel (e) as a means to quickly identify stellar variability and compare it to any possible periodicity in the PRV periodogram of panel (d).

Using the stellar and planetary parameters derived by Morton et al. (2016), we estimate that the planetary Hill radius ($R_{\mathrm{H}}$) is about 540,000 km wide. Given that prograde moons can only be stable out to about $0.5 \times R_{\mathrm{H}}$ (Domingos et al. 2006), the range of prograde and gravitationally stable moon orbits is limited to about 270,000 km, corresponding to about $9\,R_{\mathrm{p}}$ for Kepler-856 b. For comparison, Io, the innermost of the Galilean moons, orbits Jupiter at about 6.1 planetary radii, corresponding to about 8 % of Jupiter's Hill radius. So the space for orbital stability is quite narrow around Kepler-856 b compared to the orbits of the solar system moons, though a close-in sufficiently large moon could still be physically plausible.

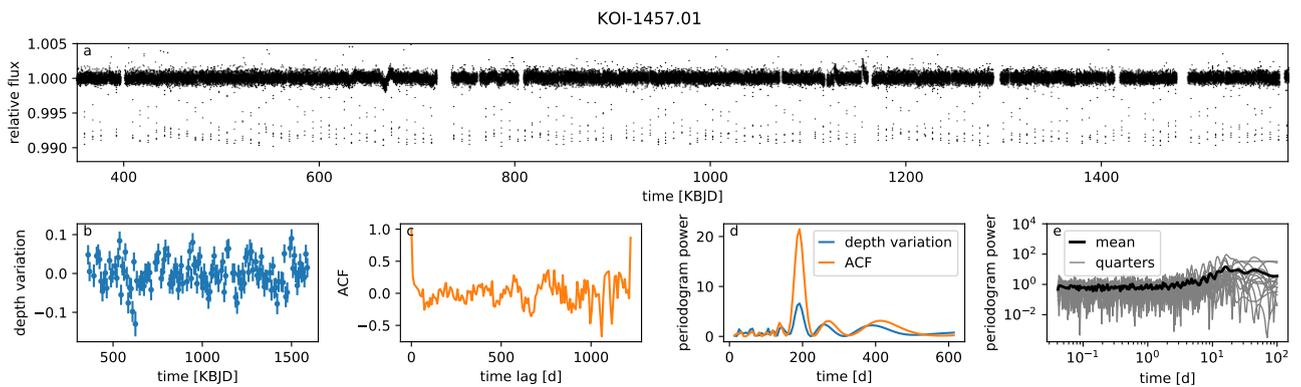

Figure 8.9: Analysis of the Kepler light curve of Kepler-856 b (KOI-1457.01). **(a)** PDCSAP Kepler light curve. **(b)** Fractional transit depths variation from Holczer et al. (2016). **(c)** Autocorrelation of the transit depth variation. **(d)** Periodogram of the transit depth variation and the autocorrelation function. **(e)** Periodograms of the PDCSAP flux of 14 Kepler quarters (gray lines) and their mean (black line) after masking out the planetary transits.

### 8.2.7 Discussion

Comparing the TTV, TDV, and PRV distributions as a function of the moon's orbital phase for systems of various orbital inclinations, we noticed that systems without occultations (and significant inclinations) exhibit more sinusoidal patterns. A comparison of the blue (with occultations) and orange (without occultations) lines in any of the upper panels of Fig. 8.5(a) - (c) serves as an illustrative example for this. Configurations without occultations thus naturally produce higher peaks in in the corresponding periodograms. An improvement to our PRV indicator, in particular for systems with occultations, could be achieved if the PRV signal could be modeled and fitted to the PRV time series, or if it could be used as a fitting function in the periodogram. This approach, however, could become as mathematically and computationally involved as implementing a planet-moon transit model to the light curve to begin with.

The detectability of moon-induced PRV signals increases with the number of transits observed and it is desirable to at least observe a full period of the PRV signal. Hence, this new exomoon indicator can preferably be detected with longterm transit surveys such as Kepler or PLATO, which deliver (or are planned to deliver) continuous stellar high-precision photometry for several years.

Localized features on a star's surface such as spots or faculae can also cause variations in the measured exoplanet radius. Hence, any search for moon-induced PRVs needs to include the analysis of stellar variations and verify that any candidate signals are not caused by the star.nents to the

observed – or in our case the fitted – TTV signal, we noticed that both components tend to cancel each other, see Fig. 8.5(b). This means that for low-density moons the TTV signal will be dominated by the photometric component, while for high-density moons the barycentric component will be more relevant. For moons with densities similar to the densities of the terrestrial or planets and moons of the solar system, which range between about $3\,\mathrm{g\,cm^{-3}}$ and $5\,\mathrm{g\,cm^{-3}}$, the overall TTV effect might actually be very hard to observe as the barycentric and photometric and contributions essentially cancel each other out.

The amplitude of the PRV signal increases with increasing orbital separation between the planet and the moon. The PRV amplitude is maximized when the orbital semimajor axis around the planet is larger than two stellar radii, at which point the two transits of the planet and the moon can occur without any overlap in the light curve. Transits without an overlap happen when the moon's orbital phase is near $\varphi_\mathrm{s} = 0.25$ or $\varphi_\mathrm{s} = 0.75$, that is, when the planet and the moon show maximum tangential deflection on the celestial plane. In these cases of a transit geometry, there is no "contamination" of the planetary transit by the moon and the lowest possible planetary radius is fitted, thereby maximizing the variation with respect to those transits of a transit sequence in which the planet and the moon transit along or near the line of sight, that is, near $\varphi_\mathrm{s} = 0$ or $\varphi_\mathrm{s} = 0.5$, at which point the maximum planetary radius (with maximum contamination by the moon) is fitted.

## 8.2.8 Conclusions

We use standard tools for the analysis of exoplanet transits to study the usefulness of TTVs, TDVs, and PRVs as exomoon indicators. Our aim is to identify patterns in the data that is now regularly being obtained by exoplanet searches, which could betray the presence of a moon around a given exoplanet without the need of dedicated star-planet-moon simulations.

We first test these indicators on simulated light curves of different stellar systems and find that the PRVs observed over multiple transits show periodic patterns that indicate the presence of a companion around the planet. The TTV and TDV signals, however, may be less useful indicators for the presence of an exomoon, depending on whether the density of the moon. The moon density determines whether the barycentric or photometric component of the respective effect dominates. The PRV signal, for comparison, is always dominated by the photometric contribution and, as a consequence, has more predictable, understandable variation.

Our search for indications of an exomoon in the 2598 series of transit mid-points, transit durations, and transit depths by Holczer et al. (2016) reveals the hot Saturn Kepler-856 b as an interesting object with strong PRVs that are likely not caused by stellar variability. While an exomoon interpretation would be a natural thing to put forward in this study, we caution that the range of dynamically stable moon orbits around Kepler-856 b is rather narrow due to its proximity to the star. The planetary Hill sphere is about $18\,R_\mathrm{p}$ wide, restricting prograde moons to obits less than about $9\,R_\mathrm{p}$ wide. At this proximity to the planet, however, tides may become important and ultimately present additional obstacles to the longterm survival of a massive moon (Barnes & O'Brien 2002; Heller 2012). Generally speaking, our investigations of the fitted transit depths from Holczer et al. (2016) and of the corresponding Kepler light curves show that rotation of spotted stars or an undersampling effect of the transit light curve can cause false positives.

The previously suggested method of finding exomoons in the TTV-TDV diagram (Heller et al. 2016b) might not be as efficient a tool to search for moons as thought. The TTV-TDV diagram only forms an ellipse in the barycentric regime, that is, when the moon has a very high density. For rocky or icy moons, however, the figure of the TTV-TDV diagram derived from fitting a planet-only model to the combined planet-moon transit light curve is much more complex than an ellipse.

We find that the threshold for a PRV-based exomoon detection around an $m_V = 8$ solar type star with the Kepler or PLATO missions lies at a minimum radius of about the size of the Earth.

# Appendix A

# Appendix − Non-Peer-Reviewed Conference Proceedings

**A.1  Constraints on the Habitability of Extrasolar Moons (Heller & Barnes 2014)**

# Constraints on the habitability of extrasolar moons

## René Heller[1] and Rory Barnes[2,3]

[1]Leibniz Institute for Astrophysics Potsdam (AIP), An der Sternwarte 16, 14482 Potsdam
email: `rheller@aip.de`

[2]University of Washington, Dept. of Astronomy, Seattle, WA 98195, USA

[3]Virtual Planetary Laboratory, NASA, USA
email: `rory@astro.washington.edu`

**Abstract.** Detections of massive extrasolar moons are shown feasible with the *Kepler* space telescope. *Kepler*'s findings of about 50 exoplanets in the stellar habitable zone naturally make us wonder about the habitability of their hypothetical moons. Illumination from the planet, eclipses, tidal heating, and tidal locking distinguish remote characterization of exomoons from that of exoplanets. We show how evaluation of an exomoon's habitability is possible based on the parameters accessible by current and near-future technology.

**Keywords.** celestial mechanics – planets and satellites: general – astrobiology – eclipses

## 1. Introduction

The possible discovery of inhabited exoplanets has motivated considerable efforts towards estimating planetary habitability. Effects of stellar radiation (Kasting et al. 1993; Selsis et al. 2007), planetary spin (Williams & Kasting 1997; Spiegel et al. 2009), tidal evolution (Jackson et al. 2008; Barnes et al. 2009; Heller et al. 2011), and composition (Raymond et al. 2006; Bond et al. 2010) have been studied.

Meanwhile, *Kepler*'s high precision has opened the possibility of detecting extrasolar moons (Kipping et al. 2009; Tusnski & Valio 2011) and the first dedicated searches for moons in the *Kepler* data are underway (Kipping et al. 2012). With the detection of an exomoon in the stellar irradiation habitable zone (IHZ) at the horizon, exomoon habitability is now drawing scientific and public attention. Yet, investigations on exomoon habitability are rare (Reynolds et al. 1987; Williams et al. 1997; Scharf 2006; Heller & Barnes 2012; Heller 2012). These studies have shown that illumination from the planet, satellite eclipses, tidal heating, and constraints from orbital stability have fundamental effects on the habitability of moons – at least as important as irradiation from the star. In this communication we review our recent findings of constraints on exomoon habitability.

## 2. Why bother about exomoon habitability?

The number of confirmed exoplanets will soon run into the thousands with only a handful being located in the IHZ. Why should we bother about the habitability of their moons when it is yet so hard to characterize even the planets? We adduce four reasons:

**(*i.*)** If they exist, then the first detected exomoons will be roughly Earth-sized, i.e. have masses $\gtrsim 0.25\,M_\oplus$ (Kipping et al. 2009).

**(*ii.*)** Expected to be tidally locked to their planets, exomoons in the IHZ have days much shorter than their stellar year. This is an advantage for their habitability compared to terrestrial planets in the IHZ of M dwarfs, which become tidally locked to the star.

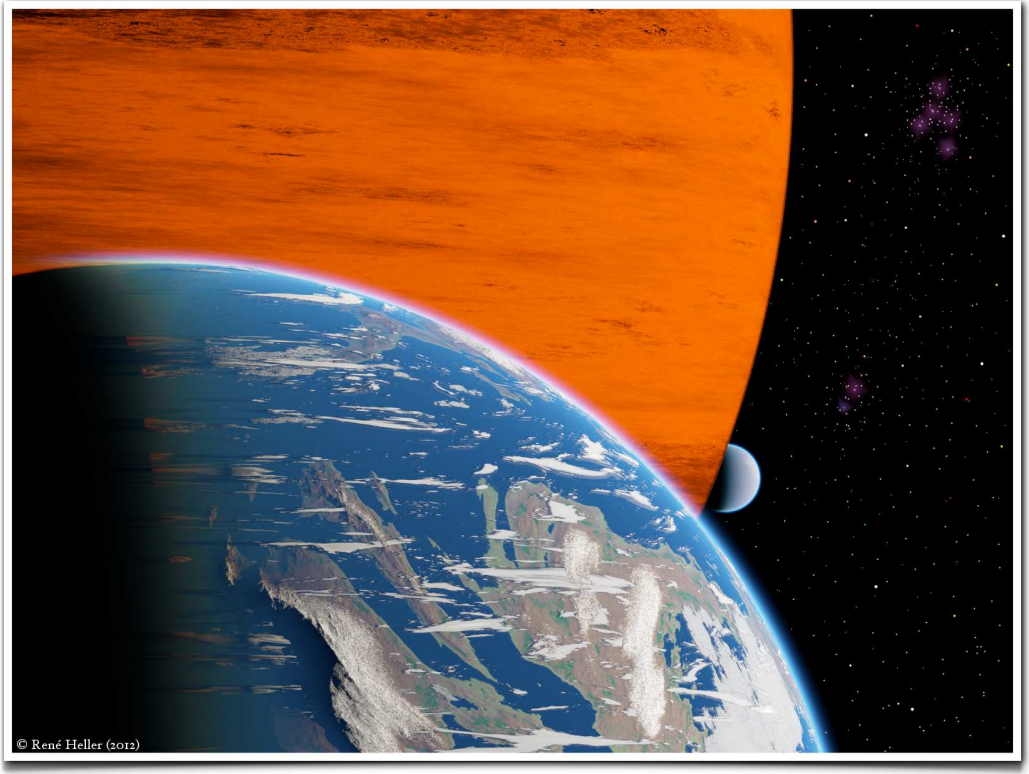

© René Heller (2012)

**Figure 1.** A hypothetical Earth-sized moon orbiting the recently discovered Neptune-sized planet Kepler-22 b in the irradiation habitable zone of a Sun-like host star. The satellite's orbit is equal to Europa's distance from Jupiter. A second moon with the size of Europa is in the background.

- **(*iii.*)** Massive host planets of satellites are more likely to maintain their primordial spin-orbit misalignment than small planets (Heller et al. 2011). Thus, an extrasolar moon in the stellar IHZ which will likely orbit any massive planet in its equatorial plane (Porter & Grundy 2011) is much more likely to experience seasons than a single terrestrial planet at the same distance from the star.
- **(*iv.*)** Extrasolar habitable moons could be much more numerous than planets. In her IAU talk on Aug. 28, Natalie Batalha has shown the "Periodic Table of Exoplanets", indicating many more Warm Neptunes and Warm Jovians (i.e. potential hosts to habitable moons) than Warm Earths in the *Kepler* data (`http://phl.upr.edu`).

   The confirmation of the Neptune-sized planet Kepler-22 b in the IHZ of a Sun-like star (Borucki et al. 2012) and the detection of Kepler-47 c in the IHZ of a stellar binary system (Orosz et al. 2012) have shown that, firstly, adequate host planets exist and, secondly, their characterization is possible. Figure 1 displays a hypothetical, inhabited Earth-sized moon about Kepler-22 b in an orbit as wide as Europa's semi-major axis about Jupiter.

## 3. Constraints on exomoon habitability

### 3.1. *Illumination*

Similar to the case of planets, where the possibility of liquid surface water defines habitability (Kasting et al. 1993), we can approach a satellite's habitability by estimating its orbit-averaged global energy flux $\bar{F}_\mathrm{s}^\mathrm{glob}$. If this top-of-the-atmosphere quantity is less than the critical flux to induce a runaway greenhouse process $F_\mathrm{RG}$ and if the planet-moon duet is in the IHZ, then the moon can be considered habitable. Of course, a planet-moon system can also orbit a star outside the IHZ and tidal heating could prevent the moon from becoming a snowball (Scharf 2006). However, the geophysical and atmospheric properties of extremely tidally-heated bodies are unknown, making habitability assessments challenging. With Io's surface tidal heating of about $2\,\mathrm{W/m}^2$ (Spencer et al. 2000) in mind, which leads to rapid reshaping of the moon's surface and global volcanism, we thus focus on moons in the IHZ for the time being.

Computation of $\bar{F}_\mathrm{s}^\mathrm{glob}$ includes phenomena that are mostly irrelevant for planets. We must consider the planet's stellar reflection and thermal emission as well as tidal heating in the moon. Only in wide circular orbits these effects will be negligible. Let us assume a hypothetical moon about Kepler-22 b, which is tidally locked to the planet. In Fig. 2 we show surface maps of its flux averaged over one stellar orbit and for two different orbital configurations. In both scenarios the satellite's semi-major axis is 20 planetary radii and eccentricity is 0.05. Tidal surface heating, assumed to be distributed uniformly over the surface, is $0.017\,\mathrm{W/m}^2$ in both panels. For reference, the Earth's outward heat flow is $0.065\,\mathrm{W/m}^2$ through the continents and $0.1\,\mathrm{W/m}^2$ through the ocean crust (Zahnle et al. 2007). Parametrization of the star-planet system follows Borucki et al. (2012). In the left panel, the moon's orbit about the planet is assumed to be coplanar with the circumstellar orbit, i.e. inclination $i = 0°$. The satellite is subject to eclipses almost once per orbit about the planet. An observer on the moon could only watch eclipses from the hemisphere which is permanently faced towards the planet, i.e. the moon's pro-planetary hemisphere. Eclipses are most prominent on the sub-planetary point and make it the coldest point on the moon in terms of average illumination (Heller & Barnes 2012). In the right panel, the moon's orbit is tilted by $45°$ against the circumstellar orbit and eclipses occur rarely (for satellite eclipses see Fig. 1 in Heller 2012). Illumination from the planet overcompensates for the small reduction of stellar illumination and makes the sub-planetary point the warmest spot on the moon.

To quantify a moon's habitability we need to know its average energy flux. In Heller & Barnes (2012) and Heller (2012) we show that

$$\bar{F}_{\mathrm{s}}^{\mathrm{glob}} = \frac{L_* \ (1 - \alpha_{\mathrm{s}})}{16\pi a_{*\mathrm{p}}^2 \sqrt{1 - e_{*\mathrm{p}}^2}} \left( x_{\mathrm{s}} + \frac{\pi R_{\mathrm{p}}^2 \alpha_{\mathrm{p}}}{2a_{\mathrm{ps}}^2} \right)$$

$$+ \frac{R_{\mathrm{p}}^2 \sigma_{\mathrm{SB}} (T_{\mathrm{p}}^{\mathrm{eq}})^4}{a_{\mathrm{ps}}^2} \frac{(1 - \alpha_{\mathrm{s}})}{4} + h_{\mathrm{s}}(e_{\mathrm{ps}}, a_{\mathrm{ps}}, R_{\mathrm{s}}) \quad , \tag{1}$$

where $L_*$ is stellar luminosity, $a_{*\mathrm{p}}$ the semi-major axis of the planet's orbit about the star, $a_{\mathrm{ps}}$ the semi-major axis of the satellite's orbit about the planet, $e_{*\mathrm{p}}$ the circumstellar orbital eccentricity, $R_{\mathrm{p}}$ the planetary radius, $\alpha_{\mathrm{p}}$ and $\alpha_{\mathrm{s}}$ are the albedos of the planet and the satellite, respectively, $T_{\mathrm{p}}^{\mathrm{eq}}$ is the planet's thermal equilibrium temperature, $h_{\mathrm{s}}$ the satellite's surface-averaged tidal heating flux, $\sigma_{\mathrm{SB}}$ the Stefan-Boltzmann constant, and $x_{\mathrm{s}}$ is the fraction of the satellite's orbit that is *not* spent in the shadow of the planet. Note that tidal heating $h_{\mathrm{s}}$ depends on the satellite's orbital eccentricity $e_{\mathrm{ps}}$, its semi-major axis $a_{\mathrm{ps}}$, and on its radius $R_{\mathrm{s}}$.

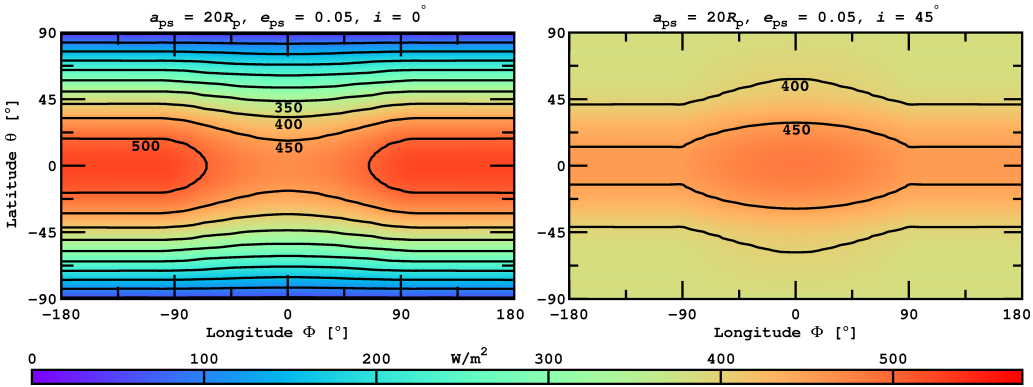

**Figure 2.** Orbit-averaged surface illumination of a hypothetical exomoon orbiting Kepler-22 b. Stellar reflection and thermal emission from the planet as well as tidal heating are included. *Left*: Orbital inclination is $0°$, i.e. the moon is subject to periodic eclipses. The subplanetary point at $(\phi = 0° = \theta)$ is the coldest location. *Right*: Same configuration as in the left panel, except for an inclination of $45°$. Eclipses are rare and the planet's illumination makes the subplanetary point the warmest location on the moon.

This formula is valid for any planetary eccentricity; it includes decrease of average stellar illumination due to eclipses; it considers stellar reflection from the planet; it accounts for the planet's thermal radiation; and it adds tidal heating. Analyses of a planet's transit timing (Sartoretti & Schneider 1999; Szabó et al. 2006) & transit duration (Kipping 2009a,b) variations in combination with direct observations of the satellite transit (Szabó et al. 2006; Simon et al. 2007; Tusnski & Valio 2011) can give reasonable constraints on Eq. (1) and thus on a moon's habitability. In principle, these data could be obtained with *Kepler* observations alone but *N*-body simulations including tidal dissipation will give stronger constraints on the satellite's eccentricity than observations.

### 3.2. *The habitable edge and Hill stability*

The range of habitable orbits about a planet in the IHZ is limited by an outer and an inner orbit. The widest possible orbit is given by the planet's sphere of gravitational dominance, i.e. Hill stability, the innermost orbit is defined by the runaway greenhouse limit $\bar{F}_{\mathrm{s}}^{\mathrm{glob}} = F_{\mathrm{RG}}$ and is called the "habitable edge" (Heller & Barnes 2012).

Space for habitable orbits about a planet decreases when the planet's Hill radius moves inward and when the habitable edge moves outward. This is what happens when we virtually move a given planet-moon binary from the IHZ of G a star to that of a K star and finally into the IHZ of an M star (Heller 2012). Shrinking the planet's Hill sphere means that any moon must orbit the planet ever closer to remain gravitationally bound. Additionally, perturbations from the star on the moon's orbit become substantial and due to the enhanced eccentricity and the accompanying tidal heating the habitable edge moves outward. Hence, the range of habitable orbits vanishes. As the planetary Hill sphere in the IHZ about an M dwarf is small and moons follow eccentric orbits, satellites in M dwarf systems become subject to catastrophic tidal heating, and this energy dissipation induces rapid evolution of their orbits.

Let us take an example: Imagine a planet-moon binary composed of a Neptune-sized host and a satellite 10 times the mass of Ganymede ($10\,M_{\mathrm{Gan}} \approx 0.25\,M_{\oplus}$, $M_{\oplus}$ being the mass of the Earth). This duet shall orbit in the center of the IHZ of a $0.5\,M_{\odot}$ star ($M_{\odot}$ being the solar mass), i.e. at a stellar distance of roughly $0.3\,\mathrm{AU}$ (Selsis et al. 2007). The outermost stable satellite orbit, which numerical simulations have shown to be generally about 1/3 the planet's Hill radius (Barnes & O'Brien 2002), then turns out at $486{,}000\,\mathrm{km}$. This means that the moon's circum-planetary orbit must be at least as tight as Io's orbit about Jupiter! Recall that Io is subject to enormous tidal heating. Yet, the true tidal heating of this hypothetical moon will depend on its eccentricity (potentially forced by the close star and/or by further satellites), and with masses and radii different from those of Jupiter and Io, tidal dissipation in that system will be different.

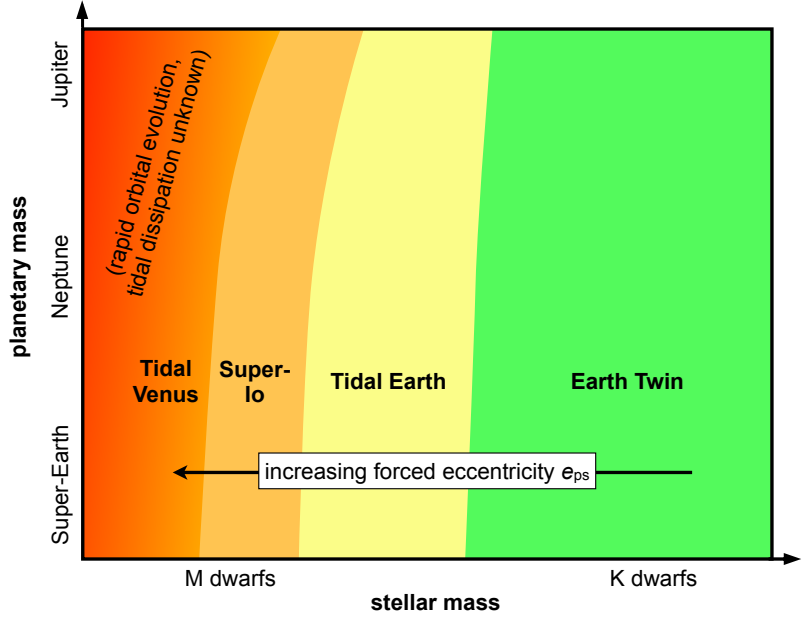

**Figure 3.** Schematic classification of hypothetical $10\,M_{\mathrm{Gan}}$-mass ($\approx 0.25\,M_{\oplus}$) moons in the widest Hill stable orbits about planets in the stellar IHZ. In the IHZ about M dwarfs a satellite's eccentricity $e_{\mathrm{ps}}$ is strongly forced by the close star, which induces strong tidal heating in the moon. A Tidal Venus moon is uninhabitable.

Moving on to a $0.25\,M_{\odot}$ star, the IHZ is now at $\approx 0.125\,\mathrm{AU}$ and the satellite's orbit must be within $255{,}000\,\mathrm{km}$ about the planet. For a $0.1\,M_{\odot}$ star with its IHZ at about $0.05\,\mathrm{AU}$ the moon's orbital semi-major axis must be $< 139{,}000\,\mathrm{km}$, i.e. almost as close as Miranda's orbit about Uranus. As we virtually decrease the stellar mass and as we move our planet-moon binary towards the star to remain in the IHZ, the star also forces the satellite's orbit to become more and more eccentric. We expect that for stellar masses below about $0.2\,M_{\odot}$ no habitable Super-Ganymede exomoon can exist in the stellar IHZ due strong tidal dissipation (Heller 2012).

Figure 3 shall illustrate our gedankenexperiment. The abscissae indicates stellar mass, the ordinate denotes planetary mass. For each star-planet system the planet-moon binary is assumed to orbit in the middle of the IHZ and the moon shall orbit at the widest possible orbit from the planet. With this conservative assumption, tidal heating is minimized. Colored areas indicate the type of planet according to our classification scheme proposed in Barnes & Heller (2012). Exomoons in K dwarf systems will hardly be subject to the dynamical constraints illustrated above (green area), thus Earth twin satellites could exist. However, if roughly Earth-mass exomoons exist in lower-mass stellar systems, then they can only occur as Tidal Earths with small but significant tidal heating (yellow area); as Super-Ios with heating $> 2\,\mathrm{W/m^2}$ but not enough to induce a runaway greenhouse process (orange area); or Tidal Venuses, i.e. with powerful tides and $\bar{F}_{\mathrm{s}}^{\mathrm{glob}} > F_{\mathrm{RG}}$ (red areas). A Tidal Venus is uninhabitable by definition. Tidal dissipation in the upper-left corner of Fig. 3 will be enormous and will lead to so far unexplored geological and orbital evolution on short timescales.

## 4. Prospects for habitable extrasolar moons

The quest of habitable moons seeks objects unknown from the solar system. Even the most massive moon, Ganymede, has a mass of only $\approx 0.025\,M_\oplus$. It is not clear whether moons as massive as Mars ($\approx 0.1\,M_\oplus$) or 10 times as massive as Ganymede ($\approx 0.25\,M_\oplus$) exist (Sasaki et al. 2010; Ogihara & Ida 2012). But given the unexpected presence of giant planets orbiting their stars in only a few days and given transiting planetary systems about binary stellar systems, clearly a Mars-sized moon about a Neptune-mass planet does not sound absurd.

Although Fig. 3 is schematic and urgently requires deeper investigations, it indicates that habitable exomoons cannot exist in the IHZ of stars with masses $\lesssim 0.2\,M_\odot$ (Heller 2012). Orbital simulations, eventually coupled with atmosphere or climate models, have yet to be done to quantify these constraints. With NASA's *James Webb Space Telescope* and ESO's *European Extremely Large Telescope* facilities capable of tracking spectral signatures from inhabited exomoon are being built (Kaltenegger 2010) and future observers will need a priority list of the most promising targets to host extraterrestrial life. Exomoons have the potential to score high if their habitability can be constrained from both high-quality observations and orbital simulations.

## A.2 Hot Moons and Cool Stars (Heller & Barnes 2013b)

# Hot Moons and Cool Stars

René Heller[1,a] and Rory Barnes[2,3,b]

[1] *Leibniz-Institut für Astrophysik Potsdam (AIP), An der Sternwarte 16, 14482 Potsdam, Germany*
[2] *University of Washington, Dept. of Astronomy, Seattle, WA 98195*
[3] *Virtual Planetary Laboratory, USA*

**Abstract.** The exquisite photometric precision of the *Kepler* space telescope now puts the detection of extrasolar moons at the horizon. Here, we firstly review observational and analytical techniques that have recently been proposed to find exomoons. Secondly, we discuss the prospects of characterizing potentially habitable extrasolar satellites. With moons being much more numerous than planets in the solar system and with most exoplanets found in the stellar habitable zone being gas giants, habitable moons could be as abundant as habitable planets. However, satellites orbiting planets in the habitable zones of cool stars will encounter strong tidal heating and likely appear as hot moons.

## 1 Introduction

The advent of high-precision photometry from space with the *CoRoT* and *Kepler* telescopes has dramatically increased the number of confirmed and putative extrasolar planets. Beyond the sheer number of detections, smaller and smaller exoplanets were found with the today record being about 0.5 Earth-radii [1]. This achievement is of paramount importance for astrobiological investigations, as roughly Earth-sized planets may be inhabited – provided many other requirements are met, of course.

Habitability of terrestrial planets can formally be defined as a planet's ability to allow for liquid surface water. The presence of liquid surface water will depend on the planet's distance to its host star, amongst others, with the adequate distance range spanning the stellar habitable zone, depending on the planet's atmospheric composition and surface pressure [2, 3]. As of today, roughly 50 extrasolar planet candidates have been confirmed in the *Kepler* data, most of which are much bigger than Earth. These planets are likely to be gaseous and to resemble Neptune or Jupiter, rather than Earth. While they are not likely to be habitable, their moons might be.

Referred to exomoons, Martin Still, Director of the Kepler Guest Observer Office, said in his talk on the extended *Kepler* project during this meeting: "They are gonna come." So how can massive extrasolar moons be detected, provided they exist in the first place? And to which extent will they possibly be characterized?

---

[a]e-mail: rheller@aip.de
[b]e-mail: rory@astro.washington.edu

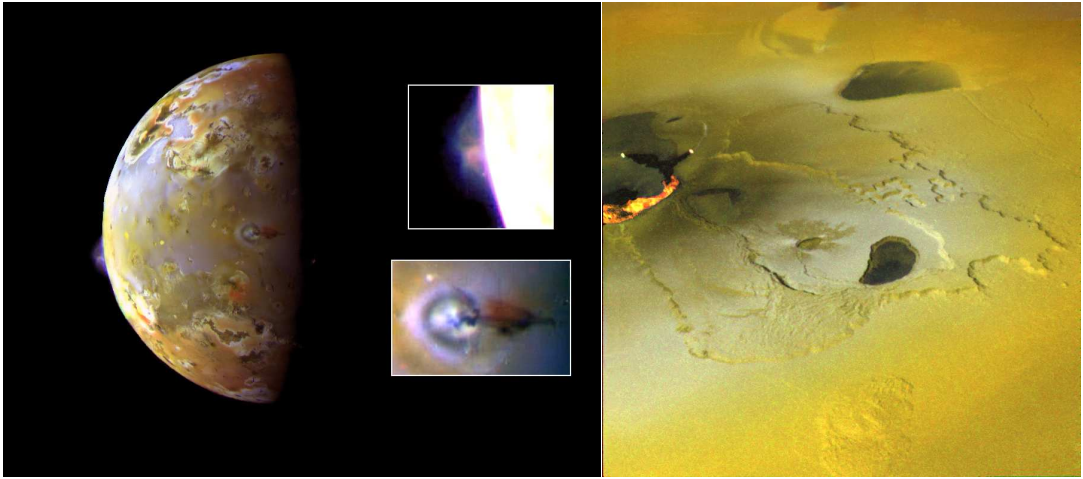

**Figure 1.** The tectonically most active body in the solar system is Jupiter's moon Io. Its enhanced volcanism is driven by tidal dissipation inside the satellite. The two inlets in the left image, taken by the Galileo Orbiter, show a sulfuric plume over a volcanic depression named Pillan Patera (upper photograph) and another eruption called the Prometheus plume (lower photograph). The right image shows the strong orange infrared emission of flowing lava in Tvashtar Catena, a chain of calderas on Io. In extrasolar moons, tidal heating may be much stronger and even make them detectable via direct imaging. (Image credits: NASA/JPL)

## 2 Detection methods

### 2.1 Transit timing and duration variations of the planet

Two of the most promising techniques proposed for finding exomoons are transit timing variations (TTVs) and transit duration variations (TDVs) of the host planet. Combination of TTV and TDV measurements can provide information about a satellite's mass, its semi-major axis around the planet [4–6], and possibly about the inclination of the satellite's orbit with respect to the orbit around the star [7]. The first dedicated hunt for exomoons in the *Kepler* data is now underway [8] and could possibly detect exomoons with masses down to 20% the mass of Earth [9]. This corresponds to roughly 10 times the mass of the two most massive moons in the solar system, Ganymede and Titan.

### 2.2 Direct observations of the moon

Observations of an exomoon transit itself [10–13] as well as planet-satellite mutual eclipses [14, 15] can provide information about the satellite's radius. Spectroscopic investigations of a moon's Rossiter-McLaughlin effect can yield information about its orbital geometry [16, 17], although relevant effects require accuracies in stellar radial velocity of the order of a few cm/s.

Moons in the stellar habitable zone of low-mass stars must orbit their host planet very closely to remain gravitationally bound [18, 19]. This will trigger enhanced tidal heating on those hypothetical moons and could make them uninhabitable. While a threat to life, enormous tidal heating in terrestrial moons around giant planets could be strong enough to make them detectable by direct imaging [20]. Tidal heating in moons has been observed in the solar system, with Jupiter's moon Io serving as the most prominent example (see Fig. 1). As tidal heating in a satellite is proportional to the host planet's mass cubed, massive planets provide the most promising targets for direct imaging detections of tidally heated exomoons.

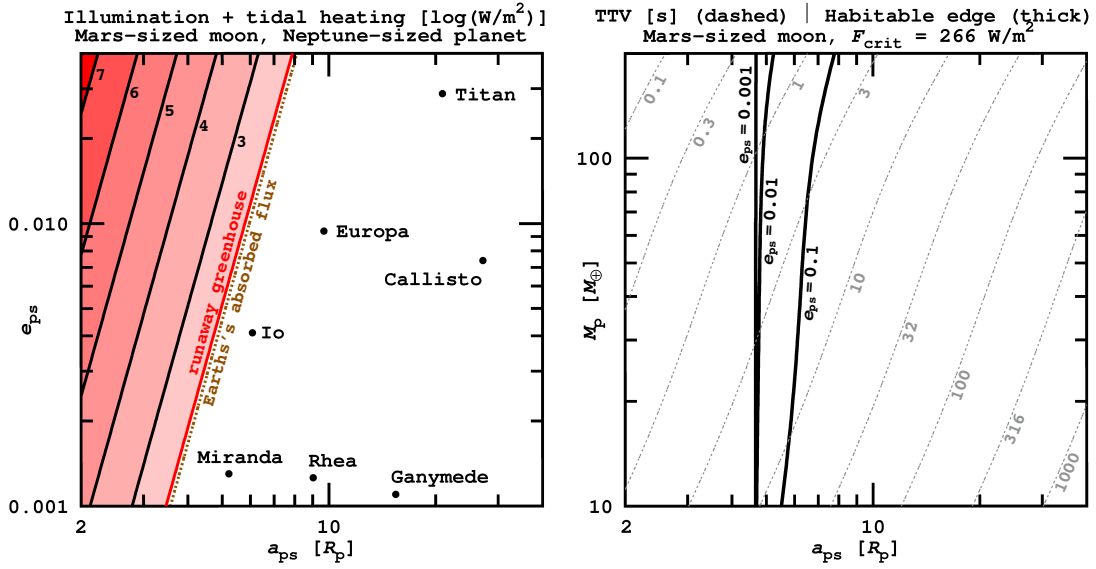

**Figure 2.** *Left:* Total top-of-the-atmosphere flux (in logarithmic units of W/m$^2$) of a Mars-sized moon about a Neptune-sized planet in the habitable zone of a 0.25 $M_\odot$ star. Tidal heating increases with decreasing semi-major axis $a_{ps}$ (abscissa) and increasing eccentricity $e_{ps}$ (ordinate). Some examples for orbital elements of solar system moons are indicated. *Right:* Amplitude of the transit timing variation (dashed lines) for a Mass-sized moon about a range of host planets. Planetary masses are shown in Earth masses on the ordinate. The habitable edge [21] is indicated for three different orbital eccentricities $e_{ps}$ of the satellite: 0.1, 0.01, and 0.001.

## 3 Characterizing exomoons in the stellar habitable zone

A massive moon in the stellar habitable zone can be subject to strong tidal heating and hence be uninhabitable. To estimate its habitability, we have set up a model that includes stellar and planetary illumination as well as tidal heating [21]. If their sum is greater than the critical flux for the moon to be subject to a runaway greenhouse effect [22], the moon will lose all its water and become uninhabitable. In the left panel of Fig. 2 we show contours of the orbit-averaged illumination plus tidal heat flux of a hypothetical Mars-sized moon orbiting a Neptune-sized planet in the habitable zone of a 0.25 $M_\odot$ star. Abscissa indicates the planet-satellite semi-major axis in planetary radii, ordinate shows orbital eccentricity. In the reddish regions this prototype satellite will be desiccated and uninhabitable.

Given sufficiently long observational coverage and high-accuracy data, the techniques described in Sect. 2 make it possible to detect and considerably characterize a sub-Earth-sized moon orbiting a giant planet. In the right panel of Fig. 2 we show the TTV amplitudes of a Mars-sized moon orbiting a range of host planets. It is assumed that the moon's orbit is circular and that both the circumstellar and the circum-planetary orbit are seen edge-on from Earth [8].

To determine a satellite's mass and orbit via TTV and TDV, many transits of the host planet need to be observed. Thus, to find moons about planets in the stellar habitable zone within the *Kepler* duty cycle of 7 years, one might be tempted to conclude that they can preferably be detected in cool star systems (i.e. around M dwarfs), because their habitable zones are close-by where a planet performs potentially many transits in a given time span. However, the amplitude of a planet's TTV is smaller in cool star systems, given a fixed semi-major axis [10], and the lack of M dwarfs in the *Kepler* sample [23] further decreases the chance of finding habitable moons in cool star systems.

Beyond that, moons of planets in the habitable zones of cool stars might not be habitable in the first place [18]. The planet's sphere of gravitational dominance, i.e. its Hill sphere, is relatively small due to the close star. Hence, any moon would have to follow a very tight orbit about the planet. Moreover, the close star will force the satellite's orbit to be non-circular. Both aspects, small orbital distance and an eccentric orbit, will cause any Earth-sized moon of a massive gaseous planet to experience enormous tidal heating. Ultimately, stellar irradiation and tidal heating will sum up to a top-of-the-atmosphere energy flux that exceeds the critical flux for the initiation of the runaway greenhouse effect [21]. Moons of planets in the habitable zones of cool stars will thus be hot rather than habitable.

To characterize potentially habitable moons, that is to say, moons in the habitable zones of K and G type stars, using the TTV and TDV techniques will take about a decade of observations, at least. With the *Kepler* mission being scheduled for a total mission cycle of 7 years, such a detection might just be at the edge of what is possible [8–10].

# Appendix B

# Appendix − Popular Science Publications

## B.1    Better than Earth (Heller 2015)

This popular science article was published as a cover story of the January 2015 issue of *Scientific American* (www.nature.com/scientificamerican/journal/v312/n1/full/scientificamerican0115-32.html).

# Better Than Earth

**Planets quite different from our own may be the best homes for life in the Universe.**

By René Heller

*Origins Institute, McMaster University, Department of Physics and Astronomy, Hamilton (ON) L8S 4M1, Canada*
rheller@physics.mcmaster.ca
www.physics.mcmaster.ca/~rheller

DO WE INHABIT THE BEST O ALL POSSIBLE WORLDS? German mathematician Gottfried Leibniz thought so, writing in 1710 that our planet, warts and all, must be the most optimal one imaginable. Leibniz's idea was roundly scorned as unscientific wishful thinking, most notably by French author Voltaire in his magnum opus, *Candide*. Yet Leibniz might find sympathy from at least one group of scientists – the astronomers who have for decades treated Earth as a golden standard as they search for worlds beyond our own solar system.

Because earthlings still know of just one living world – our own – it makes some sense to use Earth as a template in the search for life elsewhere, such as in the most Earth-like regions of Mars or Jupiter's watery moon Europa. Now, however, discoveries of potentially habitable planets orbiting stars other than our sun – exoplanets, that is – are challenging that geocentric approach.

Over the past two decades astronomers have found more than 1,800 exoplanets, and statistics suggest that our galaxy harbors at least 100 billion more. Of the worlds found to date, few closely resemble Earth. Instead they exhibit a truly enormous diversity, varying immensely in their orbits, sizes and compositions and circling a wide variety of stars, including ones significantly smaller and fainter than our sun. Diverse features of these exoplanets suggest to me and to others that Earth may not be anywhere close to the pinnacle of habitability. In fact, some exoplanets, quite different from our own, could have much higher chances of forming and maintaining stable biospheres. These "superhabitable worlds" may be the optimal targets in the search for extraterrestrial, extrasolar life.

## AN IMPERFECT PLANET

OF COURSE, our planet does possess a number of properties that, at first glance, seem ideal for life. Earth revolves around a sedate, middle-aged star that has shone steadily for billions of years, giving life plenty of time to arise and evolve. It has oceans of life-giving water, largely because it orbits within the sun's "habitable zone," a slender region where our star's light is neither too intense nor too weak. Inward of the zone, a planet's water would boil into steam; outward of the area, it would freeze into ice. Earth also has a life-friendly size: big enough to hold on to a substantial atmosphere with its gravitational field but small enough to ensure gravity does not pull a smothering, opaque shroud of gas over the planet. Earth's size and its rocky composition also give rise to other boosters of habitability, such as climate-regulating plate tectonics and a magnetic field that protects the biosphere from harmful cosmic radiation.

Yet the more closely we scientists study our own planet's habitability, the less ideal our world appears to be. These days habitability varies widely across Earth, so that large portions of its surface are relatively devoid of life – think of arid deserts, the nutrient-poor open ocean and frigid polar regions. Earth's habitability also varies over time. Consider, for instance, that during much of the Carboniferous period, from roughly 350 million to 300 million years ago, the planet's atmosphere was warmer, wetter and far more oxygen-rich than it is now. Crustaceans, fish and reef-building corals flourished in the seas, great forests blanketed the continents, and insects and other terrestrial creatures grew to gigantic sizes. The Carboniferous Earth may have supported significantly more biomass than our present-day planet, meaning that Earth today could be considered less habitable than it was at times in its ancient past.

Further, we know that Earth will become far less life-friendly in the future. About five billion years from now, our sun will have largely exhausted its hydrogen fuel and begun fusing more energetic helium in its core, causing it to swell to become a "red giant" star that will scorch Earth to a cinder. Long before that, however, life on Earth should already have come to an end. As the sun burns through its hydrogen, the temperature at its core will gradually rise, causing our star's total luminosity to slowly increase, brightening by about 10 percent every billion years. Such change means that the sun's habitable zone is not static but dynamic, so that over time, as it sweeps farther out from our brightening star, it will eventually leave Earth behind. To make matters worse, recent calculations suggest that Earth is not in the middle of the habitable

## The Evolution of the Solar Habitable Zone

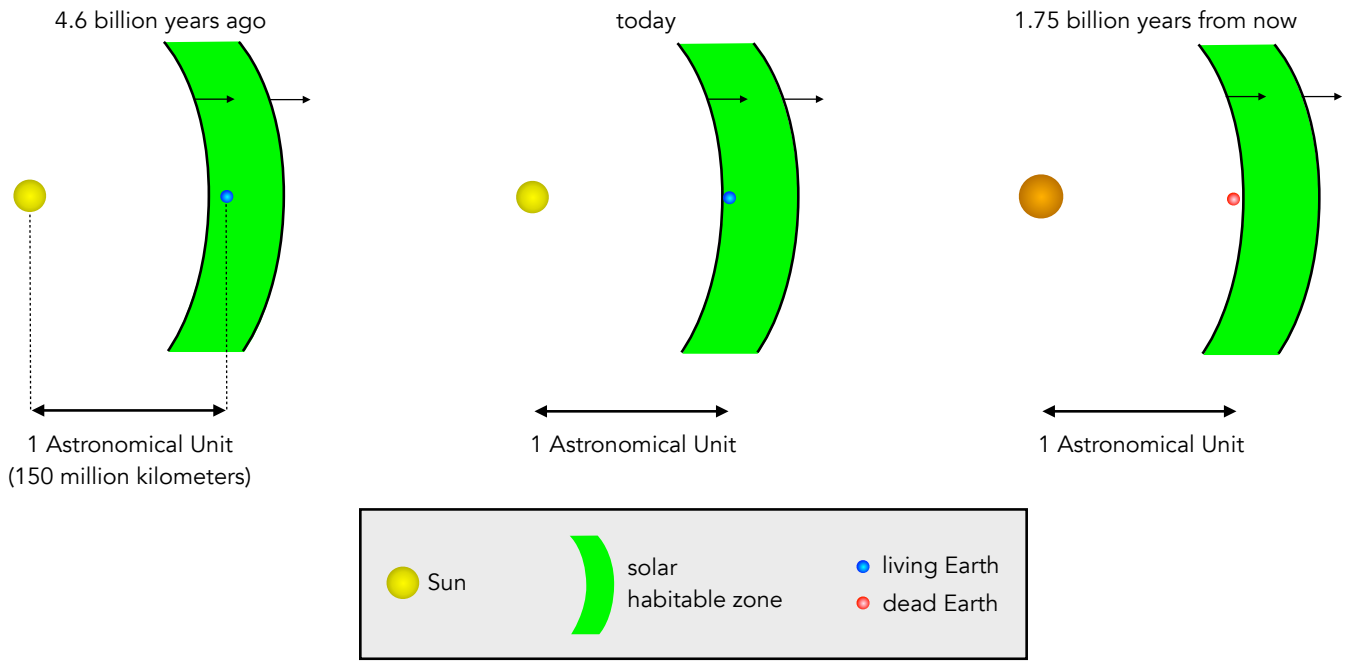

**Figure 1:** *This graphic shows the Earth's distance to the Sun and its location in the solar habitable zone (nothing is to scale here) during three different epochs of stellar evolution. In general, the habitable zone of a star is the distance range in which an Earth-like planet would have the ability to sustain liquid surface water – the essential ingredient to make a world habitable. As time goes by (see panel titles), the solar luminosity and radius increase, and so the solar habitable zone moves away from the Sun. In about 1.75 billion years, the Earth will leave the habitable zone and become a desiccated giant rock. While Earth is a nice place to live on today, it can be regarded "marginally habitable" from a cosmological point of view – both in spatial and in temporal dimensions.*

zone but rather on the zone's inner cusp, already teetering close to the edge of over-heating (see Figure 1).

Consequently, within about half a billion years our sun will be bright enough to give Earth a feverish climate that will threaten the survival of complex multicellular life. By some 1.75 billion years from now, the steadily brightening star will make our world hot enough for the oceans to evaporate, exterminating any simple life lingering on the surface. In fact, Earth is well past its habitable prime, and the biosphere is fast-approaching its denouement. All things considered, it seems reasonable to say our planet is at present only marginally habitable.

### SEEKING A SUPERHABITABLE WORLD

IN 2012 I FIRST BEGAN THINKING about what worlds more suitable to life might look like while I was researching the possible habitability of massive moons orbiting gas-giant planets. In our solar system, the biggest moon is Jupiter's Ganymede, which has a mass only 2.5 percent that of Earth – too small to easily hang on to an Earth-like atmosphere. But I realized that there are plausible ways for moons approaching the mass of Earth to form in other planetary systems, potentially around giant planets within their stars' habitable zones, where such moons could have atmospheres similar to our own planet.

Such massive "exomoons" could be superhabitable because they offer a rich diversity of energy sources to a potential biosphere. Unlike life on Earth, which is powered primarily by the sun's light, the biosphere of a super-habitable exomoon might also draw energy from the reflected light and emitted heat of its nearby giant planet or even from the giant planet's gravitational field. As a moon orbits around a giant planet, tidal forces can cause its crust to flex back and forth, creating friction that heats

**IN BRIEF**

**Astronomers are searching** for twins of Earth orbiting other sunlike stars. **Detecting Earth-like twins** remains at the edge of our technical capabilities. **Larger "super-Earths"** orbiting smaller stars are easier to detect and may be the most common type of planet. **New thinking suggests** that these systems, along with massive moons orbiting gas-giant planets, may also be superhabitable – more conducive to life than our own familiar planet.

the moon from within. This phenomenon of tidal heating is probably what creates the subsurface oceans thought to exist on Jupiter's Europa and Saturn's moon Enceladus. That said, this energetic diversity would be a double-edged sword for a massive exomoon because slight imbalances among the overlapping energy sources could easily tip a world into an uninhabitable state.

No exomoons, habitable or otherwise, have yet been detected with certainty, although some may sooner or later be revealed by archival data from observatories such as NASA's Kepler space telescope. For the time being, the existence and possible habitability of these objects remain quite speculative.

Superhabitable planets, on the other hand, may already exist within our catalogue of confirmed and candidate exoplanets. The first exoplanets found in the mid-1990s were all gas giants similar in mass to Jupiter and orbiting far too close to their stars to harbor any life. Yet as planet-hunting techniques have improved over time, astronomers have begun finding progressively smaller planets in wider, more clement orbits. Most of the planets discovered over the past few years are so-called super-Earths, planets larger than Earth by up to 10 Earth masses, with radii between that of Earth and Neptune. These planets have proved to be extremely common around other stars, yet we have nothing like them orbiting the sun, making our own solar system appear to be a somewhat atypical outlier.

Many of the bigger, more massive super-Earths have radii suggestive of thick, puffy atmospheres, making them more likely to be "mini Neptunes" than super-sized versions of Earth. But some of the smaller ones, worlds perhaps up to twice the size of Earth, probably do have Earth-like compositions of iron and rock and could have abundant liquid water on their surfaces if they orbit within their stars' habitable zones. A number of the potentially rocky super-Earths, we now know, orbit stars called M dwarfs and K dwarfs, which are smaller, dimmer and much longer-lived than our sun. In part because of the extended lives of their diminutive stars, these super-sized Earths are currently the most compelling candidates for super-habitable worlds, as I have shown in recent modeling work with my collaborator John Armstrong, a physicist at Weber State University.

### THE BENEFITS OF LONGEVITY

WE BEGAN OUR WORK with the understanding that a truly longlived host star is the most fundamental ingredient for superhabitability; after all, a planetary biosphere is unlikely to survive its sun's demise. Our sun is 4.6 billion years old, approximately halfway through its estimated 10-billion-year lifetime. If it were slightly smaller, however, it would be a much longerlived K dwarf star. K dwarfs have less total nuclear fuel to burn than more massive stars, but they use their fuel more efficiently, increasing their longevity. The middle-aged K dwarfs we observe today are billions of years older than the sun and will still be shining billions of years after our star has expired. Any potential biospheres on their planets would have much more time in which to evolve and diversify.

A K dwarf's light would appear somewhat ruddier than the sun's, as it would be shifted toward the infrared, but its spectral range could nonetheless support photosynthesis on a planet's surface. M dwarf stars are smaller and more parsimonious still and can steadily shine for hundreds of billions of years, but they shine so dimly that their habitable zones are very close-in, potentially subjecting planets there to powerful stellar flares and other dangerous effects. Being longer-lived than our sun yet not treacherously dim, K dwarfs appear to reside in the sweet spot of stellar superhabitability.

Today some of these long-living stars may harbor potentially rocky super-Earths that are already several billion years older than our own solar system. Life could have had its genesis in these planetary systems long before our sun was born, flourishing and evolving for billions of years before even the first biomolecule emerged from the primordial soup on the young Earth. I am particularly fascinated by the possibility that a biosphere on these ancient worlds might be able to modify its global environment to further enhance habitability, as life on Earth has done. One prominent example is the Great Oxygenation Event of about 2.4 billion years ago, when substantial amounts of oxygen first began to accumulate in Earth's atmosphere. The oxygen probably came from oceanic algae and eventually led to the evolution of more energy-intensive metabolisms, allowing creatures to have bigger, more durable and active bodies. This advancement was a crucial step toward life's gradual emergence from Earth's oceans to colonize the continents. If alien biospheres exhibit similar trends toward environmental enhancement, we might expect planets around long-lived stars to become somewhat more habitable as they age.

To be superhabitable, exoplanets around small, long-lived stars would need to be more massive than Earth. That extra bulk would forestall two disasters most likely to befall rocky planets as they age. If our own Earth were located in the habitable zone of a small K dwarf, the planet's interior would have grown cold long before the star expired. For example, a planet's internal heat drives volcanic eruptions and plate tectonics, processes that replenish and recycle atmospheric levels of the greenhouse gas carbon dioxide. Without those processes, a planet's atmospheric $CO_2$ would steadily decrease as rainfall washed the gas out of the air and into rocks. Ultimately the $CO_2$-dependent global greenhouse effect would grind to a halt, increasing the likelihood that an Earth-like planet would enter an uninhabitable "snowball" state in which all of its surface water freezes.

Beyond the potential breakdown of a planet-warming greenhouse effect, the cooling interior of an aging rocky world could also cause the collapse of any protective planetary magnetic field. Earth is shielded by a magnetic field generated by a spinning, convecting core of molten iron, which acts like a dynamo. The core remains liquefied because of leftover heat from the planet's formation, as well as from the decay of radioactive isotopes. Once a rocky planet's internal heat reservoir became exhausted, its core would solidify, the dynamo would cease, and the magnetic shield would fall, allowing cosmic radiation and stellar flares to erode the upper atmosphere and impinge on the surface. Consequently, old Earth-like planets would be expected to lose substantial portions of their atmospheres to space, and higher levels of damaging radiation could harm surface life.

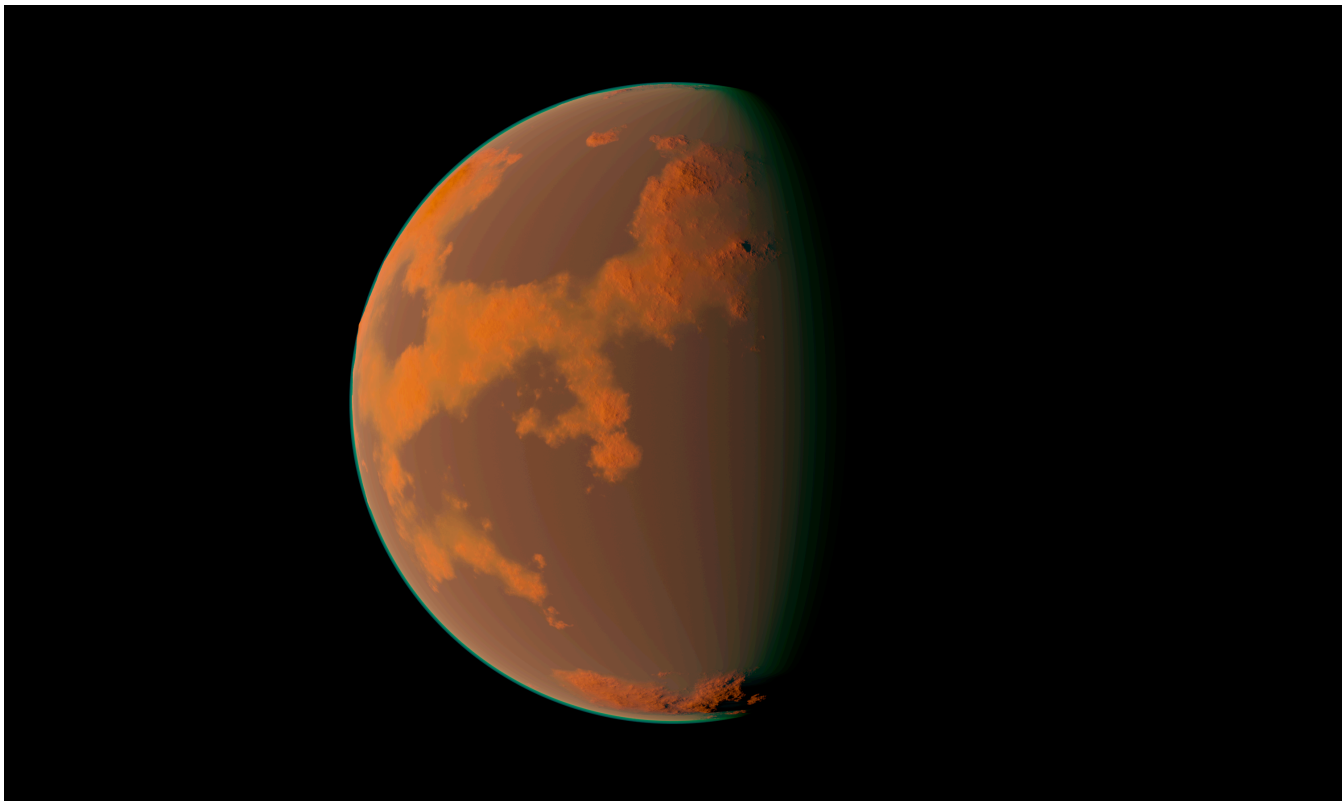

*Figure 2: This example of a hypothetical superhabitable planet visualizes a range of effects that we expect these class of planets to show. First, it appears illuminated by an orange light source, which is owed to its host star being a so-called K dwarf star, a star about 80 percent the mass of the Sun. Second, the planet shows a somewhat opaque atmosphere, which is due to its increased surface gravity compared to Earth and its ability to draw down more gas). Third, while this planet has a large amount of water, its land surface is somewhat more dispersed than on Earth. It resembles an archipelagos world rather than a planet dominated by continents, like Earth.*

Rocky super-Earths as much as twice our planet's size should age more gracefully than Earth, retaining their inner heat for much longer because of their significantly greater bulks. But planets larger than about three to five Earth masses may actually be too bulky for plate tectonics because the pressures and viscosities in their mantles become so high that they inhibit the required outward flow of heat. A rocky planet only two times the mass of Earth should still possess plate tectonics and could sustain its geologic cycles and magnetic field for several billion years longer than Earth could. Such a planet would also be about 25 percent larger in diameter than Earth, giving any organisms about 56 percent more surface area than our world on which to live.

### LIFE ON A SUPERHABITABLE SUPER-EARTH

WHAT WOULD A SUPERHABITABLE PLANET look like? Higher surface gravity would tend to give a middling super-Earth planet a slightly more substantial atmosphere than Earth's, and its mountains would erode at a faster rate. In other words, such a planet would have relatively thicker air and a flatter surface. If oceans were present, the flattened planetary landscape could cause the water to pool in large numbers of shallow seas dotted with island chains rather than in great abyssal basins broken up by a few very large continents (see Figure 2). Just as biodiversity in Earth's oceans is richest in shallow waters near coastlines, such an "archipelago world" might be enormously advantageous to life. Evolution might also proceed more quickly in isolated island ecosystems, potentially boosting biodiversity.

Of course, lacking large continents, an archipelago world would potentially offer less total area than a continental world for land-based life, which might reduce overall habitability. But not necessarily, especially given that a continent's central regions could easily become a barren desert as a result of being far from temperate, humid ocean air. Furthermore, a planet's habitable surface area can be dramatically influenced by the orientation of its spin axis with respect to its orbital plane around its star. Earth, as an example, has a spin-orbit axial tilt of about 23.4 degrees, giving rise to the seasons and smoothing out what would otherwise be extreme temperature differences between the warmer equatorial and colder polar regions. Compared with Earth, an archipelago world with a favorable spin-orbit alignment could have a warm equator as well as warm, ice-free poles and, by virtue of its larger size and larger surface area on its globe, would potentially boast even more life-suitable land than if it had large continents.

Taken together, all these thoughts about the features important to habitability suggest that superhabitable worlds are slightly larger than Earth and have host stars somewhat smaller and dimmer than the sun. If correct, this conclusion is tremendously exciting for astronomers because across interstellar distances super-Earths orbiting small stars are much easier to detect and study than twins of our own Earth-sun system. So far statistics from exoplanet surveys suggest that super-Earths around small stars are substantially more abundant throughout our galaxy than Earth-sun analogues. Astronomers seem to have many more tantalizing places to hunt for life than previously believed.

One of Kepler's prize finds, the planet Kepler-186f, comes to mind. Announced in April 2014, this world is 11 percent larger in diameter than Earth and probably rocky, orbiting in the habitable zone of its M dwarf star. It is probably several billion years old, perhaps even older than Earth. It is about 500 lightyears away, placing it beyond the reach of current and near-future observations that could better constrain predictions of its habitability, but for all we know, it could be a superhabitable archipelago world.

Closer superhabitable candidates orbiting nearby small stars could soon be discovered by various projects, most notably the European Space Agency's PLATO mission, slated to launch by 2024. Such nearby systems could become prime targets for the James Webb Space Telescope, an observatory scheduled to launch in 2018, which will seek signs of life within the atmospheres of a small number of potentially superhabitable worlds. With considerable luck, we may all soon be able to point to a place in the sky where a more perfect world exists.

**The author**: René Heller is a postdoctoral fellow at the Origins Institute at McMaster University in Ontario and a member of the Canadian Astrobiology Training Program. His research focuses on the formation, orbital evolution, detection and habitability of extrasolar moons. He is informally known as the best German rice pudding cook in the world.

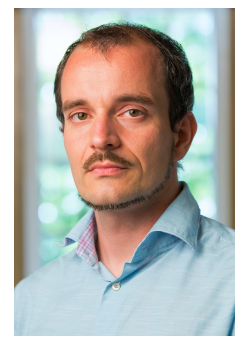

# B.2    Extrasolare Monde – schöne neue Welten? (Heller 2013)

# Extrasolare Monde – Schöne Neue Welten?

von René Heller

*Während mittlerweile knapp 800 Planeten außerhalb des Sonnensystems gefunden wurden, steht der Nachweis von extrasolaren Monden noch aus. Aktuelle Studien zeigen, dass ihre Detektion mit der heutigen Technologie zum ersten Mal möglich ist. Für diese Exomonde sagen Wissenschaftler nun bisher unbekannte astrophysikalische Phänomene voraus.*

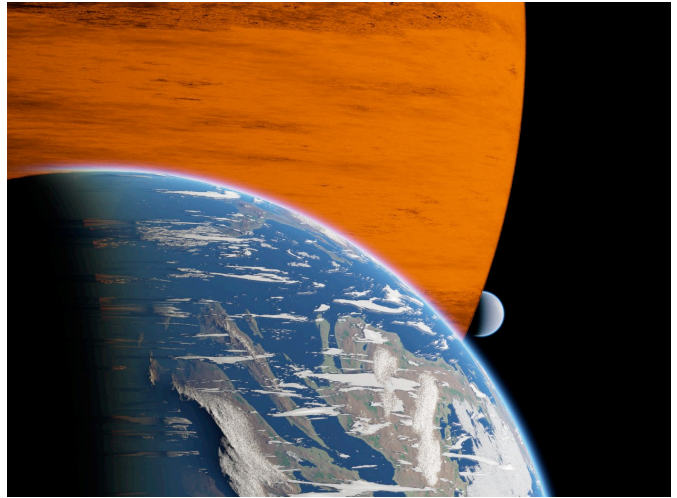

**Monde um extrasolare Planeten könnten lebensfreundliche Bedingungen aufweisen. Ihre Klimaeigenschaften werden u.a. von der Bestrahlung des Sterns und des Planeten sowie von der Gezeitenheizung abhängen.**

Der Nachweis von Monden um extrasolare Planeten, sogenannter Exomonde, ist nunmehr möglich geworden, vor allem durch das Weltraumteleskop *Kepler*. Zahlreiche Studien der vergangenen Jahre belegen, dass die Detektion von Exomonden unmittelbar bevorsteht, vorausgesetzt, Satelliten von ungefähr der Größe des Mars existieren außerhalb des Sonnensystems. Angesichts der Schwierigkeit des Unterfangens jedoch stellen sich berechtigte Fragen danach, was wir überhaupt über Exomonde lernen können. Welche Beobachtungsgrößen gibt es und auf welche Eigenschaften von Exomonden werden sie uns schließen lassen?

Noch bevor auch nur eines dieser Objekte nachgewiesen wurde, wissen wir, dass sich diese Welten grundsätzlich verschieden darstellen werden von den zahlreichen bisher erforschten Exoplaneten. Ihr Tag-und-Nacht-Muster unterscheidet sich von dem auf Planeten durch das Wechselspiel von planetarer und stellar Bestrahlung sowie durch Bedeckungen des Sterns durch den Planeten. Sowohl planetares Licht als auch Eklipsen haben Auswirkungen auf das Mondklima. Darüber hinaus wirken Gezeiten auf eine Kopplung der Rotation von Monden an die Orbitperiode um den Planeten hin. Das heißt, in der habitablen Zone um massearme Sterne, in der erdähnliche Planeten dem Stern stets die gleiche Seite zuwenden, können erdähnliche Monde durch Gezeiteneffekte verhältnismäßig kurze Tage erfahren. Ähnlich wie Titan im Orbit um Saturn kann eine signifikante Schräglage des Planeten außerdem das Jahreszeiten auf Exomonden hervorrufen, solange diese den Planeten am Äquator umrunden. Diese Effekte könnten sich positiv auf die Stabilität eventueller Atmosphären von Exomonden auswirken und ihnen sogar Vorteile gegenüber terrestrischen Exoplaneten bringen. In engen Orbits eines Satelliten um seinen Planeten spielt darüber hinaus die Gezeitenheizung eine entscheidende Rolle für das Mondklima, eventuelle Plattentektonik und Vulkanismus. Außerdem können je nach dem, ob ein Satellit einen Gasplaneten oder terrestrischen Planeten umrundet, bestimmte Entstehungs-Szenarien und somit materielle Zusammensetzungen angenommen bzw. verworfen werden.

Angesichts der über 2000 Kandidaten für Planeten in den *Kepler*-Daten, von denen sich mehr als 50 meist Neptun- bis Jupitergroße Objekte in der HZ befinden, erleben wir mit Studien zur Physik von Exomonden gerade die Geburt eines neuen Forschungszweiges. Schauen wir uns also an, was er uns bietet!

## Entstehung von Monden

Zunächst sollten wir uns darüber im Klaren sein, dass selbst die massereichsten Monde im Sonnensystem, Ganymed und Titan, Leichtgewichte sind im Vergleich zur Erde. Ihre Massen reichen nur an den vierzigsten Teil der Erdmasse. Monde, die mit *Kepler* in absehbarer Zeit nachgewiesen werden können, müssten jedoch mindestens ein Viertel der Masse der Erde haben, also um mindestens den Faktor zehn schwerer sein als Ganymed. Forscher suchen somit nach etwas, das wir aus unserem Sonnensystem nicht kennen, was kaum eine Demotivation bedeuten kann angesichts der zuvor unvorstellbaren Vielfalt an Planeten, die man mittlerweile fand.

Basierend auf den Bahnparametern, gruppieren Astronomen die Monde im Sonnensystem in zwei Klassen: reguläre und irreguläre Monde. Erstere weisen fast kreisrunde, meist enge Orbits auf und umrunden den Planeten in seiner Äquatorebene und zwar in der gleichen Richtung wie der Planet sich dreht. Man geht davon aus, dass sie sich in-situ aus einer zirkumplanetaren Scheibe aus Eis und Gesteinen geformt haben. Die meisten der irregulären Monde hingegen sind wahrscheinlich zum größten Teil eingefangen worden. Prominentes Beispiel hierfür ist Neptuns größter Mond Triton. Sein Orbit ist um 157° gegen Neptuns Äquator geneigt, er umläuft den Gasplaneten also retrograd. Wahrscheinlich hat Neptun Triton dereinst aus einem passierenden planetaren Binärsystem herausgerissen, während Tritons ehemaliger Begleiter aus dem Sonnensystem katapultiert wurde [1].

Kritisch für das Vorkommen von massiven, regulären Exomonden sind Ergebnisse zweier Forschergruppen aus Boulder (USA) und Tokio (Japan) aus den vergangenen Jahren. Erstere fand in ihren theoretischen Untersuchungen, dass die Masse des für die Bildung von Monden zur Verfügung stehenden Materials um Gasplaneten ungefähr ein Fünftausendstel der Planetenmasse ausmacht [2]. Diesem Anteil entsprechen z.B. die Summe der Massen der Galileischen Monde im Vergleich zur Masse Jupiters und Titans Masse im Vergleich zu Saturn. Auf der Suche nach einem Mond mit der Masse des Mars müsste man also Planeten untersuchen, die mindestens doppelt so schwer sind wie Jupiter. Für Monde mit der Masse der Erde müsste der Planet bereits die Masse eines Braunen Zwerges haben.

Verantwortlich für diese Massenbarriere während der Entstehung sind Konkurrenz zweier Prozesse. Während aus der zirkumstellaren Scheibe Material auf den Gasplaneten einfällt, welches das Wachstum der Trabanten fördert, werden die massivsten Monde von der Reibung mit dem zirkumplanetaren Gas in immer engere Bahnen um den Planeten getrieben, bis sie schließlich unter enormen Gezeitenkräften zerrissen werden und eventuell auf den Planeten herabbröseln.

In einer Erweiterung dieser Theorie konnte die Tokioter Gruppe allerdings nachweisen, dass auch marsähnliche Monde um jupiterähnliche Planeten entstehen können [3]. Sogar Satelliten mit der Masse der Erde könnten vorkommen, seien aber äußerst selten.

Über das Einfangen von Objekten der Masse des Mars oder der Erde in einen stabilen Orbit um einen Gasplaneten gibt es zahlreiche Untersuchungen, die belegen, dass solche Prozesse unter dynamischen Aspekten durchaus möglich sind – allein wie wahrscheinlich und üblich solche Vorgänge sind, vermögen sie nicht aufzuzeigen. Die Frage nach solchen Wahrscheinlichkeiten werden wohl erst die Detektionen oder gegebenenfalls die Nicht-Detektionen belegen.

## Detektion von Exomonden

Bisher gibt es keine bestätigte Beobachtung eines extrasolaren Mondes. Zum einen liegt das an der zu erwartenden Rarität dieser Objekte, zum anderen fehlte bis vor kurzem die dafür erforderlichen Instrumentierung. Denn die zu erwartenden beobachterischen Effekte sind nicht nur extrem selten, sondern auch so winzig, dass eine erdgebundene Suche nach Exomonden um zufällig ausgewählte Planeten aussichtslos wäre. Durch den erfolgreichen Betrieb des *Kepler*-Teleskops seit 2009 wurde diese Hürde just genommen und Astronomen begeben sich nun auf die Suche nach den möglicherweise in den bereits gesammelten Daten versteckten Hinweisen auf Exomonde.

Erst Anfang dieses Jahres initiierte ein Team um den Astrophysiker David Kipping vom Harvard-Smithsonian Center for Astrophysics die erste dezidierte Suche nach Exomonden in den *Kepler*-Daten. Ihr Programm trägt den Namen "Hunt for Exomoons with Kepler", kurz HEK [4]. In einigen Artikeln hatten Kipping und seine Kollegen zuvor die theoretischen Grundlagen für den Nachweis von Exomonden gelegt. In diesen Arbeiten und in Studien anderer Forscher konnten mittlerweile mehrere Effekte bestimmt werden, welche die Detektion von Exomonden möglich machen. Einige von ihnen werden direkt durch einen Mond hervorgerufen, andere bestehen aus kleinen Abweichungen des Planeten von seiner Bahn. Beide Kategorien jedoch vereint, dass sie nur für eine bestimmte Klasse von Planeten auftreten: die Transitplaneten. Wagen wir einen kurzen Exkurs zu diesen Objekten, auf dass wir die Effekte ihrer Monde besser verstehen können!

Transitplaneten ziehen, von der Erde aus gesehen, einmal im Laufe ihres Orbits um den Stern vor diesem vorbei und verdunkeln ihn dabei geringfügig. Da ein Planet während des Transits, also während wir seine unbeleuchtete Seite sehen, in guter Näherung schwarz ist im Vergleich zu seinem Stern, kann man die Stärke des Helligkeitsverlusts gut dadurch abschätzen, dass man die Fläche der

Planetenscheibe $\pi R_{\mathrm{p}}^2$ durch die Fläche der Sternscheibe $\pi R_{\mathrm{s}}^2$ teilt, also $R_{\mathrm{p}}^2/R_{\mathrm{s}}^2$ berechnet.

Könnte man die Verdunklung der Sonne durch Jupiter betrachten, beobachtete man also außerhalb der Bahn Jupiters um das Zentralgestirn, so würde man eine Helligkeitseinbuße der Sonne von ungefähr 0.988% beobachten. Die Verdunklung durch die Erde beträge 0.0084%, also weniger als ein Zehntausendstel.

Mittlerweile kennt man 230 bestätigte Transitplaneten in 196 Sternsystemen, nicht zuletzt Dank der bis vor kurzem unerreichten Anzahl simultan überwachter Sterne des *Kepler*-Teleskops, verbunden mit seiner enormen Präzision. Über 2000 weitere *Kepler*-Kandidaten harren übrigens ihrer Bestätigung durch weiterführende Analysen der Lichtkurven und unabhängige Beobachtungen.

Das Haupterkennungsmerkmal für die automatische Detektions-Software der Planetensucher ist die präzise Periodizität der Bedeckungen. Sie entspricht der Orbit-Periode des Transitplaneten um den Stern. Für den Fall, dass der Planet seinen Stern ohne Mond umkreist und vorausgesetzt, dass die Bahnstörungen durch etwaige andere Planeten ausreichend gering sind, ist die Transitperiode konstant. Wird der Planet jedoch von einem Mond begleitet, so verursacht die gravitative Wechselwirkung ein Torkeln des Planeten, denn beide Körper umrunden ihren gemeinsamen Schwerpunkt. Die Auslenkung von diesem Massenzentrum erfolgt für die beiden Körper in entgegengesetzter Richtung und wird durch das Hebelgesetz als

---

### Box 1: Transit Timing Variation (TTV)

Auf dem nebenstehenden Bild zeigt die blaue, lockige Kurve die Bahn eines Mondes, während der Orbit des Planeten durch die schwarze Ellipse gekennzeichnet ist. Das Duett aus Planet und Mond umrundet gemeinsam den orangefarbenen Stern. Die Länge des Vektors $\vec{s}$ vom Mond zum Planeten ist dabei stark vergrößert dargestellt, damit das Torkeln des Satelliten sichtbar wird. Der vergrößerte Ausschnitt zeigt einen Zoom in den Bereich um den Zeitpunkt, zu dem das Planet-Mond-System von der Erde aus gesehen vor dem Stern vorbeizieht. Offensichtlich befindet sich der Mond am Ende seines gemeinsamen Orbits mit dem Planeten um den Stern nicht an der gleichen Position wie zum Start der Simulation.

Einen entgegengesetzten räumlichen Versatz erfährt dabei auch der Planet. Allerdings ist dieser noch um einiges kleiner als der des Mondes und so ist er auf dieser Abbildung nicht erkennbar. Wir können dennoch aus der Abbildung erahnen (oder kompliziert genau berechnen), dass die Periode des Transitzeitpunkts des Planeten von Orbit zu Orbit kleinen Schwankungen unterworfen sein wird. Diese Abweichung bezeichnet man als Transit Timing Variation (TTV).

$$\frac{M_p}{M_m} = \frac{d_m}{d_p}$$

beschrieben, wobei $M_p$ und $M_m$ jeweils die Masse des Planeten und des Mondes bezeichnen und $d_p$ und $d_m$ die Abstände der beiden Objekte vom Massenzentrum. Die Auslenkung des Planeten wird also typischerweise viel kleiner sein als die des Mondes. Je nach dem, in welcher Konstellation das Planet-Mond-Paar aus der Erde aus gesehen vor dem Stern entlangzieht, wird der Planet mal in Richtung seiner Bewegung um den Stern ausgelenkt sein, mal in die entgegengesetzte Richtung. Im erste Fall passiert der Transit etwas früher als im Durchschnitt, im zweiten Fall etwas später. Diese Variationen sind je nach den Massen- und Abstandsverhältnissen in dem Dreikörper-System aus Stern, Planet und Mond in der Größenordnung von Sekundenbruchteilen bis zu wenigen Minuten. Der englische Ausdruck für dieses Phänomen lautet "transit timing variation" (TTV, siehe Box 1).

Bereits Ende des 1990er Jahre wurde vorhergesagt, dass für die Amplitude der zeitlichen Variation der Transitperiode

$$\Delta_{TTV} \sim M_m \times a_{pm}$$

gilt, wobei $a_{pm}$ die große Halbachse der Mondbahn um den Planeten ist. Diese Proportionalität allein lässt also nicht eindeutig auf jeweils den Abstand zwischen Planet und Mond die Masse des Mondes schließen. Eine weitere Beobachtungsgröße muss her.

Im Rahmen seiner Dissertation am University College London konnte David Kipping einen neuen Effekt ausfindig machen, den der sogenannten "transit duration variation" (TDV). Es handelt sich dabei also um die Variation der Dauer des Planetentransits. Diese Schwankung kann zweierlei Ursprung haben. Zum einen variiert neben der Auslenkung auch die tangentiale Geschwindigkeitskomponente des Planeten. Je nach dem, in welche Richtung der Mond sich während des Transits gerade um den Planeten bewegt, wird der Planet eine zusätzliche Geschwindigkeit in Richtung seines Orbits um den Stern erfahren oder ein wenig langsamer vor der stellaren Scheiben entlang ziehen. Dadurch dauert der Transit jeweils etwas kürzer oder länger als im Durchschnitt. Da die Variation der Geschwindigkeitsrichtung der Auslöser für diese Sorte von TDV verantwortlich ist, wird diese "TDV-V" abgekürzt, wobei das letzte "V" für "velocity", also die Geschwindigkeit steht.

Zum anderen kann der Orbit des Mondes um seinen Planeten relativ zum Orbit des Planeten um den Stern gekippt sein. Dadurch erfährt der Planet während seiner Transits eine Auslenkung aus der mittleren Bahnebene und zieht mal näher zur Mitte der Sternscheibe, mal eher am Rand der Sternscheibe entlang. Näher zur Mitte ist der Weg über die Sternscheibe länger, in der Mitte selbst entspricht er einfach ihrem Winkeldurchmesser. Somit entsteht die sogenannte "TDV-TIP", wobei der Suffix "TIP" für das englische "transit impact parameter" steht, also den Abstand des Planetentransits von der Sternmitte.

Die Schwankungen von beiden TDV-Effekten sind in der gleichen Größenordnung wie die TTV. Phänomenalerweise besteht für die Amplitude der TDV jedoch die Proportionalität

$$\Delta_{TDV} \sim \frac{M_m}{\sqrt{a_{pm}}},$$

so dass durch simultane Messung von $\Delta_{TTV}$ und $\Delta_{TDV}$ die mathematische Entartung von $M_m$ und $a_{pm}$ gelöst werden kann. In den kompletten mathematischen Ausdrücken für die beiden Observablen stecken die Masse des Planeten und des Sterns sowie die Orbitperiode des Planet-Mond-Duetts um den Stern. Diese Parameter sind durch zeitaufgelöste Spektroskopie allesamt bestimmbar, so dass schließlich tatsächlich die Masse des Mondes sowie sein Abstand zum Planeten bestimmt werden kann.

Die bisher beschriebenen Effekte sind allesamt indirekter Natur, insofern als nicht der Mond, sondern der Planet beobachtet und auf die Existenz des Mondes geschlossen wird. Natürlich kann man sich auch vorstellen, dass der Transit eines Mondes direkt beobachtet wird. *Kepler* wurde schließlich zu Zwecke der Detektion von erdgroßen Planeten gestartet und konnte bereits deutlich kleinere Objekte mit der Größe des Mars nachweisen. Ein marsgroßer Mond, wie er um einen jupitergroßen Planeten durchaus vorkommen mag, wäre somit direkt nachweisbar. Der entscheidende Gewinn einer solchen Beobachtung läge in der Messung des Mondradius. Zusammen mit der Masse des Mondes ließe sich dann auf seine Dichte und somit die Zusammensetzung schließen.

Mehrere Untersuchungen der Heidelberger Forscherin Lisa Kaltenegger und Kollegen zeigen zudem, dass die spektralen Signaturen von Leben auf Exomonden, sogenannter Biomarker bzw. Bioindikatoren (letztere können auch abiotisch produziert werden), mit Weltraum-Teleskopen der nächsten Generation nachgewiesen werden können. Hierzu zählen molekularer Sauerstoff ($O_2$), Ozon ($O_3$), Methan ($CH_4$) und Distickstoffmonoxid (auch bekannt als Lachgas, $N_2O$) bzw. Kohlendioxid ($CO_2$) und Wasserdampf ($H_2O$). Die dafür notwendigen Messungen jedoch wären auf Grund des notwendigen Signal-zu-Rausch-Verhältnisses nur für Monde um M-Sterne in unserer kosmischen Nachbarschaft nachweisbar. Die Nähe zur Sonne ist dabei wichtig für eine ausreichende Lichtausbeute dieser leuchtschwachen Objekte. Da sich ihre habitable Zone zudem sehr nah am Stern befindet, hat ein Planet-Mond-Duett eine kurze Umlaufperiode um den Stern und kann innerhalb weniger Erdjahre ausreichend viele Transits zeigen, deren Signal sich aufsummieren lässt.

Während direkte spektroskopische Beobachtungen von Exomonden noch die ein oder andere Dekade unzugänglich sein werden, könnten wir aus den oben aufgezählten Bahnparametern des Systems aus einem Stern, einem Planeten und einem Mond bereits eine Menge über die Bedingungen auf der Mondoberfläche erschließen. Dabei gilt es verschiedene Effekte zu berücksichtigen, welche für die Charakterisierung von Exoplaneten keine Rolle spielen.

## Einstrahlung von Stern und Planet

Ein wesentlicher Unterschied zwischen einem frei rotierenden, erdgroßen Planeten und einem Mond gleicher Größe besteht darin, dass der Mond von zwei bedeutenden Lichtquellen beschienen wird. Auf der Erde, also auf einem Planeten stehend, kennen wir das umgekehrte Phänomen, dass wir in einer klaren Nacht bei Vollmond sogar lesen können. Man stelle sich vor wie hell mit Nacht auf Jupiters Mond Europa sein mag, wenn am Mitternacht der gleißende helle Gasriese im Zenit steht!

Mein Kollege Rory Barnes vom Astrobiology Institute der University of Washington, Seattle, und ich haben uns daran gemacht, die Einstrahlungseffekte eines Planeten auf seine Monde zu ermitteln, mit Fokus auf potenzielle Monde um Exoplaneten. Dabei haben wir sowohl das vom Planeten reflektierte Sternenlicht als auch die thermische Strahlung des Planeten berücksichtigt. Durch Beobachtungen der Monde im Sonnensystem und durch die Theorie der Gezeiten gerechtfertigte Prämisse unseres Modells ist, dass der Mond sich zum Planeten im sogenannten "tidal locking" befindet. Er wendet seinem Planeten also stets die gleiche Hemisphäre zu.

Unternehmen wir nun im Geiste eine Reise auf solche einen Exomond, der einen Gasplaneten umrundet! Wir stellen uns vor, dass auf dem Mond gerade Mitternacht herrscht und dass wir am subplanetaren Punkt auf dem Mond stehen (siehe Abbildung auf Seite 4). Der Planet steht also genau im Zenit, während der Stern sich gerade unter unseren Füßen befindet, auf der Rückseite des Mondes. Stern, Mond und Planet bilden eine Linie. Zwar ist nach den Begriffen, wie wir sie auf der Erde verwenden, gerade Mitternacht auf dem Mond, doch schauen wir hoch in den Zenit, so sehen wir die voll beleuchtete

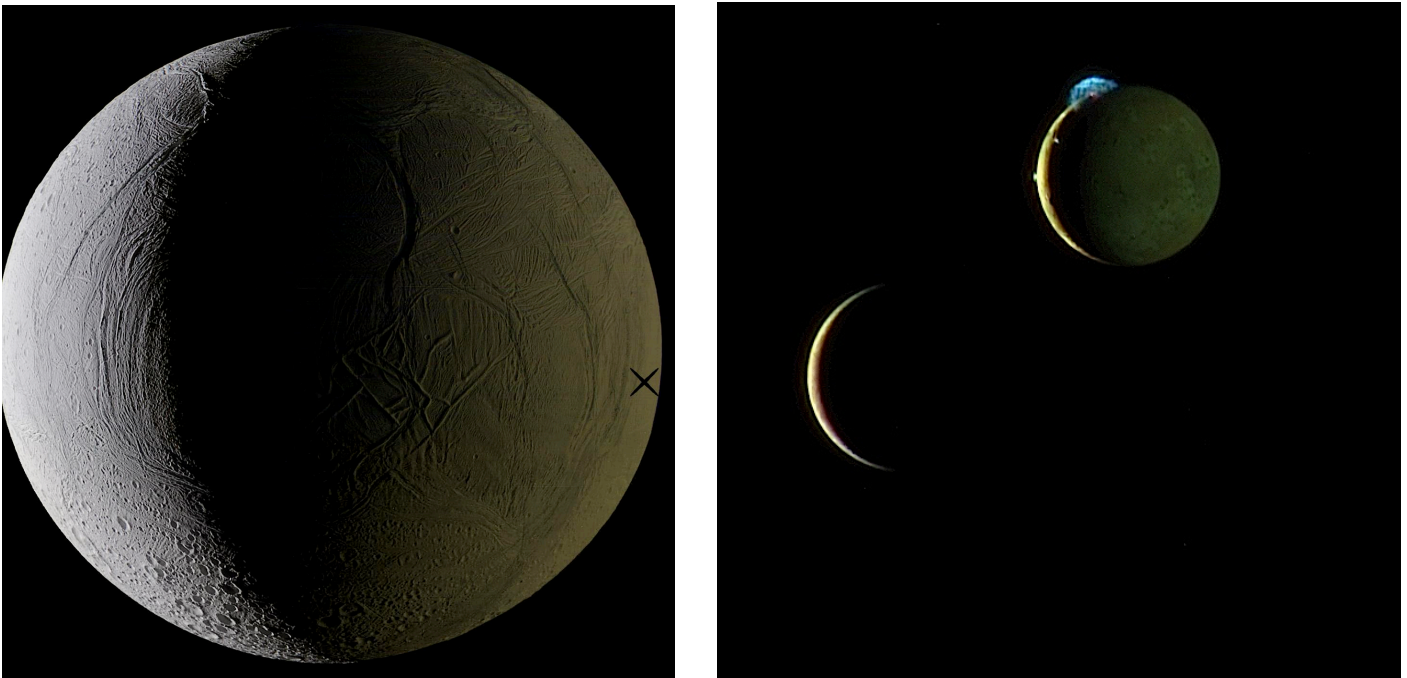

**Beispiele für das kniffelige Wechselspiel der Mond-Bestrahlung durch Stern und Planet finden wir im Sonnensystem. Im linken Bild erscheint Saturns Mond Enceladus von links beleuchtet durch die Sonne und von rechts durch das reflektierte Sonnenlicht seines Planeten. Man achte auf die verschiedenen Farben und Intensitäten der beiden Beleuchtungen! Das Kreuz deutet auf den subplanetaren Punkt des Mondes. Das rechte Bild zeigt Jupiters Monde Europa und Io. Auf Ios nördlicher Hemisphäre speit ein gewaltiger Vulkan. Beide Monde erhalten von links direkte Sonneneinstrahlung, während Io von rechts auch von Jupiters Reflexion erhellt wird. Die Monde sind ca. 800.000 km voneinander entfernt und erscheinen nur in dieser Projektion ihrer momentanen Orbitgeometrie nah beieinander. (Bildnachweis: Courtesy NASA/JPL-Caltech)**

Scheibe des Planeten. Je nach dem, wie weit unser Mond vom Planeten entfernt ist, welchen Radius der Planet hat und welchen Anteil $\alpha_p$ des einfallenden Sternenlichts der Planet reflektiert, wird die einfallende Strahlung eine Leistung zwischen ungefähr einem und hundert Watt pro Quadratmeter haben. Zum Vergleich: Die Sonne strahlt mit einer Leistung von ca. 1400 W/m² auf die Erde, während der Vollmond ungefähr 0.01 W/m² auf die Erde reflektiert. Das reflektierte Licht eines Exoplaneten mit einer Leistung von einigen zehn Watt pro Quadratmeter kann also durchaus die Nacht buchstäblich zum Tag machen! Zusätzlich zu der Spiegelung des Sternenlichts gibt der Planet thermische Strahlung an den Mond ab. Für realistische Albedos des Planeten fanden wir, dass ihr Beitrag um den Faktor zehn kleiner als das reflektierte Licht und damit meist vernachlässigbar ist. Die Abbildung oben veranschaulicht die interessante Überlagerung von stellar und planetarer Bestrahlung mit zwei Momentaufnahmen aus dem Sonnensystem. Auf dem subplanetaren Punkt des Enceladus im linken Bild – markiert durch ein Kreuz – herrscht gerade die beschriebene Mitternacht.

Noch verquerer wird es, wenn wir nun versuchen, uns den Tagesablauf dem Mond vorzustellen. Die Einstrahlung vom Planeten hängt nämlich von dessen Phase ab. Um Mitternacht scheint der Planet über dem subplanetaren Punkt auf dem Mond am stärksten. Danach nehmen seine Phase und Einstrahlungsintensität ab, bis bei Sonnenaufgang nur noch die dem Stern zugewandte Hälfte scheint, sozusagen "Halbplanet" in Anlehnung an den von der Erde aus betrachteten Halbmond. Nun nähern wir uns in Gedanken einem Spektakel. Kurz vor der Mittagszeit nämlich wird es auf einmal stockdunkel. Der Planet schiebt sich täglich zur gleichen Uhrzeit vor den Stern und da wir uns nun über der unbestrahlten Hemisphäre des Planeten befinden, ist es tatsächlich duster, denn die thermische Strahlung des Planeten kann, wie erwähnt, vernachlässigt werden. Die Rückseite des Planeten schneidet derweil einen schwarzen Kreis aus dem Himmel aus – und das zur Mittagszeit! Während dieser Minuten bis wenige Stunden dauernden Bedeckung dürften die Temperaturen auf dem Mond spürbar sinken. Kurz danach geht der gleißend helle Stern hinter dem Planeten wieder auf und bei Sonnenuntergang erscheint nun die andere Hälfte des Planeten beleuchtet und nimmt weiter zu bis Mitternacht. Es sei bemerkt, dass die Taglänge dabei genau der Orbitperiode des Mondes um seinen Planeten entspricht. Die Tage auf Jupiters Galileischen Monden entsprechen daher ungefähr 1,8 (Io), 3,6 (Europa), 7,2 (Ganymed) und 16,7 (Callisto) Erdtagen und Titans Tag dauert 15,9 Erdtage.

Während dieses hypothetischen Vorgangs haben wir angenommen, dass sich das Planet-Mond-Paar nicht nennenswert um den Stern bewegt. Das Bestrahlungsverhalten wird jedoch komplexer, wenn das Massenzentrum von Planet und Mond einen exzentrischen Orbit um den Stern beschreibt (siehe Box 1). Dann hängt die Einstrahlung von einem im Laufe eines Jahres variierenden Abstand zum Stern ab.

In unserem Artikel haben Rory Barnes und ich über das hier geschilderte Szenario hinaus Fälle erwogen, in denen der Orbit des Mondes gegen den Orbit, welchen das Planet-Mond-System um den Stern beschreibt, um den Winkel $i$ geneigt ist. Im Sonnensystem gibt das Saturn-Titan-System hierfür ein schönes Beispiel. Die Rotationsachse des Planeten ist um 26.7° gegen den Orbit um die Sonne geneigt. Saturn erfährt also über einen Zeitraum von 29½ Erdjahren Jahreszeiten. Titan umrundet Saturn in dessen Äquatorebene und erfährt somit auch Jahreszeiten. Durch diese starke Inklination kommt es von Titan aus gesehen nicht wie in unserem oben betrachteten Falle, einmal pro Planet-Mond-Orbit zu Eklipsen der Sonne hinter Saturn. Über die meiste Zeit des Jahres geht die Sonne nämlich zur Mittagszeit unter oder über Saturn hinweg. Lediglich um den Frühlings- und um den Herbstpunkt des Saturns, also wenn die Sonne seine Äquatorebene durchquert, wird sie einmal in Titans Orbit um Saturn bedeckt.

## Gezeitenheizung

Stockdunkle Mittagszeit, hell-erleuchtete Nacht, Tage mit Längen von mehreren Erdtagen – was für bizarre Welten wir uns da vorstellen! Natürlich sind wir noch nicht am Ende. Betrachten wir noch einen Aspekt, nämlich den der Gezeitenheizung! Für Monde in engen Orbits, d.h. mit Abständen von weniger als ungefähr zehn Planetenradien, wird diese Energiequelle signifikant. Prominentestes Beispiel aus dem Sonnensystem ist hierfür Io. In der Abbildung auf Seite 4 ist einer seiner zahlreichen aktiven Vulkane zu sehen. Io gilt als der geologisch aktivste Körper des Sonnensystems. Während auf der Erde ein Wärmefluss von ca. 0.08 W/m² aus dem Inneren herrscht, hauptsächlich gespeist durch radioaktive Zerfälle im Kern und nur einen kleinen Teil durch Gezeitenheizung vom Mond, so emittiert Io satte 2 W/m² aus seiner Gezeitenheizung. Die Folgen sind globaler Vulkanismus, das Ausströmen von Gasen mit einem durchschnittlichen Massenverlust von 1 Tonne pro Sekunde und wahrscheinlich ein unterirdischer, über 1200°C heißer Magma-Ozean aus diversen Schwefel- und Eisenverbindungen.

Auf Europa, der seinen Orbit um Jupiter weiter außen zieht, herrscht verhältnismäßig geringe Gezeitenheizung. Beobachtungen der NASA-Sonde Galileo aus den 1990er Jahren deuten jedoch darauf hin, dass ihre Leistung ausreicht, unter der gefrorenen Oberfläche des Mondes einen gewaltigen Ozean aus Wasser flüssig zu halten. In Gedanken an die sogenannten Schwarzen Raucher (englisch "black smoker") am Grunde der Tiefsee auf der Erde, in deren Umgebung man komplexe, von der Erdoberfläche und dem Sonnenlicht unabhängige Ökosysteme gefunden hat, lassen uns Galileos Befunde an die Möglichkeit von Leben auf Europa denken.

Extrasolare Monde mit solch einer moderaten Gezeitenheizung können bewohnbar sein oder nicht, wir würden das beim besten Willen nicht voraussagen können. Mit Sicherheit können wir jedoch sagen, dass Exomonde mit Gezeitenheizung vergleichbar der des Io nicht bewohnbar sind. Selbst wenn ein Exomond mit seinem Planeten in der habitablen Zone um den Stern zieht, würde das durch Vulkanismus freigesetzte Kohlendioxid den Mond in ein Treibhaus verwandeln, siehe Venus. Das von Rory Barnes und mir entwickelte Modell, welches die Einflüsse von stellarer sowie planetarer Einstrahlung mit der Gezeitenheizung koppelt, soll helfen, kostbare

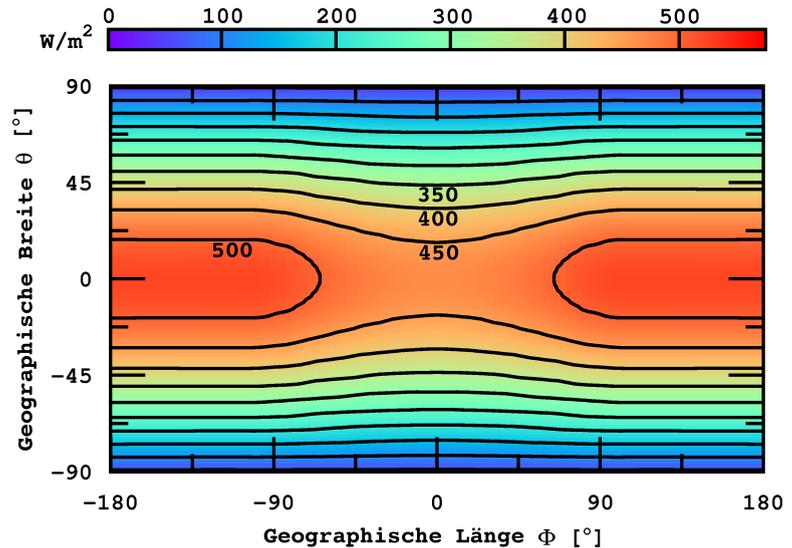

Durchschnittlicher Energiefluss über der Atmosphäre eines hypothetischen Mondes um Kepler-22b. Die Konturen geben konstante Flüsse in Einheiten von W/m² an (siehe Farbleiste). Die Einbuchtung um den subplanetaren Punkt, also bei ϕ = 0° = θ resultiert aus den Eklipsen.

Beobachtungszeit an den Großteleskopen sinnvoll einzuteilen und nur den aussichtsreichen Objekten Priorität zu geben. Schließlich werden Monde, die eventuell bewohnbare oder gar bewohnte Oberflächen haben und von einer erdähnlichen Atmosphäre umgeben sein könnten, die attraktivsten Ziele darbieten.

## Synthese von Einstrahlung und Gezeitenheizung

Unser Modell, das Bestrahlung und Gezeitenheizung auf Monden berücksichtigt, haben wir in einem Computerprogramm geschrieben und konnten damit den Energiefluss auf Monden simulieren. Betrachten wir als Beispiel einen hypothetischen Exomond um den neulich entdeckten Exoplaneten Kepler-22b, der ein sonnenähnlichen Stern in der habitablen Zone umrundet. Der Planet weist einen Radius $R_p$ von ungefähr 2,38 Erdradien auf und ist daher in Masse und Struktur vermutlich ähnlich Uranus oder Neptun.

Für unsere Simulation haben wir angenommen, dass der Mond Kepler-22b in $20R_p$ Entfernung auf einer leicht exzentrischen Bahn umrundet, nämlich mit $e = 0,05$. Dies sollte zu starker Gezeitenheizung führen. Außerdem legten wir den Orbit des Mondes in die gleiche Ebene wie den Orbit um den Stern, so dass Eklipsen auftreten.

Die Abbildung oben zeigt den über einen Orbit (also ein "Jahr" von Kepler-22b) gemittelte Energiefluss auf der Oberfläche des Mondes. Die Ordinate vermittelt geographische Breite ϕ, die Abszisse geographische Länge θ, jeweils gemessen vom subplanetaren Punkt in der Mitte. Über ϕ = 0° = θ steht also der Mond Kepler-22b im Zenit. Die Grafik ähnelt somit einer Weltkarte, nur dass hier die Kontinente, Meere und andere Oberflächeneigenschaften unterlassen sind und stattdessen der Energiefluss aufgetragen ist. Dabei beobachten wir ein spannendes Phänomen: der subplanetare Punkt ist der "kühlste" Ort entlang des Äquators. Das rührt von den Eklipsen her, denn auf der dem Planeten abgewandten Seite des Mondes sind Eklipsen nie beobachtbar und verursachen somit auch keine Reduktion der stellaren Einstrahlung. Für Orbits mit Inklinationen von wenigen Grad fanden wir übrigens das Gegenteil. Dort wurde auf Grund der zusätzlichen Einstrahlung vom Planeten der subplanetare Punkt der "heißeste" Ort der gesamten Oberfläche, denn Eklipsen traten nun sehr selten im Laufe des Jahres auf.

Auch wenn der Effekt der Gezeiten in dieser Abbildung nicht ohne Weiteres abzulesen ist, so sei erwähnt, dass Gezeitenheizung hier gleichmäßig verteilt 42W/m² beiträgt. Ohne Gezeitenheizung würden die Farben bei gleichbleibender Farbskala am Äquator also nur bis ins Orange gehen und nicht ins Rot, während sie an den Polen den Wert 0W/m² erreichen würden, was sie hier – zugegeben: schwer erkennbar – nicht tun.

## Atmosphären und die habitable Zone

Die Atmosphäre ist von grundlegender Bedeutung für die Oberflächeneigenschaften

---

### Box 2: Der Treibhauseffekt

Das thermische Gleichgewicht auf einem Planeten oder Mond ergibt sich aus der Annahme, dass er dieselbe Menge an Strahlung abgibt wie er absorbiert. Mit $T_\odot$ als der Oberflächentemperatur (genauer der "effektiven Temperatur") der Sonne, $R_\odot$ als ihrem Radius und $\alpha_\oplus$ als der Bond-Albedo, also der Oberflächenreflektivität der Erde, ergibt sich ihre Oberflächentemperatur im thermischen Äquilibrium zu

$$T^{\text{eq}} = T_\odot \left( \left[ \frac{R_\odot}{1\,\text{AU}} \right]^2 \frac{(1 - \alpha_\oplus)}{4} \right)^{1/4} = 255\,\text{K} \equiv -18°\text{C} ,$$

wobei eine Astronomische Einheit (AU) dem Abstand zwischen Sonne und Erde entspricht. Die durchschnittlich gemessen Temperatur auf der Erde beträgt jedoch +16°C. Der Unterschied von 34°C wird auf den Treibhauseffekt zurückgeführt. Dieses Phänomen tritt auf, weil Kohlendioxid, Wasserdampf und Methan in der Erdatmosphäre teilweise intransparent sind für die thermische Abstrahlung der Erde. Ein Teil der absorbierten Sonneneinstrahlung kann also nicht reemittiert werden und erwärmt die Erde.

Für die Venus in 0.723 AU Abstand zur Sonne und mit einer Albedo von 0.75 ergibt die Gleichung eine Gleichgewichtstemperatur von 232 K $\equiv$ –41°C. Jedoch bewirkt der Treibhauseffekt in der von Kohlendioxid dominierten Atmosphäre eine mittlere Oberflächentemperatur vor beeindruckenden +464°C. Zwar besteht auch die Atmosphäre des Mars vor allem aus Kohlendioxid, doch ist sie so dünn, dass ihr Treibhauseffekt verschwindend gering bleibt.

---

des Körpers, nicht zuletzt für seine eventuelle Bewohnbarkeit. Ihrer Wärmekapazität führt zu einer zeitlichen Latenz in der Abkühlung der Oberfläche bei Nacht und der Erwärmung bei Tag. Auch bestimmt sie den Wärmetransport von der bestrahlten auf die unbestrahlte Seite des Körpers, bietet ggf. Schutz gegen schädigende Strahlung aus dem All – so schützt auf der Erde die Ozon-Schicht gegen UV-Strahlung der Sonne – und ist schließlich das Medium für den globalen Gas- und Wasseraustausch. Darüber hinaus wirkt sie wie ein Treibhaus. Auf der Erde z.B. beträgt die durchschnittliche Oberflächentemperatur +16°C, obwohl sich aus der Berechnung des thermischen Gleichgewichts lediglich –18°C ergeben (siehe Box 2). Ausschlaggebend dafür ist der Treibhauseffekt, der auch zur Definition der habitablen Zone berücksichtigt wird. Dieser Abstandsbereich um einen Stern definiert einen Orbit-Gürtel, genauer eine Sphäre, innerhalb der ein terrestrischer Planet mit einer Atmosphäre ähnlich der der Erde flüssiges Wasser auf seiner Oberfläche führen würde. Da Wasser als notwendige Bedingung für Leben angesehen wird, nennt man diesen Bereich eben die bewohnbare Zone. Ihr innerer Rand beschreibt den Abstand zum Stern, in dem eine erdähnliche Atmosphäre heiß und mit Wasserdampf gesättigt wäre und außerdem ihre Tropopause verlöre, so dass das Wasser hoch in die Stratosphäre steigen könnte, wo es dann durch die energiereiche Strahlung in Wasserstoff und Sauerstoff zerlegt würde. Der Wasserstoff würde ins All entweichen, der Sauerstoff zurückbleiben und der Planet würde sein Wasservorrat verlieren, also austrocknen und unbewohnbar werden. Diesen Effekt nennen wir ein "runaway greenhouse", also einen irreversibel zunehmenden Treibhauseffekt.

Im Sonnensystem gibt es nur einen Mond innerhalb der habitablen Zone, nämlich den Erdmond. Dieser allerdings ist zu massearm, als dass seine Gravitation im Stande wäre eine signifikante Atmosphäre zu binden. Überhaupt gibt es anscheinend nur zwei Monde im Sonnensystem, die eine nennenswerte Gashülle tragen.

Auf Saturns Mond Titan, ca. 9.5 AU von der Sonne entfernt, könnte ein Mensch dank der dichten Stickstoff-Atmosphäre und der Oberfläche einen Druck von ca. 1.5 bar erzeugt, mit Flügeln ähnlich denen von Vögeln fliegen, auch dank Titans niedriger Oberflächengravitation. Auf Spaziergängen wäre ein Druckausgleich nicht nötig, man benötigte lediglich einen Temperaturanzug und eine Atemmaske. Zur Bespaßung aller Beteiligten könnte man noch kurz die Atemmaske absetzen und ins Feuerzeug ausatmen. Der expirierte Sauerstoff ergäbe mit dem Methan der Atmosphäre ein explosives Gemisch.

Am Rande des Sonnensystems wird Neptuns Mond Triton von einer hauchdünnen Atmosphäre aus Stickstoff, Kohlenmonoxid und Methan umgeben. Wie die Sonde Voyager 2 herausfand, bilden sich in ihr bei -230°C zarte Wolken aus Stickstoff.

## Was lernen wir uns zukünftigen Beobachtungen?

Der Nachweis von Wolken auf Exomonden ist allerdings noch eine Strophe in der Zukunftsmusik. Vorerst müssen wir uns damit begnügen, die grundlegenden Parameter von Exomonden zu bestimmen. Diese werden sich jedoch nur im Verhältnis zu anderen Parametern, nämlich denen des Sterns und des Planeten, ermitteln lassen, so dass die Aufgabe letztlich darin besteht, mindestens drei Körper zu charakterisieren – und eventuell weitere Planeten oder Monde in dem System.

Die stellare Leuchtkraft $L_s$ ließe sich entweder aus der Oberflächentemperatur $T_s$ des Sterns und seinem Radius $R_s$ bestimmen oder aus der Parallaxe und Magnitude des Sterns. Für die erste Methode müssten hochaufgelöste Spektren verfügbar sein und seismische Daten, z.B. aus der präzisen *Kepler*-Photometrie, welche Rückschlüsse auf den Radius erlauben. Die zweite Methode wäre nur auf Sterne in der Sonnenumgebung anwendbar und benötigte astrometrische Messungen. Aus Modellen für Sternevolution ließe sich dann die Sternmasse $M_s$ bestimmen. Unter der Annahme, dass die Masse $M_p$ des Planeten viel größer ist als die des Mondes, wäre $M_p$ aus Messungen der Radialgeschwindigkeit des Sterns berechenbar und sowohl die Orbitperiode $P_{sp}$ als auch die große Halbachse und Exzentrizität $e_{sp}$ des Planet-Mond-Systems um den Stern würden sich automatisch ergeben. Wie oben ausgeführt, ließen sich für Transitplaneten dann TTV und TDV nutzen, um die Mondmasse $M_m$, die große Halbachse des Mondorbits um seinen Planeten $a_{pm}$, und eventuell die Bahnneigung $i$ zwischen den beiden Orbits zu erlangen. Die für die Gezeitenheizung essentiell wichtige Exzentrizität $e_{pm}$ des Mondorbits um den Planeten kann lediglich simuliert werden, insbesondere ist hier die Anwesenheit weiterer Monde zu prüfen (s. Io). Die Albedos $\alpha_p$ und $\alpha_m$ des Planeten und des Mondes müssten voraussichtlich geschätzt werden. Ähnliches gilt für die Materialeigenschaften des Mondes, welche seine Gezeitenkopplung beschreiben, typischerweise ausgedrückt durch die Dissipationskonstanten $Q$ oder $\tau$, und $k_2$.

Zwar wird die direkte Charakterisierung von Exomond-Atmosphären mittelfristig nicht möglich sein, doch können uns auch die durch die derzeitige Technologie und Theorie zugänglichen Parameter bereits viel über die zu erwartenden Oberflächenbedingungen verraten. Aus Masse und Radius lässt sich die Dichte bestimmen, welche wiederum Schlüsse auf die Zusammensetzung zulässt. Und mit der abschätzbaren Einstrahlung, die sich aus unserem Modell ergibt, zusammen mit der aus Masse und Radius ebenfalls berechenbaren Oberflächengravitation können sich gewisse atmosphärische Kompositionen als wahrscheinlich herausstellen.

Der neue Beitrag von Rory Barnes und mir bestand in der Synthese von stellarer und planetarer Einstrahlung mit der Gezeitenheizung auf potentiellen Exomonden zur Beschreibung des Klimas auf solchen Welten. Als Pointe haben wir die Bewohnbarkeit von Exomonden dadurch definiert, dass die Summe aller beteiligter Energieflüsse gering genug sein muss, so dass ein Mond mit erdähnlicher Masse und Atmosphäre nicht einen irreversiblen Treibhauseffekt erfährt. So konnten wir nachweisen, dass etwaige erdgroße Monde um den neulich entdeckten Planeten Kepler-22b, welcher seinen sonnenähnlichen Stern in der HZ umrundet, bewohnbar sein könnten [5], vorausgesetzt, diese Monde hätten ein TTV-Signal $\Delta_{\mathrm{TDV}} \leq 10$ Sekunden.

Noch gibt es auch auf der theoretischen Seite einiges zu tun, z.B. bei der Simulation und Auswertung der Transit-Effekte für den Fall von Mehrfach-Mond-Systemen. Eine analytisch geschlossene Theorie für die TTV und TDV in Systemen mit mehr als einem Mond existiert noch nicht. Numerische Simulationen können derweil eine Stütze bieten. Eine grundlegende Arbeit hierzu könnte aus $N$-Körper-Simulationen erstellt werden, wobei $N$ die Anzahl aller beteiligten Körper und in diesem Fall größer als drei ist. $N$-Körper-Simulationen sind auch nötig um die langfristige Evolution der Geometrie des Mondsystems kennenzulernen, welche das Bestrahlungsmuster bestimmt. Untersuchungen der Stabilität solcher Systeme sind darüber hinaus notwendig um etwaige Funde von Exomonden auf Konsistenz zu prüfen und Exomond-Interpretationen von beobachteten TTVs und TDVs zu bekräftigen oder auszuschließen.

Die Anfang Mai 2012 von der ESA ernannte große Mission "*Jupiter Icy Moons Explorer*" (*JUICE*) mit Start 2022, wird ab 2030 Jupiters große Monde untersuchen. Eines der Hauptziele des Projekts ist es, das Potenzial der Monde Europa, Ganymed und Callisto als Hort von Leben zu erkunden. Dazu wird *JUICE* deren Topographie zentimetergenau vermessen und somit Rückschlüsse auf Verformungen von Gezeiten zulassen. Die präzise Vermessung von Nutation und Libration sowie der Rotationsschieflagen der Monde gegen den Orbit werden weitere interessante Verbesserungen der Modellierung dynamischer Eigenschaften von Satelliten-Systemen erlauben. Die Erkundung der Oberflächenchemie, die Vermessung der strukturellen Zusammensetzung, die Suche nach Wasservorkommen, die Vermessung von Ganymeds Magnetfeld und das Überwachen von Ios vulkanischer Aktivität lassen fundamental neue Einsichten in die Planetologie dieser Monde erwarten. Schließlich wird *JUICE* Schlüsse auf die Physik von Exomonden zulassen, deren Detektion mit *Kepler* uns bald bevorstehen mag.

## Literaturhinweise

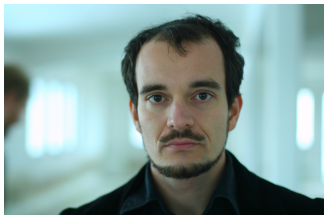

*Dr. René Heller befasst sich am Leibniz-Institut für Astrophysik Potsdam (AIP) sowohl mit der Spektralanalyse von Weißen Zwergen sowie mit der Bewohnbarkeit von extrasolaren Planeten und Monden.*

## B.3  Ist der erste Mond außerhalb des Sonnensystems entdeckt? (Heller 2018b)

# Hinweise auf die erste Entdeckung eines Mondes außerhalb des Sonnensystems

von René Heller

*Während um die acht Planeten des Sonnensystems bisher insgesamt fast 200 Monde entdeckt wurden, konnte noch kein einziger Mond um die annähernd 4000 bekannten Planeten außerhalb des Sonnensystems nachgewiesen werden. Doch neue Beobachtungen des Hubble Weltraumteleskops weisen nun auf die erste Entdeckung eines Exomondes hin.*

Wie heißt es doch in einem populären Popsong der deutschen Hip-Hop-Band Die Fantastischen Vier? "Es könnte so einfach sein, isses aber nicht."

Seit beinahe 20 Jahren nun machen sich Astrophysiker ernsthafte Gedanken über die Möglichkeiten, Monde außerhalb des Sonnensystems zu entdecken. Die einen favorisieren die Transit-Methode als den verheißungsvollsten Ansatz (siehe Box 1), andere bevorzugen die Suche mittels Microlensing, wieder andere Astronomen haben vorgeschlagen, durch Gezeiten extrem aufgeheizte Monde zu suchen, die im infraroten Licht sichtbar sein könnten – sozusagen, super-Ios. Und so schön die theoretischen Ansätze zur Exomond-Detektion auch in mathematischer Form oder in Computersimulationen auch aussehen, eines haben sie fast alle gemeinsam: sie basierend entweder auf stark vereinfachten Annahmen über das Rauschen in den Daten echter Beobachtungen oder sie vernachlässigen das Rauschen gleich ganz.

Und so verwundert es nicht, dass um keinen der beinahe 4000 mittlerweile bekannten Exoplaneten bisher eindeutig ein Exomond gefunden werden konnte.

Neue Beobachtungen des Hubble-Weltraumteleskops vom Stern Kepler-1625 jedoch, welcher von einem jupitergroßen Exoplaneten begleitet wird, liefern starke Hinweise auf den ersten Fund eines solchen Exomondes [1]. Die ersten Hinweise hatten sich bereits in den Daten des Kepler-Teleskops gefunden [2]. In drei Transits, die Kepler zwischen 2009 und 2013 beobachtete, zeigte der Stern nicht nur die charakteristische Verdunklung während des Transits seines Planeten Kepler-1625b, sondern es zeigten sich in der gemittelten Lichtkurve aus den drei Transits zusätzliche Hinweise auf eine leichte Verdunklung sowohl vor als auch nach dem Transit des Planeten. Dieser Effekt wurde 2014 vom Autor dieses Artikels vorhergesagt [3] und ist darauf zurückzuführen, dass der Mond im statistischen Mittel über mehrere Transits mal vor und mal nach dem Planeten die Sternscheibe von der Erde aus gesehen passiert.

Die detaillierte Untersuchung der drei Transits von Kepler-1625b ergab dann Hinweise auf einen Mond, der ungleich allen Monden des Sonnensystems wäre. Dieser Kandidat wäre so groß wie Neptun und würde seinen jupitergroßen Planeten in einer Entfernung von 5 bis 10 Planetenradien umrunden. Zum Vergleich, die Galileischen Monde um Jupiter befinden sich in Entfernungen von ca. 6 (Io) bis 27 (Kallisto) Jupiterradien. Der schwerste Mond des Sonnensystems jedoch, Ganymed, ist nur halb so schwer wie der leichteste Planet des Sonnensystems, Merkur. Die Exomond-Kandidat um Kepler-1625b aber wäre ca. zehnmal so schwer wie alle Gesteinsplaneten und alle Monde des Sonnensystems zusammengenommen.

## Entstehung von Riesenmonden

Wie ein solcher Riesenmond entstehen kann, ist den Astrophysikern derzeit ein Rätsel [4]. Für die Monde des Sonnensystems haben sich drei Entstehungsszenarien als möglich erwiesen. Erstens können Monde nach Einschlägen auf erdähnliche Gesteinsplaneten entstehen. Dieses Szenario erklärt z.B. am besten die beobachteten Eigenschaften des Erde-Mond-Systems. Zweitens können Monde in den Akkretionsscheiben um junge Gasplaneten entstehen. Dieser Mechanismus wird

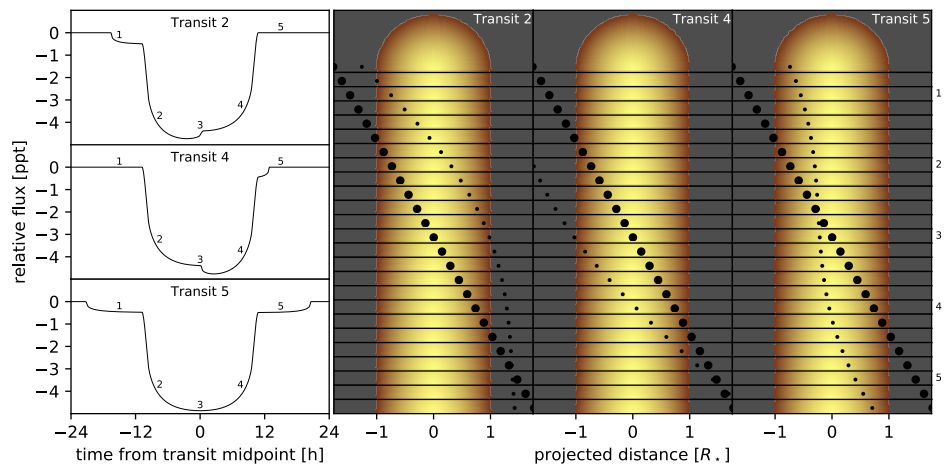

**Zieht ein Exoplanet von der Erde aus gesehen vor seinem Heimatstern entlang, so kommt es zu einer charakteristischen Verdunklung des Sterns. Solch ein Transit dauert typischerweise mehrere Stunden. Wird der Exoplanet zusätzlich von einem Exomond umkreist, so kann auch der Begleiter einen Transiteffekt verursachen und den Stern zusätzlich verdunkeln. Die drei links gezeigten Transit-Lichtkurven sind das Ergebnis aus Simulationen der drei Transits von Kepler-1625 b mit seinem Exomond-Kandidaten, die mit dem Kepler-Teleskop von 2009 bis 2013 beobachtet wurden.**

**Eine Animation zu dieser Abbildung ist online verfügbar: https://goo.gl/kucjUw.**

verantwortlich gemacht für die Existenz der beeindruckenden Mondsysteme um Jupiter, Saturn, Uranus und Neptun. Und drittens können Monde eingefangen werden. Prominentestes Beispiel hierfür ist sicherlich Neptuns größter Mond Triton, aber auch viele der kleineren natürlichen Satelliten um die Gasplaneten sind wahrscheinlich eingefangene Asteroiden. Für Kepler-1625b-i, wie der Exomond-Kandidat vorläufig provisorisch genannt wird, kommt wahrscheinlich keiner dieser Prozesse in Frage.

## Stochern im Datennebel

Die entscheidende Frage, die die neuen Hubble-Daten liefern sollten, nämlich ob das zusätzliche Exomond-Signal in den Kepler-Daten durch einen Mond oder vielleicht durch stellare Variabilität oder gar instrumentelle Effekte verursacht wird, wurde auch im Angesicht der neuen Beobachtungen nicht eindeutig beantwortet. Zwar sind die Hubble-Daten ungefähr viermal genauer als die Kepler-Daten, allerdings beinhalten sie nur einen Transit, während in den Kepler-Daten drei verfügbar sind. Auch findet sich das vermutete Exomond-Signal erneut in den Daten, allerdings ist das Signal nach einem Update der Kepler-Datenreduktions-Software kurioserweise aus den Kepler-Daten fast verschwunden. In der Gesamtschau der Daten ist damit die Signifikanz des Signals immerhin leicht erhöht gegenüber der vorherigen Analyse der Kepler-Daten.

In der Gesamtschau bleibt zu sagen, dass dieses System der bei weitem aussichtsreichste Kandidat für einen Exomond-Fund ist, den es derzeit gibt. Und nicht nur das – würde sich die Existenz dieses Systems mit zukünftigen Beobachtungen bestätigen, dann würde Kepler-1625b und sein Riesenmond ein neues Kapitel in der Forschung der Planetenentstehung öffnen.

## Literaturhinweise

*Dr. René Heller ist Projektwissenschaftler für die PLATO-Mission der ESA, die ab 2026 erdähnliche Planeten mittels der Transitmethode finden soll. Er hat mehr als ein Dutzend Forschungsartikel zu neuen Detektionsmethoden von Exomonden, zu ihren möglichen Entstehungsmechanismen und zu möglichem Leben auf Exomonden in Fachzeitschriften publiziert.*

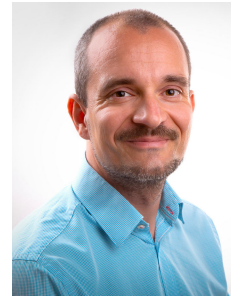

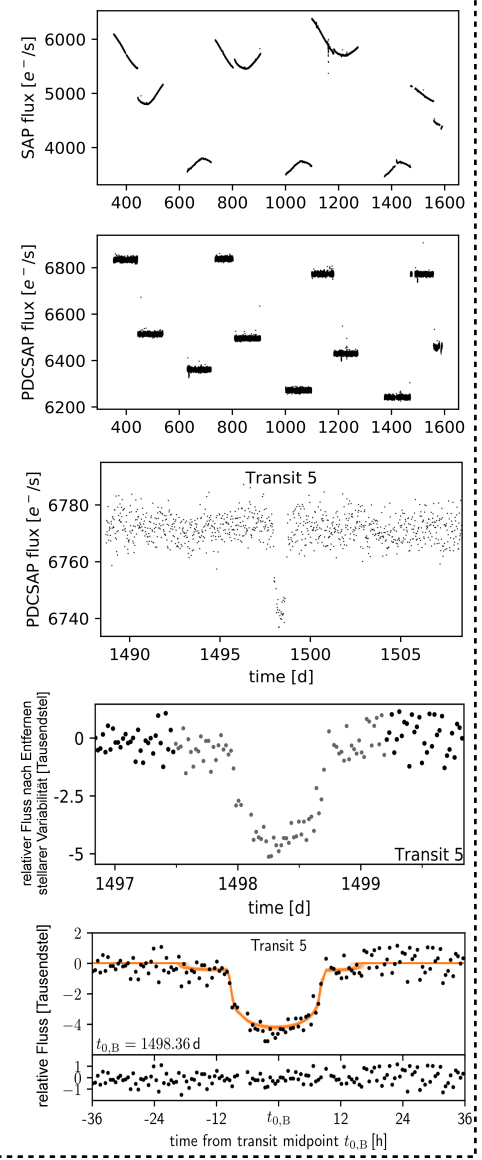

### Box 1: Datenaufbereitung von Transit-Lichtkurven

Die Crux in der Interpretation des Exomond-Signals in der Transit-Lichtkurve des Sterns Kepler-1625 liegt in dem Wissen um die Effekte der verschiedenen Zwischenschritte in der Datenaufbereitung. Beispielhaft ist hier die Aufbereitung der Kepler-Daten gezeigt [5]. Das obere Panel zeigt die unbehandelten Rohdaten des Kepler-Teleskops. Jeder Schnipsel der Lichtkurve ist knapp 100 Tage lang. Die starken Helligkeitsschwankungen sind hervorgerufen durch die Drift des Kepler-Teleskops und die damit verbundene Bewegung des Objekts Kepler-1625 über CCD-Pixel verschiedener Sensitivität. Im zweiten Panel wurden diese instrumentellen Effekte korrigiert. Das mittlere Panel zeigt den letzten der drei Transits aus den Kepler-Daten bei ca. 1498 Tagen. Er trägt die Nummer 5, da zwei weitere Transits in Beobachtungslücken unglücklicherweise nicht aufgenommen wurden. In diesem mittleren Panel kann die stellare Variabilität auf einer Zeitskala von mehreren Tagen erkannt werden. Im vorletzten Panel wurden Variationen in der Lichtkurve auf einer Zeitskala größer als die Dauer des Transits entfernt. Im unteren Panel schließlich sehen wir die Beobachtungsdaten nach dem Detrending zusammen mit einer Schar von 100 Transit-Modellen eines Planet-Mond-Systems, das aus einer Sequenz von Markov-Chain-Monte-Carlo-Simulationen ermittelt wurde [5].

# Acknowledgements

Firstly, I express my sincere gratitude to my most longterm collaborator and reference for astrobiology and exoplanet research, Rory Barnes of the Department of Astronomy at the University of Washington, Seattle. Our discussions about the textual and graphical presentations of our work were crucial sources of inspiration for me and they still serve as a guidelines for me today. Ralph Pudritz of the Department of Physics and Astronomy at McMaster University and founding director of the Origins Institute at McMaster made key contributions to my theoretical work on moon formation. Interaction with him as my supervisor during my two years as a postdoctoral fellow of the Canadian Astrobiology Training Program was a major stimulation for me to grow not only as a scientist. I am particularly grateful to professors Stefan Dreizler, Laurent Gizon, and Ansgar Reiners for giving me the opportunities to teach at the Georg-August-University of Göttingen. My special thanks go to computer scientist and data analyst Michael Hippke for many disruptive conversations and for our highly efficient collaboration on several topics related to the science of extrasolar planets and moons.

I thank Kai Rodenbeck, former PhD student at the Institute for Astrophysics at the University of Göttingen and the Max Planck Institute for Solar System Research, for his permission to present preliminary results from our ongoing collaboration and for providing me with Figs. 8.3 to 8.9. I thank Anina Timmermann, master student at the Institute for Astrophysics of the University of Göttingen, for her cooperation on the analysis of the CARMENES spectra of Kepler-1625 and for providing Table 8.1 and Fig. 8.2.

I also wish to thank the Hans and Clara Lenze Foundation in Aerzen for their financial support of this thesis with a "Habilitations-Stipendium".

The original source of inspiration for me to work in the field of exomoon research is in the 2009 movie Avatar that was written, directed, and produced by James Cameron.